\documentclass[preprint]{aastex63}
\usepackage{apjfonts}
\usepackage{epsfig}
\usepackage{graphics}
\usepackage{verbatim}
\usepackage{natbib,graphicx}
\begin{document}

\title{Following up the Kepler field: \\ Masses of Targets for transit timing and atmospheric characterization}

\suppressAffiliations

\correspondingauthor{Daniel Jontof-Hutter}
\email{djontofhutter@pacific.edu}

\author[0000-0002-6227-7510]{Daniel Jontof-Hutter}
\affil{Department of Physics\\
University of the Pacific\\ 
601 Pacific Avenue\\
Stockton, CA 95211, USA}

\affil{Department of Astronomy \& Astrophysics\\ 
525 Davey Laboratory\\ 
The Pennsylvania State University\\ 
University Park, PA 16802, USA}

\affil{Center for Exoplanets \& Habitable Worlds\\
525 Davey Laboratory\\
The Pennsylvania State University\\
University Park, PA 16802, USA}

\author[0000-0003-2862-6278]{Angie Wolfgang}
\affil{Department of Astronomy \& Astrophysics\\ 
525 Davey Laboratory\\ 
The Pennsylvania State University\\ 
University Park, PA 16802, USA}

\affil{Center for Exoplanets \& Habitable Worlds\\
525 Davey Laboratory\\
The Pennsylvania State University\\
University Park, PA 16802, USA}

\author[0000-0001-6545-639X]{Eric B. Ford}
\affil{Department of Astronomy \& Astrophysics\\
525 Davey Laboratory\\
The Pennsylvania State University\\ 
University Park, PA 16802, USA}

\affil{Center for Exoplanets \& Habitable Worlds\\
525 Davey Laboratory\\
The Pennsylvania State University\\
University Park, PA 16802, USA}

\affil{Institute for Computational and Data Sciences\\
The Pennsylvania State University\\
University Park, PA 16803, USA}

\affil{Center for Astrostatistics\\
525 Davey Laboratory\\
The Pennsylvania State University\\
University Park, PA 16803, USA}

\affil{Institute for Advanced Study\\
Einstein Dr\\
Princeton, NJ 08540, USA}

\author[0000-0001-6513-1659]{Jack J. Lissauer}
\affil{Space Science \& Astrobiology Division\\
MS 245-3\\
NASA Ames Research Center \\
Moffett Field, CA 94035, USA}

\author[0000-0003-3750-0183]{Daniel C. Fabrycky}
\affil{Department of Astronomy and Astrophysics\\
University of Chicago\\
5640 South Ellis Avenue\\
Chicago, IL 60637, USA}

\author[0000-0002-5904-1865]{Jason F. Rowe}
\affil{Department of Physics and Astronomy\\
Bishops University \\
2600 Rue College \\
Sherbrooke, QC J1M 1Z7,
Canada}

\begin{abstract}
We identify a set of planetary systems observed by \textit{Kepler} that merit transit-timing variation (TTV) analysis given the orbital periods of transiting planets, the uncertainties for their transit times, and the number of transits observed during the \textit{Kepler} mission. We confirm the planetary nature of four Kepler Objects of Interest within multicandidate systems. We forward-model each of the planetary systems identified to determine which systems are likely to yield mass constraints that may be significantly improved upon with follow-up transit observations. We find projected TTVs diverge by more than 90 minutes after 6000 days in 27 systems, including 22 planets with orbital periods exceeding 25 days. Such targets would benefit the most from additional transit-timing data. TTV follow-up could push exoplanet characterization to lower masses, at greater orbital periods and at cooler equilibrium temperatures than is currently possible from the \textit{Kepler} dataset alone. Combining TTVs and recently revised stellar parameters, we characterize an ensemble of homogeneously selected planets and identify planets in the \textit{Kepler} field with large-enough estimated transmission annuli for atmospheric characterization with the James Webb Space Telescope. 
\end{abstract}

\section{Introduction}
Characterizing the masses of transiting planets is of particular value in comparative planetary science. Measuring both planetary sizes and masses permits estimates of bulk density and hence a planet's composition. Characterizing the host-star properties and the orbital period permits an estimate of a planetary equilibrium temperature, which combined with planetary mass and radius measurements constrains the atmospheric scale height. In some cases, though not yet among \textit{Kepler} planets, transmission spectroscopy has permitted the retrieval of atmospheric atomic or molecular species (see summaries: \citealt{Stevenson2016,Crossfield2017}). 

The \textit{Kepler} mission led to the discovery of thousands of exoplanets in a wide variety of planetary systems and characterized their physical radii and orbital periods. A small subset of these exoplanets, $\approx 30$, have measured masses based on radial velocity spectroscopy (RV). In addition, from a set of multiplanet systems with measurably interacting neighbors, transit-timing variations (TTVs) have permitted mass measurements of $\approx 73$ transiting planets, including planets as small as Mars (\citealt{Jontof-Hutter2015,Mills2017}) and at orbital periods up to 191 days \citep{ofir14}. 

TTVs have proven to be complementary to RV mass characterizations. Both techniques have observational biases which lead them to probe different regions of parameter space in orbital period, mass and radius, with RV dominating at short periods. For given masses and orbital period ratio, TTVs increase in signal strength with orbital period and have enabled low-mass characterizations at orbital distances beyond the reach of RV surveys (\citealt{Steffen2016,Mills2017b}). Fortunately, there is some overlap in the two techniques, and several systems in the \textit{Kepler} field benefit from both RV and TTV detections (e.g. Kepler-9, \citealt{Borsato2019}; Kepler-18, \citealt{coch11}; Kepler-89, \citealt{mas13,weis13}).

The most precise TTV characterizations are among systems where the timescales of TTVs are less than the time span of the photometric data. For example, planet pairs near first-order mean motion resonances with TTV periodicities from several hundred days to roughly the four-year photometric baseline of \textit{Kepler} have been studied by several authors and have productively populated the planetary "mass-radius diagram" (e.g., \citealt{xie14,Jontof-Hutter2016,Hadden2017}). In addition, in some cases, TTVs are detectable at period ratios far from first-order resonances due to higher-order resonances (e.g., Kepler-29, \citealt{Jontof-Hutter2016}), or synodic encounters (Kepler-36, \citealt{Deck2012}). While TTV signal strength increases with orbital period, this is tempered by the limited number of transits for long-period systems, where there are too few data points to uniquely invert the TTV signal for planet masses and eccentricities, even with precisely measured transit times. 

It is therefore imperative to monitor future transits in multiplanet systems by ground-based or space-based observatories. \citet{VonEssen2018} and \citet{Vissapragada2020} have begun this process for select systems in the \textit{Kepler} field. In some cases, follow-up transit data will not meaningfully improve the TTV mass constraints for decades; such is the pristine quality of the \textit{Kepler} data. The first aim of this study is to identify which targets in the \textit{Kepler} field have their TTVs tightly constrained well into the future from the Kepler dataset, and which planets have uncertain future transit times. The latter group may have their masses meaningfully constrained by additional transit-timing data. Identifying these candidates optimizes the efficiency of follow-up campaigns.

The second aim of this study is to characterize the masses and densities of a homogeneously selected sample of candidate TTV planets independently of whether TTVs have been clearly detected or analyzed in prior studies. The majority of our sample has been studied by other authors, and in many cases, detailed TTV modeling by prior authors has not been improved upon. Nevertheless, since the first data releases of \textit{Gaia} \citep{Andrae2018A&A}, improved precision on stellar parameters for \textit{Kepler} hosts has enabled more precise planetary characterizations. Furthermore, 11 systems with previously unreported mass constraints are presented here for the first time. We envision this sample will prove valuable for future detailed population studies of TTV systems.
 
The third aim of this study is to estimate the atmospheric scale heights and transmission annuli of our planet sample, and hence identify which targets in the \textit{Kepler} field likely have detectable atmospheres in the limit of cloud-free or haze-free transmission annuli. 

\section{System Selection}
There have been several studies of \textit{Kepler} planetary systems observed to have significant TTVs. Near-resonant planet pairs have a TTV period and amplitude that can be measured to infer planet masses and constrain free orbital eccentricity components \citep{Lith2012}. Transit-timing catalogs of \citet{maz13}, \citet{rowe15a}, and \citet{Holczer2016} identified several dozen planet pairs with measurable TTV periodicities and amplitudes appropriate for this solution, enabling TTV systems to yield estimates of the planetary mass-radius relation \citep{wu13} and eccentricity distribution \citep{had14}. This approach has the advantage of an analytical solution that can be computed quickly. However, it neglects information that remains in the TTVs from nonresonant interactions and higher-order resonances. More detailed analytical solutions (e.g. \citealt{Agol2016a, Agol2016b}) more closely approach the results of dynamical models to transit-timing data \citep{Jontof-Hutter2016}. 

Prior studies share the bias of reporting masses and eccentricities among systems with strongly detected TTVs. This causes difficulty in analyzing such systems as a population. One solution is to perform TTV models on all systems, which would be numerically expensive. Most of the numerical expense would be wasted on planet pairs that are unlikely to have any detectable interactions. We address this by selecting our sample of systems purely on their photometric properties and not on the measured transit times or properties inferred from TTVs. This makes our sample more amenable to detailed population modeling. In this paper, we identify a set of systems with planets where there is an expectation of detectable TTVs \textit{prior to} any TTV analysis.

We estimated the minimum expected TTVs among planets in the \textit{Kepler} dataset given their orbital periods, and a conservative estimate of their minimum mass given their radii. These of course depend on the properties of host stars. For the masses of the host stars, we took the nominal values from \citet{Fulton2018} who used \textit{Gaia} parallaxes to constrain stellar parameters, with ground-based spectral classification as part of the California Kepler Survey, and for Kepler Objects of Interest (KOIs) missing from that catalog, we rely on \citet{Berger2020}. For planet sizes we rely on \citet{Fulton2018}. However, several planet candidates within multitransiting systems are missing from that catalog, and for these we rely on an updated, unified catalog of planet transit parameters in preparation (Lissauer et al.). 

We used the following simple mass-radius relation to estimate a minimum mass for every planet:
\begin{equation}
M_{p,min} = min(4,R_{p}^3), 
\label{m-r-simple}
\end{equation}
where $M_p$ and $R_{p}$ are in units of $M_{\oplus}$ and $R_{\oplus}$, respectively. With this simple relation, we ensure that planets that are Earth-sized have a density equal to Earth, but we neglect the effects of compression for larger planets. We put a cap on planet masses, to allow for large gaseous envelopes around 4 M$_{\oplus}$ cores.

Given the wide range of densities observed among planets larger than 1.6 $R_{\oplus}$ (\citealt{Rogers2015, Jontof-Hutter2016, Wolfgang2016}), this criterion underestimates the mass of many planets but ensures that all planets included in our sample have an expectation of detectable TTVs.

We estimated the minimum expected TTVs among planet pairs in our sample with two analytical solutions for TTVs, for systems of up to four transiting planets. The solution of \citet{Agol2016a, Agol2016b} as implemented by \textit{TTVfaster} analytically calculates transit times that are accurate to first order in the planet-star mass ratios and in the orbital eccentricities. We found a simple empirical expression for the amplitude of TTVs for a pair of planets on circular orbits with no mutual inclination. For each model pair, we numerically measured the difference between the maximum and minimum deviation in transit time from a linear ephemeris (in minutes) expected from the transits. We fitted empirical models for the minimum expected nonresonant TTVs of an inner planet (TTV) based on the orbital periods ($P_{inner}$ and $P_{outer}$ measured in days, with ratio ($\mathcal{P} = \frac{P_{outer}}{P_{inner}}$), and the mass ratio of an outer planet to the star ($\mu_{outer}$) is approximated as:
\begin{equation}
\rm{TTV} (mins) \gtrsim C' \left(\frac{A}{\mathcal{P}^{10}}+\frac{B}{\mathcal{P} ^{4}} \right)
\label{empiricalTTV1}
\end{equation}
where $A = 50\times P_{inner}$, $B = 6\times P_{inner}$ and $C' = 10^{4} \mu_{outer}$. 

A similar exercise on the TTVs of an outer planet (TTV') based on the perturbations caused by the inner planet yields
\begin{equation}
\rm{TTV'} (mins) \gtrsim C \left(\frac{A'}{\mathcal{P}^{13}}+\frac{B'}{\mathcal{P} ^{7}} \right)
\label{empiricalTTV2}
\end{equation}
where $A' = 100\times P_{outer}$, $B' = 16\times P_{outer}$ and $C = 10^{4} \mu_{inner}$. 

\begin{figure}[ht!]
\begin{center}
\includegraphics [height = 2.2 in]{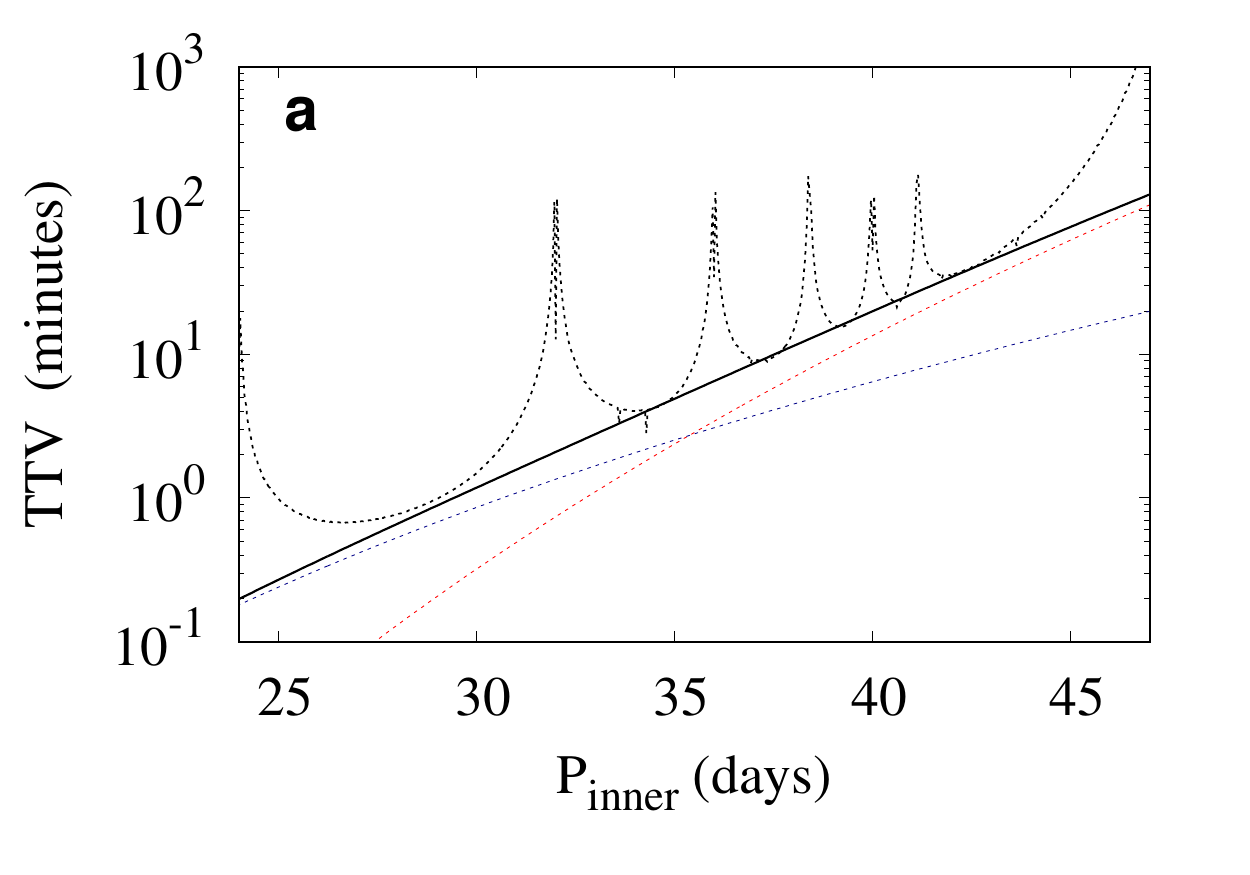}
\includegraphics [height = 2.2 in]{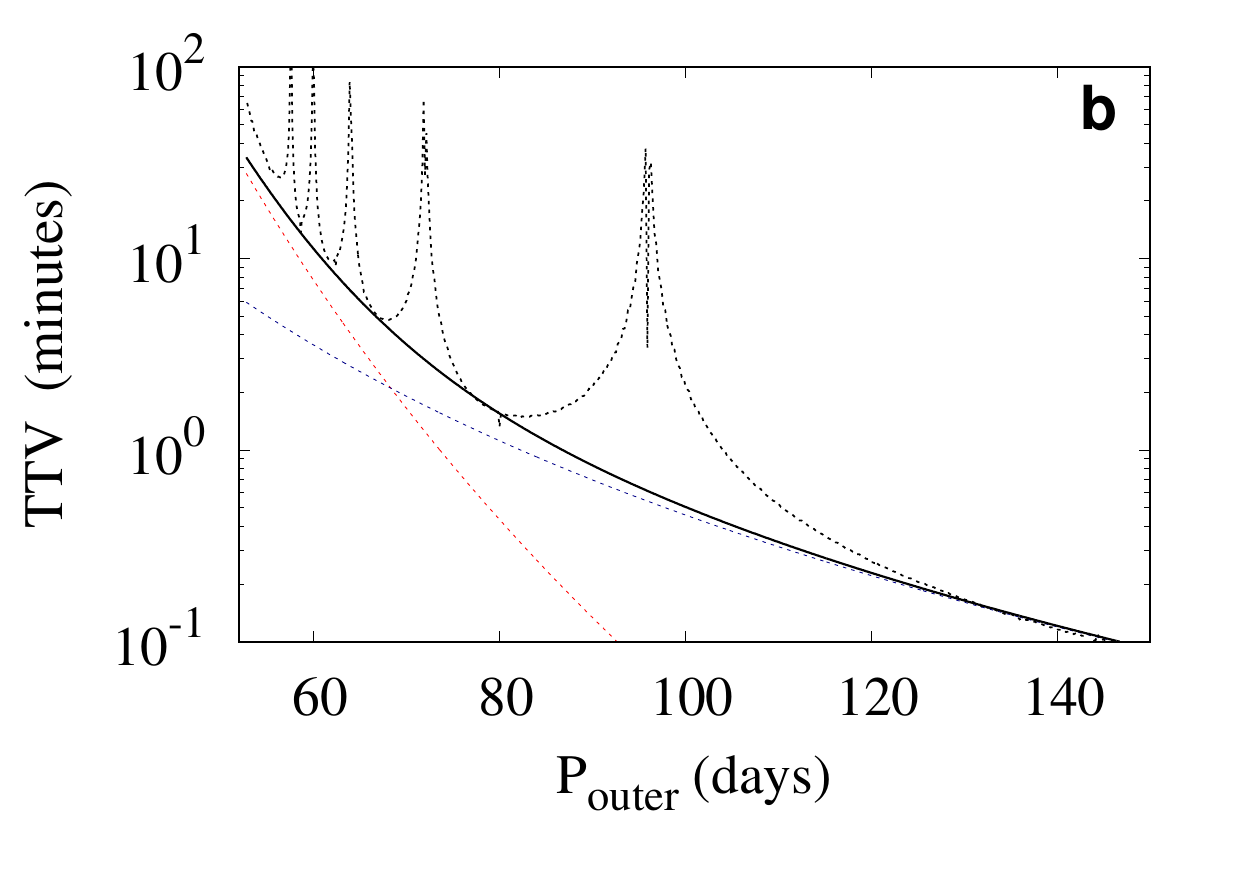}
\caption{Empirical model for the TTVs induced on an outer planet at 48 days by an inner perturber (a), anf the TTVs induced on an inner planet at 48 days by an outer perturber (b) where both bodies have dynamical mass  $\mu_{outer} = 3.00304\times 10^{-6}$ and are on circular orbits, over 15,000 days. The dashed curve marks the max - min TTV found over a four-year baseline given the detailed analytical solution of \textit{TTVfaster} \citet{Agol2016b}. The blue and red curves mark the two components of the power laws in Equations~\ref{empiricalTTV1} and ~\ref{empiricalTTV2}, and their sums are marked in solid black.}
\label{fig:TTVemp} 
\end{center}
\end{figure}

As shown in Figure~\ref{fig:TTVemp}, these empirical fits neglect the sharp peaks in TTV amplitudes near resonance, but they trace the overall trend in expected TTVs for a planet perturbed by a neighbor over a wide range of period ratios. The sharp features including drops in TTV amplitude precisely at resonance in Figure~\ref{fig:TTVemp} are due to the timescale for planetary perturbations to accumulate to detectable TTVs becoming greater than the 15,000 days used to model the TTVs in \textit{TTVfaster}. 

We used this empirical model to identify systems from \textit{Kepler} that ought to have significant TTVs even if they are not in or near resonance. 

For planet pairs near (but not in) first-order mean motion resonance, TTVs are observed as a cycle over the coherence time of two orbital periods, the so-called TTV ``superperiod''. We estimated the minimum amplitude of such TTVs for two interacting planets ($V$ and $V'$) using the solution of \citet{Lith2012} and again neglecting eccentricities:

\begin{equation}
V = \frac{P \mu' f}{\pi j^{\frac23} \left( j-1 \right)^{\frac13} \Delta}, \\
\label{lithwick1}
\end{equation}
and 
\begin{equation}
V' = \frac{P' \mu g}{\pi j \Delta},
\label{lithwick2}
\end{equation}
where $j$ is an integer that identifies the nearest first-order mean motion resonance such that the ratio of orbital periods is close to $j:j-1$, $\Delta = \frac{P}{P'}\frac{j-1}{j}-1$ is the distance of the planet pair from exact resonance, and $f$ and $g$ are sums of Laplace coefficients that depend on the orbital period ratio \citep{Lith2012}. We defined the region in period ratios near first-order resonance as bounded by the nearest third-order mean motion resonances. For example, we estimated the resonant TTV score of all planet pairs near 2:1 such that the ratio of periods lies in the range [7:4, 5:2]. We included near-first-order resonances up to $j = 5$ and only included planet pairs for which $\Delta < 0.1$. 

For each planet in our sample, we estimated the minimum signal-to-noise ratio (S/N) of the TTVs in the non-resonant regime (using Equations~\ref{empiricalTTV1} and ~\ref{empiricalTTV2}), and the resonant regime (using Equations~\ref{lithwick1}, and \ref{lithwick2}). For both, we divided the TTV amplitude by the median transit-timing uncertainty for the planet as reported in the TTV catalog of \citet{rowe15a} and multiplied by the square root of the number of transits. Henceforth, we call these criteria for the expected TTVs of a system the ``nonresonant detectability score'' and ``resonant detectability score'' respectively. 

In assessing the expectation of near-resonant TTVs, we included all adjacent pairs among \textit{Kepler}'s multis as well as pairs with one intermediate neighbor. For the nonresonant pairs, we estimated the detectability score of adjacent pairs only. We include only multitransiting planetary systems of four or fewer planetary candidates, since higher-multiplicity systems require significantly more computational time for each simulation and have more parameters to explore.

\begin{figure}[!h]
\begin{center}
\includegraphics [height = 2.5 in]{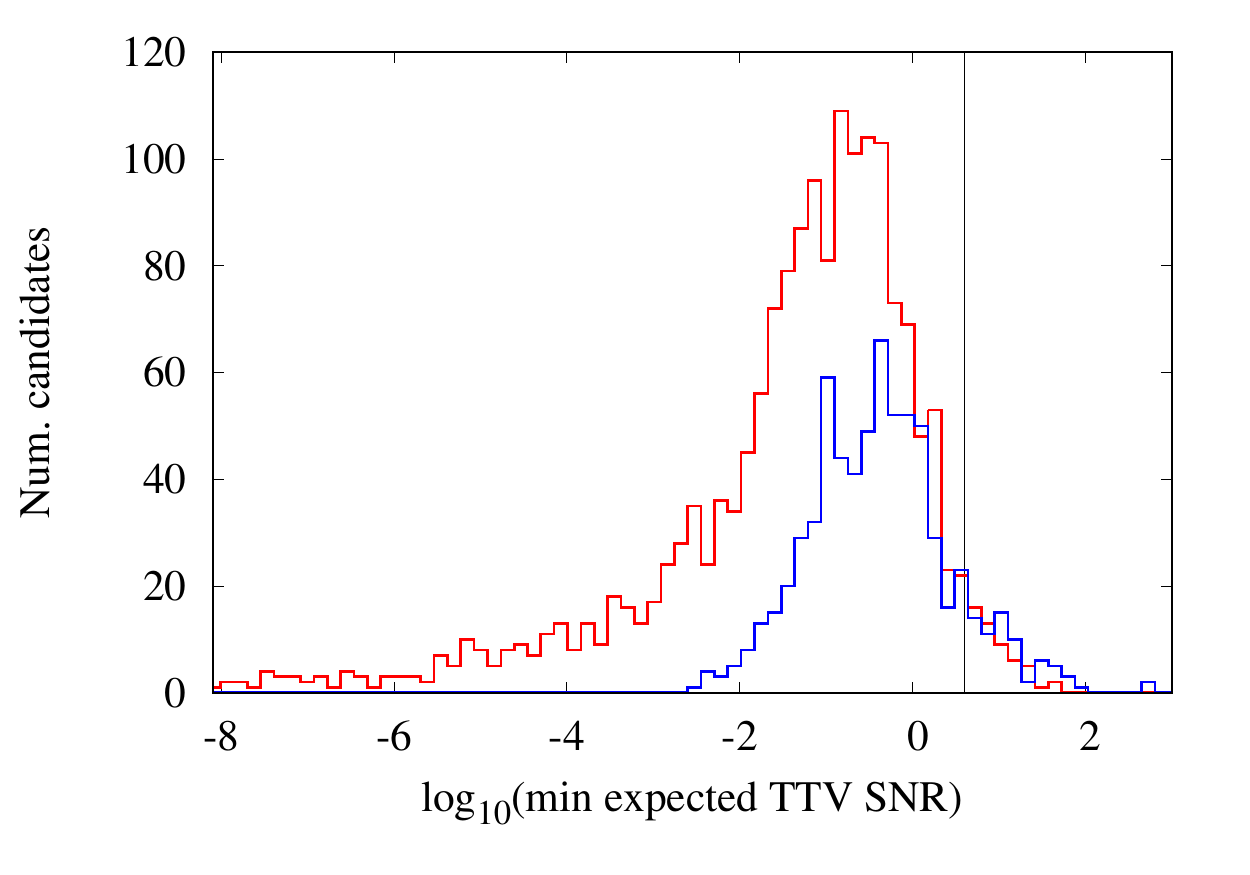}
\caption{Histogram of minimum expected TTV S/N for nonresonant (red) and resonant TTVs (blue) in multiples with up to four candidates, with known periods and radii, assuming circular orbits and the mass-radius relation of Equation~\ref{m-r-simple}. Note the logarithmic scale. Planet candidates with an expected S/N $\geq$4 (marked as a black line on the histogram) were selected for further analysis. 
\label{fig:histo-snr}} 
\end{center}
\end{figure}

In cases where the transit times of a particular planet candidate were not available in the \citet{rowe15a} catalog, we did not calculate an expected TTV score, with the exception of Kepler-289 (KOI-1353), which has a confirmed planet without a KOI number \citep{Schmitt2014b}. All multitransiting systems where at least one planet scores $>$4 on either test are included in our sample. The bar is set low for this criterion to ensure that we include nondetections in our target list. Note that our list excludes several systems with detected TTVs since the planetary masses may well be higher than our expected minimum masses, and our criterion is chosen to ensure that our list of systems is not overwhelmingly dominated by nondetections. Figure~\ref{fig:histo-snr} shows a histogram of the expected minimum TTV S/N for all candidates, and it highlights how unlikely TTVs are among planet pairs within \textit{Kepler}'s multiplanet systems. 

Our search for candidates with an expectation of TTVs was from a sample that includes 680 planetary candidates in near-resonant pairs and 849 planetary candidates in nonresonant pairs, with substantial overlap between these two lists. From these lists, we identified 57 planetary candidates with a nonresonant detectability score greater than 4, and 72 planetary candidates with a resonant detectability score above 4. Among these, some were discarded if no satisfactory preliminary fits were found, as explained in detail in the next section. The scores for systems that were ultimately included for further analysis are listed in Tables~\ref{tbl-SystemSelection1} and~\ref{tbl-SystemSelection2}.

\section{Preliminary TTV model fitting}
We performed preliminary dynamical fits against cataloged transit-timing data for our selected systems. Our dynamical models included five parameters for each planet: the orbital period, the time of the first model transit after epoch BJD-2,455,680.0, the planet-star mass ratio, and the eccentricity vector components $e\cos\omega$ and $e\sin\omega$. We assumed orbits are coplanar since mutual inclinations that are significant enough to cause detectable TTVs are unlikely in multitransiting systems (\citealt{fab14,nes14}).

In many systems, one or more planets were on short orbital periods or had a large period ratio with their interacting neighbors and hence likely contribute little to the TTVs. For consistency, we included these planets in our TTV models if their transit times were available, since their mass upper limits may still be informative. 

In some cases, the transit-timing catalogs of \citet{rowe15a} and \citet{Holczer2016} lack data on candidates that are potentially interacting. For KOI-1574 (Kepler-87), we relied on the measured transit times of \citet{ofir14}, treating it as a three-planet system with candidates orbiting at 5.8, 114.7 and 191.2 days. A fourth candidate discovered in \textit{Kepler} DR 24 at 8.98 days is potentially interacting with the innermost planet, but is unlikely to affect the transit times of the outer two planets. For this paper, we measured long-cadence transit times for KOIs: 520, 750, 1353 and 3503. We have only measured long cadence times for consistency with other systems studied in this paper, and because long cadence transit-timing uncertainties were used in the selection of our sample. We leave more detailed studies of individual systems of interest that we identify in this paper to future authors.

\textit{Planet Hunters} discovered a transiting planet at KOI-1353 (PH3 c), which is confirmed as Kepler-289 d but has no candidate number \citep{Schmitt2014b}. A fourth candidate designated as KOI-1353.03 is an alias of Kepler-289 d and hence a false positive. 

We excluded the potentially false multiplanet system KOI-284 which has transiting planets orbiting at 6.2 and 6.4 days. This may be a binary system with planets orbiting separate stars with similar orbital periods \citep{liss14}. KOI-521 and KOI-3444 are flagged as potentially binaries in \textit{ExoFOP} and we exclude them from our analysis. KOI-750.04 is marked as a false positive in \textit{Kepler} DR 25. We exclude it and treat the remaining candidates as a three-planet system.

To identify models that closely match the data, we initialized orbital periods and phases assuming a linear ephemeris, the stellar mass at 1 M$_{\odot}$ and planetary masses at 6 M$_\oplus$. We initialized eccentricities at 0.001 and performed a grid-search in eccentricity vector components, with periapses initialized at $\pm 45^{\circ}$ and  $\pm 135^{\circ}$ for each planet. We performed Levenberg-Marquardt minimization of the goodness-of-fit parameter, $\chi^2$ and noted the best-fit model for each system. We identified systems for which we could not find a satisfactory preliminary model. We excluded systems where the best-fit model had a reduced $\chi^2 > 2.6$, where the reduced $\chi^2$ is the goodness-of-fit divided by the degrees of freedom in the model, (number of transits minus the number of free model parameters). The excluded systems were KOI-94 (red. $\chi^2$ = 3.3), KOI-262 (red. $\chi^2$ = 2.8);  KOI-312 (red. $\chi^2$ = 2.7); KOI-880 (red. $\chi^2$ = 7.1),  KOI-1236 (red. $\chi^2$ = 5.5),  KOI-1426 (red. $\chi^2$ = 16), KOI-1525 (red. $\chi^2$ = 10), KOI-1858 (red. $\chi^2$ = 4.0), KOI-2038 (red. $\chi^2$ = 3.1), KOI-2173 (red. $\chi^2$ = 3.8) and KOI-2672 (red. $\chi^2$ = 6.0). 

KOI-3791 (Kepler-460) has two planets and just 10 measured transit times, leaving zero degrees of freedom for our model fits. We exclude this system from our sample.  

\begin{table}[ht!]
  \begin{center}
  \tiny
    \begin{tabular}{|c|c|c|c|c|}
      \hline
     name &  Non-resonant TTV Score & Resonant TTV Score &  $\min \chi^2_{red}$   \\
 \hline 
  KOI- 85.01,  Kepler-65 c     &  7.3    &   --- &  1.7   \\ 
  KOI- 115.01, Kepler-105 b   &  8.6  &  5.0  &  1.5   \\ 
  KOI- 115.02, Kepler-105 c   &  4.4  &  2.7  &  1.5   \\ 
  KOI- 137.01, Kepler-18 c & 3.2  & 8.6 & 1.5  \\  
  KOI- 137.02, Kepler-18 d & 2.1 & 5.5 & 1.5 \\  
  KOI-152.01, Kepler-79 d    &  15.1 &  7.8  &  1.2  \\ 
   KOI-156.01, Kepler-114 c &  4.9 &  3.3  &  2.2    \\ 
   KOI- 156.03, Kepler- 114 d  &  10.2  & 9.5  &   2.2   \\ 
   KOI-168.01, Kepler-23 c  &  2.9  & 4.7 &  1.4   \\ 
   KOI- 222.01, Kepler-120 b &  0.9 & 5.7 &  1.4    \\ 
  KOI-  244.01, Kepler-25 c   &  1.8  & 12.5  &  1.1  \\ 
  KOI-  244.02, Kepler-25 b   & 1.2  & 5.6  &  1.1  \\ 
  KOI-  248.01, Kepler- 49 b &  5.8  & 10.0 &  1.7  \\ 
  KOI-  248.02, Kepler- 49 c &  2.2  & 8.5  &  1.7  \\
   KOI- 250.01, Kepler-26 b    &  21.6  &  2.9  & 1.5  \\
   KOI-255.01, Kepler-505 b &    0.2  & 11.0 &  1.4  \\ 
   KOI- 277.02, Kepler-36 c   & 50.2  & --- & 2.6  \\ 
  KOI- 277.02, Kepler-36 b    & 6.1  & --- & 2.6  \\ 
 KOI-314.01, Kepler-138 c &  24.2   &  8.6 &  1.9 \\
 KOI-314.02, Kepler-138 d &  4.2   &  --- &  1.9 \\ 
  KOI-314.03, Kepler-138 b &  4.5   & 15.6 &  1.9 \\ 
  KOI-  377.01,  Kepler-9 b   &   6.6   &  45.7   & 1.9 \\
   KOI-  377.02,  Kepler-9 c   &   2.2   &  12.7   & 1.9 \\
 KOI- 401.01,  Kepler-149 b   & 3.9  & 5.1 & 1.0   \\ 
KOI-430.01, Kepler-551 b  &  3.2 & 17.9 &   1.5  \\ 
KOI-457.01, Kepler-161 b & 4.0 & --- & 1.4 \\ 
 KOI- 520.01, Kepler- 176 c    & 0.76  &  6.8 &  1.3  \\ 
 KOI- 520.03, Kepler- 176 d    & 0.56  &  7.8 &  1.3  \\
KOI-  523.01, Kepler- 177 c & 36.3 & 40.6  &  1.5  \\ 
KOI-  523.02, Kepler- 177 b  & 8.5 & 12.1  &  1.5  \\ 
 KOI-  567.02, Kepler- 184 c  & 4.7 &  1.4 & 1.4  \\ 
  KOI-  567.03, Kepler- 184 d  & 5.0 &  2.0 & 1.4  \\ 
  KOI-  620.01,  Kepler-51 b  &  10.6  &  13.0  &  1.9  \\ 
 KOI-  620.02,  Kepler-51 d  &  44.5  &  75.6  &  1.9  \\ 
 KOI-  620.03,  Kepler-51 c  &  11.9  &  12.6  &  1.9 \\
KOI-  654.01, Kepler-200 b   &   6.9  &  --- &  2.1   \\ 
KOI-  654.02, Kepler-200 c   &   9.0  &  --- &  2.1   \\ 
   KOI- 730.01, Kepler- 223 d &  1.8 & 29.2  & 1.5  \\ 
   KOI- 730.02, Kepler- 223 c &  0.4 & 27.0  & 1.5    \\
   KOI- 730.03, Kepler- 223 e &  1.3 & 26.2  & 1.5   \\
   KOI- 730.04, Kepler- 223 b &  0.7 & 20.1  & 1.5   \\
  \hline
     \end{tabular}
    \caption{Identifying systems for TTV modeling: Part 1 of 2. The first column identifies the transiting planet with the highest expected TTV signal in its system, by KOI number and Kepler number if available. The second and third column respectively list the best scores for expected chopping  ``non-resonant" TTV signal  and scores for expected TTV signals from near first-order resonances ``resonant signal'' given the planetary orbital periods, sizes, stellar parameters, median transit-timing uncertainties and number of measured transit times. The next column lists the reduced $\chi^2$ for the best fit model we have found, including all transiting exoplanets with cataloged transit times. Systems with a best fit reduced $\chi^2 > 2.5$ are excluded from this table but are listed in the text.}\label{tbl-SystemSelection1}
  \end{center}
 \end{table}

 \begin{table}[ht!]
  \begin{center}
  \tiny
    \begin{tabular}{|c|c|c|c|}
      \hline
      KOI &  Non-resonant TTV Score & Resonant TTV Score &  $\min \chi^2_{red}$ \\
 \hline
 KOI- 738.01, Kepler-29 b  & 8.0  & ---  &  2.1    \\ 
  KOI- 738.02, Kepler-29 c  & 8.0  & ---  &  2.1    \\ 
 KOI-750.01, Kepler-662 b & 2.7 & 15.0 & 1.2 \\  
 KOI- 806 .02,  Kepler- 30 c   &  6.1   &  6.0 &  2.1  \\
    KOI- 877.01 , Kepler- 81 b  &  1.1  & 7.2 & 1.5   \\ 
      KOI- 886.01,  Kepler- 54 b & 1.1  & 6.8  & 1.6   \\ 
      KOI- 886.02,  Kepler- 54 c & 0.3  & 5.6  & 1.6   \\ 
   KOI- 934.03, Kepler- 254 d  &  0.4  & 5.0 &  1.5    \\ 
      KOI-1070.03 & 11.8 & --- & 1.7  \\ 
  KOI-1279.01, Kepler-804 b & 0.8 & 4.1  & 1.2 \\ 
  KOI-1338 .02 & 0.06  & 12.8 & 1.6  \\ 
  KOI-1338 .03 & 0.15 & 33.3 &  1.6 \\ 
   Kepler-289 d & 3.2 & 4.9 & 1.2  \\ 
 KOI-1353.01, Kepler-289 c & 9.4 & 14.2 & 1.2  \\ 
KOI-1574.01, Kepler-87 b & 19.1  &  17.1 & 1.5  \\  
KOI-1576.01, Kepler-307 b & 14.2 & 13.4 & 2.2  \\ 
KOI-1576.02, Kepler-307 c & 14.1 & 9.4 & 2.2  \\ 
 KOI- 1598.01, Kepler- 310 c    & 6.1  & ---  &  1.2  \\ 
 KOI- 1599.01, Kepler-1659 c  &  0.5  & 9.7 & 2.2   \\  
 KOI- 1599.02, Kepler-1659  b & 0.5  & 17.2 & 2.2    \\  
 KOI- 1783.01  &  8.0  & 13.3 & 0.7   \\ 
 KOI- 1831.01, Kepler- 324 c   &  2.1  & 6.0  & 1.5     \\ 
  KOI- 1833.02  & 10.4  & 9.8  &  1.6    \\  
 KOI- 1833.03   & 5.3  & 4.4  &  1.6    \\ 
KOI- 1955.02, Kepler- 342 c & 1.3  & 12.4 &  1.3    \\ 
KOI- 1955.04, Kepler- 342 d & 1.8  & 10.6 &  1.3    \\
KOI- 2086.01,    Kepler-60 b  & 2.4  &  32.2 &  1.6  \\ 
KOI- 2086.02,    Kepler-60 c  & 2.2  &  50.2 &  1.6   \\
KOI- 2086.03,    Kepler-60 d  & 0.0  &  24.9 &  1.6   \\
KOI- 2092.01, Kepler-359 c  &  17.6  &   60.6 & 1.6   \\ 
KOI- 2092.03, Kepler-359 d  &  10.5  &   40.1 & 1.6   \\ 
KOI- 2113.01, Kepler- 417 c &  8.0 & --- &  1.1   \\ 
KOI- 2113.02, Kepler- 417 b &  6.1 & --- &  1.1   \\ 
KOI- 2174.01  & 16.4 &  --- & 1.2    \\  
 KOI- 2195.01,   Kepler- 372 c      & 1.1  & 62.8  & 1.7    \\ 
 KOI- 2195.02,   Kepler- 372 d      & 0.5  & 50.6  & 1.7    \\ 
 KOI- 2414.01,   Kepler- 384 b    & 0.3 & 7.0  &  1.5   \\ 
 KOI- 2414.02,   Kepler- 384 c    & 0.2  & 5.6  &   1.5  \\ 
KOI- 3503.01   & 0.3 & 4.3  &   1.3  \\ 
\hline
     \end{tabular}
    \caption{Identifying systems for TTV modeling: Part 2 of 2. The first column identifies the transiting planet with the highest expected TTV signal in its system, and the second and third column respectively list the best non-resonant TTV scores and resonant TTV scores within that system given the planetary orbital periods, sizes, stellar parameters, median transit-timing uncertainties and number of measured transit times. The next column list the reduced $\chi^2$ for the best fit model we have found for each system, including all transiting exoplanets with cataloged transit times. }\label{tbl-SystemSelection2}
  \end{center}
\end{table}
\clearpage
To sample the mass posteriors of these planets, we used a differential evolution Markov Chain Monte Carlo (MCMC) algorithm (\citealt{ter06, Jontof-Hutter2015, Jontof-Hutter2016}), beginning the chains close to the best-fit model found from our preliminary fitting. We adopted a uniform prior on orbital periods, initial orbital phases, and (positive, definite) planet-star mass ratios, and a Gaussian prior on eccentricity vector components with a standard deviation of 0.1. This was chosen to be wide enough to include most eccentricities typical in multitransiting systems ($e$ $\sim$ 0.032, \citealt{fab14}) but disfavors eccentricities much higher than 0.1. Higher eccentricities are unlikely since many of the models would be unstable. However, due to a degeneracy in TTVs, in many cases high eccentricities fit the data well and permit long-term stability if the orbits are apsidally aligned (\citealt{Jontof-Hutter2016,Gratia2021}). Our prior in eccentricity vector components reduces the number of high-eccentricity models in our posterior samples (compared to a uniform prior), but does not eliminate them. After burn-in, we drew samples and from our MCMC chains measured the autocorrelation length in dynamical masses to ensure 10,000 effective random samples were taken after thinning. In some cases, the sampling was slower to converge and the effective number of samples was less than 10,000; KOI-137 (effectively 2400 samples), KOI-250 (effectively 3500 samples), KOI-1338 (effectively 3000 samples), KOI-2174 (effectively 7000 samples), and KOI-2195 (effectively 2500 samples).

Our posterior samples from TTV modeling are available in electronic format with 5 columns per planet listed in the order: $\frac{m_{p}}{M_{\star}} \frac{M_{\odot}}{m_{\oplus}}$, $P$ (days), $e\sin\omega$, $e\cos\omega$, $T_{0}$, and first model transit after BJD-2455680 calculated as (BJD-2454900). The samples are made available on Zenodo: doi:10.5281/zenodo.4422053.

\section{Projected Transit Times}
From our posteriors, we measured the 15.9th and the 84.1th percentiles, and then we drew 1000 additional samples with the mass of each planet within, above and below these approximate 1$\sigma$ bounds. For each of these three subsamples, we ran dynamical models integrating the planetary motions for a total of 6000 days (from just before the start of the \textit{Kepler} mission through 2025 August 12) and took the mean and standard deviation of our model transit times at every projected transit date. This gives us an uncertainty on future transit times for our samples, which we use to determine whether projected TTVs are rapidly diverging. Systems where future transit times diverge by an amount that exceeds the transit-timing uncertainty of a follow-up observation are targets where additional data can improve mass constraints. In addition, our projections of subsamples where the masses are below or above the 1$\sigma$ credible intervals in some cases highlight where projected transit times are separated by uncertainties in dynamical masses. These are displayed in Appendix A.

\begin{figure}
\begin{center}
\figurenum{9}
\includegraphics[height = 1.45 in]{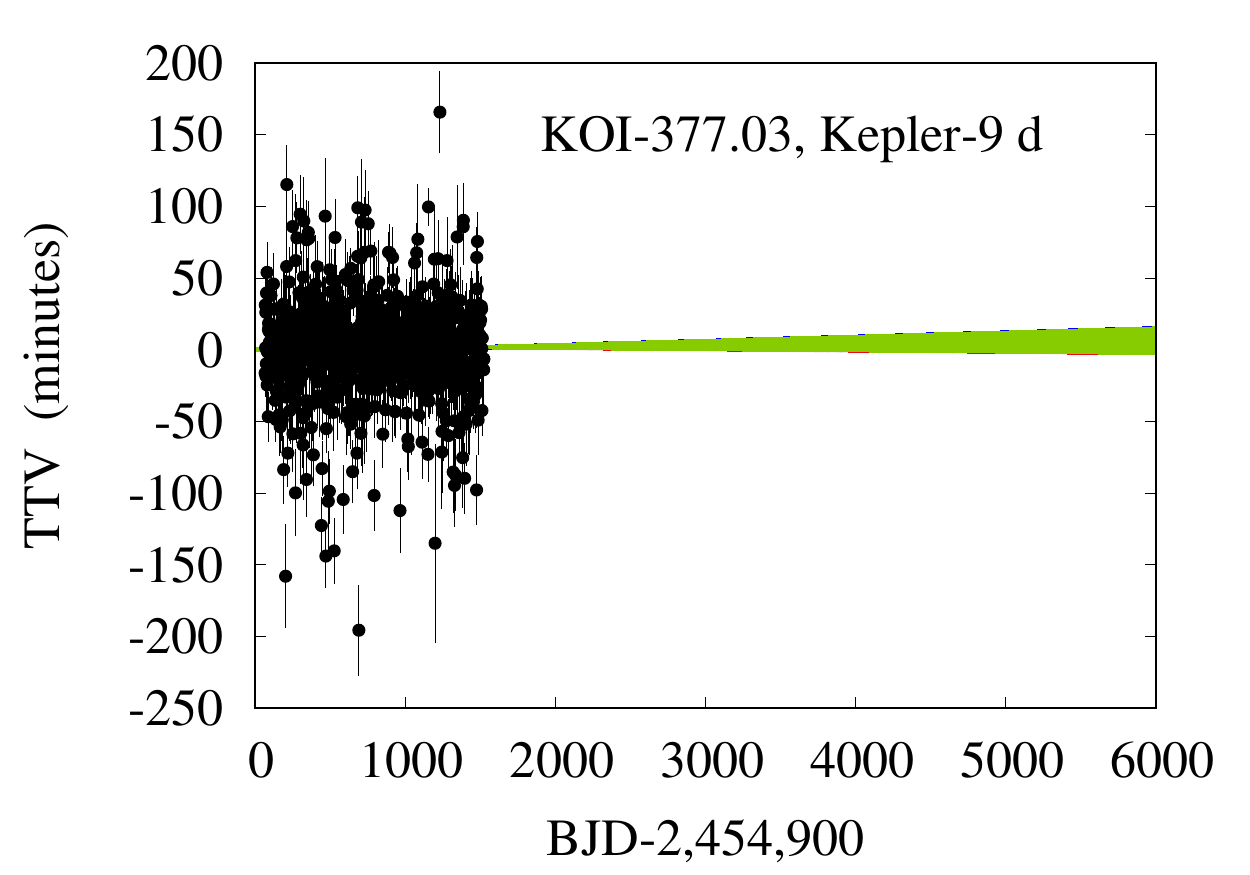}
\includegraphics[height = 1.45 in]{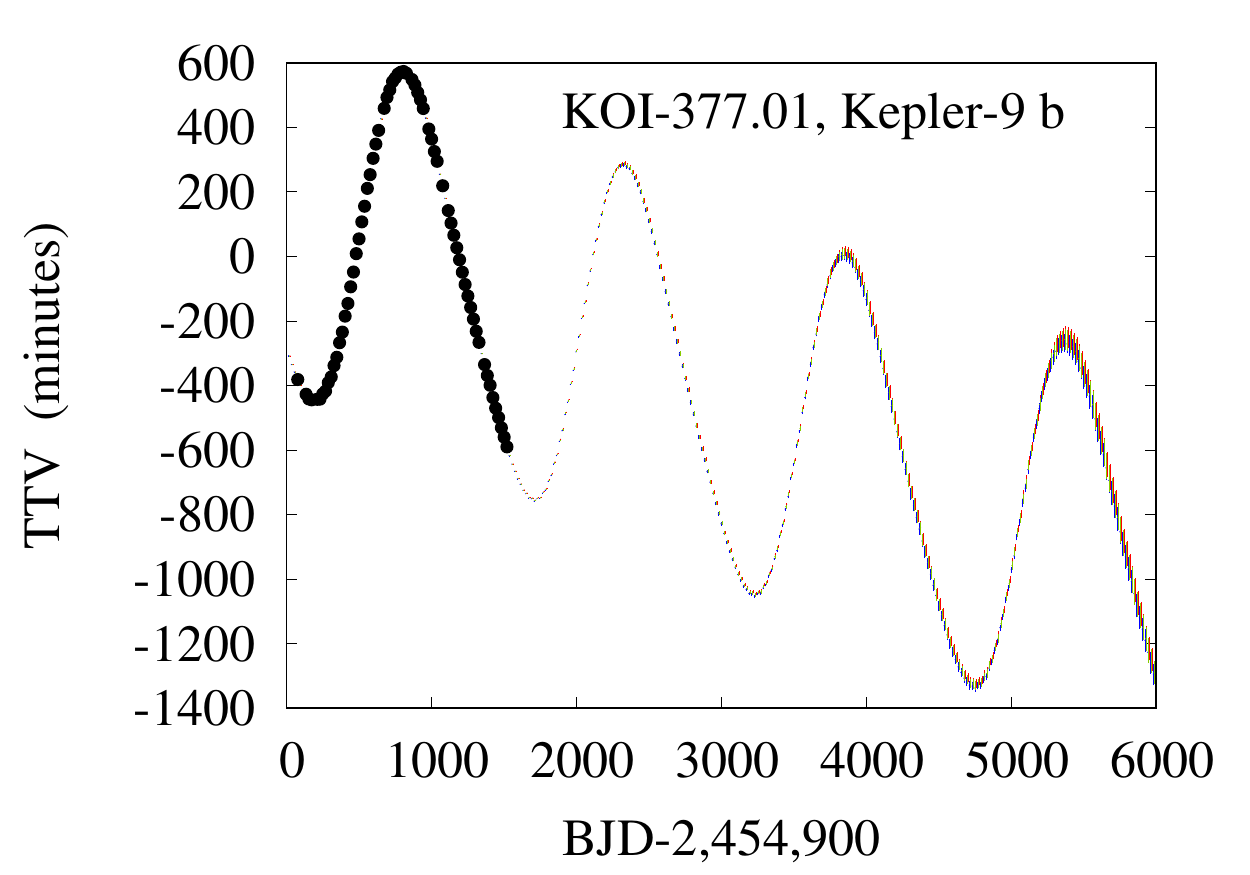}
\includegraphics[height = 1.45 in]{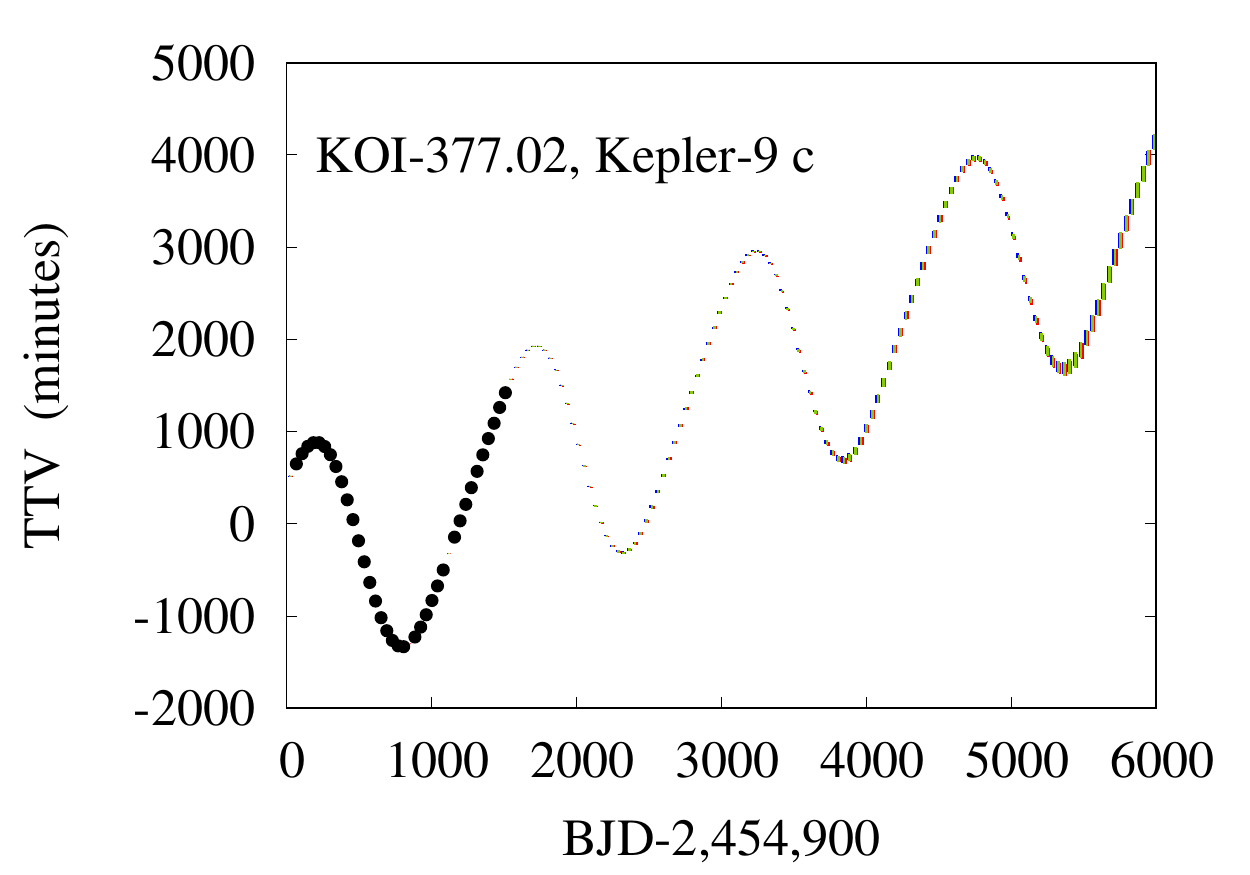}
\includegraphics[height = 1.45 in]{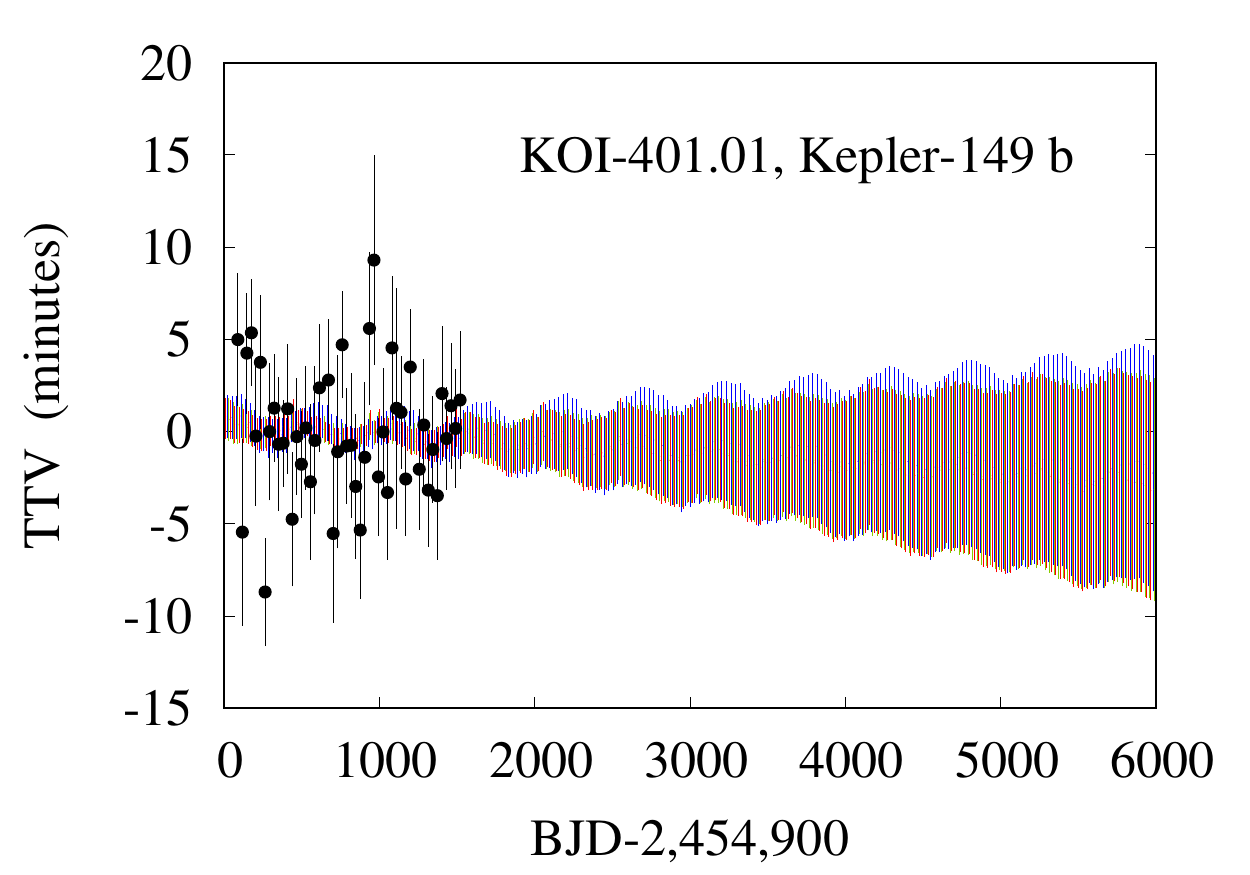}
\includegraphics[height = 1.45 in]{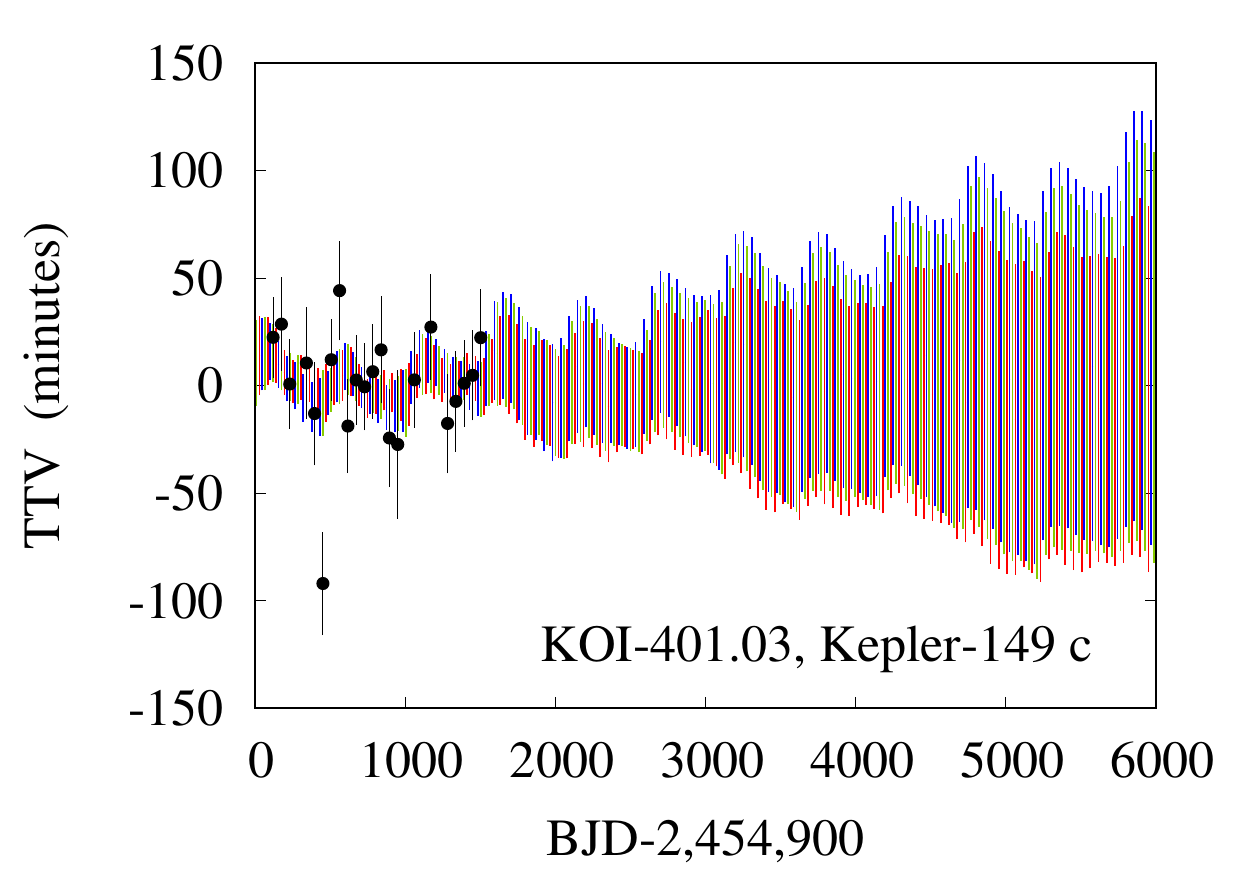}
\includegraphics[height = 1.45 in]{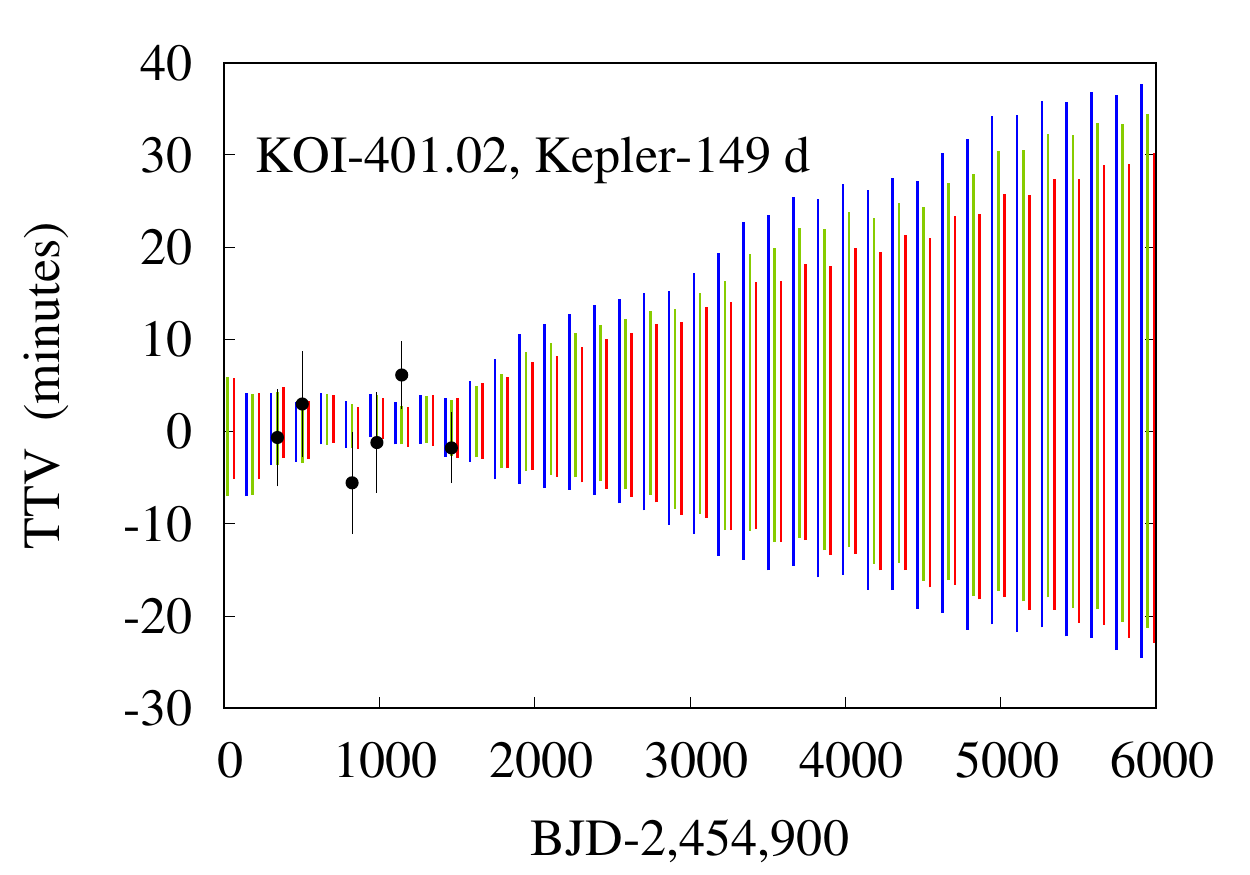}
\includegraphics[height = 1.05 in]{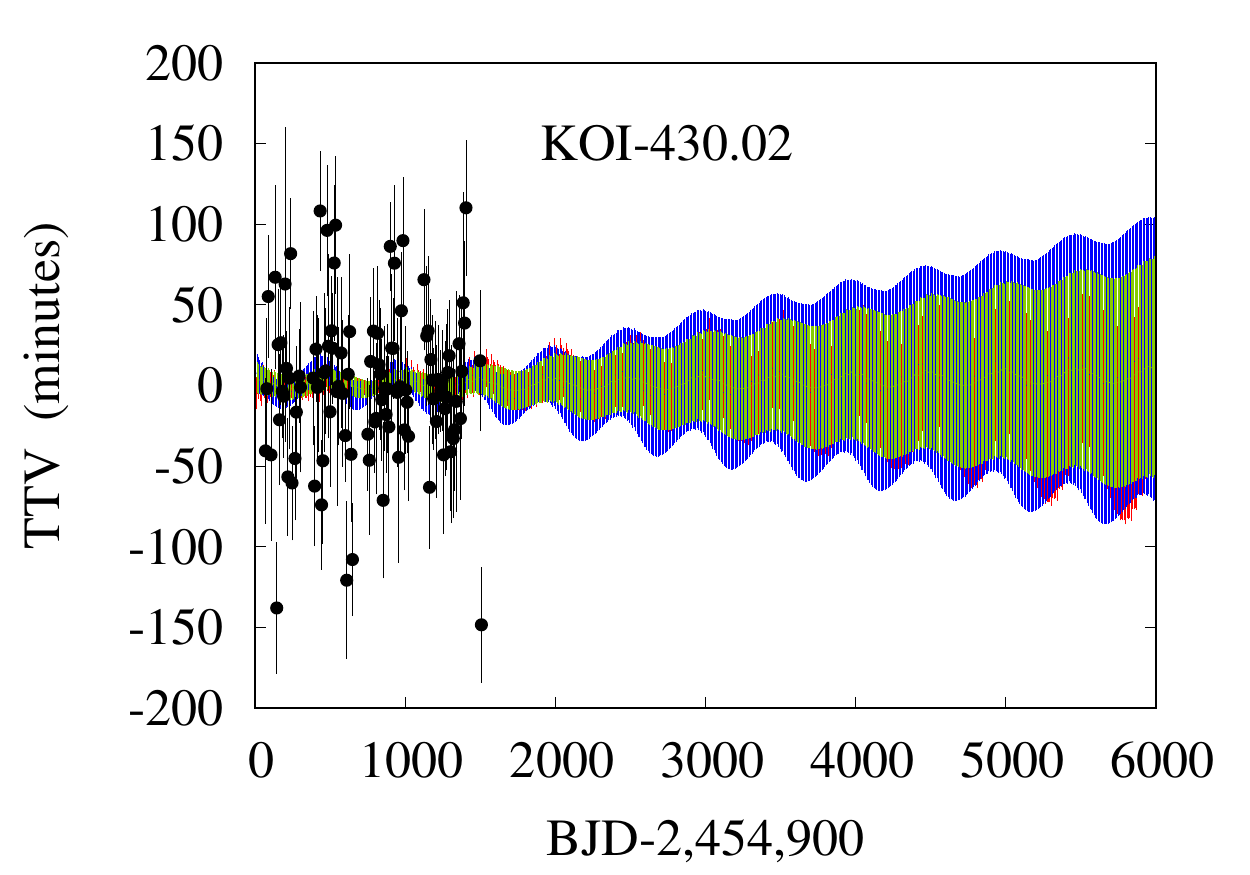}
\includegraphics[height = 1.05 in]{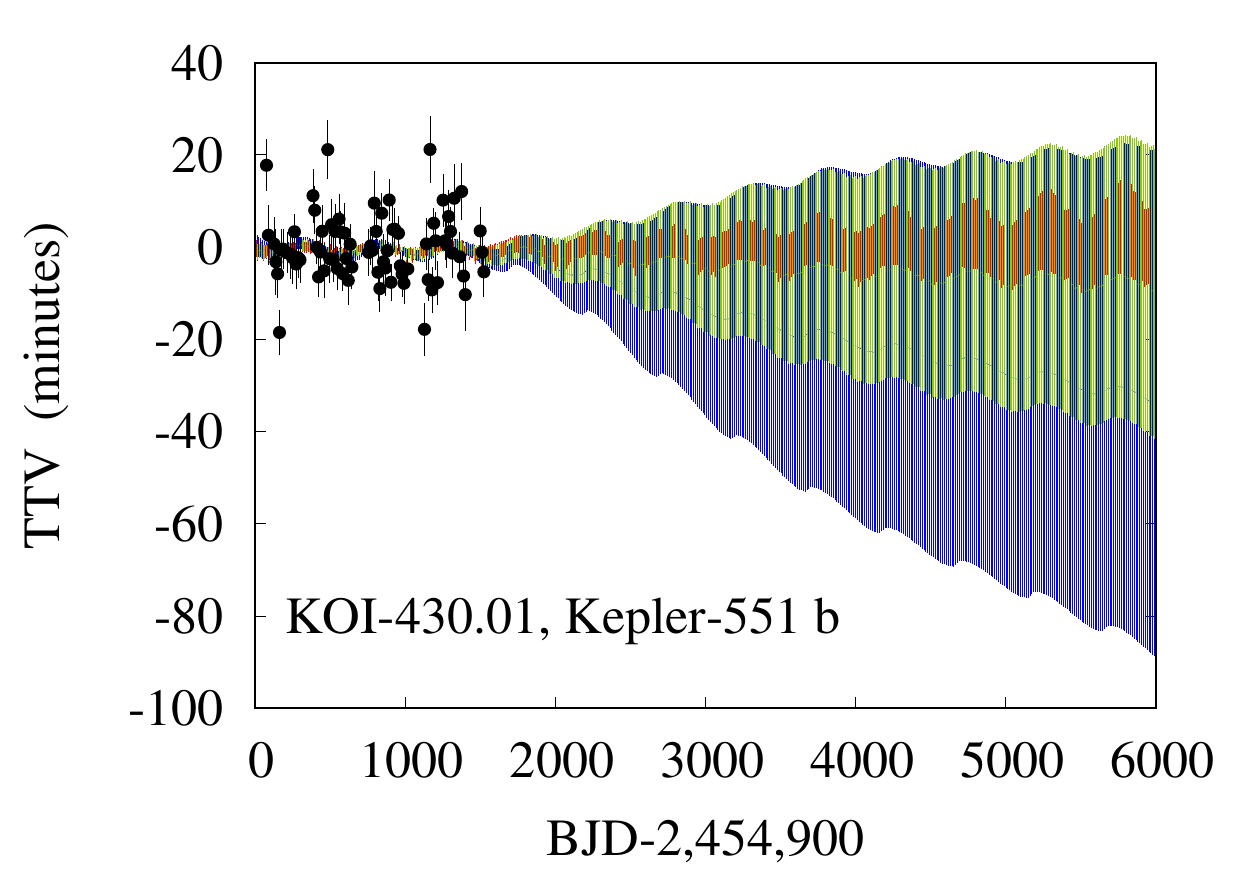}
\includegraphics[height = 1.05 in]{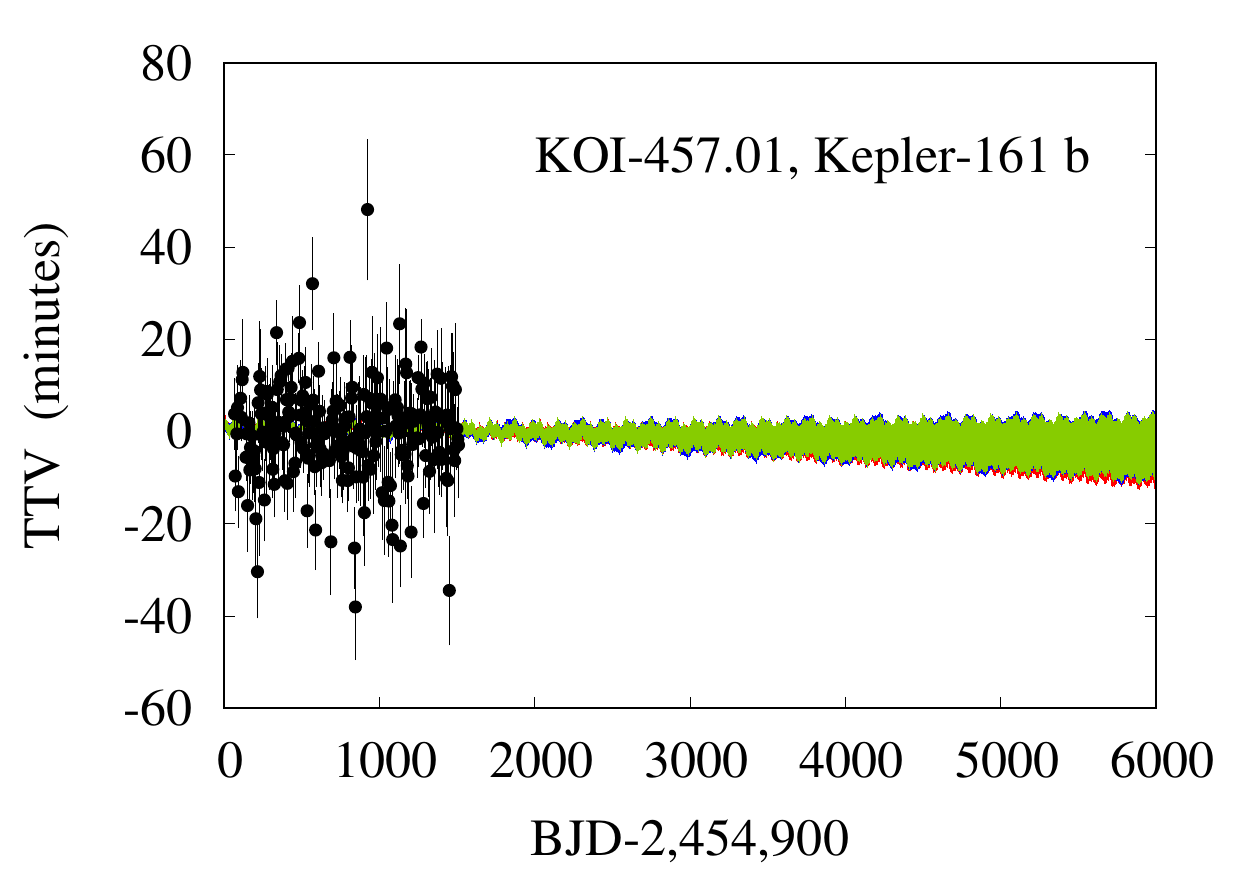}
\includegraphics[height = 1.05 in]{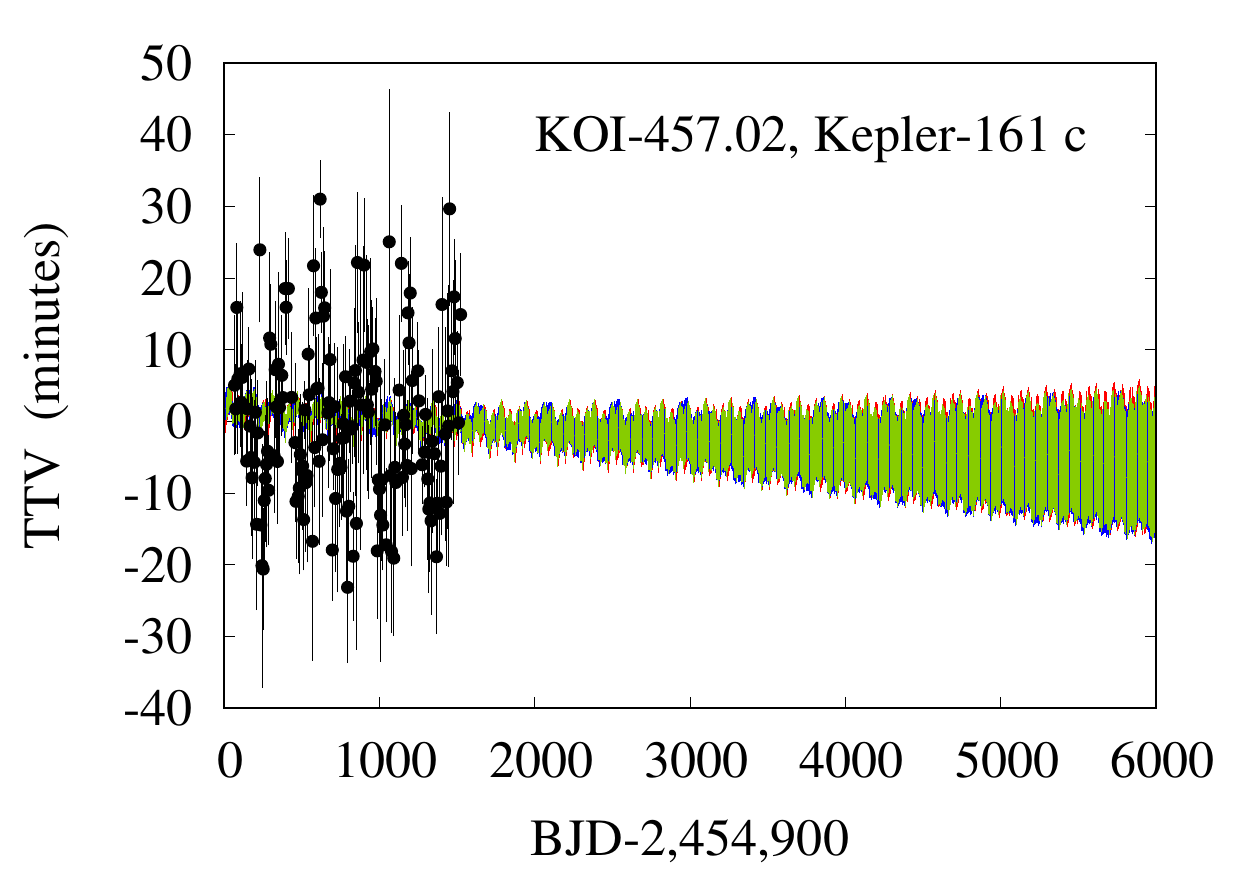}
\includegraphics[height = 1.05 in]{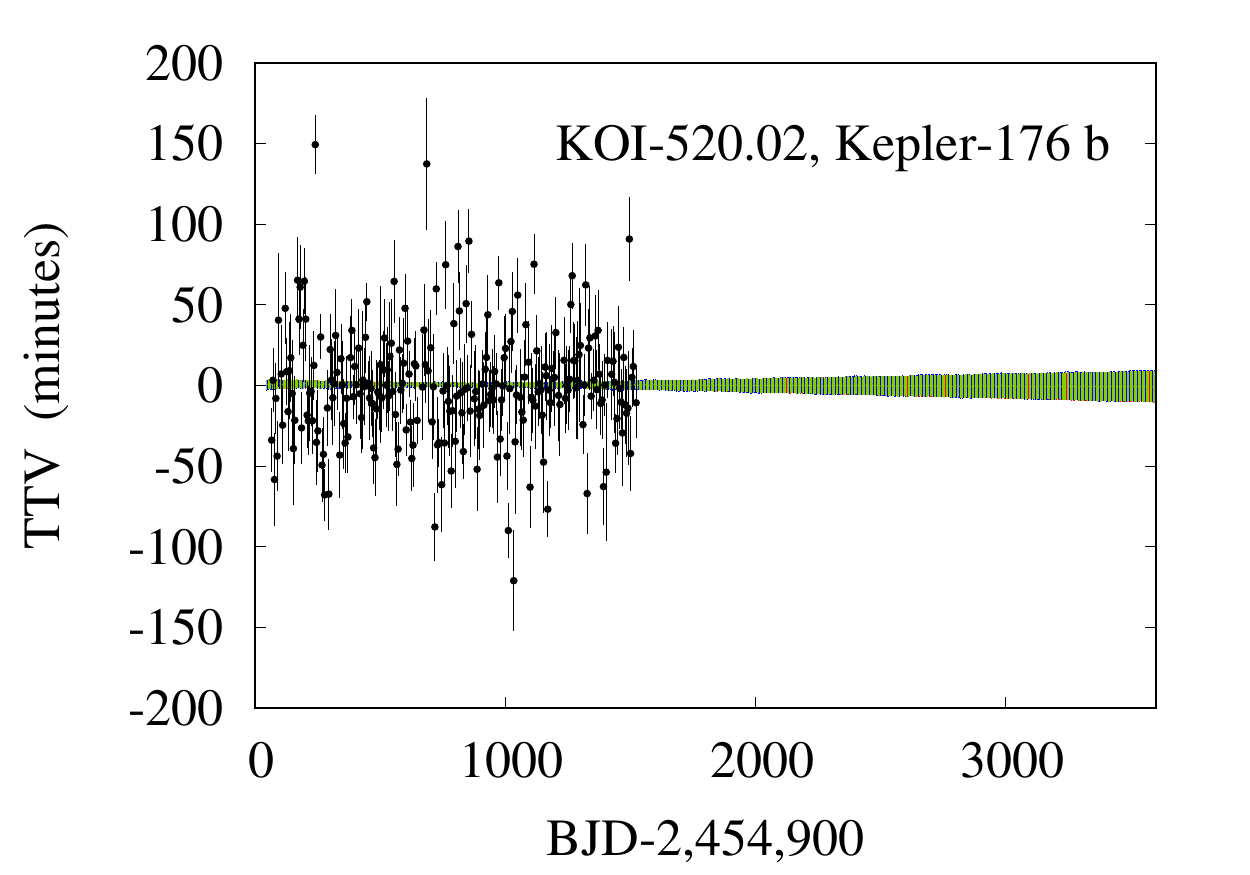}
\includegraphics[height = 1.05 in]{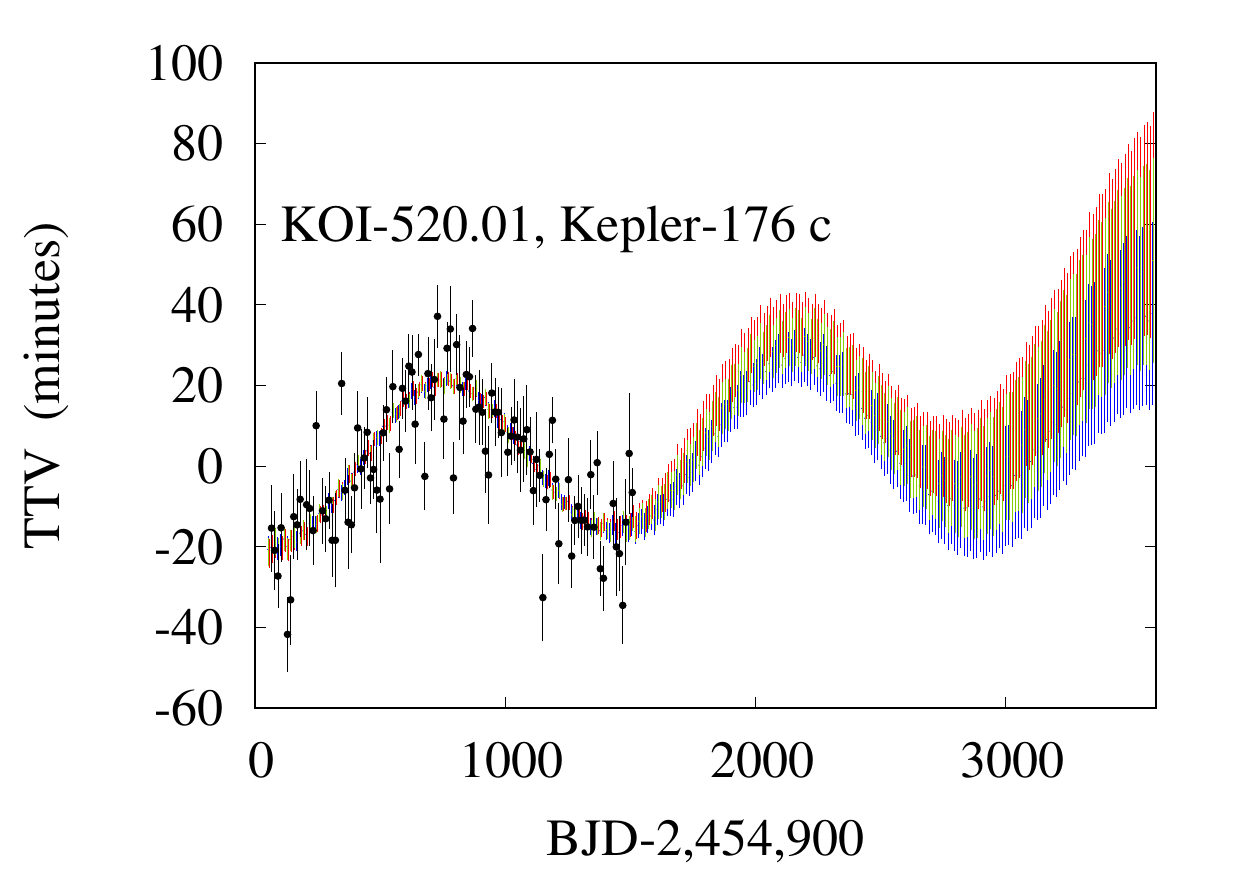}
\includegraphics[height = 1.05 in]{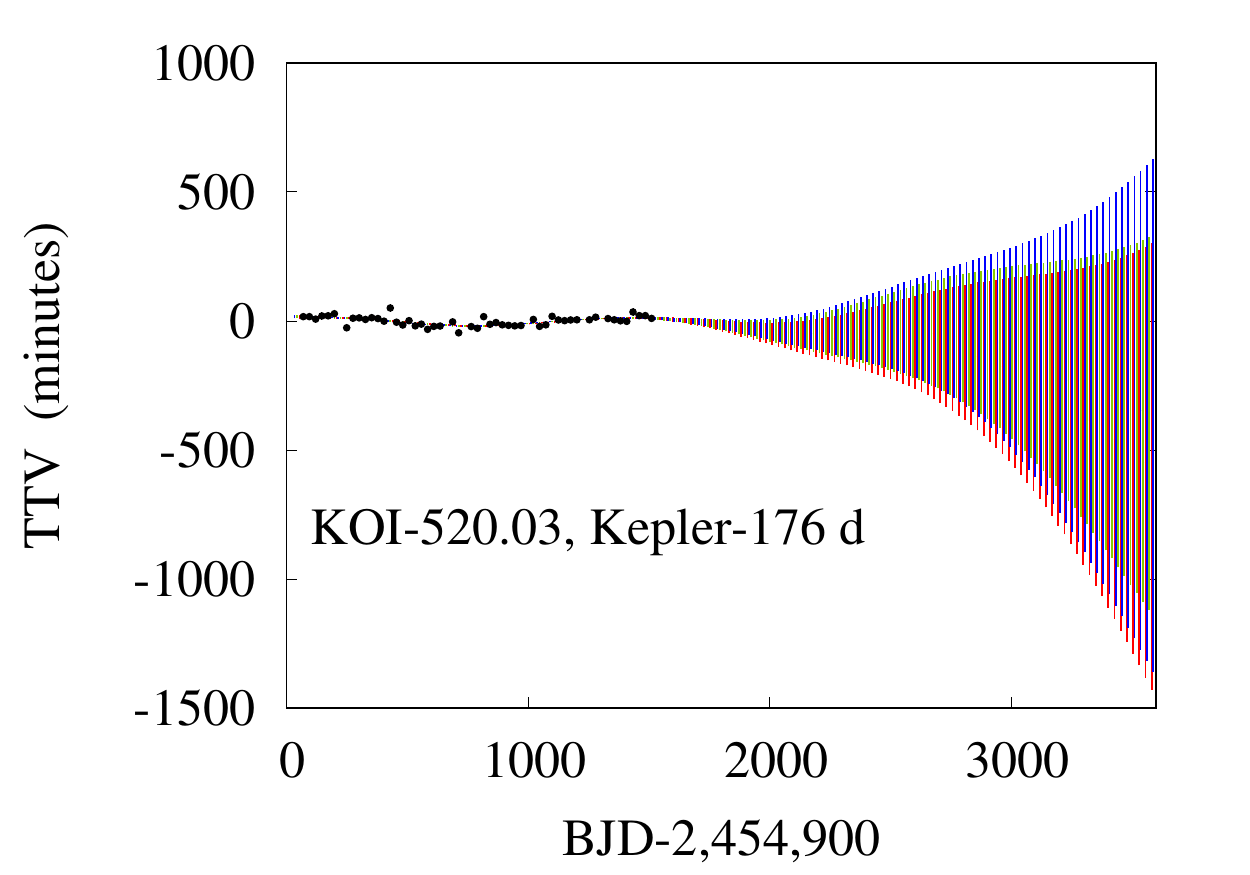}
\includegraphics[height = 1.05 in]{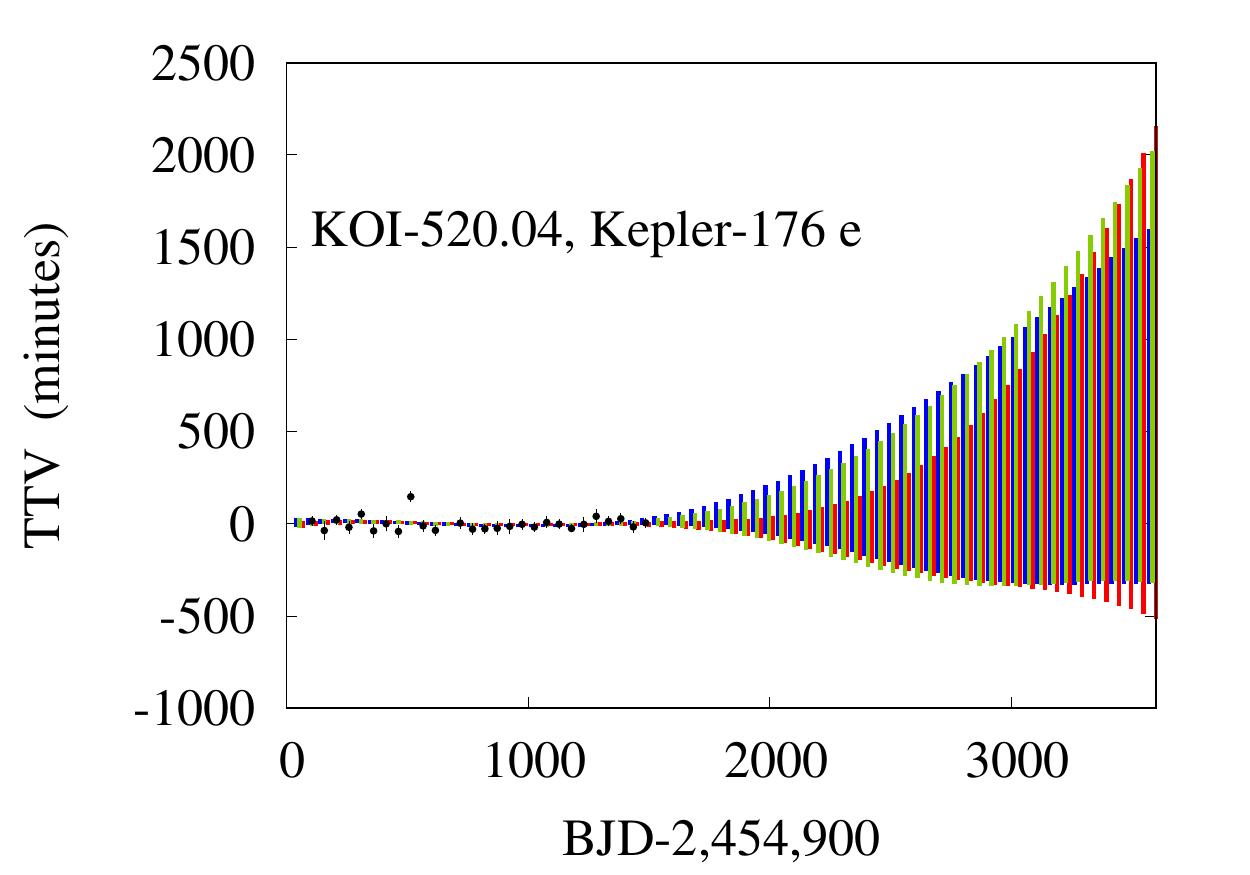}
\caption{Distribution of projected transit times as a function of time for the planet labelled in each panel (part 3). Black points mark transit times in the catalog of \citet{rowe15a} with 1$\sigma$ error bars. In green are 68.3\% confidence intervals of simulated transit times from posterior sampling. In blue (red), are a subset of samples with dynamical masses below (above) the 15.9th (84.1th) percentile. \label{fig:KOI-377fut}}
\end{center}
\end{figure}

\begin{figure}
\begin{center}
\figurenum{12}
\includegraphics [height = 1.45 in]{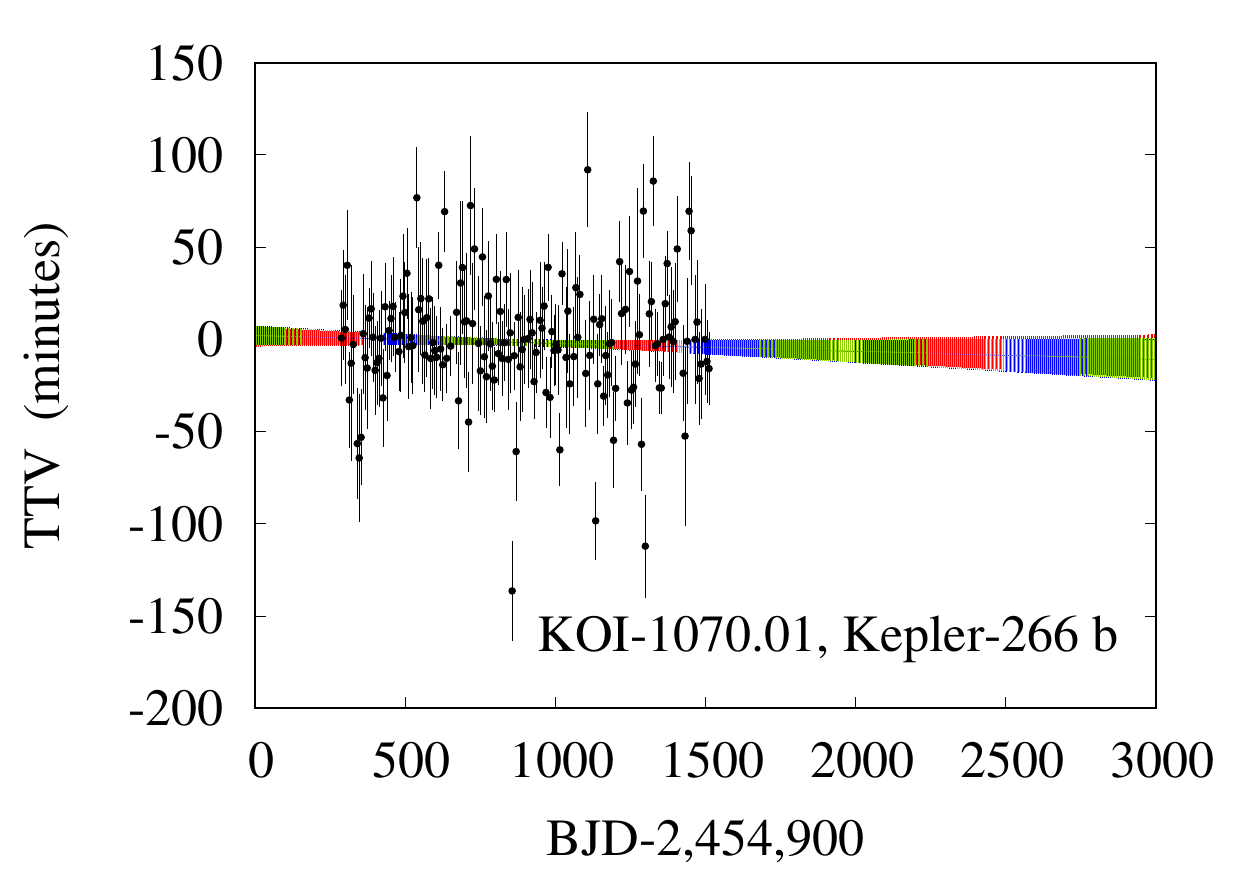}
\includegraphics [height = 1.45 in]{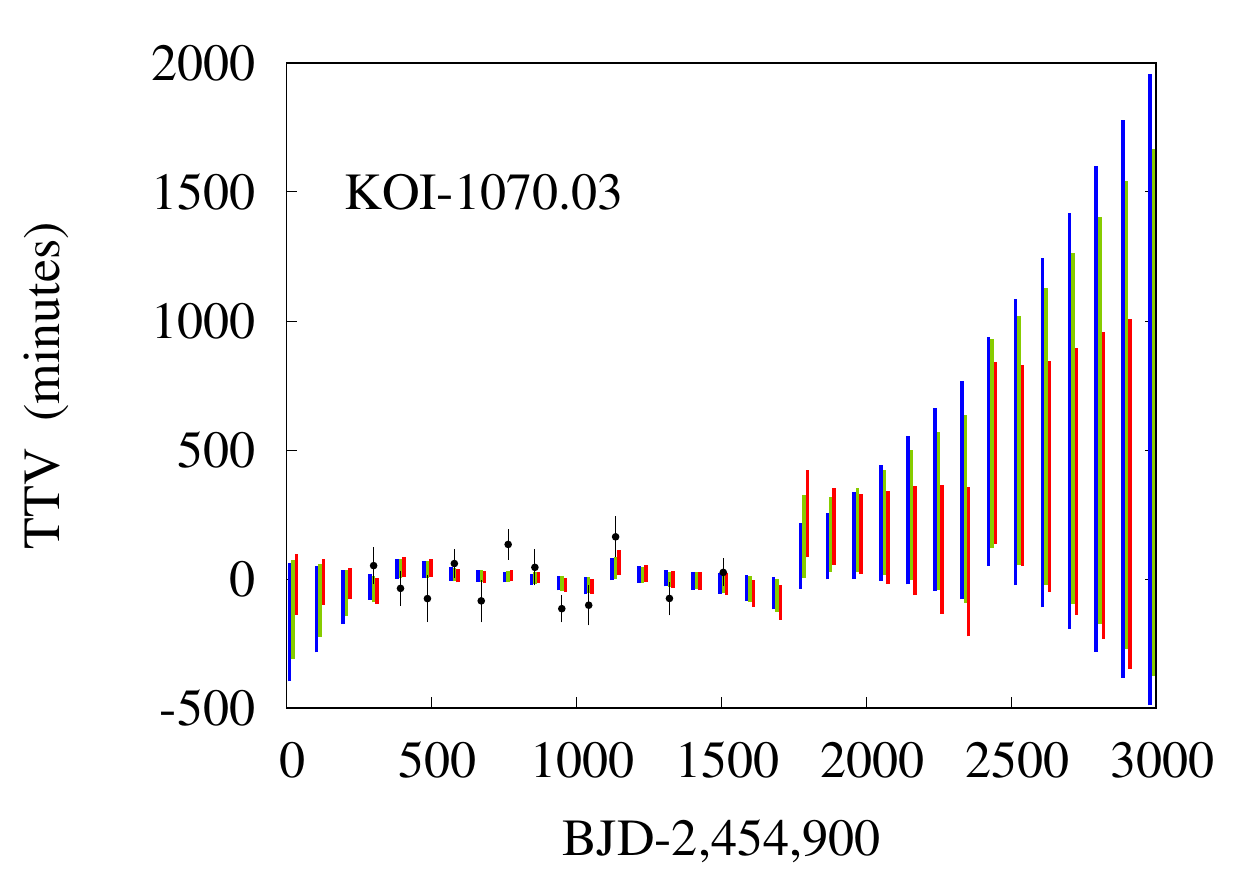}
\includegraphics [height = 1.45 in]{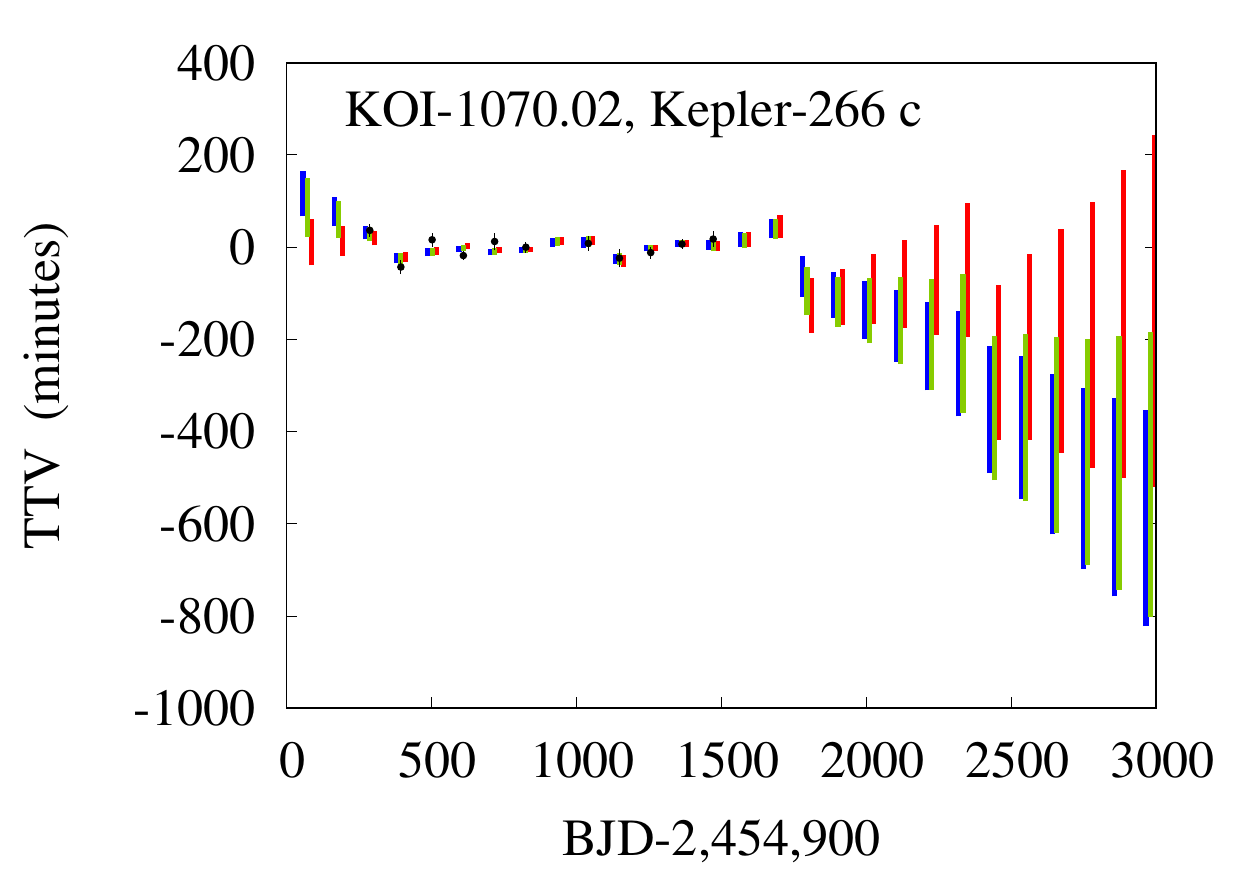}
\includegraphics[height = 1.45 in]{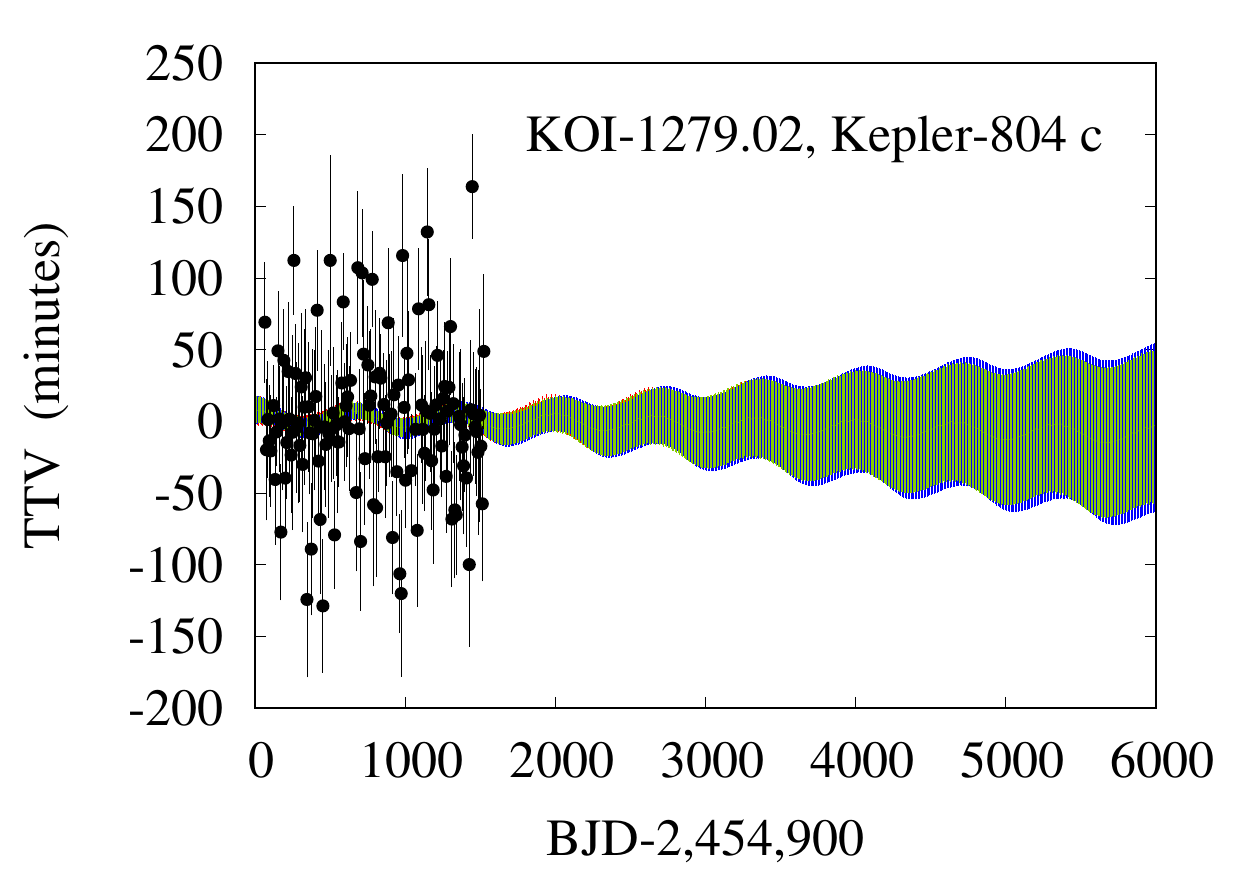}
\includegraphics[height = 1.45 in]{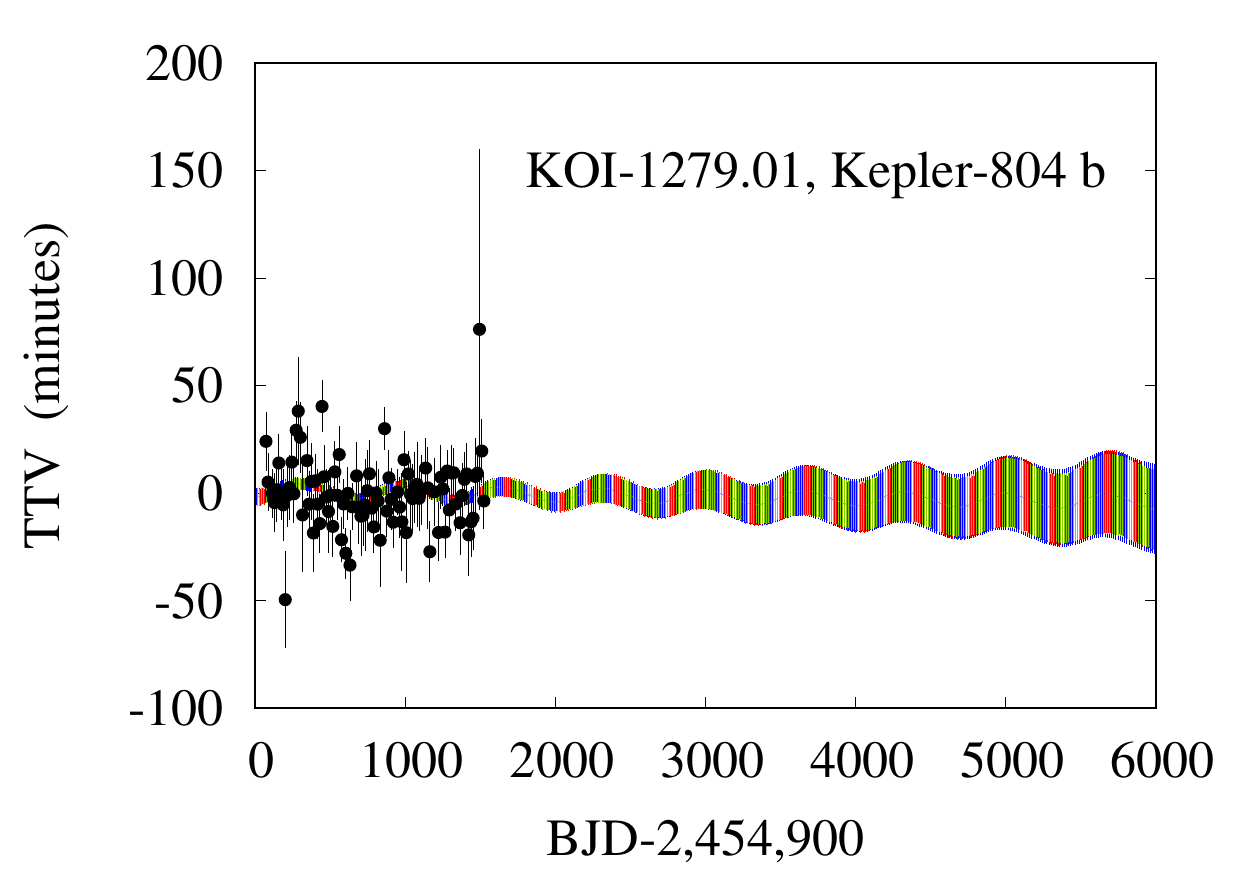} \\
\includegraphics[height = 1.45 in]{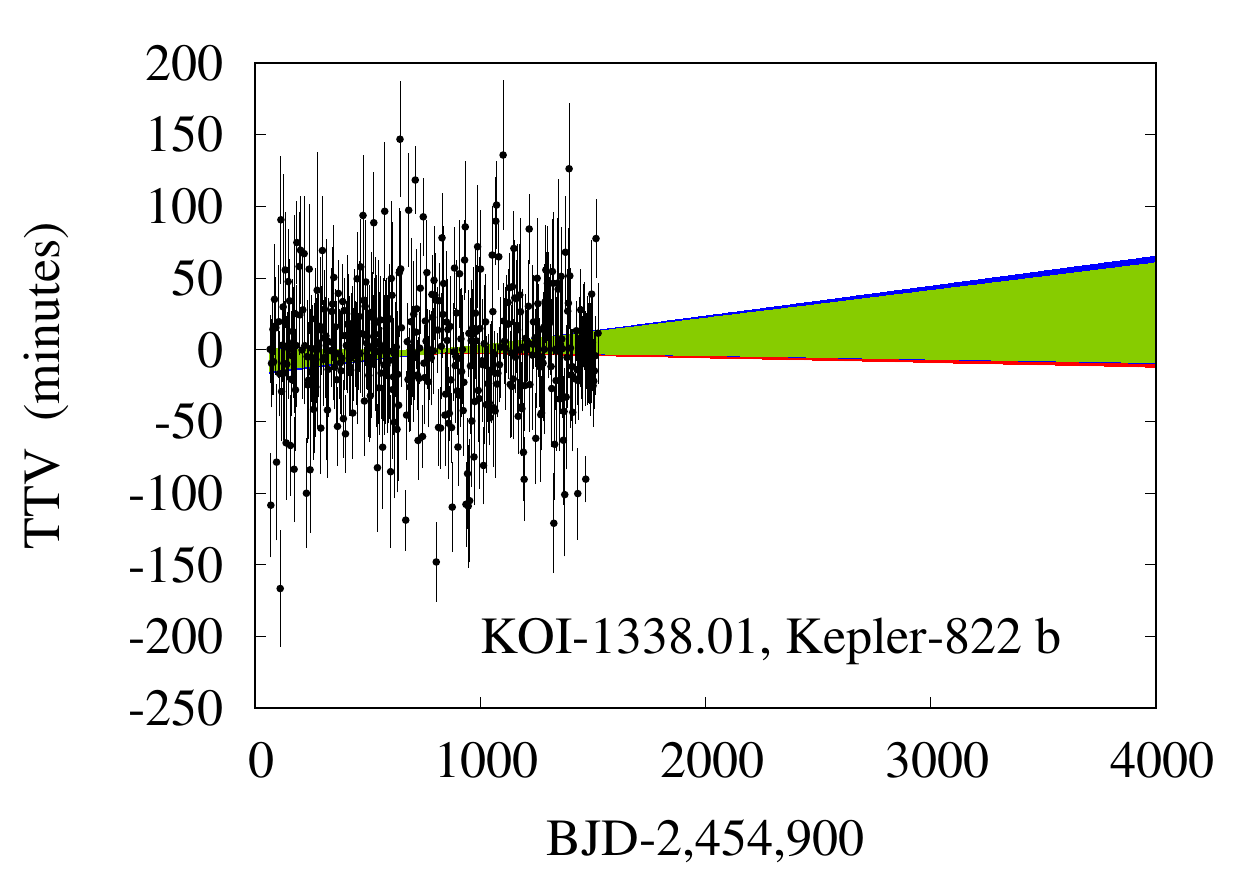}
\includegraphics[height = 1.45 in]{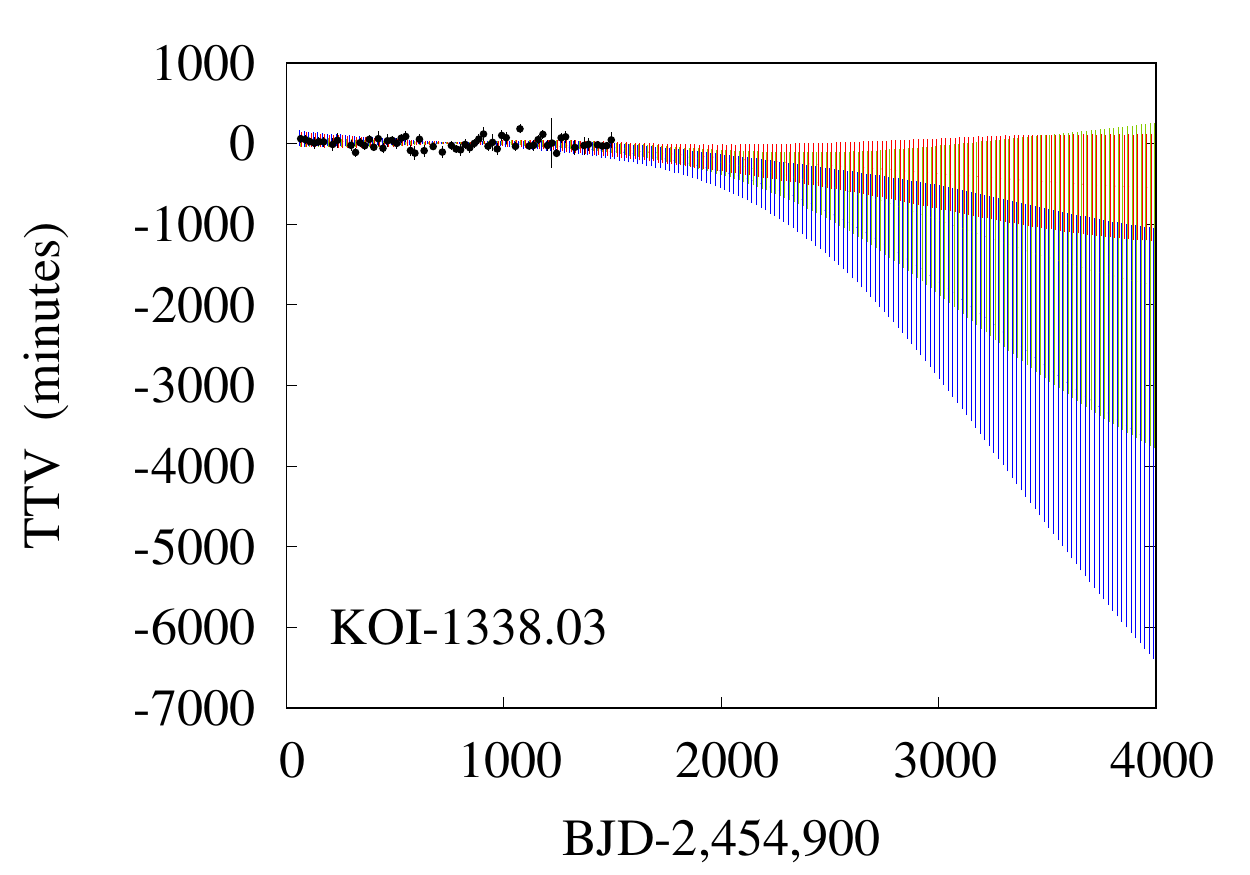}
\includegraphics[height = 1.45 in]{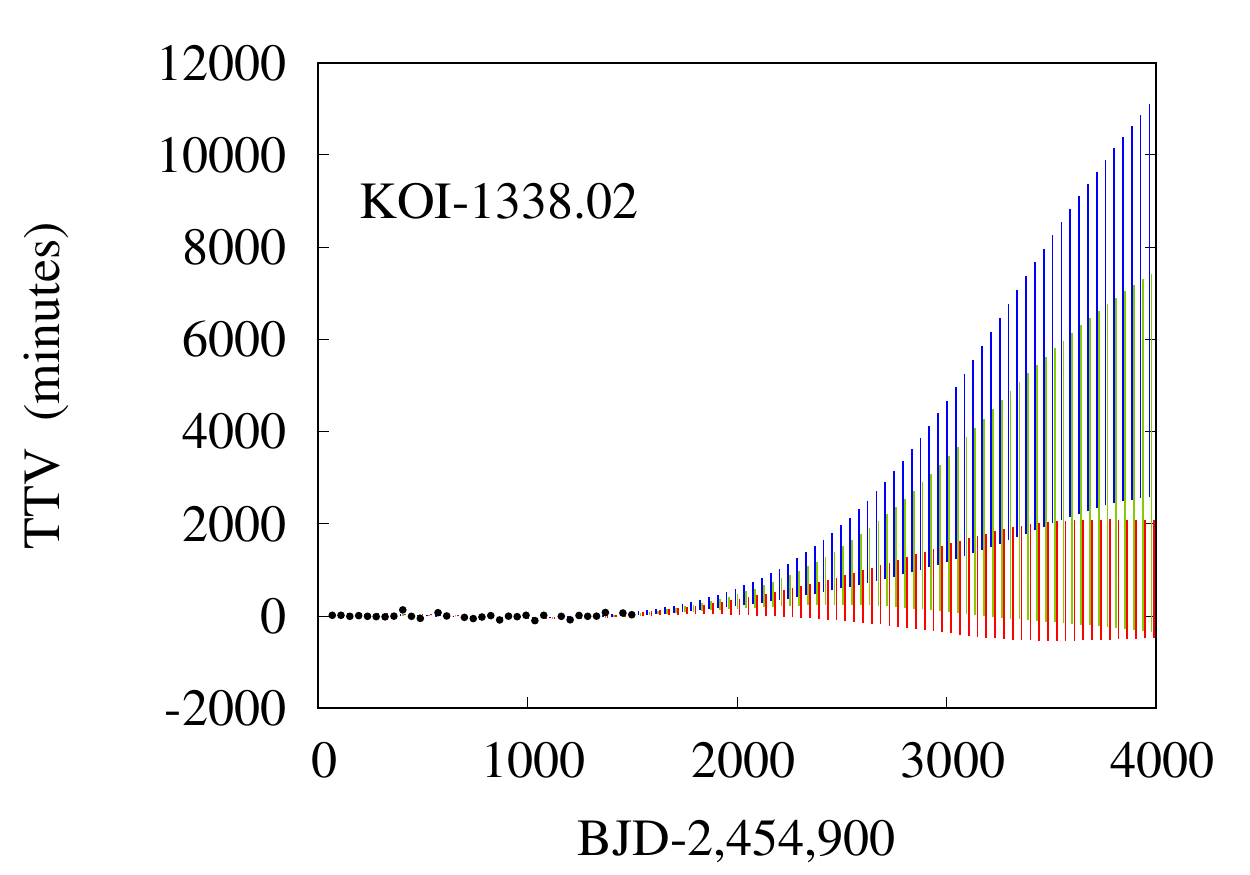}
\includegraphics[height = 1.45 in]{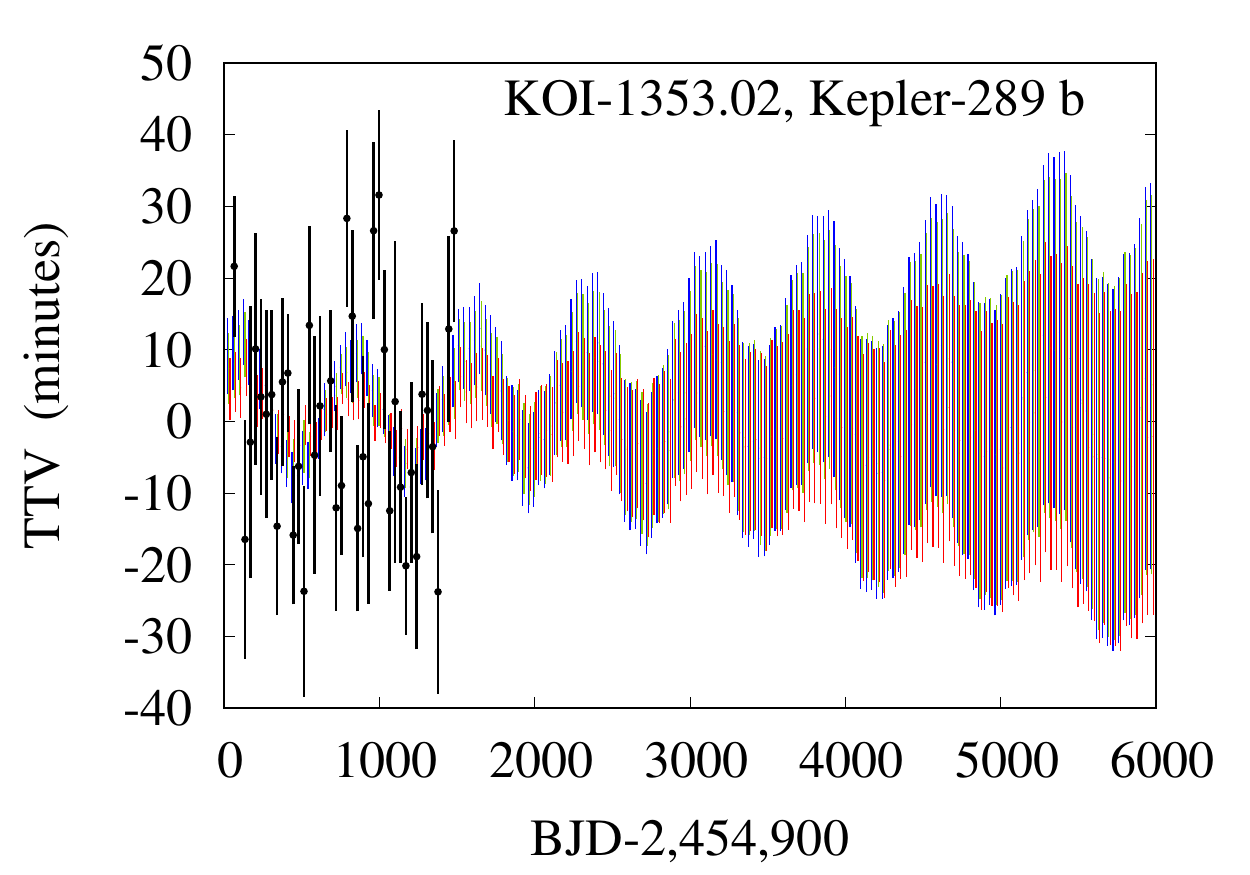}
\includegraphics[height = 1.45 in]{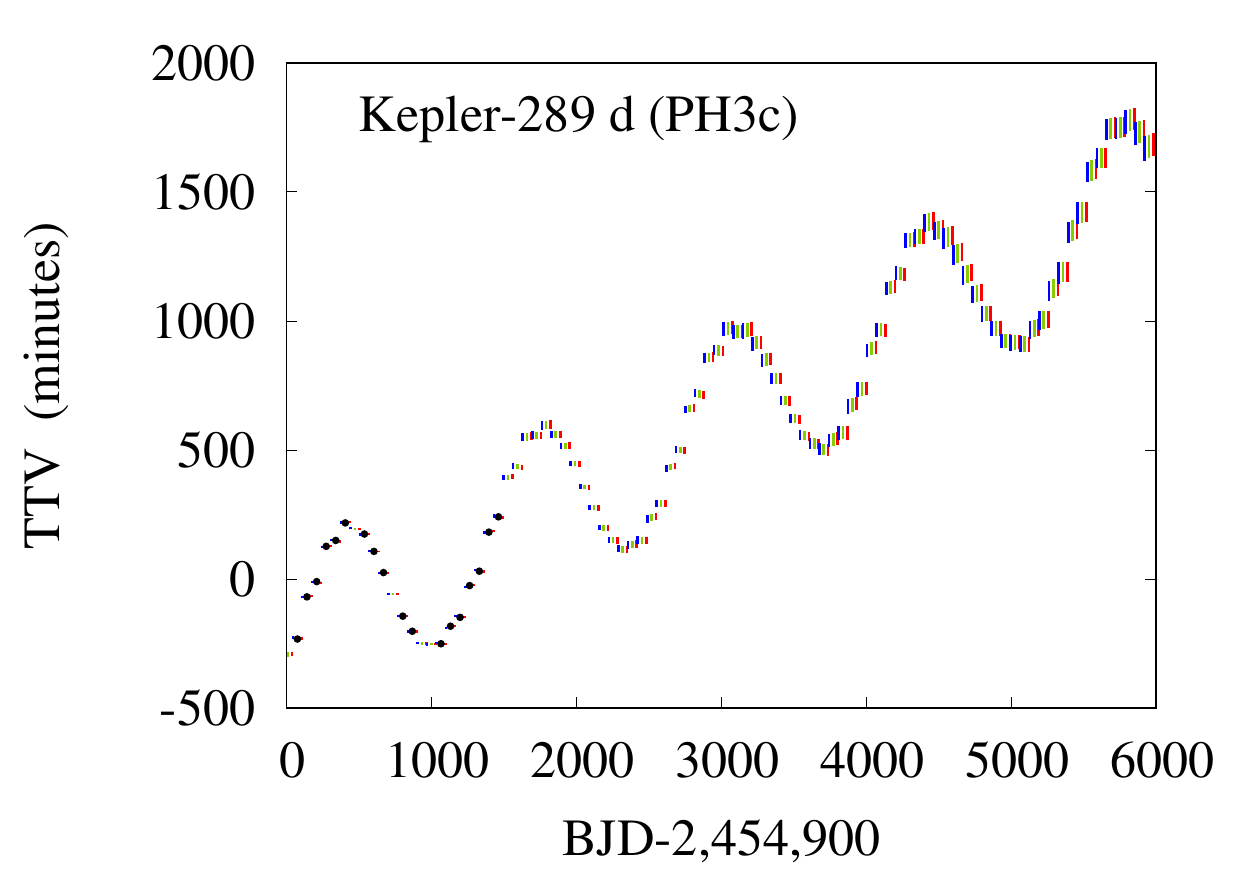} 
\includegraphics[height = 1.45 in]{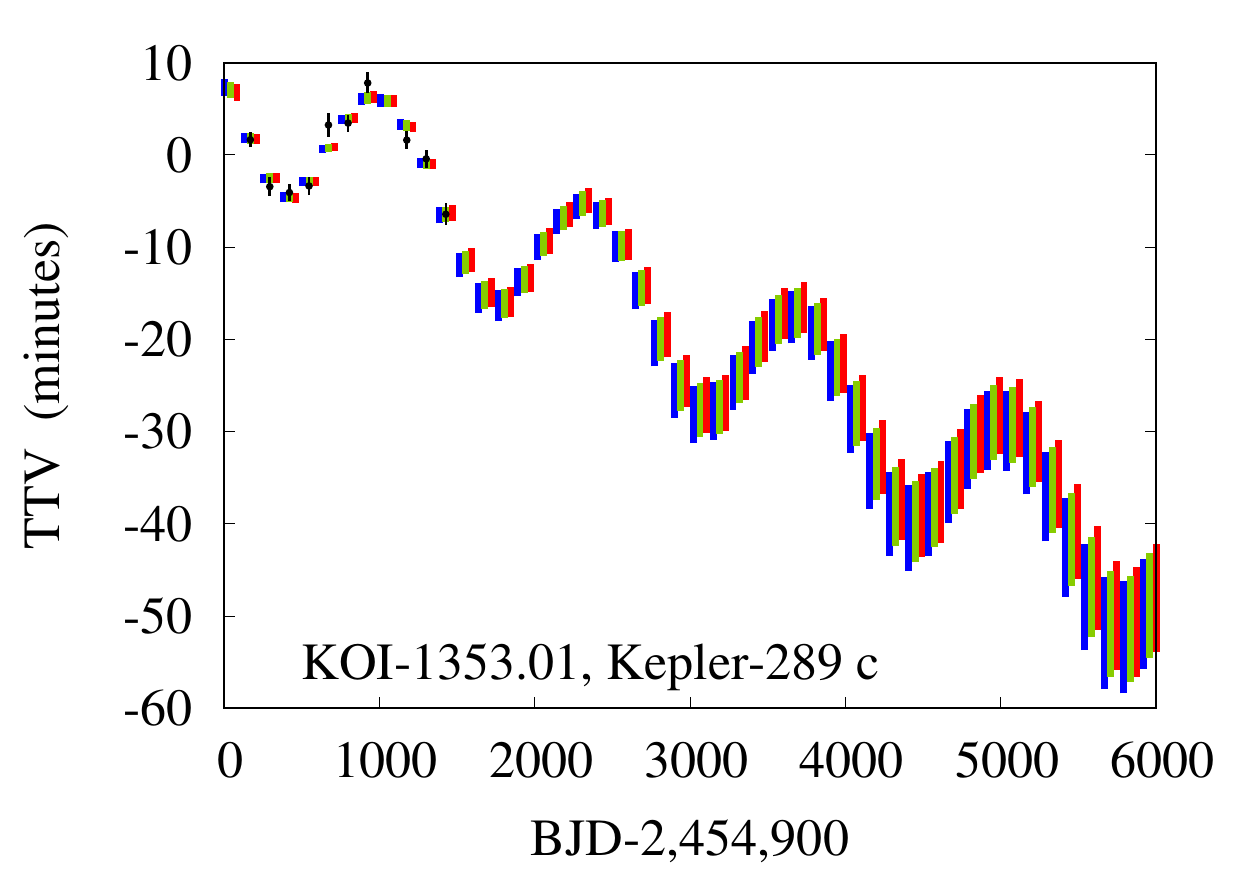}
\includegraphics[height = 1.45 in]{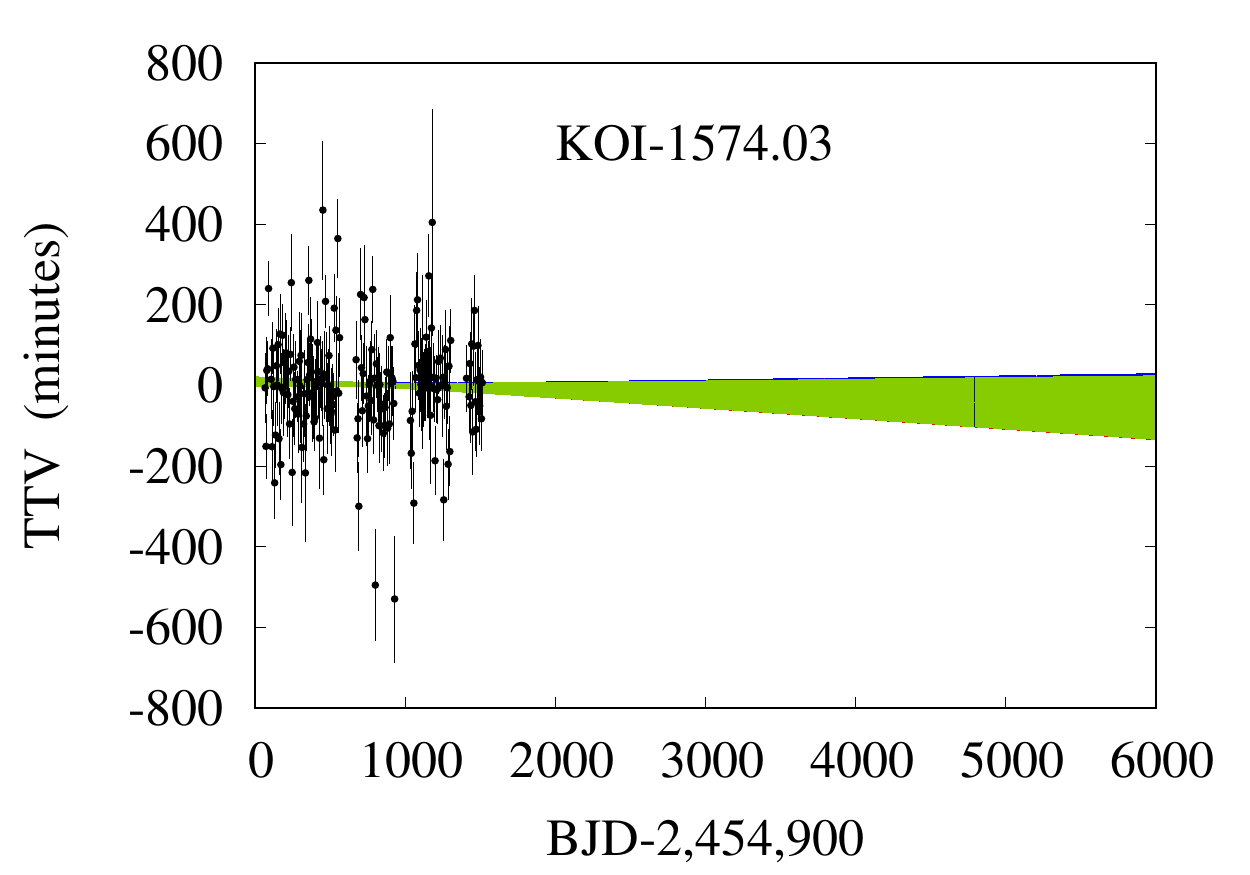}
\includegraphics[height = 1.45 in]{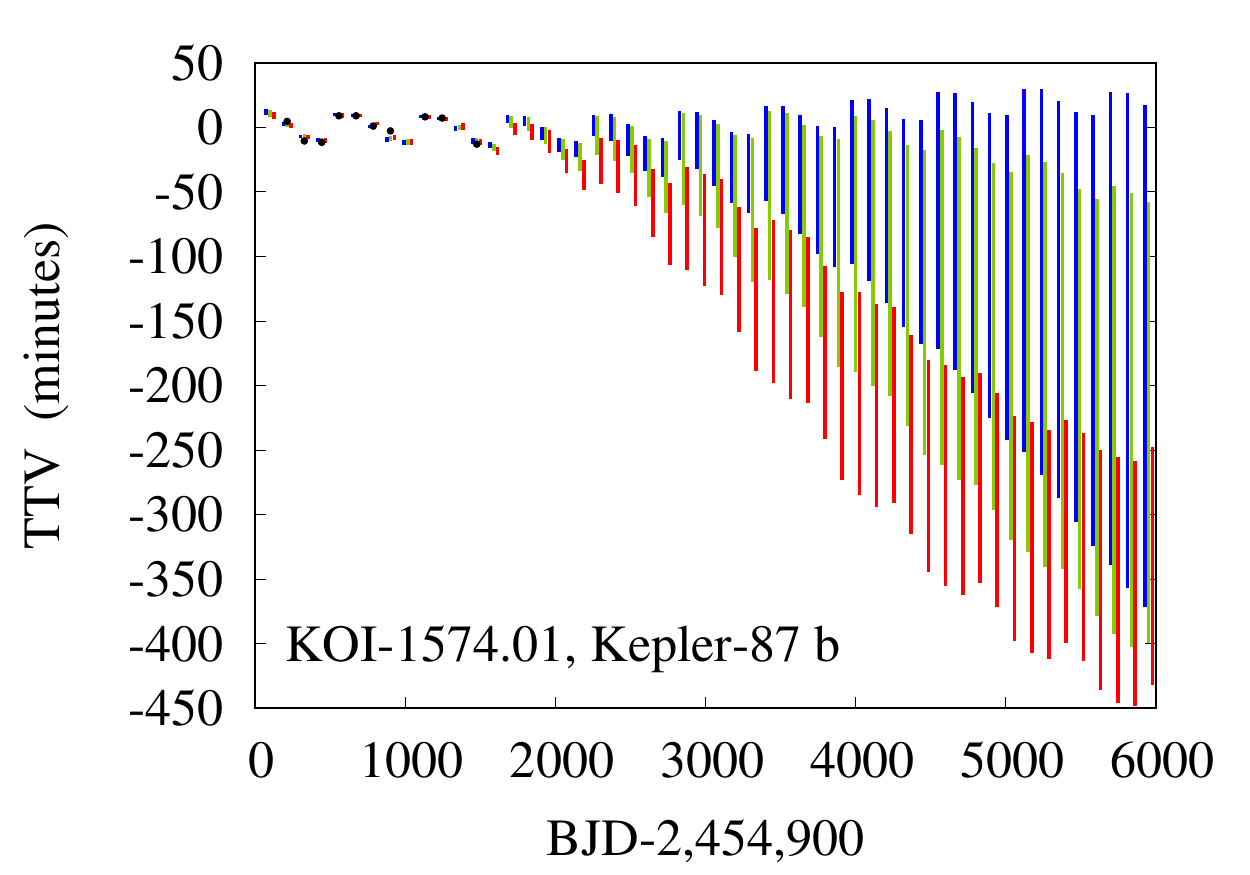}
\includegraphics[height = 1.45 in]{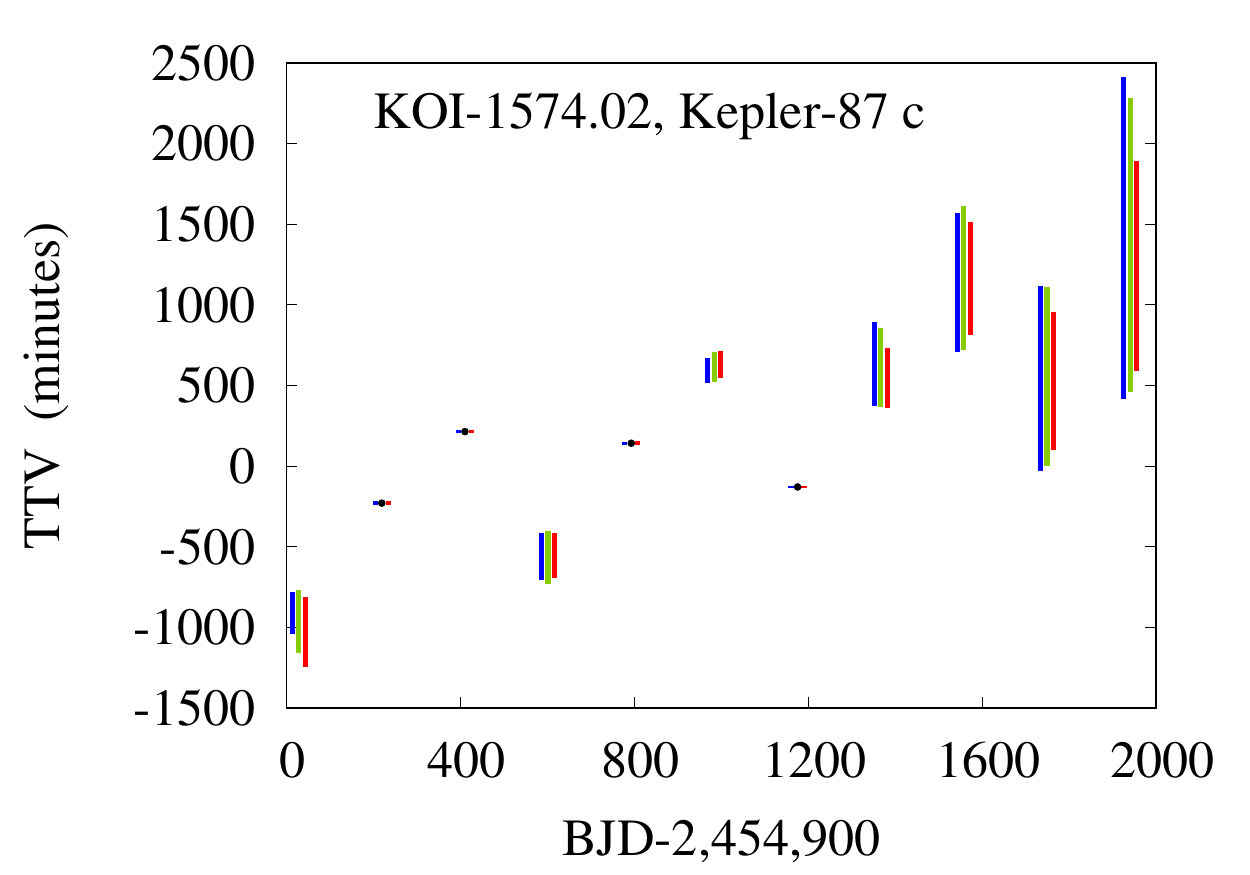}
\caption{Distribution of projected transit times as a function of time for the planet labelled in each panel (part 6). Black points mark transit times in the catalog of \citet{rowe15a} with 1$\sigma$ error bars. In green are 68.3\% confidence intervals of simulated transit times from posterior sampling. In blue (red), are a subset of samples with dynamical masses below (above) the 15.9th (84.1th) percentile. \label{fig:KOI-1070fut}} 
\end{center}
\end{figure}

\begin{figure}
\begin{center}
\figurenum{14}
\includegraphics [height = 1.05 in]{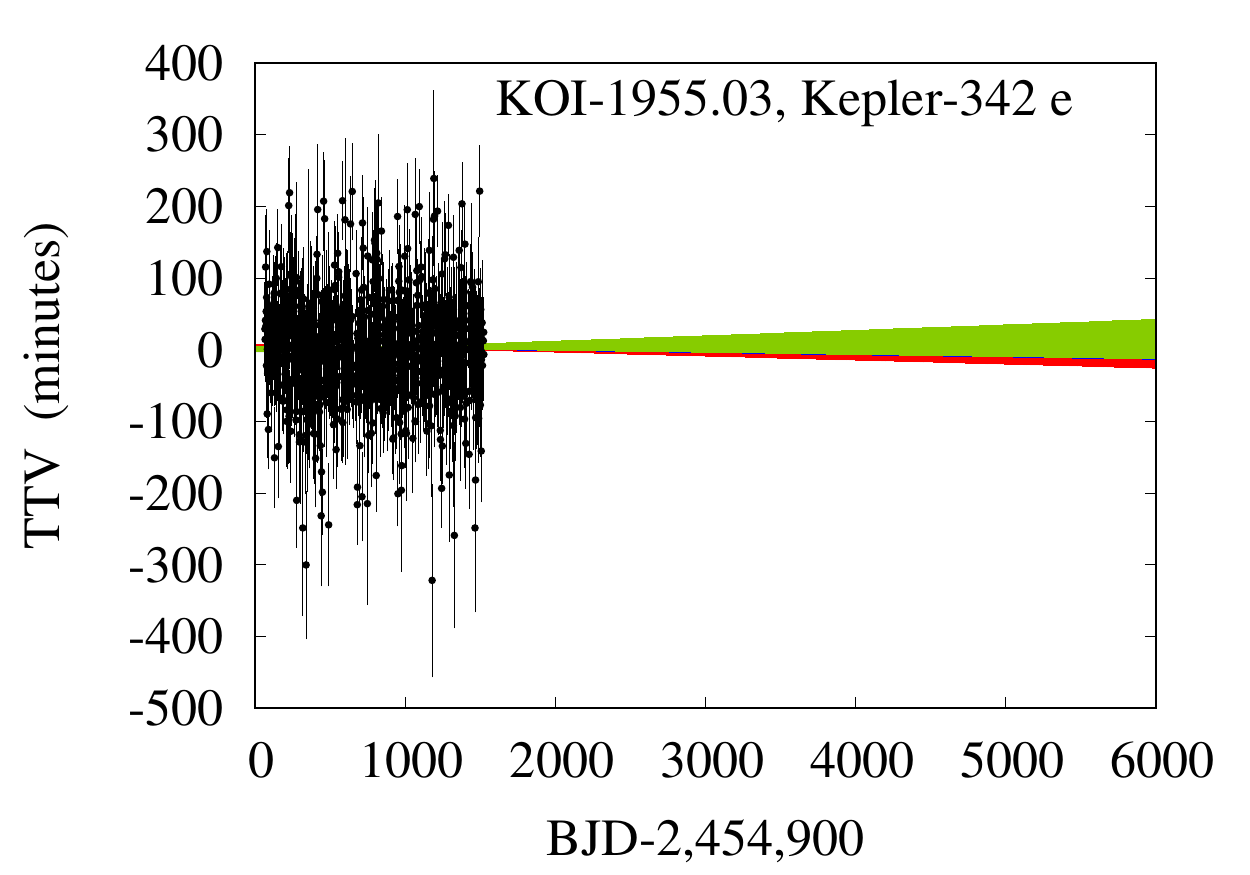}
\includegraphics [height = 1.05 in]{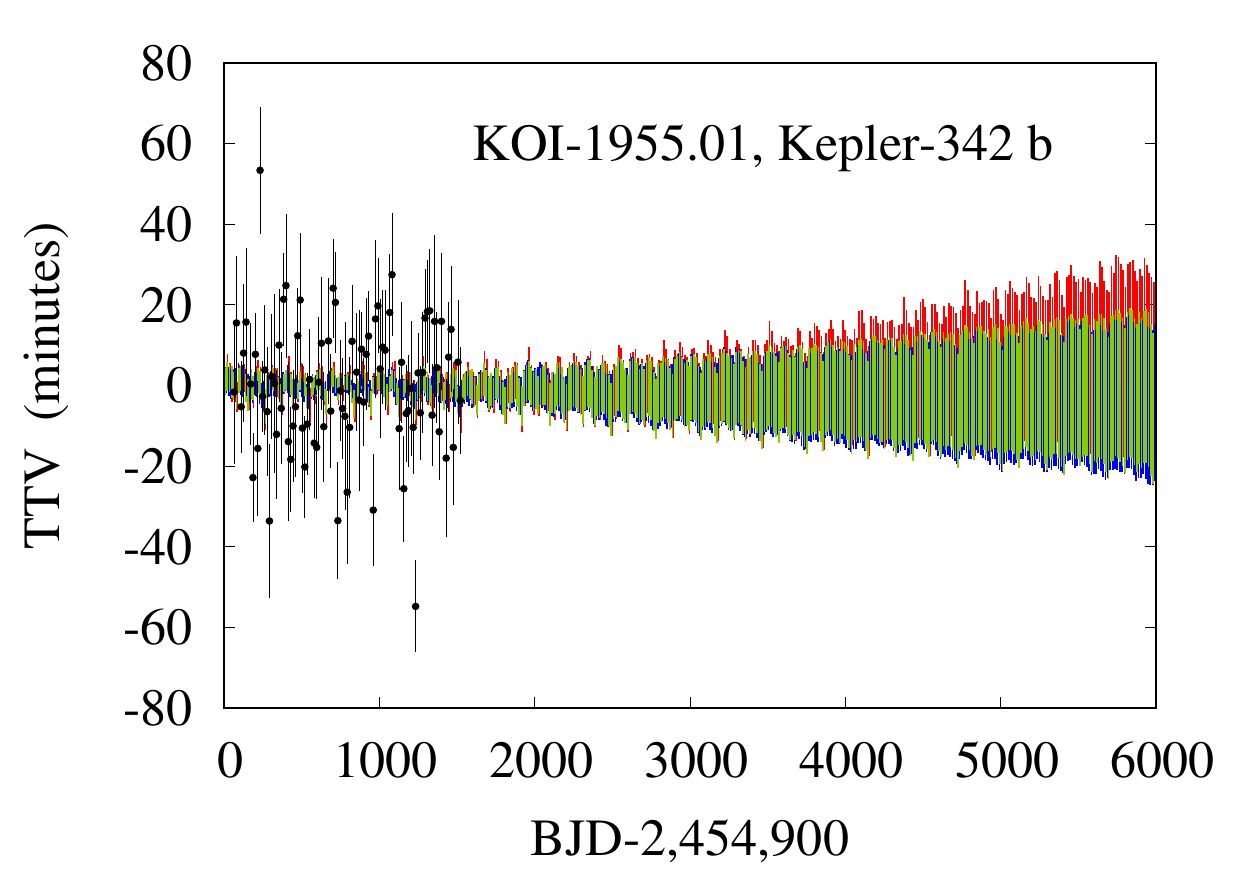}
\includegraphics [height = 1.05 in]{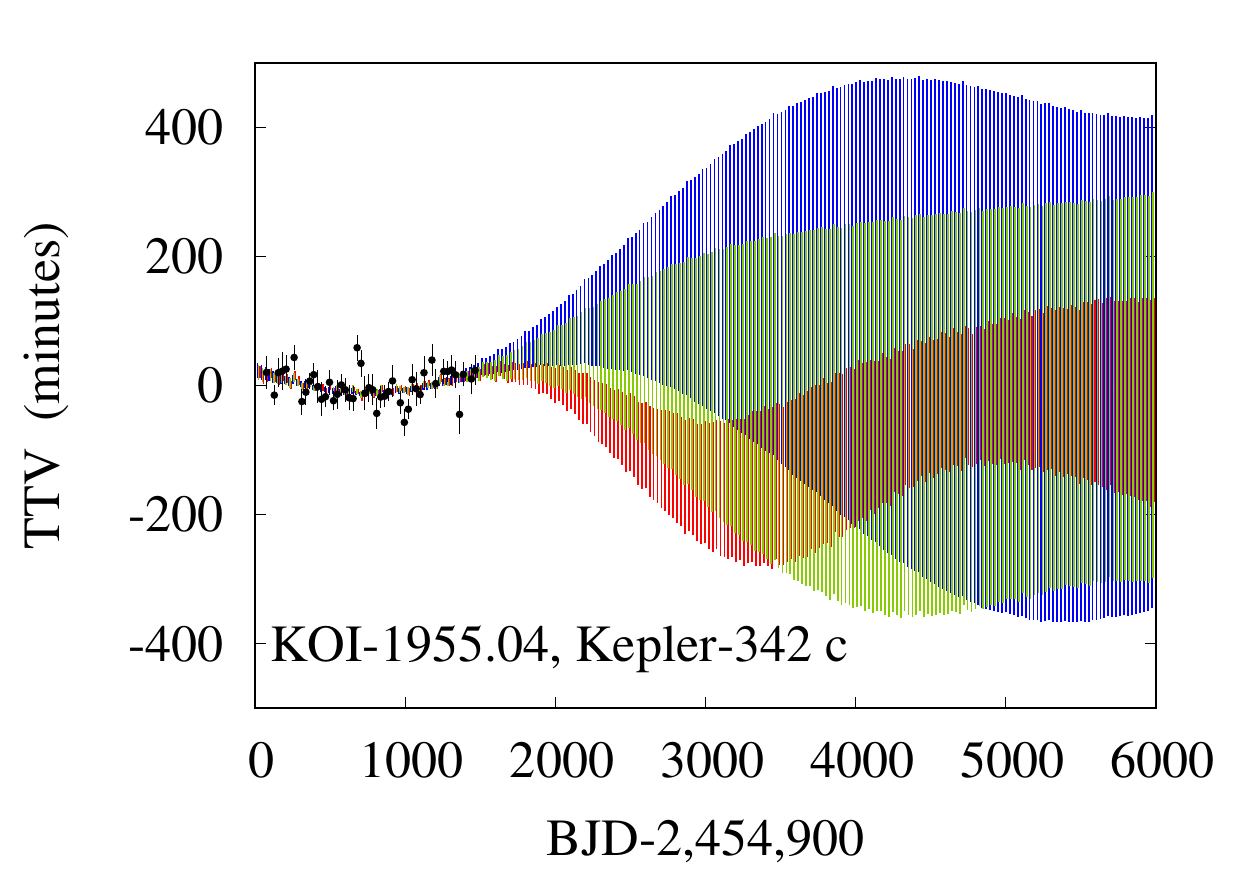}
\includegraphics [height = 1.05 in]{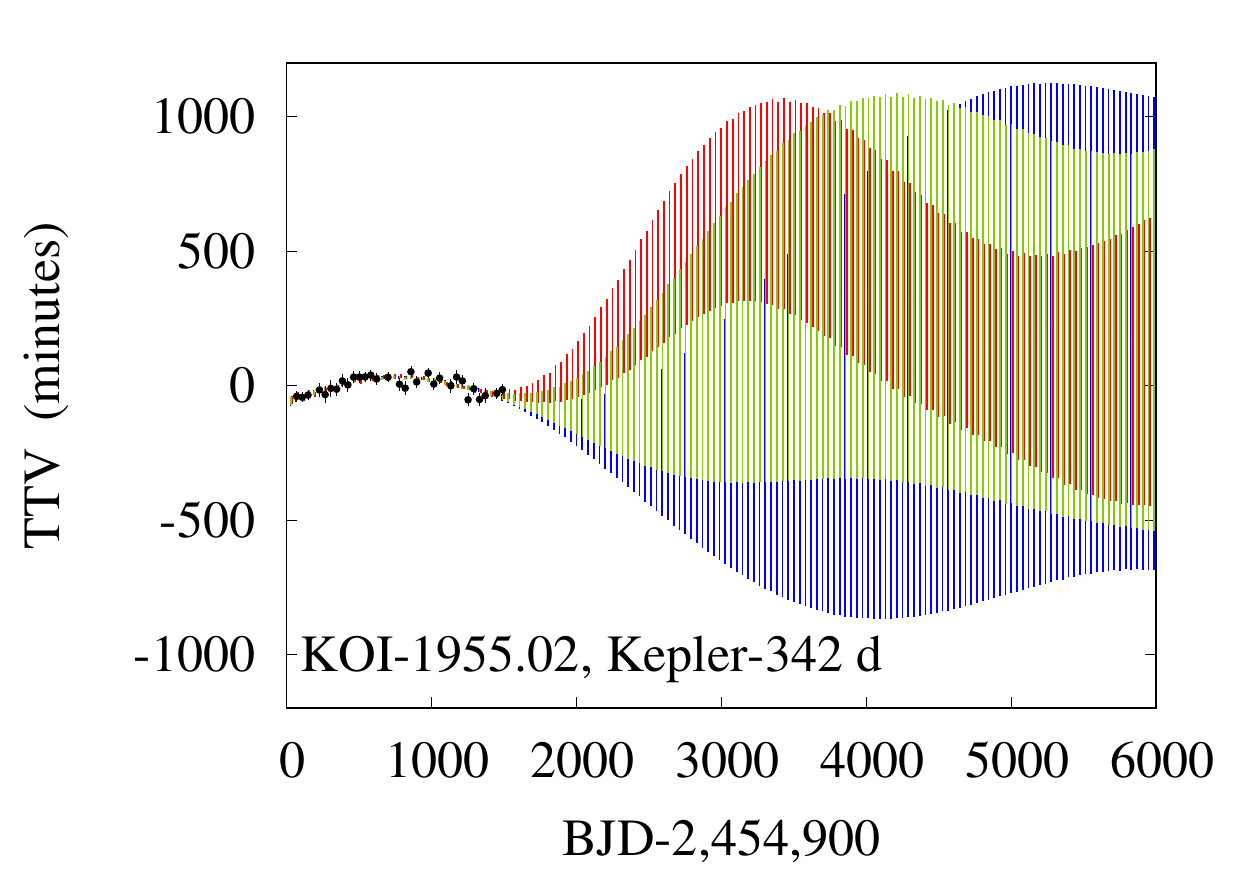}
\includegraphics[height = 1.45 in]{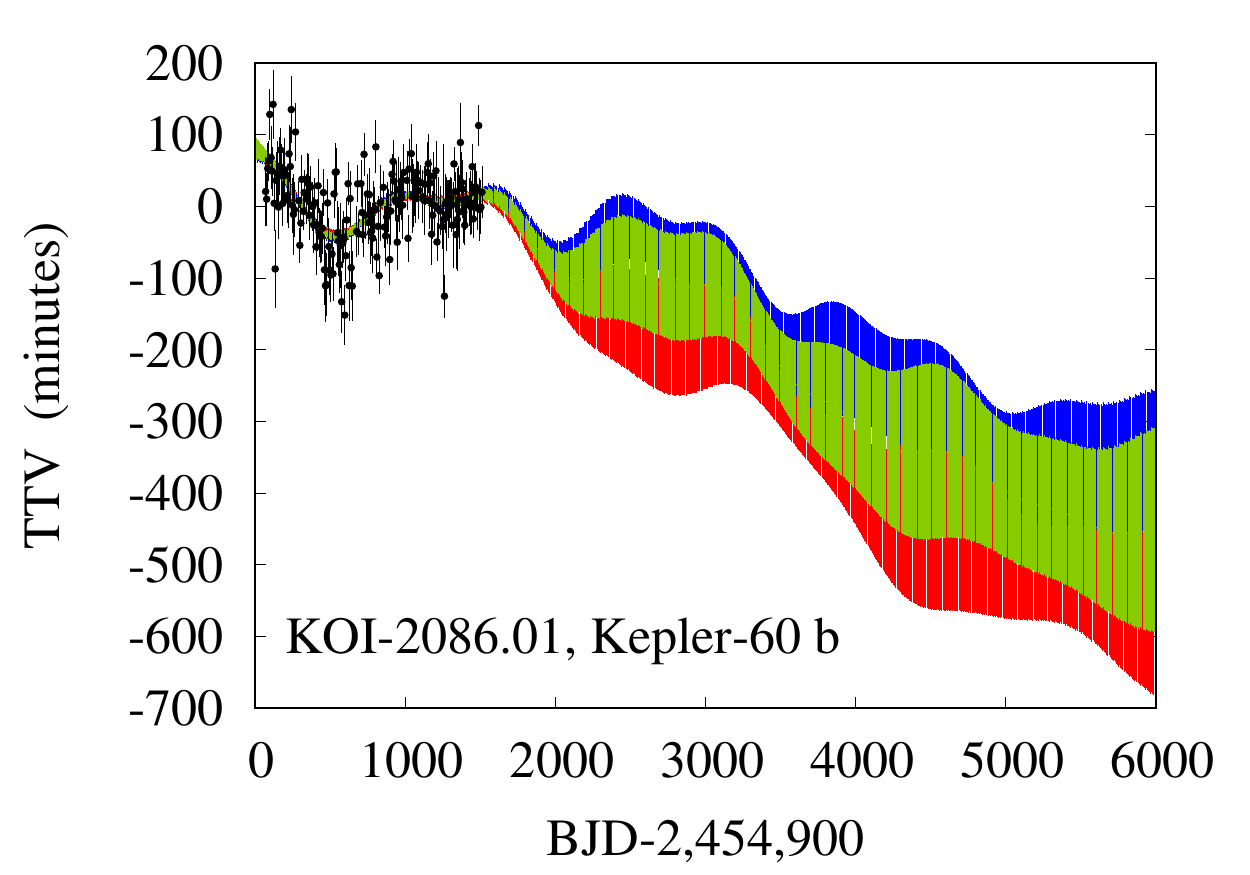} 
\includegraphics[height = 1.45 in]{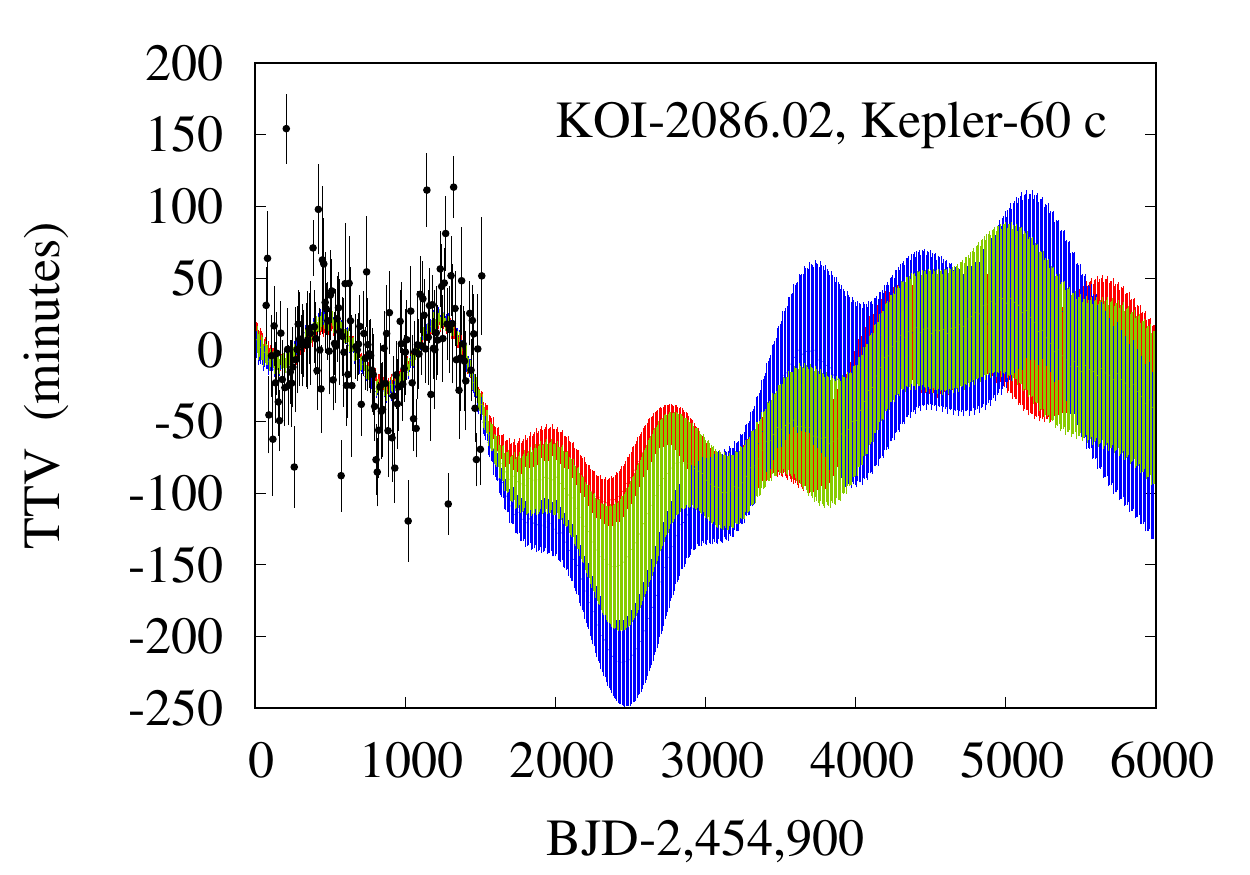}
\includegraphics[height = 1.45 in]{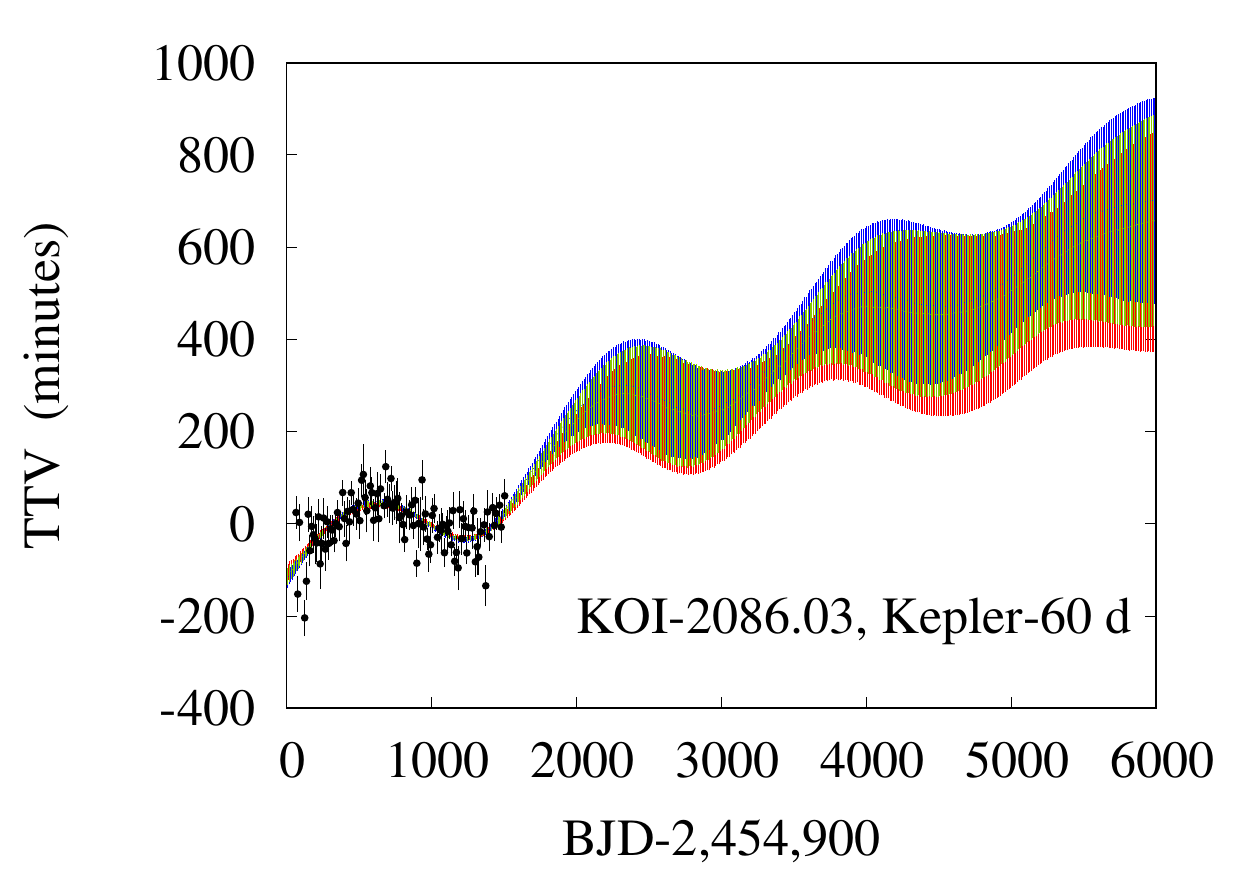}
\includegraphics[height = 1.45 in]{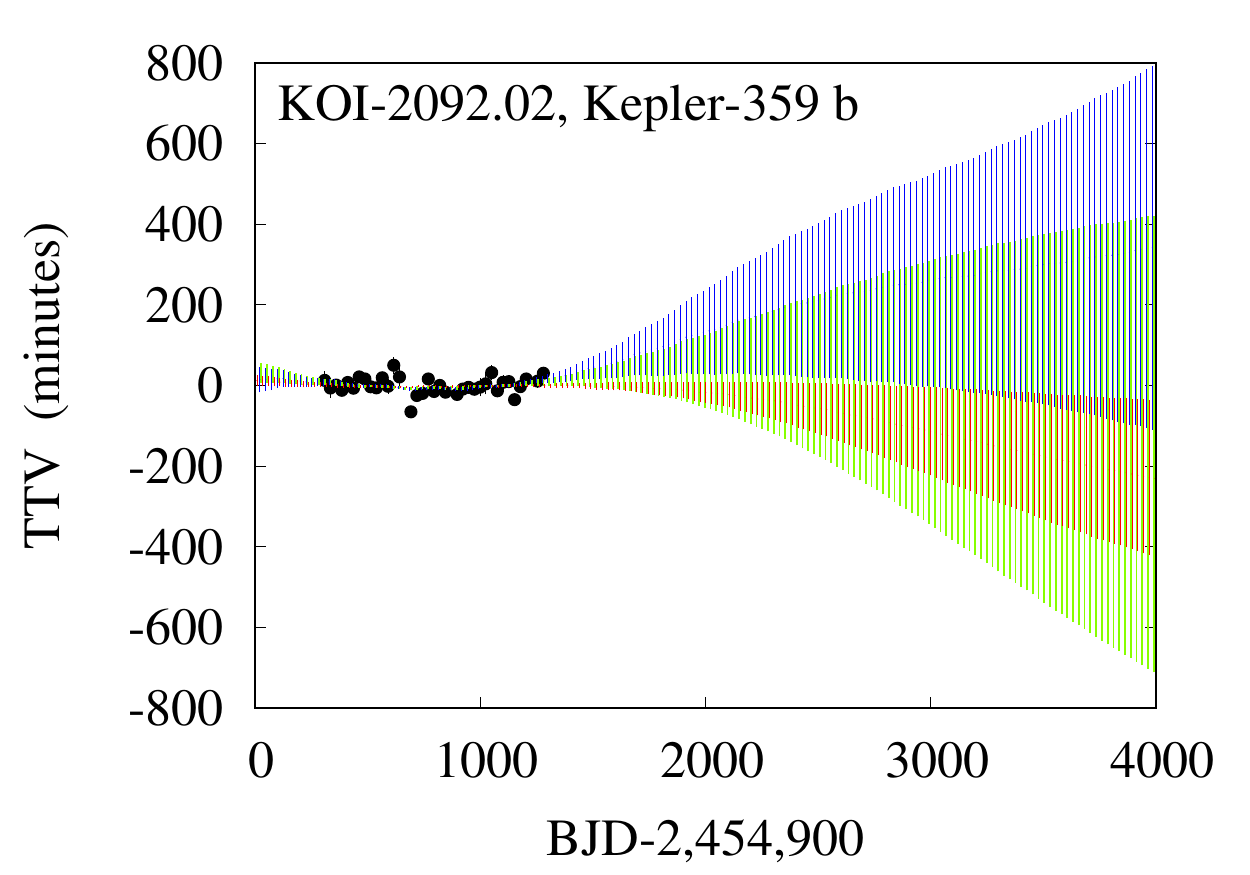} 
\includegraphics[height = 1.45 in]{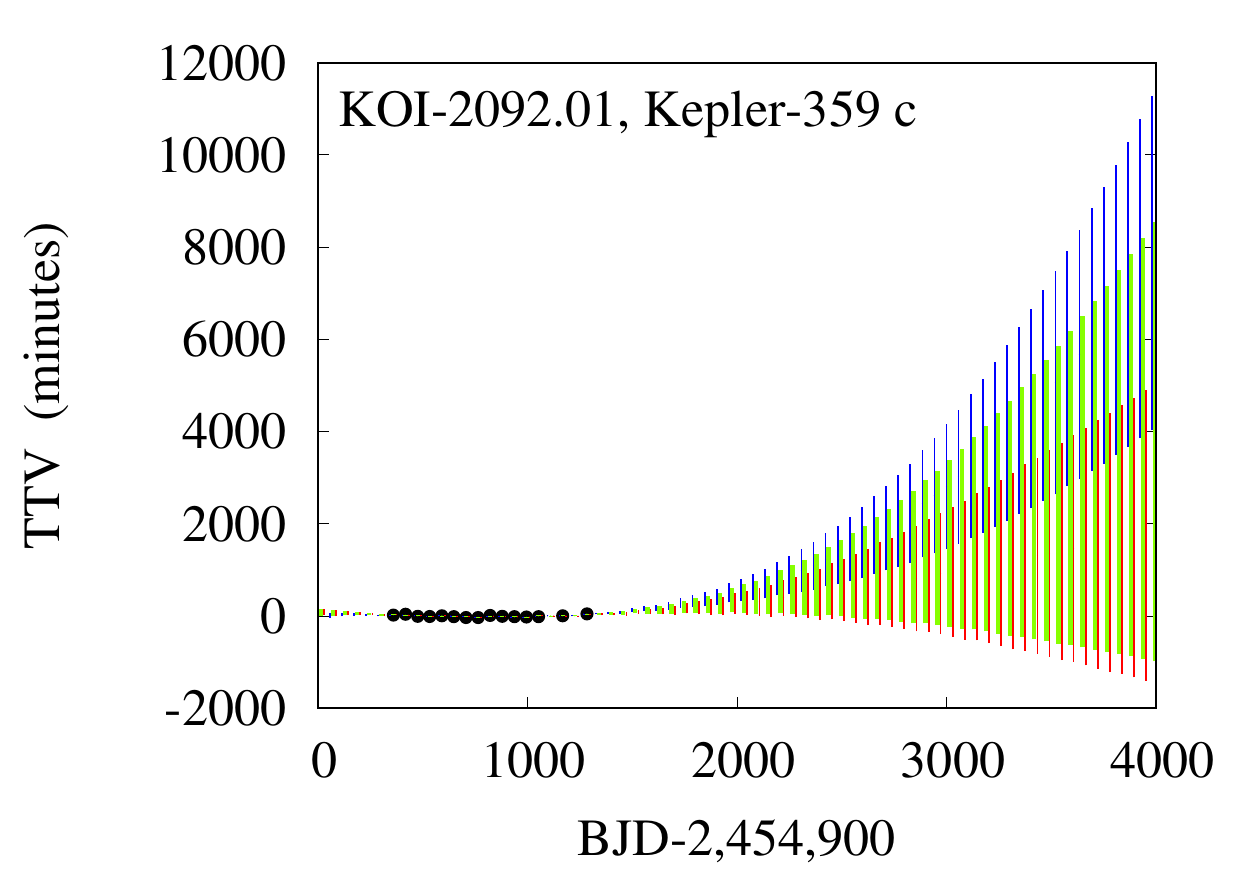}
\includegraphics[height = 1.45 in]{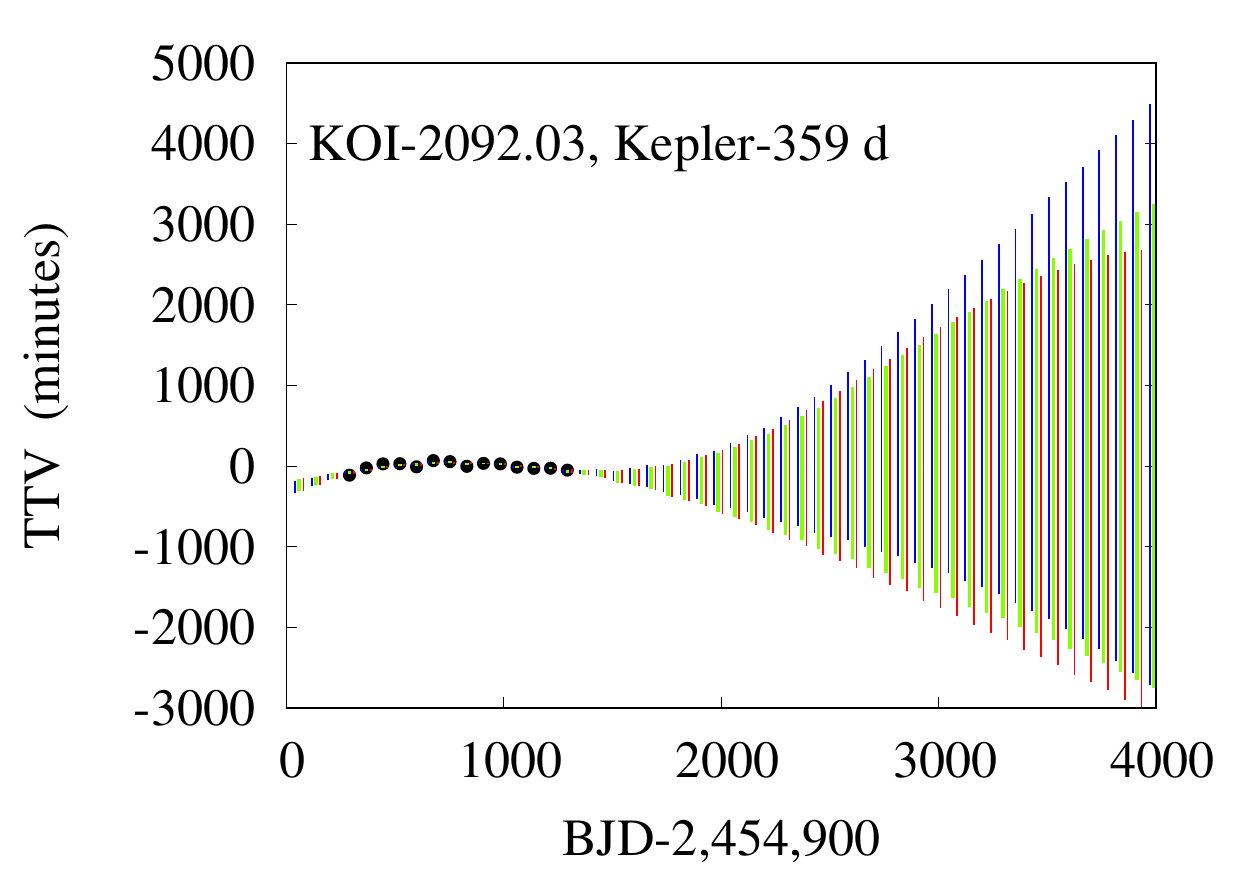} 
\includegraphics[height = 1.45 in]{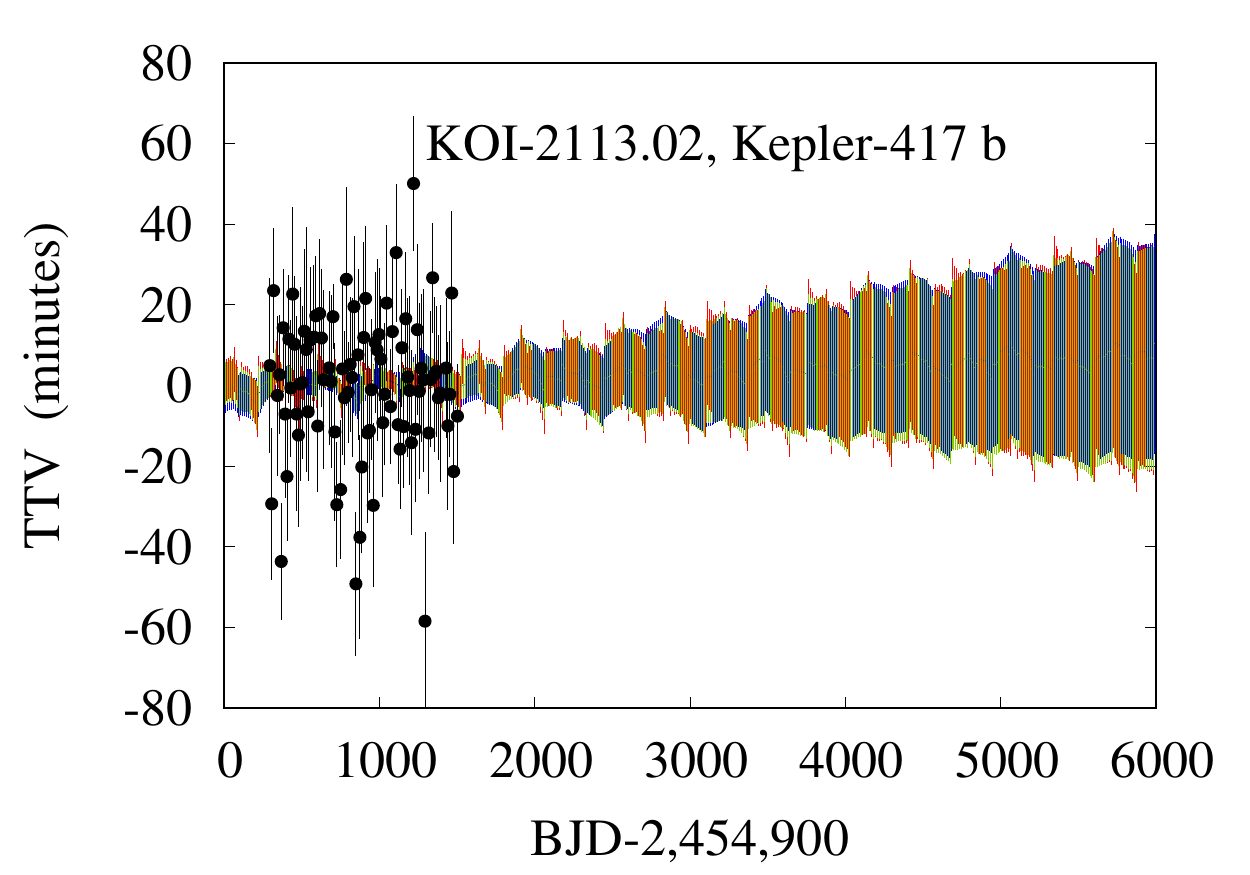}
\includegraphics[height = 1.45 in]{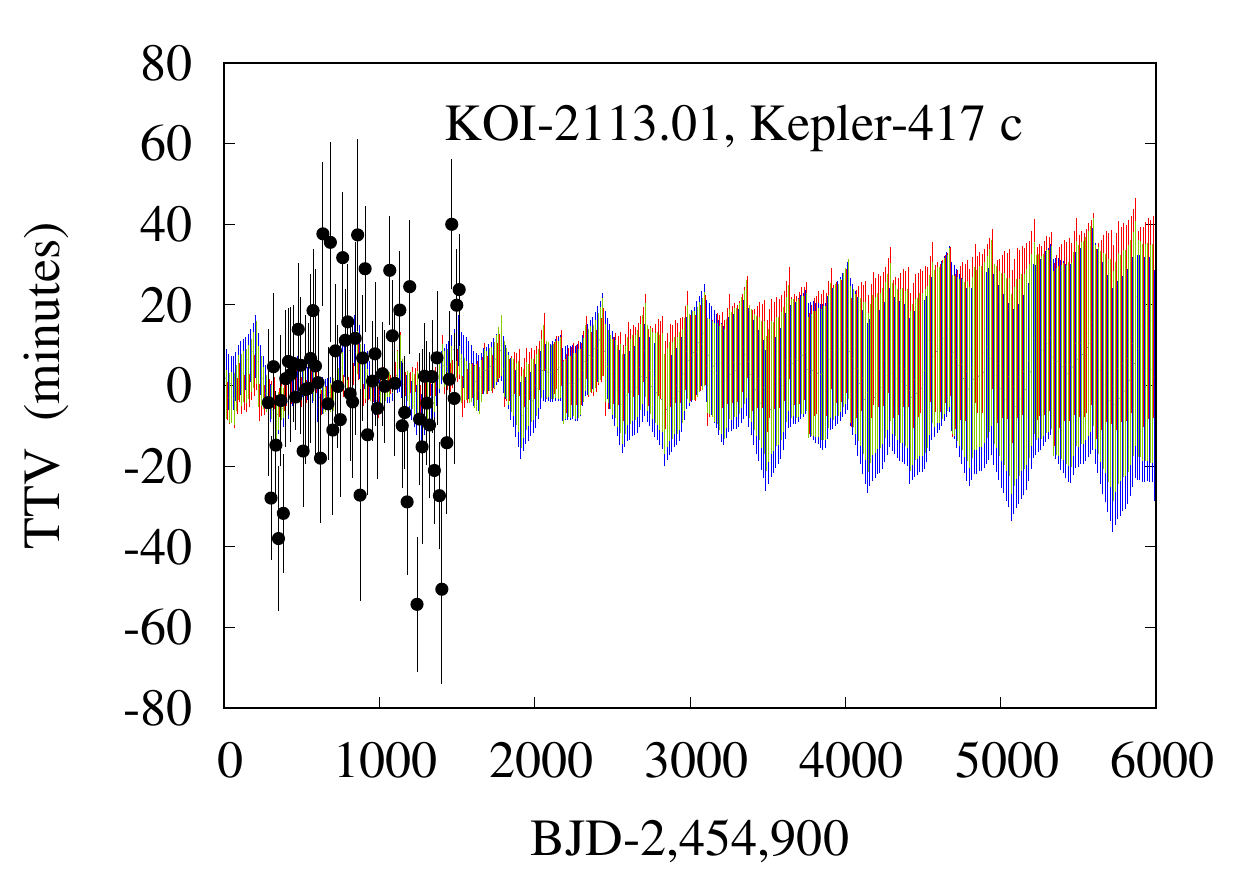} \\
\includegraphics[height = 1.05 in]{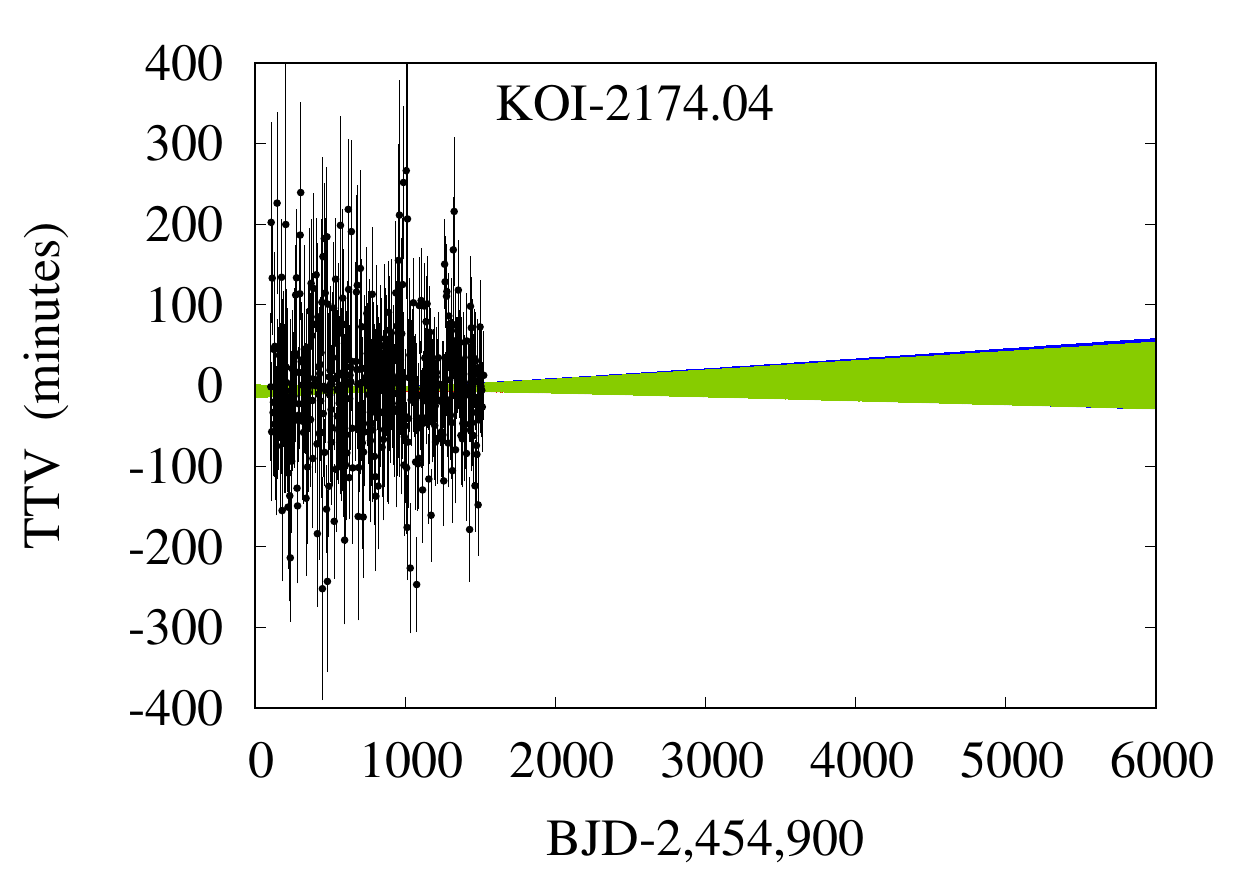}
\includegraphics[height = 1.05 in]{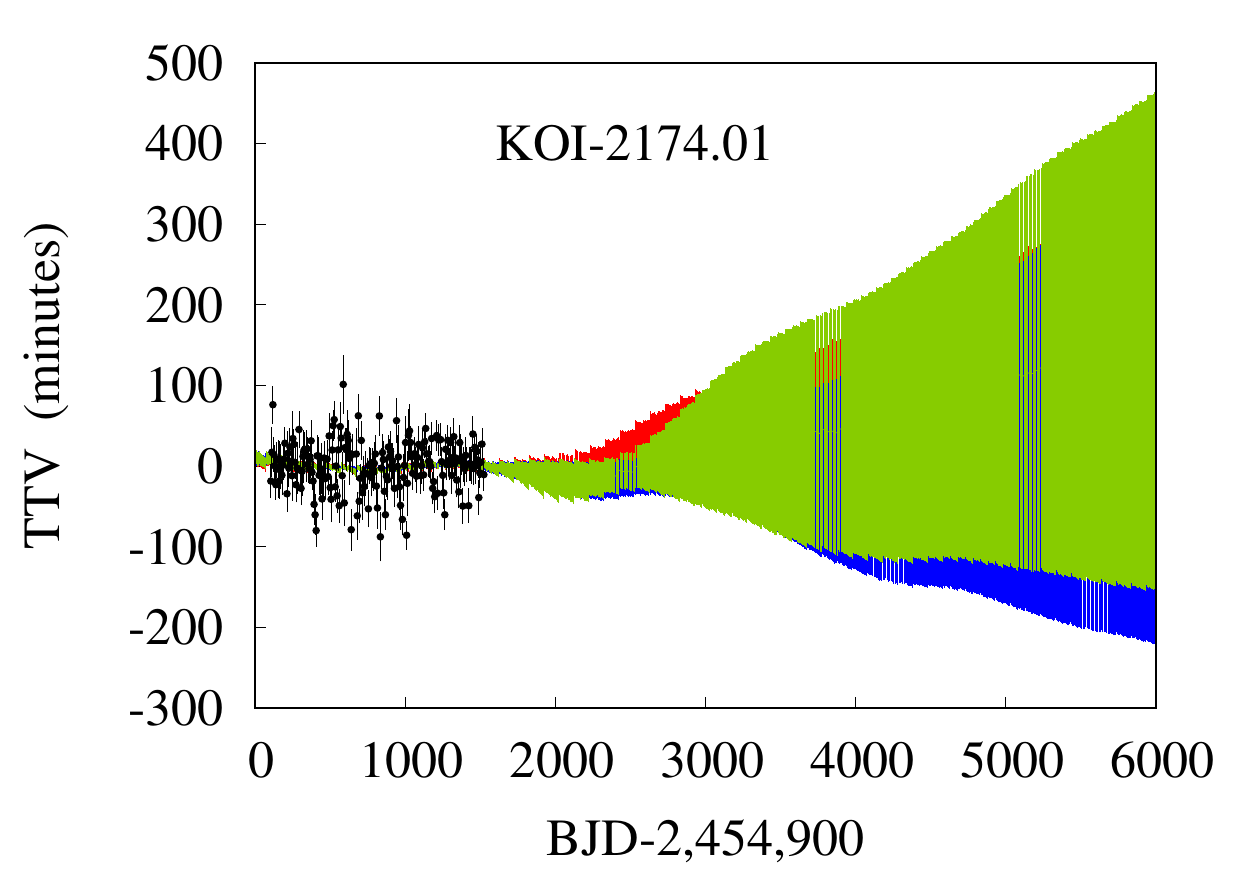}
\includegraphics[height = 1.05 in]{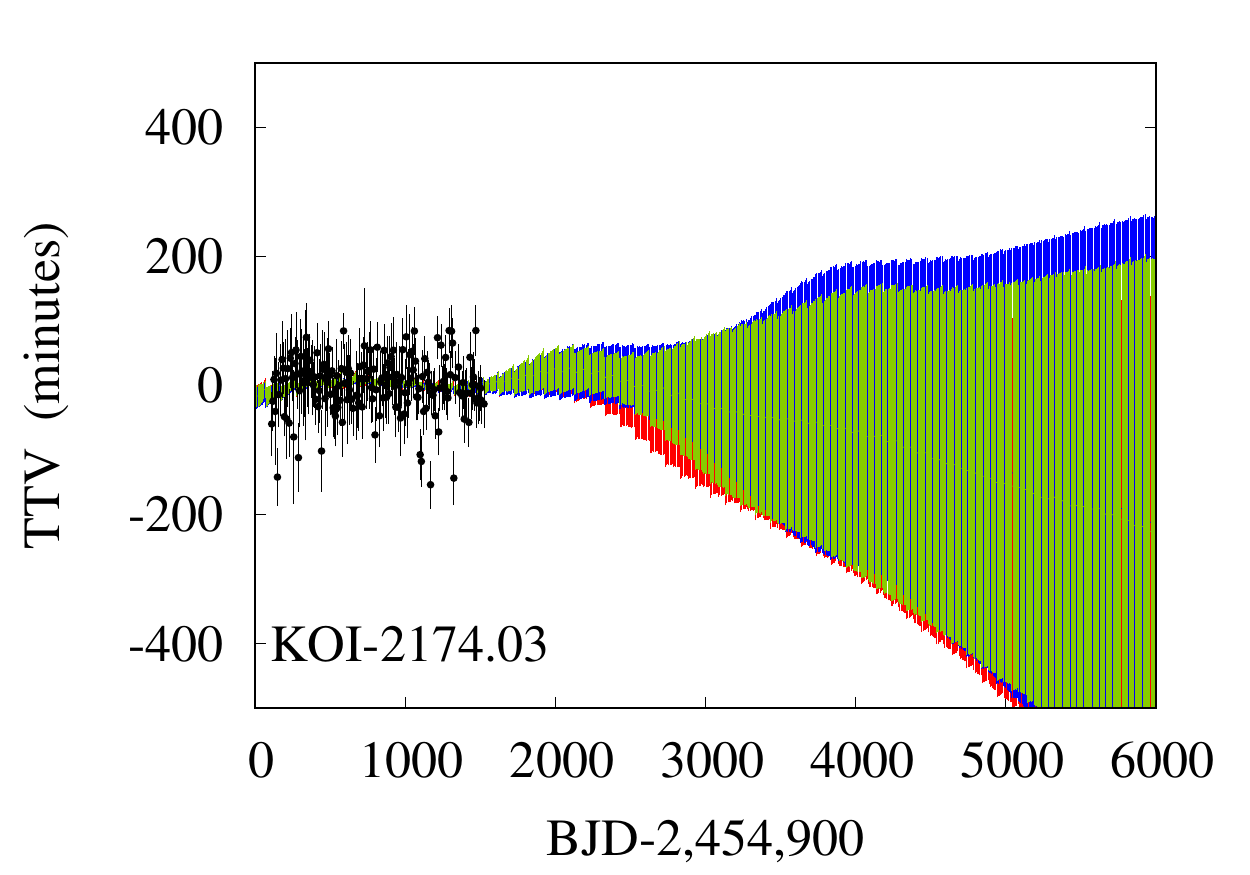} 
\includegraphics[height = 1.05 in]{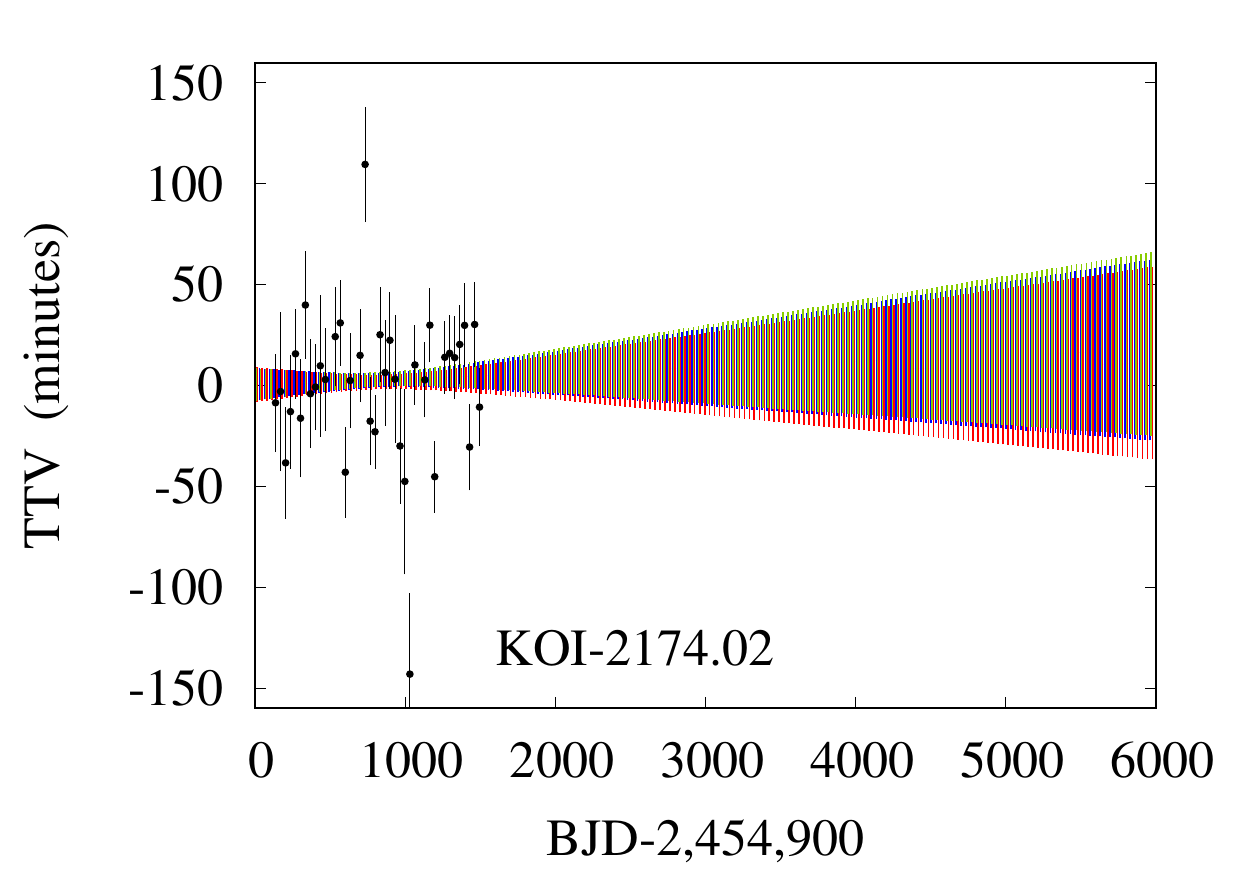} \\
\includegraphics[height = 1.45 in]{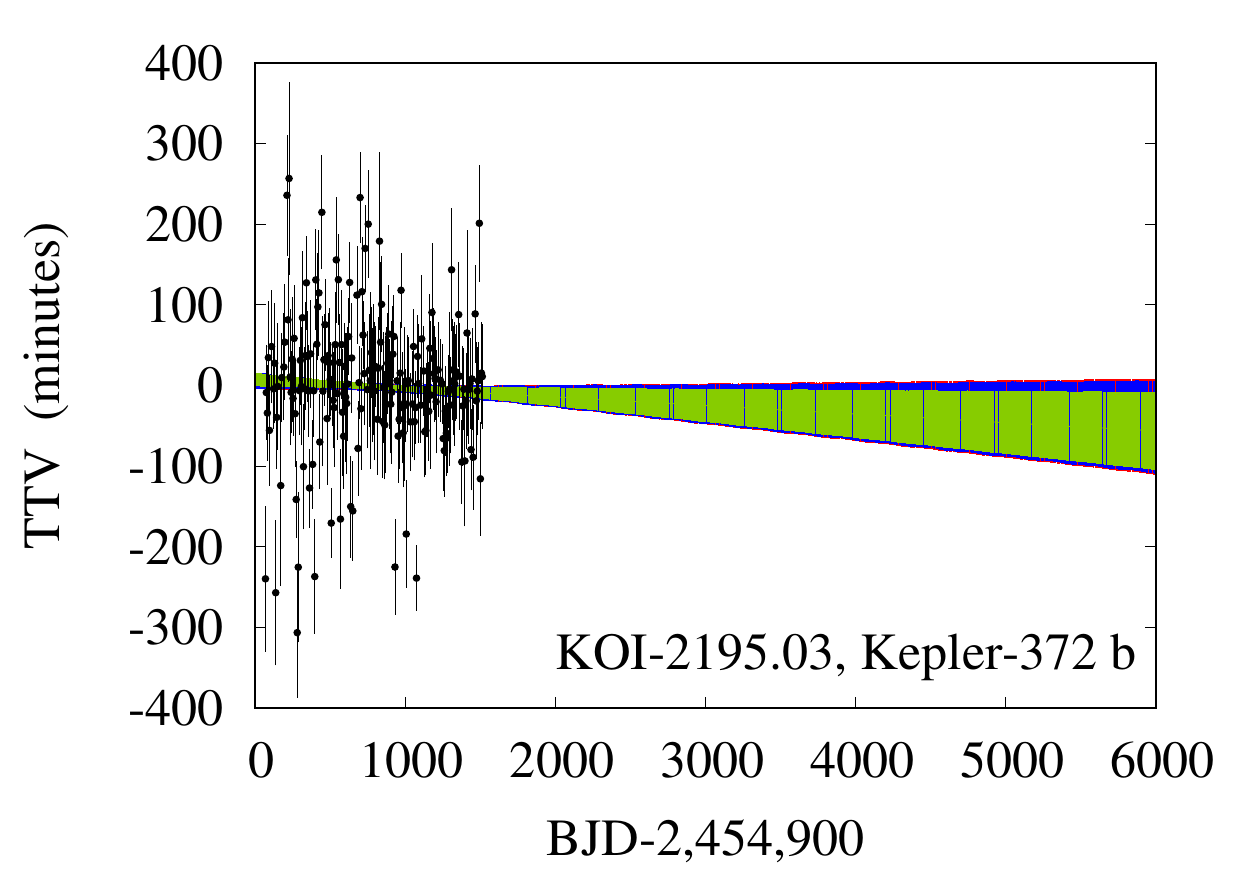}
\includegraphics[height = 1.45 in]{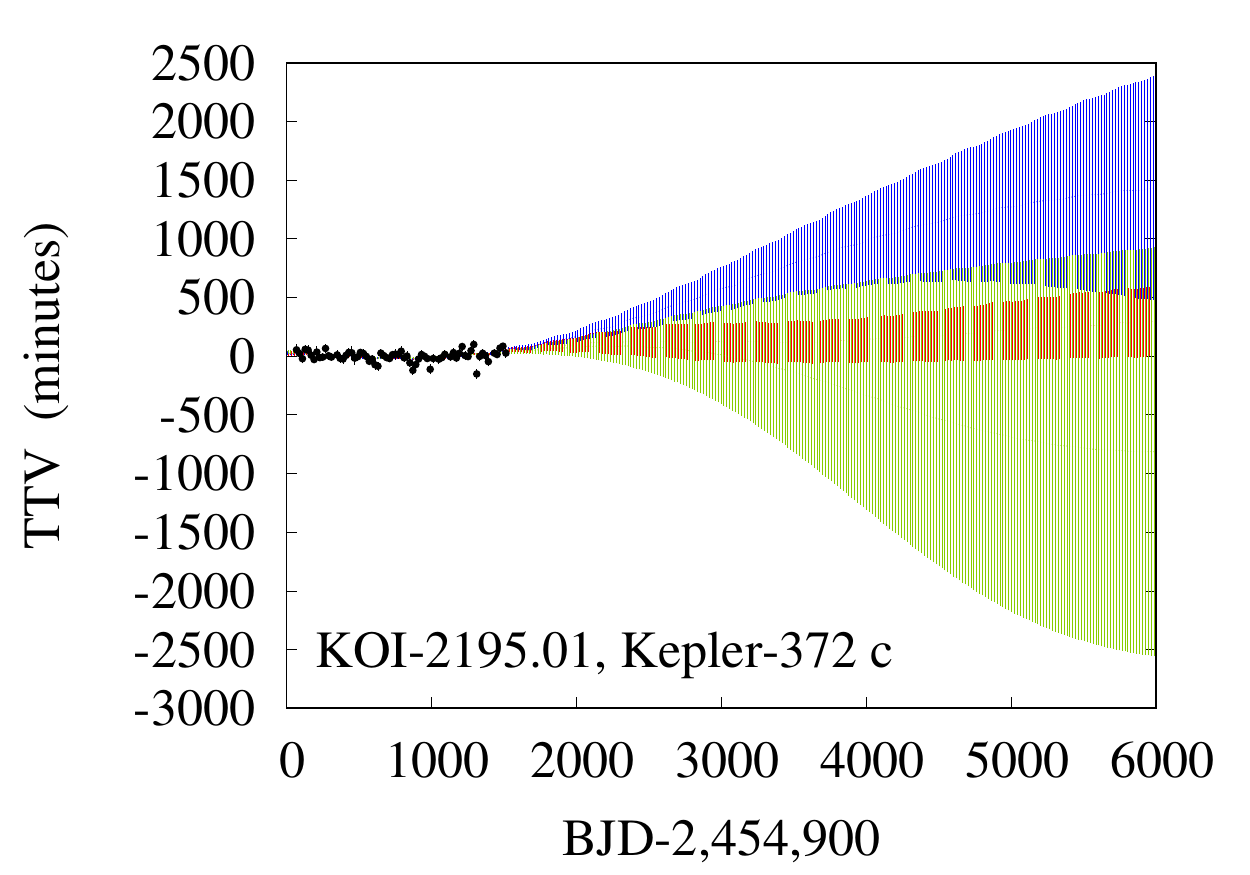}
\includegraphics[height = 1.45 in]{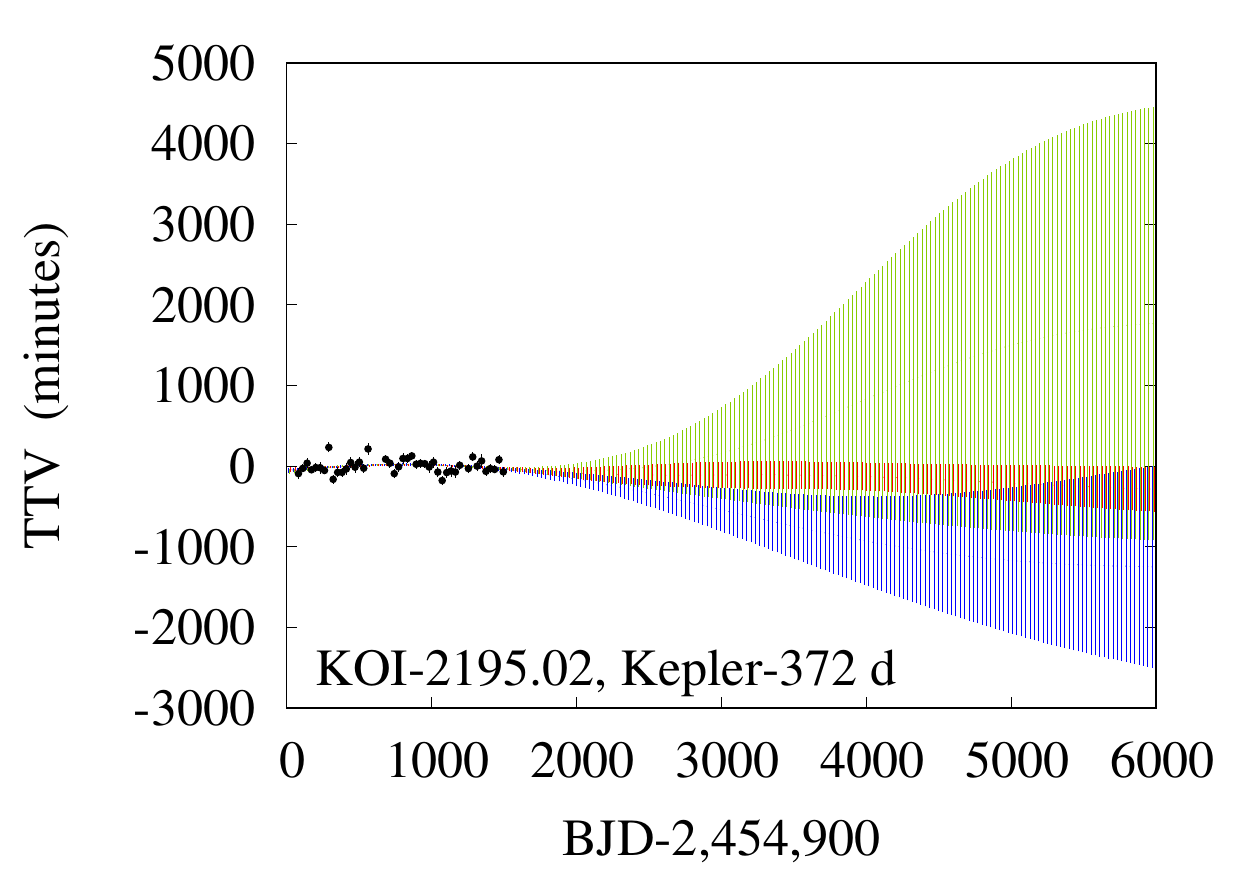}
\caption{Distribution of projected transit times as a function of time for the planet labelled in each panel (part 8). Black points mark transit times in the catalog of \citet{rowe15a} with 1$\sigma$ error bars. In green are 68.3\% confidence intervals of simulated transit times from posterior sampling. In blue (red), are a subset of samples with dynamical masses below (above) the 15.9th (84.1th) percentile. 
\label{fig:KOI-1955fut}} 
\end{center}
\end{figure}

In many cases, the projected transit times are known precisely for decades beyond the \textit{Kepler} mission. Figures~\ref{fig:KOI-85fut}--\ref{fig:KOI-2414fut} show the divergence of projected transit times and uncertainties for all of the planets in our final sample. They are displayed in order of the KOI number of the host, but within each set, the candidate's TTVs are plotted in order of their orbital periods. These figures include candidates with rapidly diverging projected TTVs, candidates with slowly diverging TTVs, and candidates with no TTVs.

Tables~\ref{tbl-ObsProperties1a}--\ref{tbl-ObsProperties3} summarize the photometric properties of each target and the divergence of their TTVs after 4000 days and after 6000 days or until the projected uncertainty in TTVs diverges by more than one day. In systems where TTVs of at least one planet are readily detected and the uncertainties diverge by at least 90 minutes, future transit-timing data will likely improve constraints on planetary and orbital parameters. We consider these the most urgent candidates for follow-up transit timing, and these are listed in Table~\ref{tbl-ObsProperties1a} and ~\ref{tbl-ObsProperties1b}. In other cases, TTVs that are strongly detected and were identified by \citet{Holczer2016} but diverge slowly are in Table~\ref{tbl-ObsProperties2}. A third category in Table~\ref{tbl-ObsProperties3} includes systems with weakly detected TTVs that also diverge slowly. 
 \begin{table}[ht!]
  \begin{center}
  \tiny
    \begin{tabular}{|c|c|c|c|c|c|c|c|}
      \hline
      KOI &  Depth (ppm) & Dur (hours) & Kep-mag & J-mag   & $\delta$ TTV (mins) (a) & $\delta$ TTV (mins) (b) & Fig. \\
 \hline
248.01 & 1766  & 2.54 & 15.264  & 13.184  & 9 & 18  & ~\ref{fig:KOI-168fut}\\
248.02 & 1387  & 2.18 &  &   & 12  & 33 & \\
248.03 & 853  & 1.60 &  &   & 12  & 24  & \\
248.04 & 814  & 2.31 &   &   & 136  &  278 & \\
\hline
277.01 & 502 & 7.47   & 11.866  & 11.124 & 89 &  144  & ~\ref{fig:KOI-168fut}  \\ 
277.02 & 85 &   7.28  &  &  & 162  &  256  & \\ 
\hline
401.01 & 2047 & 5.12   & 14.001  & 12.694 & 4 &  6   & ~\ref{fig:KOI-377fut}  \\ 
401.02 & 1553 & 5.47   &  &  & 18 & 28    &   \\ 
401.03 & 326 & 6.92   &   &  & 50 &  96    &  \\ 
\hline
520.01 & 882  &  3.59 & 14.550  & 13.097  & 11 &  18 & ~\ref{fig:KOI-377fut} \\
520.02 & 328  & 2.40  &  &  & 33  & 61   & \\
520.03 & 739  & 2.77  &  &  &  > 1  day &  > 1 day & \\
520.04 &  259 &  4.64 &  &  &  >  1 day  & > 1 day  & \\
\hline
523.01 & 3184  & 5.28  & 15.000  & 13.858  & 30 & 52  & ~\ref{fig:KOI-523fut} \\
523.02 &  714 & 7.50  &  &  & 53  & 97  & \\
\hline
567.01 &  778 & 3.38  & 14.338  & 13.100  & 25 & 54   &  ~\ref{fig:KOI-523fut} \\ 
567.02 &  533 & 4.50  &   &   & 168 &  275  & \\ 
567.03 &  638 & 4.03  &   &   & 135 &  369   & \\ 
\hline
730.01 & 782  & 6.12  & 15.344  & 14.095  & 411 & 691  &  ~\ref{fig:KOI-523fut}  \\ 
730.02 &  437 & 5.75  &  &   & 253 & 643  & \\ 
730.03 &  578 &  7.20 &   &   & 915 & 1423  & \\ 
730.04 &  350 & 5.74  &  &   & 233 &  449  & \\ 
\hline
738.01 & 1185 &  3.08 & 15.282  & 14.131 & 102 &  238  &   ~\ref{fig:KOI-523fut}  \\ 
738.02 & 1040 &  3.28 &   &  & 138 &  310  & \\ 
\hline
750.01 & 857  & 3.46  &  15.377 & 13.810  & > 1 day  & > 1 day &  ~\ref{fig:KOI-750fut}  \\ 
750.02 &  172 &  1.62 &   &   & > 1 day  & > 1 day&    \\ 
750.03 &  227 & 2.71  &   &   & 26 & 43 &    \\ 
\hline
877.01 & 1411  &  2.39 &  15.019  & 13.168 & 27 & 74  &  ~\ref{fig:KOI-750fut}   \\ 
877.02 &  1251 & 2.74  &    &  & 14 &  31  & \\ 
877.03 & 439  & 2.37  &   &  & 154 & 364  & \\ 
\hline
886.01 & 1334  & 2.31 &  15.847  & 13.508 & 32 &  76   &  ~\ref{fig:KOI-750fut}  \\ 
886.02 & 759  & 4.32 &    &  & 71 &  131  & \\ 
886.03 & 767  & 2.63 &   &  & 31 &  49  & \\ 
\hline
934.01 &  1539 & 2.92 &   15.843  & 14.675  & 5 & 9   &  ~\ref{fig:KOI-750fut}  \\ 
934.02 &  611 & 3.74 &     &   & 33 & 94  & \\ 
934.03 &  797 & 3.90 &    &   & 27 &  43  & \\ 
\hline
1070.01 & 549  & 3.68  &   15.590  & 14.300 & 17 &  27  & ~\ref{fig:KOI-1070fut}   \\ 
1070.02 &  1396 & 8.04 &     &   & > 1 day & > 1 day   & \\ 
1070.03 & 359  & 6.64 &    &   & > 1 day & > 1 day   & \\ 
\hline
     \end{tabular}
    \caption{Nominal parameters for prime targets for follow-up transit photometry (Part 1 of 2). Projected TTVs of at least one planet in each system diverge by at least 90 minutes. The columns denote KOI, transit depth (ppm), nominal duration (hours), apparent magnitudes in the Kepler band and J band, and projected transit-timing uncertainty on the first transit after (a) BJD: 2458900 and the last transit before (b) BJD: 2460900 respectively. The final column references the figure where the projected TTVs are illustrated. 
    }\label{tbl-ObsProperties1a}
  \end{center}
\end{table}

 \begin{table}[ht!]
  \begin{center}
  \tiny
    \begin{tabular}{|c|c|c|c|c|c|c|c|}
      \hline
      KOI &  Depth (ppm) & Dur (hours) & Kep-mag & J-mag   & $\delta$ TTV (mins) (a) & $\delta$ TTV (mins) (b) & Fig. \\
\hline
1338.01 & 233  & 2.92  &   15.590  & 14.609 & 13 &  21  &  ~\ref{fig:KOI-1070fut}    \\ 
1338.02 &  267 & 6.28 &     &   & > 1 day & > 1 day   & \\ 
1338.03 & 164  & 4.51 &     &   & > 1 day & > 1 day   & \\ 
\hline
1574.01 & 4850  & 12.04 &   14.600  & 13.389 & 99 & 170  &   ~\ref{fig:KOI-1070fut}   \\ 
1574.02 & 1061  & 16.94 &     &  & > 1 day  &  < 1 day   & \\ 
1574.03 & 134  & 4.58 &     &  & 50  &  79  & \\ 
\hline
1576.01 & 825  & 2.77 &   14.072  & 12.833 & 6 & 10  &   ~\ref{fig:KOI-1576fut}  \\ 
1576.02 & 662  & 3.04 &    &  & 11 &  19  & \\ 
1576.03 & 125  & 1.92 &     &  & 69 & 108  & \\ 
      \hline
1599.01 & 363.4  & 8.84 &   14.802  & 13.361  & 845  & 1434  & ~\ref{fig:KOI-1576fut}  \\
1599.02 & 376.1  & 5.18 &     &   & 713 & 1132  & \\
\hline
1783.01 & 4034 & 5.93 &   13.929  & 12.917 & 48 &  95  & ~\ref{fig:KOI-1576fut} \\ 
1783.02 & 1628 & 8.30 &     &  & 319 &  562  & \\ 
\hline
1831.01 &  1025 & 6.34 &   14.122  & 12.783  & 99 & 152  & ~\ref{fig:KOI-1576fut}  \\ 
1831.02 & 199  & 2.51  &    &   & 21 & 41 & \\ 
1831.03 & 204  & 4.49  &     &   & 120 & 218  & \\ 
1831.04 &  356 & 1.23  &     &   & 19 & 34 & \\ 
\hline
1955.01 & 225  & 4.33 &   13.147  & 12.220  & 12 & 19   & ~\ref{fig:KOI-1955fut}  \\ 
1955.02 & 245  & 9.56 &     &   & 707 & 711 & \\ 
1955.03 & 48  & 2.91 &     &   & 17 & 28  & \\ 
1955.04 & 197  & 3.54 &     &   & 297 & 299  & \\ 
\hline
2086.01 & 145  & 4.37 &   13.959  & 12.804 & 94 &  143  & ~\ref{fig:KOI-1955fut}  \\ 
2086.02 &  181 & 4.14 &     &  & 39 &  53  & \\ 
2086.03 & 131  & 3.02  &     &  & 142 & 230  & \\ 
\hline
2092.01 & 1627  & 5.12 &   15.886  & 14.736  & > 1 day &  > 1 day  & ~\ref{fig:KOI-1955fut}   \\
2092.02 & 1125  & 4.69 &    &   & 498 & 759  & \\
2092.03 &  943 & 4.77 &     &   & > 1 day &  > 1 day  & \\
\hline
2174.01 &  782 & 2.31  &   15.673  & 13.732  & 158  & 309   & ~\ref{fig:KOI-1955fut}  \\
2174.02 &  825 & 4.16 &     &   &  28 &  45   & \\
2174.03 &  382 & 2.04 &     &   &  216 & 426   & \\
2174.04 &  196 & 1.57 &    &  & 25  &   42 & \\
\hline
2195.01 & 305  & 7.17 &   14.881  & 13.878  & 975 &  > 1 day  & ~\ref{fig:KOI-1955fut}  \\ 
2195.02 & 243  & 6.65 &  &   & > 1 day &  > 1 day  & \\ 
2195.03 & 139  & 4.92 &    &  & 30 & 48  & \\ 
\hline 
2414.01 & 143  & 6.18 &   13.584  & 12.419 & > 1 day &  > 1 day  & ~\ref{fig:KOI-2414fut}  \\
2414.02 & 163  & 4.86 &     &  & > 1 day  &  > 1 day  & \\
\hline 
3503.01 & 72  & 3.69 &   13.827  & 12.807  & 292 &  390  & ~\ref{fig:KOI-2414fut}  \\
3503.02 & 79  & 4.17 &     &  &  351 & 432   & \\
\hline
     \end{tabular}
    \caption{Nominal parameters for prime targets for follow-up transit photometry (Part 2 of 2). Projected TTVs of at least one planet in each system diverge by at least 90 minutes. The columns denote KOI, transit depth (ppm), nominal duration (hours), apparent magnitudes in the Kepler band and J band, and projected transit-timing uncertainty on the first transit after (a) BJD: 2458900 and the last transit before (b) BJD: 2460900 respectively. The final column references the figure where the projected TTVs are illustrated. 
    }\label{tbl-ObsProperties1b}
  \end{center}
\end{table}

 \begin{table}[ht!]
  \begin{center}
  \tiny
    \begin{tabular}{|c|c|c|c|c|c|c|c|}
      \hline
          KOI &  Depth (ppm) & Dur (hours) & Kep-mag & J-mag   & $\delta$ TTV (mins) (a) & $\delta$ TTV (mins) (b) & Fig. \\
\hline
137.01 & 2270 & 3.41   &13.549  & 12.189  & 1 &  2  &  ~\ref{fig:KOI-85fut}  \\ 
137.02 & 3270 & 3.53  &   &   & 1 &  2  &   \\ 
137.03 & 315 & 1.98 &   &   & 5  &   6  &  \\ 
\hline
 152.01 &  2894 & 8.64  & 13.914  & 12.913  & 5 &   9 &  ~\ref{fig:KOI-85fut} \\ 
152.02 & 748  & 6.85 &   &   & 15 &  24 & \\ 
152.03 & 654 & 5.02  &  &   & 8 &  13  &  \\ 
152.04 & 476 & 3.00 &   &   & 34  & 55   &  \\ 
\hline
156.01 & 591  & 2.63 & 13.738  & 12.035  & 19 &  35  & ~\ref{fig:KOI-85fut}  \\ 
156.02 &  359 & 2.37  &    &   & 40 &  70  &   \\ 
156.03 & 1431  & 2.90  &   &   & 3 &  5   &   \\ 
\hline
168.01 & 415 & 6.06 & 13.438  & 12.353  & 10 & 18   & ~\ref{fig:KOI-168fut} \\ 
168.02 & 207 & 6.08 &   &   & 19  &  30  &   \\ 
168.03 & 131 & 5.15 &    &   & 36 &  57   &   \\ 
\hline
250.01 & 2795 & 2.71  &  15.473  & 13.408  & 46 & 74  & ~\ref{fig:KOI-168fut} \\ 
250.02 & 2055  & 2.04 &    &   & 44 & 64  & \\ 
250.03 &  507 & 1.87 &    &   & 13 &  22 & \\ 
250.04 & 1538  & 1.90 &   &   & 13 & 23 & \\ 
\hline
314.01 & 756 & 2.32 & 12.925 & 10.293 & 4 &  5  &  ~\ref{fig:KOI-168fut}   \\
314.02 & 598 & 1.70 &  &  & 5 & 8  & \\
314.03 & 138 & 2.01 &  &  & 32 & 43  & \\
\hline
377.01 & 6661  & 4.13  & 13.803  & 12.710  & 21 & 37  &  ~\ref{fig:KOI-377fut}  \\ 
377.02 & 6159  & 4.53  &   &   & 47 & 84  & \\ 
377.03 &  248 & 1.93  &   &   & 6 &  10  & \\ 
\hline
620.01 &  6209 & 5.78  & 14.669  & 13.562 & 4 &  5  &  ~\ref{fig:KOI-523fut} \\ 
620.02 & 11571 & 8.45  &  &  & 30 &  56  & \\ 
620.03 & 1903 &  2.74 &   &  & 38 &  71  & \\ 
\hline
806.01 & 10566  & 8.93 &  15.403  & 13.997 & 6 &  9  &  ~\ref{fig:KOI-750fut}  \\ 
806.02 & 20280  & 6.62 &    & & 2 & 3  &   \\ 
806.03 &  488 & 4.80 &    &  & 22 &  48 &   \\ 
\hline
1353.01 & 12389 & 9.02  &  13.956   & 12.861  & 3  &  6  &  ~\ref{fig:KOI-1070fut}    \\ 
1353.02 & 430  & 3.48 &     &   & 17 &   26 & \\ 
Kep-289d & 617 & 4.30  &     &   & 26  &  45  & \\ 
\hline
     \end{tabular}
    \caption{Nominal target parameters for TTV systems (identified in \citealt{Holczer2016}) with slowly diverging projected TTVs. Columns: transit depth (ppm), nominal duration (hours), apparent magnitudes in the Kepler band and J band, and projected transit-timing uncertainty after (a) BJD: 2458900 and (b) BJD: 2460900 respectively. The final column references the figure where the projected TTVs are illustrated. 
    }\label{tbl-ObsProperties2}
  \end{center}
\end{table}

 \begin{table}[ht!]
  \begin{center}
  \tiny
    \begin{tabular}{|c|c|c|c|c|c|c|c|}
      \hline
          KOI &  Depth (ppm) & Dur (hours) & Kep-mag & J-mag   & $\delta$ TTV (mins) (a) & $\delta$ TTV (mins) (b) & Fig. \\
 \hline
85.01 &  323  & 4.02  &  11.018  & 10.066   & 19 & 30  & ~\ref{fig:KOI-85fut}   \\ 
85.02 &  100 &  3.13 &  &    & 5 & 7  &      \\ 
85.03 &  112  & 4.24  &  &   & 13 &  18   &   \\ 
\hline
115.01 & 602  &  2.95  &12.791  & 11.811  & 2 &  4  & ~\ref{fig:KOI-85fut}   \\ 
115.02 & 192 &  3.00 &   &   & 5 &  9  &   \\ 
115.03 &  23  & 3.13 &   &   & 19 &  31  &   \\ 
\hline
222.01 & 1286  & 2.75 &  14.735  & 13.019  & 4 & 6   & ~\ref{fig:KOI-168fut}  \\ 
222.02 &  817 & 3.42 &   &   & 9 & 13   &   \\ 
\hline
244.01 & 1180  & 2.73 & 10.734  & 9.764  & 1 &  1  & ~\ref{fig:KOI-168fut} \\ 
244.02 &  402 & 3.53 &   &   & 2 &  4  &   \\ 
\hline
255.01 &  2313 & 4.07 &  15.108  & 12.912  & 7 & 10  & ~\ref{fig:KOI-168fut} \\ 
255.02 &  181 & 2.59 &  &   & 50 & 73  &   \\ 
\hline
430.01  &  1709 & 2.71 &  14.897 & 12.991 & 22 & 32 & ~\ref{fig:KOI-377fut} \\  
430.02  &  201 &  2.57   &   &  & 41 & 68 & \\ 
\hline
457.01  &  759 & 1.88 &  14.196 & 12.767 & 4 & 7 & ~\ref{fig:KOI-377fut} \\  
457.02  &  732 &  1.35   &   &  & 6 & 9 & \\ 
\hline
654.01 & 337  & 3.25  & 13.984  & 12.871  & 16 & 28   & ~\ref{fig:KOI-523fut}  \\ 
654.02 & 238  & 1.19  &   &   & 18 & 30   &   \\ 
\hline
1279.01 & 336  & 4.79 &  13.749  & 12.631 & 11 & 18   & ~\ref{fig:KOI-1070fut}   \\ 
1279.02 & 103  & 4.30 &    &  & 34 & 53   &  \\ 
\hline
1598.01 & 1092 & 5.96 &   14.279  & 13.056  & 16  & 32  &   ~\ref{fig:KOI-1070fut}  \\ 
1598.02 & 659 & 7.47  &     &  & 31 &  56 & \\ 
1598.03 & 187 & 2.31  &     &  & 22 & 35  & \\ 
\hline
1833.01 & 763  & 1.77  &   14.265  & 12.518  & 5 &  9  & ~\ref{fig:KOI-1576fut}  \\ 
1833.02 &  1175 & 1.44 &     &   & 5 & 9   & \\ 
1833.03 & 587  & 1.62 &     &   & 9 & 27   & \\ 
\hline
2113.01 & 1094  & 3.49 &   15.886  & 14.426  & 17  & 27  & ~\ref{fig:KOI-1955fut}  \\ 
2113.02 &  872 & 3.07 &     &   & 17 & 28 &   \\ 
\hline
     \end{tabular}
    \caption{Nominal target parameters for TTV targets unreported by \citet{Holczer2016} and with no projected TTVs diverging by 90 minutes. Columns: transit depth (ppm), nominal duration (hours), apparent magnitudes in the Kepler band and J band, and projected transit-timing uncertainty after (a) BJD: 2458900 and (b) BJD: 2460900 respectively. The final column references the figure where the projected TTVs are illustrated. 
    }\label{tbl-ObsProperties3}
  \end{center}
\end{table}
\clearpage

\section{Individual systems: TTV results and the value of follow-up transit timing}
Joint posteriors of dynamical masses between adjacent planets and eccentricity vector components are displayed in Appendix B. 

We chose our systems for analysis based on the expectation of TTVs given the periods of the planets and the median uncertainty of transit times using two analytical models for TTVs. Here, we consider the value of the existing transit-timing dataset and future data to group the systems as follows:  (1) systems with an expectation of both types of TTV signal, with a score in Tables~\ref{tbl-SystemSelection1} and ~\ref{tbl-SystemSelection2} above 7;  (2), systems with an expected near-resonant TTV score above 7 but an expected nonresonant signal below 4; (3) systems that had an expected nonresonant TTV score above 7 but no expected near-resonant signal;  and (4) systems with no expectation of strongly interacting planets such that the highest expected resonant or nonresonant TTV scores are between 4 and 7. We expect weak upper limits on the masses of most of these planets. However, planets that are significantly more massive than 4 $M_{\oplus}$ may be strongly detected within this category. When characterizing the value of future data for any system, we make the assumption that if dynamical constraints are expected to improve for any particular planet, then they would likely improve for all interacting neighbors within that that system.  
 
 \subsection{Resonant and Nonresonant Interactions Expected}
The extreme low-density planet KOI-152.01 (Kepler-79 d) has strongly detected TTVs induced by two neighboring planets. All four planets at Kepler-79 are well characterized in both dynamical mass and orbital eccentricity. The projected TTVs diverge slowly for the Kepler-79 d, and the range of projected transit times diverges by less than one hour. However, the TTVs of KOI-152.04 (Kepler-79 e) diverge to roughly one hour of total range. Hence, transit-timing precision of one hour or less may further constrain planetary masses and orbital parameters.

KOI-523 (Kepler-177) has strongly detected near-resonant and nonresonant TTVs. The projected TTVs diverge quickly, and hence follow-up transit-timing (e.g., \citealt{Vissapragada2020}) would further improve the planet masses and eccentricities. 

KOI-620 (Kepler-51) has strongly detected near-resonant TTVs, and nonresonant TTVs in KOI-620.03 (Kepler-51 c). The projected transit times diverge quickly, and hence additional transit-timing data (e.g. \citealt{LibbyRoberts2020}) would be valuable. 

KOI-1353.01 and its inner neighbor Kepler-289 d (PH3 c) show near-resonant TTVs at a superperiod of $\sim$1370 days. Both planets have moderately increasing uncertainties on their projected transit times (see Table~\ref{tbl-ObsProperties2}). However, the outermost planet has a very high transit S/N and its transit timing uncertainties in the Kepler data are $\sim$1 minute. Hence, if a transit-timing uncertainty of a few minutes is achievable for KOI-1353.01, or a few tens of minutes for Kepler-289 d or KOI-1353.02, with follow-up transit photometry, the TTV constraints on the planet masses would likely improve.

The inner pair of planets at KOI-1576 (Kepler-307) have strongly detected resonant TTVs as expected. There is some divergence in the future transit times, with a separation in future times based on the planet masses, making this a strong candidate for follow-up transit timing. 

KOI-1783 (Kepler-1662) has two long-period transiting planets with relatively few transits in the Kepler dataset. The future transit times diverge rapidly, and the dynamical mass measurements would benefit from additional transit-timing data (e.g., \citealt{Vissapragada2020}). 

KOI-2092.01 (Kepler-359 c) has a strong expectation of near-resonant and nonresonant TTVs. However, the superperiod is much longer than the Kepler baseline, and the TTVs provide weak constraints on the planetary masses. The projected TTVs diverge rapidly, and additional transit-timing data is likely to significantly improve the planet characterizations in this system. 

 \subsection{Resonant Interactions Expected}
 KOI-137 (Kepler-18) has strongly detected resonant TTVs with a superperiod of $\sim$300 days which is well sampled by the Kepler baseline. The TTVs diverge extremely slowly.  Similarly, KOI-156.03 (Kepler-114 d) has strongly detected near-resonant TTVs with a superperiod that is much shorter than the Kepler baseline. In these cases, follow-up transit timing is unlikely to further constrain the planetary masses for decades. 

This is also the case for the two planets of KOI-244 (Kepler-25), which have strongly detected near-resonant TTVs with a super-period that is much shorter than the Kepler baseline. In this system, low-mass solutions with eccentricities $\lesssim 0.5$ provide a good fit to the TTVs that are inconsistent with the larger masses suggested by RV observations \citep{mar14} However, there are low-eccentricity solutions are closely consistent with the RV \citep{Hadden2017}. The projected TTVs diverge slowly, and there is little evidence that future transit times will further constrain the TTV model, although additional RV data may be of value. Kepler-25 has a third planet found by RV at 123 days weighing 160 $M_{\oplus}$. We estimate that the maximum TTVs induced by Kepler-25 d on Kepler-25 c are of order 10 s, too small to have any effect on our two-planet TTV model.

KOI-255.01 (Kepler-505 b) has a weakly detected signal. The TTVs provide upper limits only on the planetary masses, and the projected transit times diverge with some evidence that additional transit-timing data will further constrain the phase of the TTV signal as well as the planetary masses.  

KOI-314.01 (Kepler-138 c) and KOI-314.03 (Kepler-138 b) have strongly detected TTVs from the orbital periods being near the 4:3 resonance. The outer pair, KOI-314.01 (Kepler-138 c)  and KOI-314.04 (Kepler-138 d), is near the second-order 5:3 resonance and interact strongly with a superperiod of $\sim$880 days. Their TTVs diverge slowly, while the innermost planet has a TTV period of $\sim$1540 days and has rapidly diverging projected TTVs. 

The two outer planets of KOI-377 (Kepler-9) have masses that far exceed our system selection criteria, and the TTVs have the highest S/N among the interacting pairs included in our sample. The masses and orbital eccentricities are very tightly constrained from both near-resonant and nonresonant interactions. The projected TTVs of KOI-377.02 (Kepler-9 c) diverge by over an hour after 6000 days, making this system of some value for follow-up transit timing.

KOI-430.01 (Kepler-551 b) has a strong expectation of near-resonant TTVs at a superperiod of $\sim$500 days. It is clear that there are excessive outlying transit times caused by underestimated timing uncertainties or transit-timing error. It appears likely that underestimating the transit-timing measurement uncertainties was largely responsible for our model predicting that near-resonant TTVs would have a higher S/N than observed. Nevertheless, we clearly detect a periodic signal, although the phase of the signal in KOI-430.02 is poorly constrained. The projected TTVs diverge quickly, and mass constraints would benefit from additional transit-timing data.

The middle pair of the four-planet system KOI-520 (Kepler-176) is near the 2:1 resonance with a superperiod of $\sim$1400 days. The outer pair is near the 2:1 resonance  with a superperiod of $\sim$3900 days, far longer than the Kepler baseline. The TTV amplitude and phase are well constrained for KOI-520.01 (Kepler-176 c) and its projected TTVs diverge slowly. However, its outer neighbors KOI-520.03 (Kepler-176 d) and KOI-520.04 (Kepler-176 e) have a poorly constrained TTV model and rapidly diverging projected transit times. 

KOI-730 (Kepler-223) has four planets in a series of resonant chains, with TTV periods exceeding the Kepler baseline. The projected TTVs diverge rapidly and separate by mass. Hence, further characterization of this system would benefit from additional transit-timing data.

KOI-750.01 (Kepler-662 b) is near a 3:2 resonance with its inner neighbor, with a superperiod of $\sim$1600 days. The TTV model is a poor fit to KOI-750.01, with excessive outlying transit times. The projected TTVs diverge rapidly. 

The inner pair of KOI-877 (Kepler-81) has strongly detected near-resonant TTVs. The projected TTVs diverge quickly, and additional transit-timing data would further constrain the dynamical masses.

KOI-1338 (Kepler-822) has three planet candidates. The orbital periods of KOI-1338.02 and KOI-1338.03 indicate a superperiod of $\sim$180 yr, and the two candidates are likely in a 2:1 mean motion resonance. It is clear that the transit times vary little from a constant period over the Kepler baseline. The projected TTVs diverge rapidly. 

KOI-1599 (Kepler-1659) has an interacting pair with a near-resonant superperiod longer than the Kepler baseline. Projected TTVs diverge rapidly, and additional measurements would be of value in constraining the phase and the amplitude of the TTVs, and hence the planetary masses.

KOI-1833.03 (Kepler-968 c) and KOI-1833.02 are near the 4:3 mean motion resonance, with a relatively short superperiod of $\sim$200 days. Projected TTVs diverge slowly. 

KOI-1955.02 (Kepler-342 c) and KOI-1955.02 (Kepler-342 d) have strongly detected near-resonant TTVs with a superperiod of $\sim$5000 days, which is longer than the Kepler baseline. The projected TTVs diverge quickly with separation based on the planetary masses.

KOI-2086 (Kepler-60) has strongly detected TTVs, due to libration within a three-body resonance chain \citep{Gozdziewski2016}. The projected TTVs diverge, and additional transit-timing measurements would enable tighter constraints on the planetary masses. 

KOI-2195 (Kepler-372) has three transiting planets. The interacting pair have a near-resonant superperiod of $\sim$90 yr and is likely in a 3:2 mean motion resonance. With little evidence of TTVs over the four-year Kepler baseline, the planetary masses are poorly constrained. However, divergence in the projected transit times makes this system a strong candidate for follow-up transit timing.

KOI-2414.01 (Kepler-384 b) has an expectation of detectable TTVs caused by near-resonant interactions with Kepler-384 c. However, the super-period of $\sim$6700 days exceeds the Kepler baseline, and the upper limits on the masses of the planets are weak. There are clearly non-Gaussian residuals to the model fits, casting some doubt on the model. However, the projected transit times rapidly diverge, and future transit data will further constrain the model.

\subsection{Nonresonant Interactions Expected}
KOI-85.01 (Kepler-65 c) and KOI-85.02 (Kepler-65 d) interact with no TTVs discernible by eye. Nevertheless, it is illustrative to compare the projected TTVs of KOI-85.01 and its isolated inner neighbor KOI-85.02, which has a precise linear ephemeris. The weakly interacting pair has divergent projected TTVs and well-constrained dynamical mass upper limits, even from the nondetection of TTVs. It is unclear how valuable additional transit data would be for this system. 

KOI-115.01 (Kepler-105 b), has TTVs primarily caused by its outer neighbor Kepler-105 c (KOI-115.02) with a periodicity of $\sim$140 days \citep{Jontof-Hutter2016}. This is caused by a near-resonance, which, as shown in Table~\ref{tbl-SystemSelection1} was expected to cause a moderate signal. However, this system was selected because of the expectation of nonresonant TTVs, and the posteriors show relatively tight constraints on the mass of the outermost planet. The TTVs diverge slowly, and it is unlikely that mass constraints will improve with additional transit-timing data in the near future. 

KOI-250.01 (Kepler-26 b) was expected to have a strong nonresonant TTV signal. The TTVs are in fact dominated by a near-second-order resonance (7:5) and its $\sim$770 day superperiod. Nevertheless, nonresonant components are also readily detected. Both have well-characterized masses, while the other two planets have useful upper limits only. The projected TTVs of Kepler-26 b and Kepler 26 c diverge, and hence this system would benefit from follow-up transit-timing data. 

KOI-277 (Kepler-36) shows strongly detected nonresonant TTVs. The dynamical masses and eccentricities are among the most precisely characterized for low-mass exoplanets \citep{Deck2012}. We note that the future times diverge and that additional transit-timing data would further constrain the TTV models. \citet{Vissapragada2020} report follow-up transit times that further constrain the planetary masses and orbital eccentricities.

KOI-654.01 (Kepler-200 b) and KOI-654.02 (Kepler-200 c) have an orbital period ratio of just 1.19 and interact with nonresonant TTVs, although the dynamical masses and eccentricities are weakly constrained. The projected TTVs diverge slowly following the Kepler baseline.

KOI-738 (Kepler-29) has strongly detected nonresonant TTVs and a superperiod associated with the second-order 9:7 resonance, that exceeds the Kepler baseline. While it is not apparent by eye, synodic periodicities are detected in the TTVs, and the dynamical masses are well constrained even though the baseline is too short to measure the amplitude of the near-resonant TTVs. Projected TTVs diverge quickly, motivating follow-up transit timing, as has been done by \citet{Vissapragada2020}.

KOI-1070.03, an unconfirmed candidate of Kepler-266 with a period of 92.8 days, has an expected nonresonant TTV signal due to its proximity to its outer neighbor KOI-1070.02 (Kepler-266 c) with a period of 107.7 days. The projected TTVs diverge rapidly to beyond one day within a few years after the Kepler mission.  

KOI-1574.01 (Kepler-87 b) is an extreme low-density planet with a strong nonresonant TTV signal. With relatively long orbital periods, the outer pair at Kepler-87 has few measured transit times, and additional data should tighten mass constraints significantly. The projected TTVs diverge rapidly.

KOI-2174.01 and KOI-2174.03 have orbital periods around 6.7 and 7.7 days respectively, and an expectation of nonresonant TTVs. The $\sim$2000 day periodicity seen in the model fits to the TTVs is caused by the near-second-order 15:13 resonance, and future transit data may enable tighter mass constraints on these candidates. 

 \subsection{No Strongly Detectable Interactions Expected}
 
 KOI-168.01 (Kepler-23 c) has the expectation of a weak detection of near-resonant TTVs, and the TTVs are strongly detected. The projected TTVs diverge slowly, although the separation between high-mass and low-mass solutions indicates that follow-up transit timing may be of value in constraining the planetary masses. 
 
 KOI-222.01 (Kepler-120 b) has weakly detected near-resonant TTVs that are projected to diverge slowly. 
 
KOI-248.01 (Kepler-49 b) has strongly detected TTVs at the near-resonant superperiod, although the higher frequency nonresonant TTVs are not apparent. The near-resonant TTVs are also detected as expected in Kepler-49 c. More detailed modeling using short-cadence transit-timing data by \citet{Jontof-Hutter2016} provides tighter constraints on the strongly interacting middle pair of planets in this four-planet system, although we find an upper limit on the mass of the outermost planet KOI-248.04 (Kepler-49 e). However, the poor fit to the transit times of Kepler-49 e, where no strong TTV signal is expected, hints at poorly characterized transit-timing uncertainties for this small planet. The low-mass solutions are a better fit to the data than the higher mass solutions and show a wider posterior distribution in eccentricity components. These may be distinguished with additional transit timing. The projected TTVs of Kepler-49 b and Kepler-49 c diverge slowly.

KOI-255.01 (Kepler-505 b) has TTVs at the superperiod associated with the 2:1 resonance, although it appears that the phases of the TTVs are uncertain, and there is some divergence in projected transit times. Hence, additional transit-timing data, particularly of KOI-255.02 should be of value in further characterizing this system. 

The inner pair at KOI-401 (Kepler-149) has a weakly detected near-resonant signal . Diverging projected TTVs imply that follow-up transit times would be of some value, although the uncertain future transit times for the isolated outermost planet, KOI-401.02 (Kepler 149 d), are most likely due to a poorly characterized linear ephemeris, given how few transits Kepler observed.
 
KOI-457.01 (Kepler-161 b) has an expectation of weakly detected nonresonant TTVs. Nevertheless, there are strong upper limits on the dynamical masses. There are no resonances near the orbital period ratio ($\frac{P'}{P} = 1.44$), and projected TTVs diverge slowly.  
 
KOI-567.02 (Kepler-184 c) has an expectation of weakly detectable nonresonant TTVs. These are seen by eye in the model fits but not in the data. We find a strong upper limit to the dynamical mass of planet KOI-567.02 (Kepler-184 c). The projected TTVs diverge rapidly.
 
\begin{figure}
\begin{center}
\figurenum{11}
\includegraphics [height = 1.45 in]{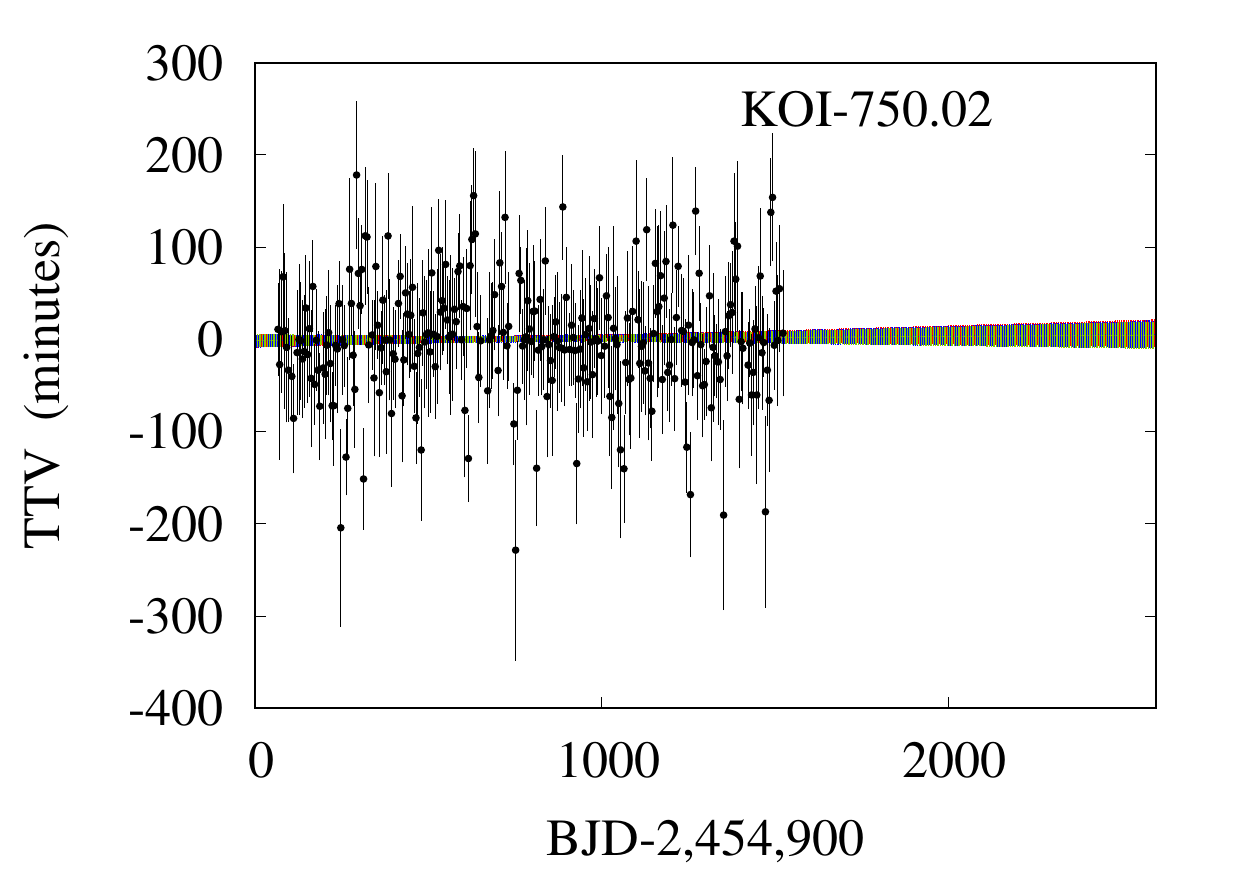}
\includegraphics [height = 1.45 in]{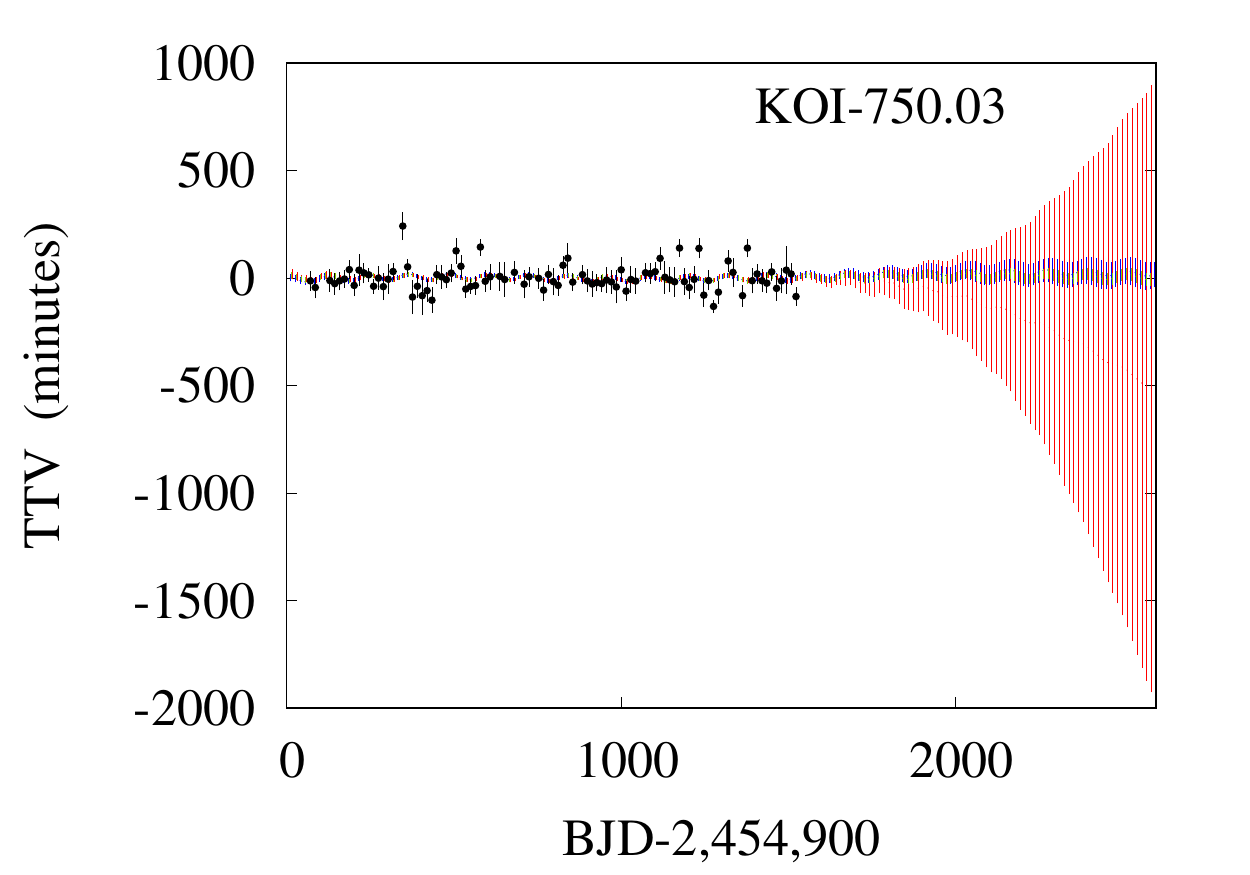}
\includegraphics [height = 1.45 in]{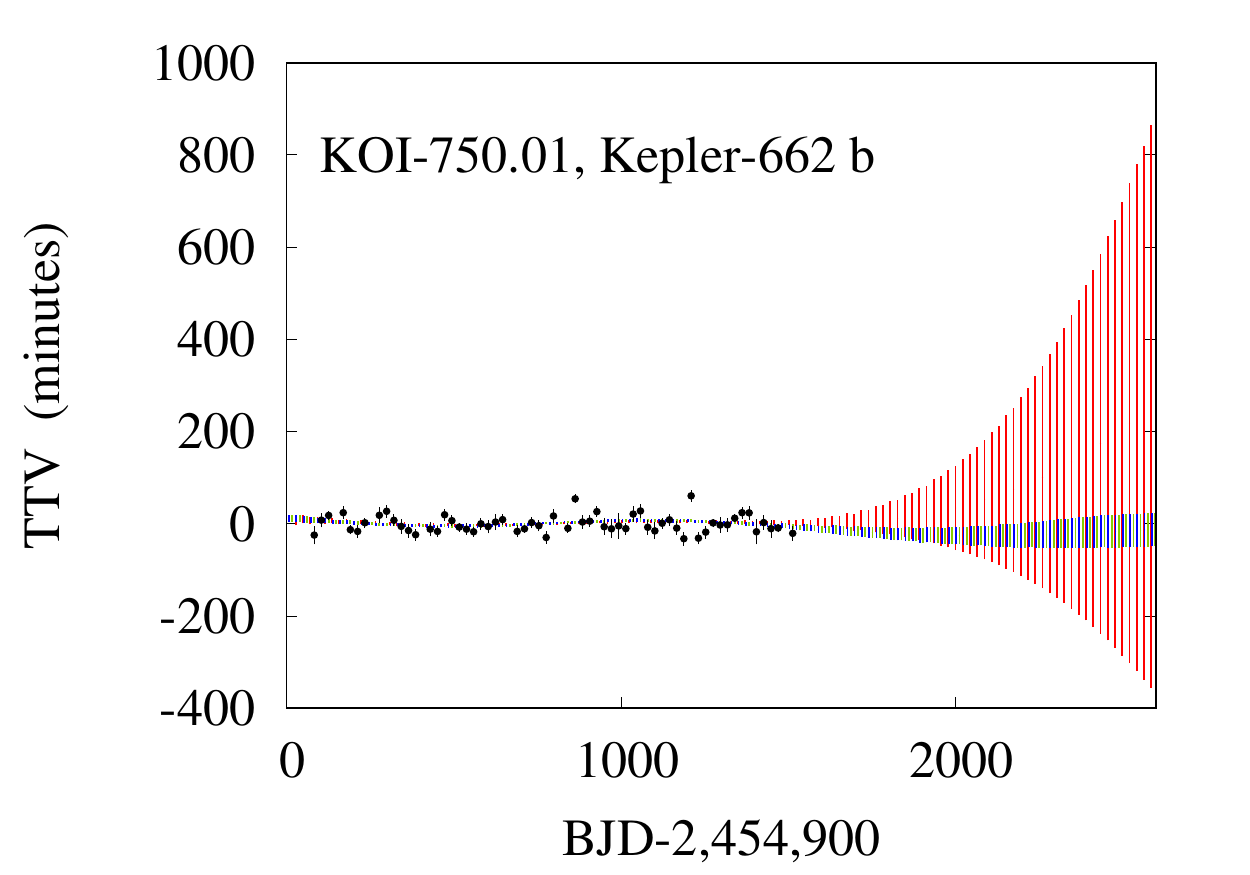}
\includegraphics[height = 1.45 in]{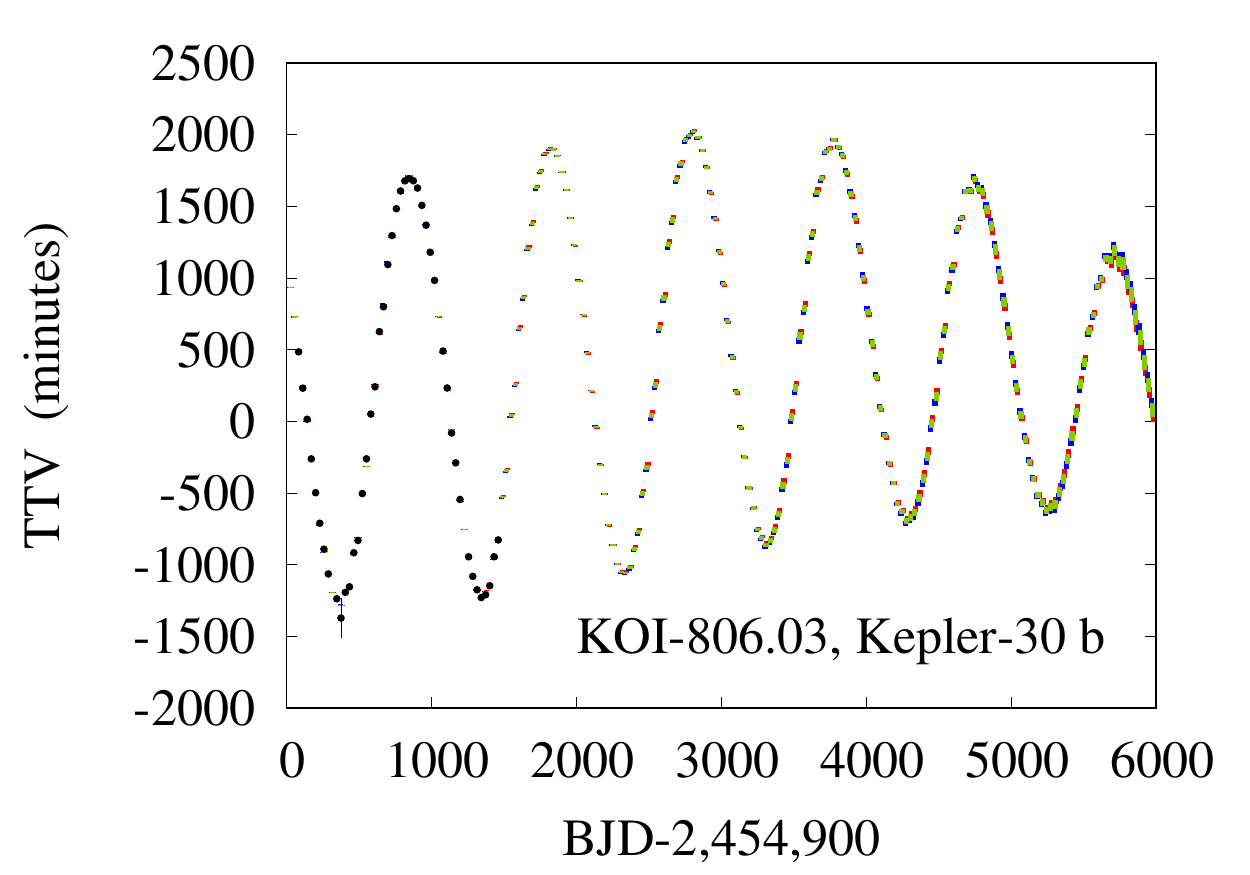}
\includegraphics[height = 1.45 in]{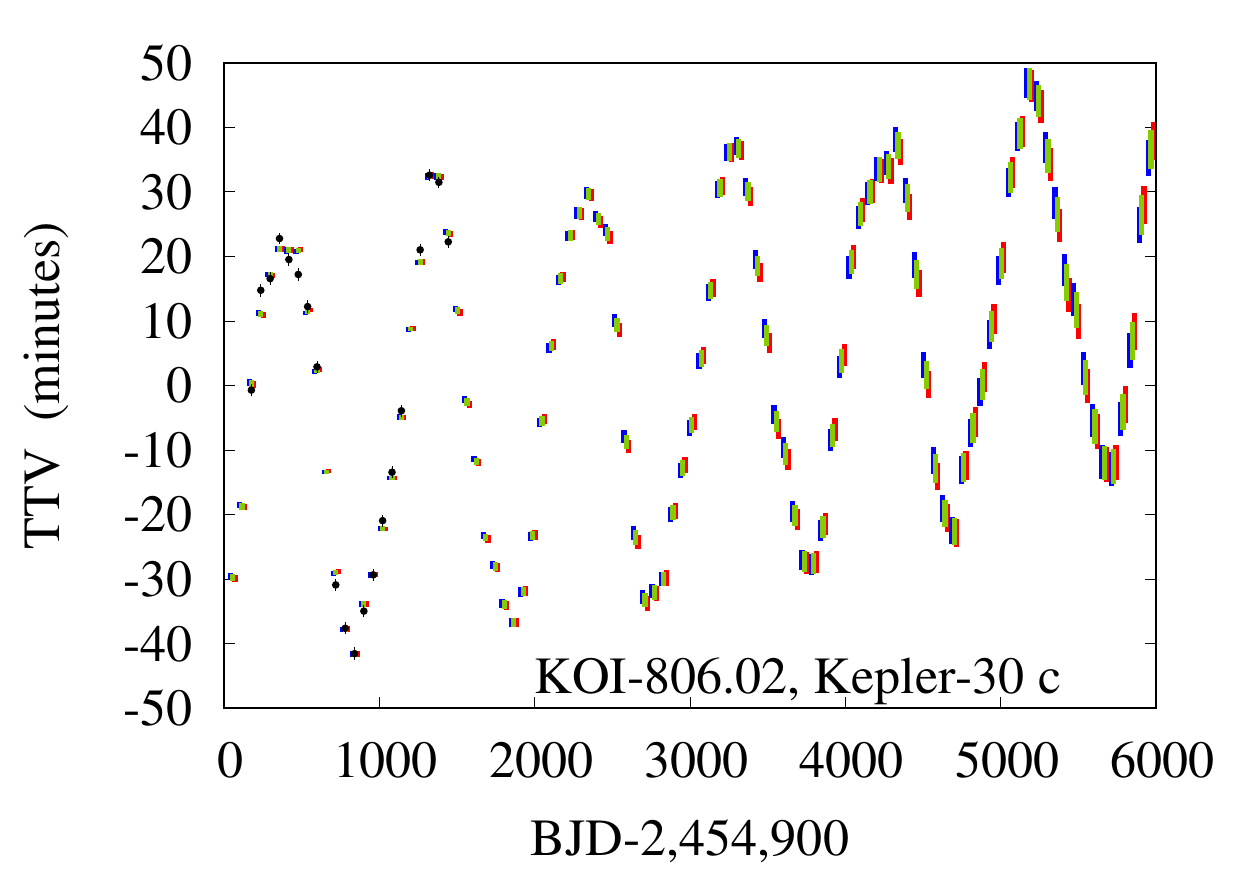}
\includegraphics[height = 1.45 in]{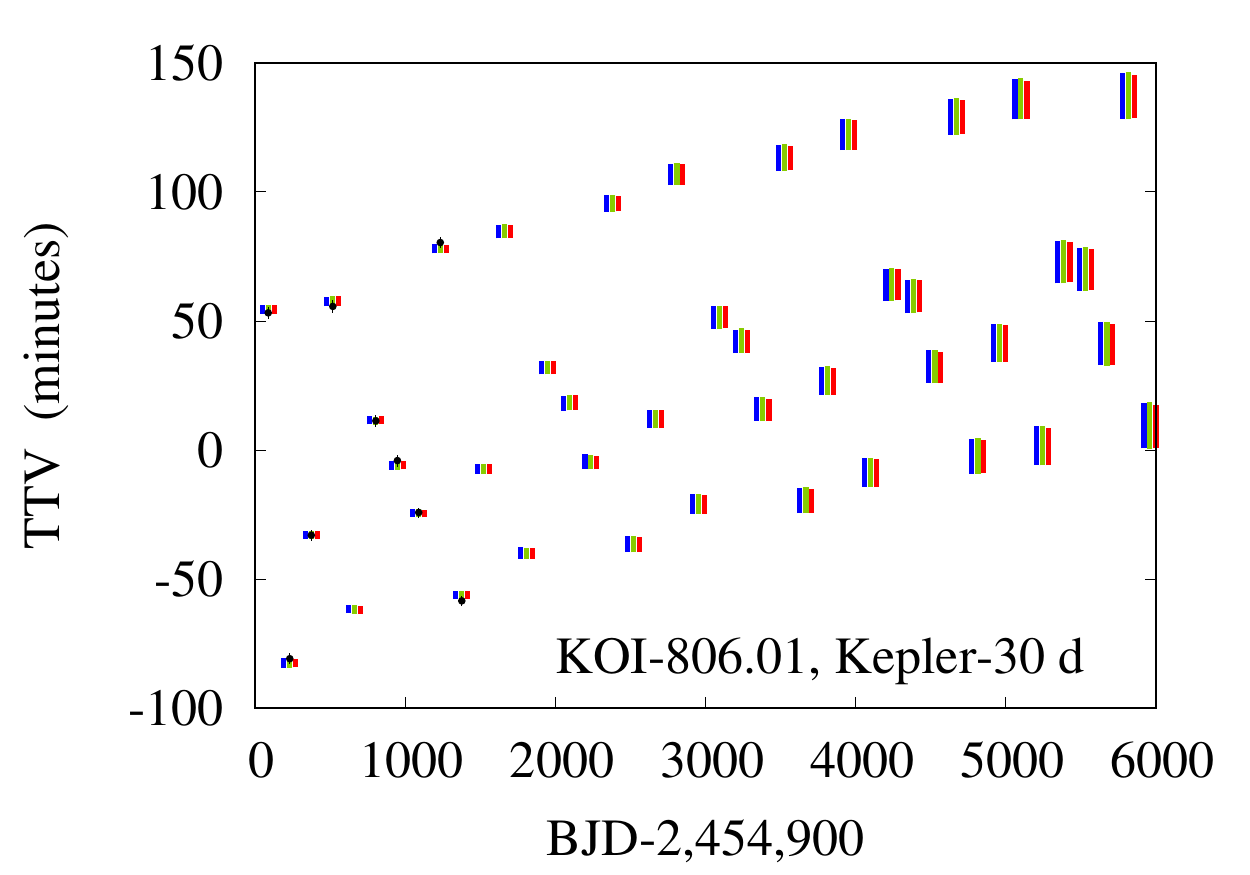}
\includegraphics[height = 1.45 in]{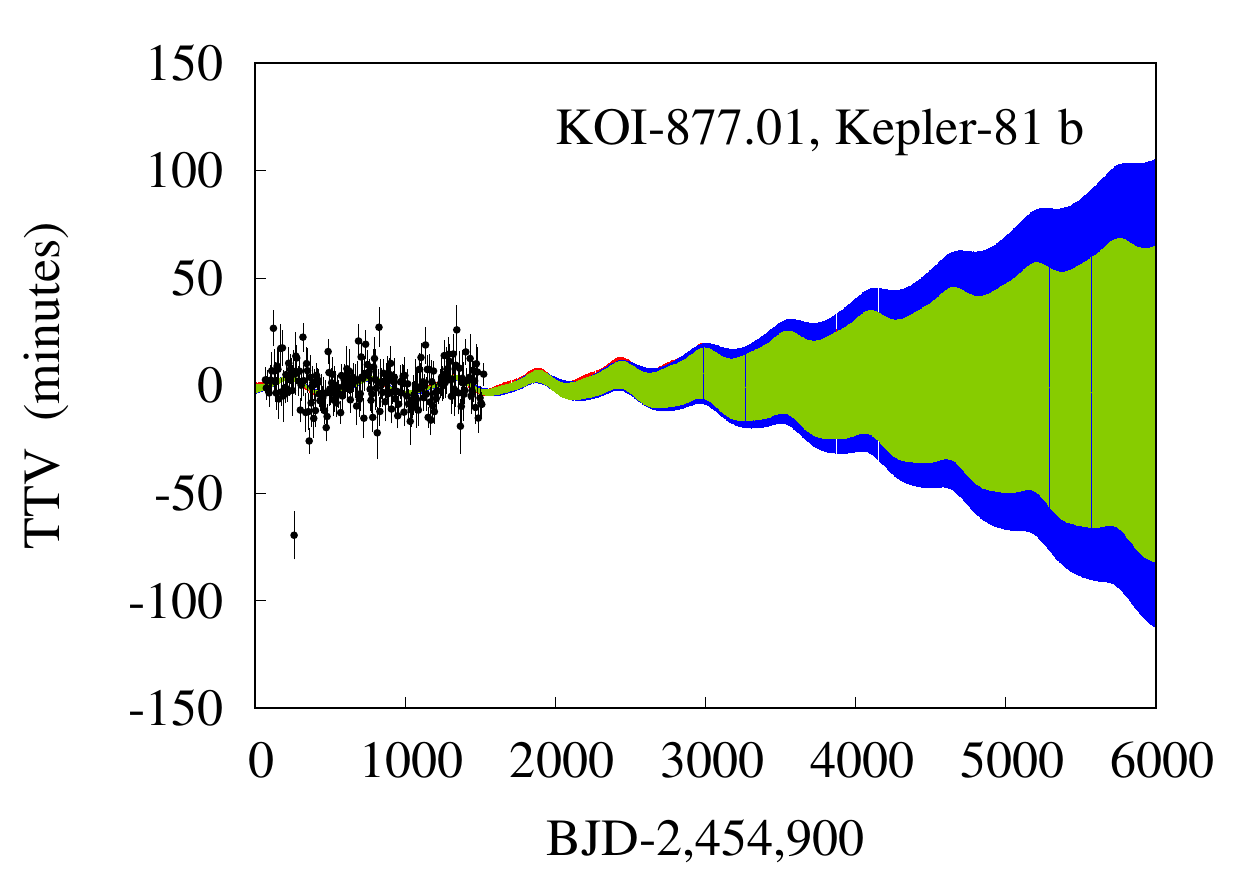}
\includegraphics[height = 1.45 in]{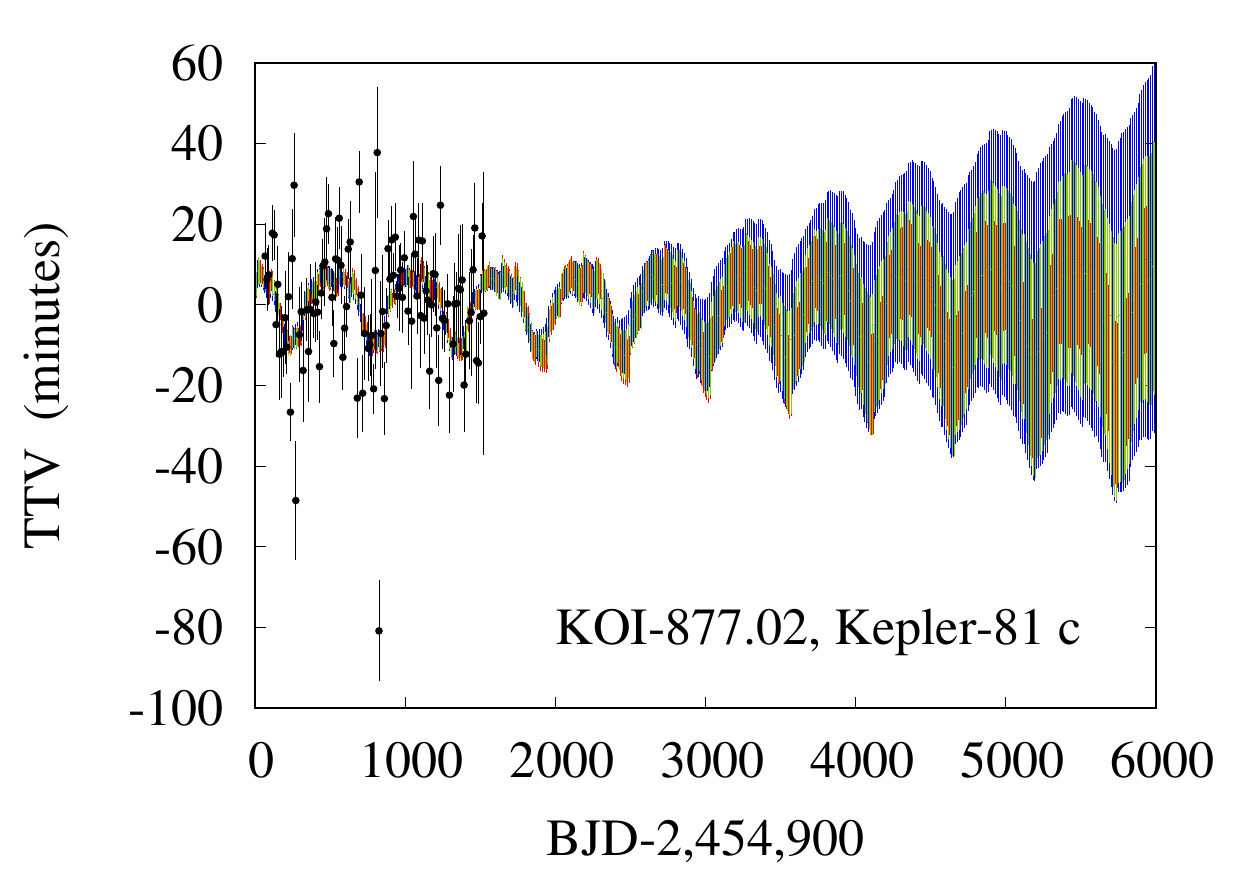}
\includegraphics[height = 1.45 in]{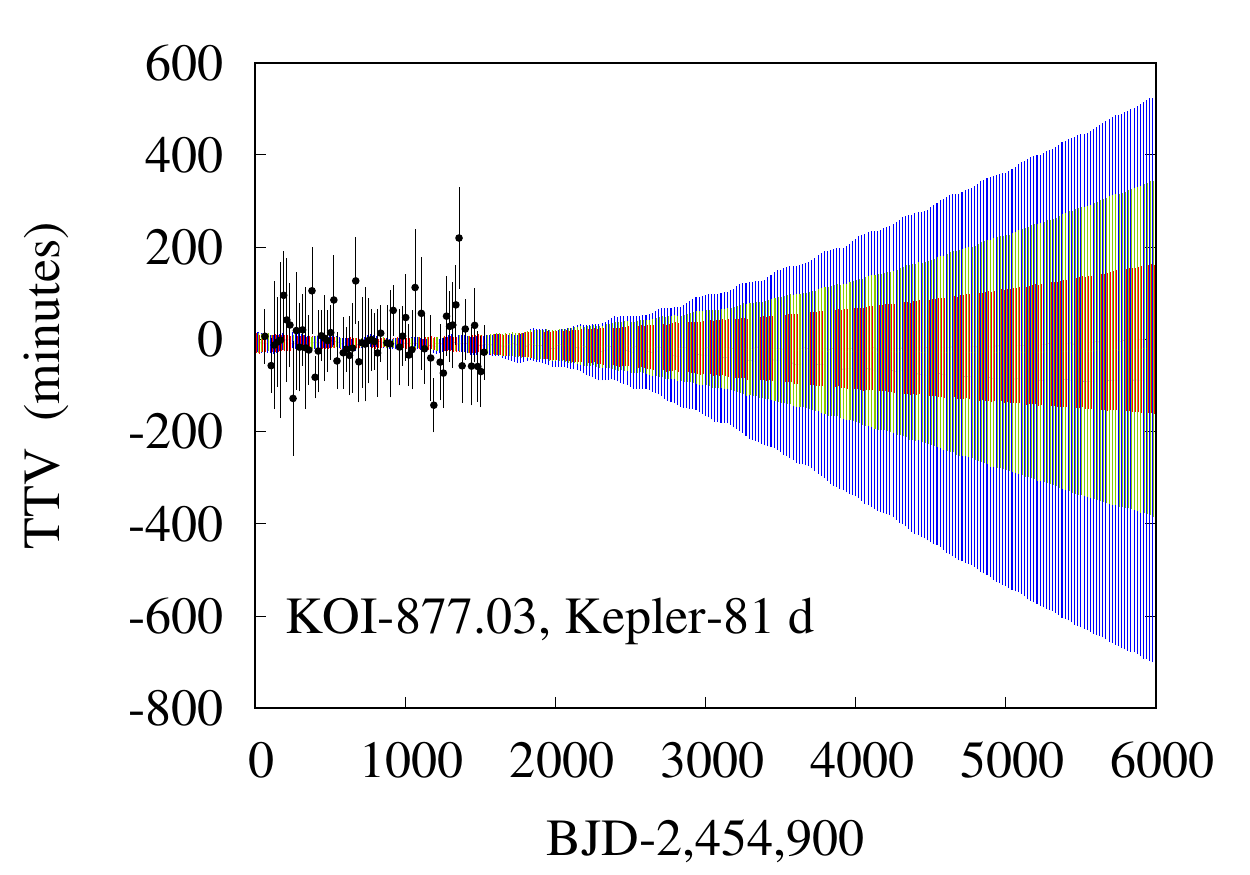}
\includegraphics[height = 1.45 in]{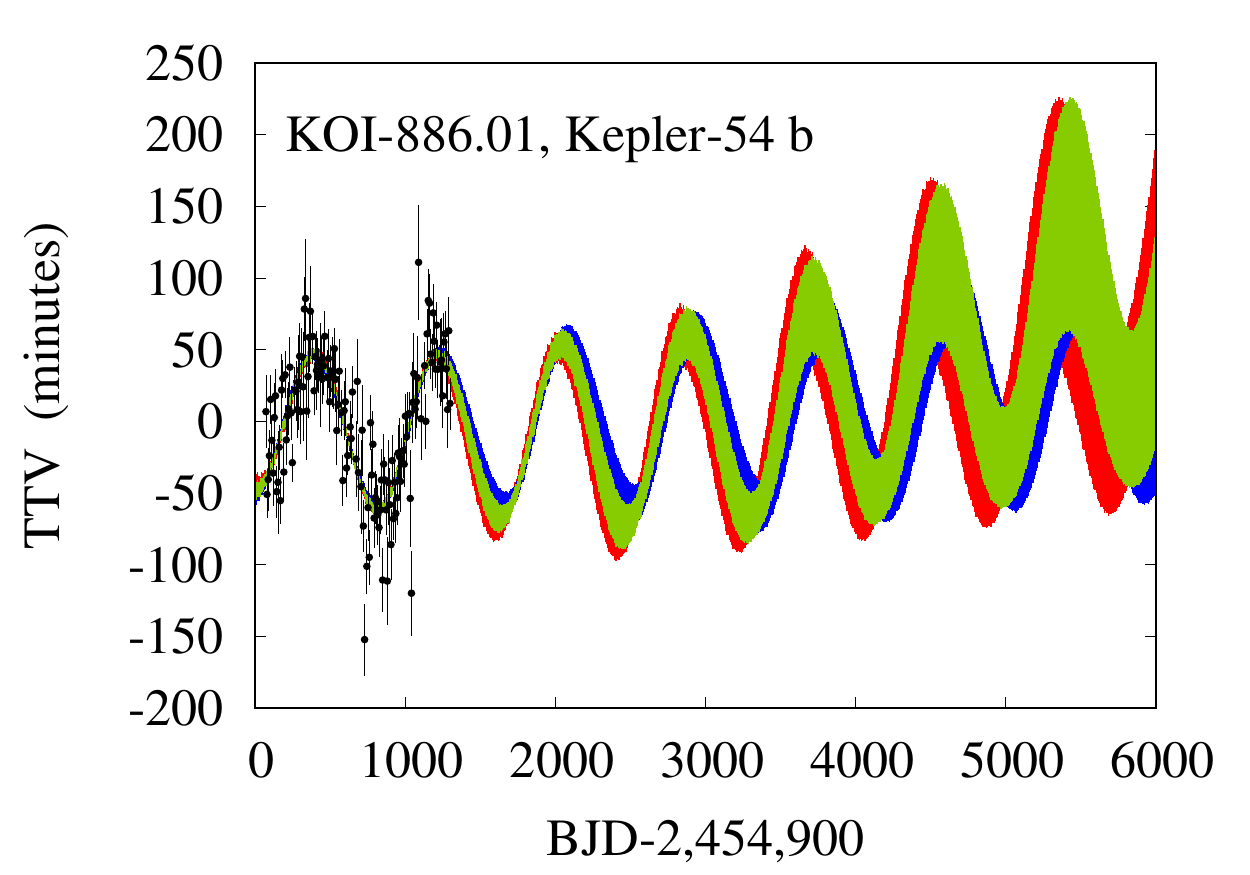} 
\includegraphics[height = 1.45 in]{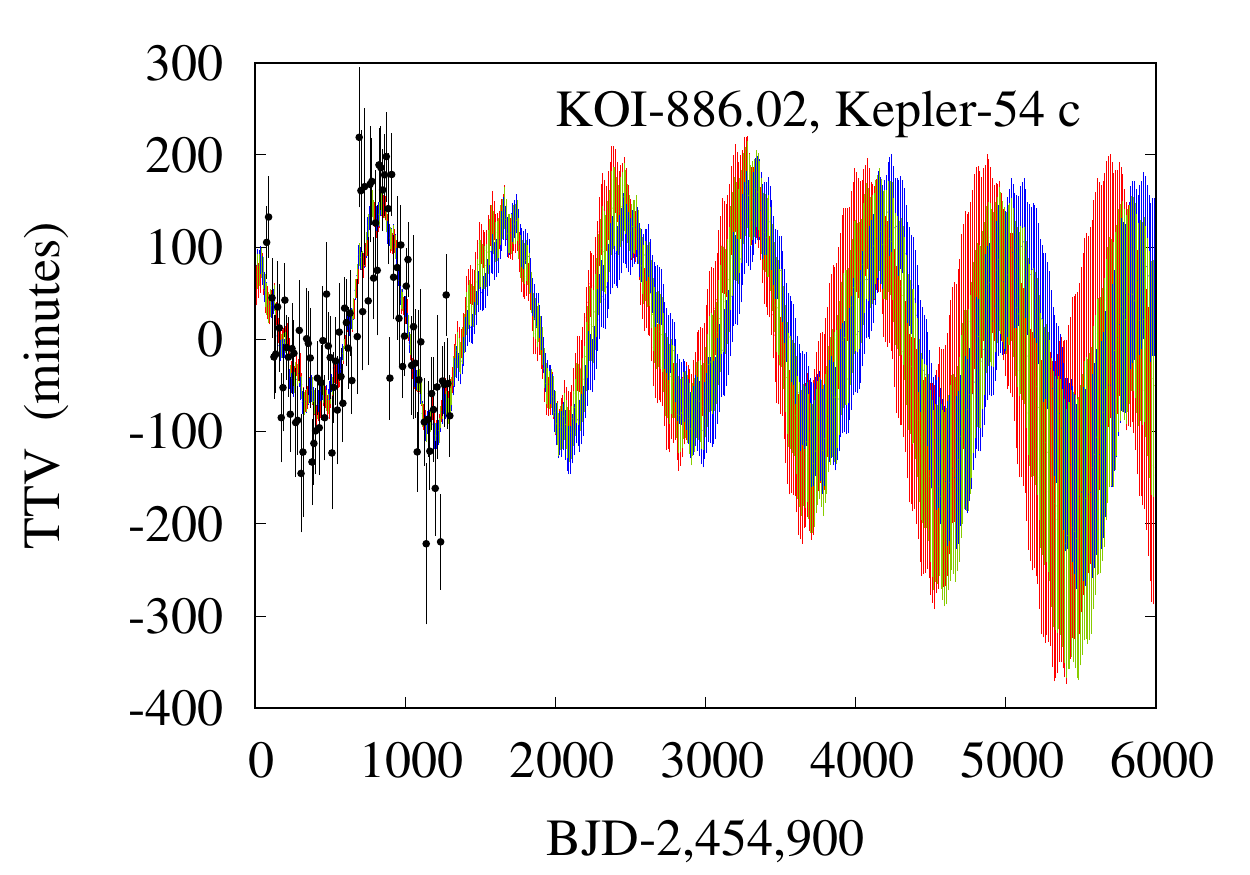}
\includegraphics[height = 1.45 in]{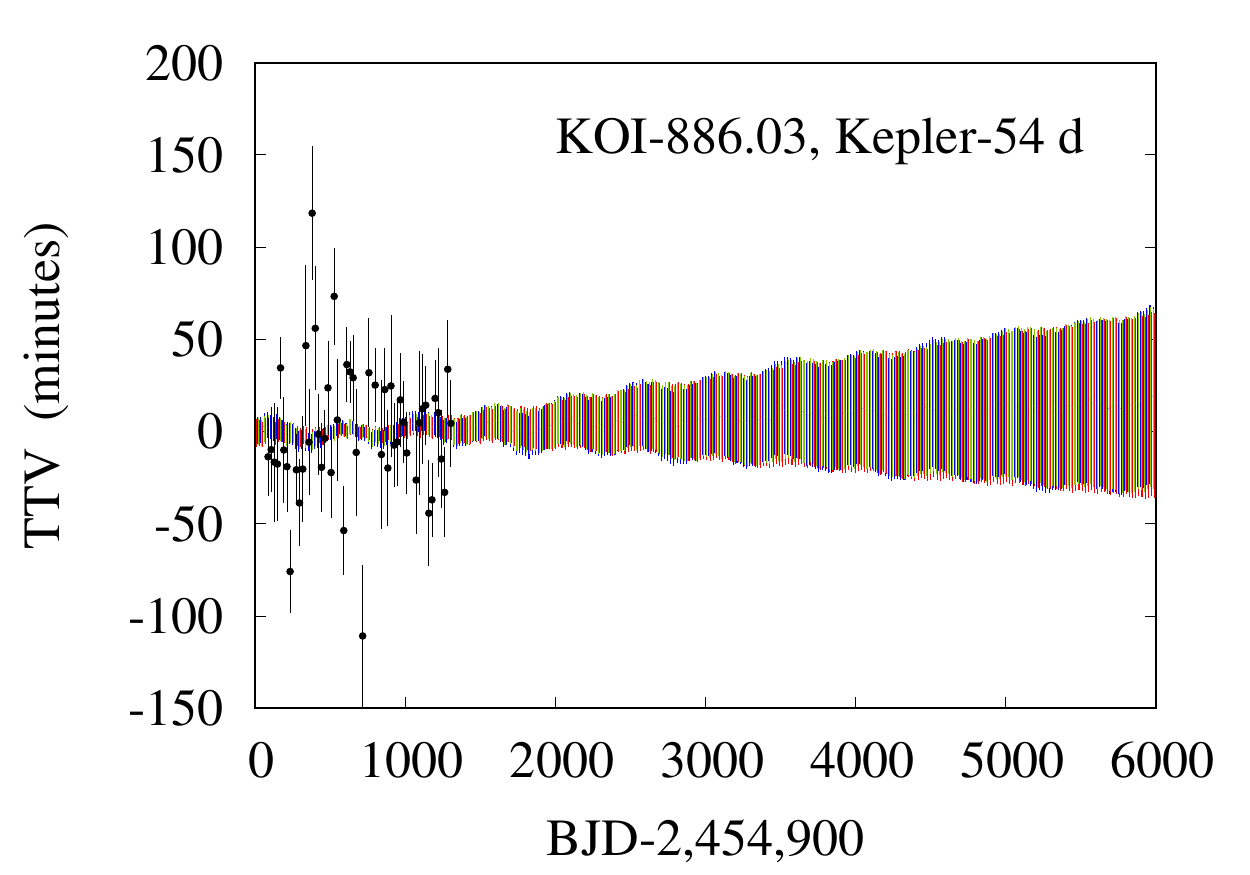}
\includegraphics[height = 1.45 in]{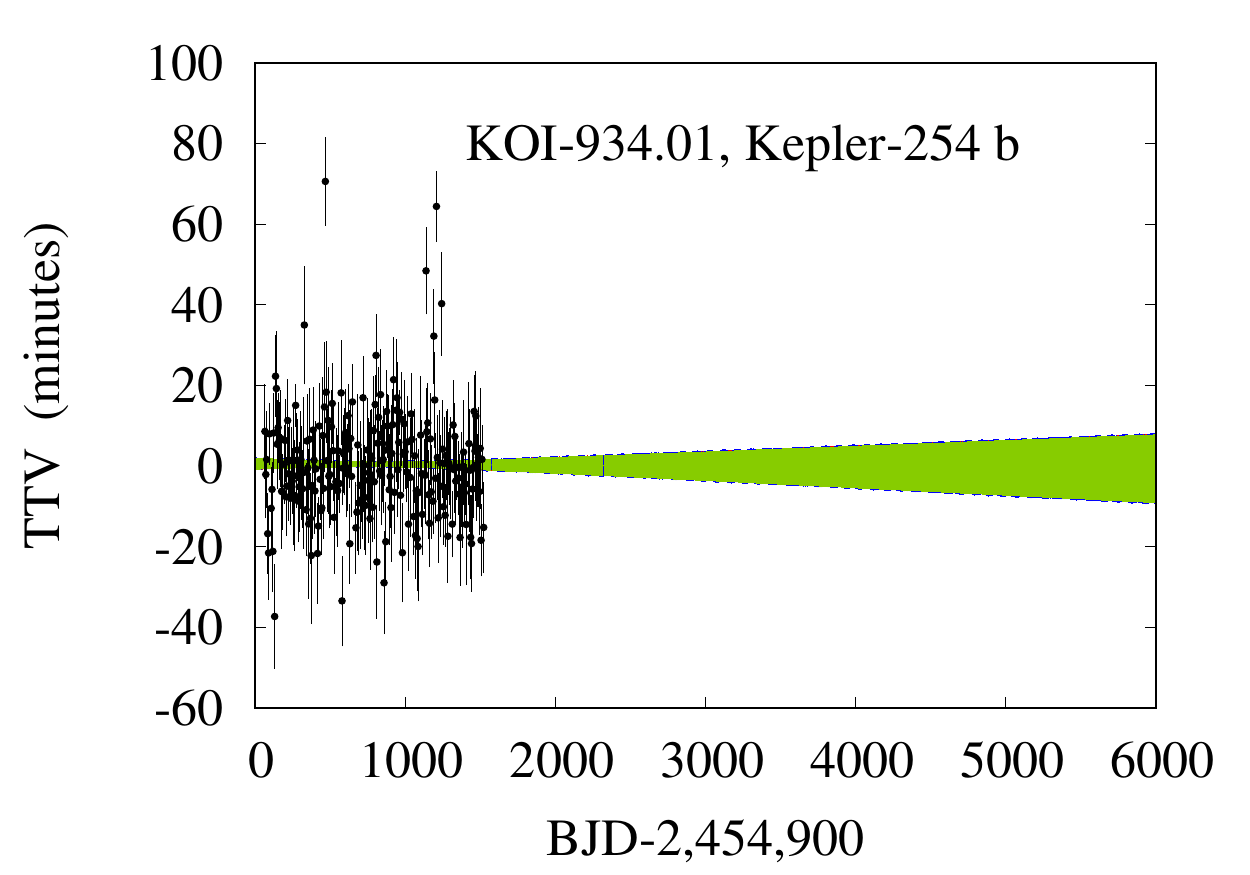}
\includegraphics[height = 1.45 in]{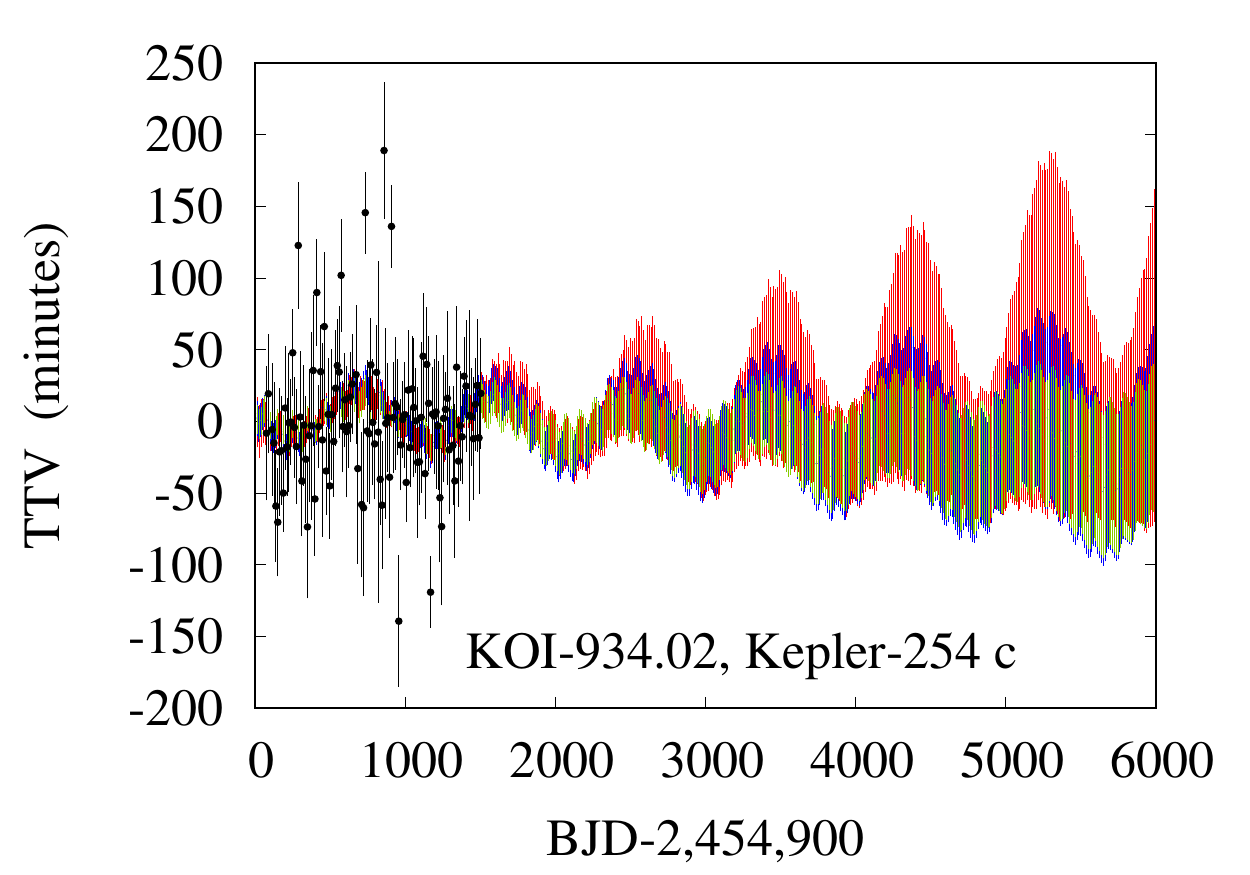}
\includegraphics[height = 1.45 in]{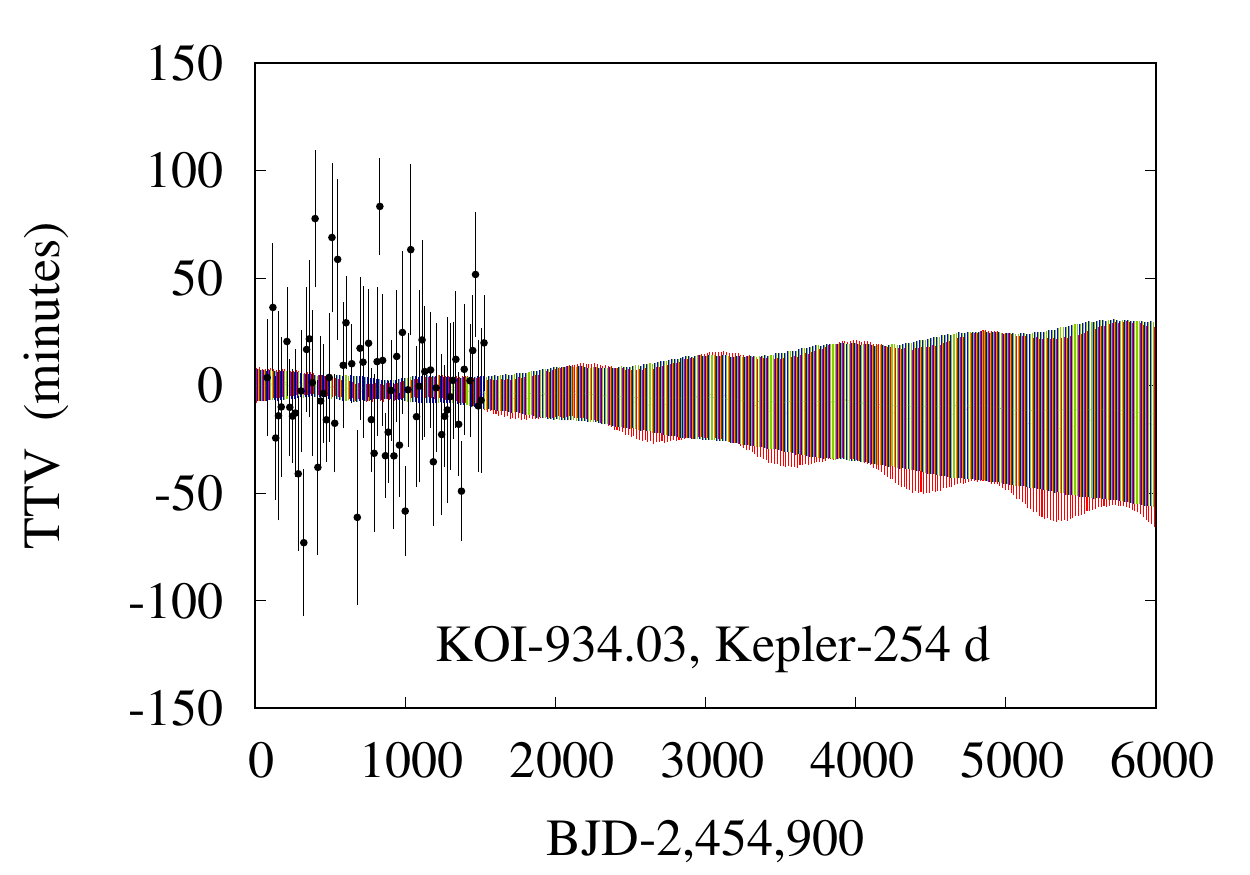}
\caption{Distribution of projected transit times as a function of time for the planet labelled in each panel (part 5). Black points mark transit times in the catalog of \citet{rowe15a} with 1$\sigma$ error bars. In green are 68.3\% confidence intervals of simulated transit times from posterior sampling. In blue (red), are a subset of samples with dynamical masses below (above) the 15.9th (84.1th) percentile.
\label{fig:KOI-750fut}} 
\end{center}
\end{figure}

KOI-806 (Kepler-30) has stronger detected TTVs than expected in Table~\ref{tbl-SystemSelection2}, and the masses were initially characterized by \citet{Sanchis2012}. There is some divergence in the projected TTVs, particularly in KOI-806.01 (Kepler-30 d), as shown in Figure~\ref{fig:KOI-750fut}, and hence additional transit-timing data may be of some value in further characterizing the planets. 

KOI-886 (Kepler-54) also has much stronger TTVs than expected in Table~\ref{tbl-SystemSelection2}. The projected transit times diverge in phase after several thousand days. This divergence appears correlated with dynamical masses, and hence additional transit-timing data would be useful in further characterizing this system.

KOI-934.03 (Kepler-254 d) is near a 3:2 resonance with its inner neighbor KOI-934.02 (Kepler-254 c), which has a strong mass upper limit. The projected TTVs of both planets diverge, and additional transit timing data would further characterize this system.

KOI-1279 (Kepler-804) has two small planets with a weakly detected near-resonant TTV signal with weak constraints on the TTV amplitude. The projected TTVs diverge slowly.

KOI-1598 (Kepler-310) has strongly detected TTVs from the second-order 5:3 near-resonance. There is also an expectation of detectable nonresonant interactions, which led to its inclusion in our sample. The outer planet, KOI-1598.02 (Kepler-310 d), has an apparent signal at the expected superperiod for the second order resonance of $\sim$1400 days. Shorter periodicities are detected in KOI-1598.02 (Kepler-310 c). The projected TTVs diverge rapidly, although there is little difference between the low-mass and high-mass solutions. 

KOI-1831.01 (Kepler-324 c) has strongly detected TTVs with a superperiod longer than the Kepler baseline. The projected TTVs diverge quickly, and follow-up transit-timing data would improve constraints on planetary masses. 

Both KOI-2113.02 (Kepler-417 b) and KOI-2113.01 (Kepler-417 c) have mass upper limits imposed by nonresonant TTVs. The projected TTVs diverge slowly.

KOI-3503 has two planet candidates with strongly detected near-resonant TTVs, despite an expectation of just weakly detectable TTVs, with a superperiod of $\sim$7800 days. The projected TTVs diverge rapidly, and this system is an ideal candidate for follow-up transit timing.

\subsection{Evaluating Our Sample}
\begin{figure}[!h]
\begin{center}
\figurenum{7}
\includegraphics[height = 1.45 in]{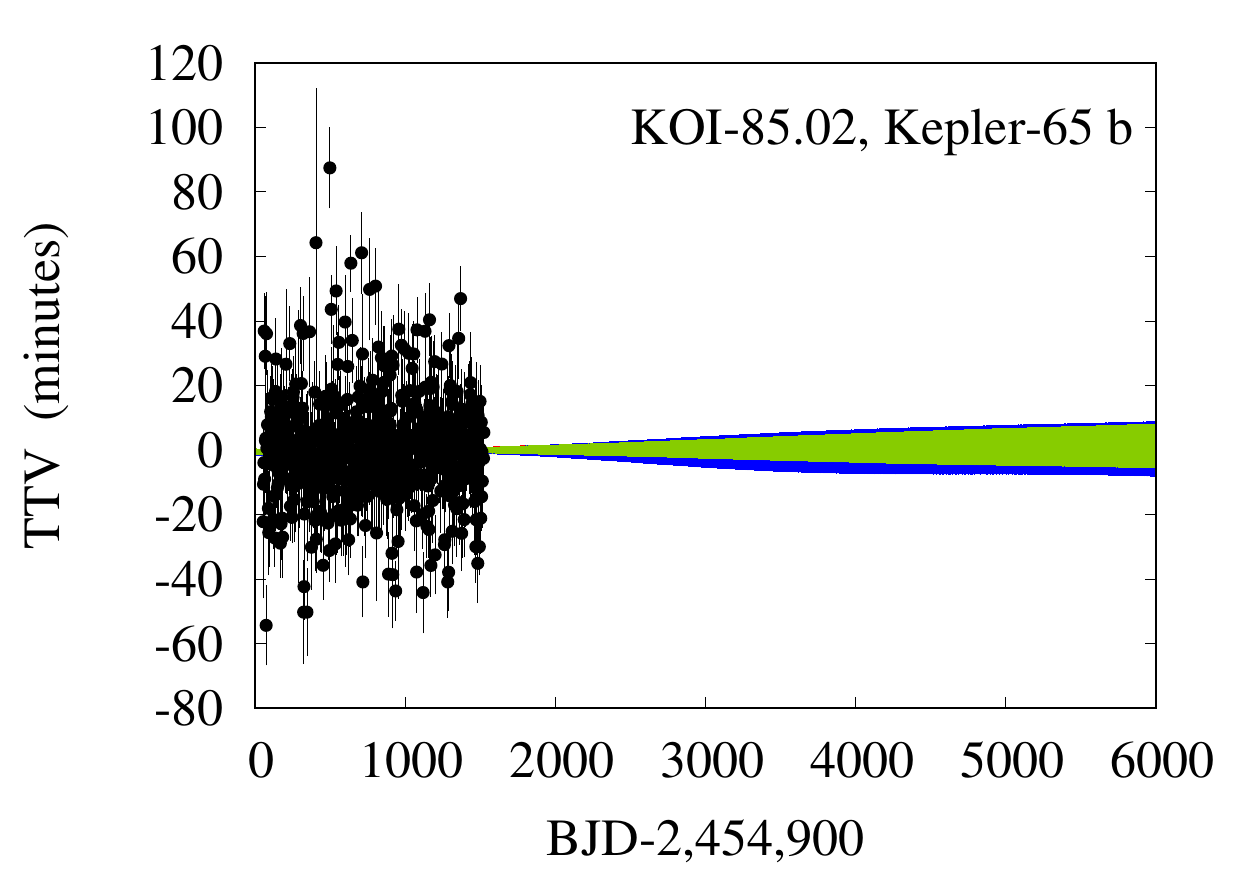}
\includegraphics[height = 1.45 in]{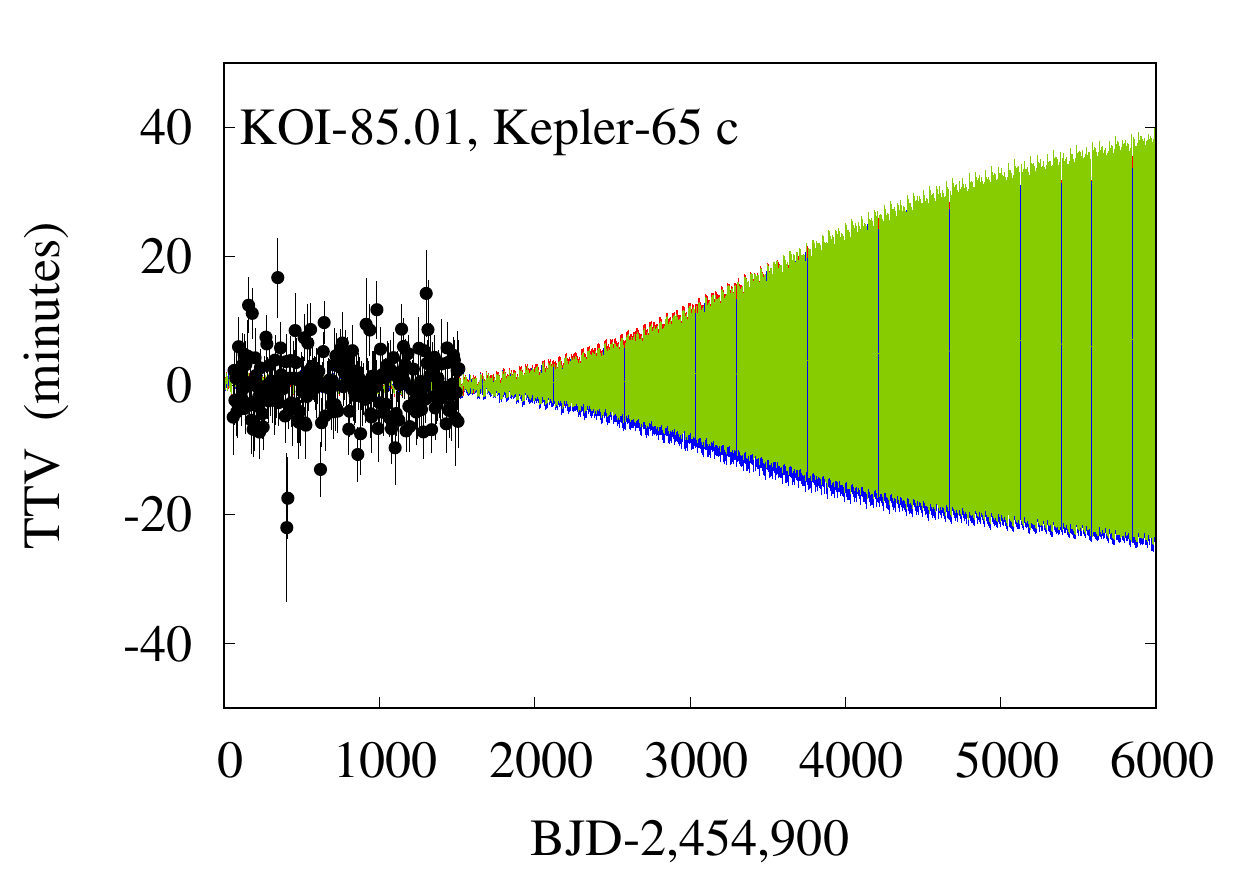}
\includegraphics[height = 1.45 in]{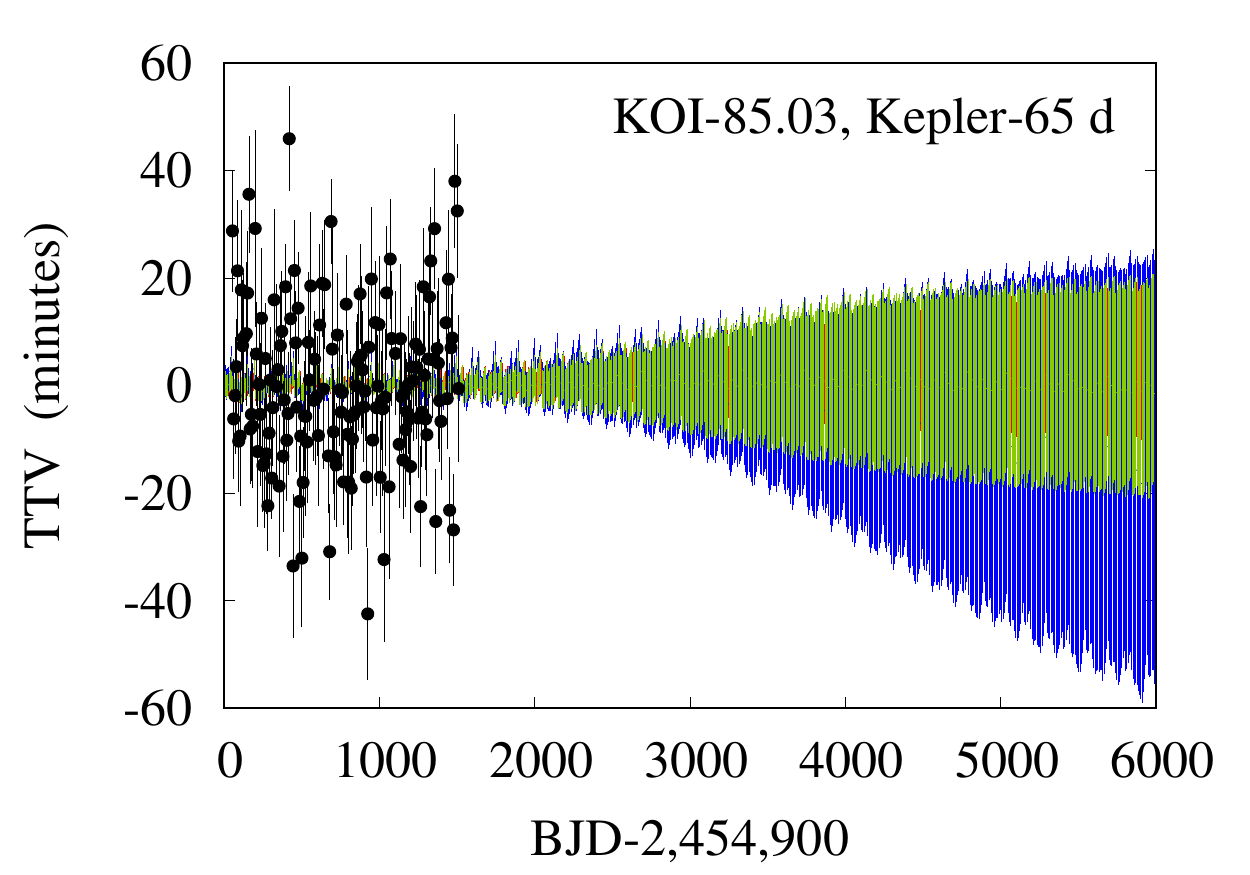}
\includegraphics[height = 1.45 in]{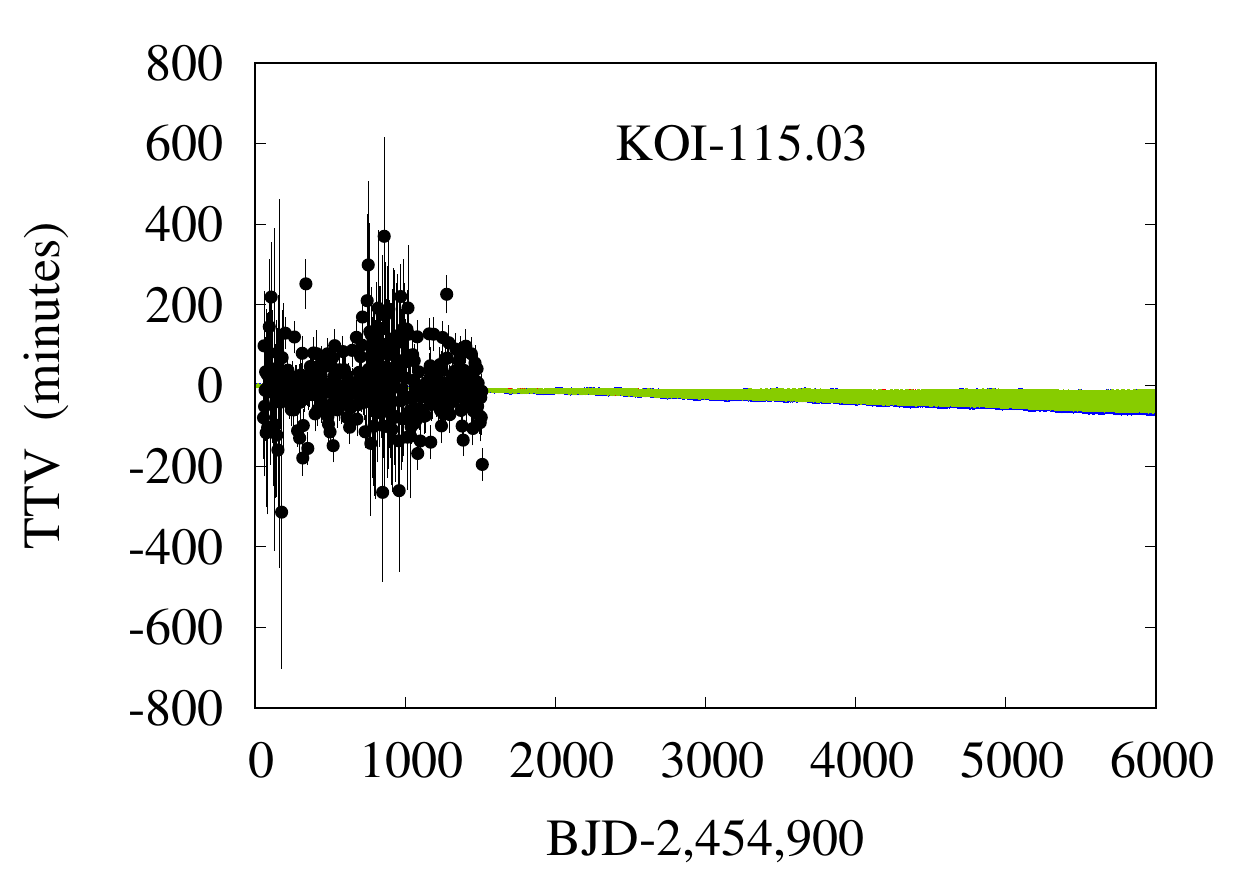}
\includegraphics[height = 1.45 in]{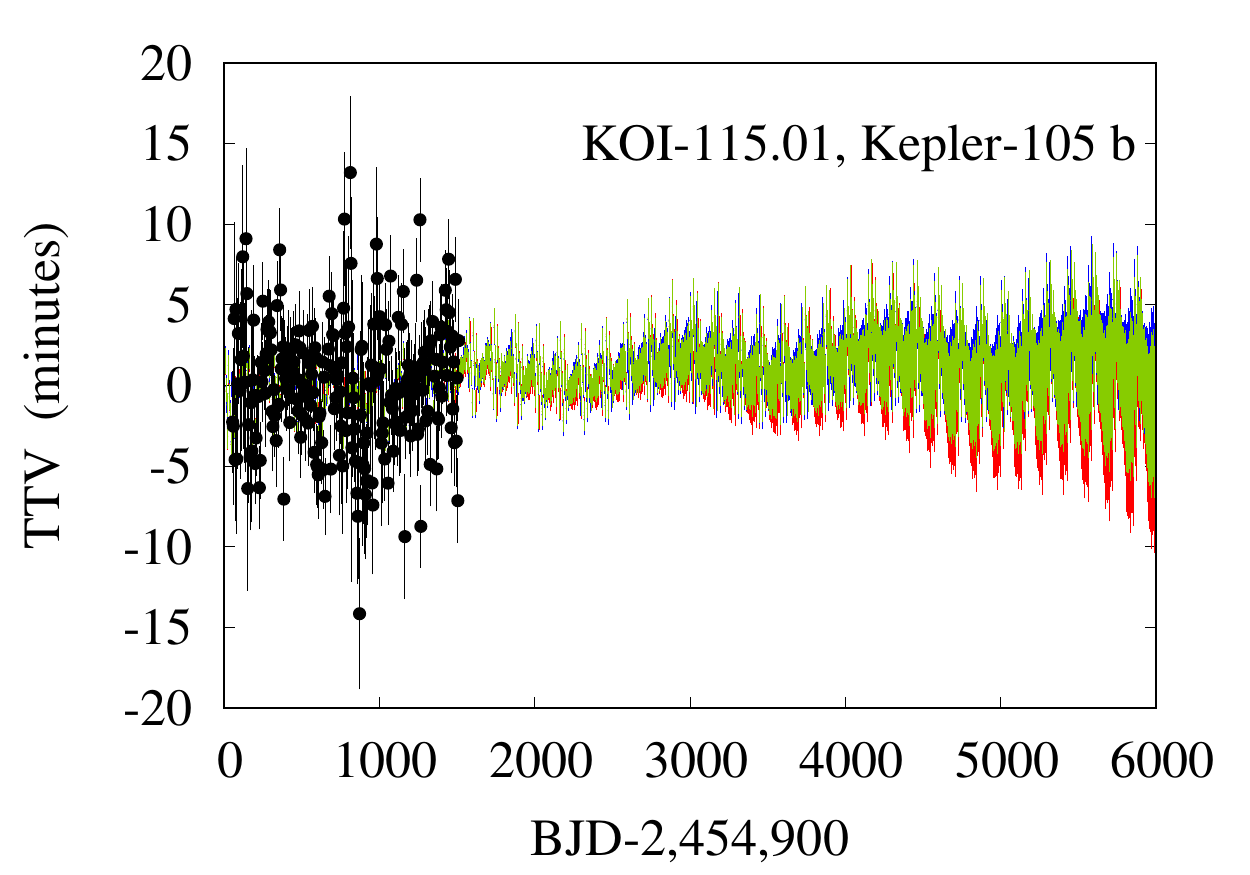}
\includegraphics[height = 1.45 in]{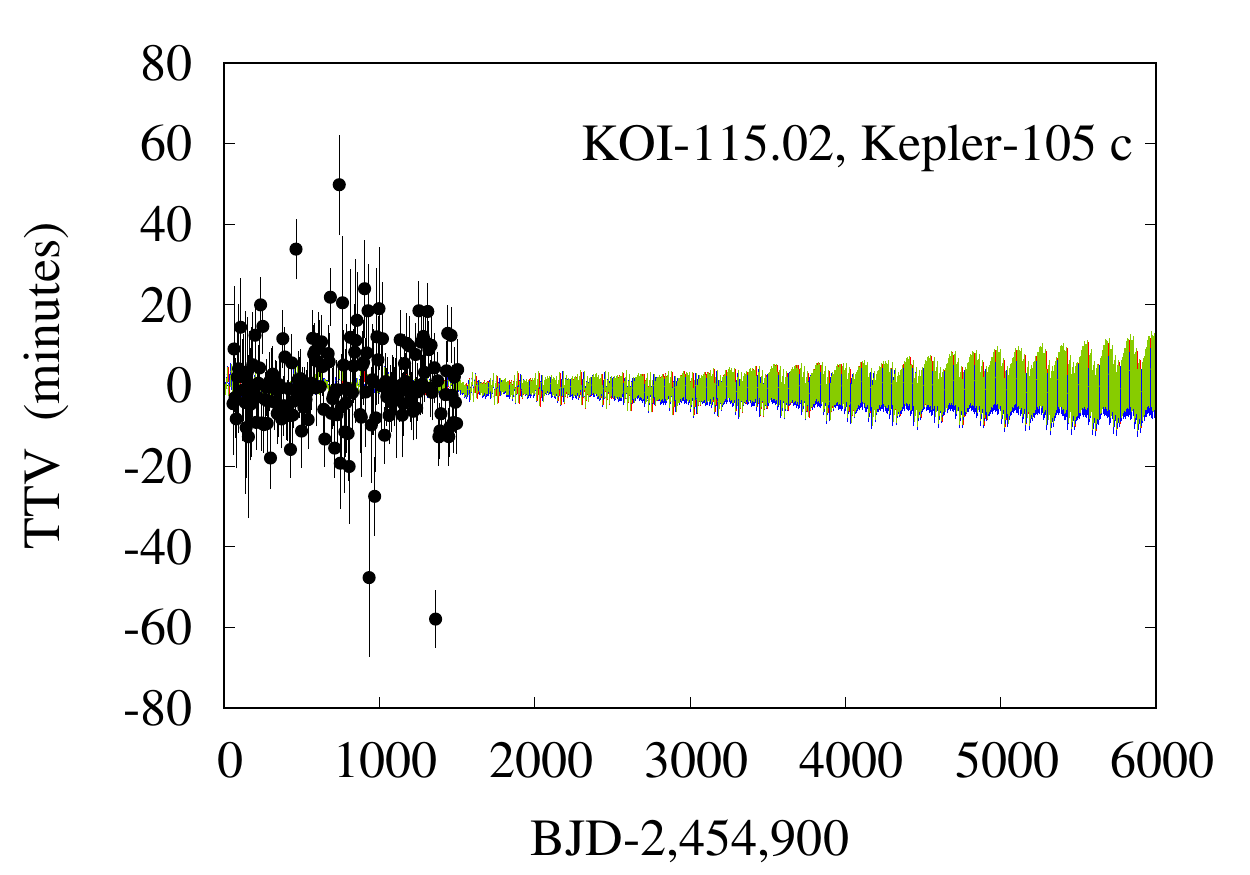}
\includegraphics[height = 1.45 in]{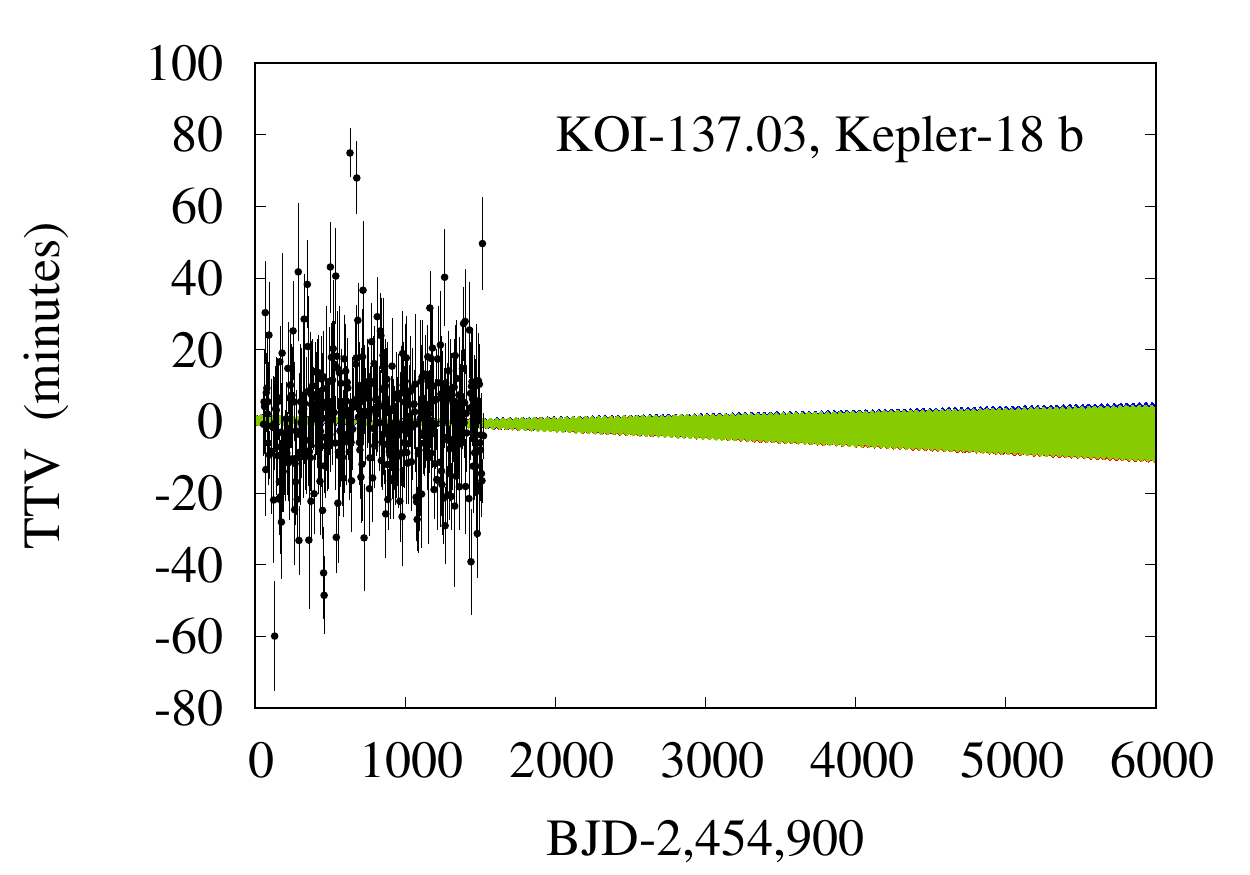}
\includegraphics[height = 1.45 in]{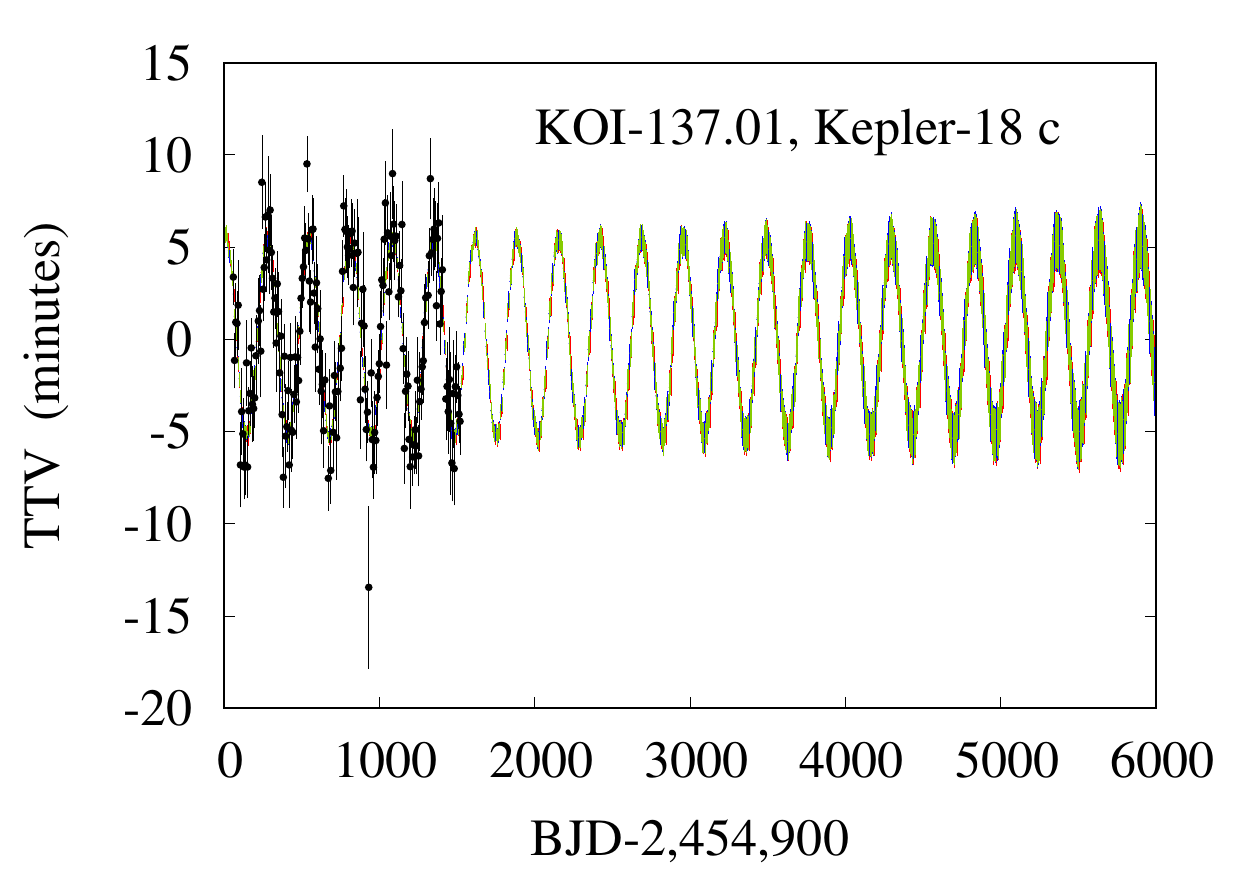}
\includegraphics[height = 1.45 in]{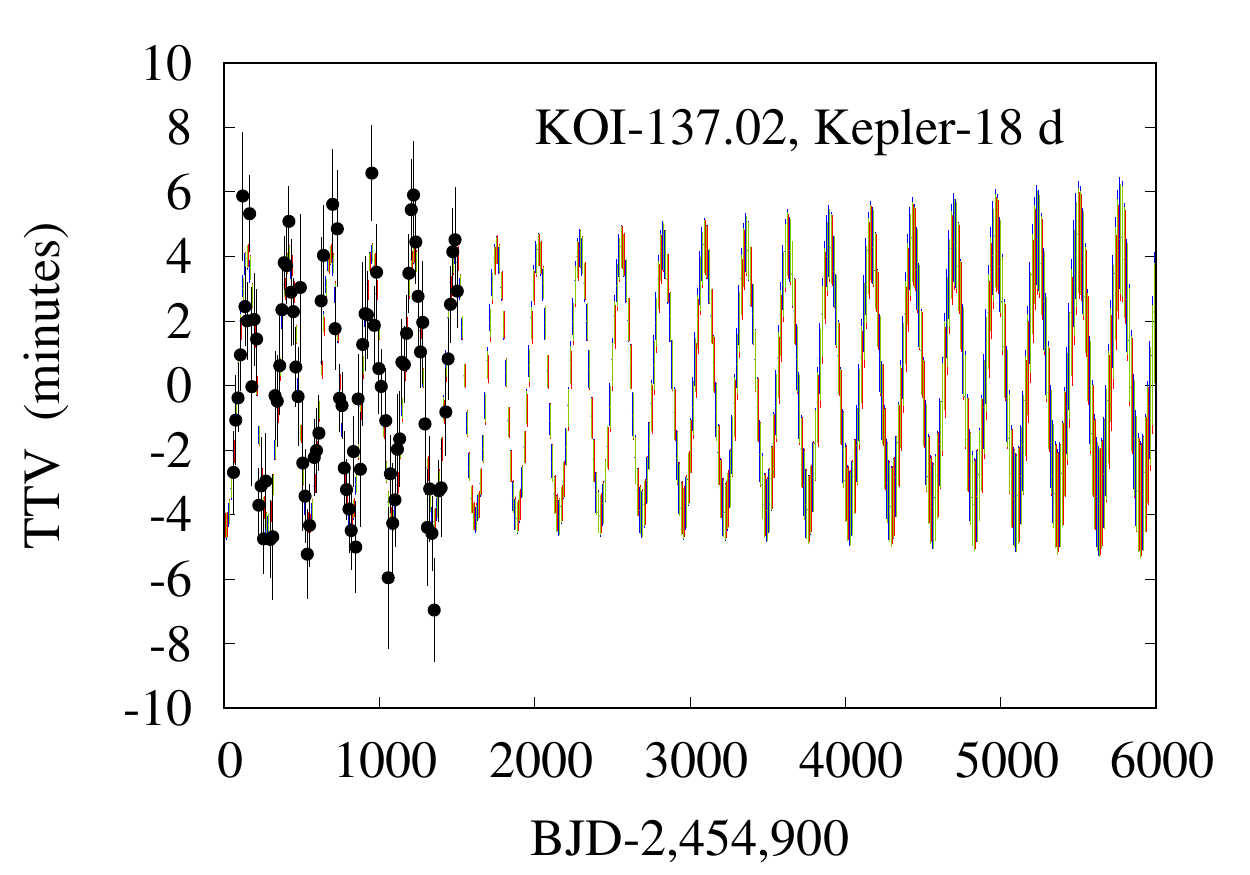}
\includegraphics[height = 1.05 in]{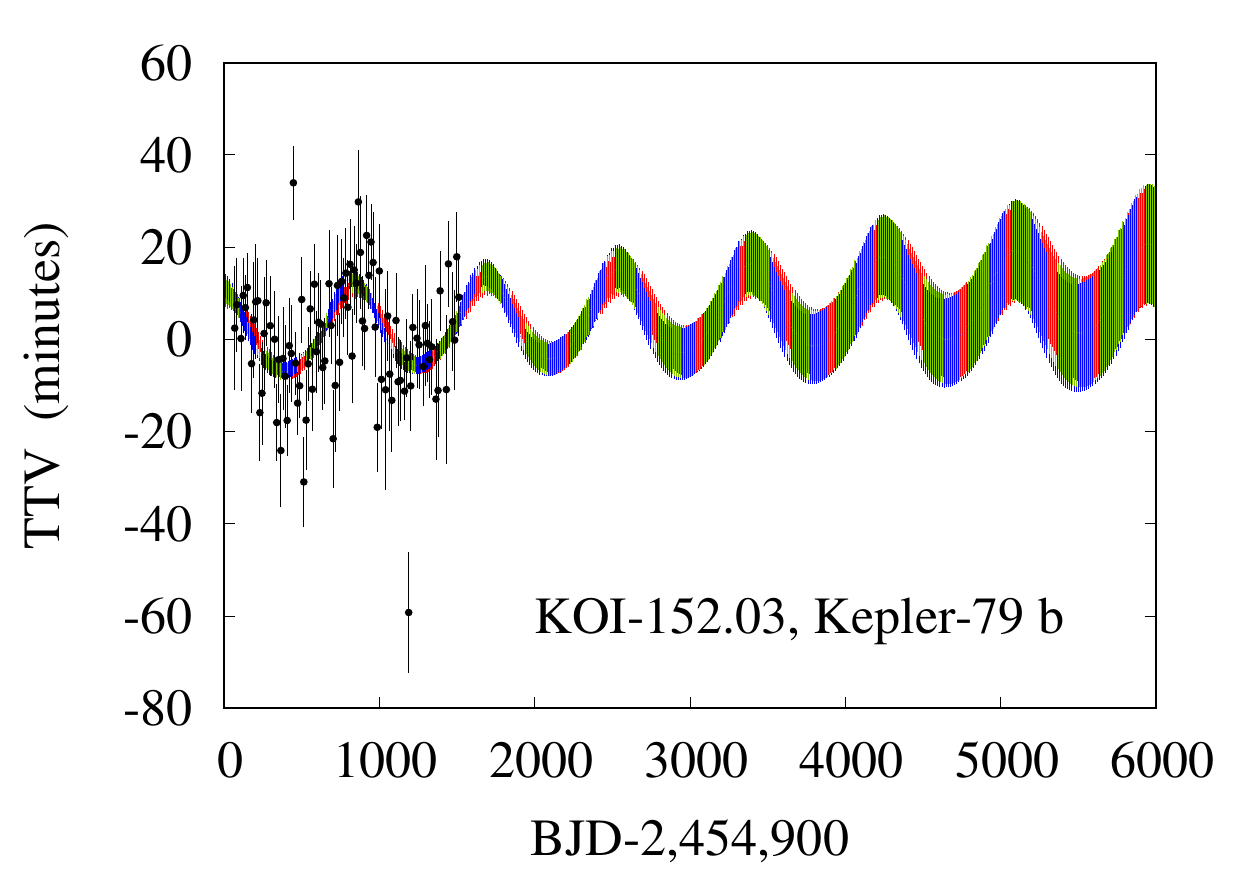}
\includegraphics[height = 1.05 in]{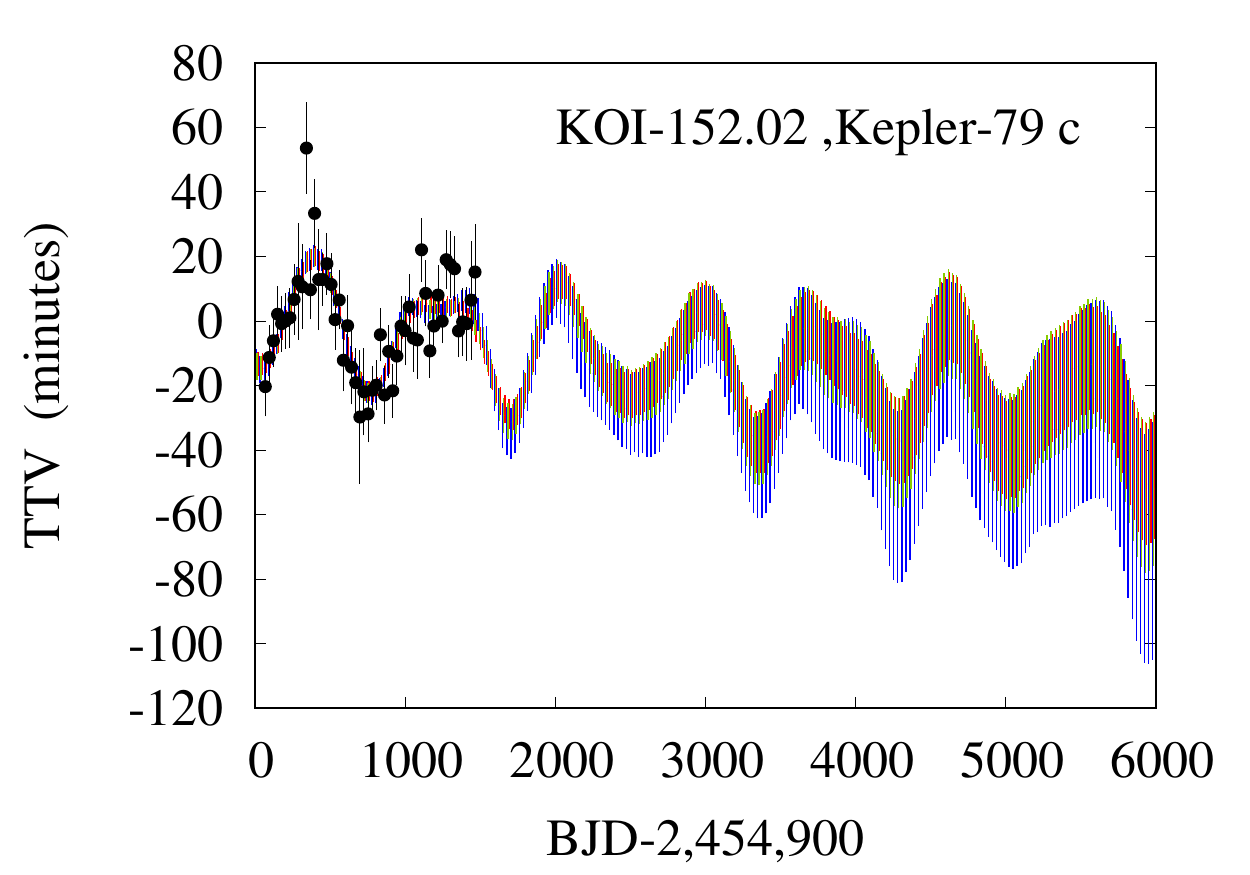}
\includegraphics[height = 1.05 in]{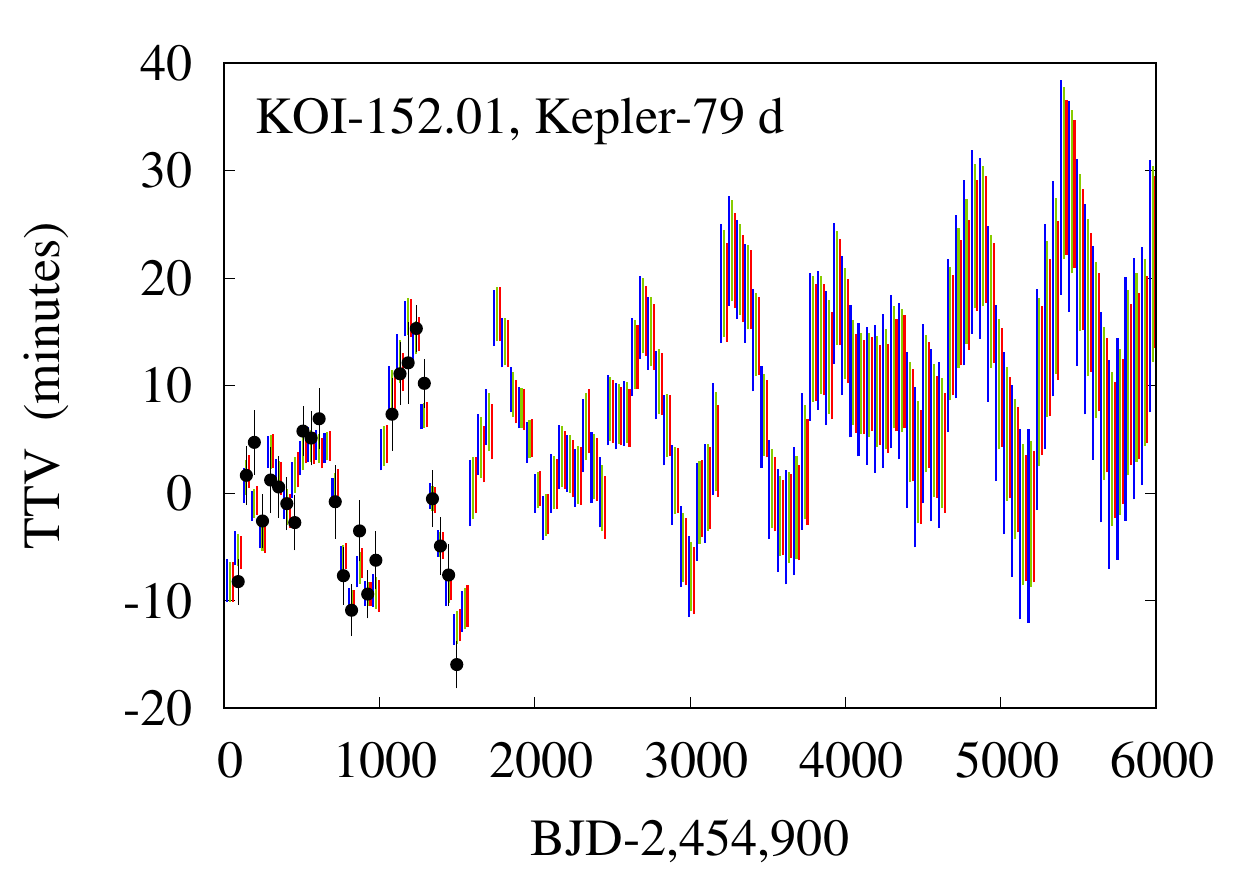}
\includegraphics[height = 1.05 in]{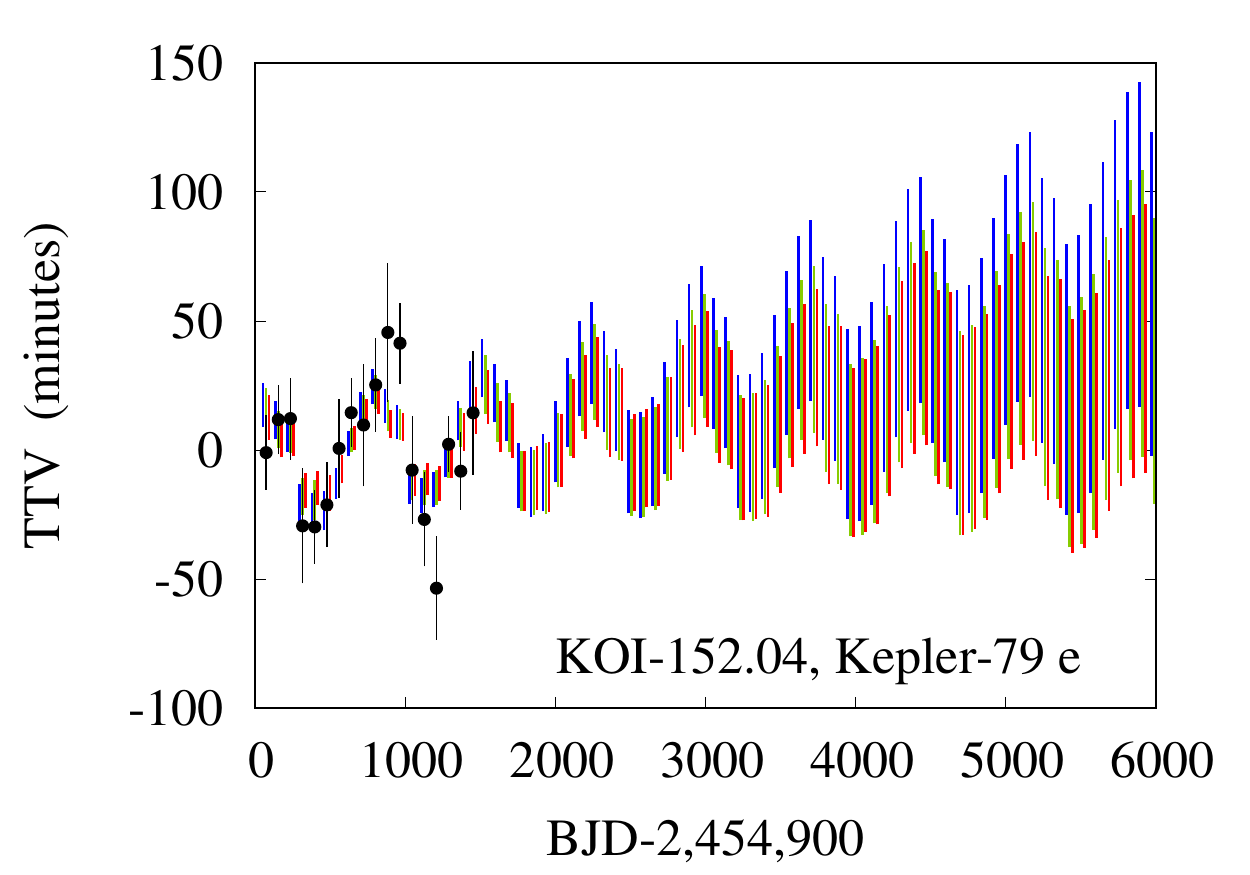}
\includegraphics[height = 1.45 in]{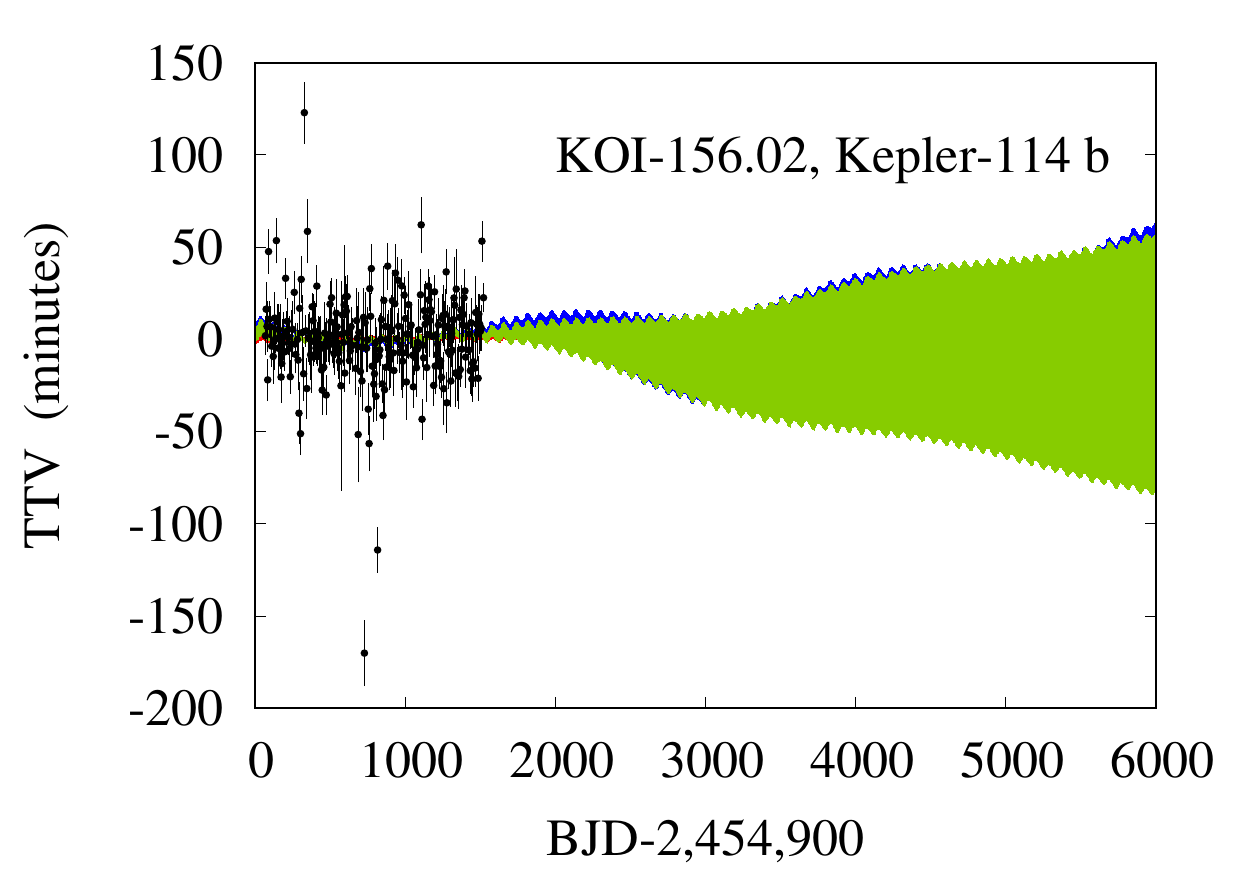}
\includegraphics[height = 1.45 in]{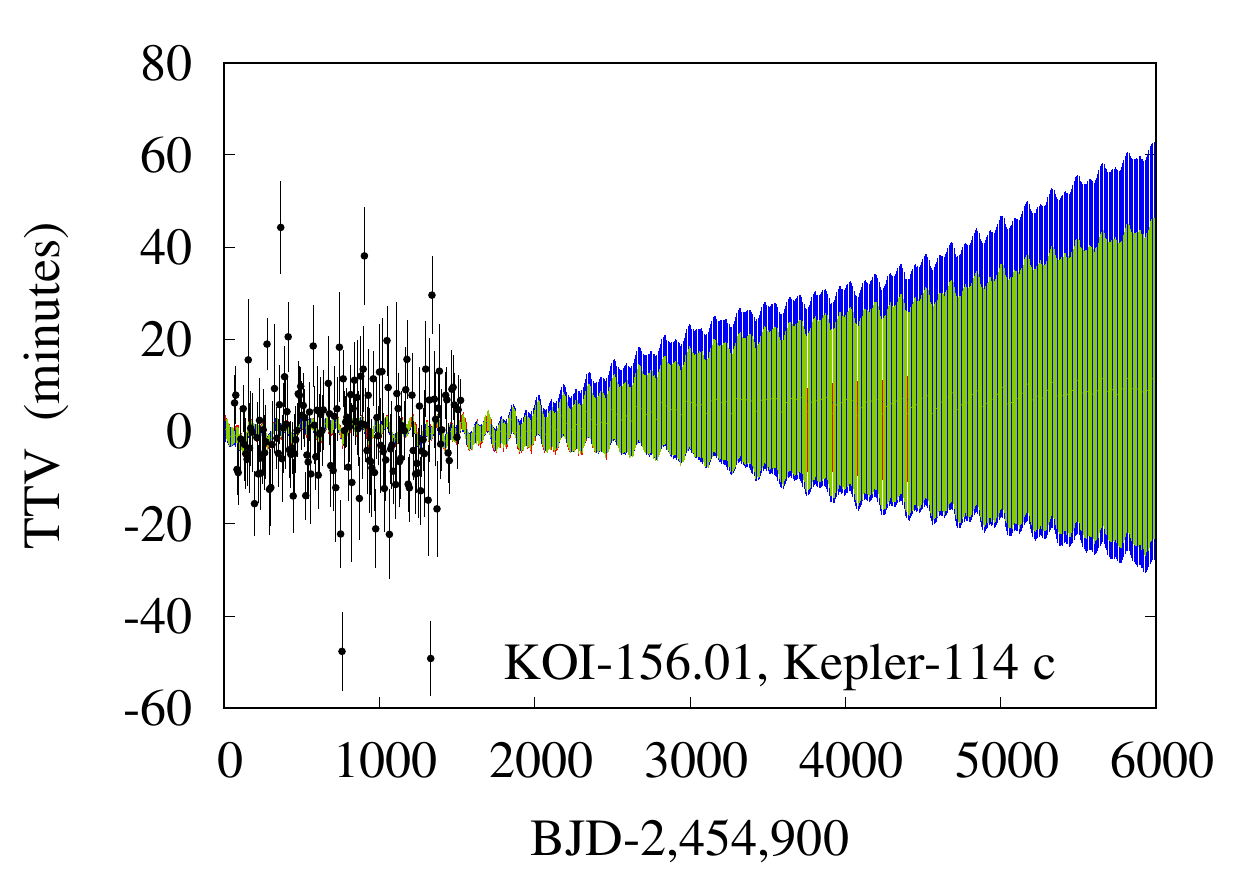}
\includegraphics[height = 1.45 in]{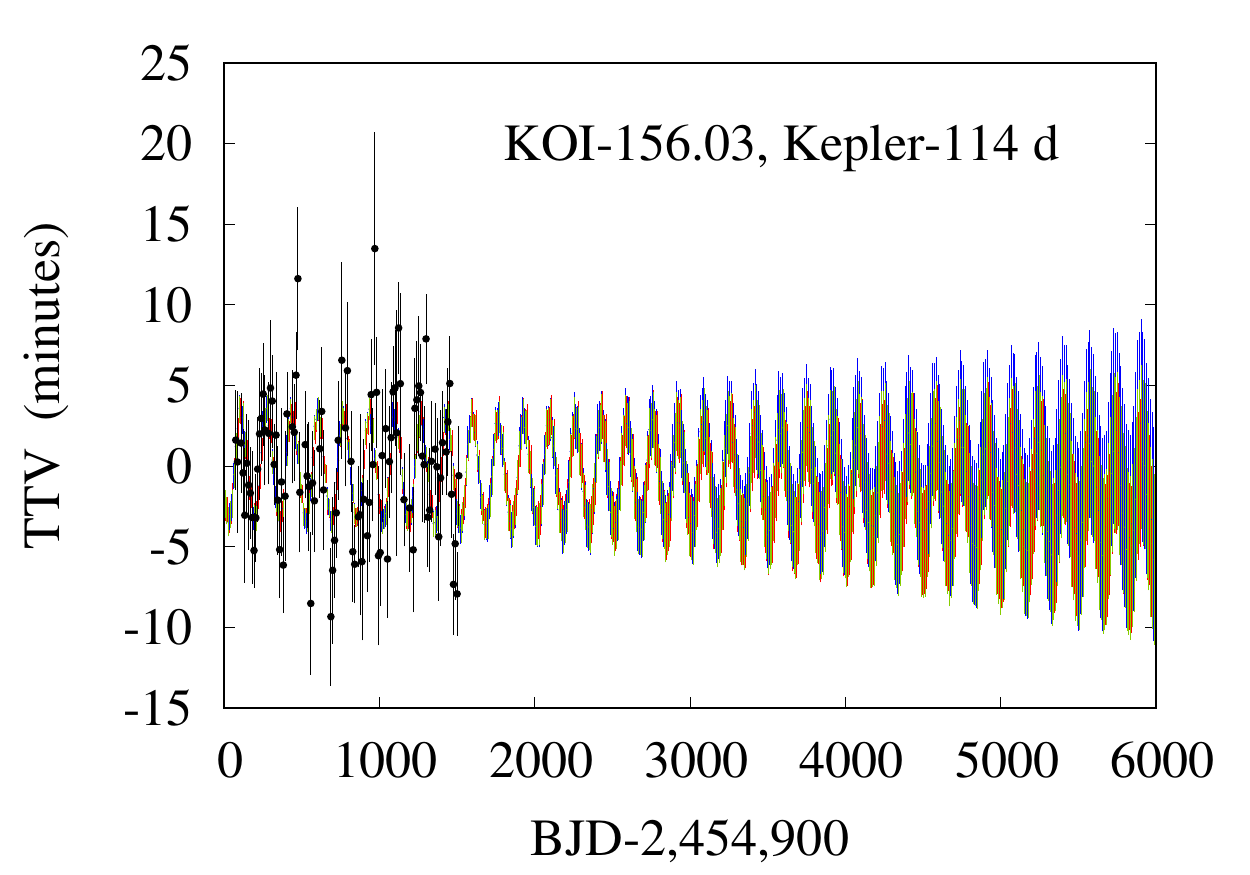}
\caption{Distribution of projected transit times as a function of time for the planet labelled in each panel (part 1). Black points mark transit times in the catalog of \citet{rowe15a} with 1$\sigma$ error bars. In green are 68.3\% confidence intervals of simulated transit times from posterior sampling. In blue (red), are a subset of samples with dynamical masses below (above) the 15.9th (84.1th) percentile. \label{fig:KOI-85fut}}
\end{center}
\end{figure}

As we have seen in this section, the future transit times of some planets diverge more quickly than others. We do not apply a quantitative test for whether TTVs are classified as rapidly or slowly diverging. Nevertheless, the projected TTVs are valuable here in identifying even weak TTV signals. The inclusion of isolated neighbors within the systems of interest in our sample highlights this. The first few examples in Figure~\ref{fig:KOI-85fut} include KOI-85.02 (Kepler-65 b), KOI-155.03, and KOI-137.03 (Kepler-18 b). For these isolated candidates, the uncertainty in the projected transit times grows linearly and is primarily due to the uncertainty in the measured linear ephemeris from the transit data. This is in contrast to candidates with an expected nonresonant TTV signal where projected transit times diverge rapidly, even though by eye, there is no obvious detection of TTVs. In most of these cases, the existing transit data may not provide useful upper limits on the masses, although there are exceptions like KOI-85.01 (Kepler-65 b), KOI-156.01 (Kepler-114 c), and both planets at KOI-2113 (Kepler-417). 

\begin{figure}[!h]
\begin{center}
\figurenum{8}
\includegraphics[height = 1.45 in]{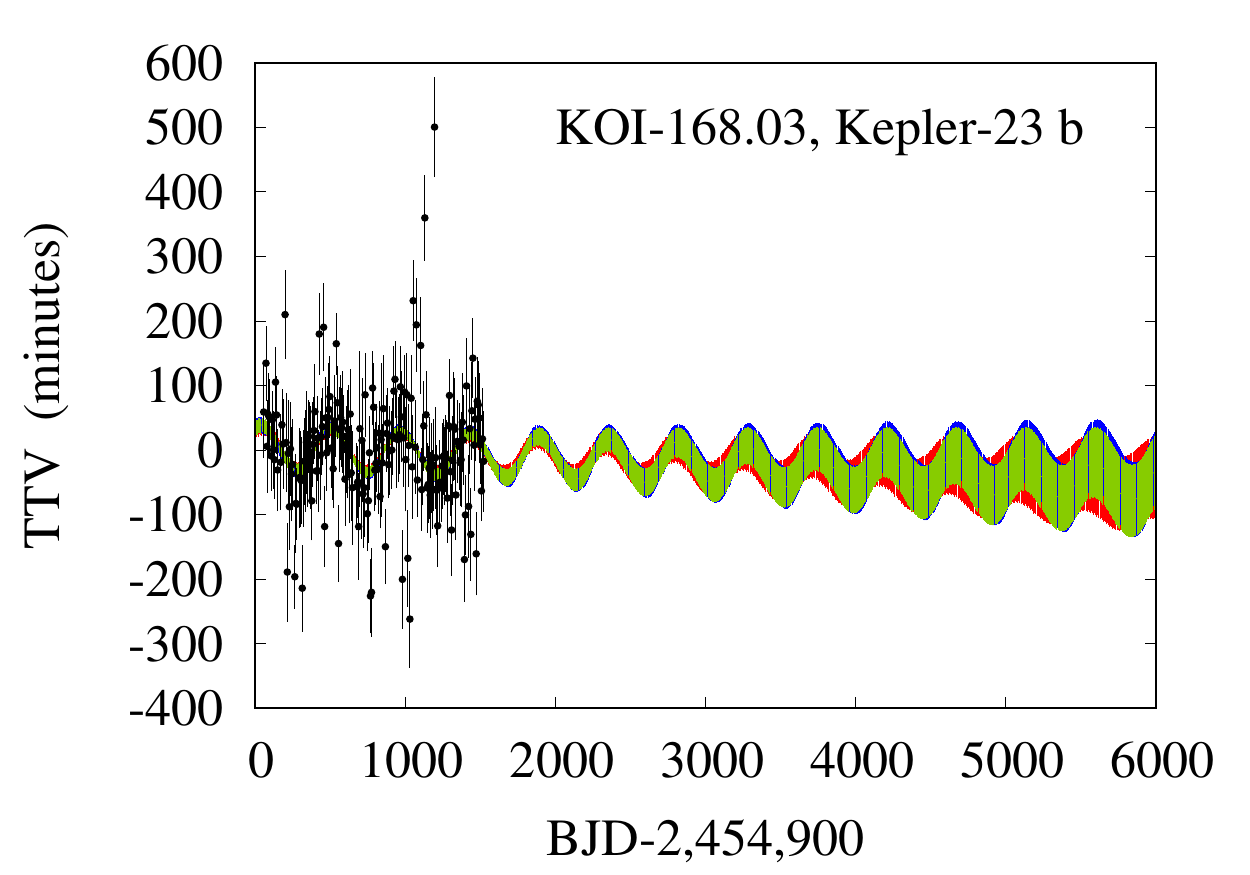}
\includegraphics[height = 1.45 in]{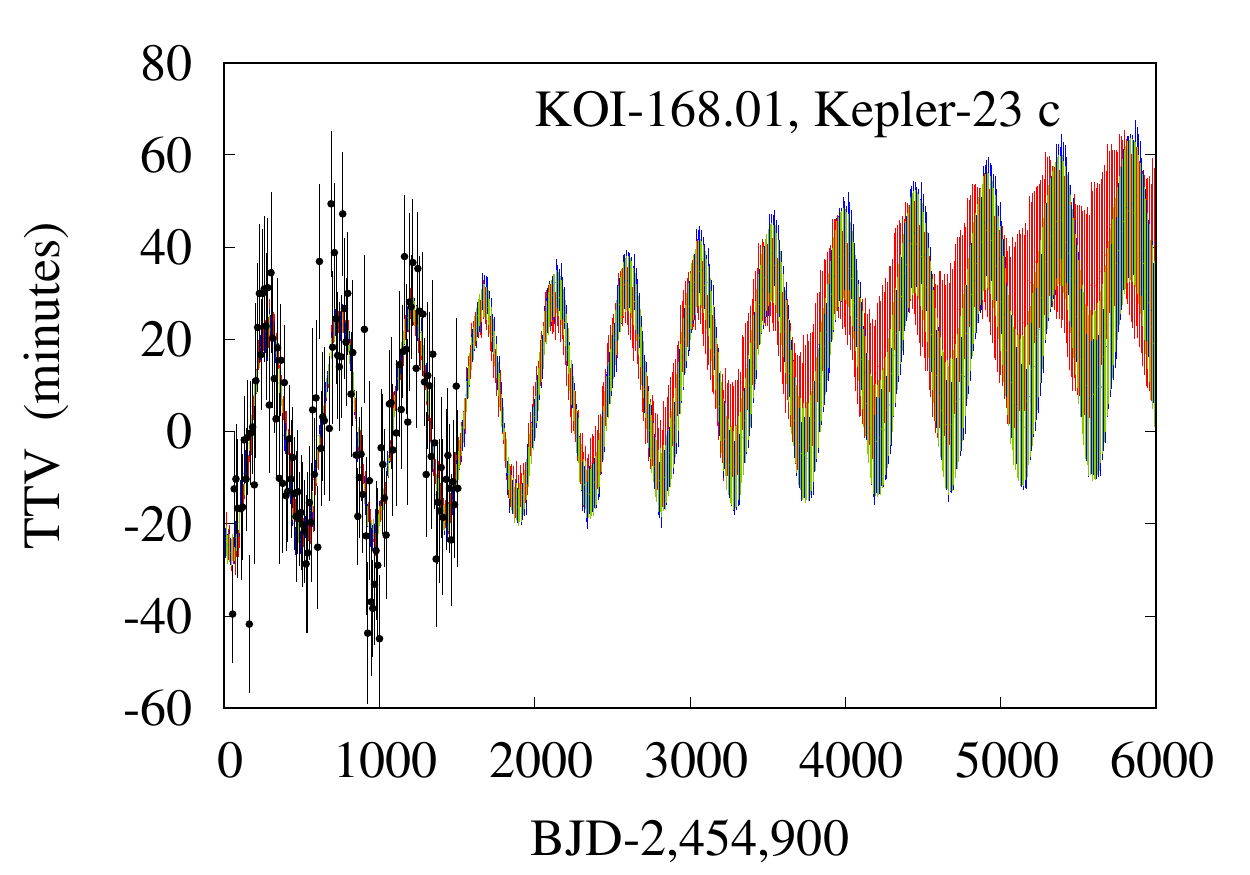}
\includegraphics[height = 1.45 in]{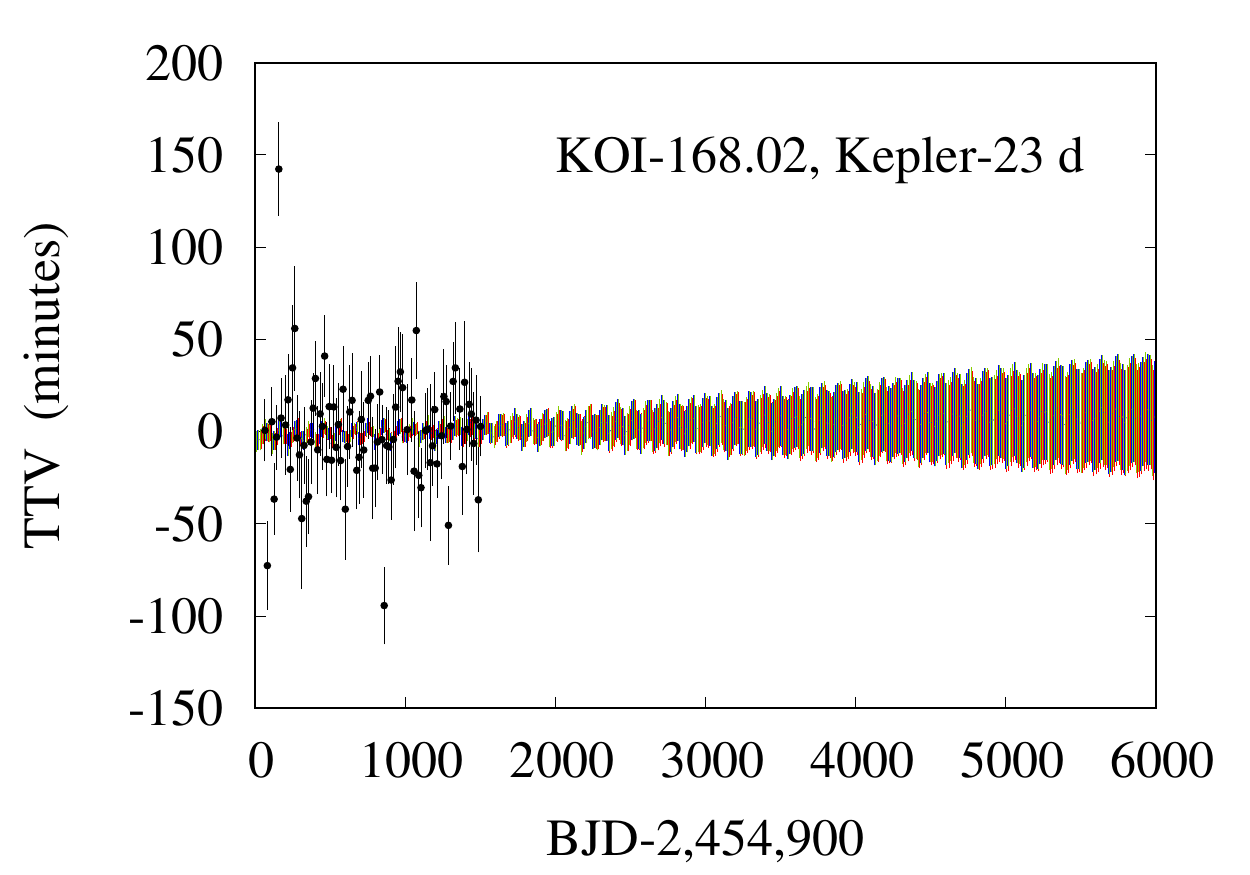}
\includegraphics[height = 1.05 in]{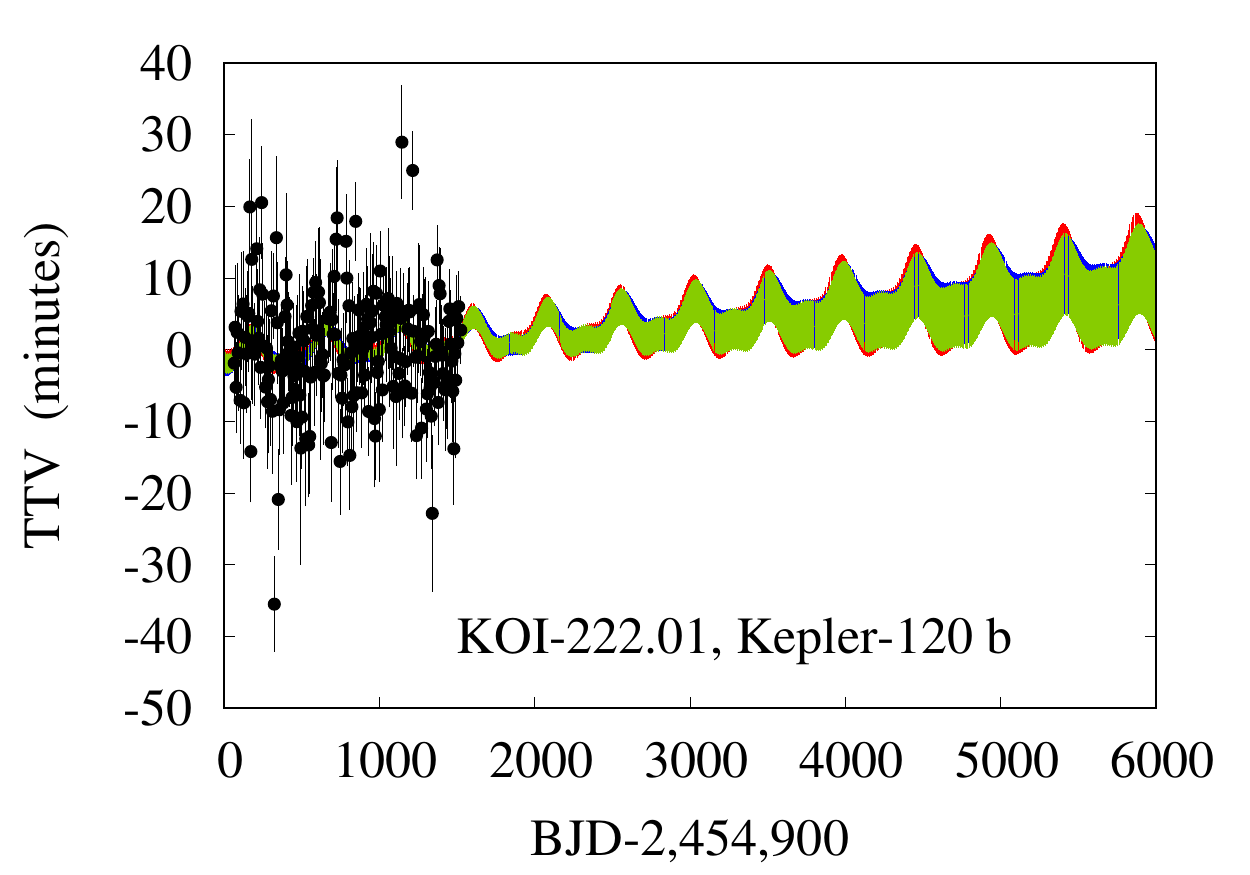}
\includegraphics[height = 1.05 in]{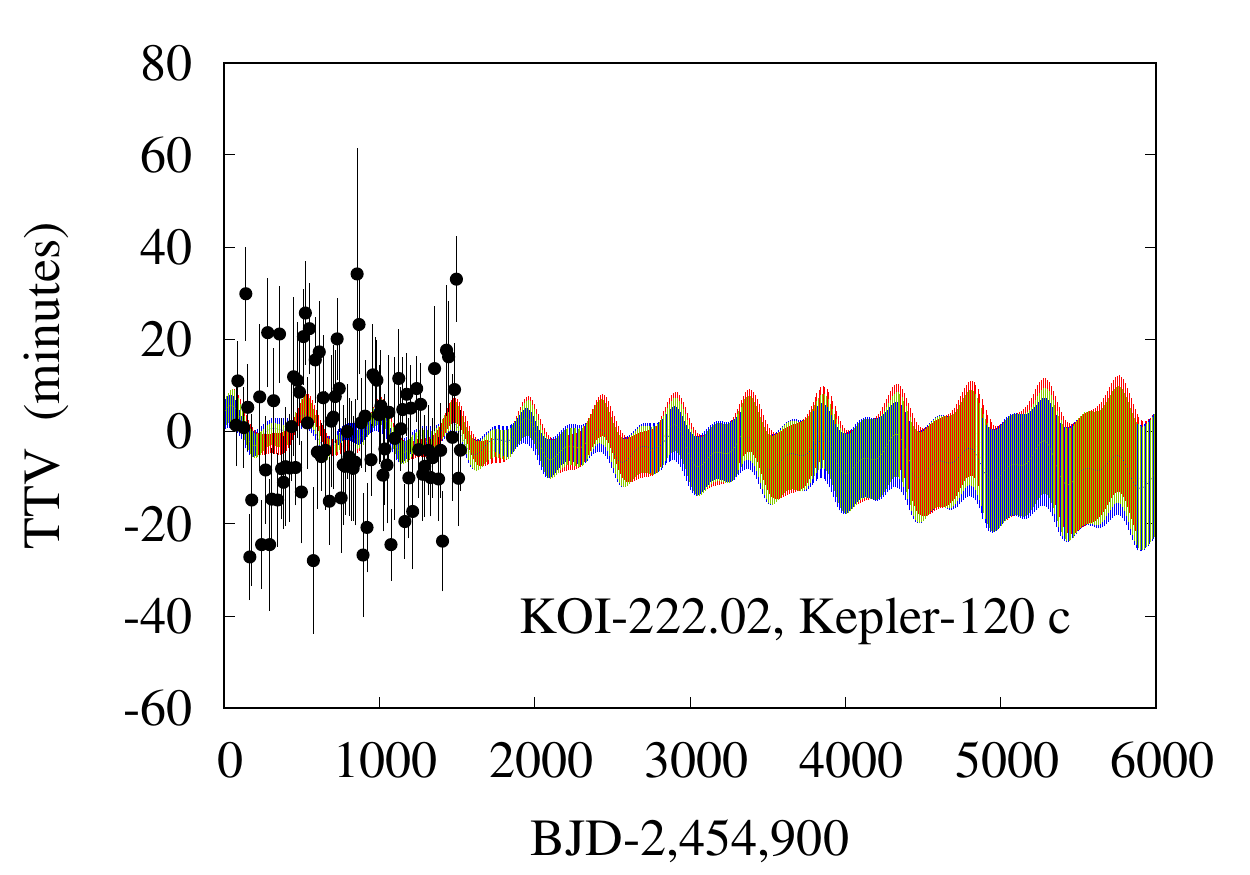}
\includegraphics[height = 1.05 in]{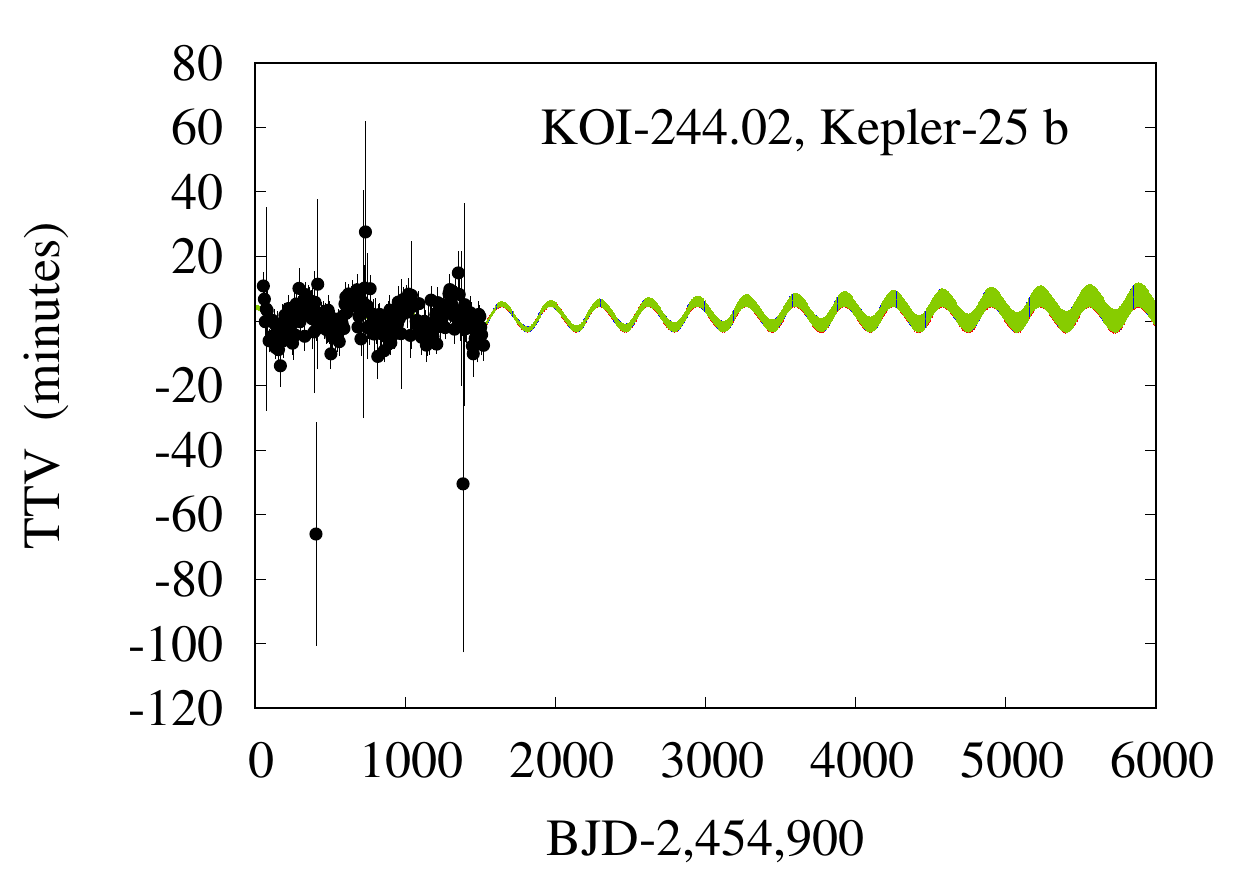}
\includegraphics[height = 1.05 in]{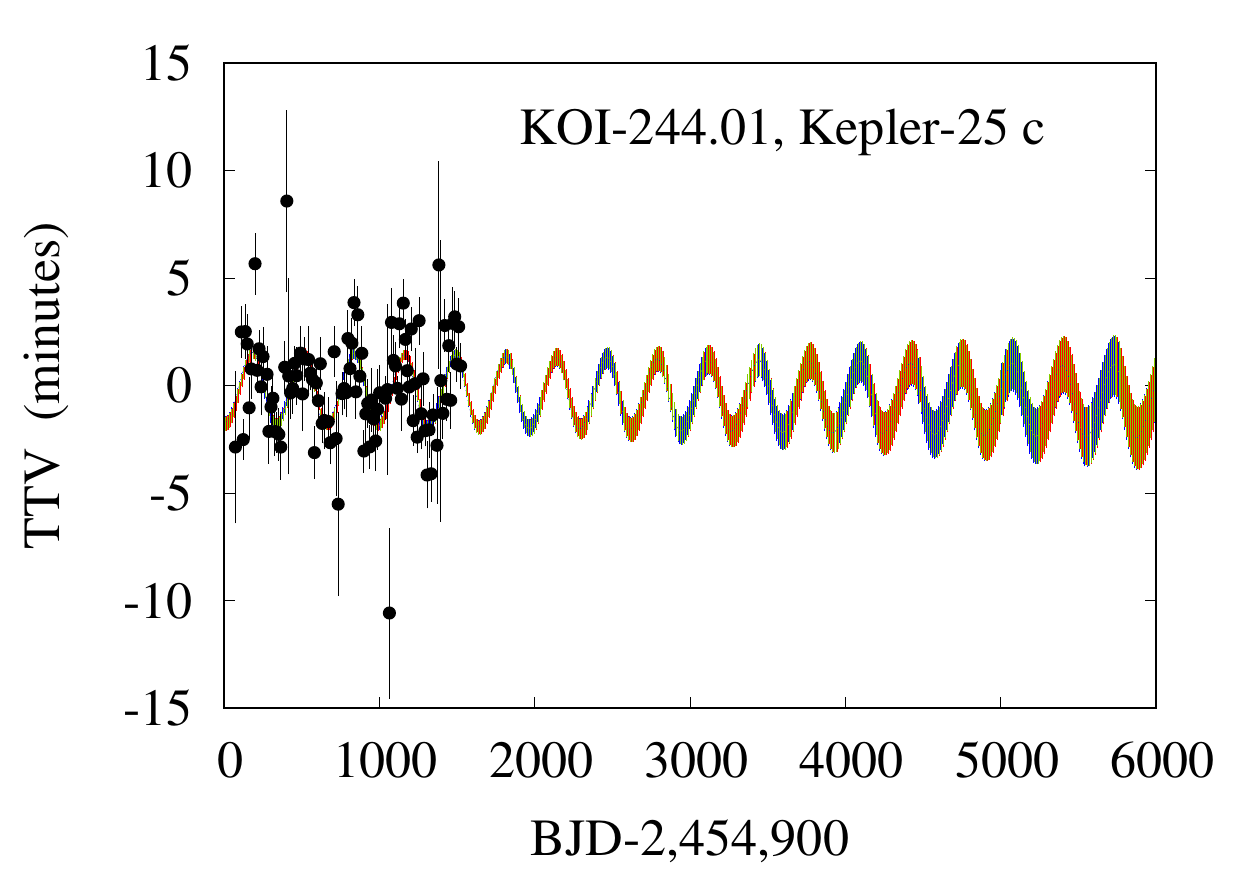}
\includegraphics[height = 1.05 in]{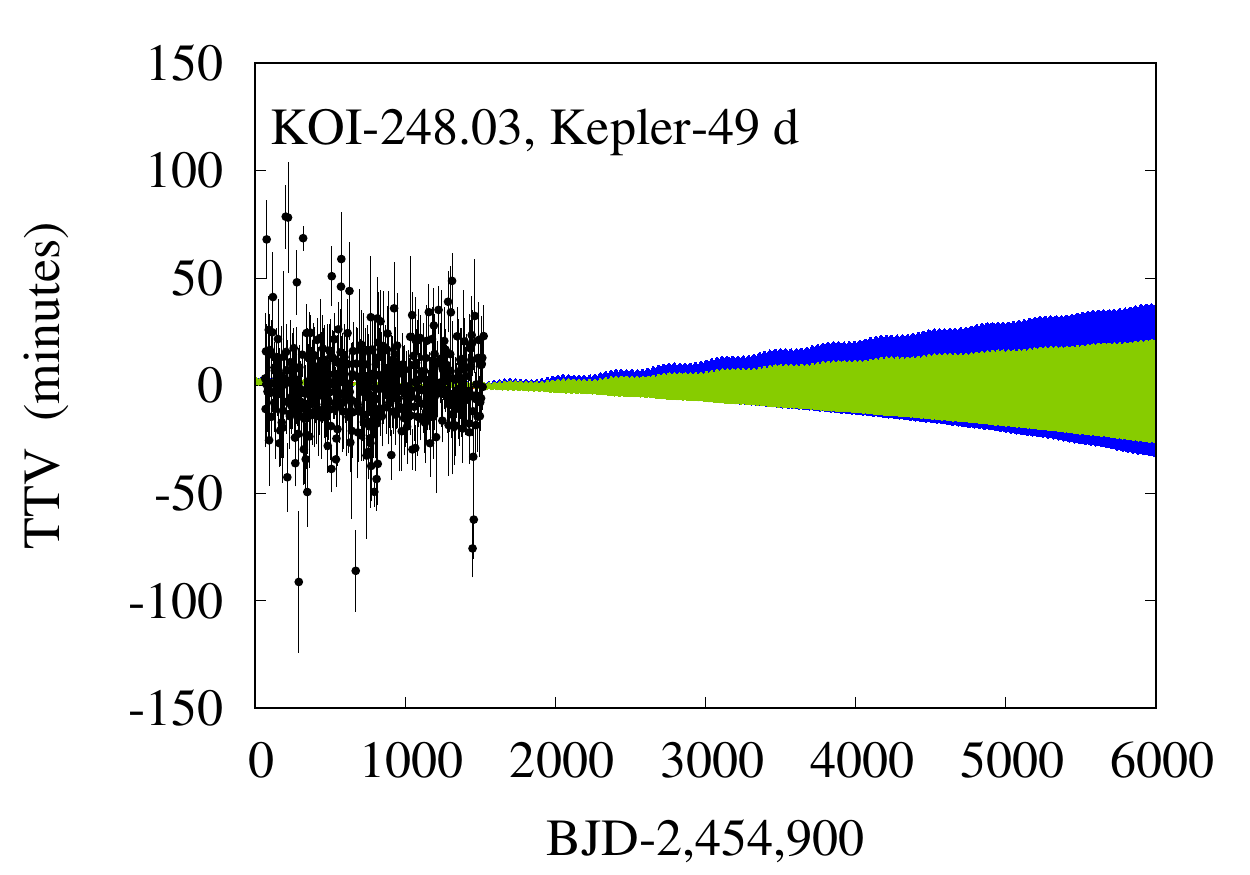}
\includegraphics[height = 1.05 in]{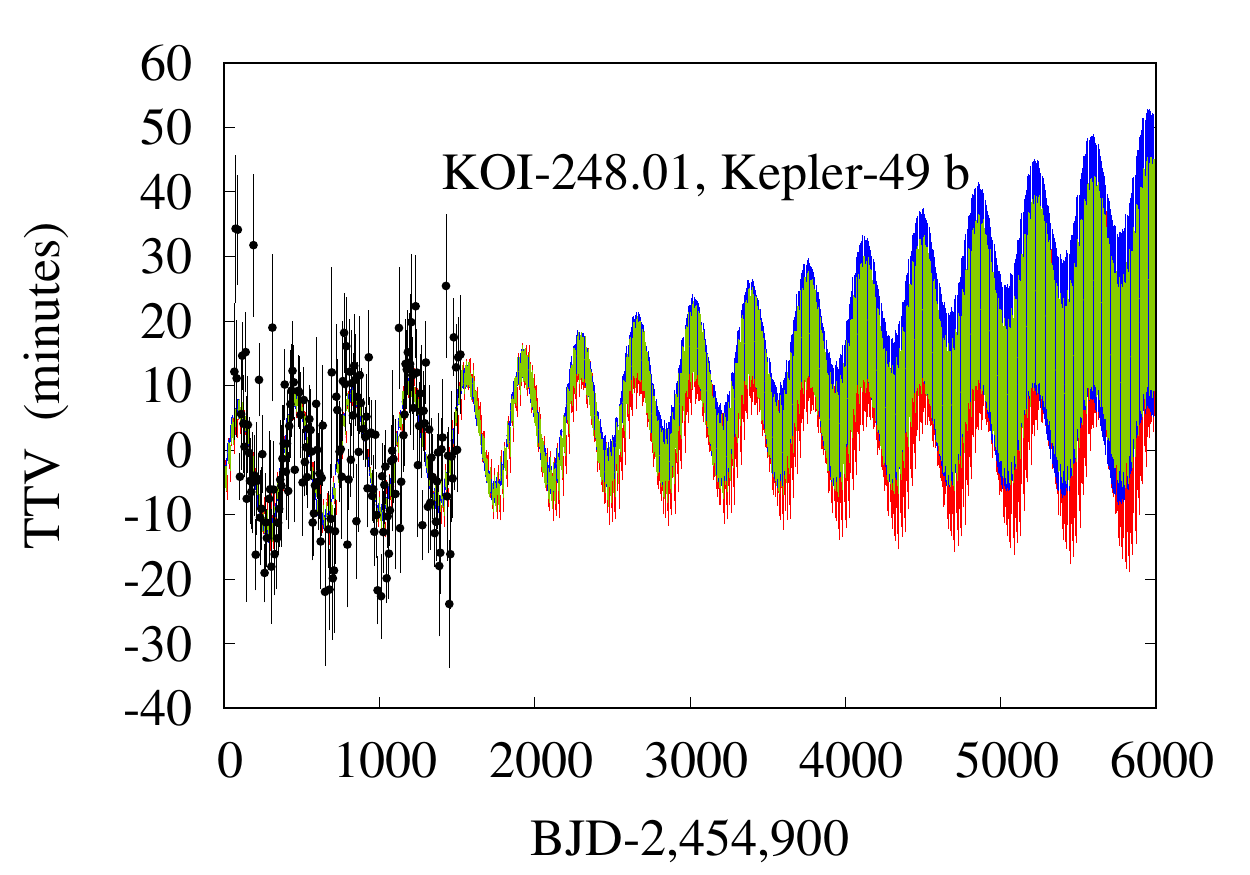}
\includegraphics[height = 1.05 in]{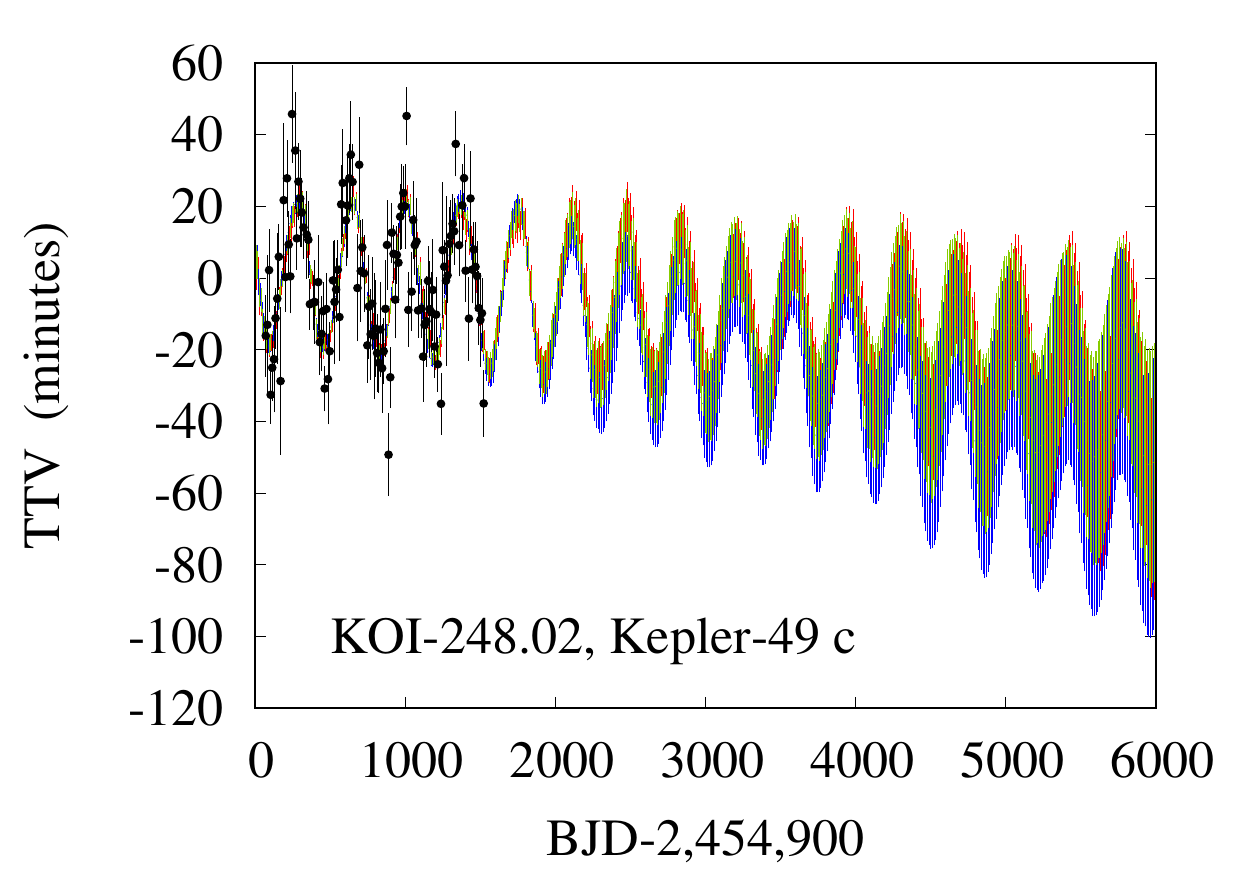}
\includegraphics[height = 1.05 in]{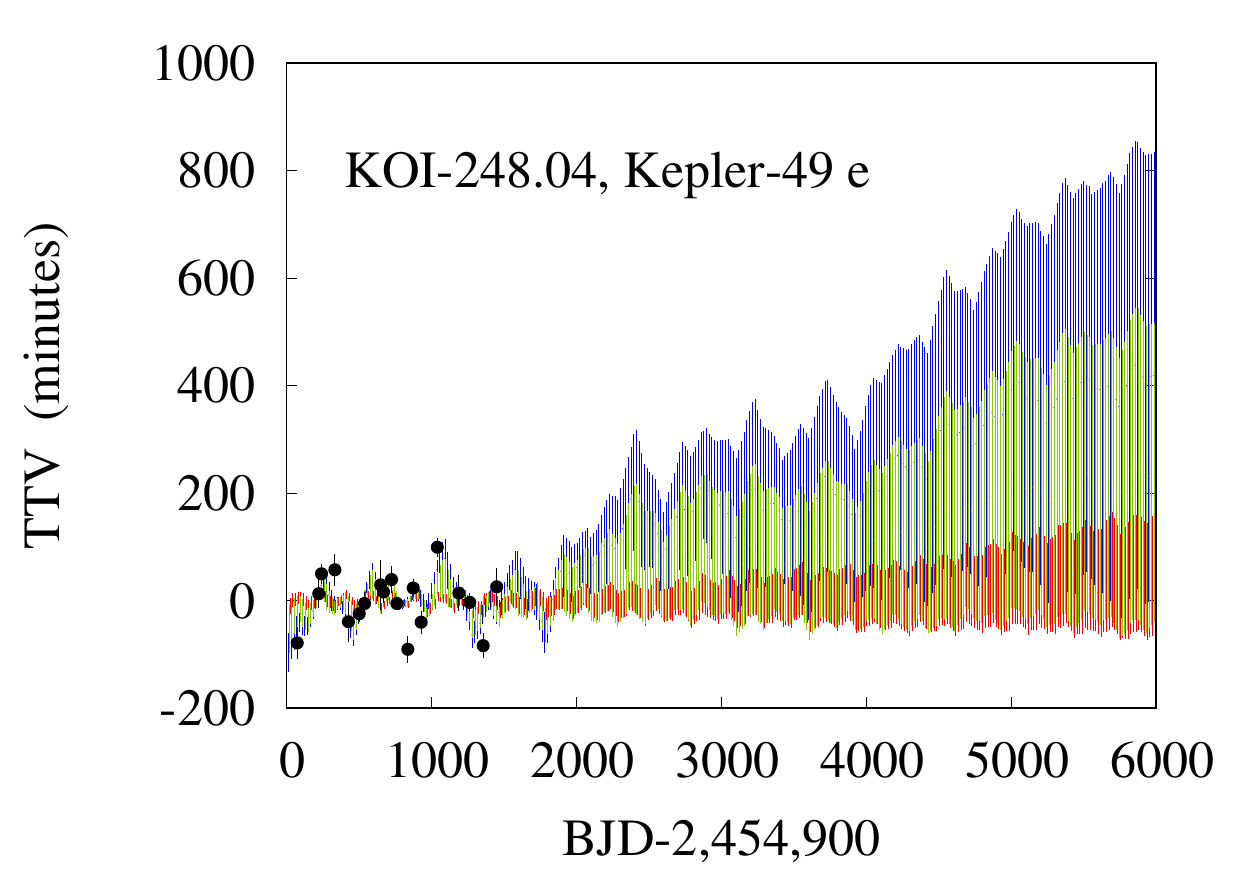}
\includegraphics[height = 1.05 in]{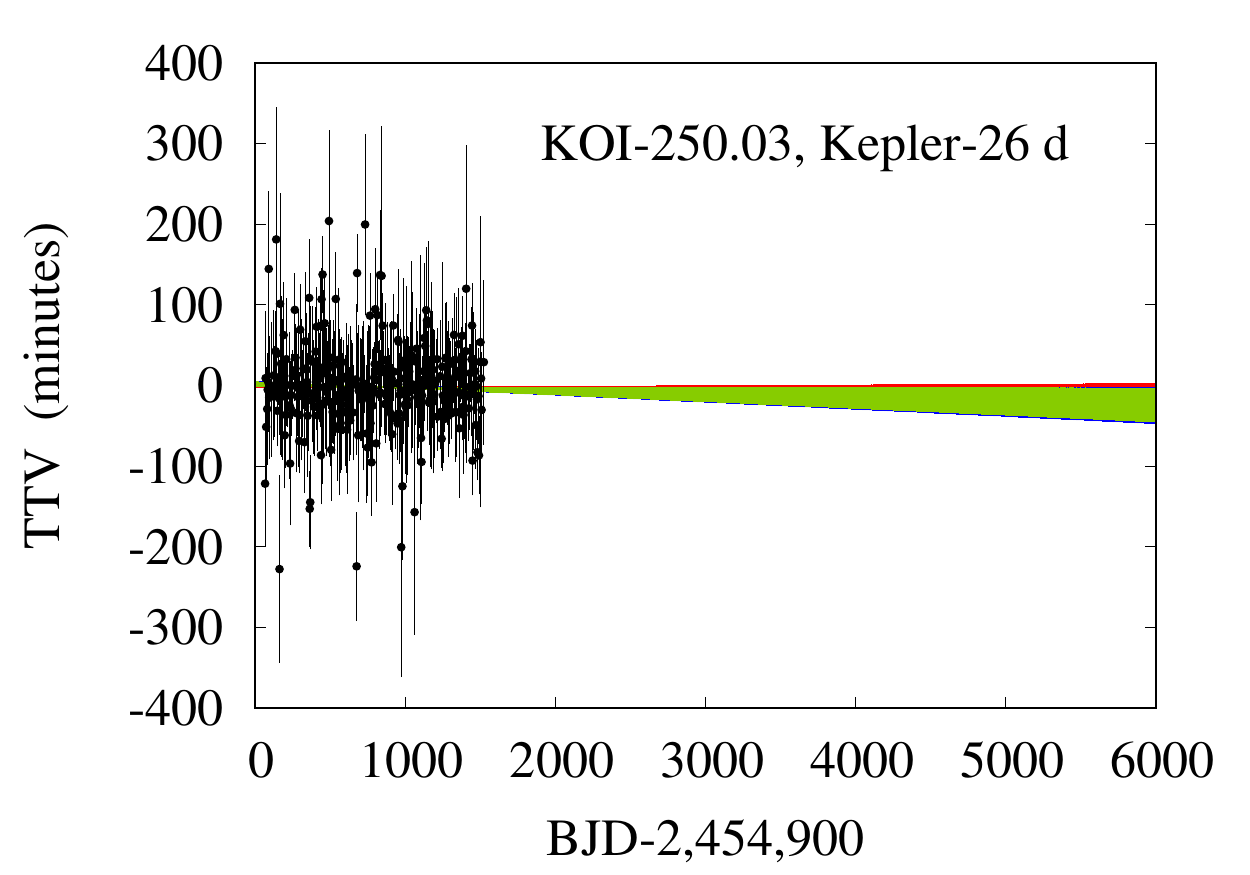}
\includegraphics[height = 1.05 in]{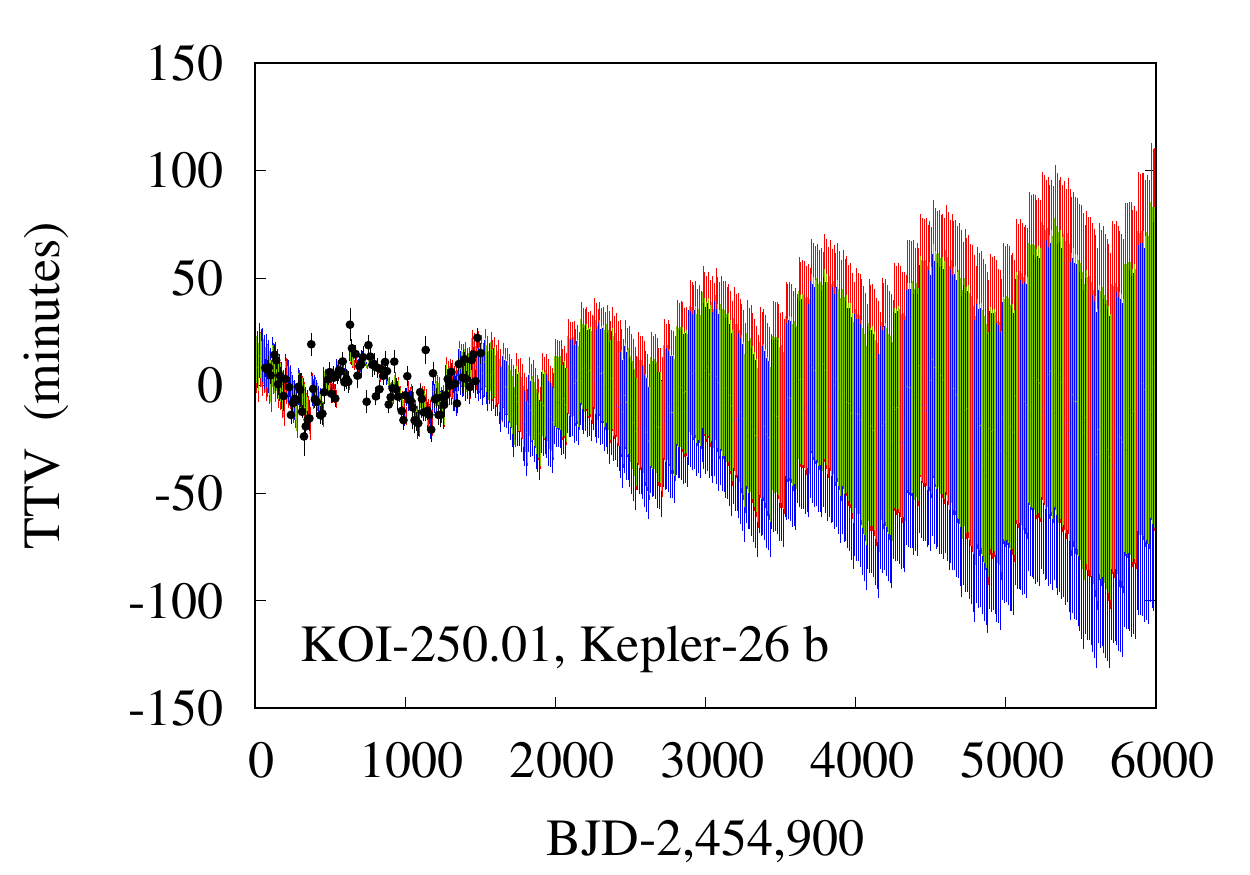}
\includegraphics[height = 1.05 in]{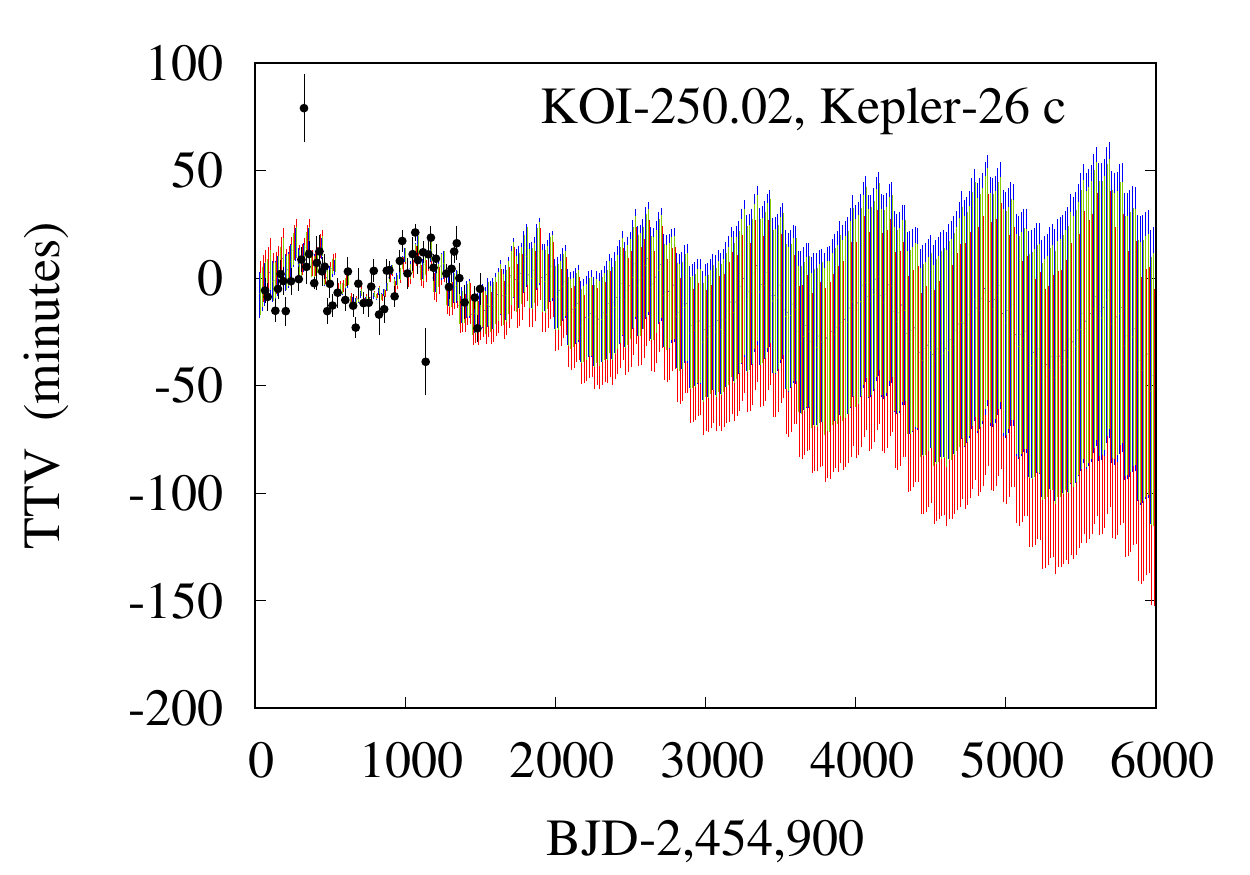}
\includegraphics[height = 1.05 in]{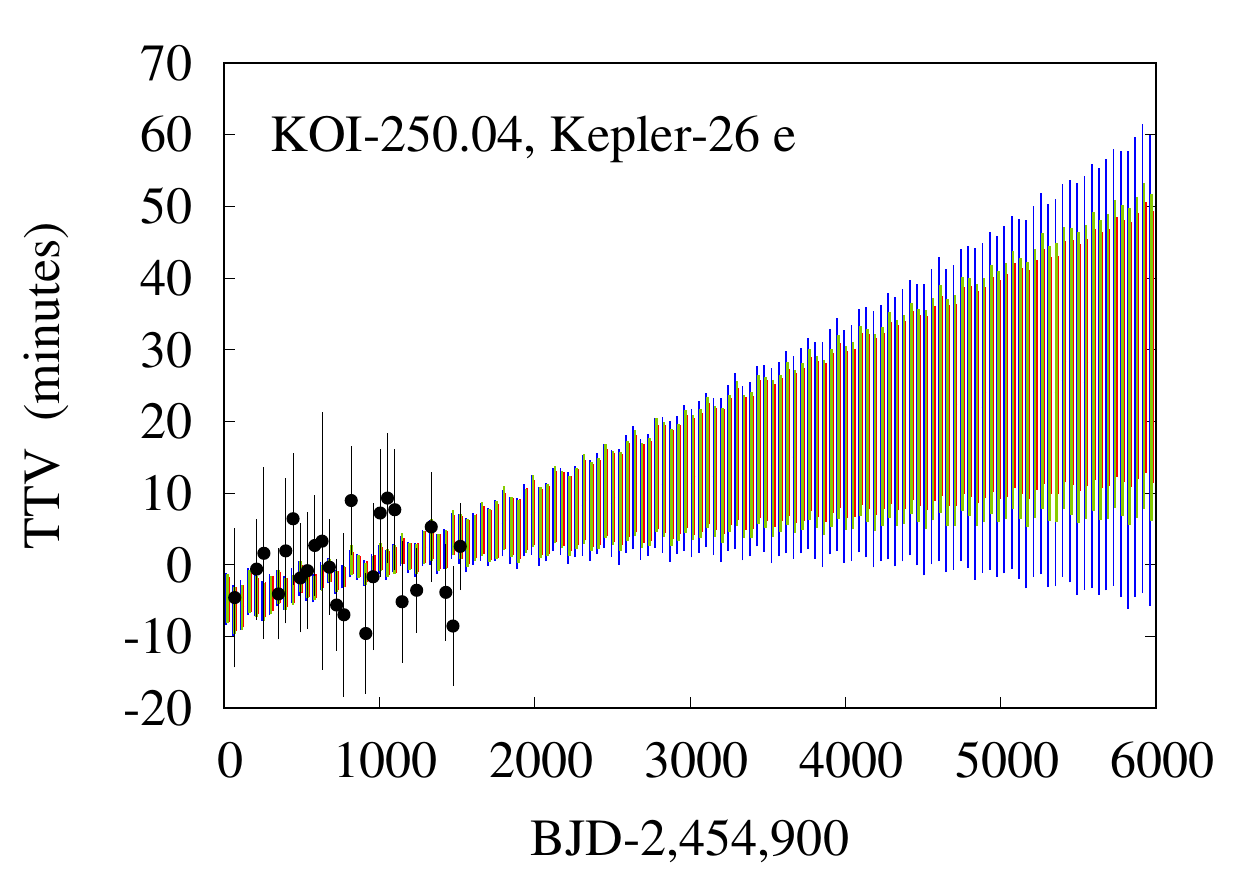}
\includegraphics[height = 1.05 in]{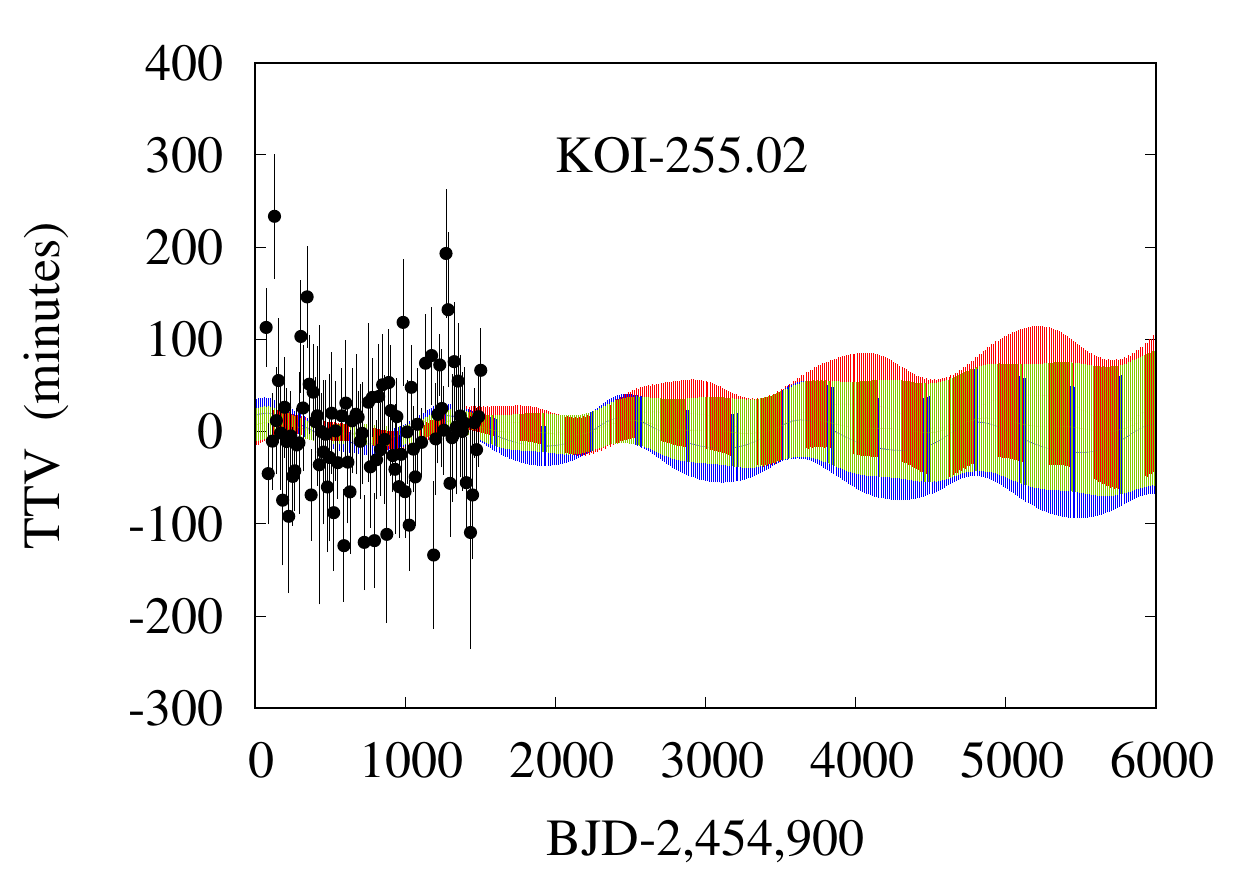}
\includegraphics[height = 1.05 in]{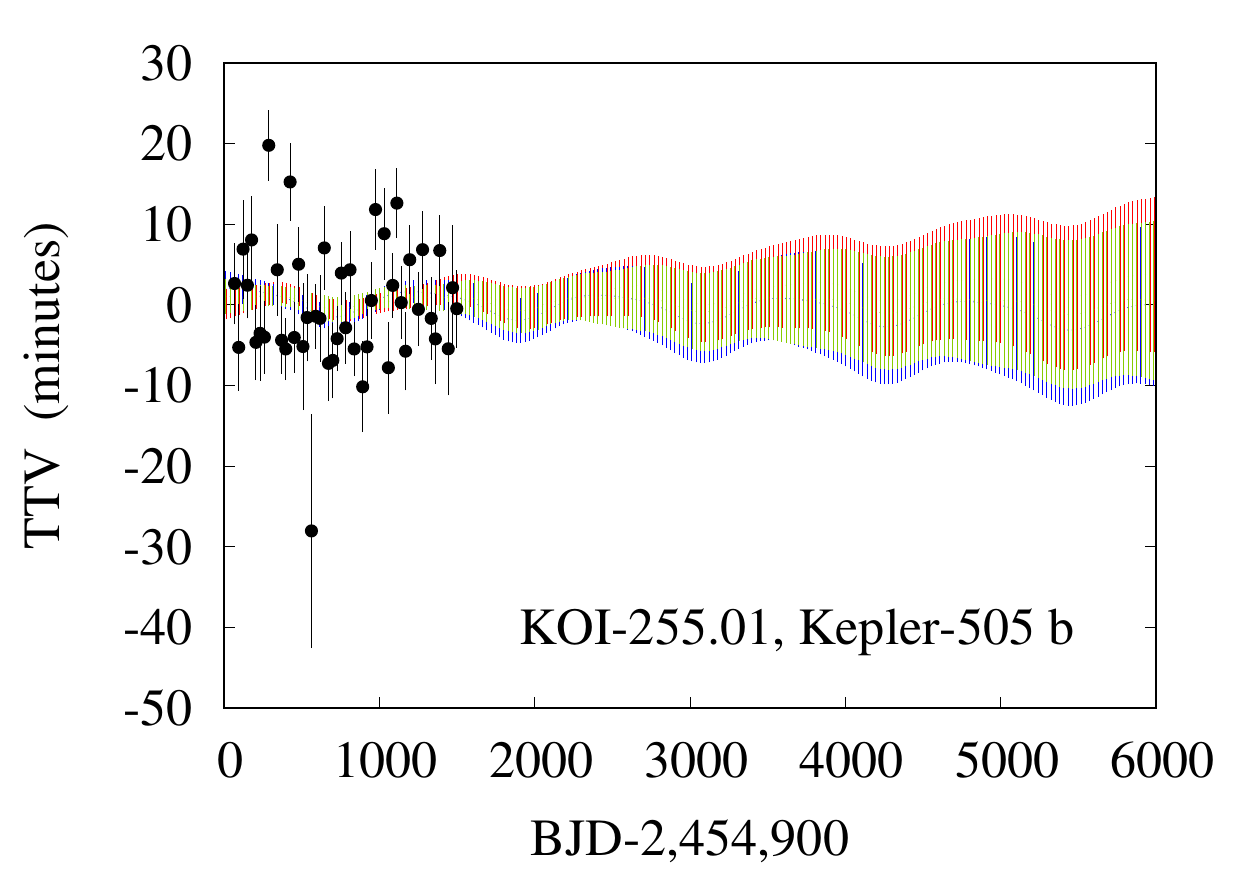}
\includegraphics[height = 1.05 in]{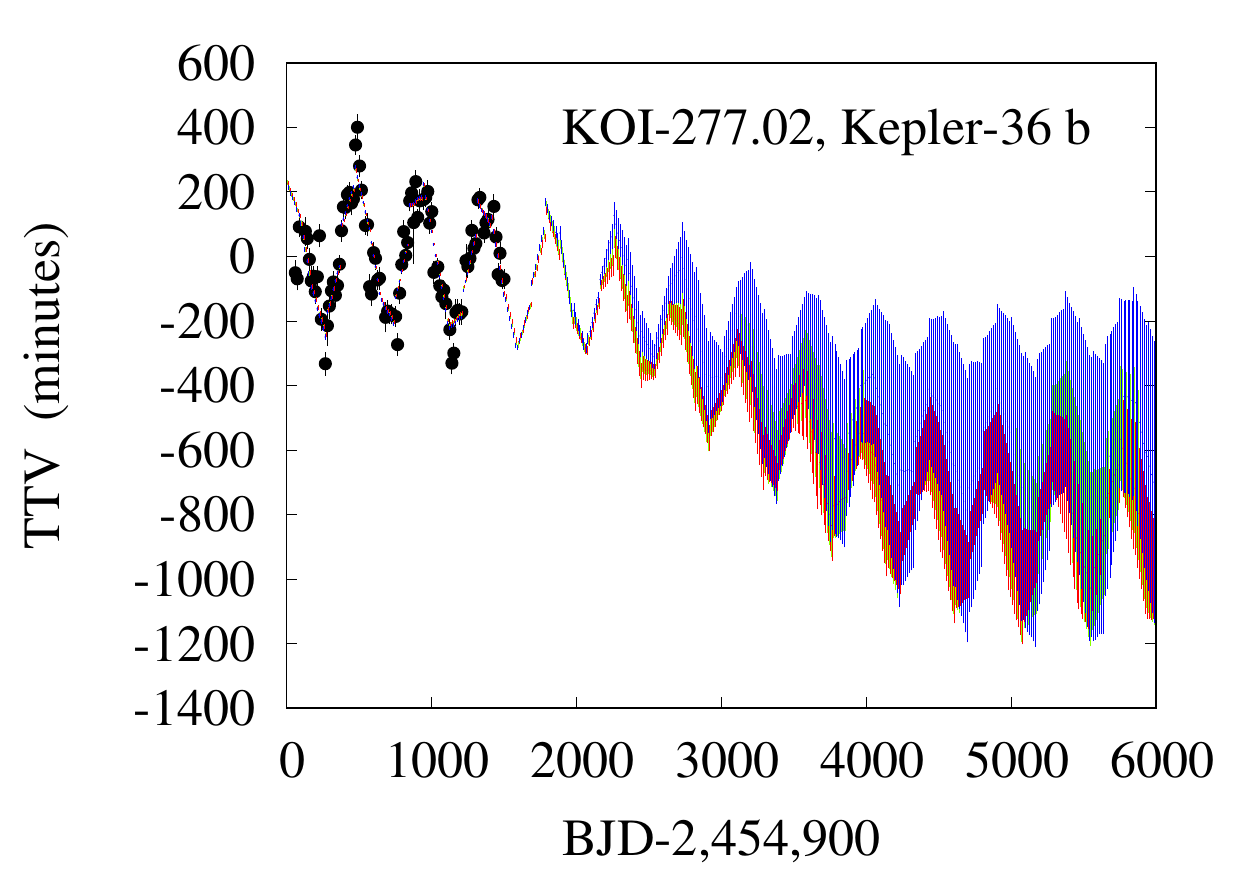}
\includegraphics[height = 1.05 in]{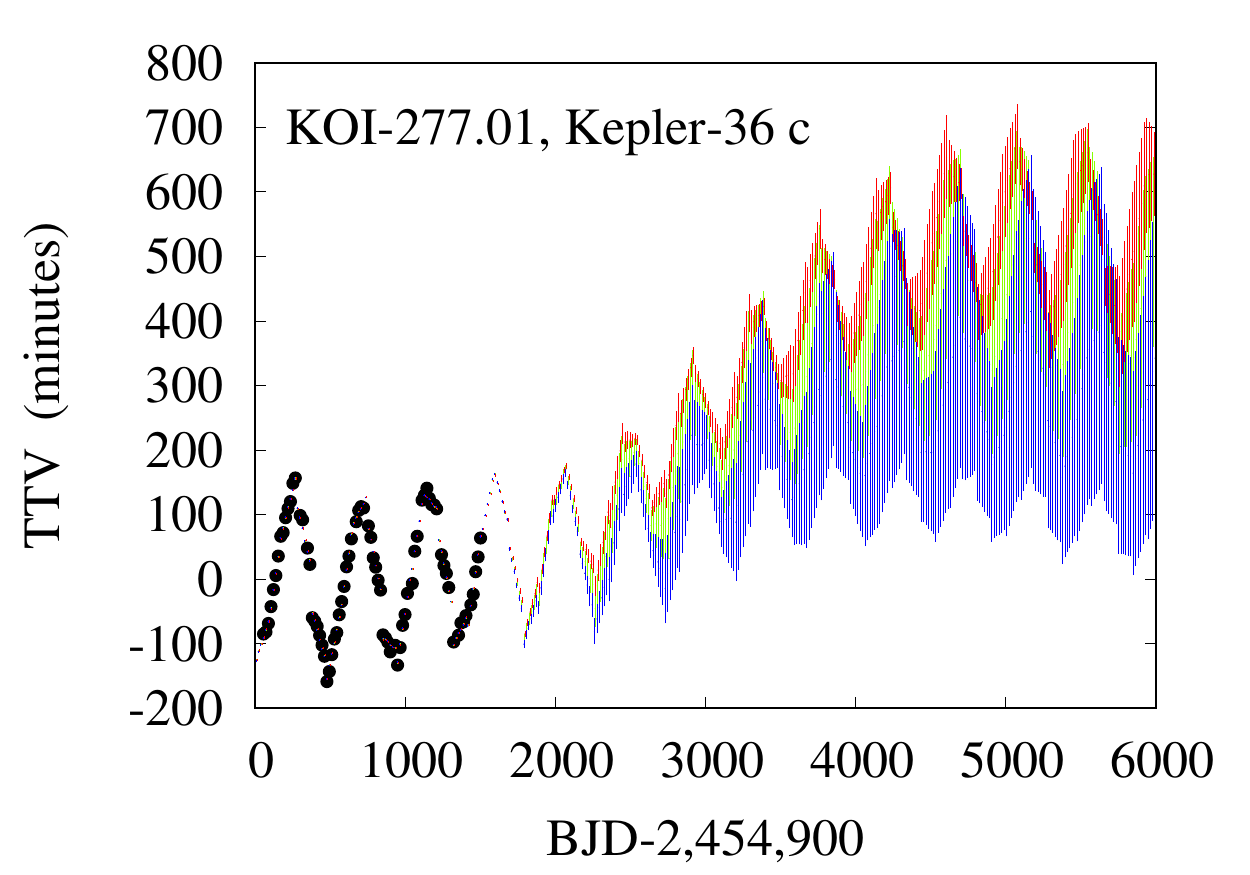}
\includegraphics[height = 1.45 in]{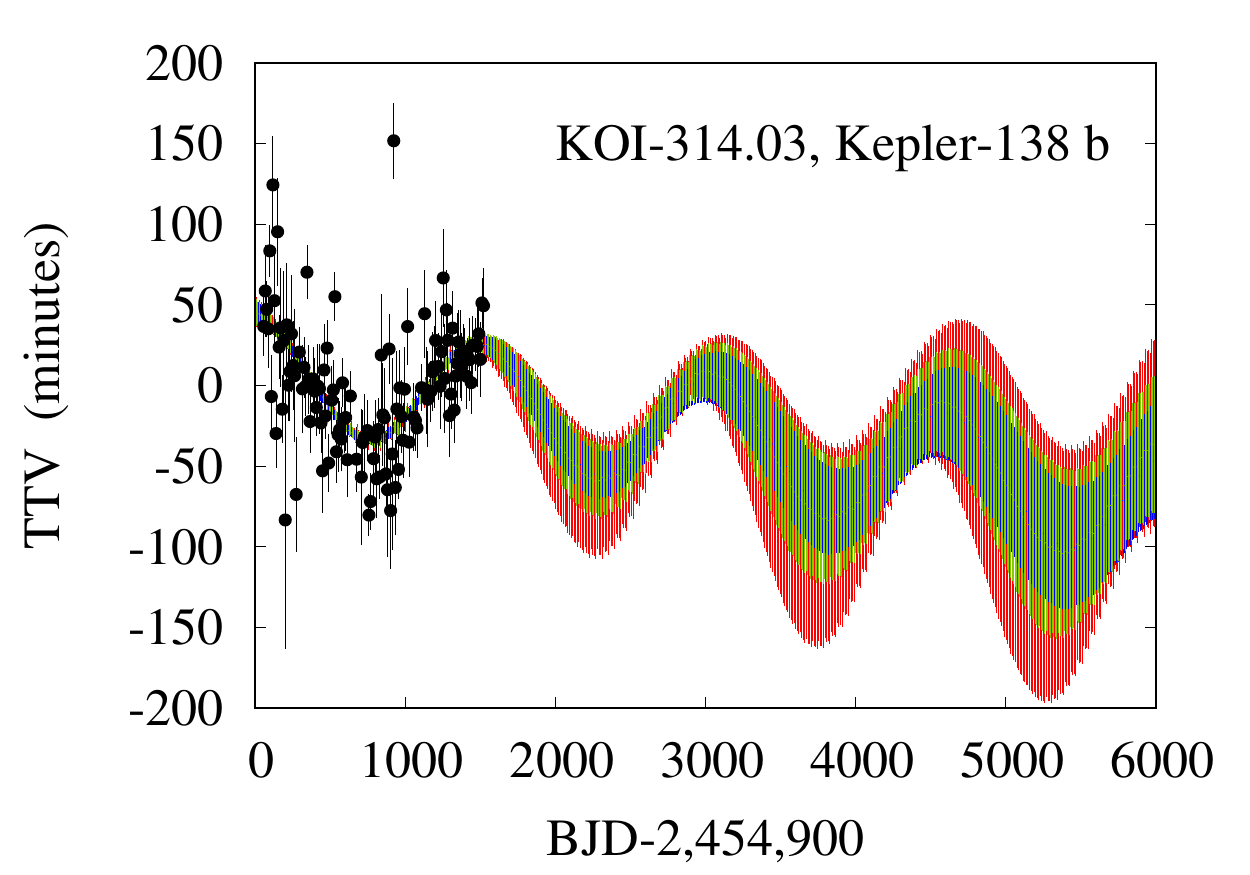}
\includegraphics[height = 1.45 in]{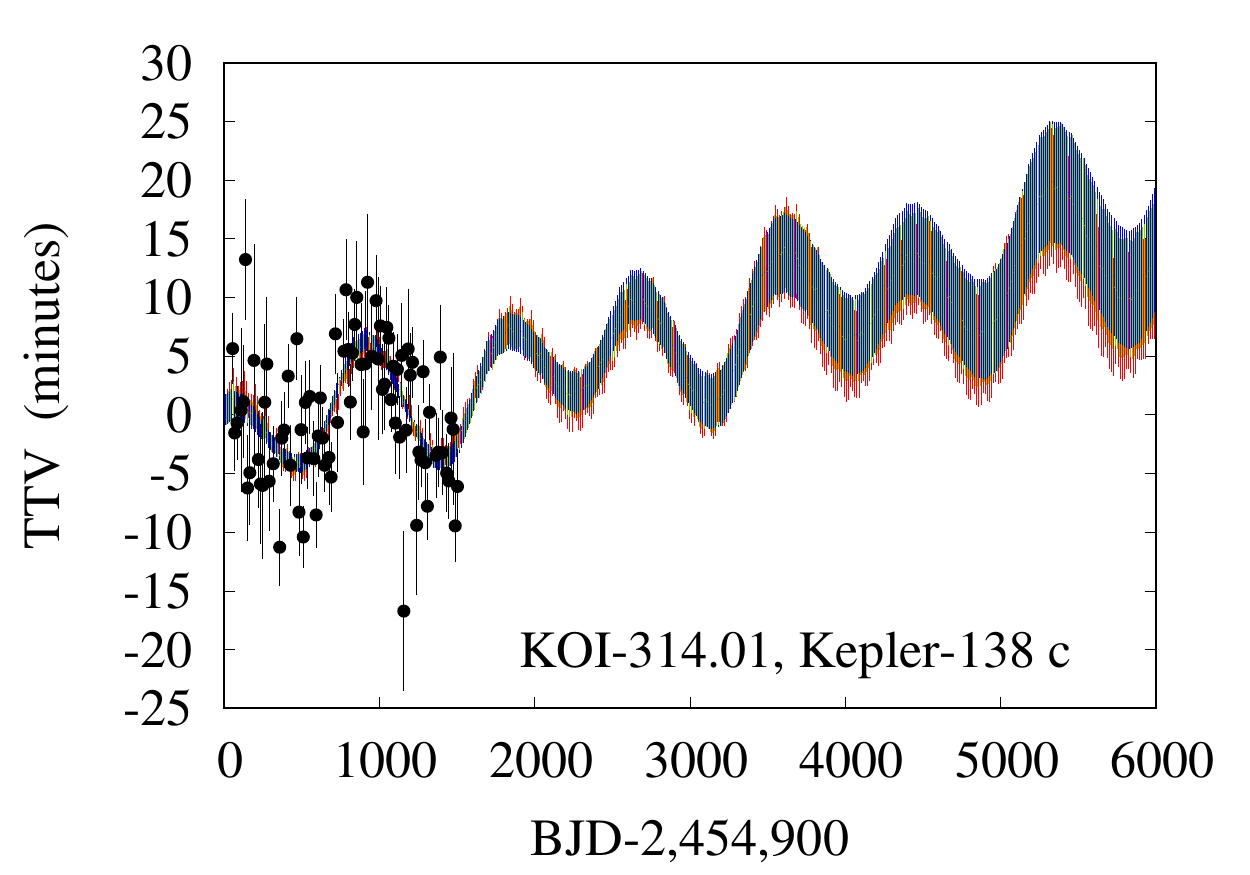}
\includegraphics[height = 1.45 in]{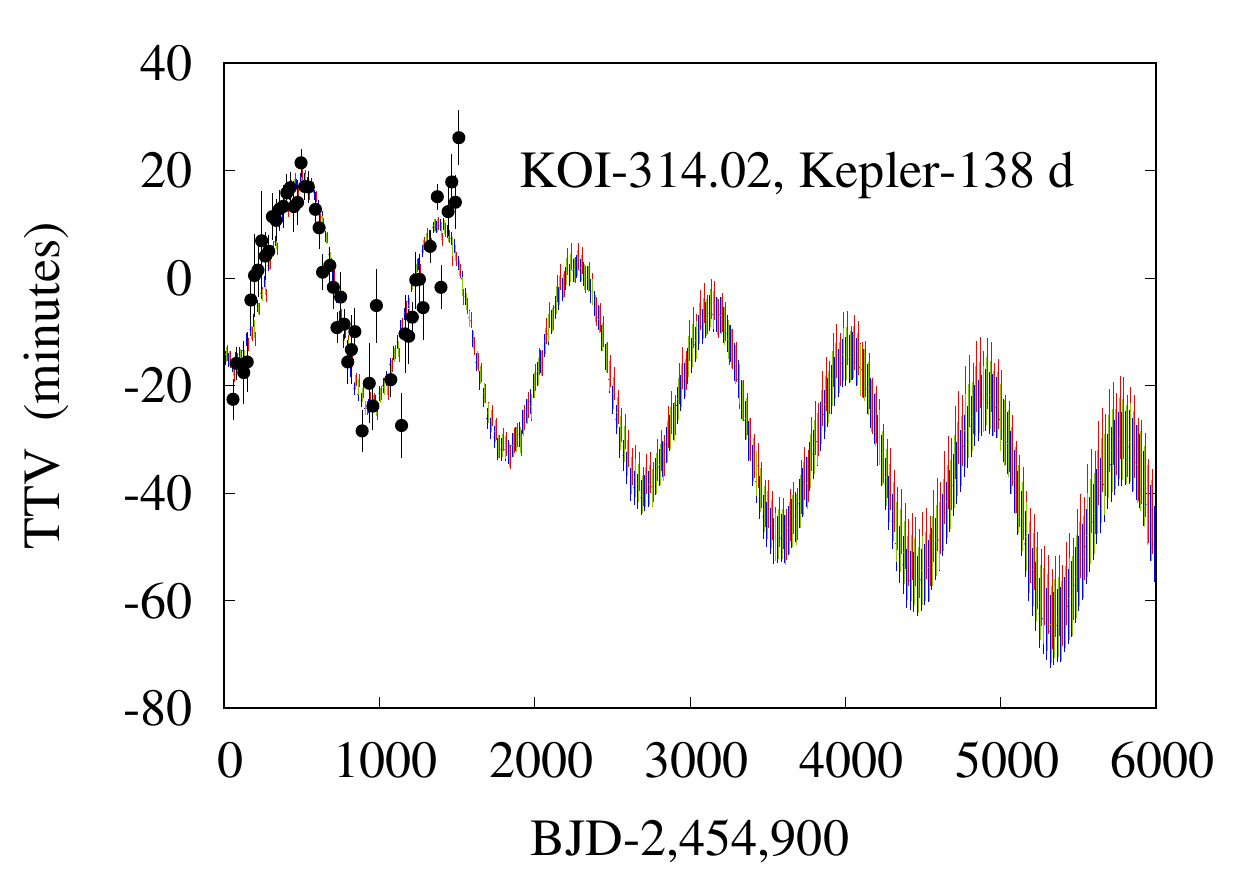}
\caption{Distribution of projected transit times as a function of time for the planet labelled in each panel (part 2). Black points mark transit times in the catalog of \citet{rowe15a} with 1$\sigma$ error bars. In green are 68.3\% confidence intervals of simulated transit times from posterior sampling. In blue (red), are a subset of samples with dynamical masses below (above) the 15.9th (84.1th) percentile. \label{fig:KOI-168fut}}
\end{center}
\end{figure}

\begin{figure}
\begin{center}
\figurenum{13}
\includegraphics [height = 1.45 in]{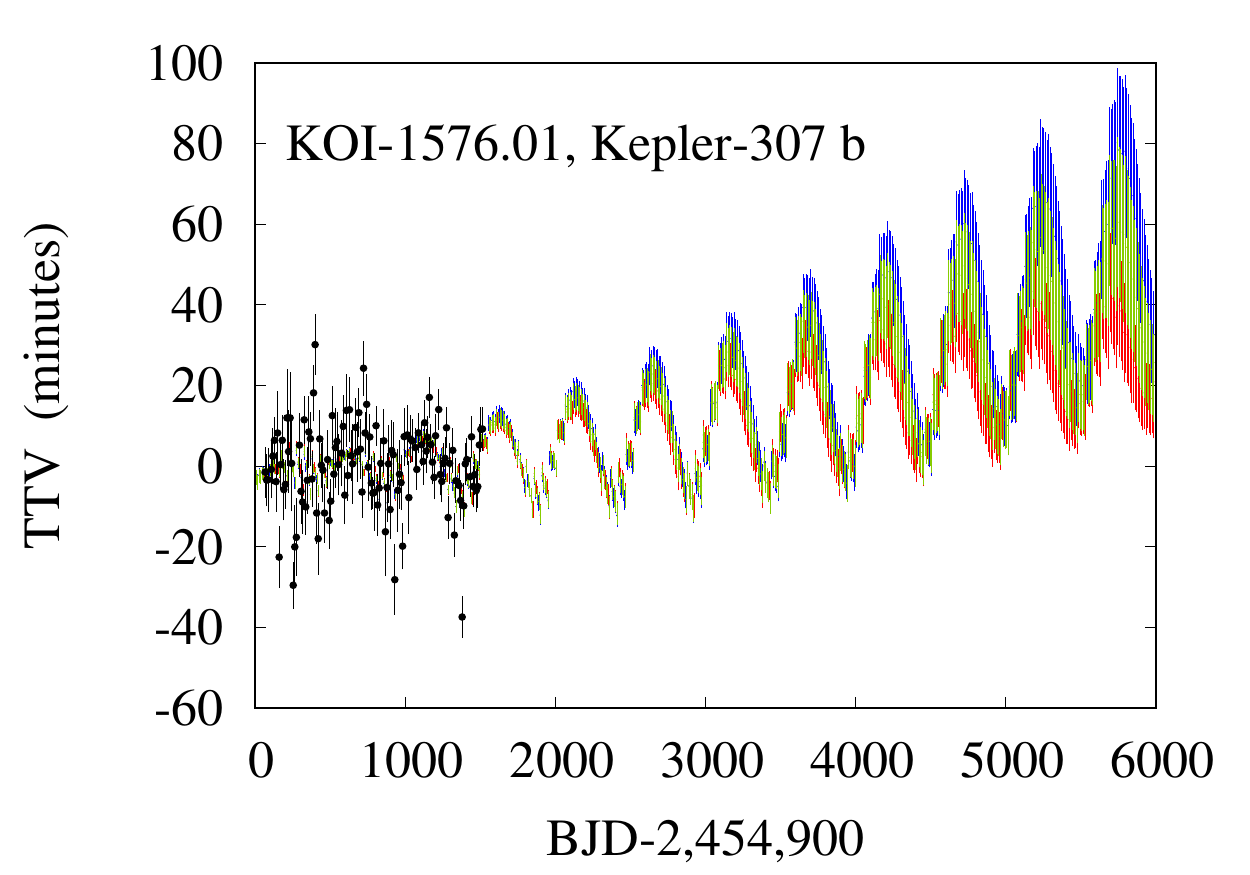}
\includegraphics [height = 1.45 in]{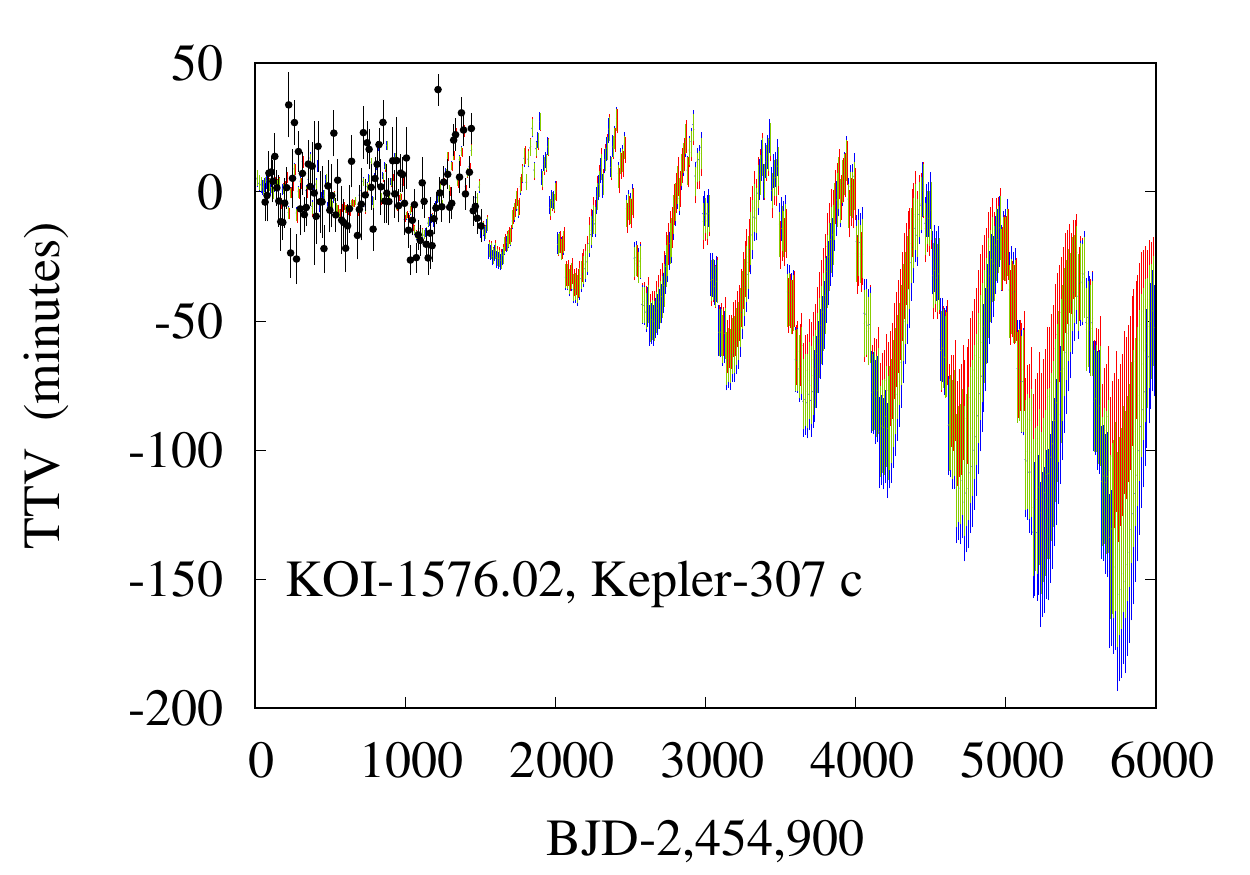}
\includegraphics [height = 1.45 in]{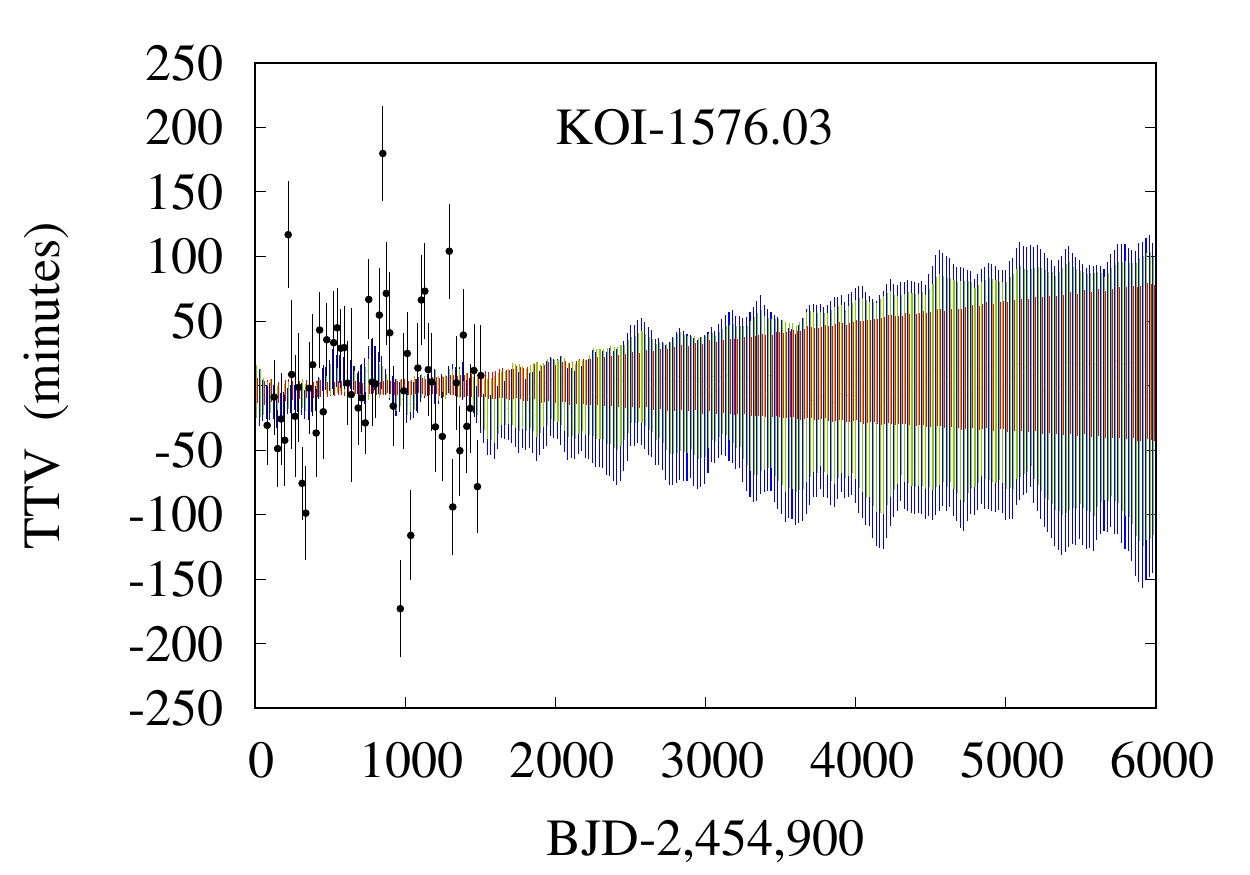}
\includegraphics[height = 1.45 in]{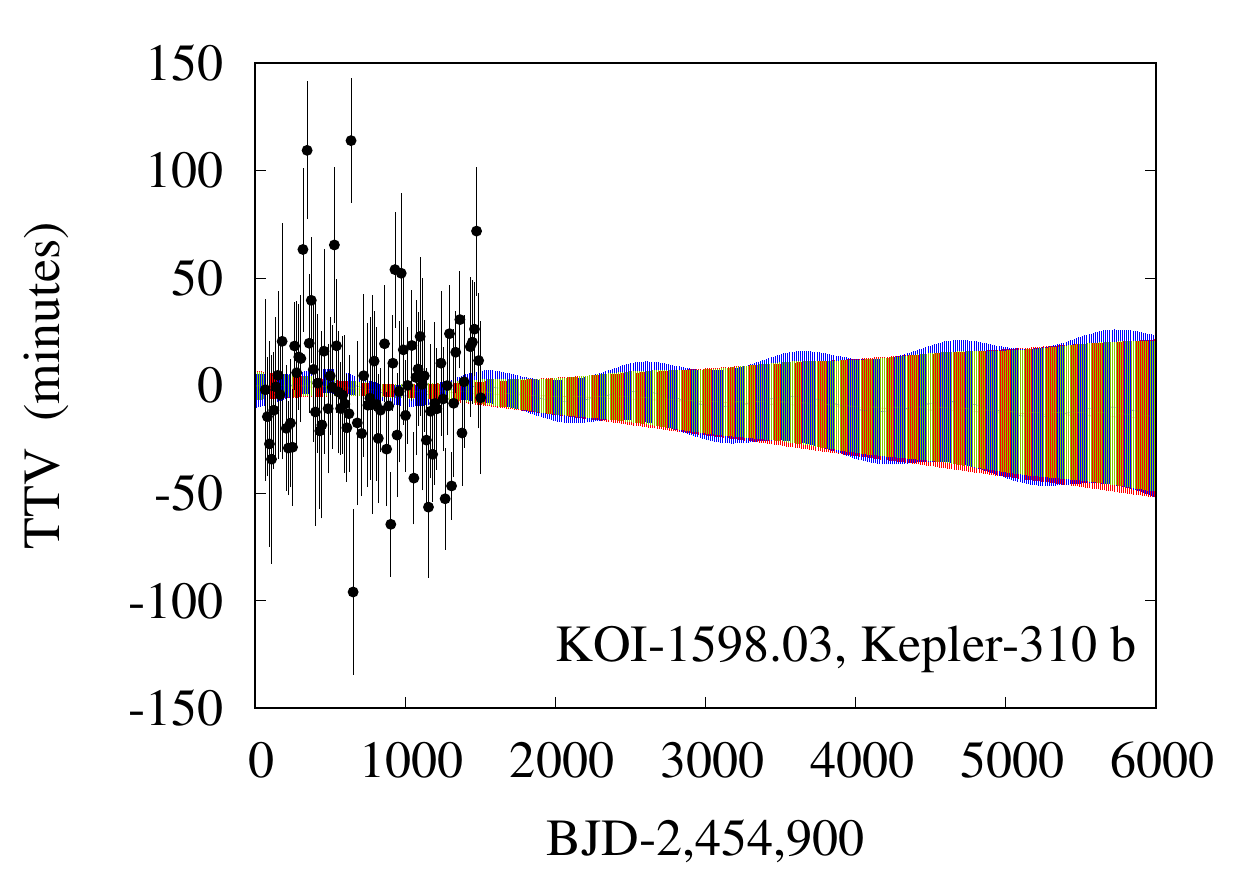}
\includegraphics[height = 1.45 in]{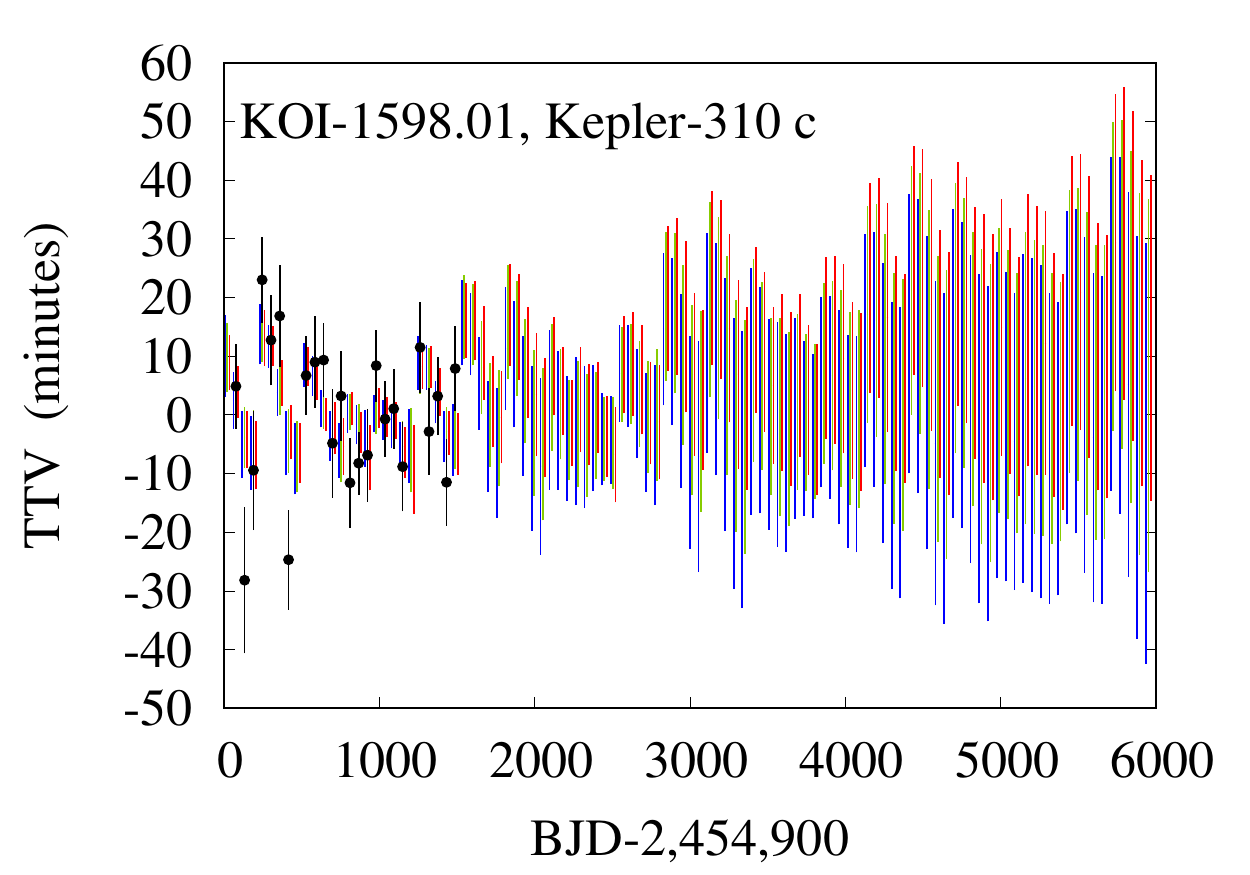} 
\includegraphics[height = 1.45 in]{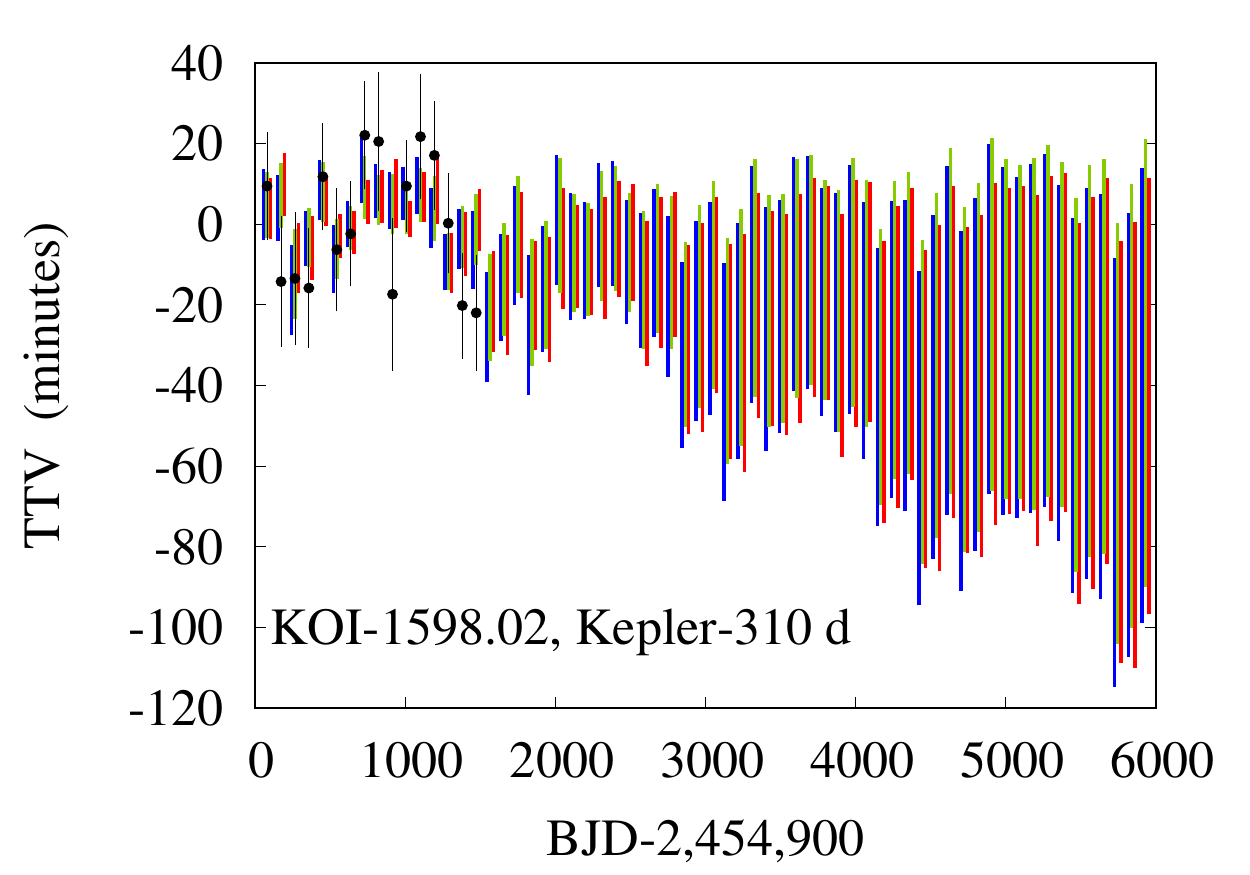}
\includegraphics[height = 1.45 in]{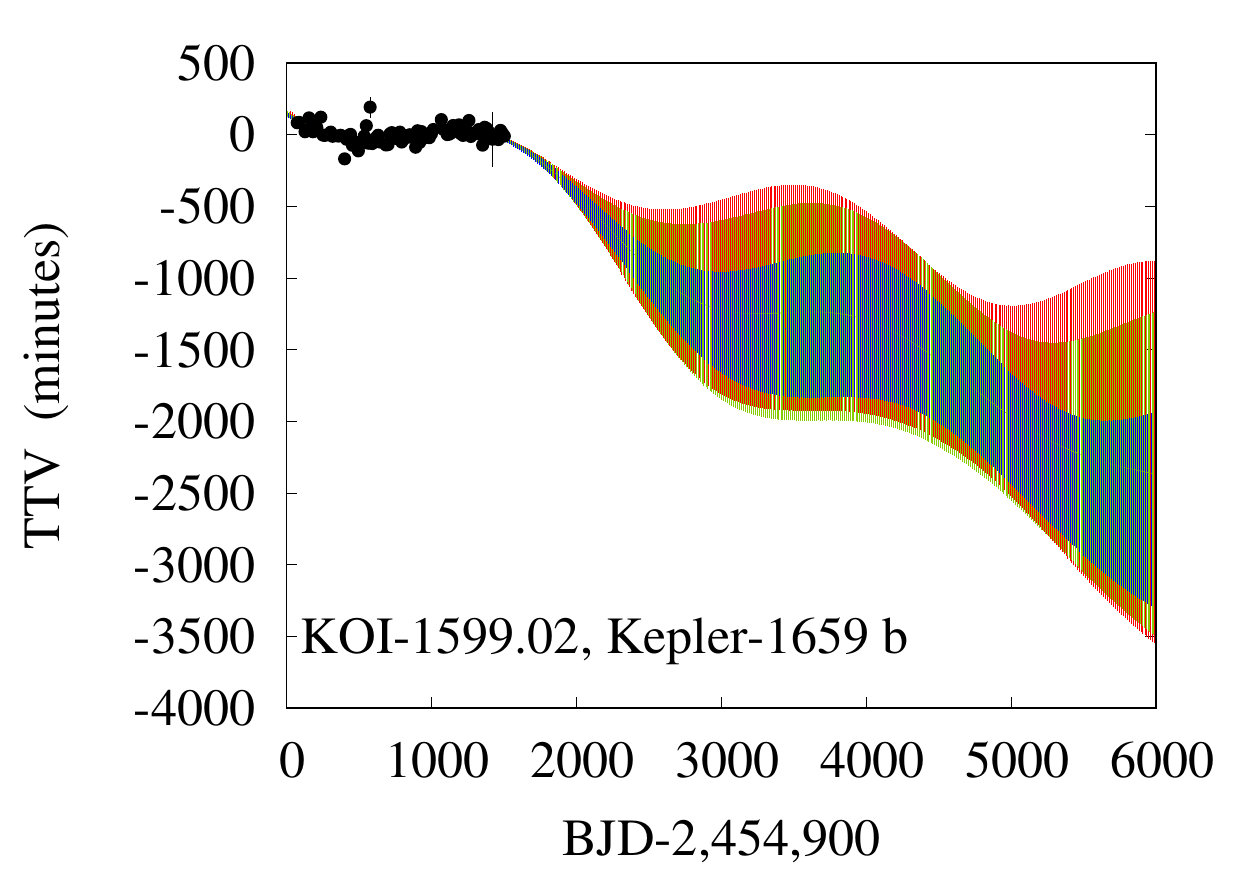}
\includegraphics[height = 1.45 in]{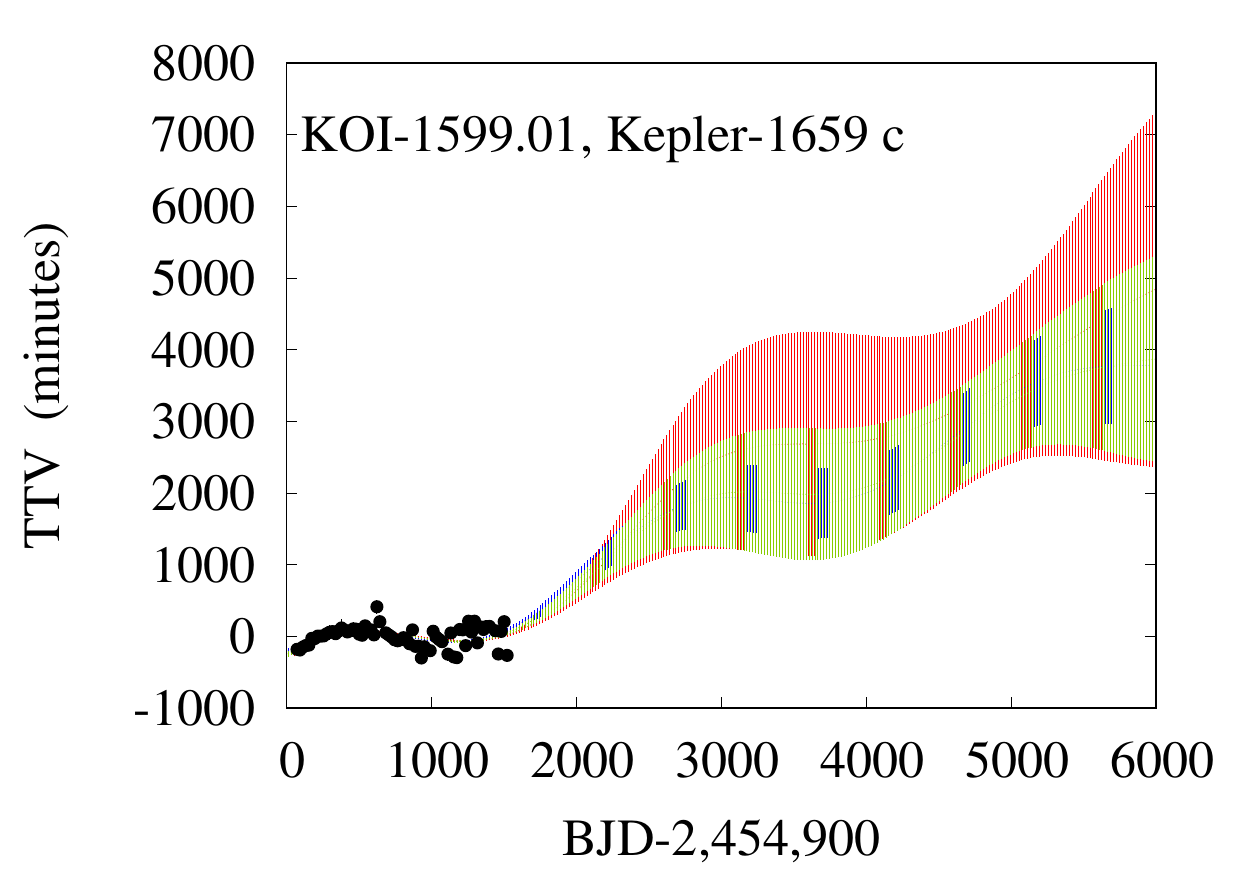} \\
\includegraphics[height = 1.45 in]{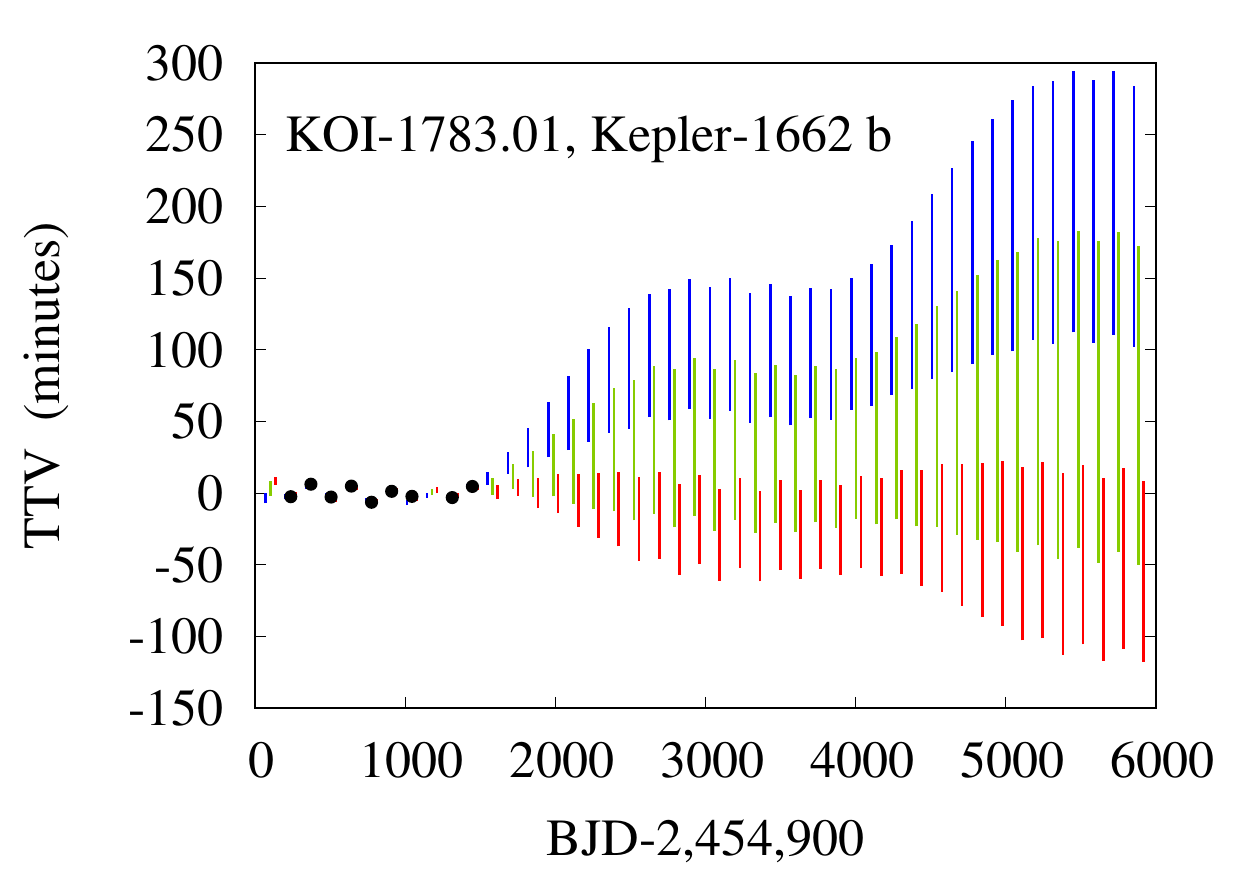}
\includegraphics[height = 1.45 in]{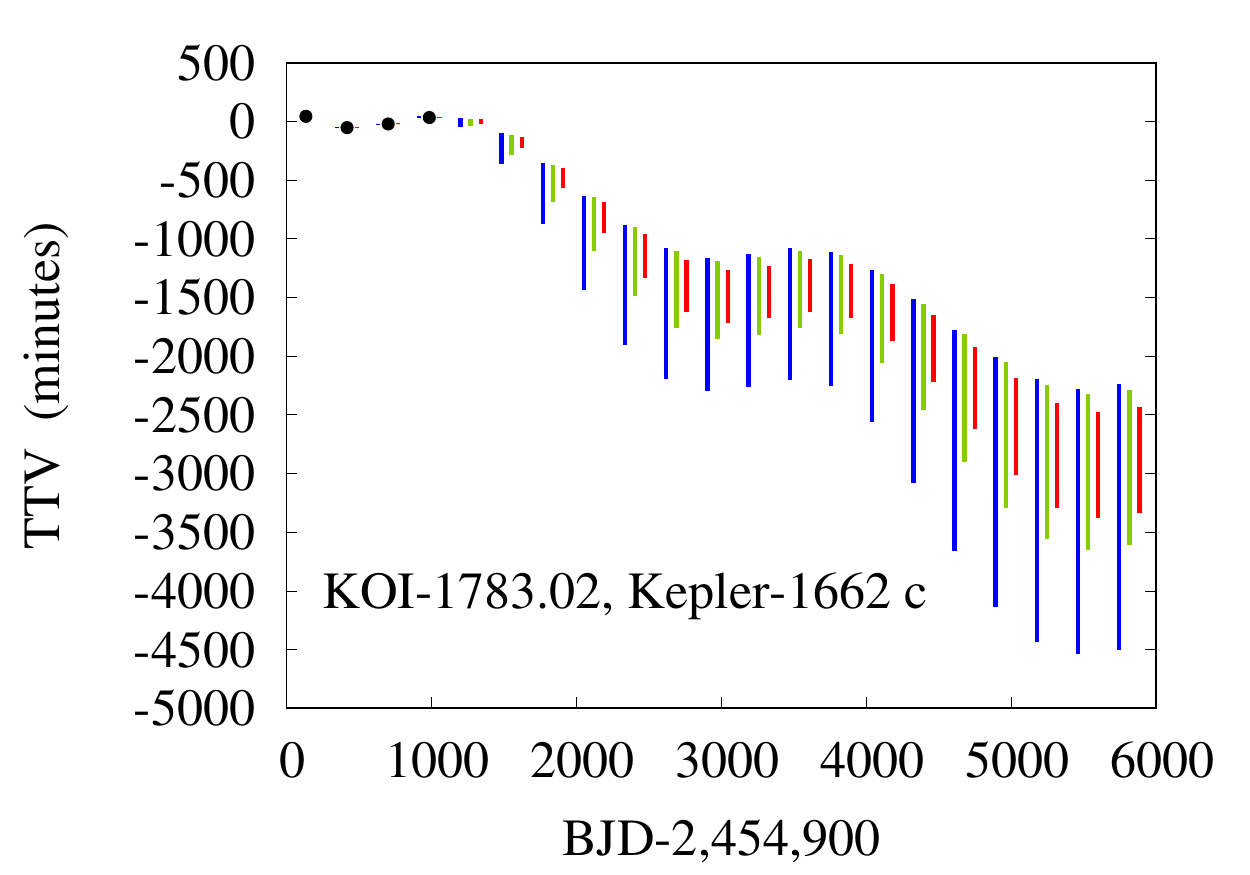} \\
\includegraphics[height = 1.05 in]{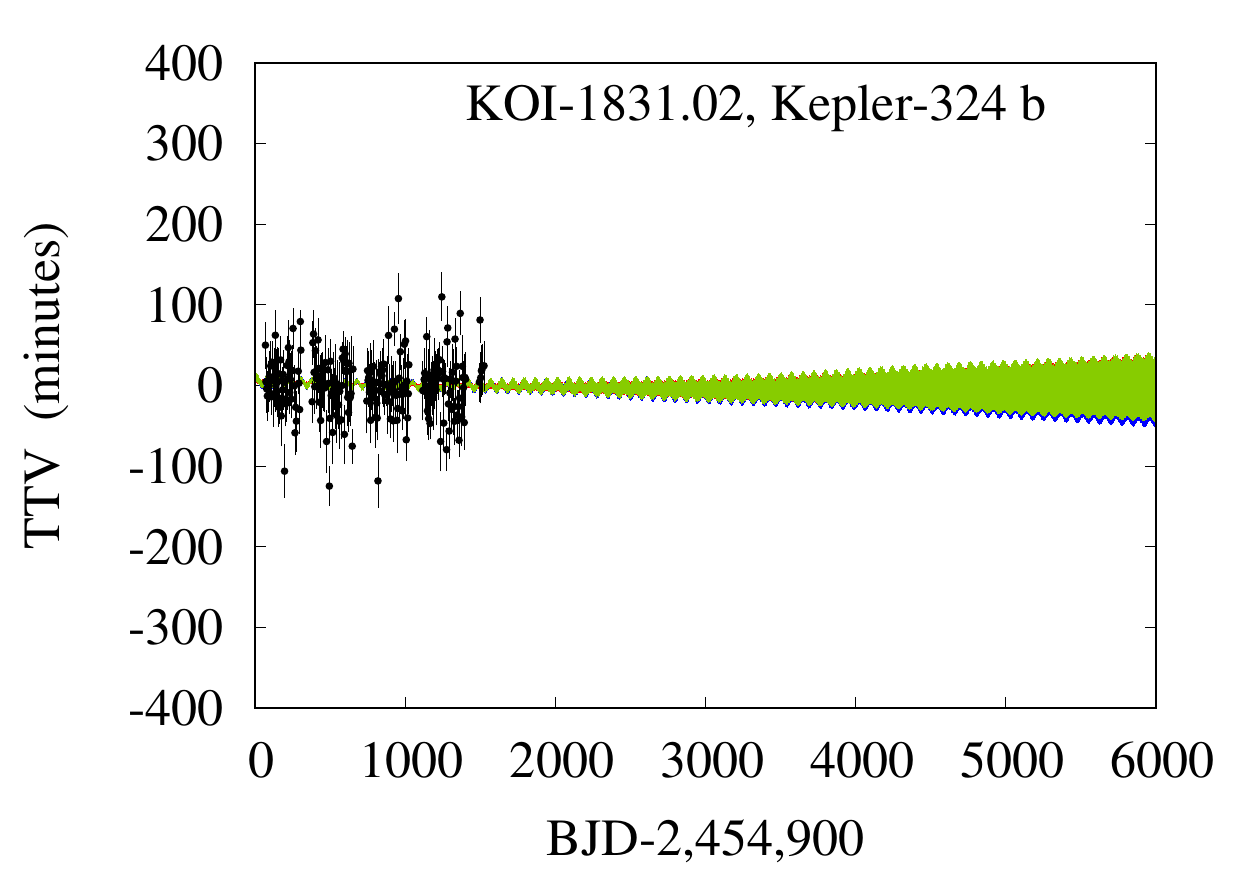}
\includegraphics[height = 1.05 in]{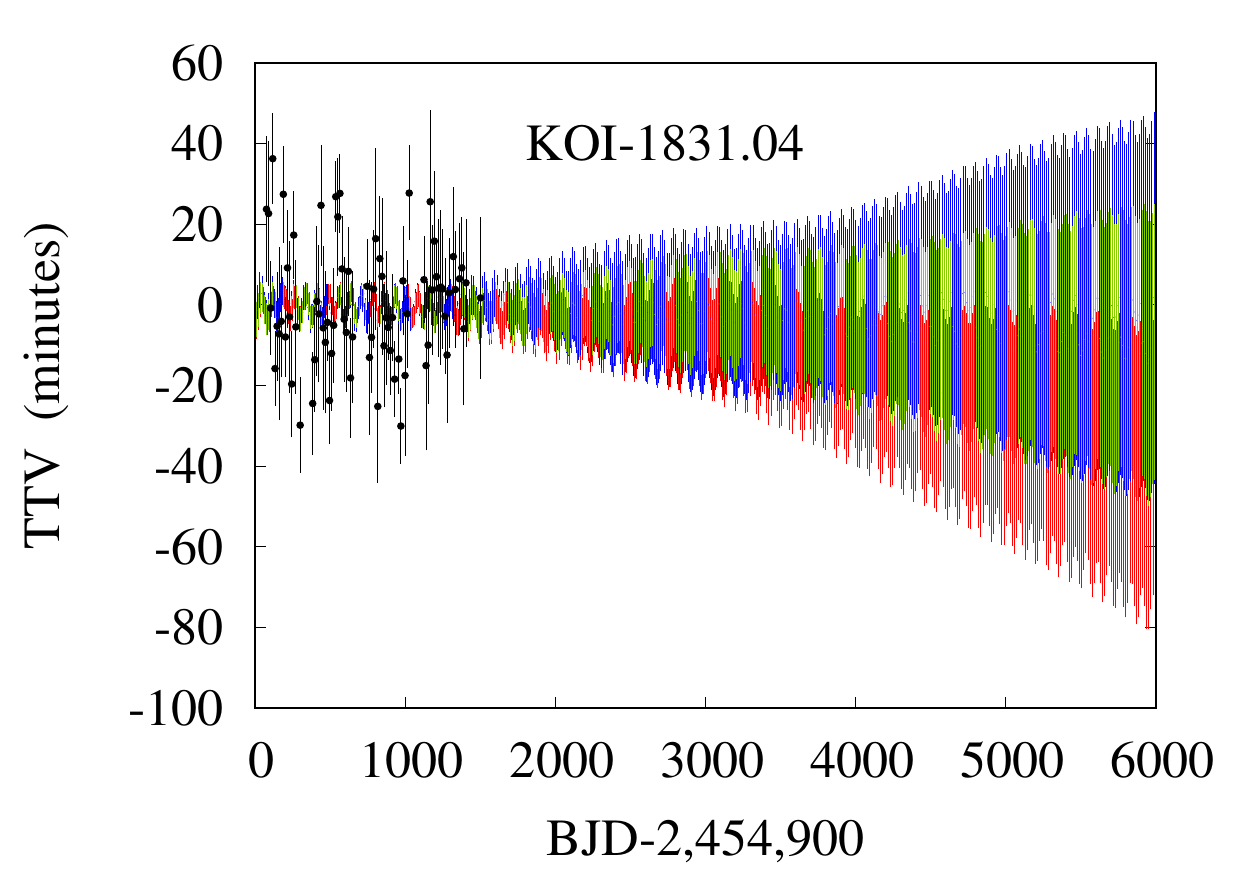}
\includegraphics[height = 1.05 in]{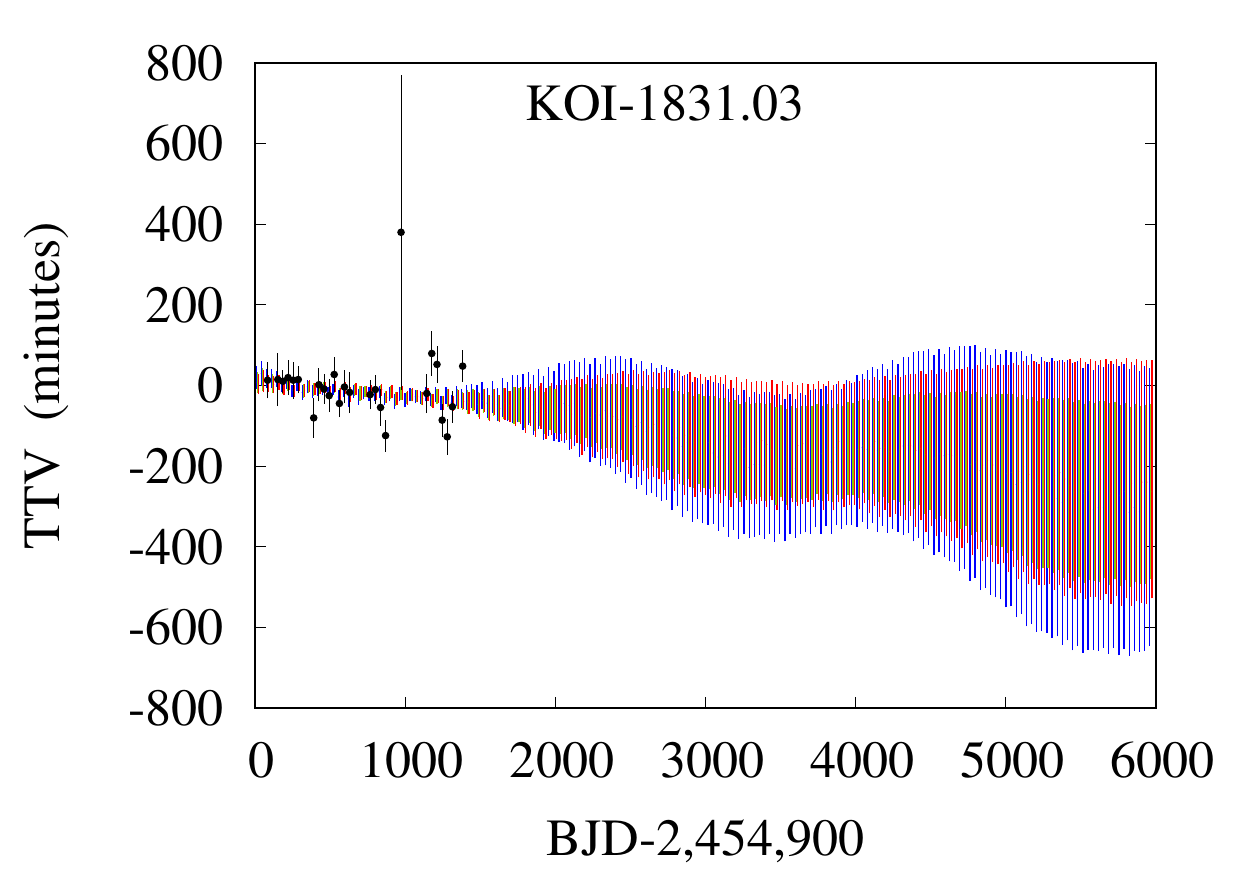}
\includegraphics[height = 1.05 in]{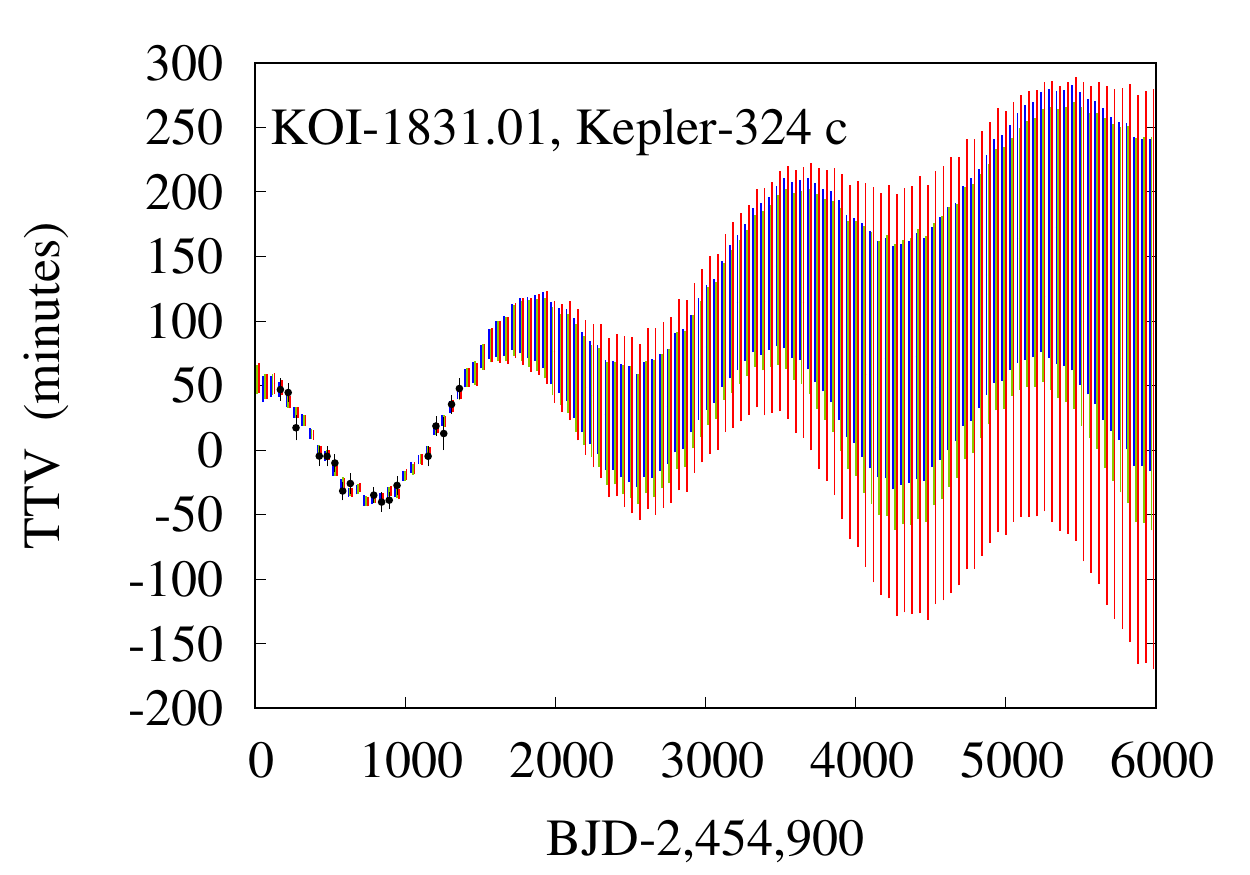}
\includegraphics[height = 1.45 in]{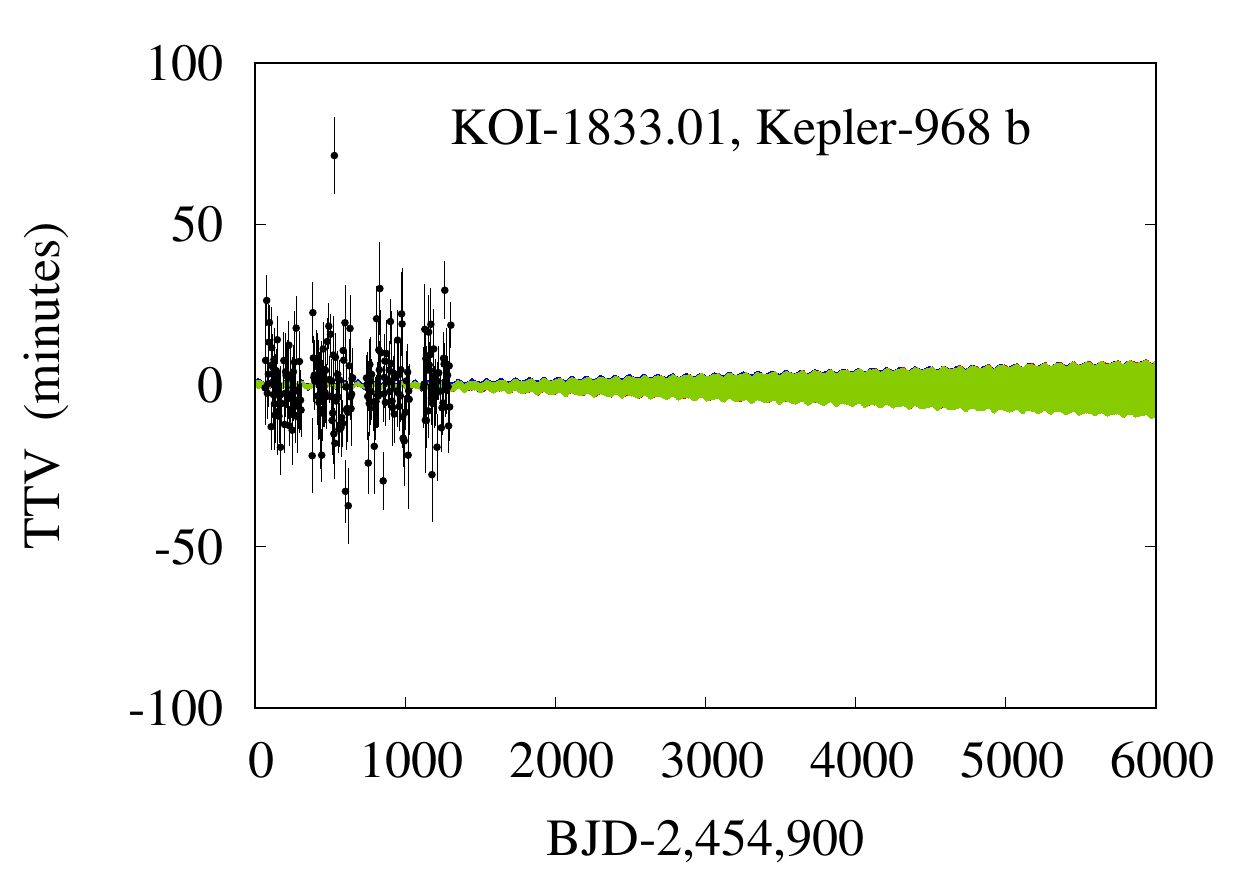} 
\includegraphics[height = 1.45 in]{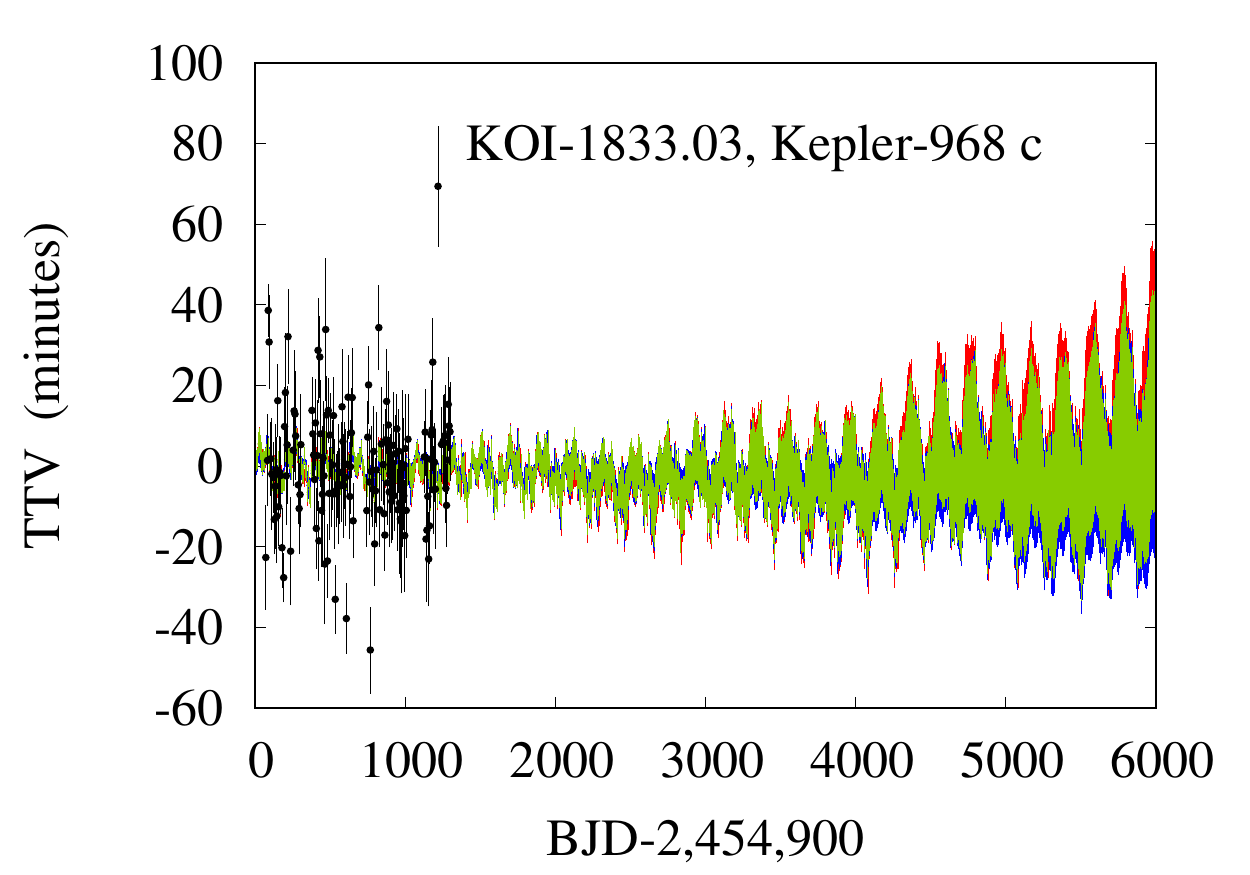}
\includegraphics[height = 1.45 in]{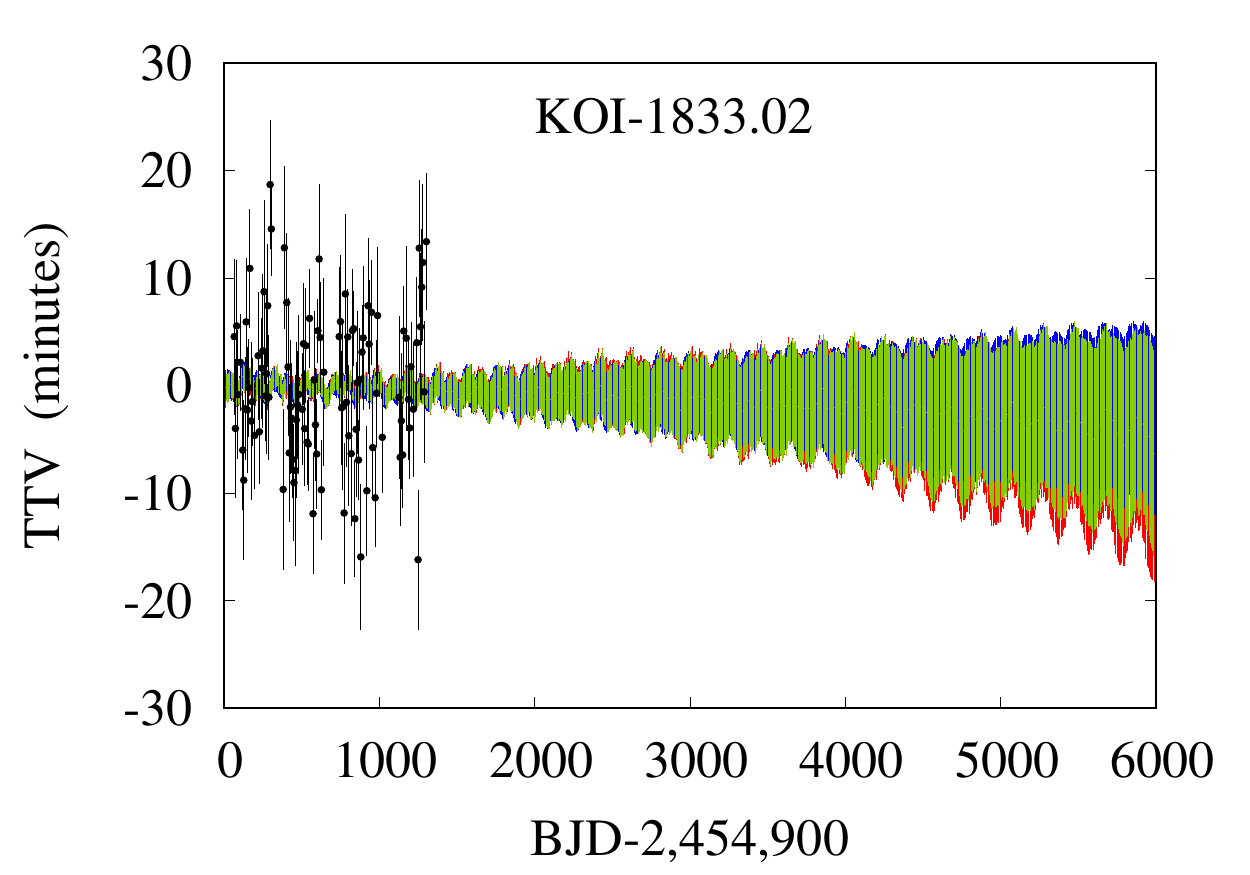}
\caption{Distribution of projected transit times as a function of time for the planet labelled in each panel (part 7). Black points mark transit times in the catalog of \citet{rowe15a} with 1$\sigma$ error bars. In green are 68.3\% confidence intervals of simulated transit times from posterior sampling. In blue (red), are a subset of samples with dynamical masses below (above) the 15.9th (84.1th) percentile.
\label{fig:KOI-1576fut}} 
\end{center}
\end{figure}

In some cases with strongly detected TTVs, projected TTVs diverge exponentially, most likely due to poorly constrained parameters within the TTV model (e.g. KOI-314 (Kepler-138) in Figure~\ref{fig:KOI-168fut}). We also observe some projected TTVs with pulsating uncertainties, like KOI-1599 (Kepler-1659) in Figure~\ref{fig:KOI-1576fut}, where the uncertainties increase overall with some modulation at the superperiod. 

\begin{figure}
\begin{center}
\figurenum{10}
\includegraphics [height = 1.45 in]{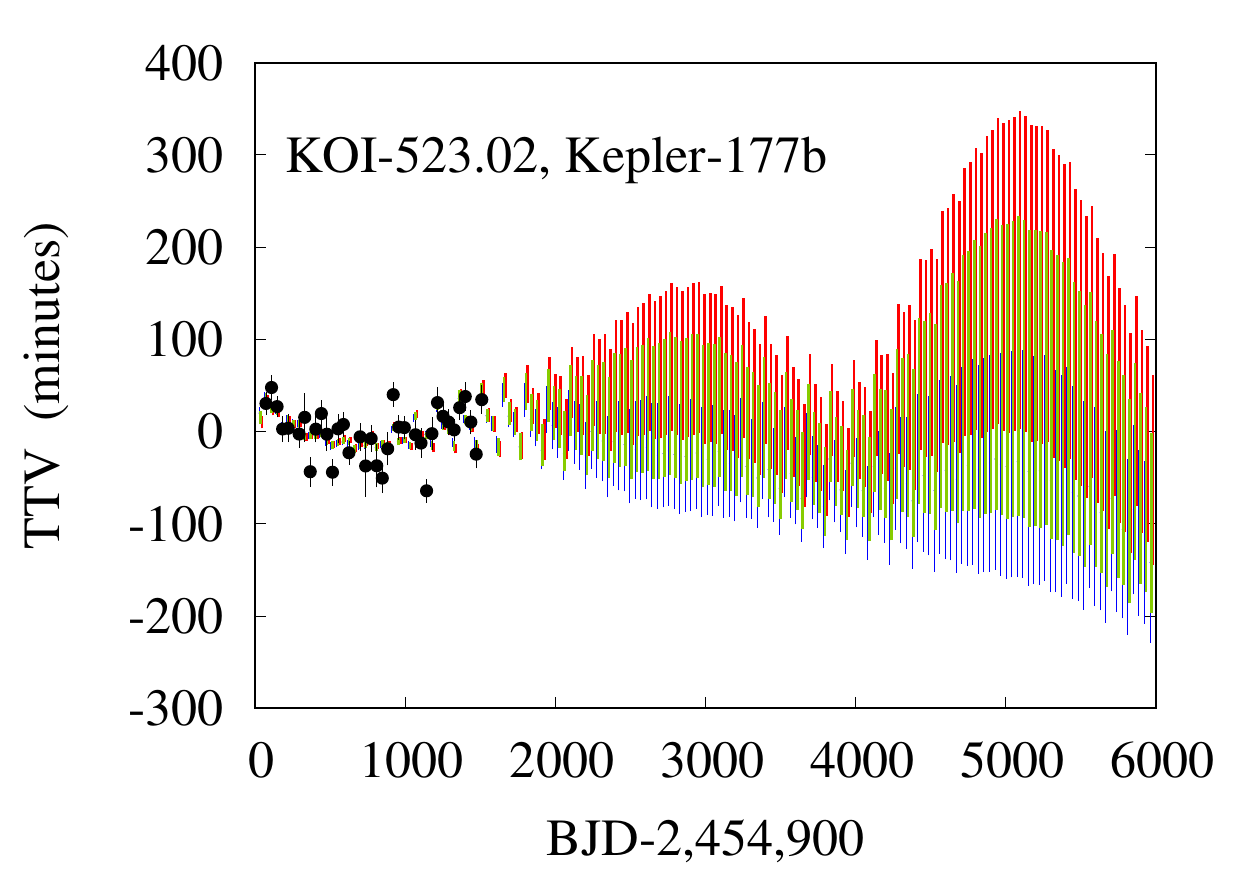}
\includegraphics [height = 1.45 in]{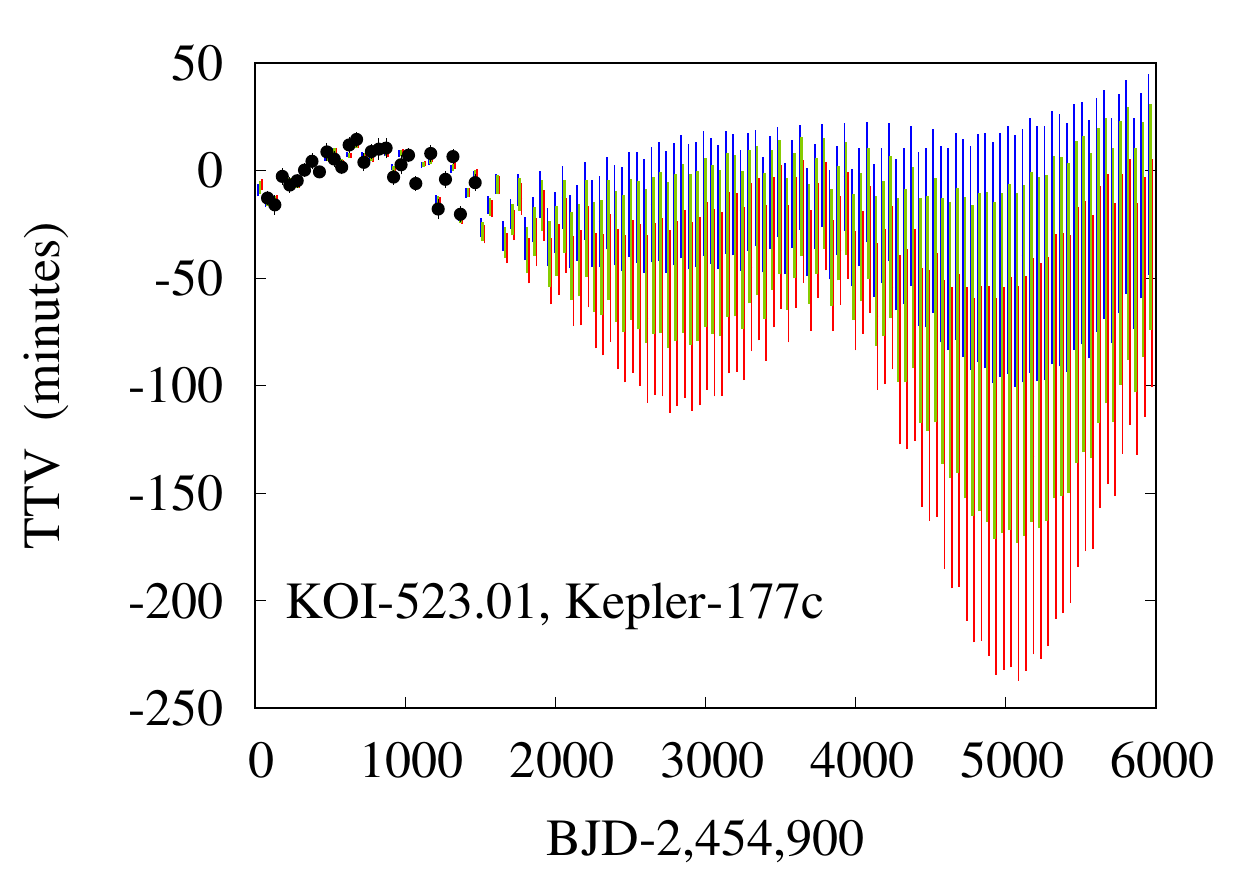} \\
\includegraphics[height = 1.45 in]{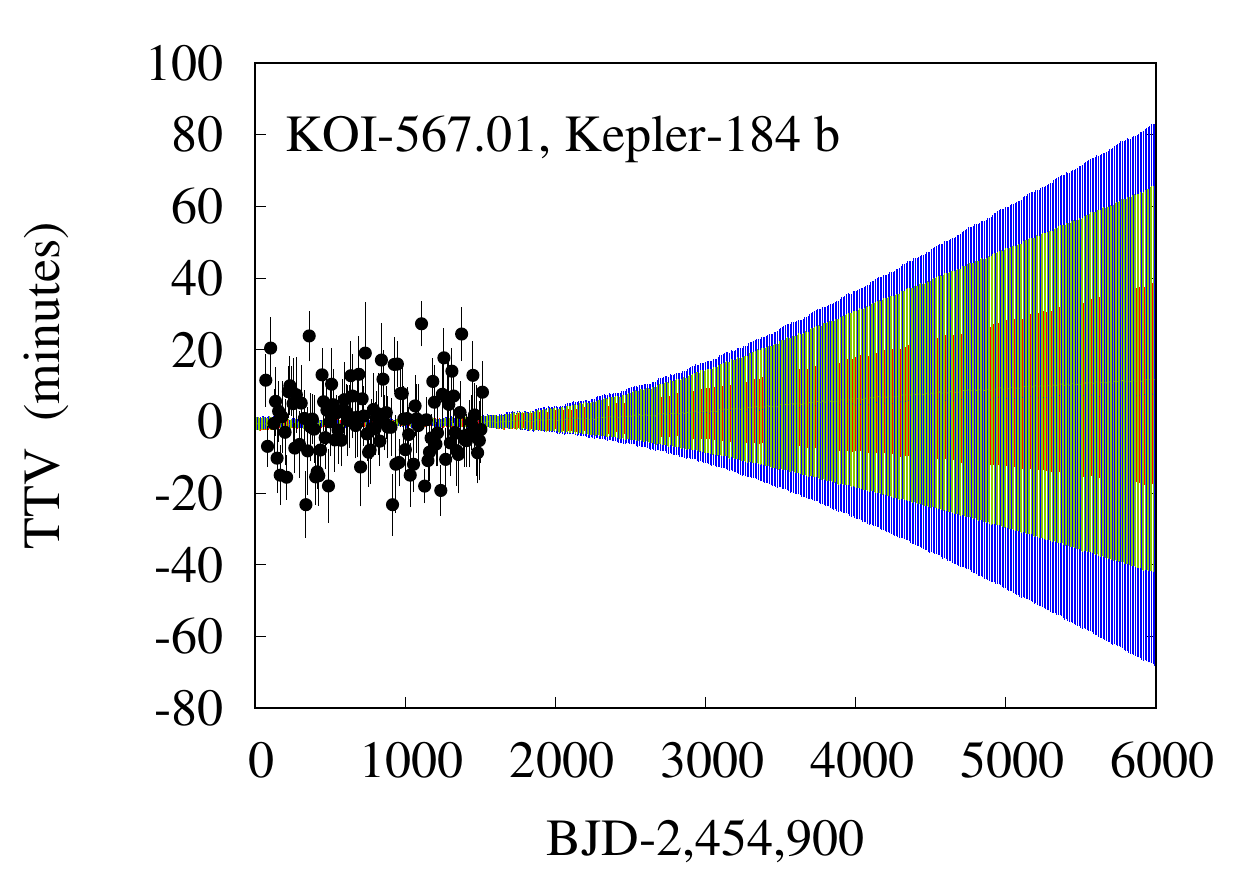}
\includegraphics[height = 1.45 in]{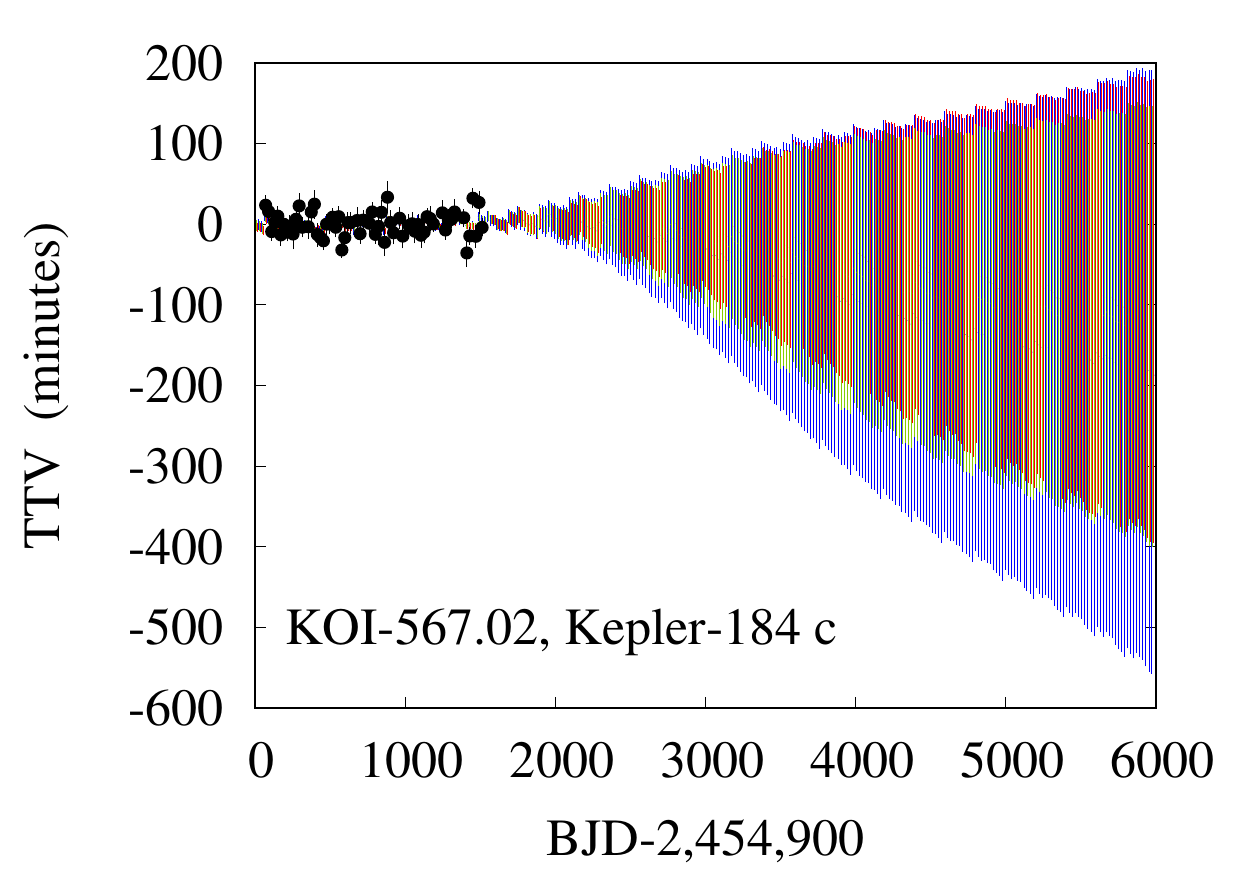}
\includegraphics[height = 1.45 in]{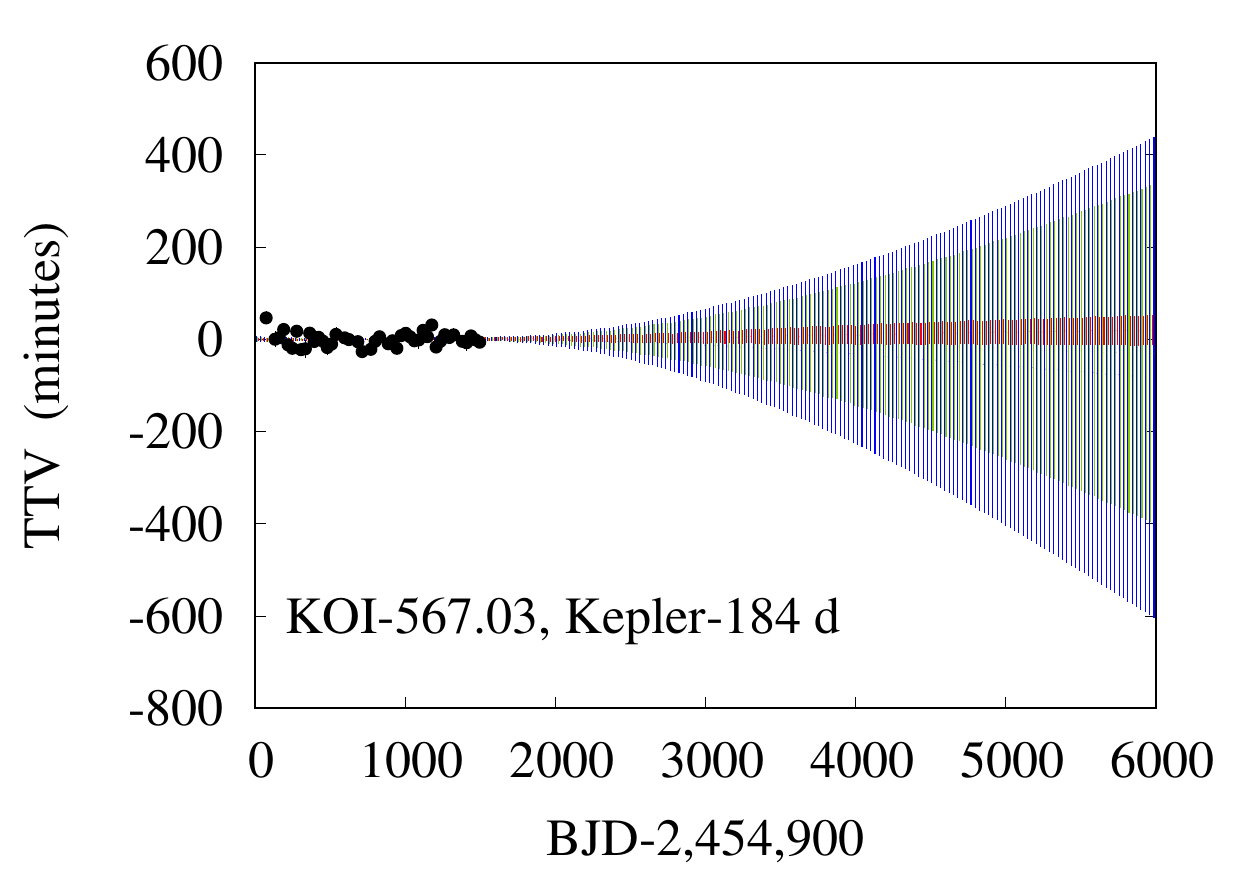}
\includegraphics[height = 1.45 in]{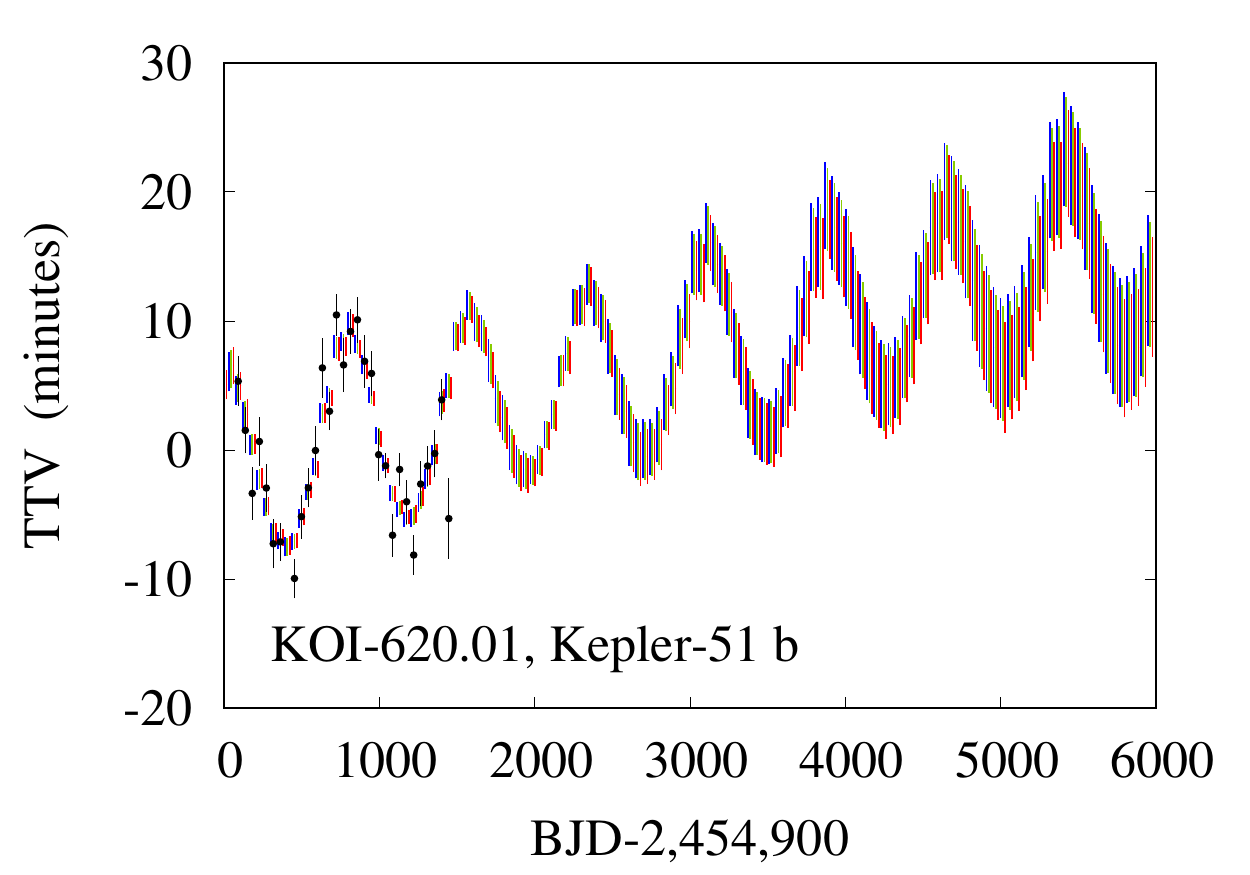}
\includegraphics[height = 1.45 in]{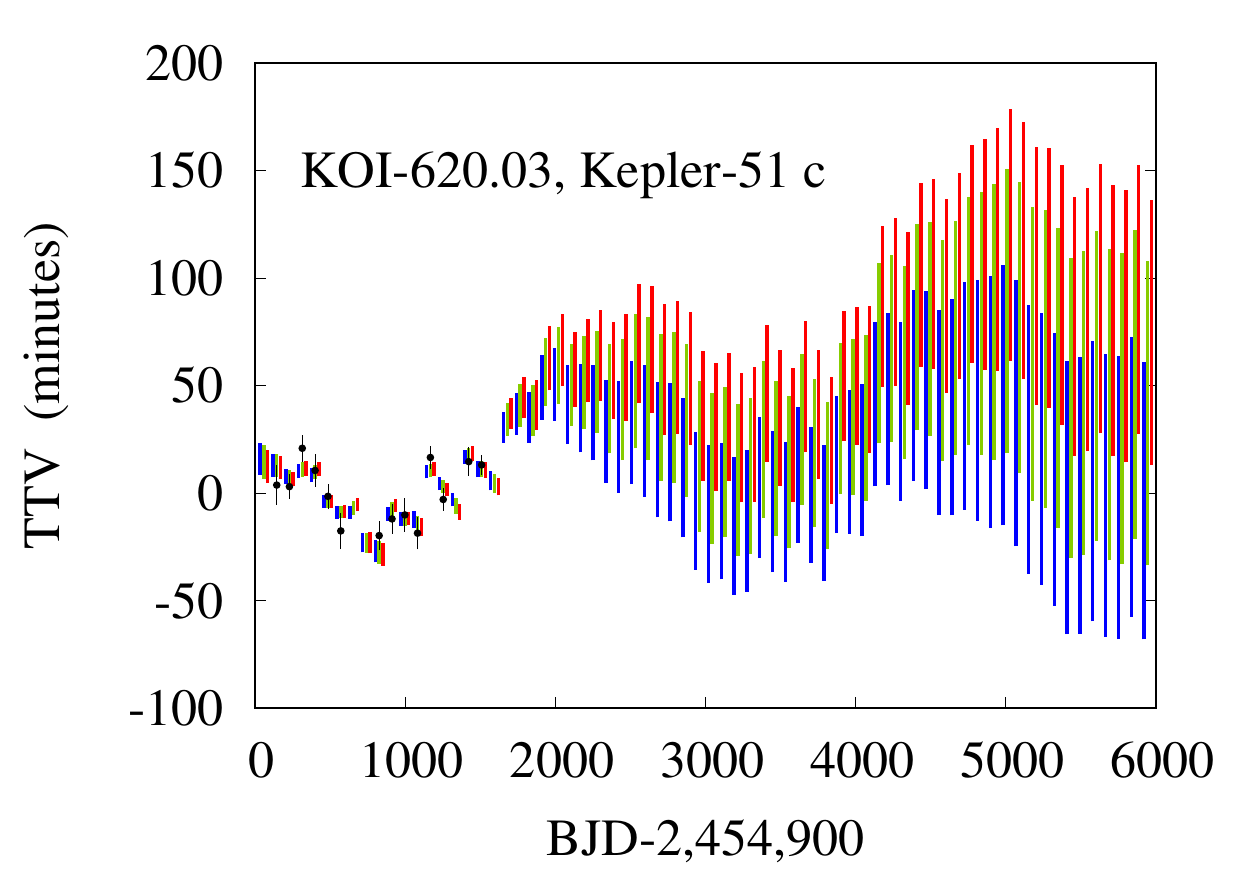}
\includegraphics[height = 1.45 in]{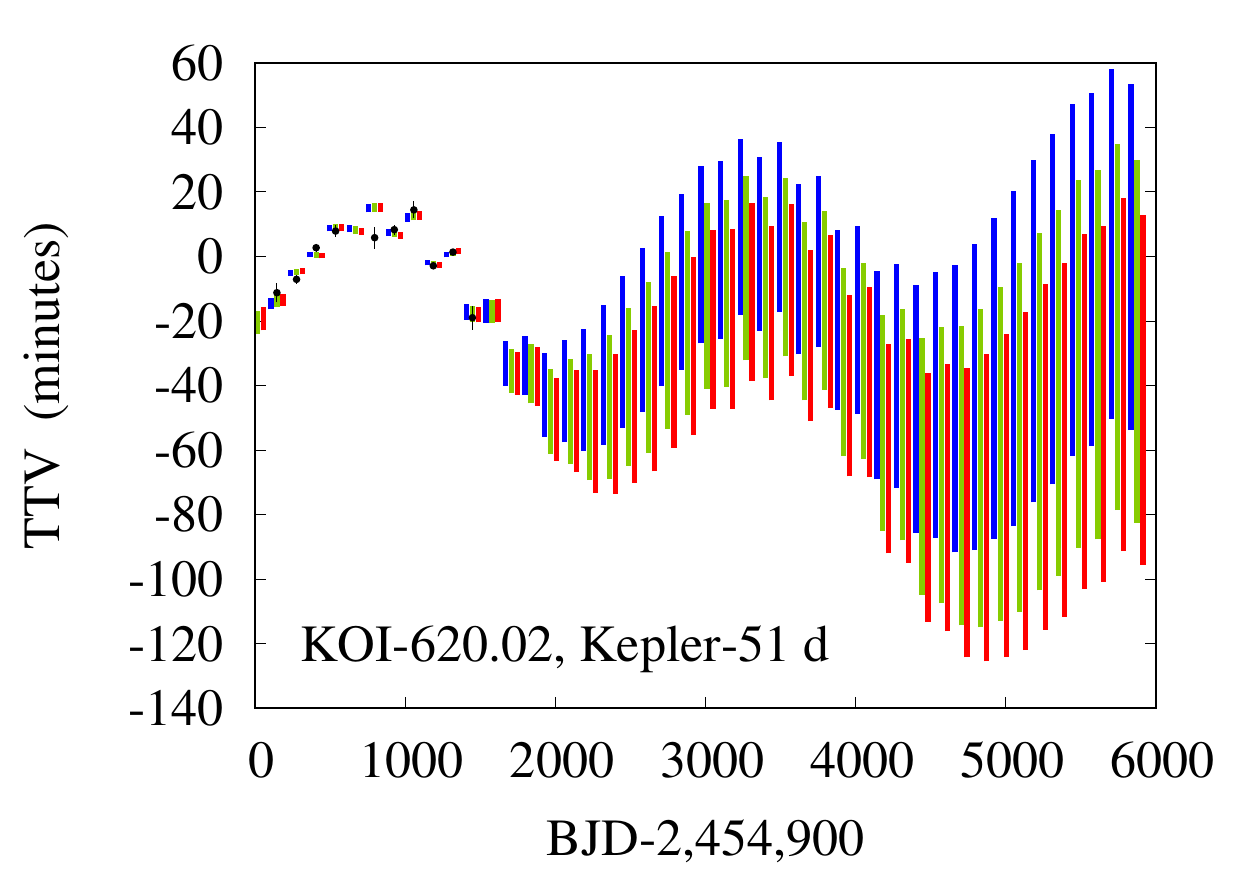}
\includegraphics[height = 1.45 in]{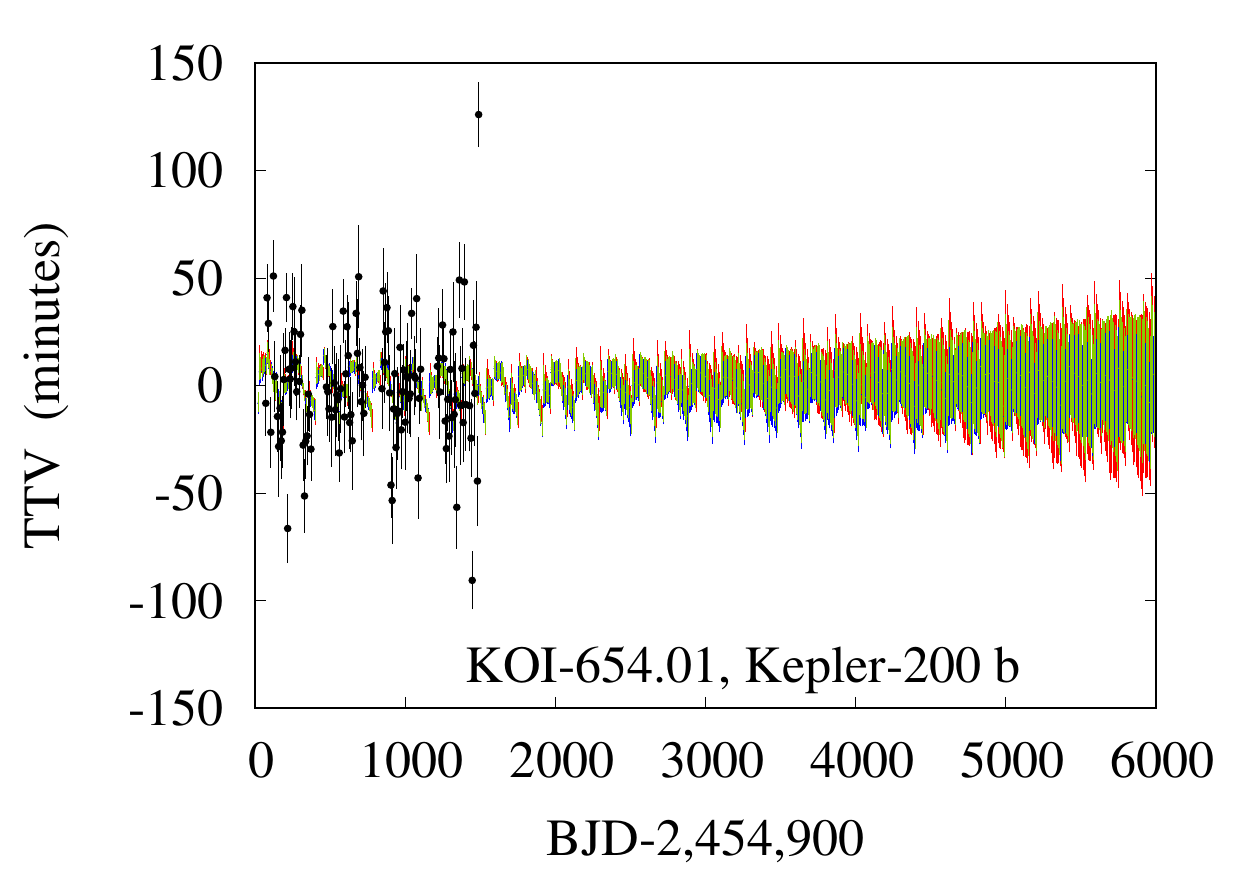}
\includegraphics[height = 1.45 in]{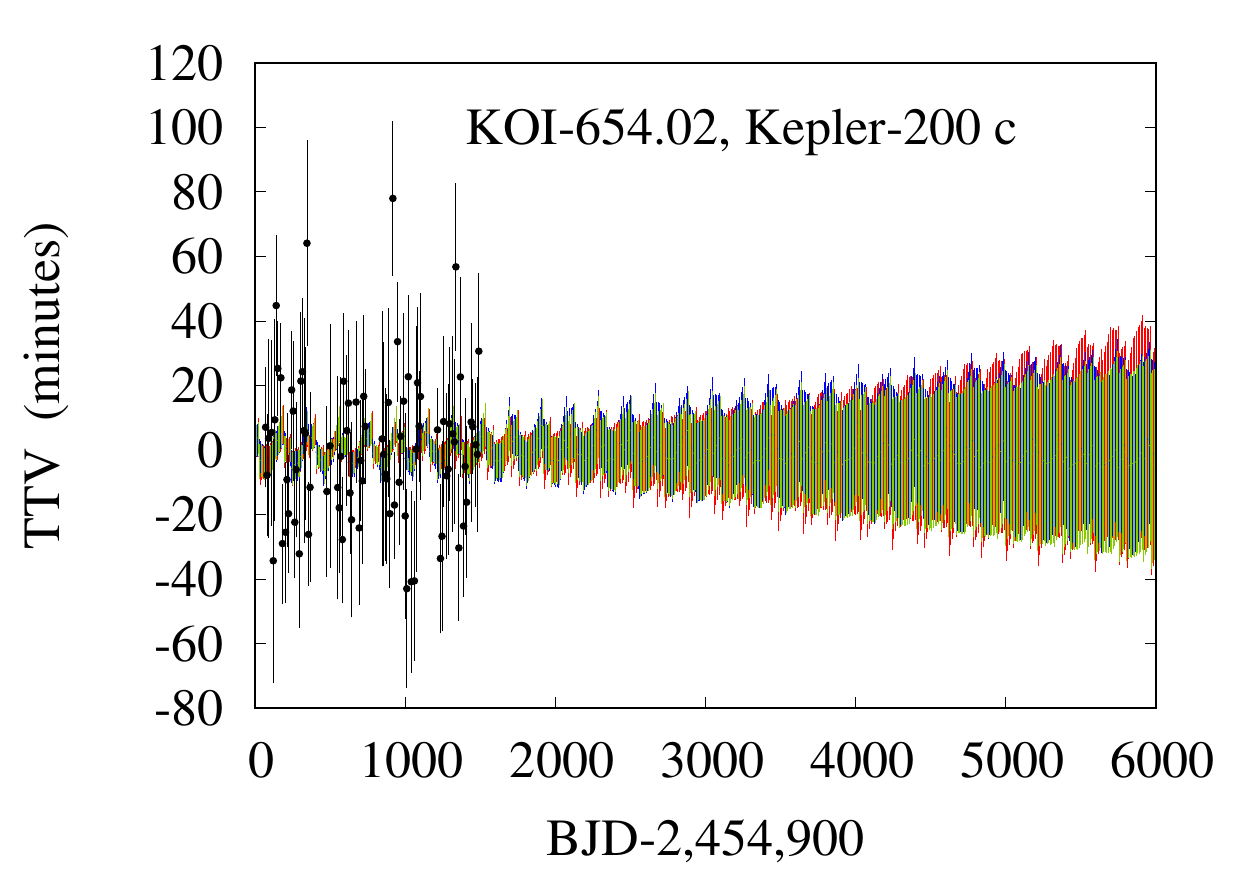} \\
\includegraphics[height = 1.05 in]{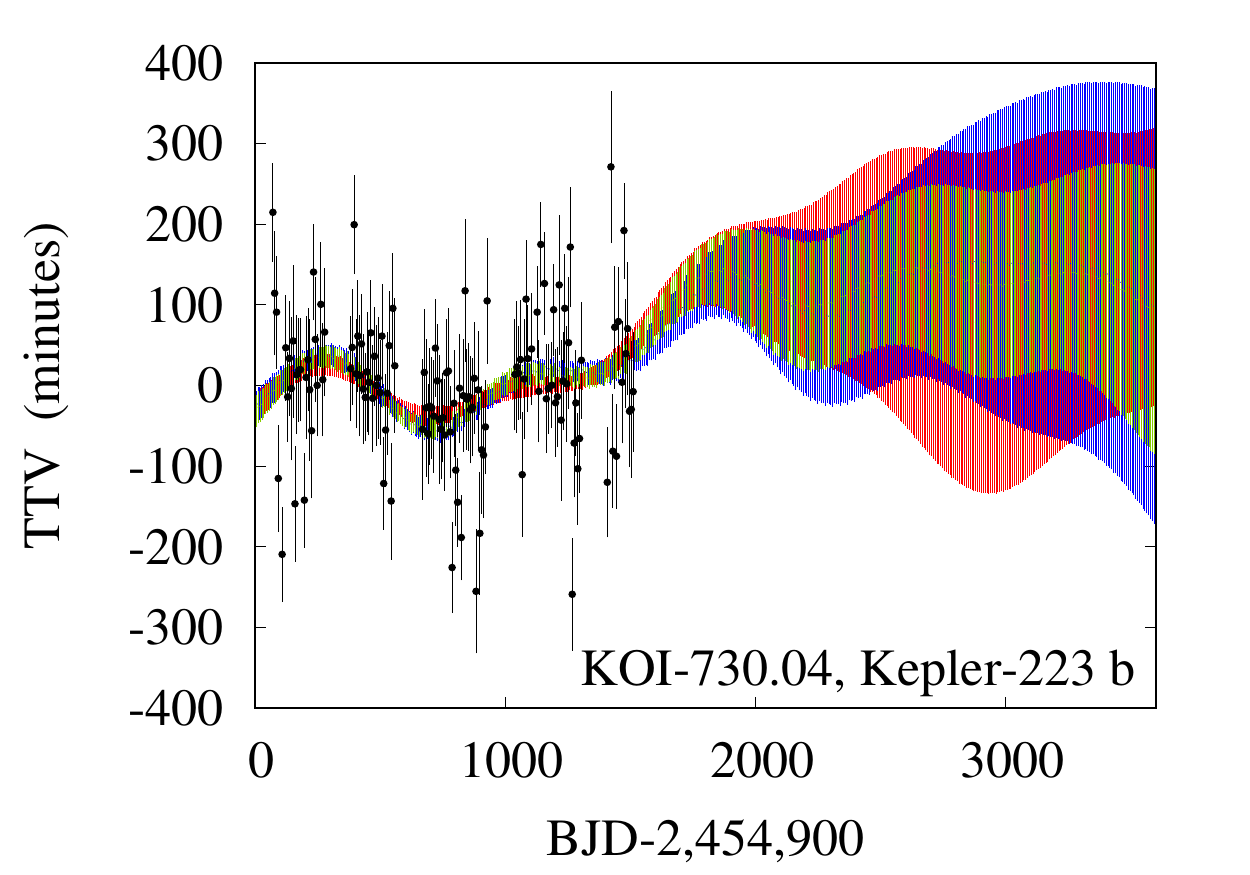}
\includegraphics[height = 1.05 in]{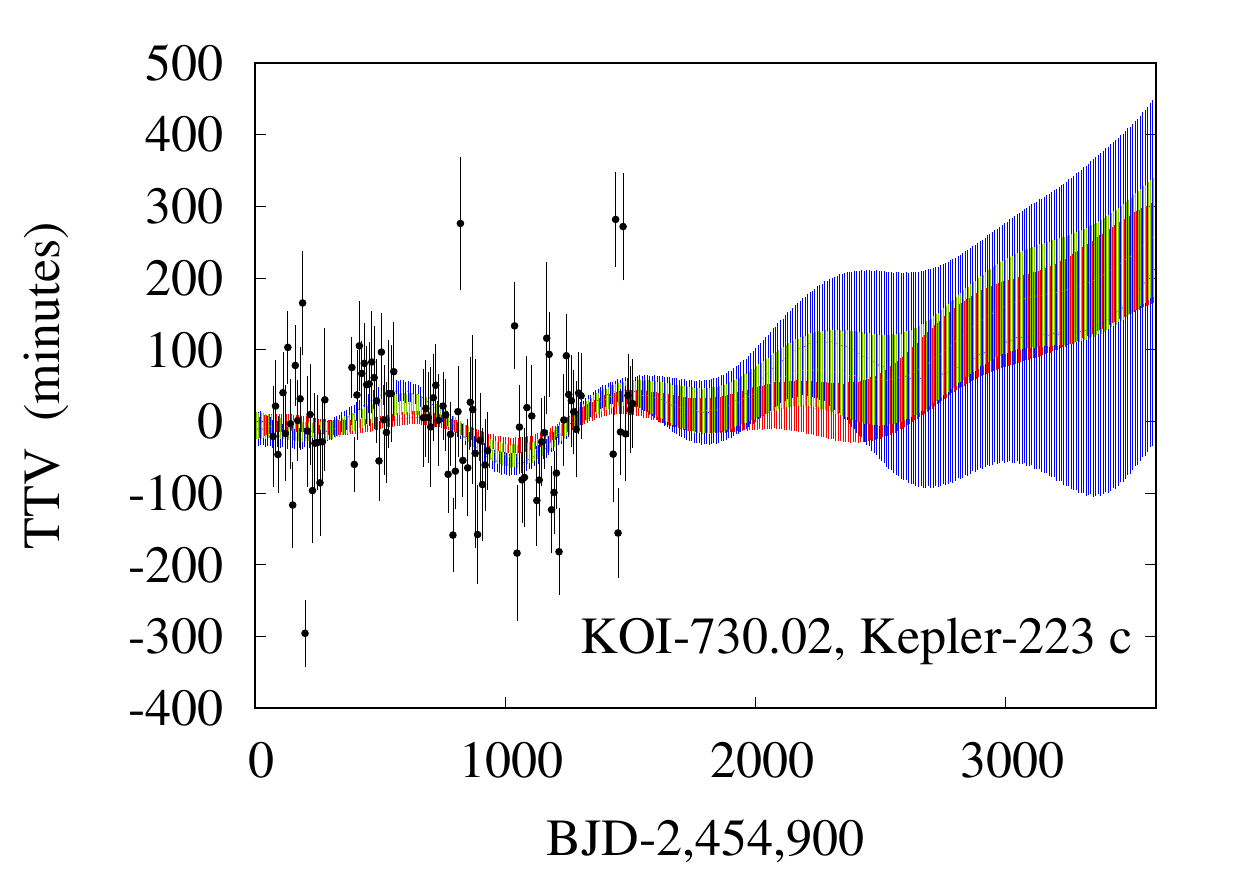}
\includegraphics[height = 1.05 in]{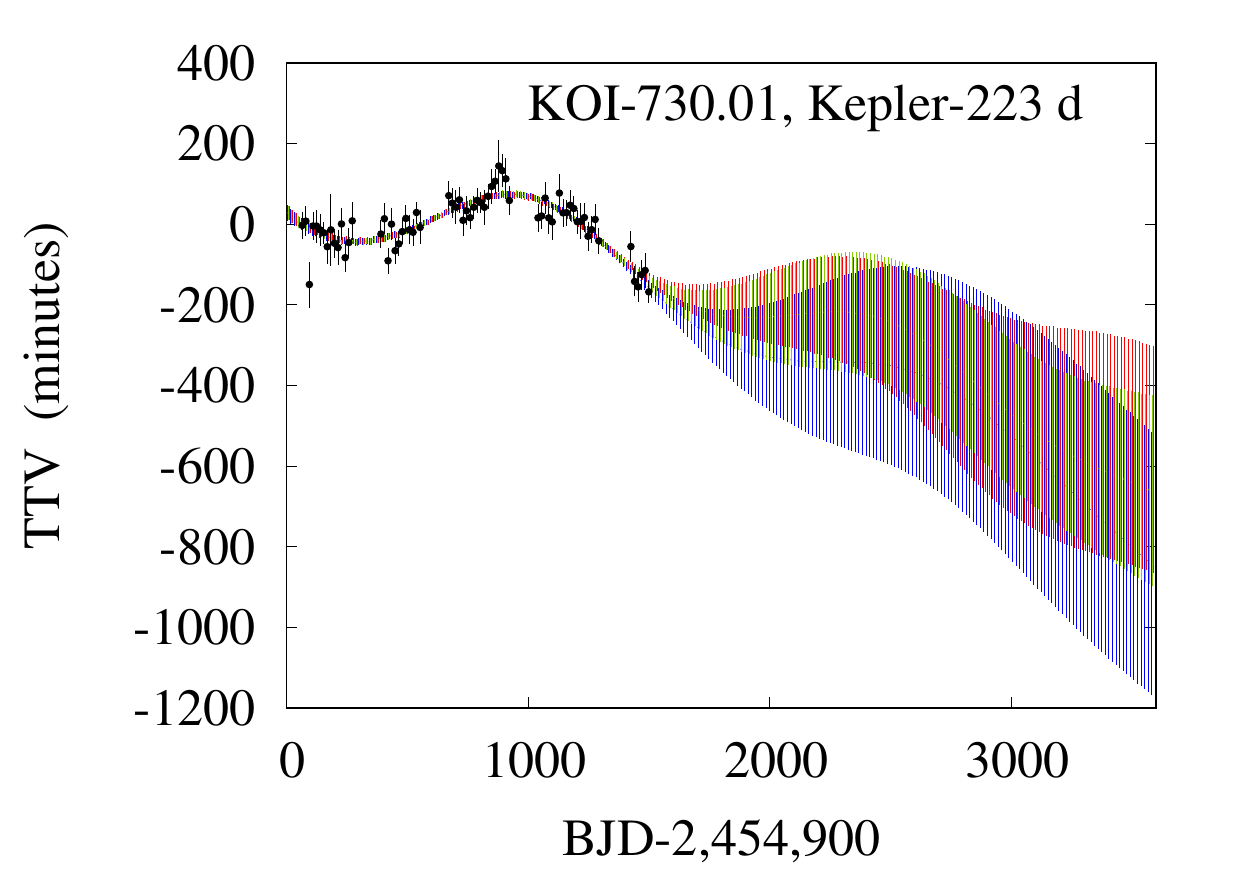}
\includegraphics[height = 1.05 in]{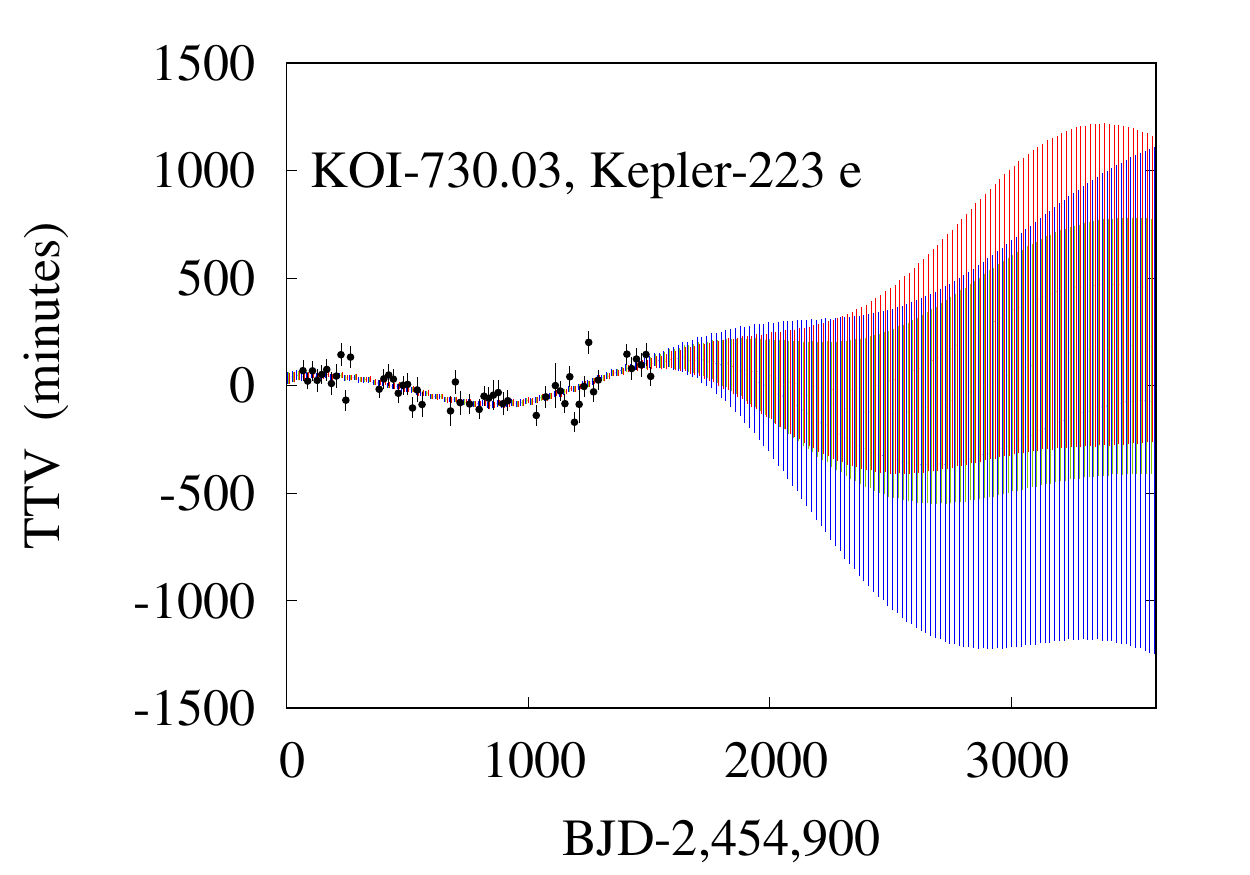}
\includegraphics[height = 1.45 in]{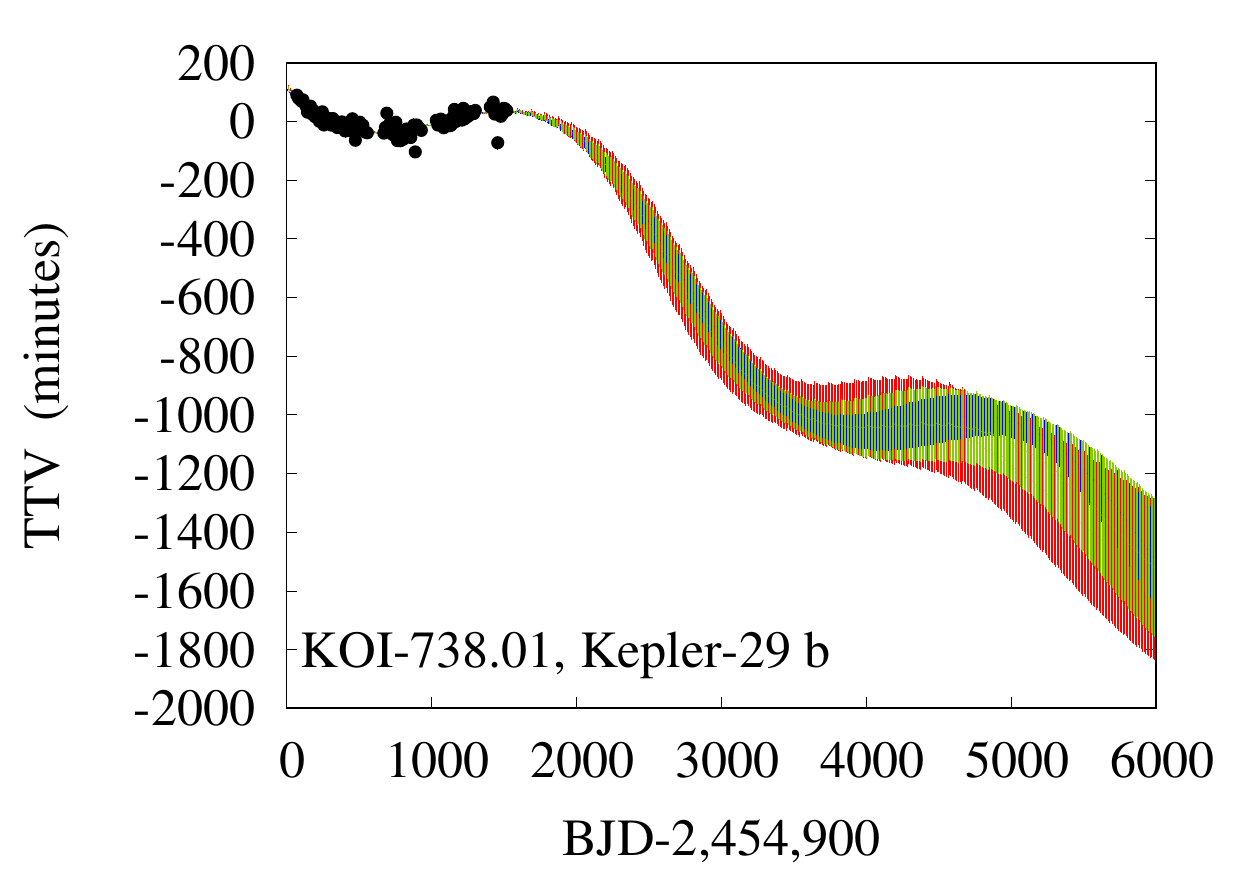}
\includegraphics[height = 1.45 in]{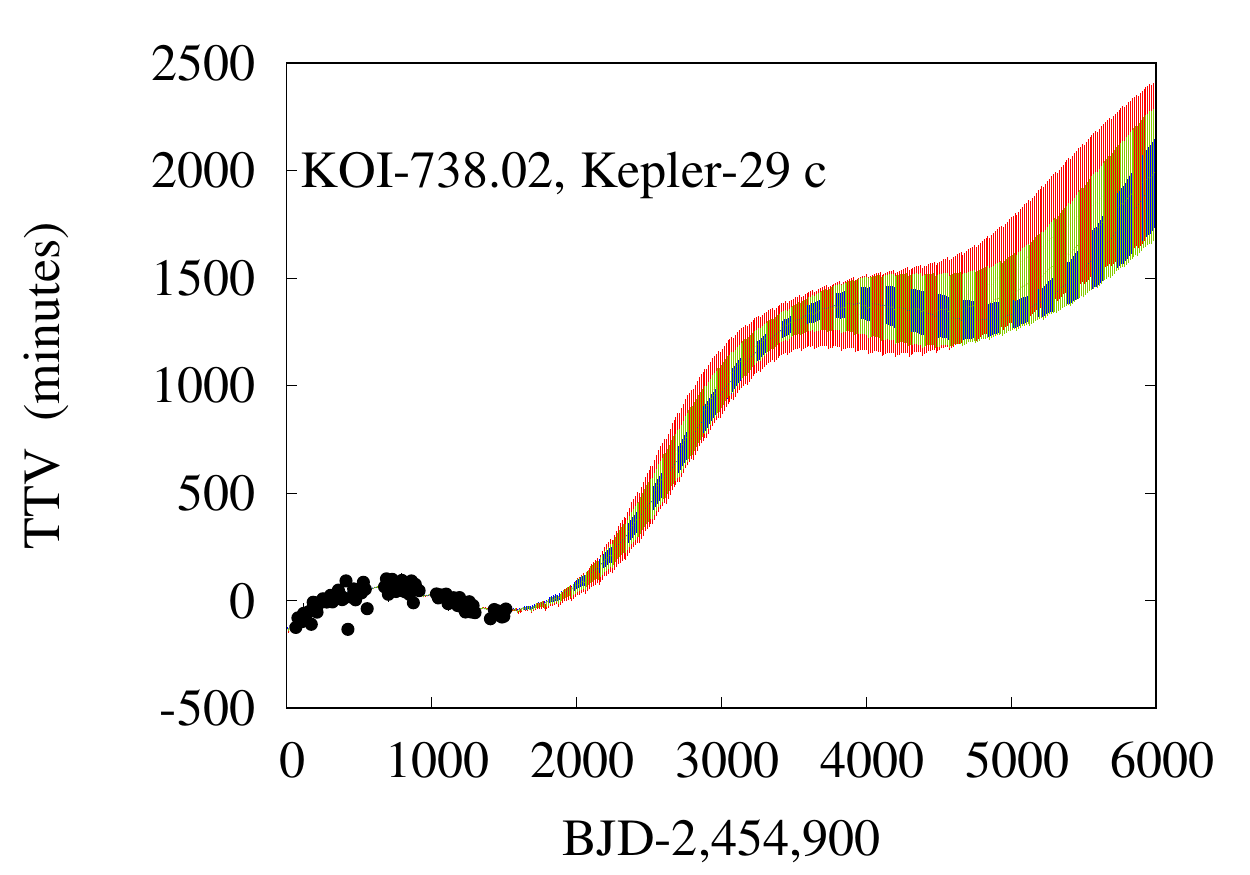}
\caption{
Distribution of projected transit times as a function of time for the planet labelled in each panel (part 4). Black points mark transit times in the catalog of \citet{rowe15a} with 1$\sigma$ error bars. In green are 68.3\% confidence intervals of simulated transit times from posterior sampling. In blue (red), are a subset of samples with dynamical masses below (above) the 15.9th (84.1th) percentile.
\label{fig:KOI-523fut}} 
\end{center}
\end{figure}

There are some common aspects to the systems that we judge to have rapidly diverging TTVs. Several candidates with an expectation of detectable resonant and nonresonant TTVs have slowly diverging TTVs, presumably because the data provides strong constraints on the TTV periodicities (e.g. KOI-115.01, KOI-152.01 in Figure~\ref{fig:KOI-85fut}). On the other hand, systems with longer TTV superperiods tend to have more rapidly diverging TTVs. In some cases (e.g. KOI-620.03 (Kepler-51 c) in Figure~\ref{fig:KOI-523fut}), this is likely due to the resonant TTV superperiod being comparable to or exceeding the Kepler baseline. In such cases, even a single future transit-timing measurement may place valuable constraints on the phase or amplitude of the TTVs. In other cases, even if the superperiod is well sampled by the Kepler baseline, the data may be noisy enough to leave some uncertainty on the amplitude or phase of the TTVs (e.g., KOI-1576, Kepler-307 in Figure~\ref{fig:KOI-1576fut}). Otherwise, most of the systems with an expectation of nearresonant TTVs where the superperiod is much shorter than the Kepler baseline tend to have slowly diverging projected TTVs (e.g., KOI-137 (Kepler-18), KOI-156 (Kepler-114) in Figure~\ref{fig:KOI-85fut}). The three-planet system KOI-314 (Kepler-138) provides an illustrative example; the coherence time of the inner pair is longer than the Kepler baseline and the projected TTVs of the innermost planet diverge rapidly, while the outer and more strongly interacting pair has a TTV period shorter than the Kepler baseline, and for both planets the projected TTVs diverge slowly (see Figure~\ref{fig:KOI-168fut}). 

\begin{figure}[ht!]
\begin{center}
\includegraphics [height = 2.3 in]{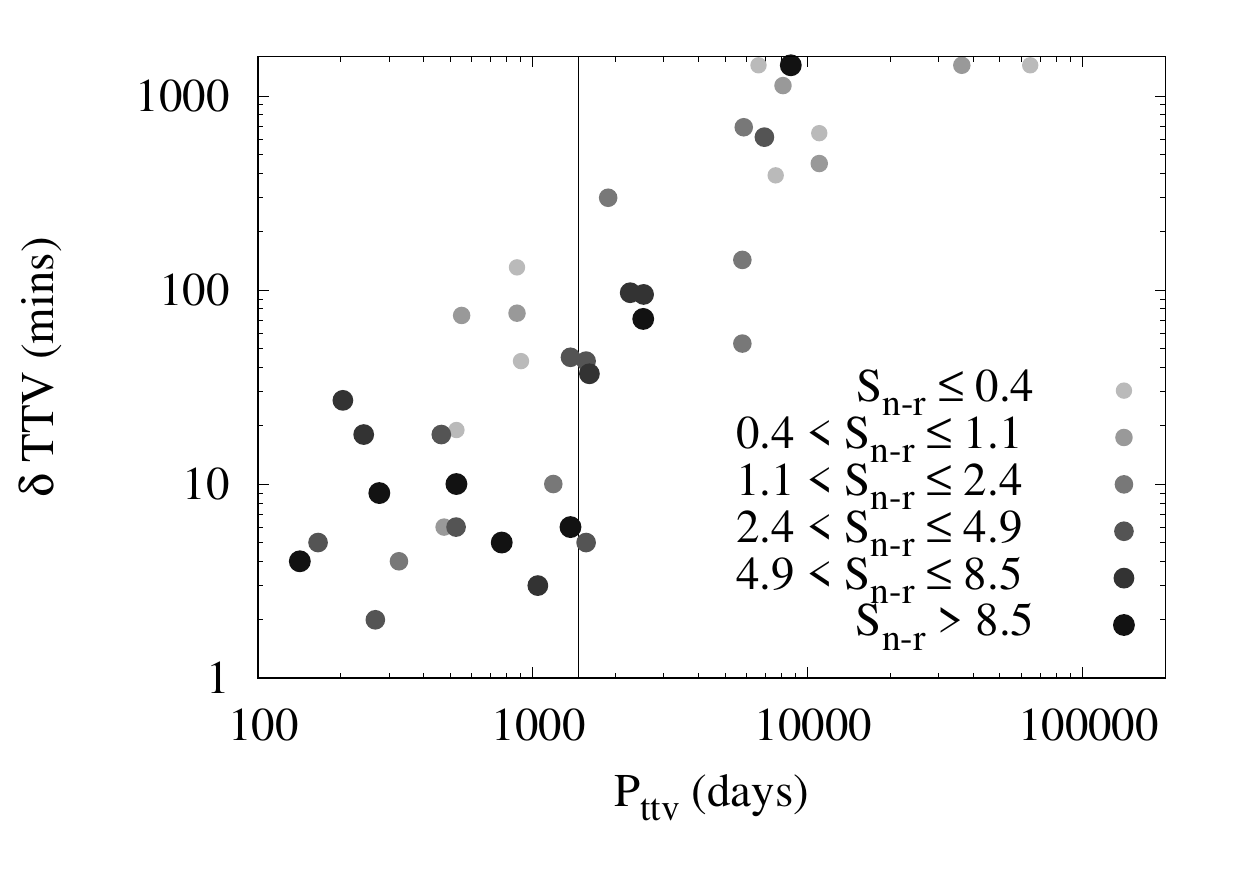}
\caption{Uncertainty of projected TTVs after 6000 days (in minutes) for near-resonance candidates with superperiods marked on the horizontal axis. The black line marks the duration of the Kepler mission. Candidates whose projected timing posteriors diverge by more than 1 day (1440 minutes) are plotted at 1440 minutes. The point sizes and grey scale mark the expected nonresonant TTV scores (S$_{n-r}$) from Tables~\ref{tbl-SystemSelection1}--\ref{tbl-SystemSelection2}, with the lowest scores in light grey, and highest scores in dark grey. We see that the divergence in projected transit times is strongly correlated with the near-resonant TTV superperiod, and weakly anticorrelated with the expectation of detectable nonresonant TTV.
\label{fig:EvalSample} }
\end{center}
\end{figure}

We consider two possible effects that determine how rapidly the projected transit times diverge. Firstly, we would expect that periodicities in the TTVs that are longer than the Kepler baseline have poorly constrained amplitudes and hence that such planets have rapidly diverging projected transit times. An additional factor could be how information-rich the data are. Where both near-resonant and nonresonant frequencies are detected in the TTV, the TTV model may be more constrained and the projected times more certain. Figure~\ref{fig:EvalSample} compares the divergence of projected TTVs with the near-resonant superperiods of relevant candidates. Note that as the orbital period ratio approaches mean motion resonance, the near-resonant superperiod increases to infinity. Such pairs have a finite TTV periodicity due to libration, which we neglect in Figure~\ref{fig:EvalSample}. The figure highlights the correlation between the divergence of the projected TTVs and the superperiod--- typically reaching hours if the superperiod exceeds the Kepler baseline of $\sim$1460 days. We also note a correlation between the expectation of nonresonant TTVs (using the scores in Tables~\ref{tbl-SystemSelection1}--\ref{tbl-SystemSelection2}) and the divergence of the projected transit times, such that planets with a lower expected signal for the nonresonant TTVs (among planets that also have an expectation of near-resonant TTVs) typically have more rapidly diverging TTVs. Hence, two factors become relevant in observing campaigns to follow up on TTV targets: the baseline of observations compared to the superperiod, and the prospect of detecting nonresonant frequencies in the TTVs. We calculated the Pearson correlation coefficient of the logarithms of TTV divergence and the superperiod to be 0.99. By comparison, the correlation coefficient between the logarithm of the TTV divergence and the nonresonant TTV expectation score among planets near first order resonances with a neighbor was -0.23, a much weaker correlation. 

\subsection{Post-Kepler Transit Timing Observations}
Some transiting planets on our list have had follow-up transit-timing observations since the Kepler mission ended. 
We compare our predicted transit times, with their 68.3\% confidence intervals, to the observed transit times in Table~\ref{tbl-comparepredictions} below. In all cases, our predictions agree closely with follow-up observations, hence we find no evidence that our multiplanet TTV models given the known planets are inadequate to explain the follow-up data.

 \begin{table}[!h]
  \begin{center}
  \small
    \begin{tabular}{|c|c|c|c|c|}
      \hline
      KOI &  Prediction &  Observation &  $D$ &  Reference \\
\hline
152.01 (Kepler-79 d) &  3321.3898 $\pm$0.0029 & 3321.3863$^{+0.0009}_{-0.0009}$ & 1.15  &  \citet{Chachan2020} \\
152.01  (Kepler-79 d) & 3529.7389$\pm$0.0025 & 3529.7253$^{+0.0074}_{-0.0066}$ & 1.74  &  \citet{Chachan2020} \\
277.01  (Kepler-36 c) & 3123.9116$\pm$0.0349 & 3123.8991 $^{+ 0.0082}_{- 0.0033}$ & 0.34  &  \citet{Vissapragada2020} \\
523.01   (Kepler-177 c)  & 3342.9484 $\pm$ 0.0240 & 3342.9695$^{+0.0060 }_{- 0.0049}$ & 0.86  &  \citet{Vissapragada2020} \\
620.01 (Kepler-51 b) & 2395.0343 $\pm$  0.0012 & 2395.0319 $^{+ 0.0017}_{- 0.0029 }$ & 1.15  &   \citet{LibbyRoberts2020} \\
620.01 (Kepler-51 b) & 2665.9580  $\pm$ 0.0015 & 2665.9525  $\pm$ 0.0040 & 1.29 &   \citet{LibbyRoberts2020} \\
620.02 (Kepler-51 d) & 2488.1946 $\pm$ 0.0170 & 2488.2006 $^{+0.0010}_{- 0.0014}$ & 0.35  &   \citet{LibbyRoberts2020} \\
620.02 (Kepler-51 d) & 2878.7397 $\pm$ 0.0198 & 2878.7534 $\pm$ 0.0014 & 0.69 &   \citet{LibbyRoberts2020} \\
738.01 (Kepler-29 b) & 3090.8608 $\pm$ 0.0506&  3090.8421$^{+ 0.0116}_{-0.0019}$ & 0.36  &  \citet{Vissapragada2020} \\
1783.01 (Kepler-1662 b) & 3329.7882 $\pm$ 0.0332 & 3329.8082$^{+ 0.0069}_{- 0.0076}$ & 0.59  &  \citet{Vissapragada2020} \\
\hline
     \end{tabular}
    \caption{Predicted (Pred) transit times from this study and observed (Obs) transit times (BJD-2454900) for post-Kepler transit observations. The fourth column provides the goodness-of-fit statistic, $D =|Pred-Obs|/\sqrt{\sigma_{pred}^2+\sigma_{obs}^2}$, that quantifies the deviations between the predicted and observed post-Kepler transit times relative to the width of a Gaussian approximation to the posterior predictive distribution.  
    }\label{tbl-comparepredictions}
  \end{center}
\end{table}
\subsection{Newly Confirmed Kepler Planets}
Several candidate KOIs remain in our sample of systems that have not been named as verified Kepler planets. We tested whether our best-fit models with free dynamical masses are better than best-fit models with the mass fixed at zero for these candidates, and we include the results in Table~\ref{tbl-confirmplanets} below. Since the mass-less models are nested within the parameter space of our standard models, we also include the $F$-statistic for model comparison.

 \begin{table}[!h]
  \begin{center}
  \small
    \begin{tabular}{|c|c|c|c|c|c|c|c|}
      \hline
      KOI & Transit S/N & Num parameters &  Num data & $\chi^2$ (free mass)  & $\chi^2$ (mass = 0) & $\Delta$BIC   & $F$-statistic \\
\hline
115.03 & 8.70 & 15 & 807 & 1187.4 & 1192.8 & -1.3 & 5.4 \\ 
255.02 & 9.40 & 10 & 143 & 191.1 & 195.0 & -1.1 & 3.9 \\
430.02 & 11.10 & 10 & 184  & 267.8 & 268.7 & -4.3 & 0.9 \\
750.02 & 10.2 & 15 & 420 & 502.6 & 505.6 & -3.4 & 3.0 \\
750.03 & 10.10 & 15 & 420 & 502.6 & 507.6 & -1.0 & 5.0 \\  
1338.02 & 14.70 &  15 & 497 &  781.6   & 783.0  & -4.8  & 1.4  \\
1338.03 & 10.20 & 15 & 497 & 781.6 & 789.0 & 1.2 & 7.4   \\
1576.03 & 8.9* & 15 & 283 & 588.5 & 591.7 & -2.4 & 3.2 \\
1831.03 & 11.60  & 20 & 353 & 502.8 & 722.1 & 213.4 & 219.3 \\ 
1831.04 & 17.10 & 20 & 353 & 502.8 & 512.1 & 3.4  & 9.3 \\ 
1833.02 & 28.50 & 15 & 459 & 723.7 & 740.1 & 10.3 & 16.3 \\
2174.01 & 23.40 & 20 & 829 & 973.5 & 988.3 & 8.1 & 14.8\\
2174.02 & 16.30 & 20 & 829 & 973.5  & 981.2 & 1.0 & 7.7 \\
2174.03 & 10.50 & 20 & 829 & 973.5  & 989.5 & 9.2 & 16.0 \\
2174.04 & 8.10 & 20 & 829 & 973.5  & 983.6 & 3.4 & 10.1 \\
3503.01 & 12.2* & 10 & 100 &  118.1 & 130.9 & 8.2 & 12.8 \\
3503.02 & 9.6* & 10 & 100 &  118.1 & 133.7 & 11.0 & 15.6 \\
\hline
     \end{tabular}
    \caption{Evidence for planet confirmation with TTV. The second column lists the transit S/N  from Kepler DR 25, or quoted here with an asterisk following our light-curve fits with free mid-transit times.. The remaining columns are used to compare best-fit models with free masses (five parameters per candidate for each system) and models with the unconfirmed candidate mass fixed at zero, with the Bayes Information Criterion (BIC) and $F$-statistic for unconfirmed Kepler candidates in our TTV models. The transit S/Ns of KOI-1576.03, KOI-3503.01, and KOI-3503.02 improve measurably when allowing for TTVs.}\label{tbl-confirmplanets}
  \end{center}
\end{table}

We exclude KOI-1574.03 (Kepler-87 from Table~\ref{tbl-confirmplanets}) since it is potentially interacting with a fourth candidate identified in Kepler DR 24 whose transit times are unavailable. We also exclude KOI-1070.03 from confirmation since in our light-curve analysis it has a transit S/N of just 7.2.  

Of the remaining unverified KOIs, we find improvement with nonzero mass where $\Delta$BIC> 3 and where $F$-test > 9  (the $3-\sigma$ level) for KOIs: 1831.03, 1833.02, 2174.01, 2174.03, 2174.04, 3503.01, and 3503.02. 

Of these, KOI-1831.03 and KOI-1831.04 are both high S/N candidates from the light curve. KOI-1831.03 is close to but not in a first-order mean motion resonance with its outer neighbor, with a period ratio of 1.515, and it is strongly detected and hence confirmed in the TTV model. KOI-1831.04, however, is not near resonance with any of the others, and the evidence for it in the TTVs is much weaker. We do not confirm it at this time. 

We confirm the planetary nature of KOI-1833.02, given the improvement in the TTV model we find with a nonzero mass. Its orbital period lies just outside the 4:3 mean motion resonance with its inner neighbor, and TTVs are expected. 

KOI-2174 has four candidates, and the expected TTVs are due to the unusually proximate orbits of the middle pair (KOI-2174.01 and KOI-2174.03), at 6.7 and 7.7 days. KOI-2174.03 has a relatively low transit S/N and may be a false alarm. Furthermore, there is a possibility that these are not false planets but perhaps a blend of two planet-hosting stars with a single light curve. However, both candidates show moderate improvement with nonzero masses over a model with zero-mass models, which provides evidence that the candidates are interacting planets. The inner-most candidate (KOI-2174.04) has a moderate improvement in the TTV model with nonzero mass, but given that there is no expectation of TTV for this candidate and the S/N < 10, a confirmation appears unjustified. The outermost planet, KOI-2174.02 is strongly detected in the light curve, but the improvement it offers the TTV model is too moderate to warrant confirmation. We leave all four candidates unconfirmed. 

\begin{figure}
\begin{center}
\figurenum{15}
\includegraphics [height = 1.45 in]{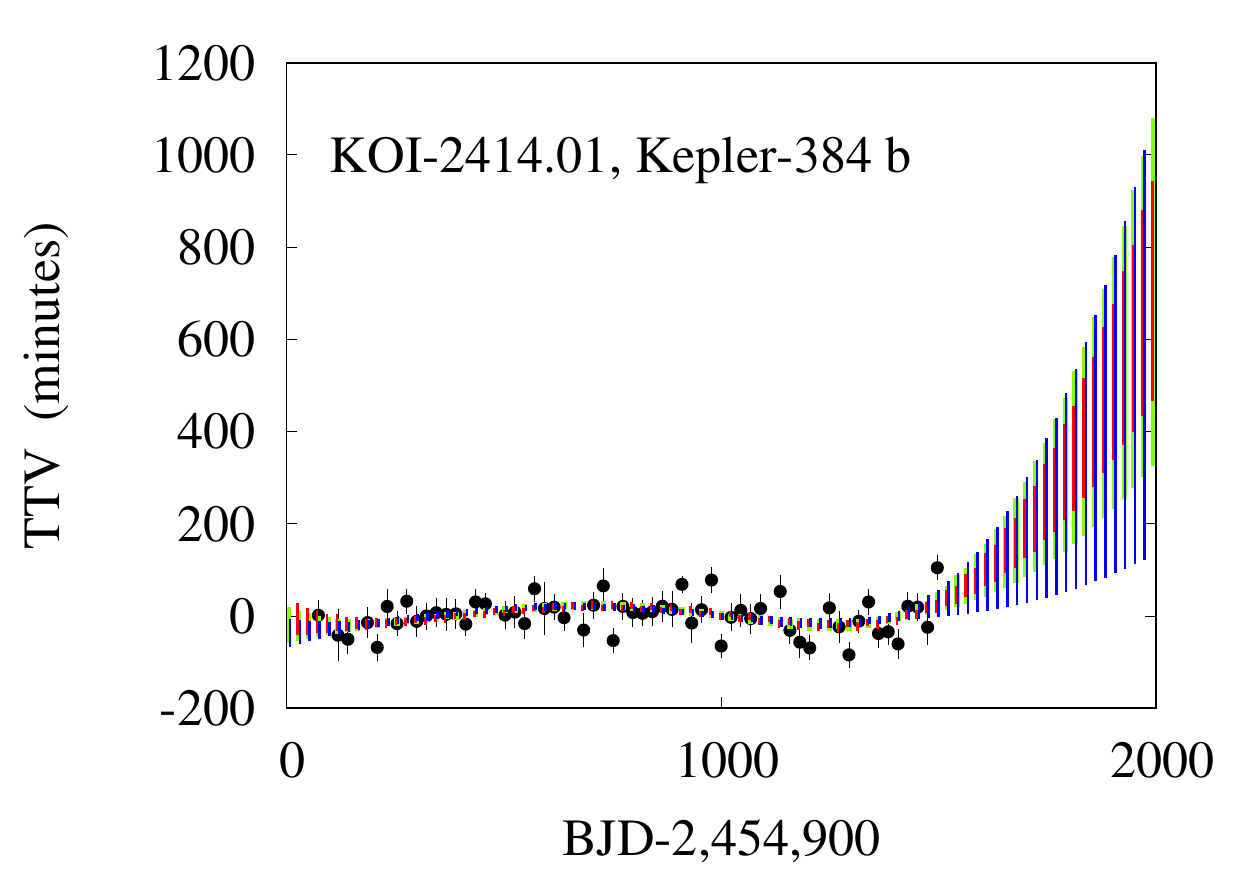}
\includegraphics [height = 1.45 in]{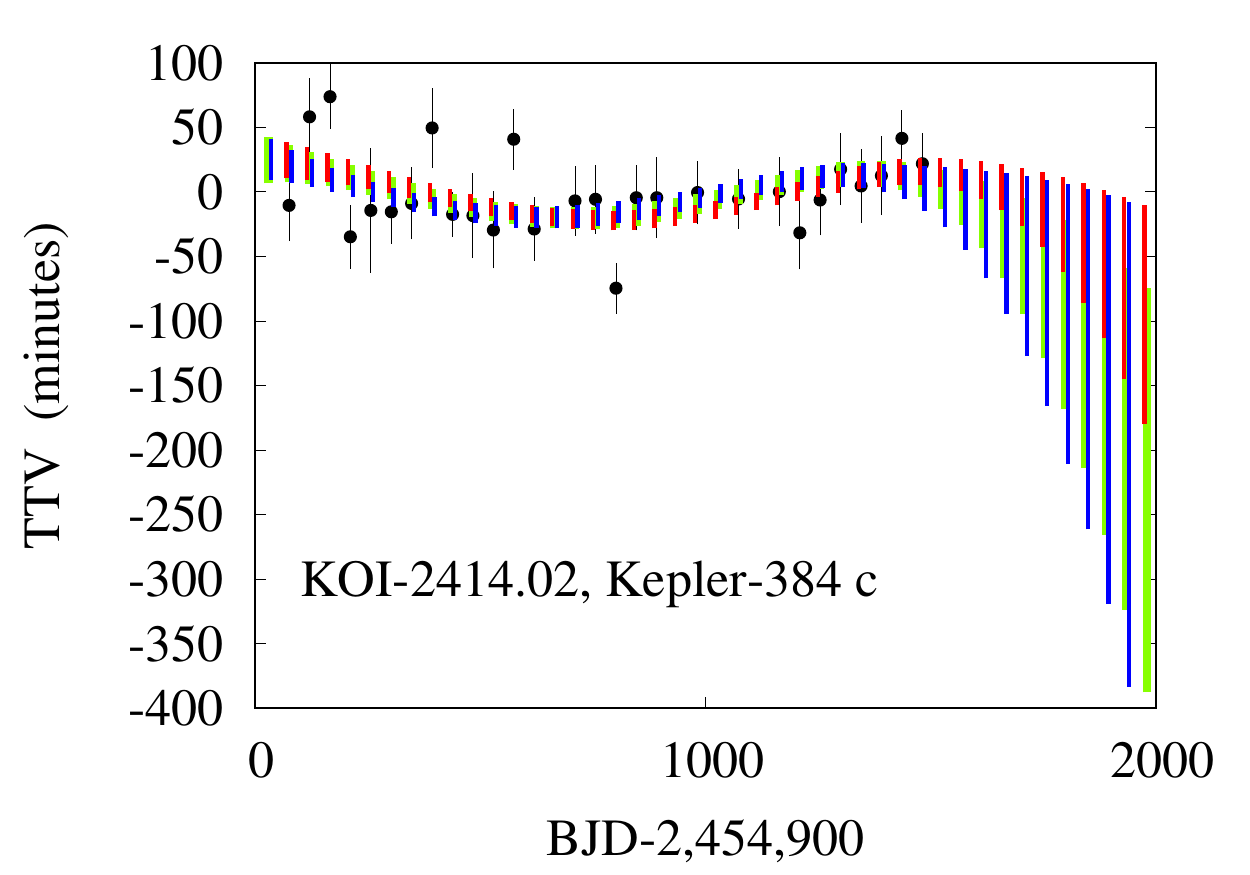} \\
\includegraphics [height = 1.45 in]{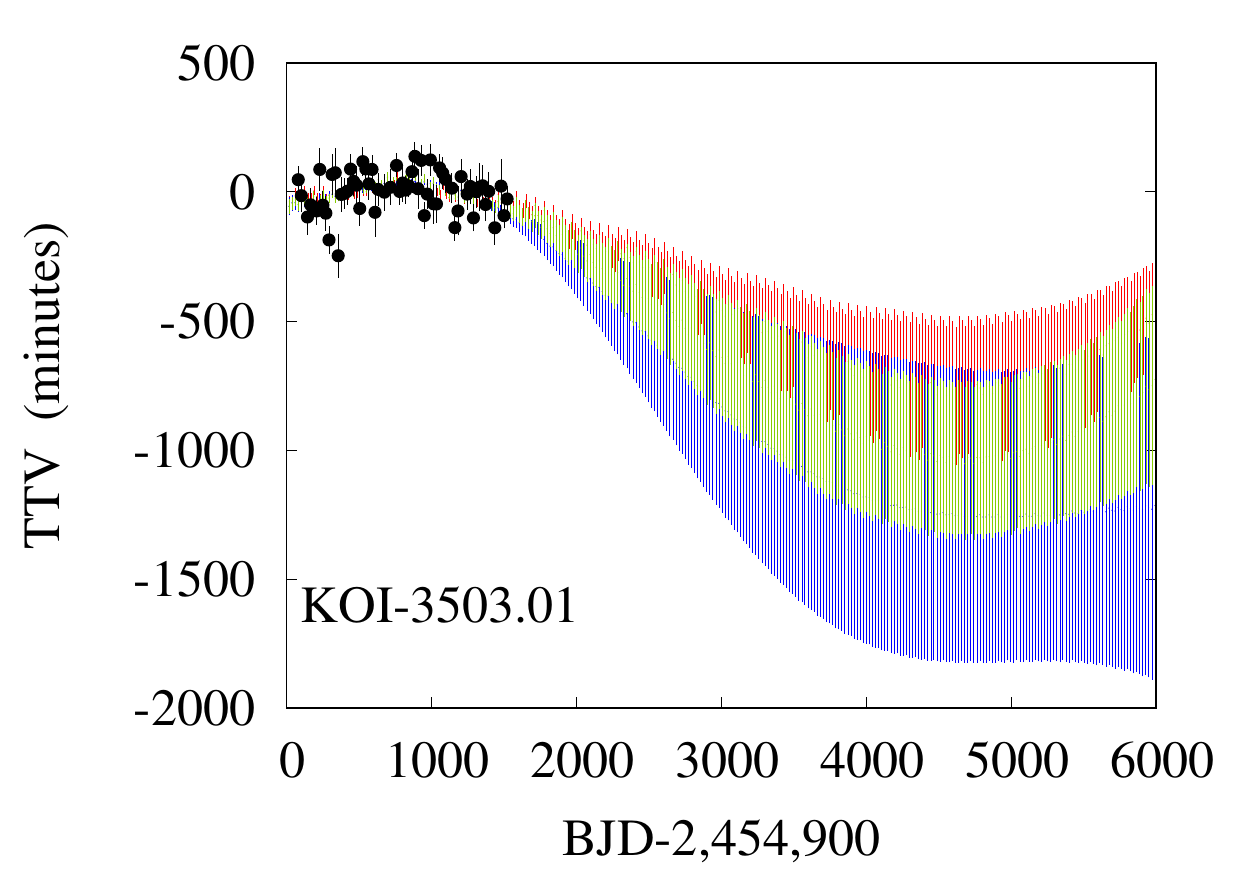}
\includegraphics [height = 1.45 in]{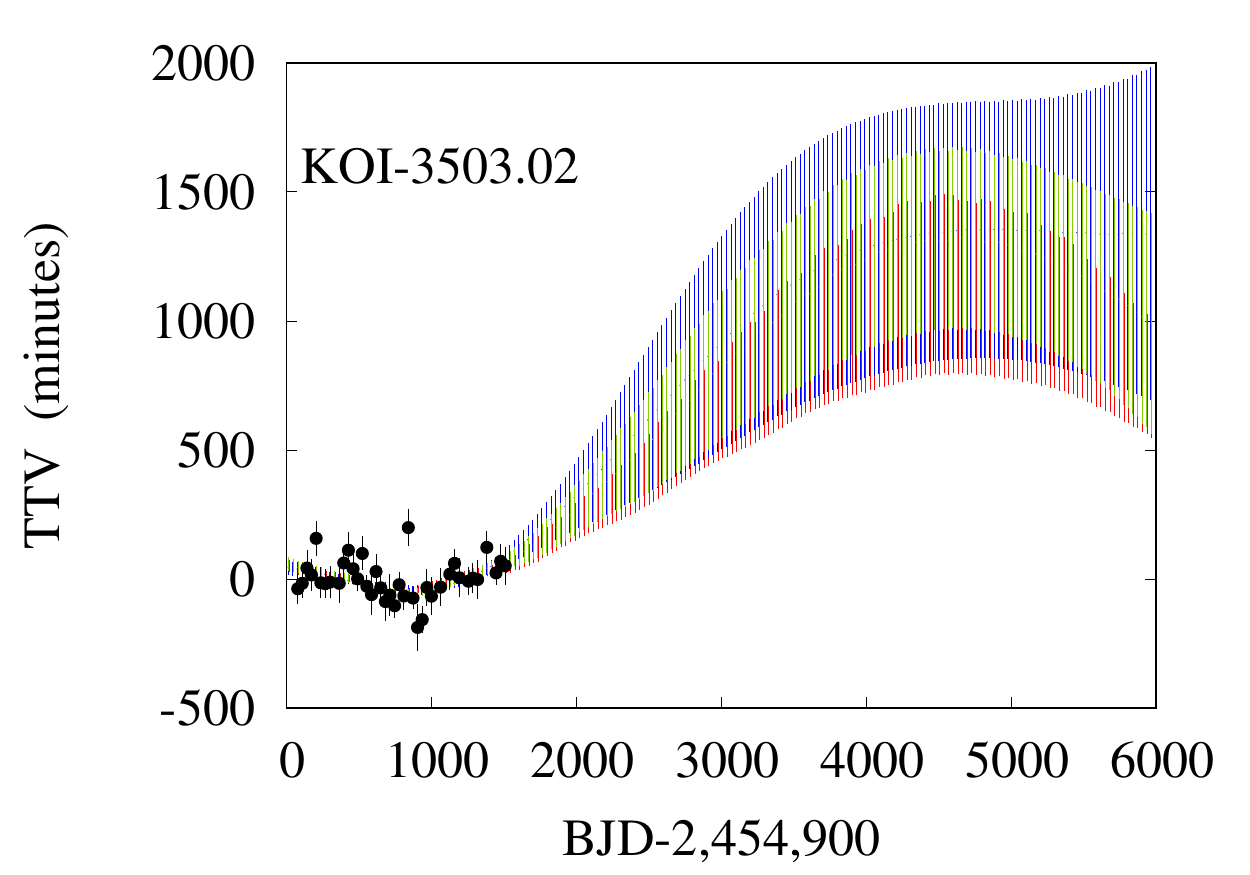}
\caption{Distribution of projected transit times as a function of time for the planet labelled in each panel (part 9). Black points mark transit times in the catalog of \citet{rowe15a} with 1$\sigma$ error bars. In green are 68.3\% confidence intervals of simulated transit times from posterior sampling. In blue (red), are a subset of samples with dynamical masses below (above) the 15.9th (84.1th) percentile. 
\label{fig:KOI-2414fut}} 
\end{center}
\end{figure}

KOI-3503.01 has S/N = 8.50 from Kepler DR 25, which is too low for validation on its own. Our fit to the light curve allowing free transit times increases the S/N to 12.2, as quoted in Table~\ref{tbl-confirmplanets}. The improvement is consistent with the expected TTVs caused by its neighbor, KOI-3503.02, given the period ratio of 1.5021, and the expected TTV period >10 yr. For KOI-3503.02, we find a weaker S/N of 9.6. Nevertheless, an astrophysical false positive seems unlikely given the anticorrelated polynomial TTV signals that can be seen by eye in the data in Figure~\ref{fig:KOI-2414fut}. Hence, we confirm these two candidates as planets. 

In summary, we confirm the planetary nature of KOI-1831.03, KOI-1833.02, KOI-3503.01, and KOI-3503.02.

\subsection{Eccentricity}
\begin{figure}[h!]
\begin{center}
\includegraphics [height = 2.2 in]{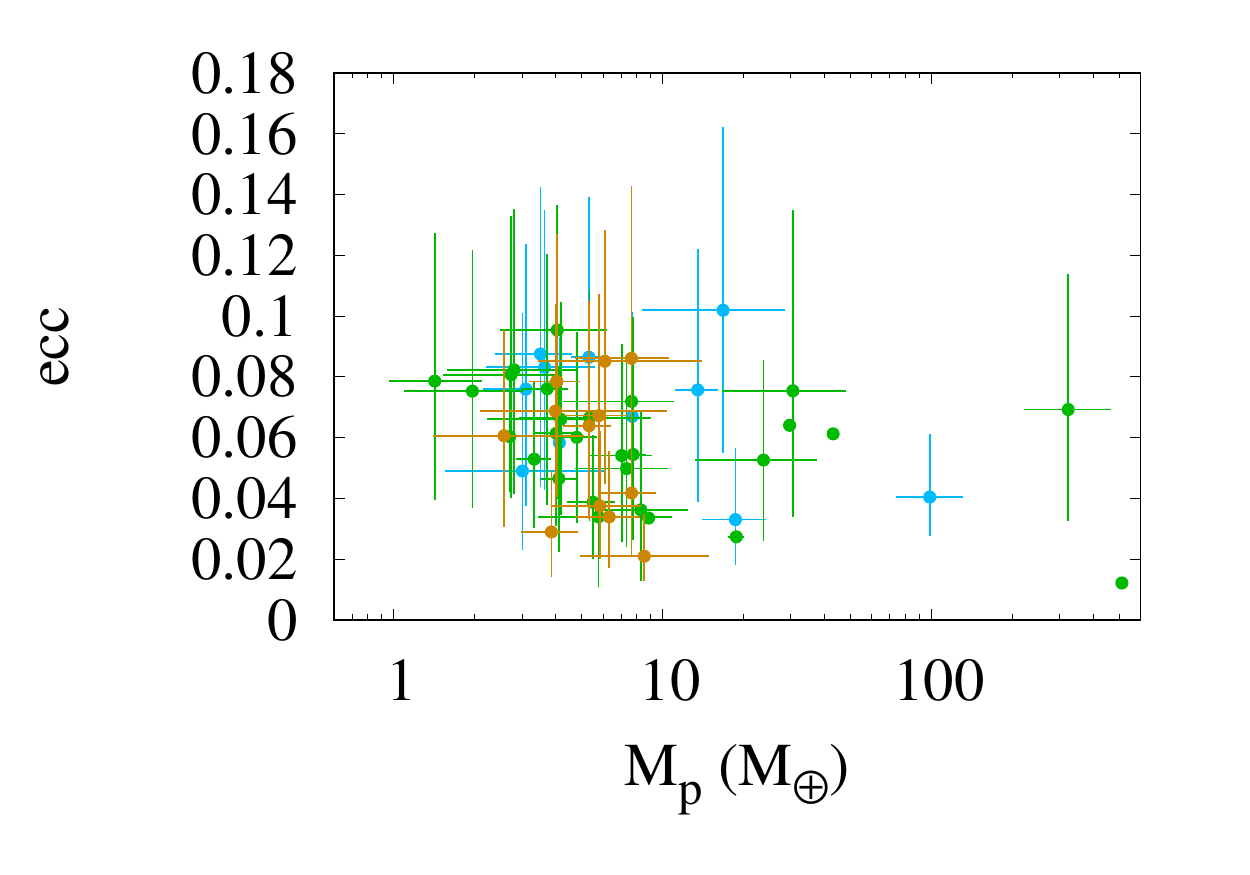}
\includegraphics [height = 2.2 in]{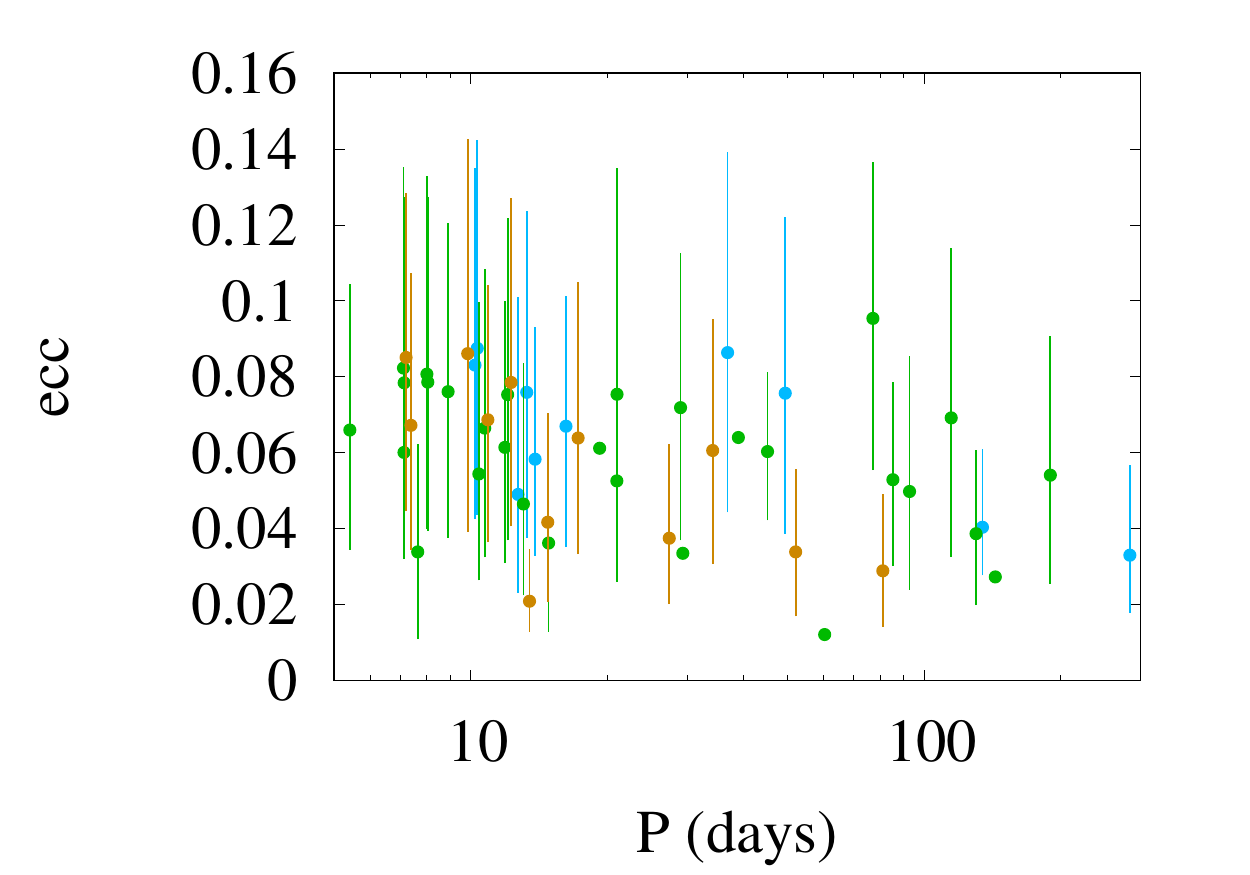}
\caption{Eccentricity posteriors for a range of planetary masses with colors signifying the number of transiting planets in the system: two (blue), three (green), and four (brown). The left panel plots eccentricities and mass, and the right panel plots eccentricity and orbital period. Note that eccentricities $\gtrsim$ 0.1 are disfavored by our prior.
\label{fig:ecc}} 
\end{center}
\end{figure}

In Figure~\ref{fig:ecc} we plot the eccentricity posteriors of planets with strongly detected dynamical masses such that the 15.9th percentile of the posterior is more than half of its median. Eccentricities are weakly anticorrelated with planetary masses, consistent with the expectation that multi-planet systems are more likely to be unstable if planets have both high masses and high eccentricities. Furthermore, we see fewer planets with both low masses and eccentricities since for Figure~\ref{fig:ecc} we have selected for strong TTV detections. There is no discernible correlation between eccentricities and orbital period.

Individual planets with strongly constrained eccentricities include planets within the systems of KOI-377 (Kepler-9), KOI-806 (Kepler-30), and KOI-1353. These systems have planets that are among the most massive of our sample, with strong near-resonant and nonresonant components to their TTVs. In these systems, dynamical masses are detected to high significance, such that the ratio of the median to the length of the interval between the 15.9th percentile and the median is greater than 9. KOI-620 (Kepler-51) and (KOI-152) Kepler-79 also have eccentricity constraints significantly narrower than our prior. For the planets in these systems, the ratio of the median to the length of the interval between the 15.9th percentile and the median ranges from 2.3 to 7.2.
 
\begin{figure}
\begin{center}
\figurenum{20}
\includegraphics [height = 1.1 in]{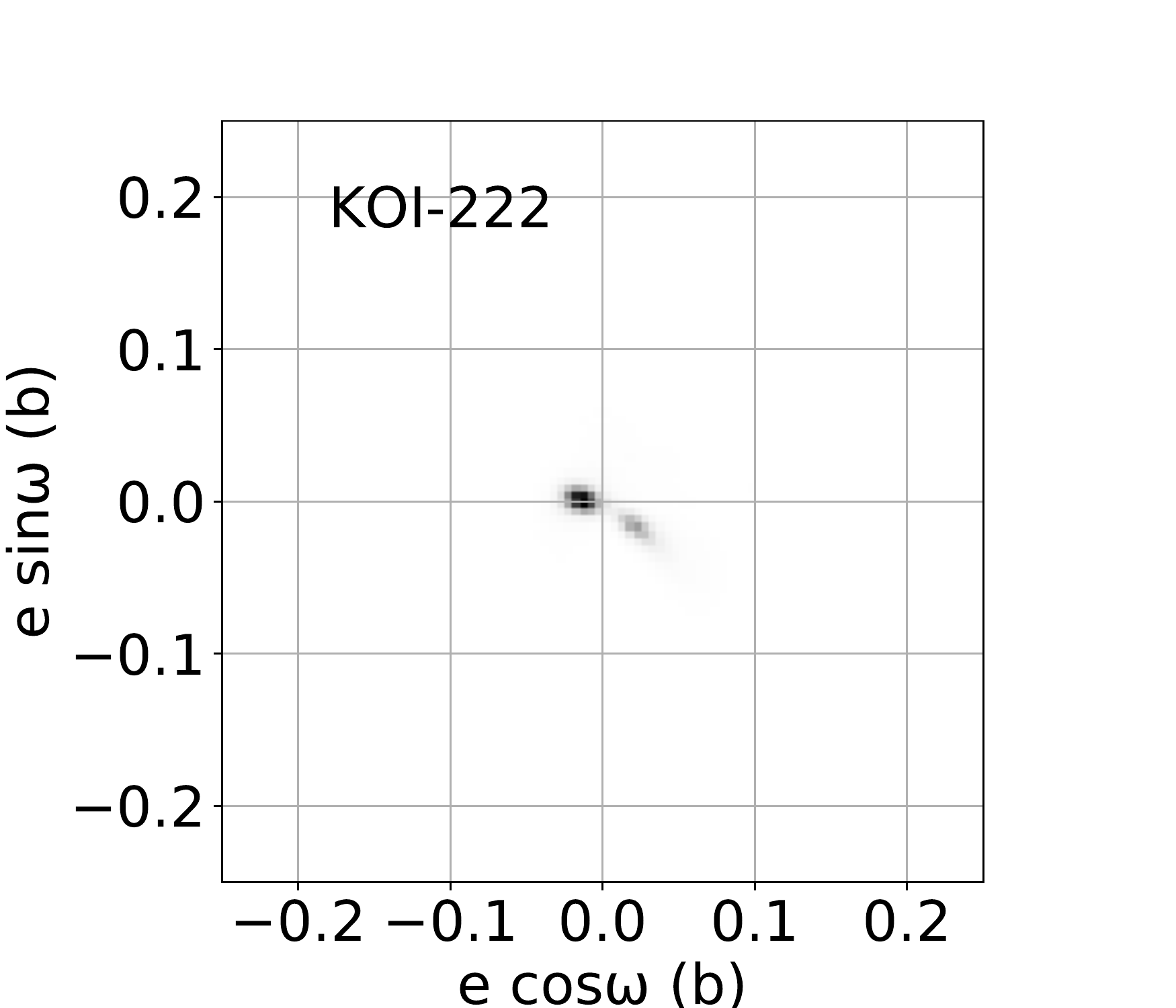}
\includegraphics [height = 1.1 in]{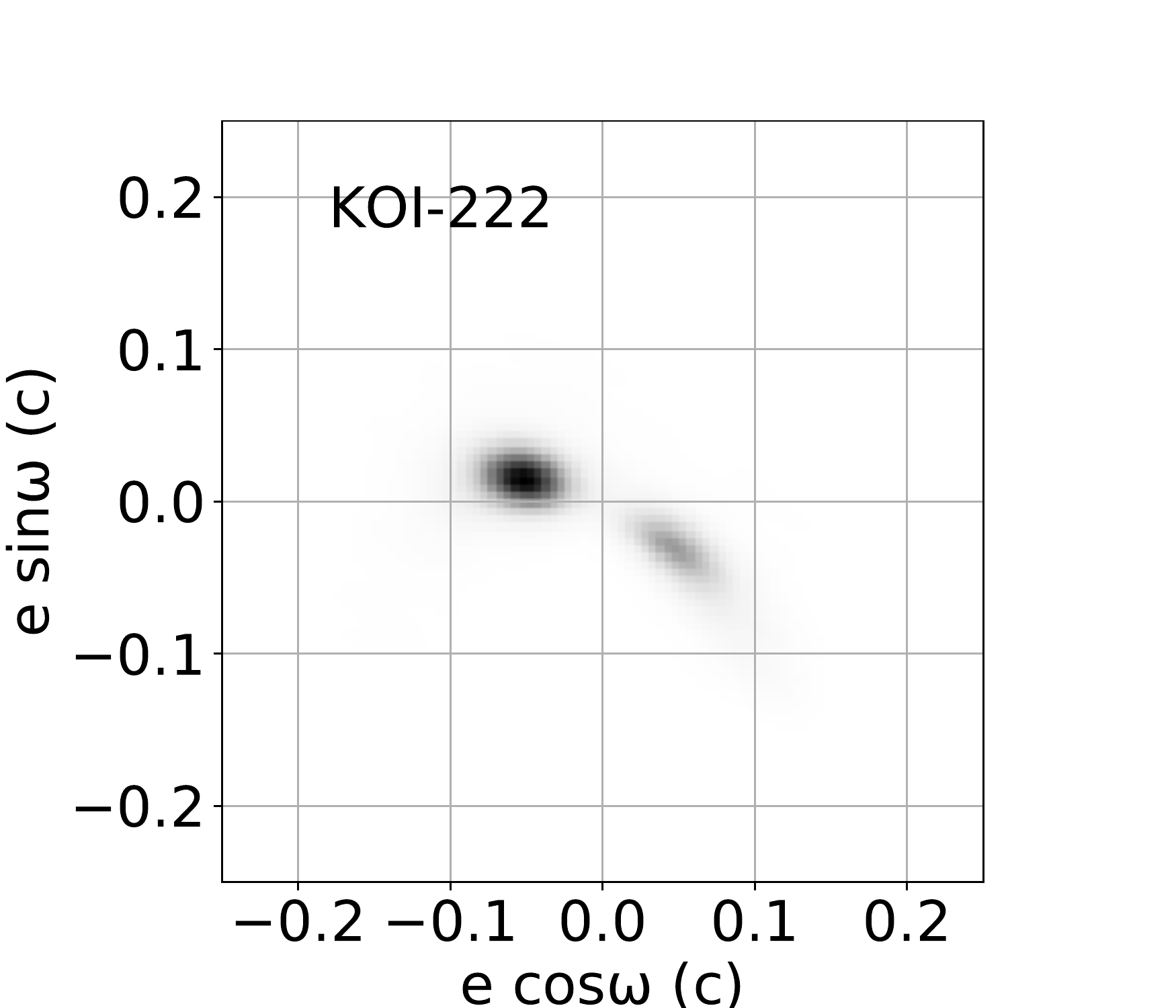}
\includegraphics [height = 1.1 in]{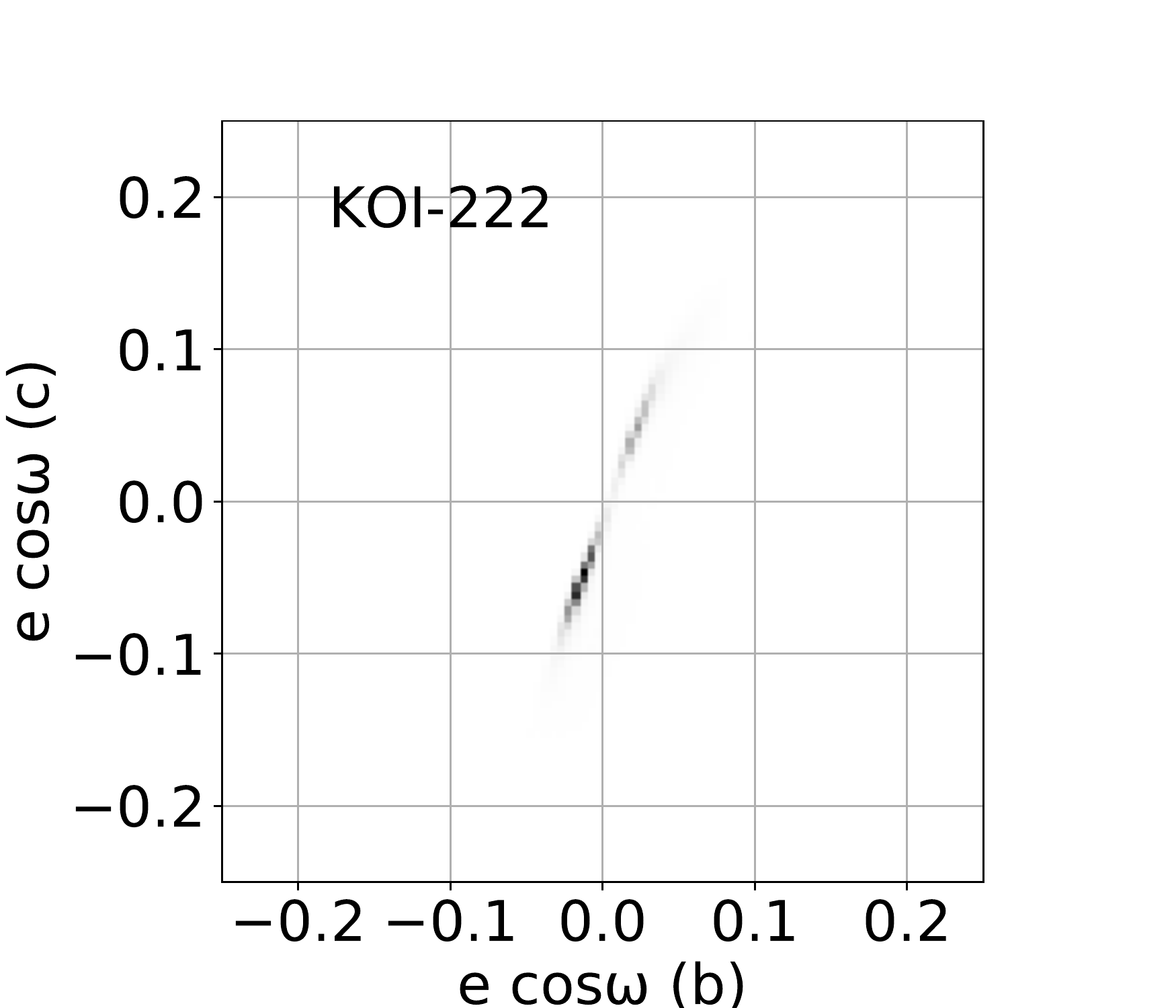}
\includegraphics [height = 1.1 in]{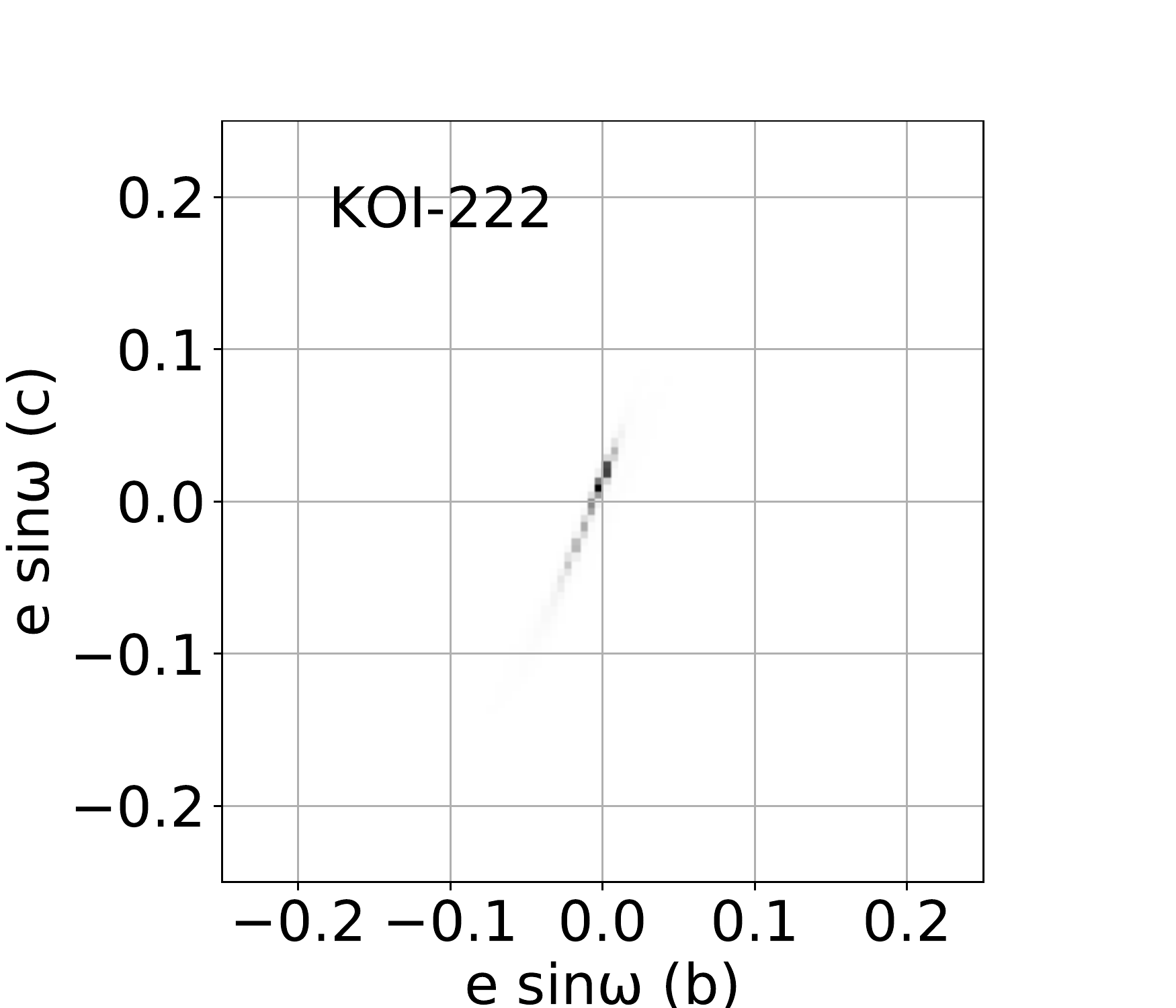} \\
\includegraphics [height = 1.1 in]{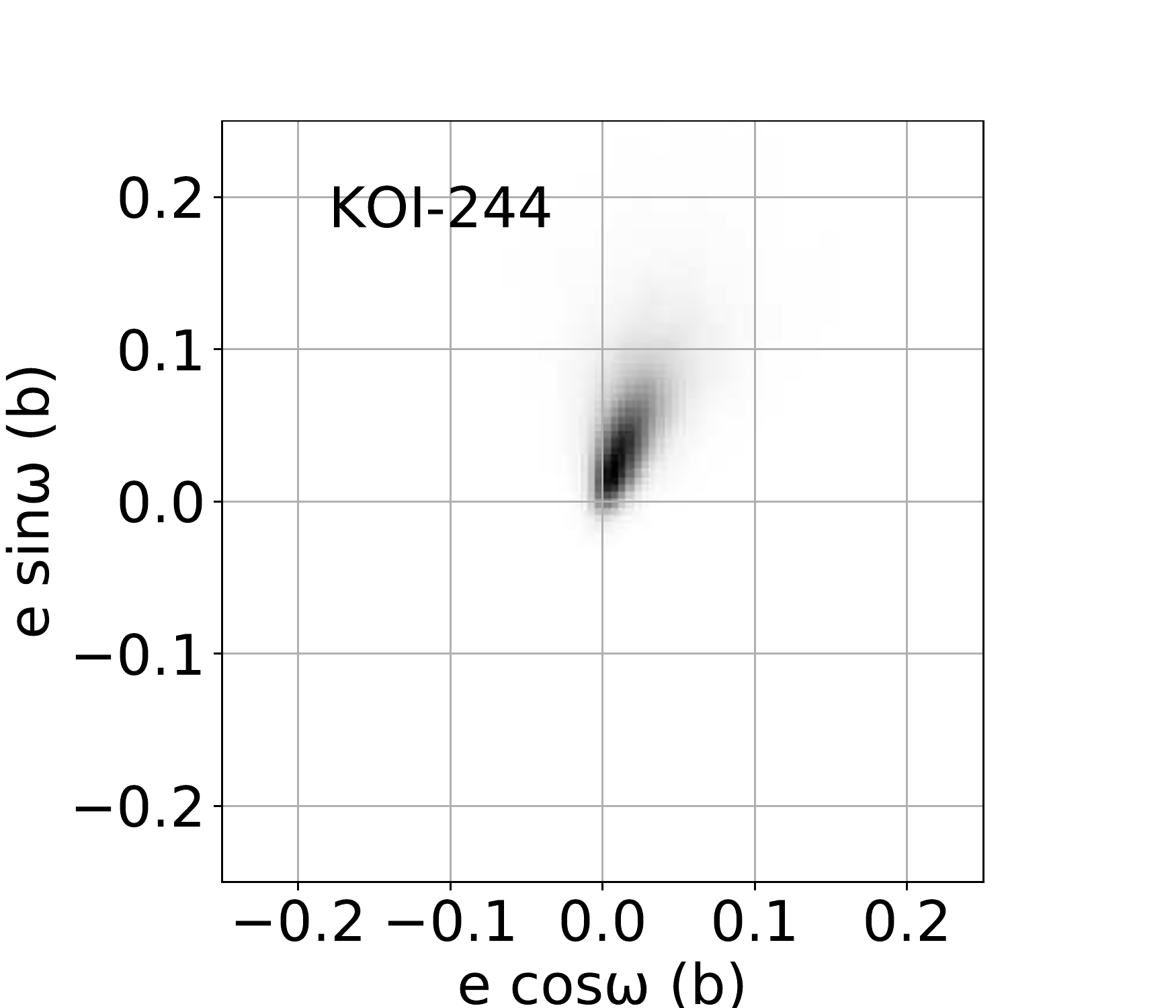}
\includegraphics [height = 1.1 in]{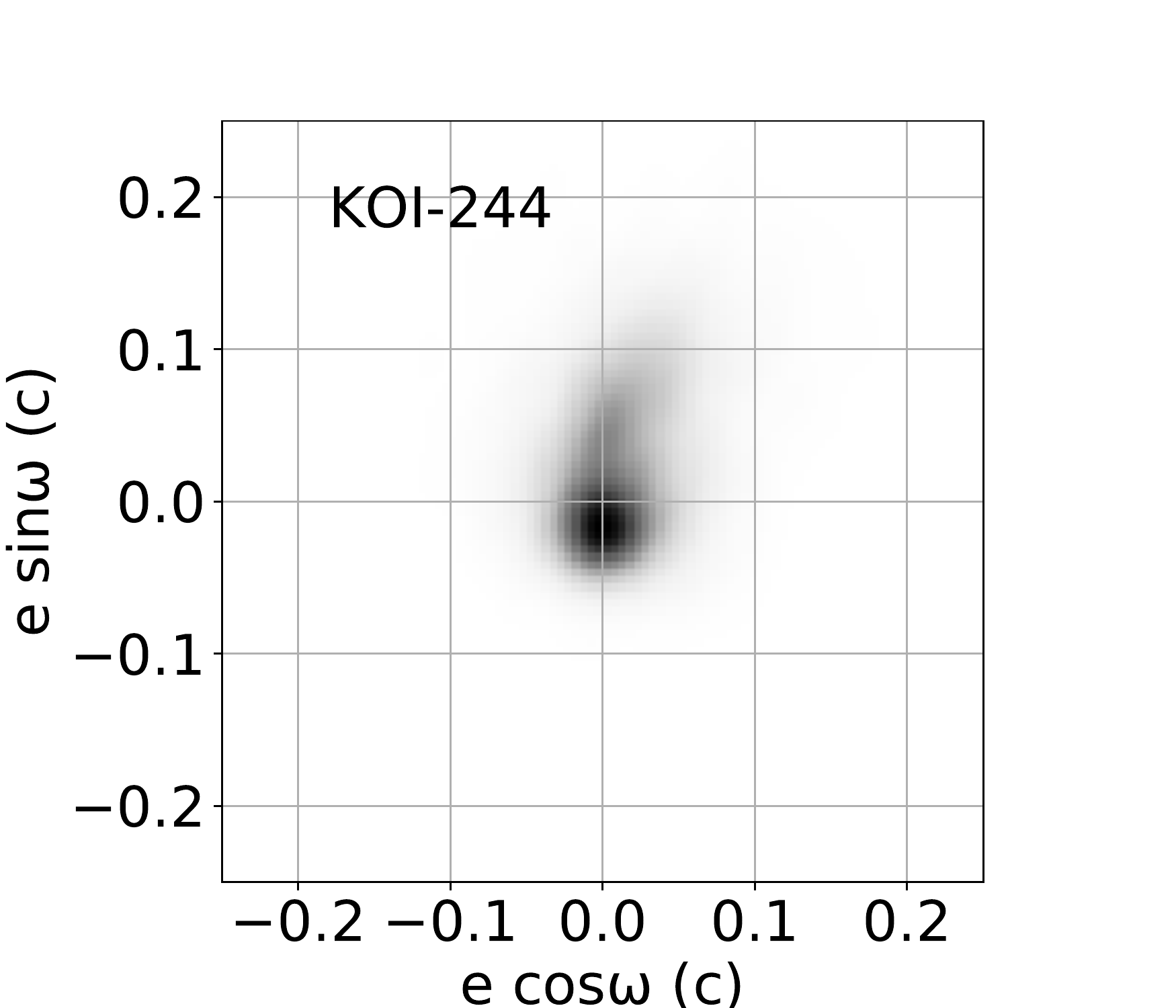}
\includegraphics [height = 1.1 in]{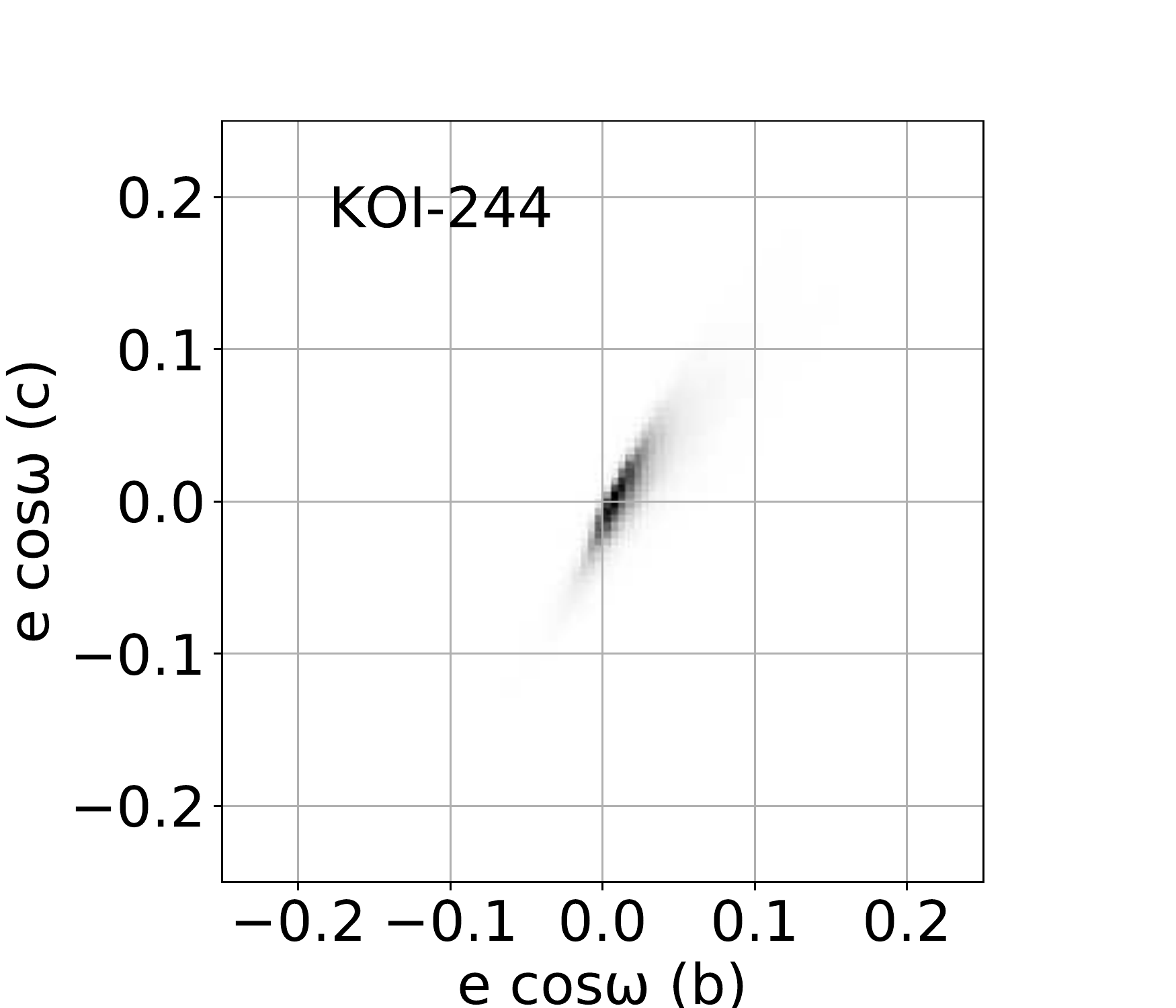}
\includegraphics [height = 1.1 in]{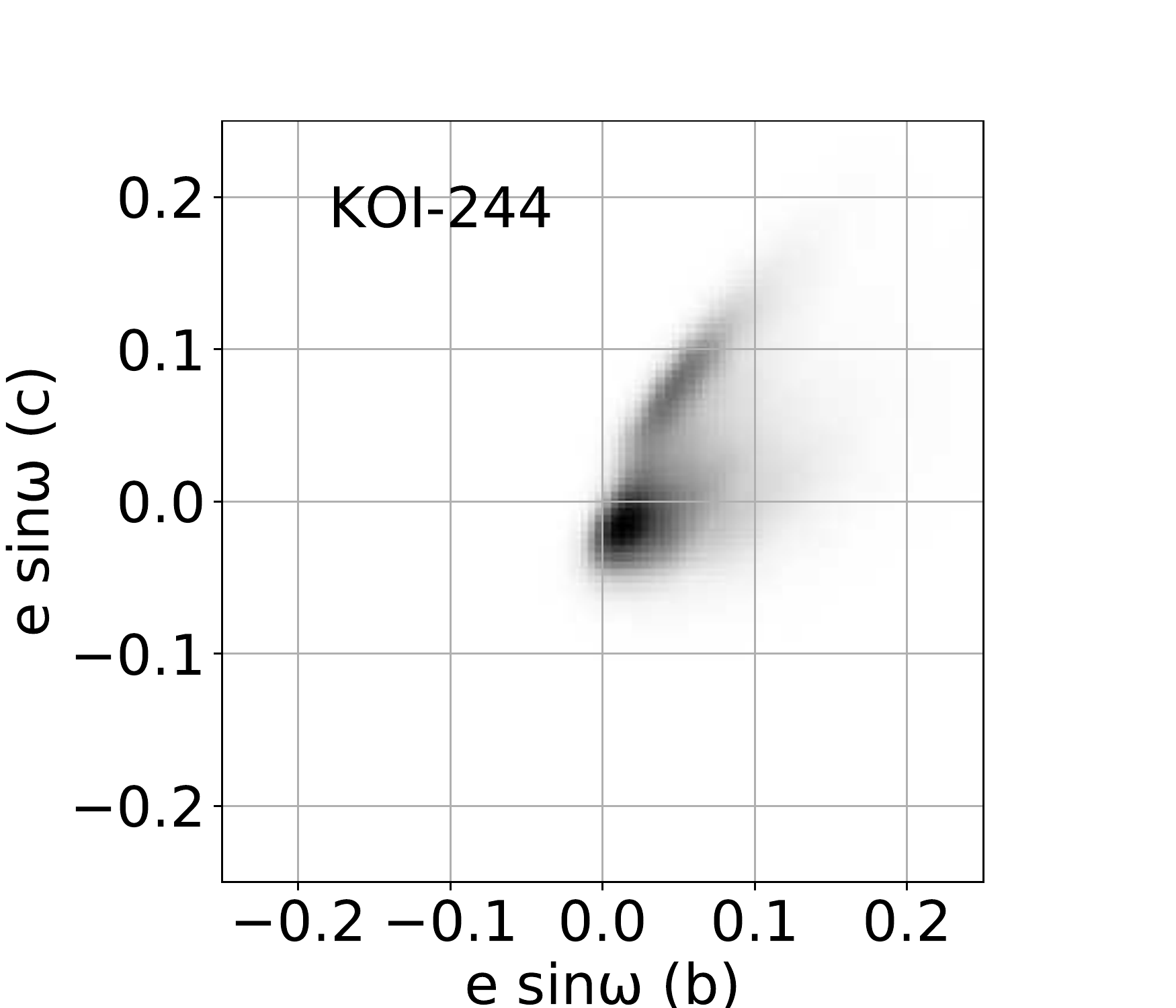} \\
\includegraphics [height = 1.1 in]{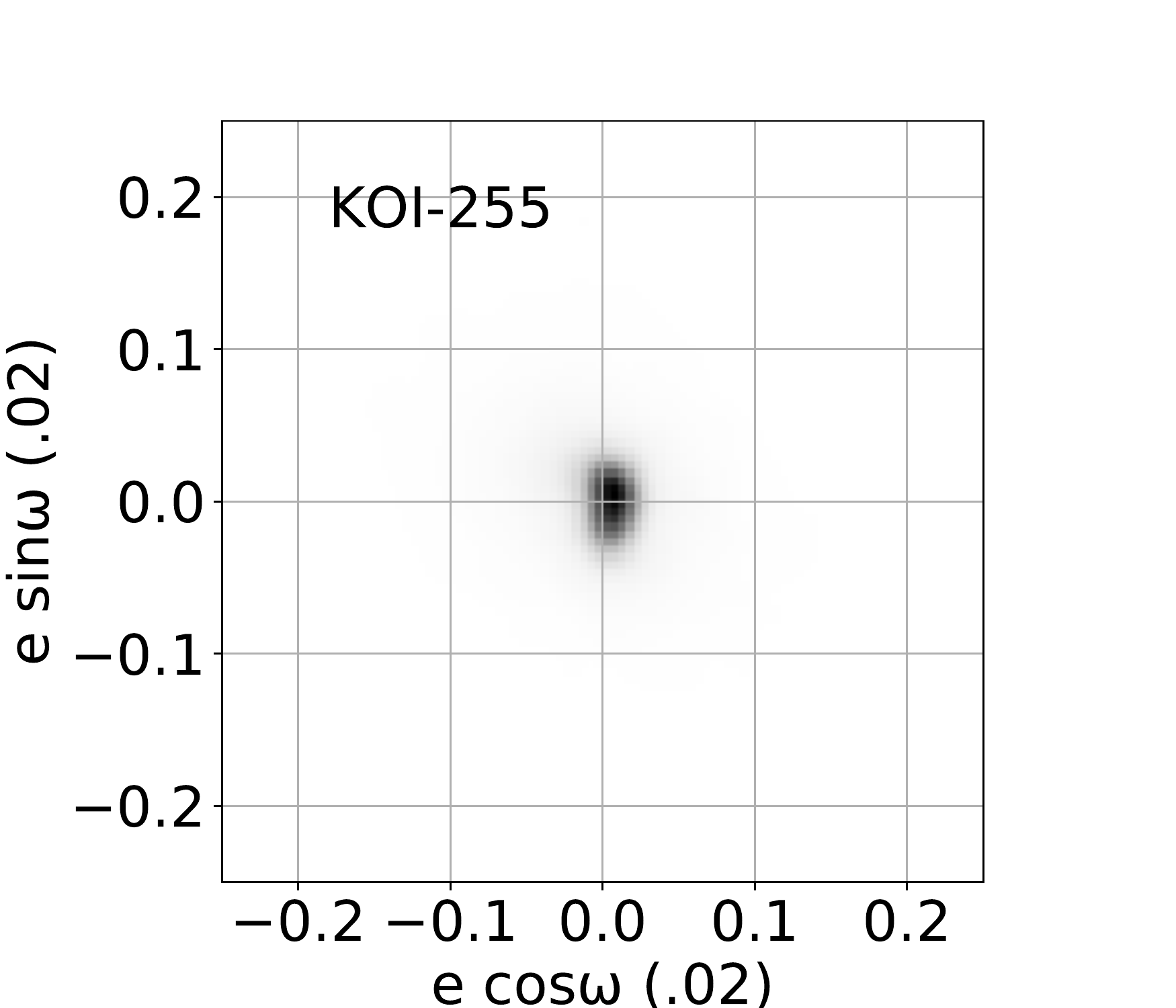}
\includegraphics [height = 1.1 in]{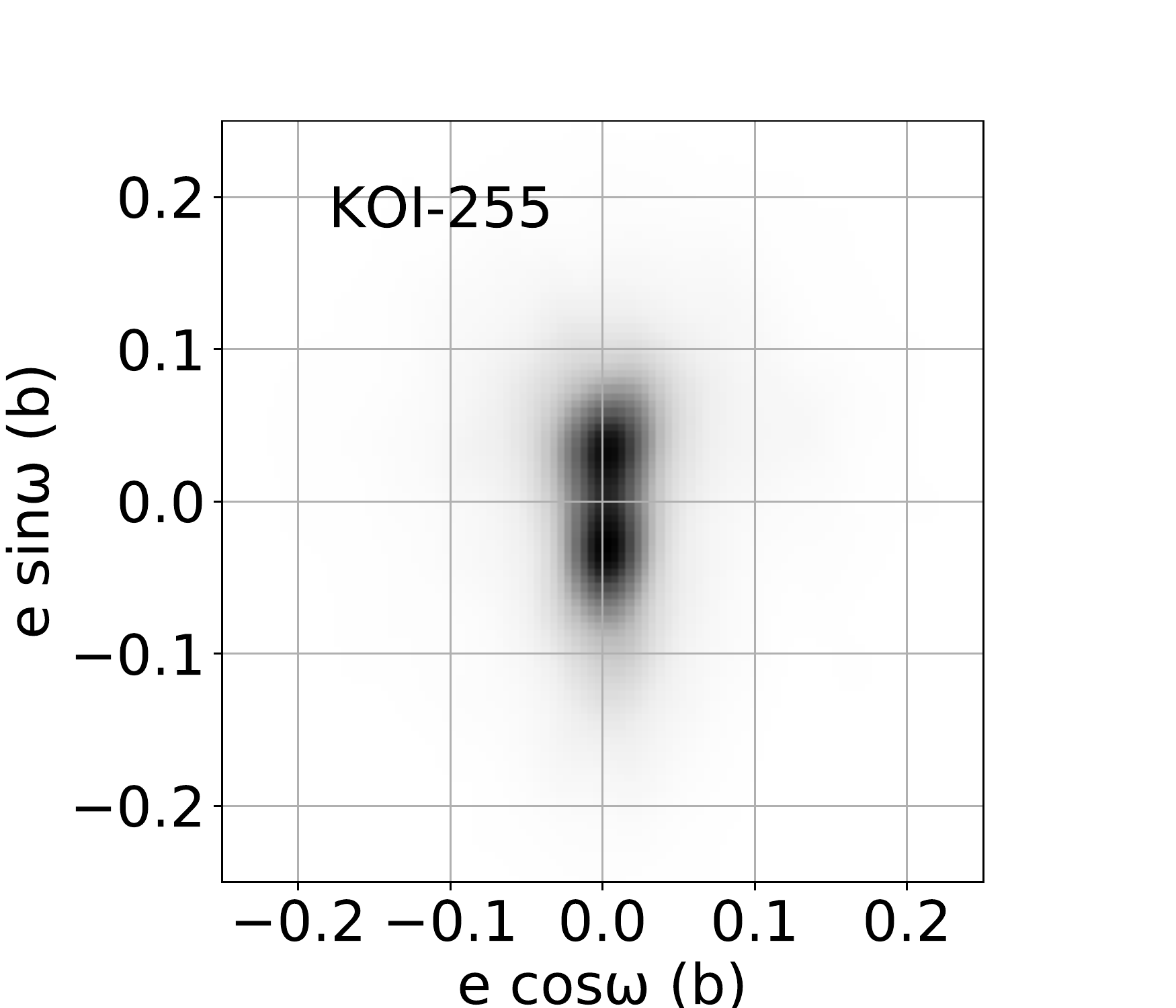}
\includegraphics [height = 1.1 in]{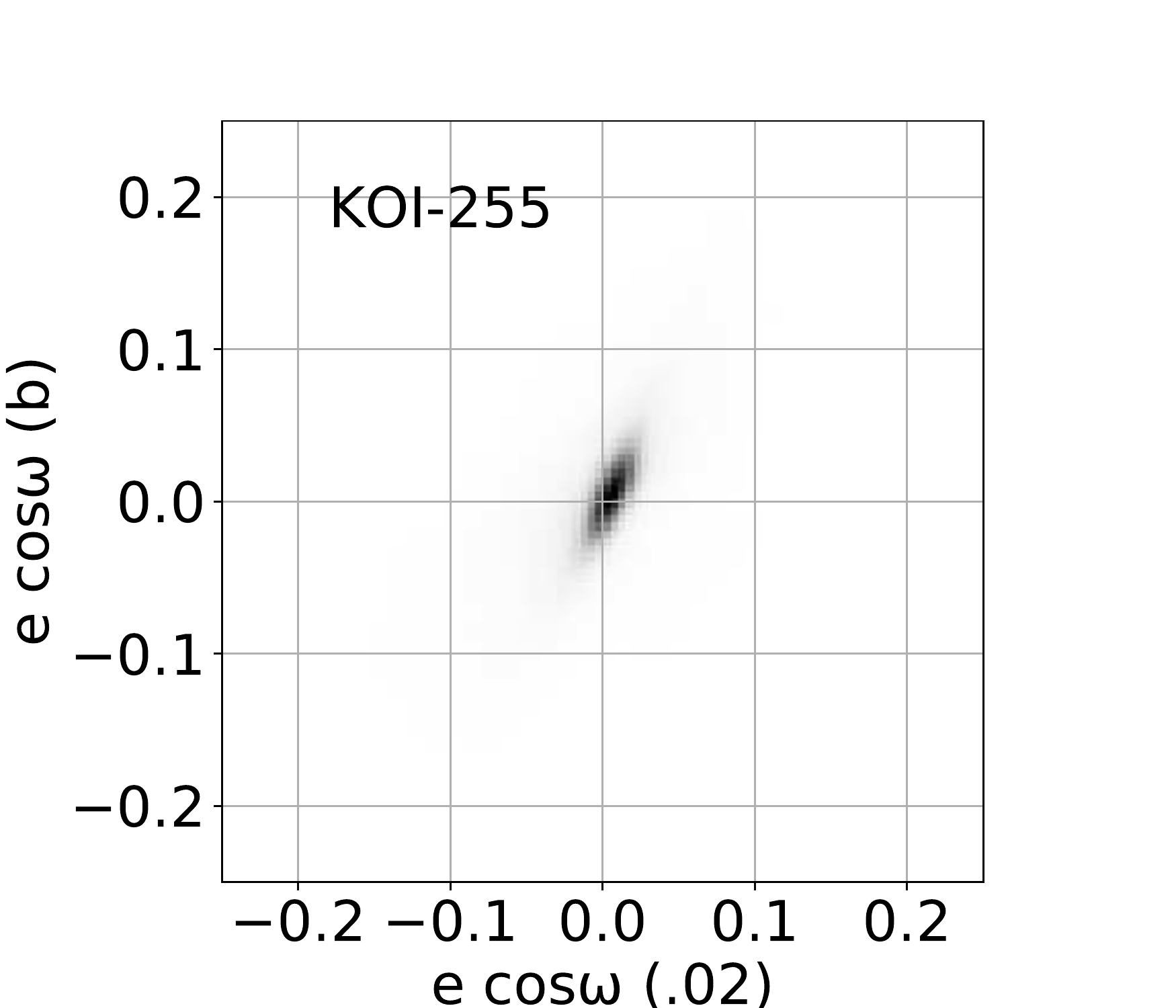}
\includegraphics [height = 1.1 in]{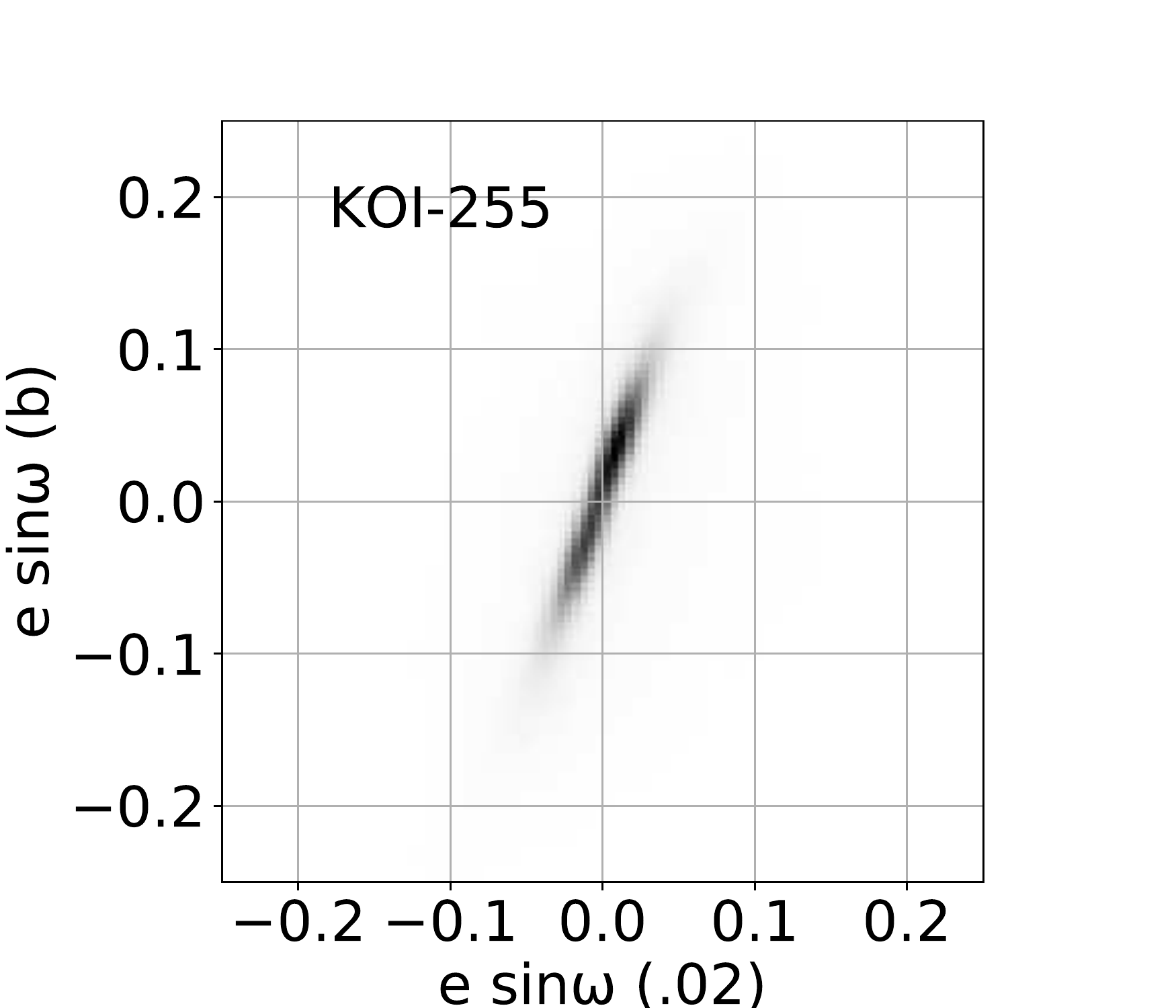} \\
\includegraphics [height = 1.1 in]{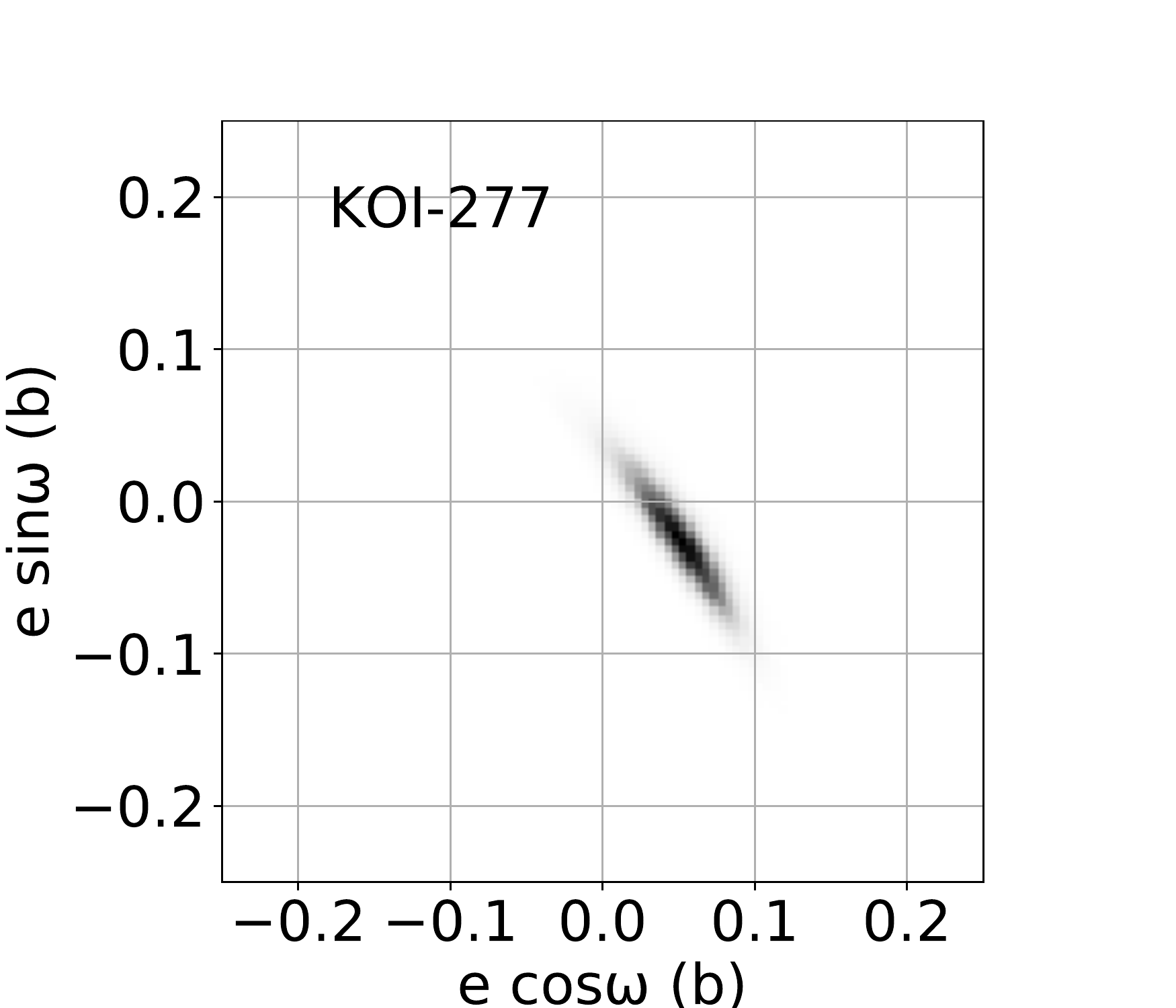}
\includegraphics [height = 1.1 in]{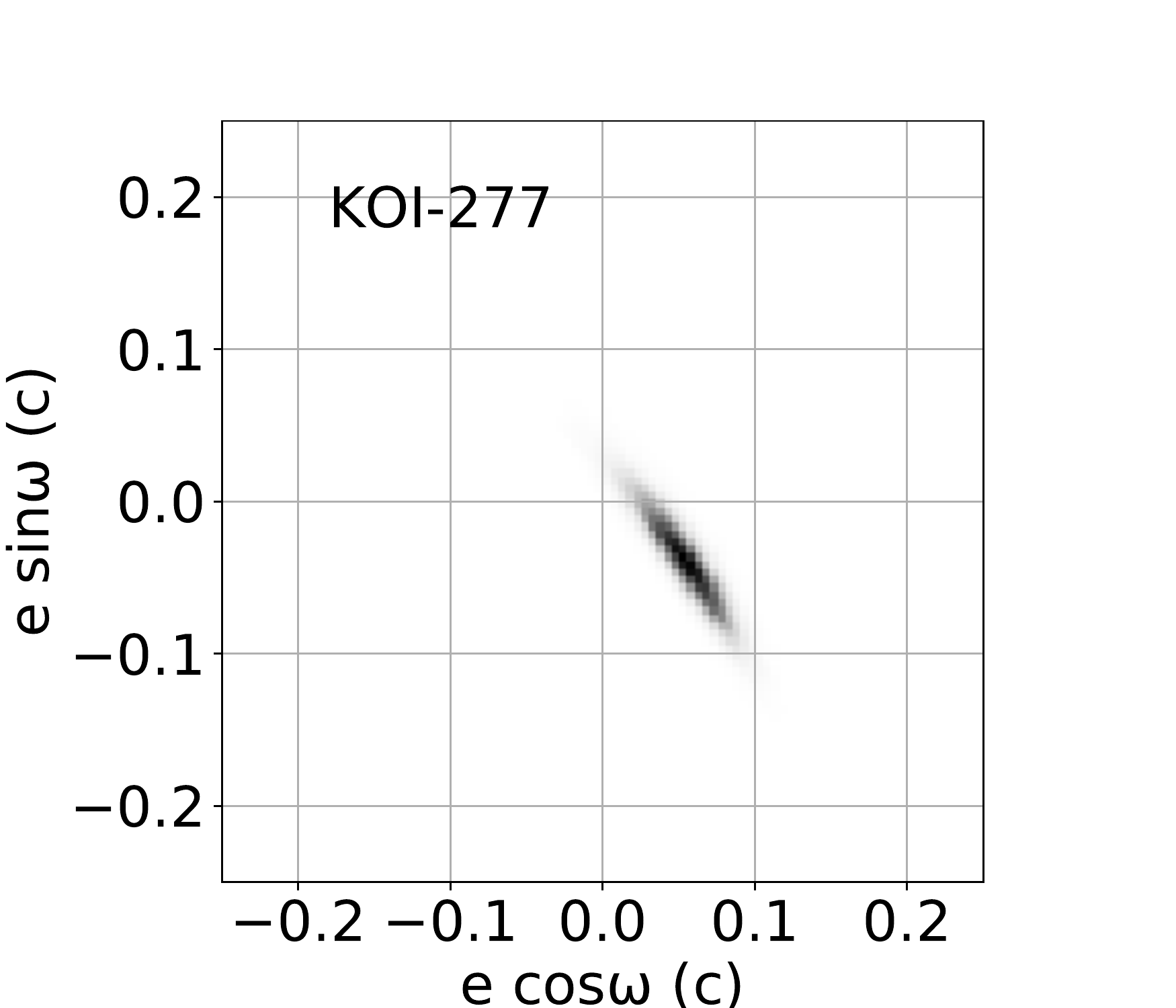}
\includegraphics [height = 1.1 in]{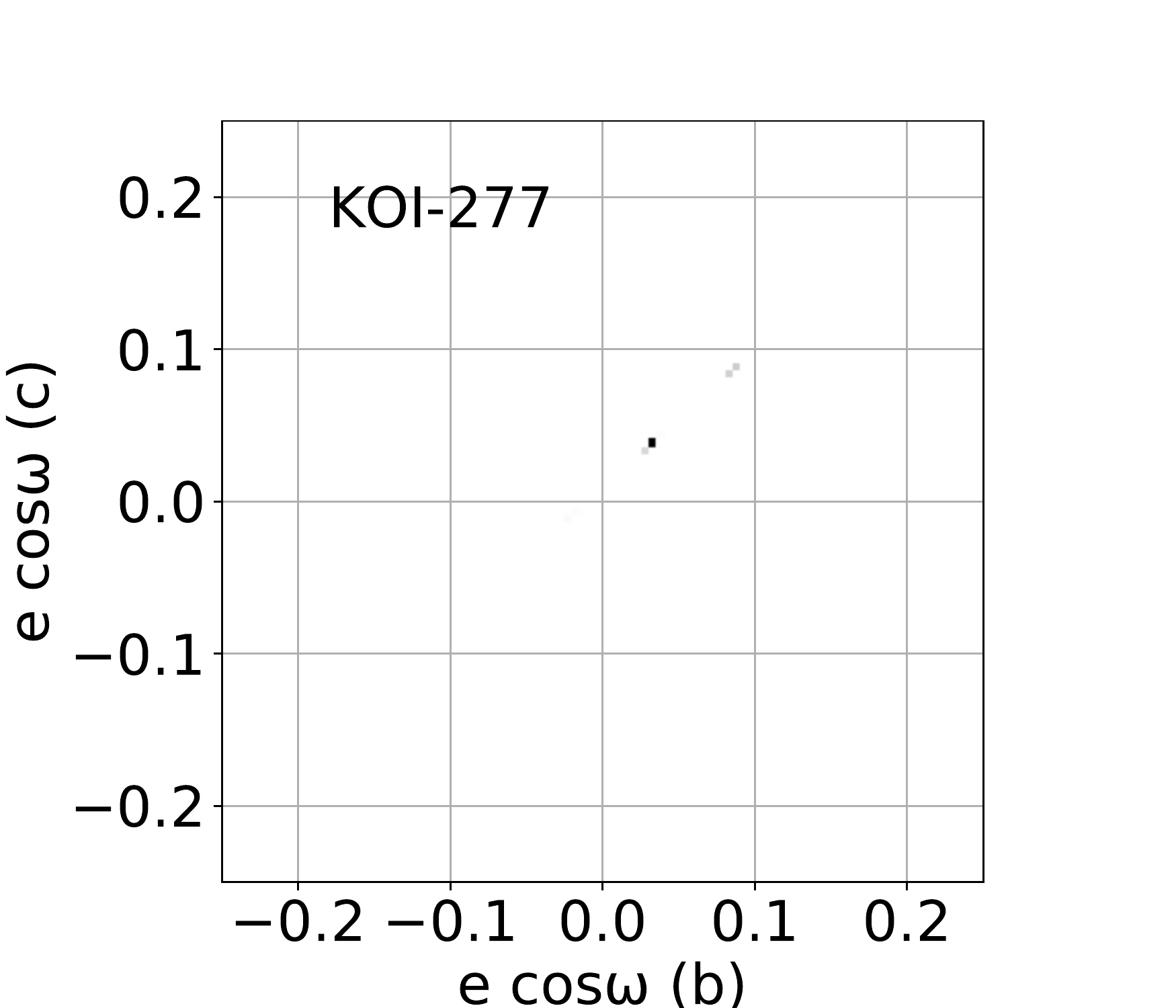}
\includegraphics [height = 1.1 in]{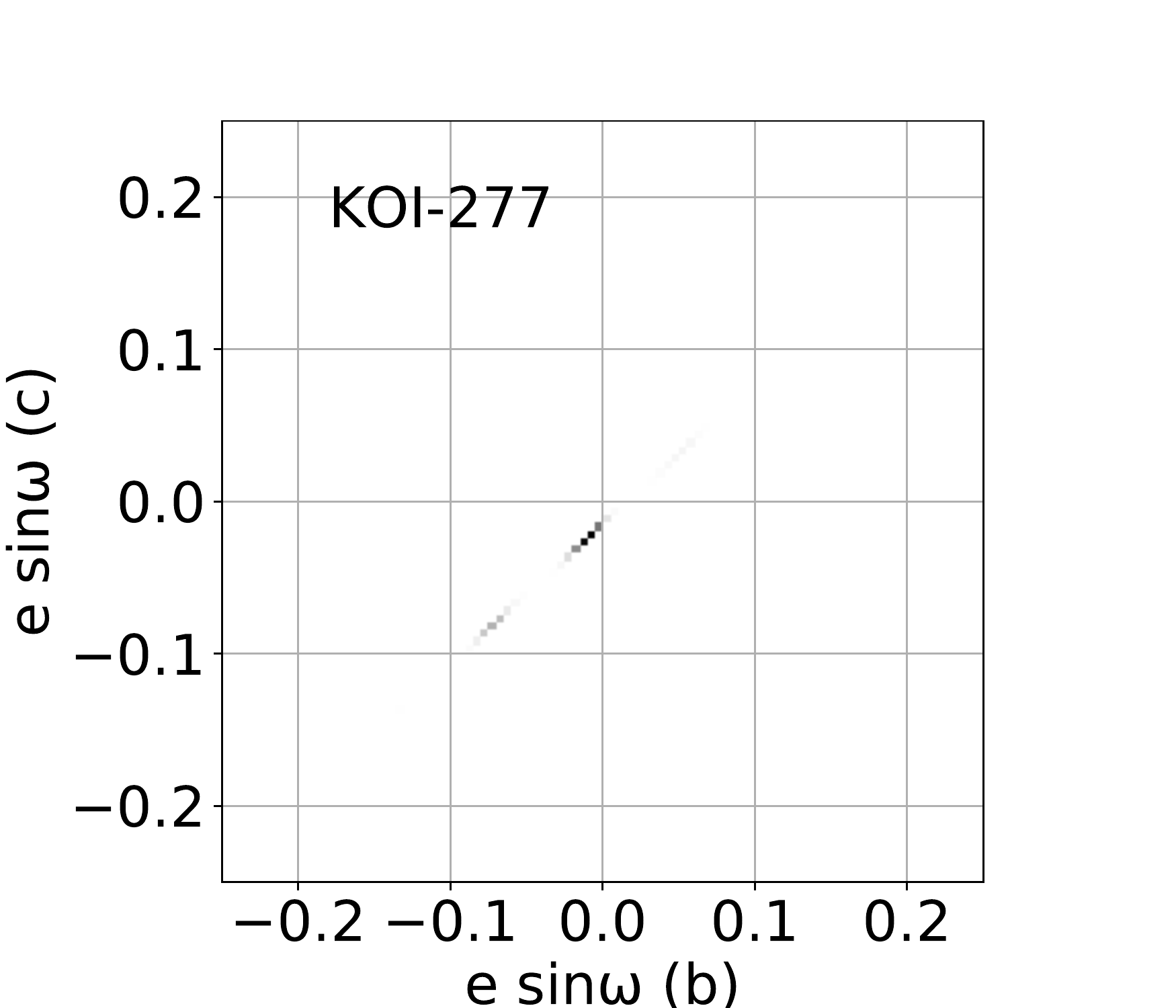} \\
\includegraphics [height = 1.1 in]{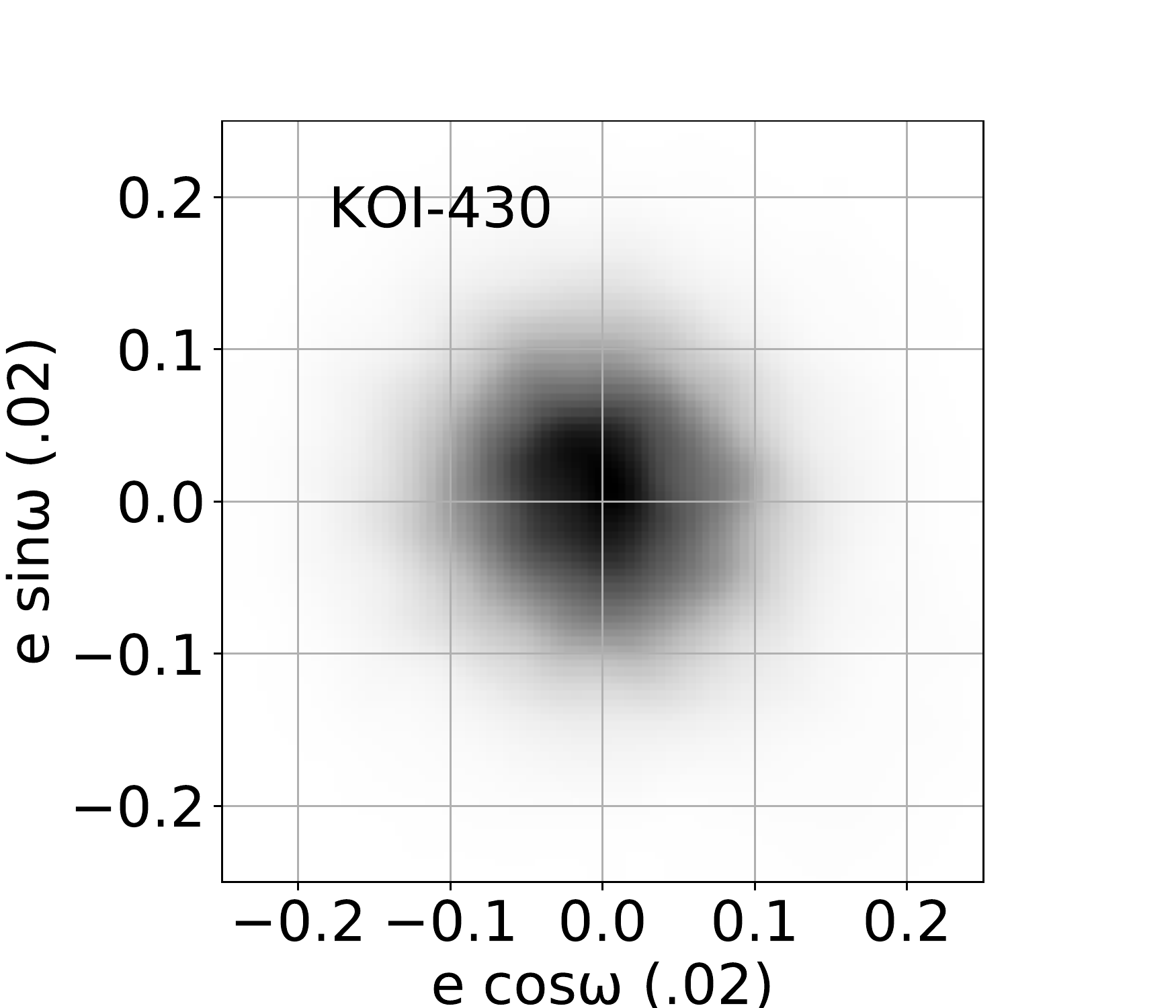}
\includegraphics [height = 1.1 in]{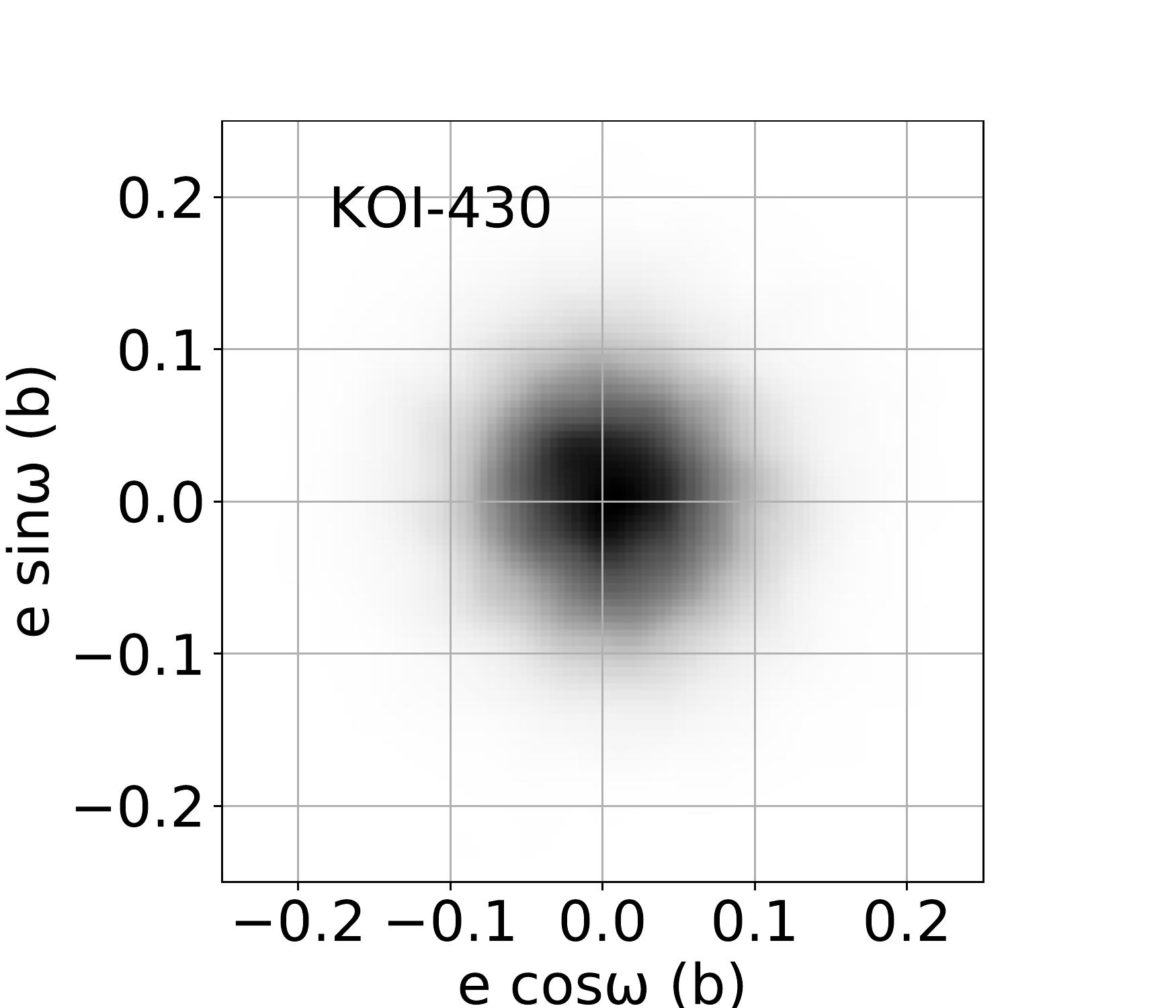} 
\includegraphics [height = 1.1 in]{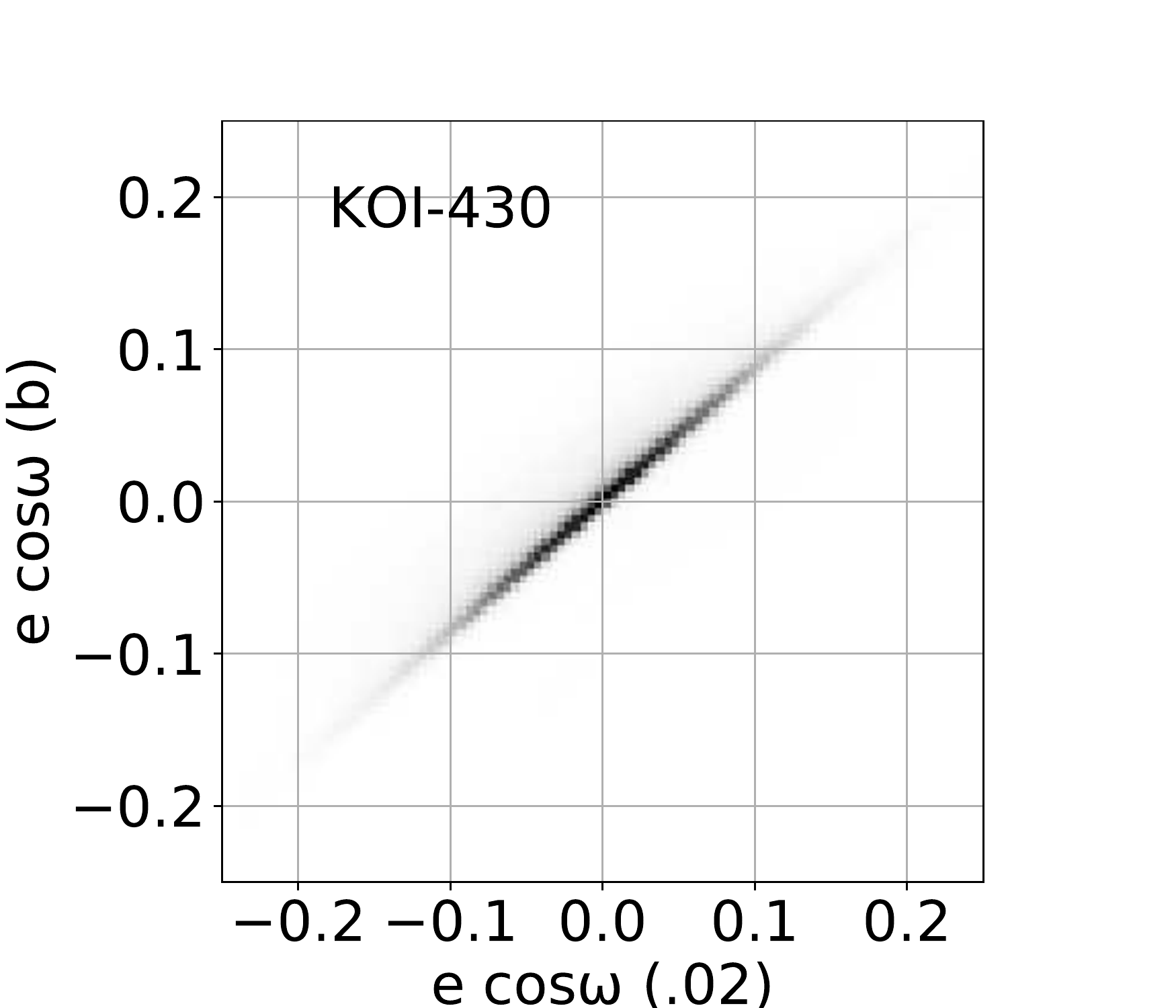}
\includegraphics [height = 1.1 in]{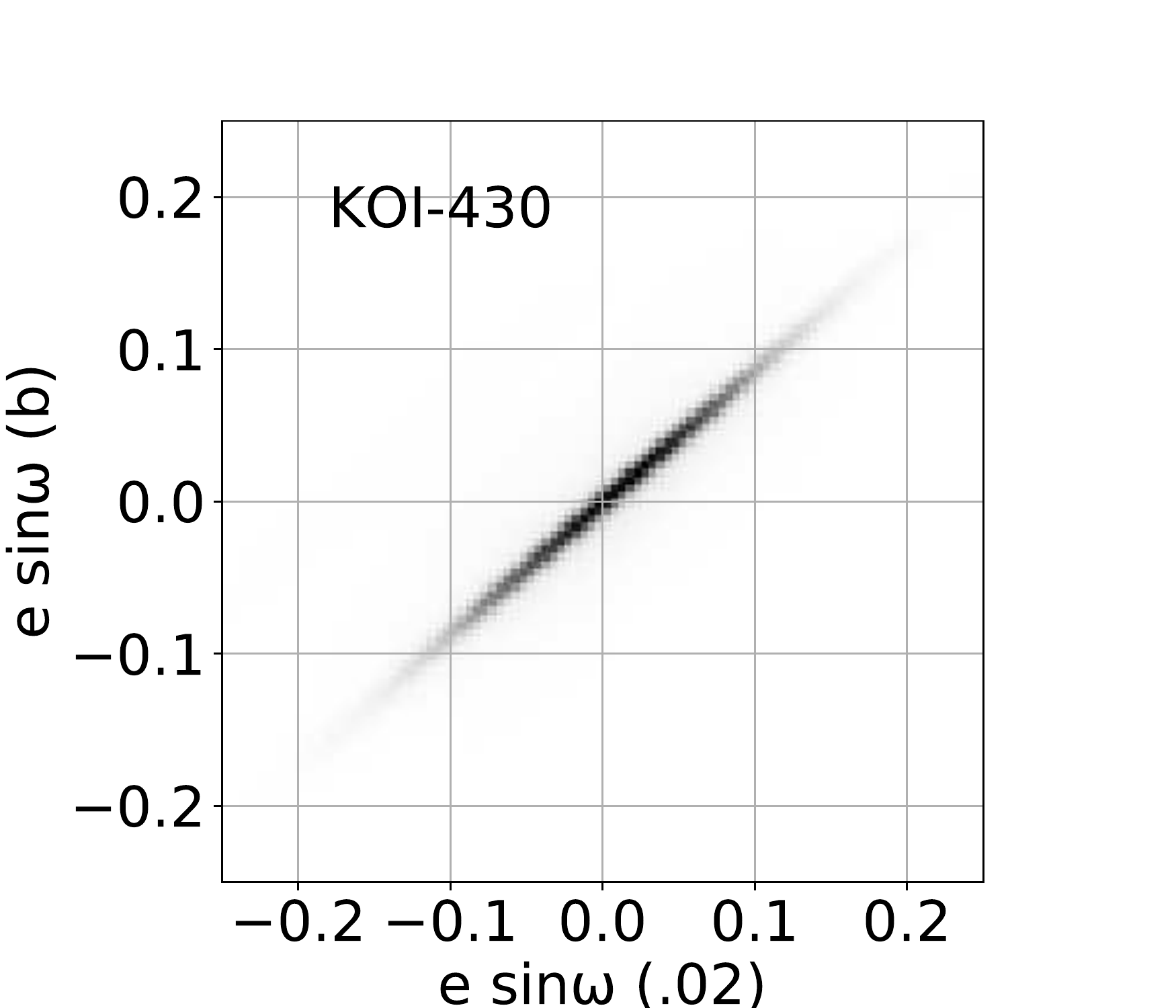} \\
\includegraphics [height = 1.1 in]{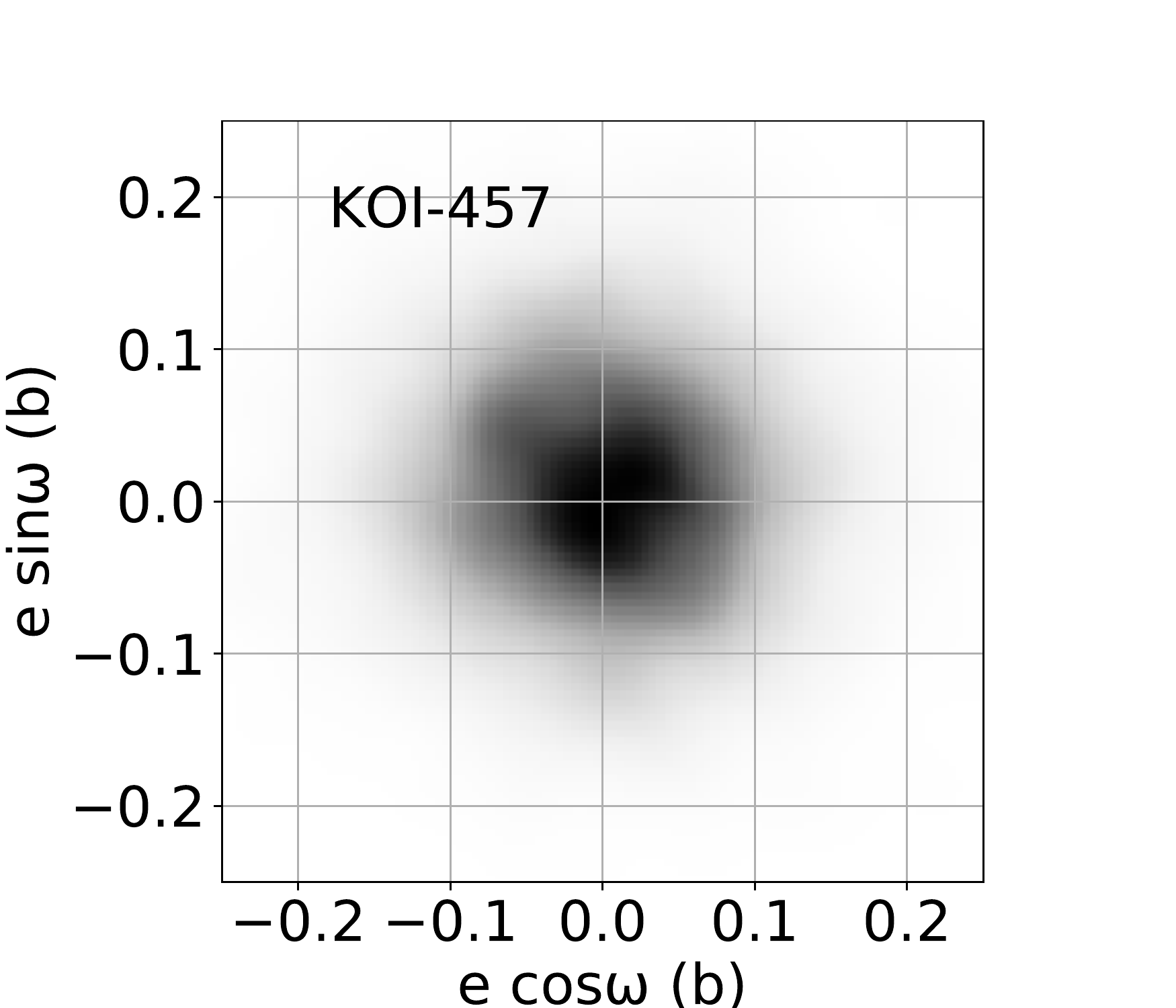}
\includegraphics [height = 1.1 in]{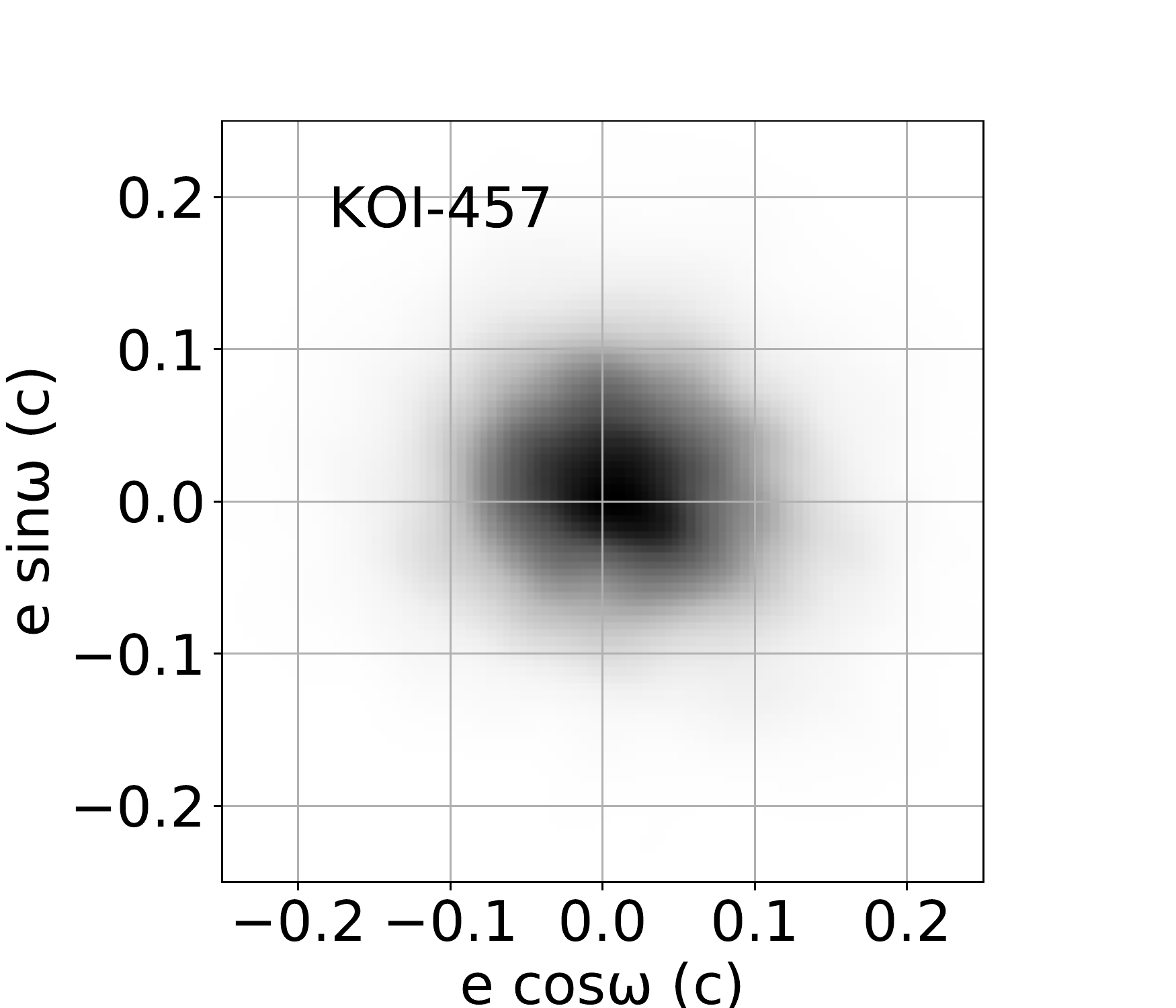} 
\includegraphics [height = 1.1 in]{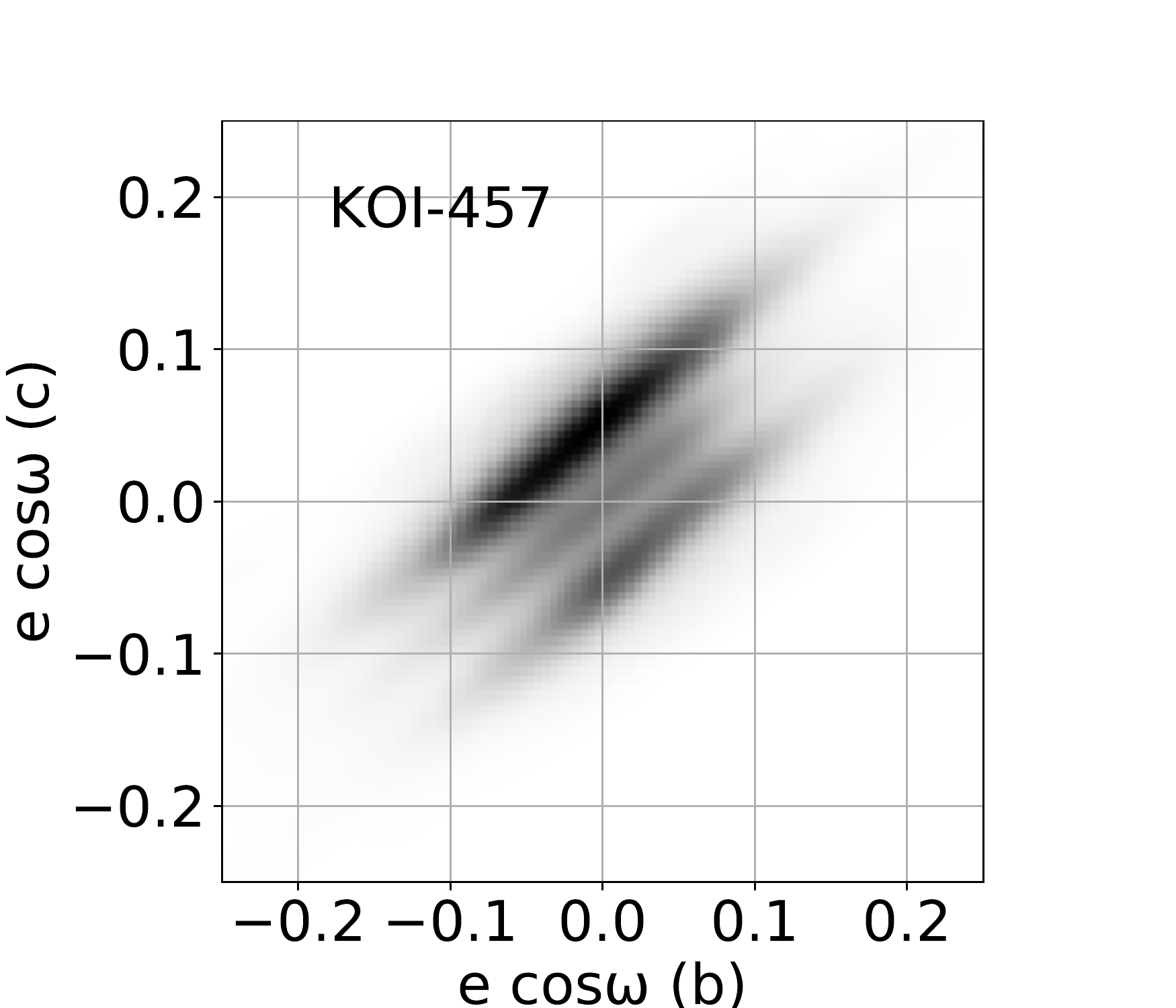}
\includegraphics [height = 1.1 in]{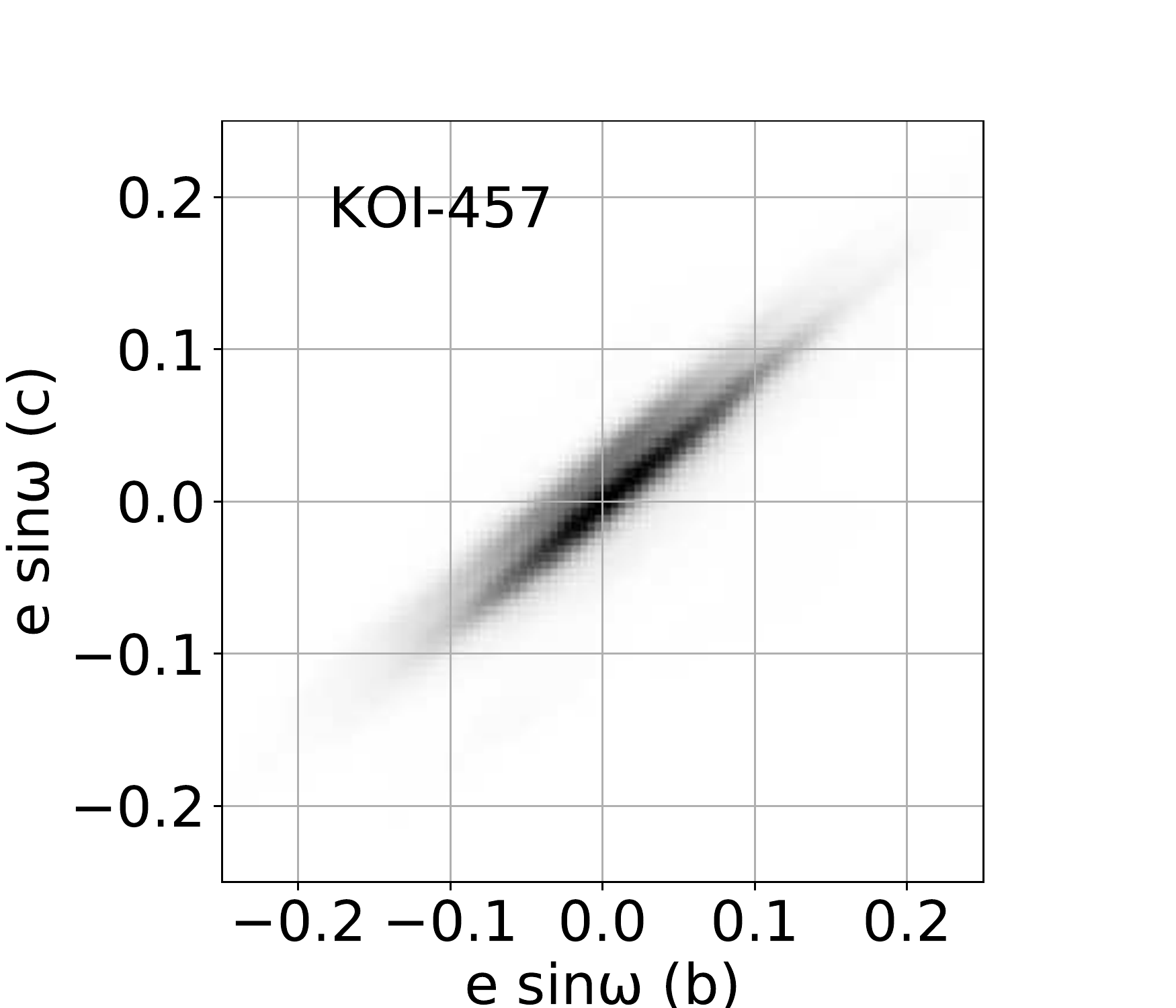} 
\caption{Two-dimensional kernel density estimators on joint posteriors of eccentricity vector components: two-planet systems. (Part 1 of 3.) \label{fig:ecc2a} }
\end{center}
\end{figure}

\begin{figure}
\begin{center}
\figurenum{21}
\includegraphics [height = 1.1 in]{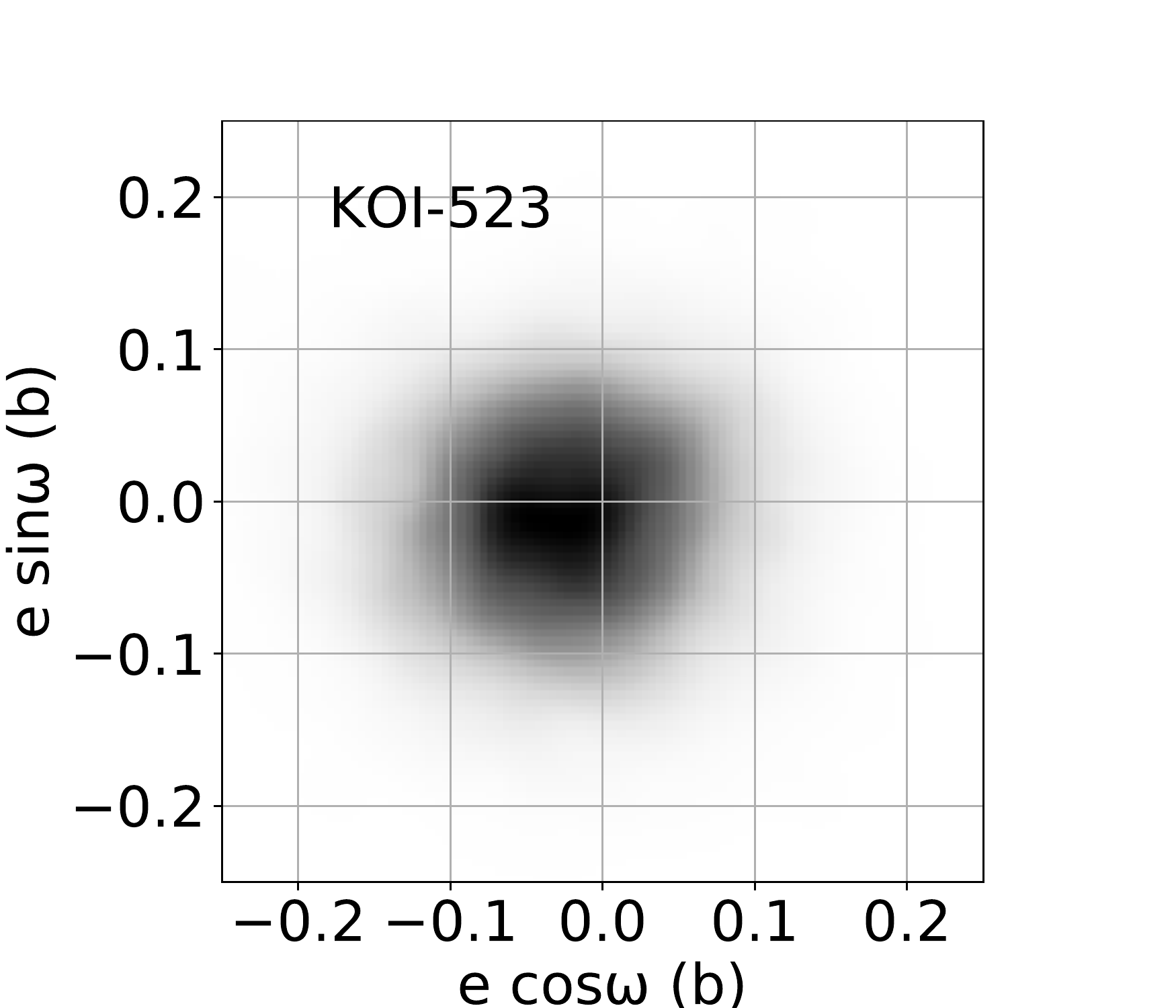}
\includegraphics [height = 1.1 in]{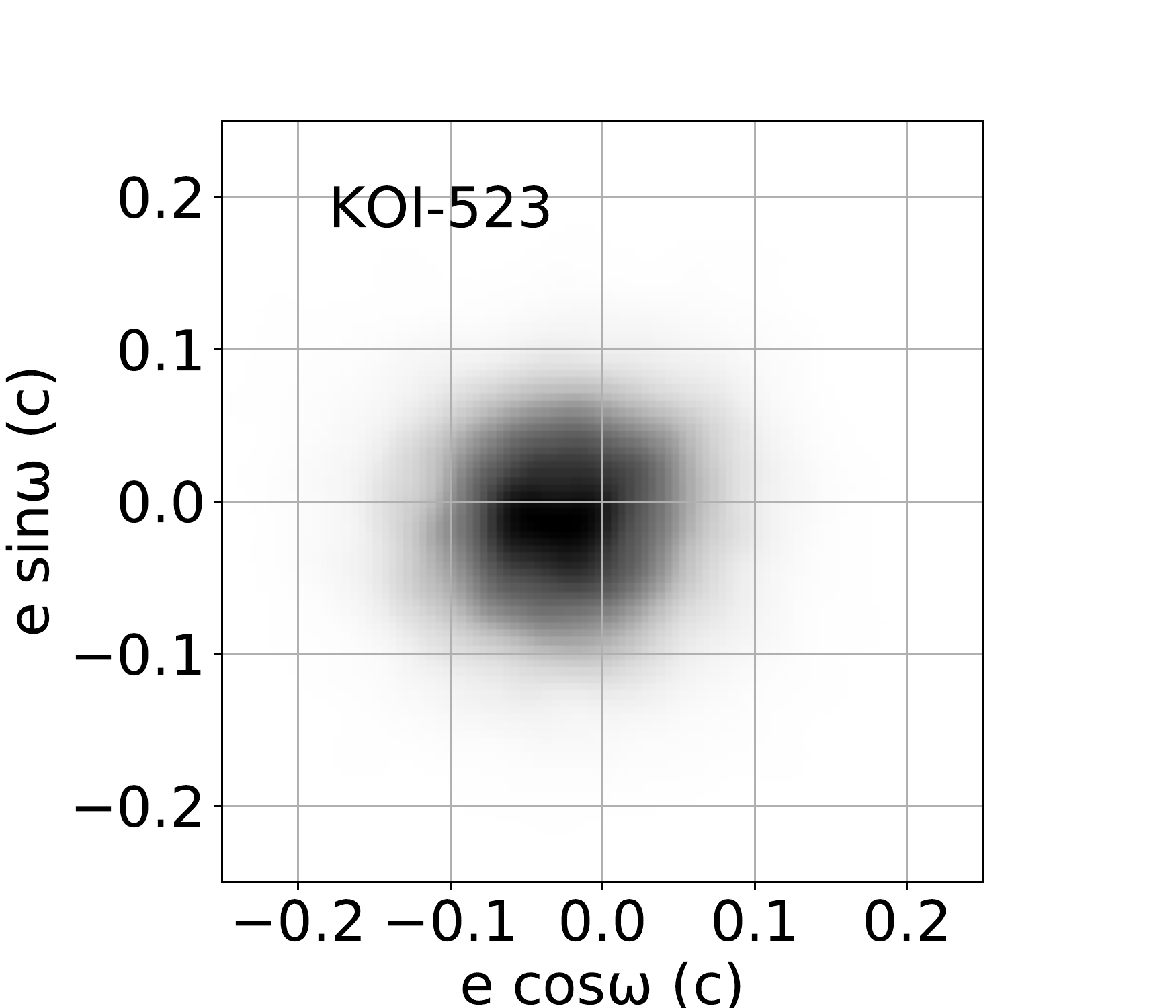}
\includegraphics [height = 1.1 in]{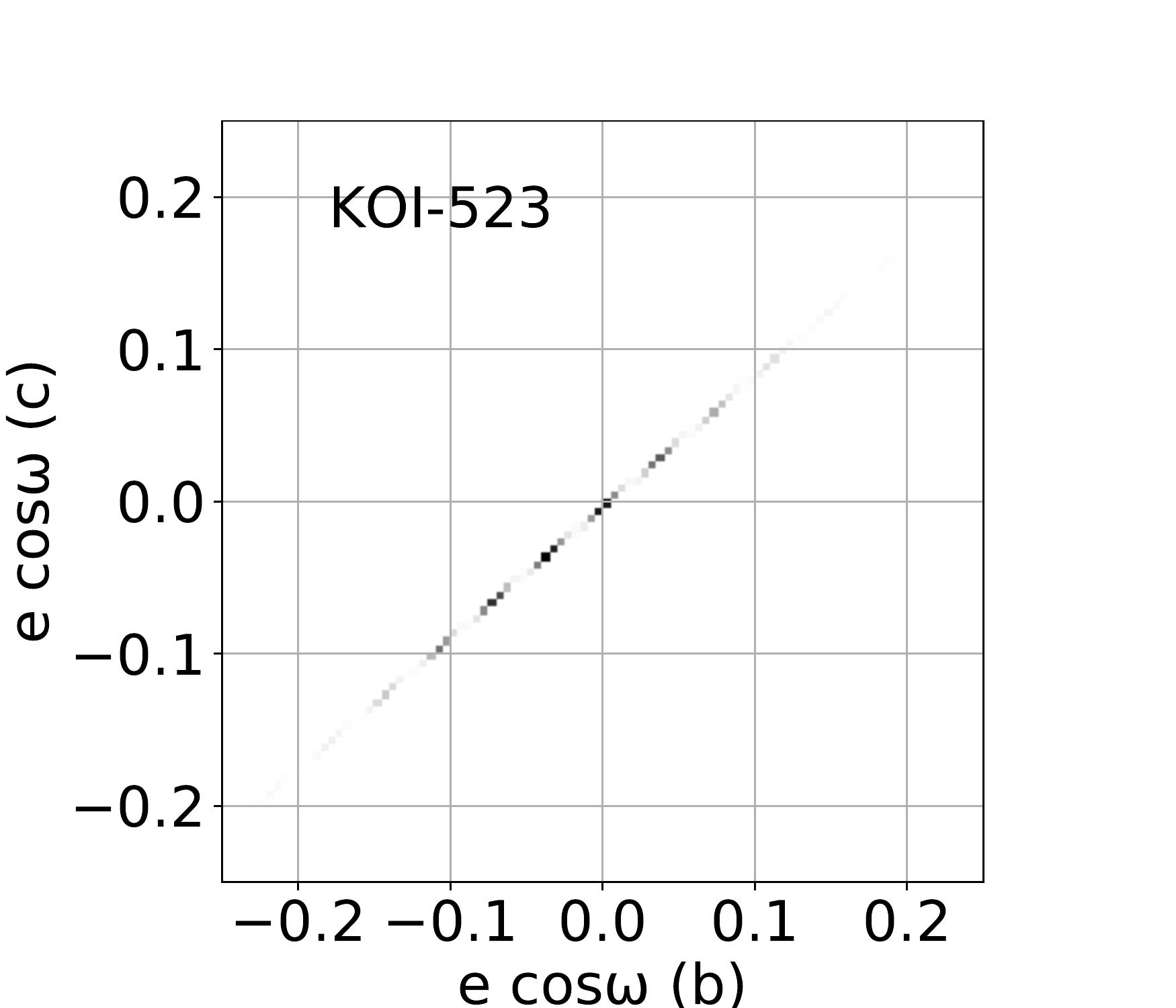}
\includegraphics [height = 1.1 in]{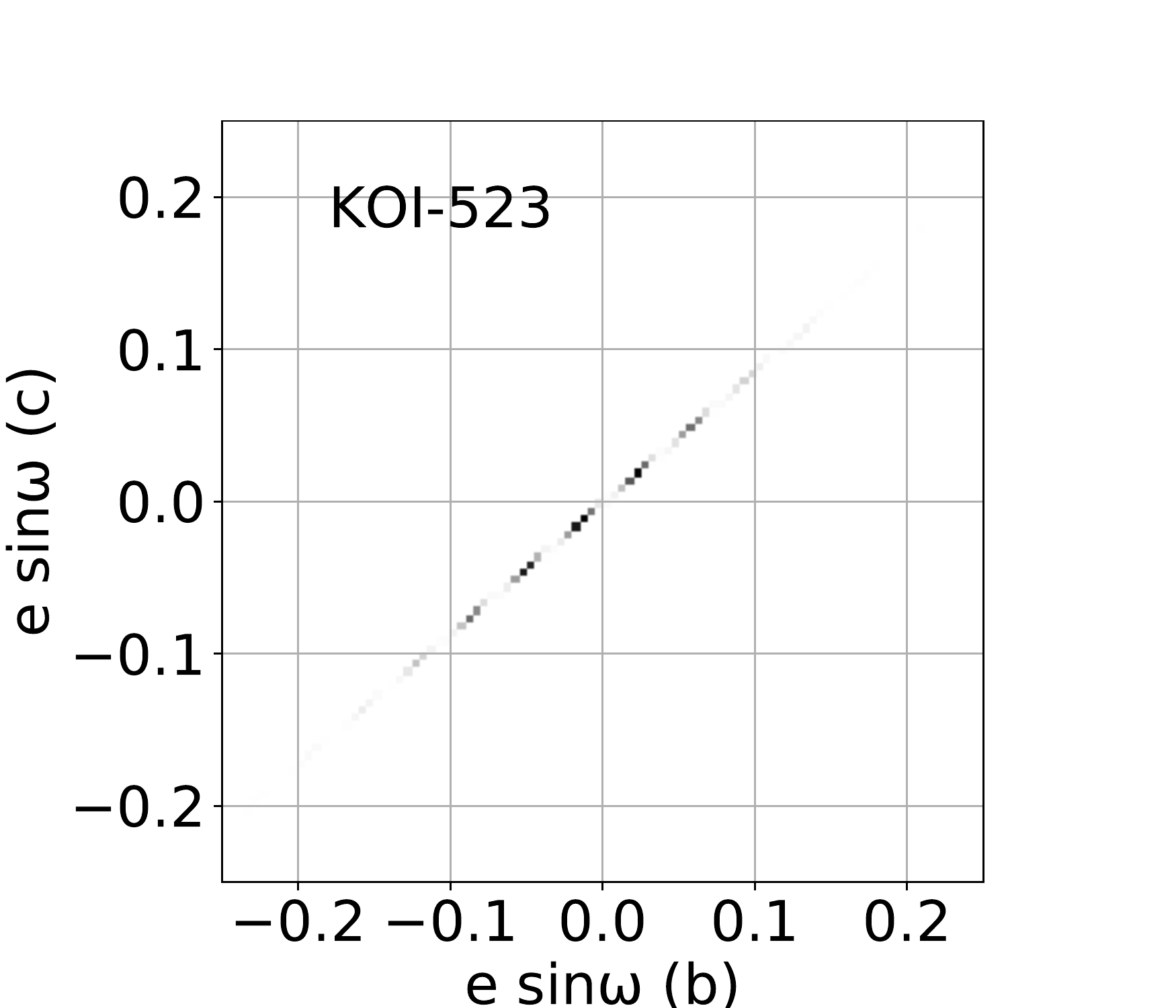} \\
\includegraphics [height = 1.1 in]{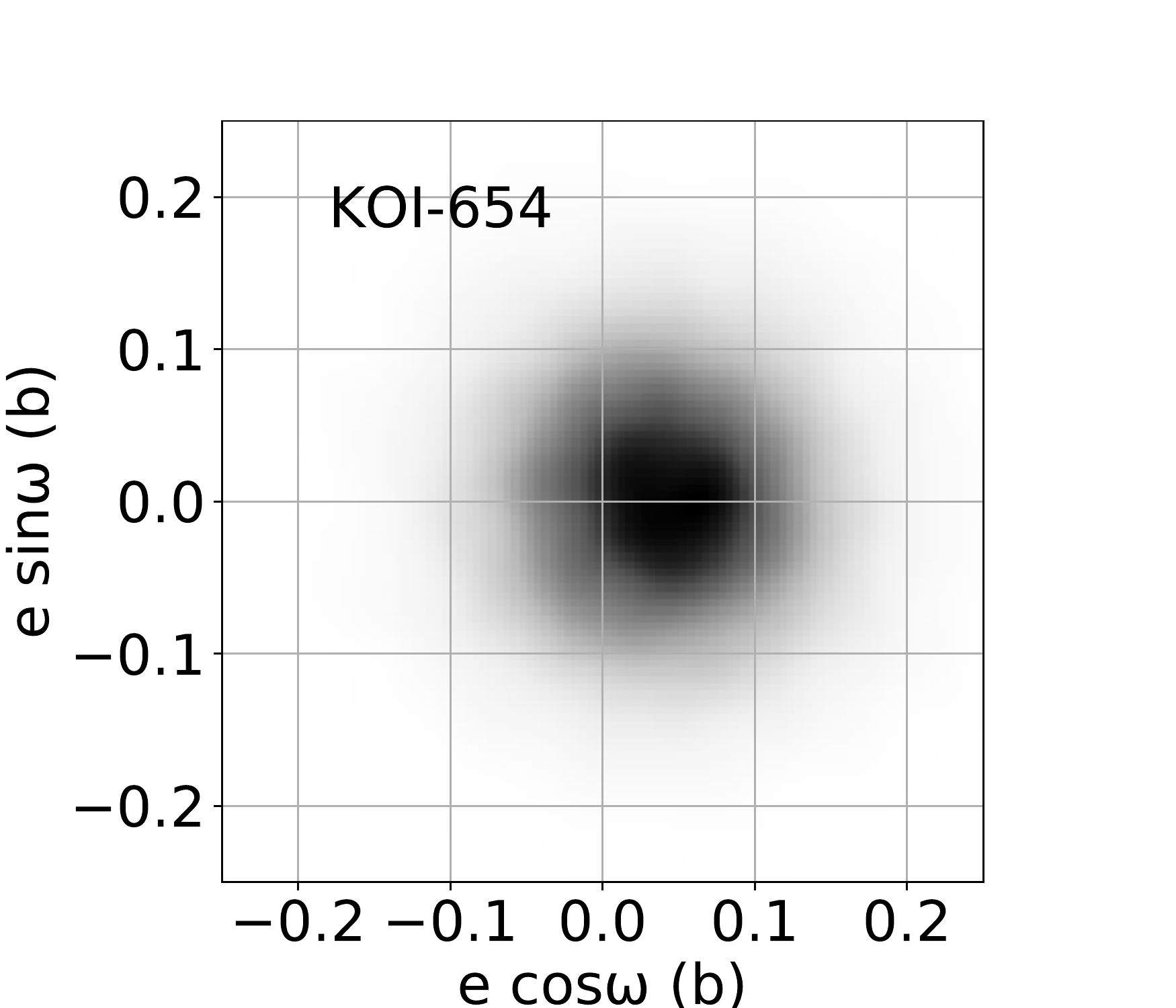}
\includegraphics [height = 1.1 in]{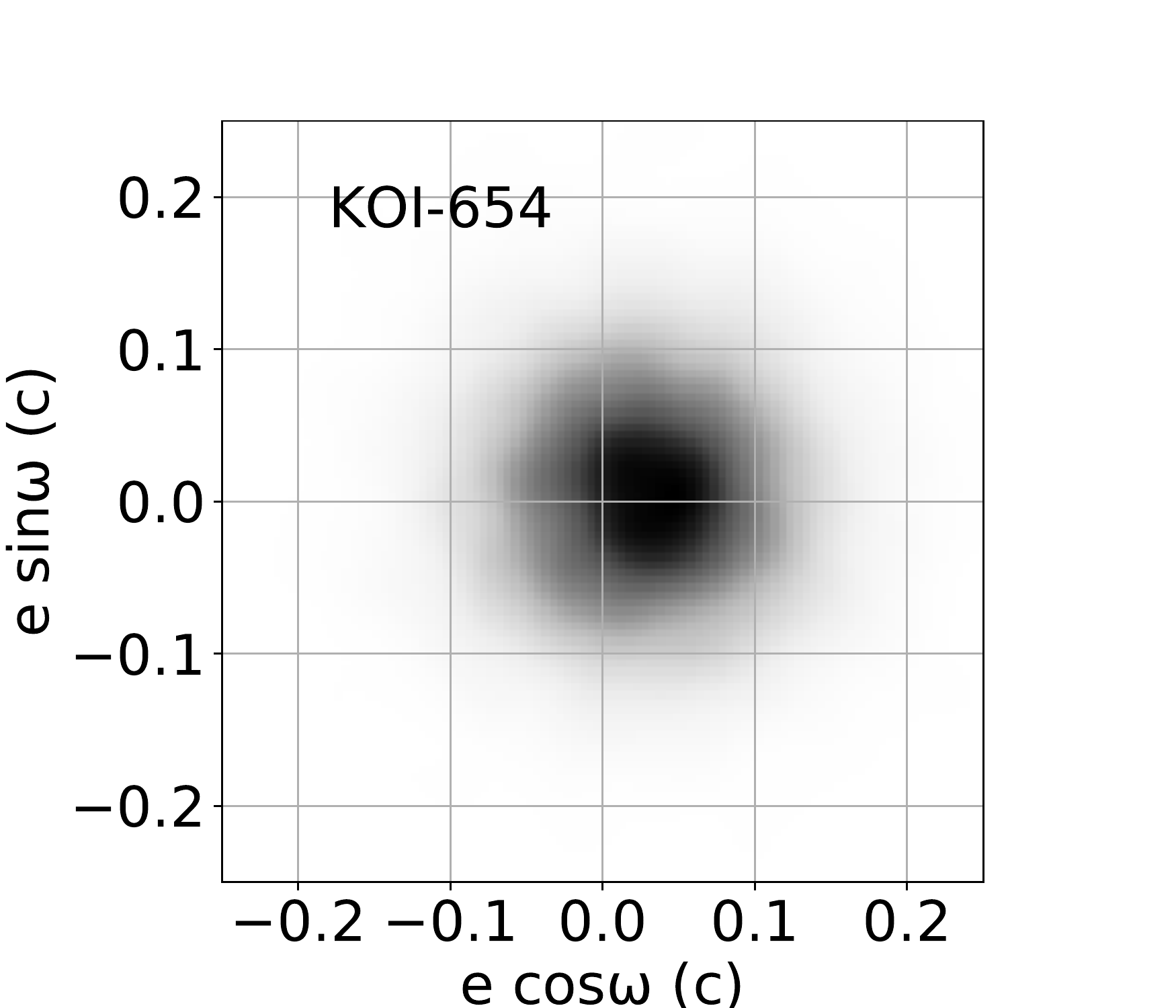}
\includegraphics [height = 1.1 in]{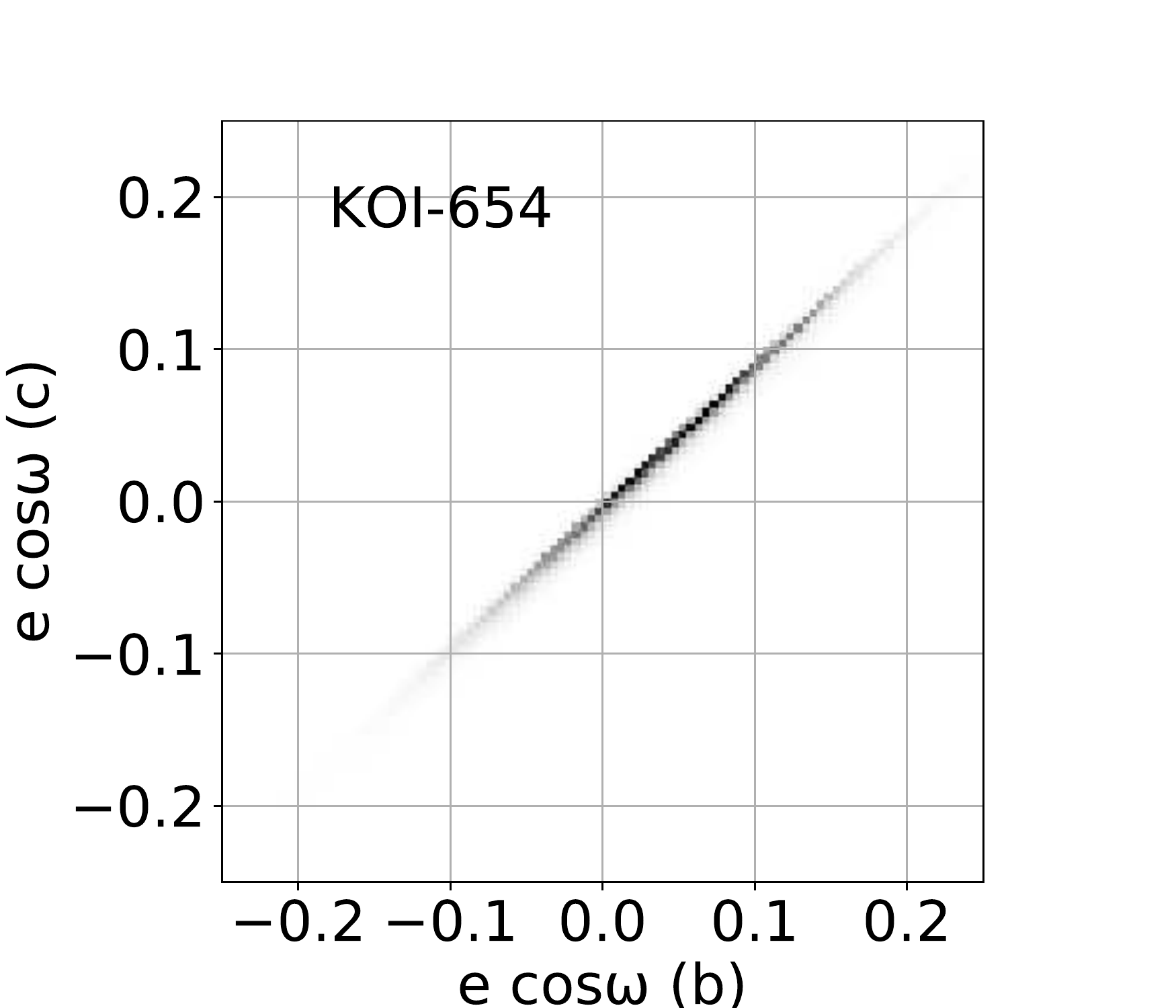}
\includegraphics [height = 1.1 in]{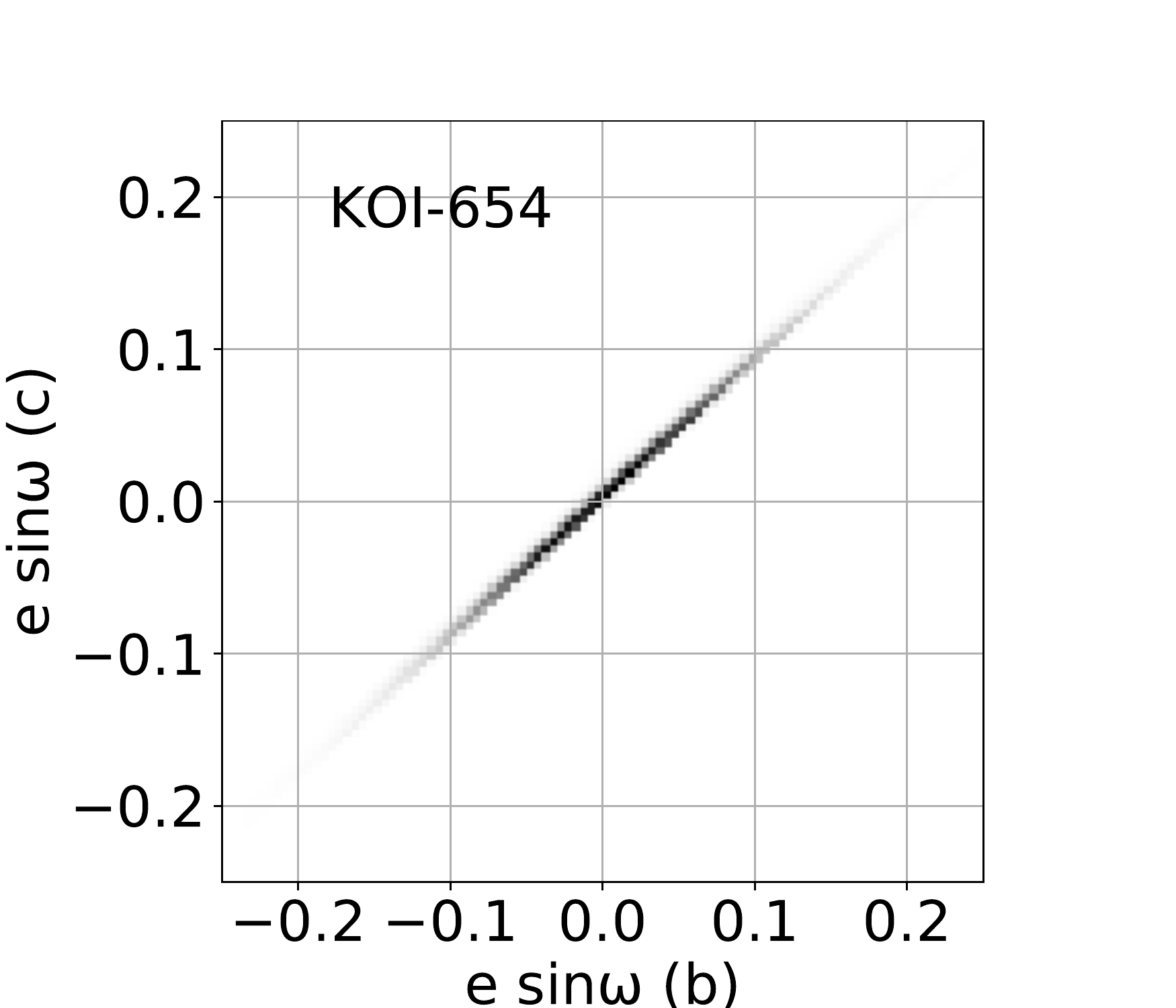} \\
\includegraphics [height = 1.1 in]{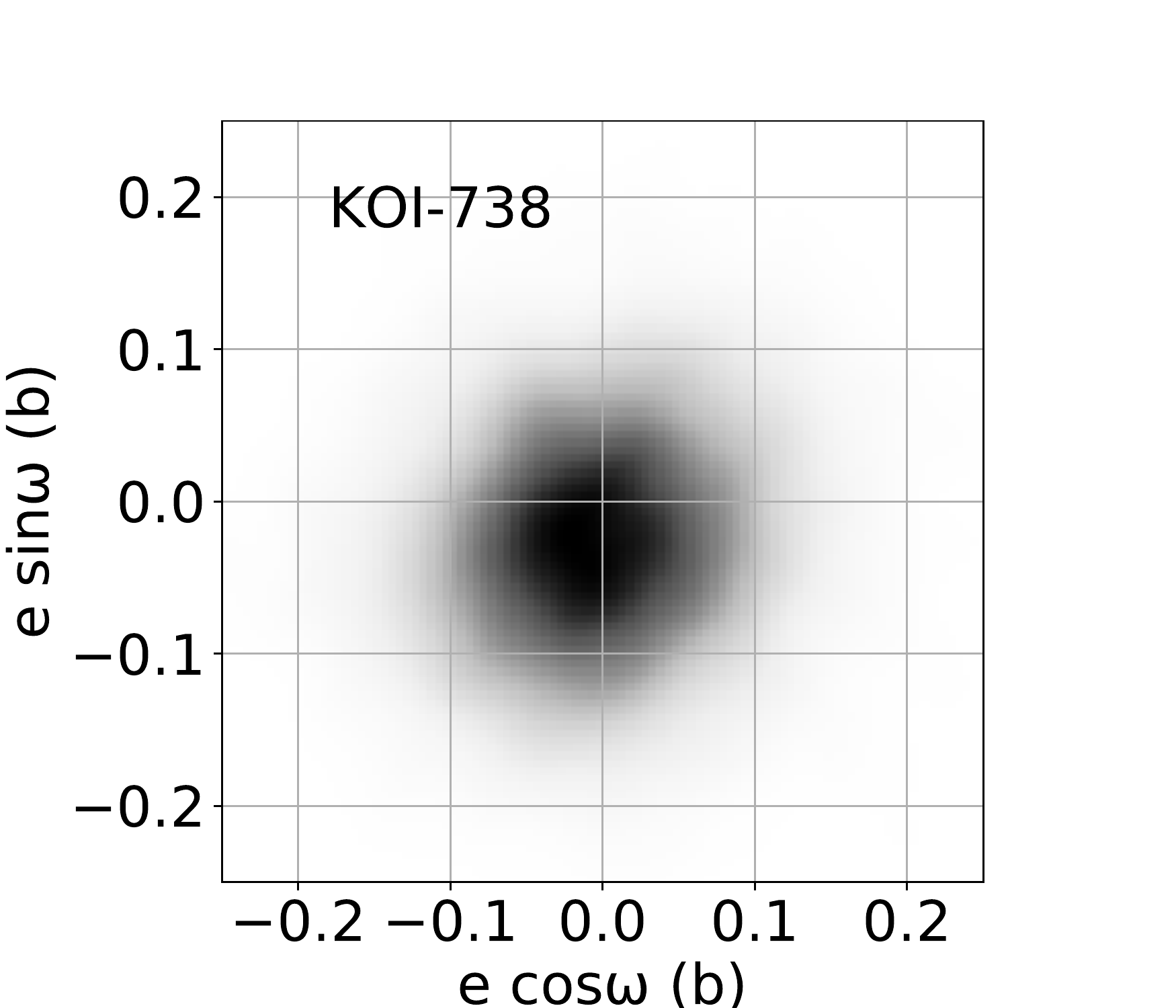}
\includegraphics [height = 1.1 in]{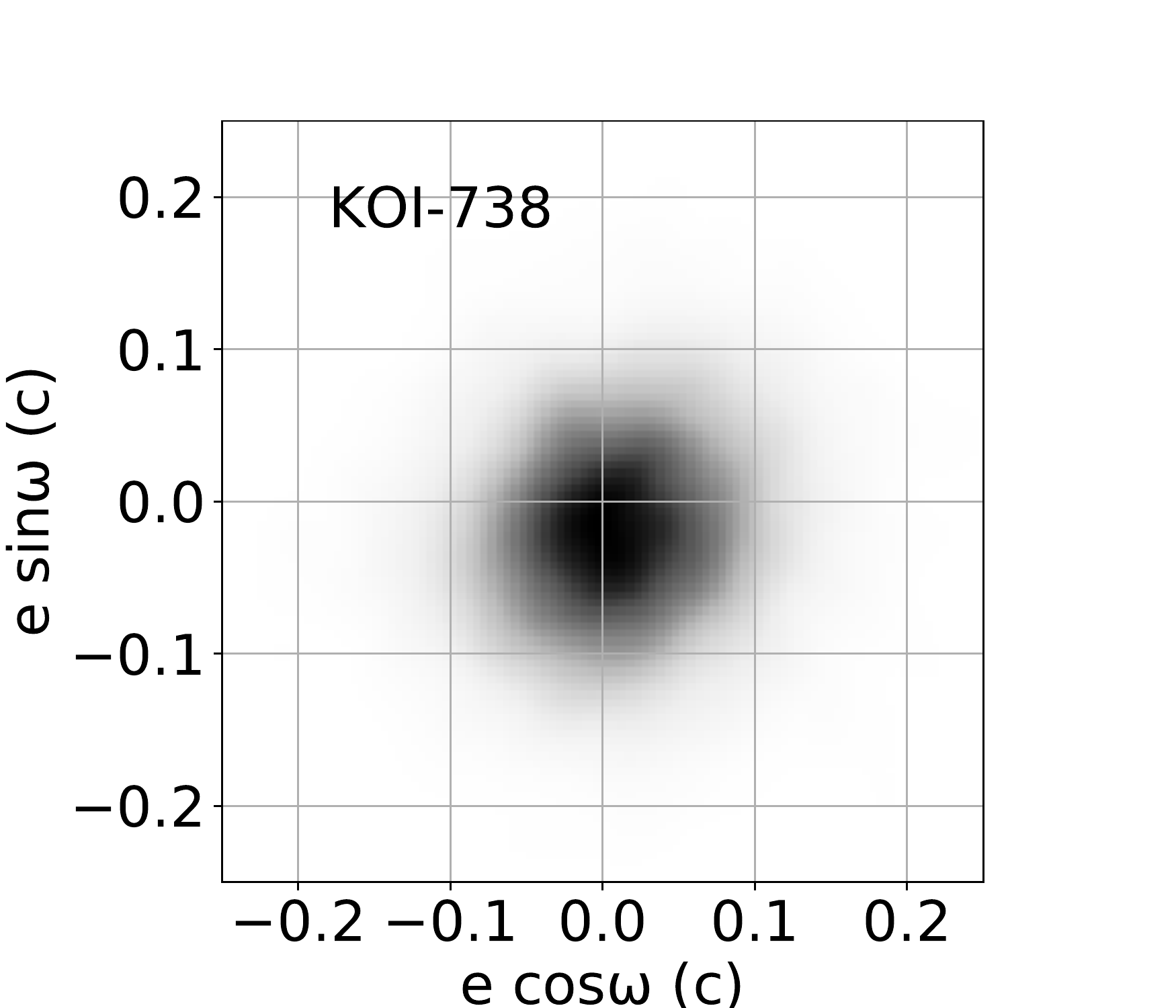}
\includegraphics [height = 1.1 in]{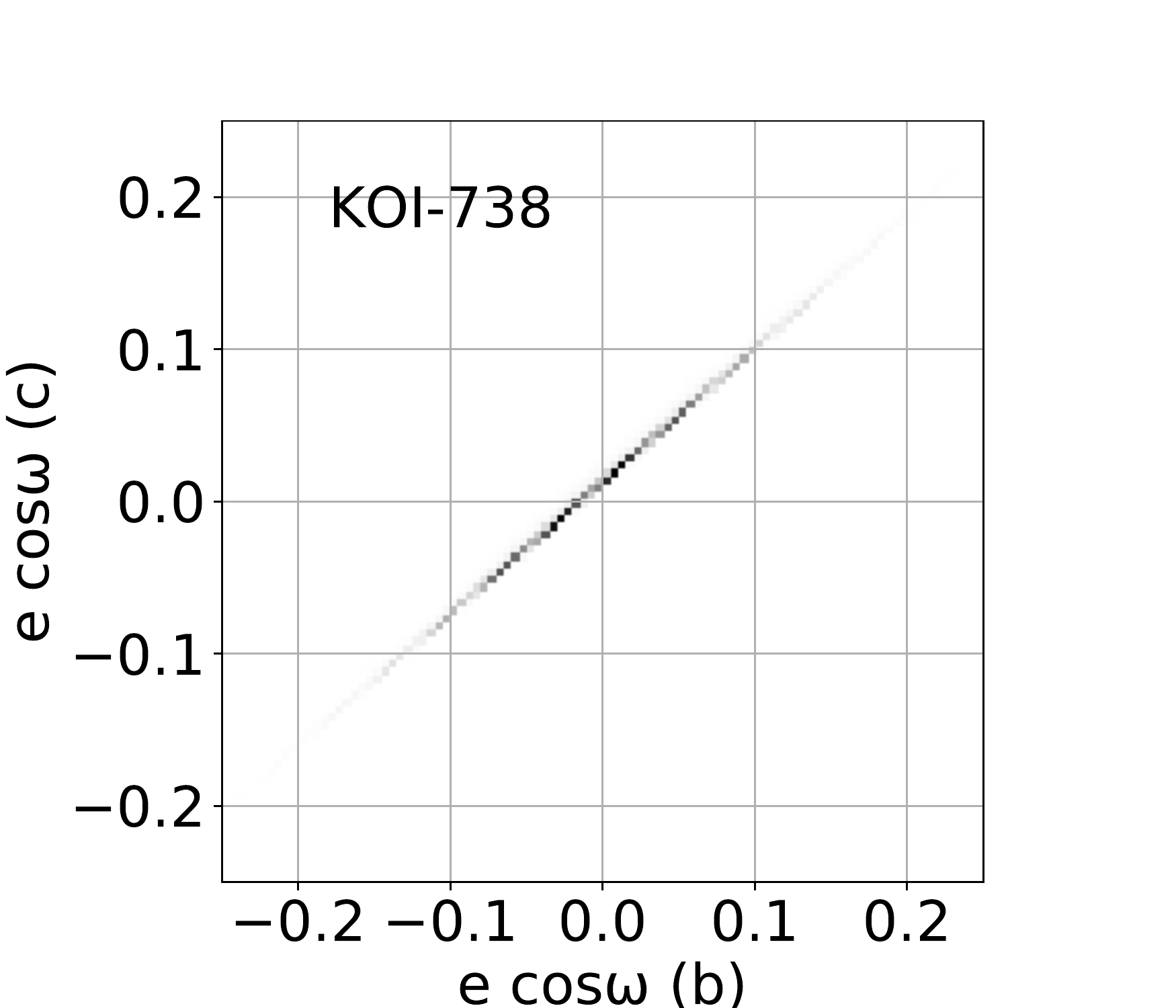}
\includegraphics [height = 1.1 in]{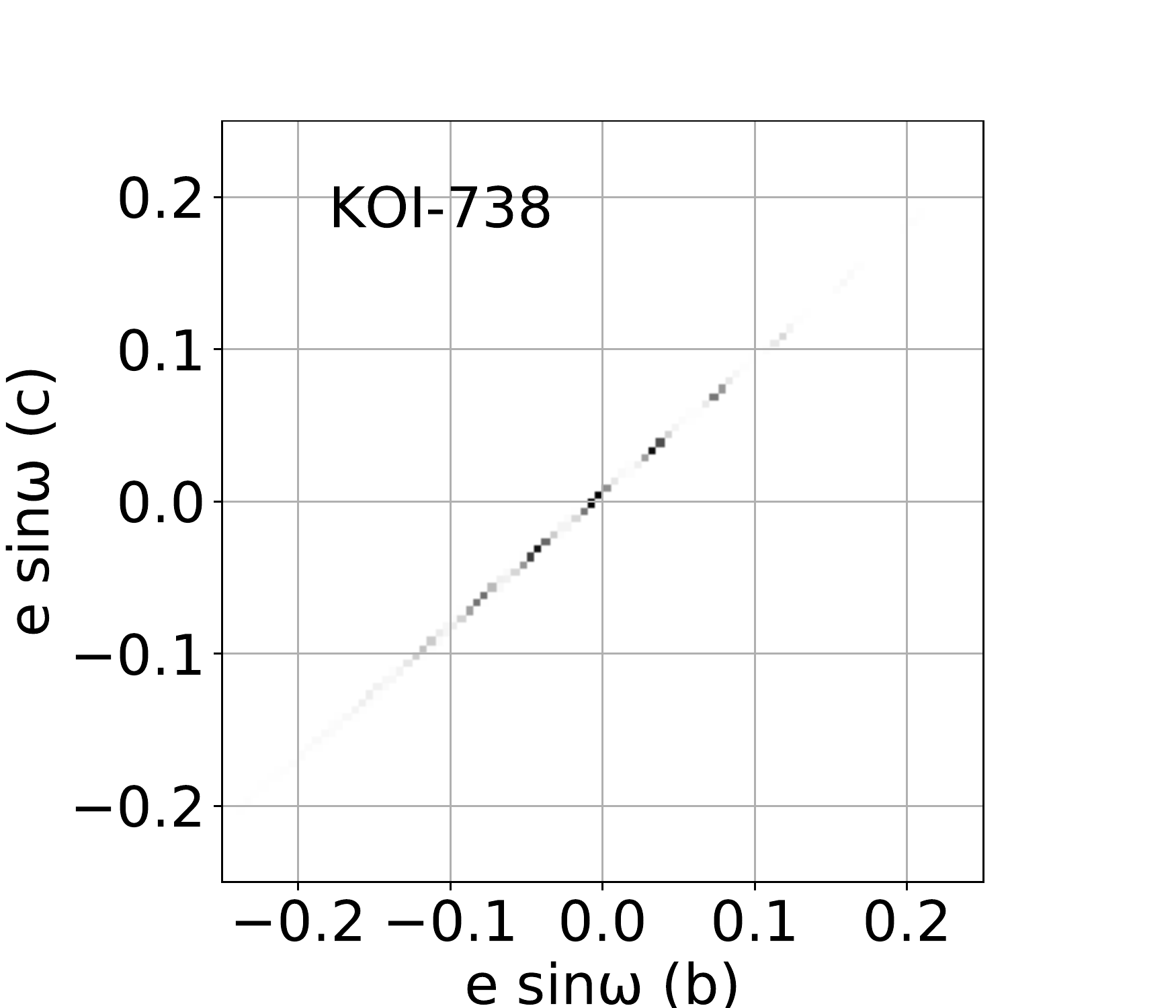} \\
\includegraphics [height = 1.1 in]{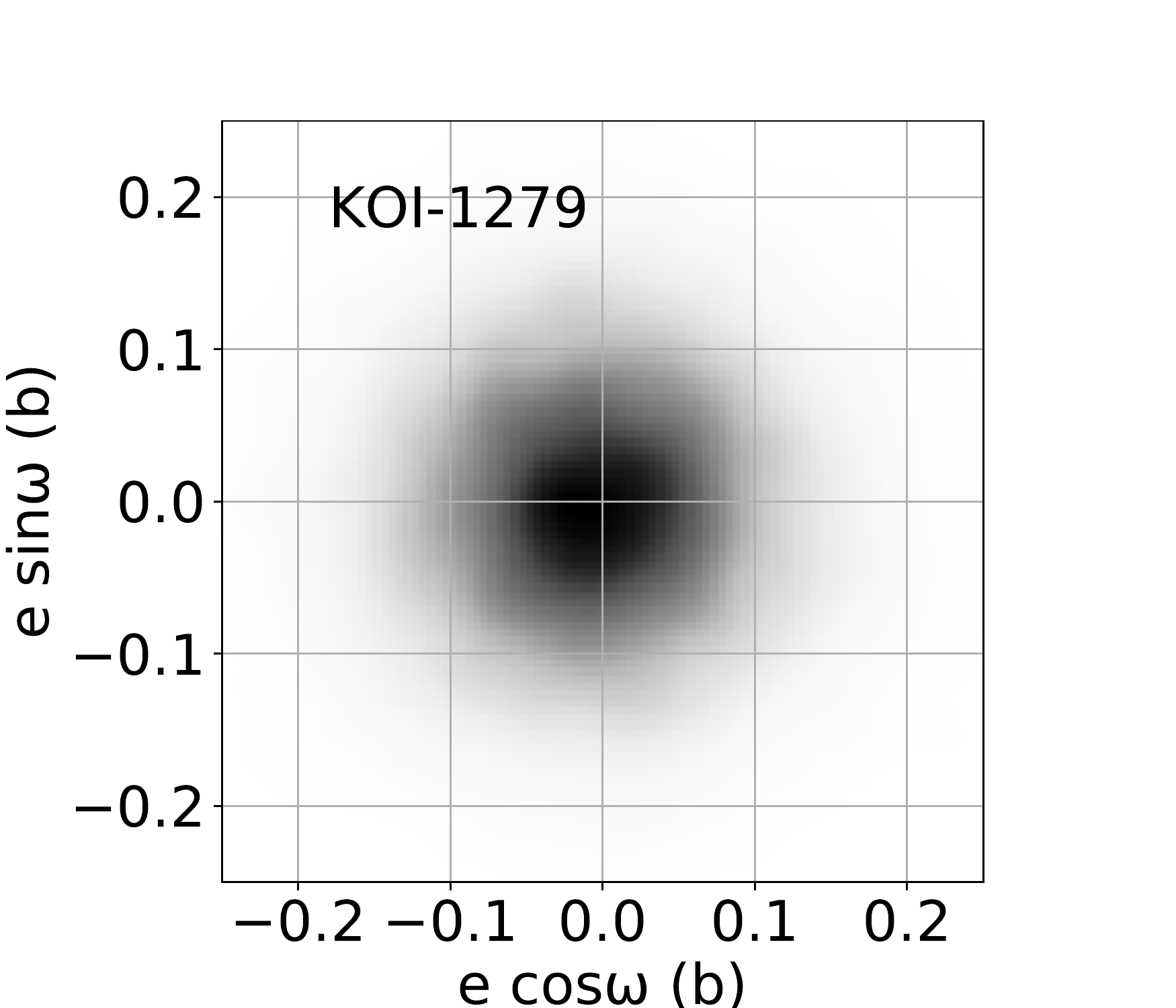}
\includegraphics [height = 1.1 in]{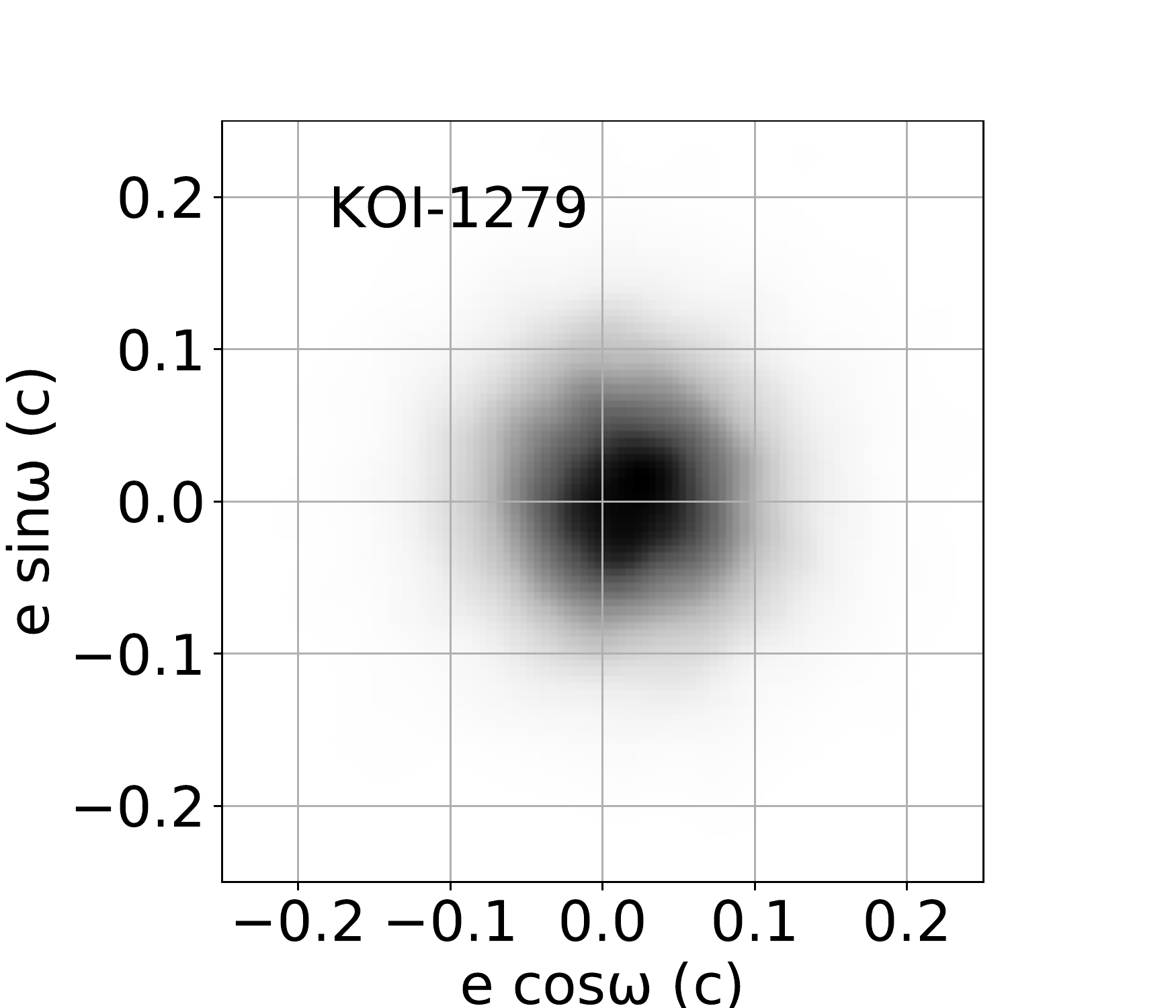}
\includegraphics [height = 1.1 in]{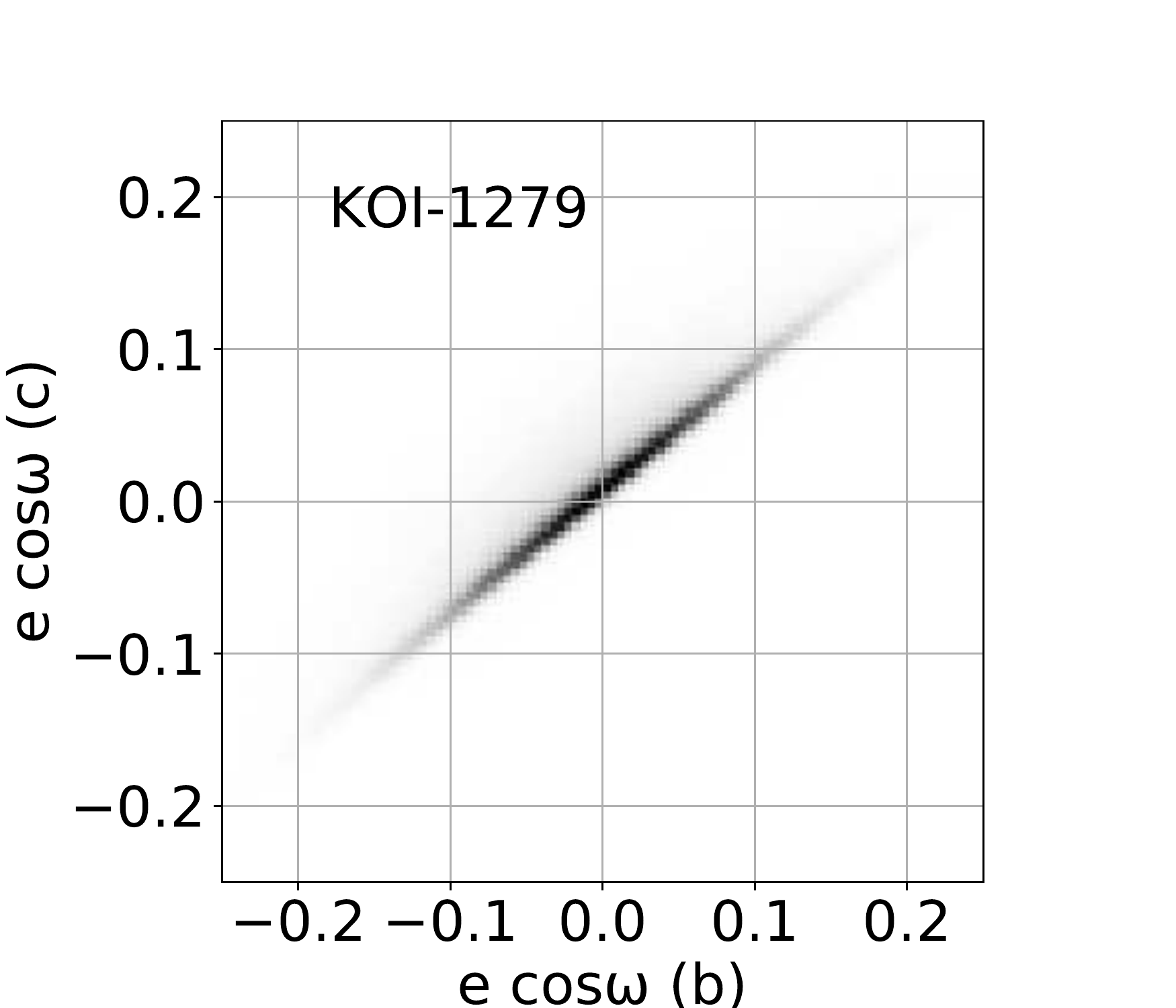}
\includegraphics [height = 1.1 in]{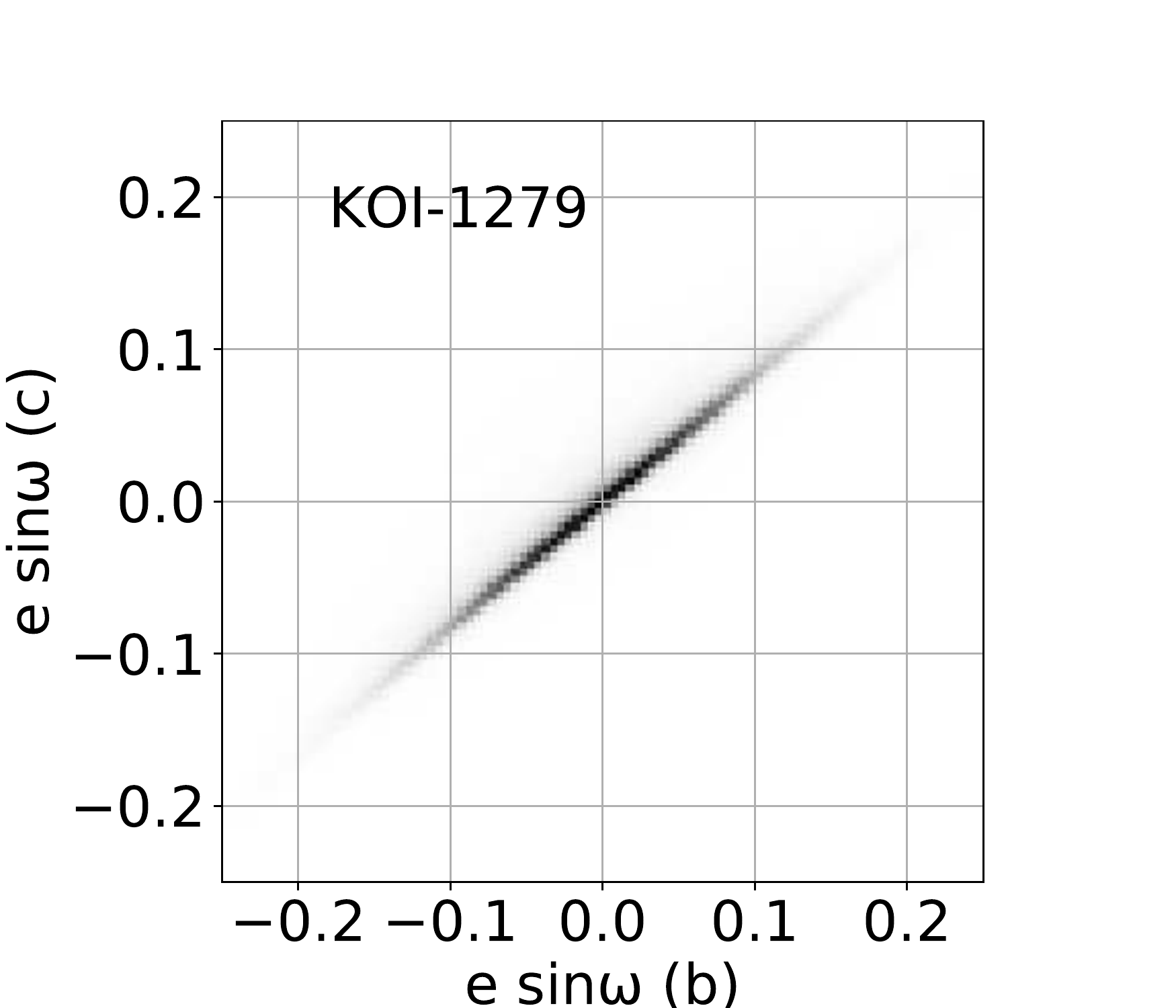} \\
\includegraphics [height = 1.1 in]{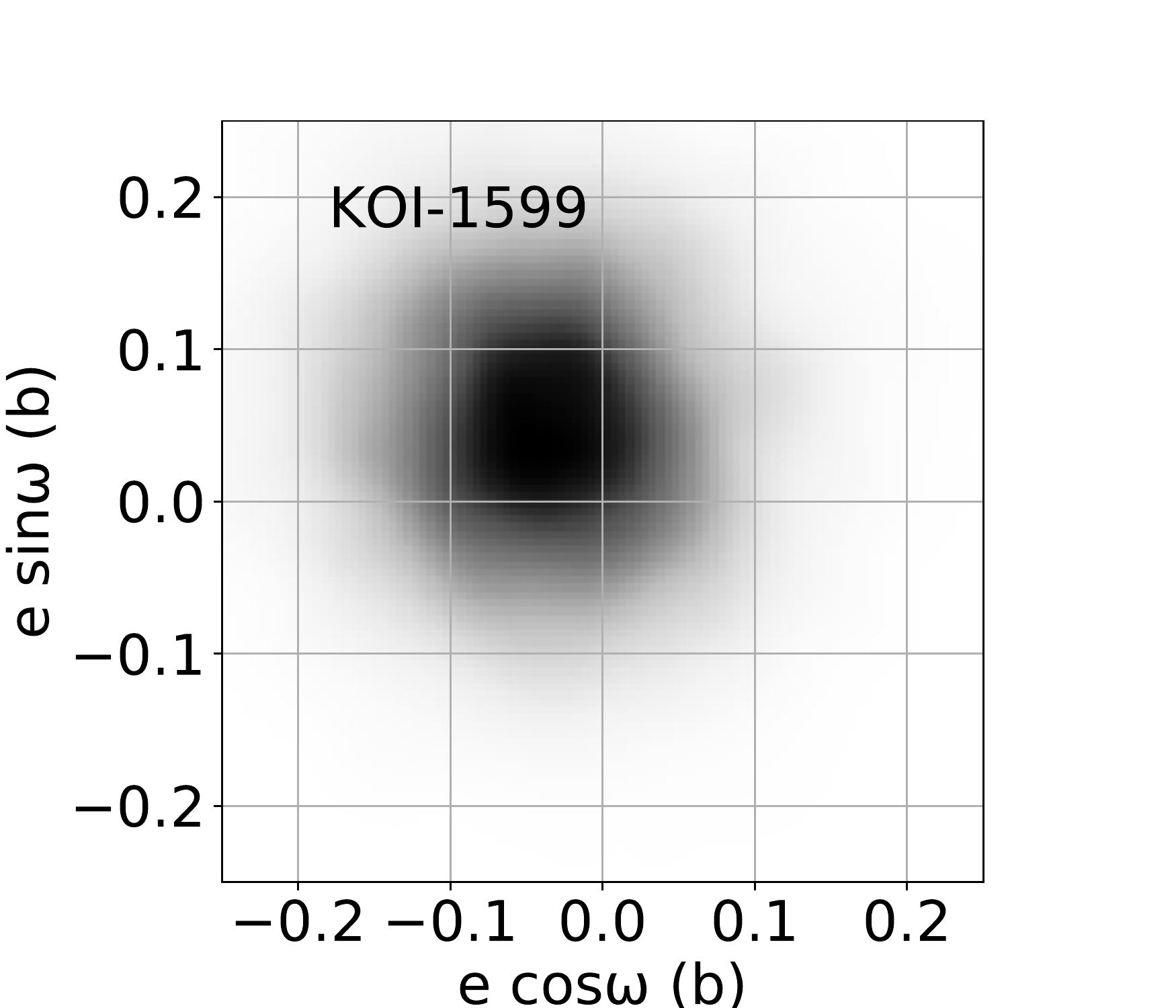}
\includegraphics [height = 1.1 in]{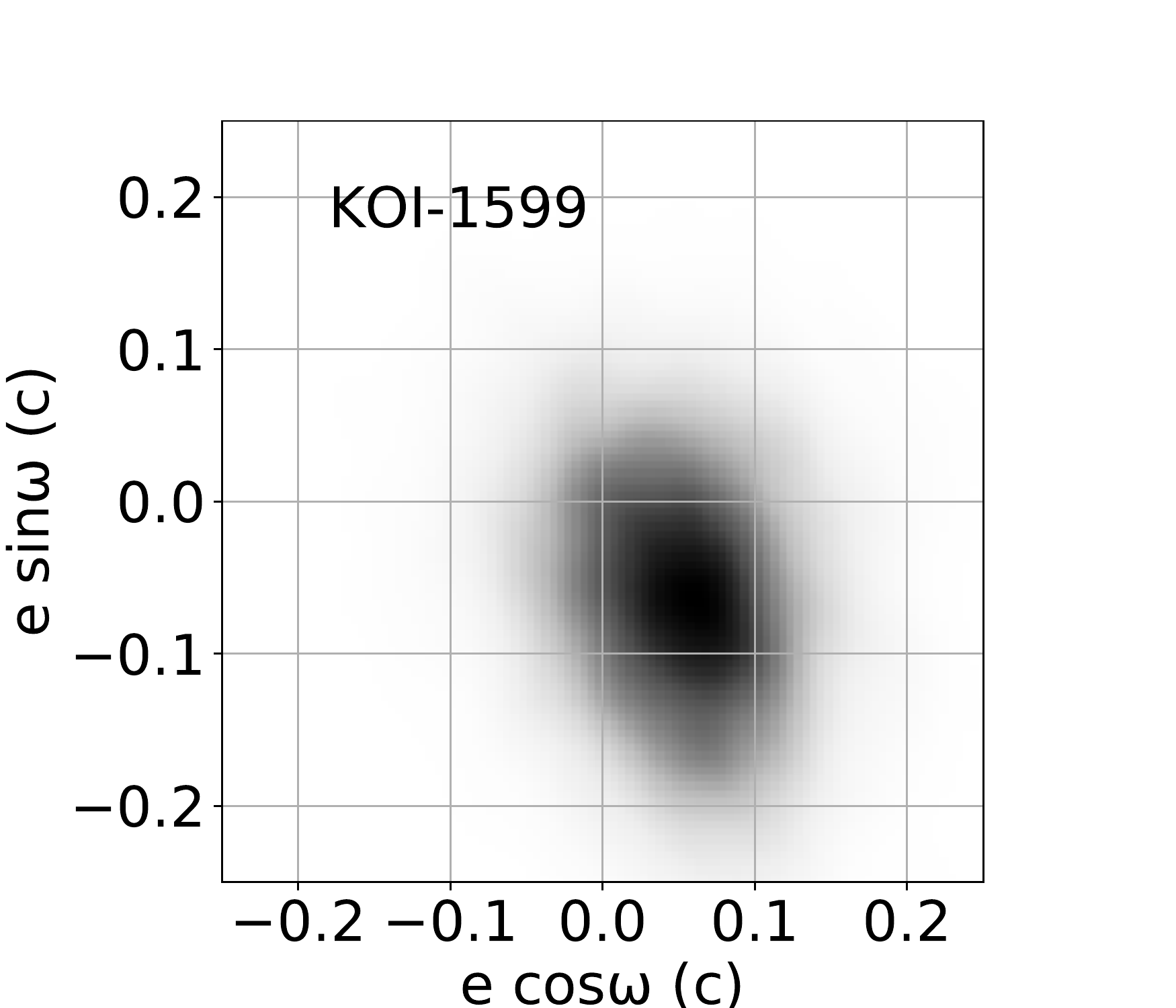} 
\includegraphics [height = 1.1 in]{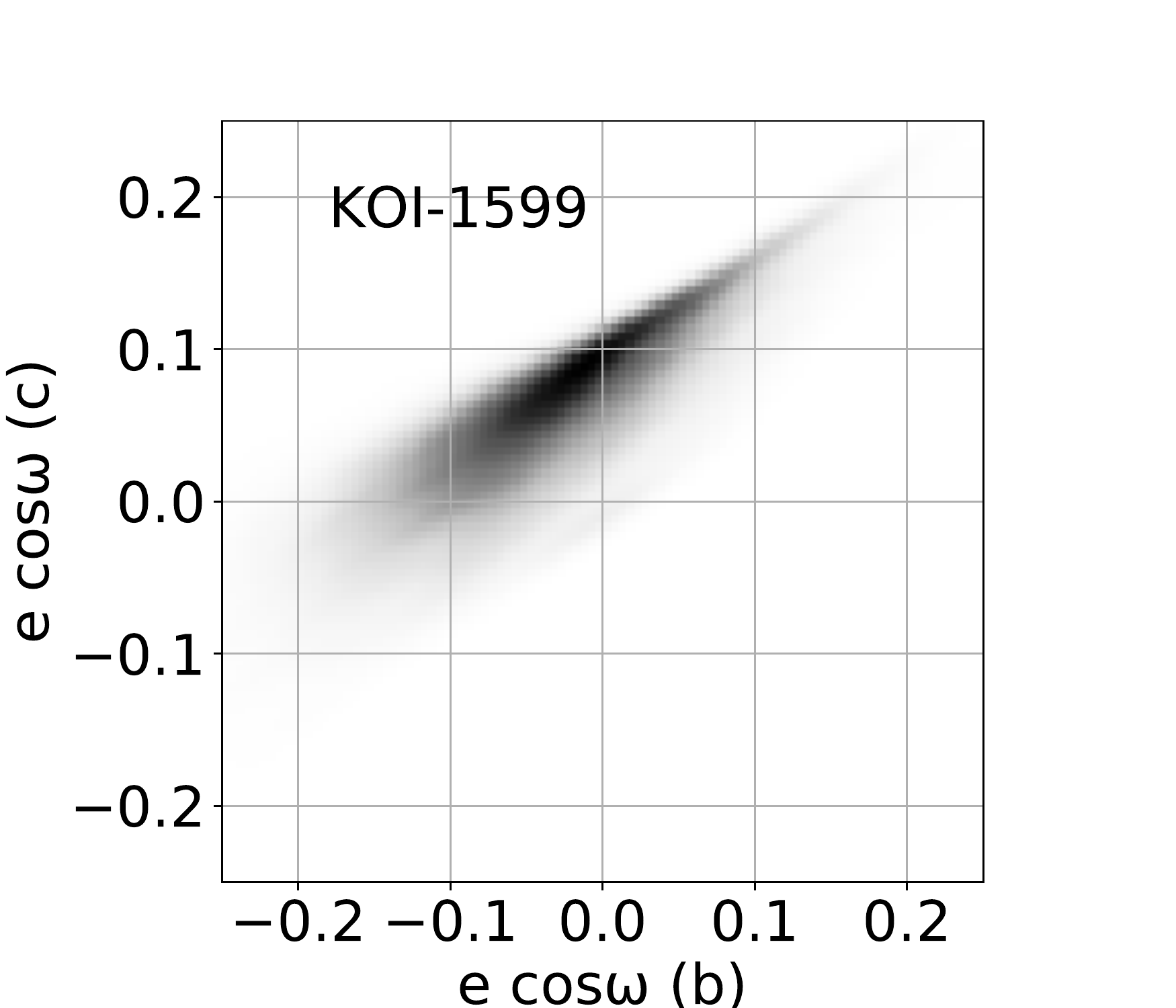}
\includegraphics [height = 1.1 in]{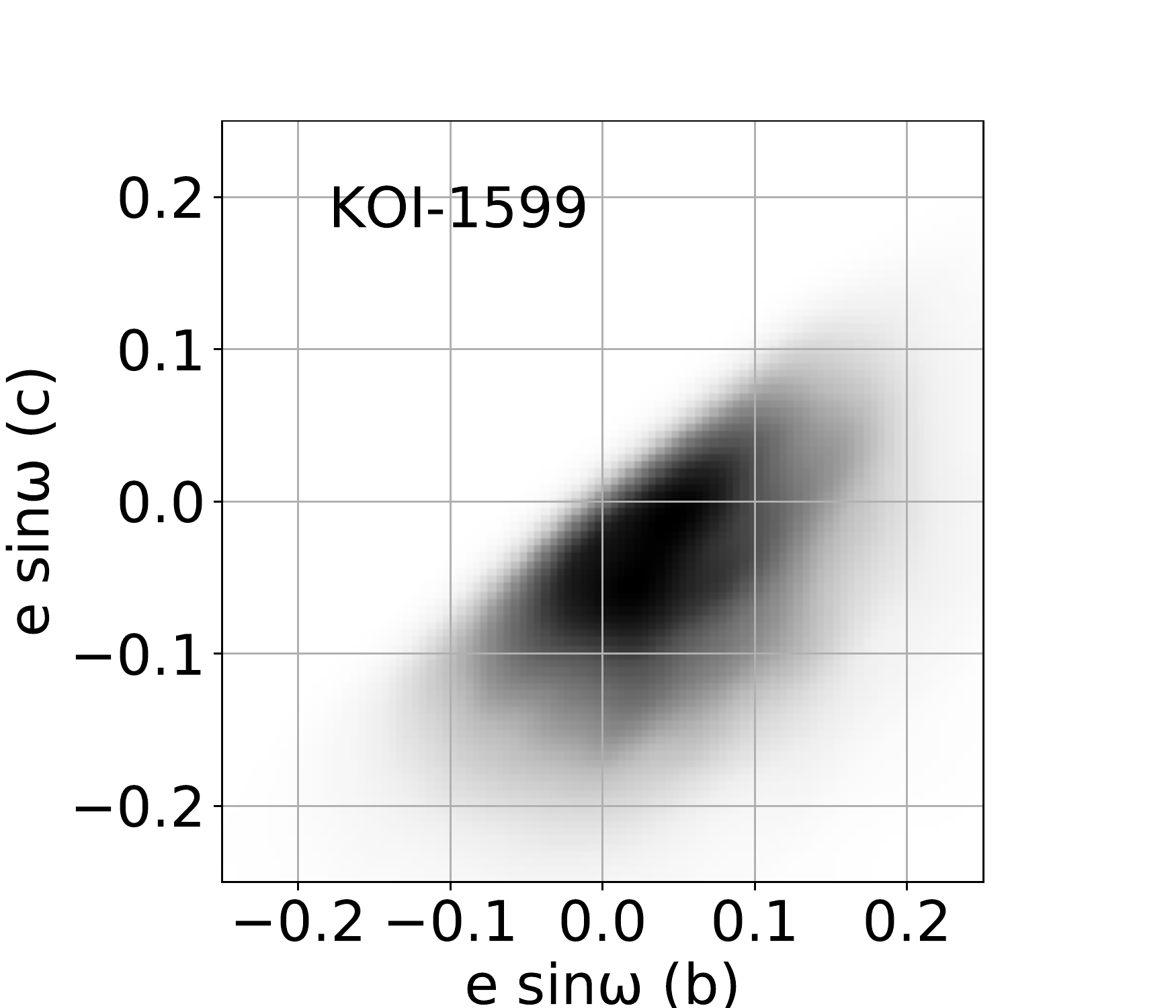} \\
\includegraphics [height = 1.1 in]{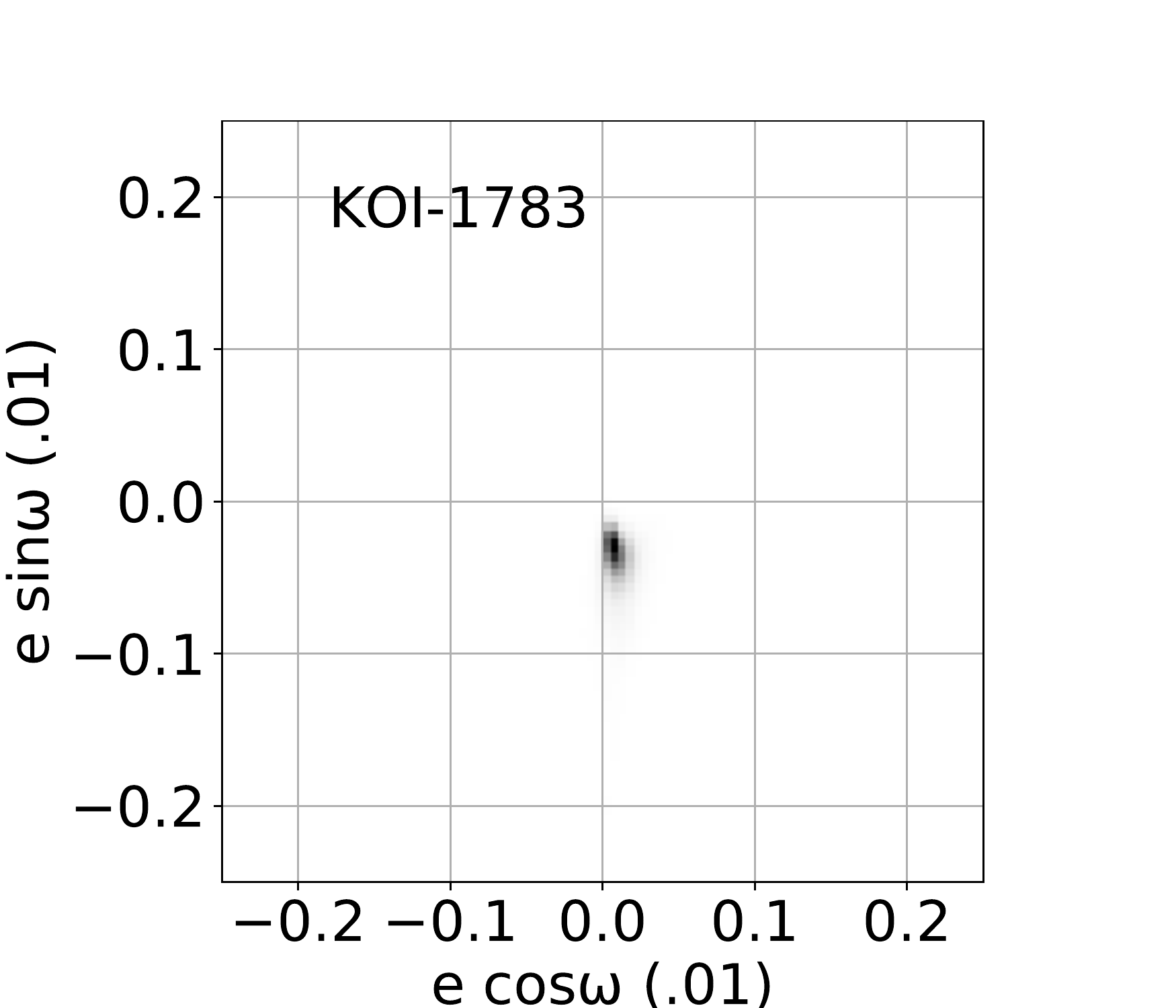}
\includegraphics [height = 1.1 in]{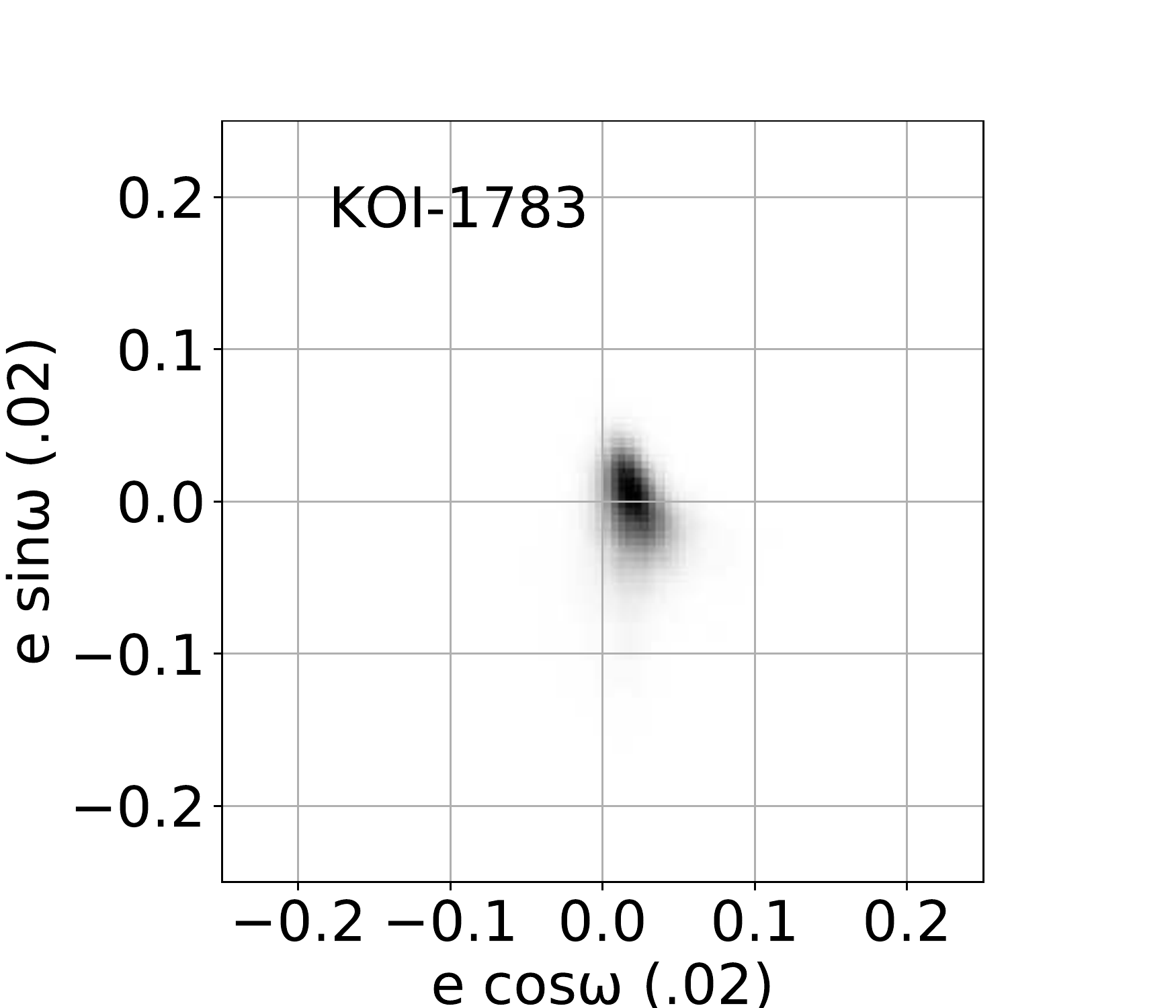} 
\includegraphics [height = 1.1 in]{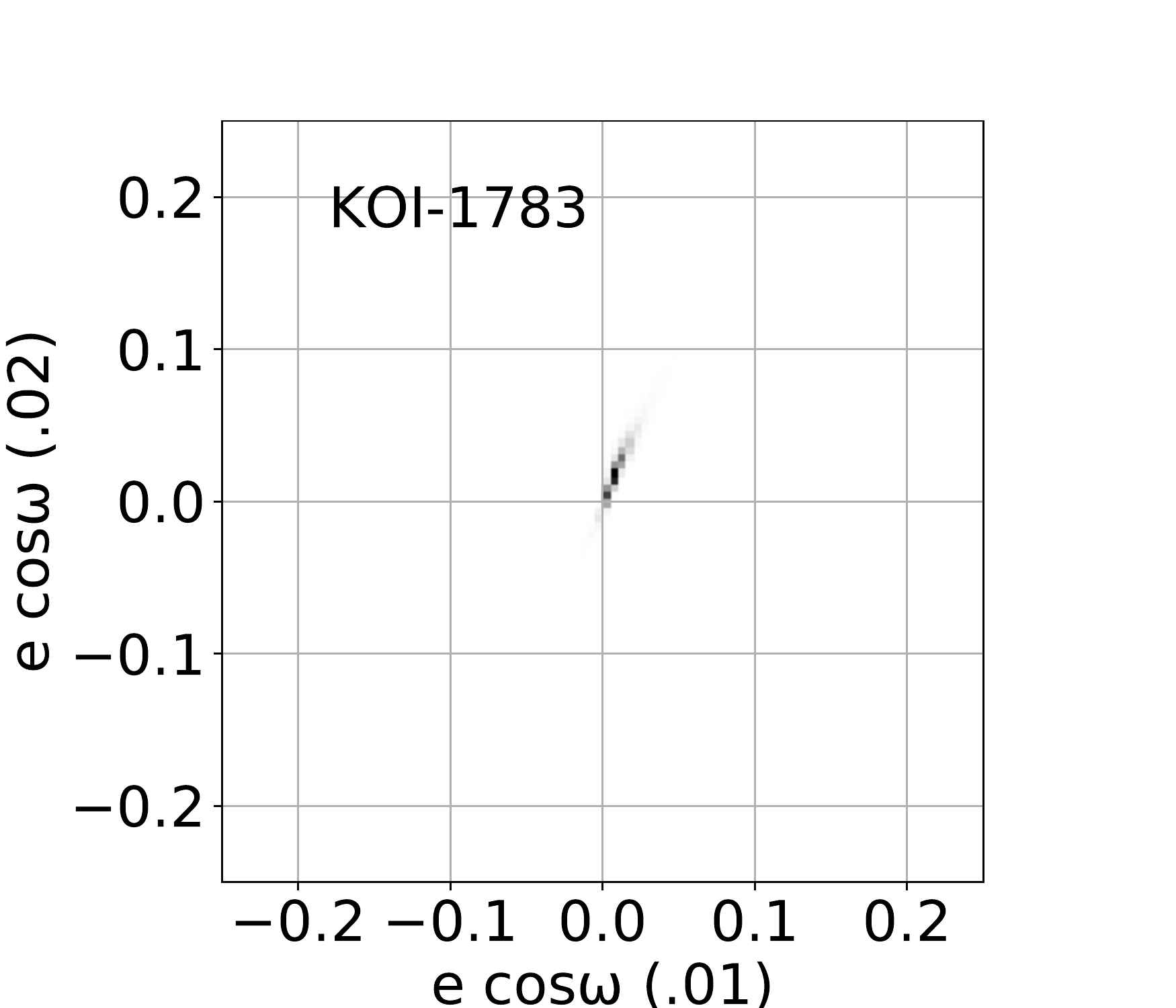}
\includegraphics [height = 1.1 in]{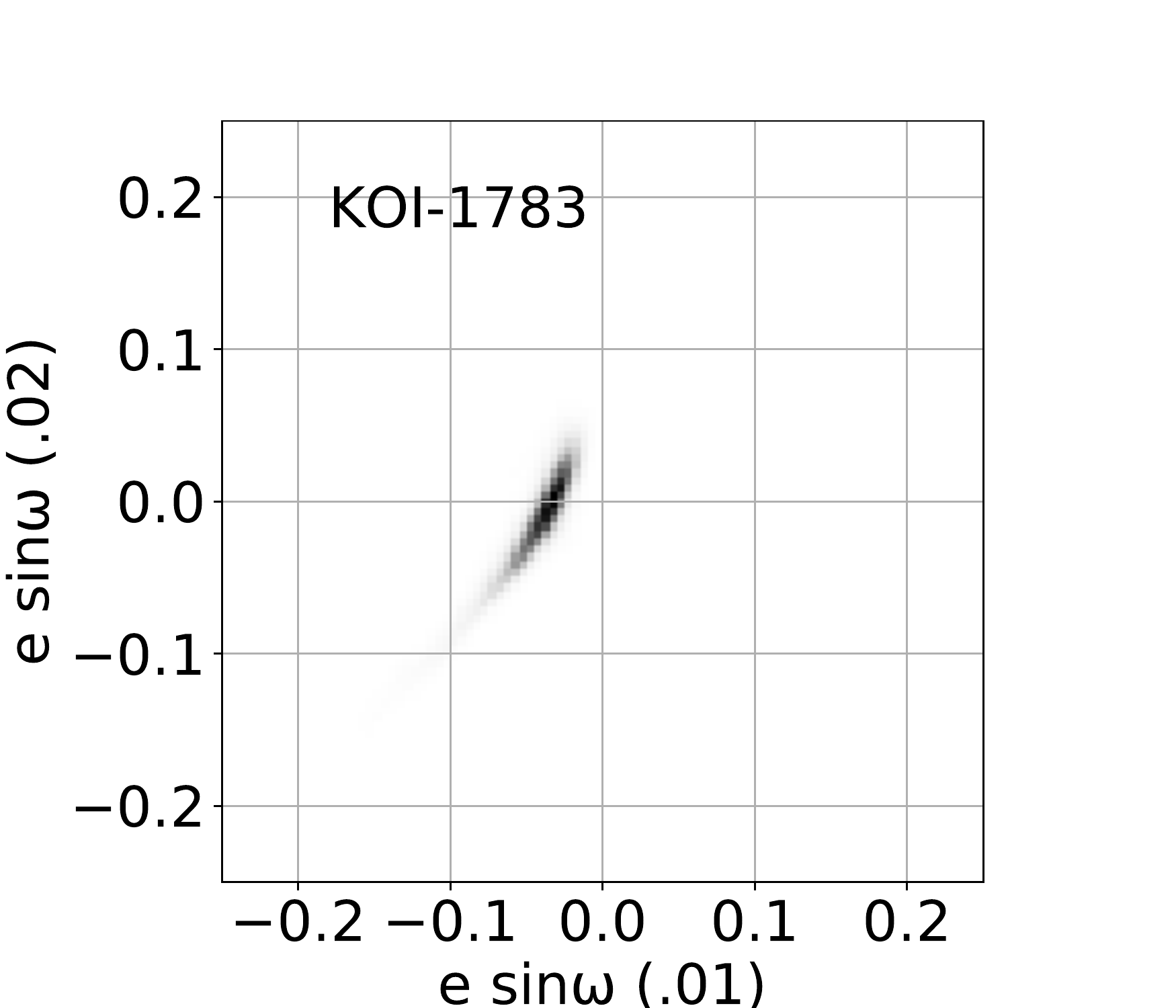} 
\caption{Two-dimensional kernel density estimators on joint posteriors of eccentricity vector components: two-planet systems. (Part 2 of 3.) 
}
\label{fig:ecc2b} 
\end{center}
\end{figure}

\begin{figure}
\begin{center}
\figurenum{22}
\includegraphics [height = 1.1 in]{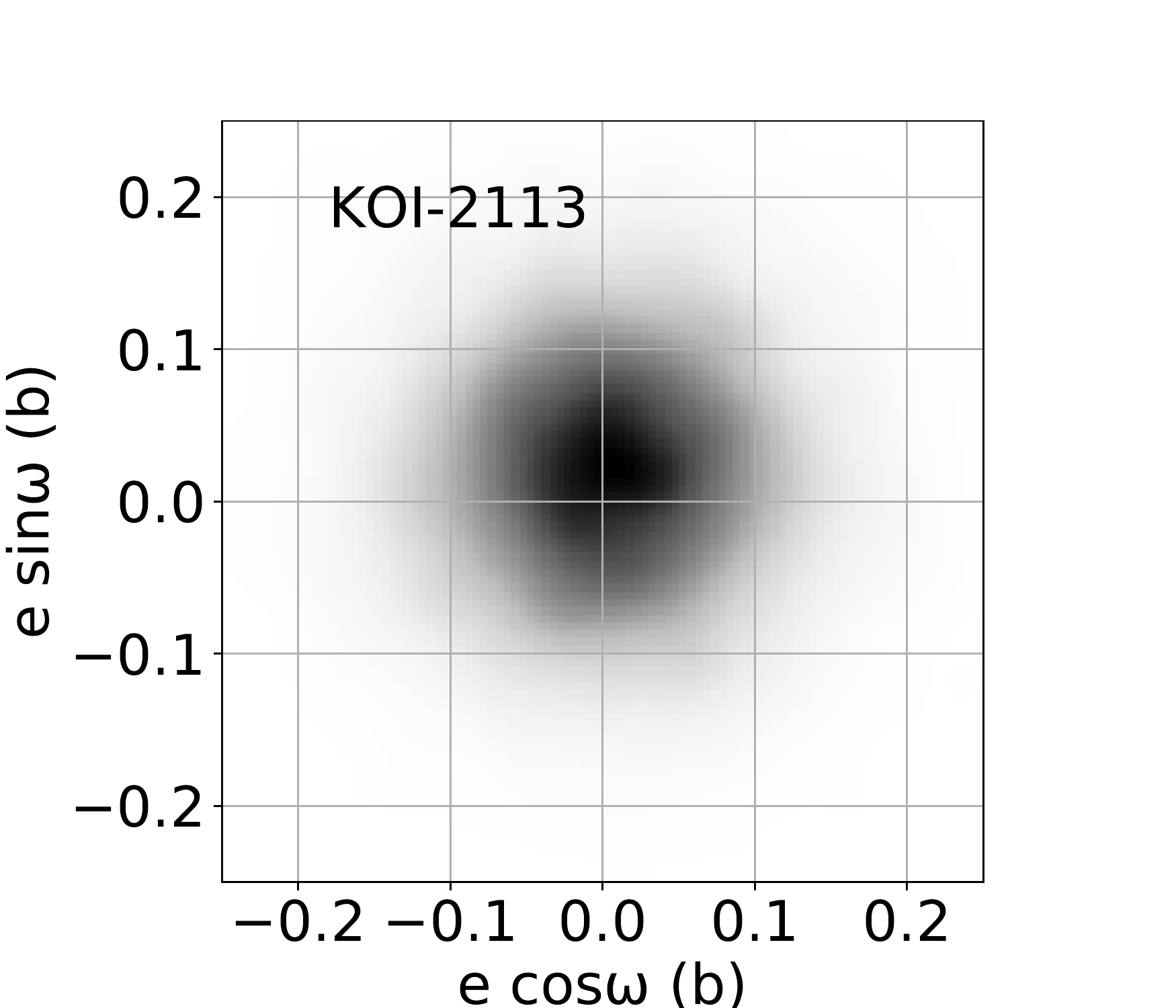}
\includegraphics [height = 1.1 in]{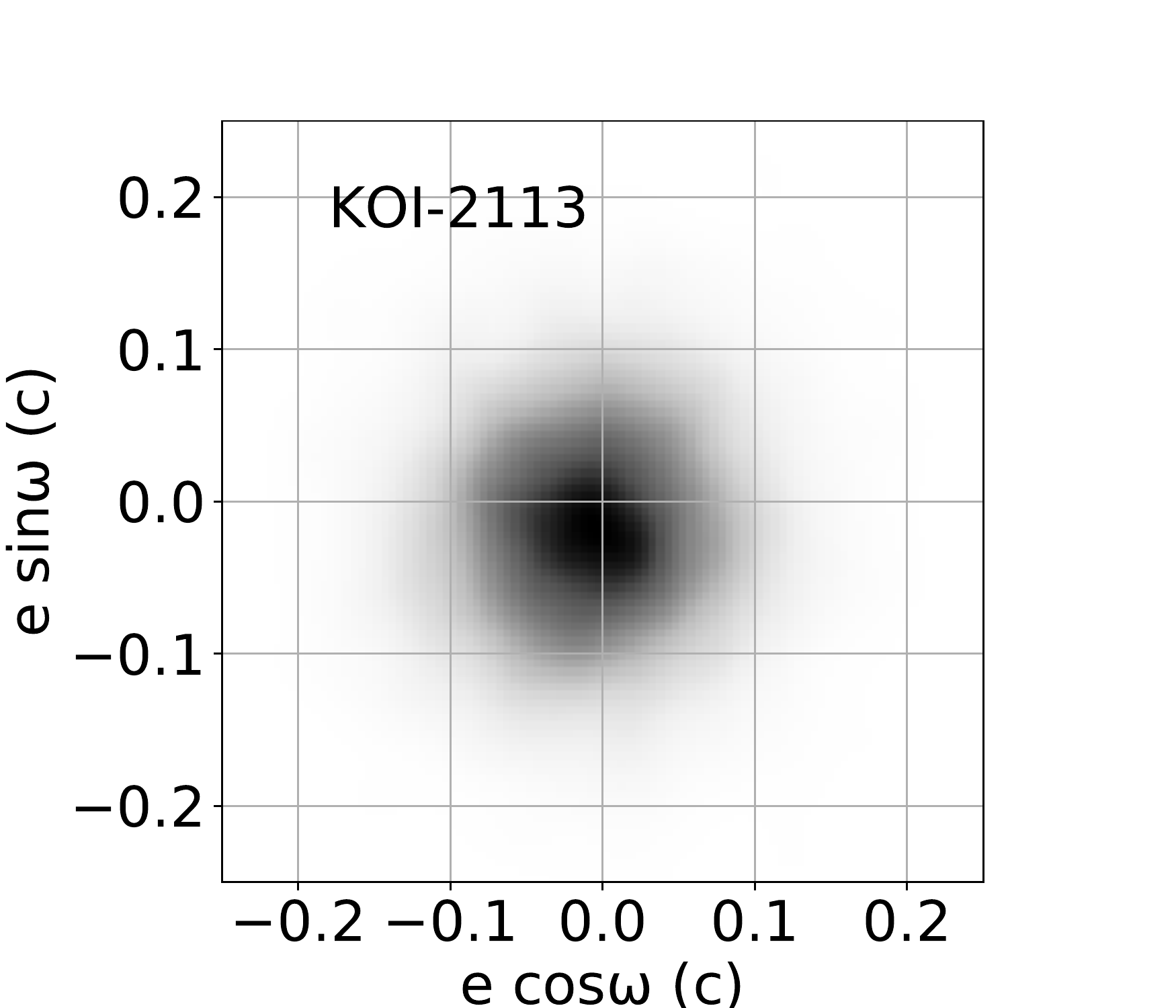}
\includegraphics [height = 1.1 in]{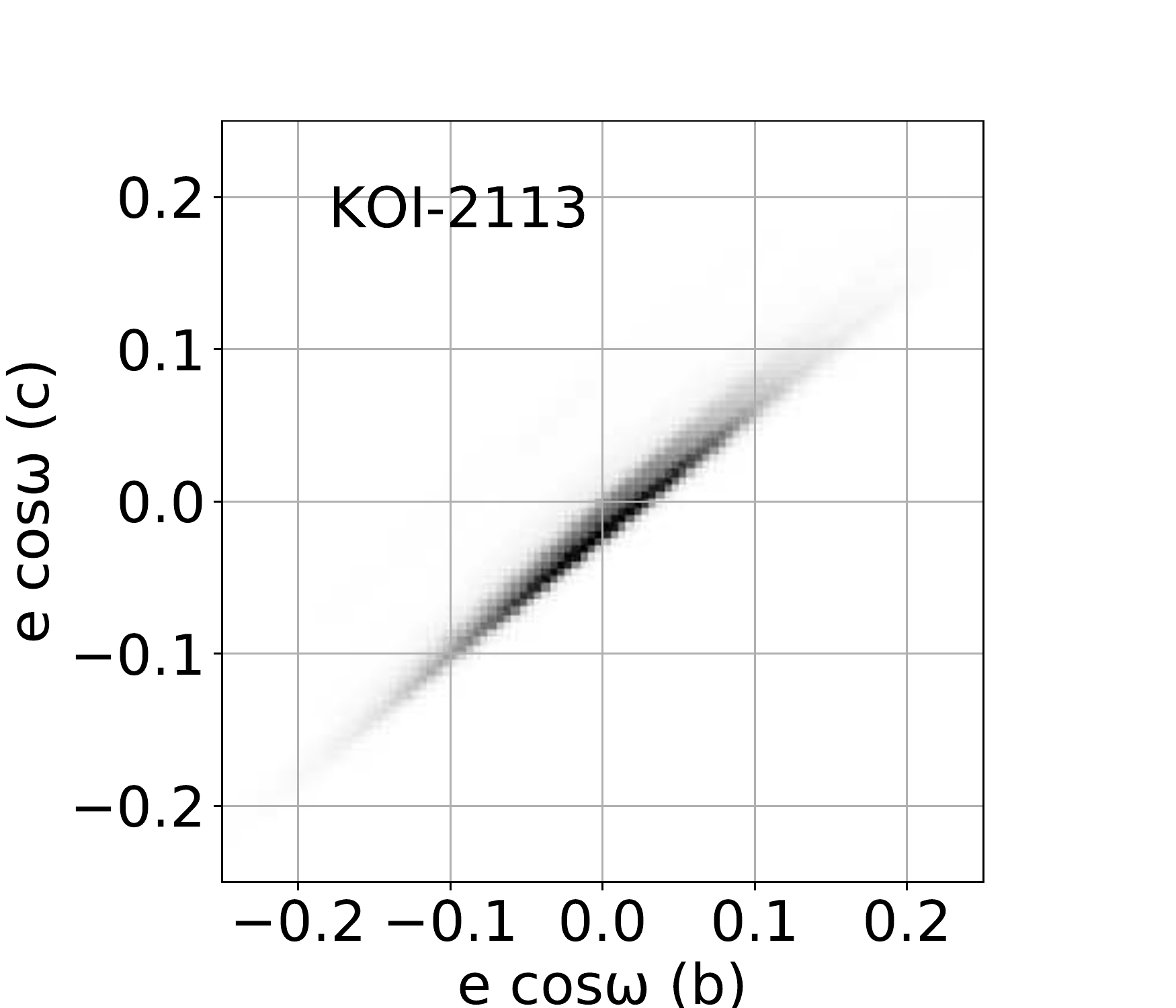}
\includegraphics [height = 1.1 in]{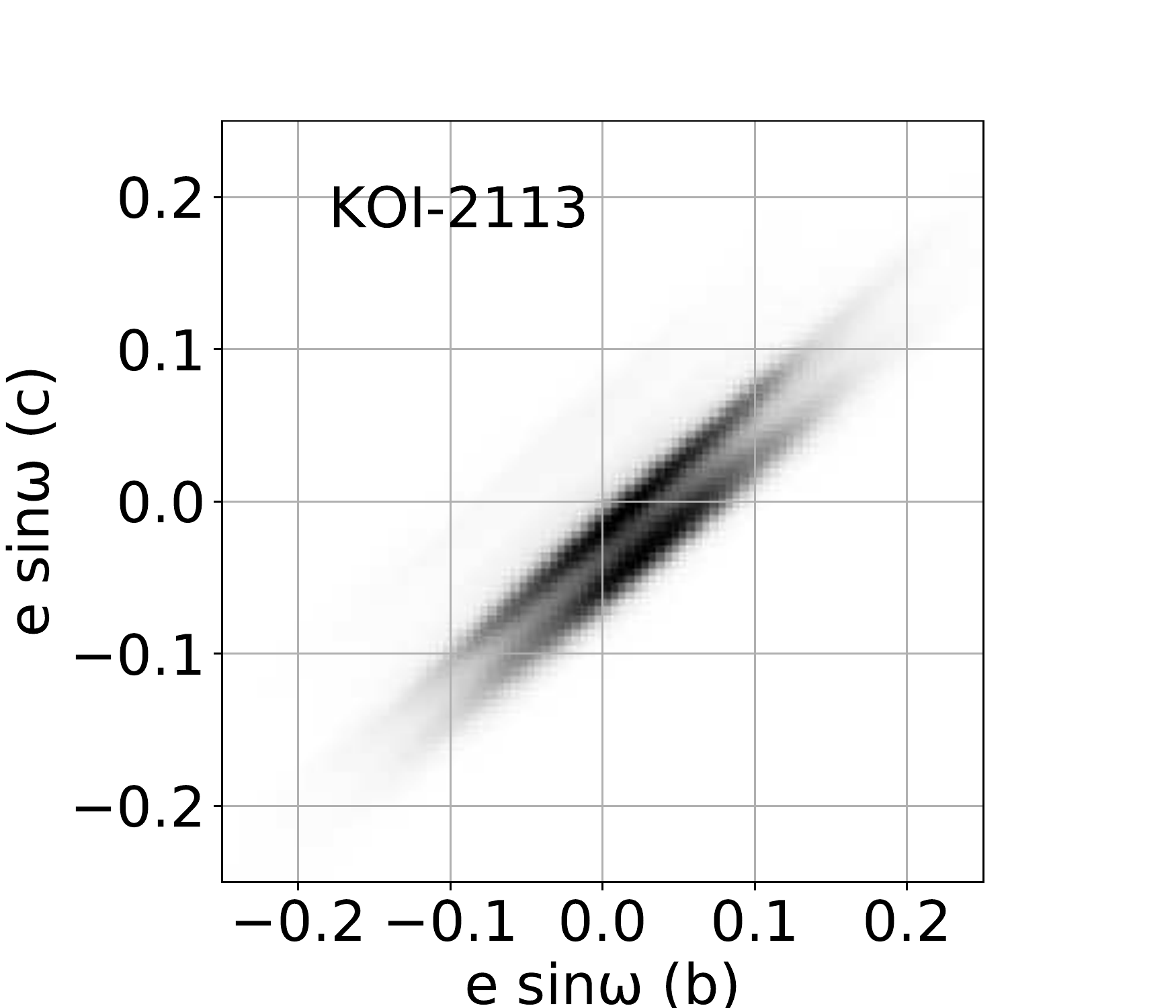}\\
\includegraphics [height = 1.1 in]{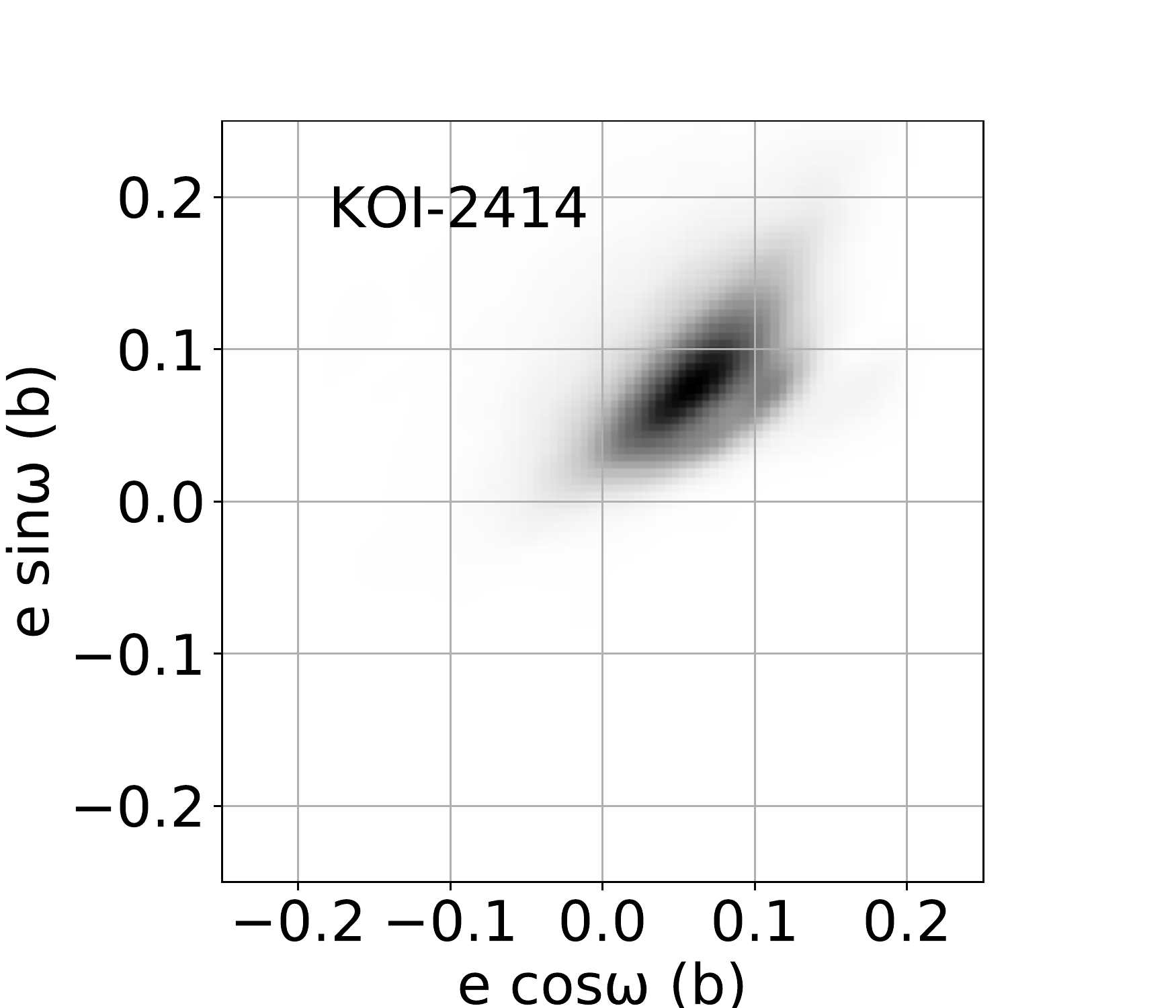}
\includegraphics [height = 1.1 in]{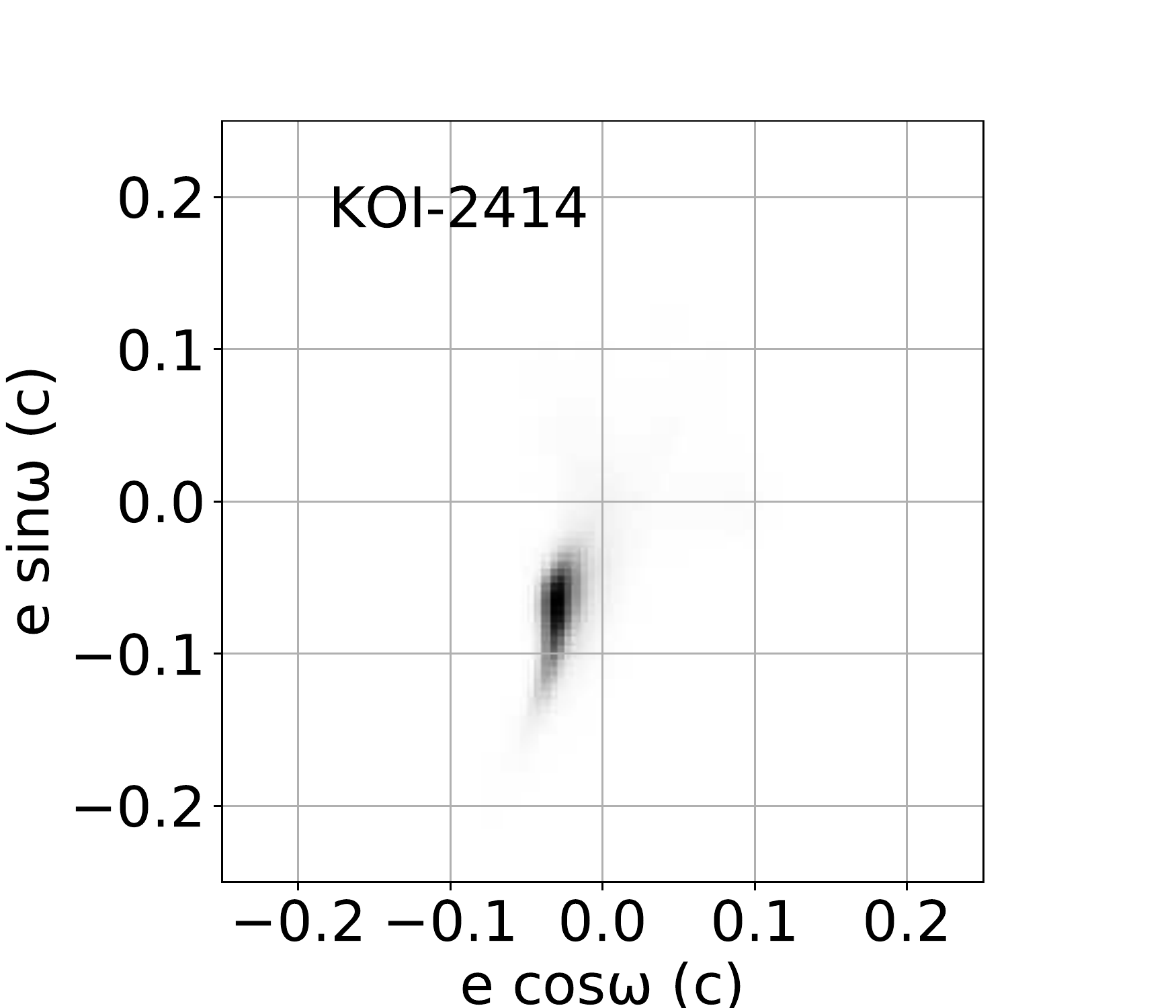}
\includegraphics [height = 1.1 in]{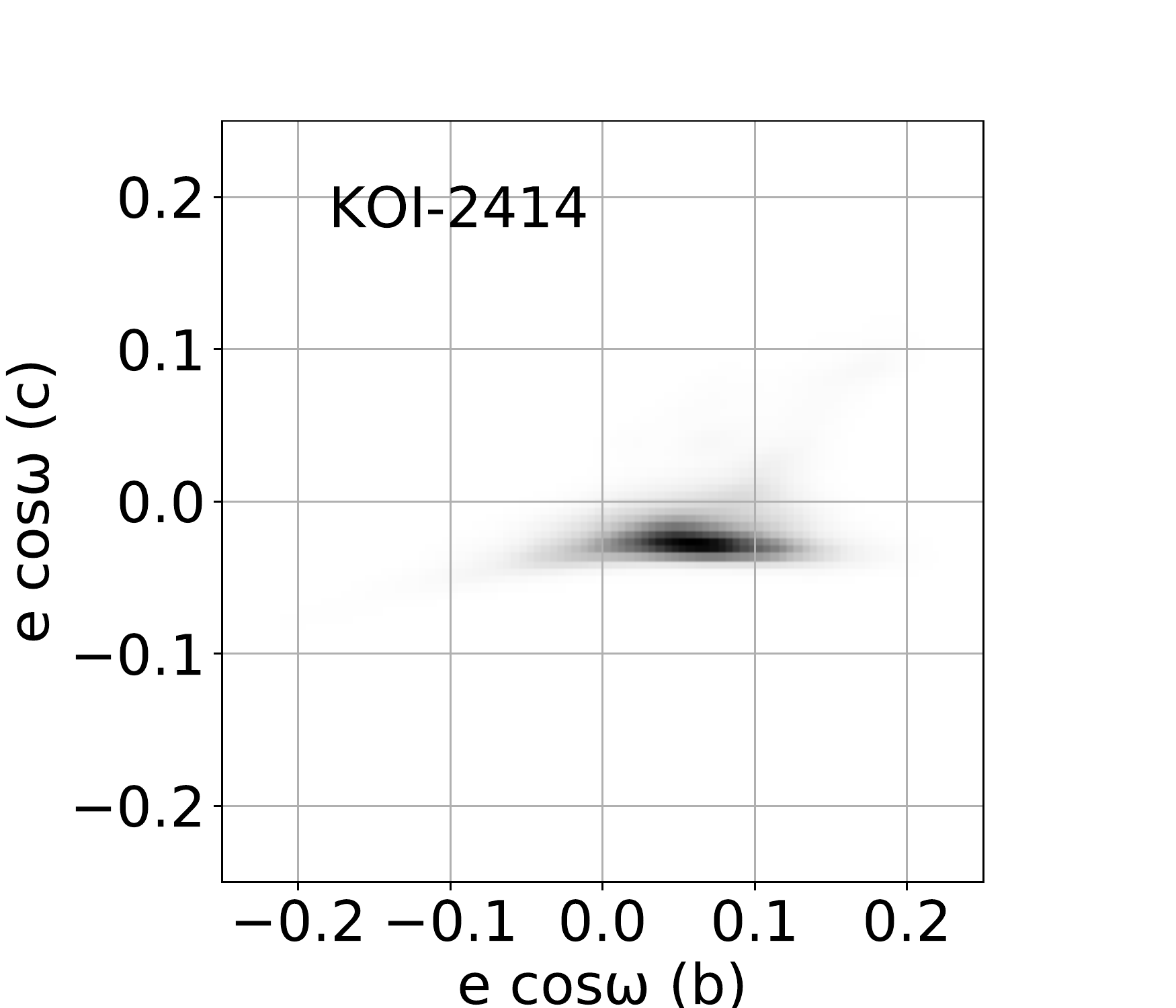}
\includegraphics [height = 1.1 in]{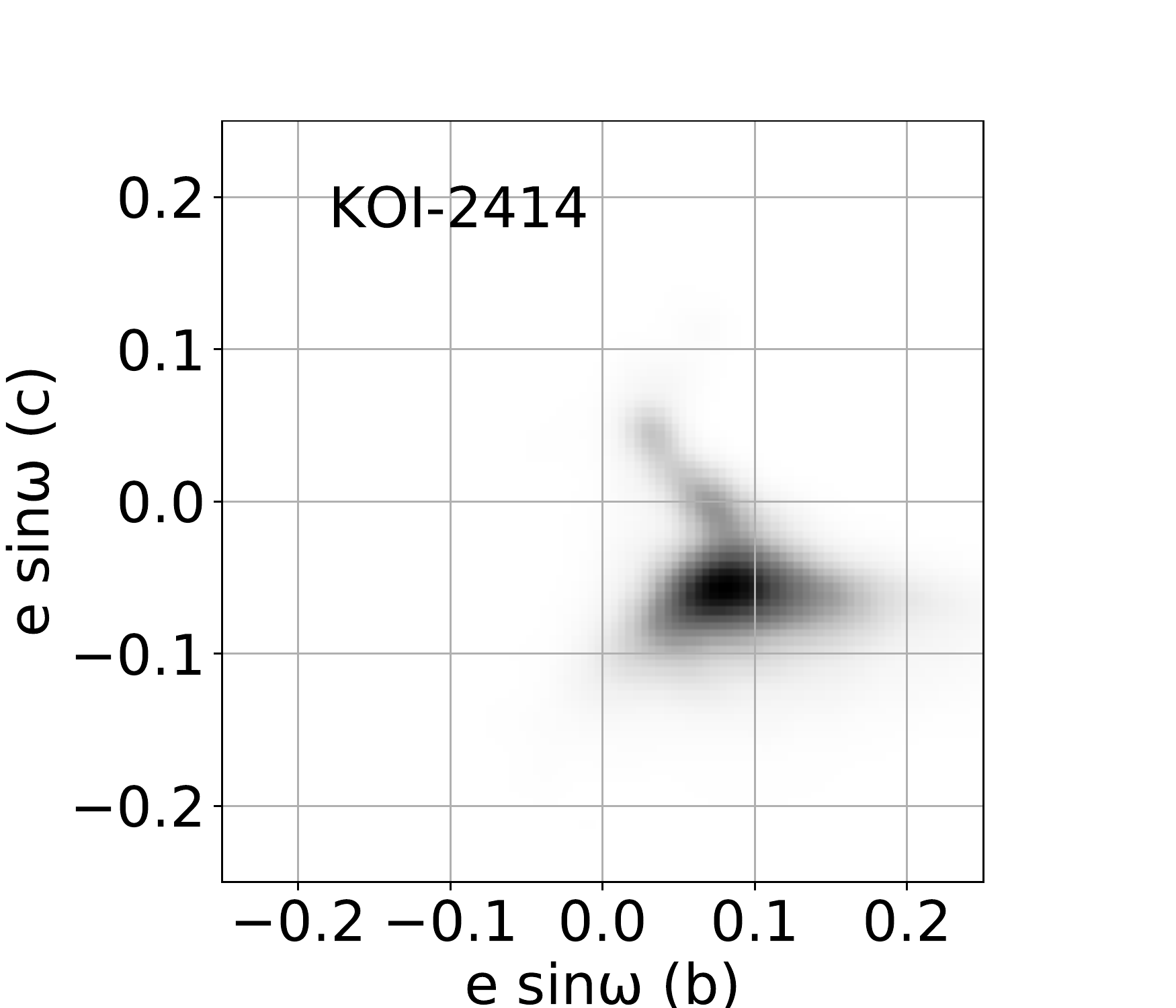} \\
\includegraphics [height = 1.1 in]{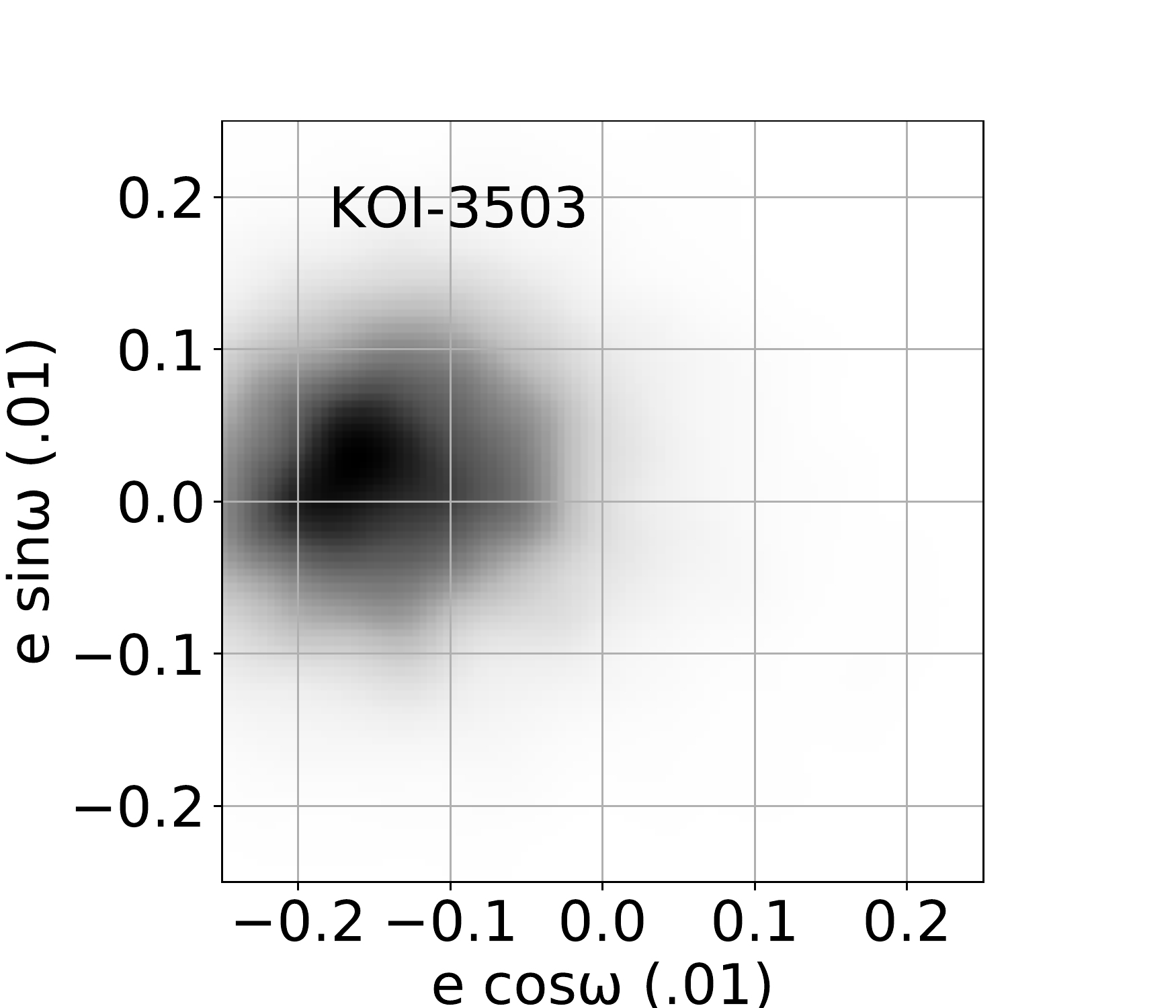}
\includegraphics [height = 1.1 in]{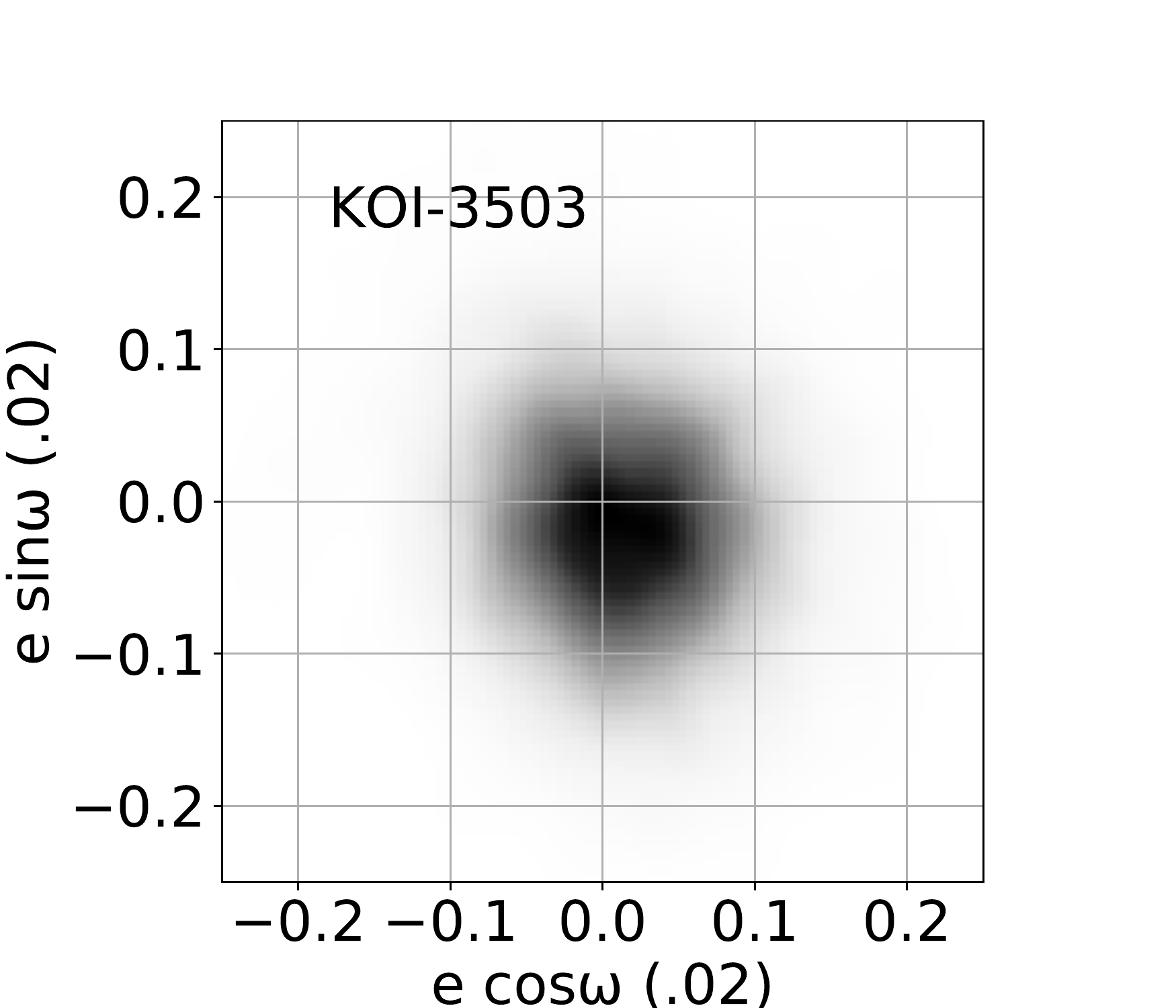}
\includegraphics [height = 1.1 in]{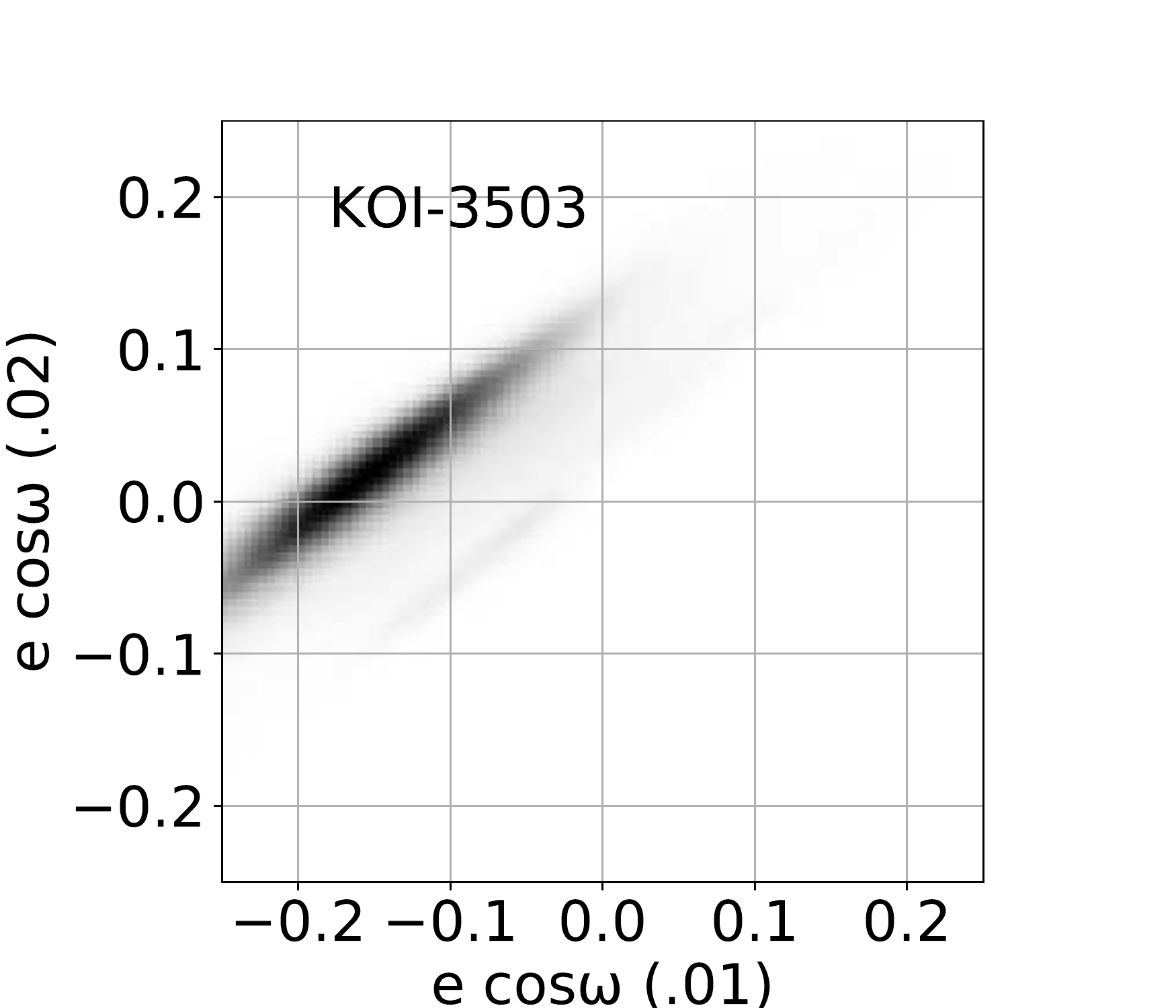}
\includegraphics [height = 1.1 in]{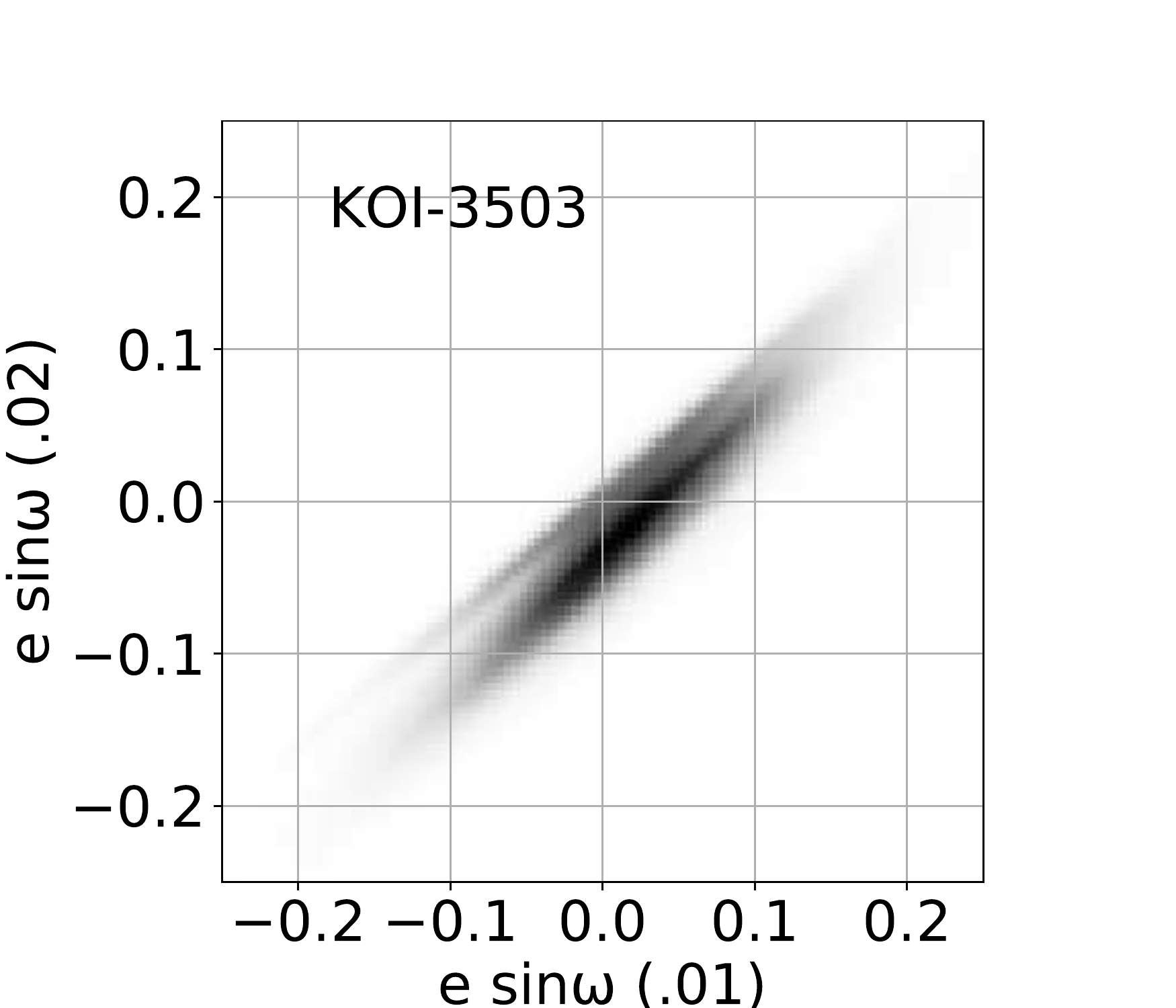} 
\caption{Two-dimensional kernel density estimators on joint posteriors of eccentricity vector components: two-planet systems. (Part 3 of 3.) 
}
\label{fig:ecc2c} 
\end{center}
\end{figure}

\begin{figure}
\begin{center}
\figurenum{23}
\includegraphics [height = 1.1 in]{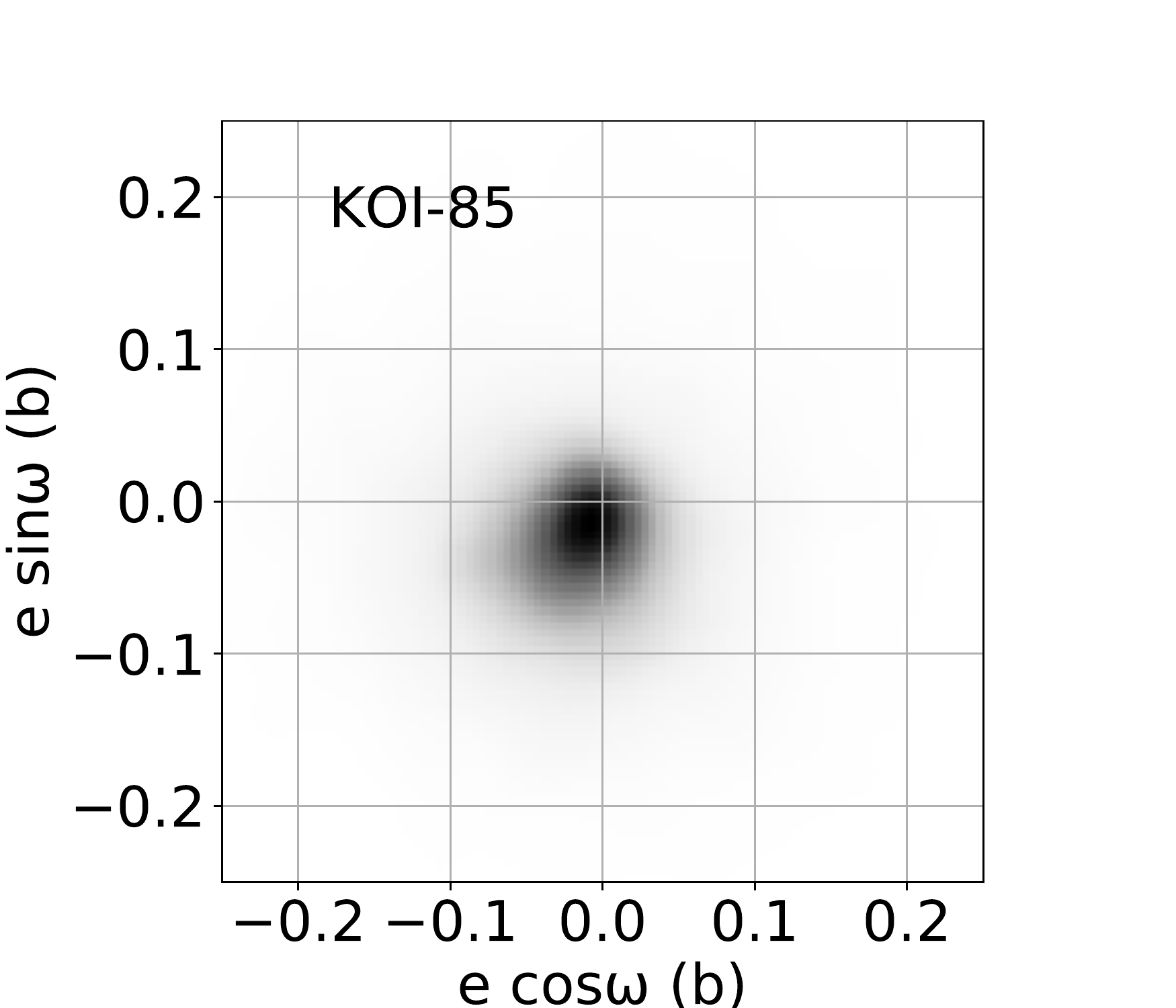}
\includegraphics [height = 1.1 in]{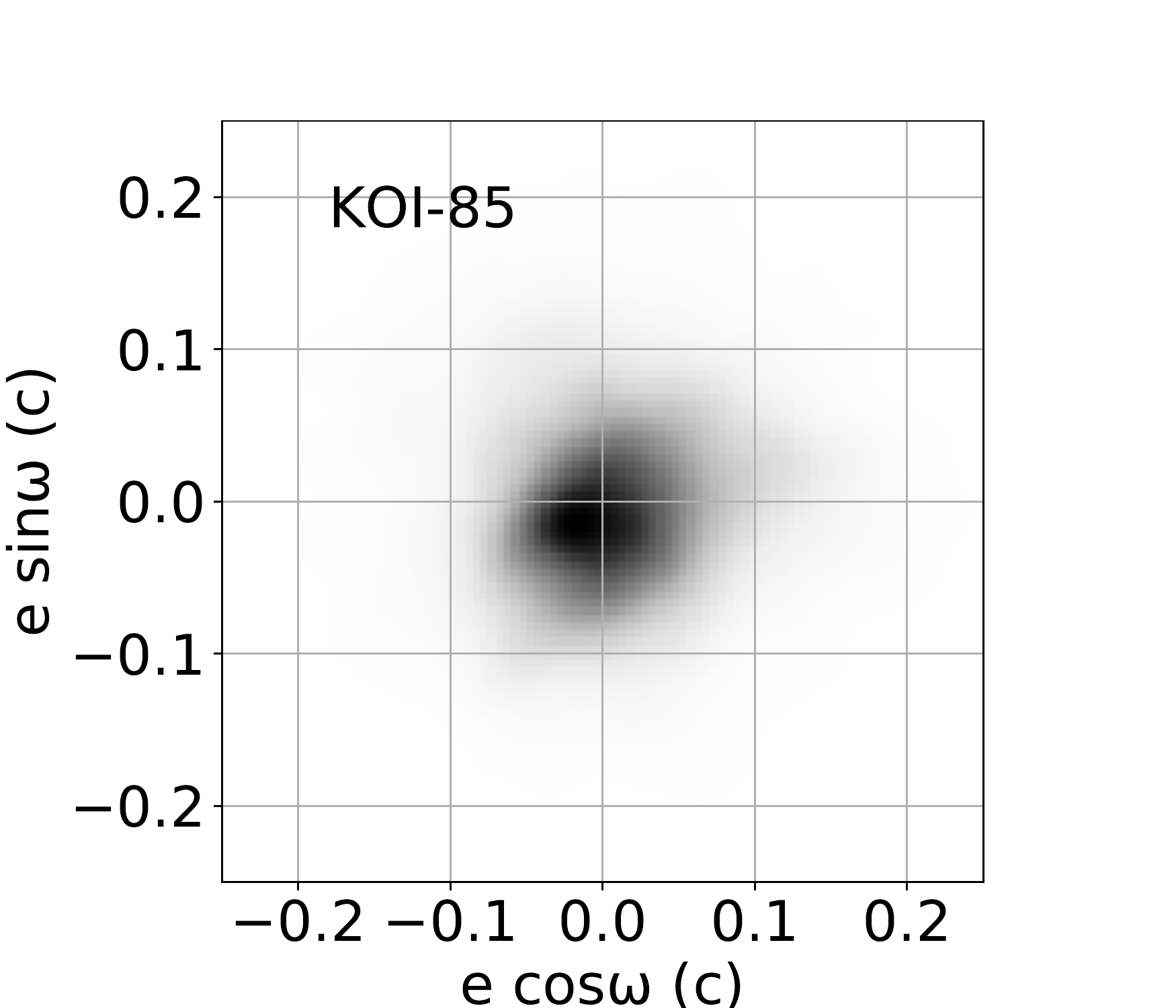}
\includegraphics [height = 1.1 in]{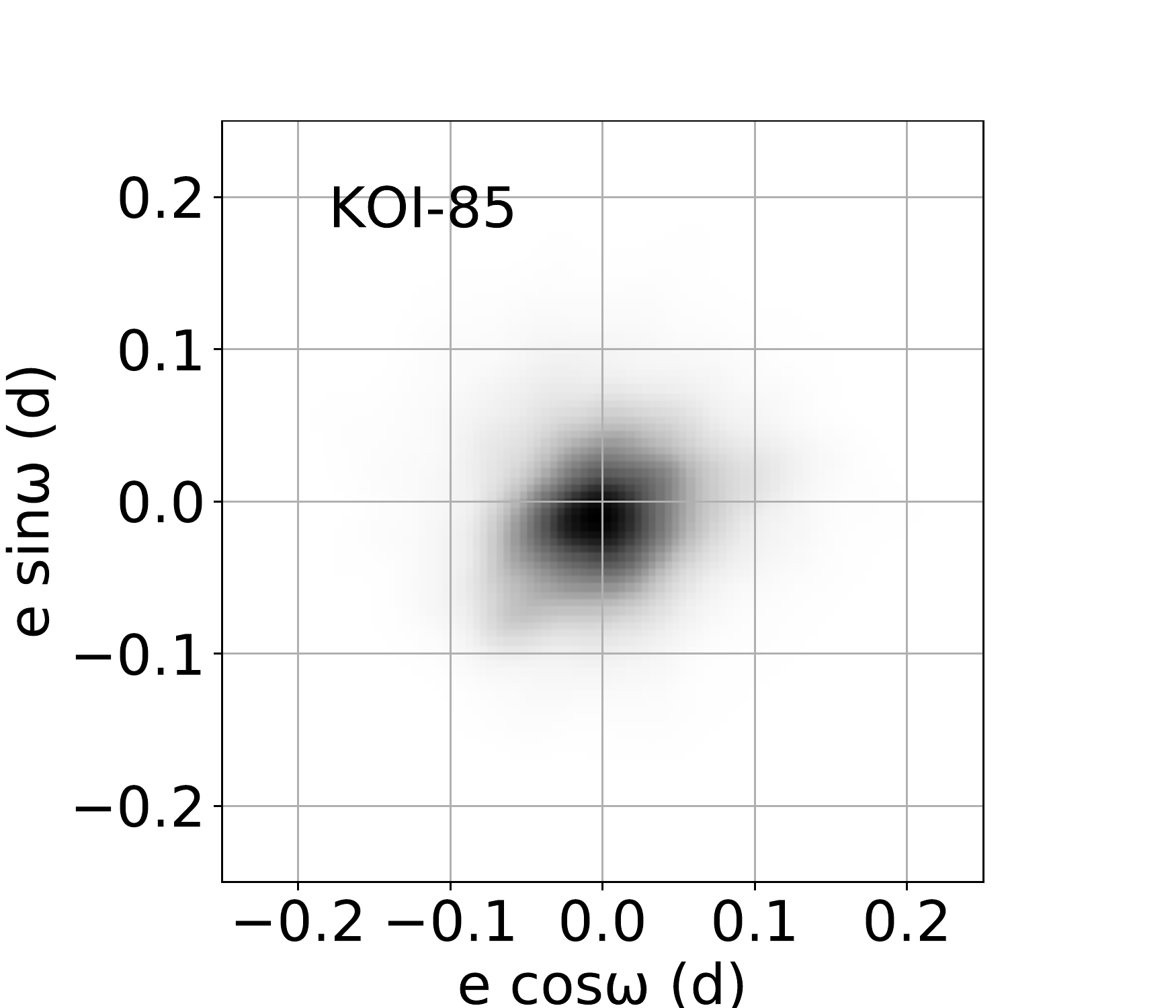}
\includegraphics [height = 1.1 in]{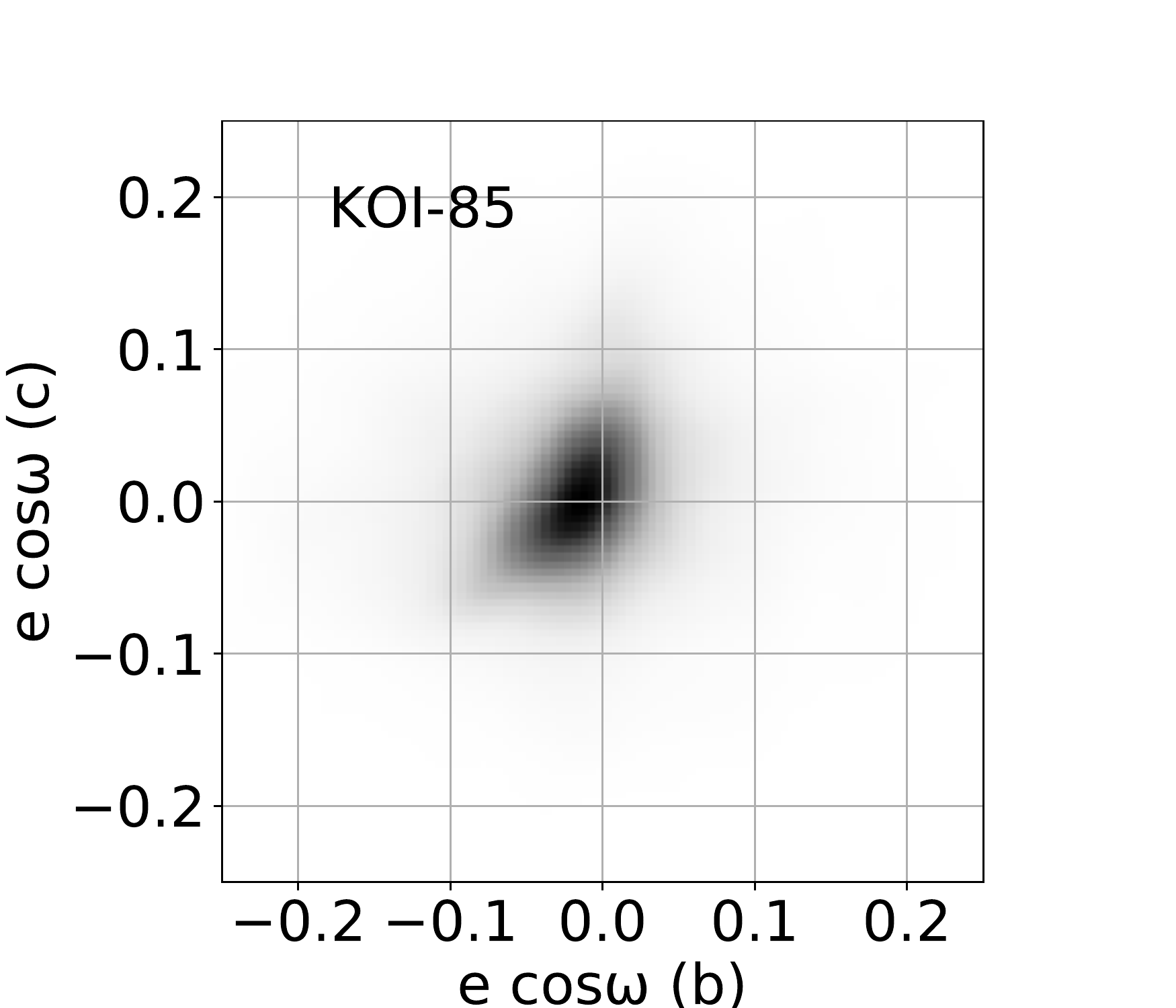} \\
\includegraphics [height = 1.1 in]{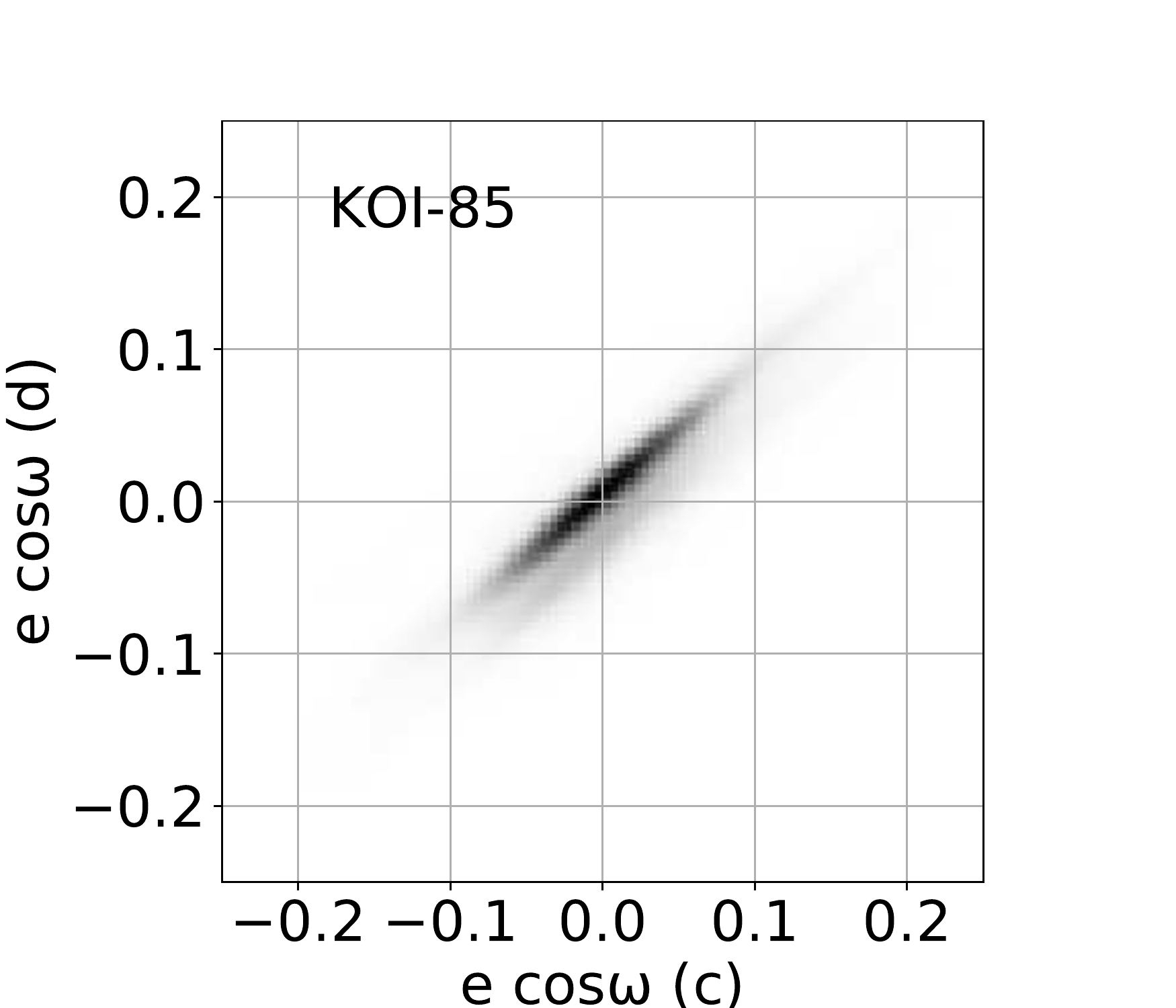}
\includegraphics [height = 1.1 in]{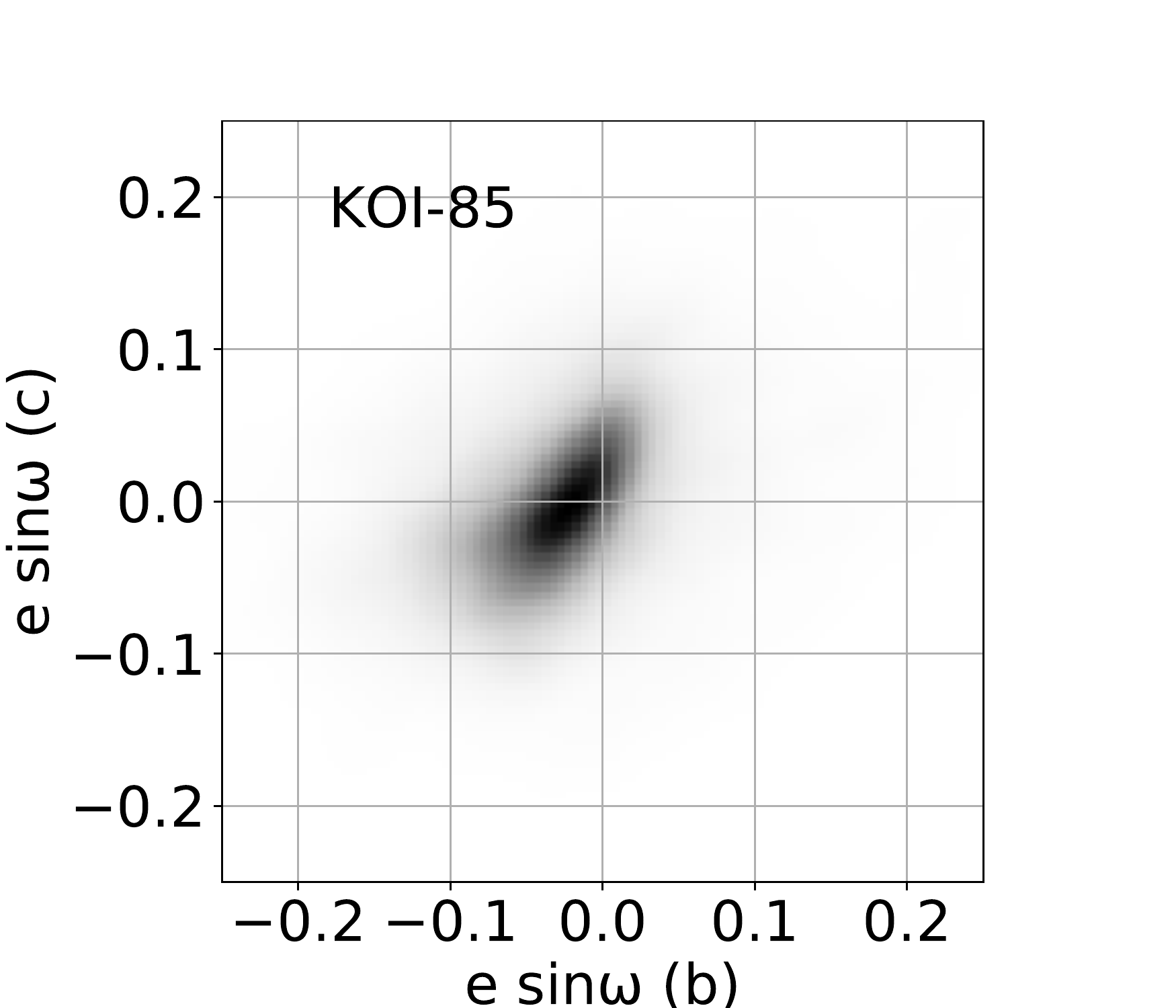}
\includegraphics [height = 1.1 in]{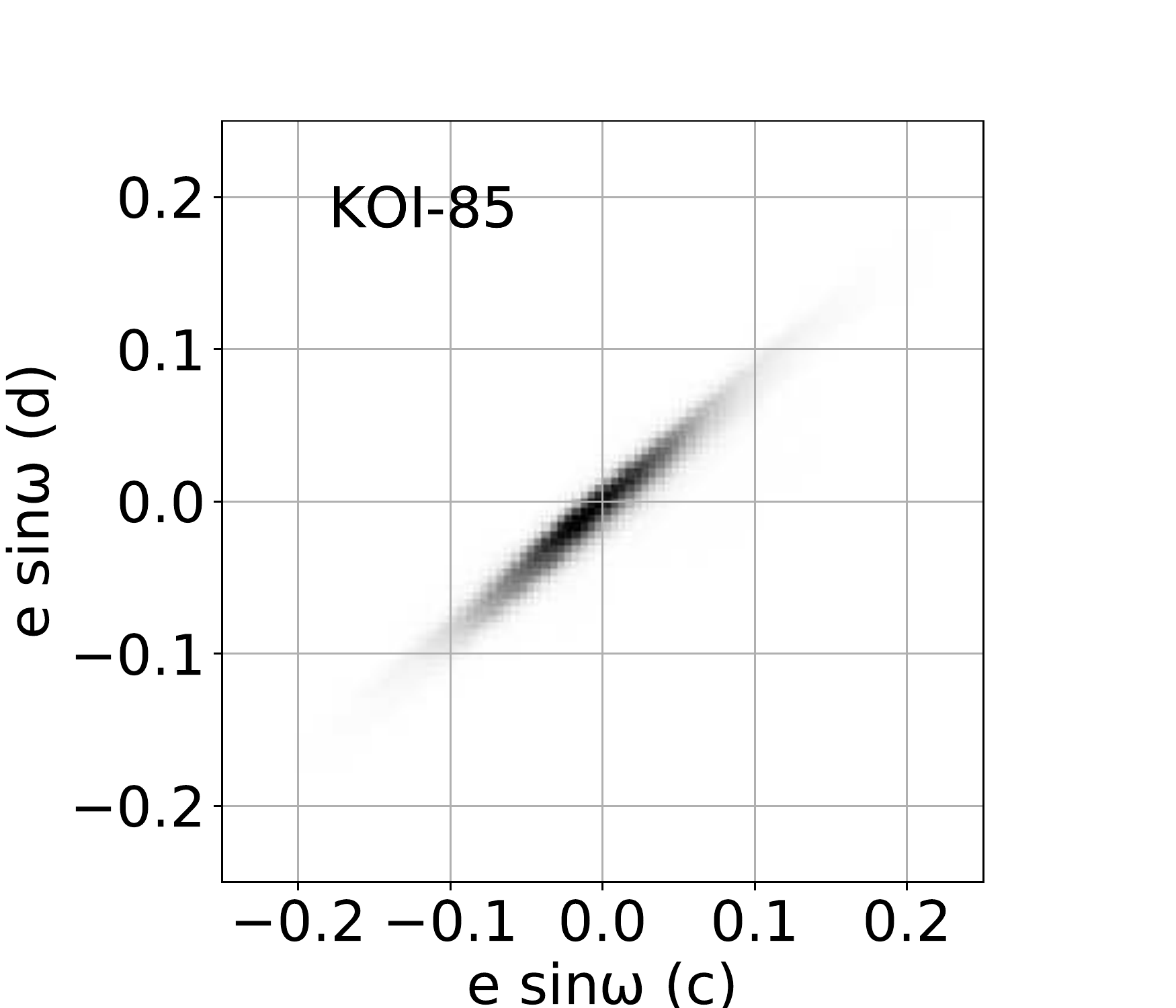}
\includegraphics [height = 1.1 in]{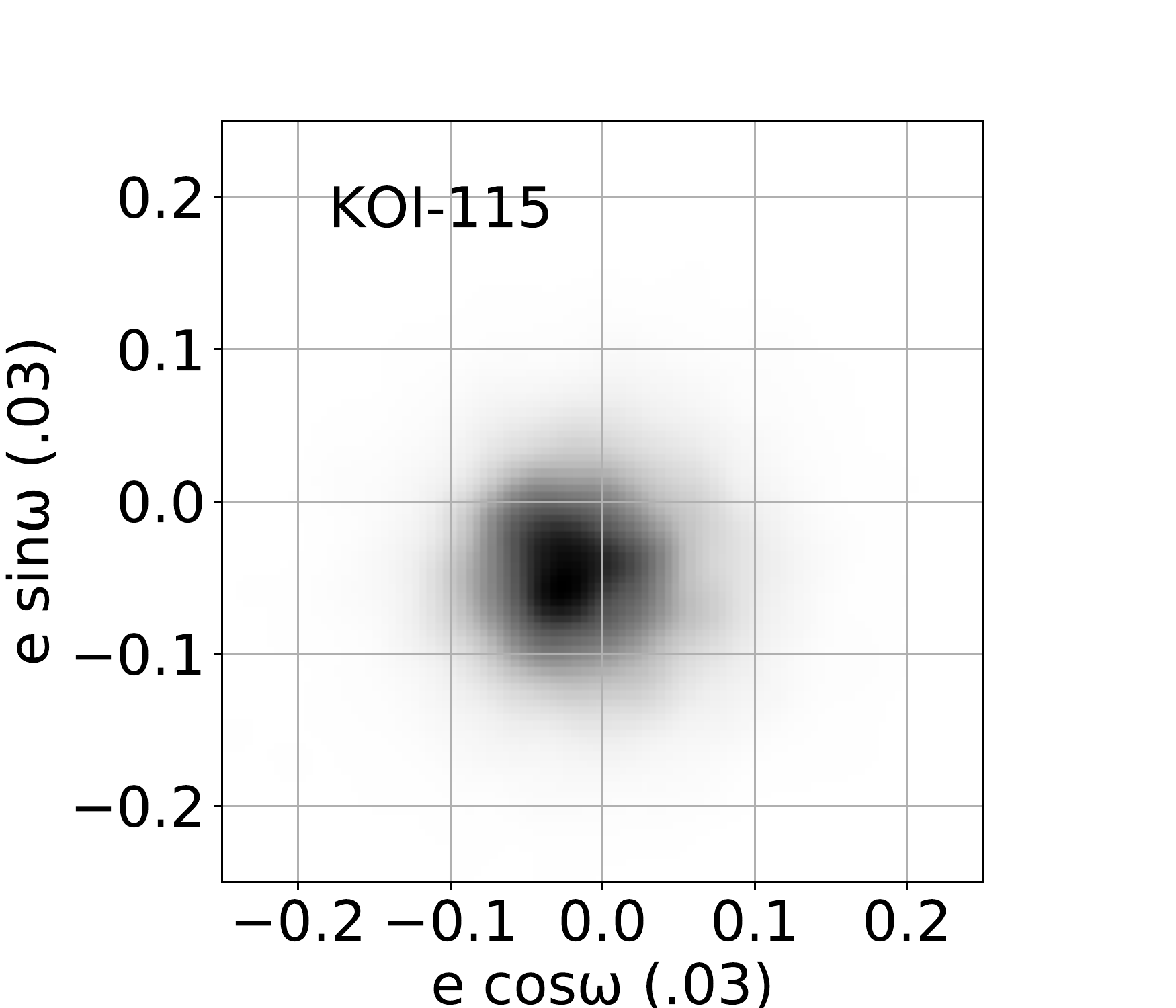} \\
\includegraphics [height = 1.1 in]{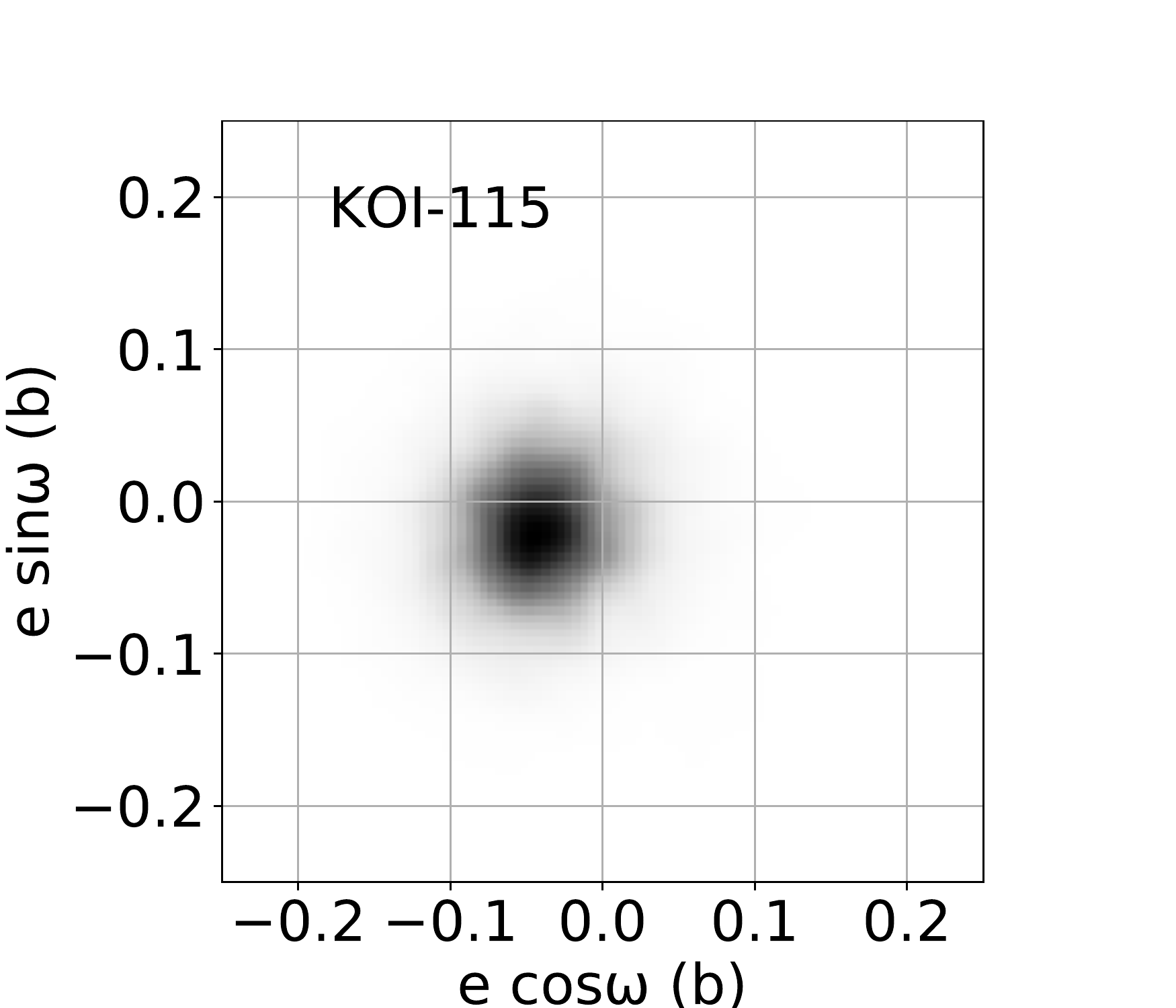}
\includegraphics [height = 1.1 in]{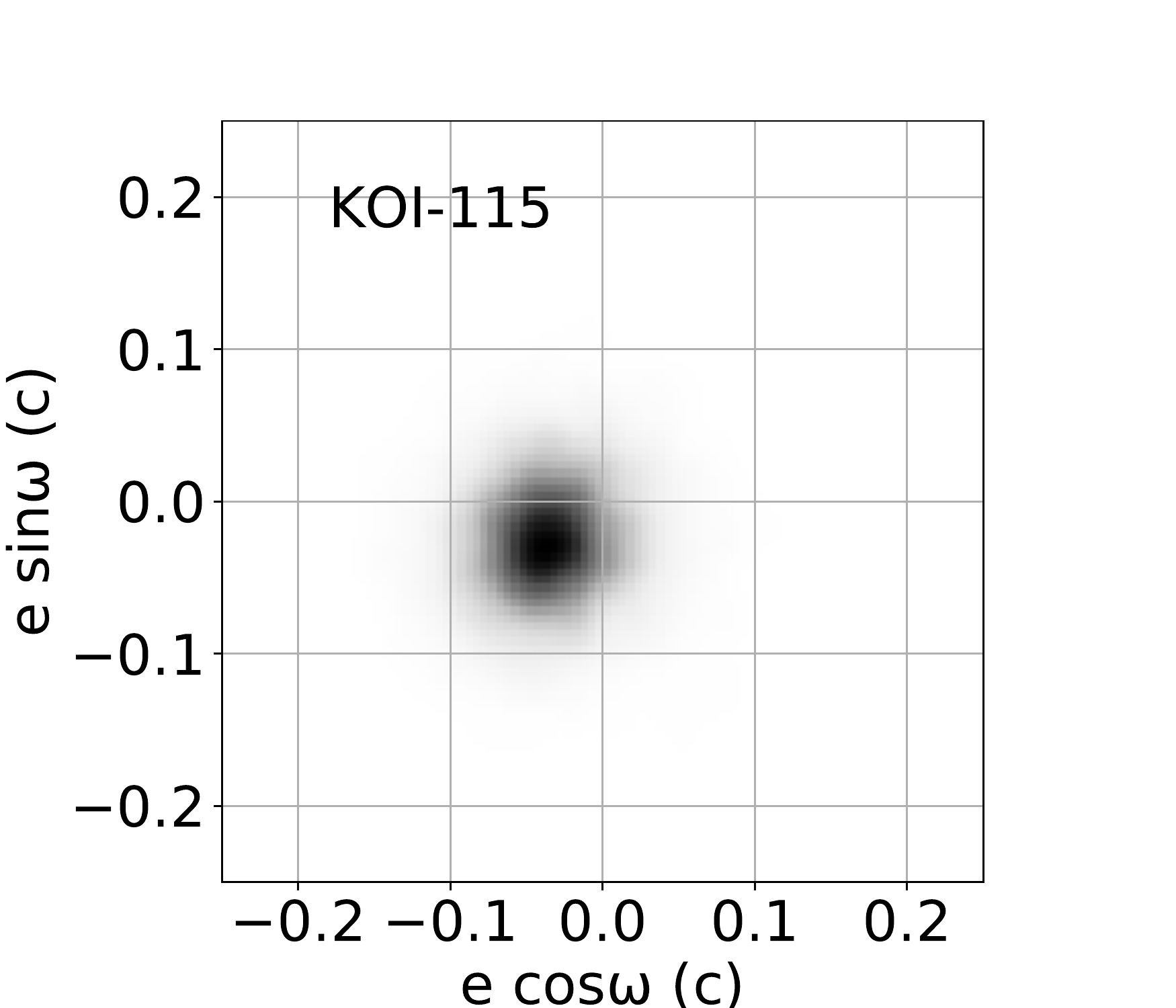}
\includegraphics [height = 1.1 in]{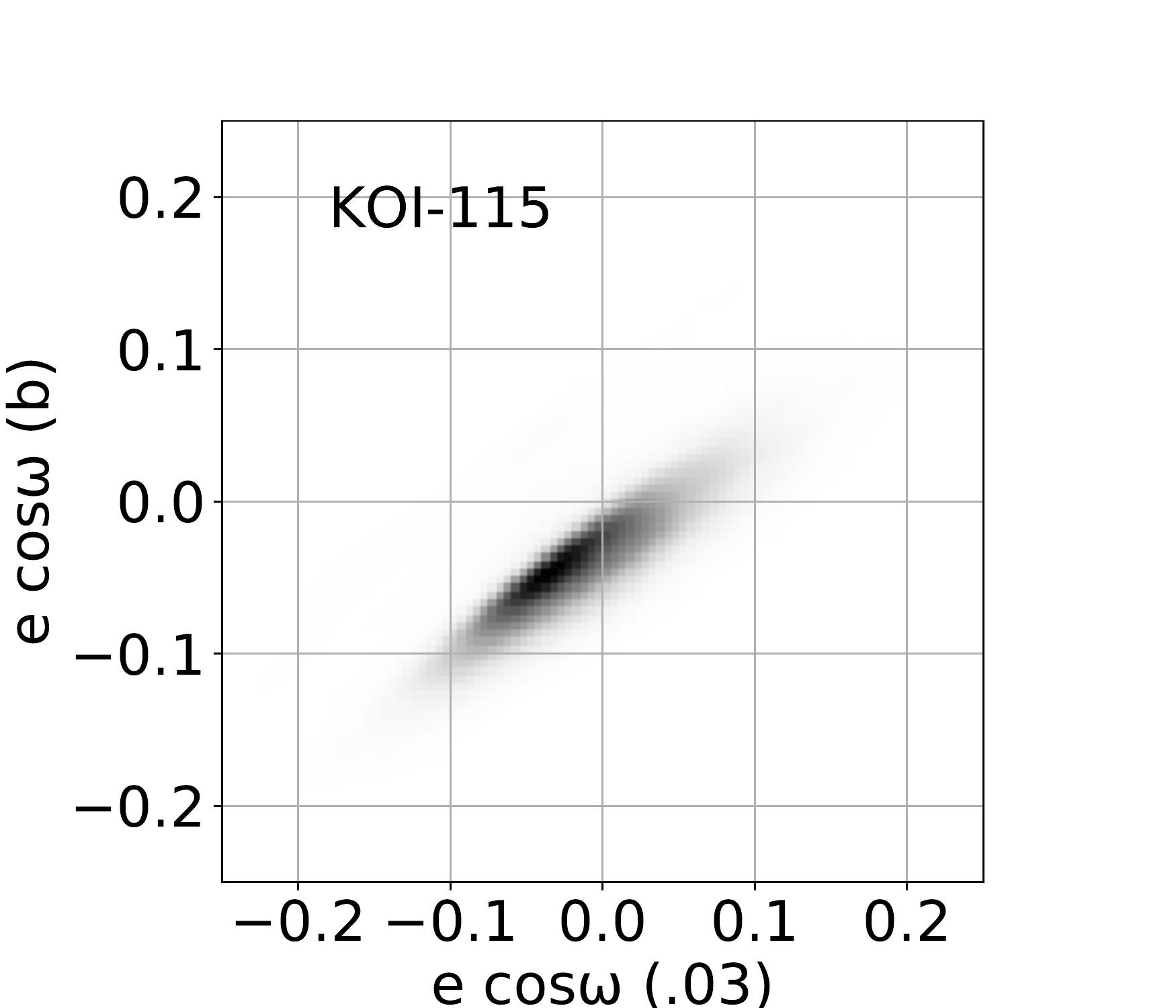}
\includegraphics [height = 1.1 in]{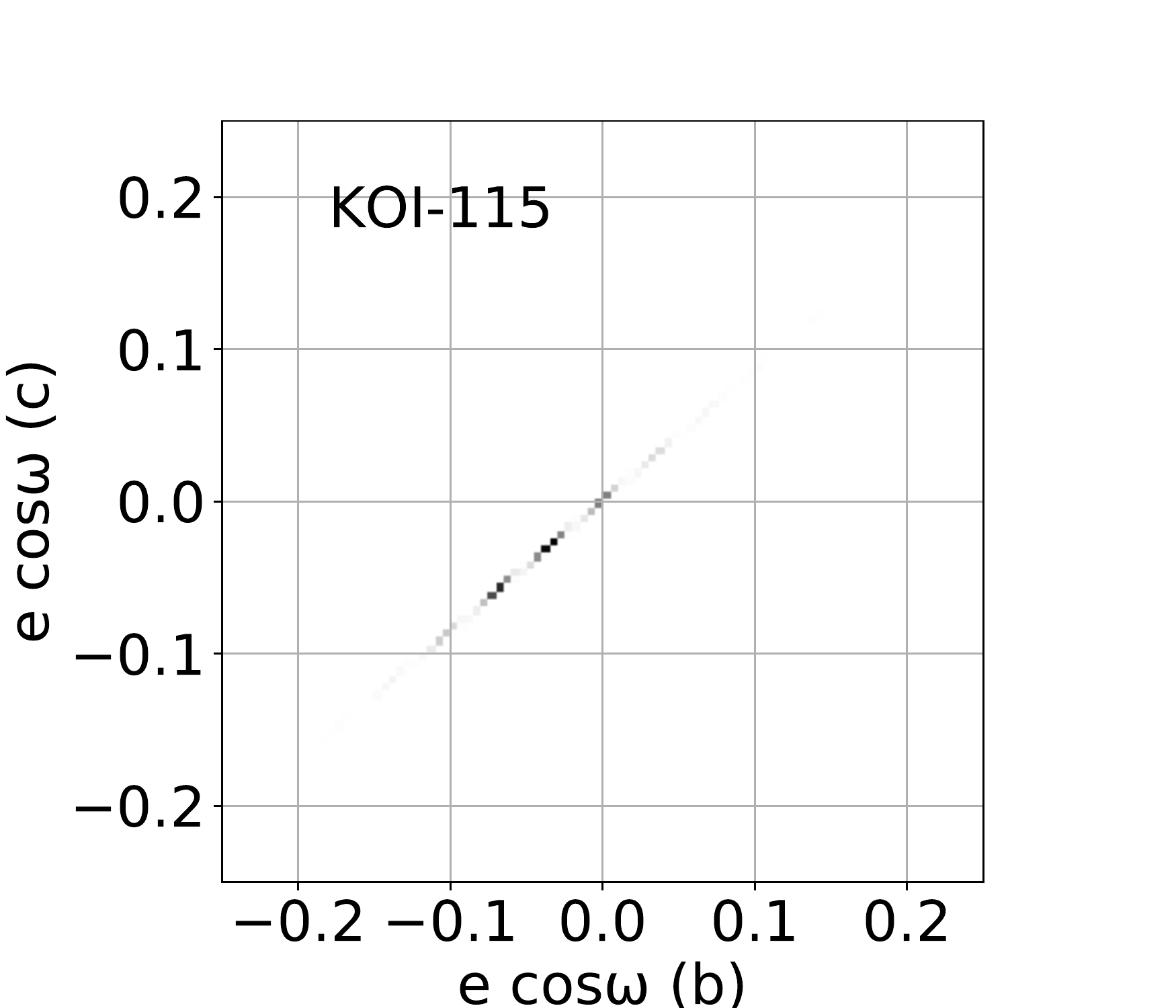} \\
\includegraphics [height = 1.1 in]{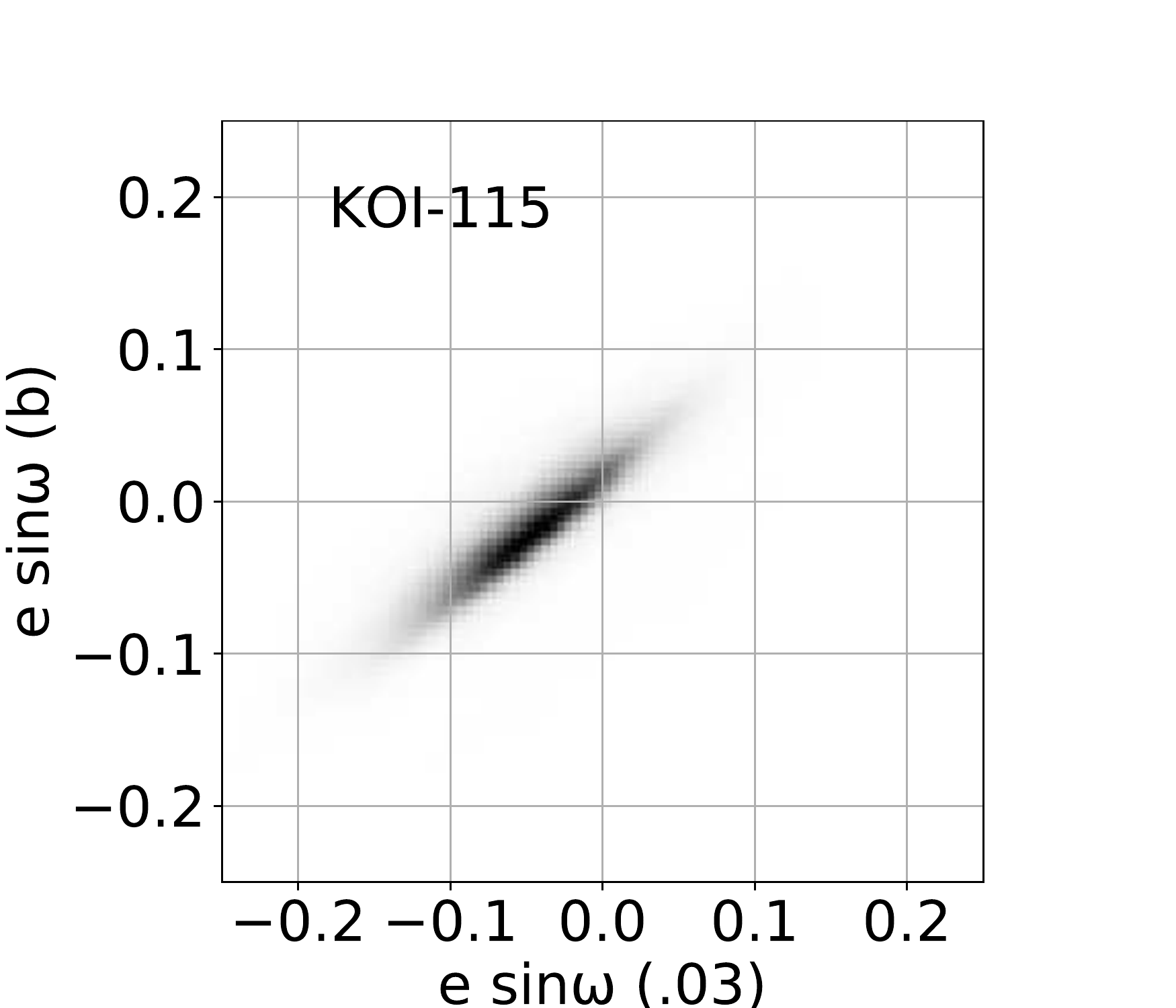}
\includegraphics [height = 1.1 in]{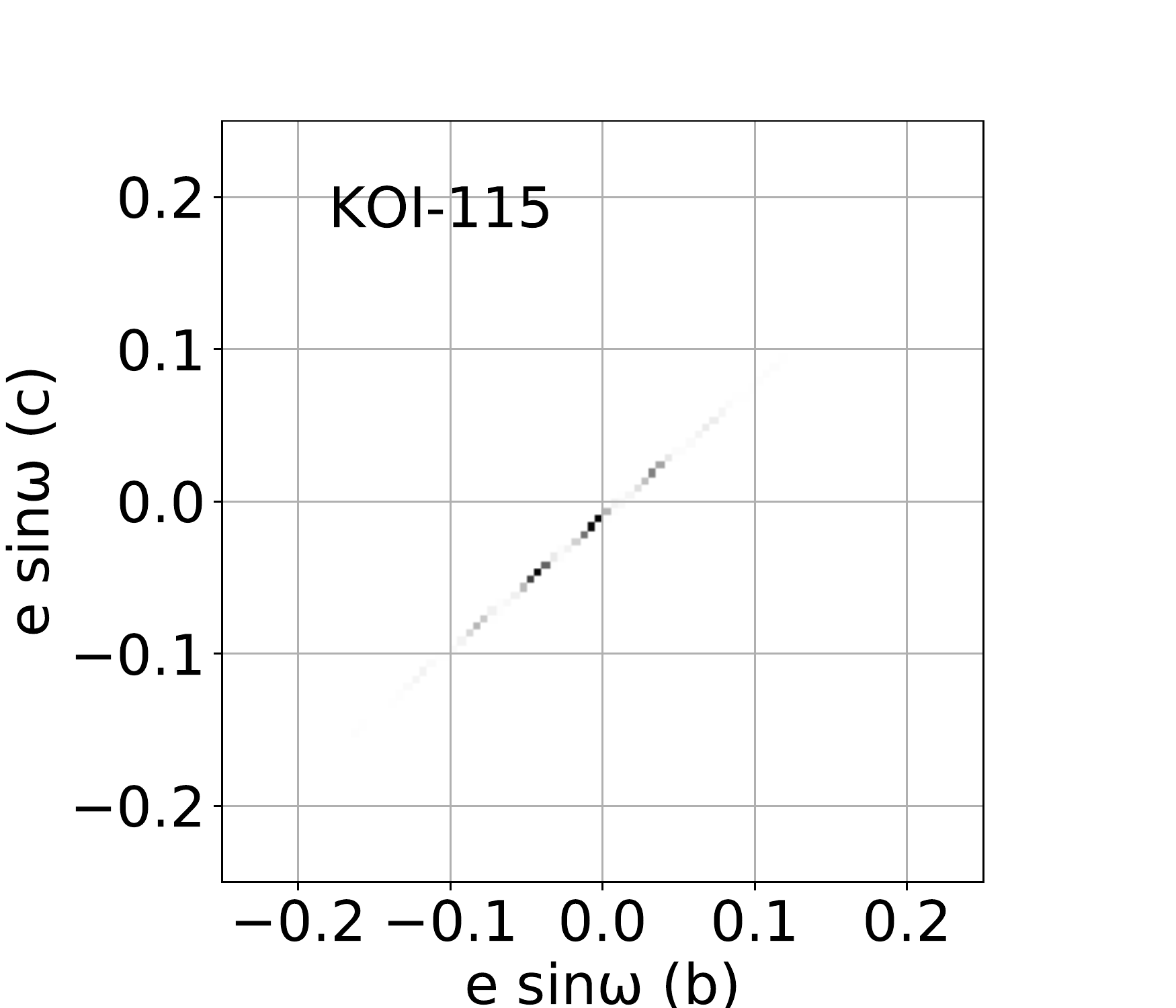}
\includegraphics [height = 1.1 in]{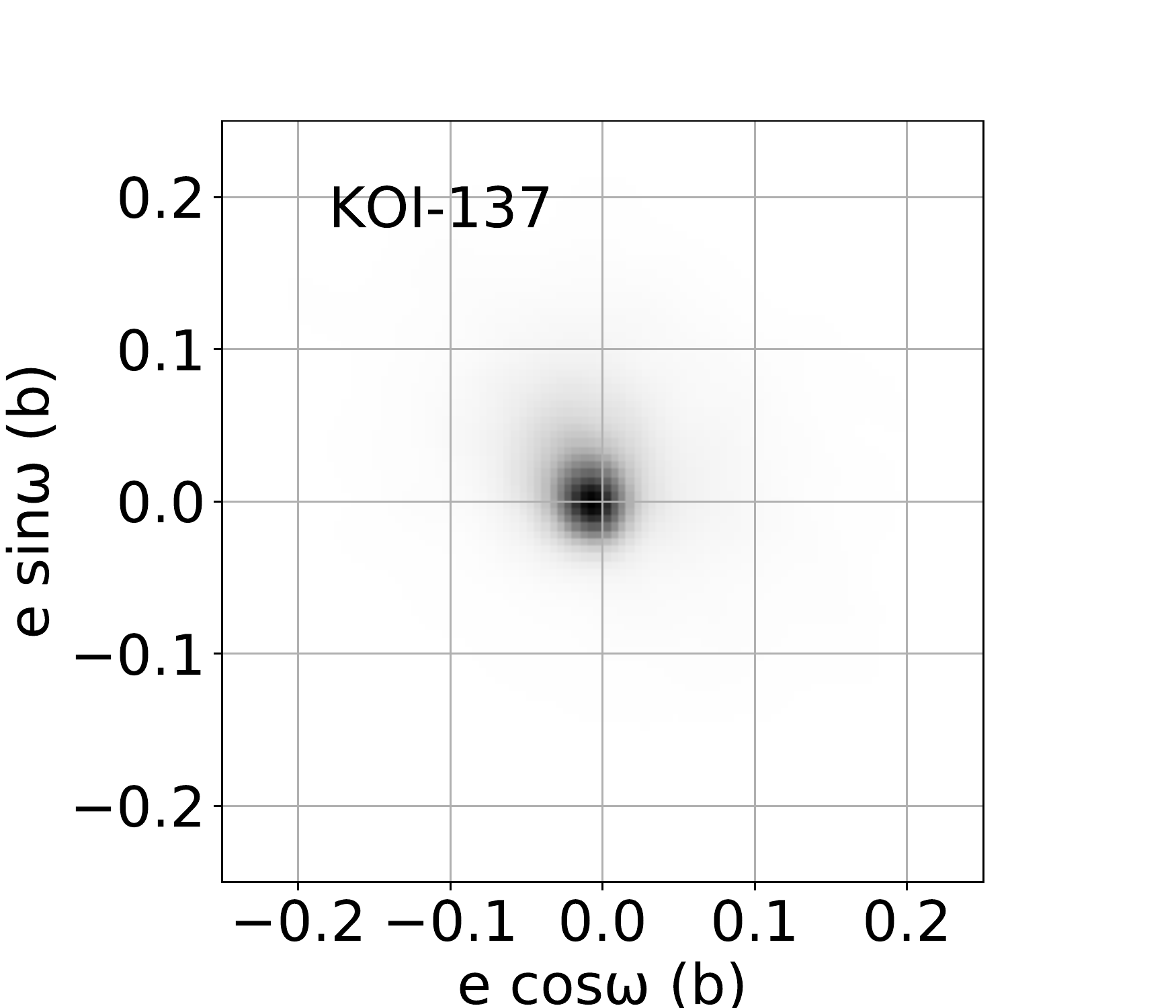}
\includegraphics [height = 1.1 in]{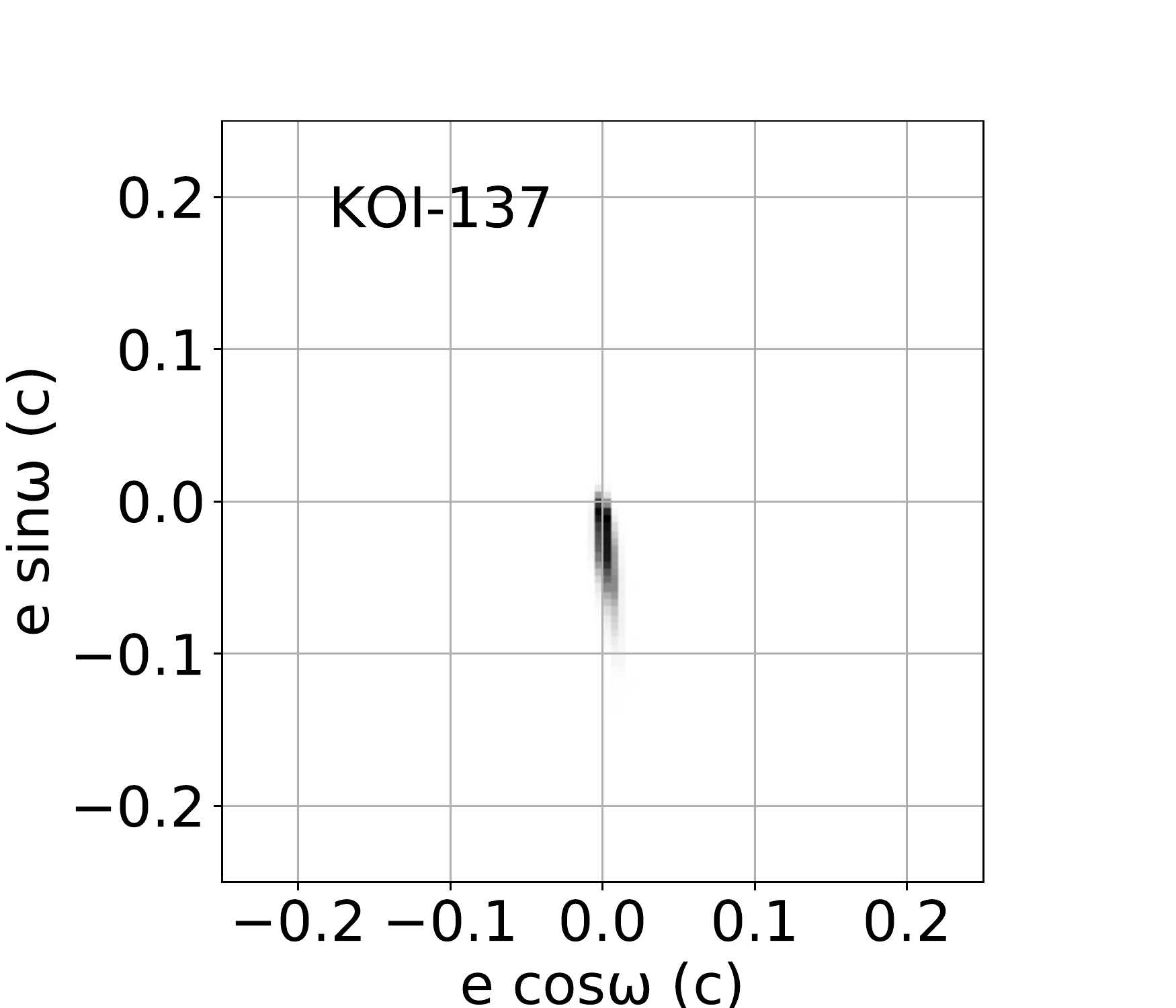} \\
\includegraphics [height = 1.1 in]{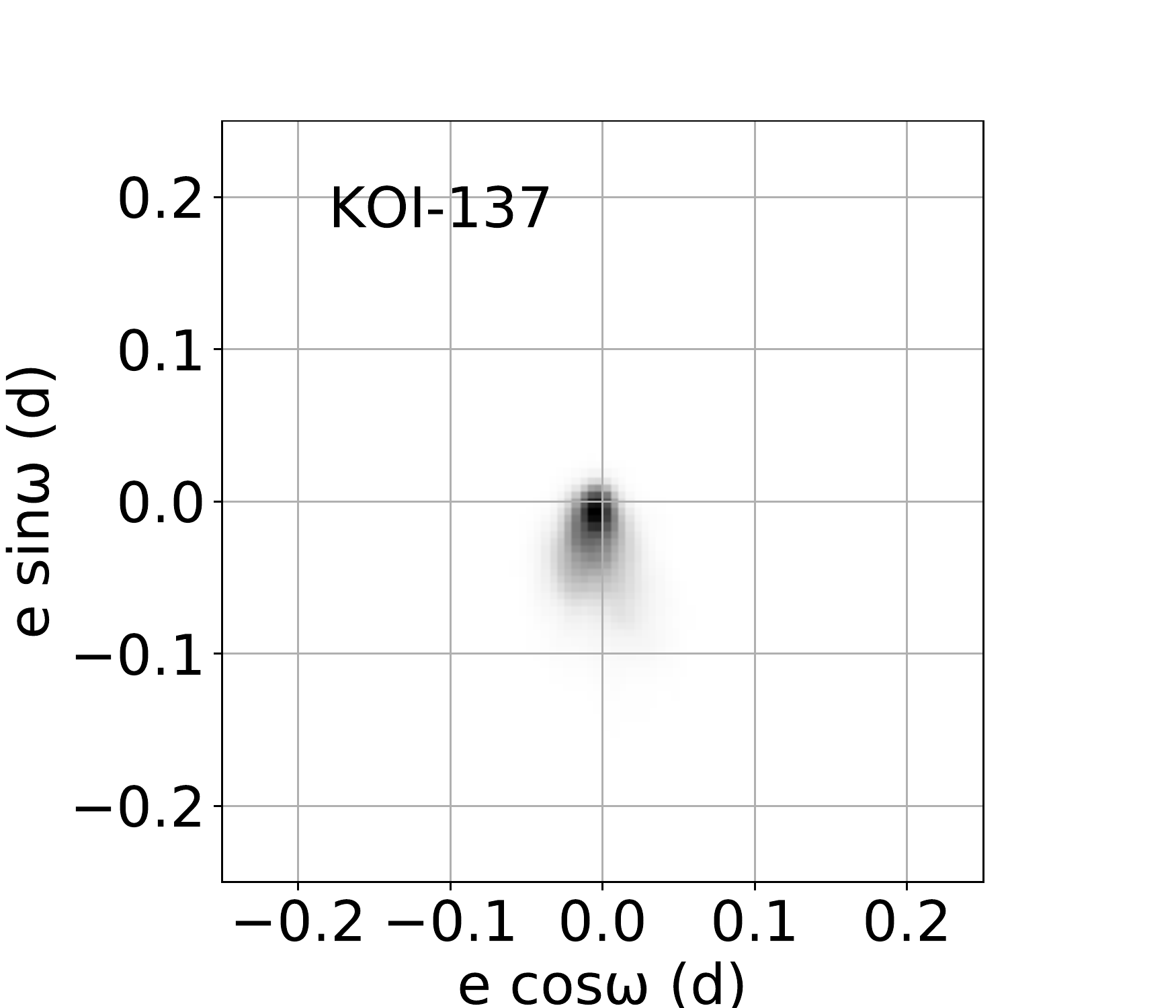}
\includegraphics [height = 1.1 in]{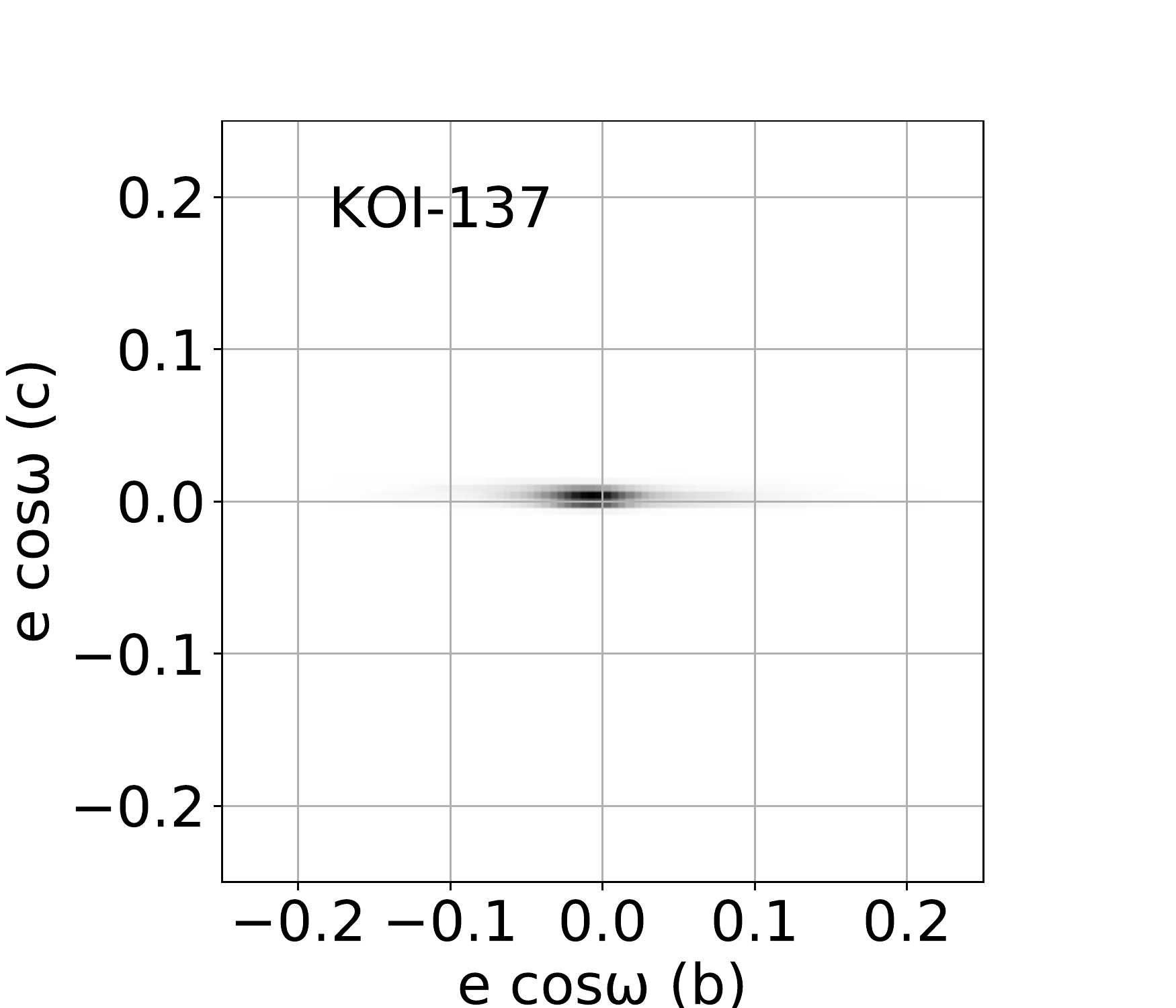} 
\includegraphics [height = 1.1 in]{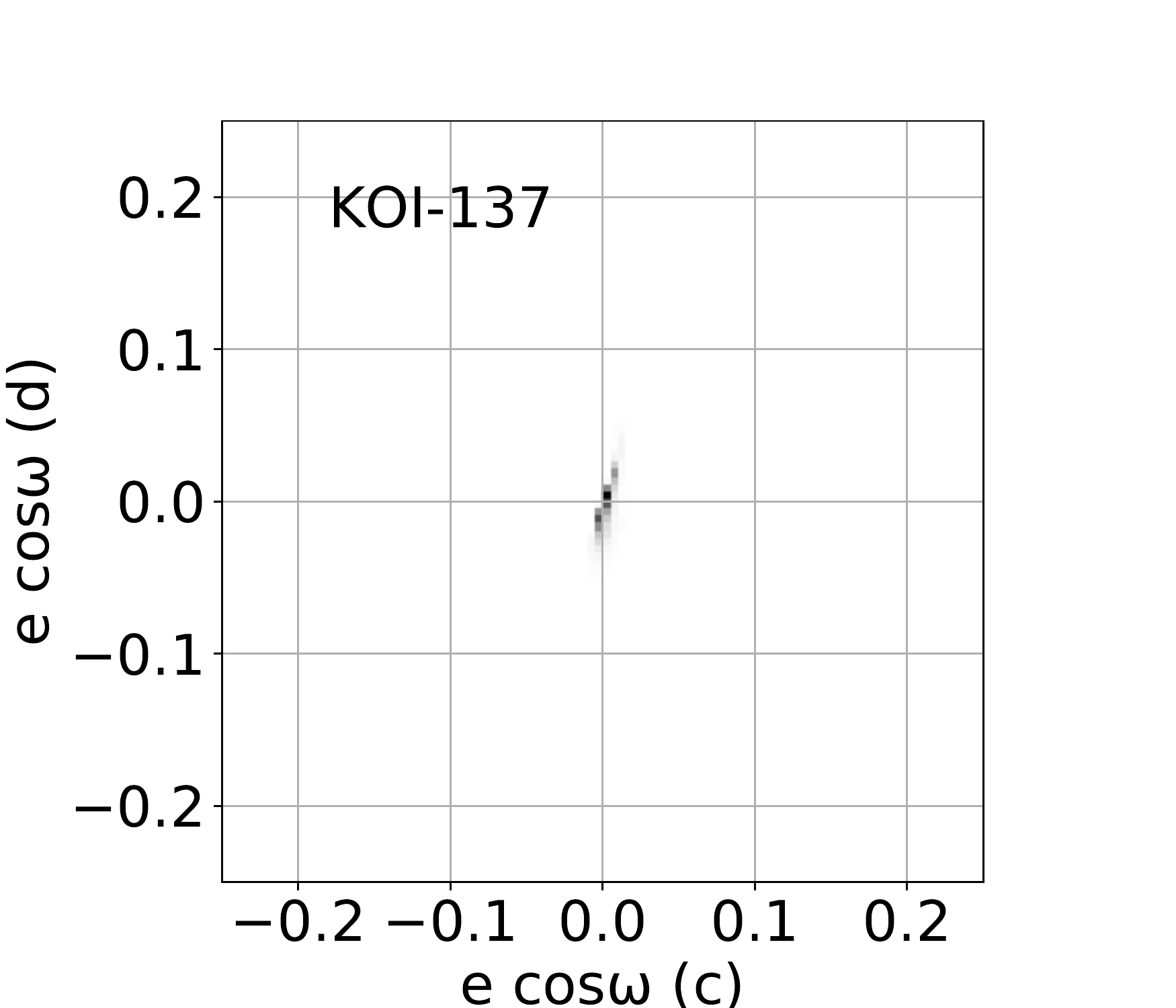}
\includegraphics [height = 1.1 in]{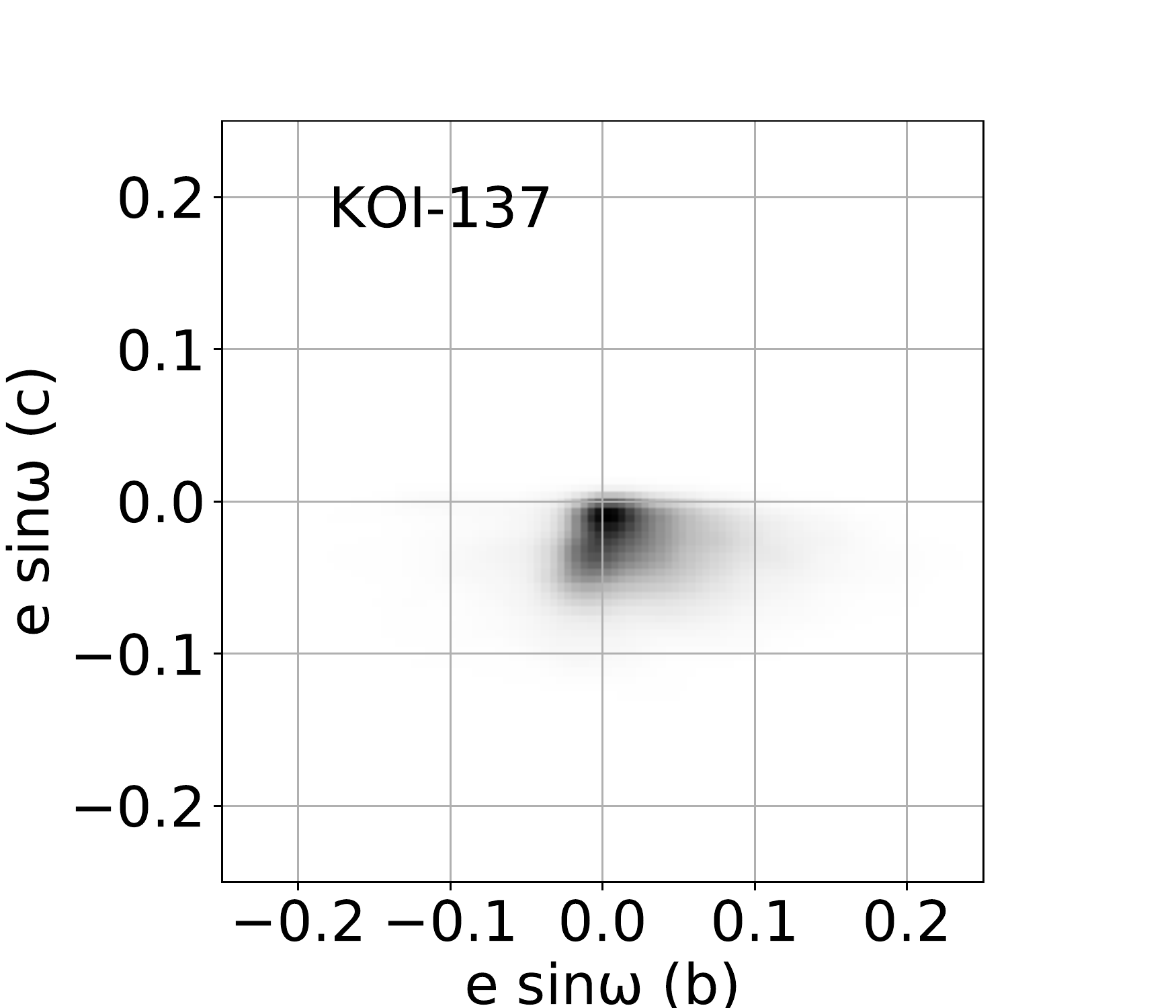} \\
\includegraphics [height = 1.1 in]{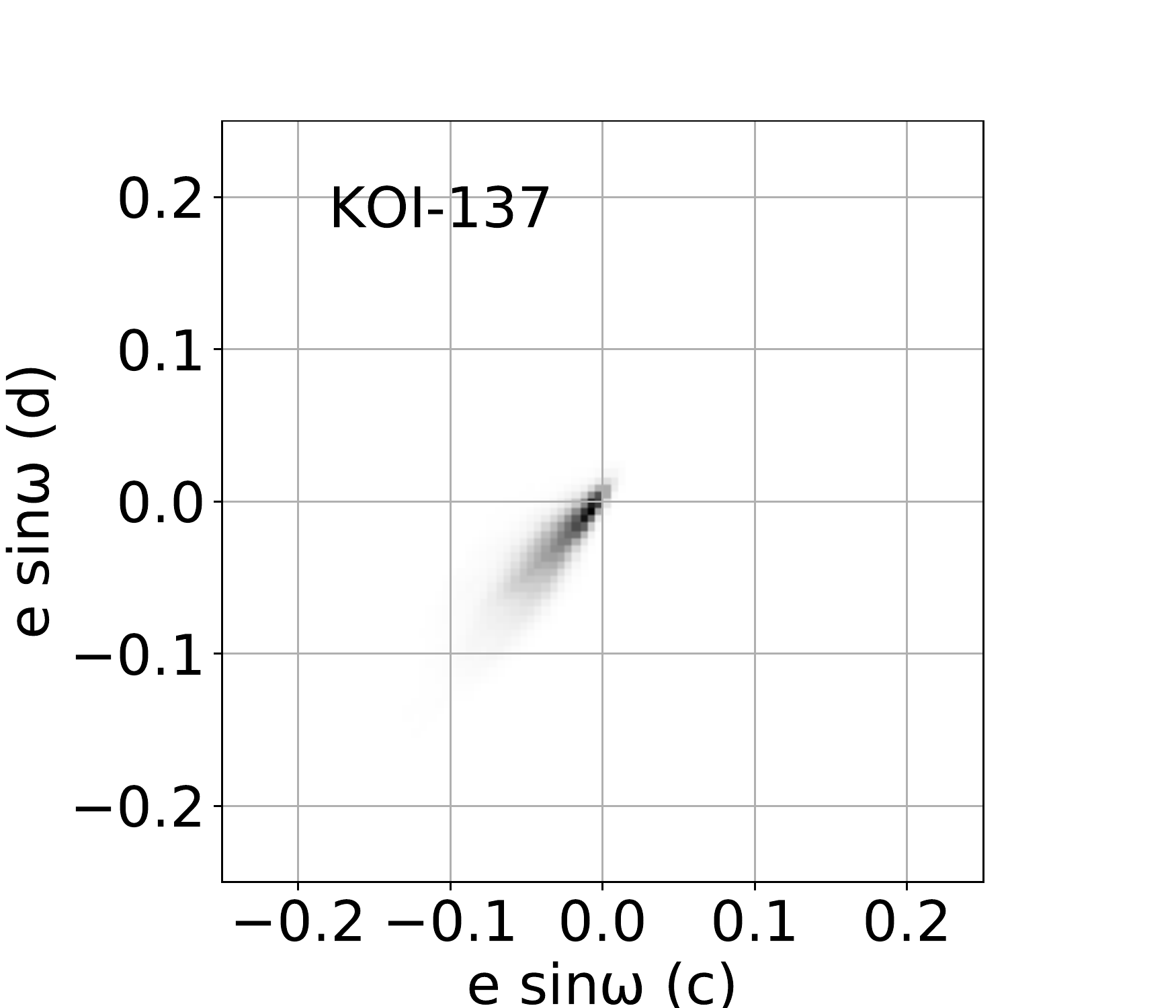}
\includegraphics [height = 1.1 in]{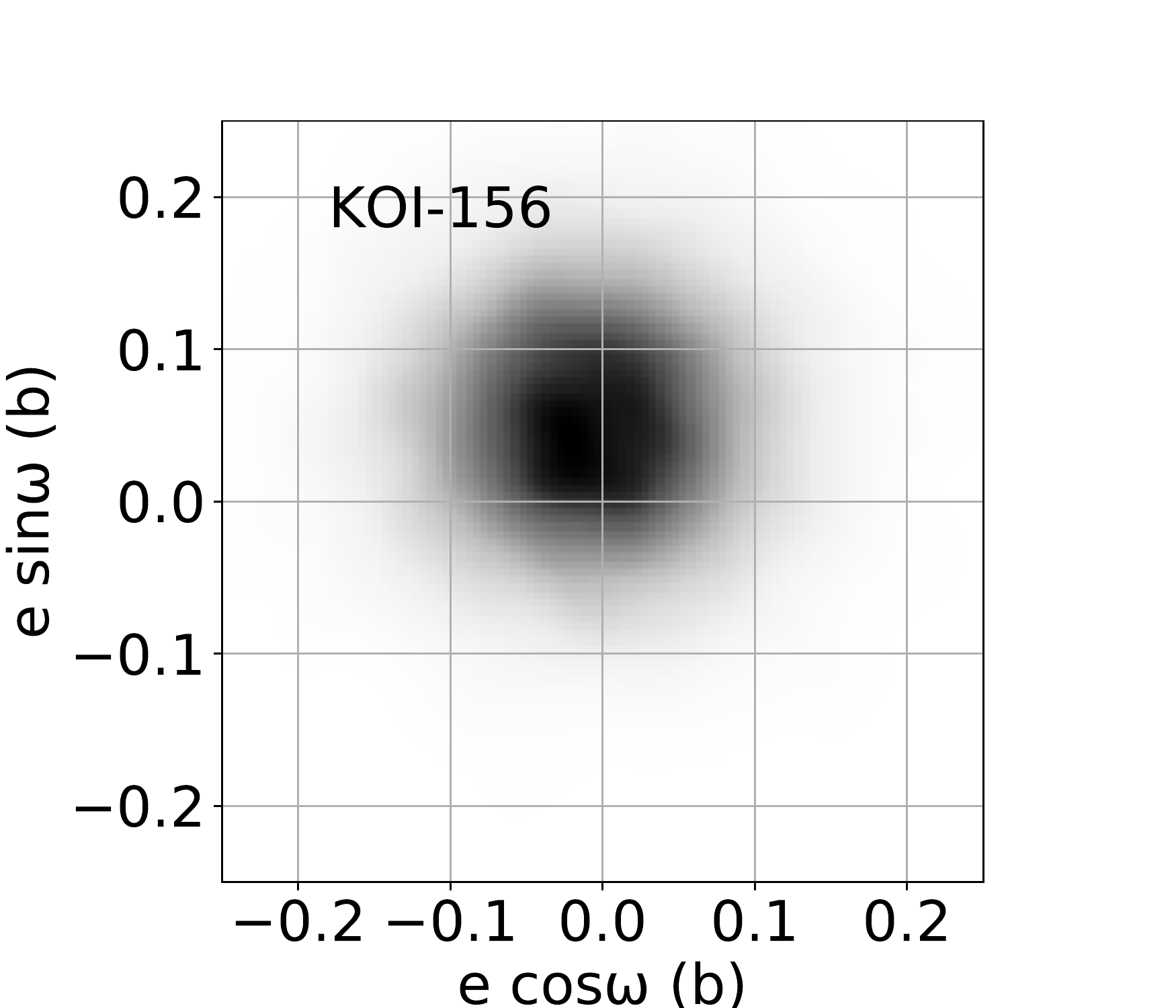} 
\includegraphics [height = 1.1 in]{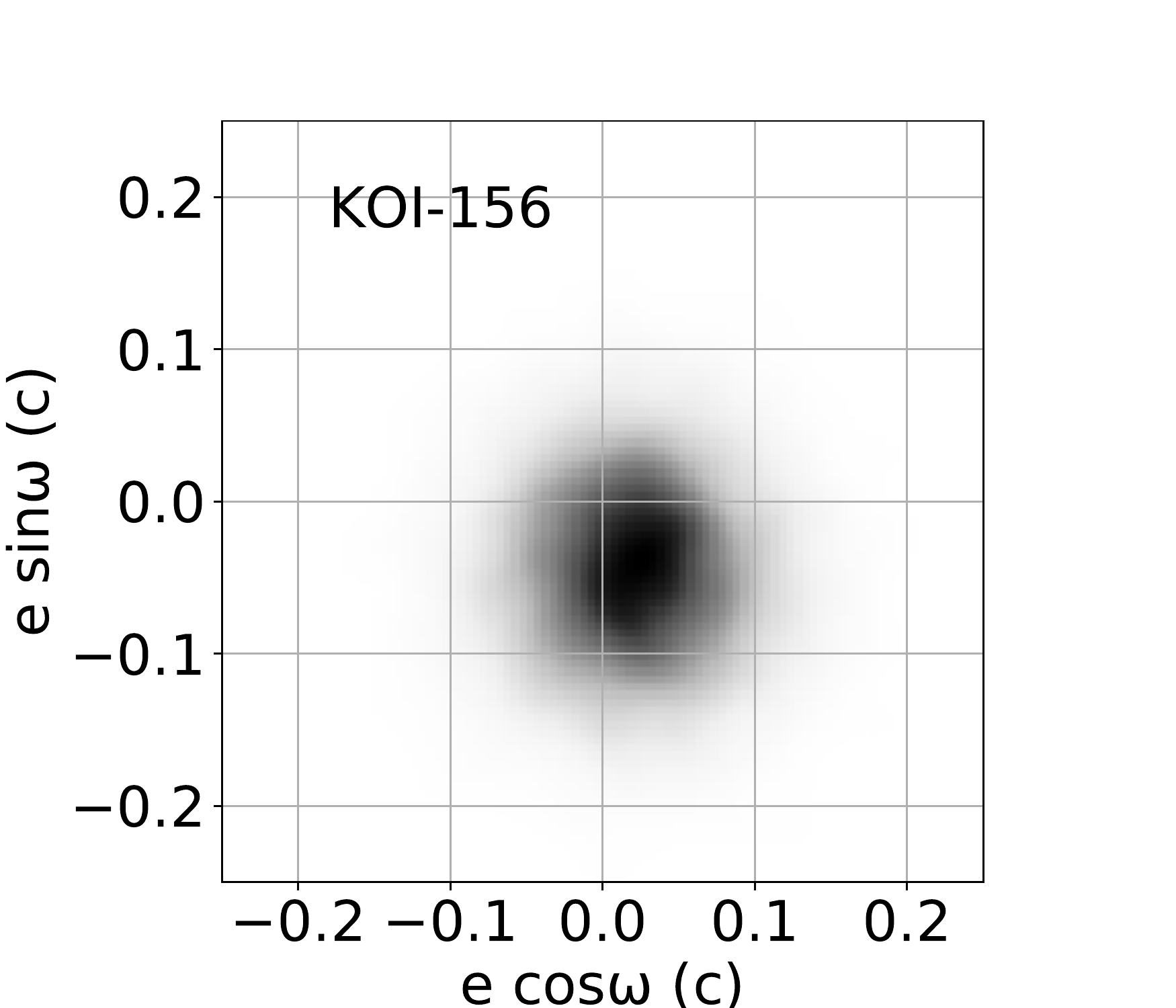}
\includegraphics [height = 1.1 in]{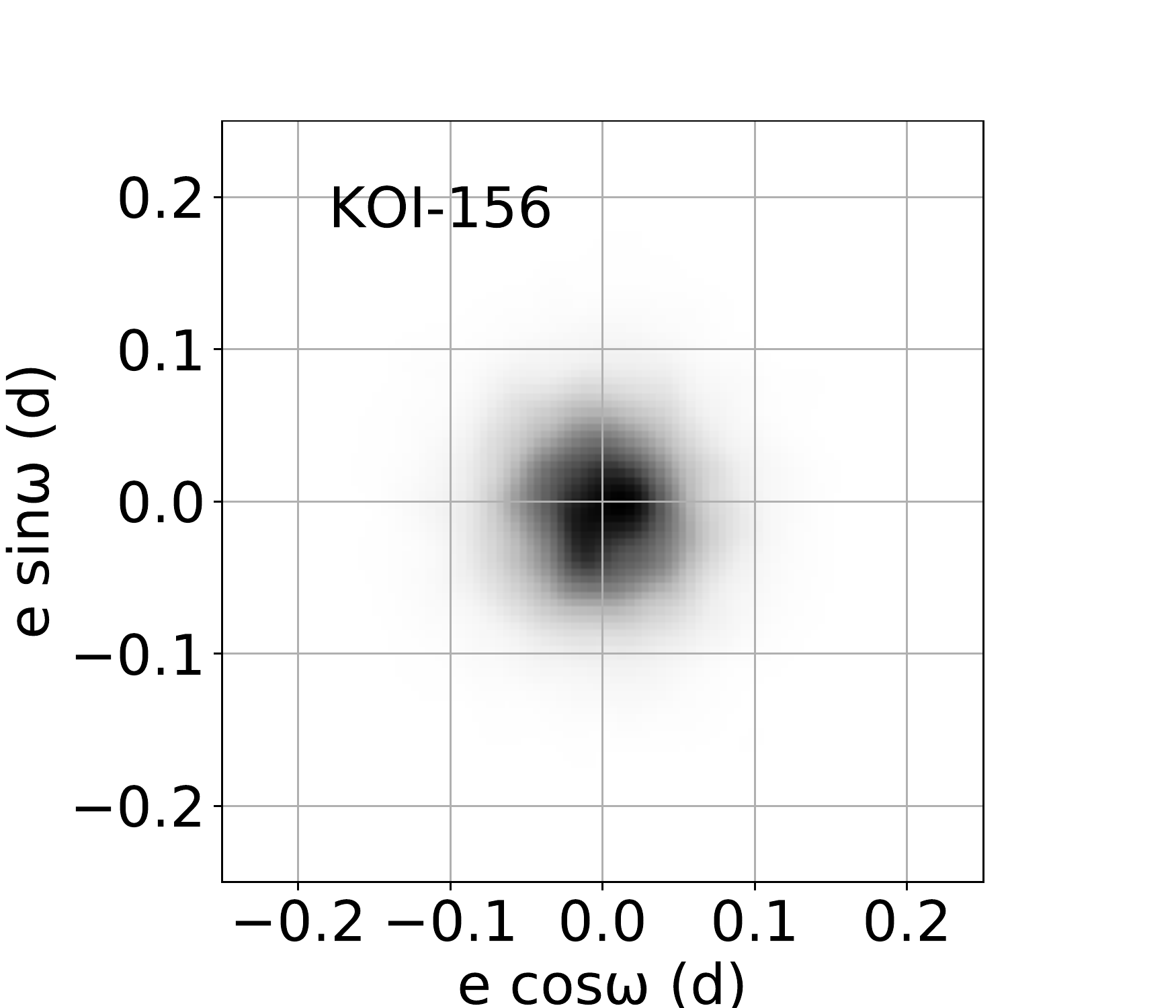} \\
\includegraphics [height = 1.1 in]{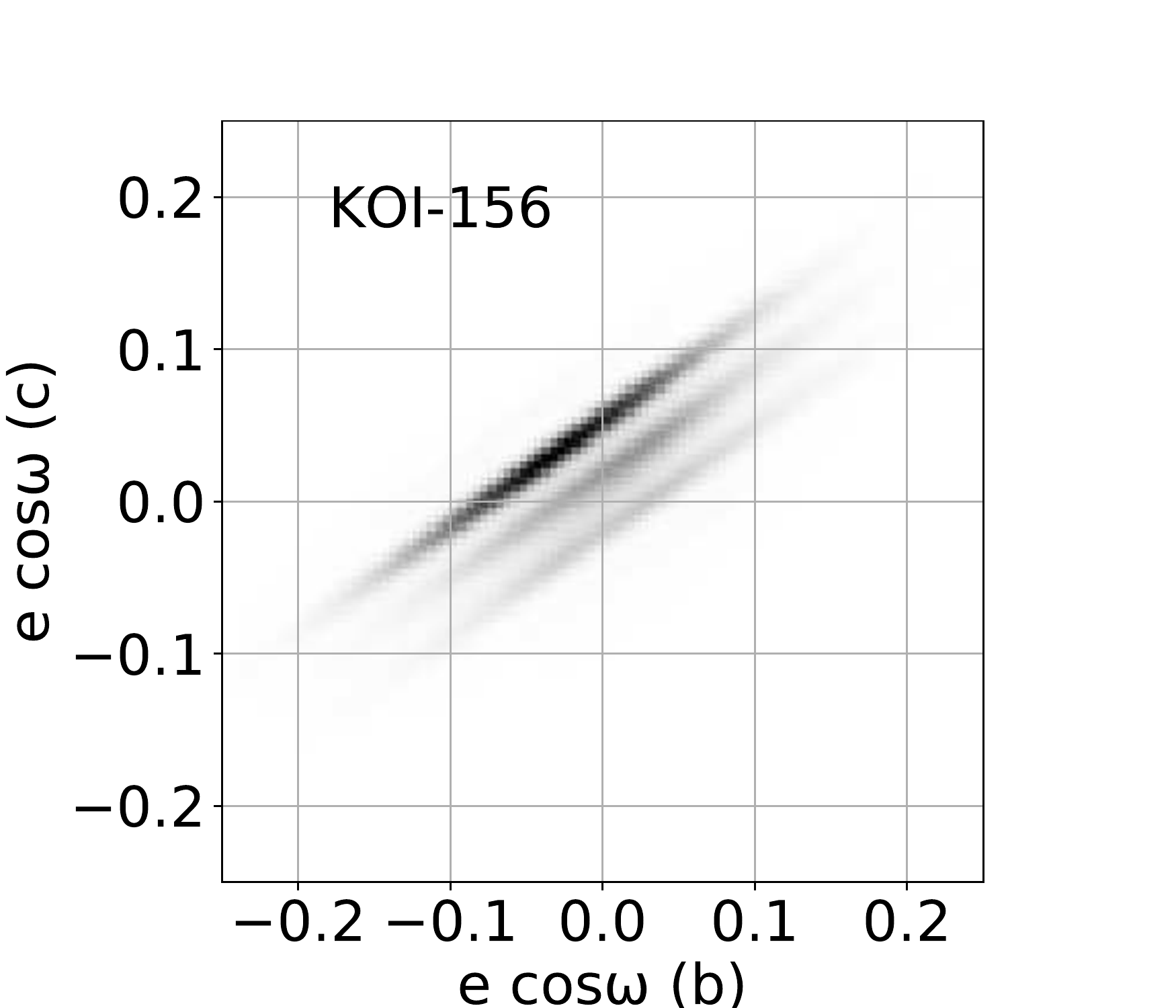} 
\includegraphics [height = 1.1 in]{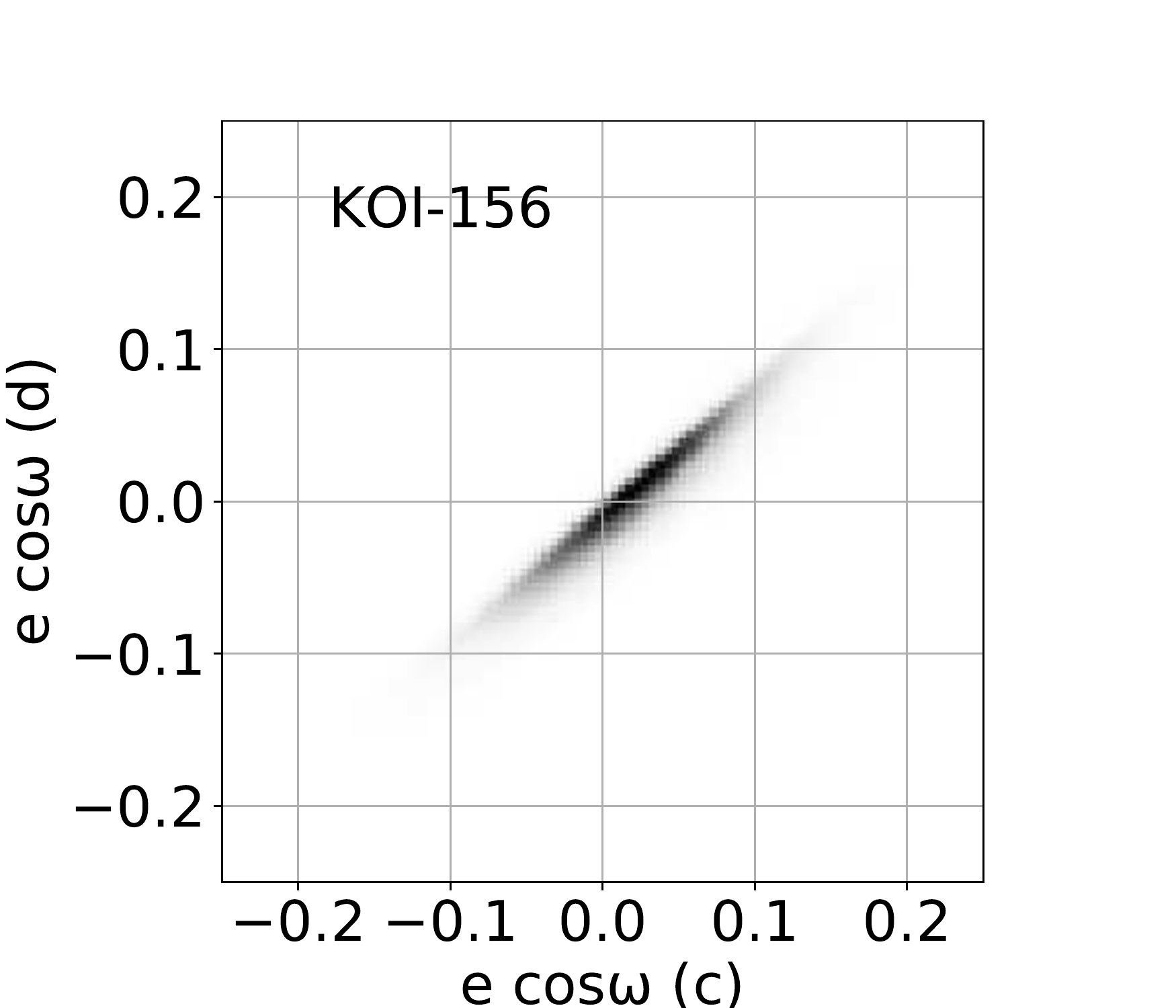}
\includegraphics [height = 1.1 in]{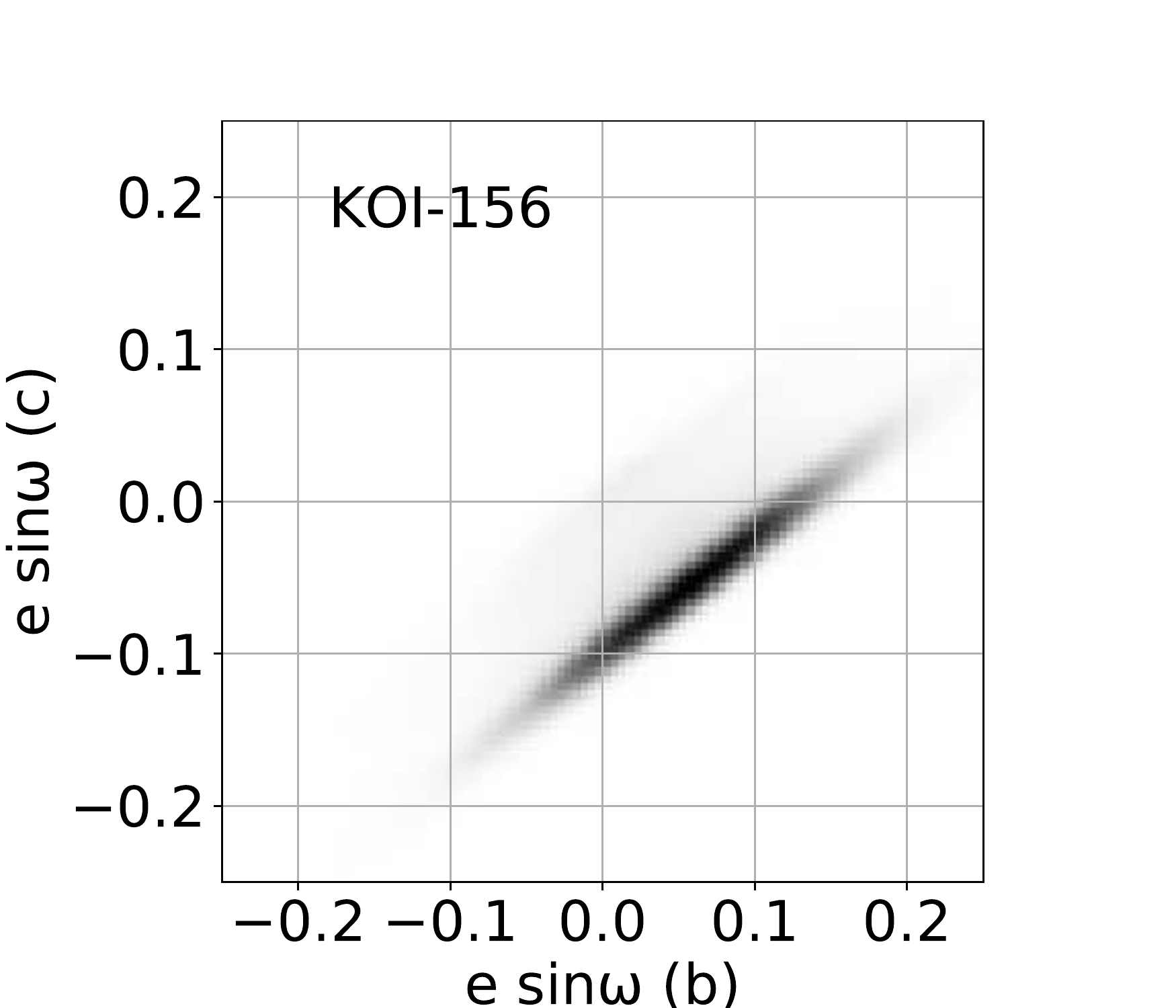} 
\includegraphics [height = 1.1 in]{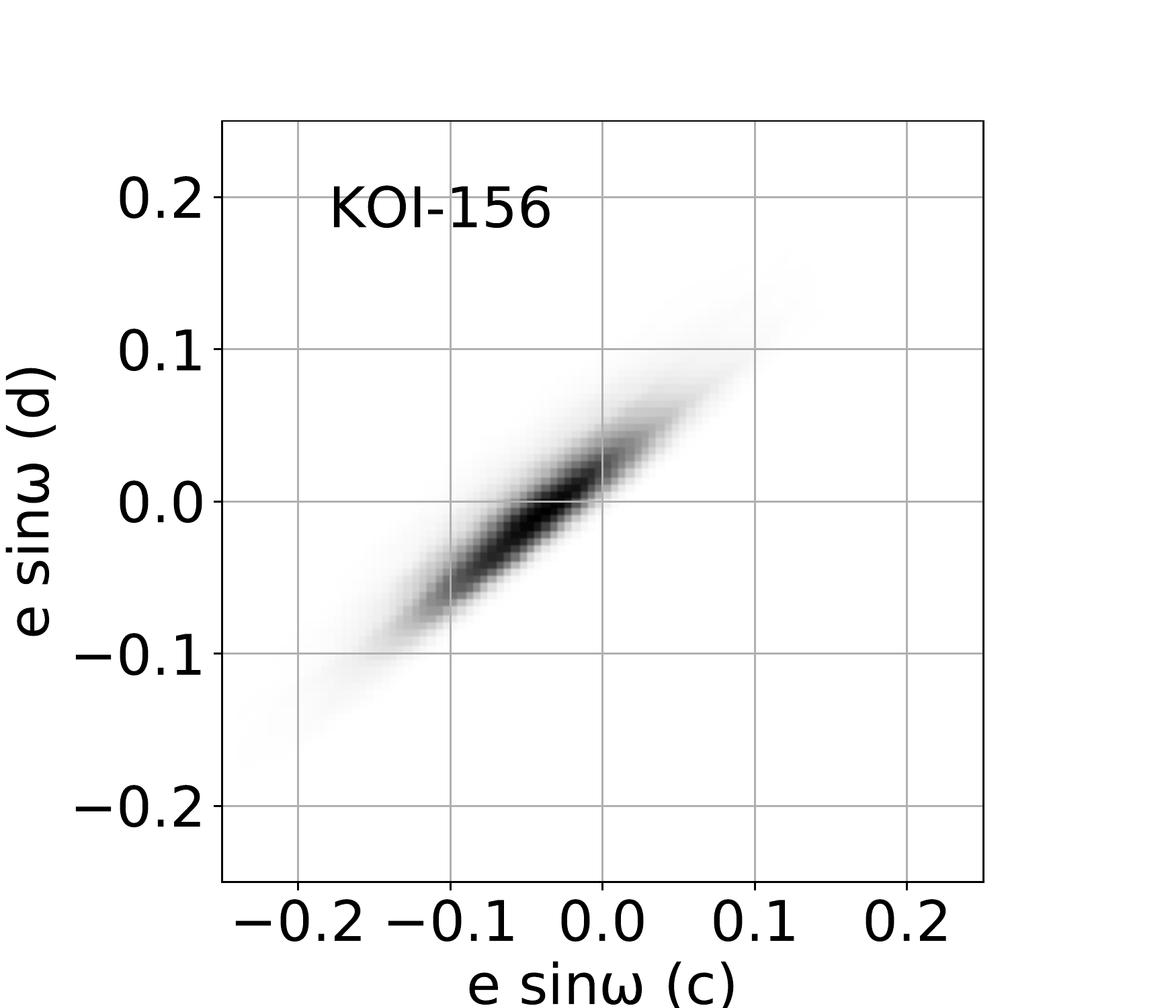}
\caption{Two-dimensional kernel density estimators on joint posteriors of eccentricity vector components: three-planet systems. (Part 1 of 7). 
}
\label{fig:ecc3a} 
\end{center}
\end{figure}

\begin{figure}
\begin{center}
\figurenum{24}
\includegraphics [height = 1.1 in]{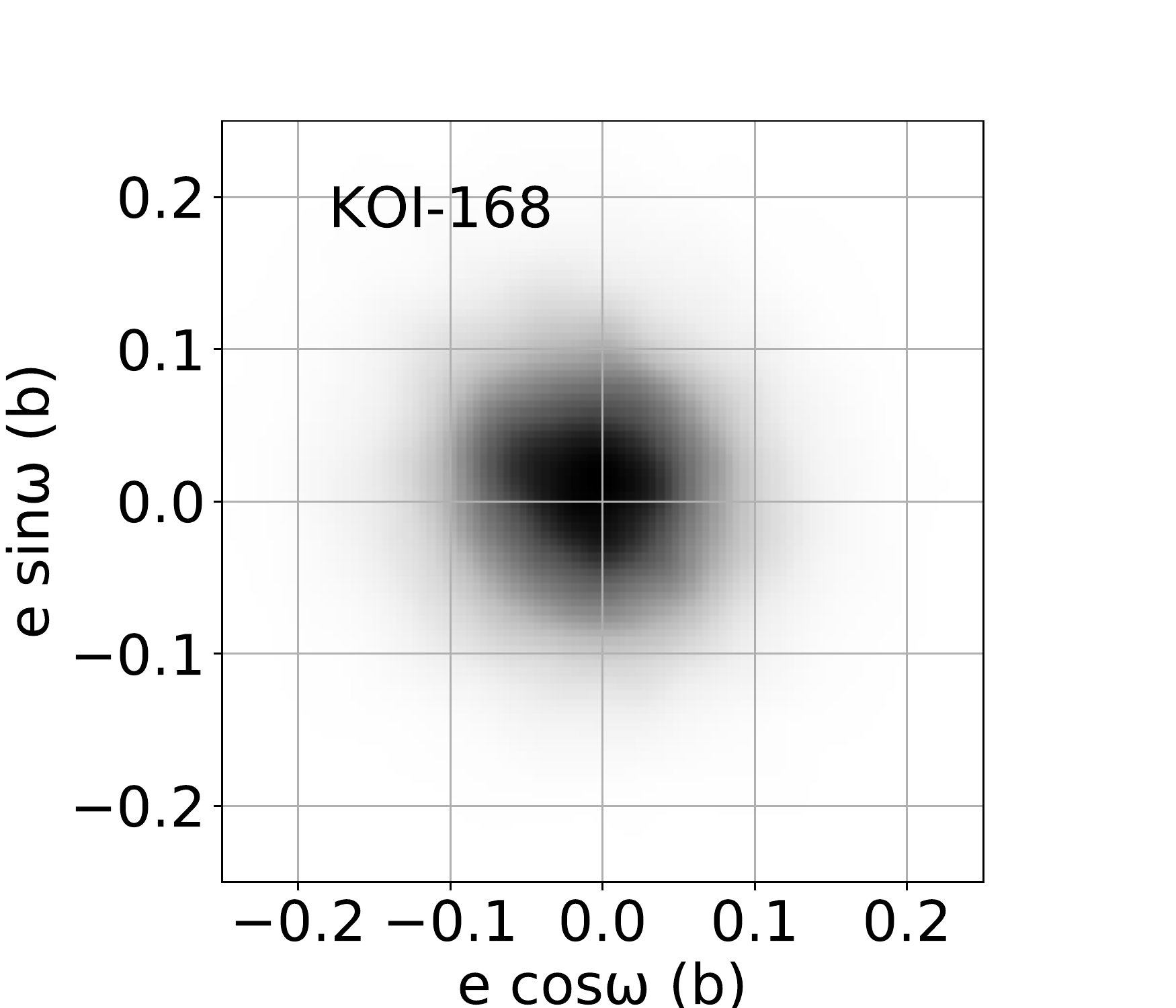}
\includegraphics [height = 1.1 in]{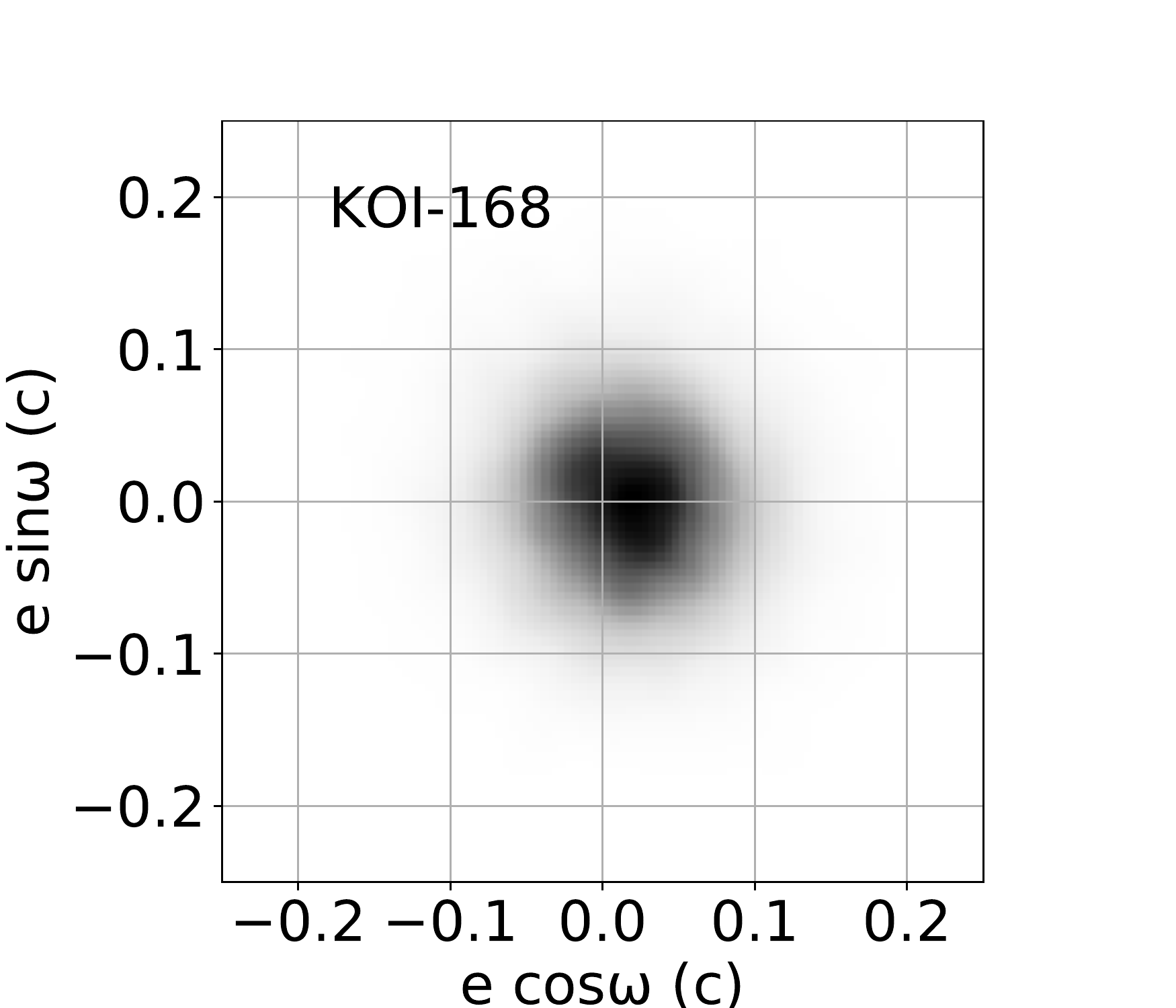}
\includegraphics [height = 1.1 in]{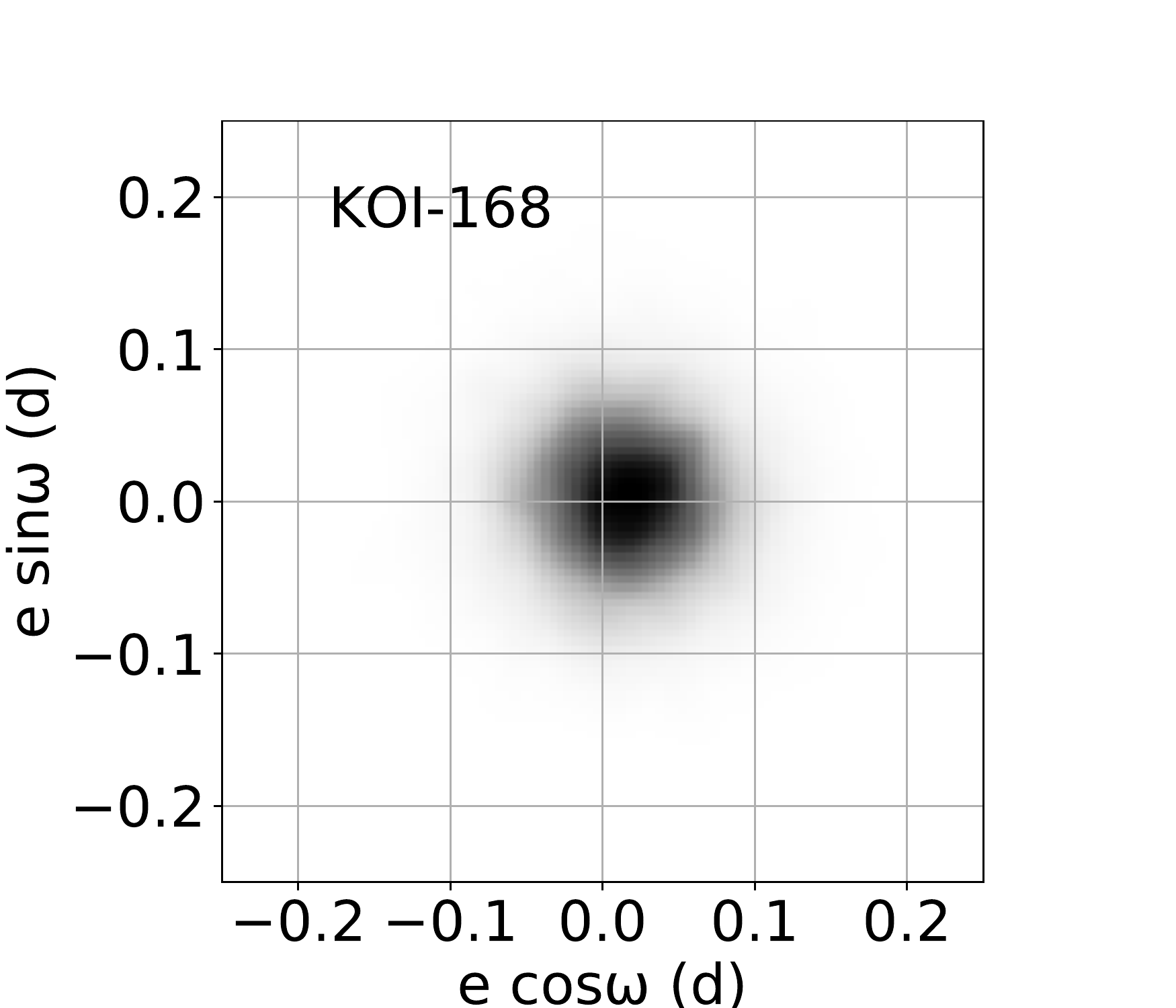}
\includegraphics [height = 1.1 in]{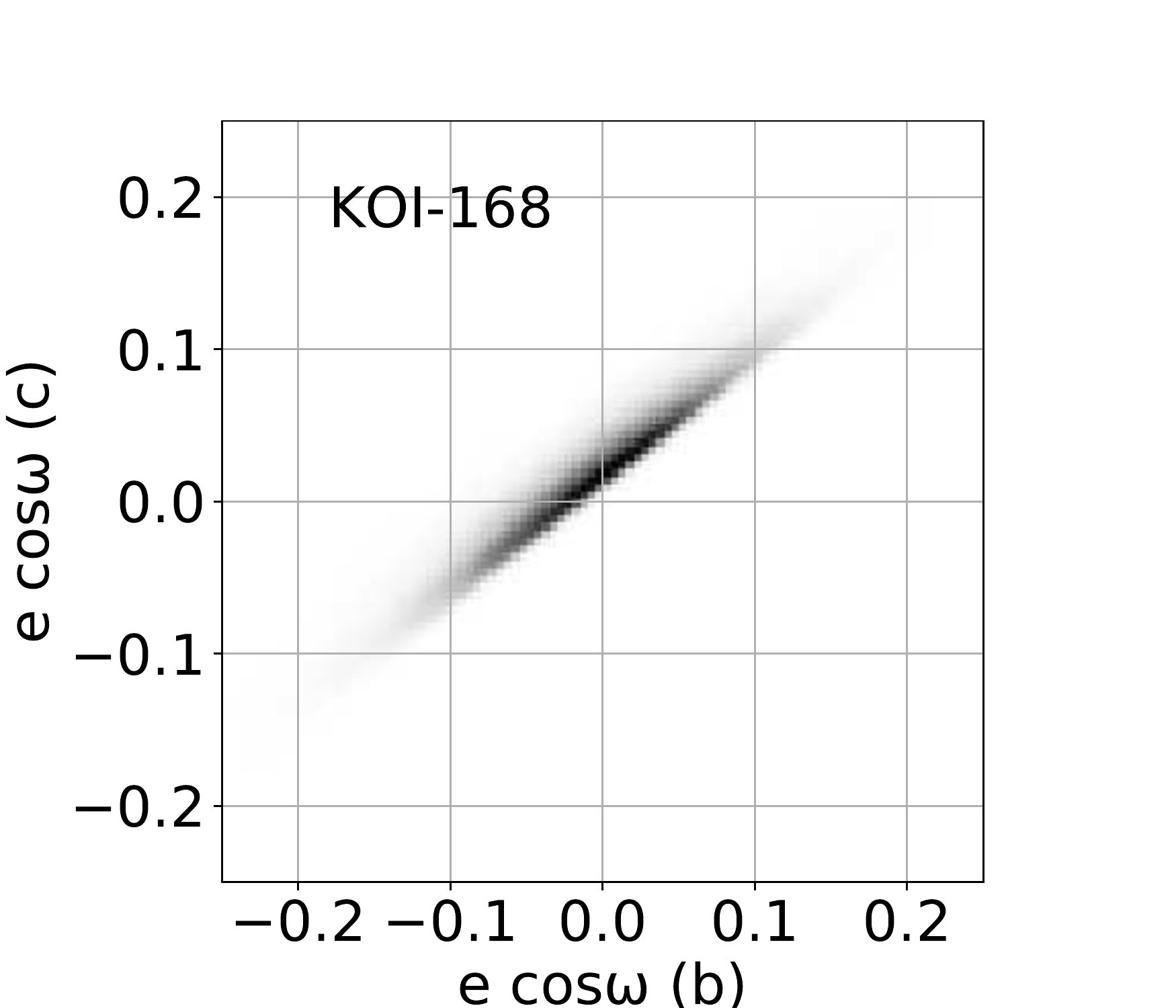} \\
\includegraphics [height = 1.1 in]{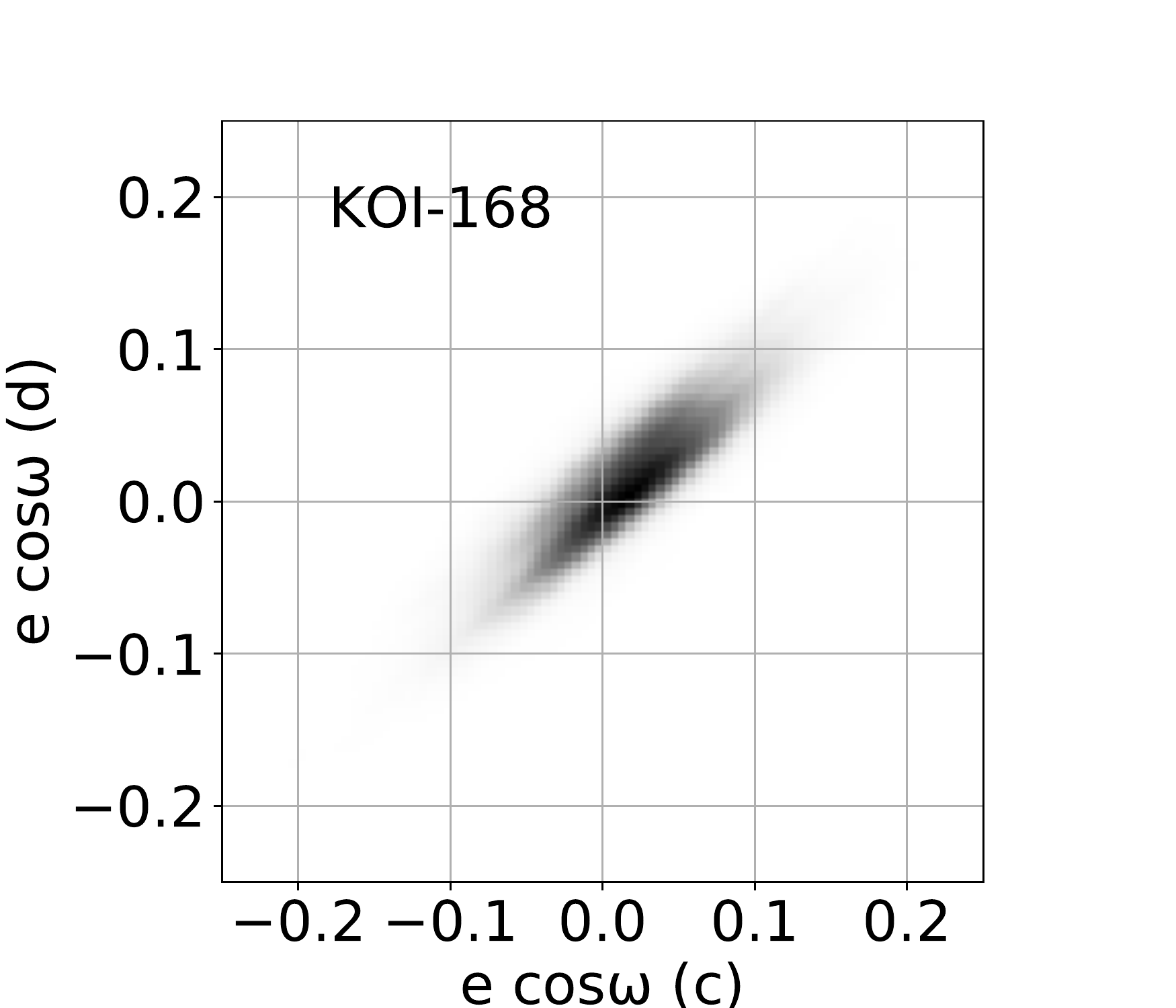}
\includegraphics [height = 1.1 in]{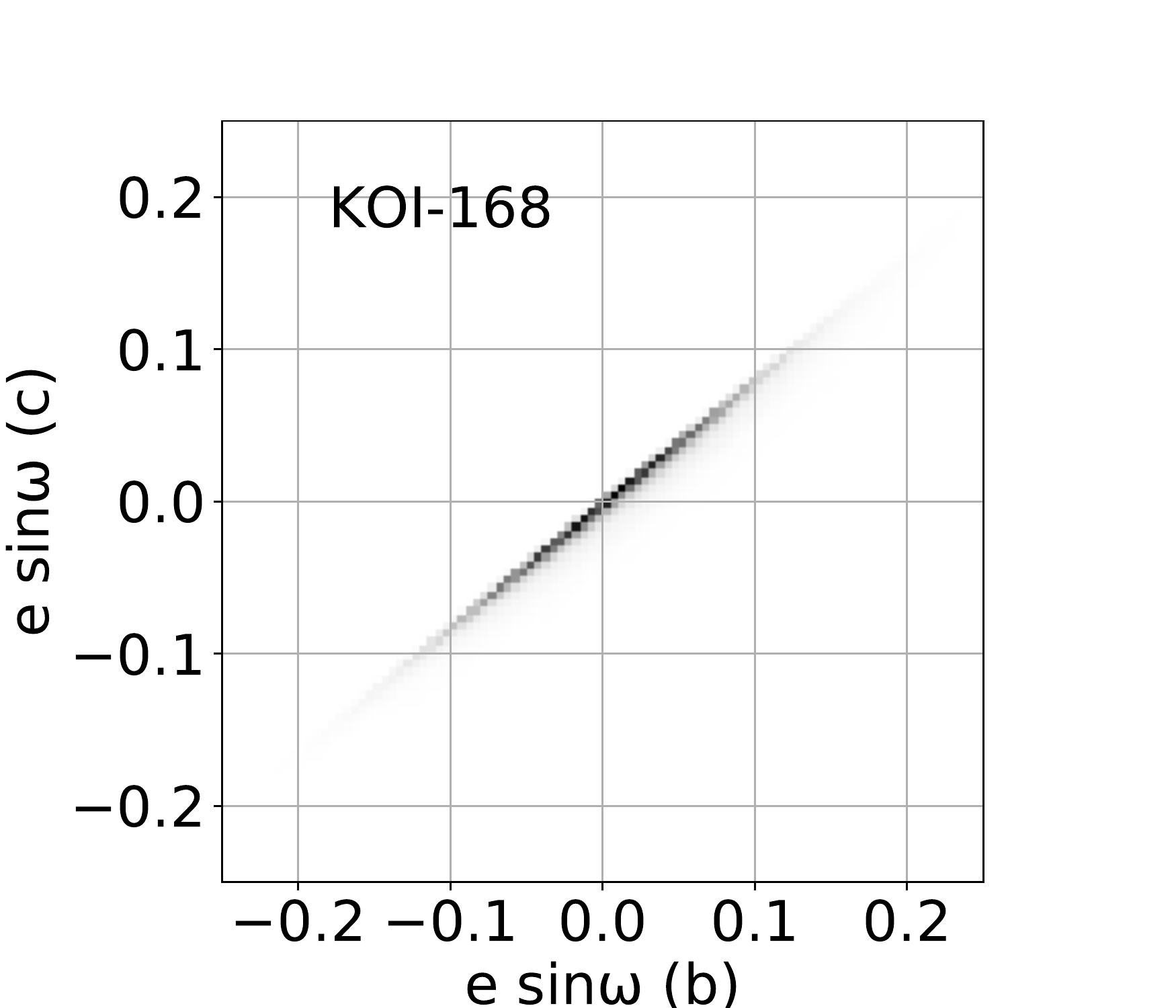}
\includegraphics [height = 1.1 in]{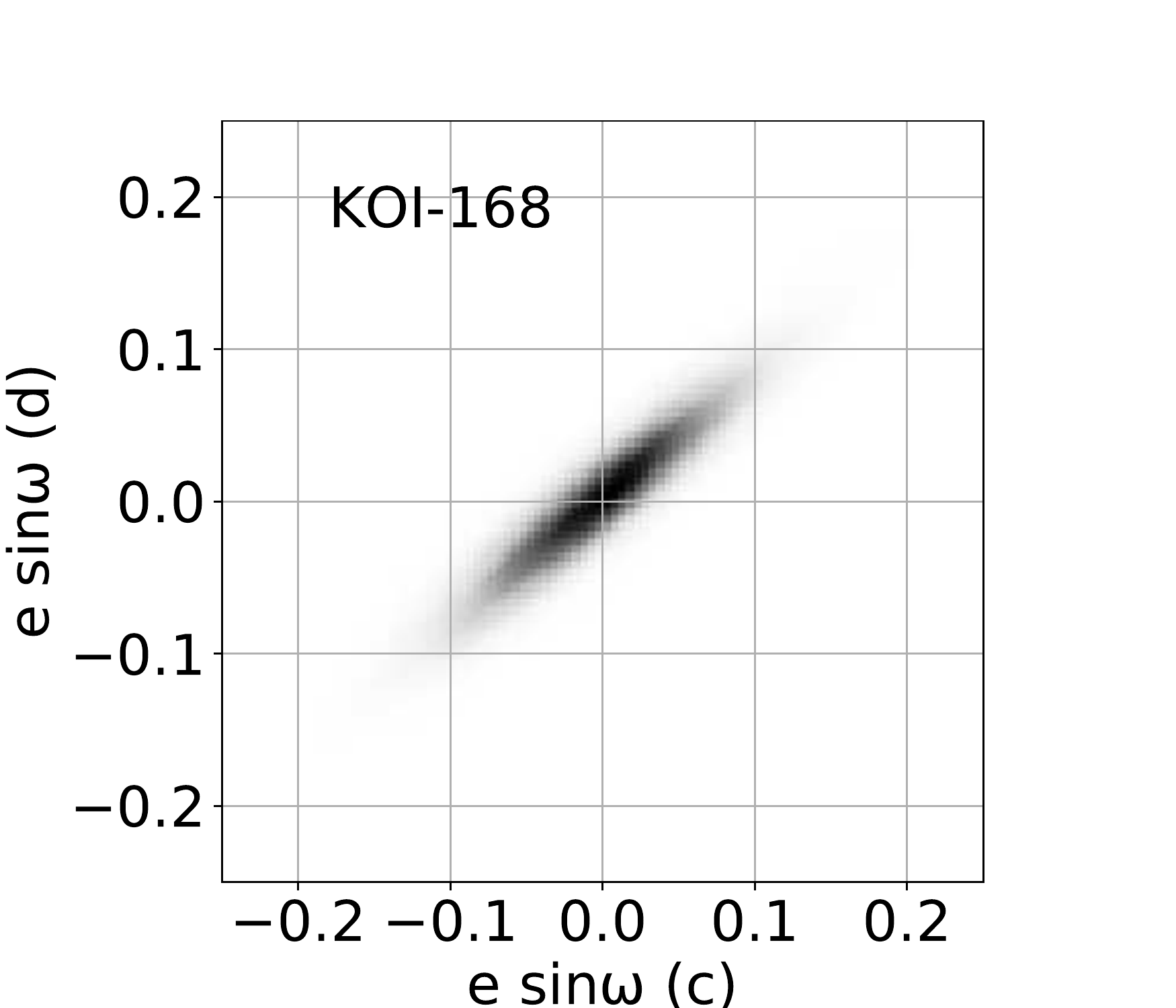}
\includegraphics [height = 1.1 in]{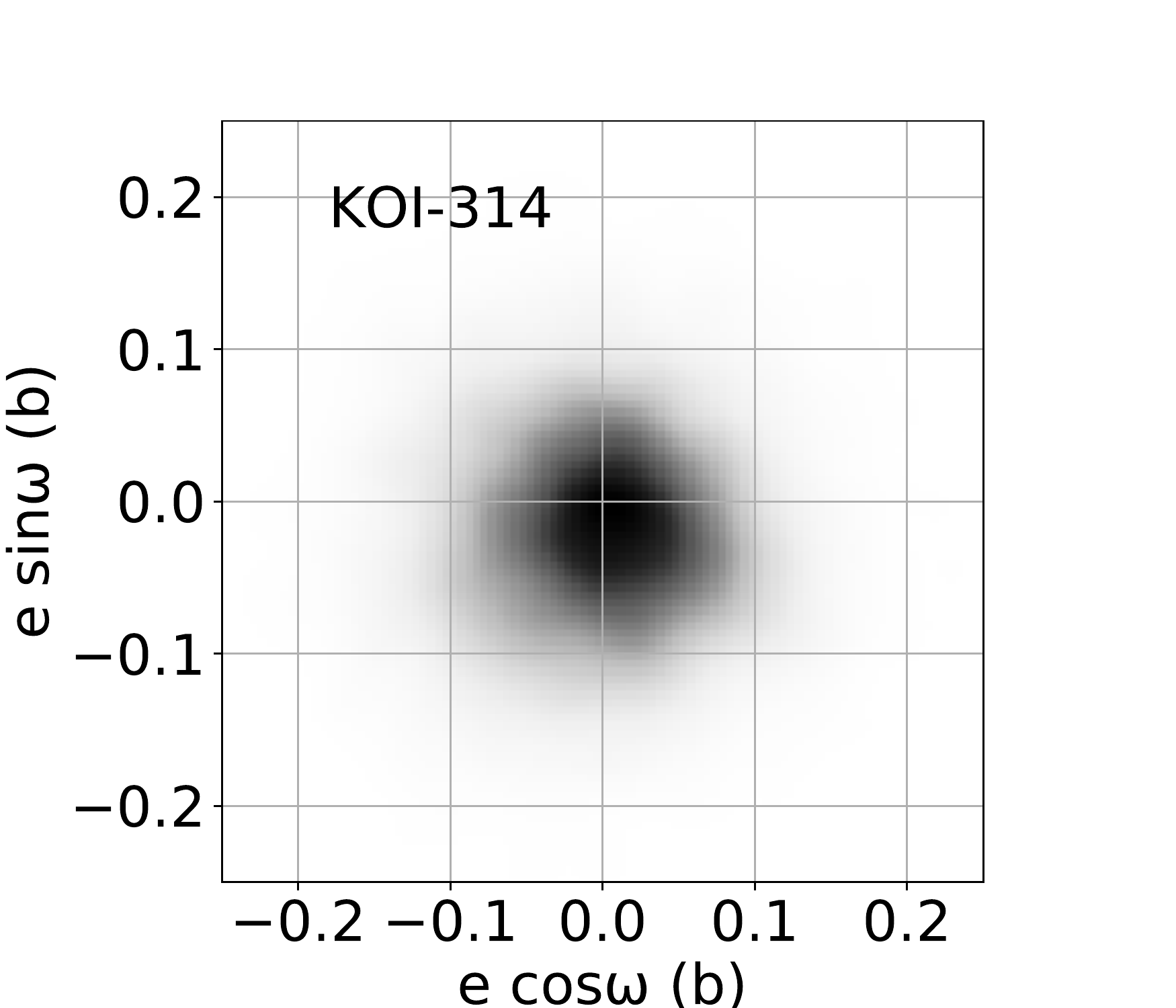} \\
\includegraphics [height = 1.1 in]{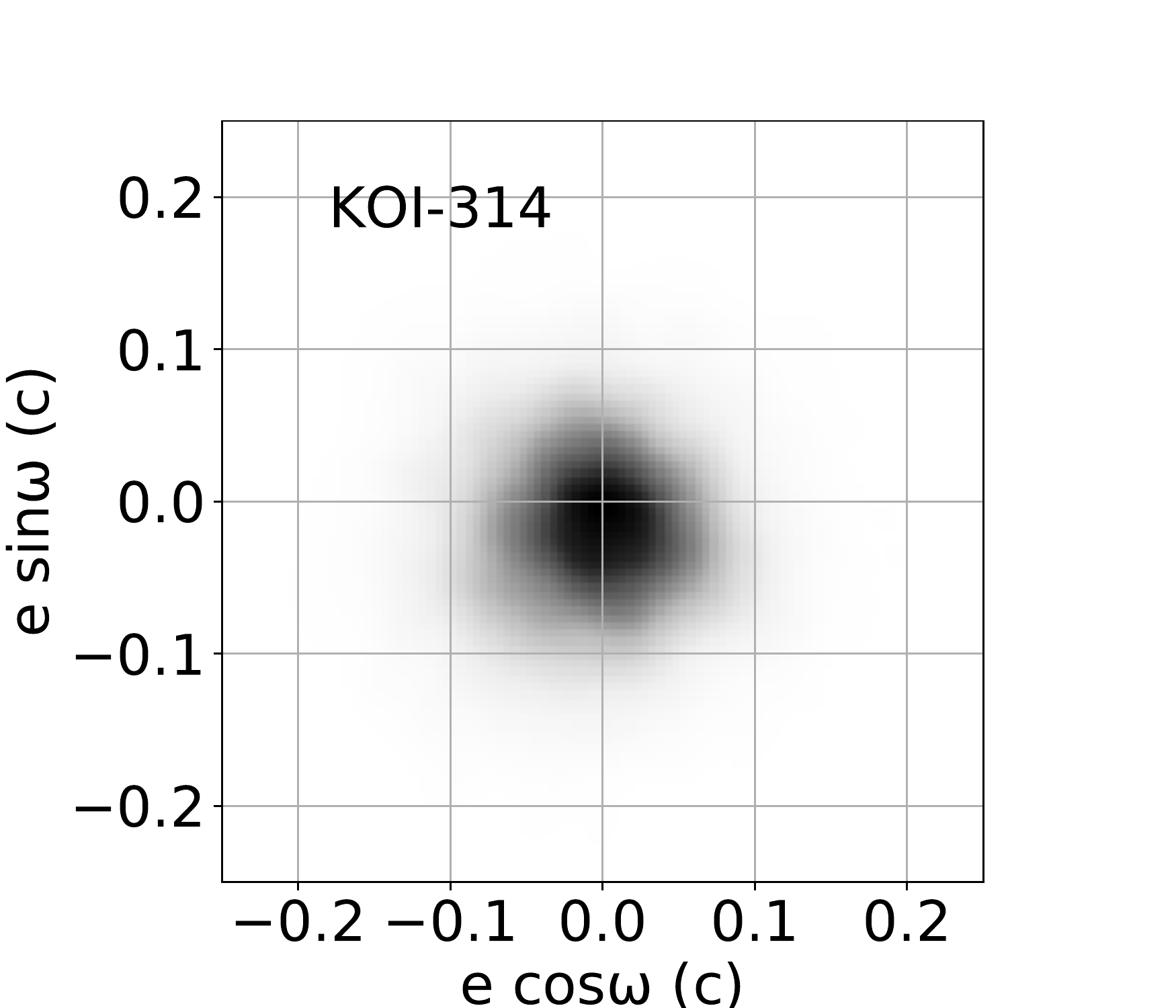}
\includegraphics [height = 1.1 in]{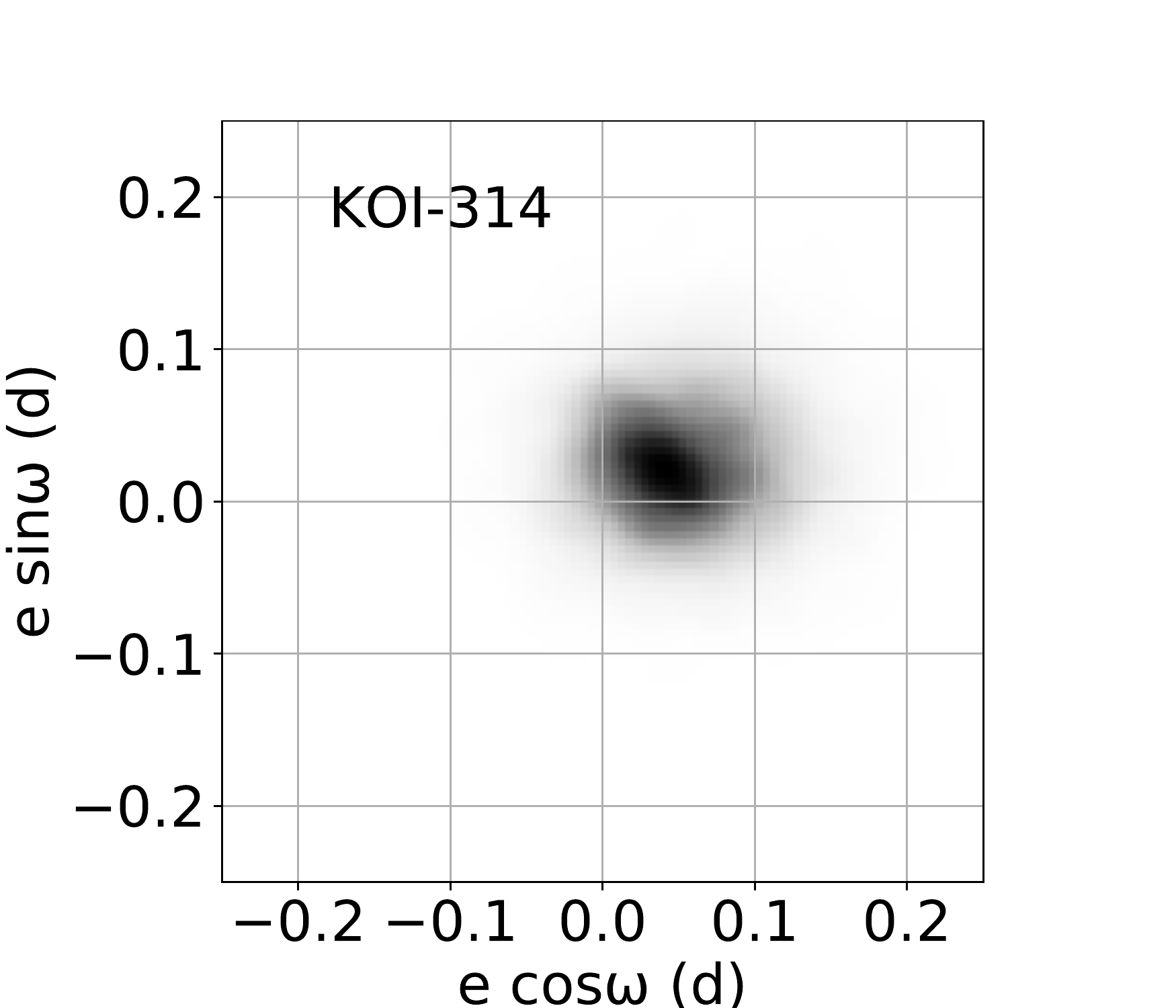}
\includegraphics [height = 1.1 in]{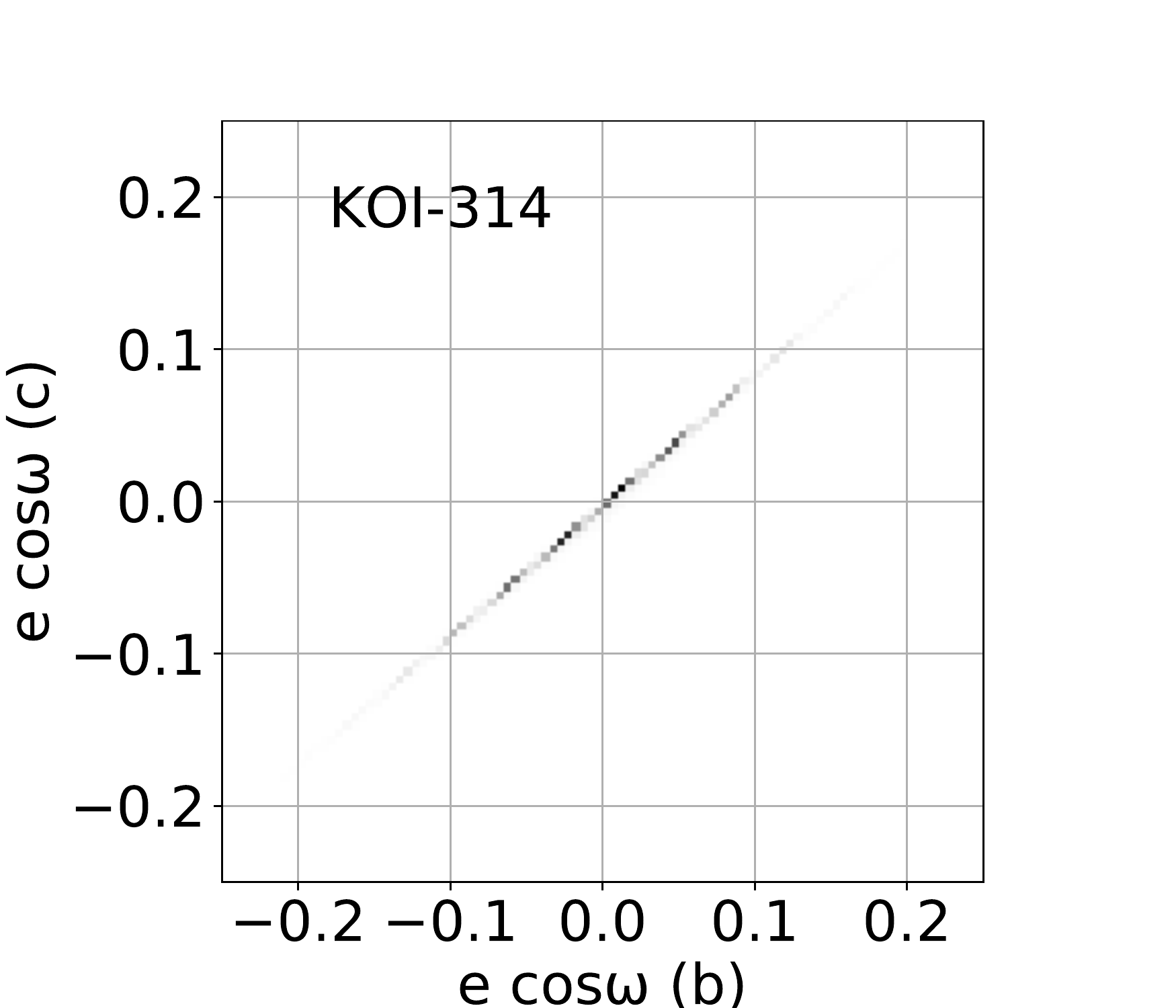}
\includegraphics [height = 1.1 in]{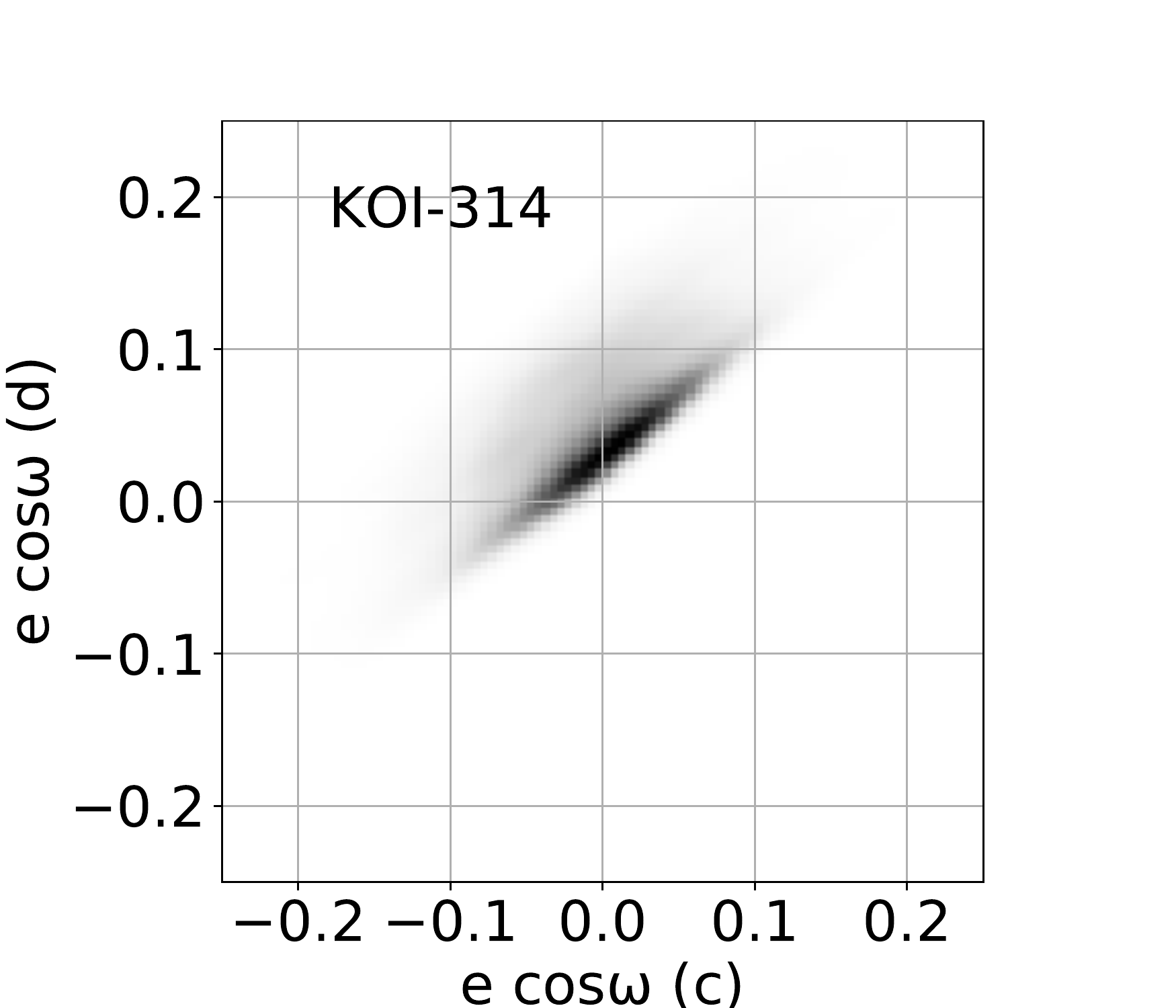} \\
\includegraphics [height = 1.1 in]{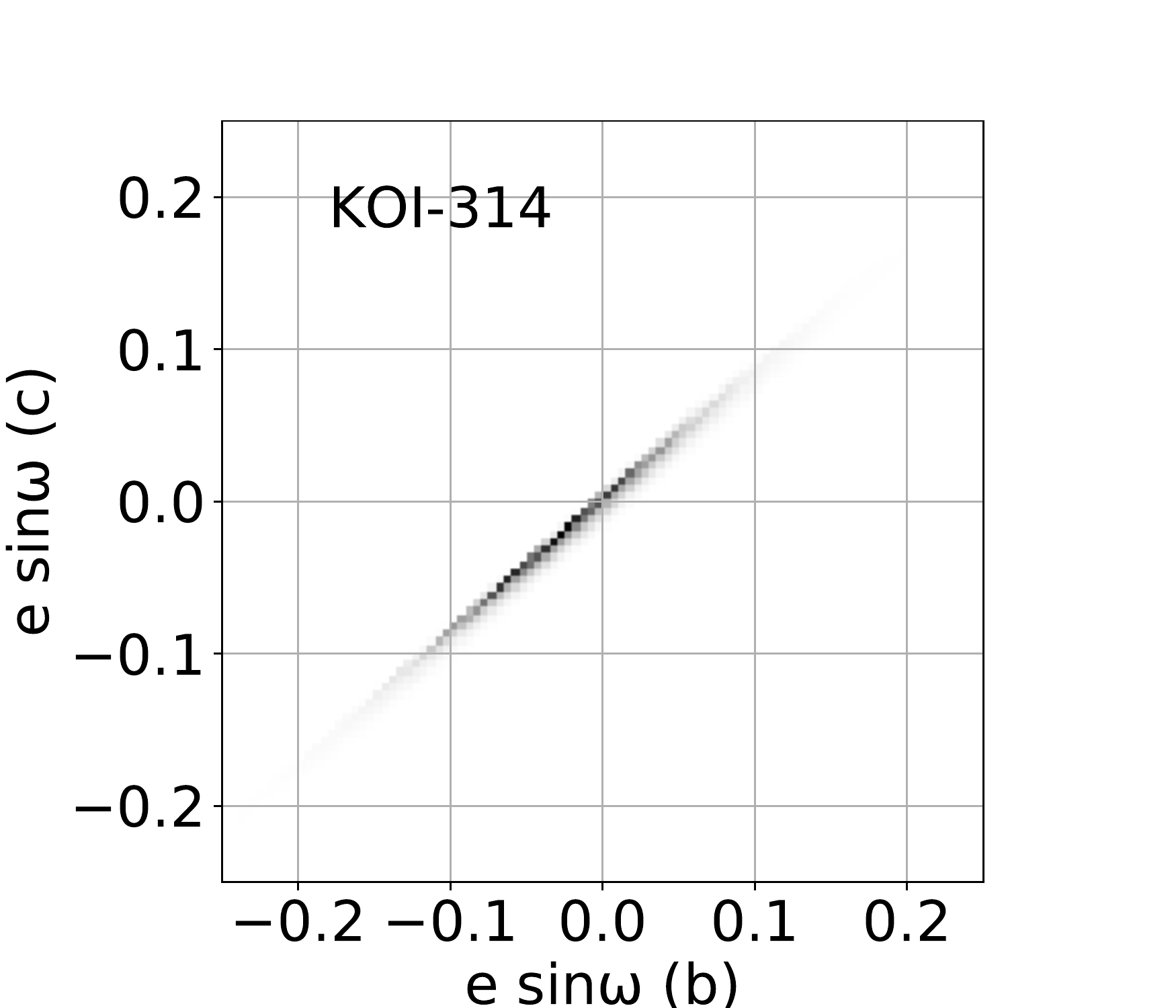}
\includegraphics [height = 1.1 in]{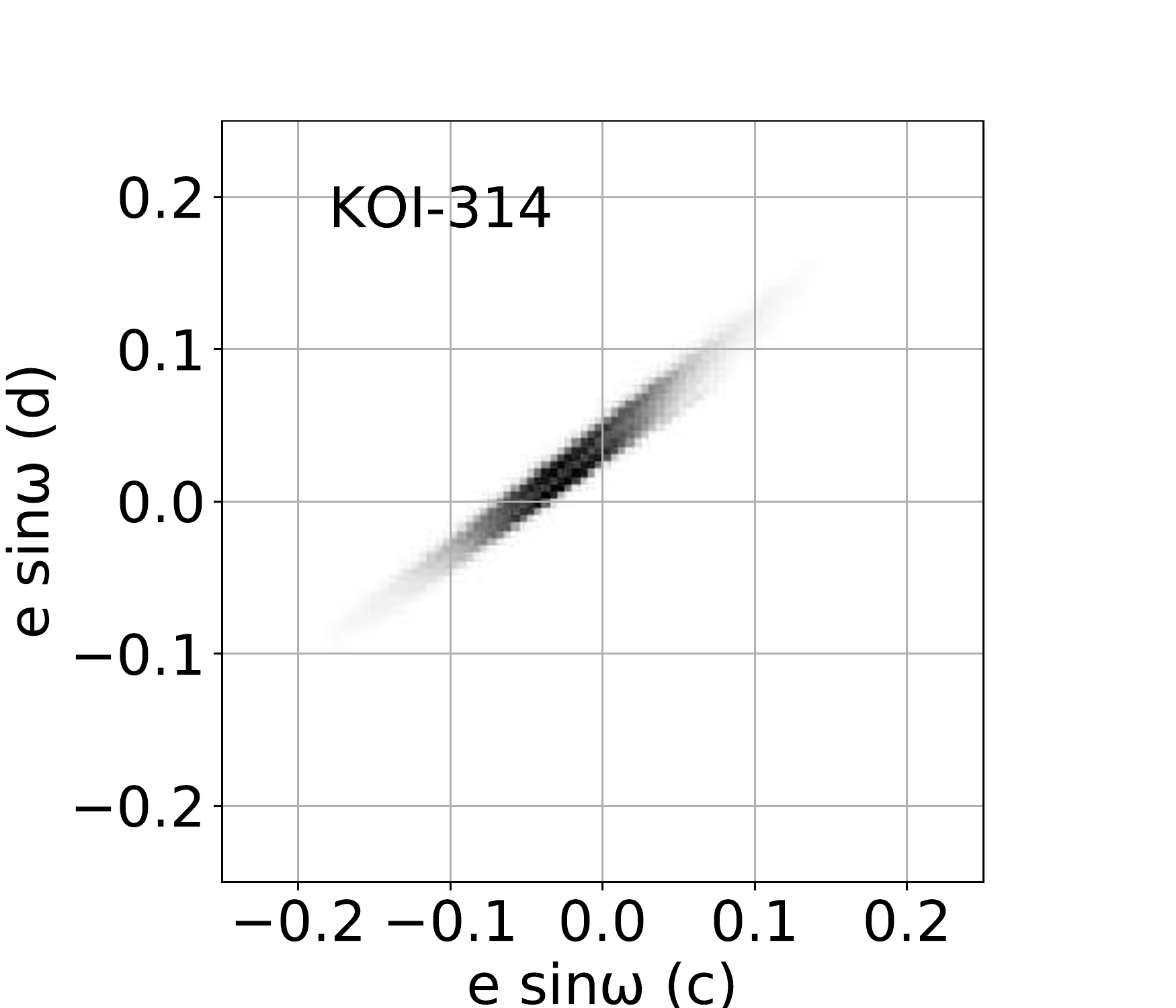}
\includegraphics [height = 1.1 in]{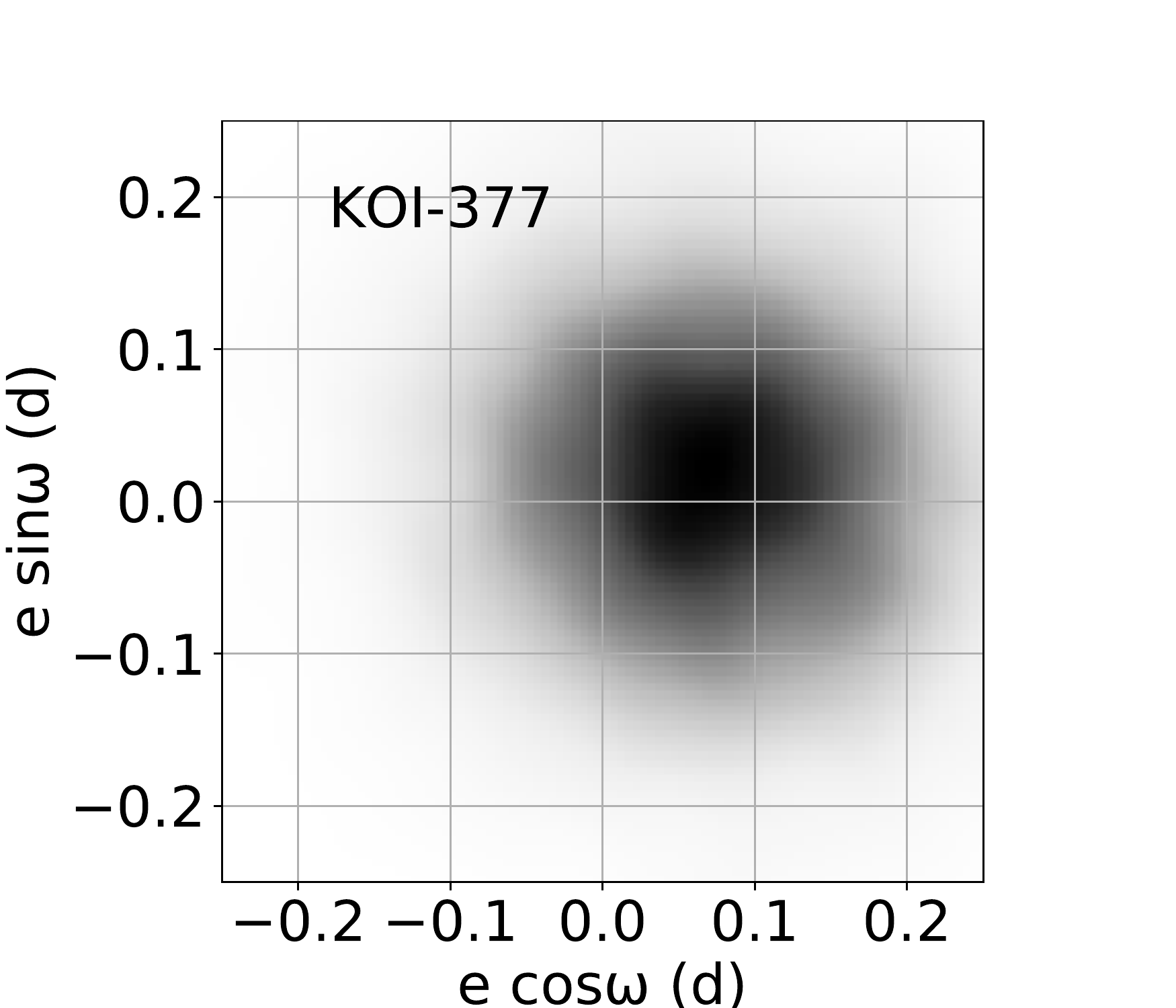}
\includegraphics [height = 1.1 in]{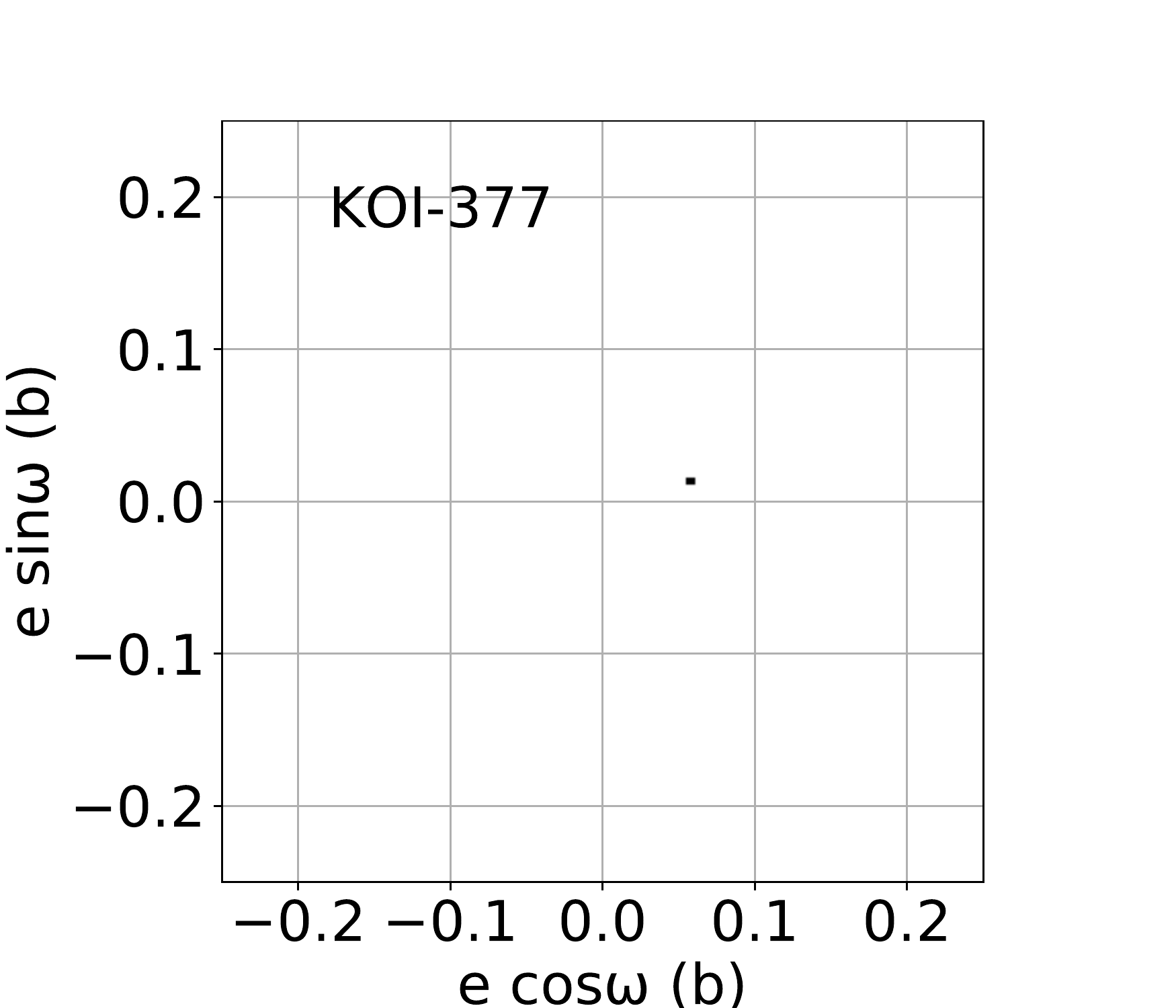} \\
\includegraphics [height = 1.1 in]{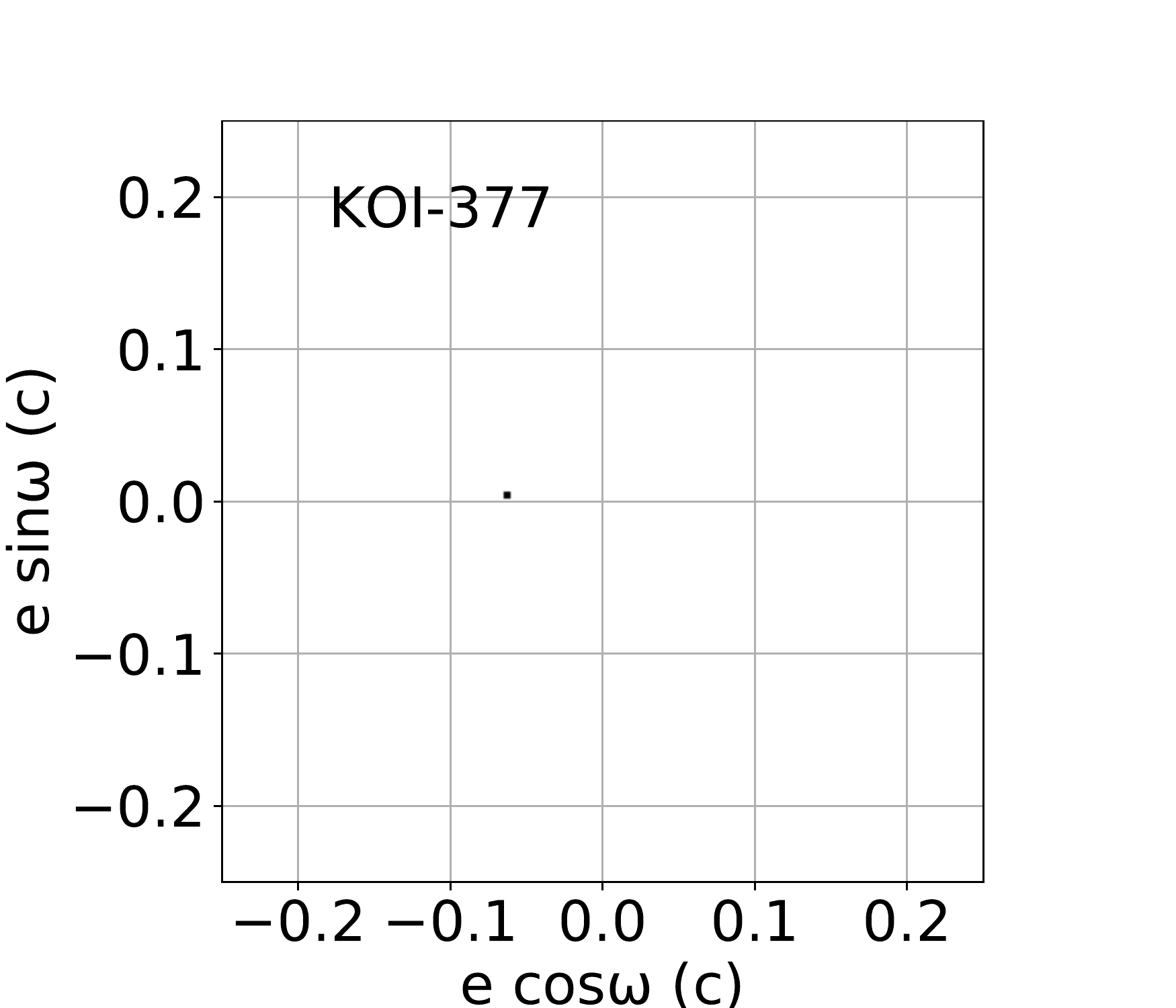}
\includegraphics [height = 1.1 in]{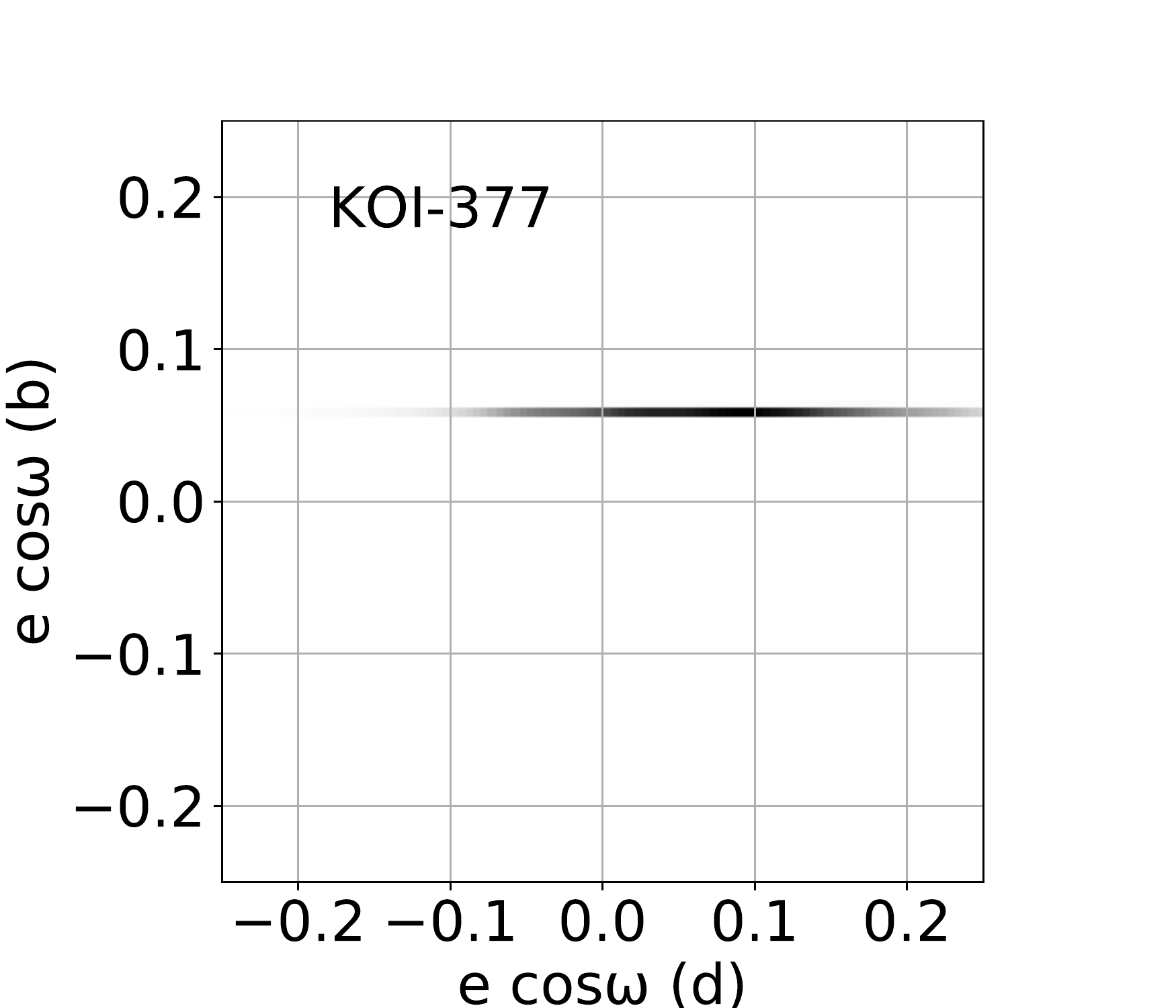} 
\includegraphics [height = 1.1 in]{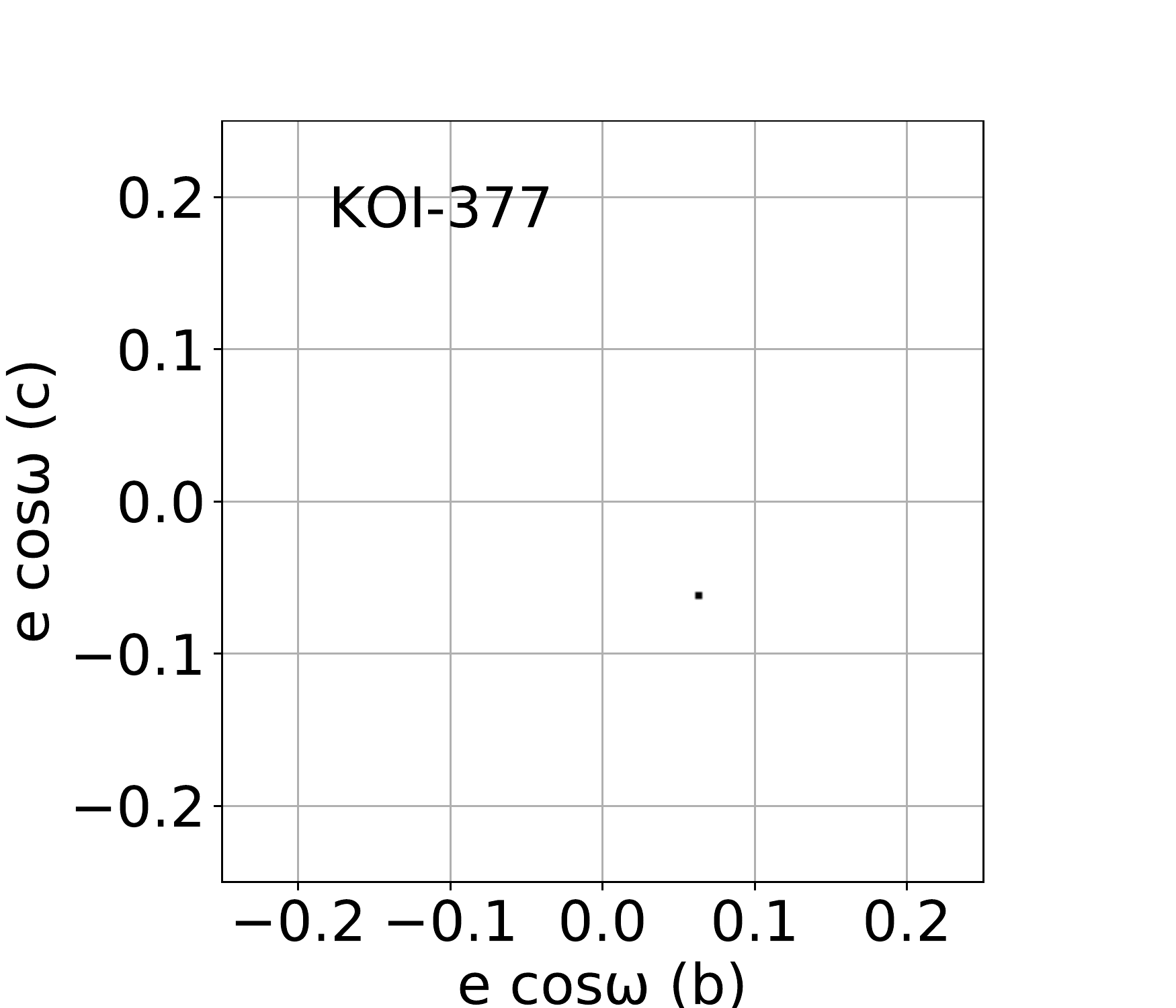}
\includegraphics [height = 1.1 in]{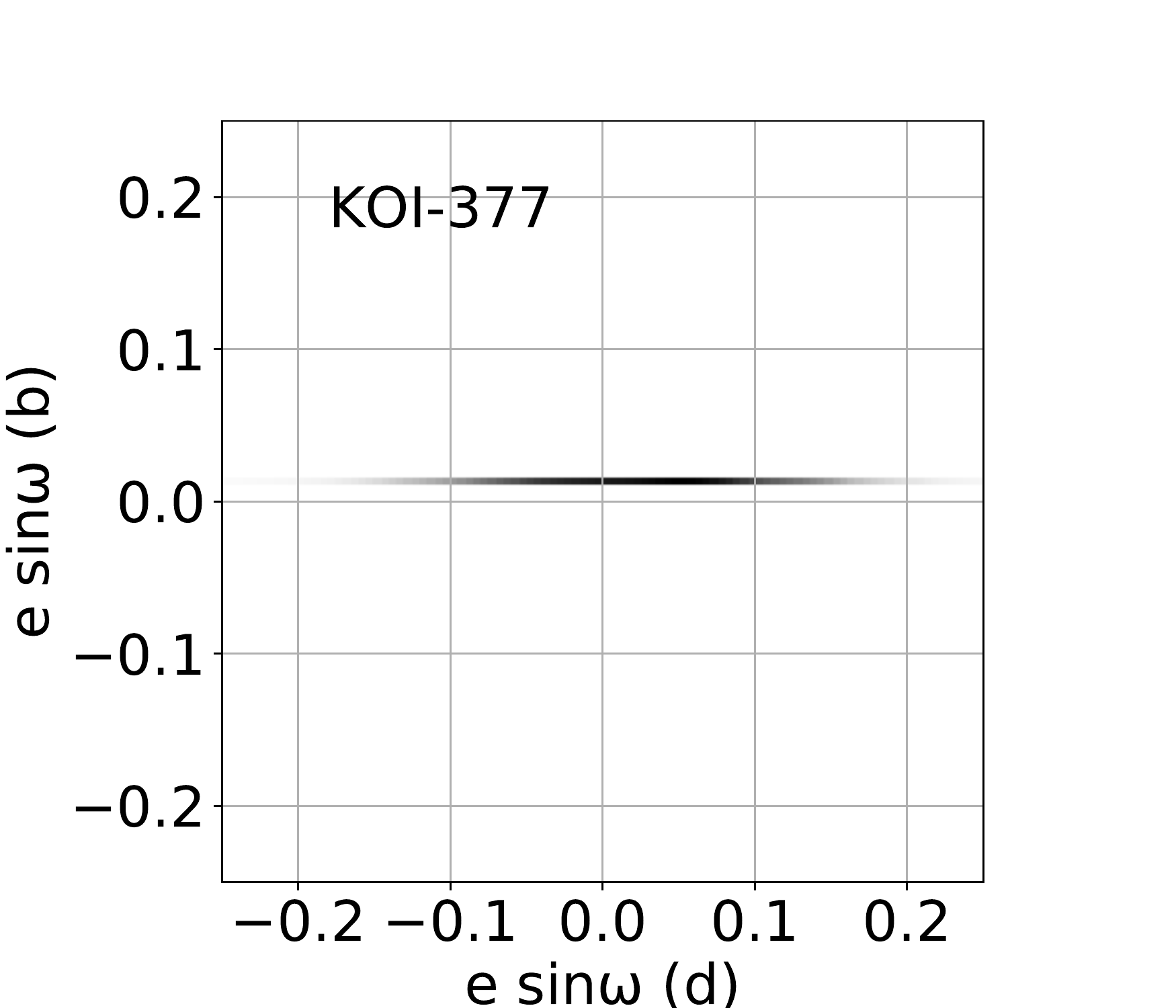} \\
\includegraphics [height = 1.1 in]{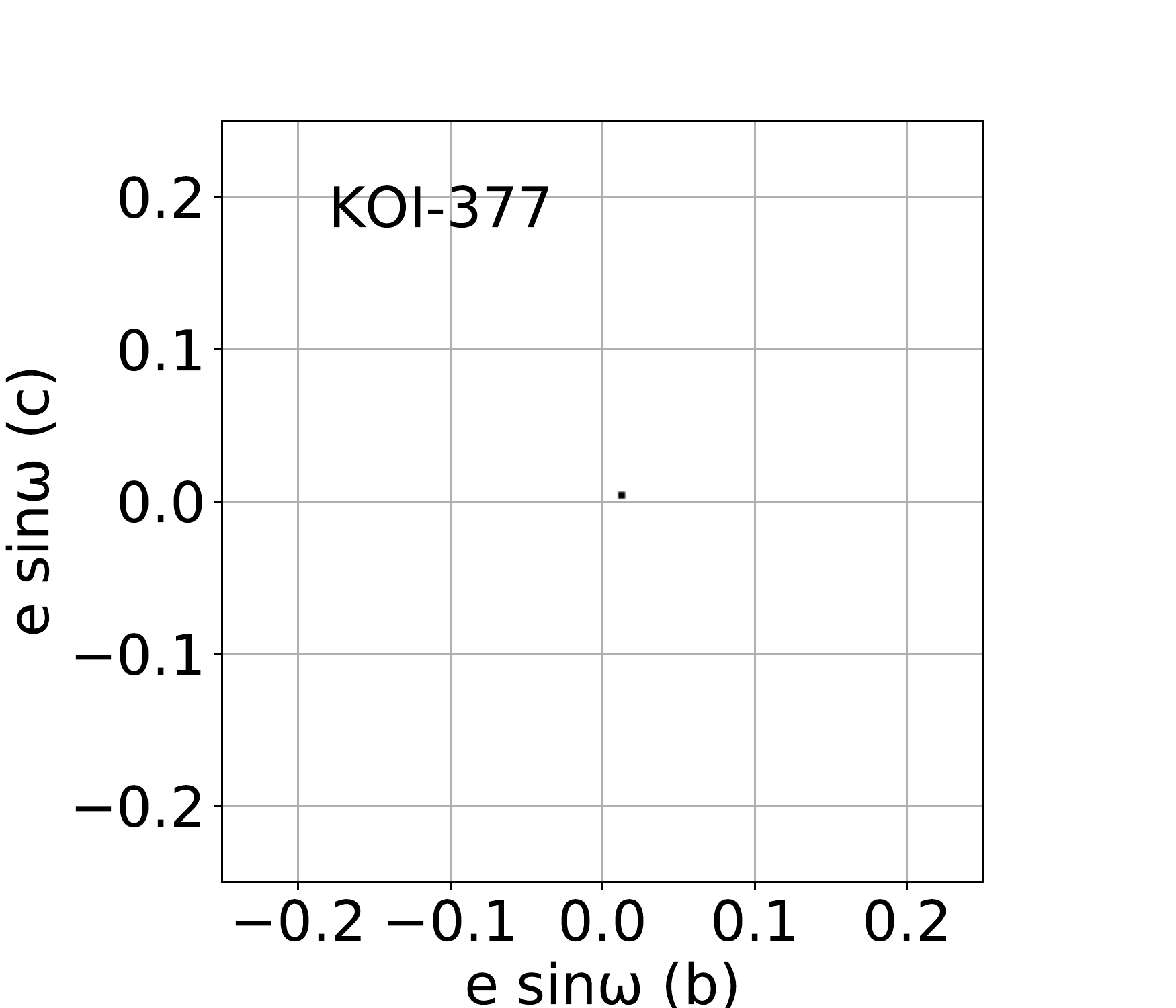}
\includegraphics [height = 1.1 in]{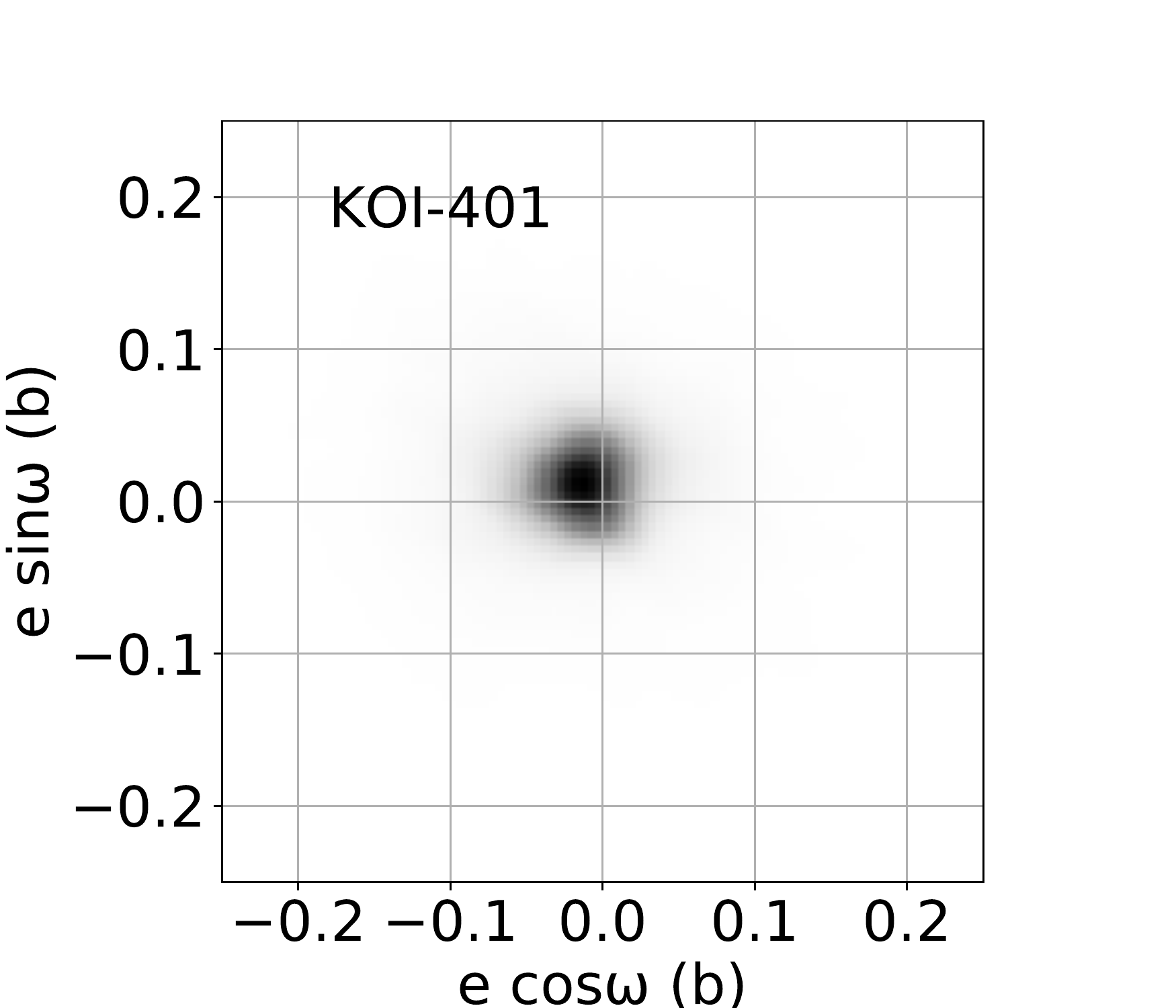} 
\includegraphics [height = 1.1 in]{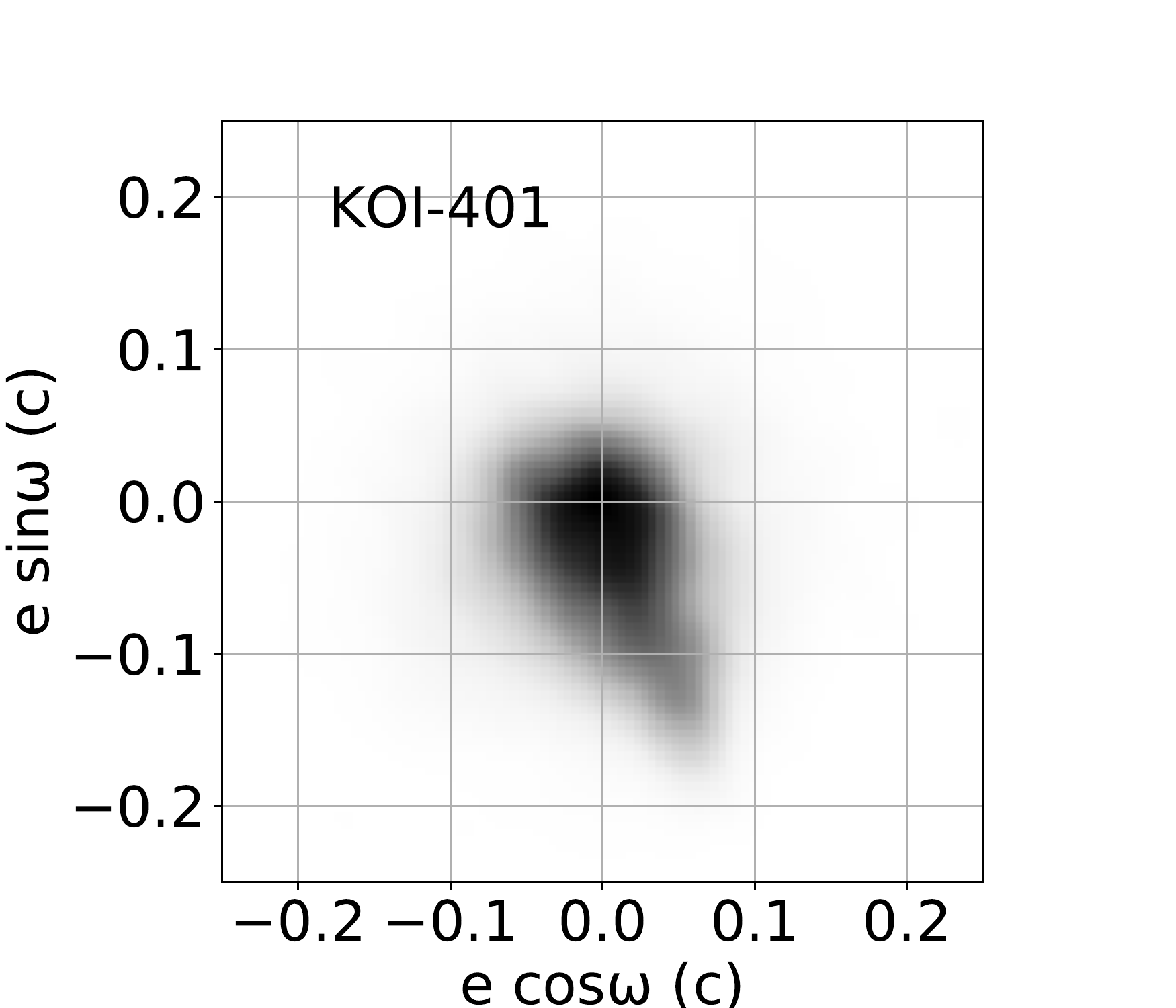}
\includegraphics [height = 1.1 in]{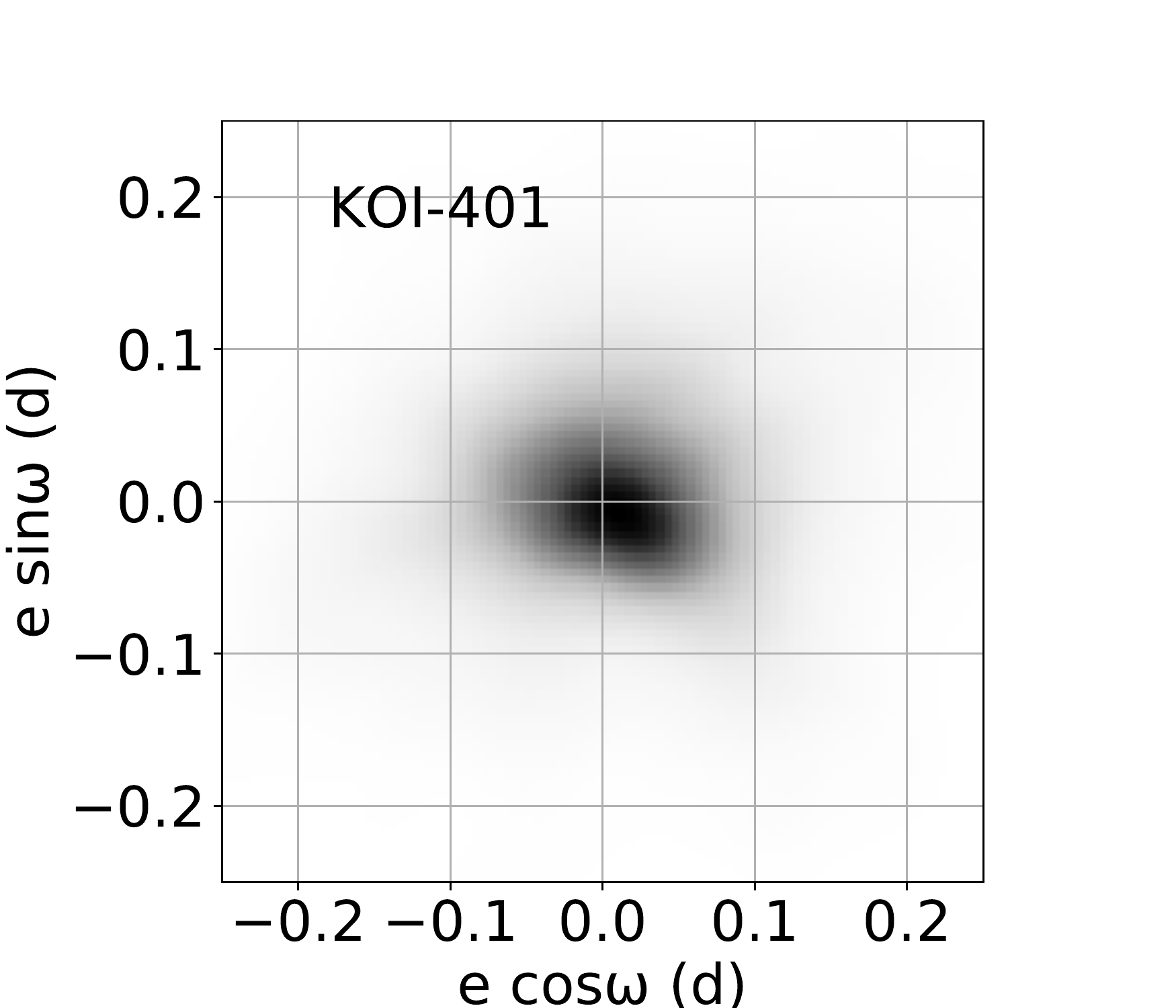} \\
\includegraphics [height = 1.1 in]{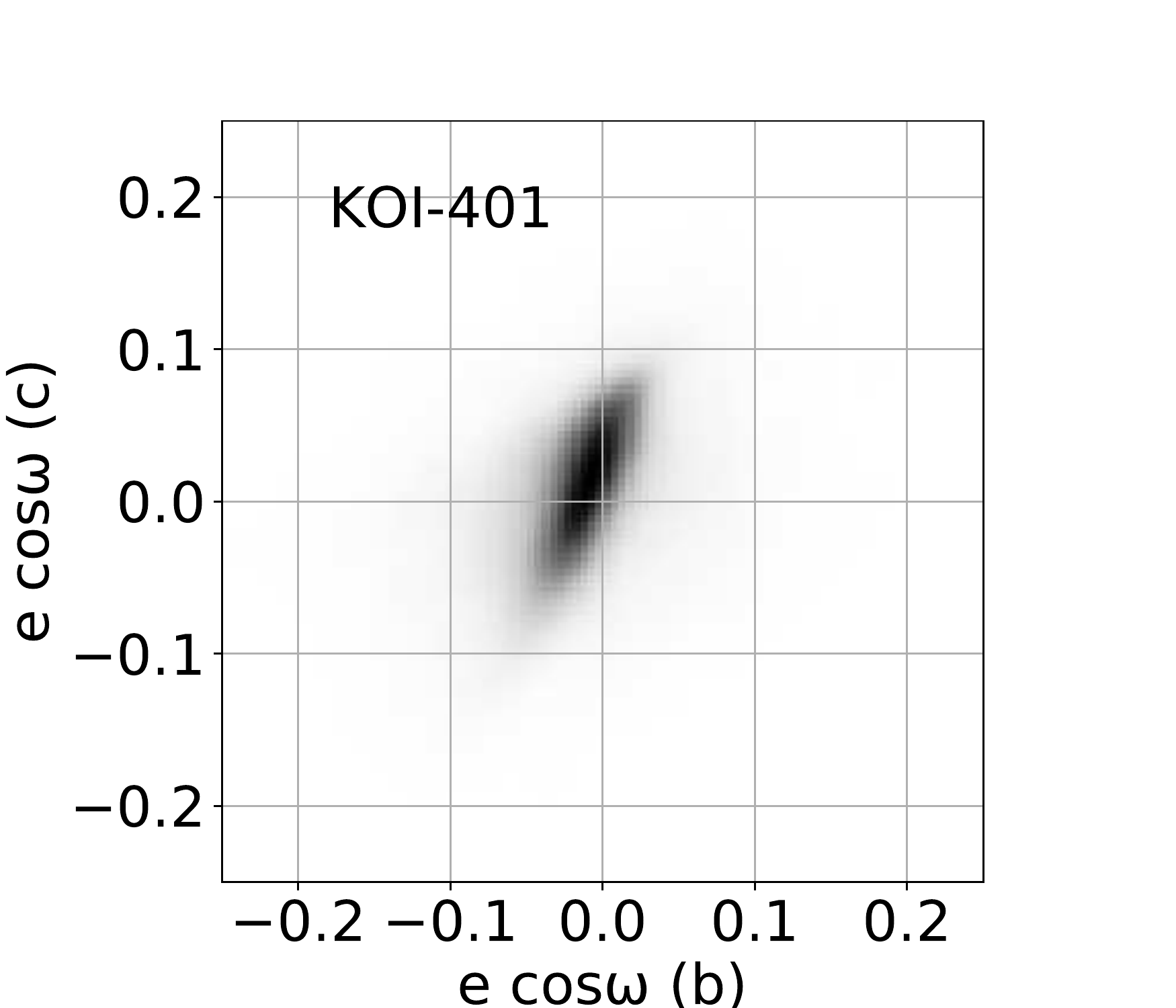} 
\includegraphics [height = 1.1 in]{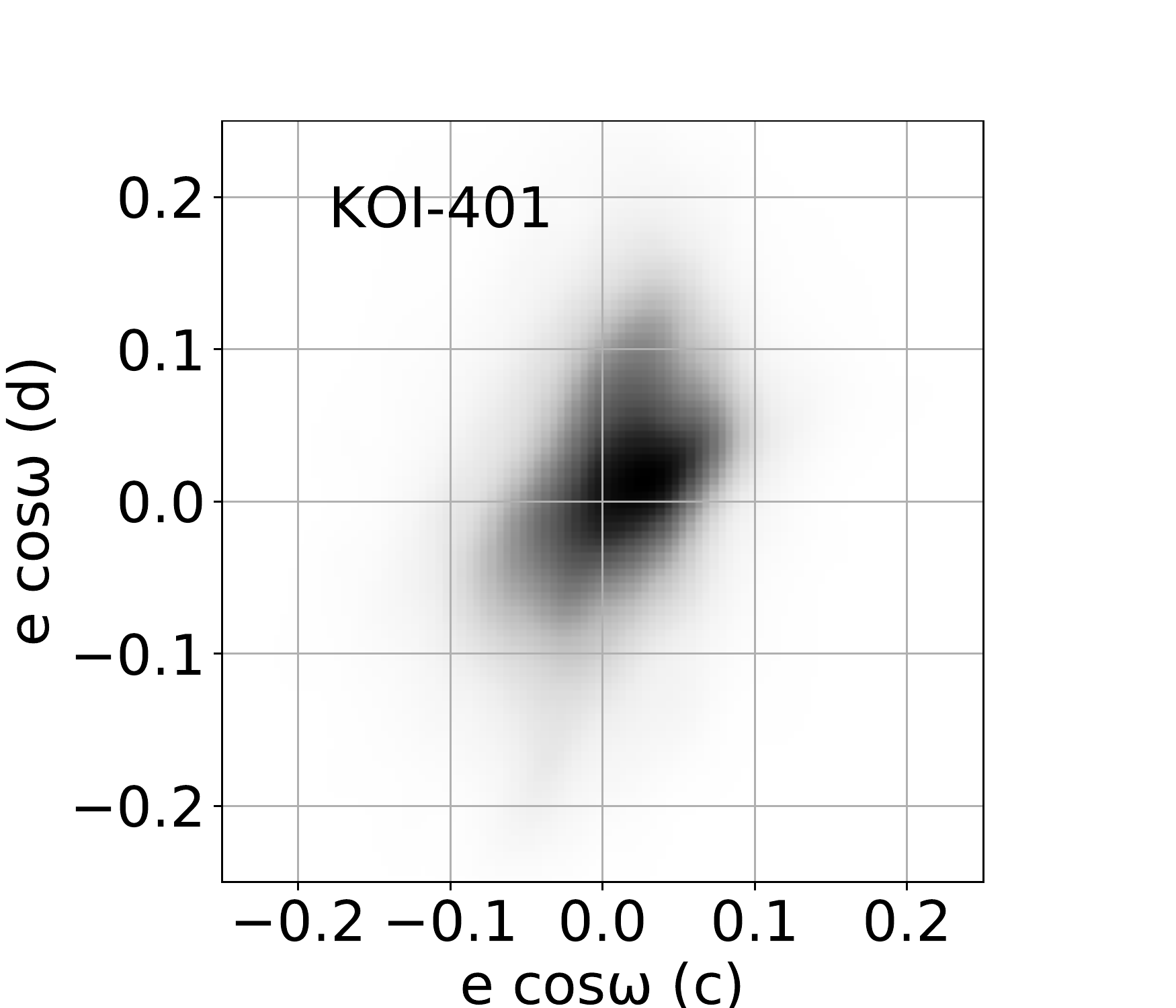}
\includegraphics [height = 1.1 in]{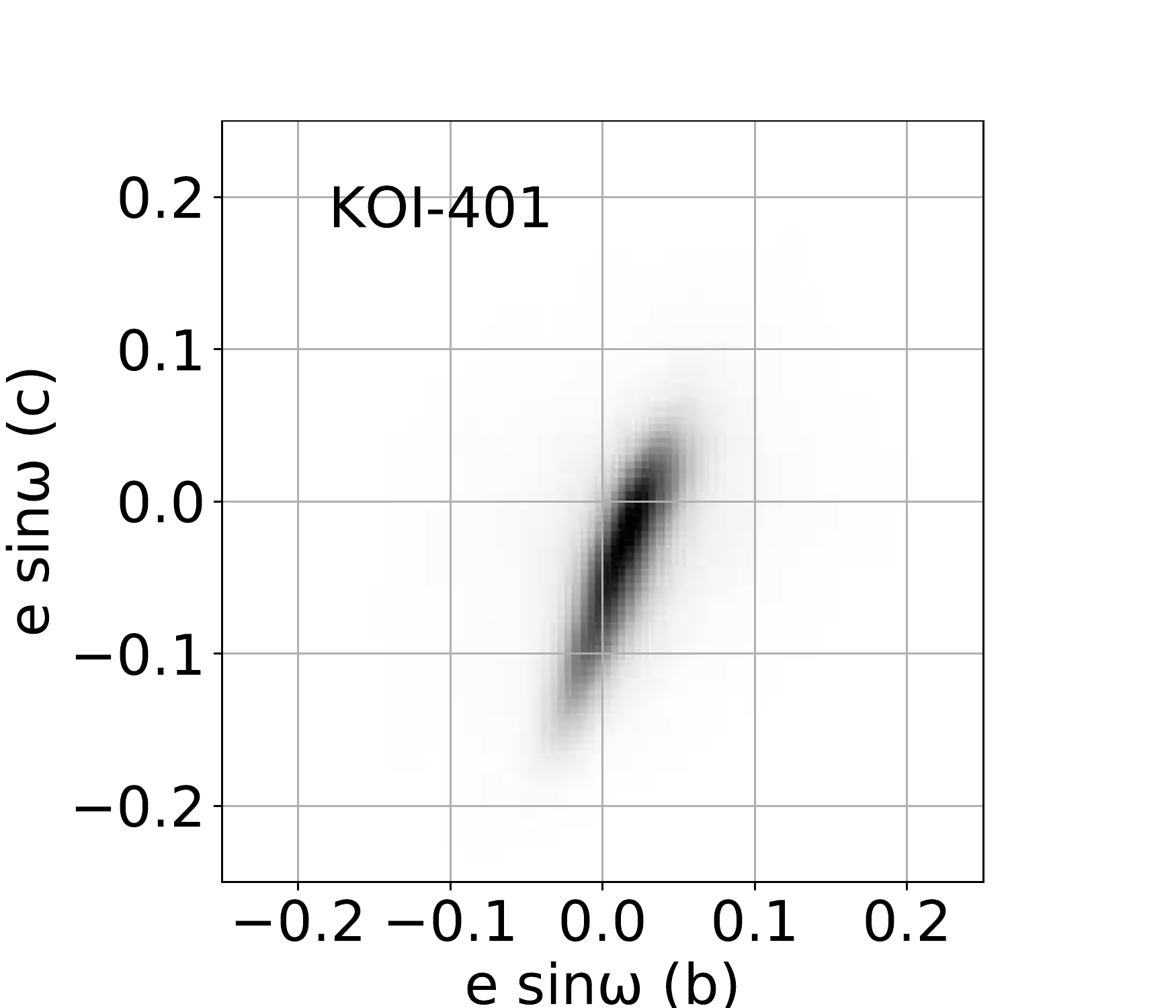} 
\includegraphics [height = 1.1 in]{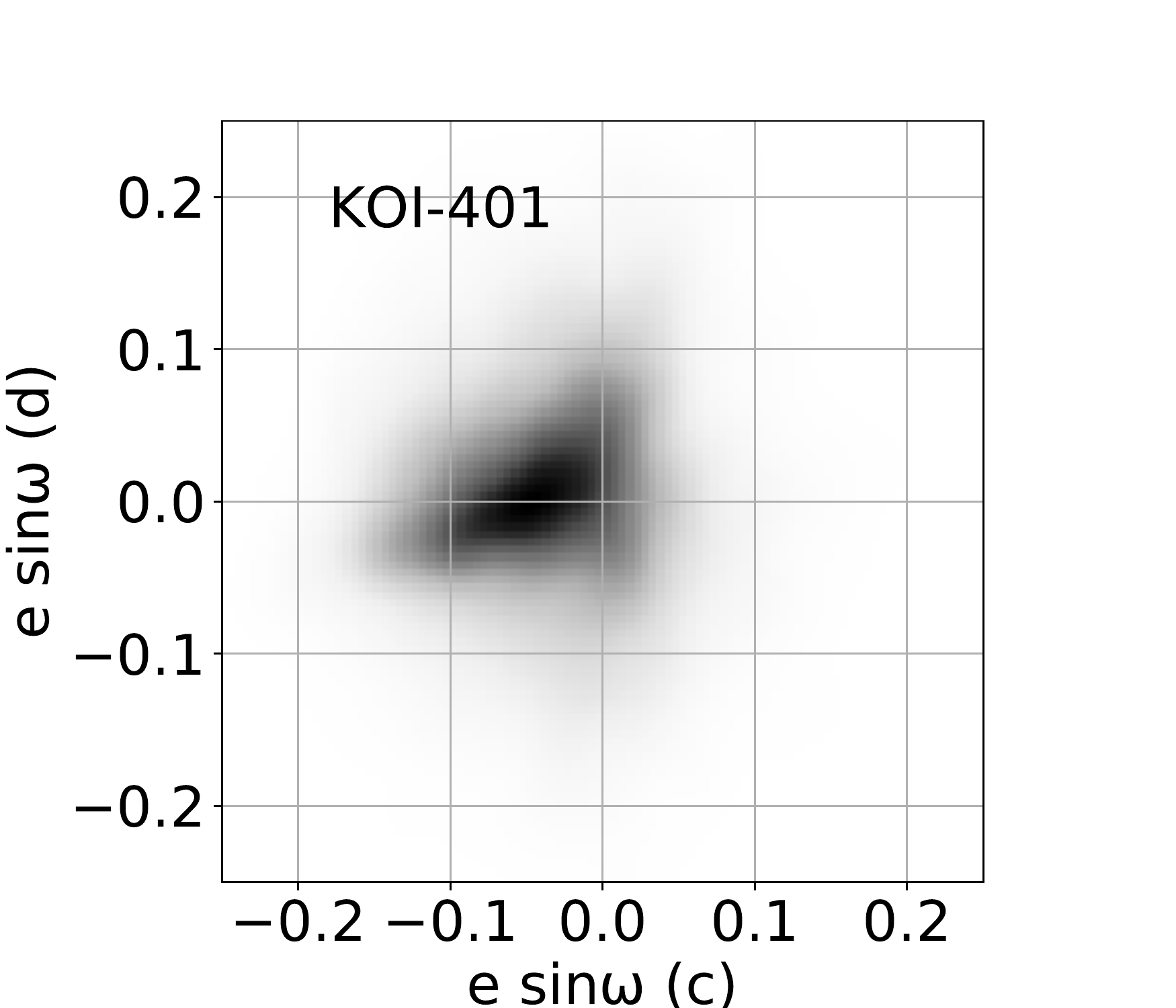}
\caption{Two-dimensional kernel density estimators on joint posteriors of eccentricity vector components: three-planet systems (Part 2 of 7). 
}
\label{fig:ecc3b} 
\end{center}
\end{figure}

\begin{figure}
\begin{center}
\figurenum{25}
\includegraphics [height = 1.1 in]{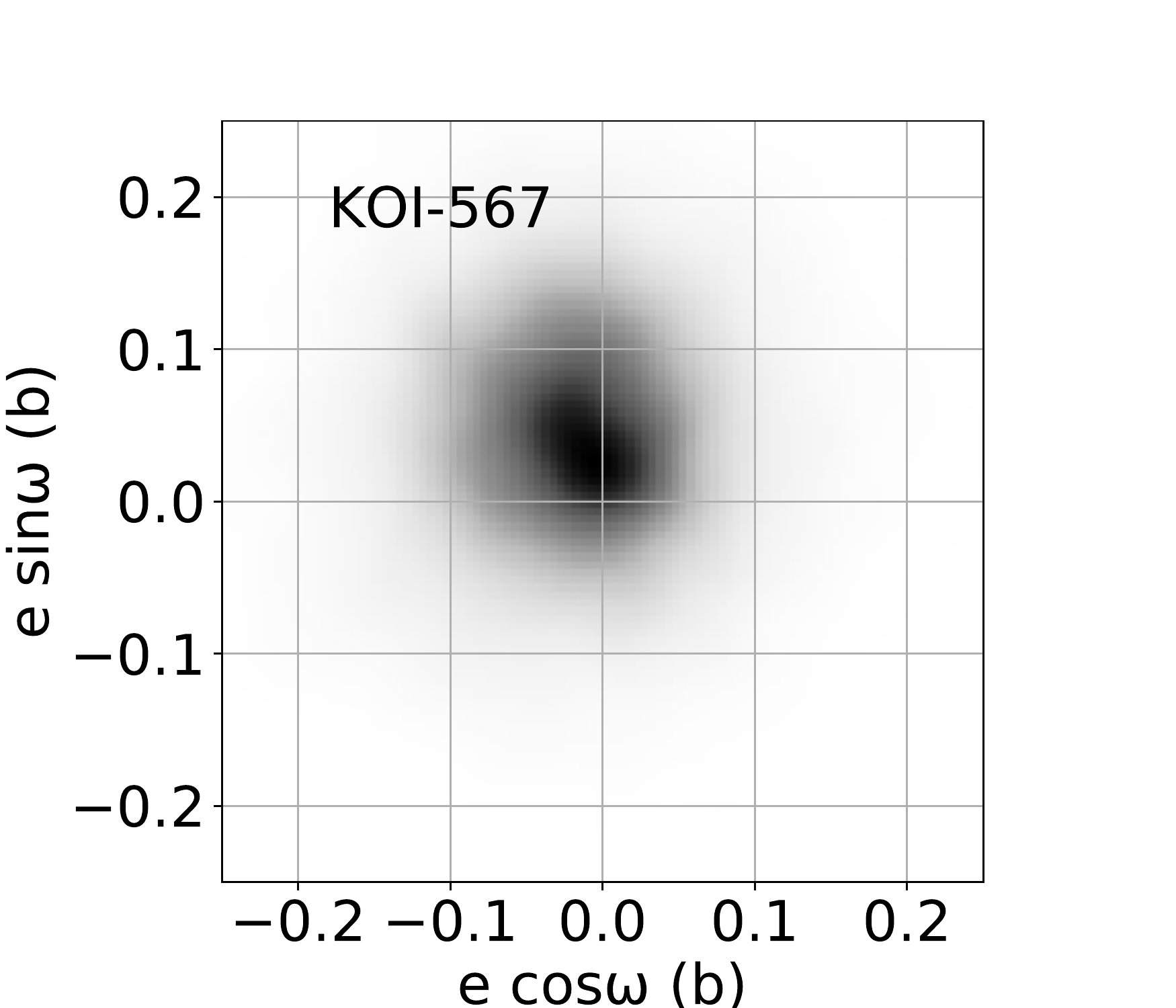}
\includegraphics [height = 1.1 in]{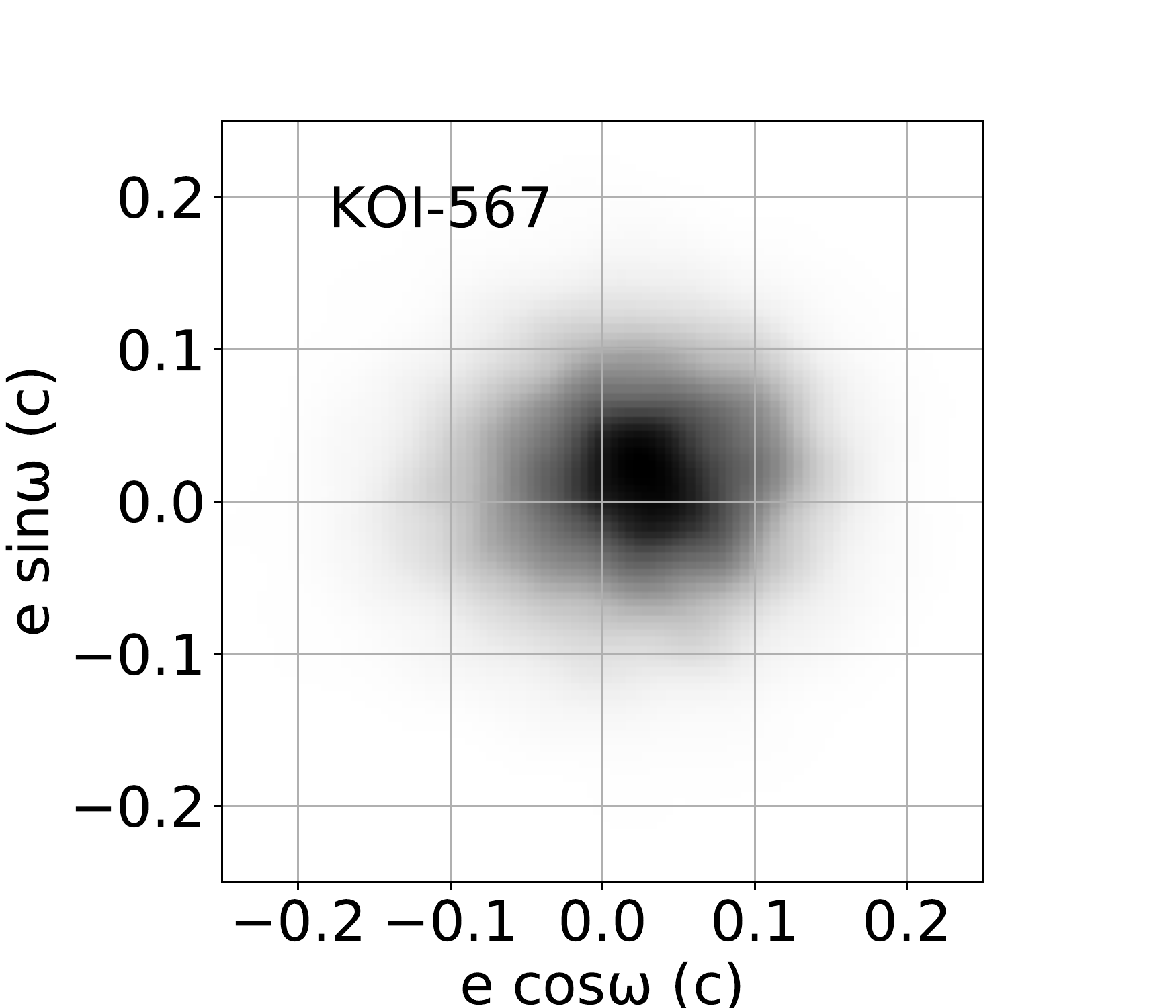}
\includegraphics [height = 1.1 in]{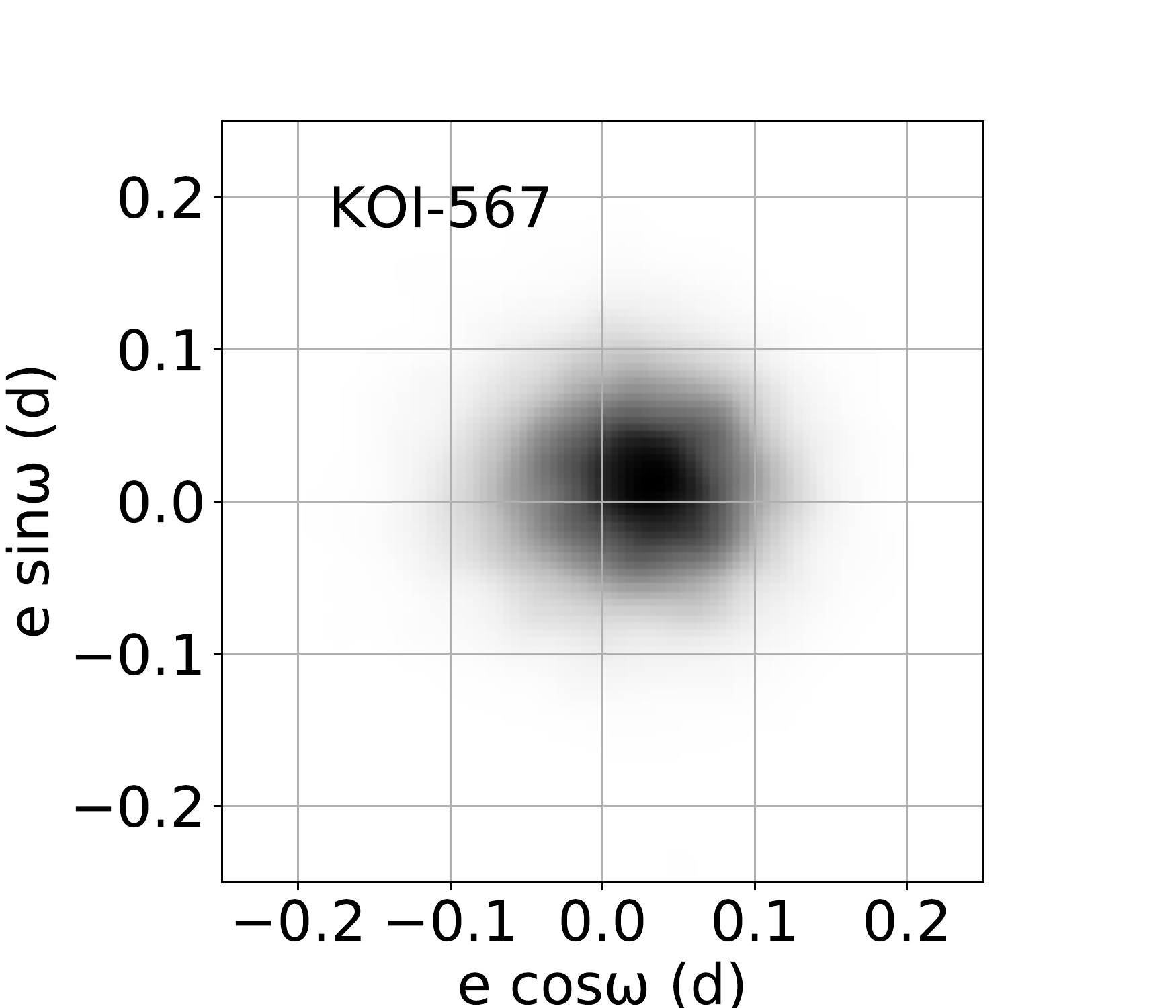}
\includegraphics [height = 1.1 in]{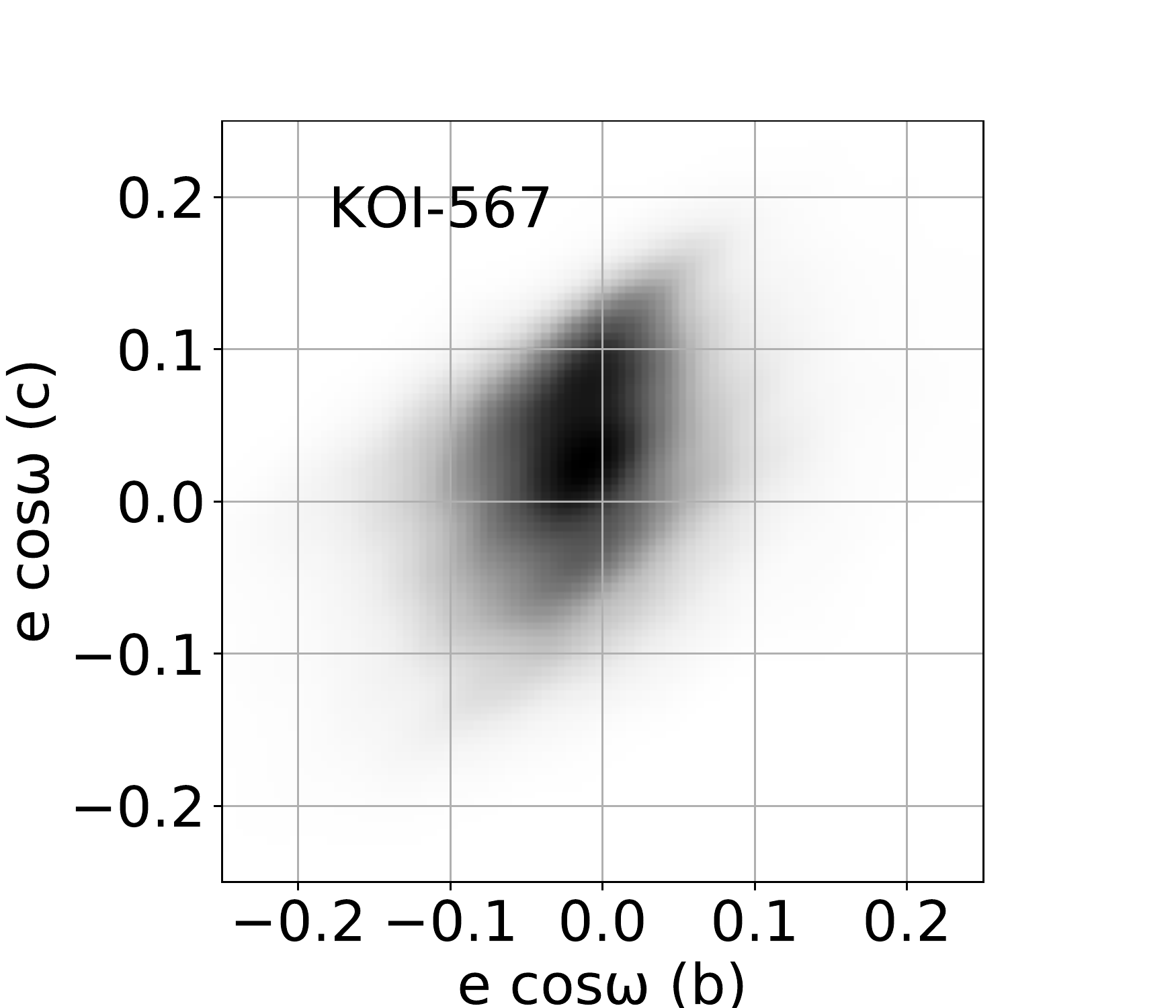} \\
\includegraphics [height = 1.1 in]{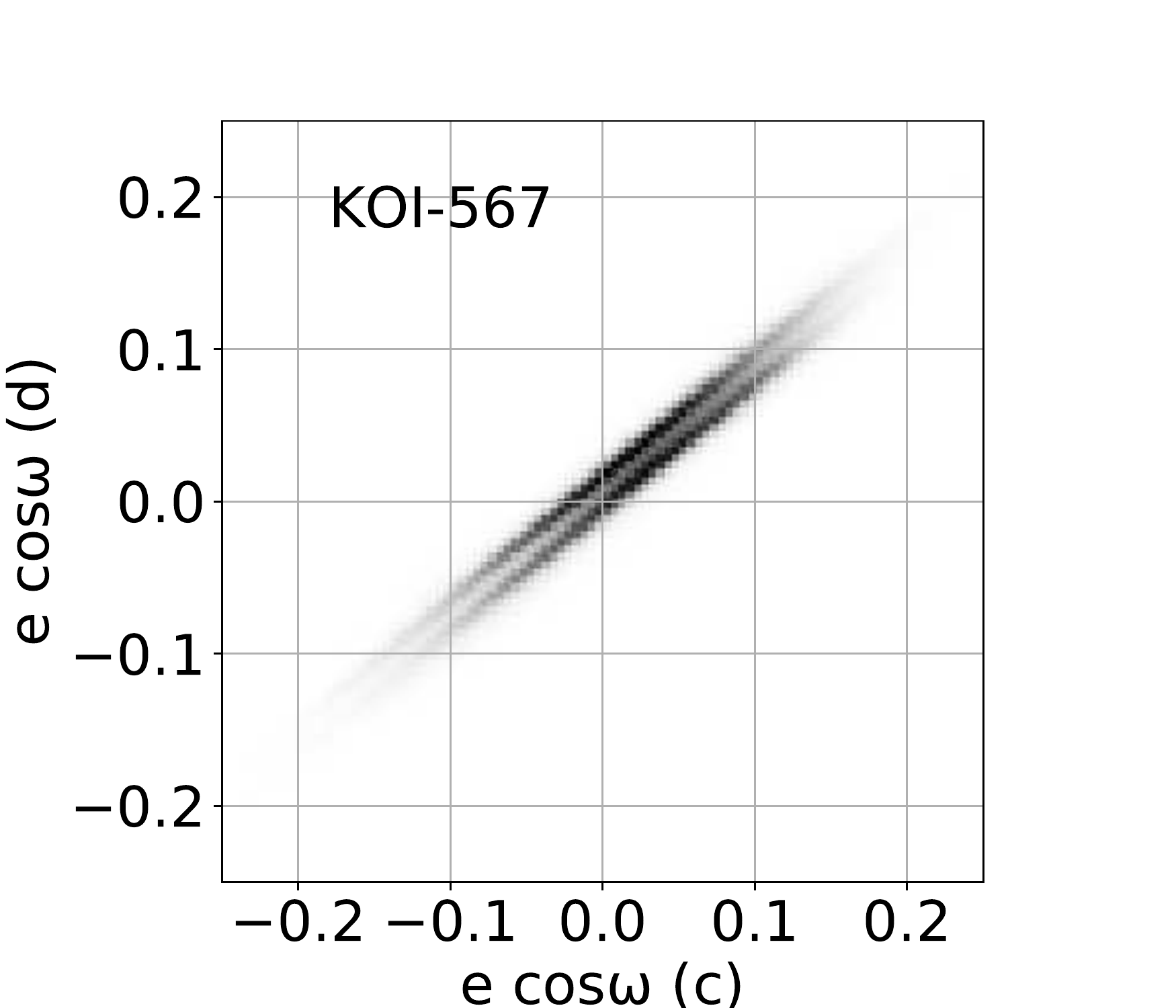}
\includegraphics [height = 1.1 in]{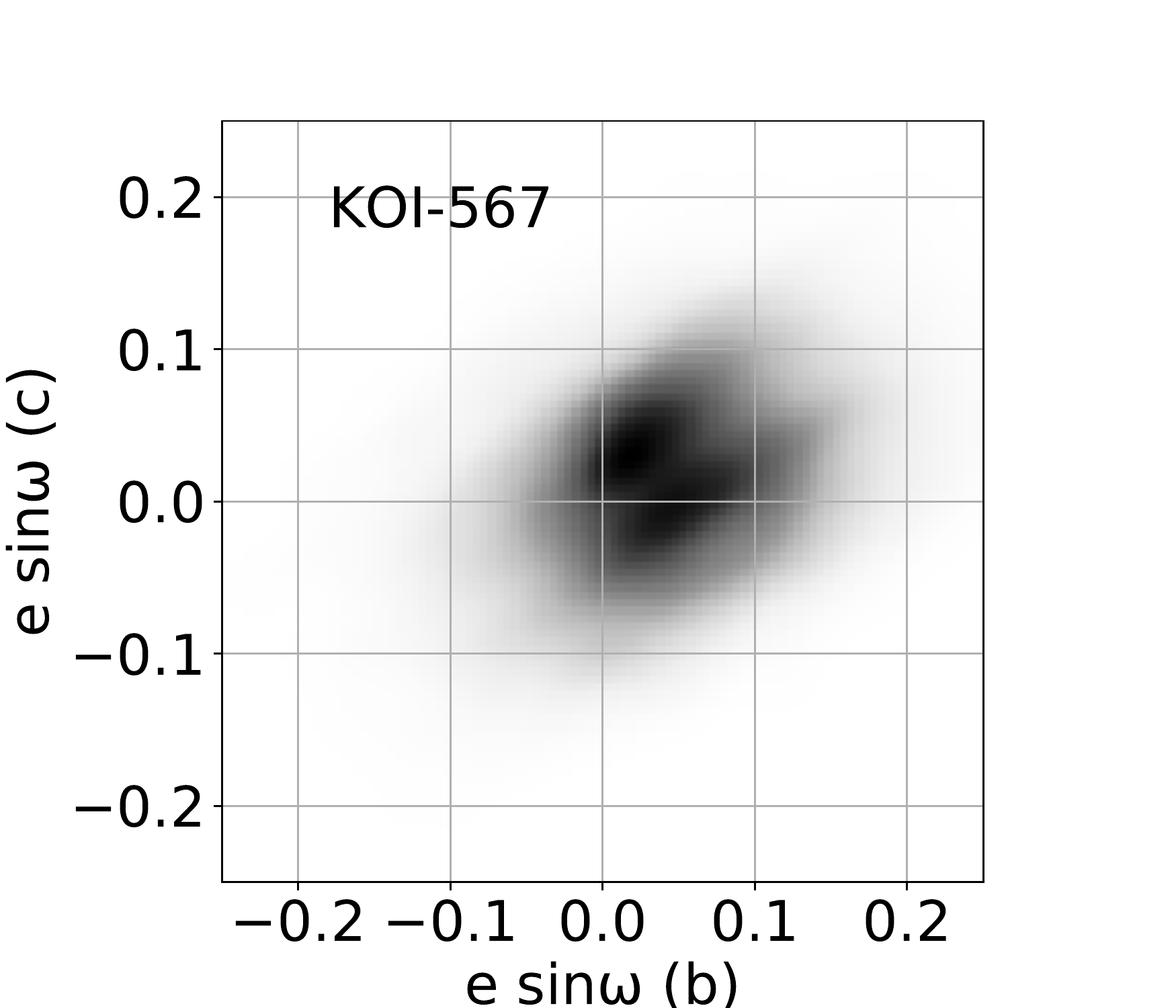}
\includegraphics [height = 1.1 in]{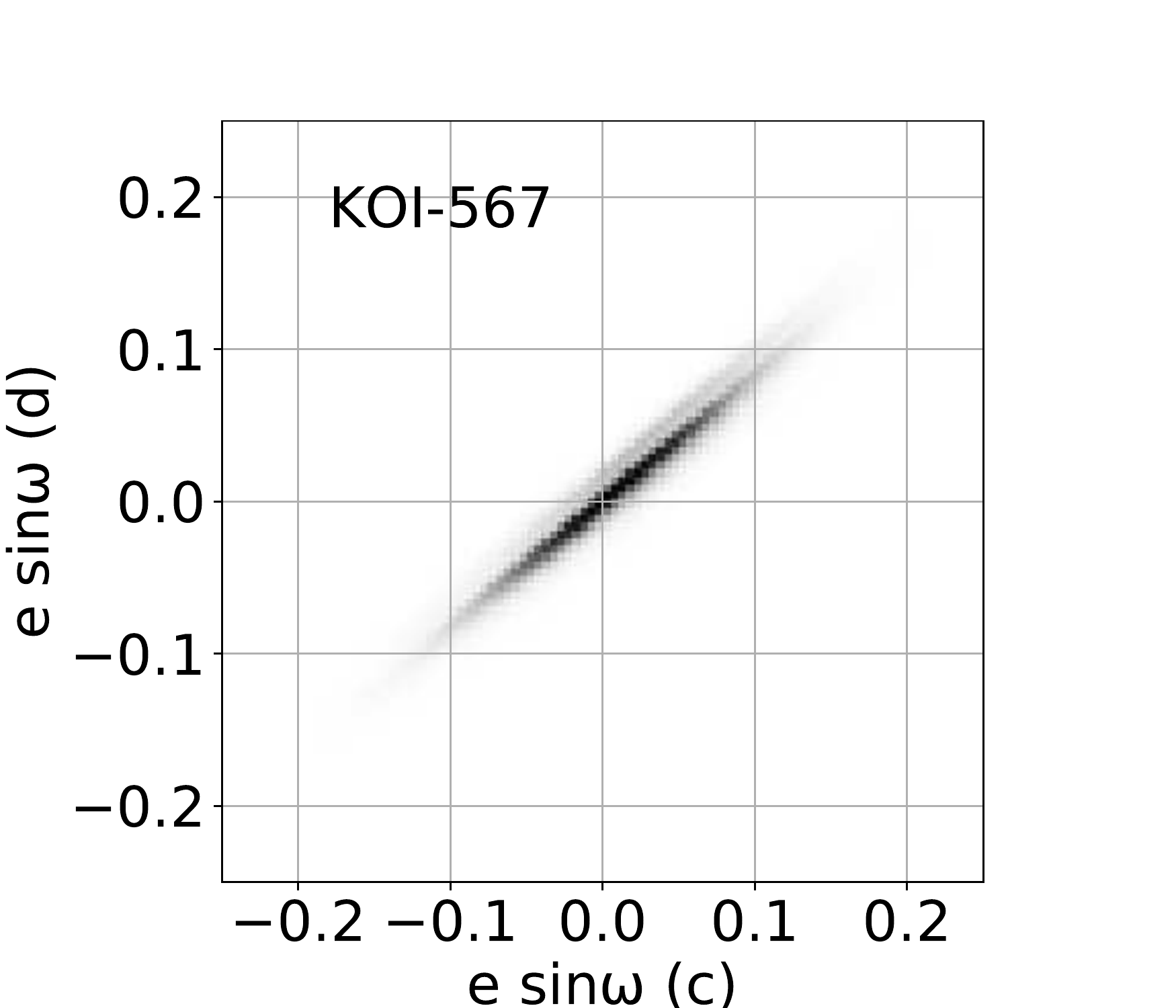}
\includegraphics [height = 1.1 in]{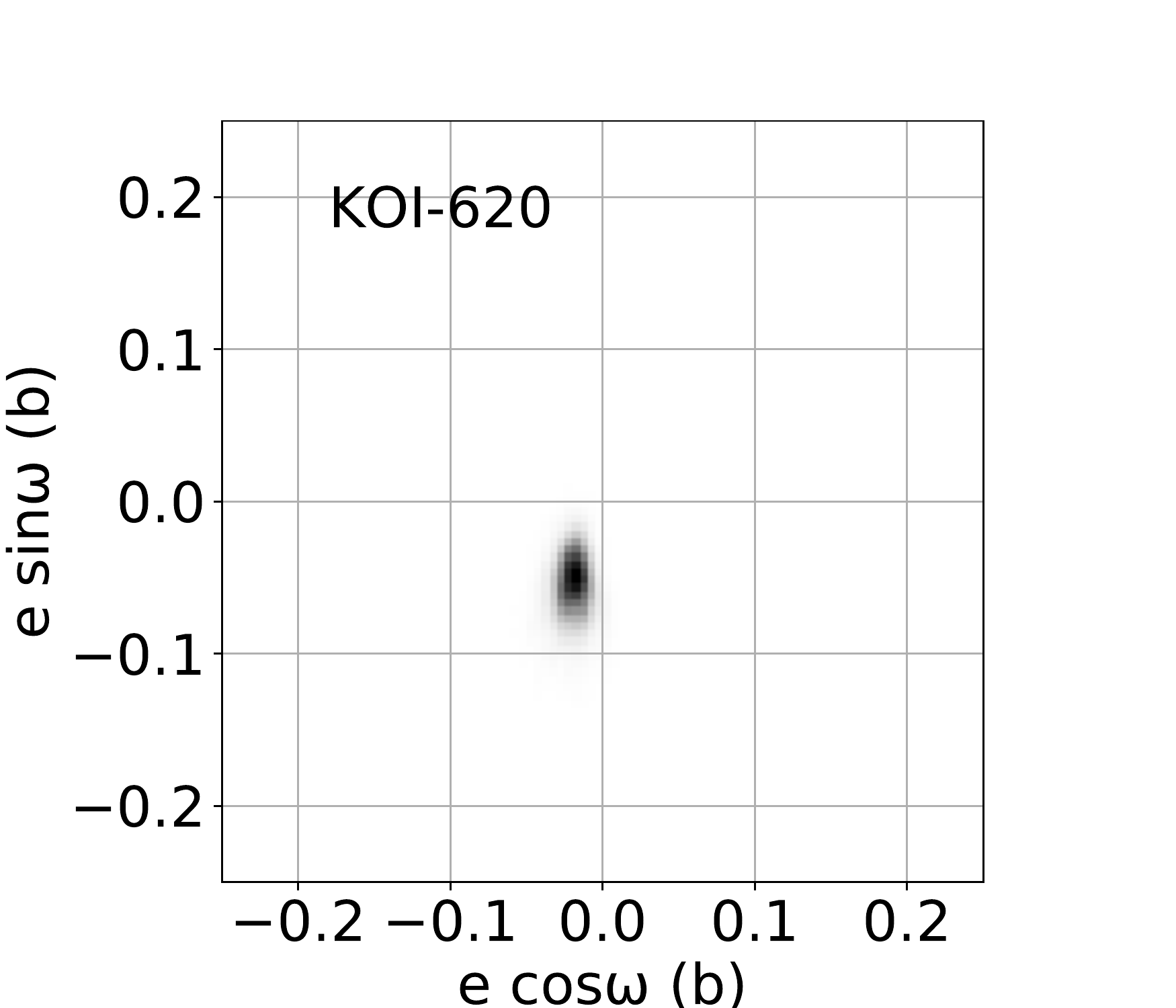} \\
\includegraphics [height = 1.1 in]{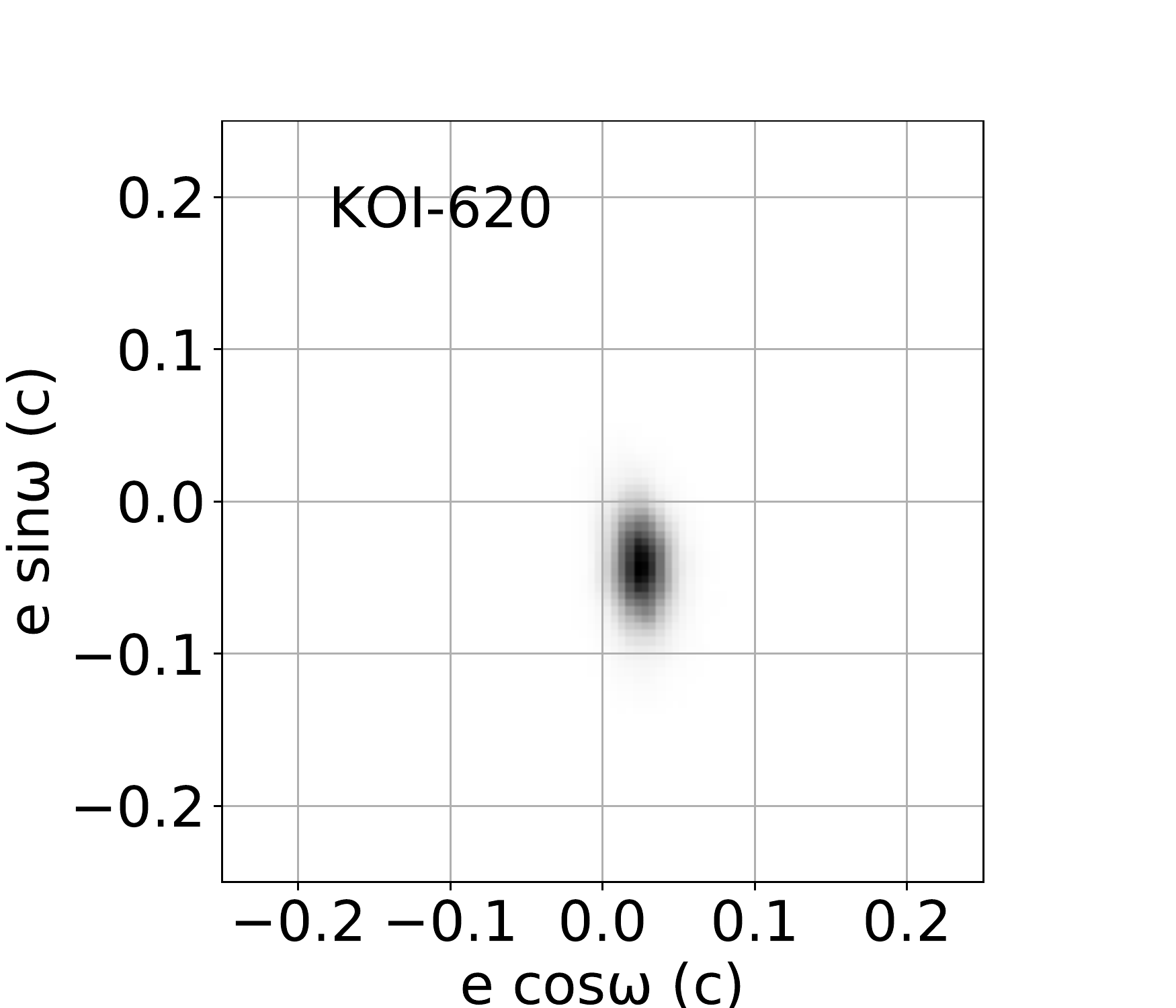}
\includegraphics [height = 1.1 in]{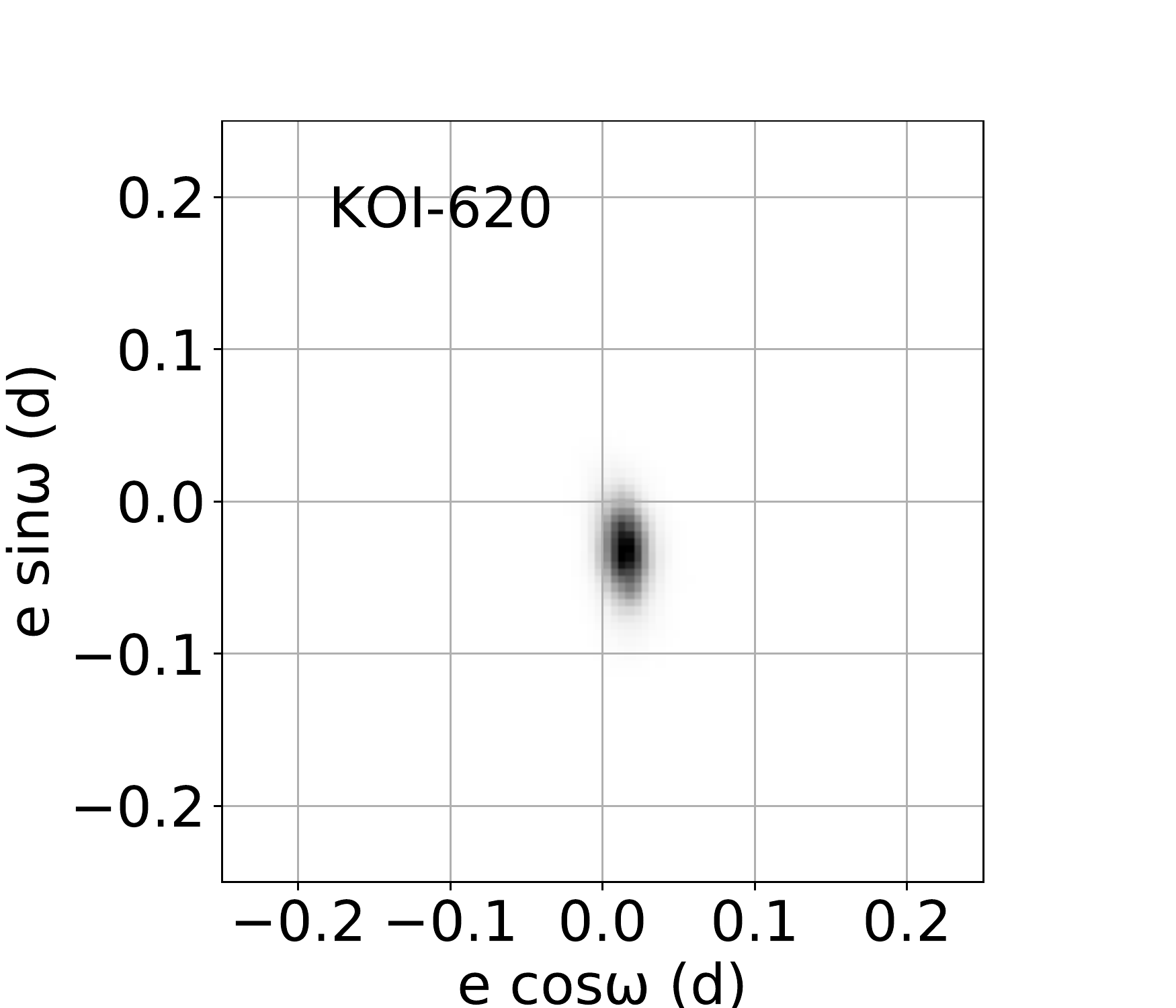}
\includegraphics [height = 1.1 in]{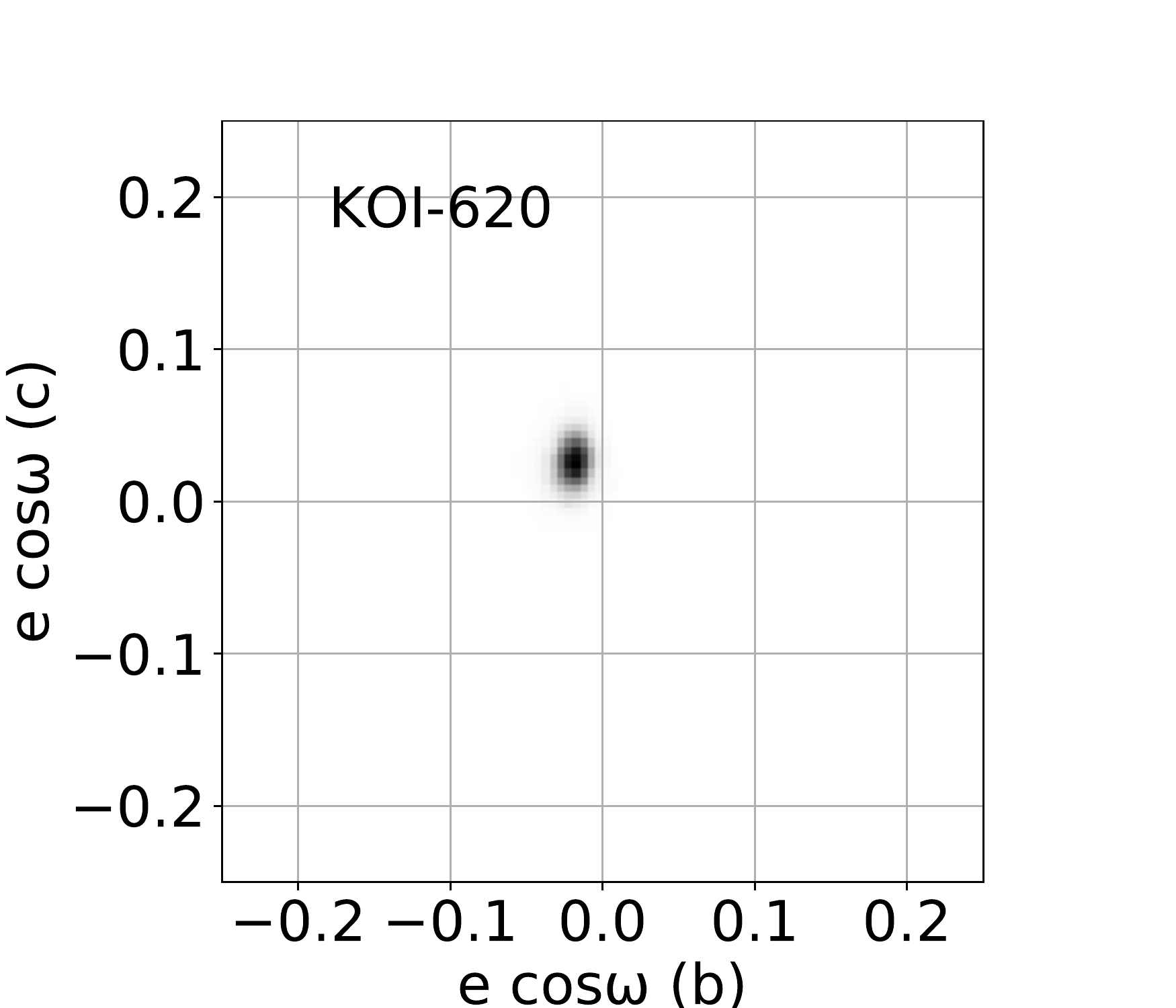}
\includegraphics [height = 1.1 in]{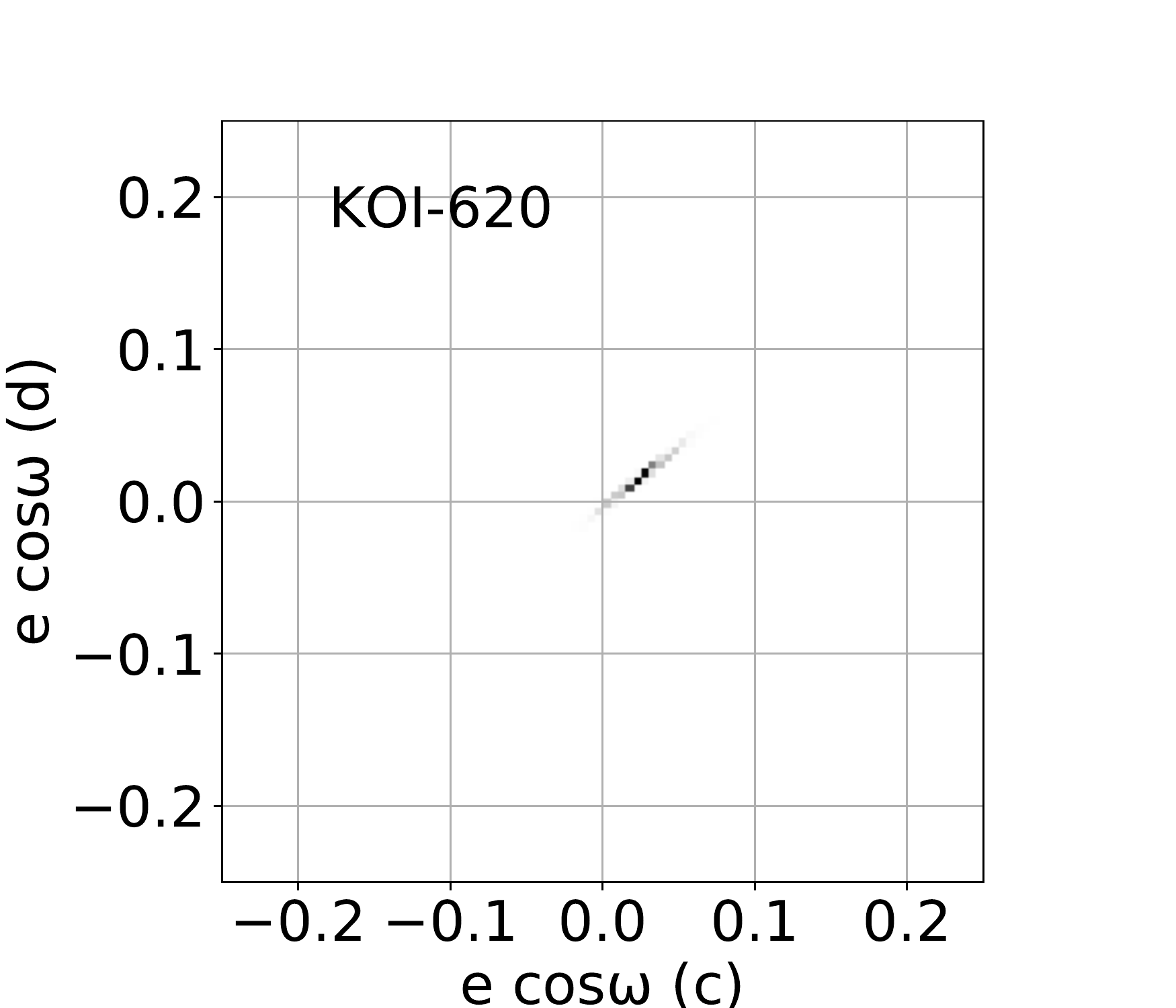} \\
\includegraphics [height = 1.1 in]{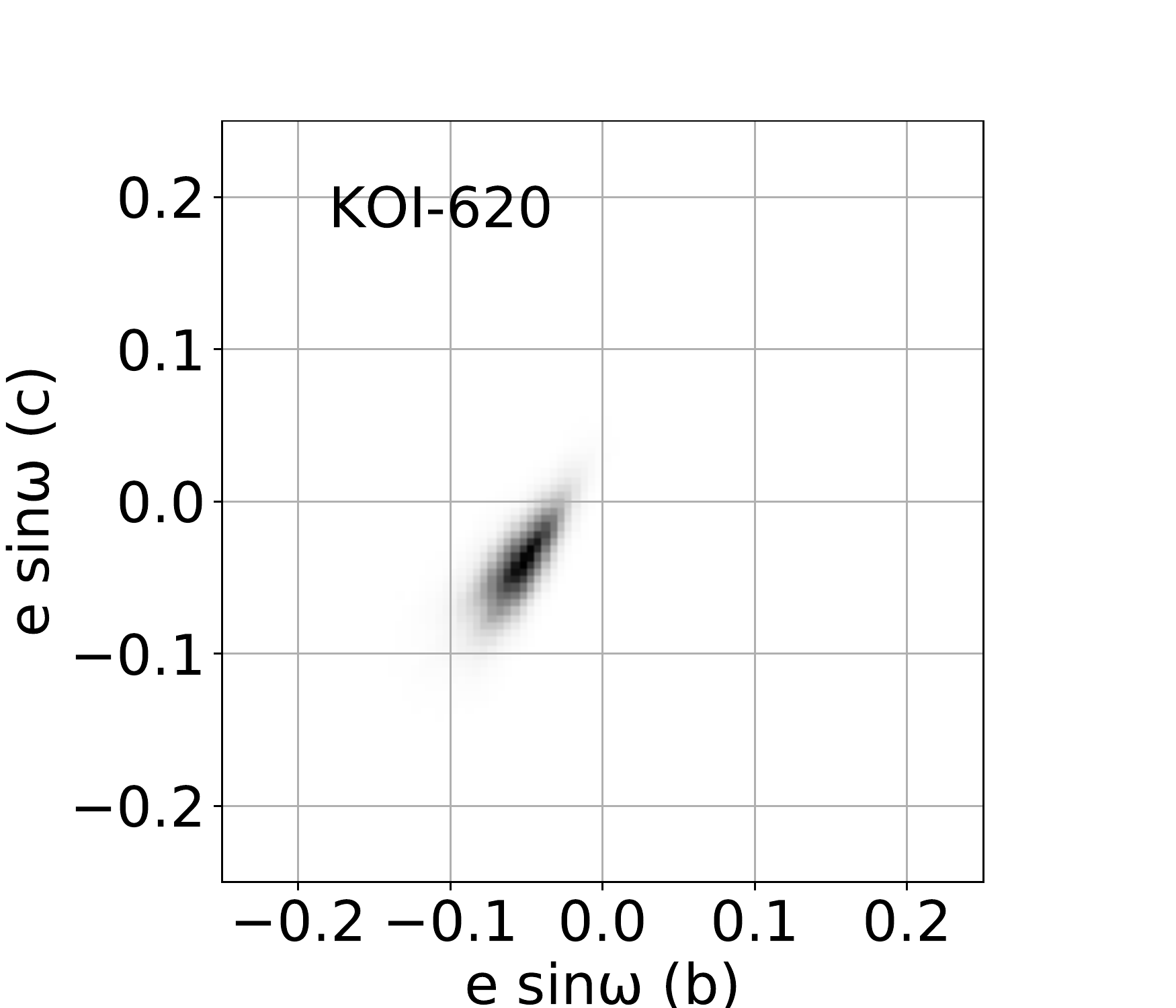}
\includegraphics [height = 1.1 in]{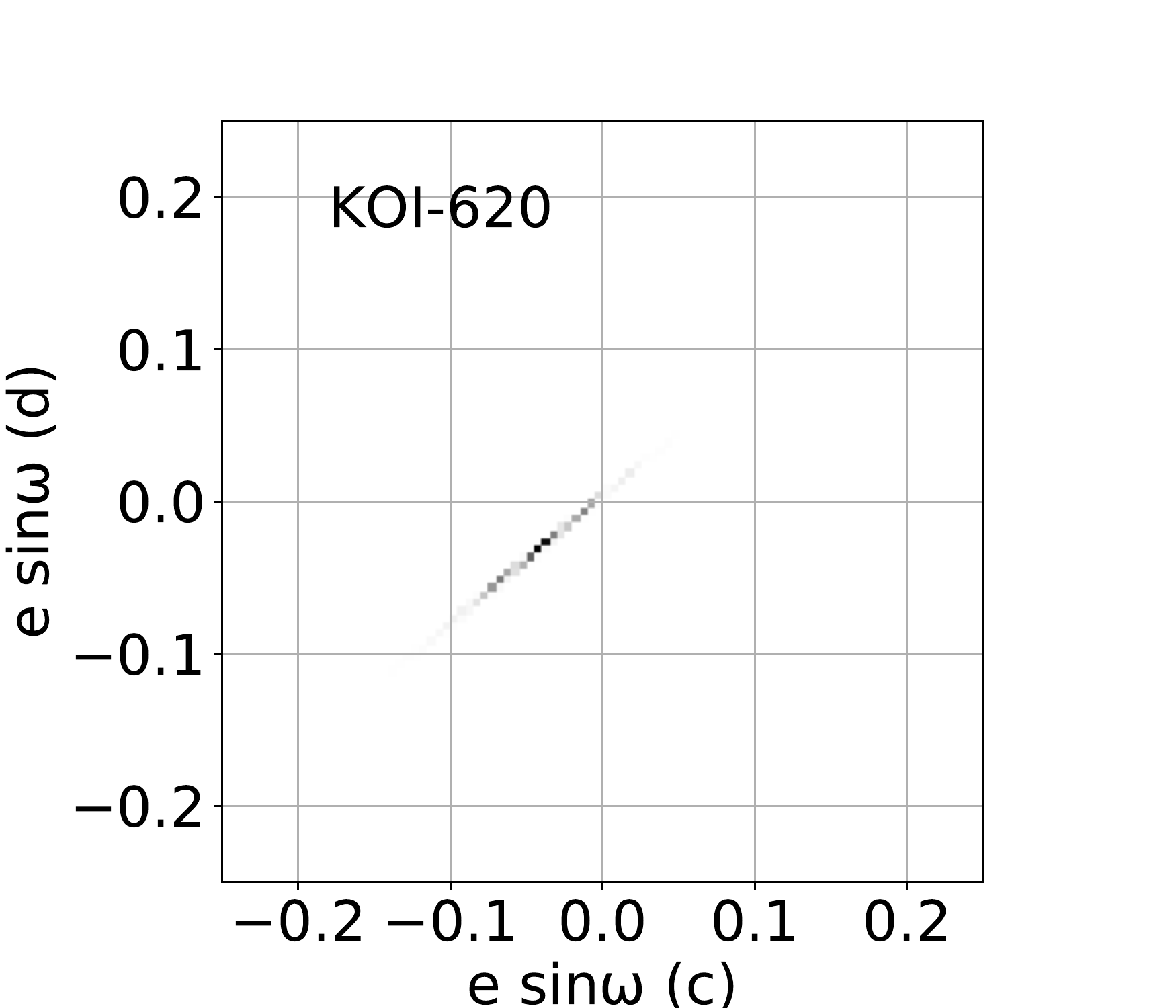}
\includegraphics [height = 1.1 in]{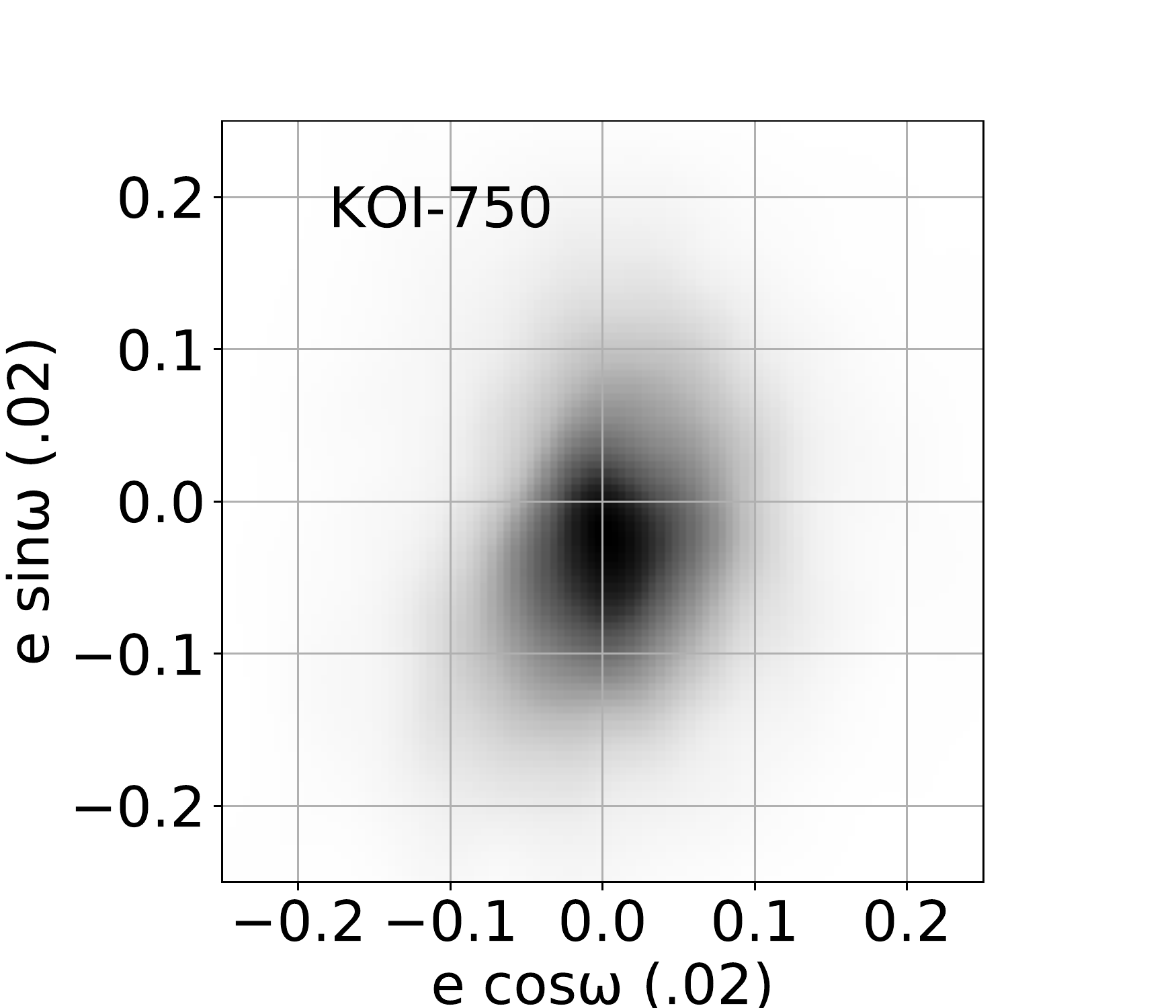}
\includegraphics [height = 1.1 in]{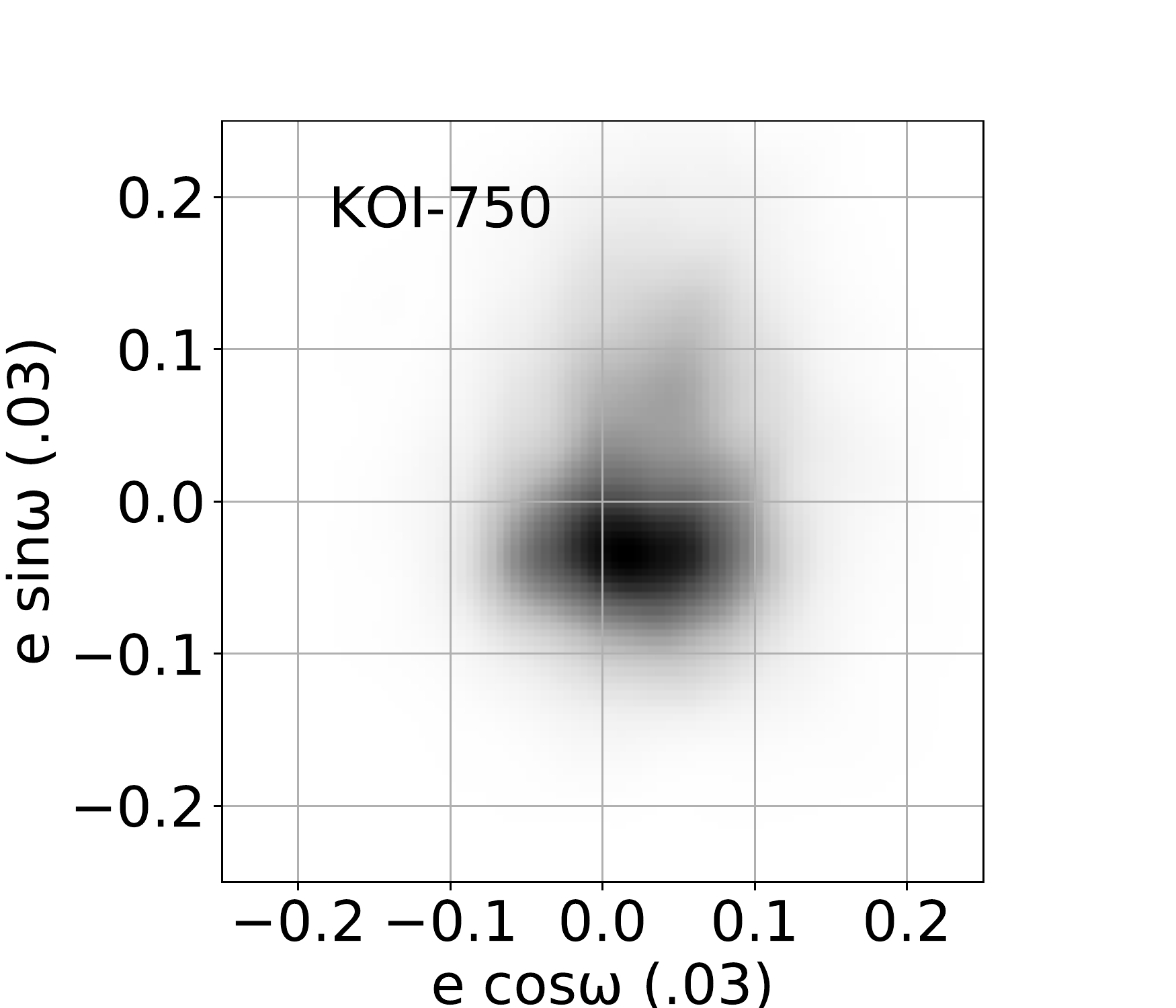} \\
\includegraphics [height = 1.1 in]{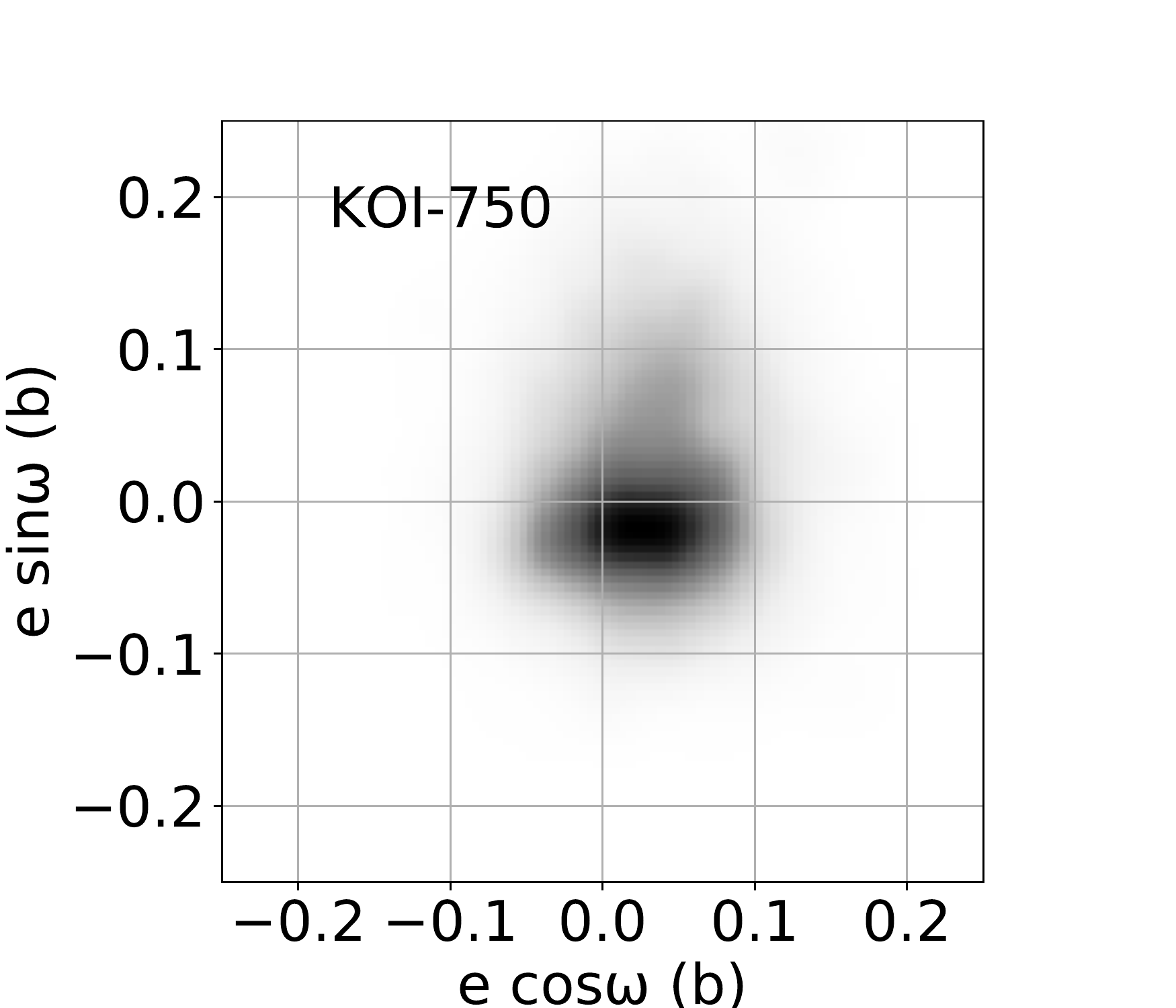}
\includegraphics [height = 1.1 in]{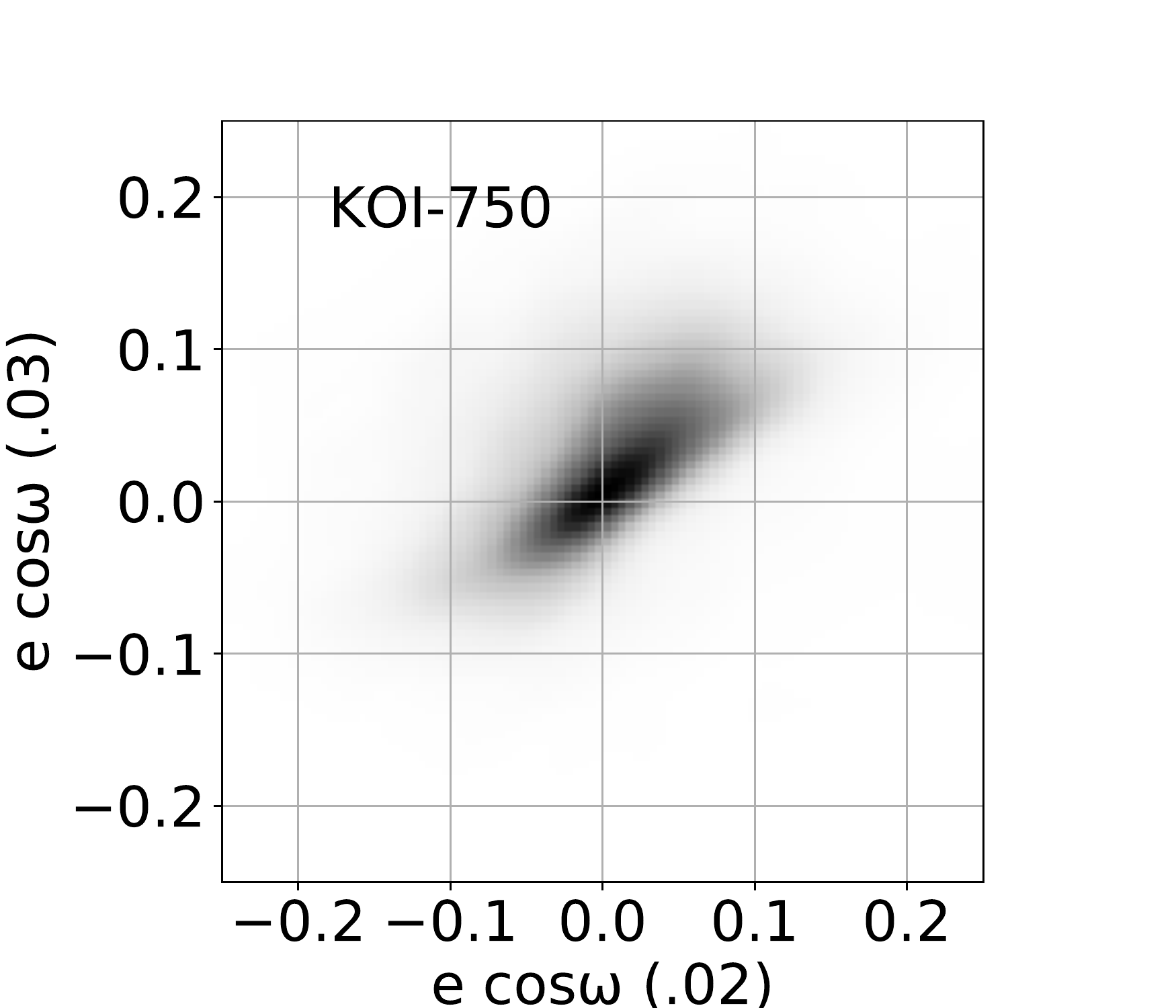} 
\includegraphics [height = 1.1 in]{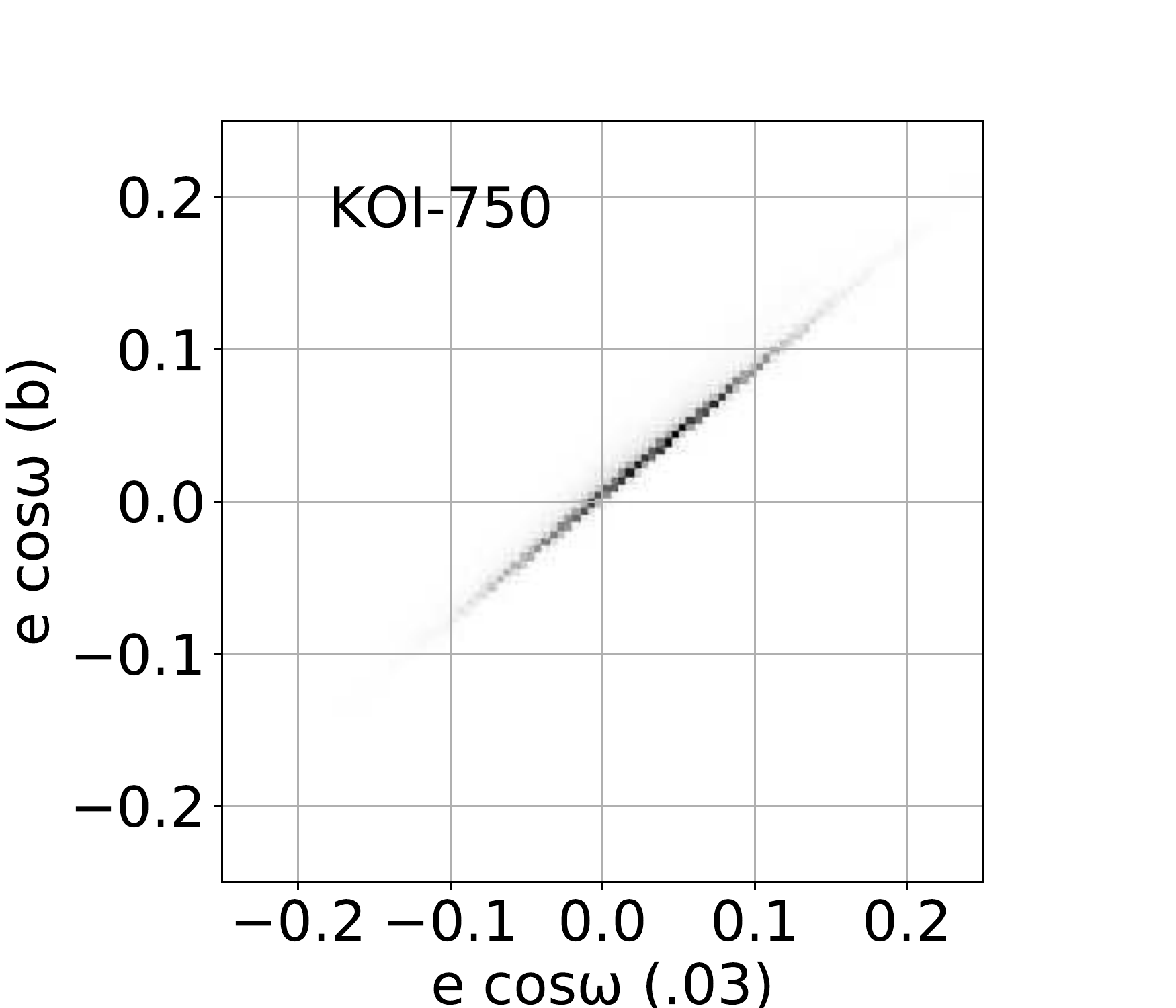}
\includegraphics [height = 1.1 in]{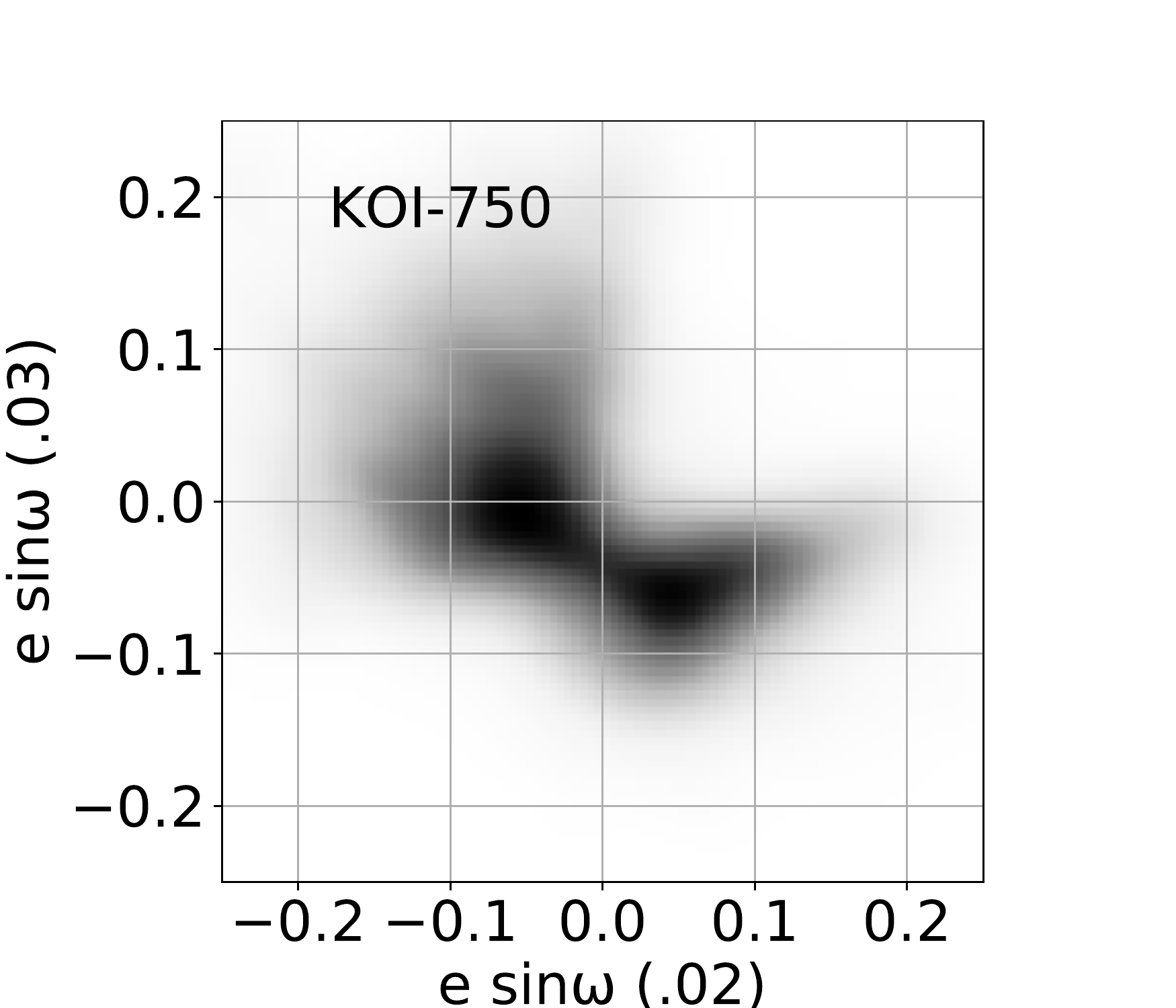} \\
\includegraphics [height = 1.1 in]{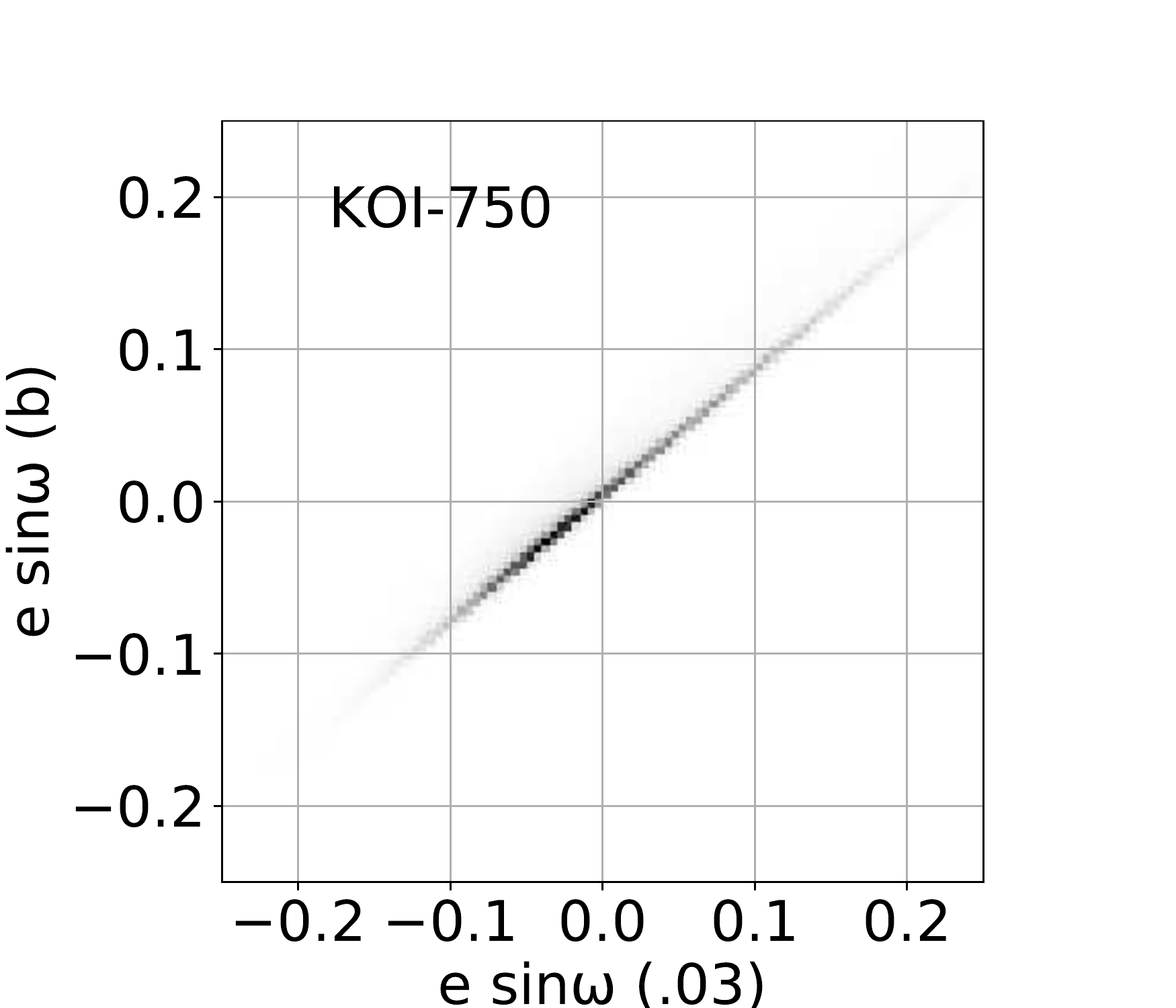}
\includegraphics [height = 1.1 in]{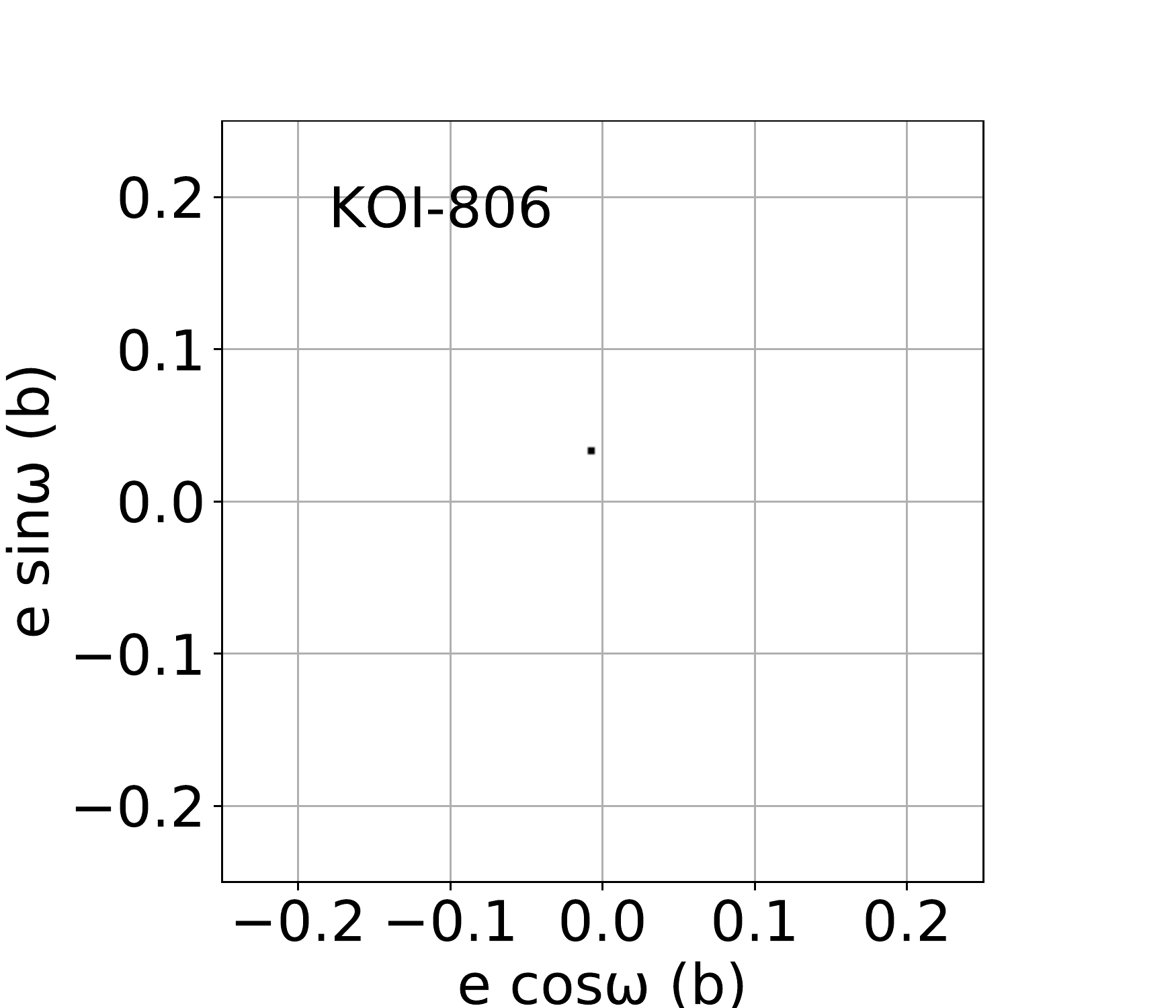} 
\includegraphics [height = 1.1 in]{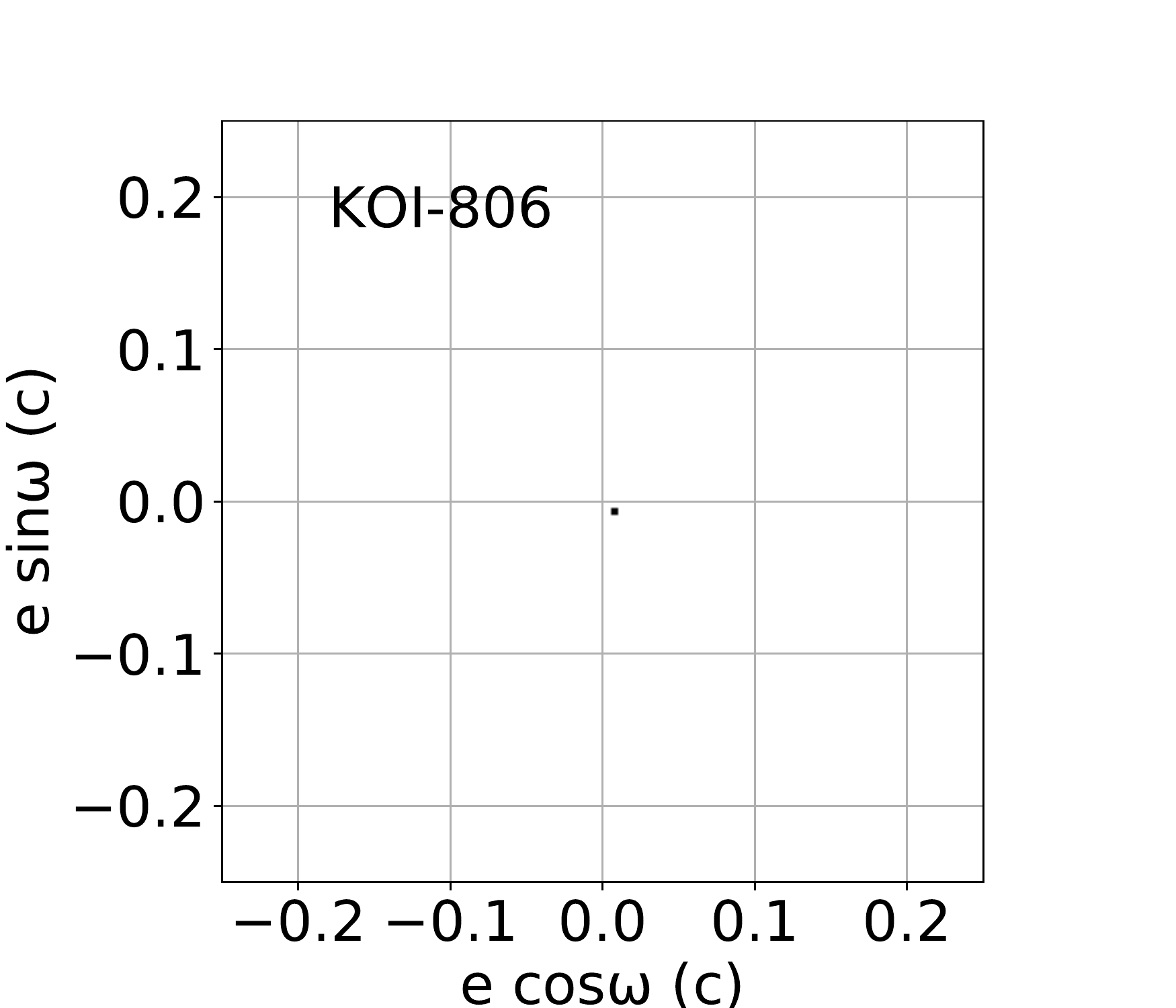}
\includegraphics [height = 1.1 in]{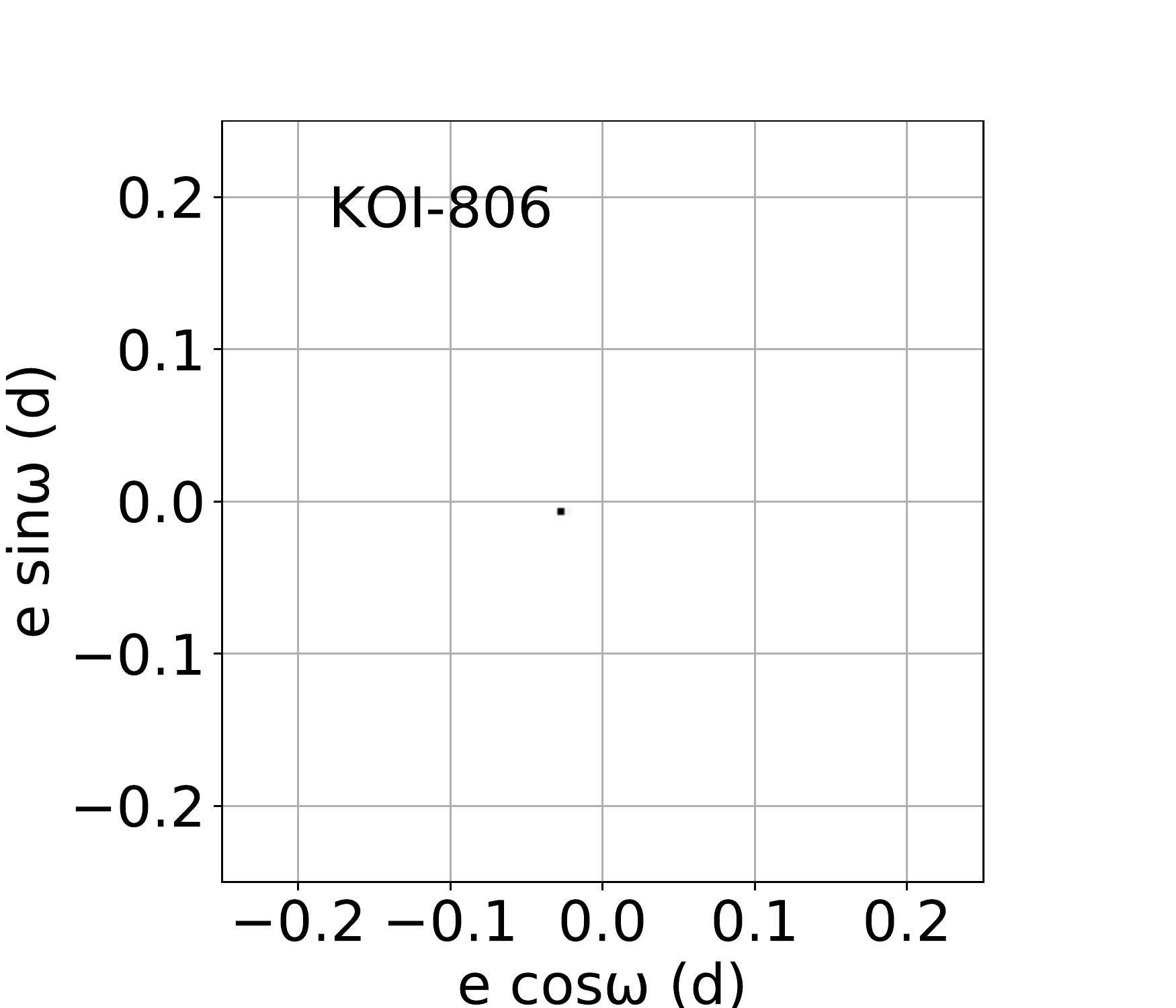} \\
\includegraphics [height = 1.1 in]{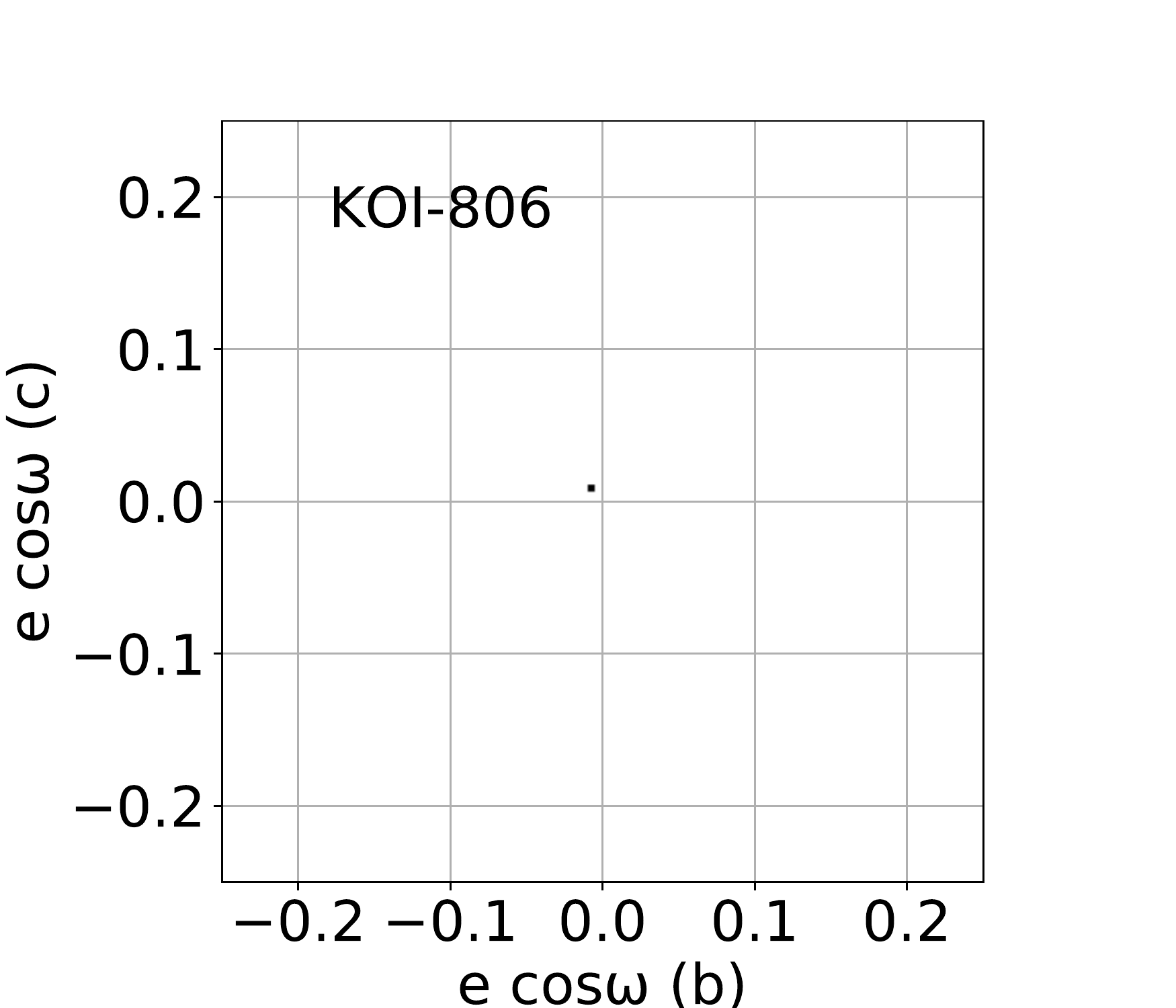} 
\includegraphics [height = 1.1 in]{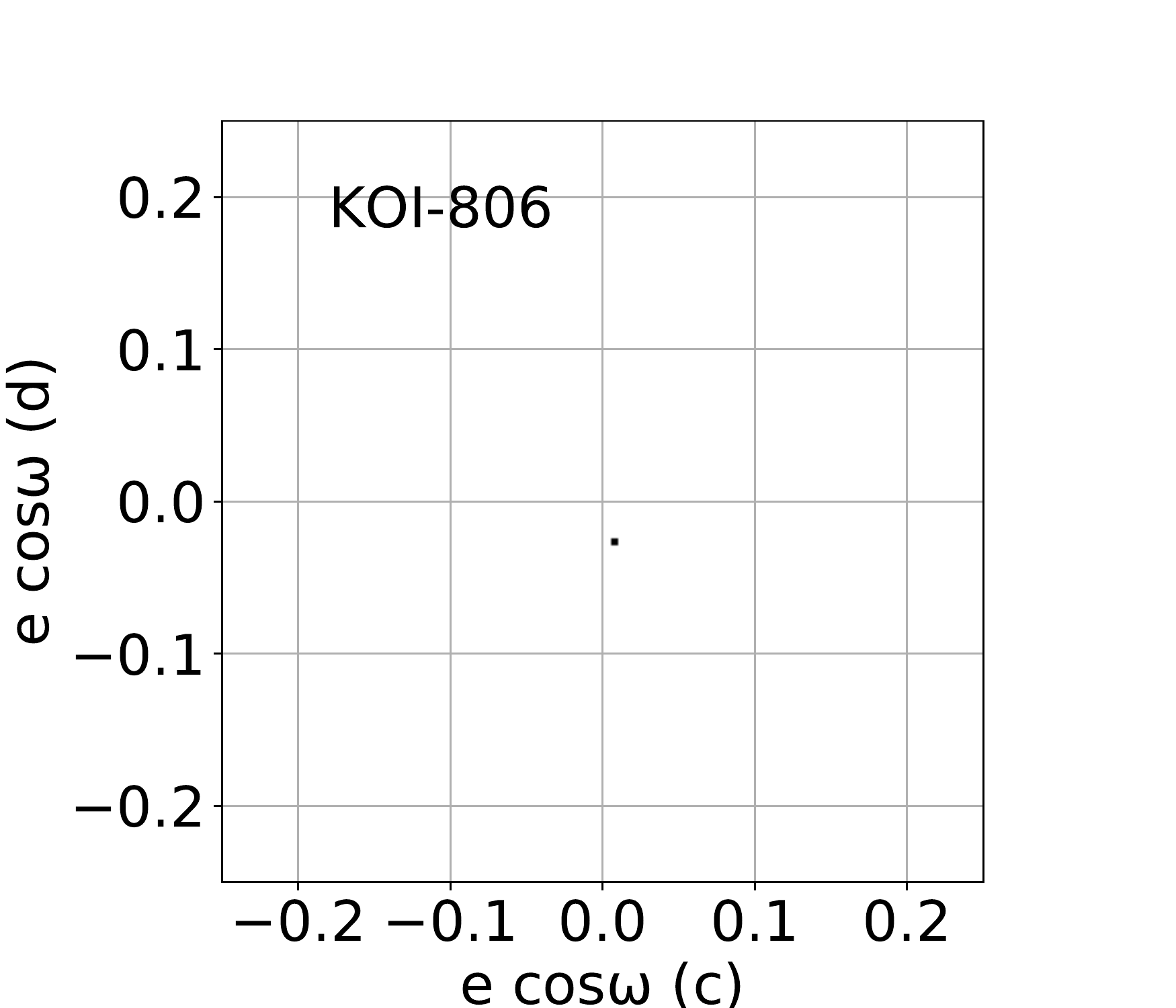}
\includegraphics [height = 1.1 in]{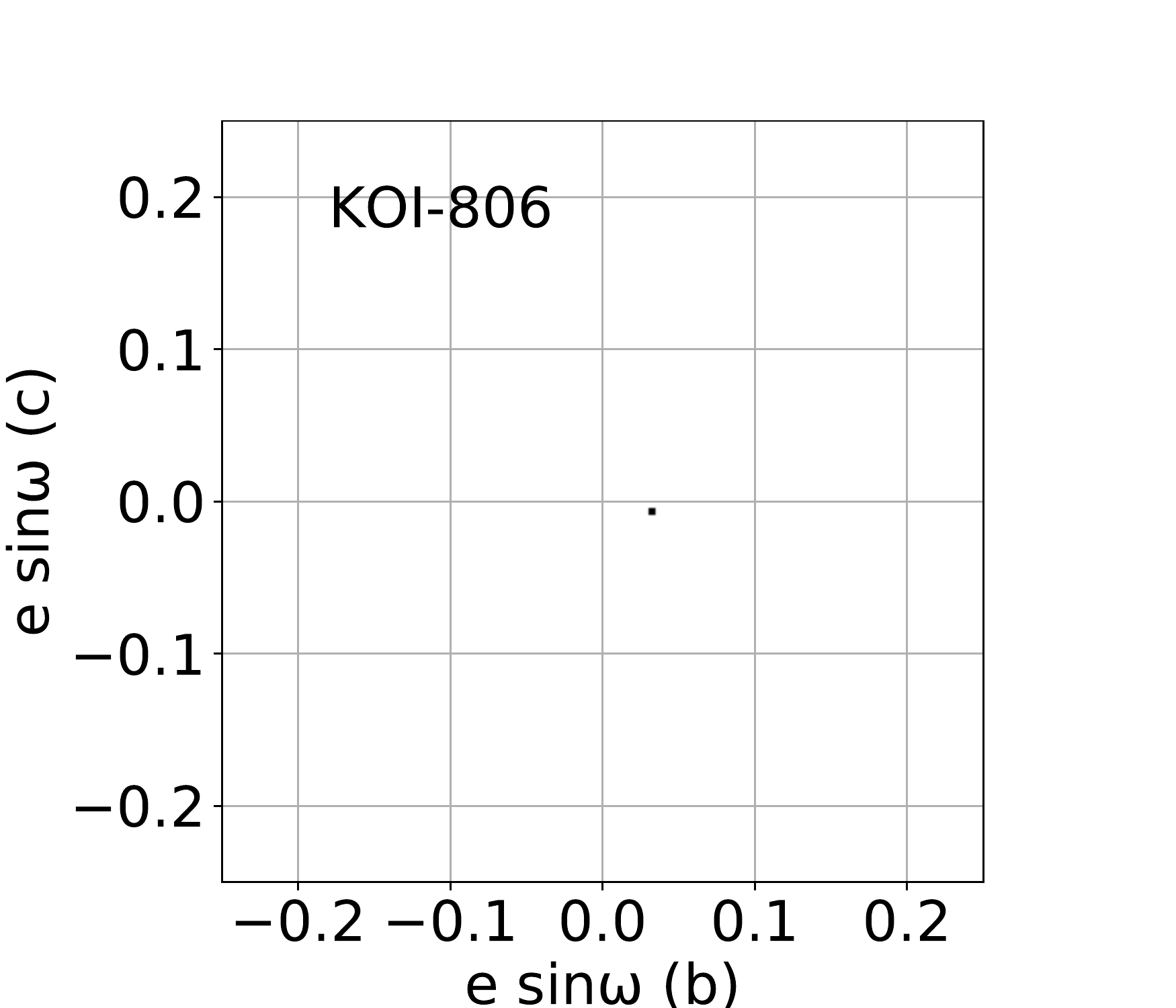} 
\includegraphics [height = 1.1 in]{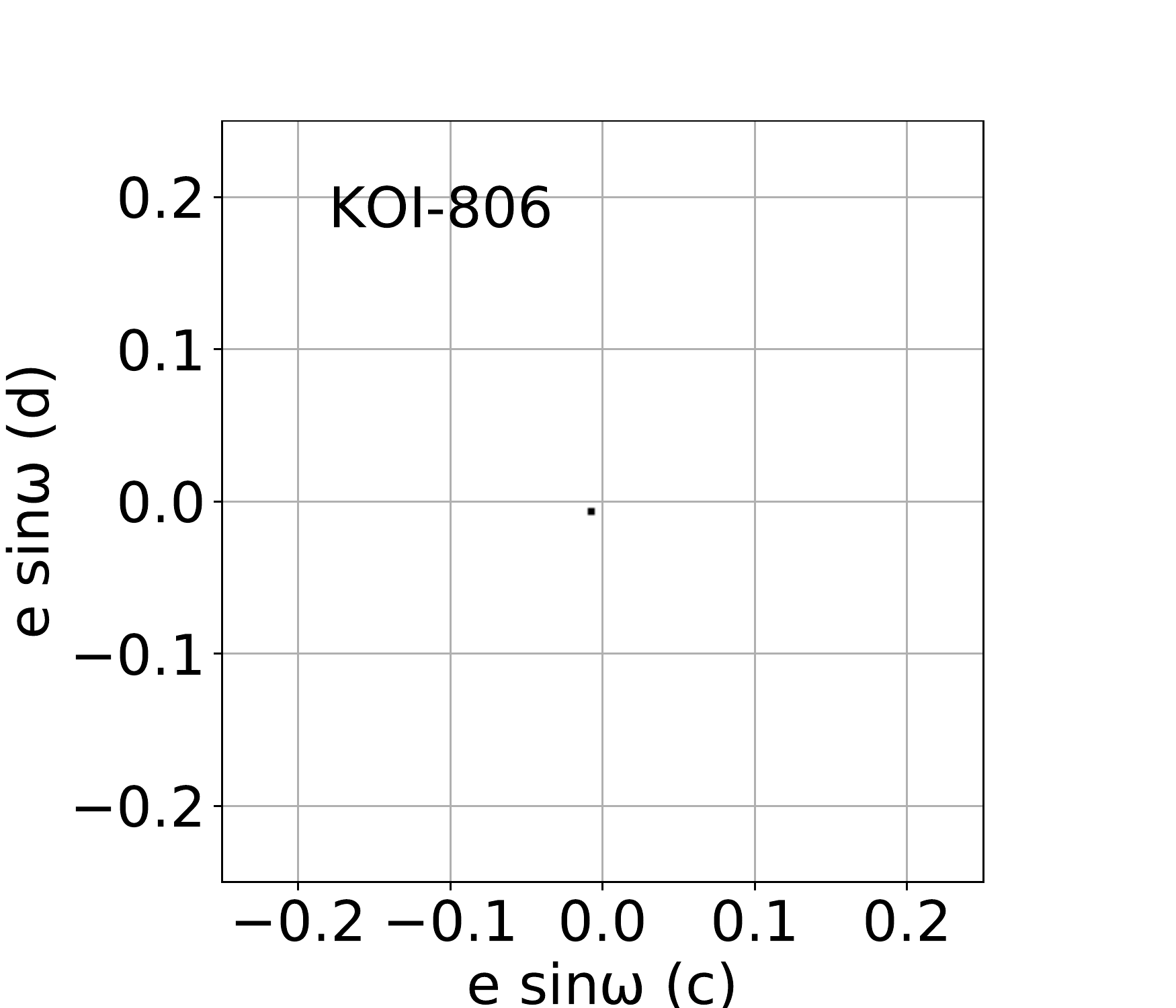}
\caption{Two-dimensional kernel density estimators on joint posteriors of eccentricity vector components: three-planet systems (Part 3 of 7). 
}
\label{fig:ecc3c} 
\end{center}
\end{figure}

\begin{figure}
\begin{center}
\figurenum{26}
\includegraphics [height = 1.1 in]{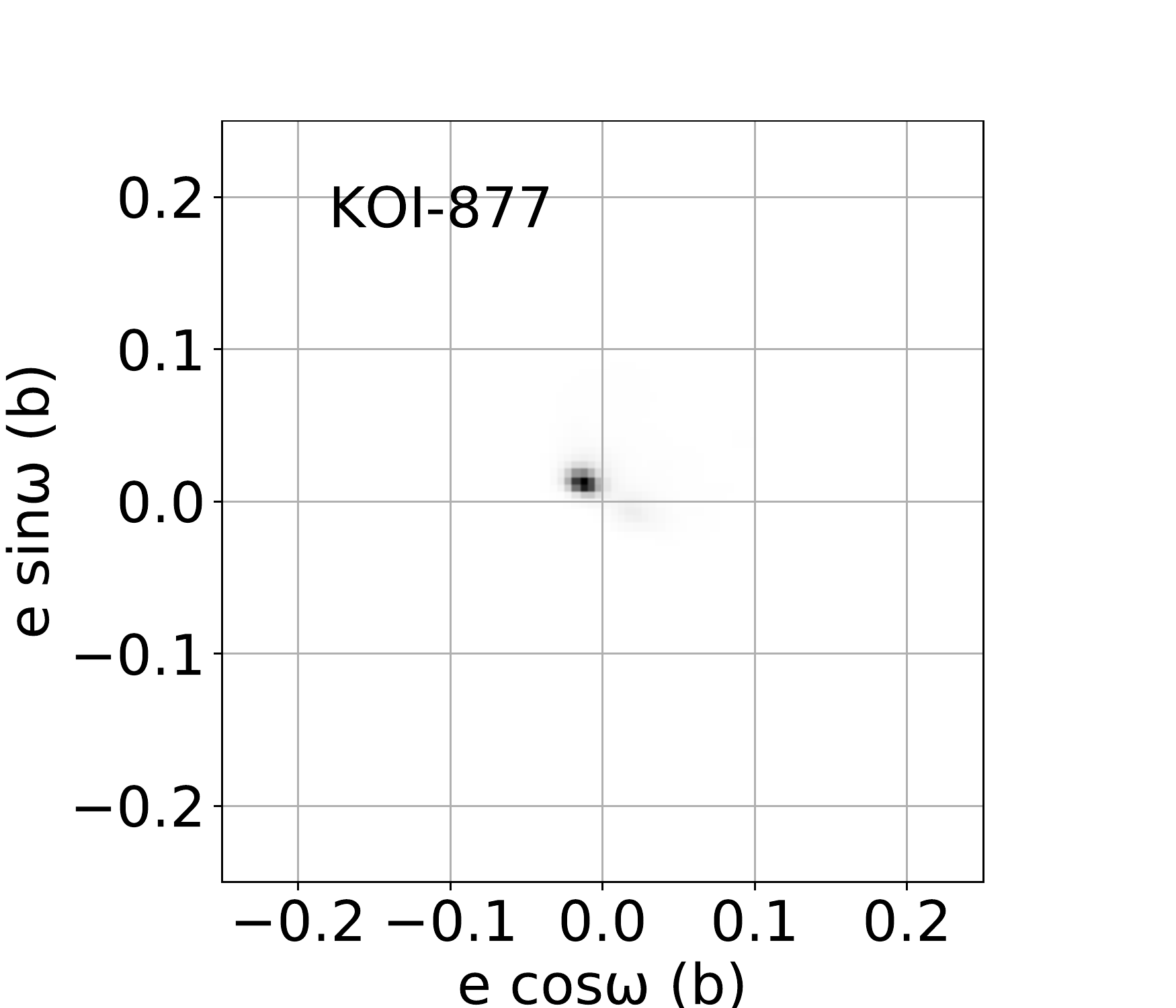}
\includegraphics [height = 1.1 in]{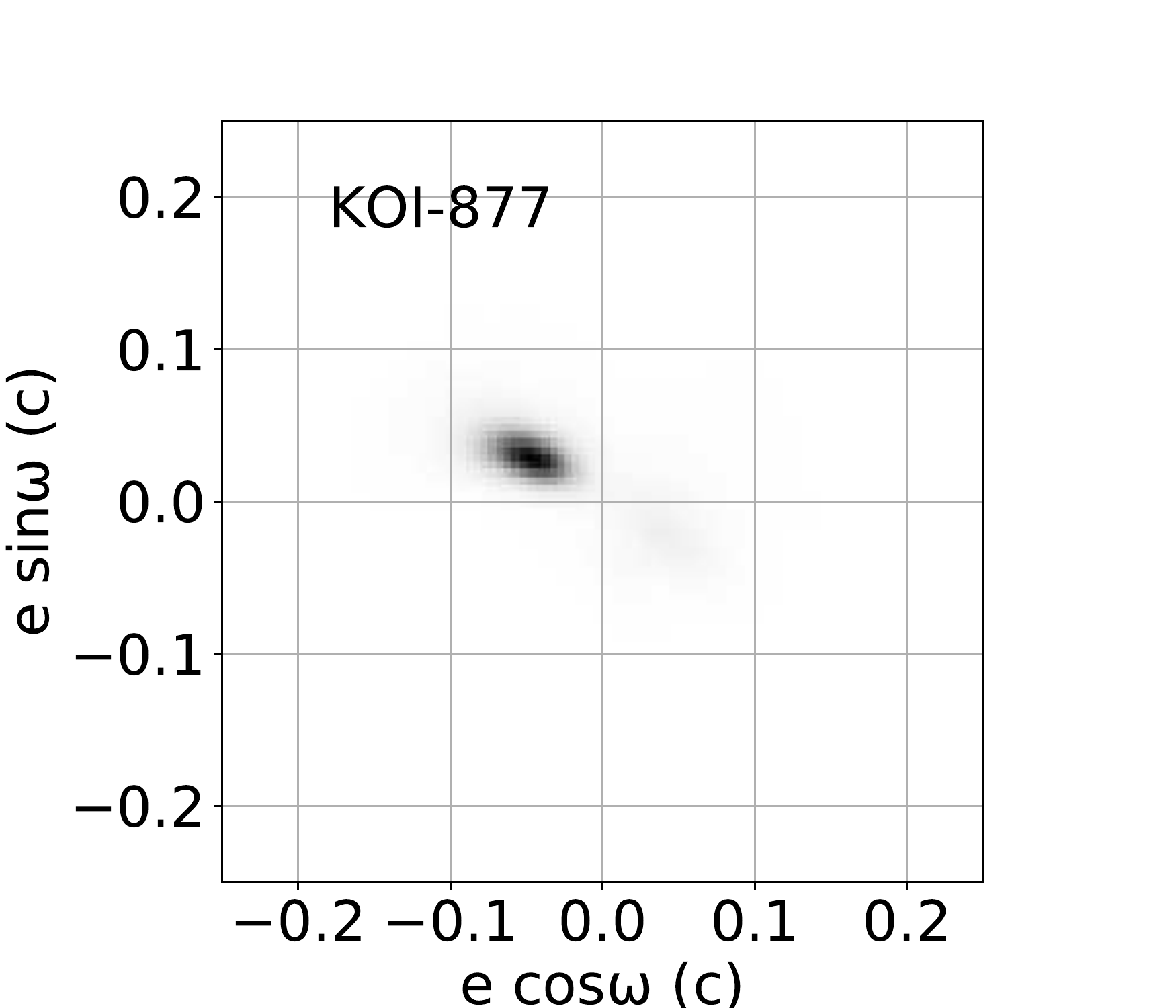}
\includegraphics [height = 1.1 in]{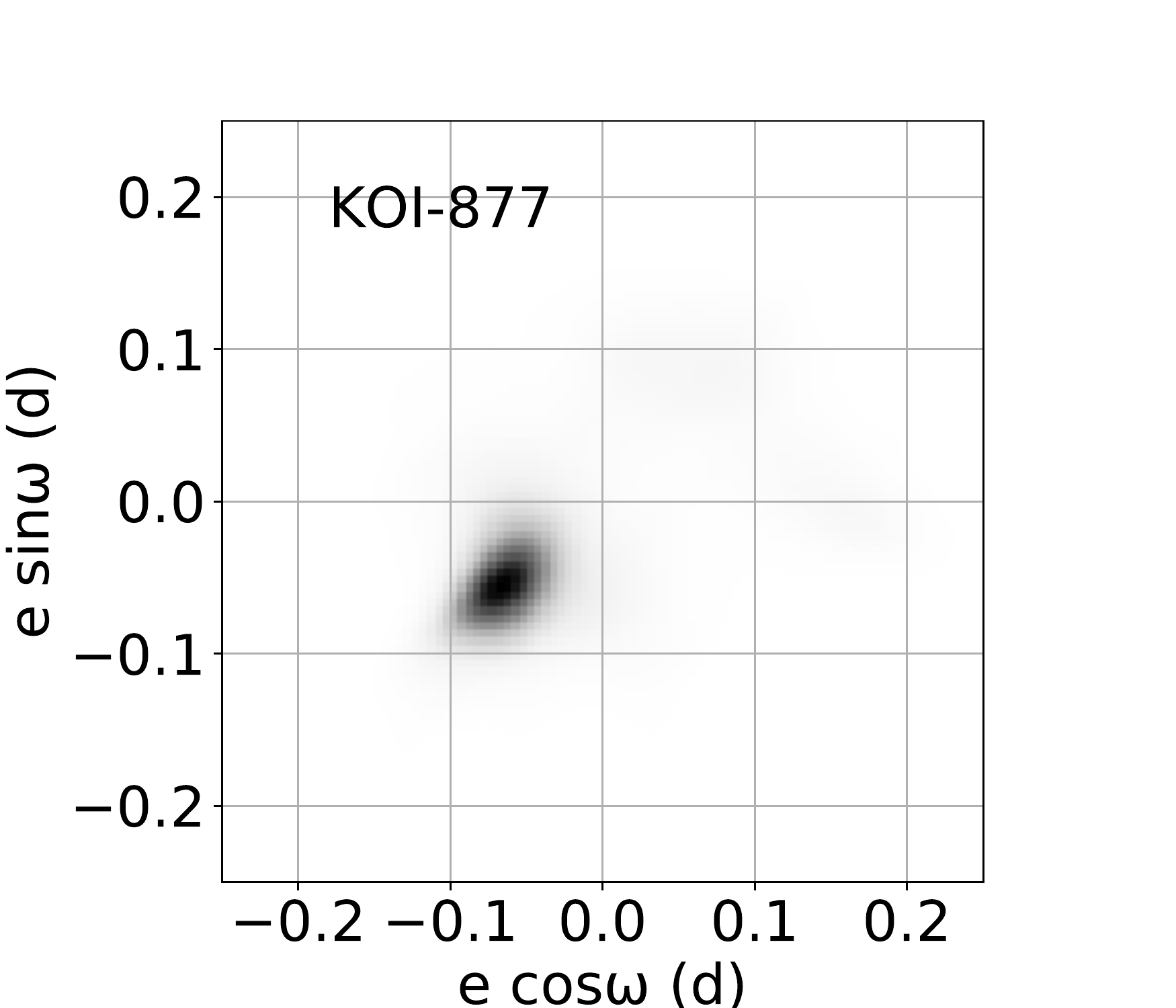}
\includegraphics [height = 1.1 in]{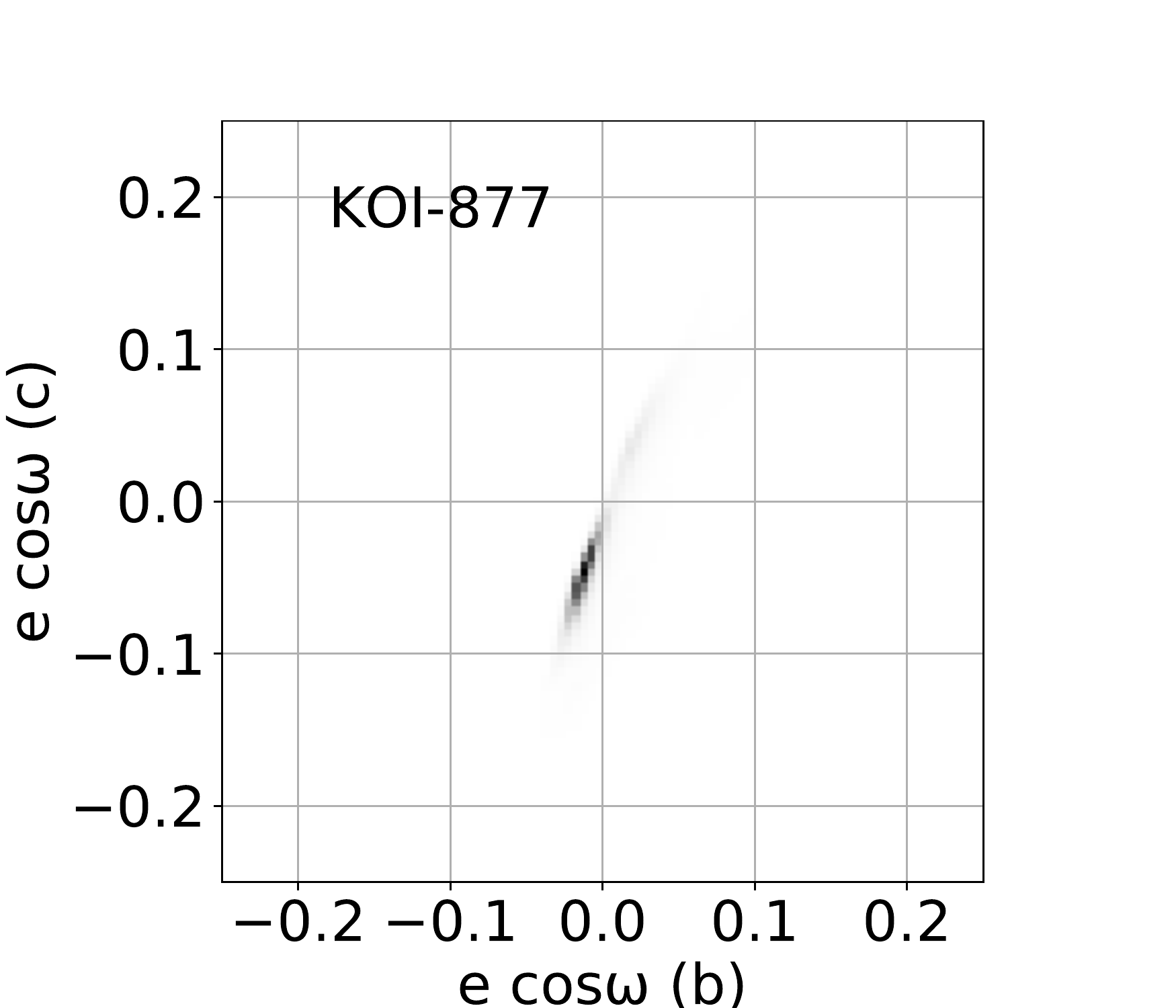}\\
\includegraphics [height = 1.1 in]{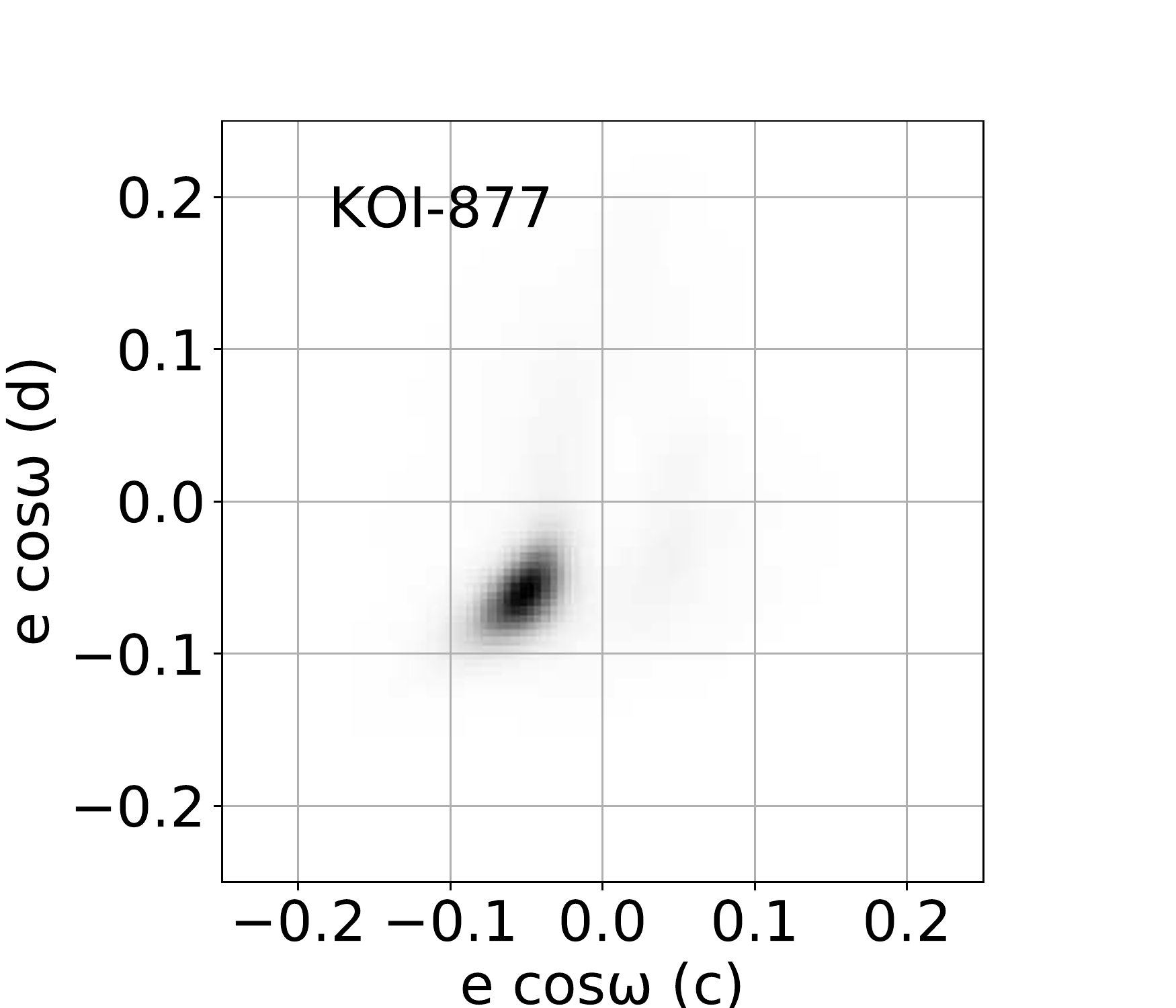}
\includegraphics [height = 1.1 in]{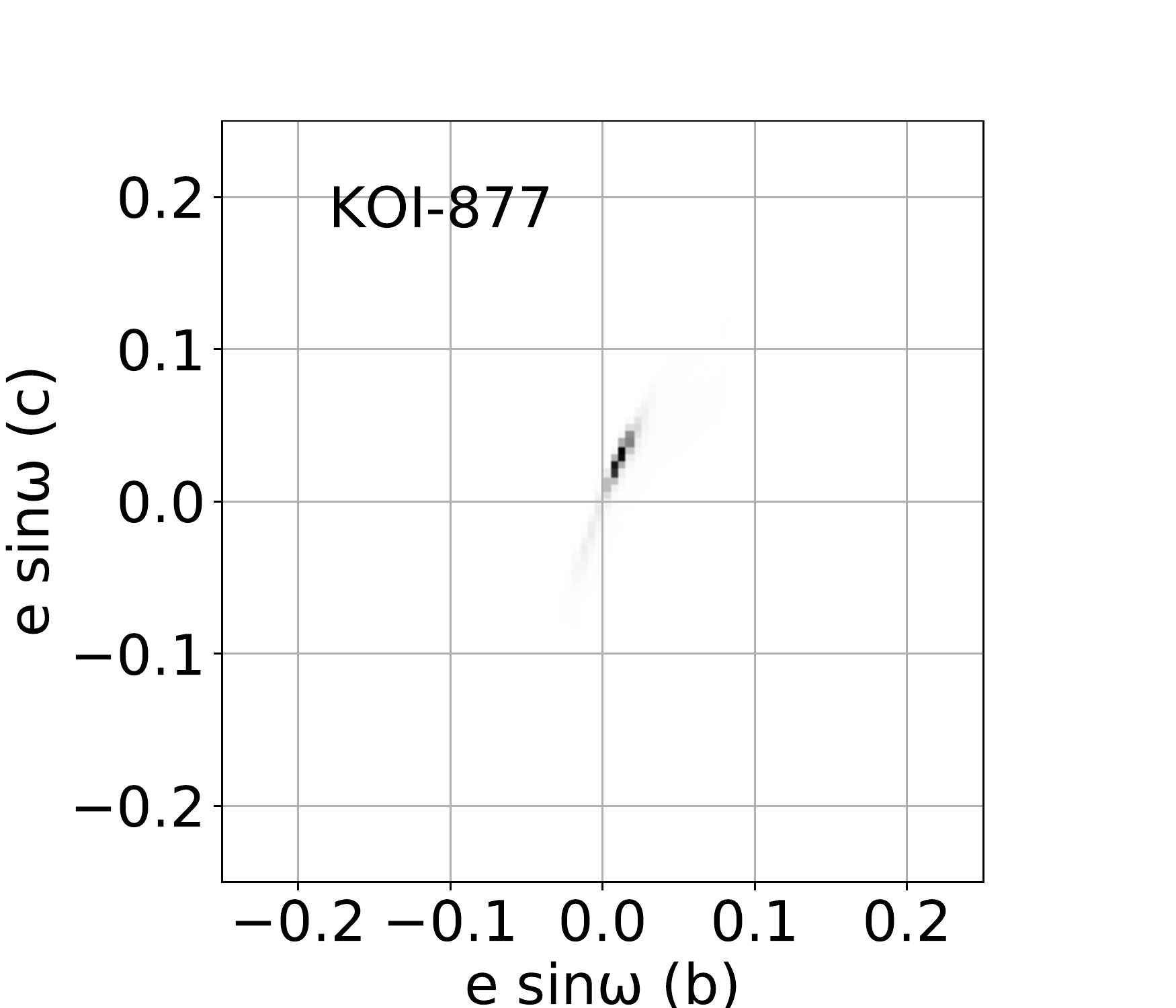}
\includegraphics [height = 1.1 in]{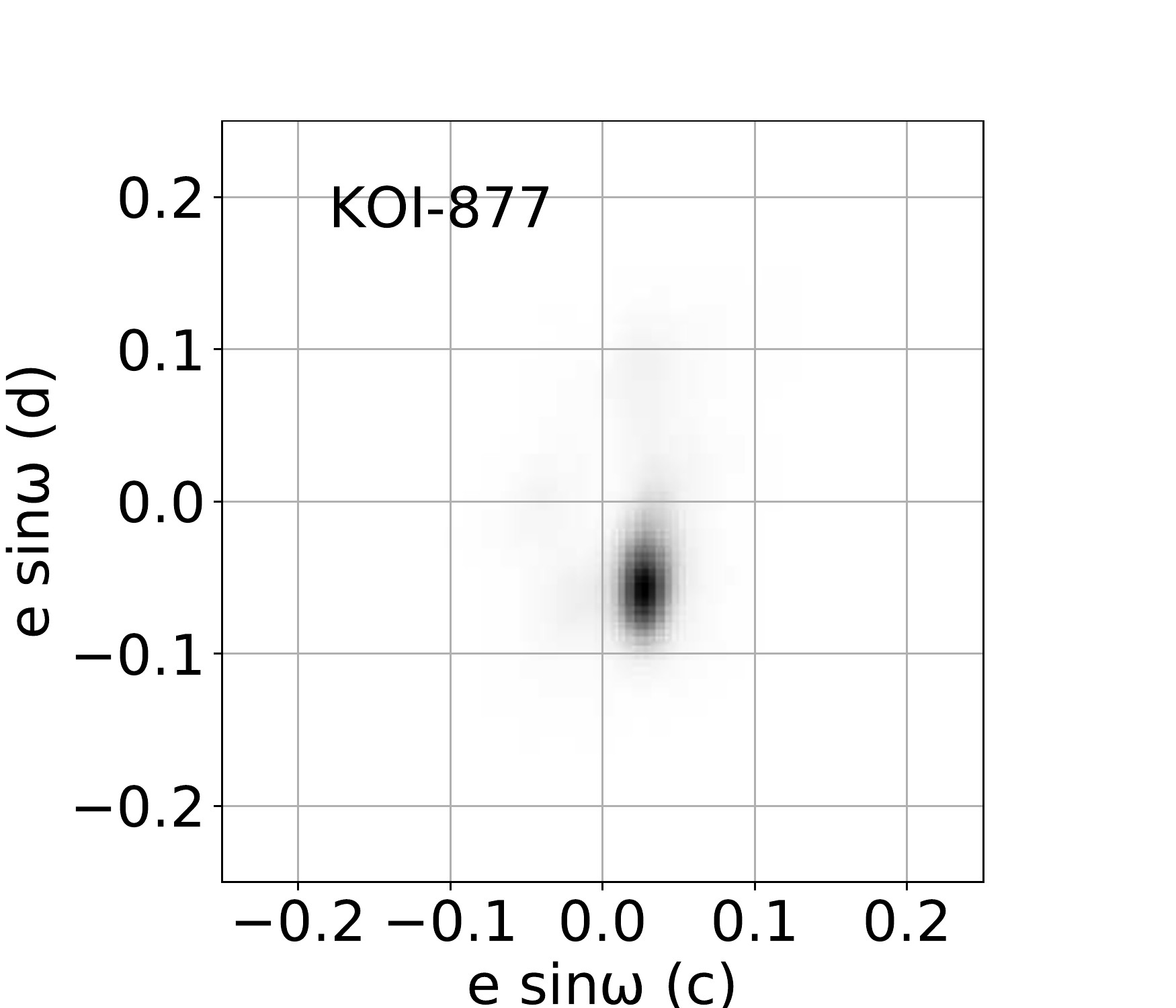}
\includegraphics [height = 1.1 in]{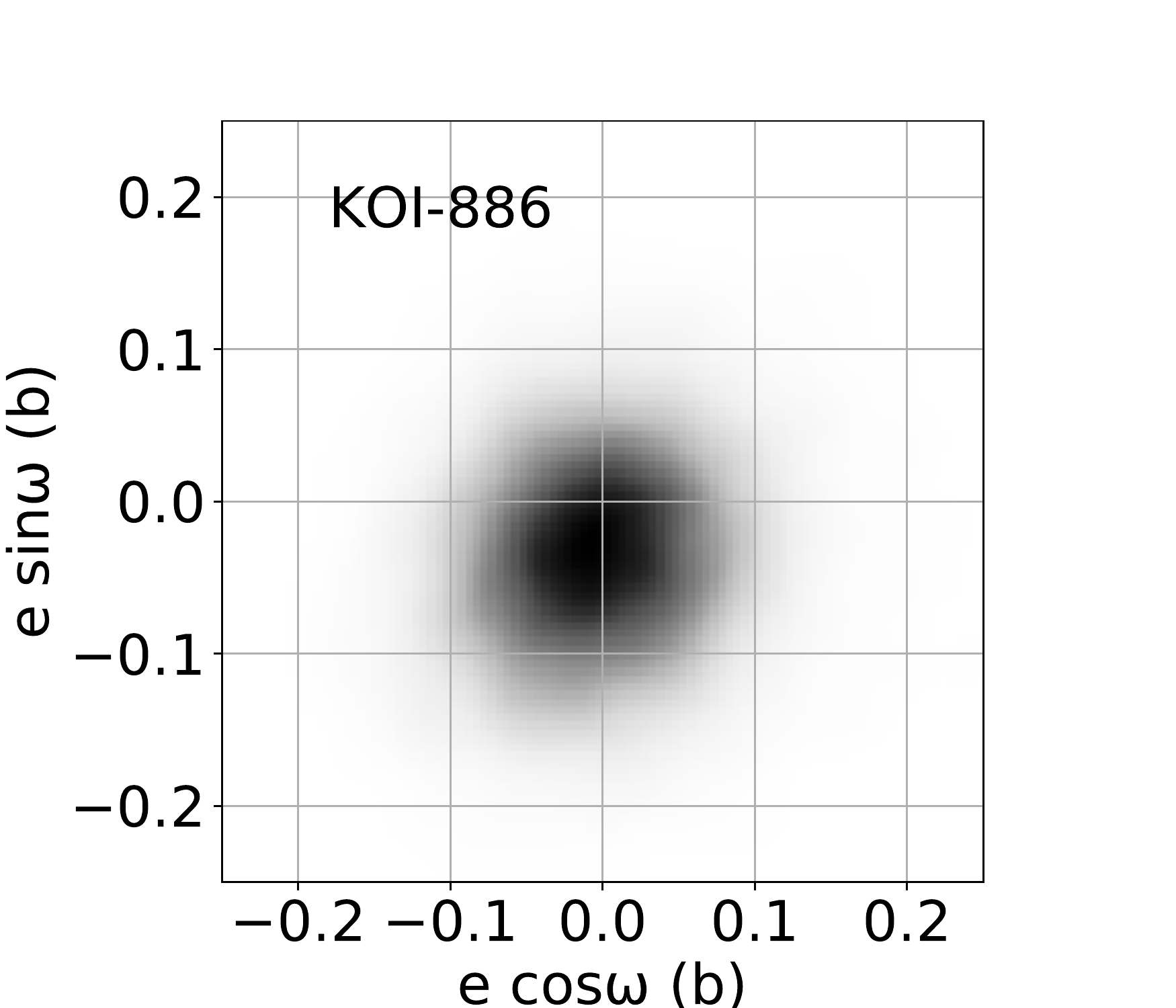} \\
\includegraphics [height = 1.1 in]{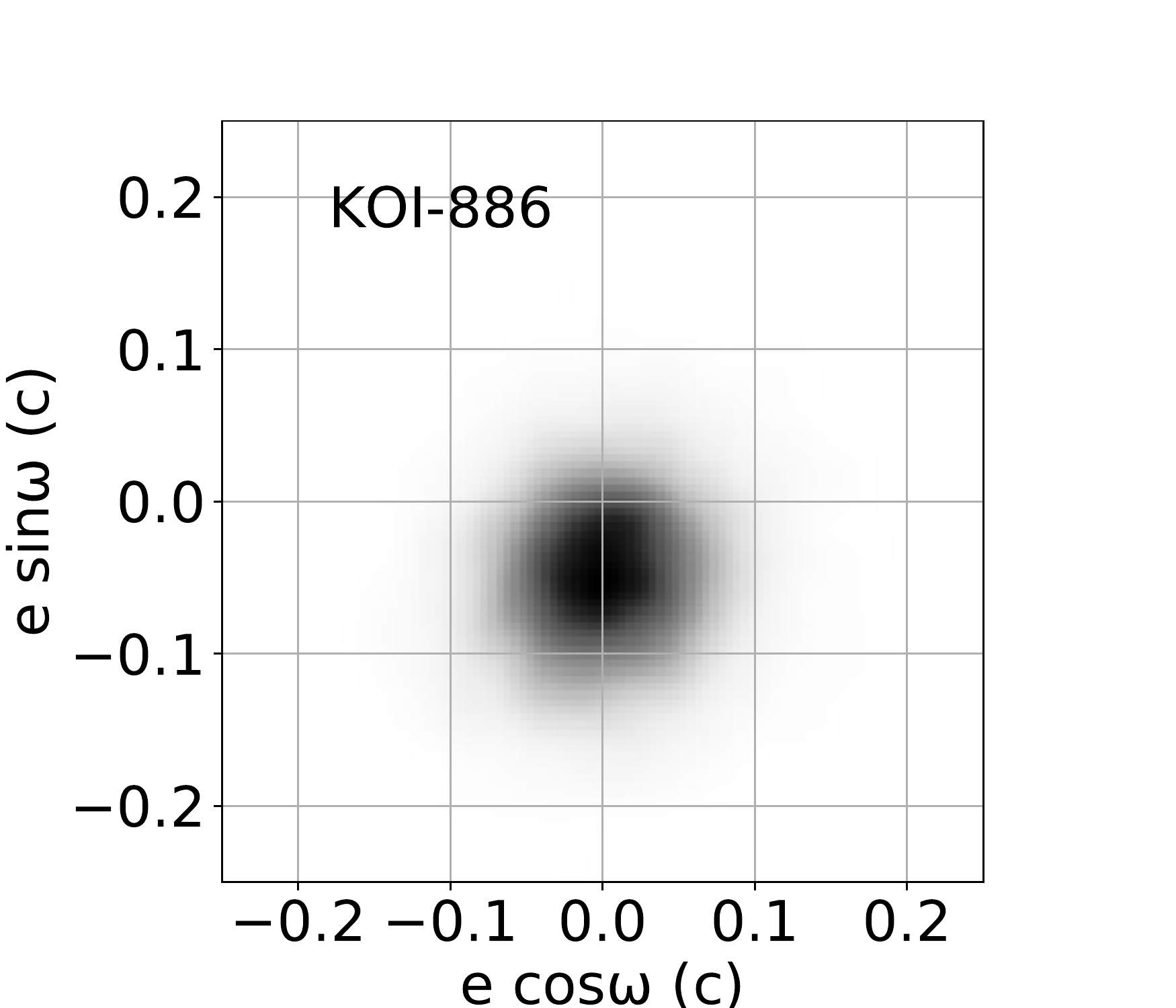}
\includegraphics [height = 1.1 in]{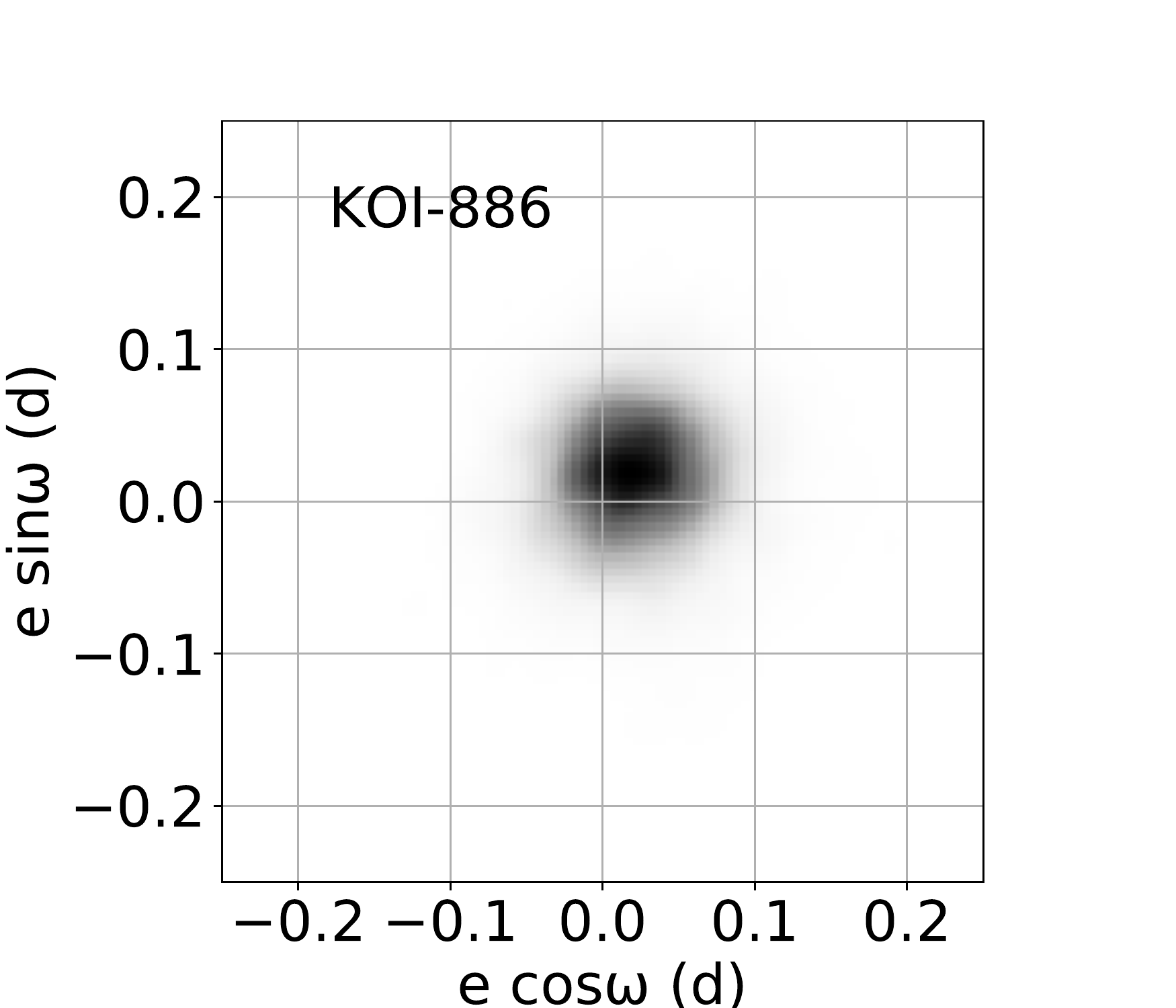}
\includegraphics [height = 1.1 in]{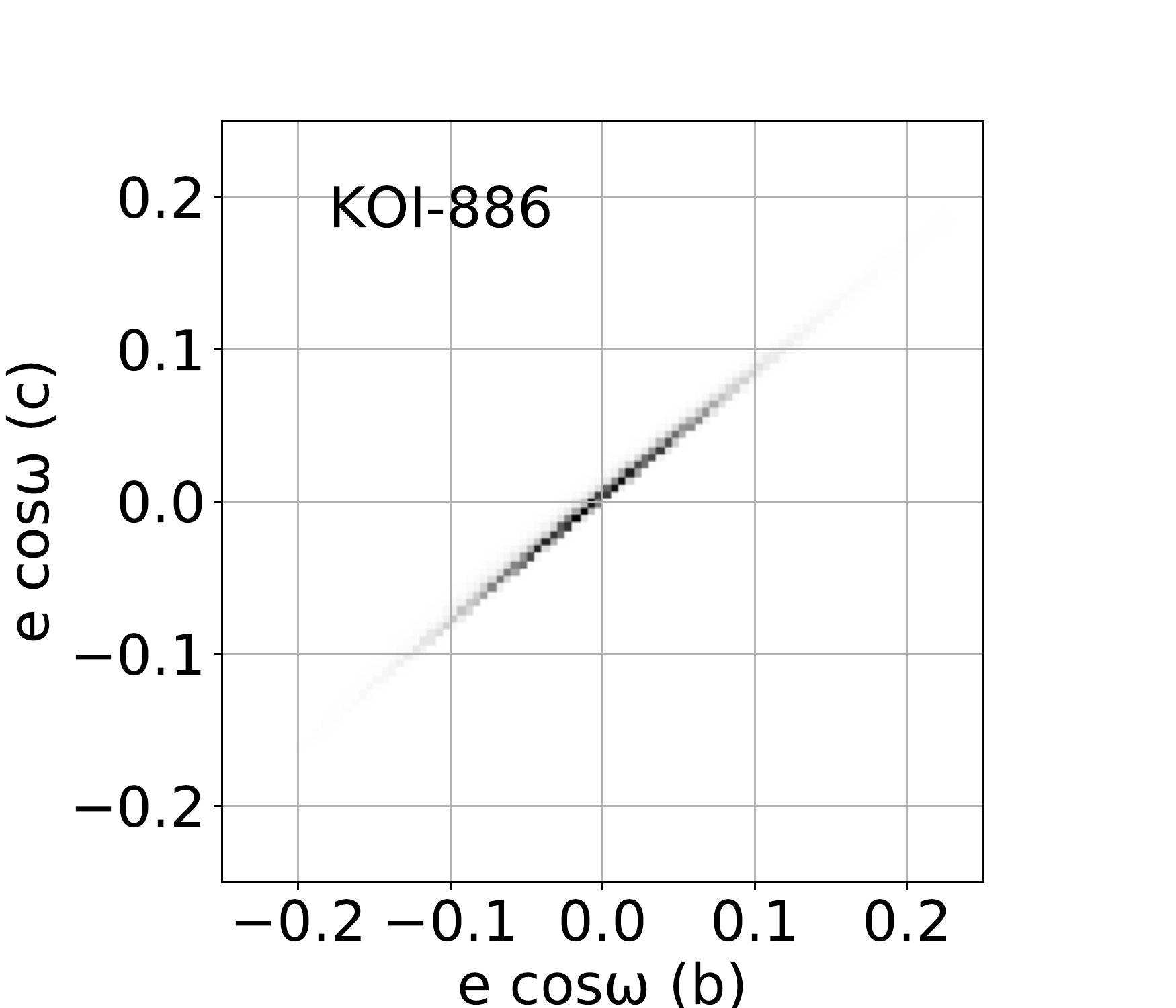}
\includegraphics [height = 1.1 in]{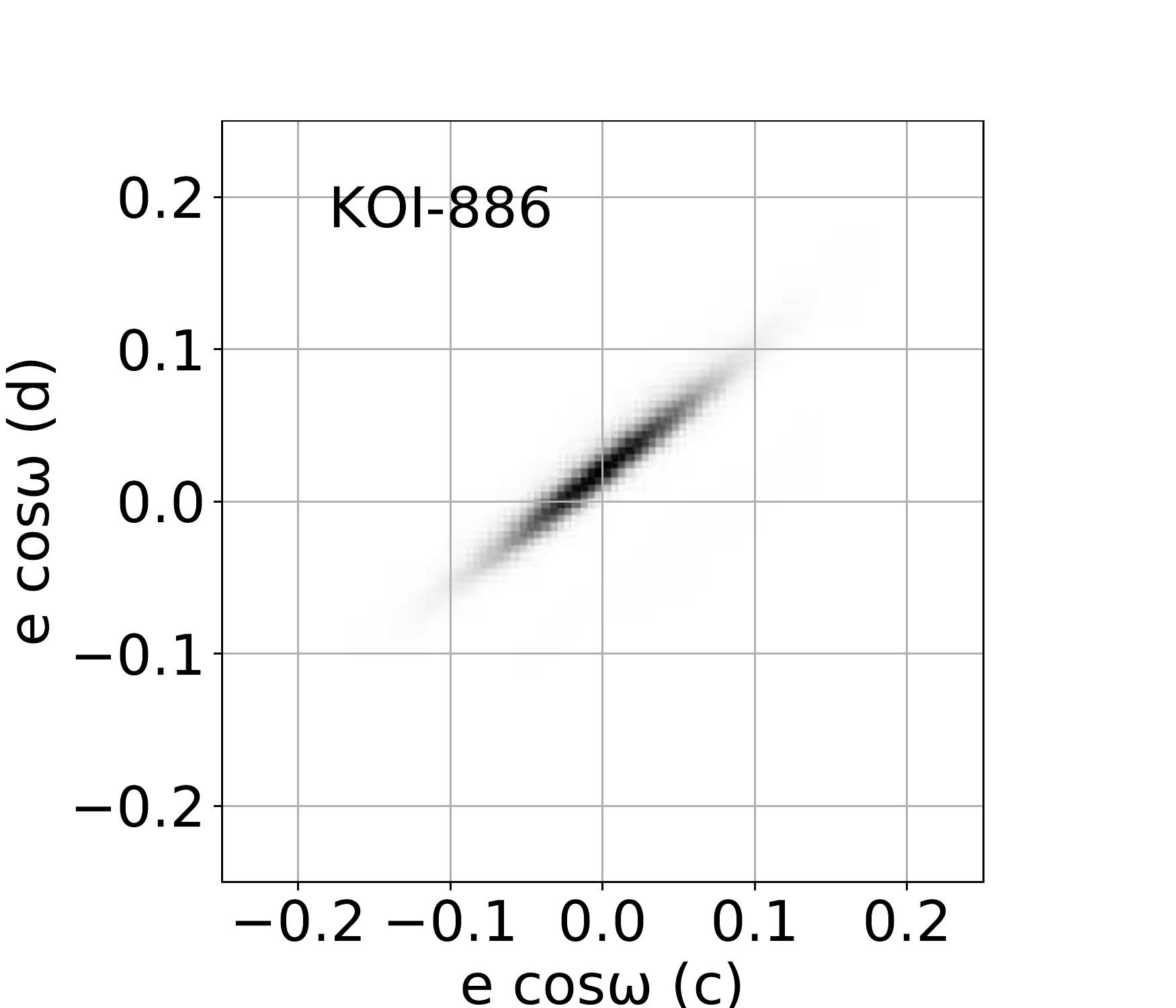} \\
\includegraphics [height = 1.1 in]{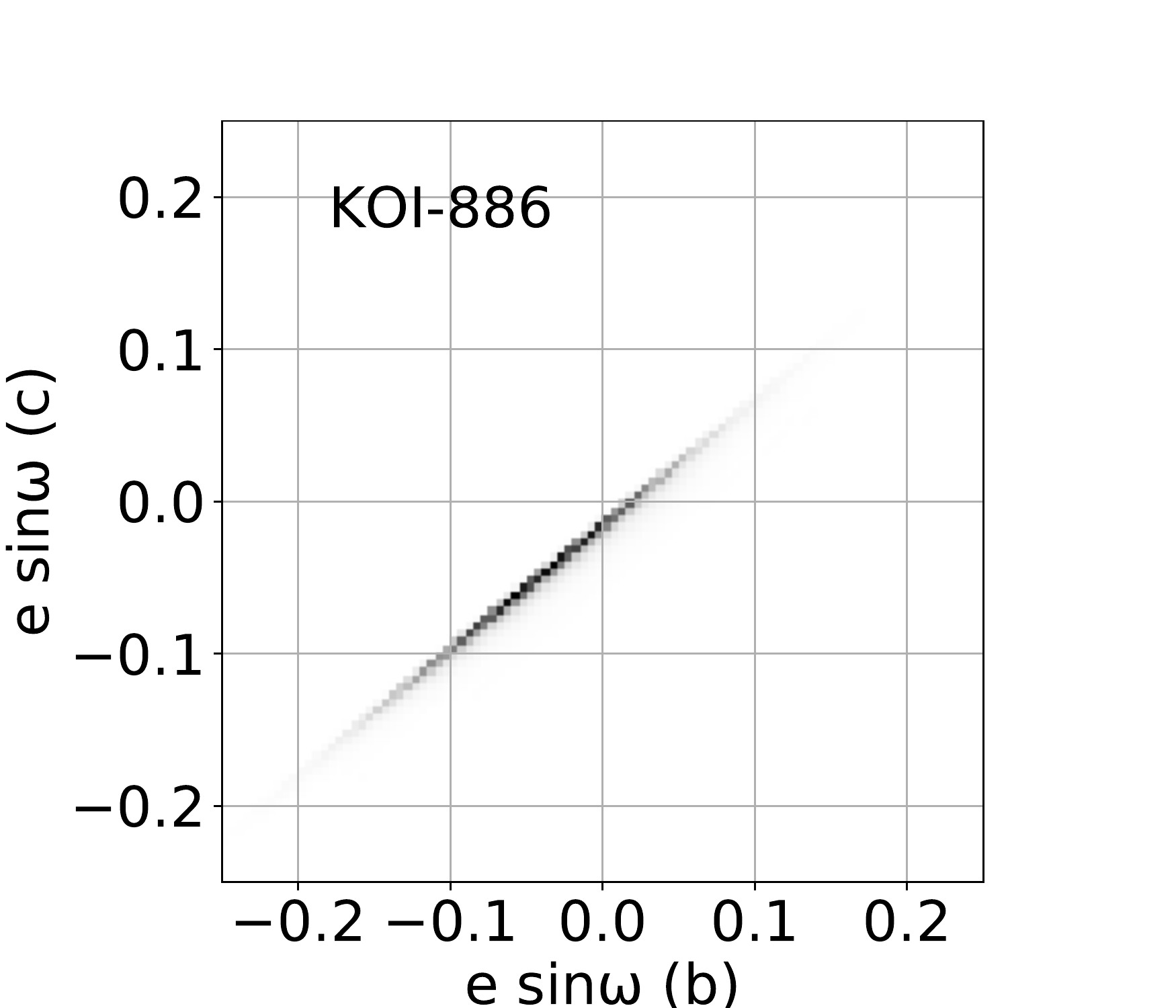}
\includegraphics [height = 1.1 in]{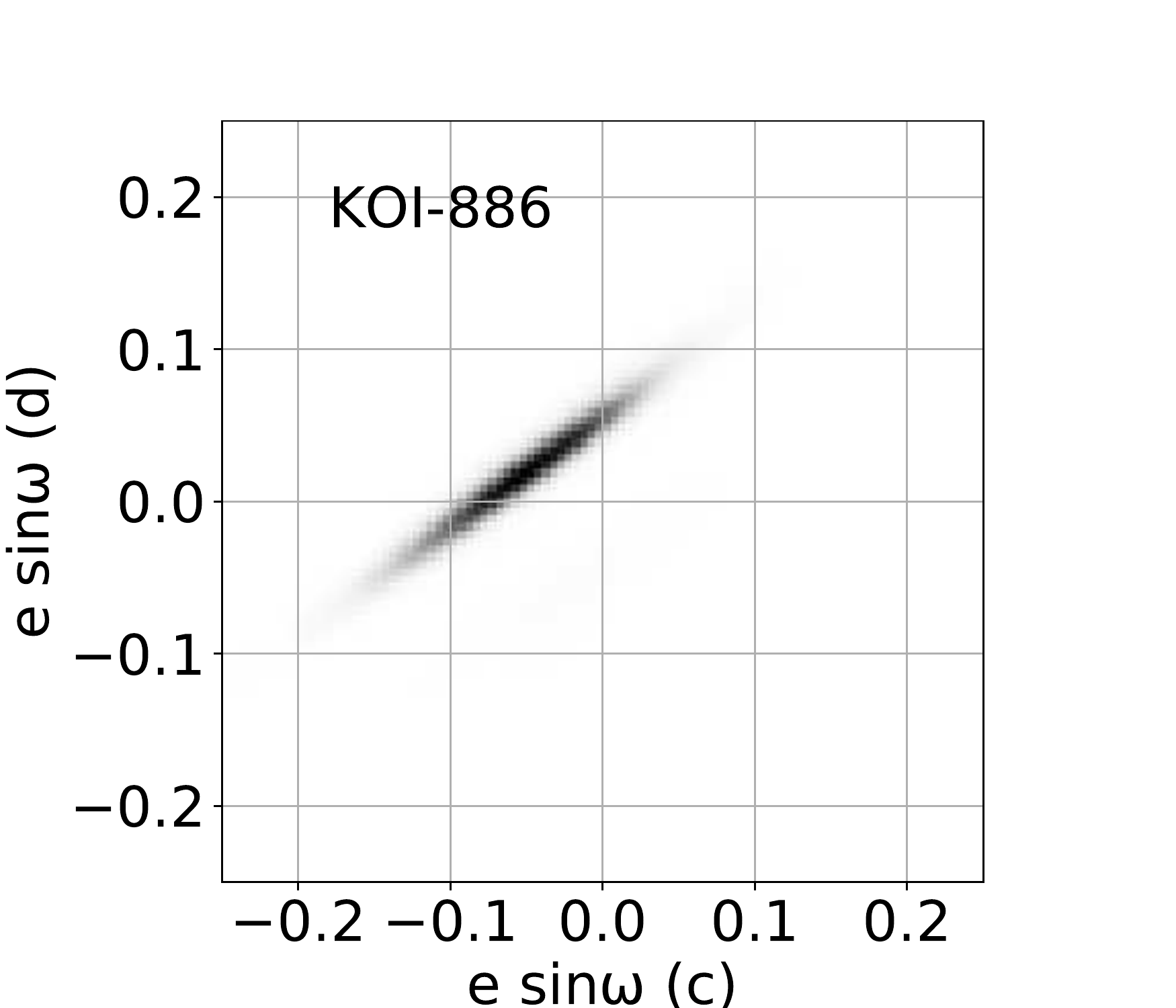}
\includegraphics [height = 1.1 in]{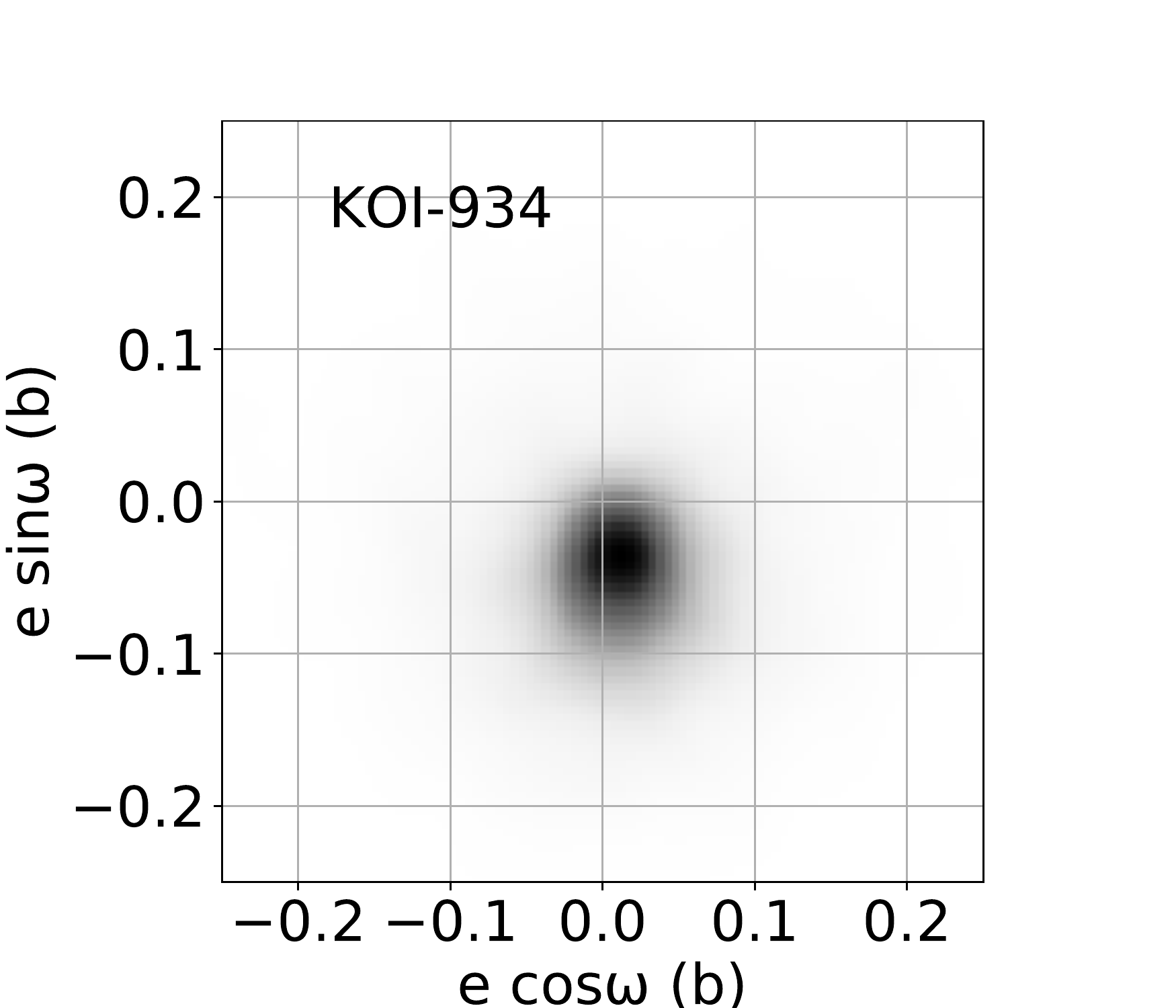}
\includegraphics [height = 1.1 in]{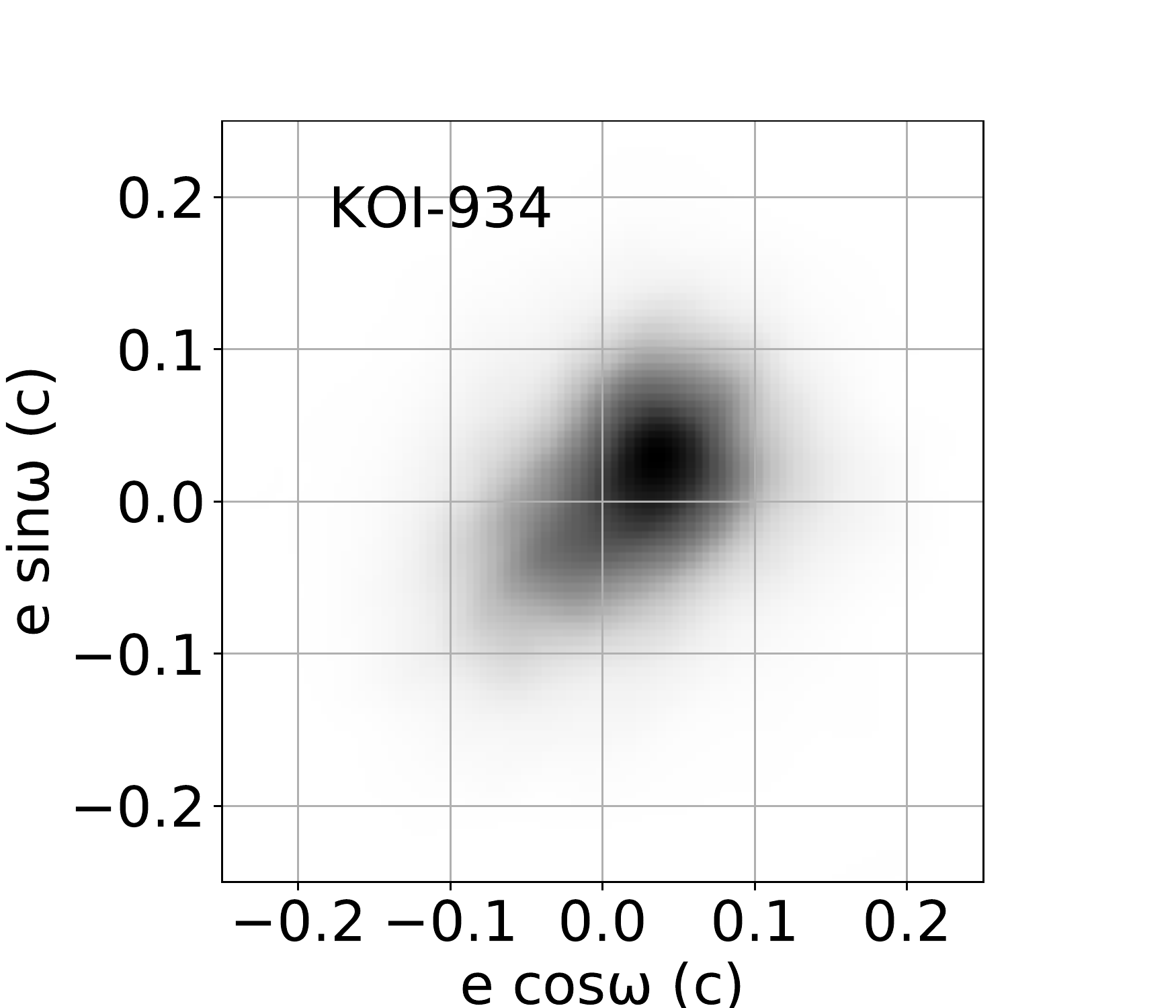} \\
\includegraphics [height = 1.1 in]{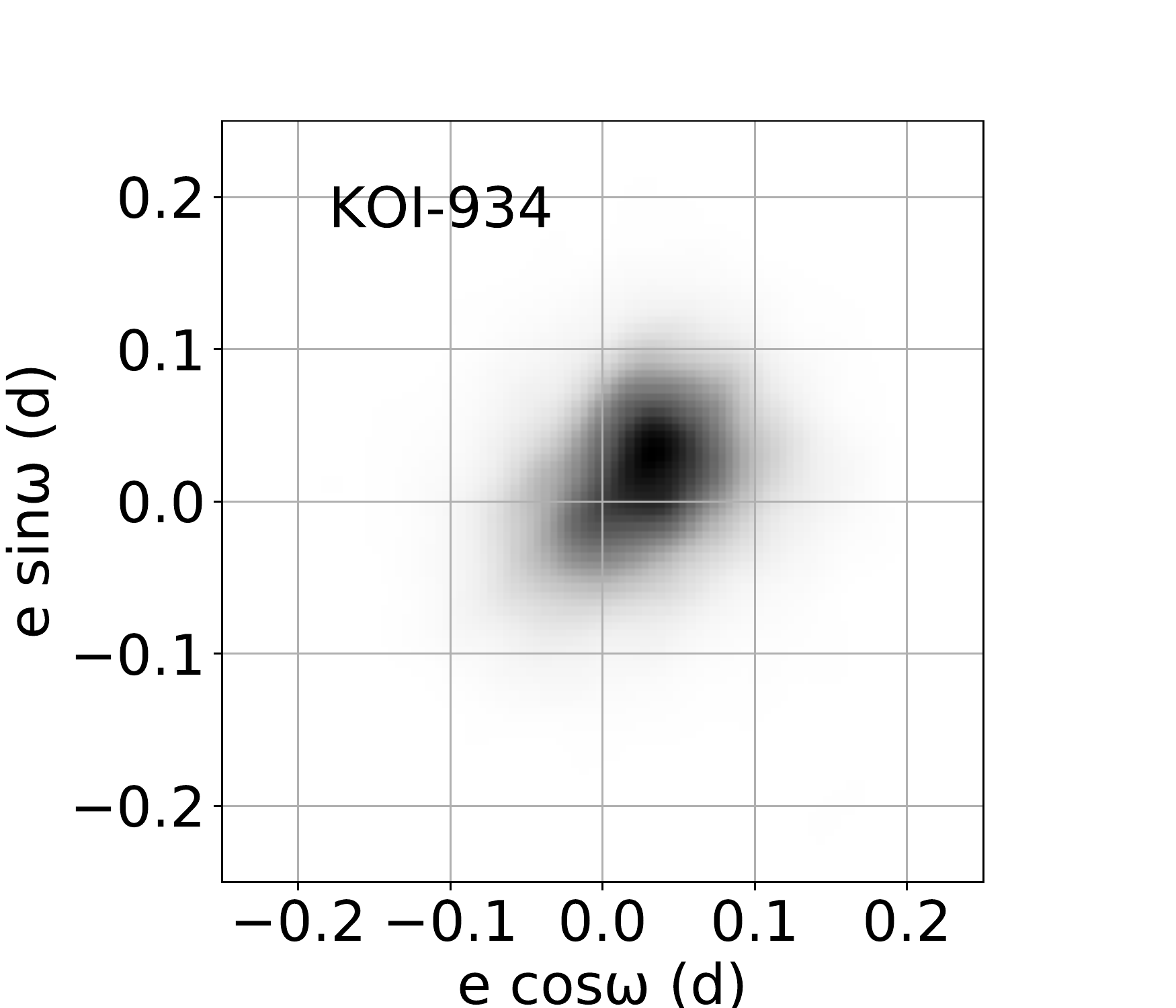}
\includegraphics [height = 1.1 in]{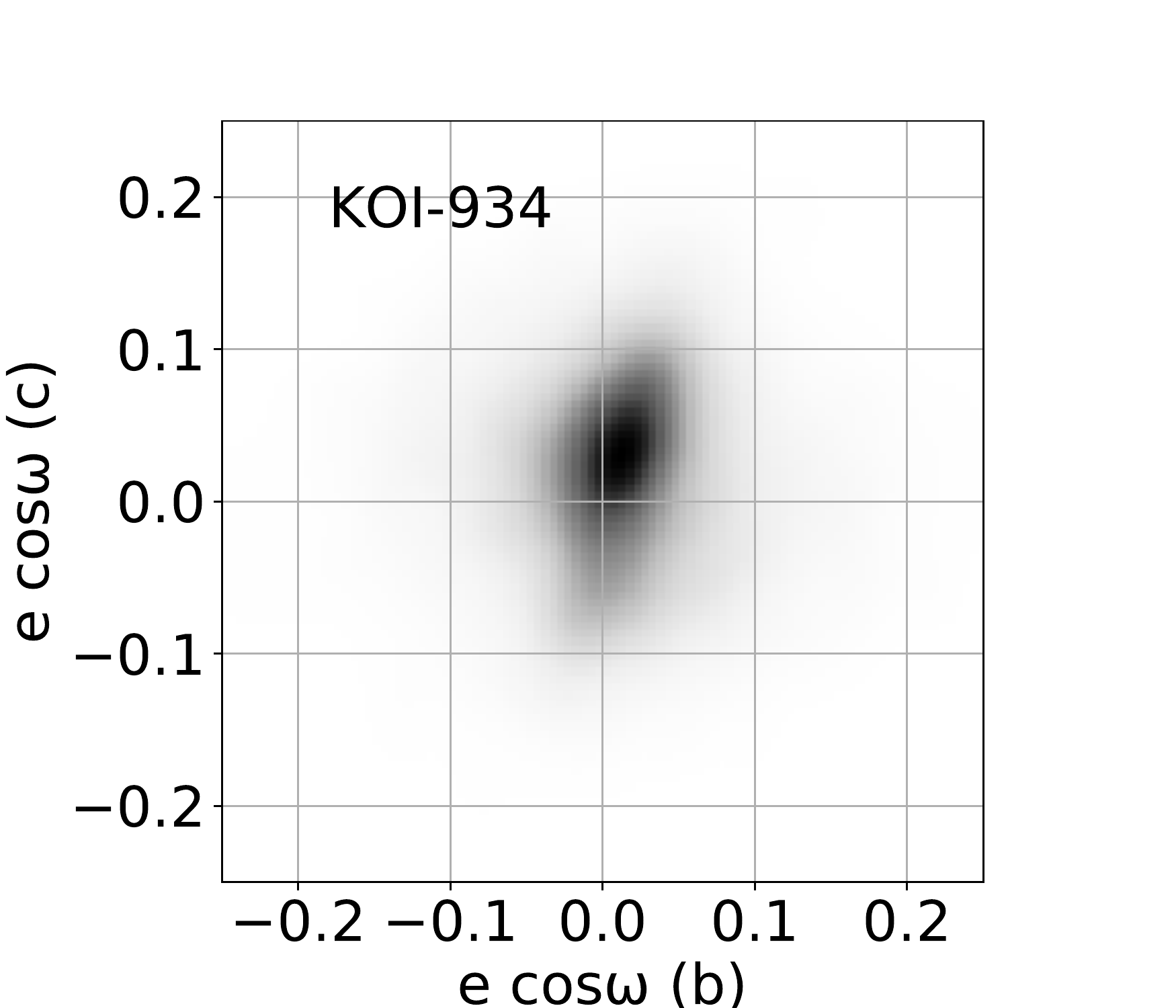} 
\includegraphics [height = 1.1 in]{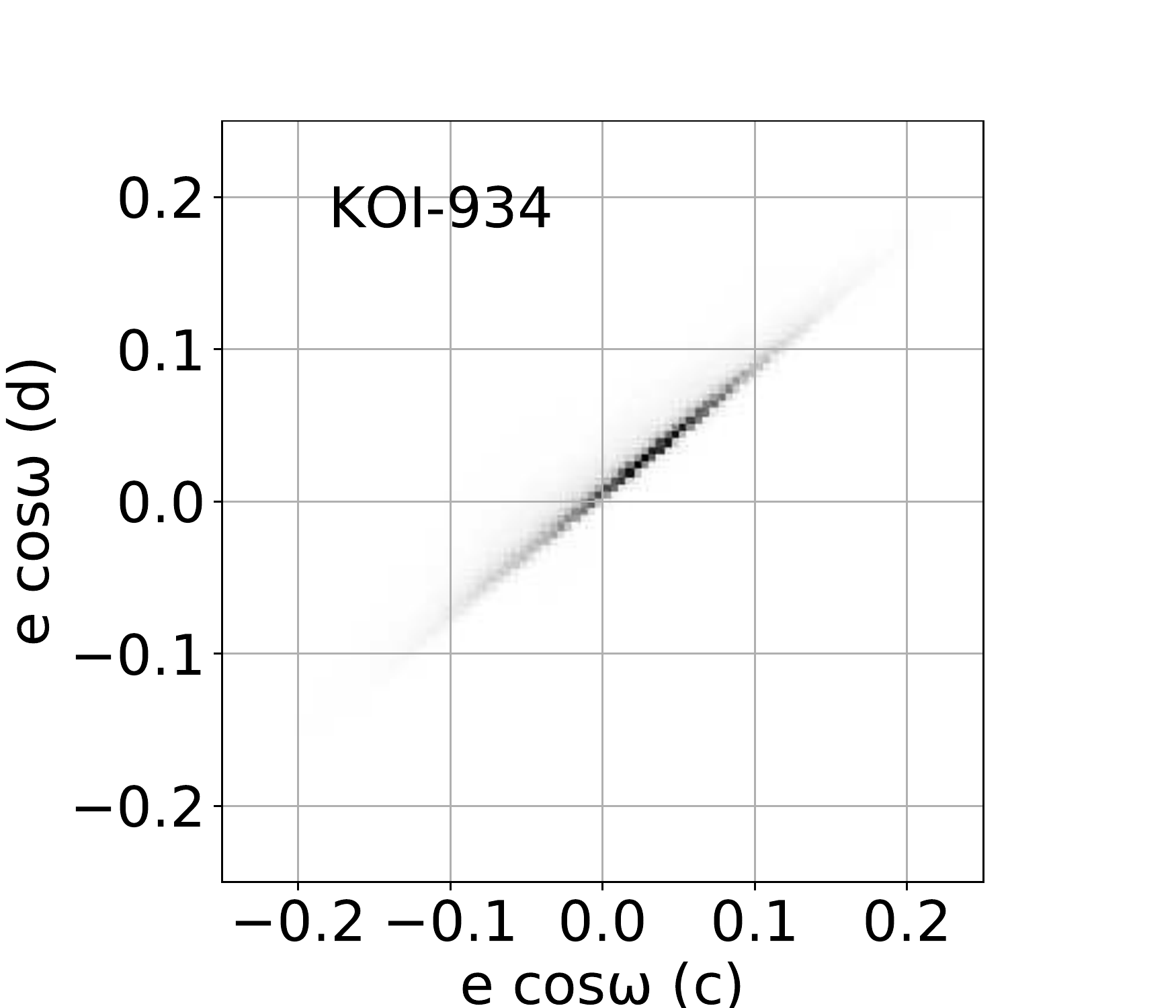}
\includegraphics [height = 1.1 in]{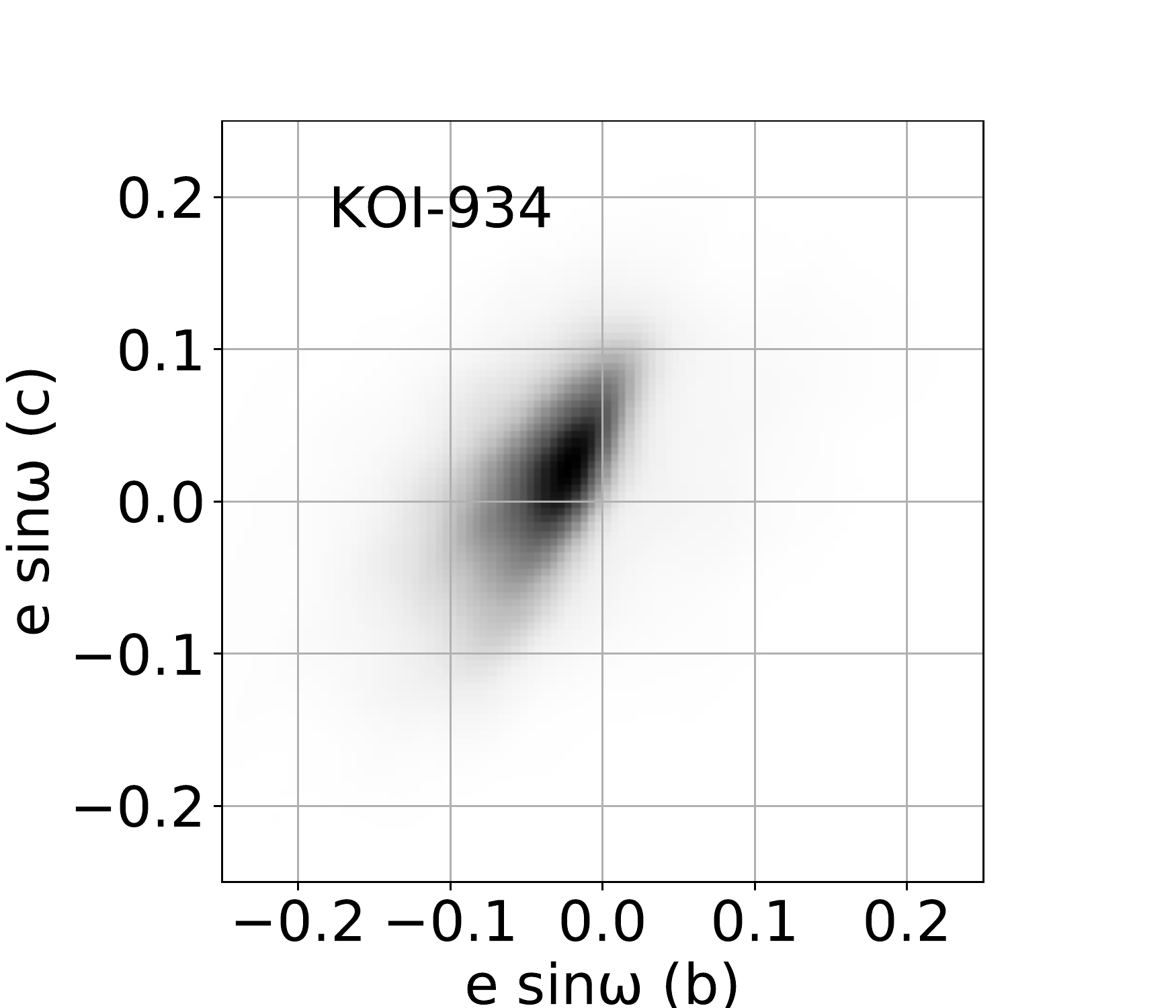} \\
\includegraphics [height = 1.1 in]{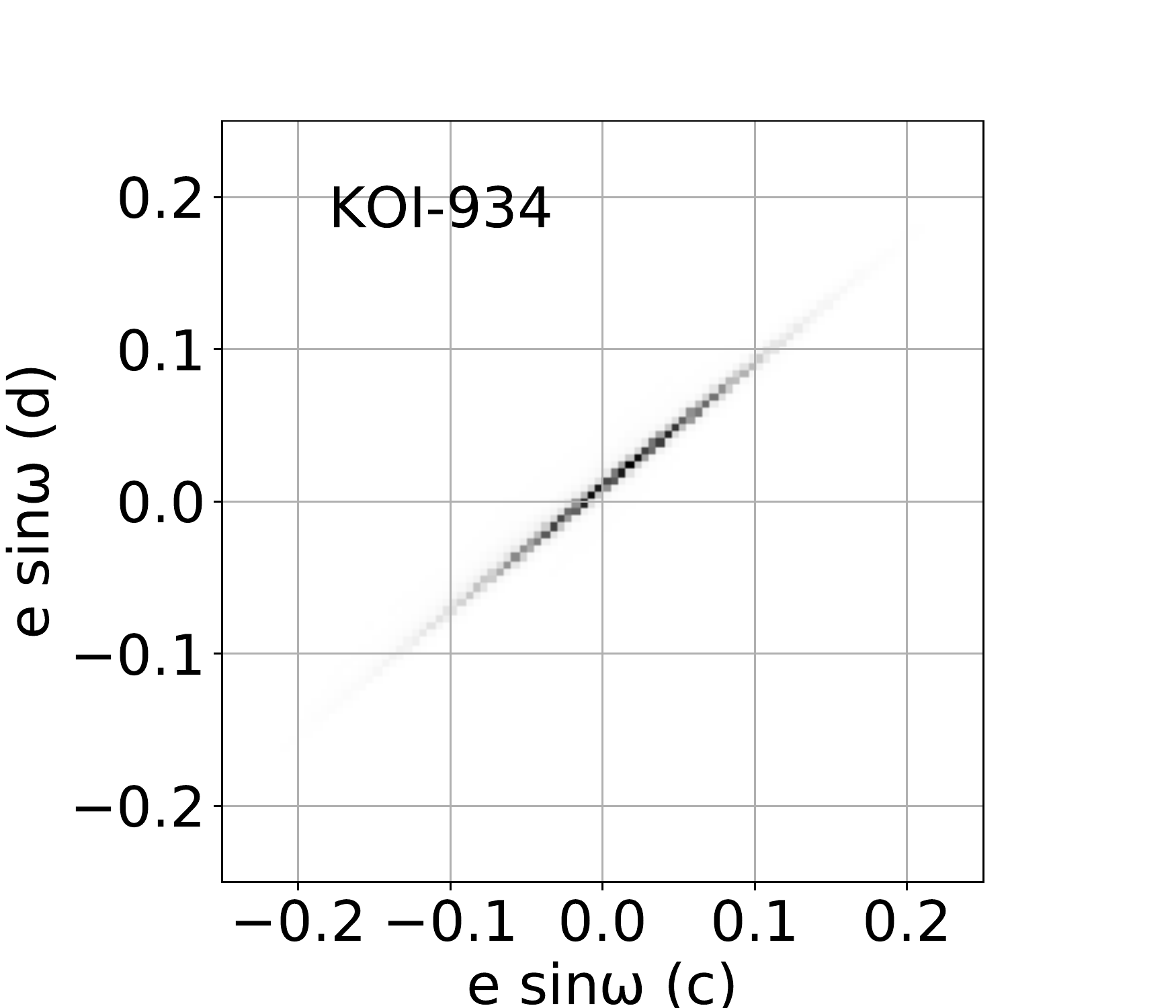}
\includegraphics [height = 1.1 in]{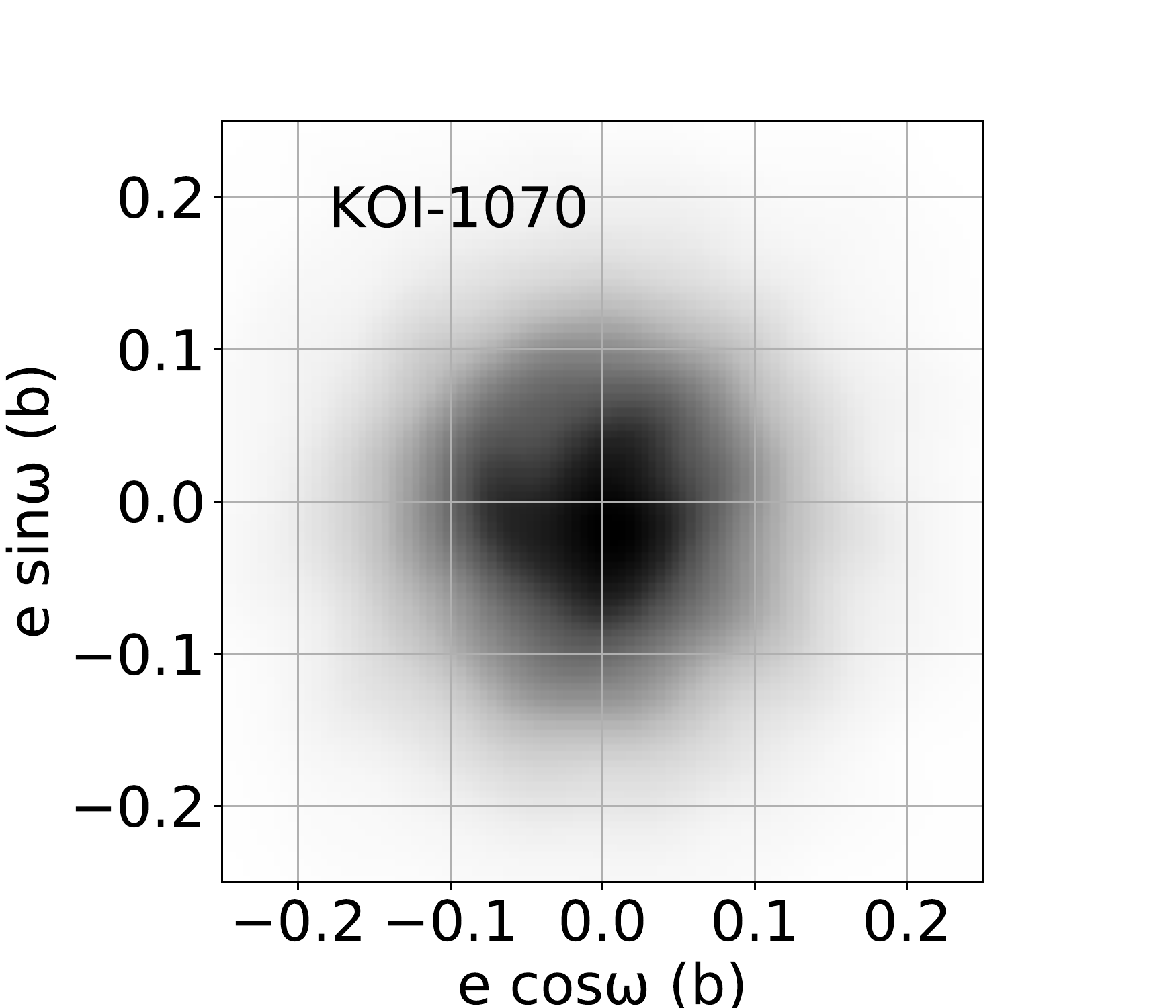} 
\includegraphics [height = 1.1 in]{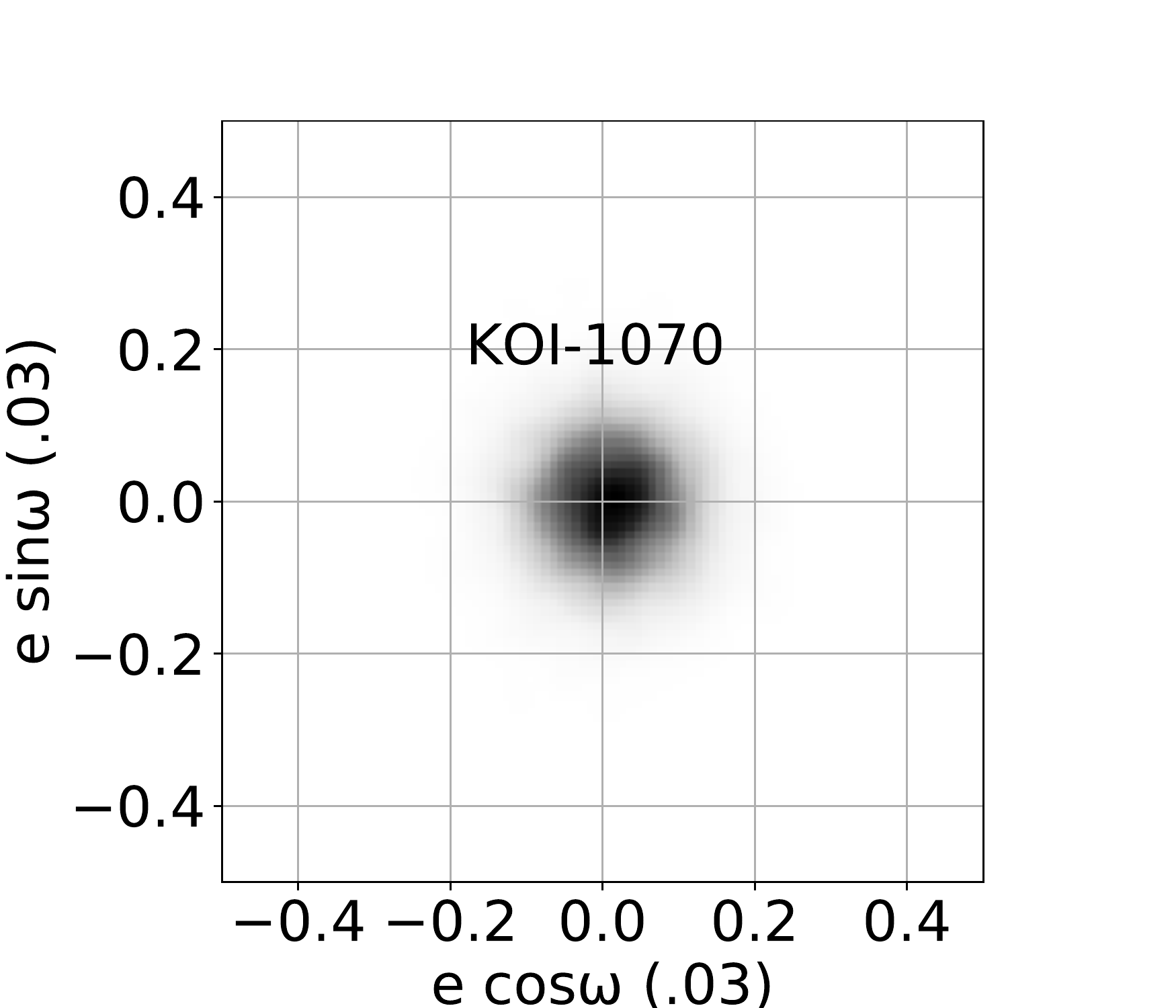}
\includegraphics [height = 1.1 in]{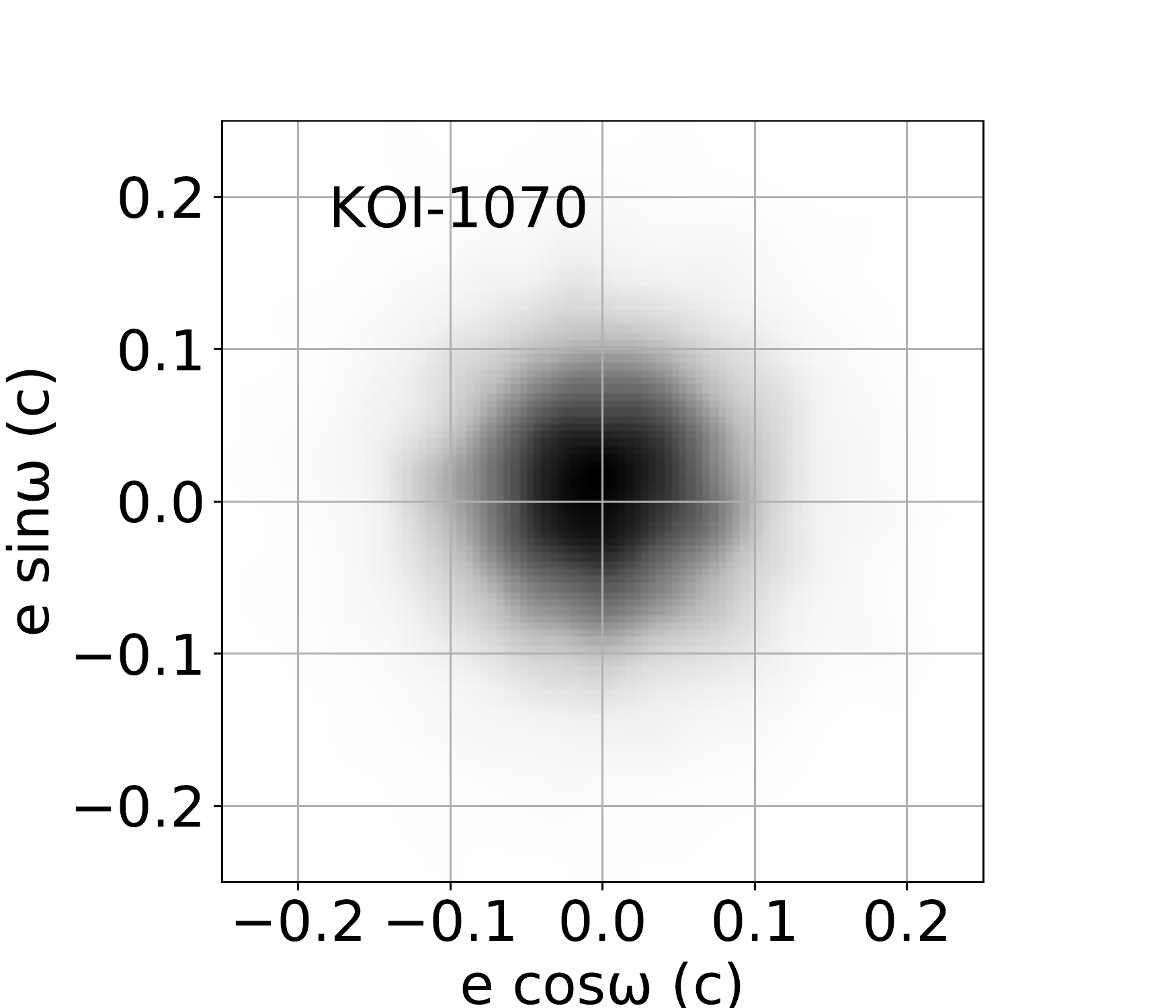} \\
\includegraphics [height = 1.1 in]{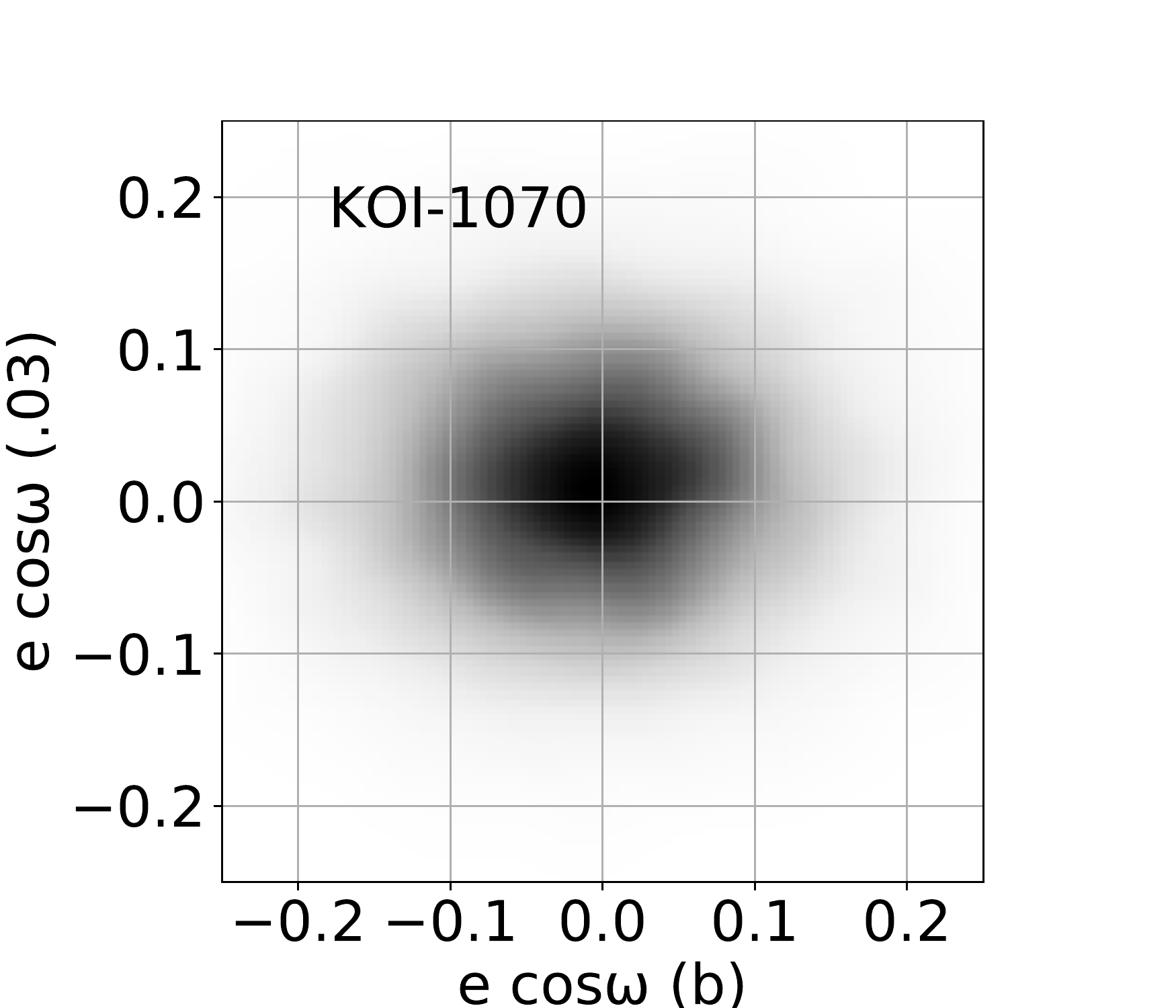} 
\includegraphics [height = 1.1 in]{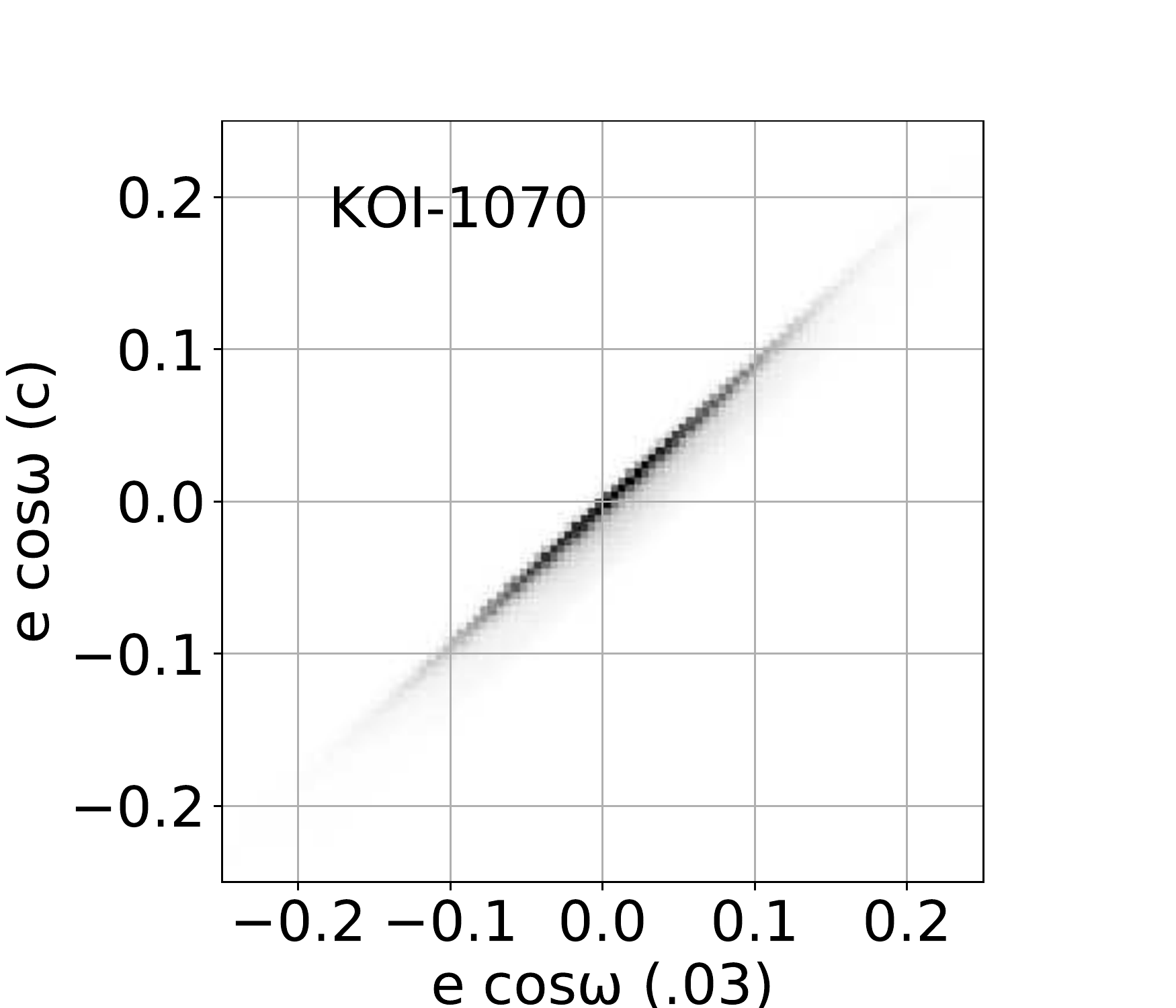}
\includegraphics [height = 1.1 in]{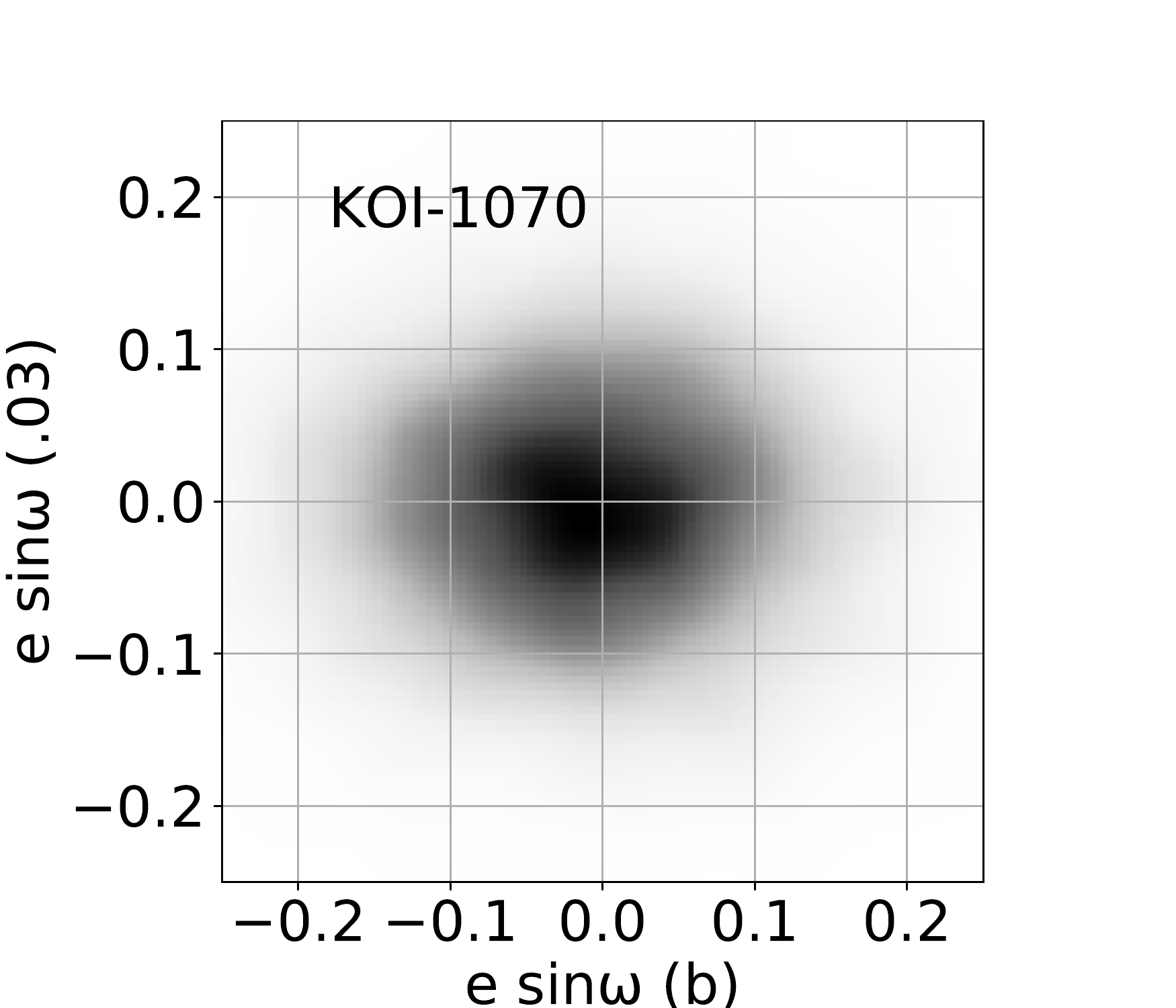} 
\includegraphics [height = 1.1 in]{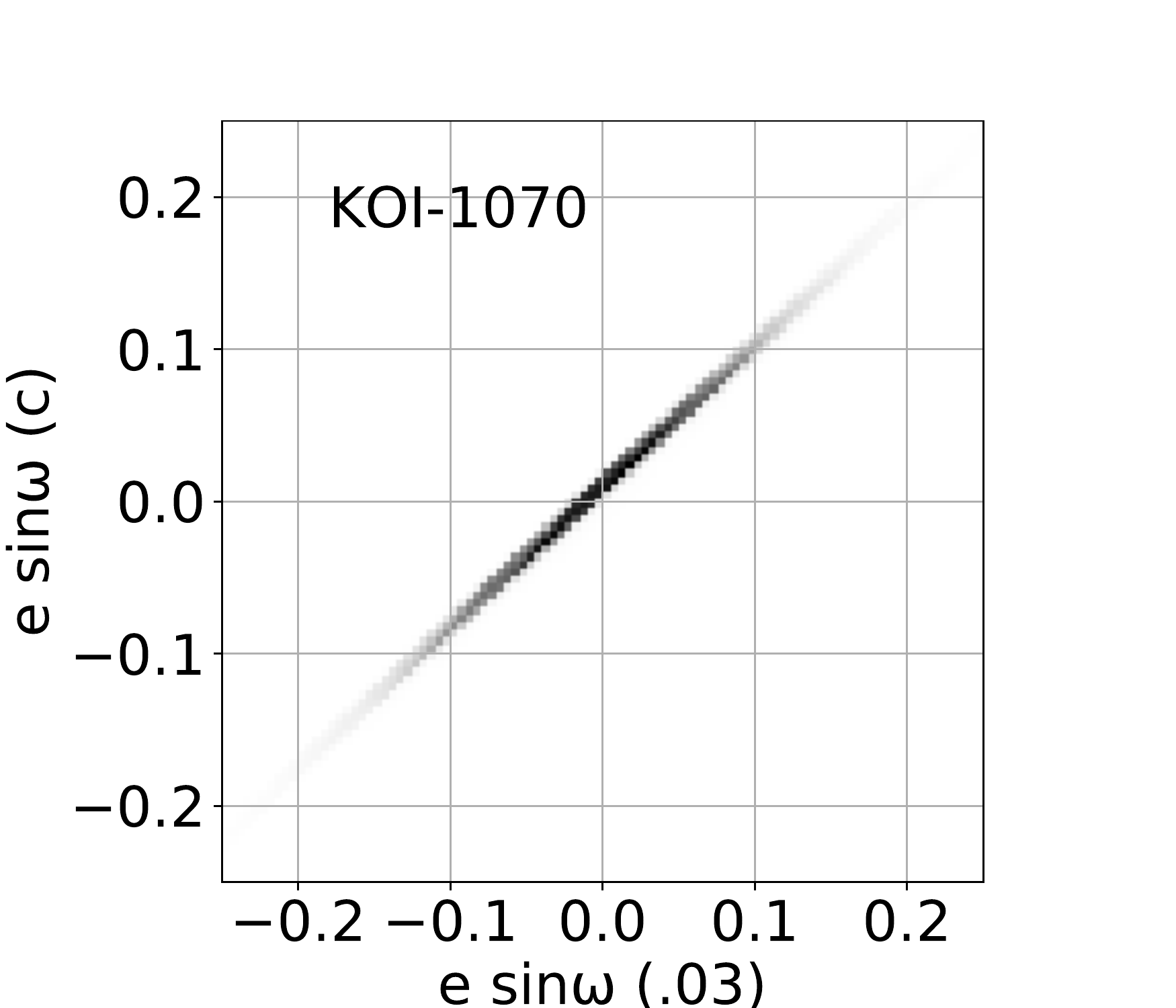}
\caption{Two-dimensional kernel density estimators on joint posteriors of eccentricity vector components: three-planet systems (Part 4 of 7). 
\label{fig:ecc3d}}
\end{center}
\end{figure}

\begin{figure}
\begin{center}
\figurenum{27}
\includegraphics [height = 1.1 in]{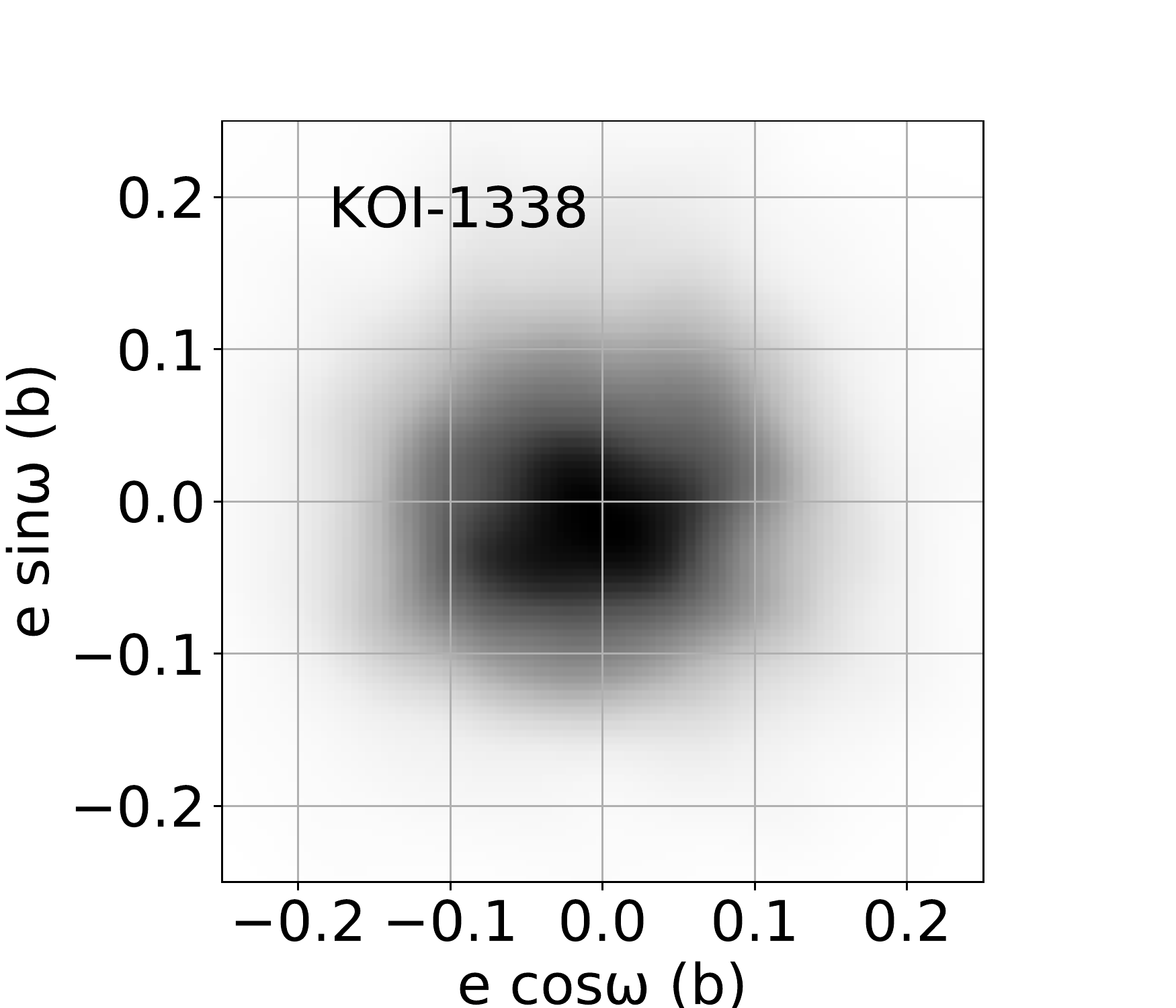}
\includegraphics [height = 1.1 in]{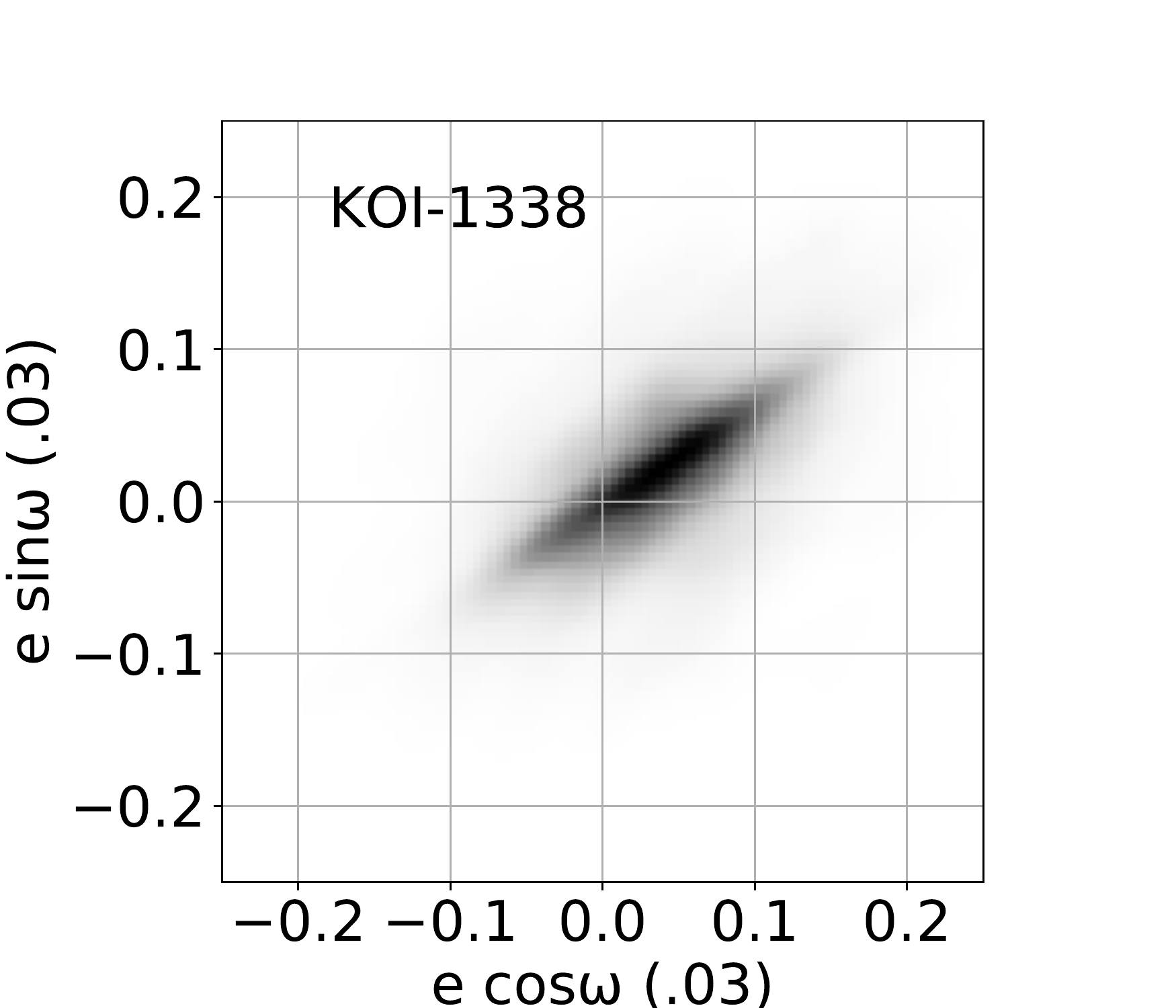}
\includegraphics [height = 1.1 in]{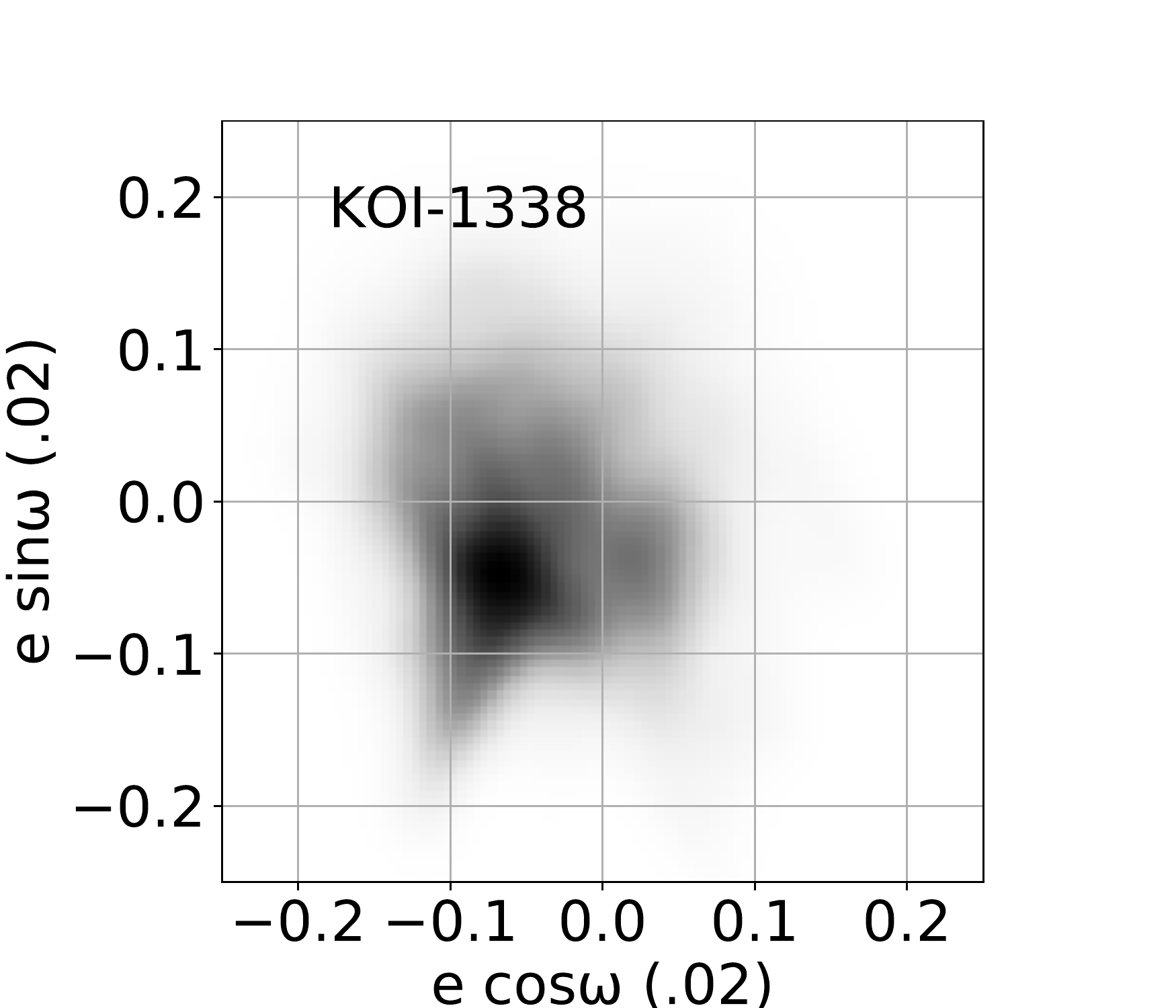}
\includegraphics [height = 1.1 in]{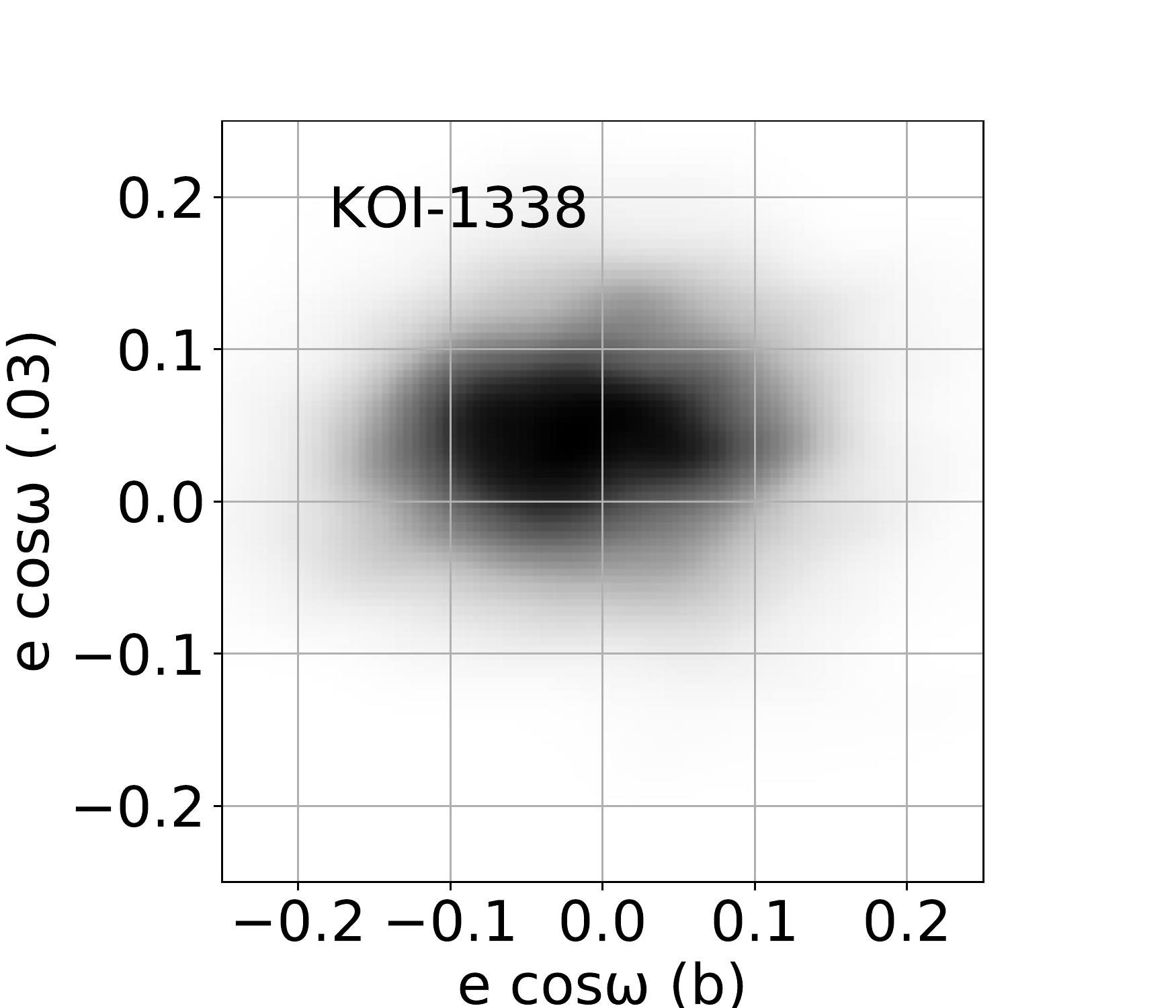} \\
\includegraphics [height = 1.1 in]{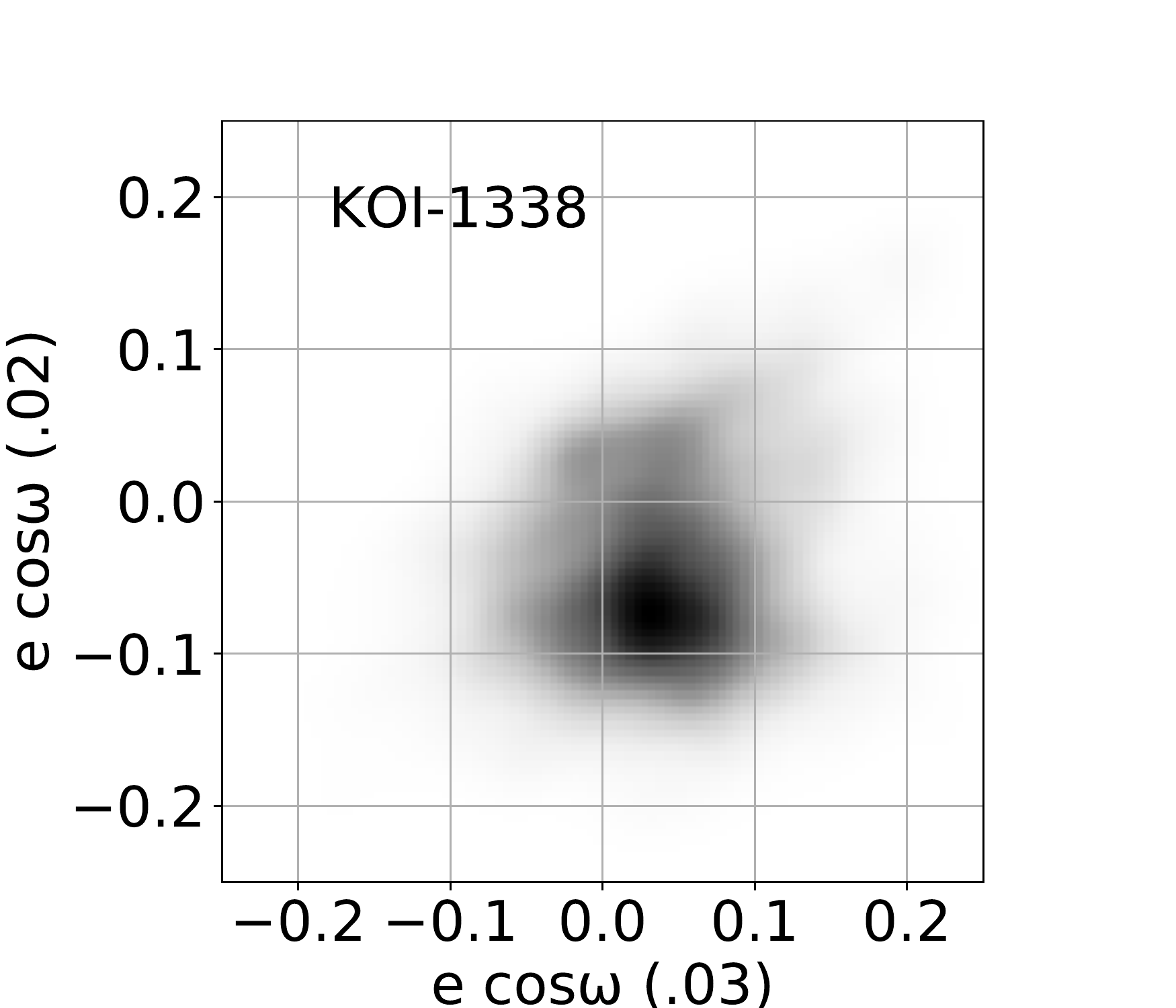}
\includegraphics [height = 1.1 in]{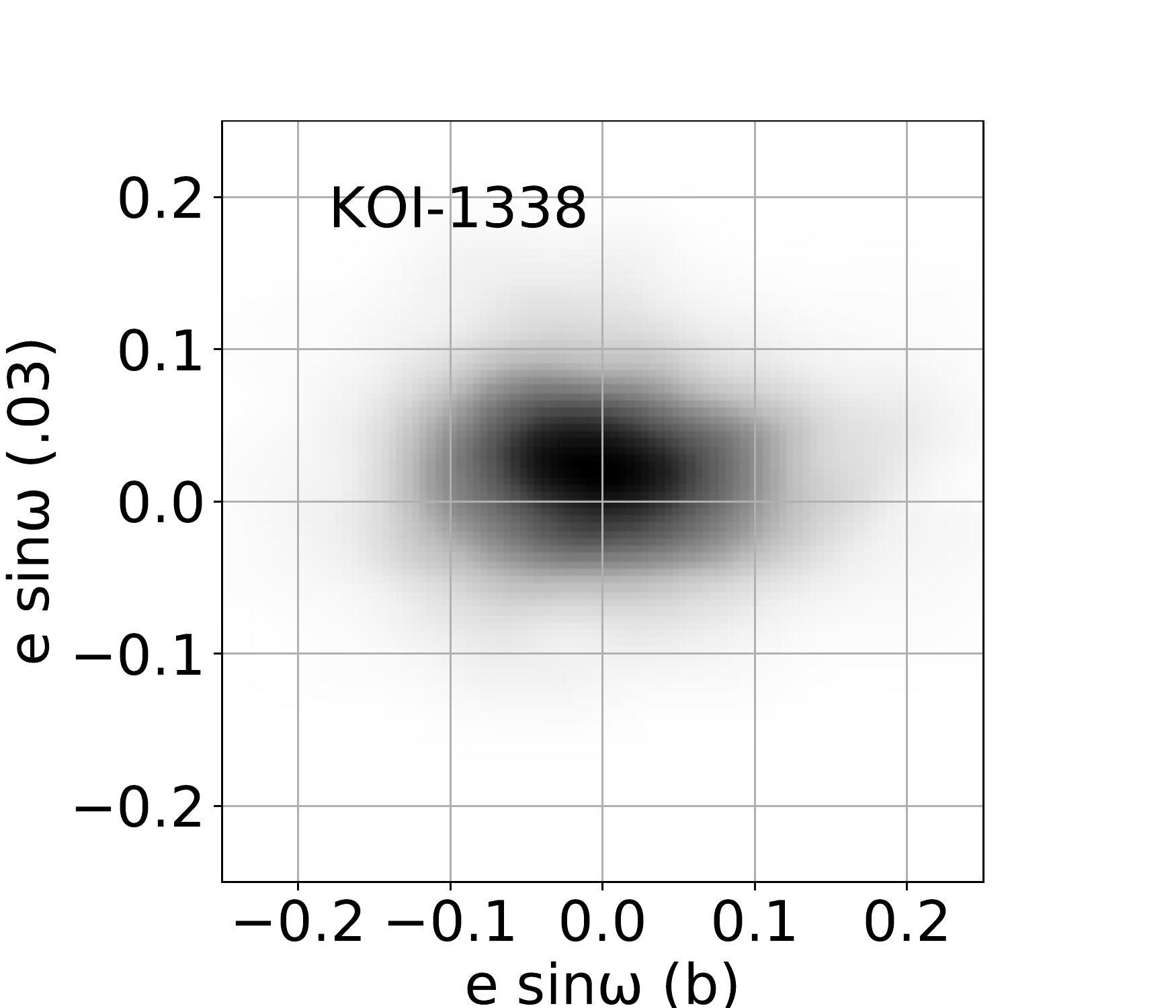}
\includegraphics [height = 1.1 in]{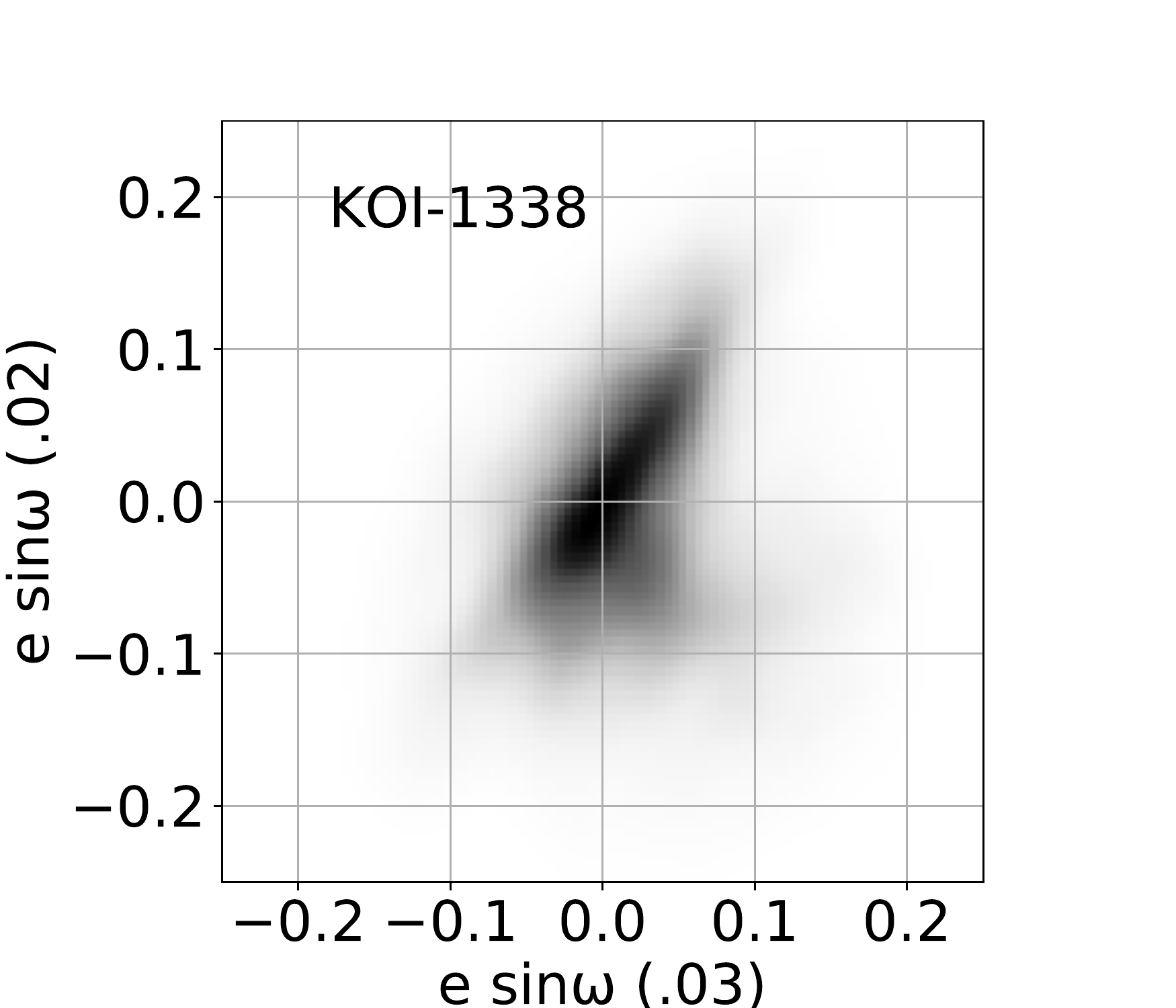}
\includegraphics [height = 1.1 in]{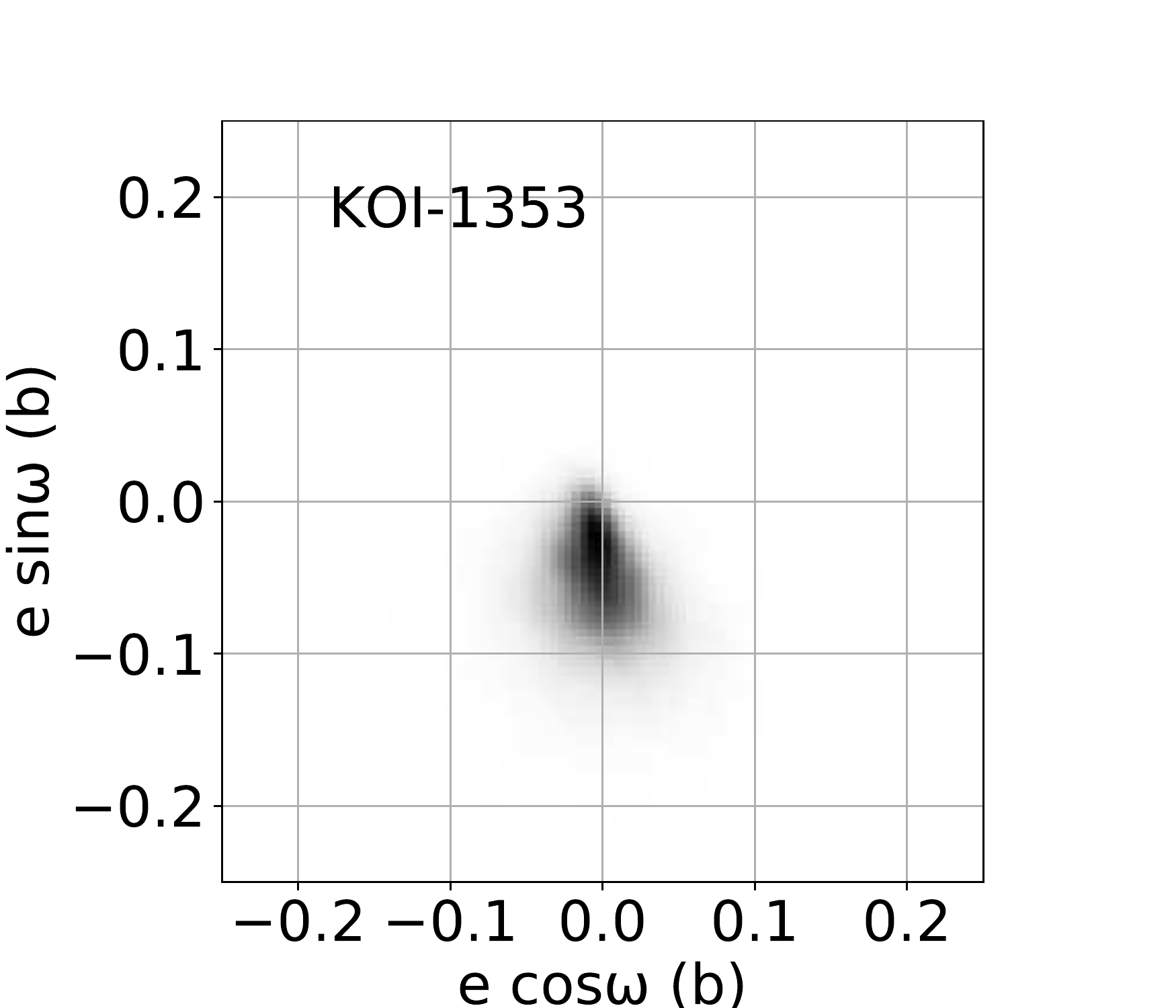} \\
\includegraphics [height = 1.1 in]{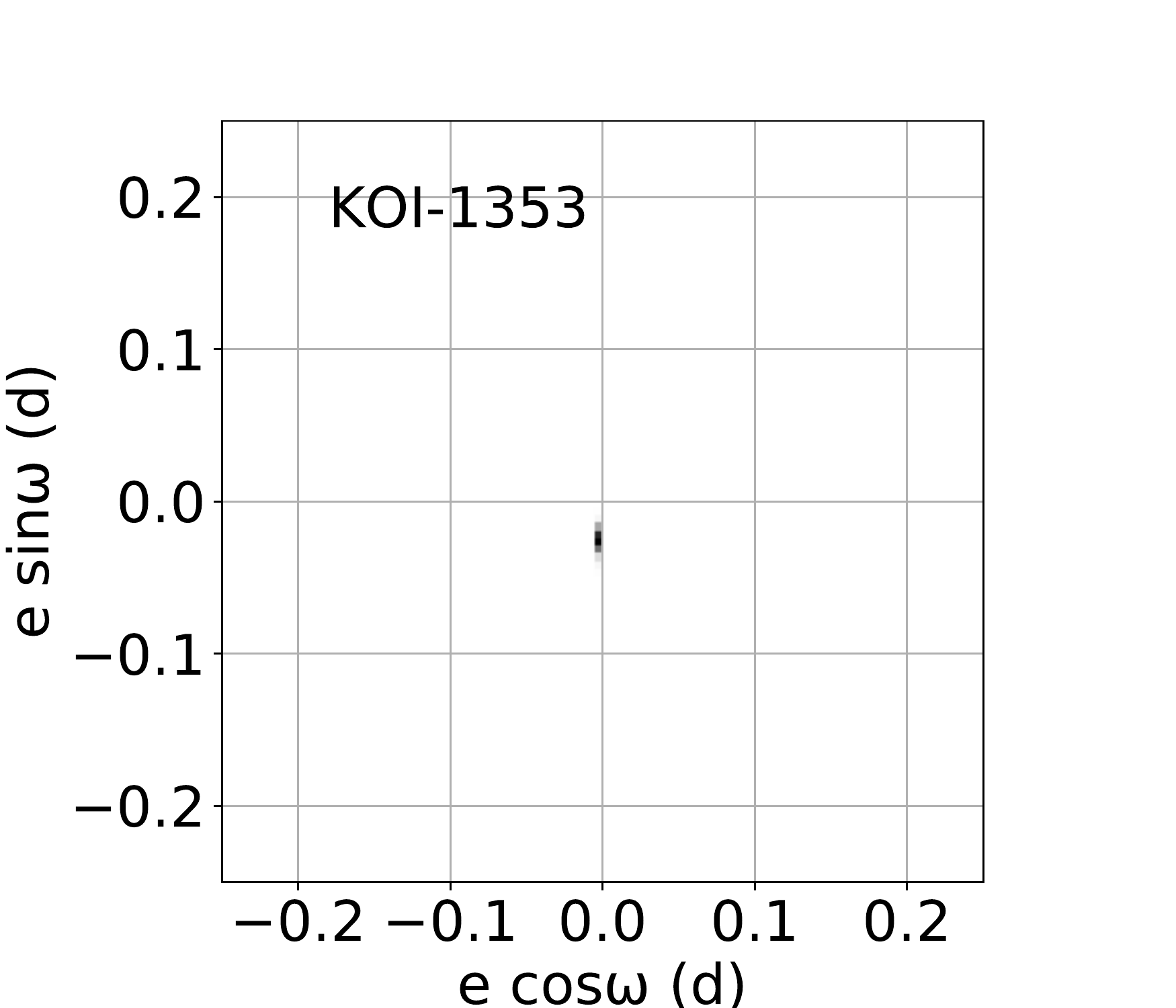}
\includegraphics [height = 1.1 in]{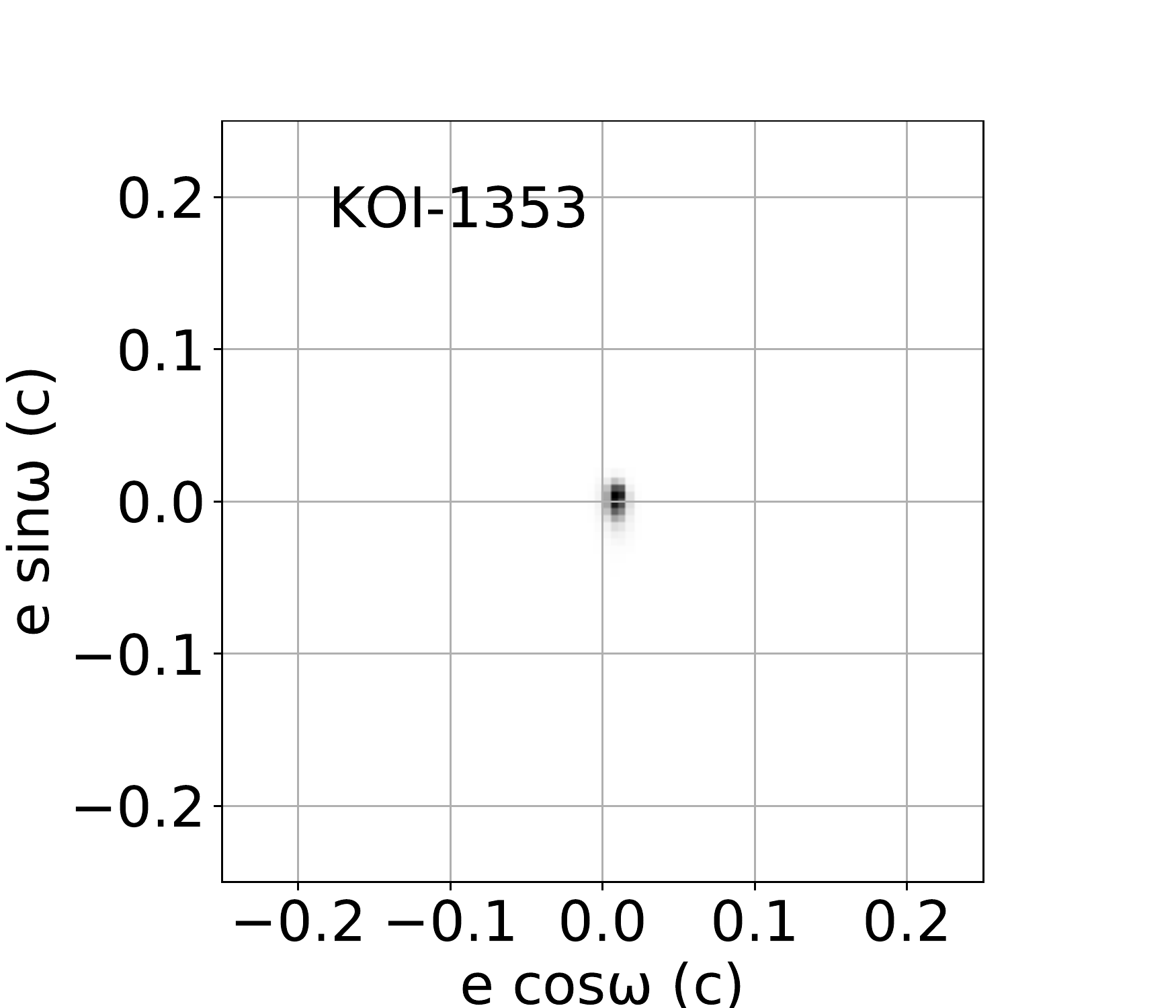}
\includegraphics [height = 1.1 in]{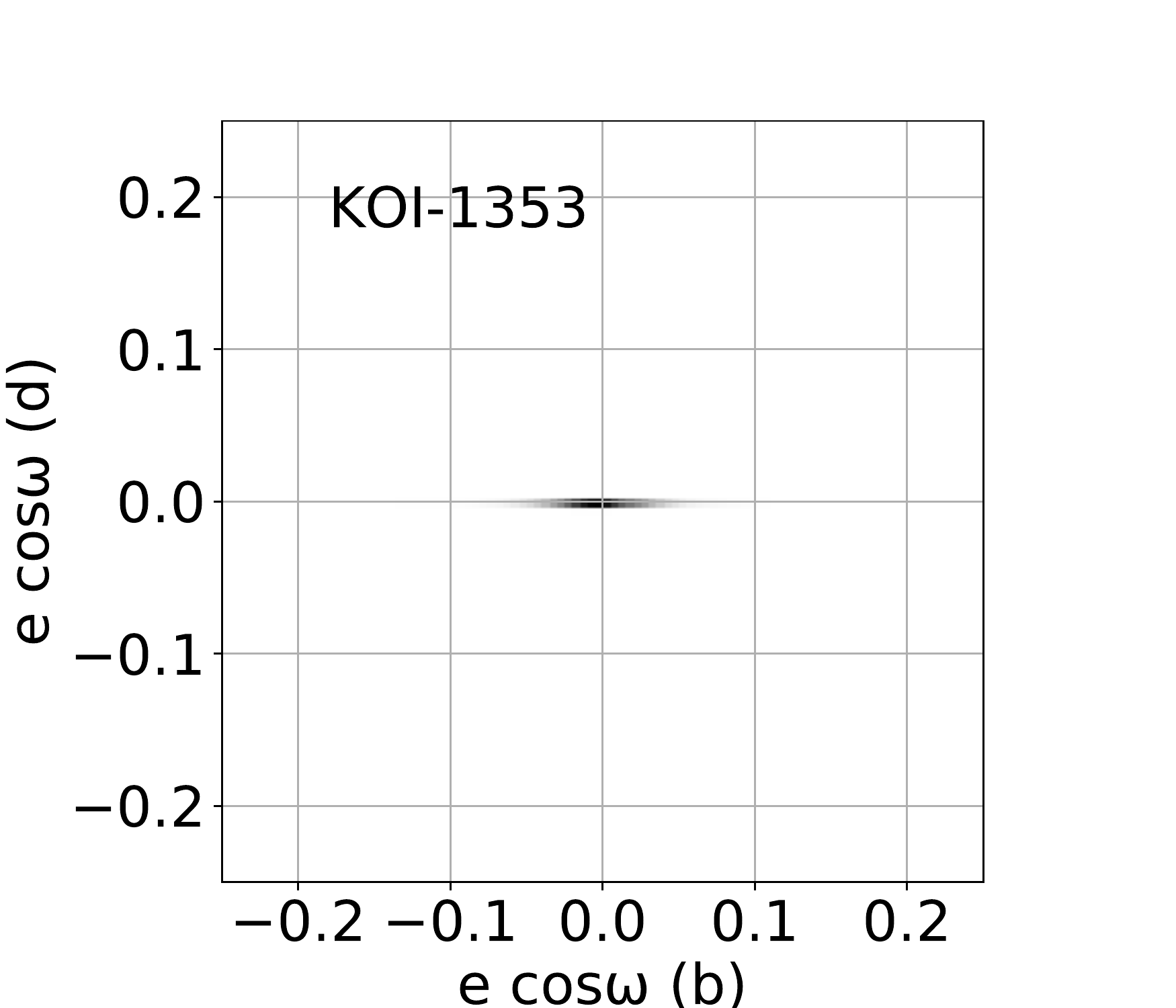}
\includegraphics [height = 1.1 in]{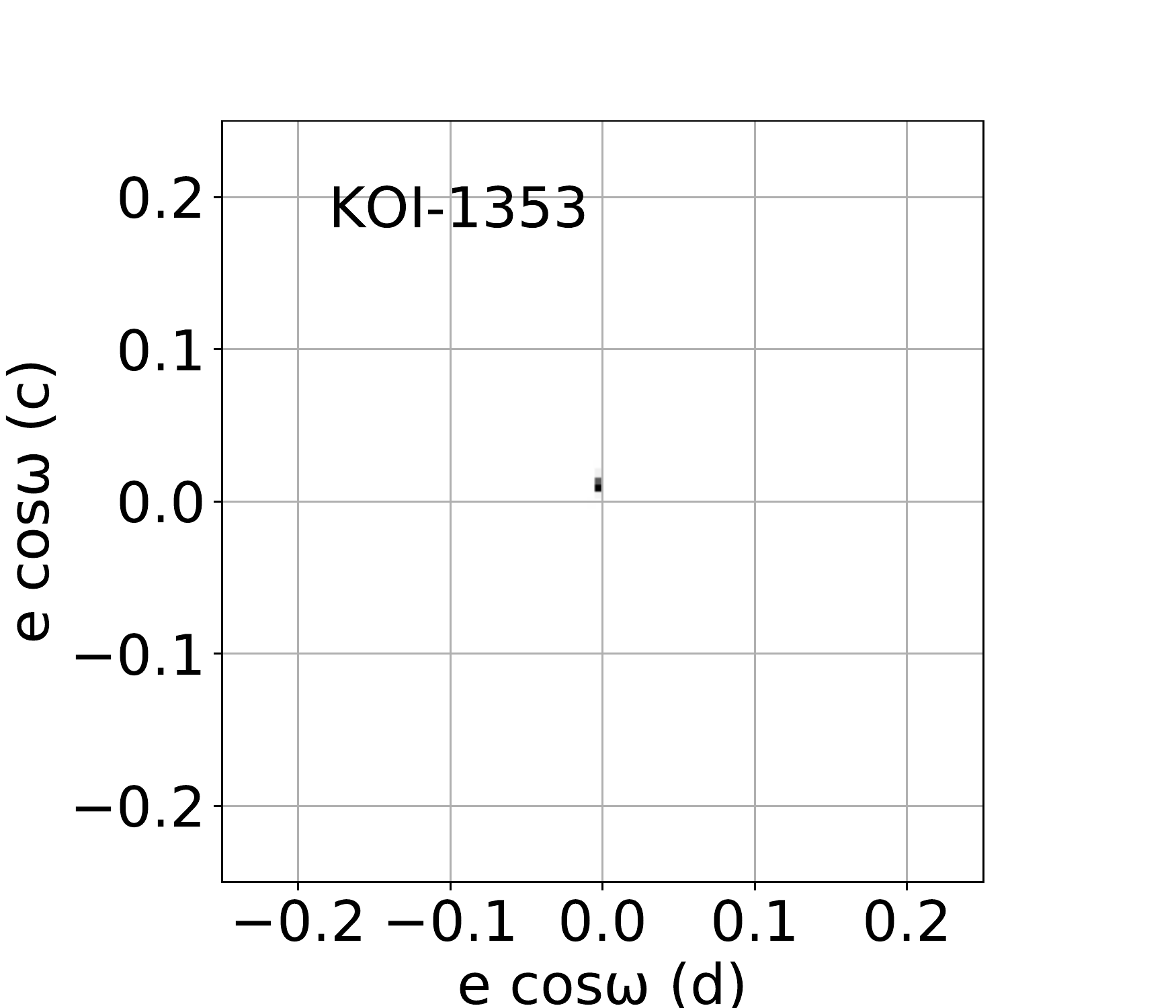} \\
\includegraphics [height = 1.1 in]{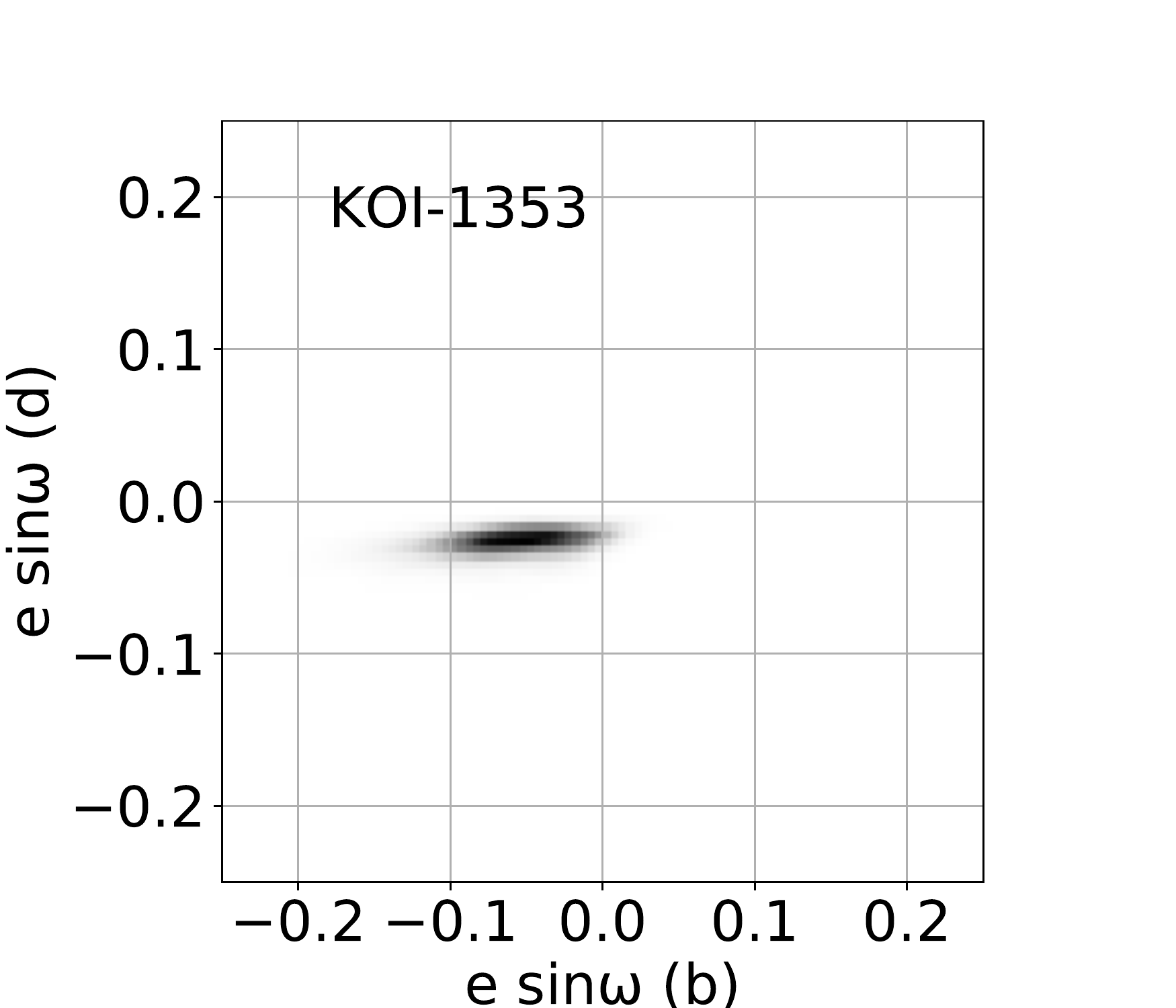}
\includegraphics [height = 1.1 in]{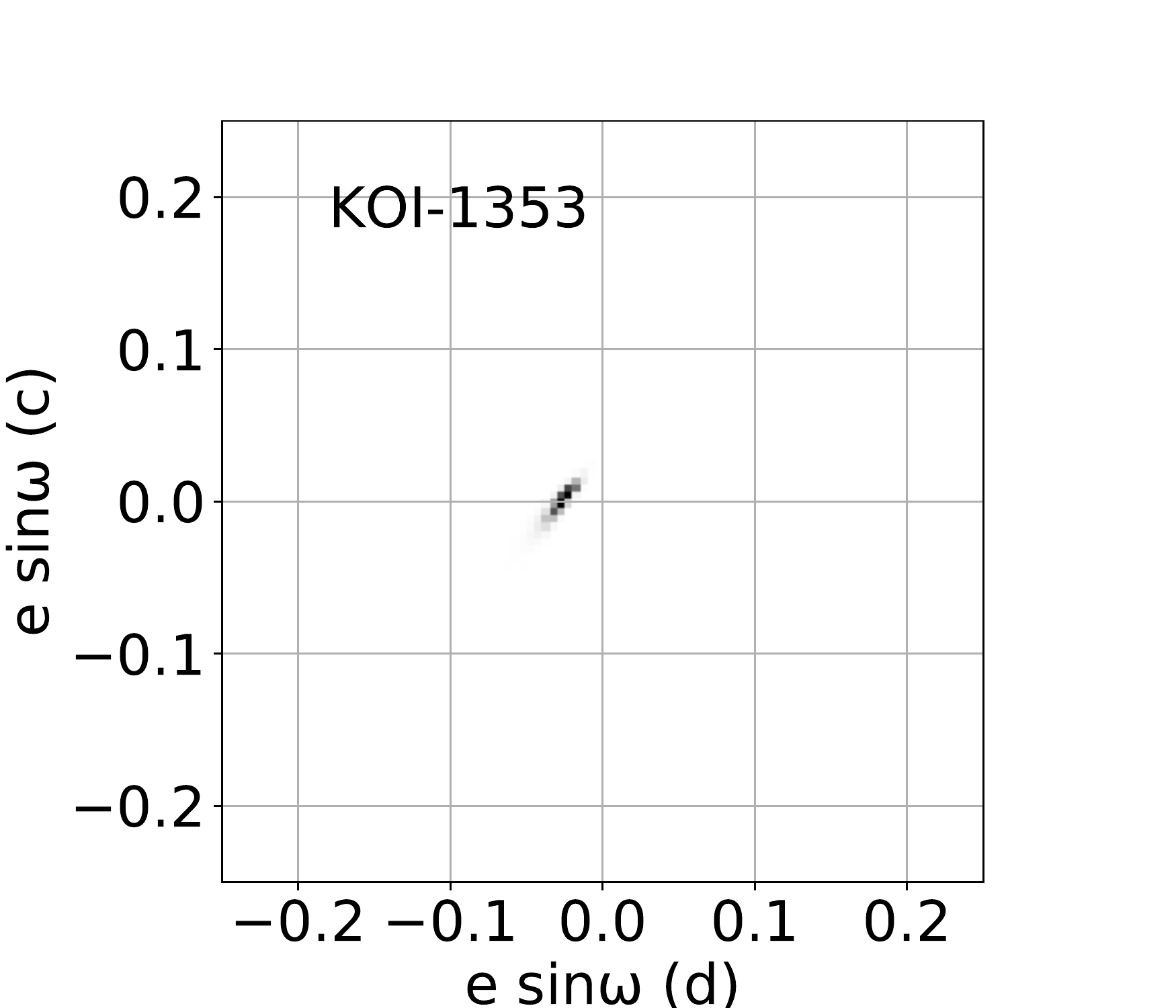}
\includegraphics [height = 1.1 in]{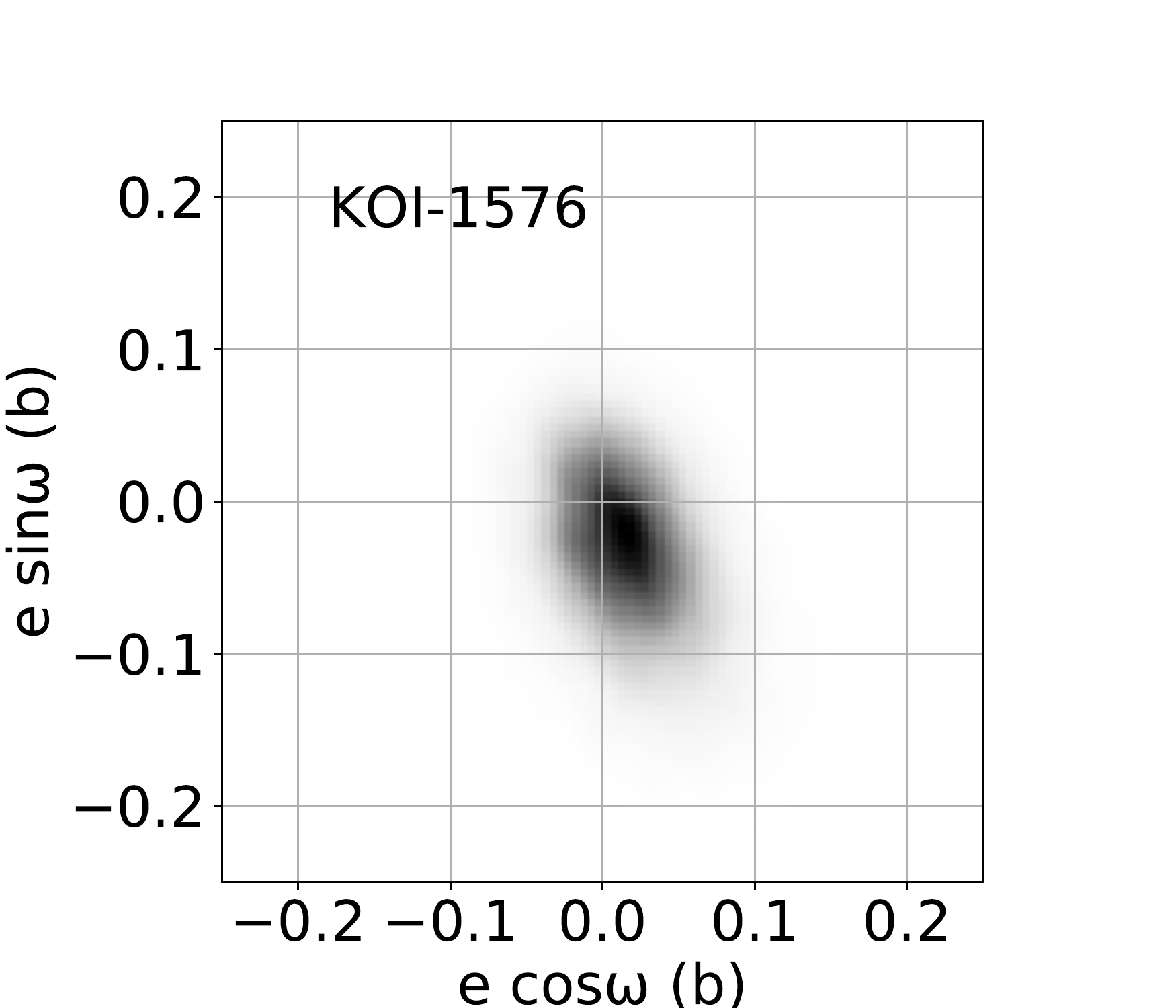}
\includegraphics [height = 1.1 in]{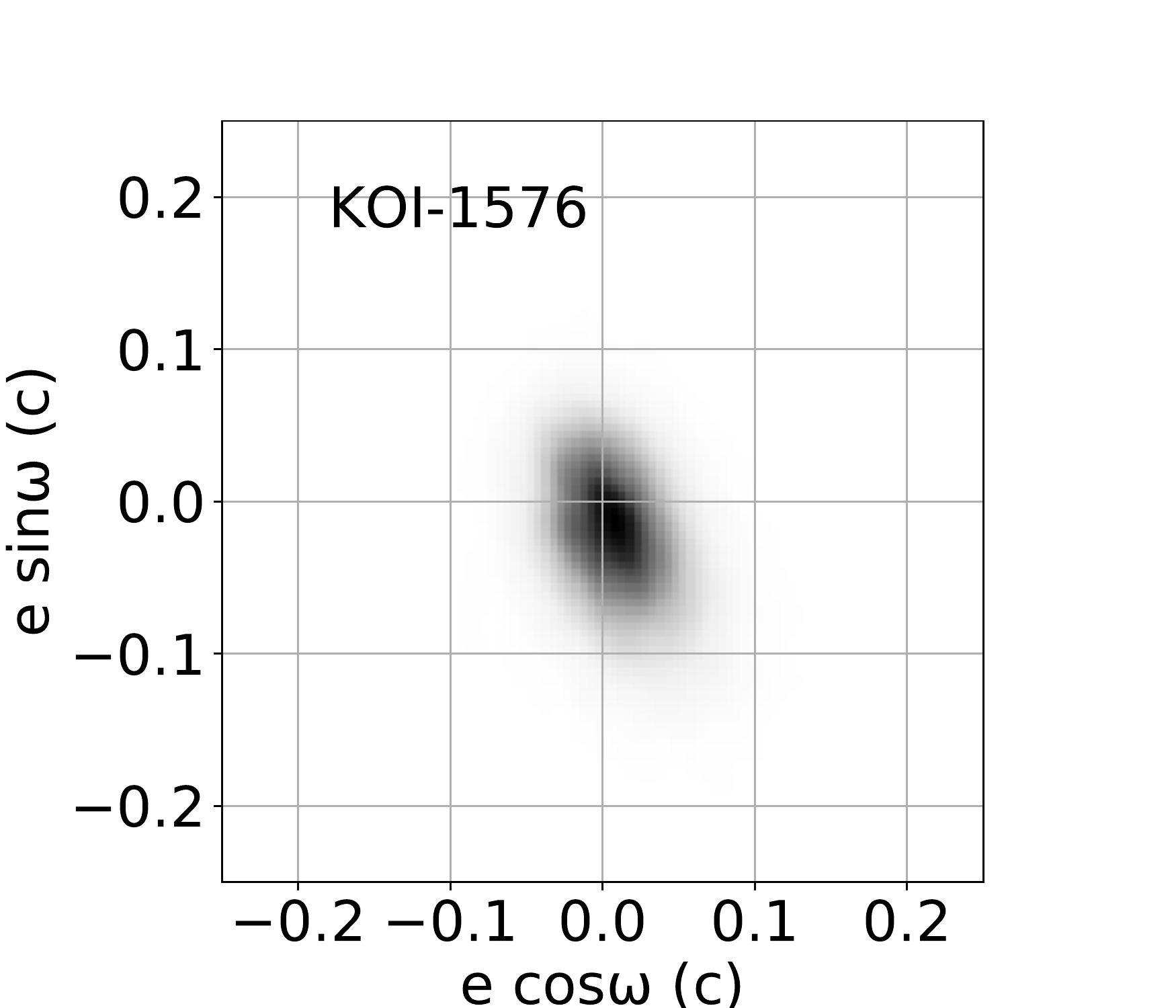} \\
\includegraphics [height = 1.1 in]{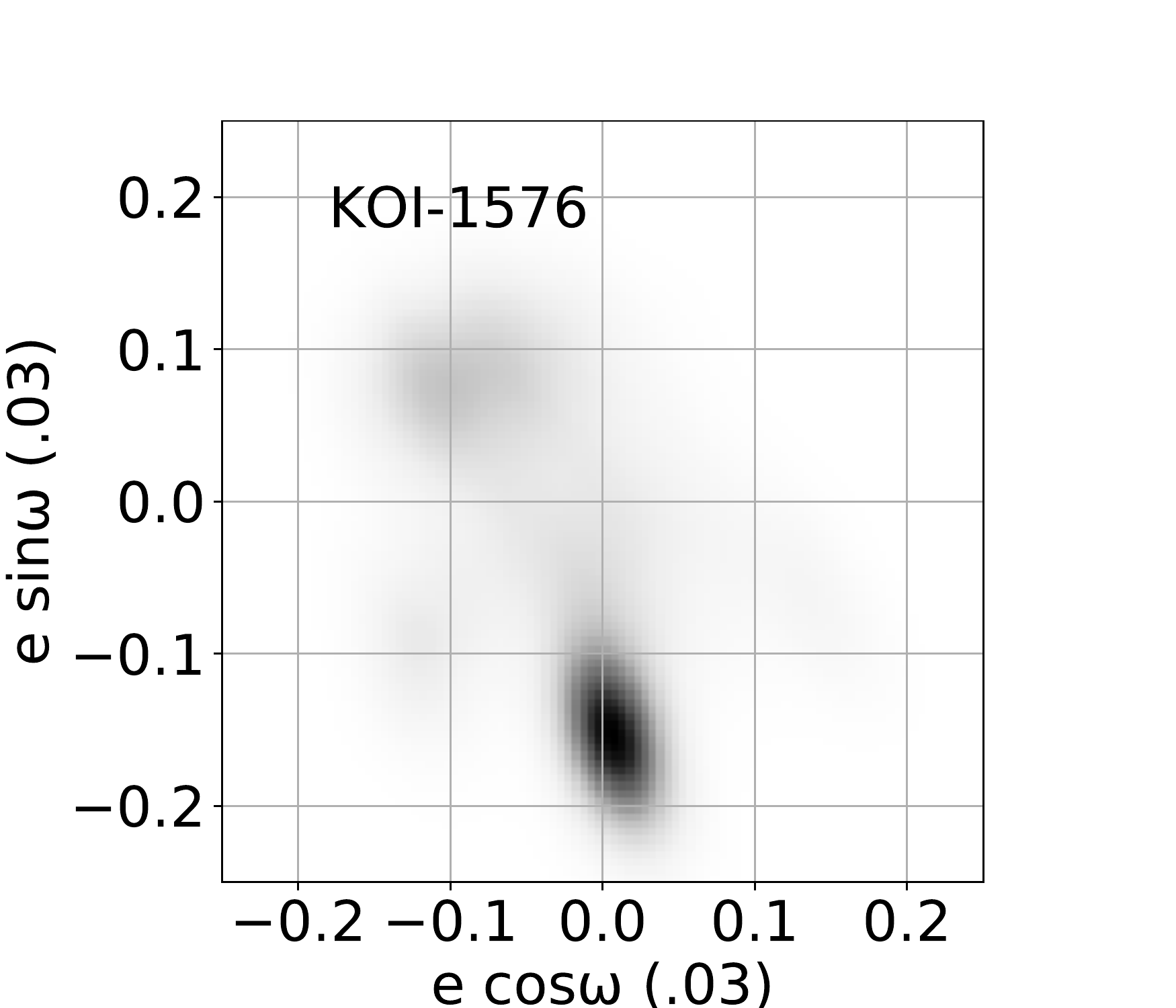}
\includegraphics [height = 1.1 in]{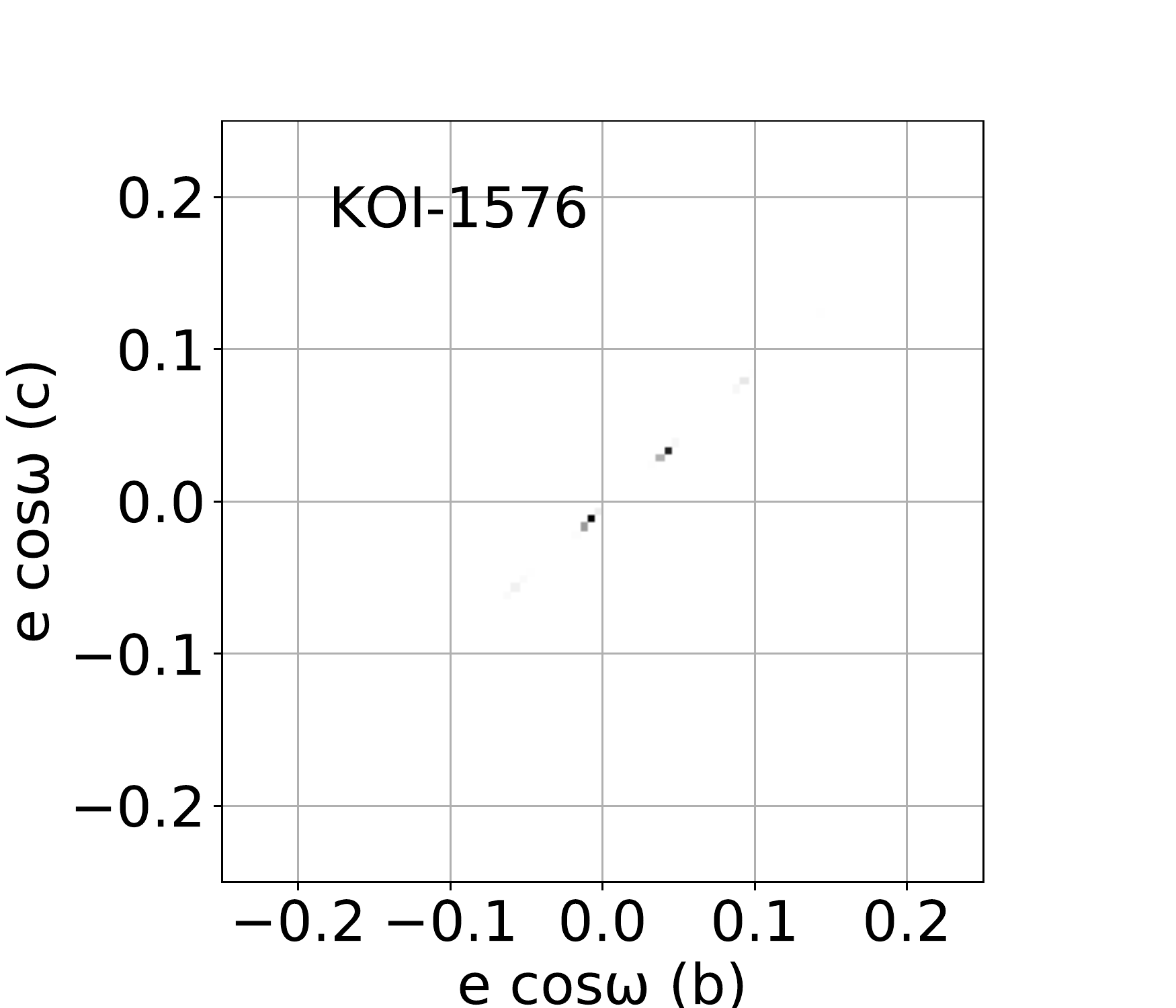} 
\includegraphics [height = 1.1 in]{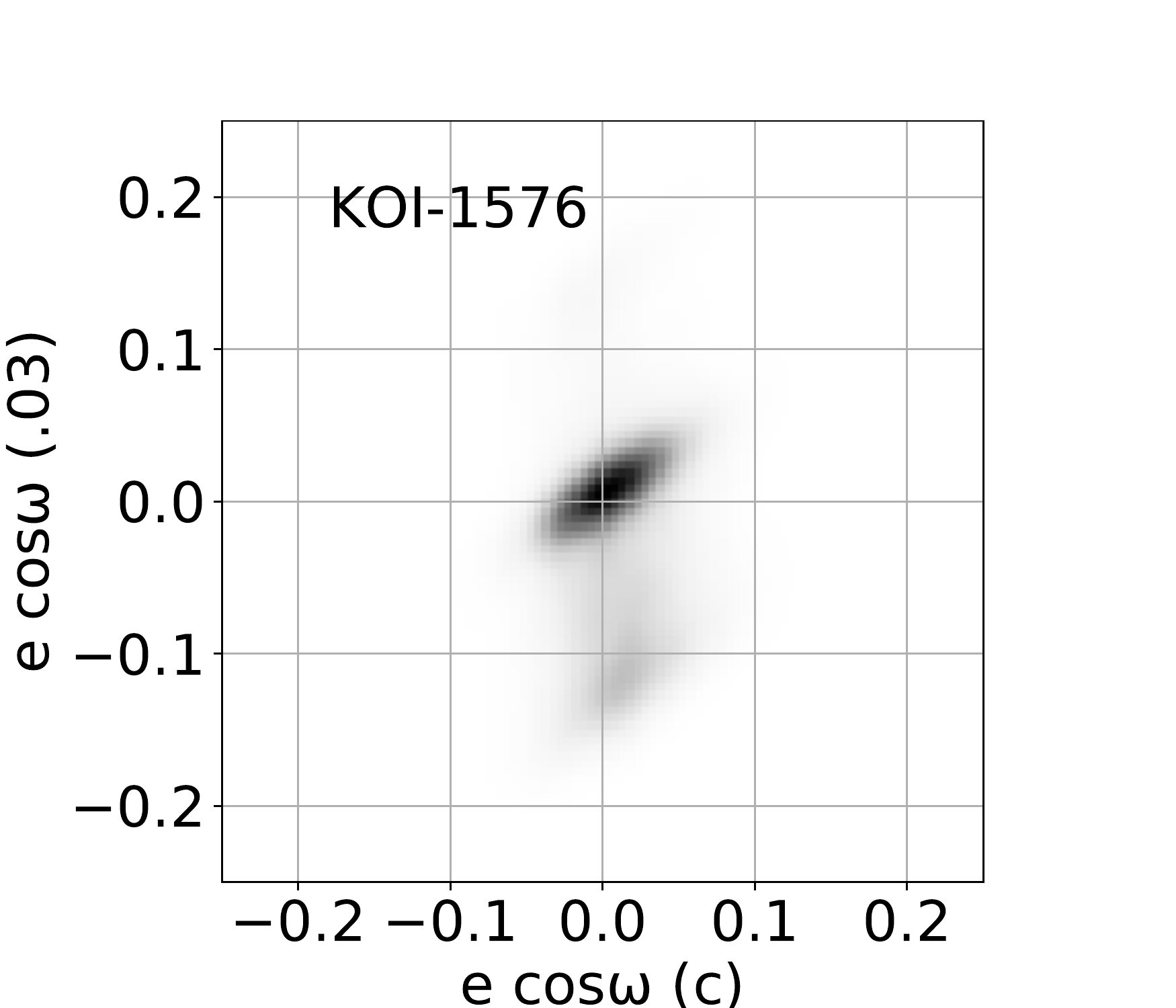}
\includegraphics [height = 1.1 in]{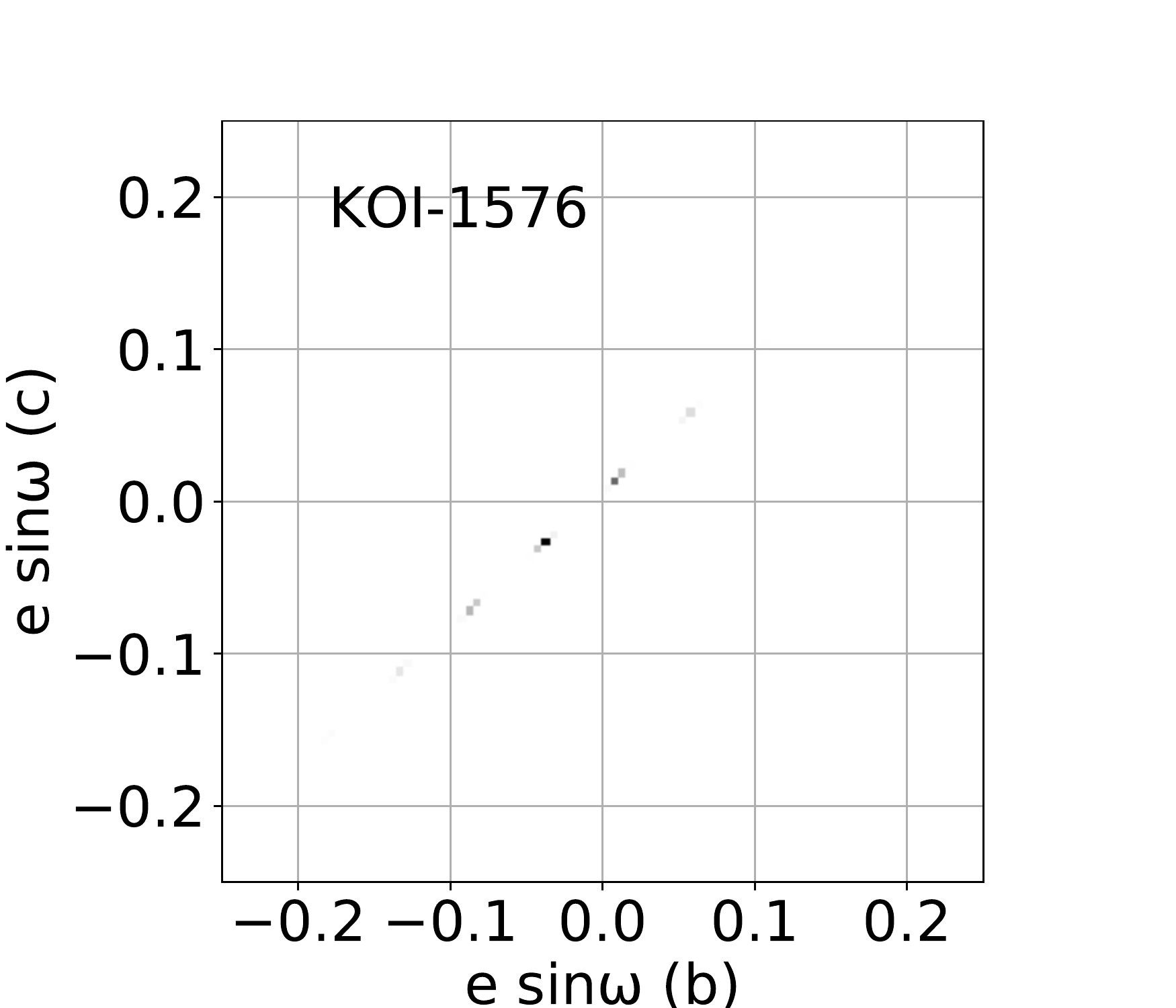} \\
\includegraphics [height = 1.1 in]{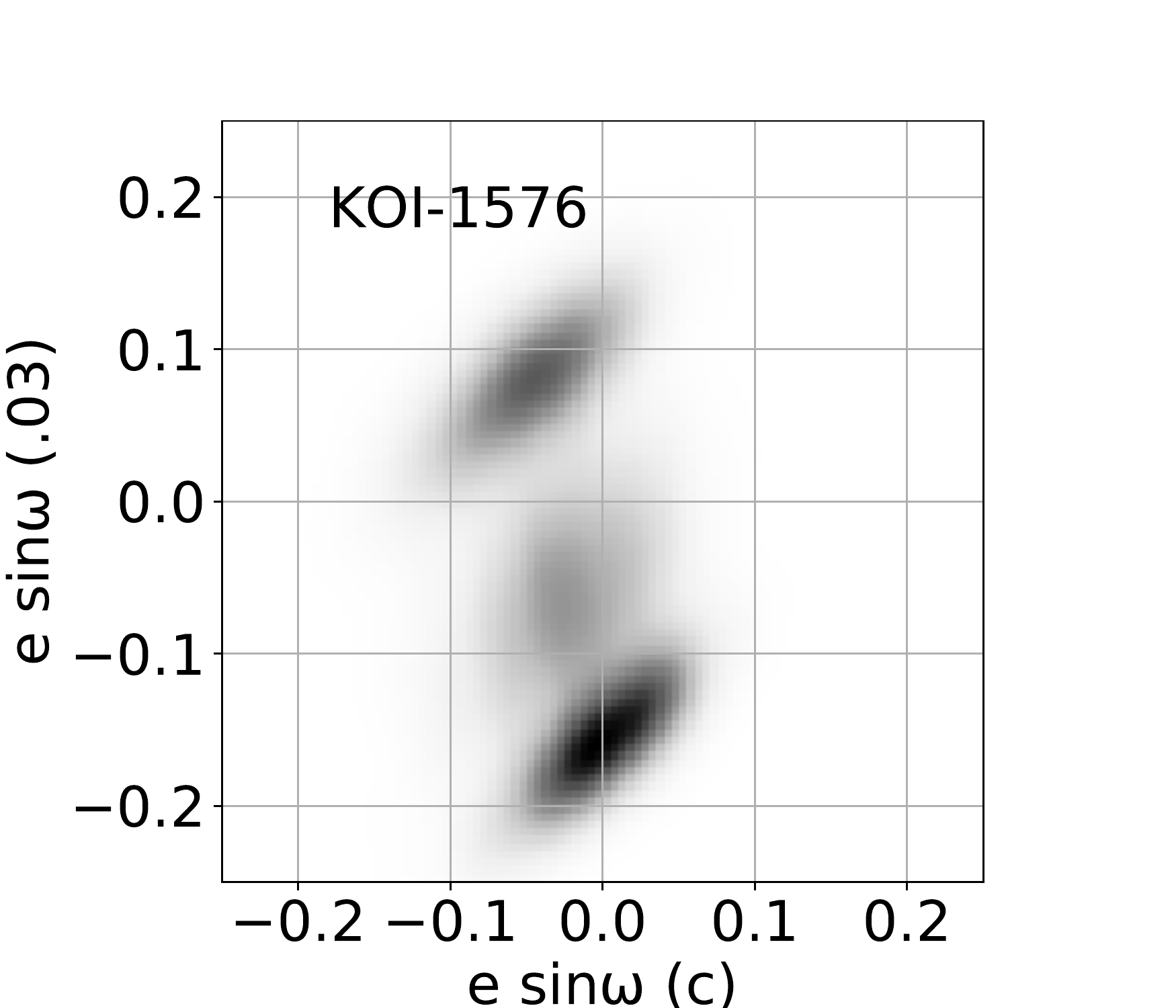}
\includegraphics [height = 1.1 in]{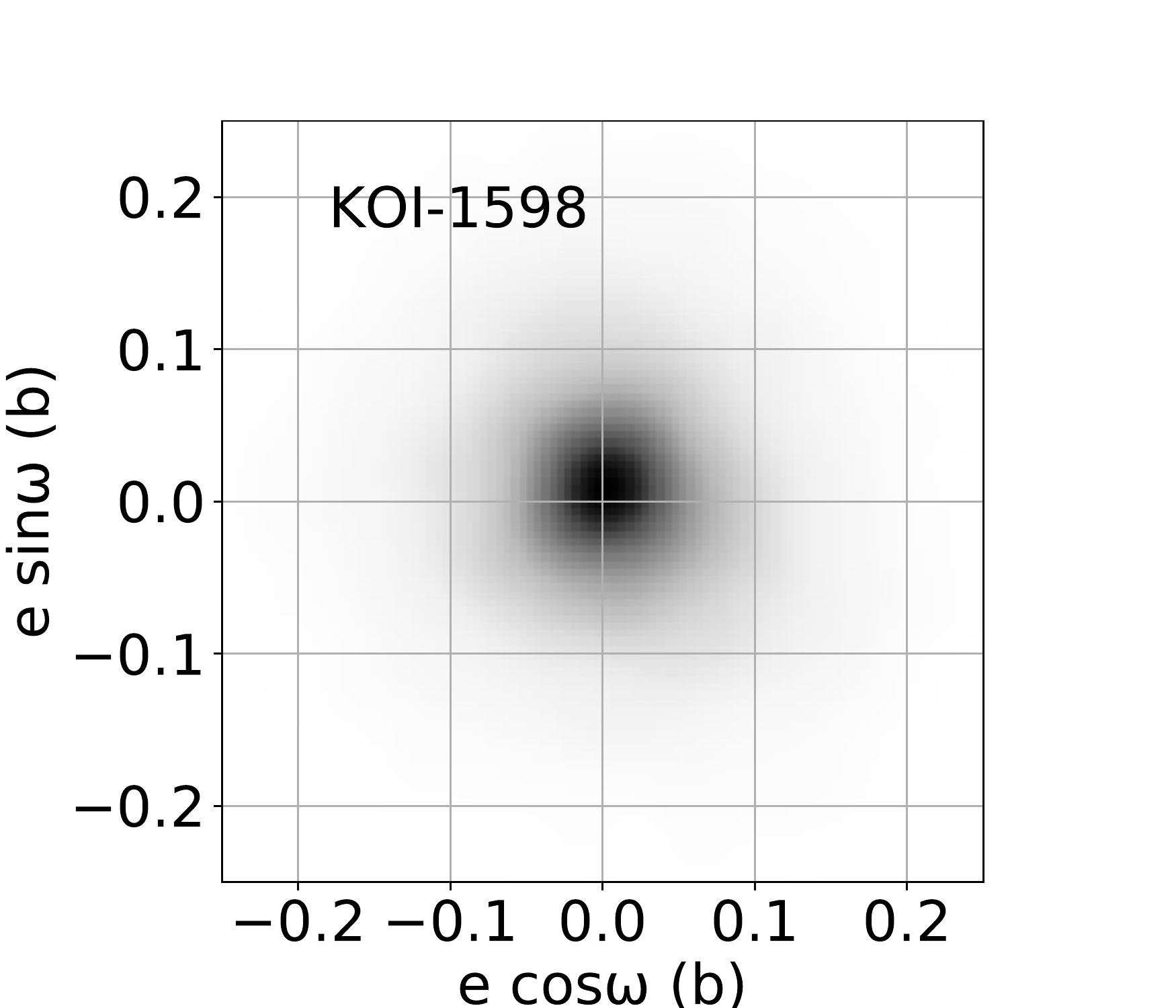} 
\includegraphics [height = 1.1 in]{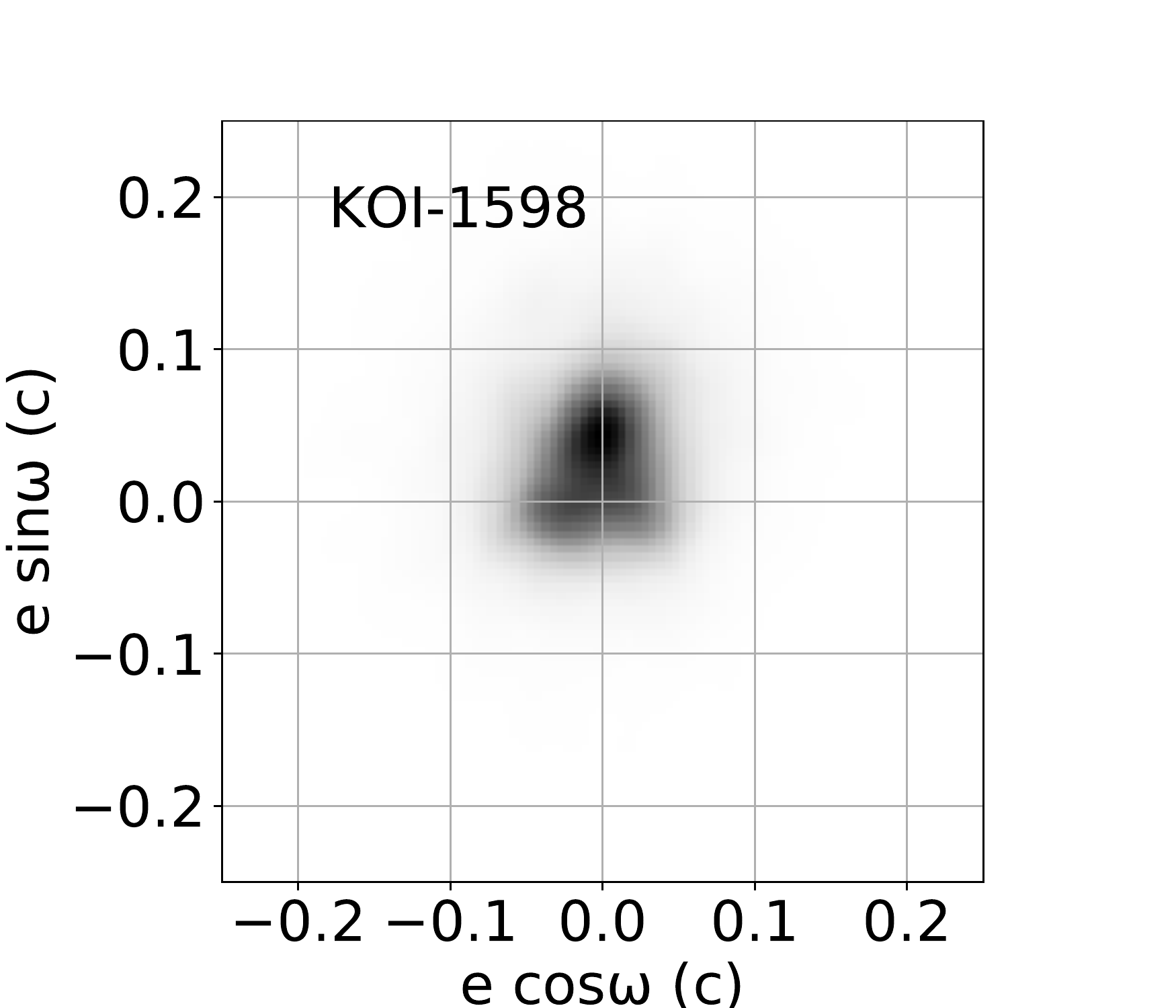}
\includegraphics [height = 1.1 in]{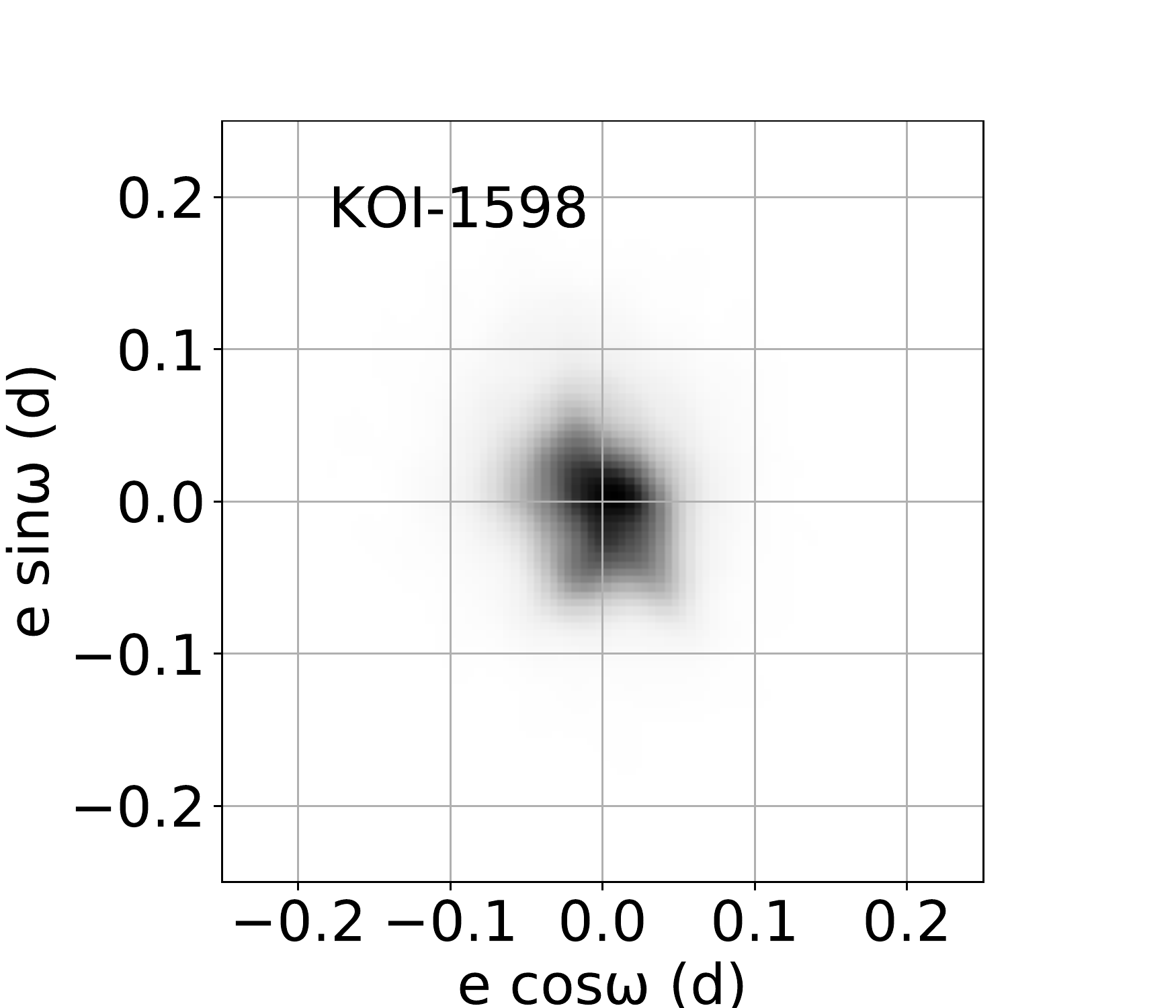} \\
\includegraphics [height = 1.1 in]{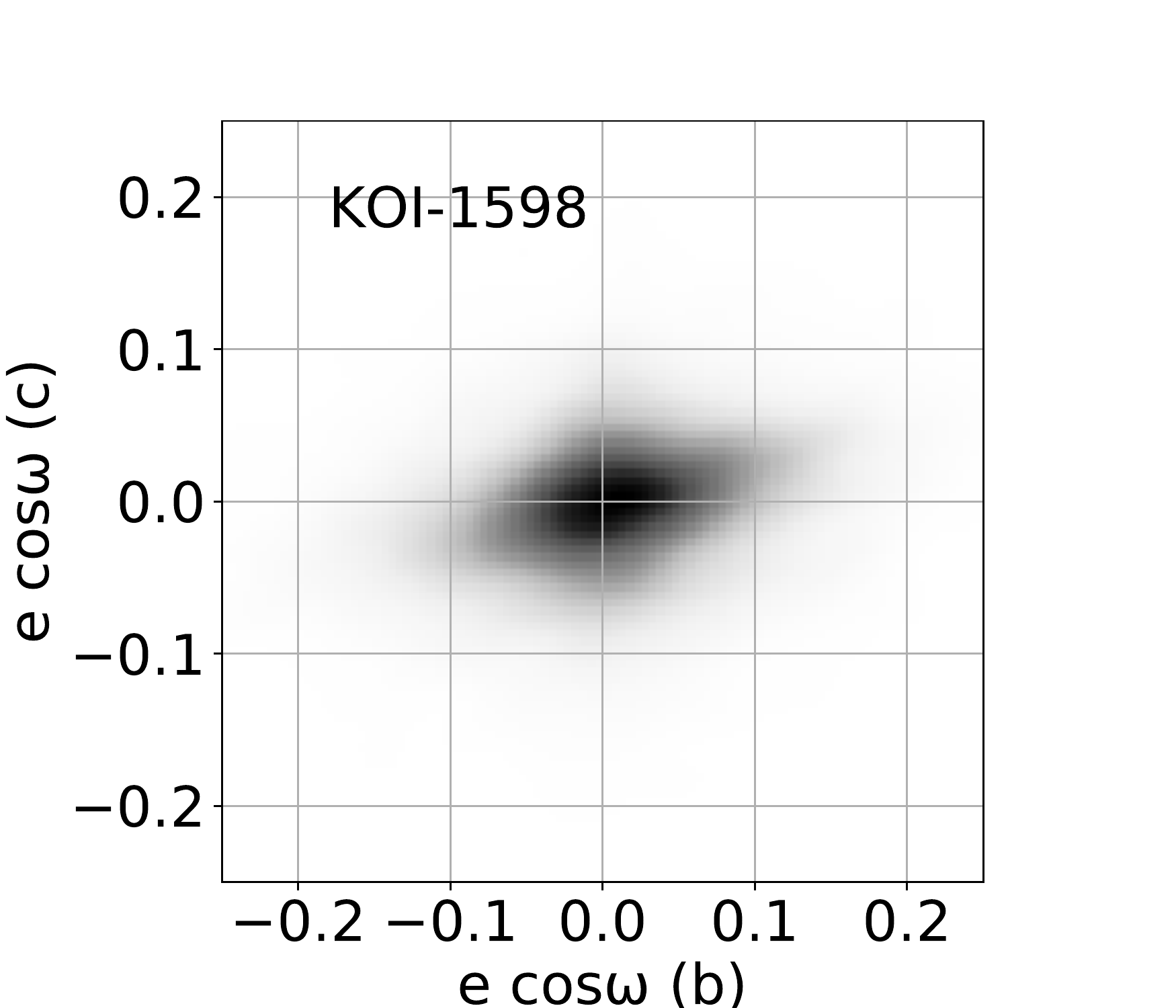} 
\includegraphics [height = 1.1 in]{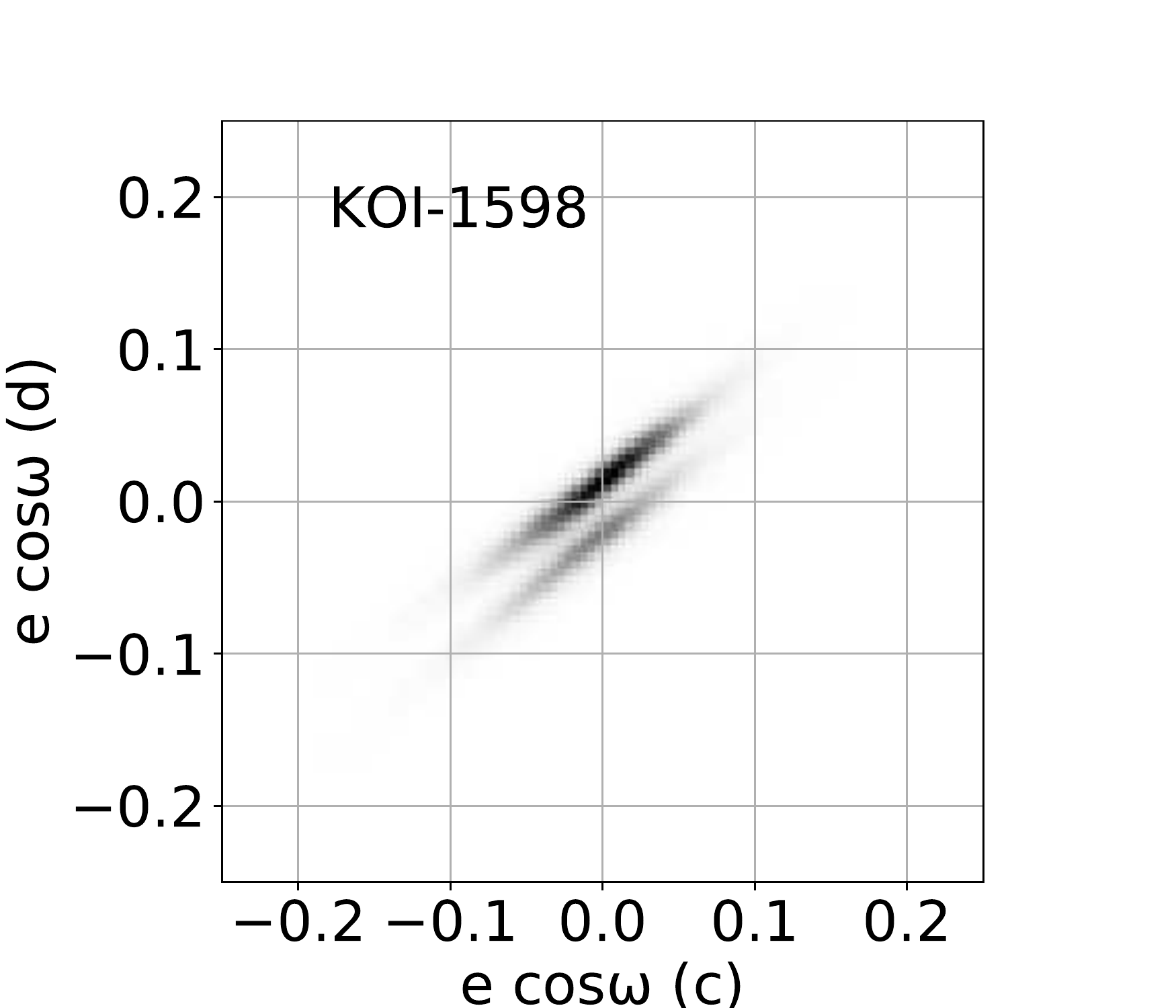}
\includegraphics [height = 1.1 in]{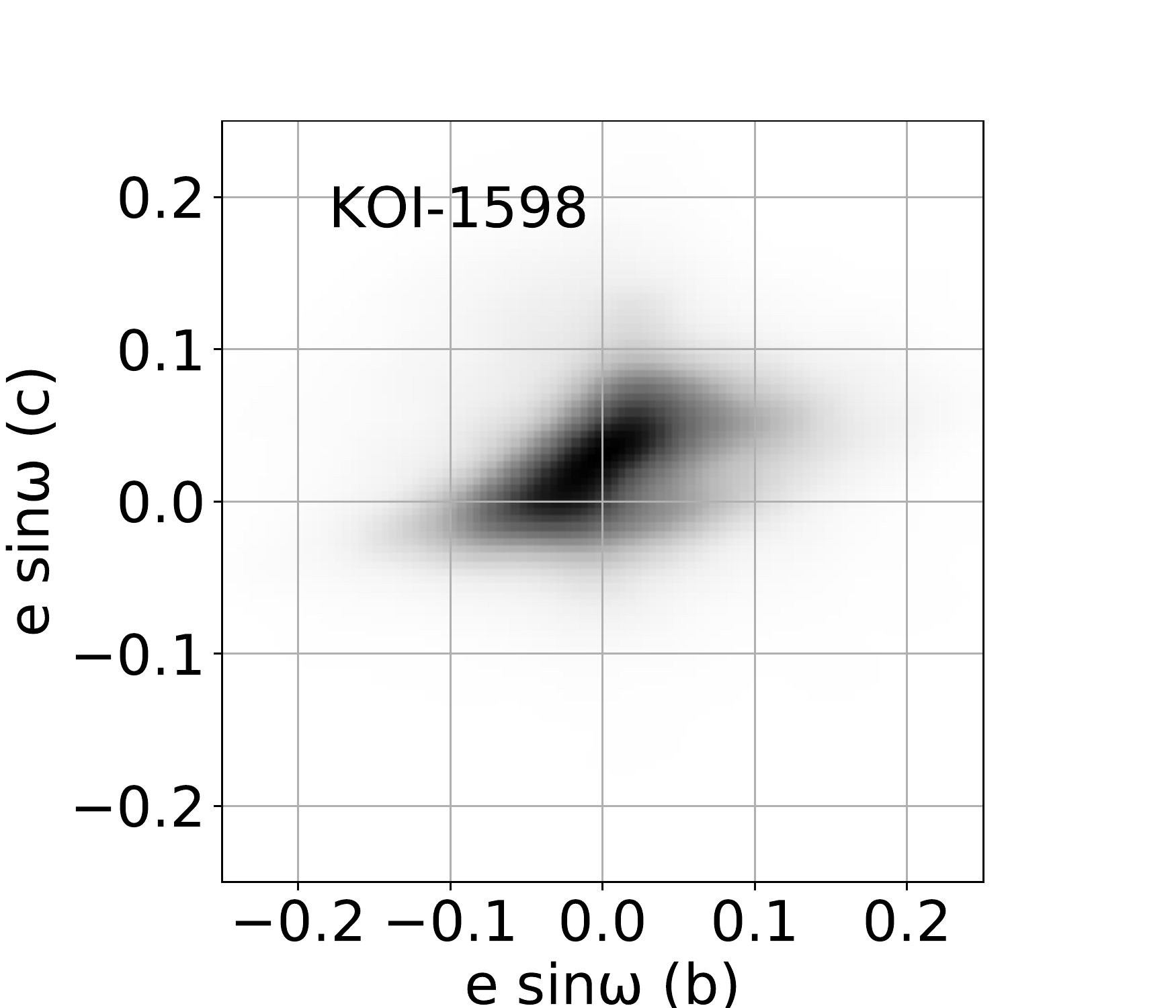} 
\includegraphics [height = 1.1 in]{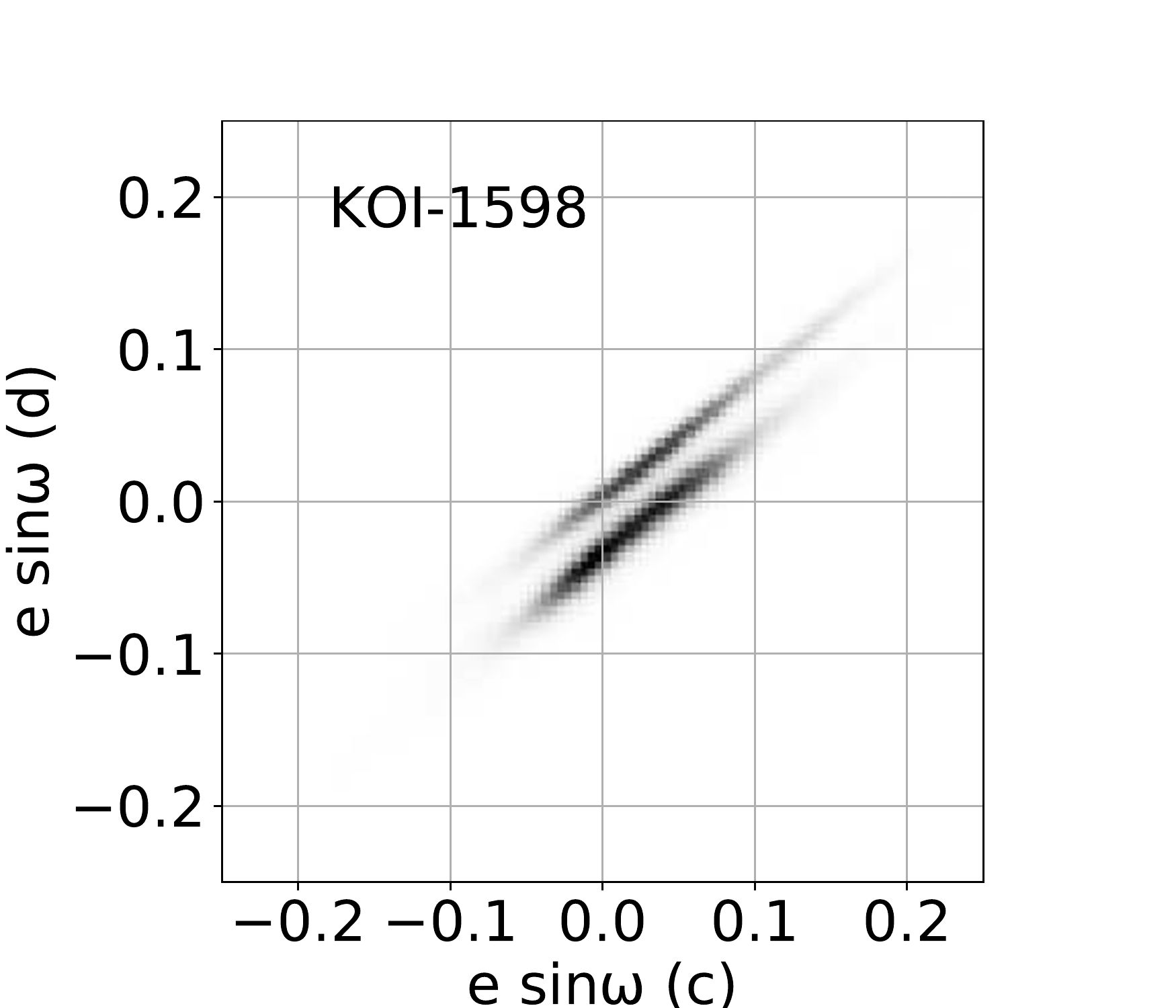}
\caption{Two-dimensional kernel density estimators on joint posteriors of eccentricity vector components: three-planet systems (Part 5 of 7). 
\label{fig:ecc3e}}
\end{center}
\end{figure}

\begin{figure}
\begin{center}
\figurenum{28}
\includegraphics [height = 1.1 in]{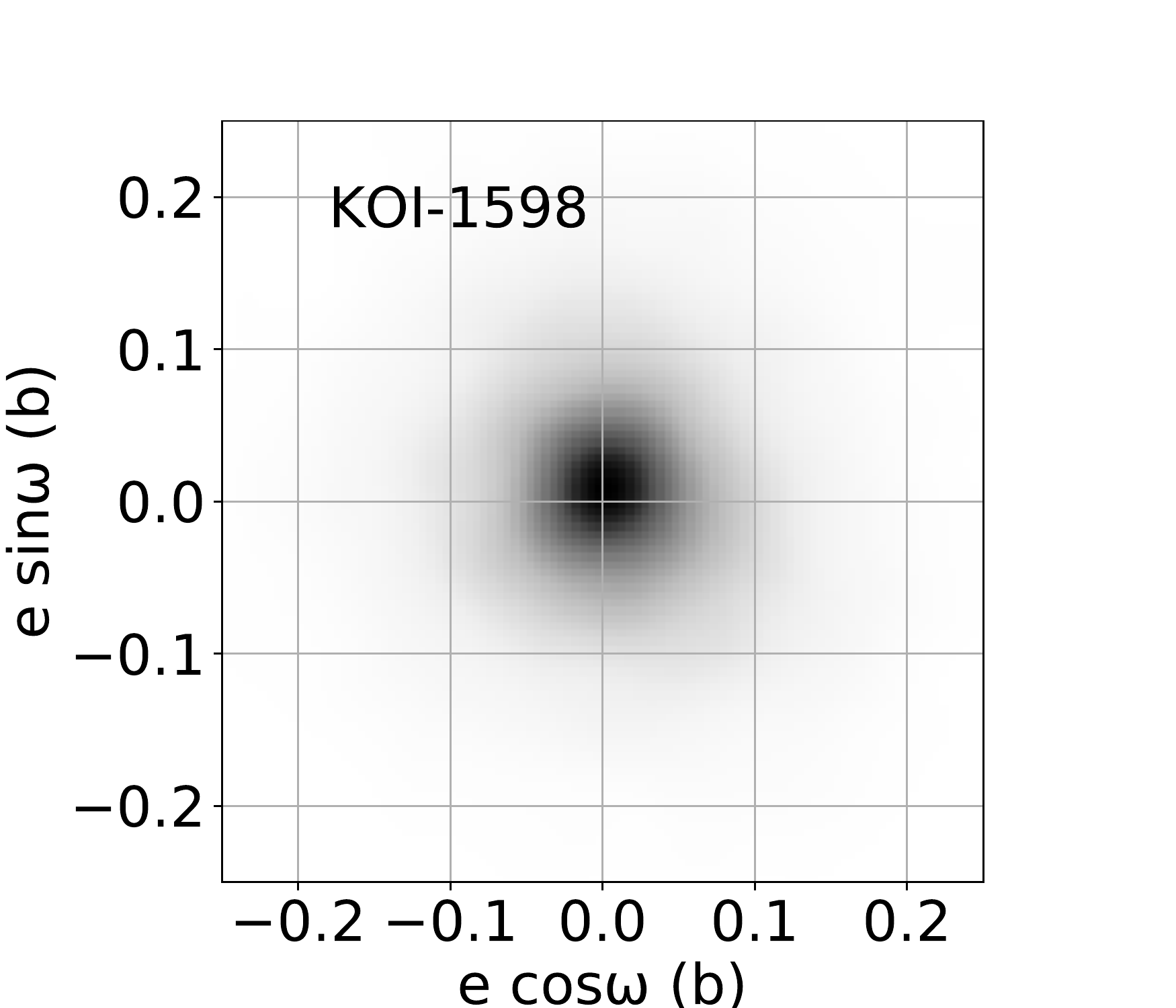}
\includegraphics [height = 1.1 in]{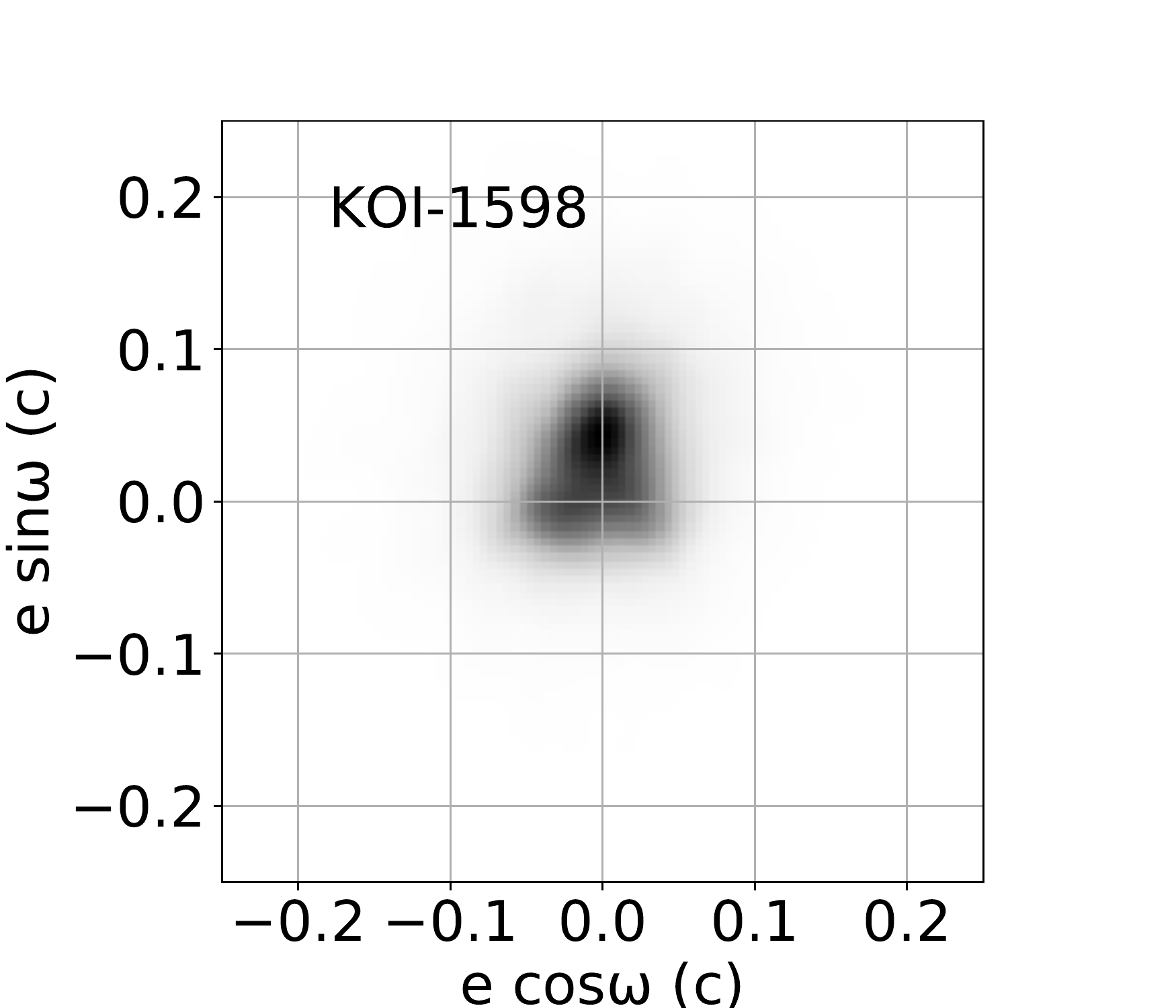}
\includegraphics [height = 1.1 in]{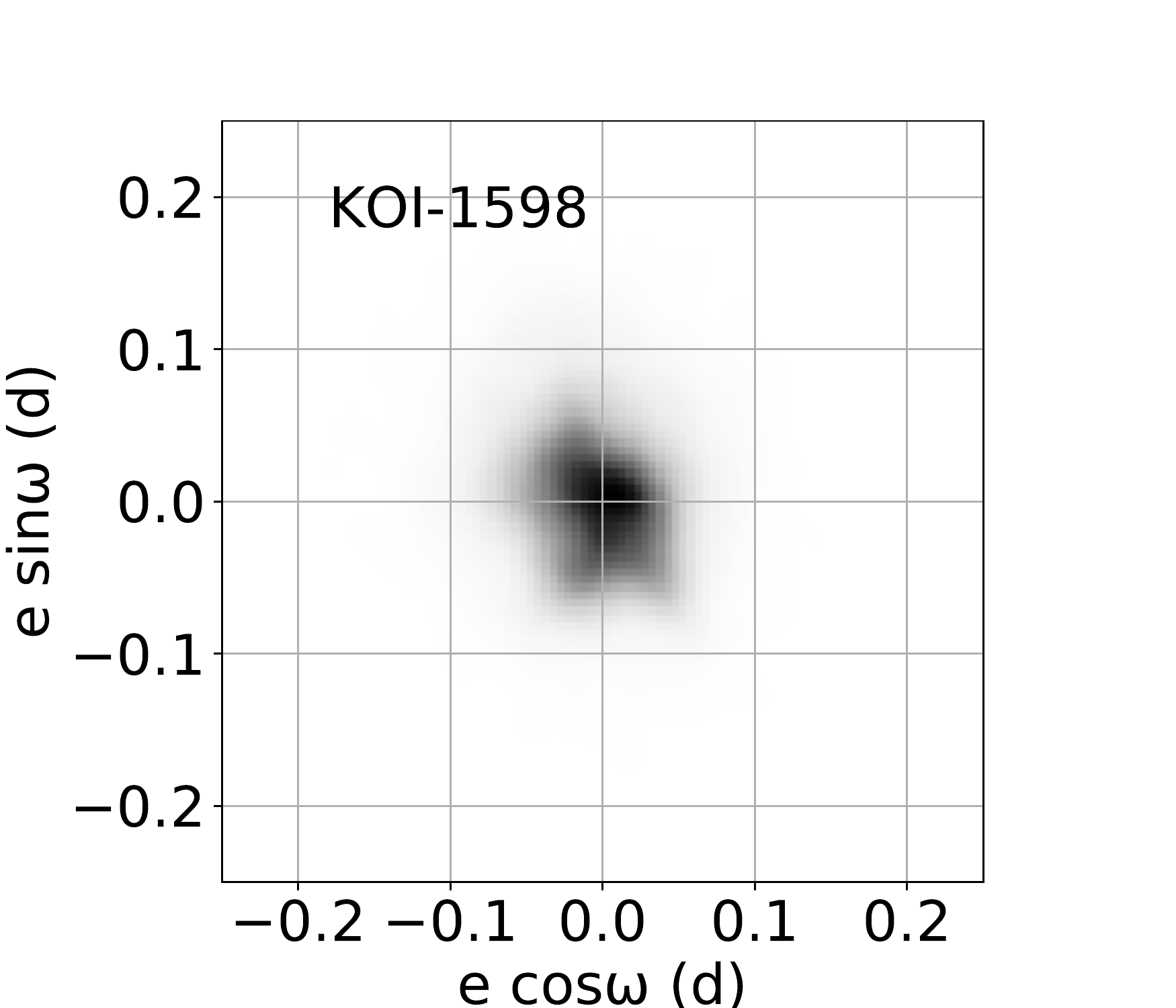}
\includegraphics [height = 1.1 in]{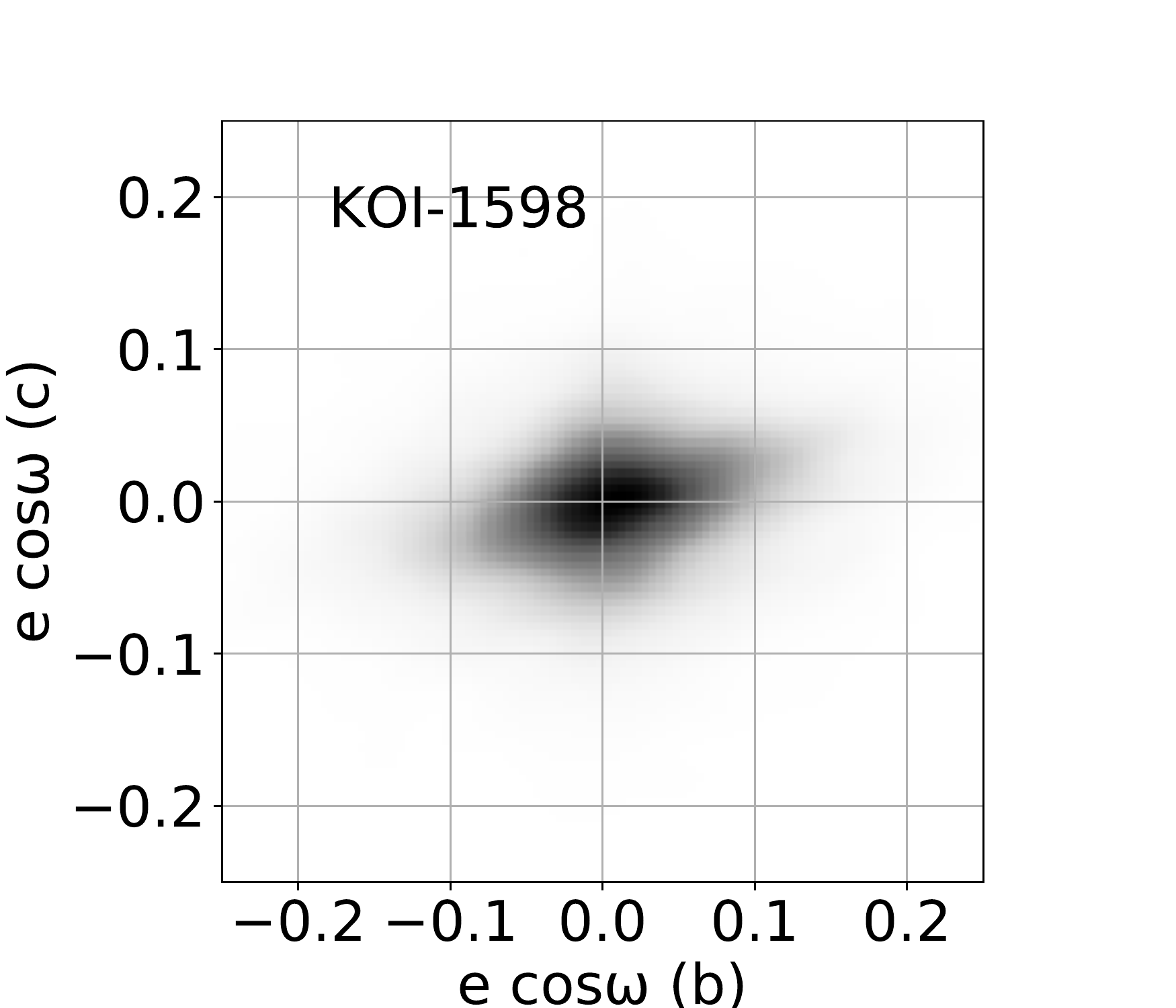}\\
\includegraphics [height = 1.1 in]{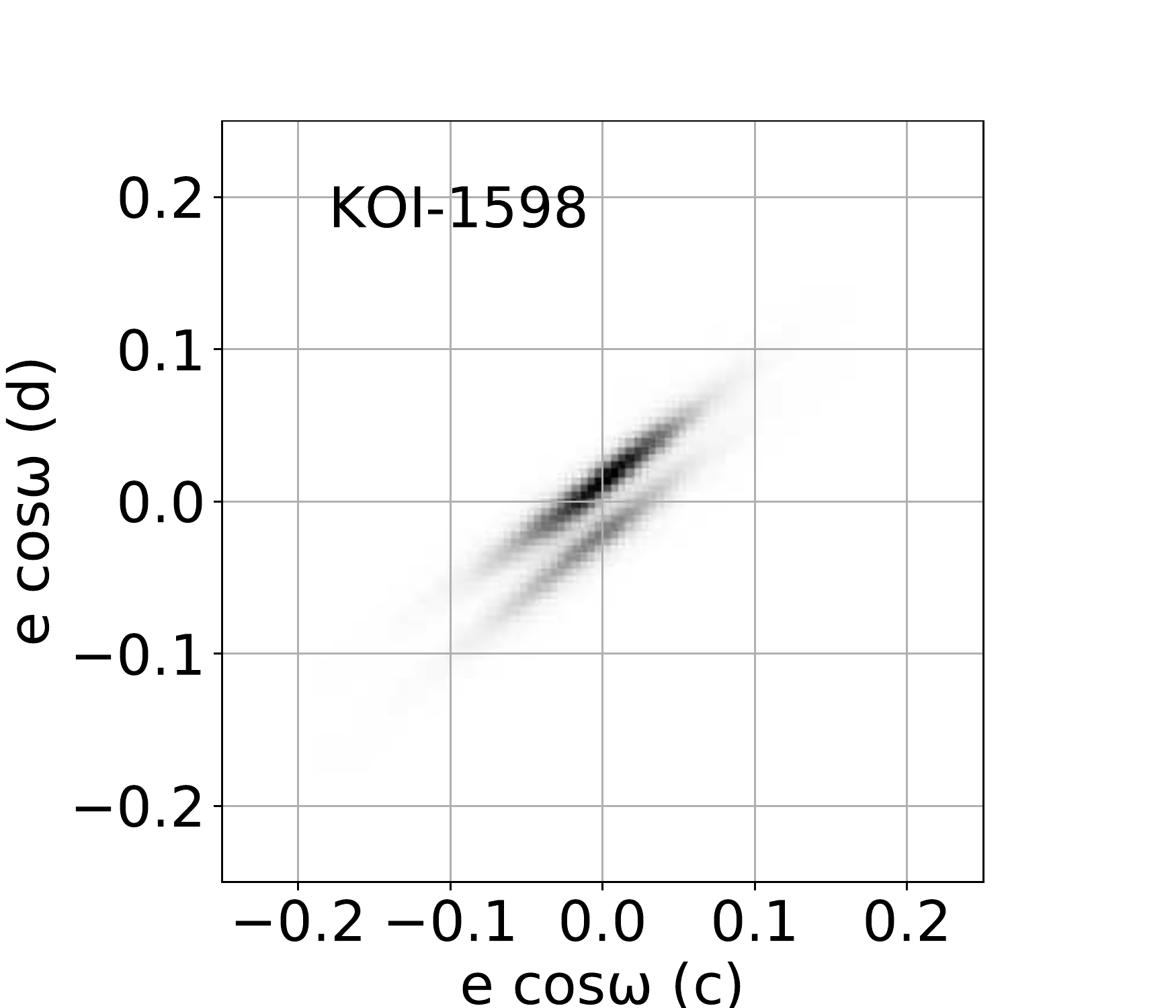}
\includegraphics [height = 1.1 in]{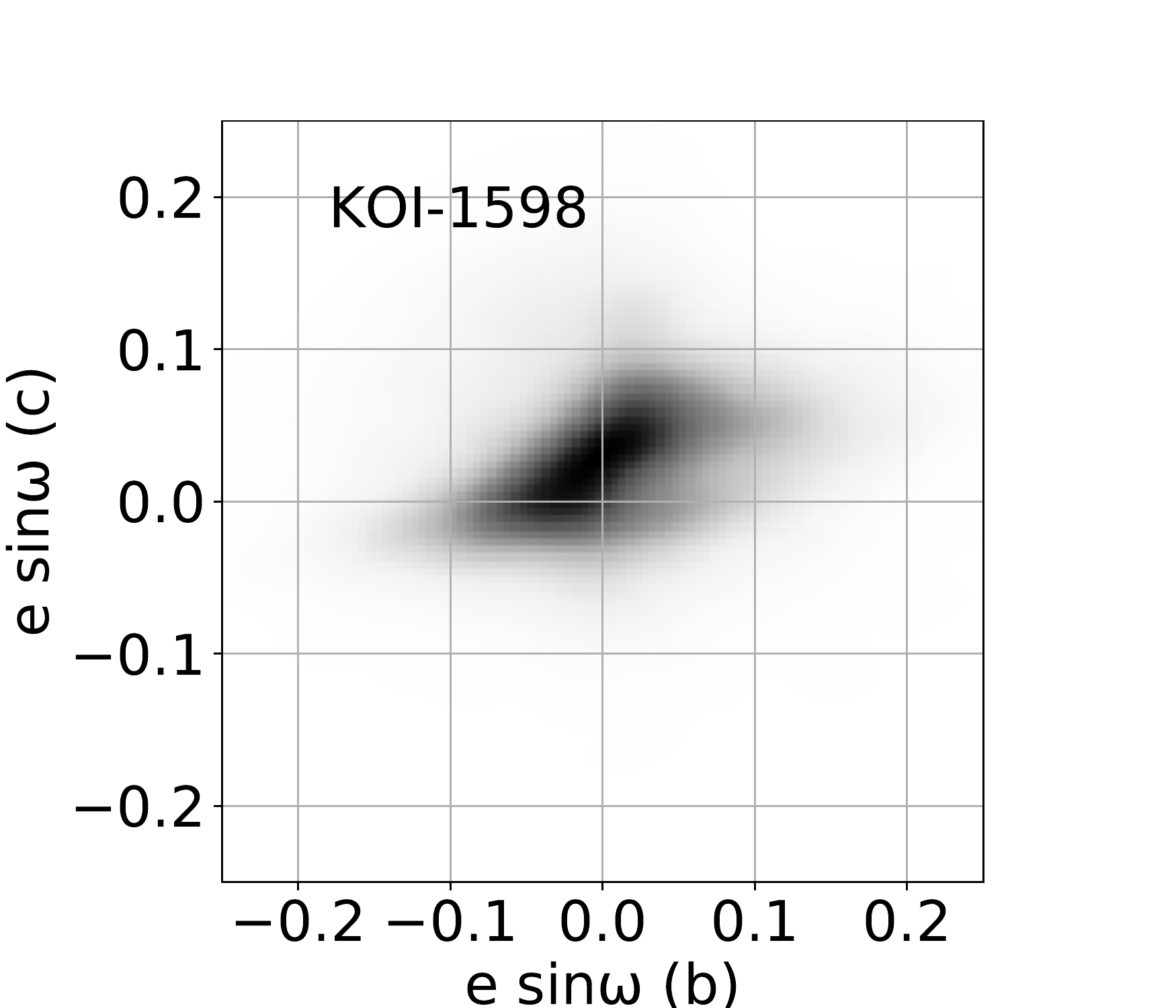}
\includegraphics [height = 1.1 in]{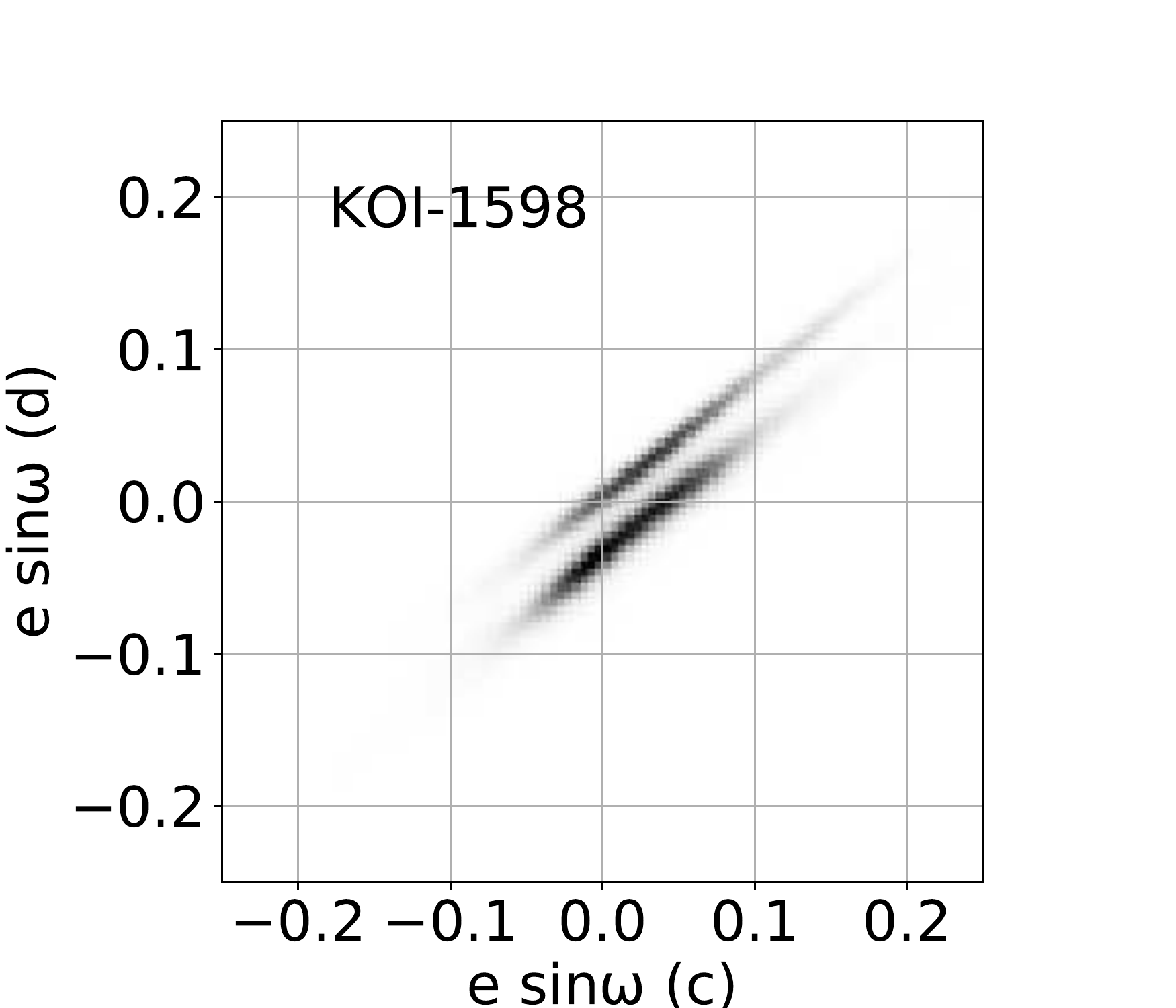}
\includegraphics [height = 1.1 in]{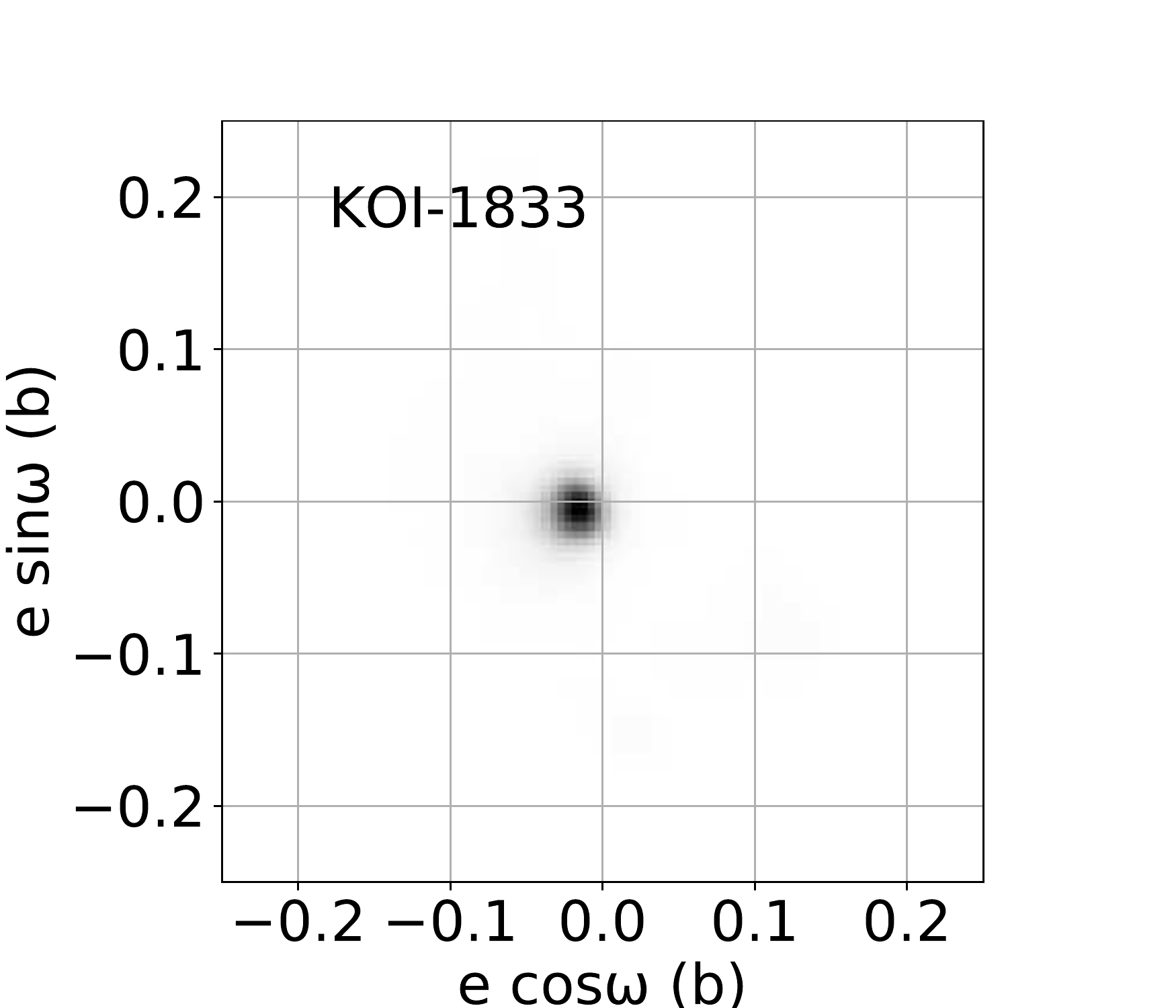} \\
\includegraphics [height = 1.1 in]{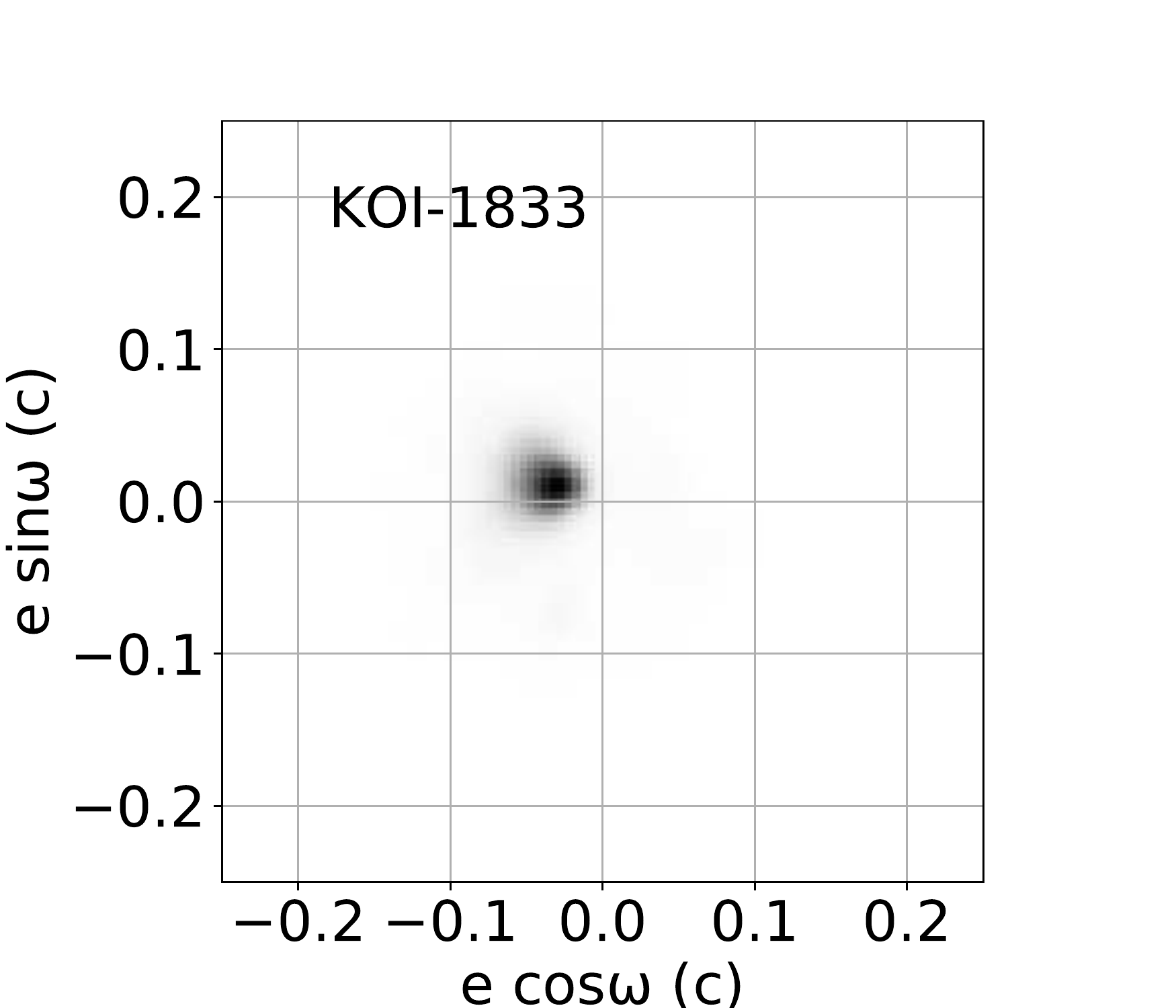}
\includegraphics [height = 1.1 in]{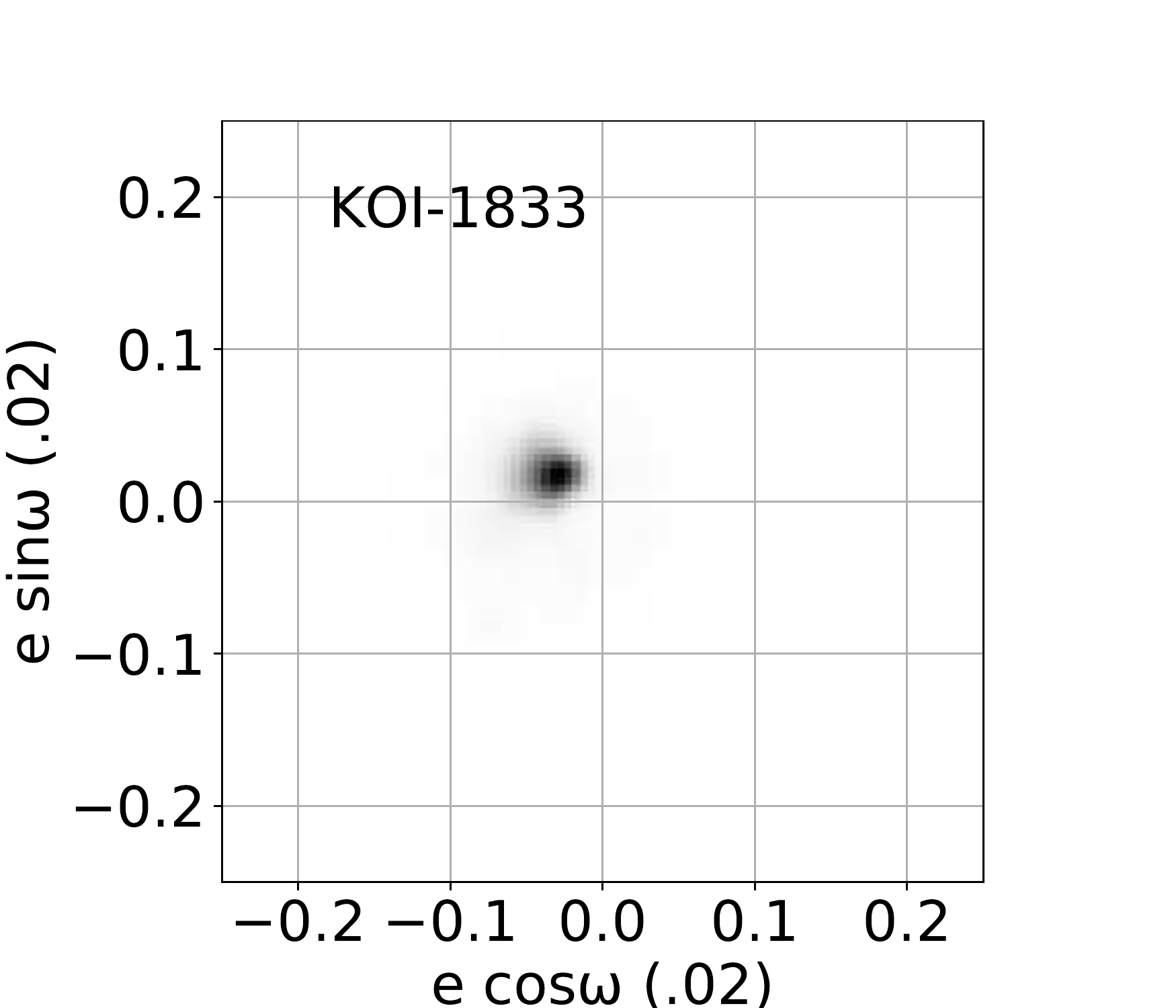}
\includegraphics [height = 1.1 in]{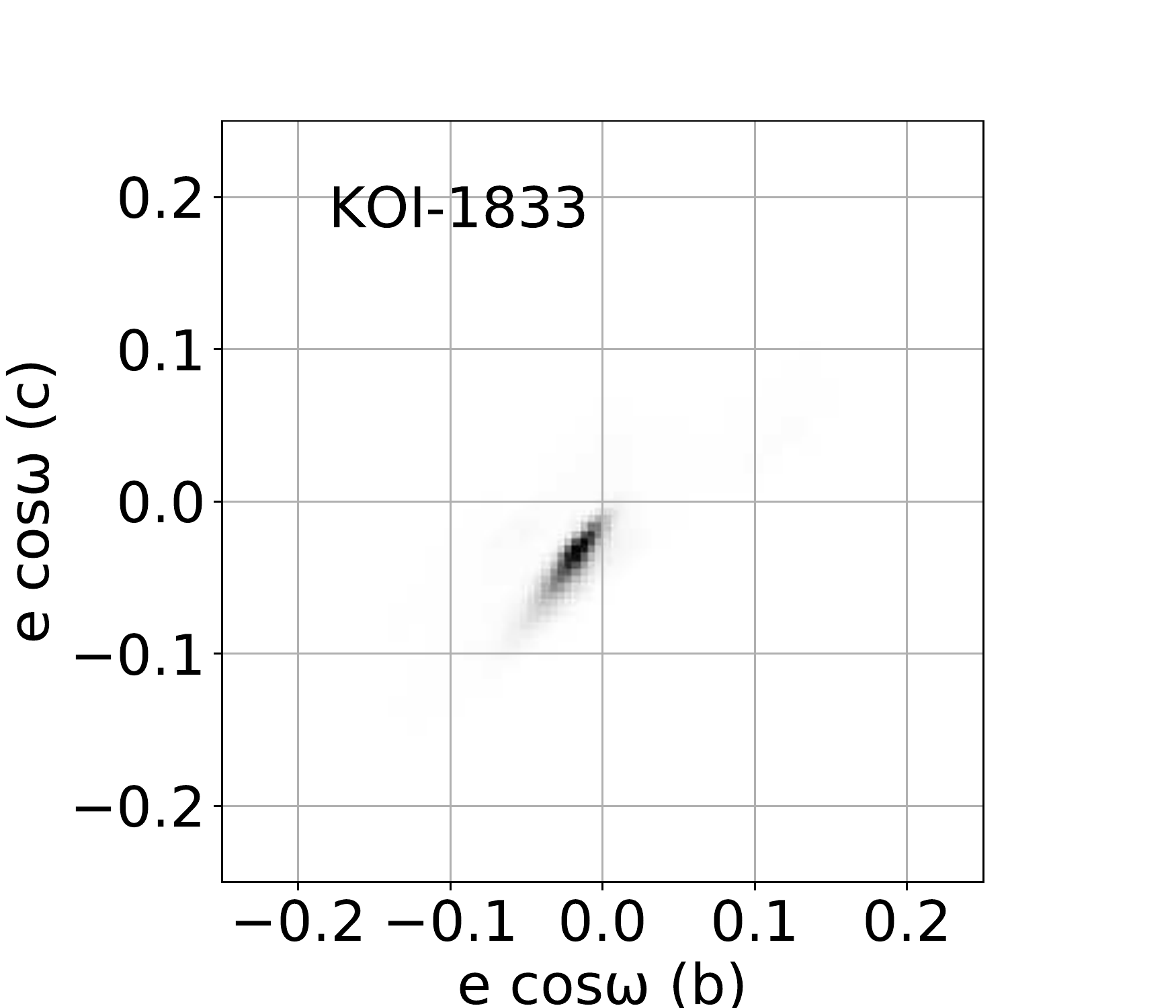}
\includegraphics [height = 1.1 in]{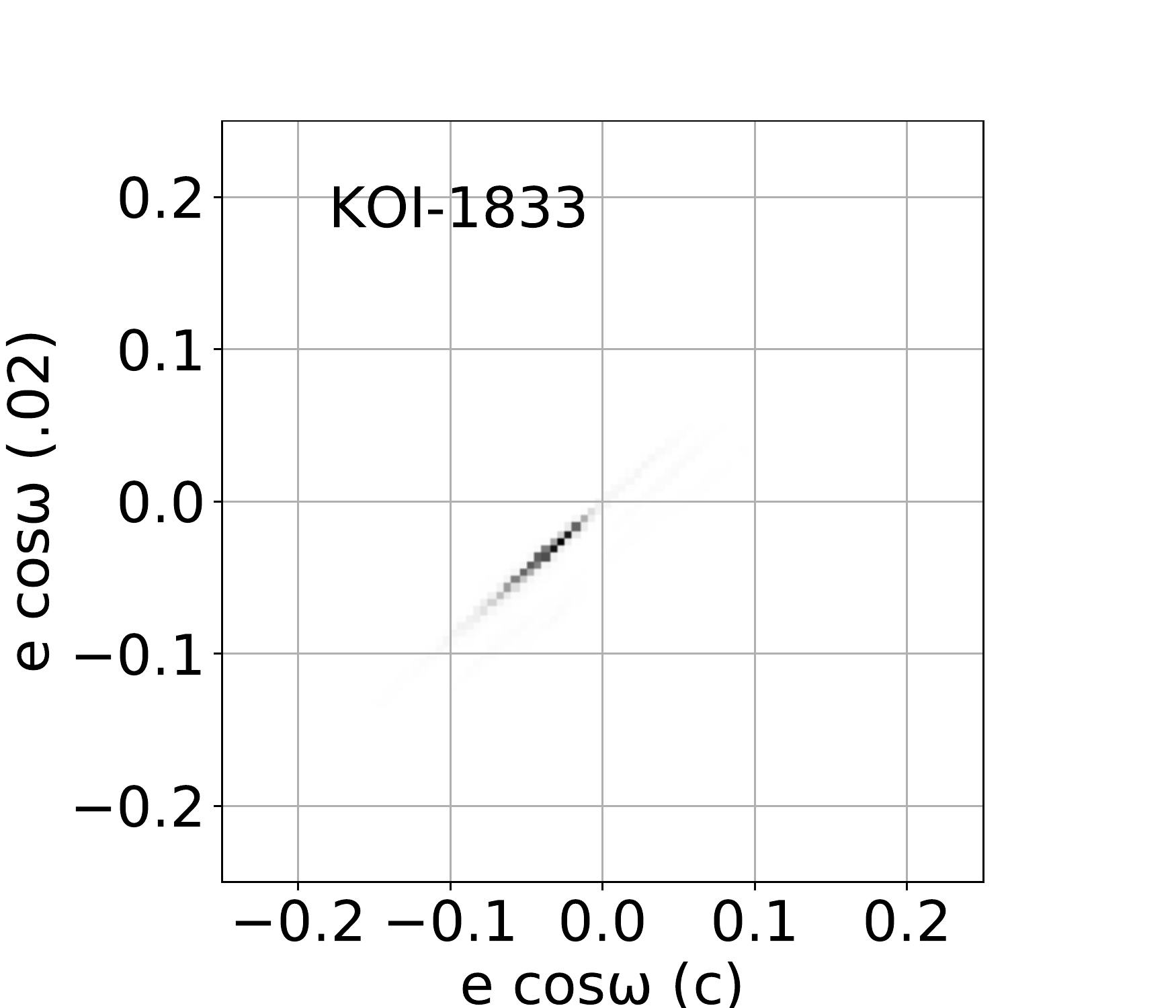} \\
\includegraphics [height = 1.1 in]{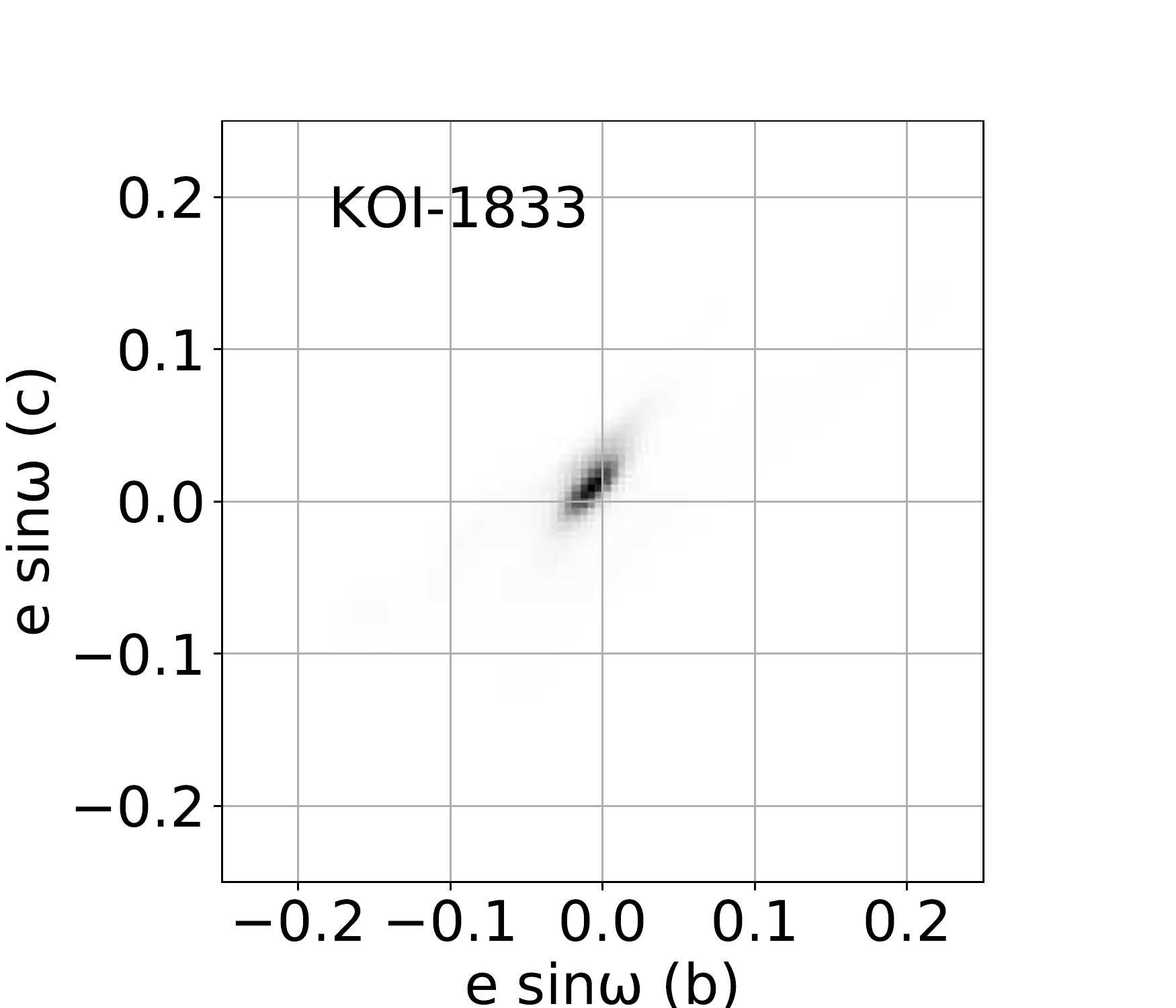}
\includegraphics [height = 1.1 in]{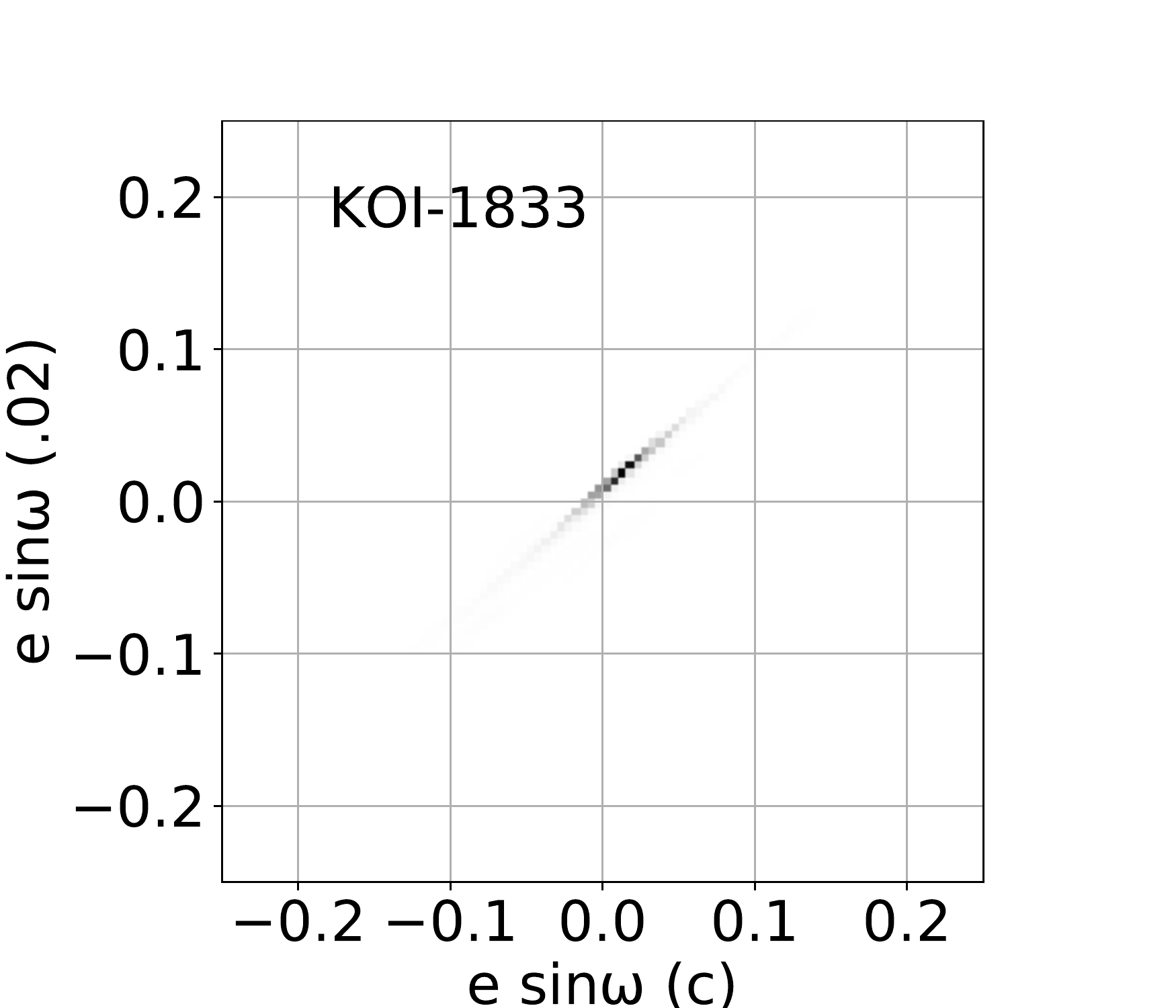}
\includegraphics [height = 1.1 in]{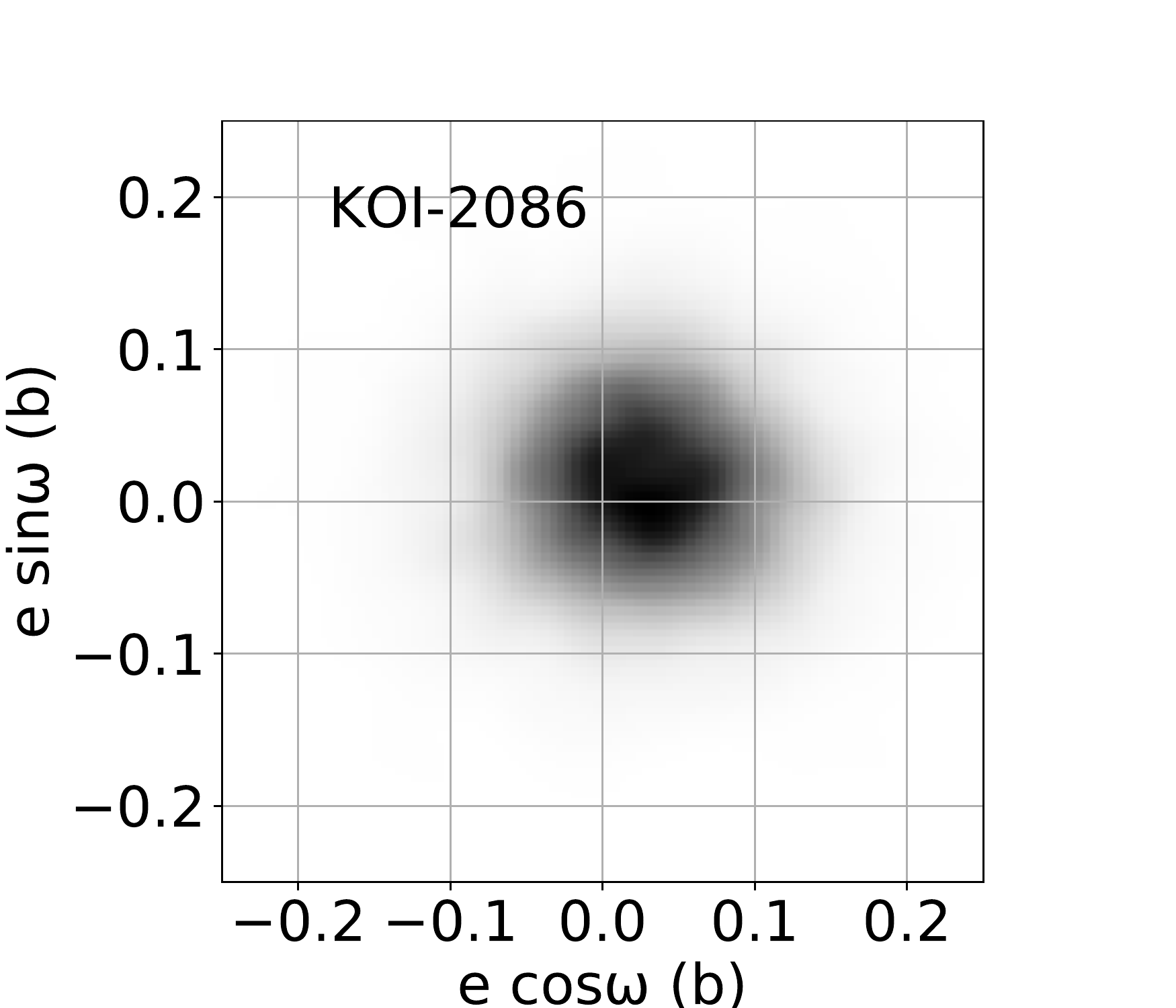}
\includegraphics [height = 1.1 in]{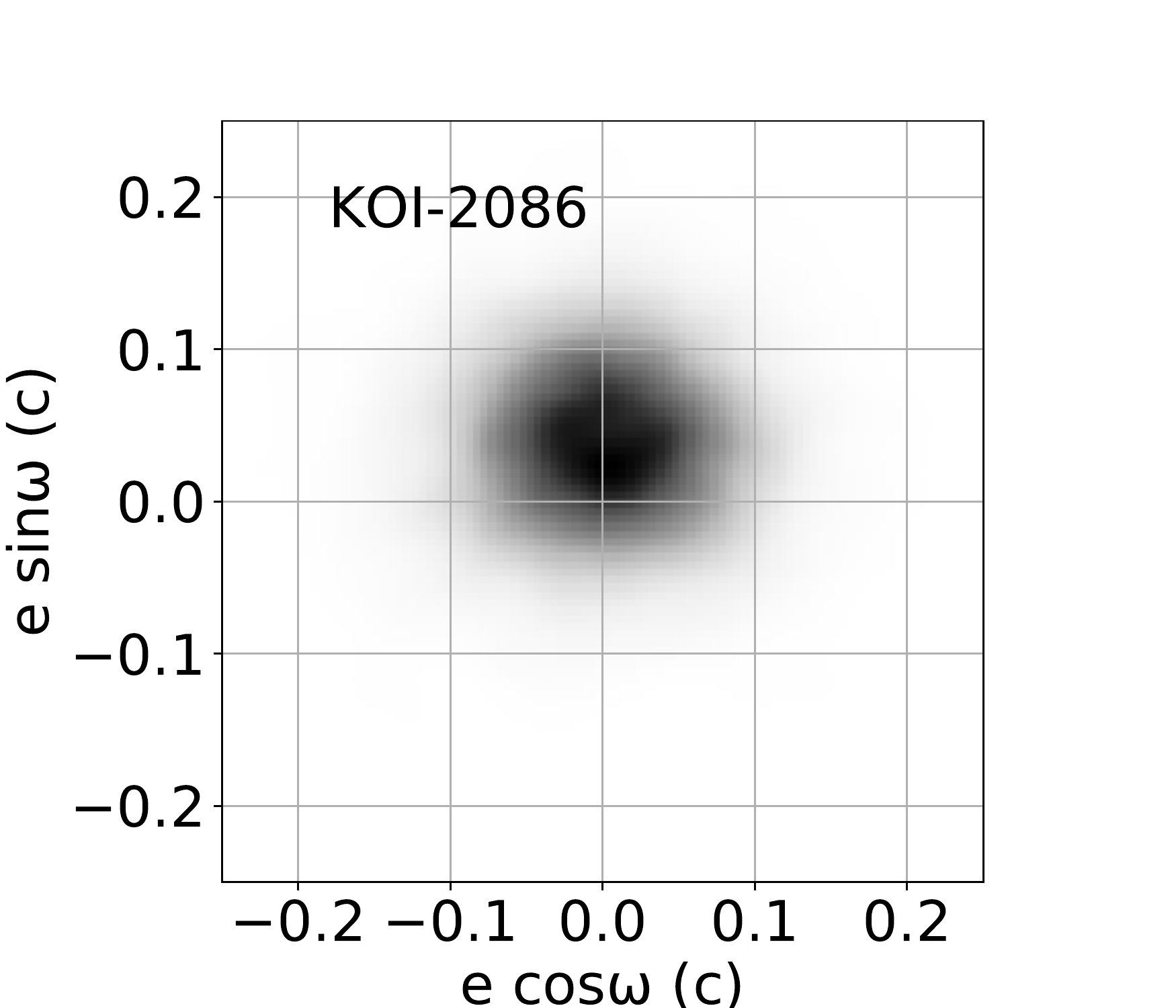} \\
\includegraphics [height = 1.1 in]{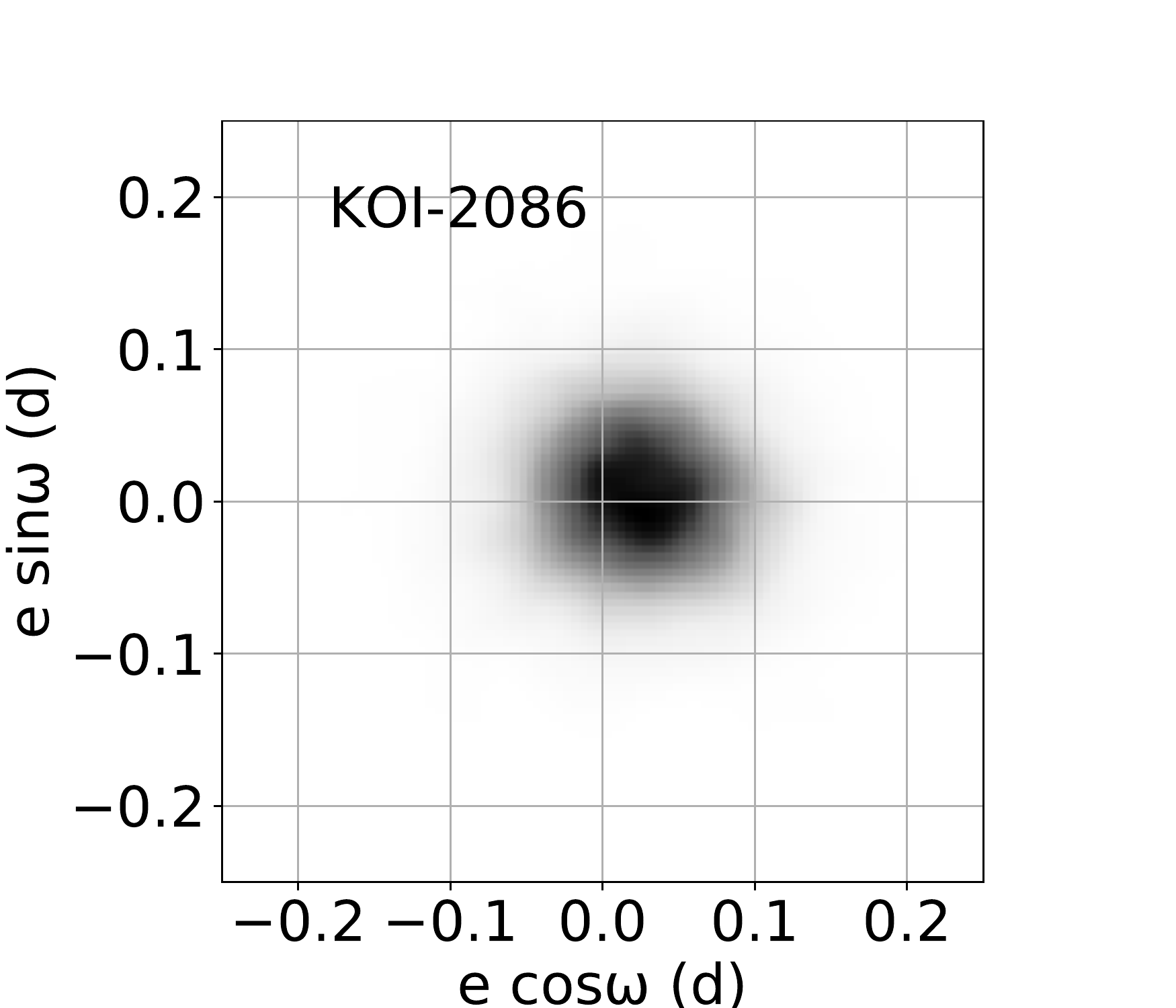}
\includegraphics [height = 1.1 in]{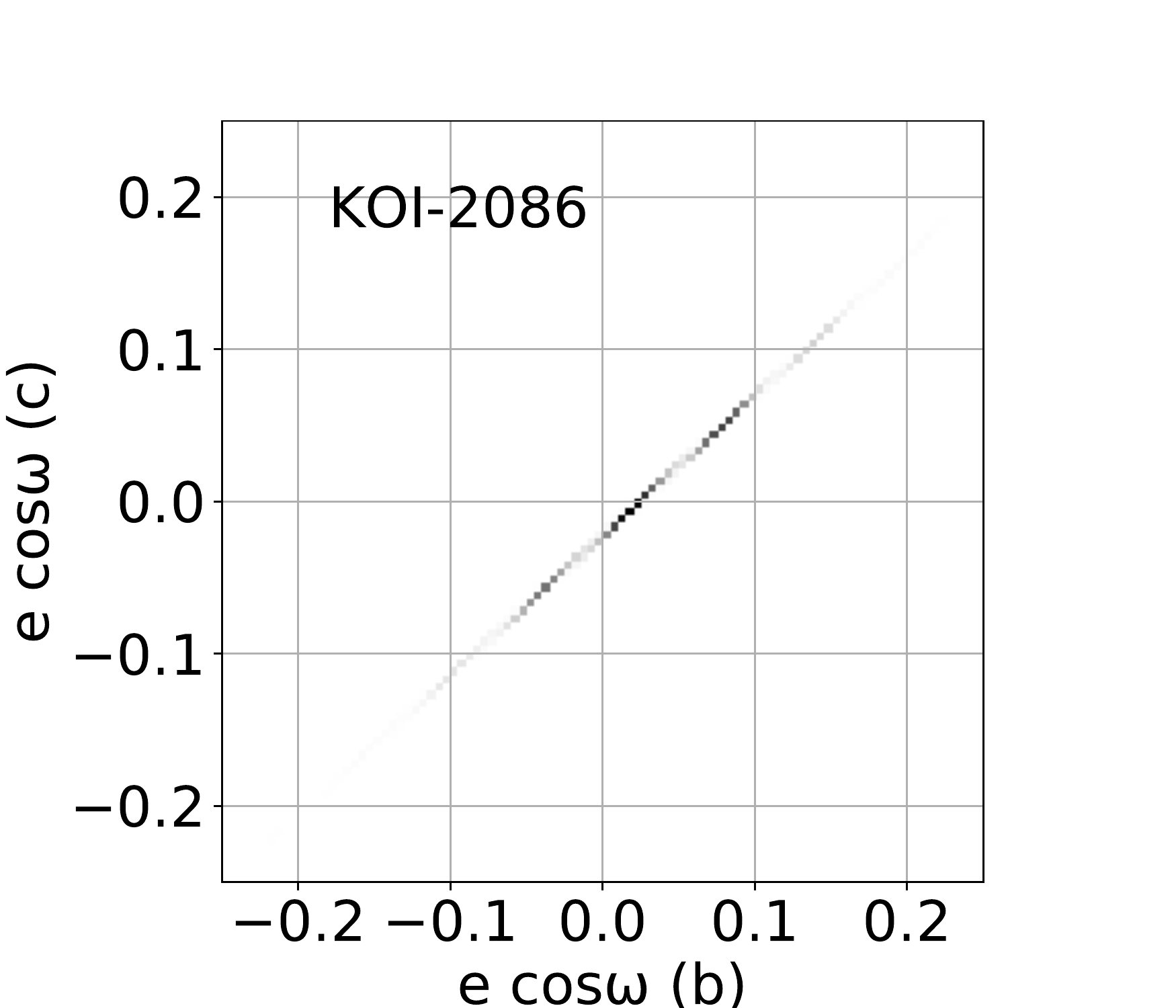} 
\includegraphics [height = 1.1 in]{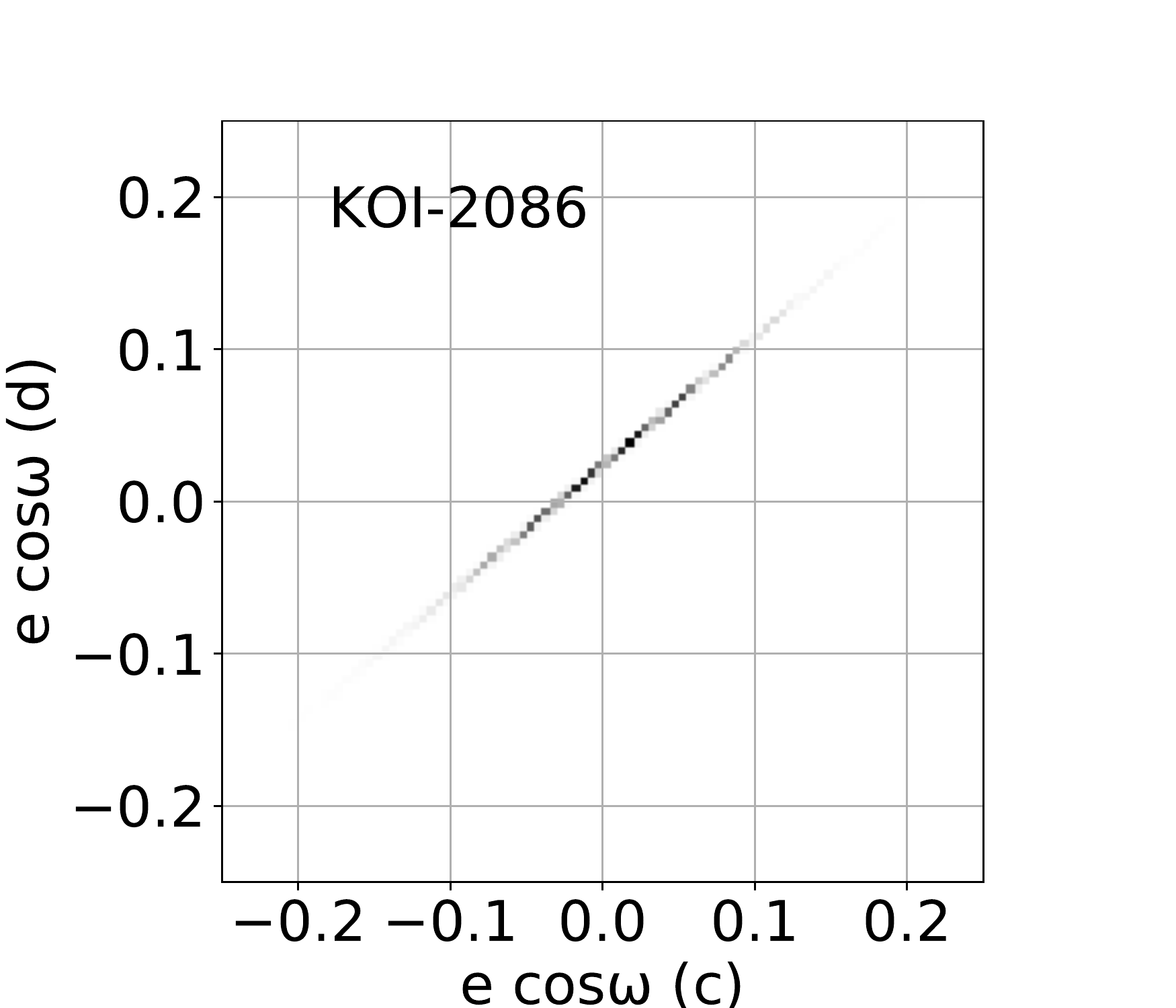}
\includegraphics [height = 1.1 in]{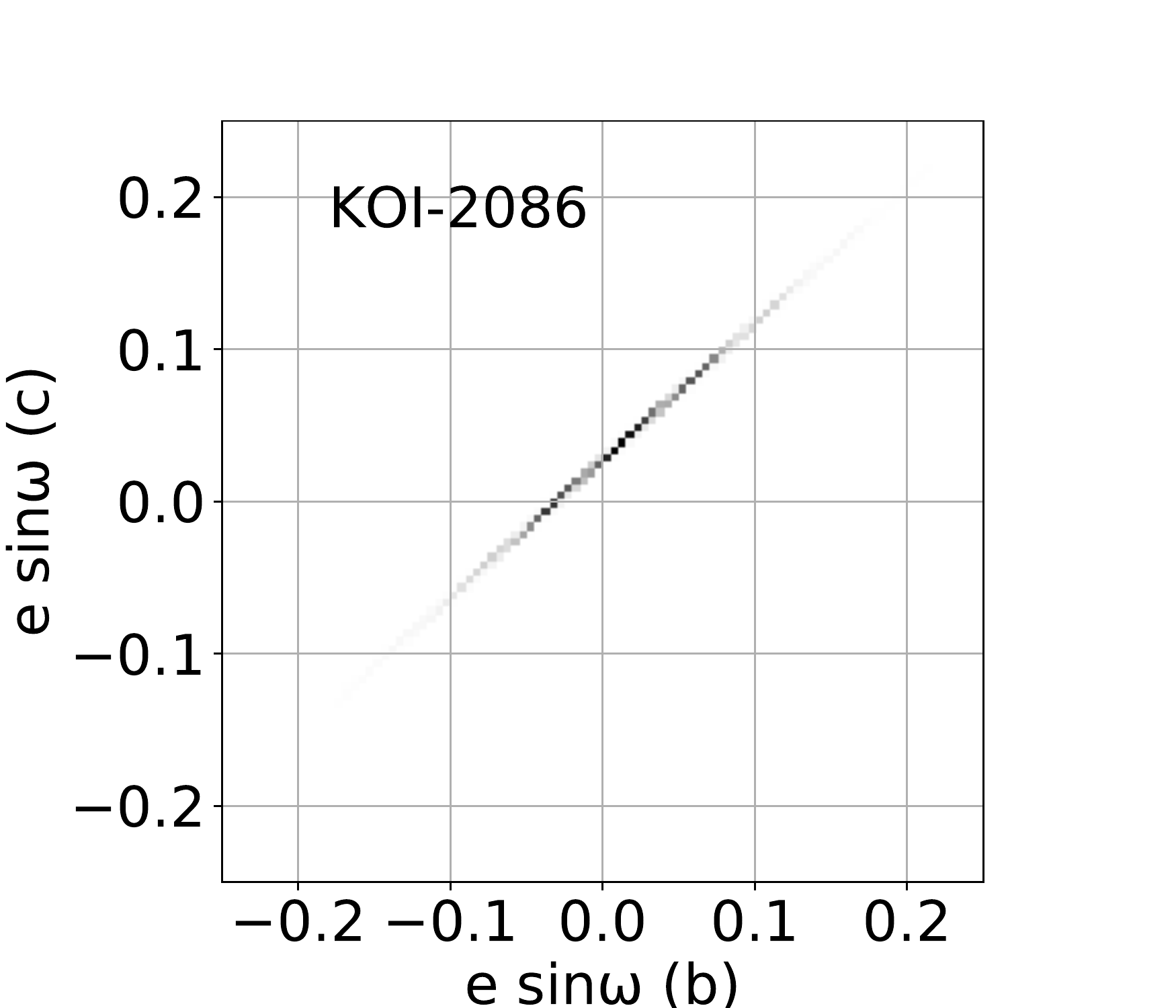} \\
\includegraphics [height = 1.1 in]{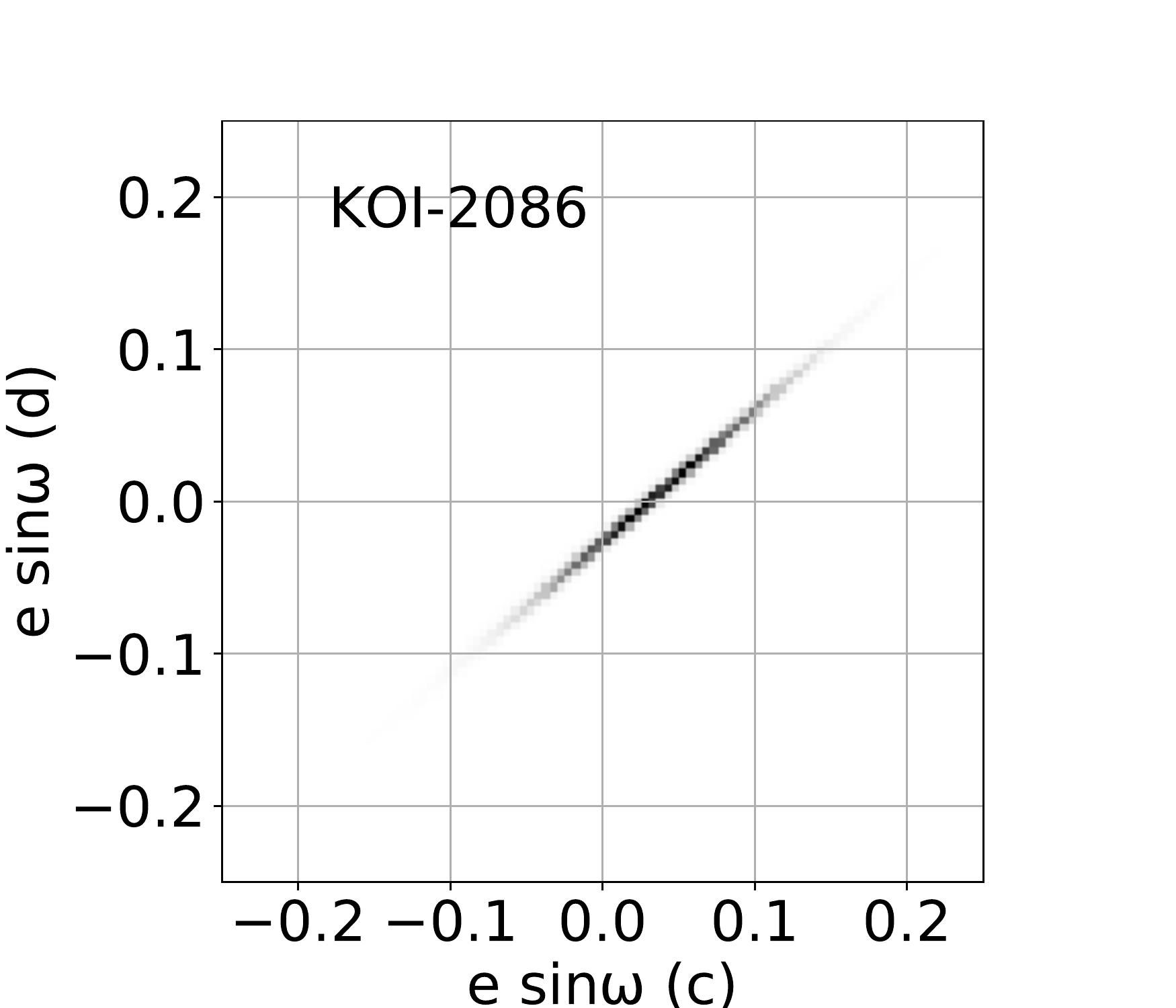}
\includegraphics [height = 1.1 in]{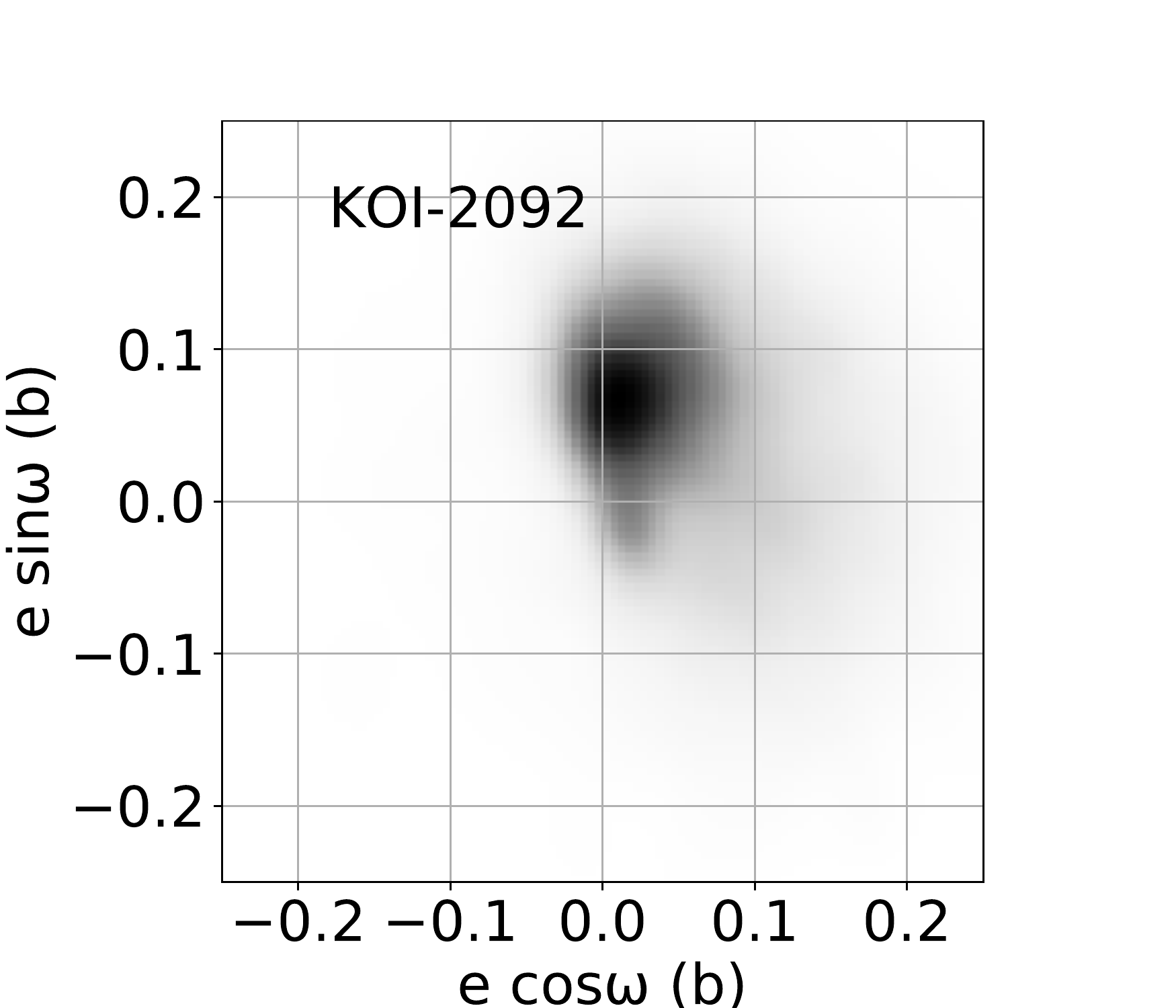} 
\includegraphics [height = 1.1 in]{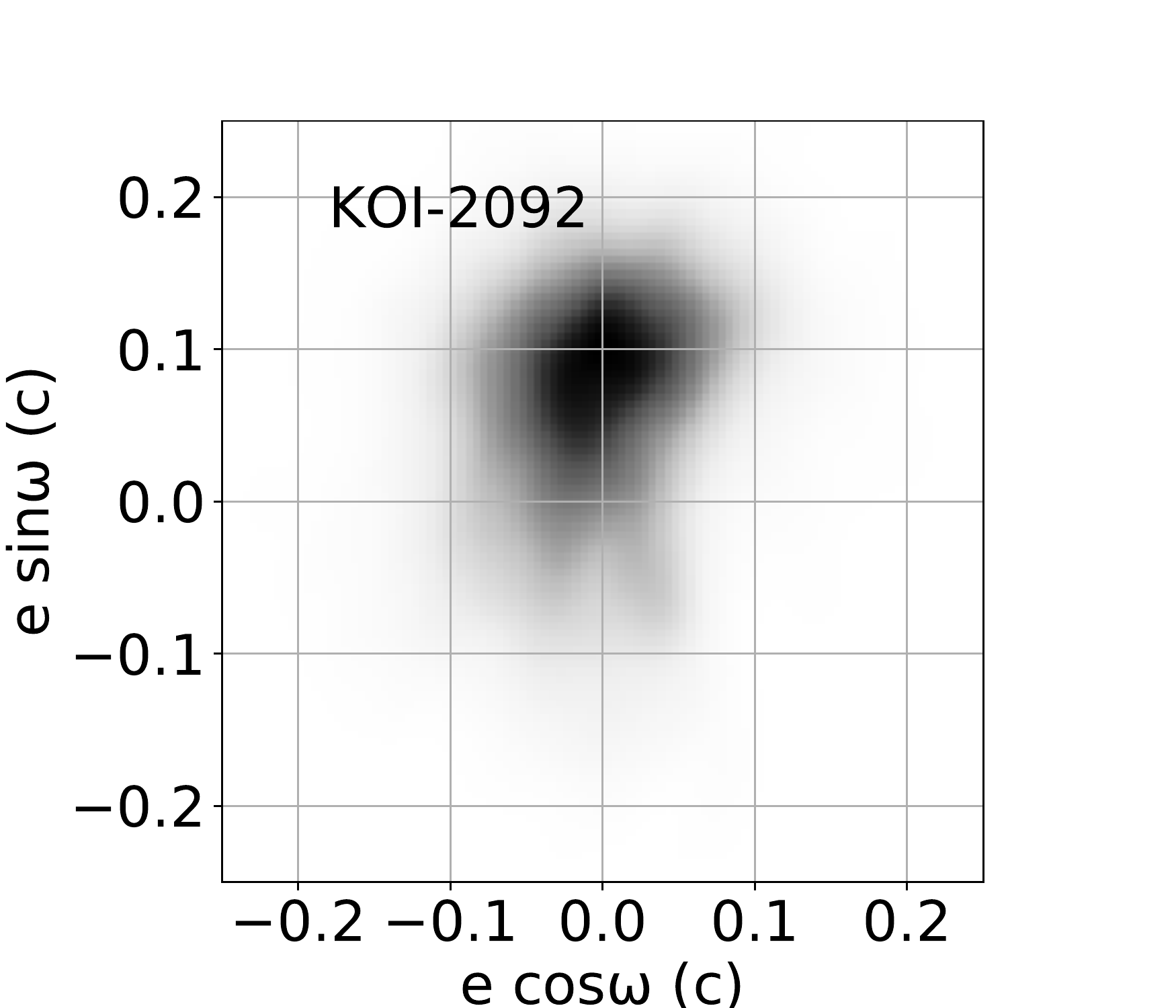}
\includegraphics [height = 1.1 in]{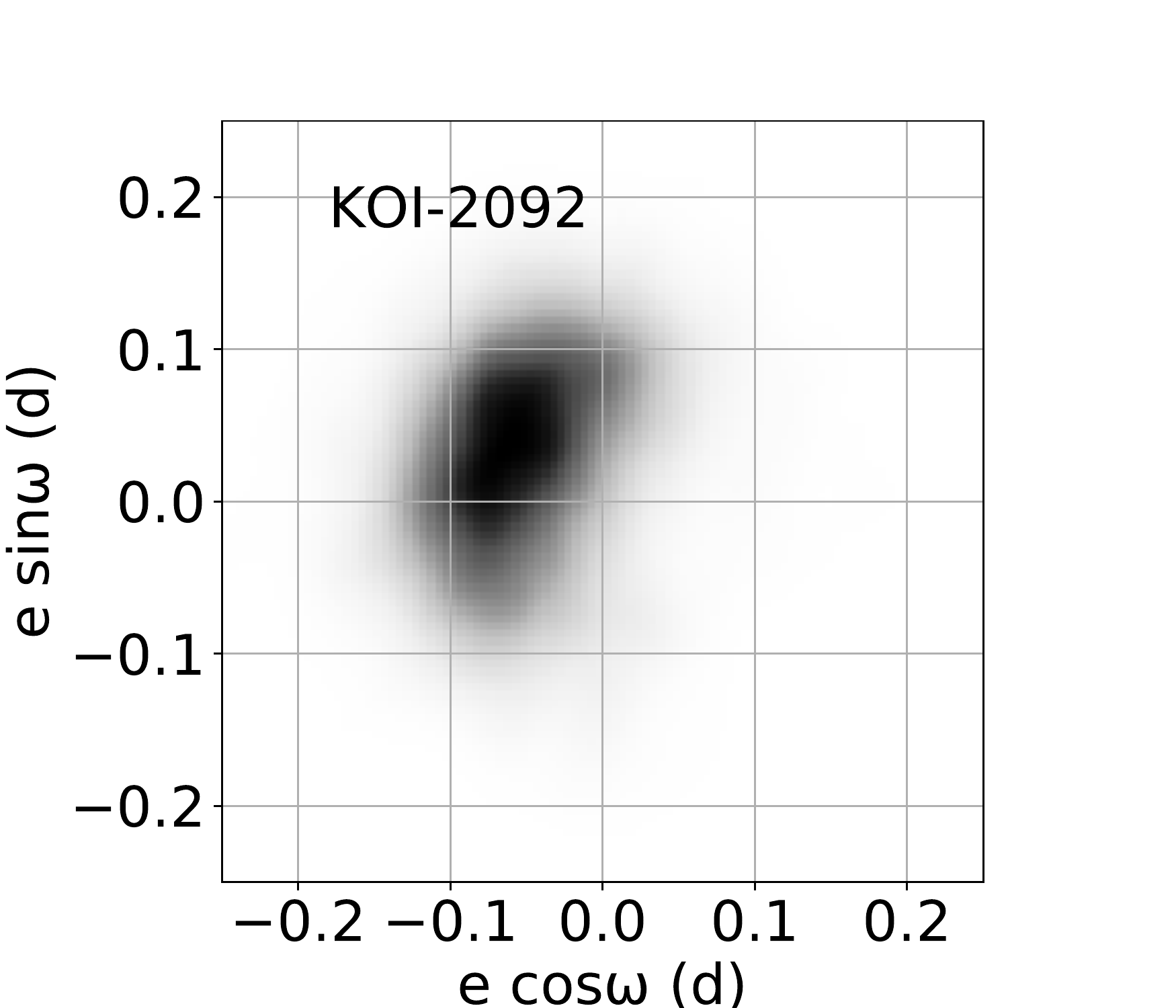} \\
\includegraphics [height = 1.1 in]{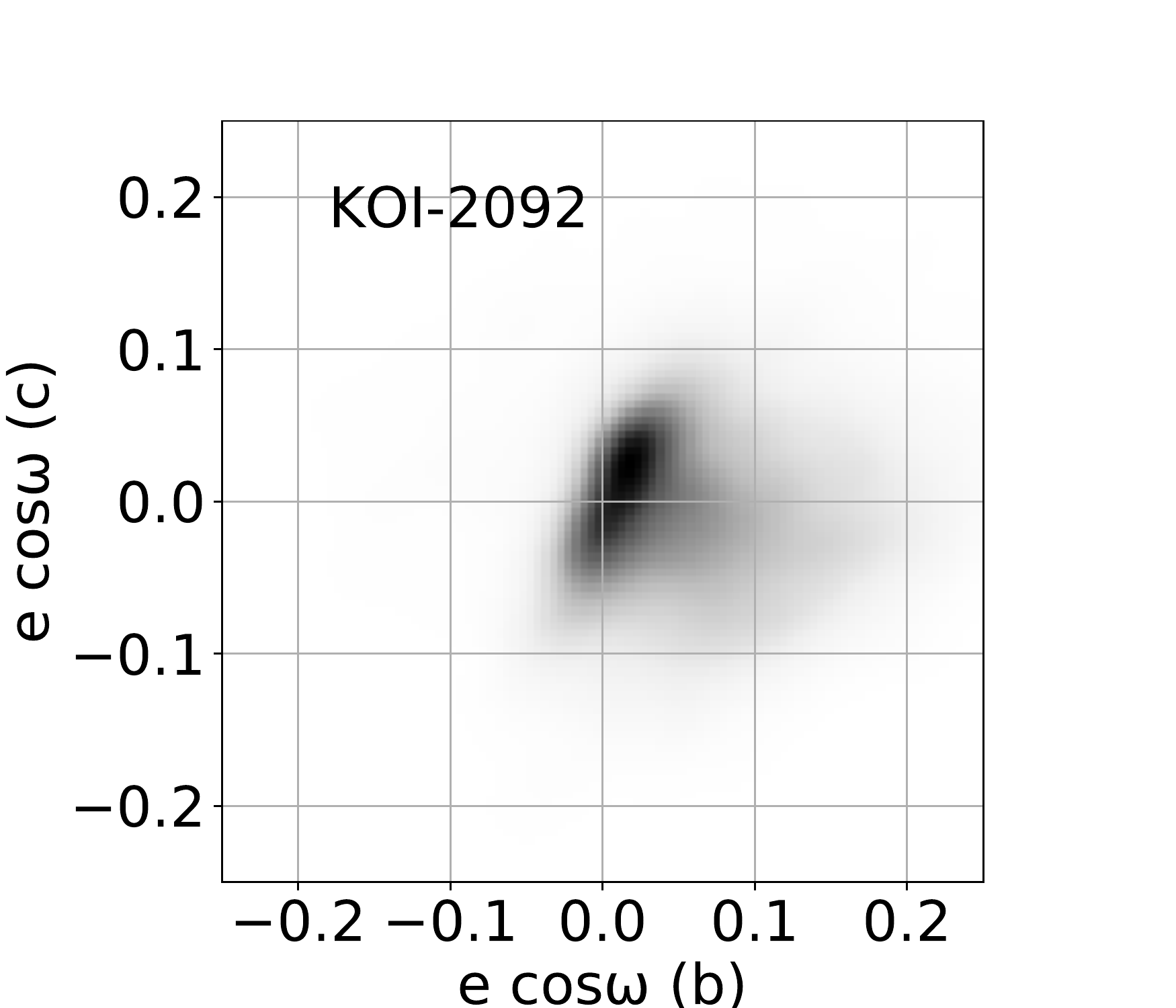} 
\includegraphics [height = 1.1 in]{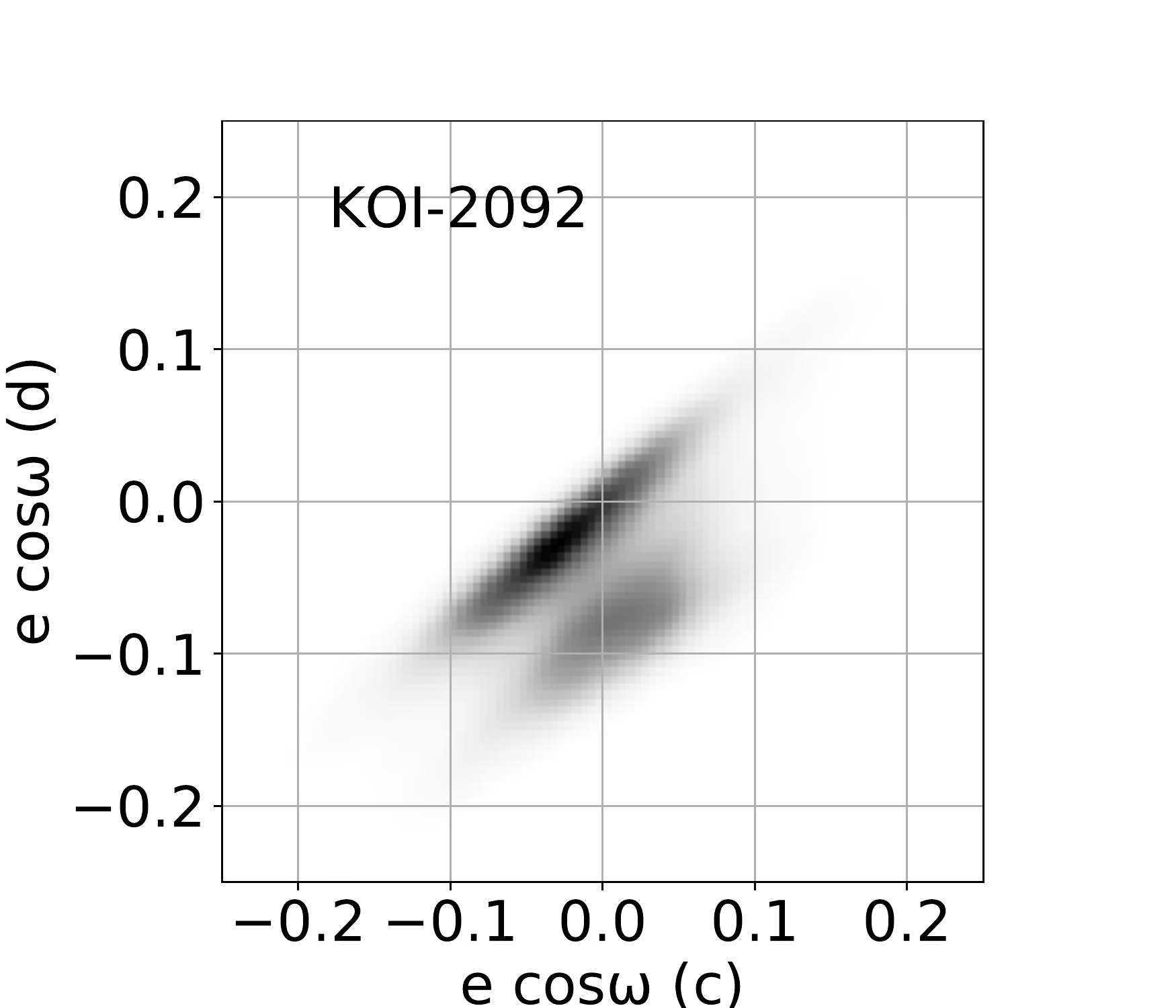}
\includegraphics [height = 1.1 in]{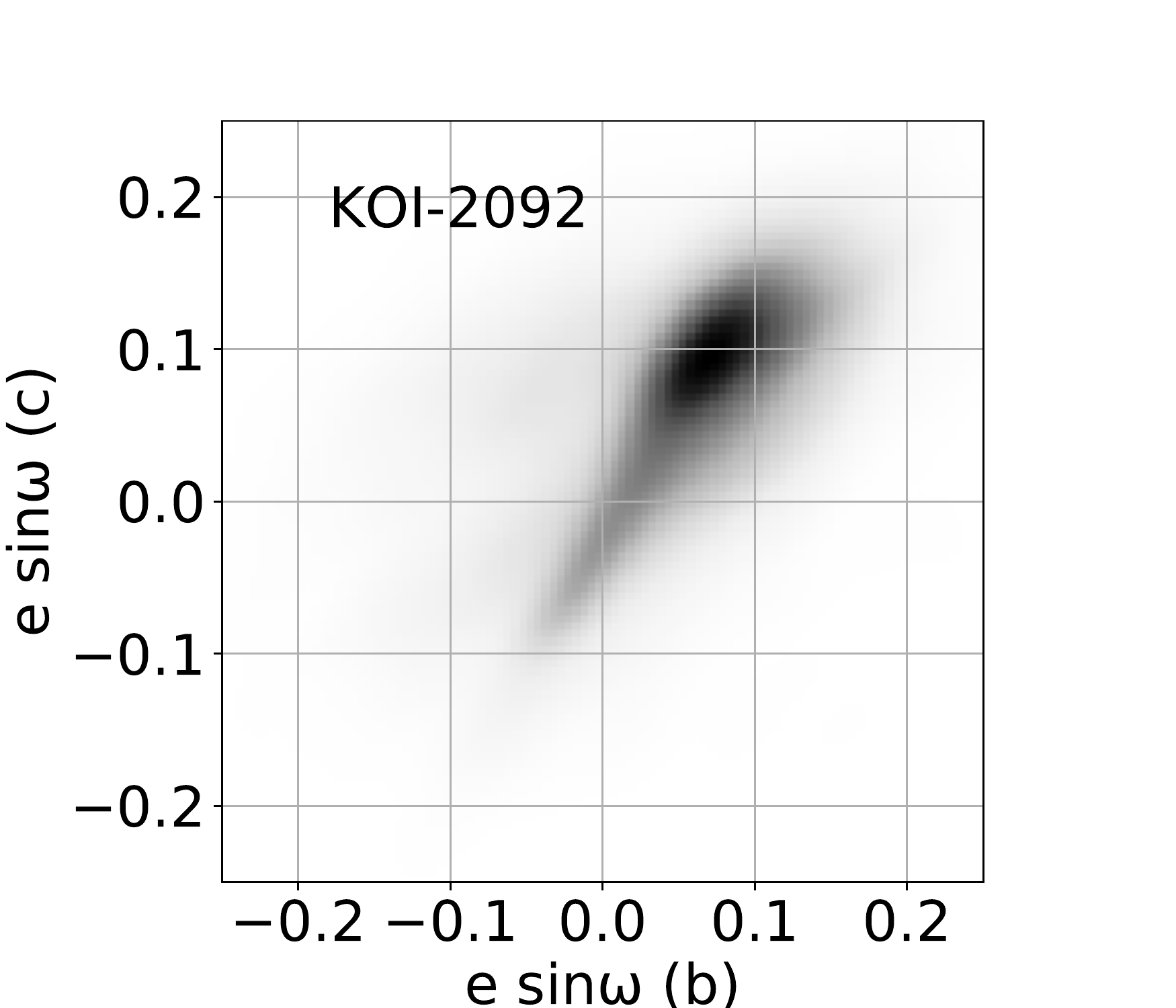} 
\includegraphics [height = 1.1 in]{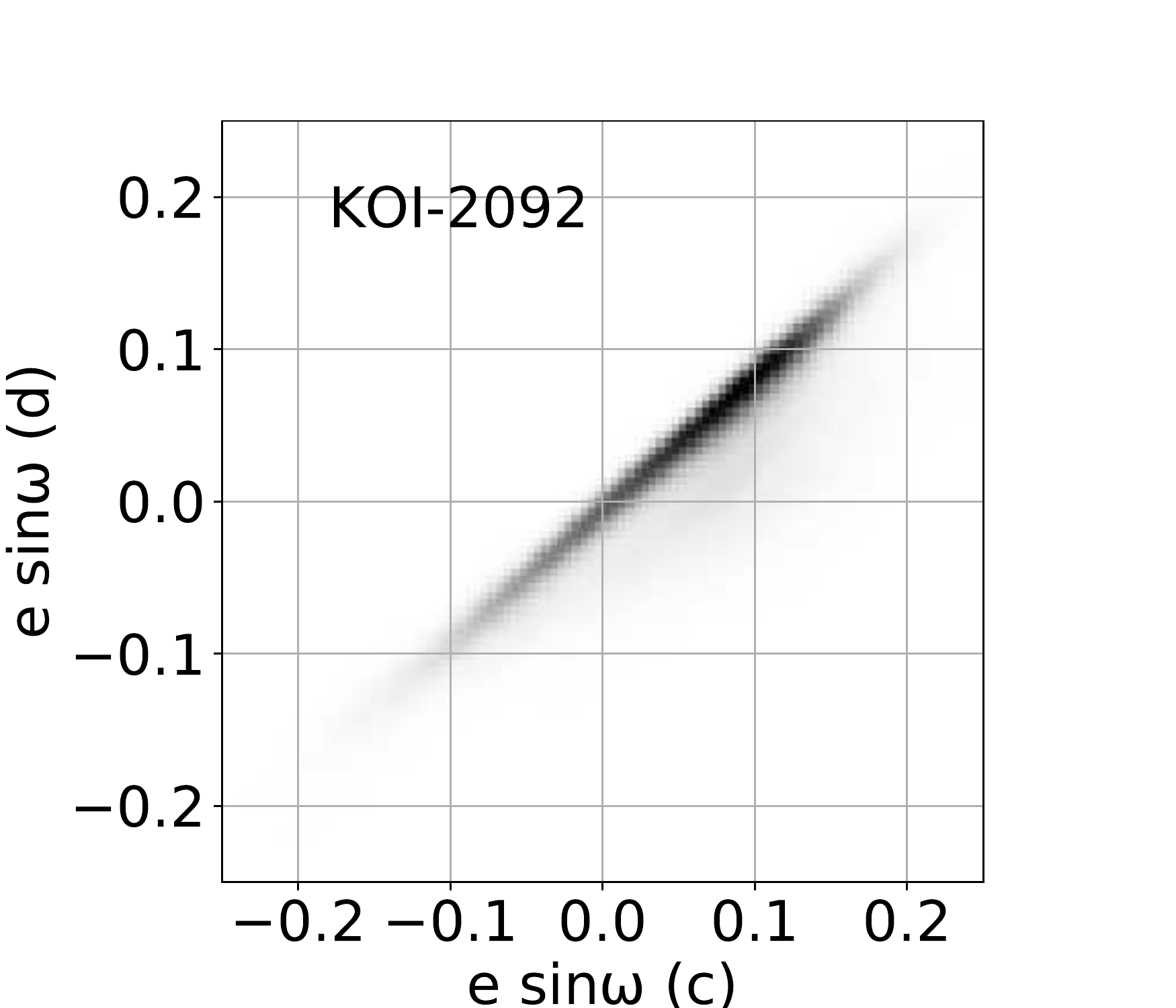}
\caption{Two-dimensional kernel density estimators on joint posteriors of eccentricity vector components: three-planet systems (Part 6 of 7). 
\label{fig:ecc3f} }
\end{center}
\end{figure}

\begin{figure}
\begin{center}
\figurenum{29}
\includegraphics [height = 1.1 in]{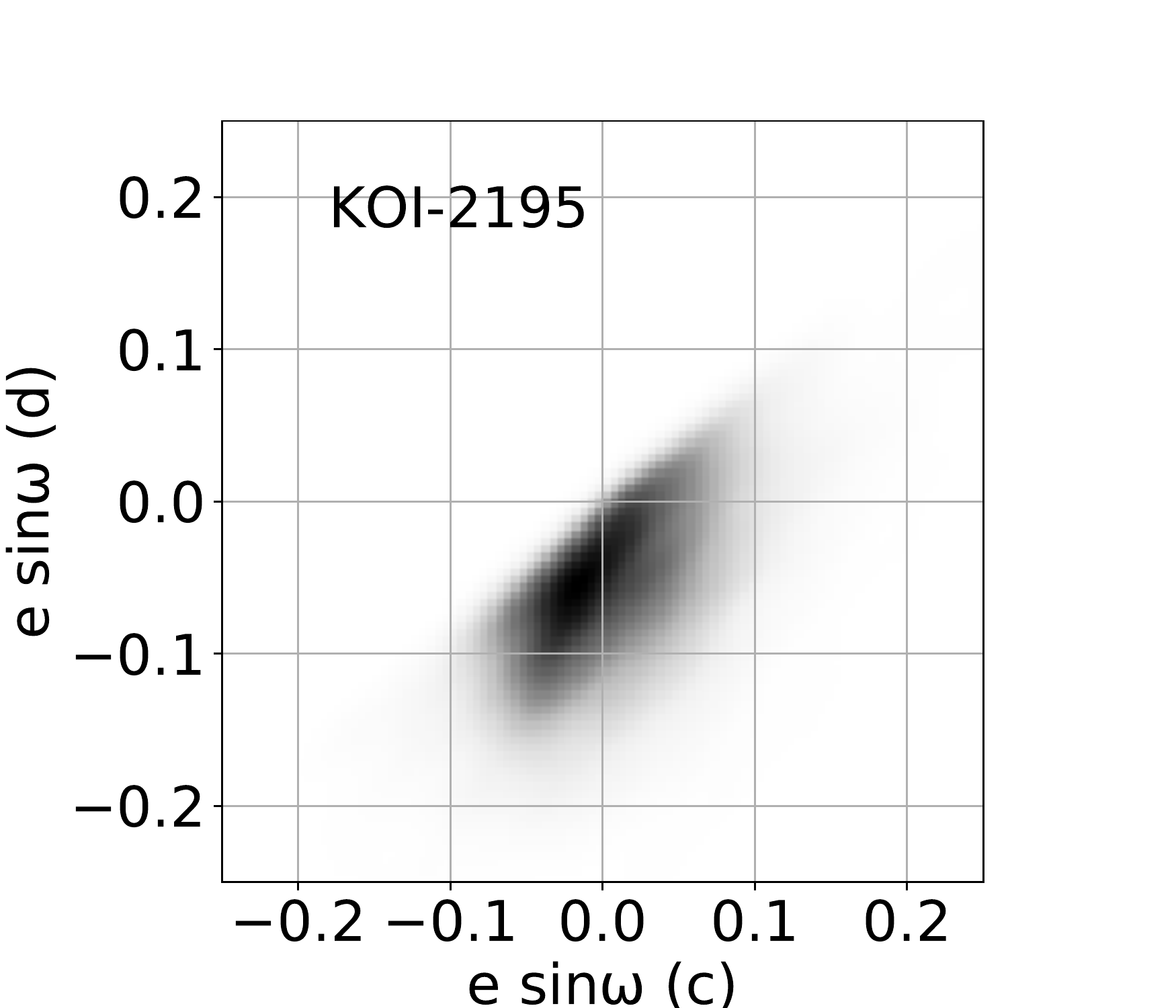}
\includegraphics [height = 1.1 in]{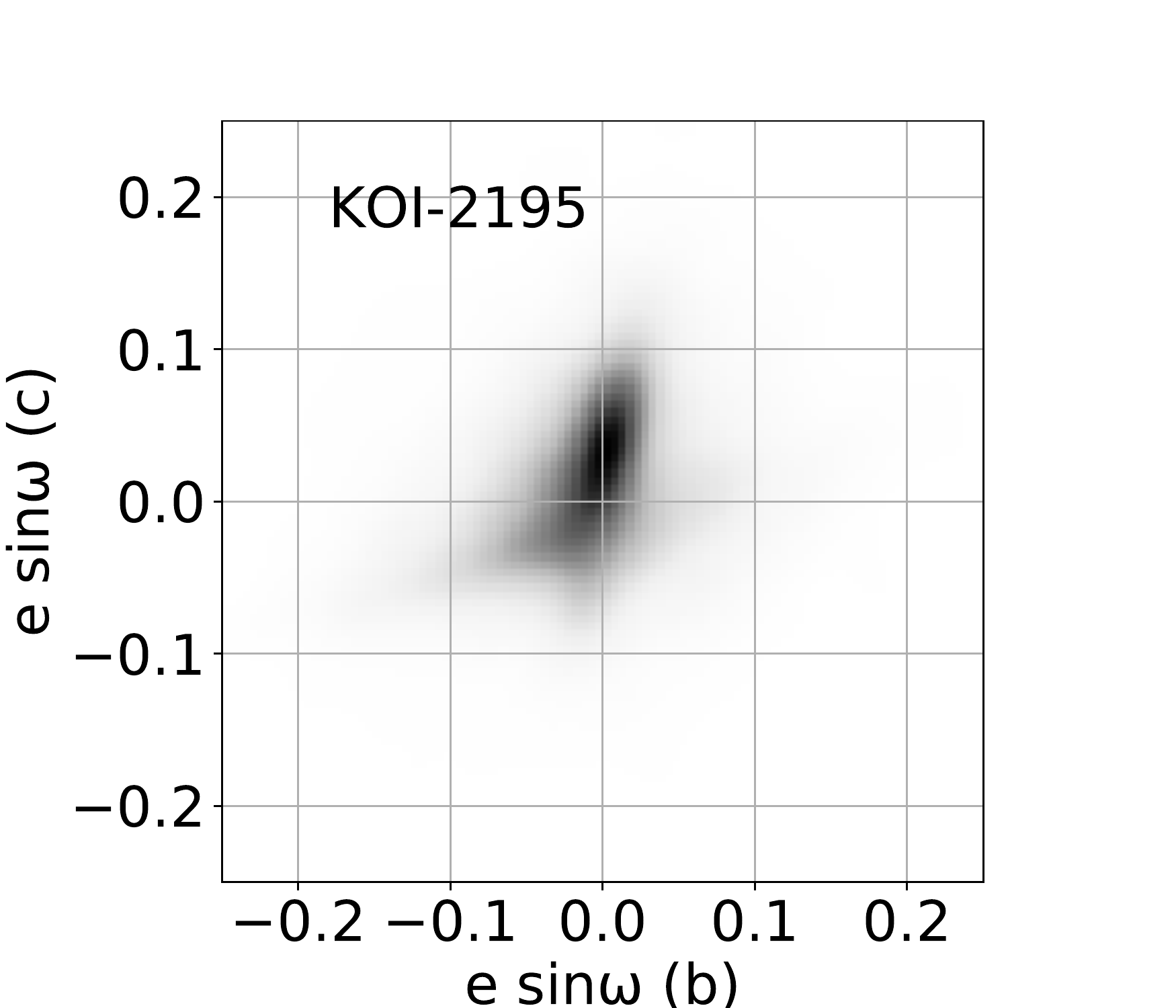}
\includegraphics [height = 1.1 in]{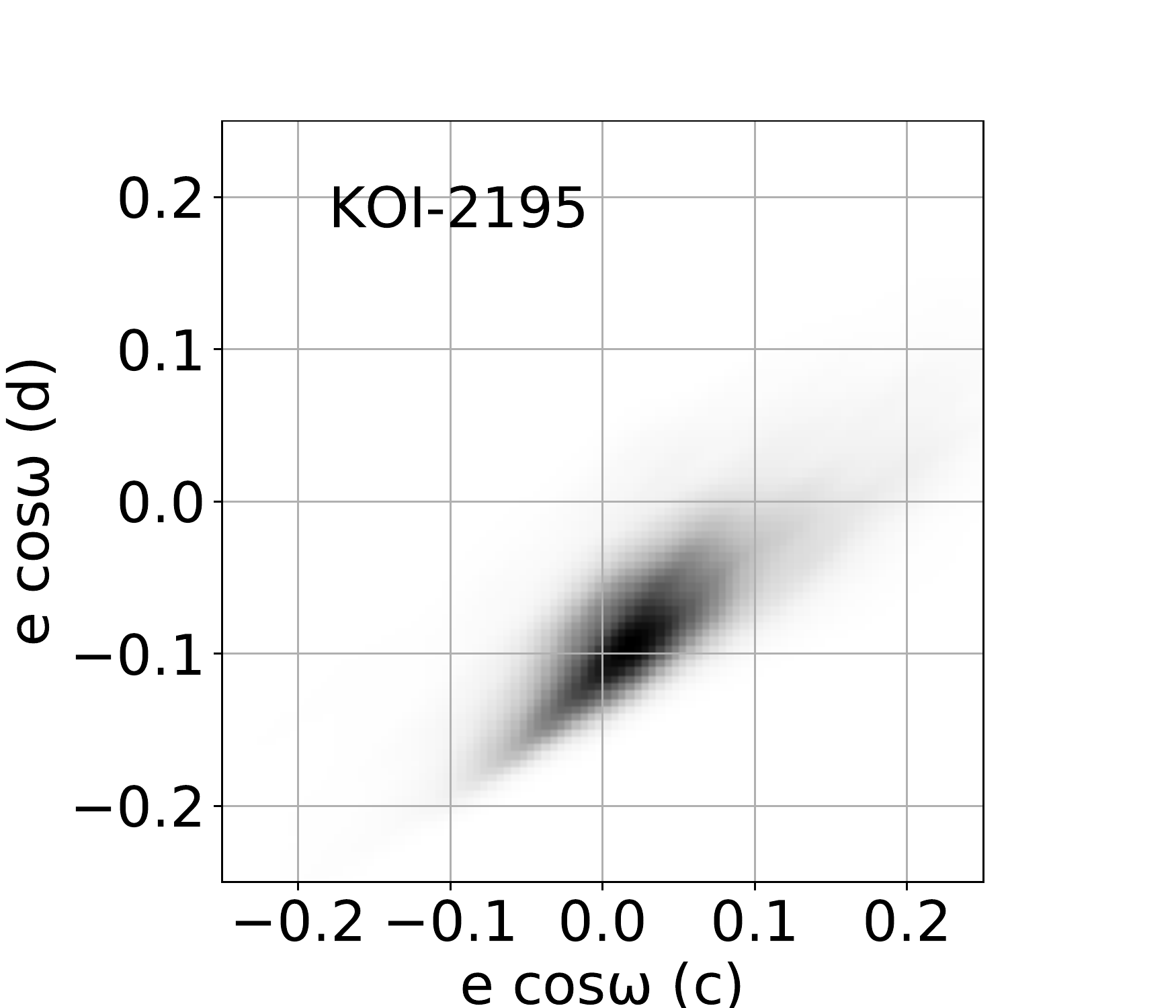}
\includegraphics [height = 1.1 in]{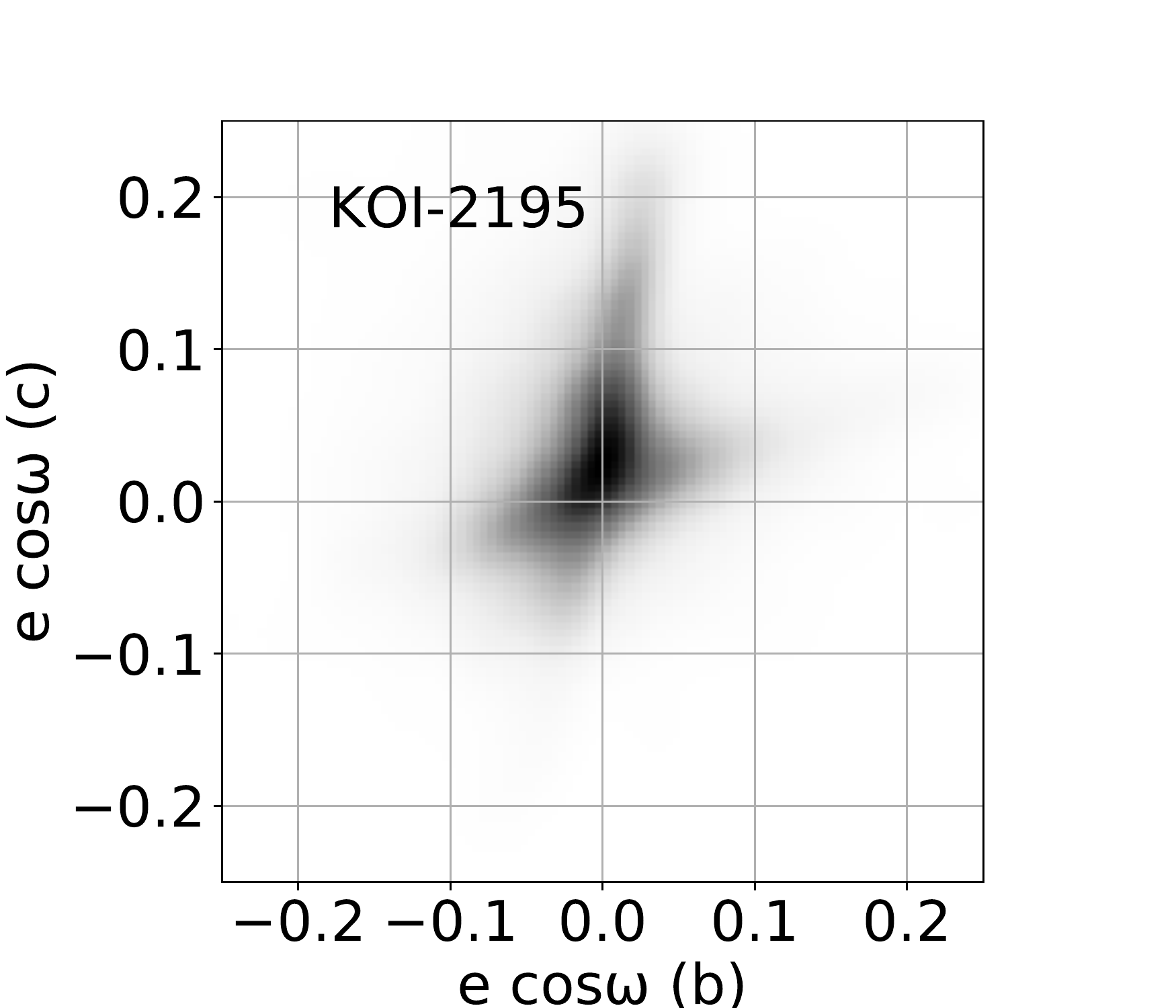} \\
\includegraphics [height = 1.1 in]{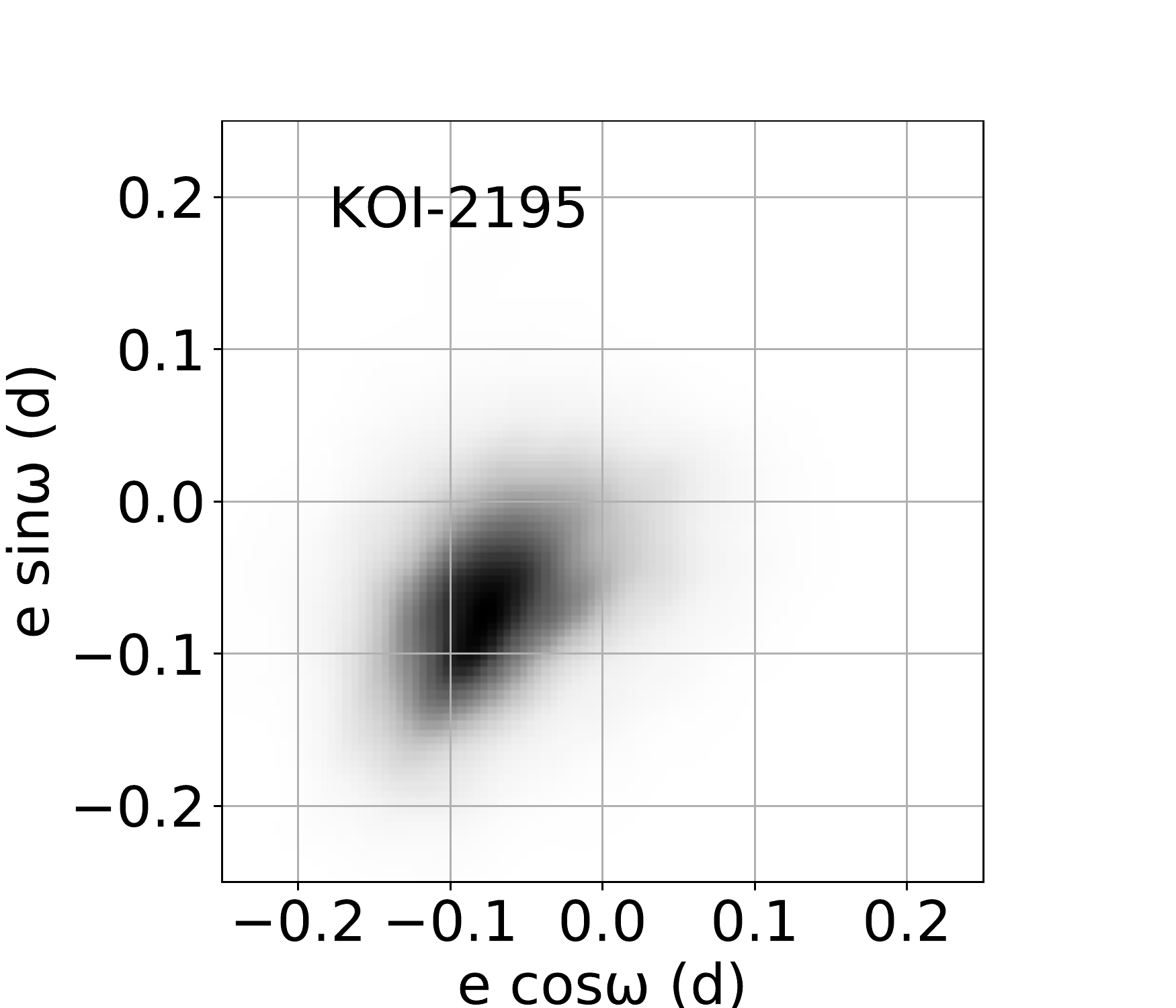}
\includegraphics [height = 1.1 in]{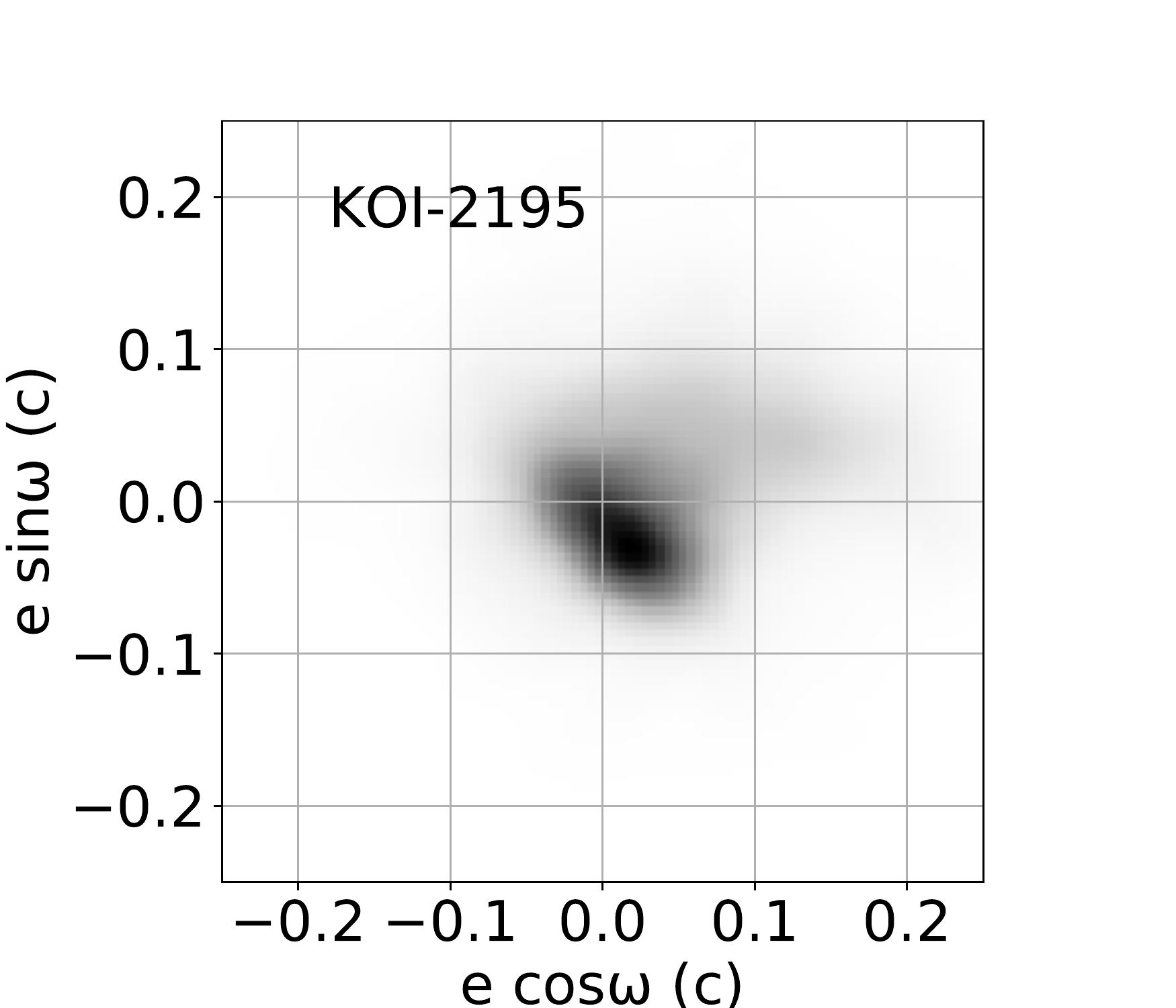}
\includegraphics [height = 1.1 in]{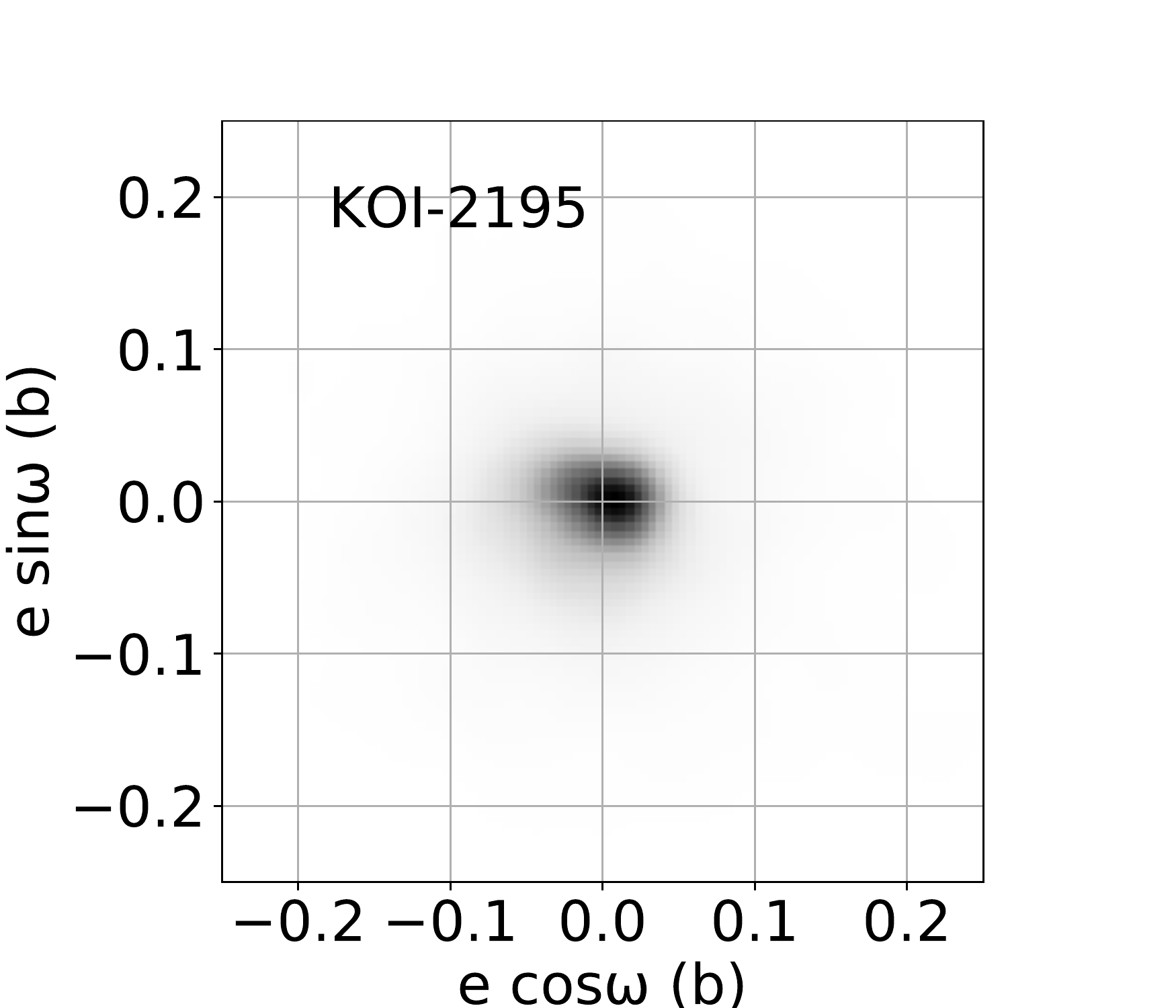}
\caption{Two-dimensional kernel density estimators on joint posteriors of eccentricity vector components: three-planet systems (Part 7 of 7). 
\label{fig:ecc3g} }
\end{center}
\end{figure}

\begin{figure}
\begin{center}
\figurenum{30}
\includegraphics [height = 1.1 in]{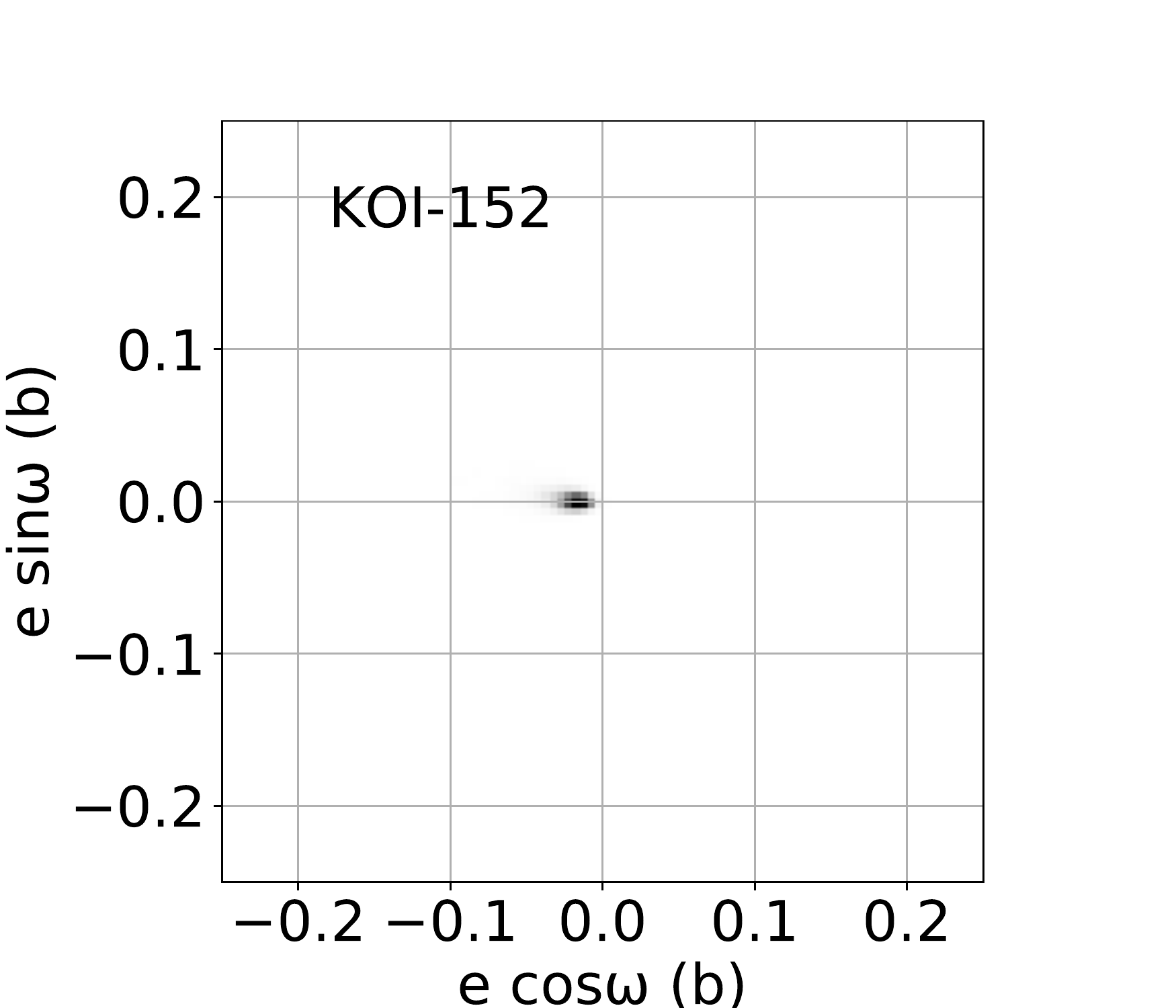}
\includegraphics [height = 1.1 in]{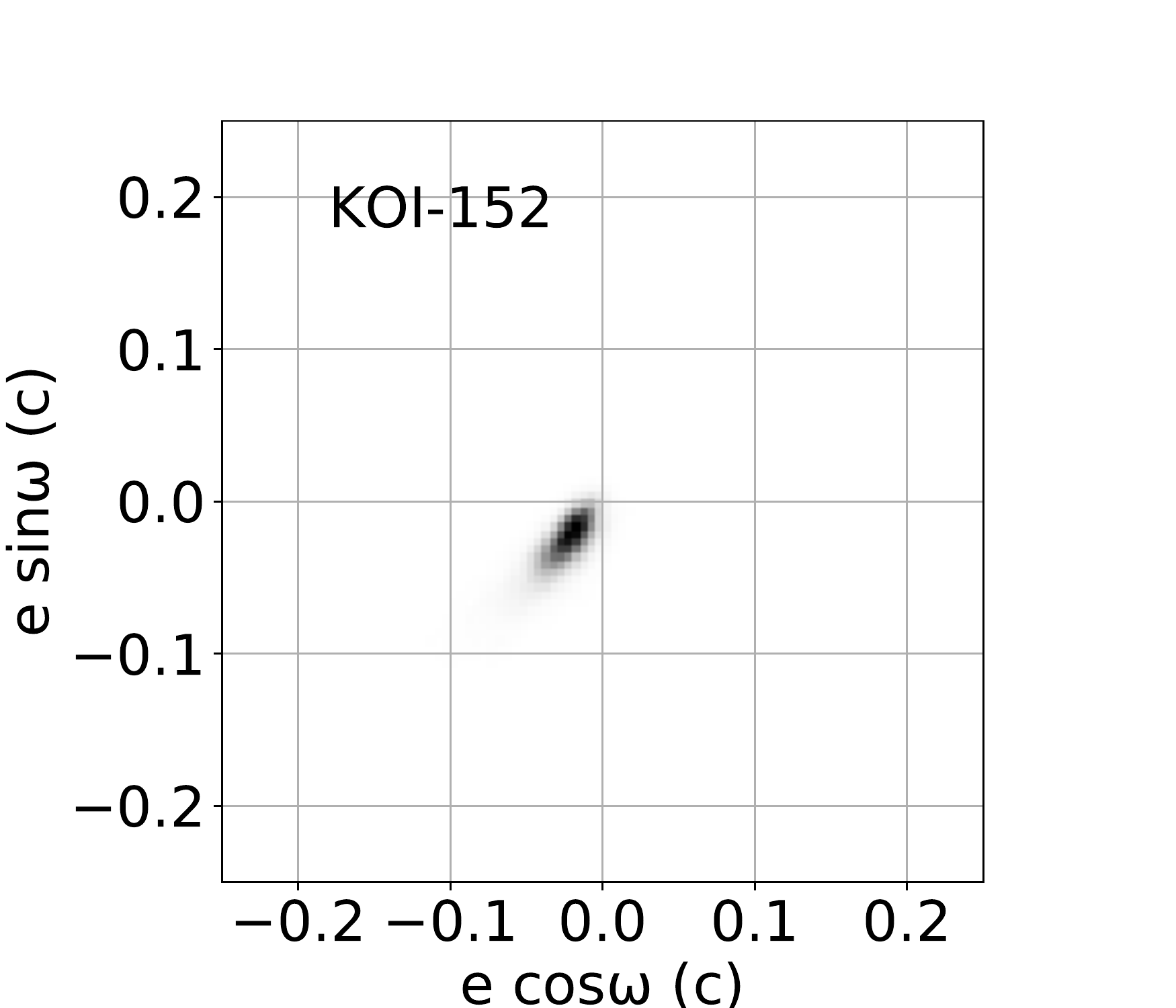}
\includegraphics [height = 1.1 in]{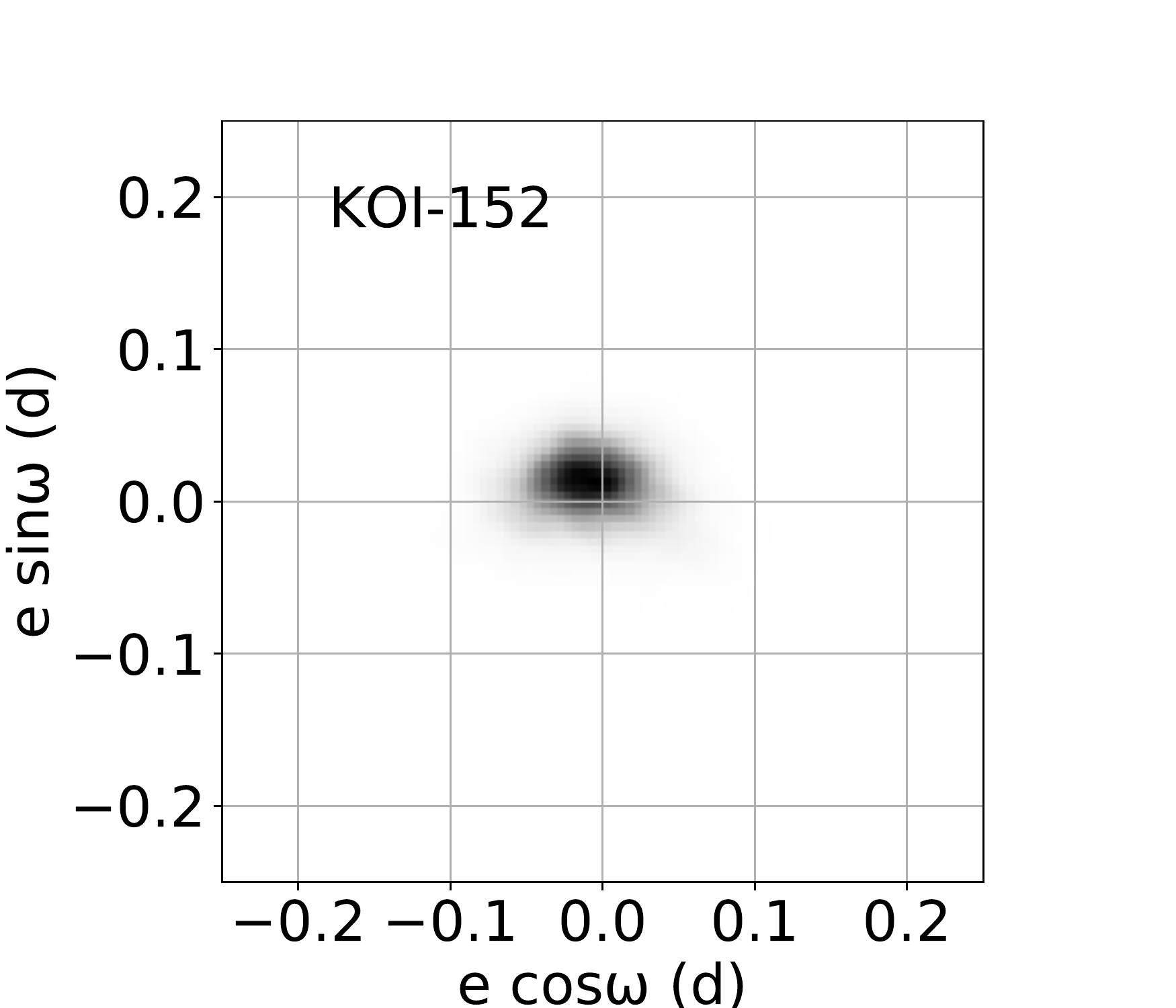}
\includegraphics [height = 1.1 in]{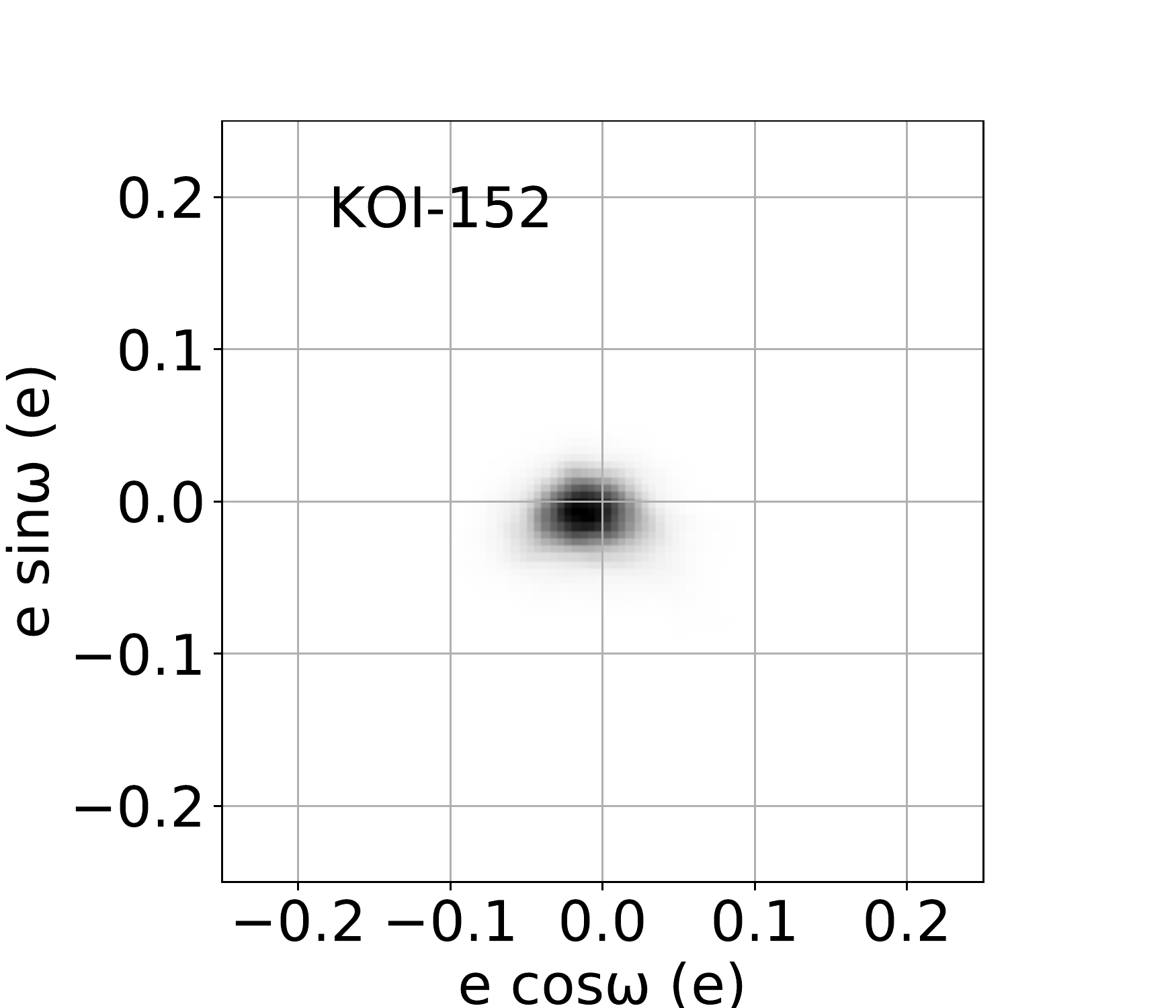} \\
\includegraphics [height = 1.1 in]{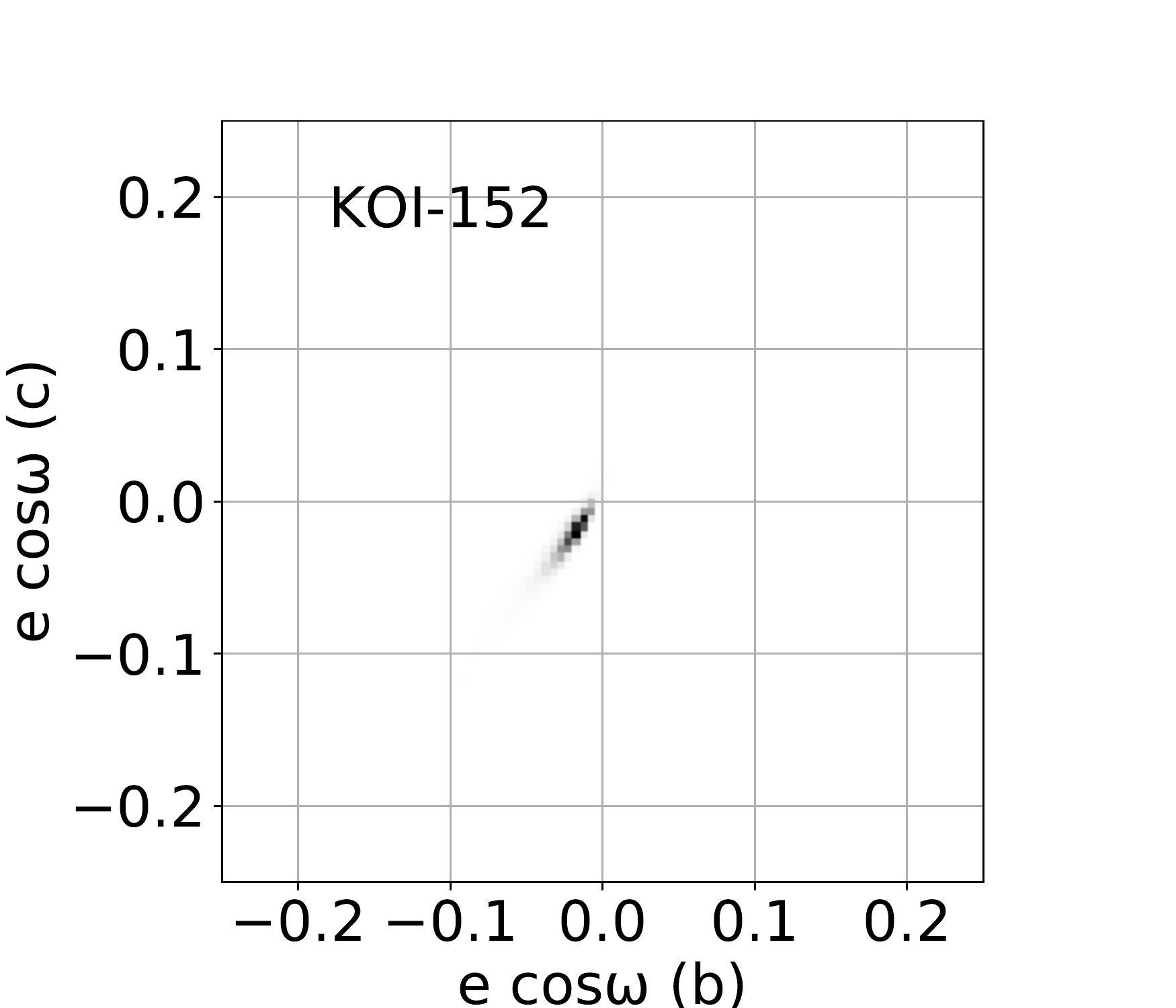}
\includegraphics [height = 1.1 in]{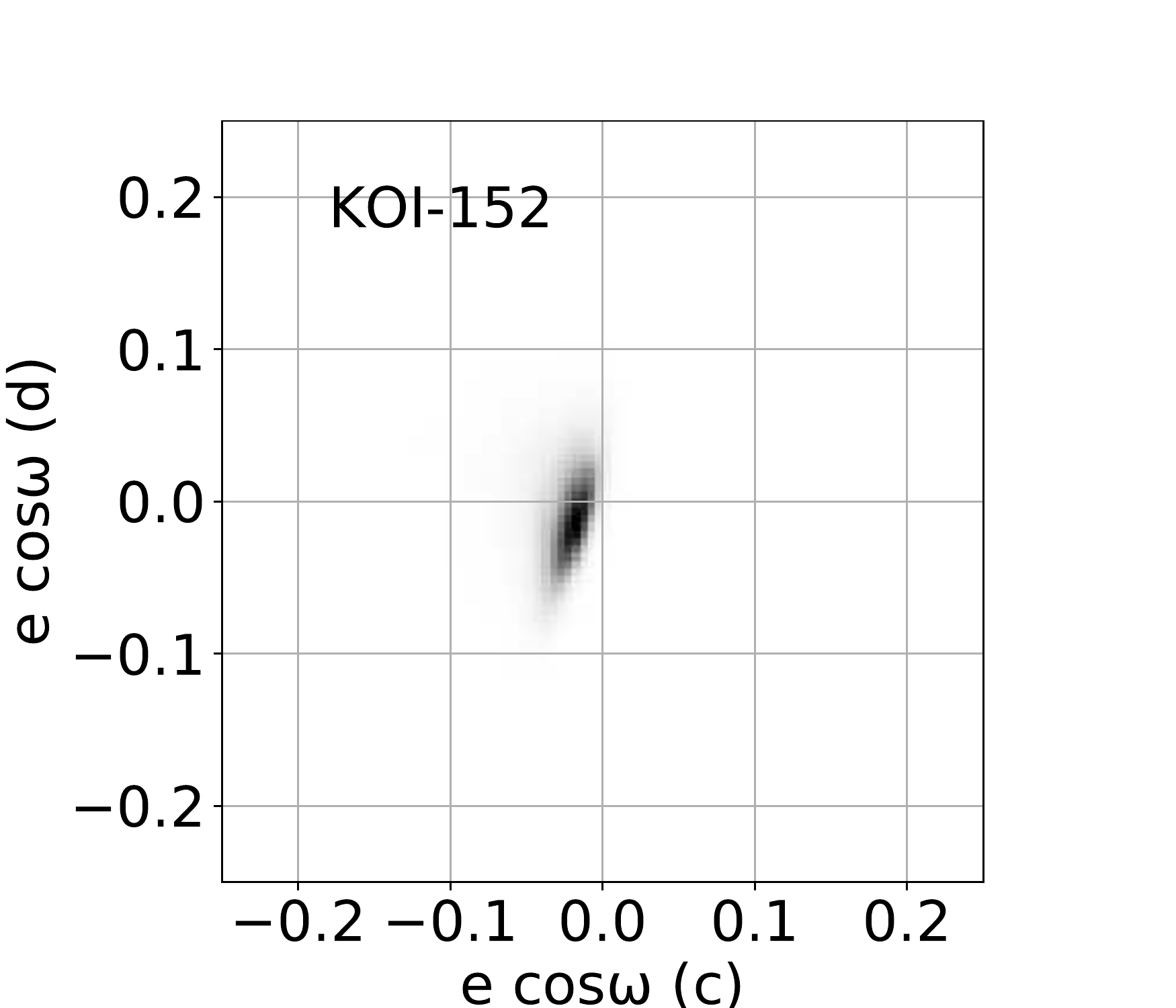}
\includegraphics [height = 1.1 in]{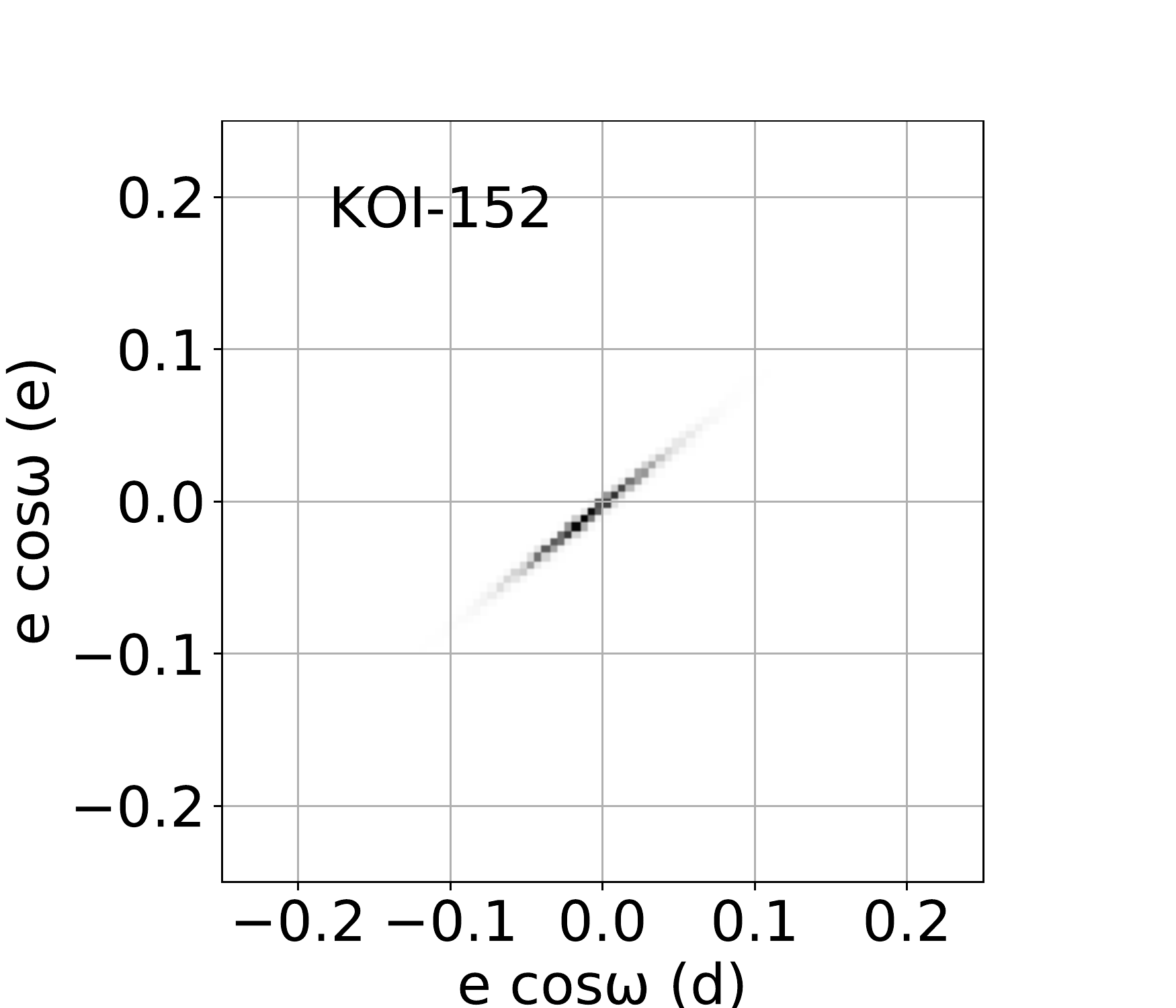}
\includegraphics [height = 1.1 in]{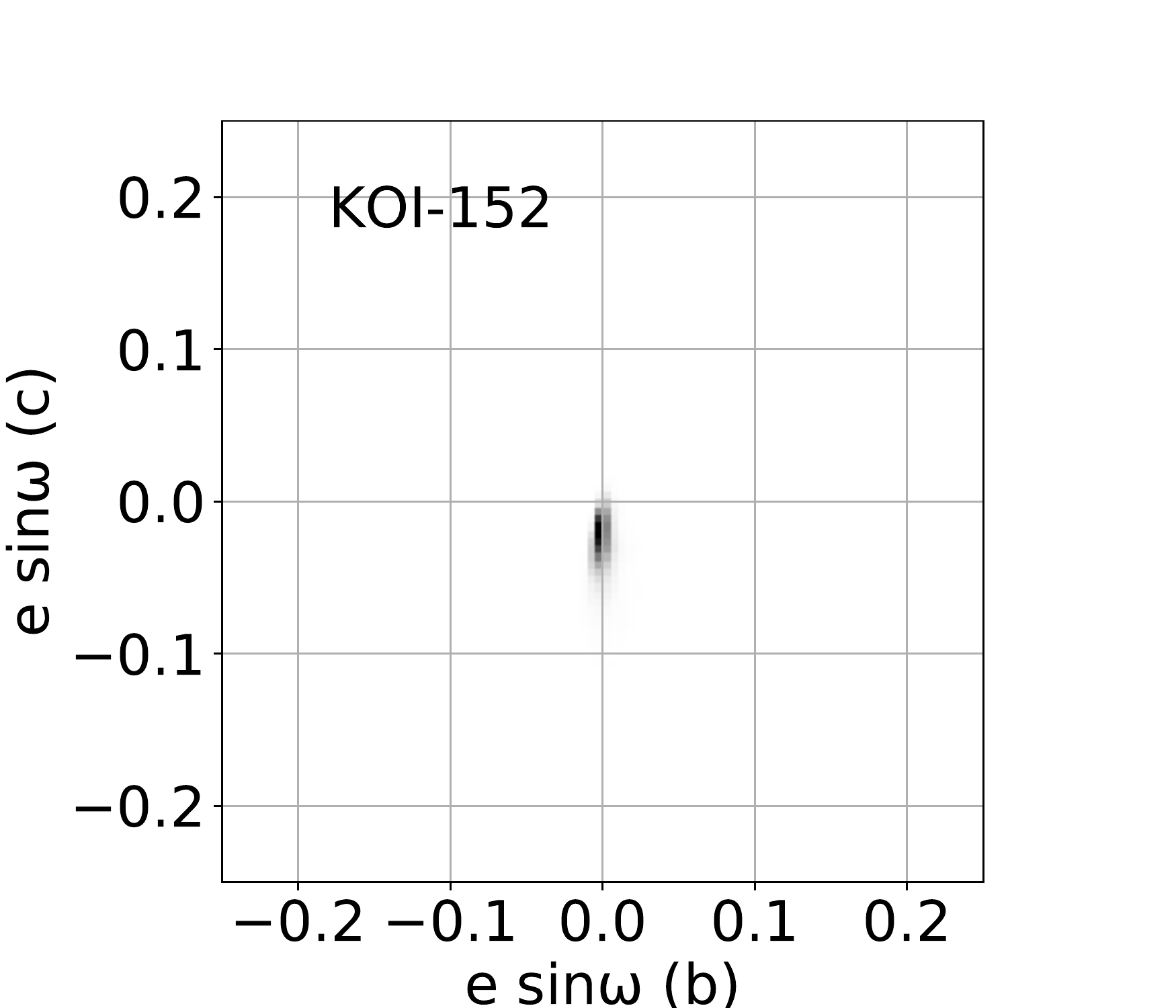} \\
\includegraphics [height = 1.1 in]{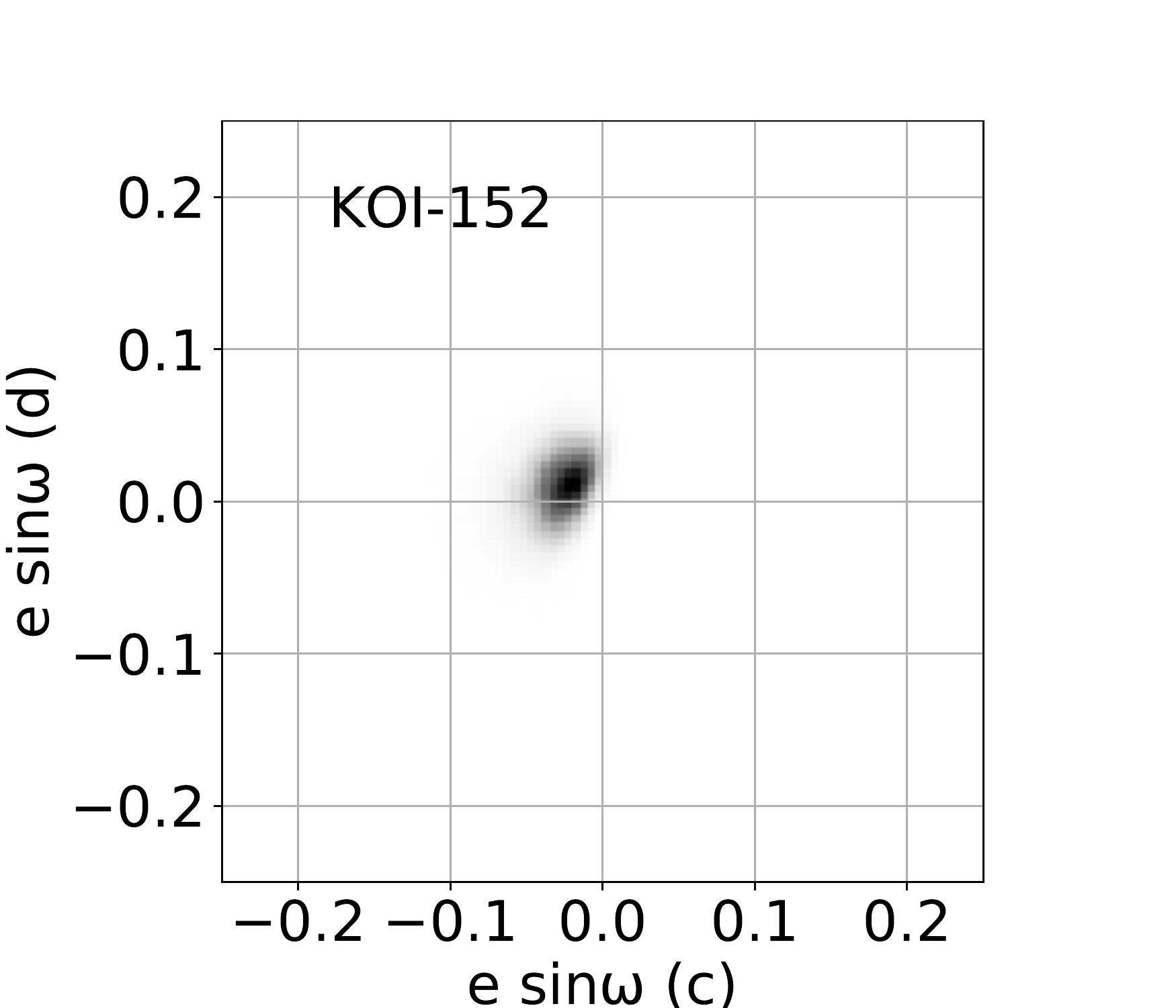}
\includegraphics [height = 1.1 in]{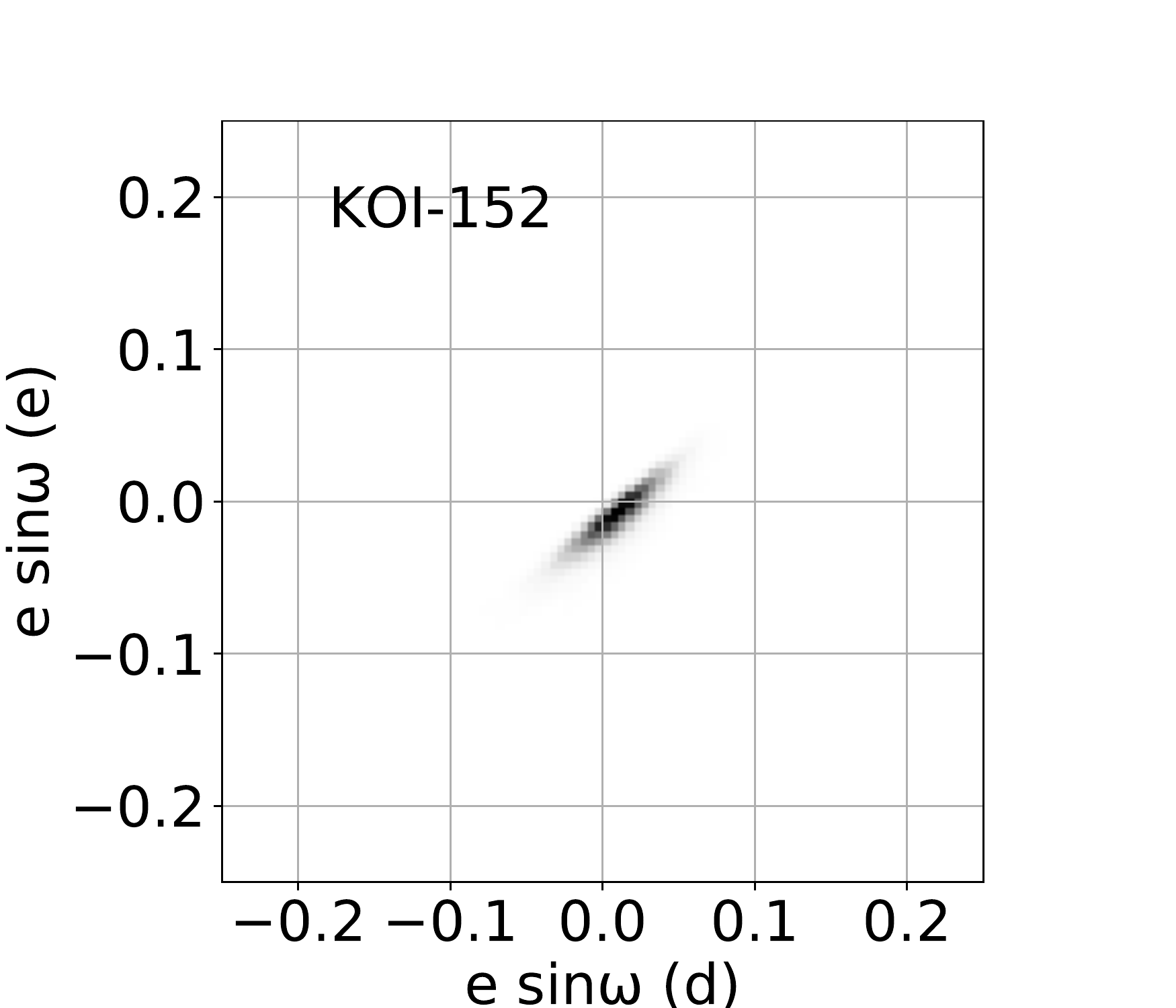}
\includegraphics [height = 1.1 in]{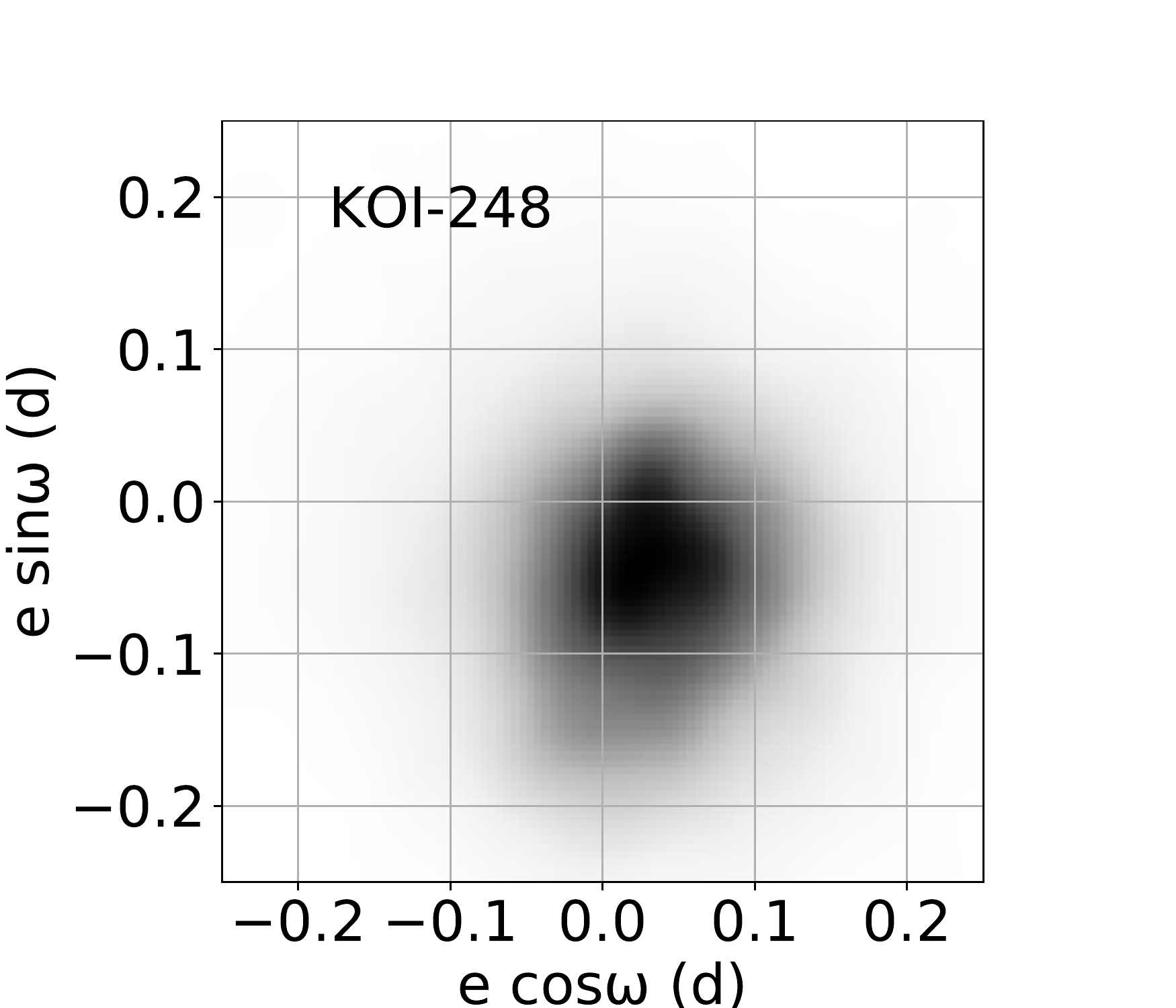}
\includegraphics [height = 1.1 in]{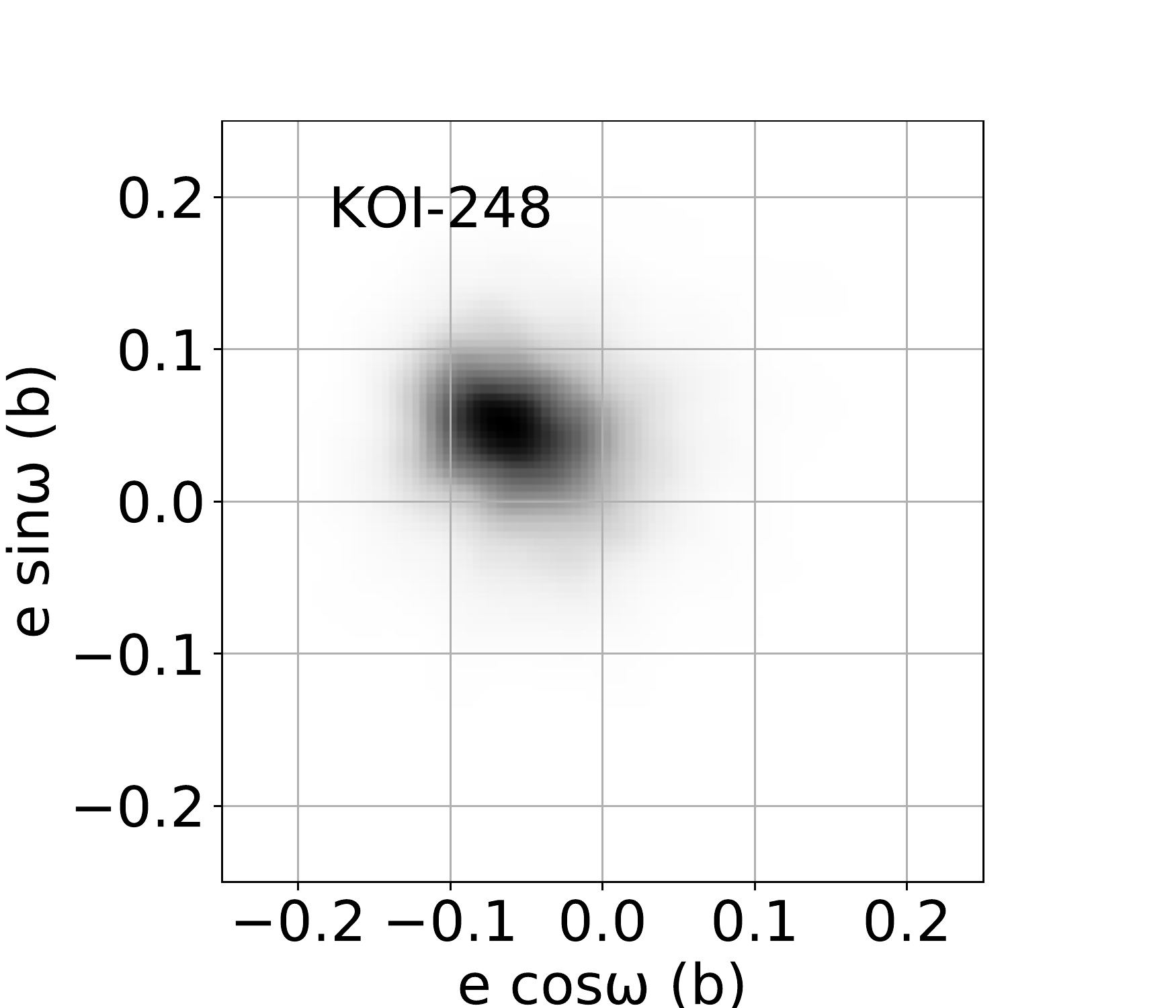} \\
\includegraphics [height = 1.1 in]{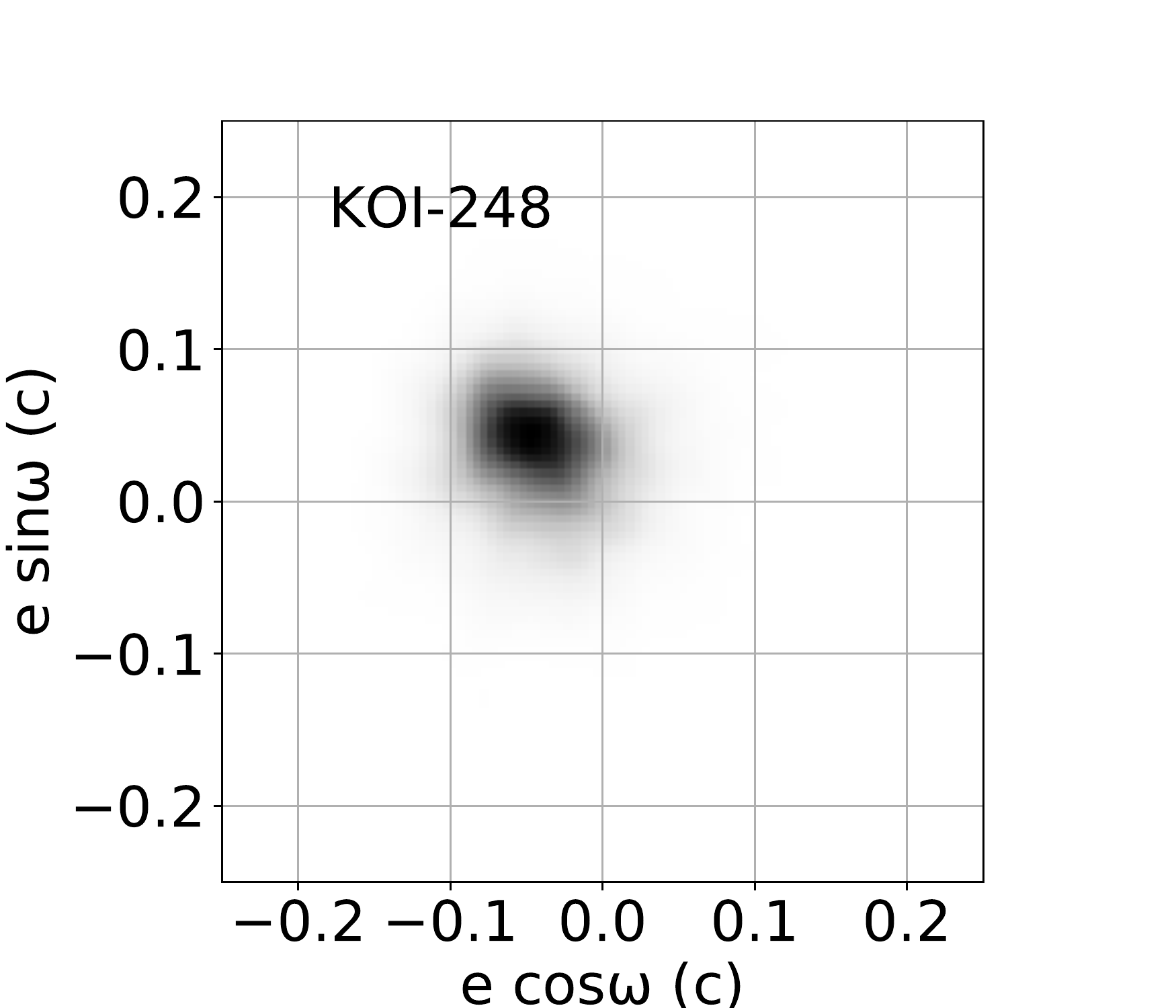}
\includegraphics [height = 1.1 in]{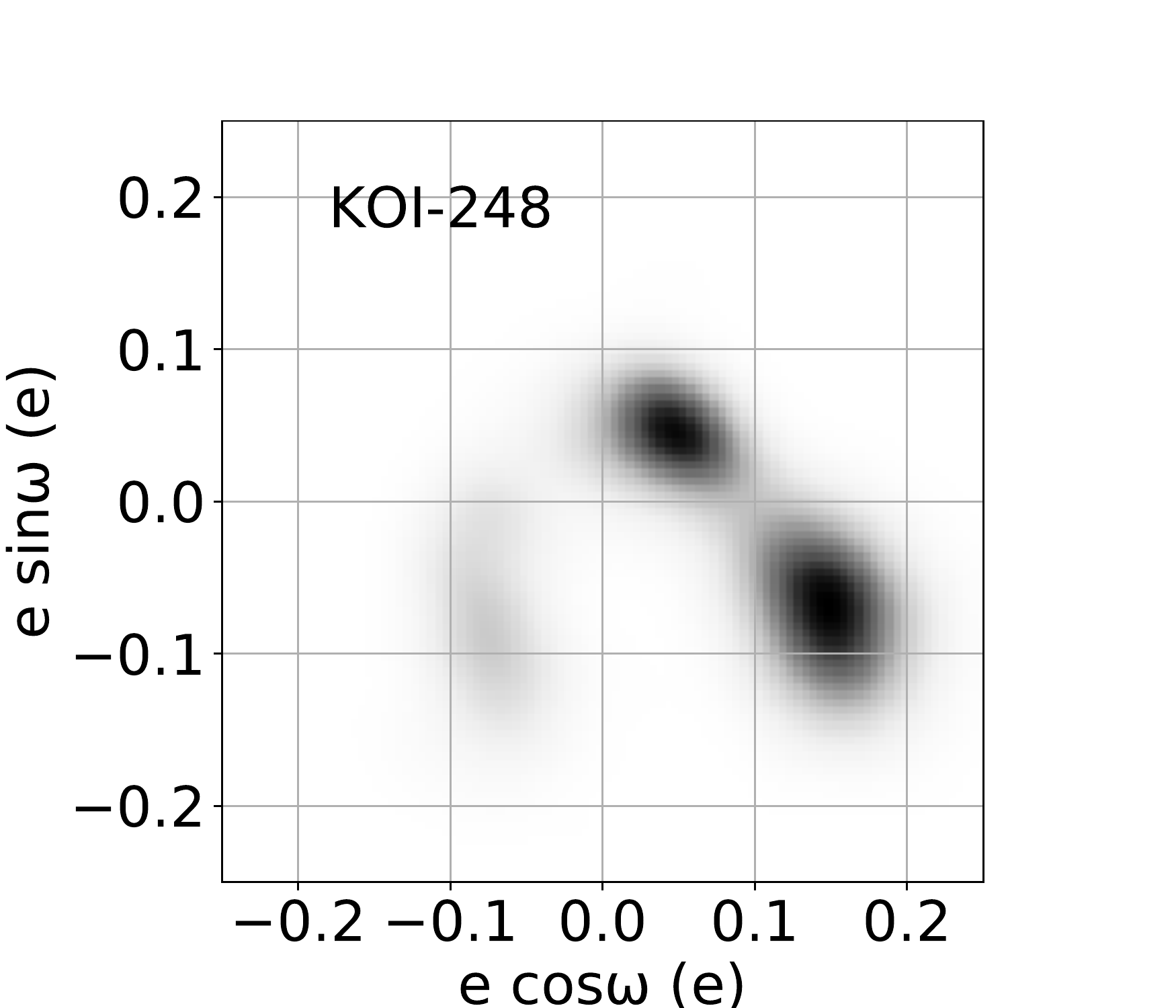}
\includegraphics [height = 1.1 in]{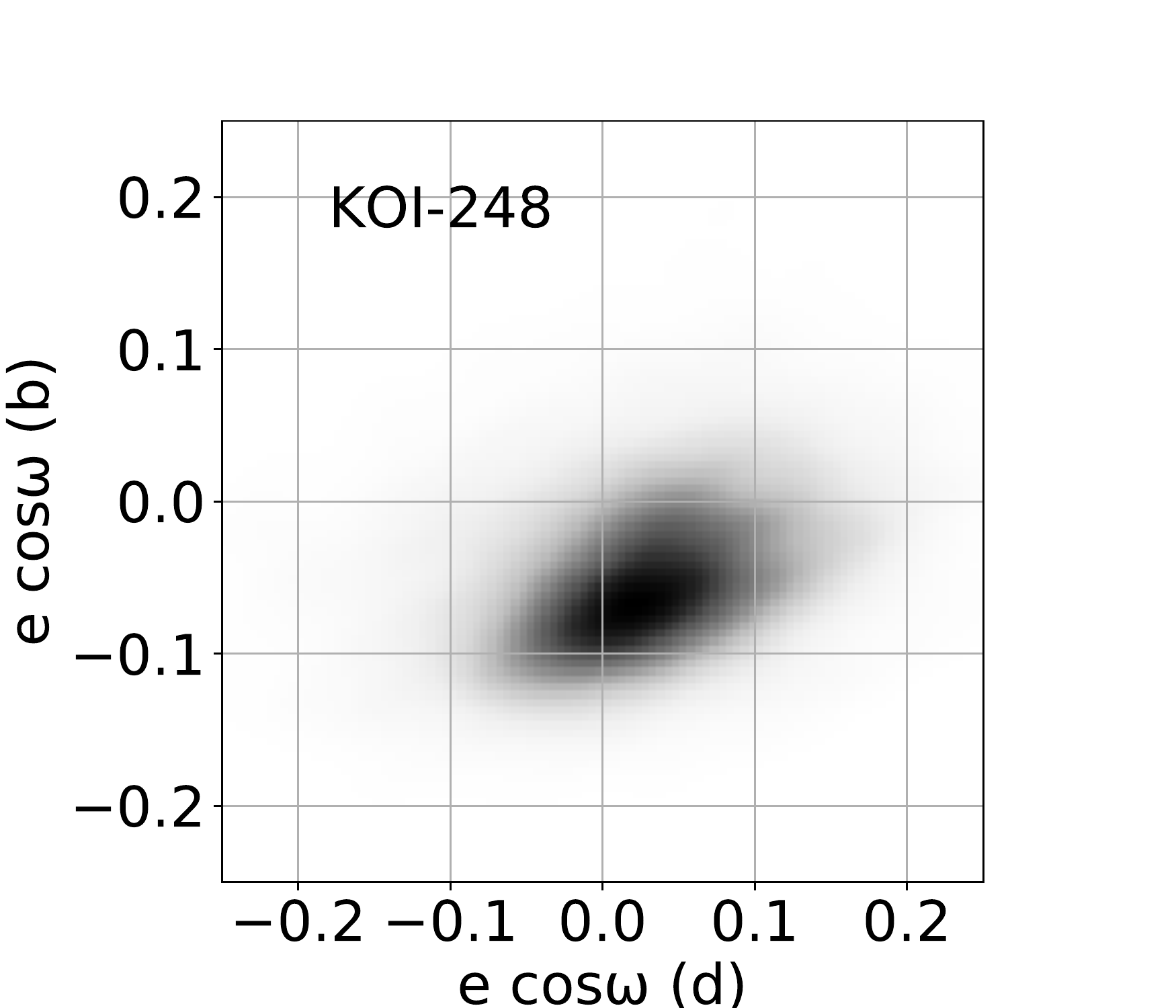}
\includegraphics [height = 1.1 in]{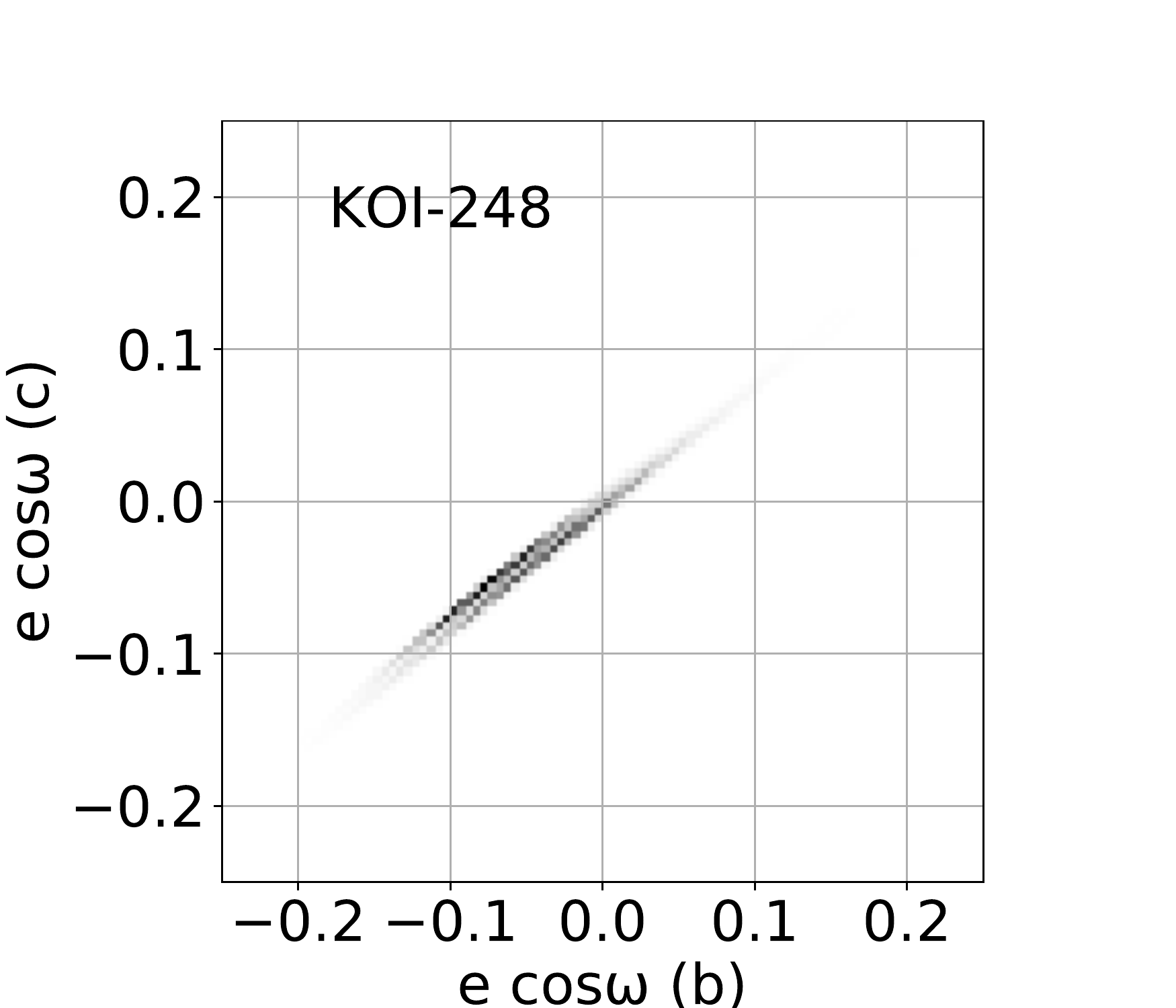} \\
\includegraphics [height = 1.1 in]{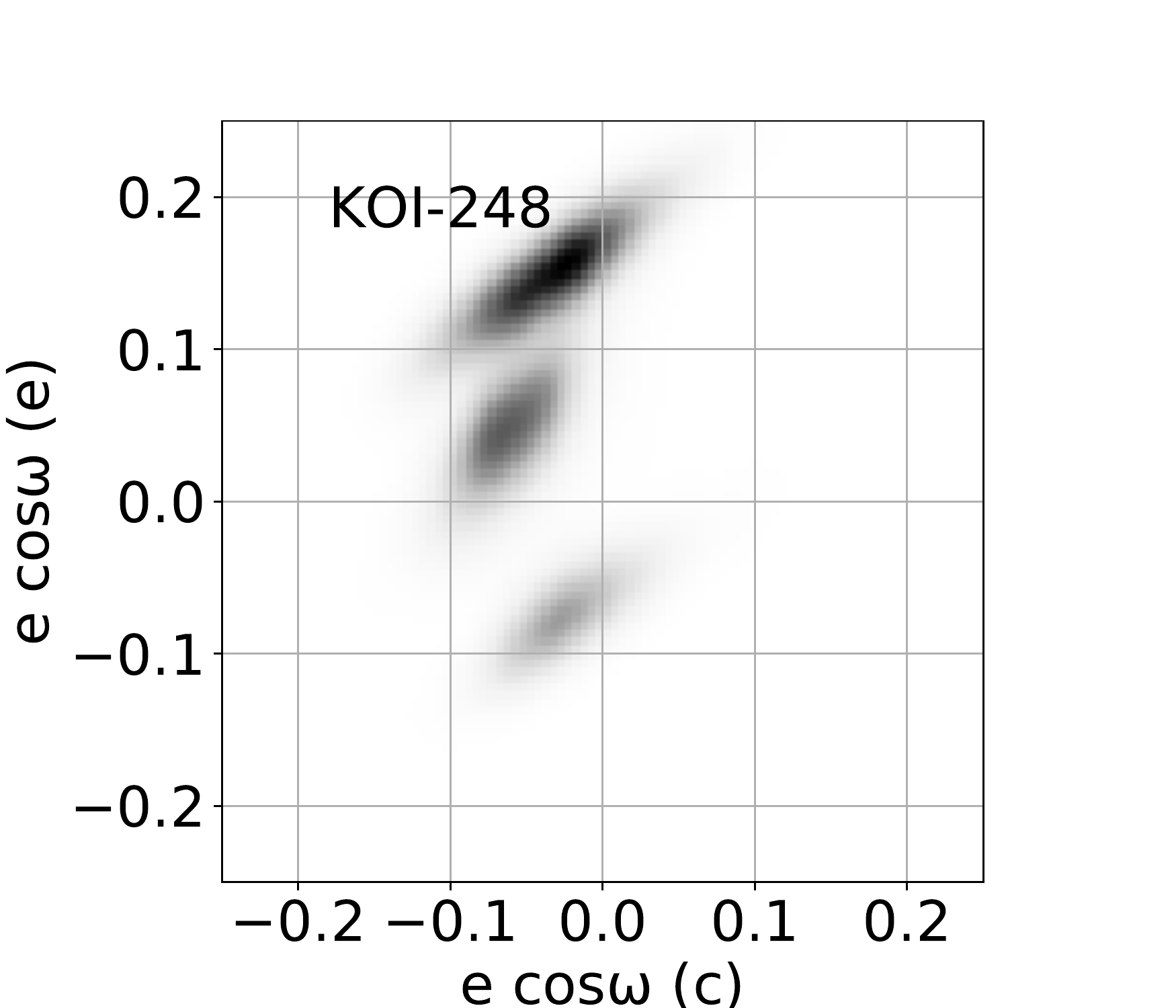}
\includegraphics [height = 1.1 in]{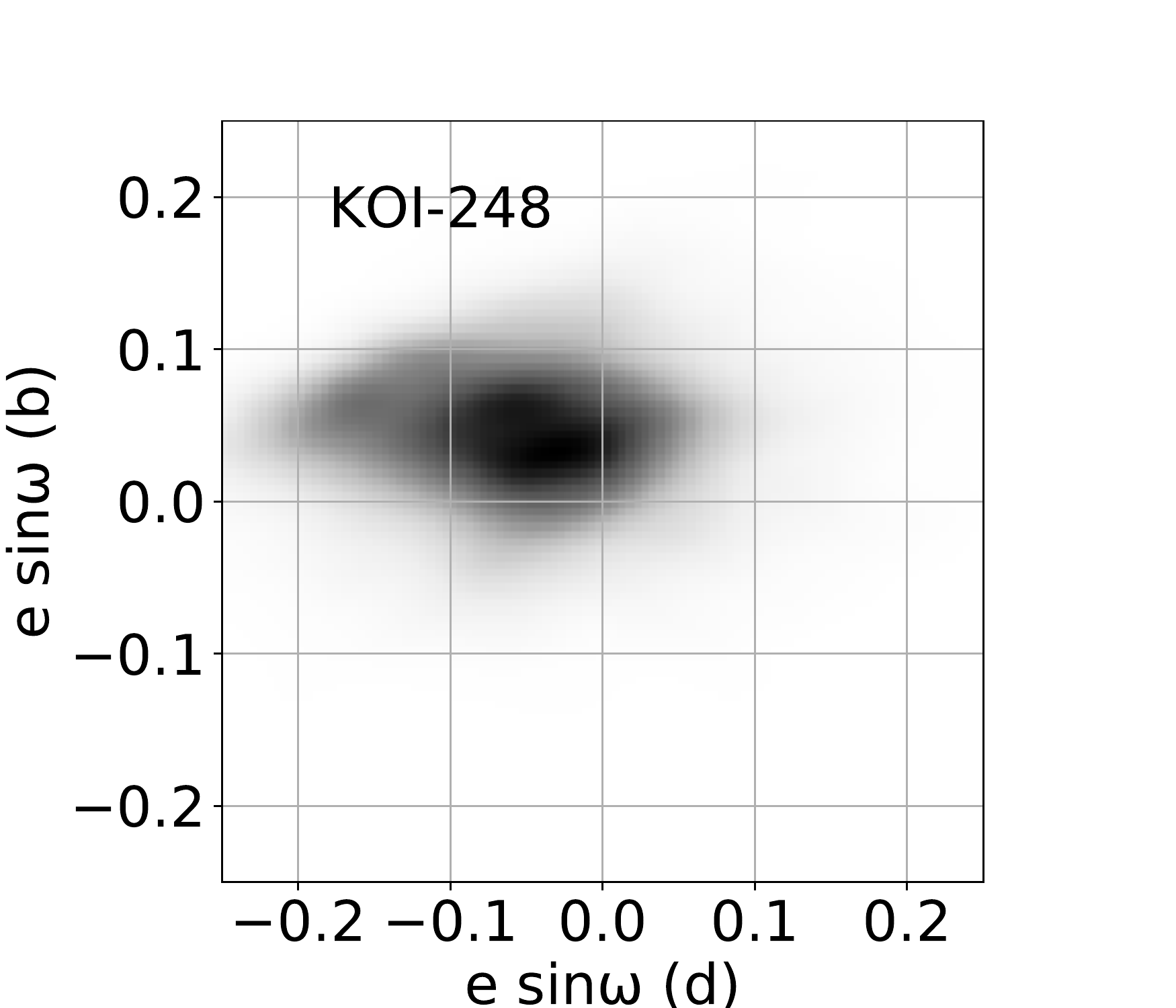} 
\includegraphics [height = 1.1 in]{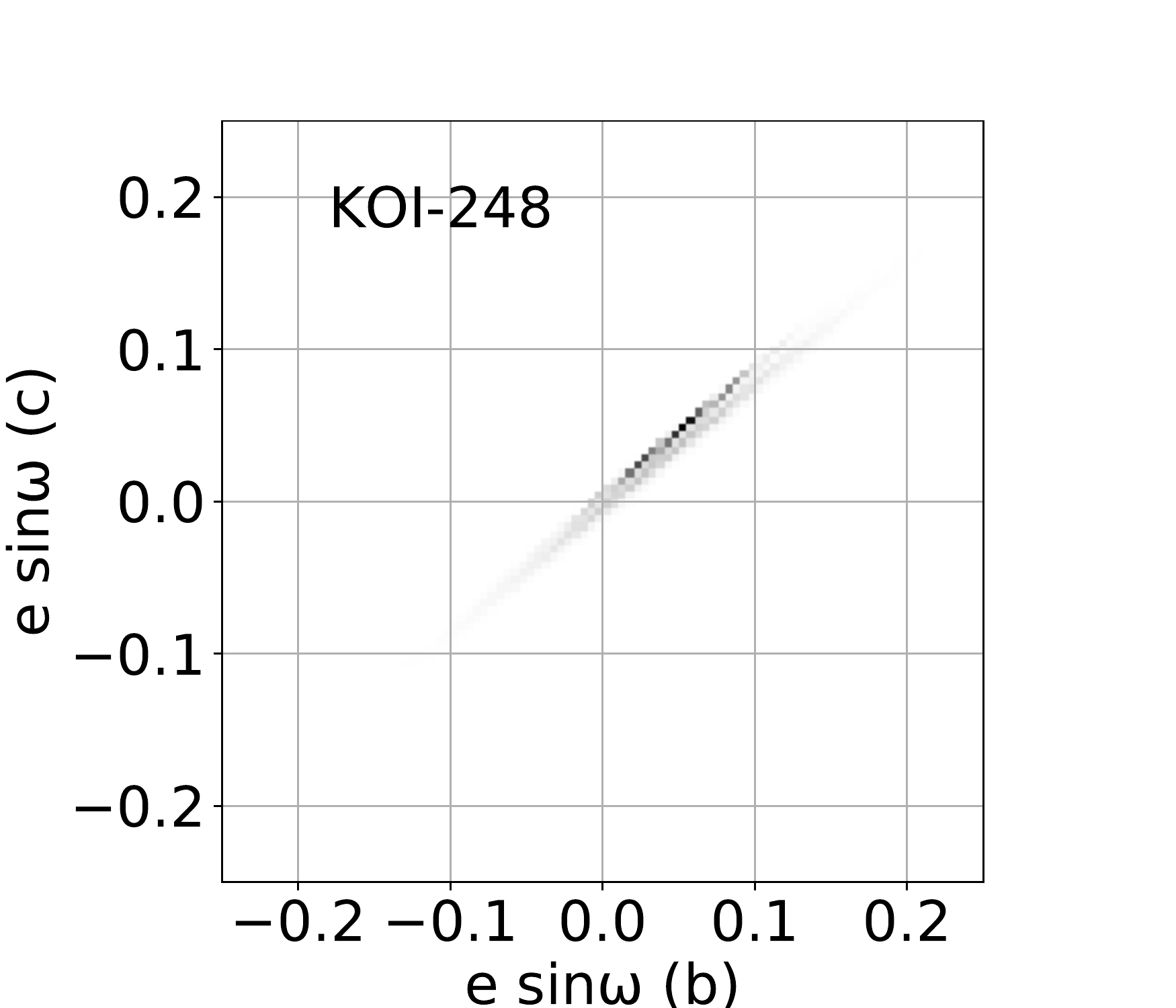}
\includegraphics [height = 1.1 in]{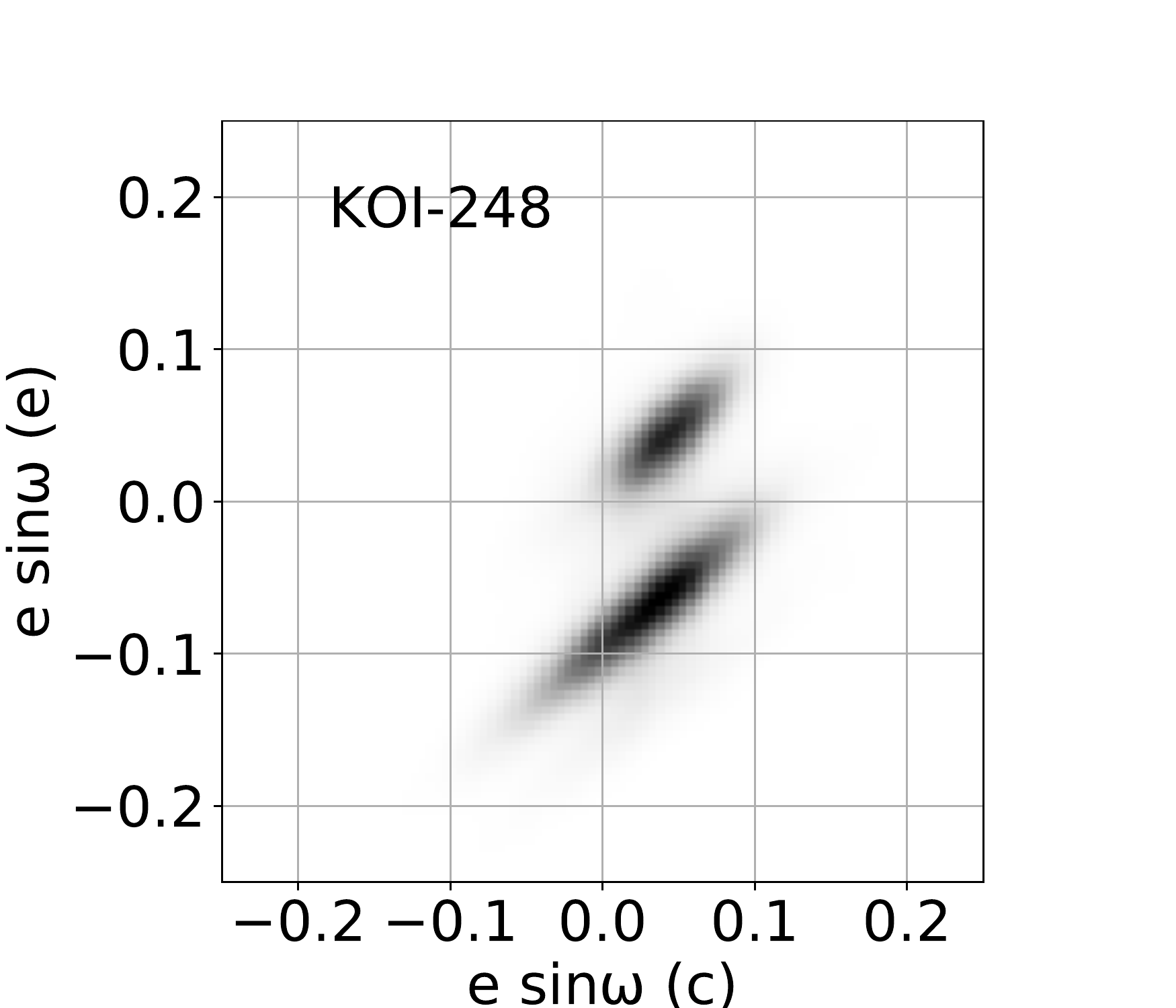} \\
\caption{Two-dimensional kernel density estimators on joint posteriors of eccentricity vector components: four-planet systems. (Part 1 of 4). 
\label{fig:ecc4a} }
\end{center}
\end{figure}

\begin{figure}
\begin{center}
\figurenum{31}
\includegraphics [height = 1.1 in]{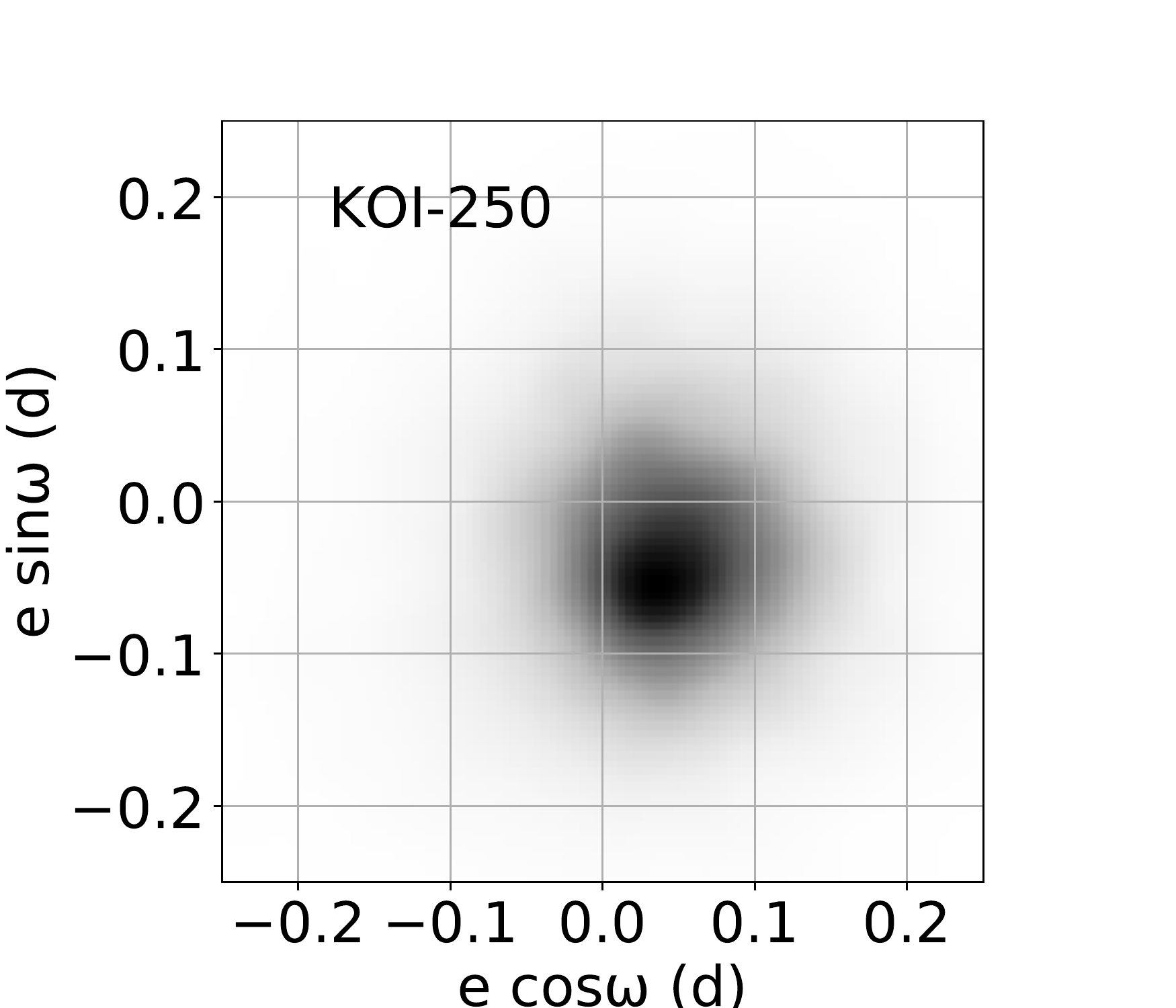}
\includegraphics [height = 1.1 in]{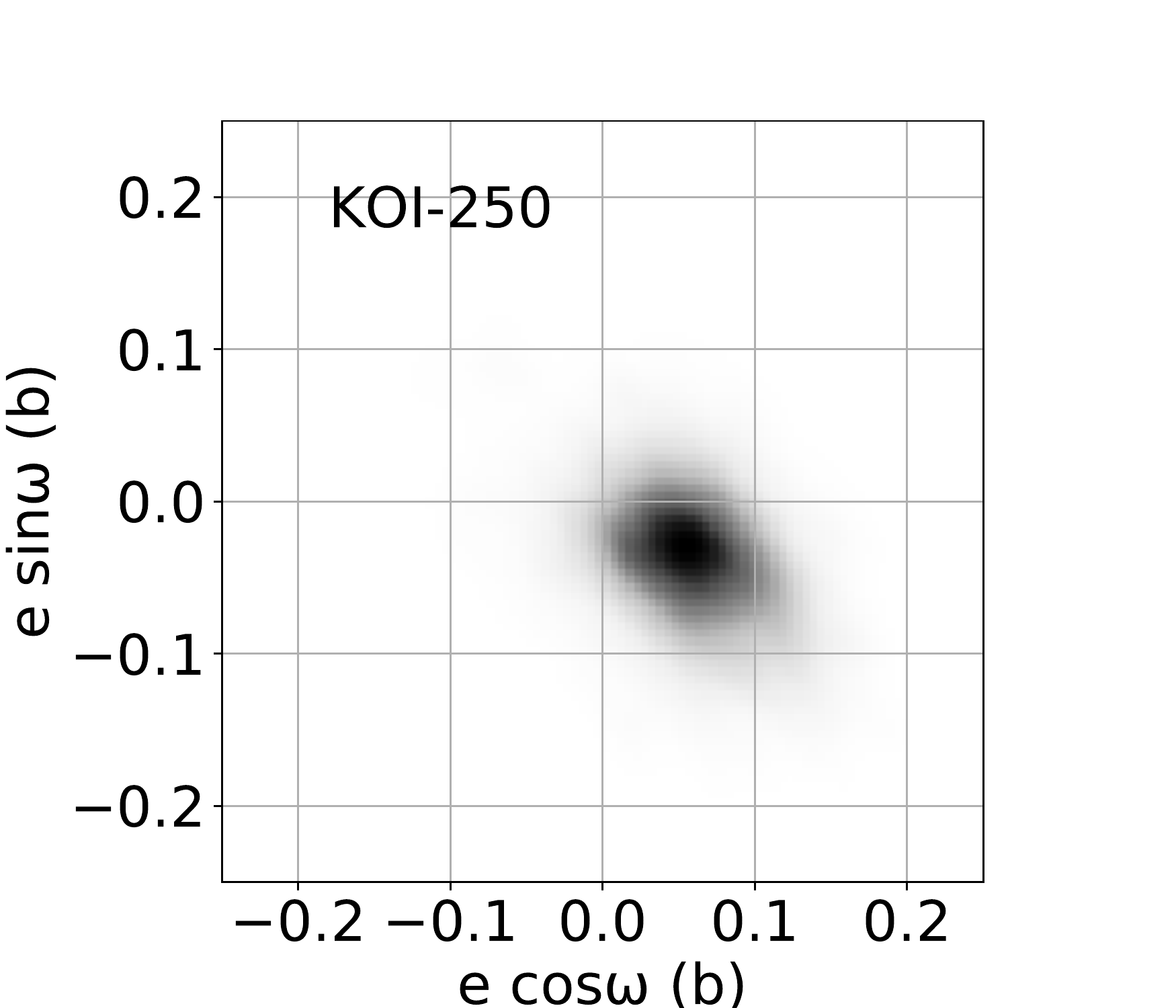}
\includegraphics [height = 1.1 in]{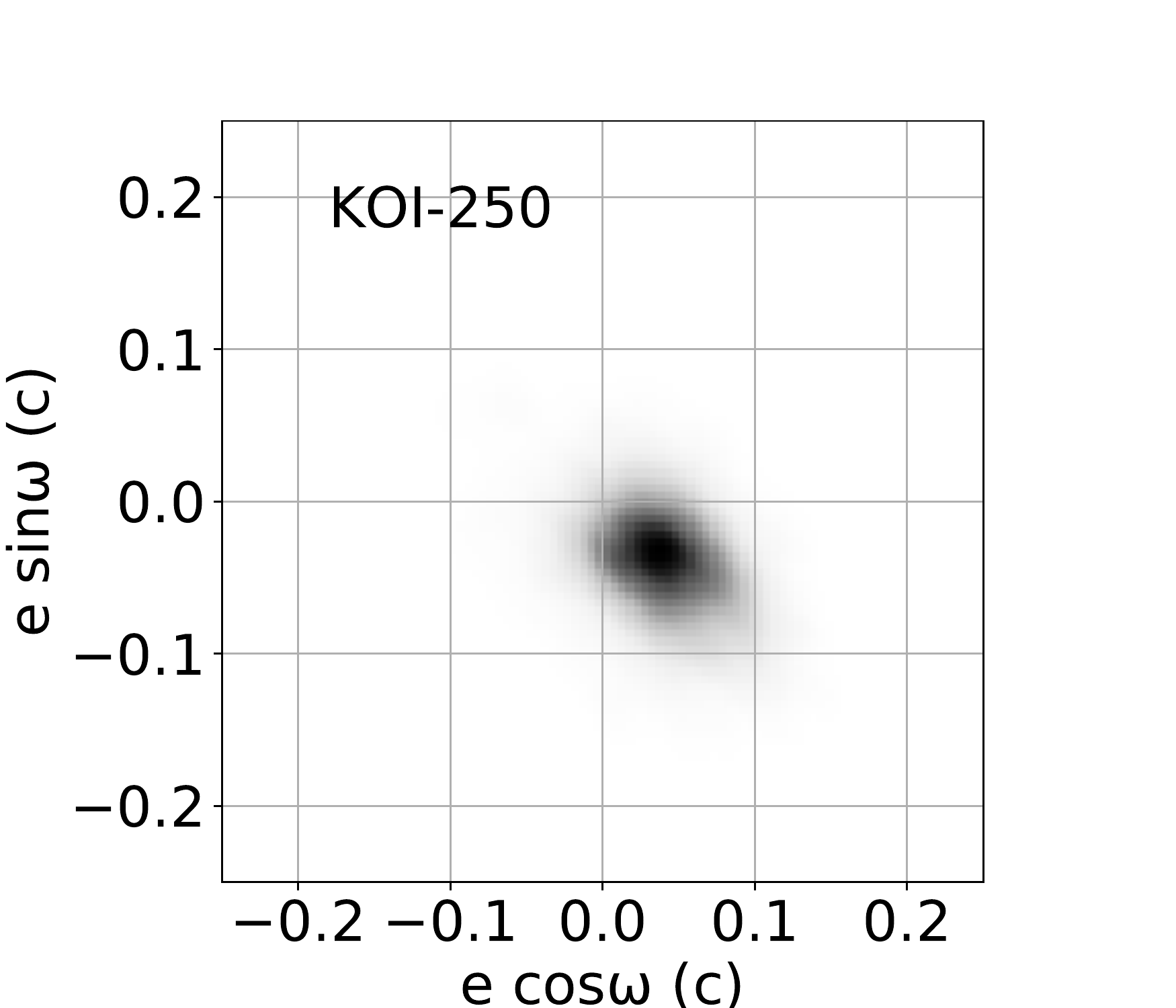}
\includegraphics [height = 1.1 in]{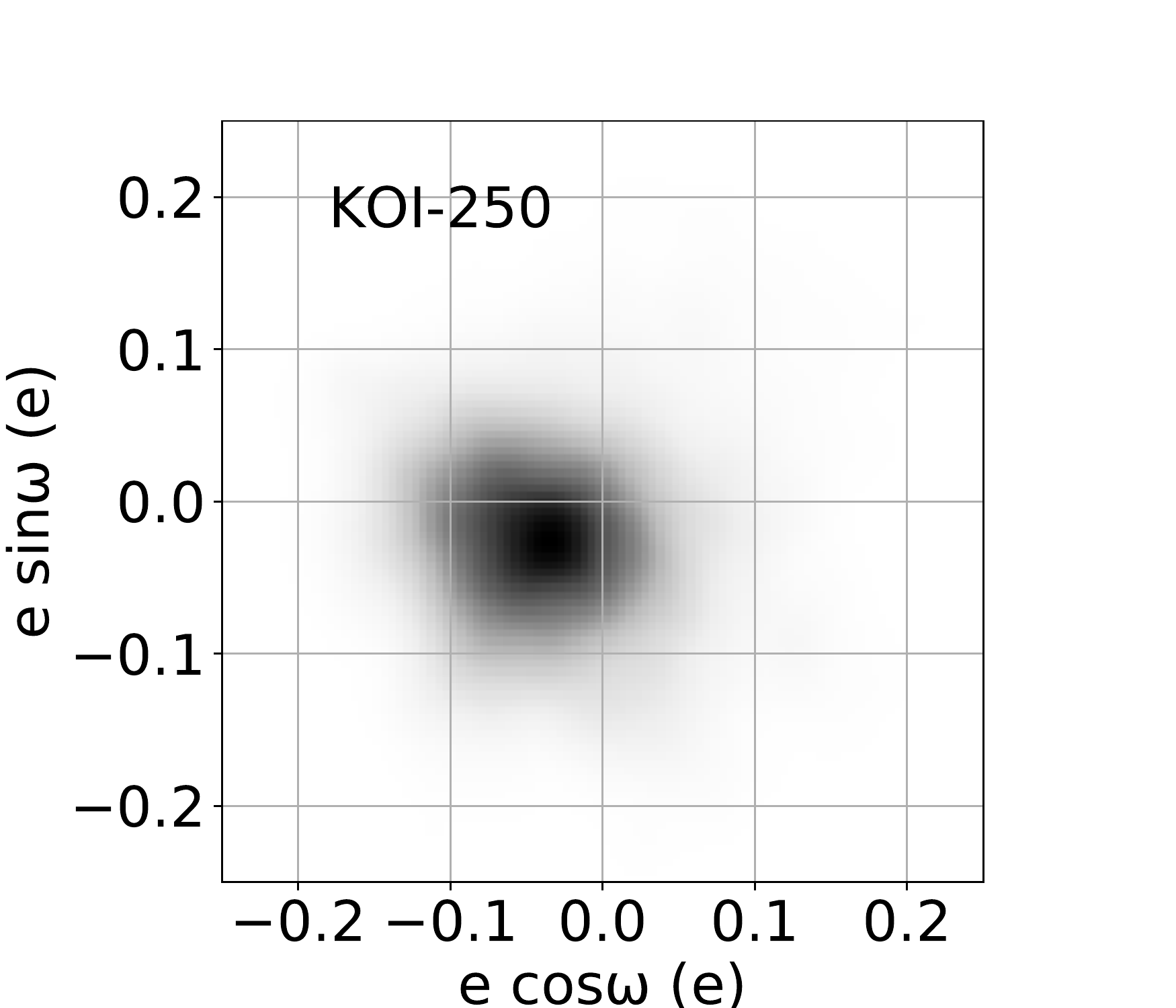} \\
\includegraphics [height = 1.1 in]{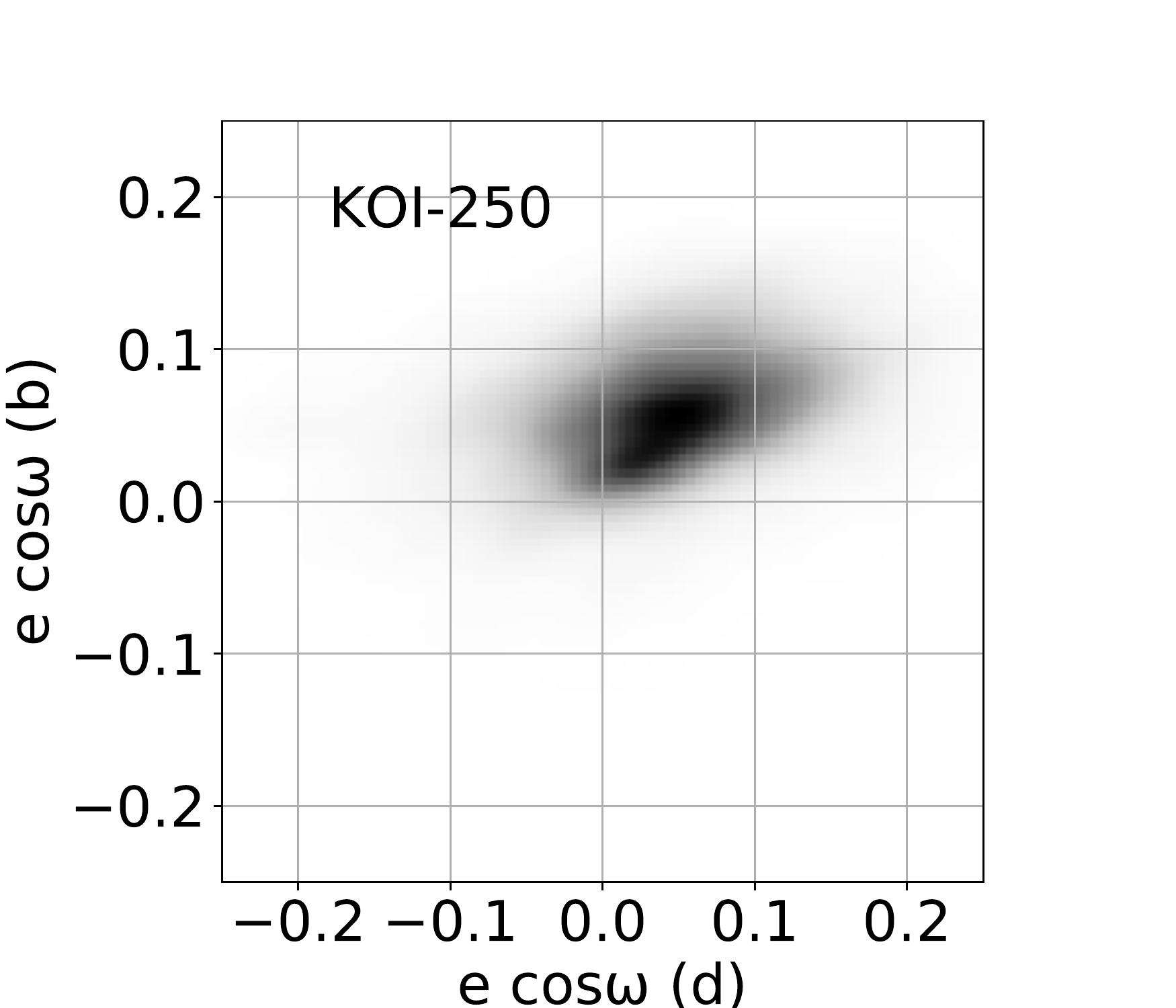}
\includegraphics [height = 1.1 in]{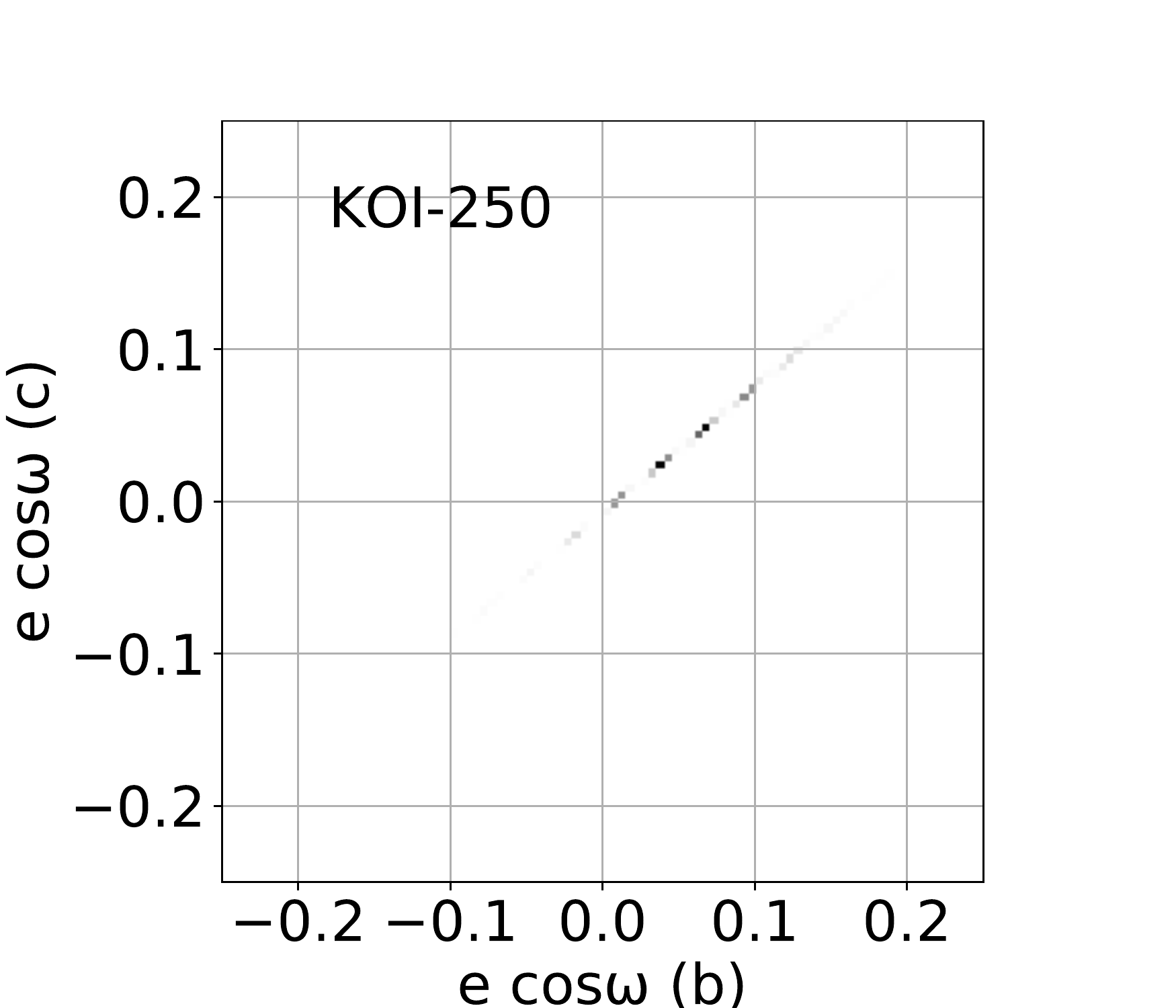}
\includegraphics [height = 1.1 in]{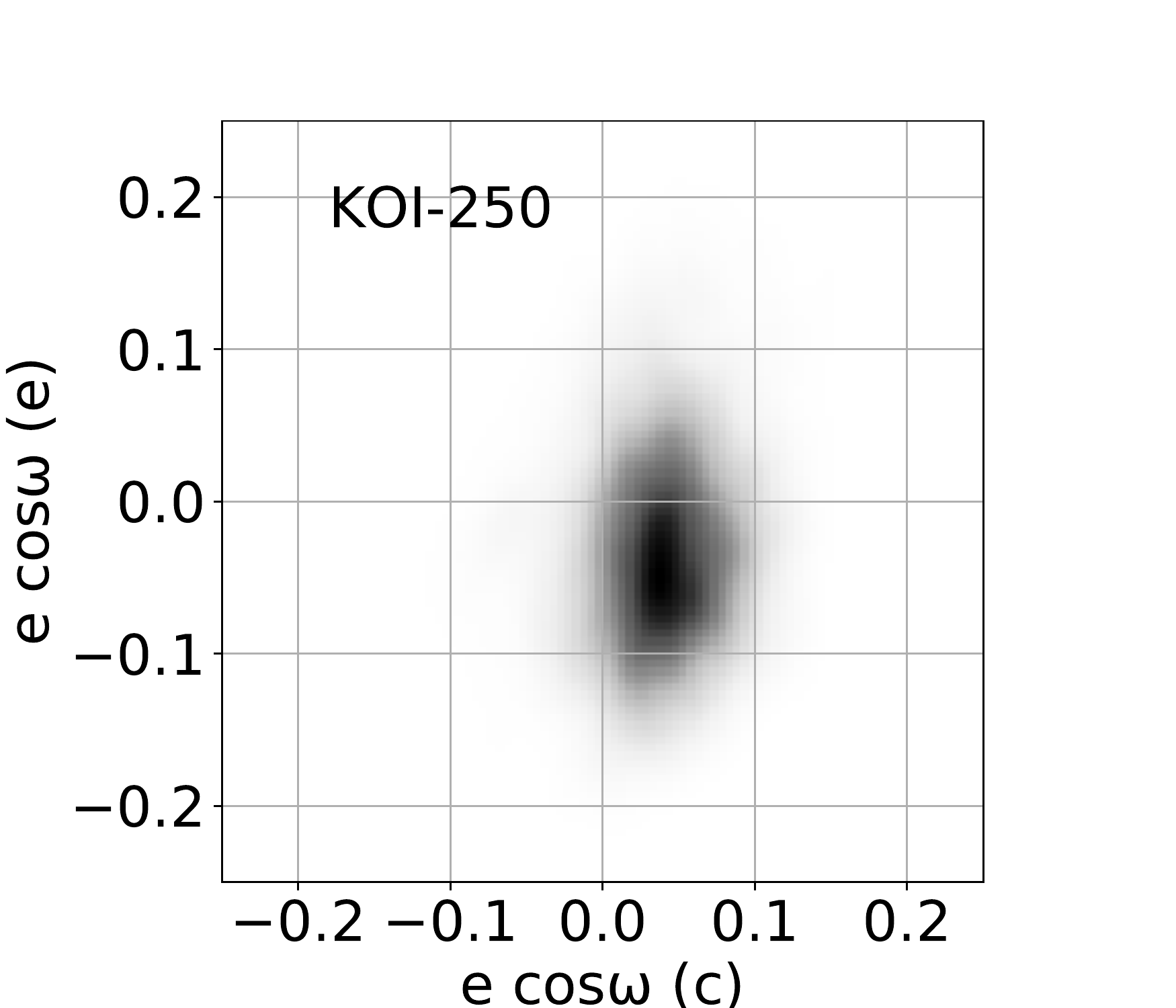}
\includegraphics [height = 1.1 in]{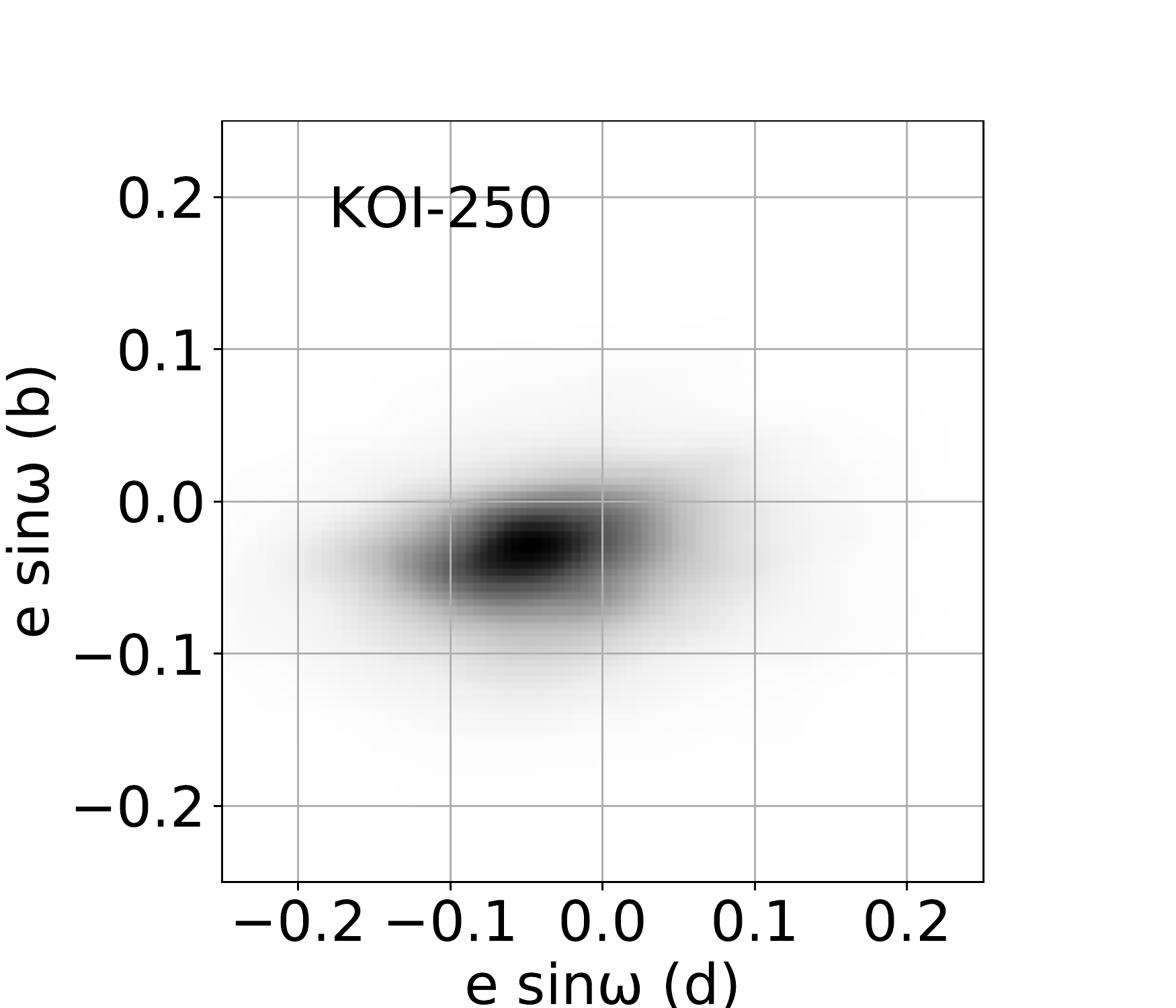} \\
\includegraphics [height = 1.1 in]{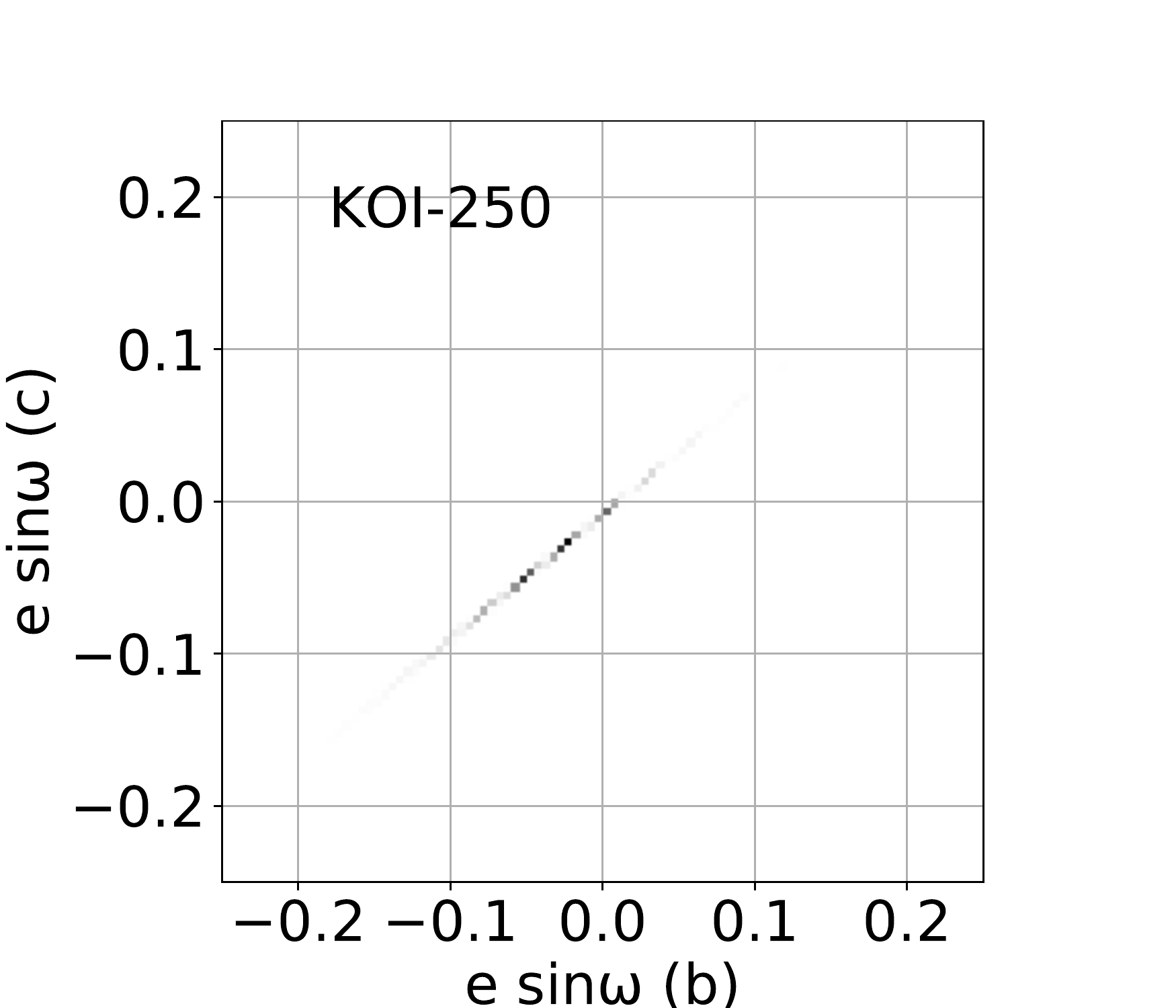}
\includegraphics [height = 1.1 in]{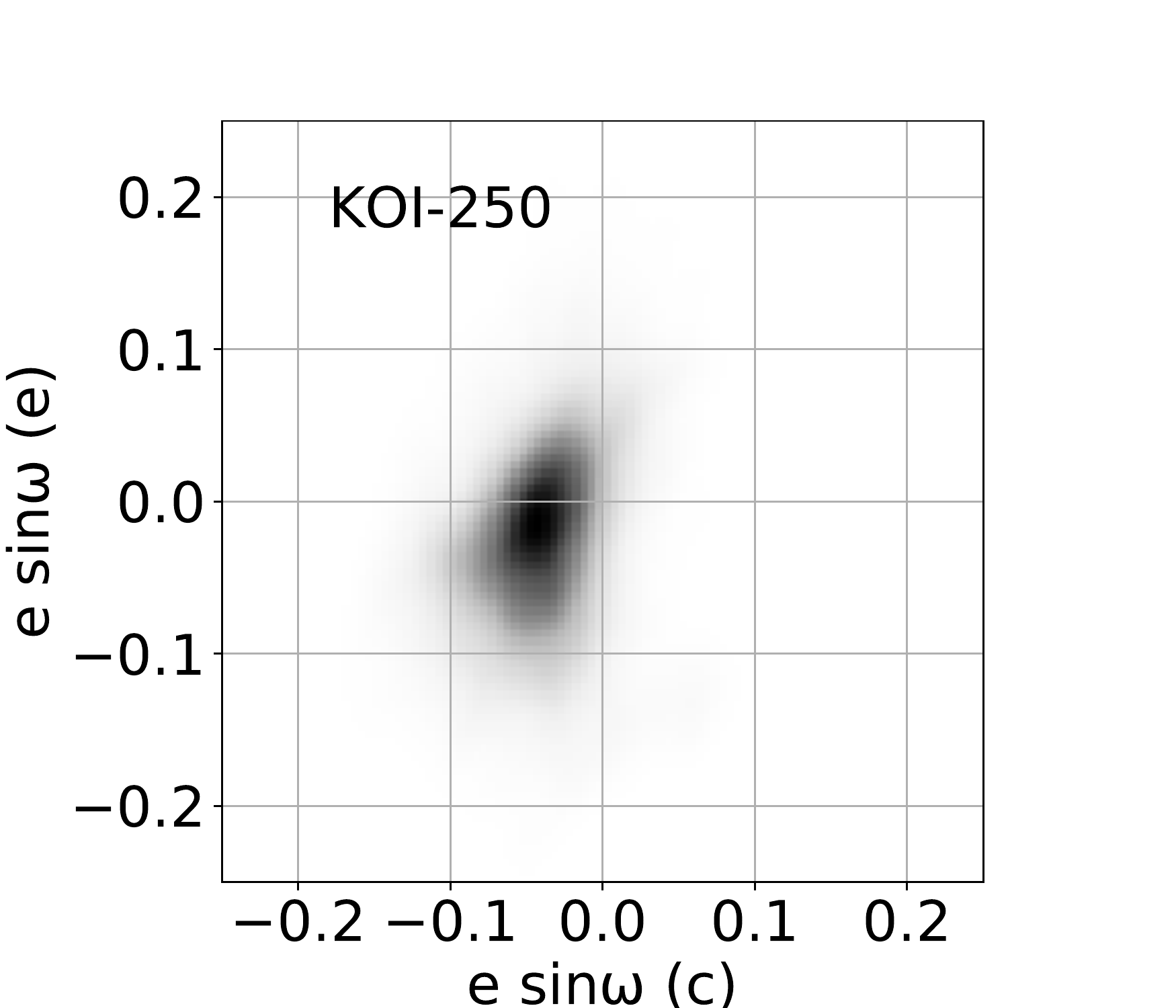}
\includegraphics [height = 1.1 in]{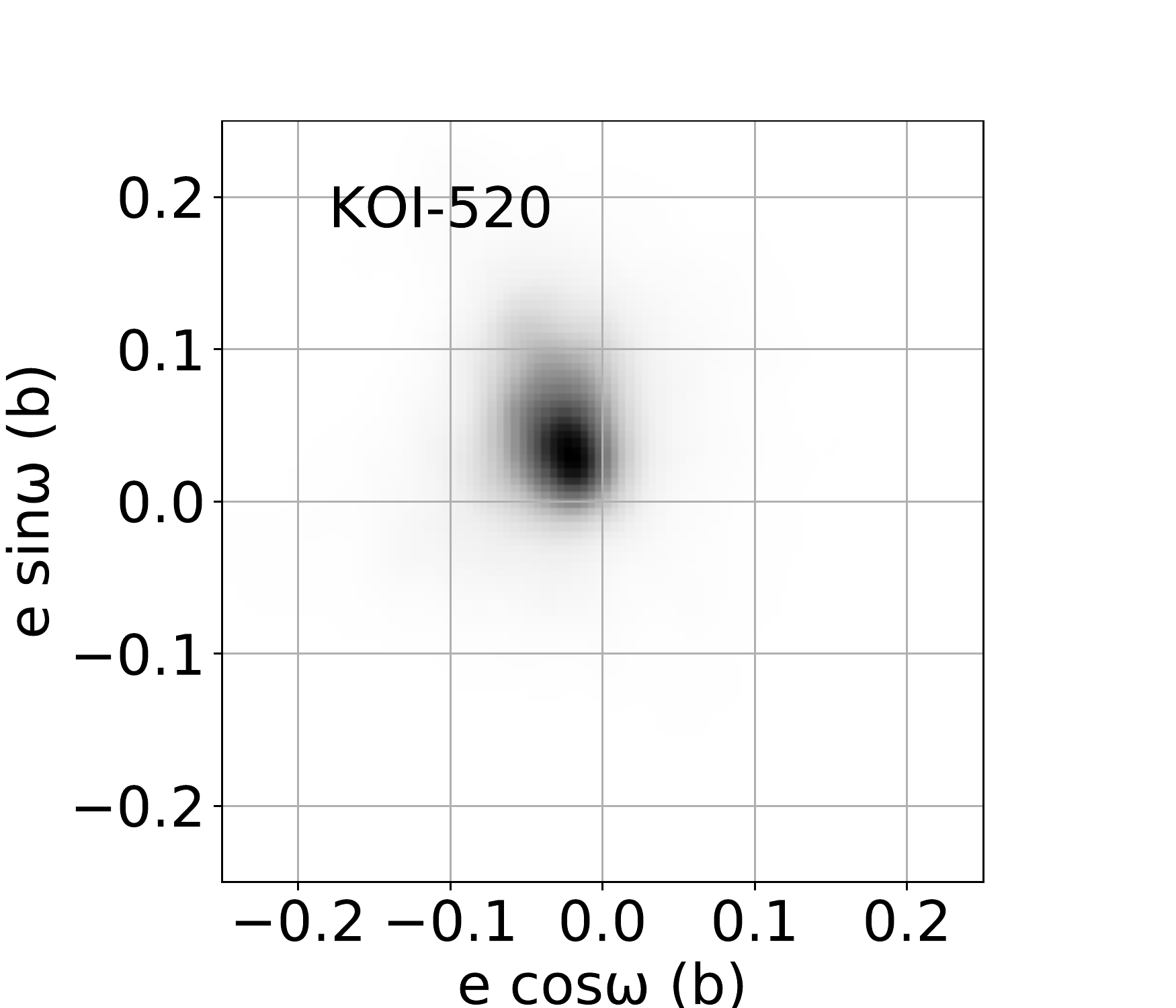}
\includegraphics [height = 1.1 in]{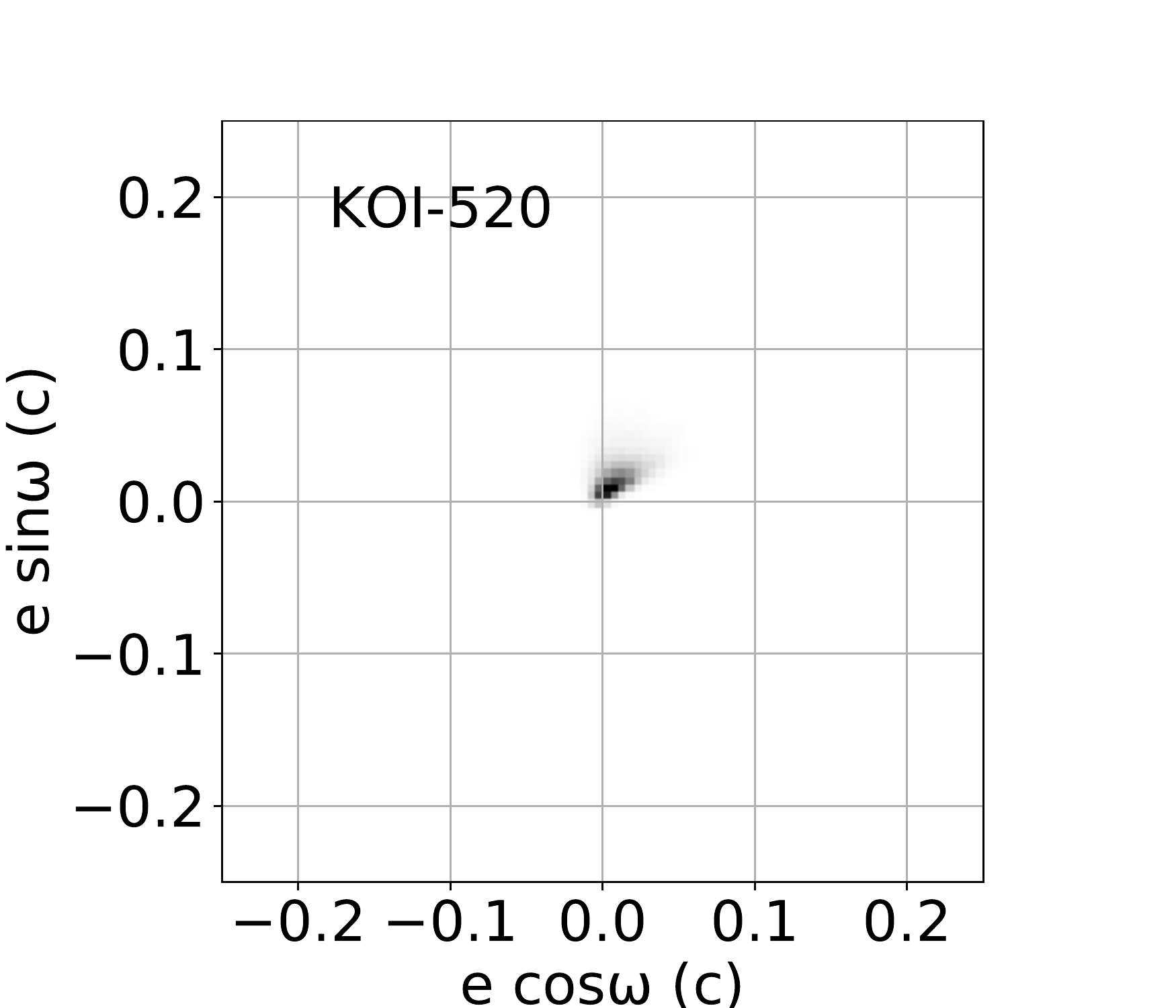} \\
\includegraphics [height = 1.1 in]{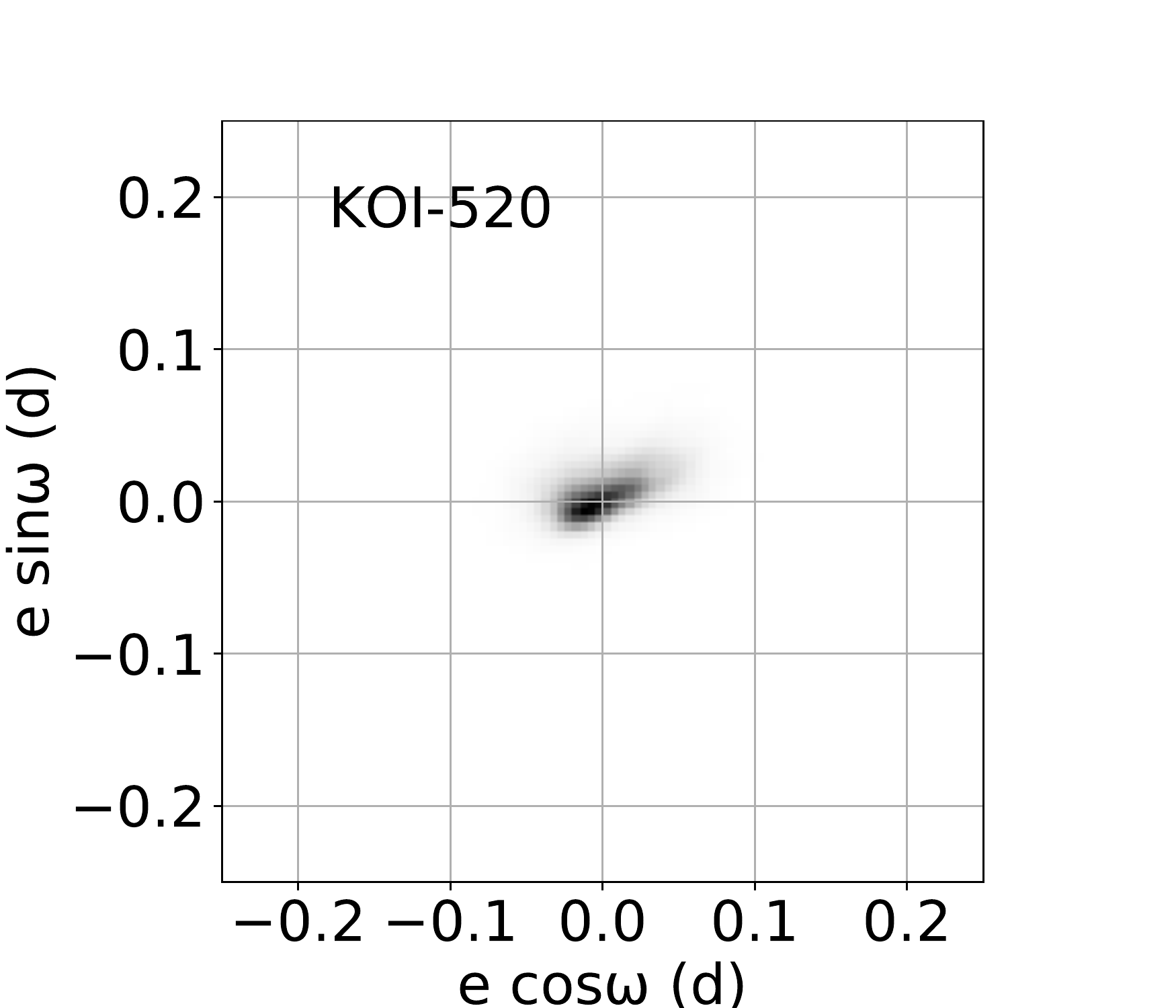}
\includegraphics [height = 1.1 in]{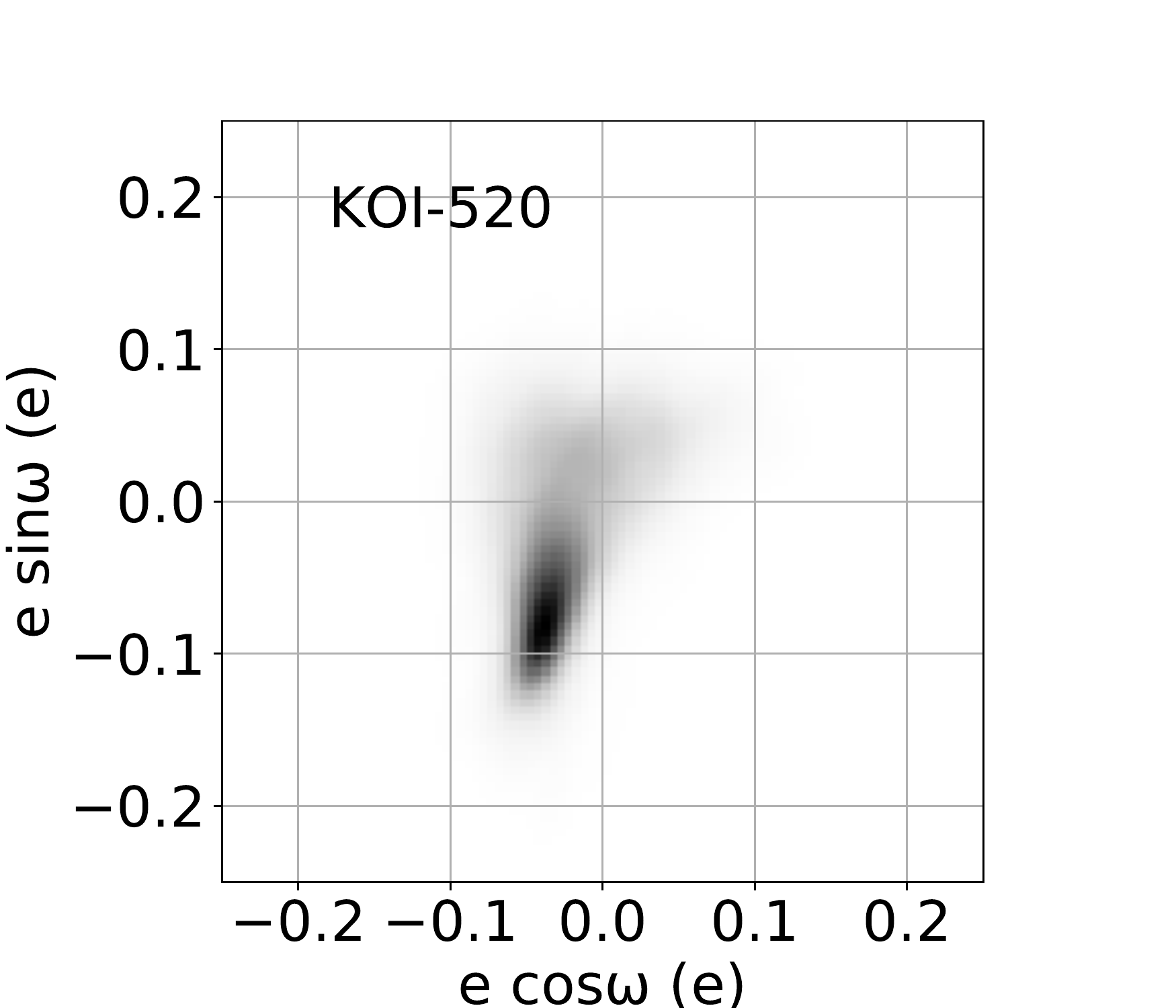}
\includegraphics [height = 1.1 in]{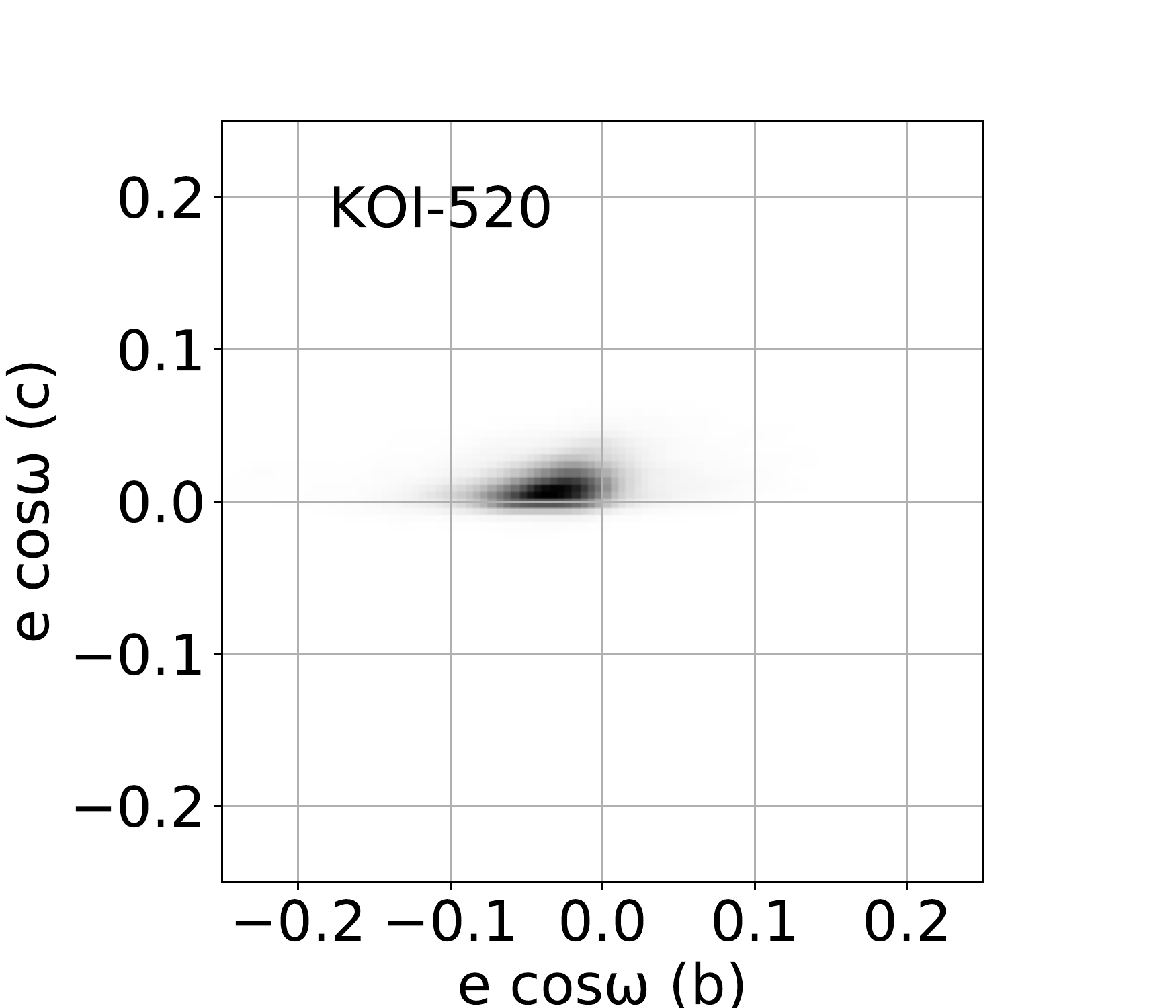}
\includegraphics [height = 1.1 in]{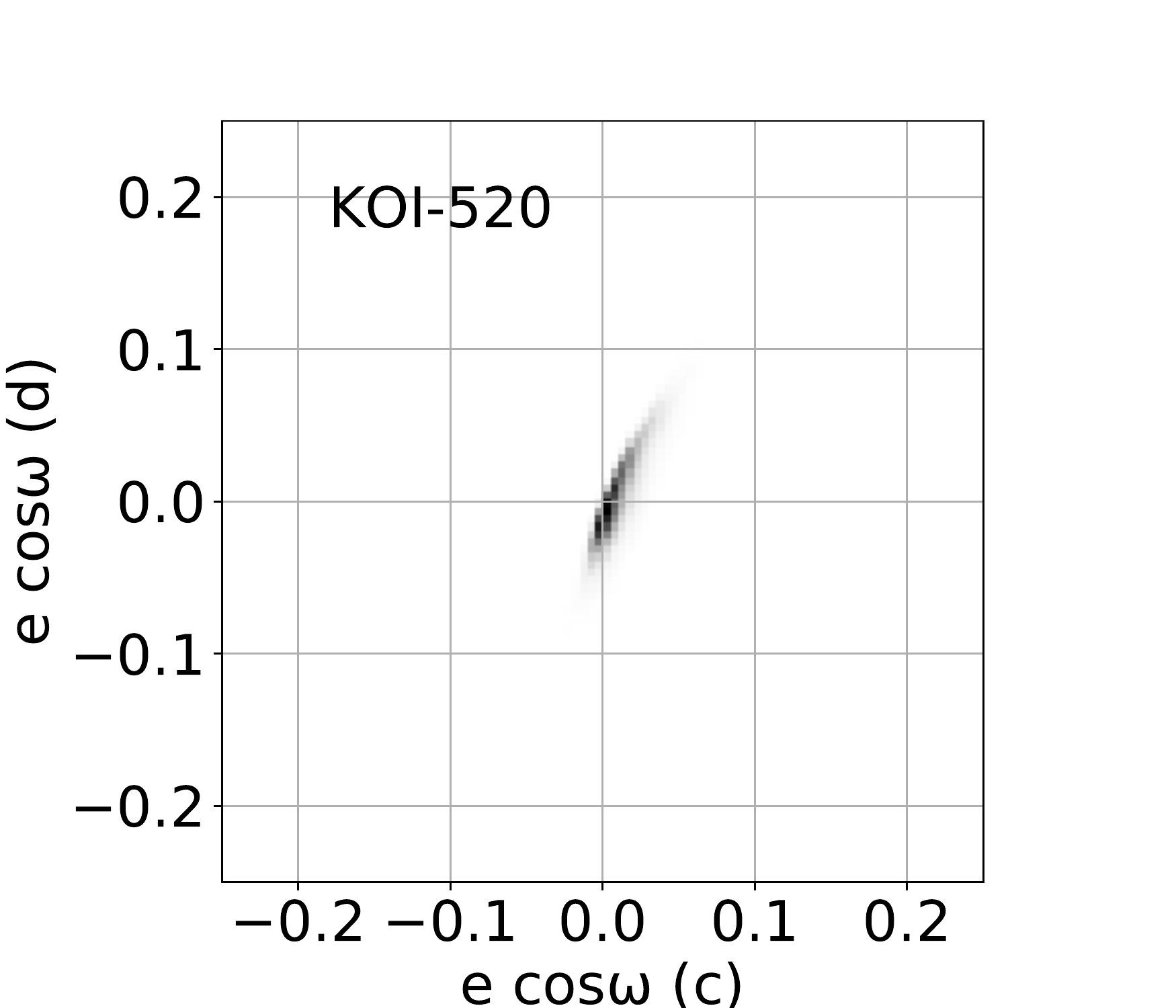} \\
\includegraphics [height = 1.1 in]{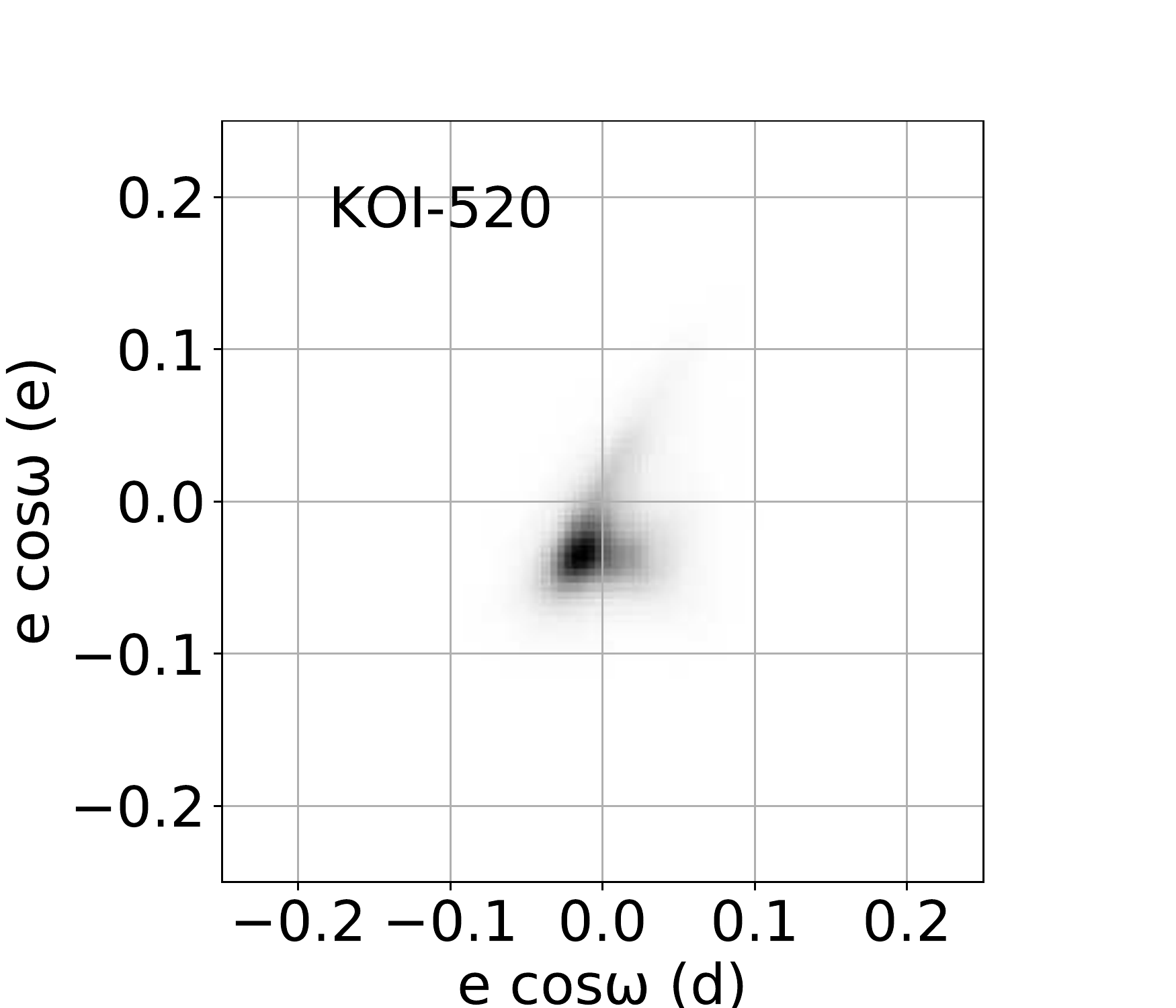}
\includegraphics [height = 1.1 in]{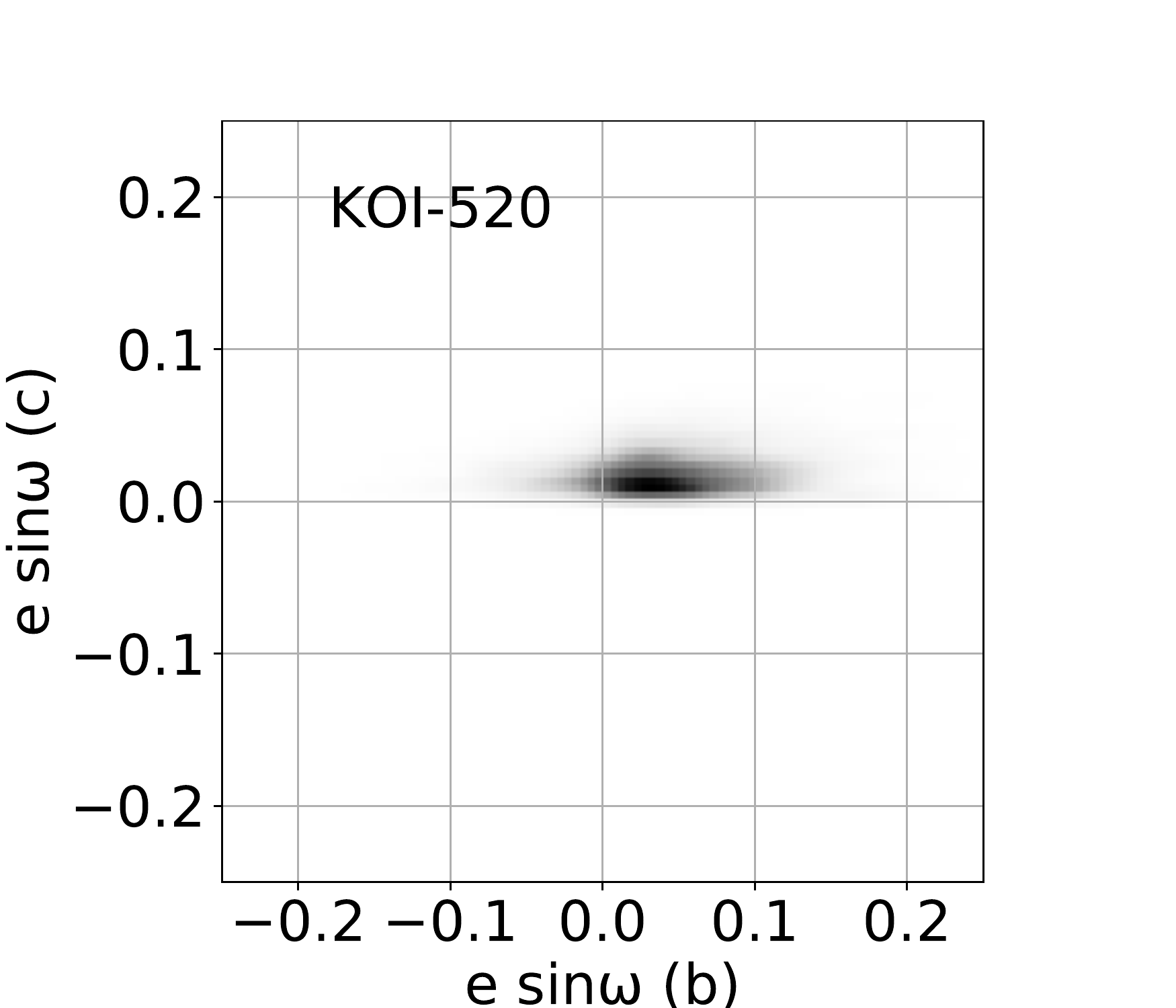} 
\includegraphics [height = 1.1 in]{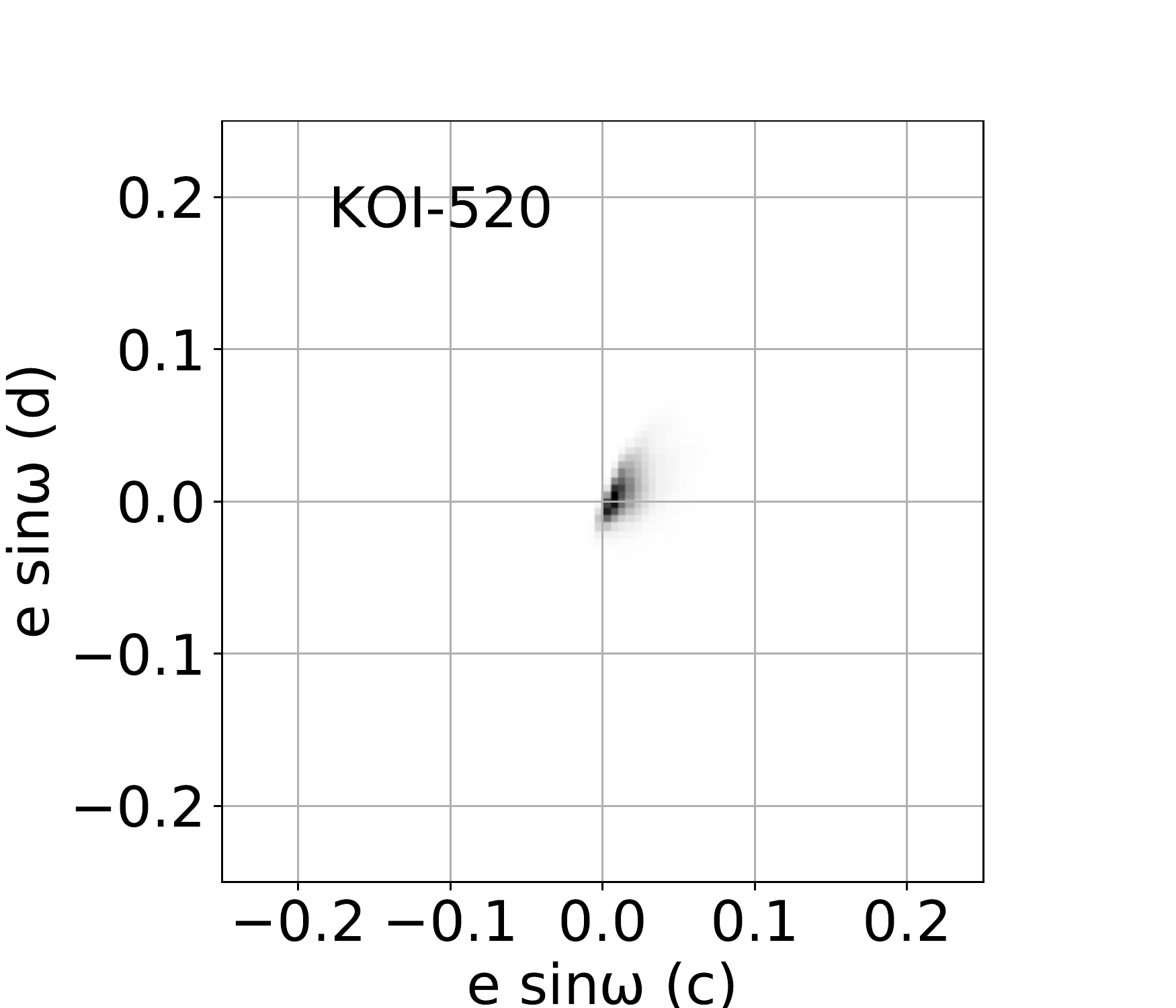}
\includegraphics [height = 1.1 in]{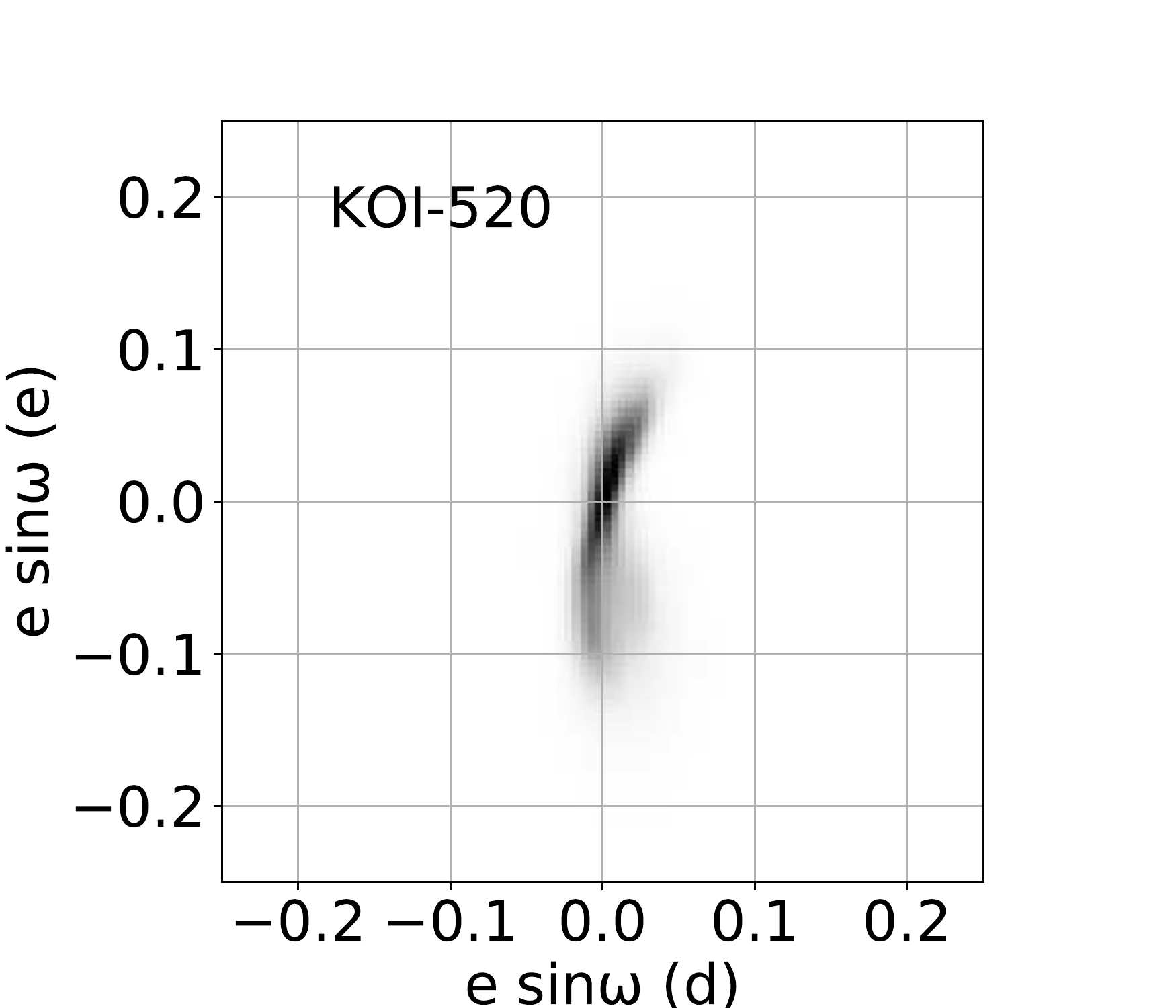} \\
\caption{Two-dimensional kernel density estimators on joint posteriors of eccentricity vector components: four-planet systems (Part 2 of 4). 
\label{fig:ecc4b} }
\end{center}
\end{figure}

\begin{figure}
\begin{center}
\figurenum{32}
\includegraphics [height = 1.1 in]{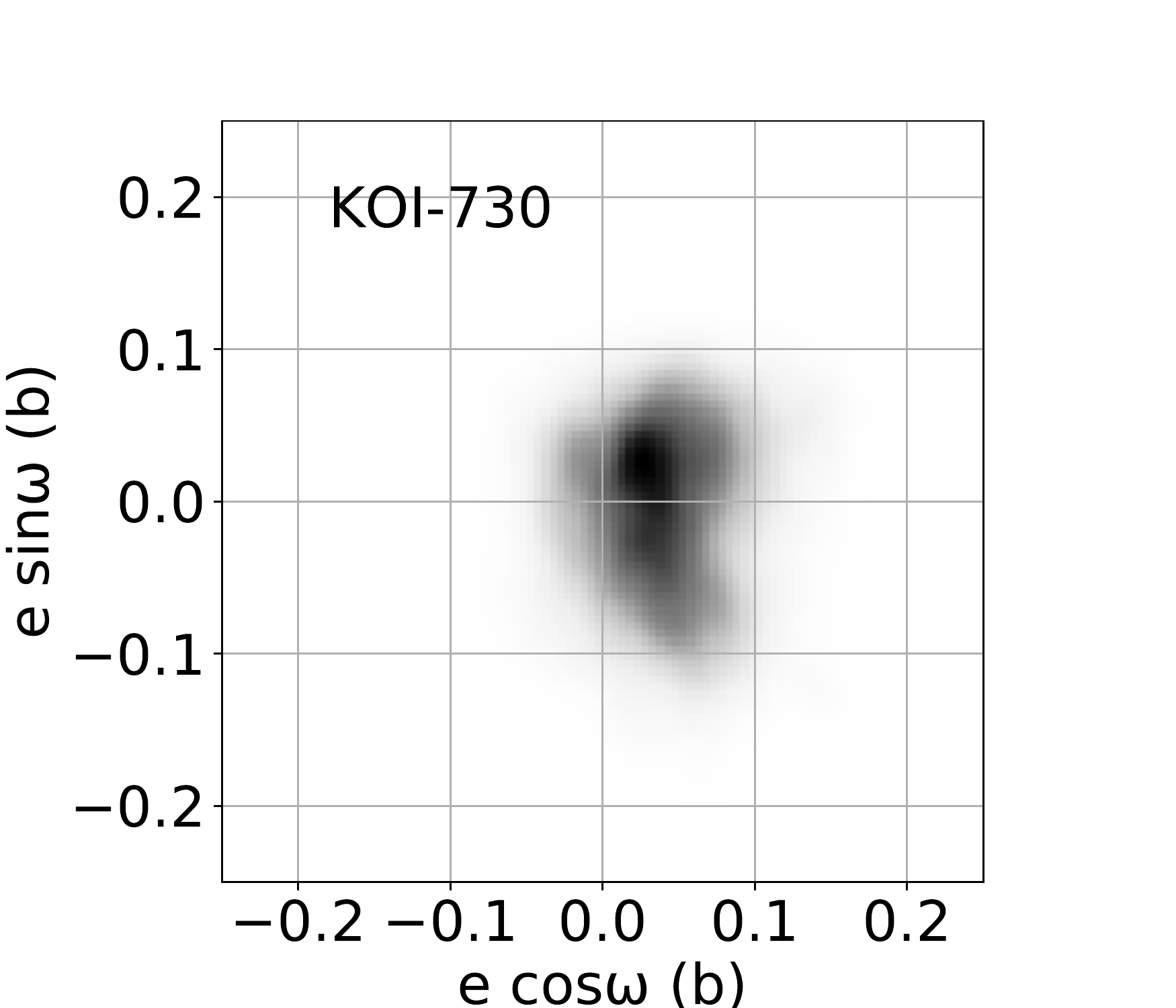}
\includegraphics [height = 1.1 in]{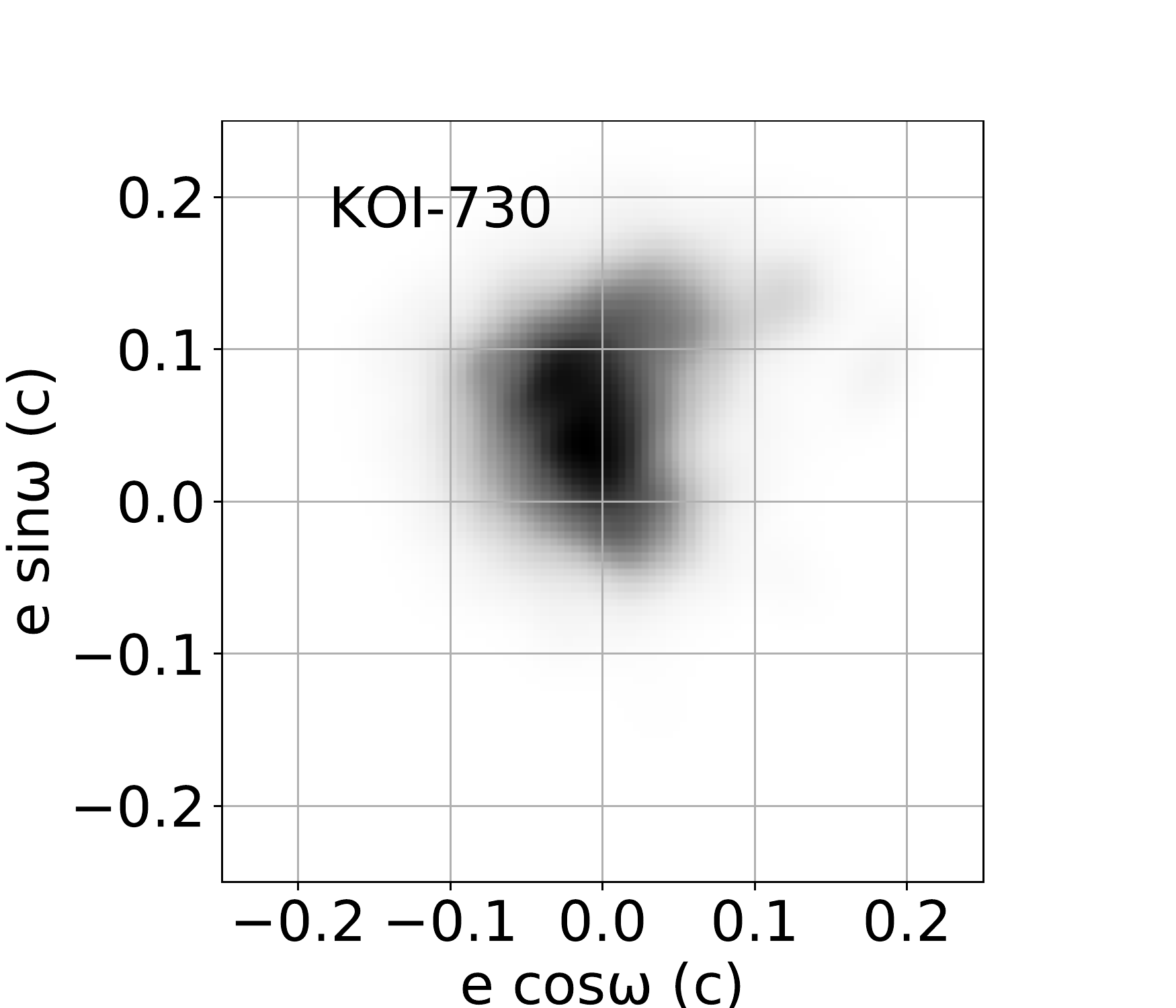}
\includegraphics [height = 1.1 in]{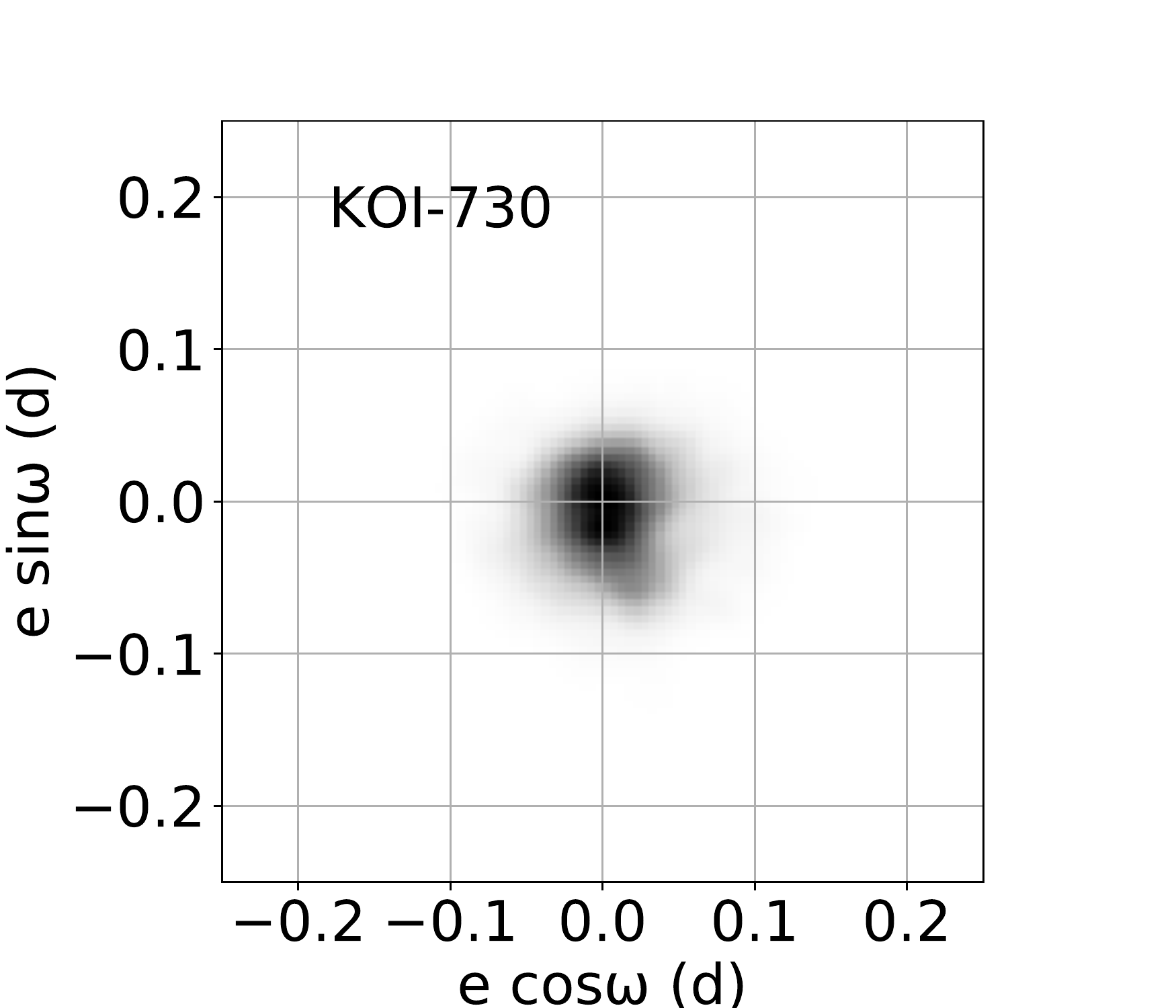}
\includegraphics [height = 1.1 in]{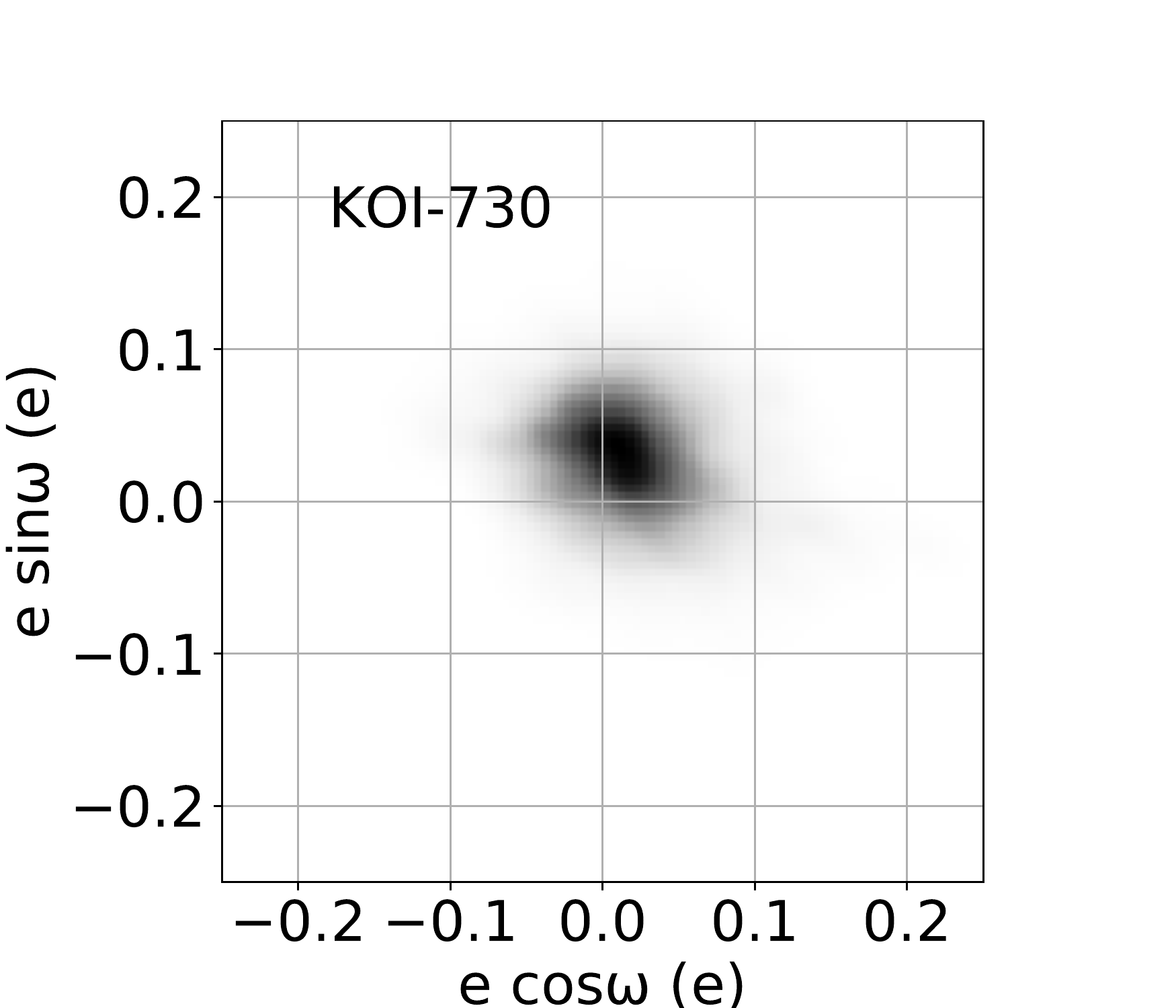} \\ 
\includegraphics [height = 1.1 in]{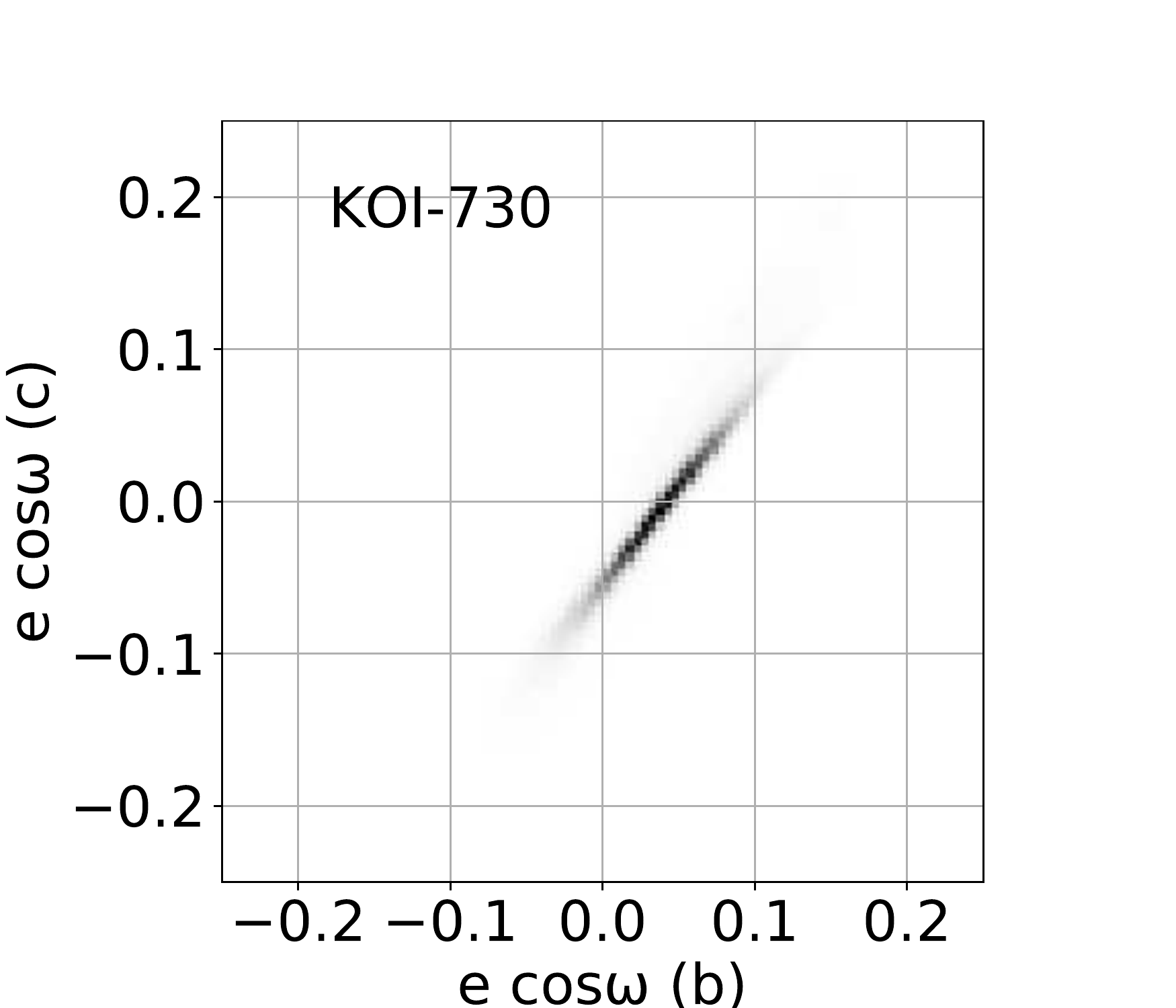}
\includegraphics [height = 1.1 in]{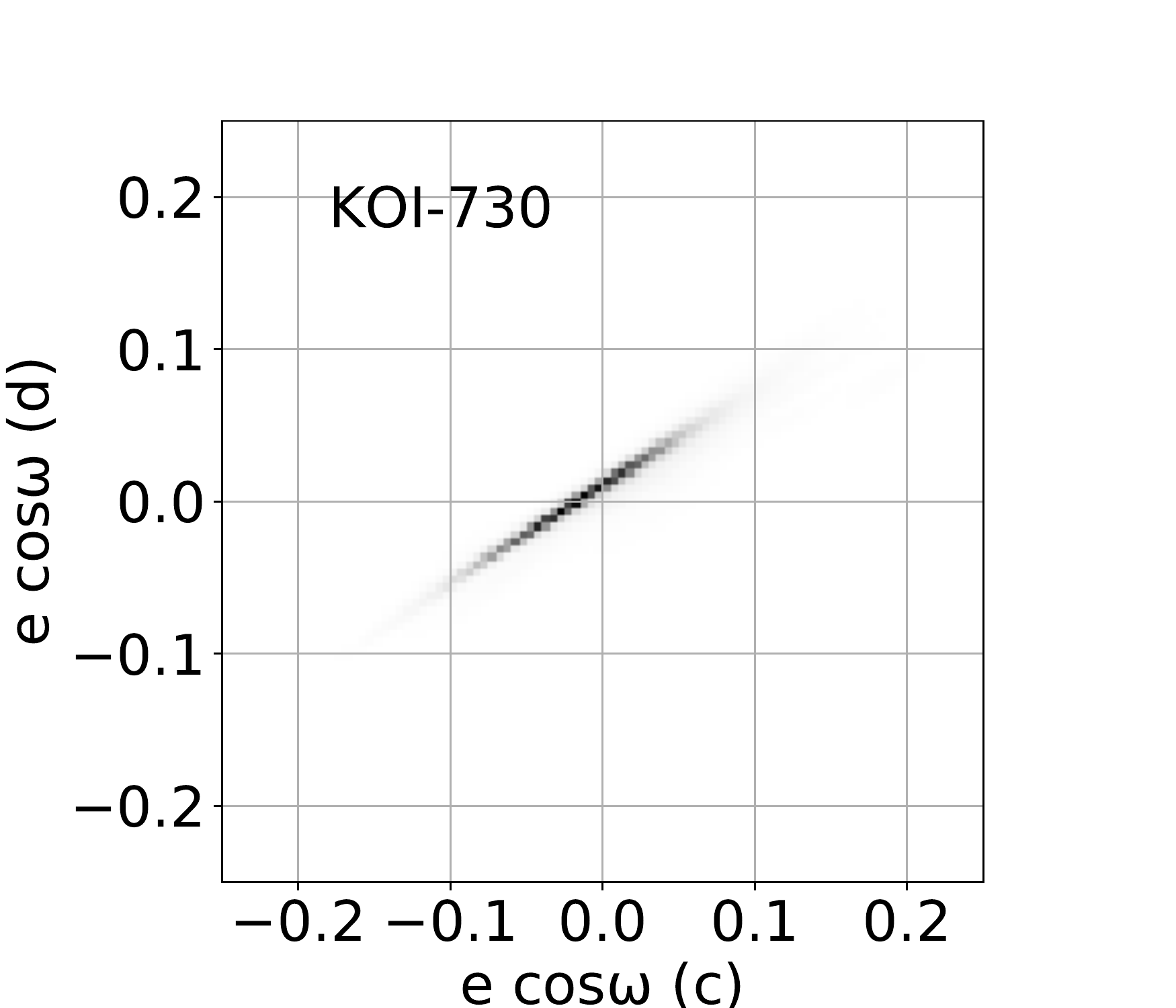}
\includegraphics [height = 1.1 in]{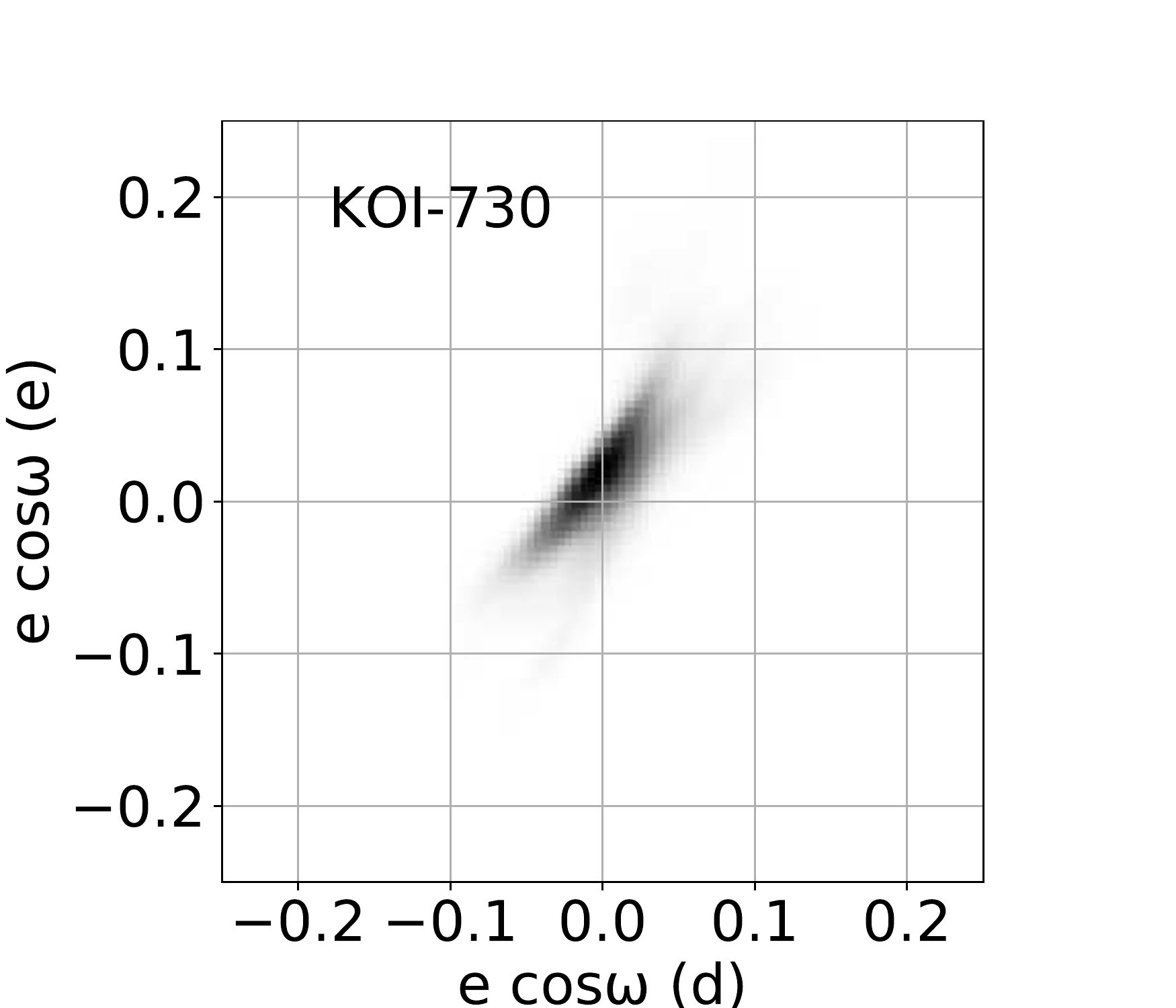}
\includegraphics [height = 1.1 in]{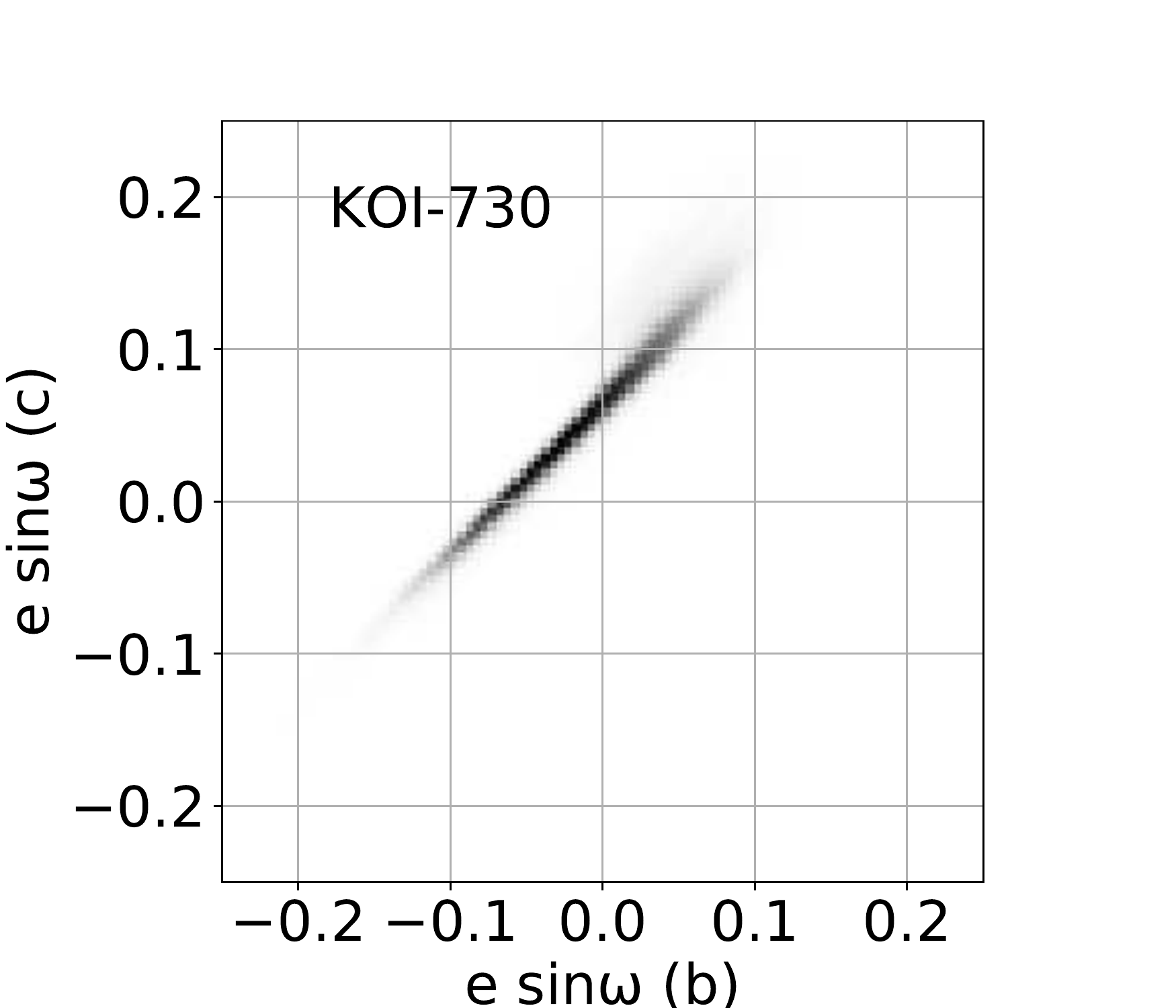} \\
\includegraphics [height = 1.1 in]{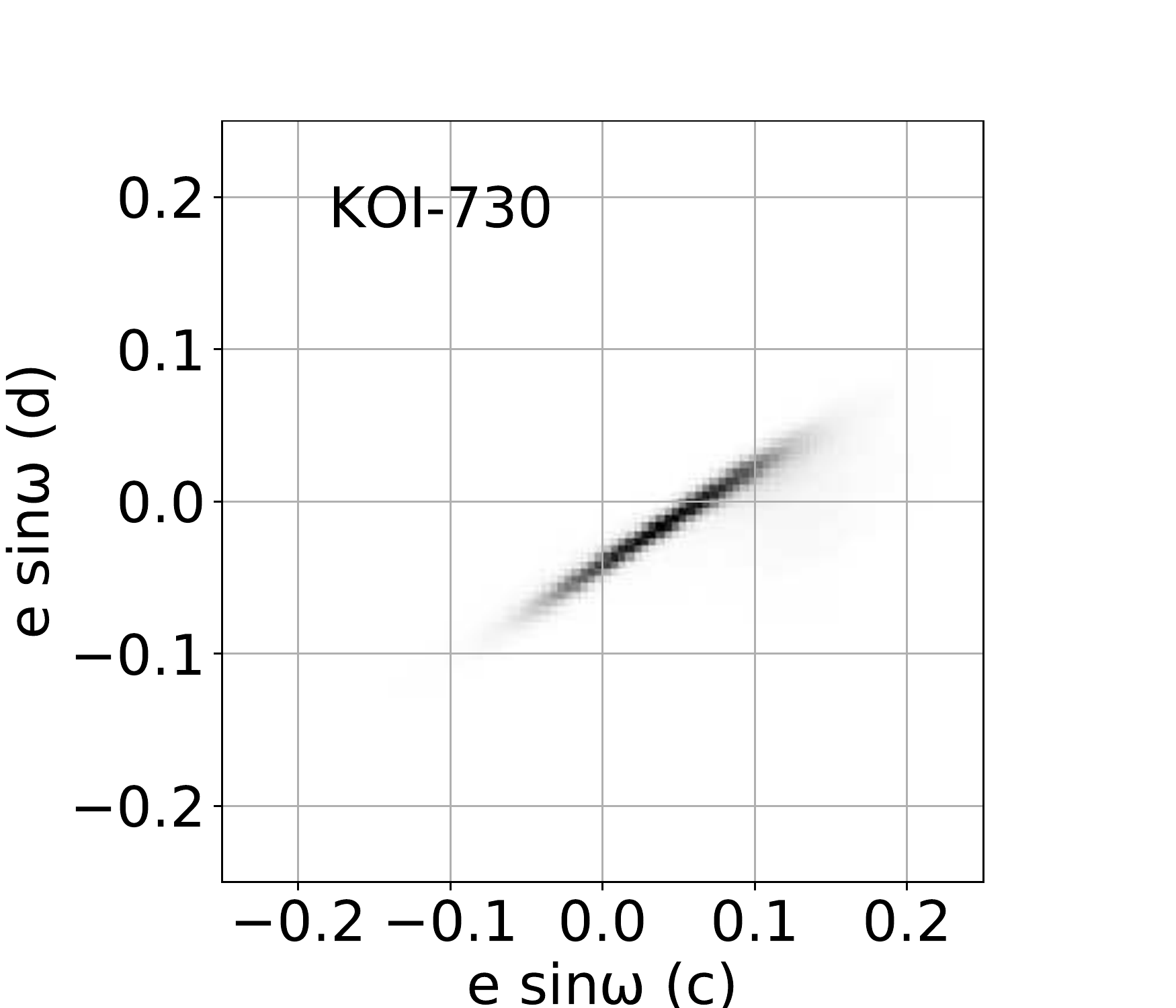}
\includegraphics [height = 1.1 in]{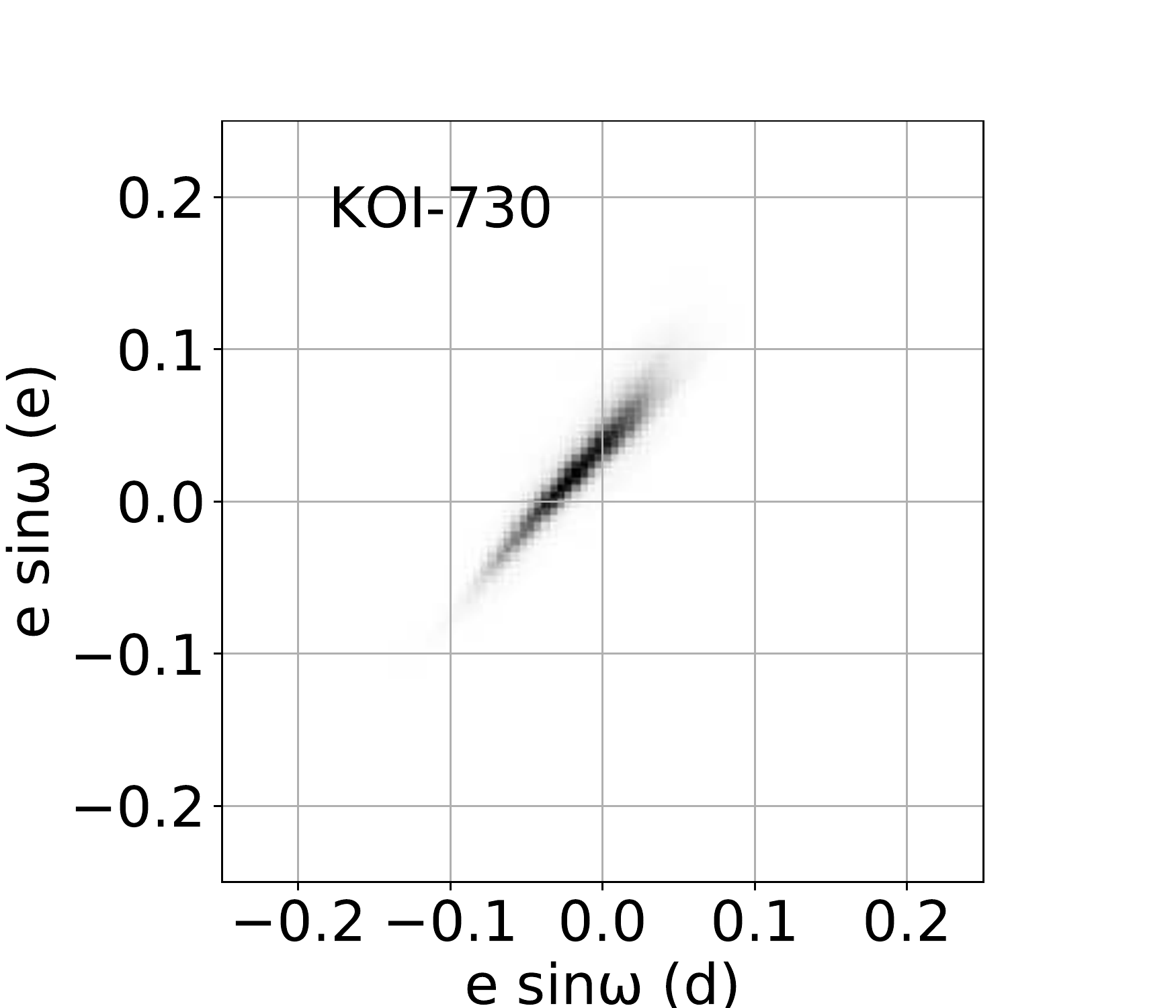}
\includegraphics [height = 1.1 in]{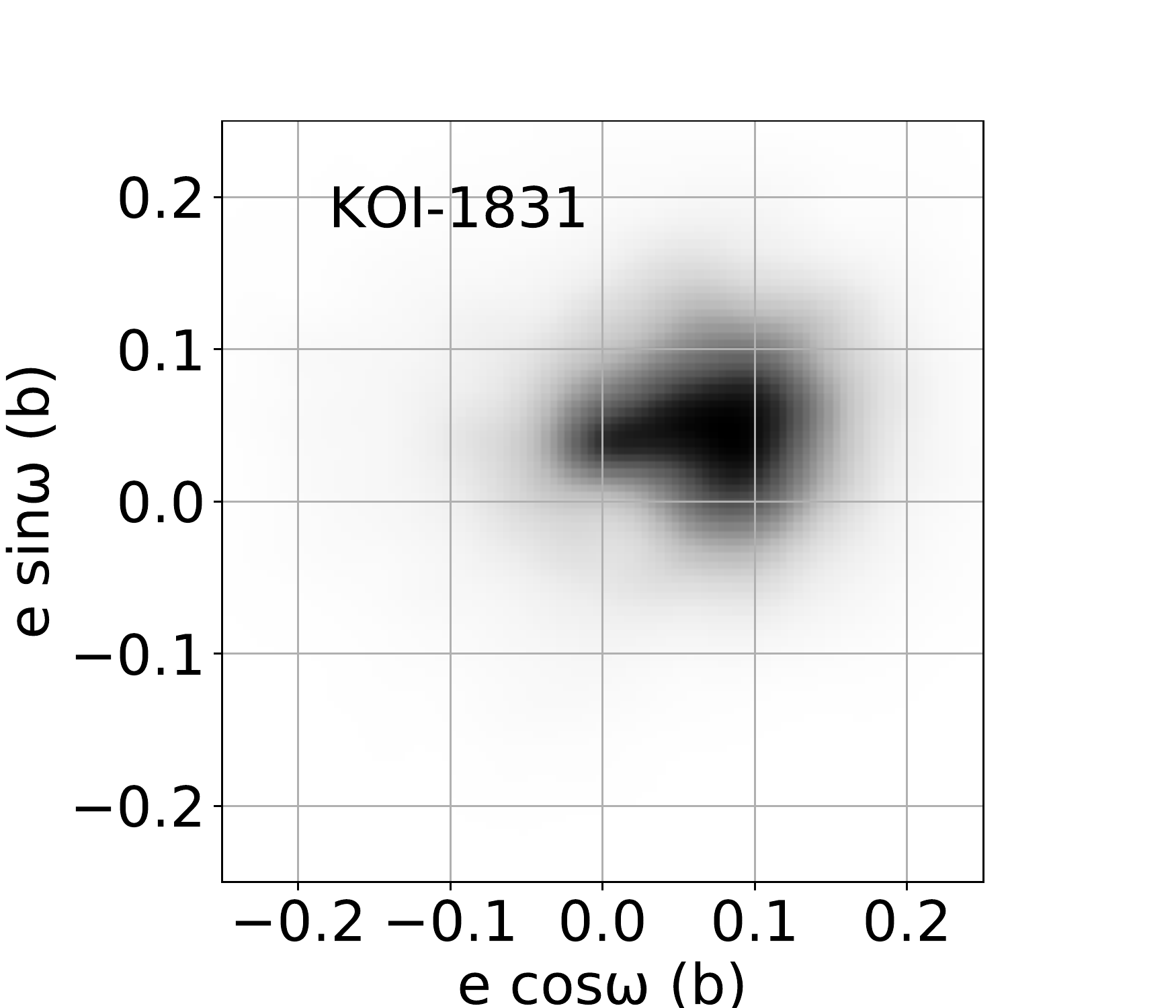}
\includegraphics [height = 1.1 in]{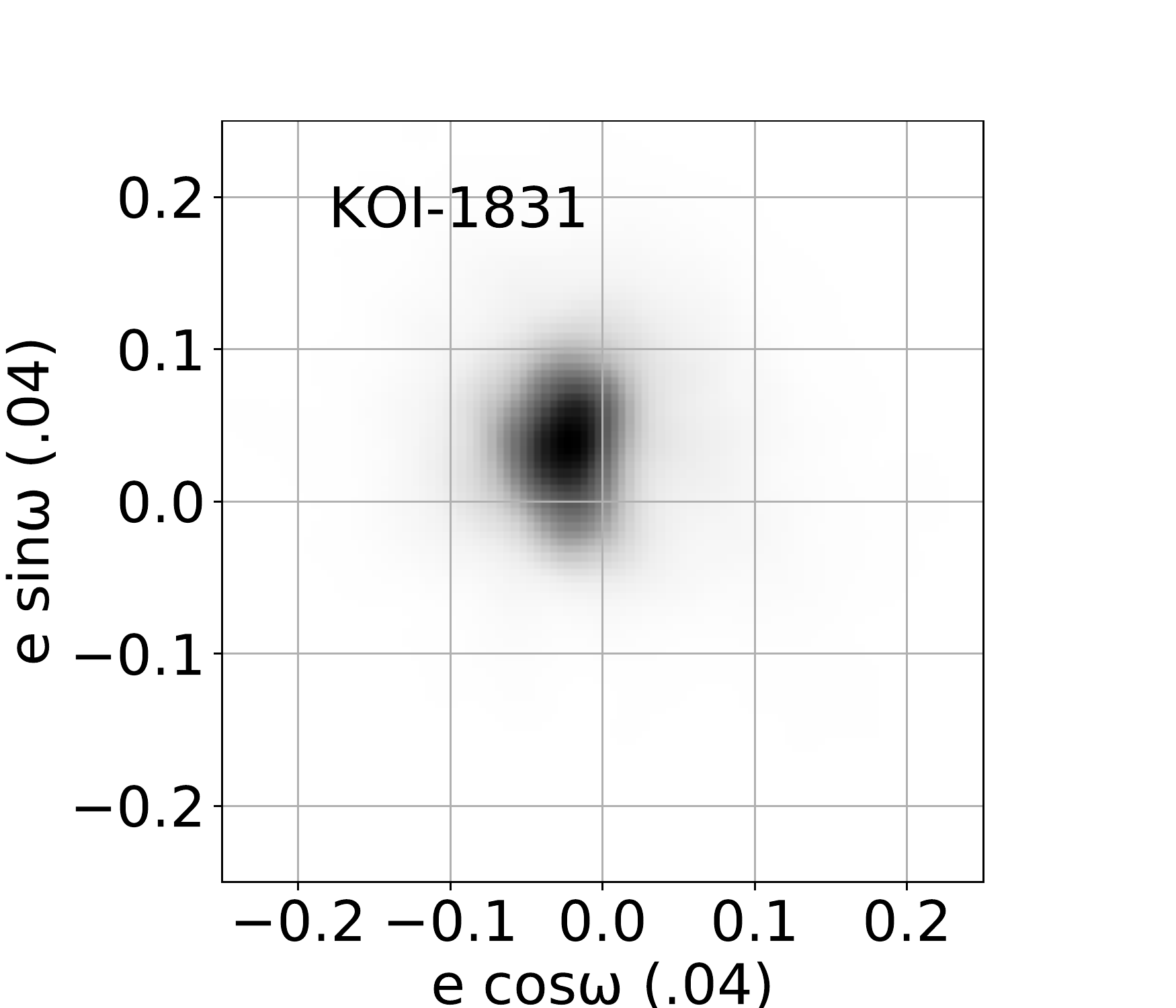} \\
\includegraphics [height = 1.1 in]{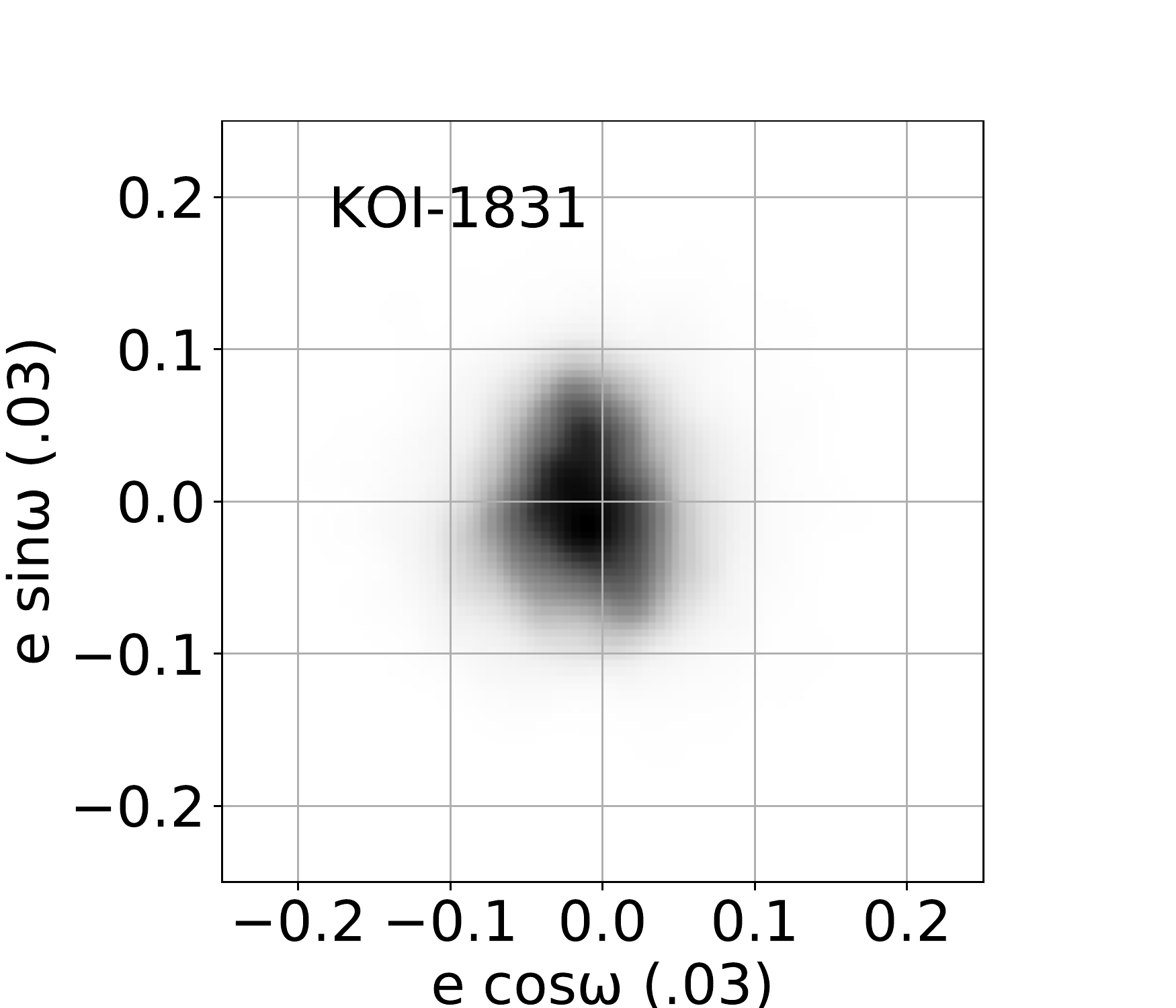}
\includegraphics [height = 1.1 in]{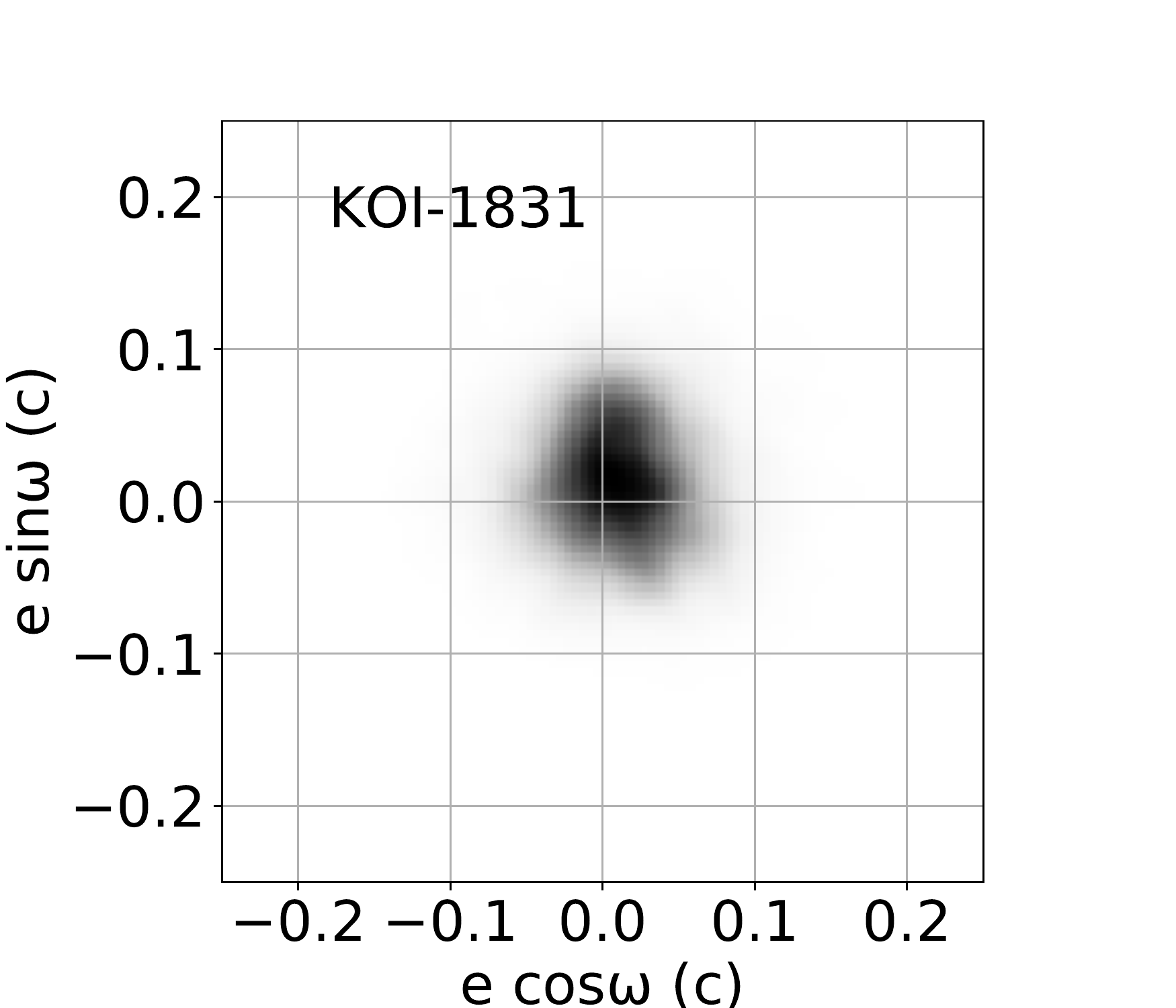}
\includegraphics [height = 1.1 in]{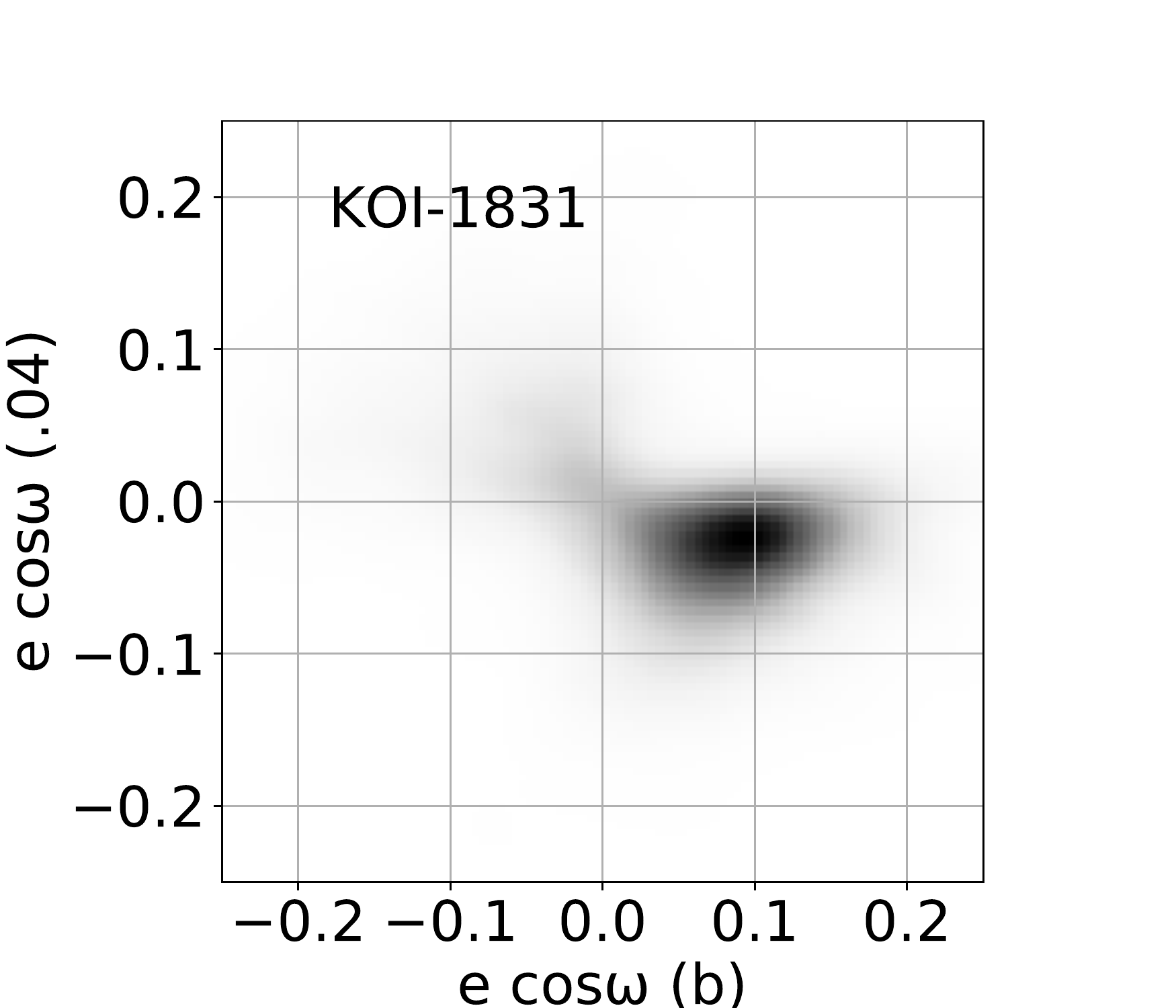}
\includegraphics [height = 1.1 in]{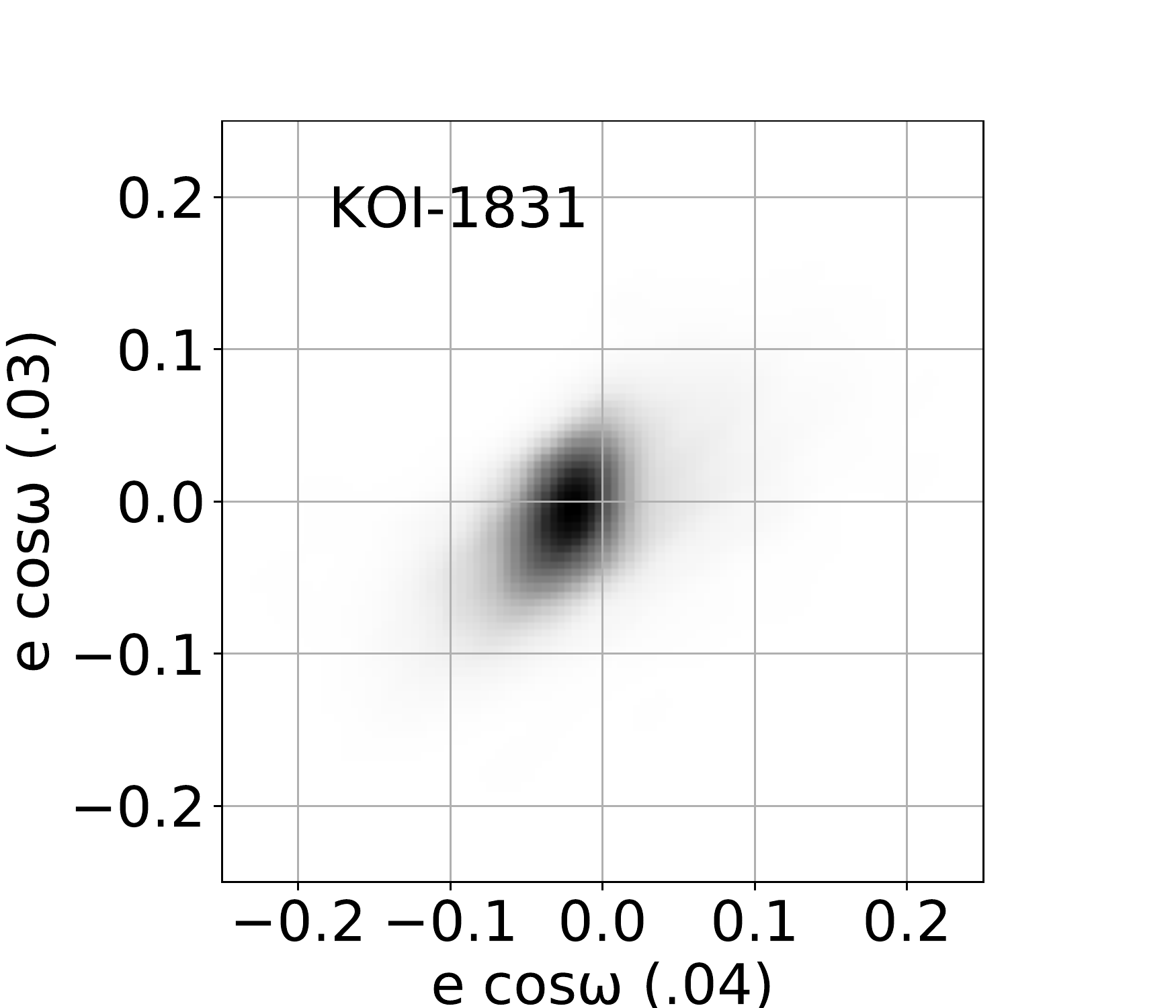} \\
\includegraphics [height = 1.1 in]{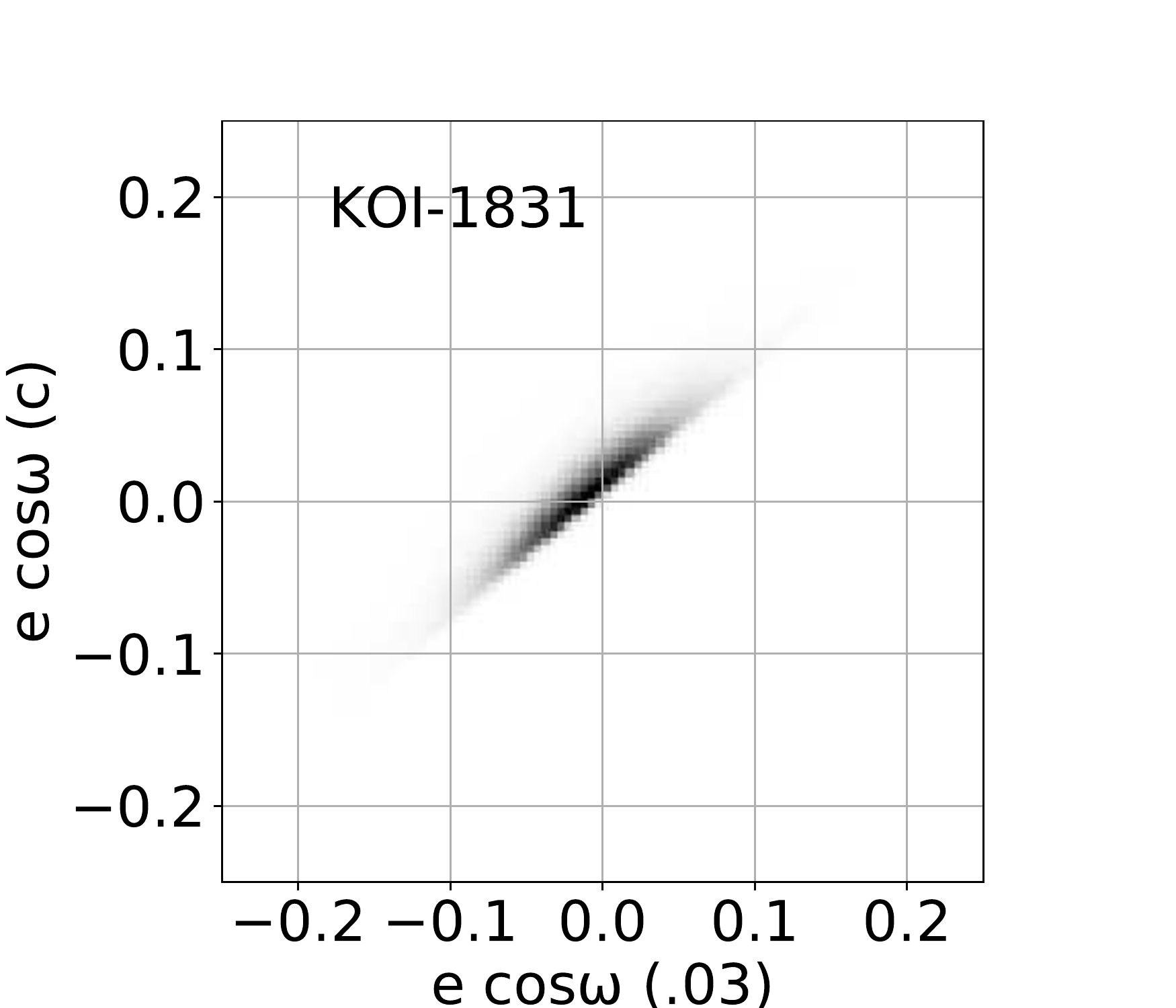}
\includegraphics [height = 1.1 in]{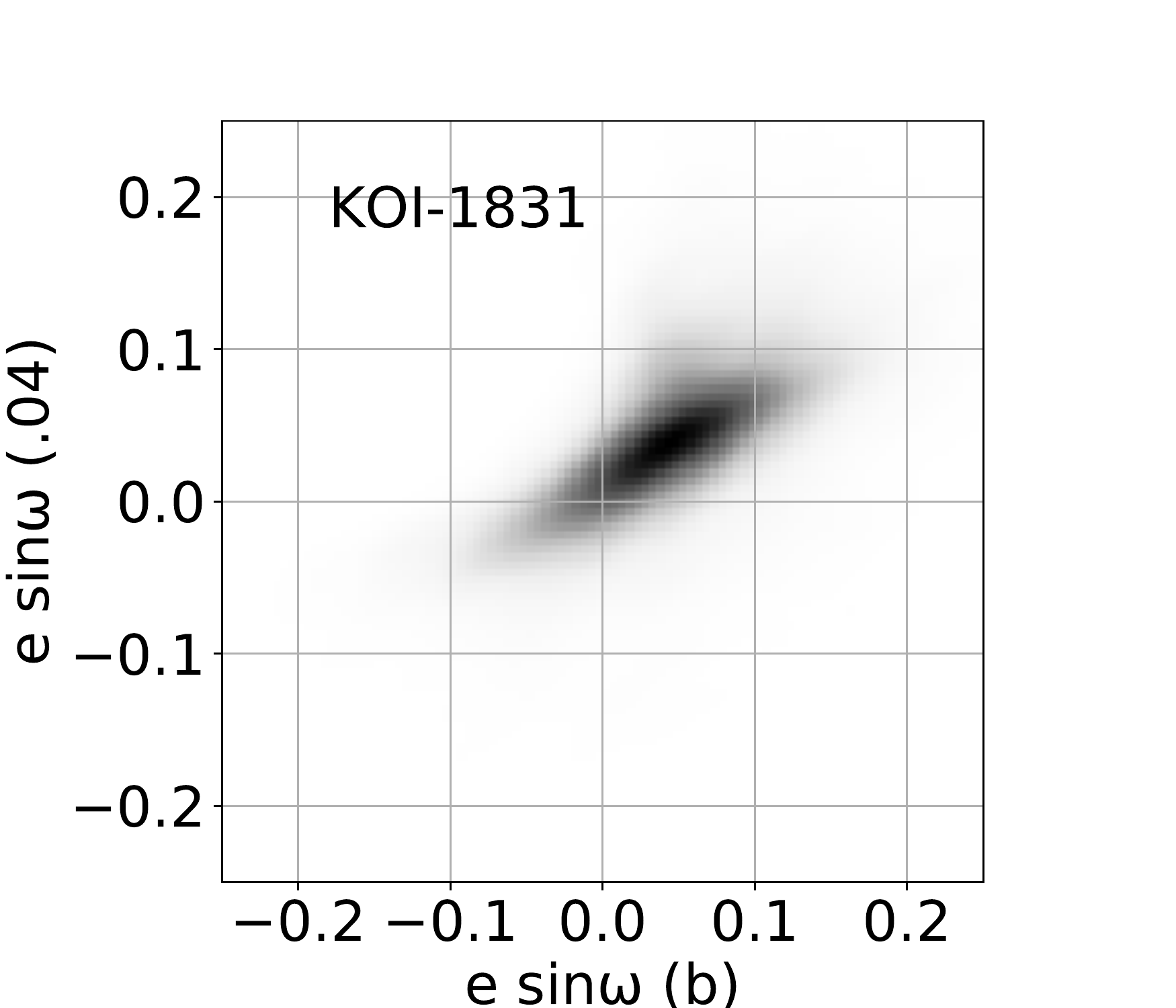} 
\includegraphics [height = 1.1 in]{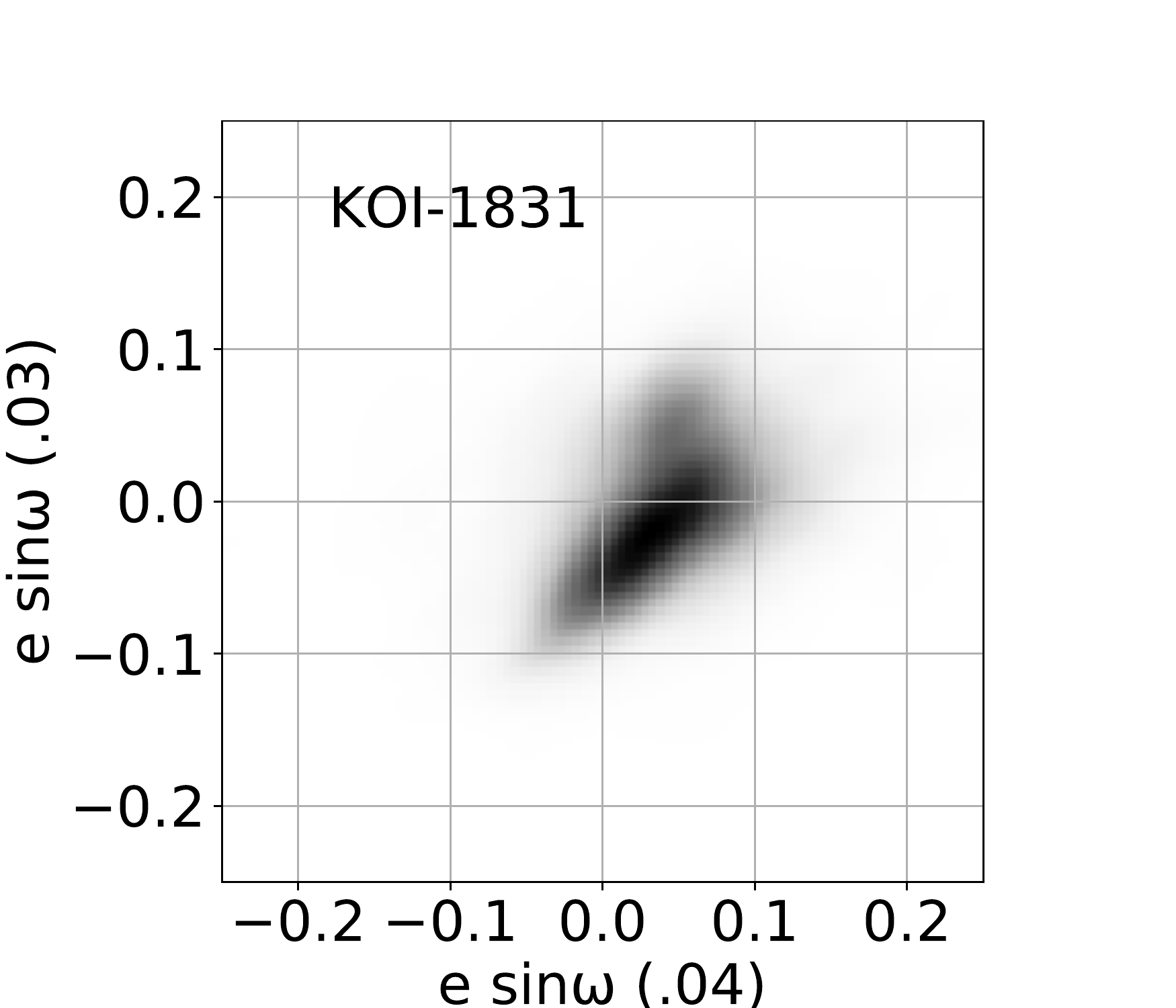}
\includegraphics [height = 1.1 in]{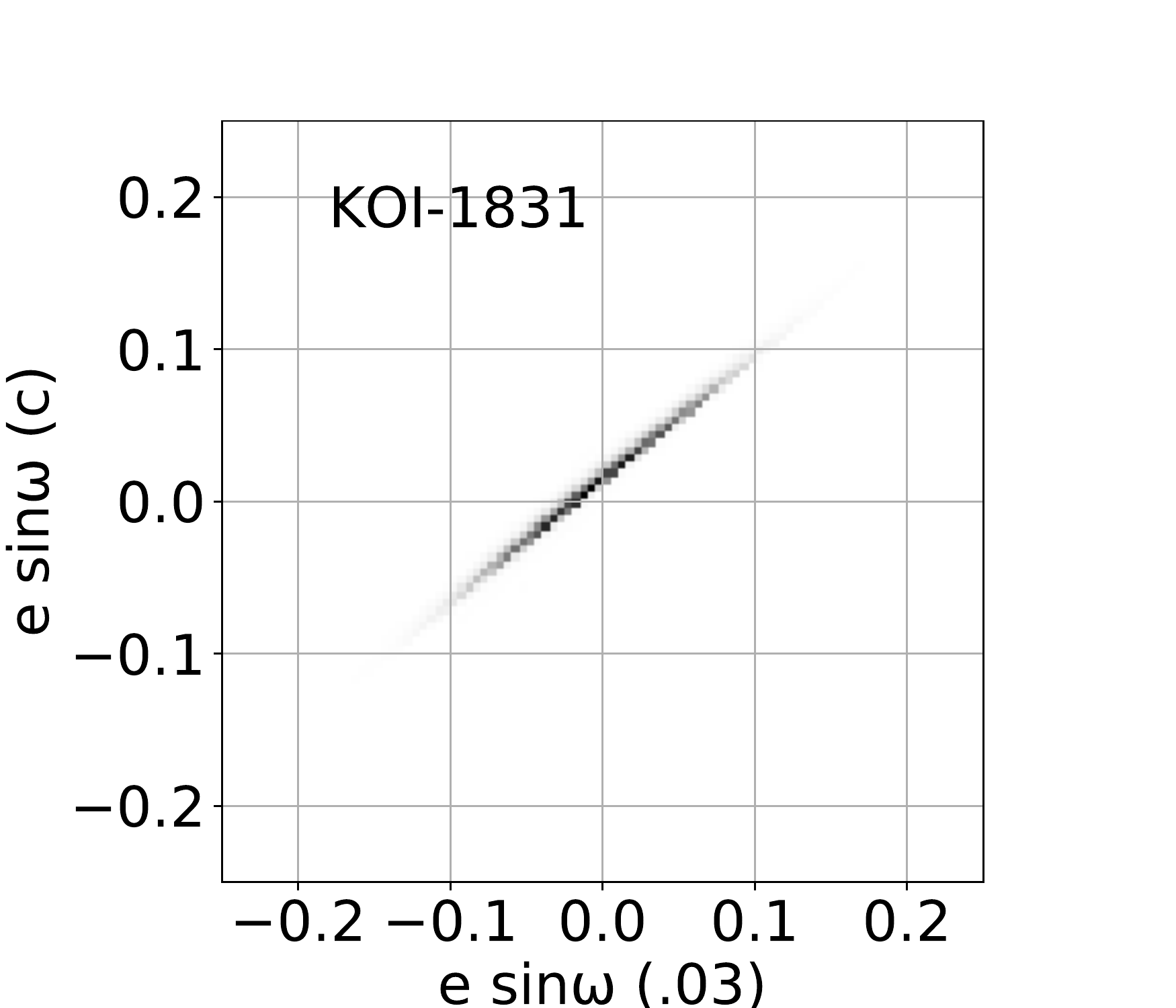} \\
\caption{Two-dimensional kernel density estimators on joint posteriors of eccentricity vector components: four-planet systems (Part 3 of 4). 
\label{fig:ecc4c} }
\end{center}
\end{figure}

\begin{figure}
\begin{center}
\figurenum{33}
\includegraphics [height = 1.1 in]{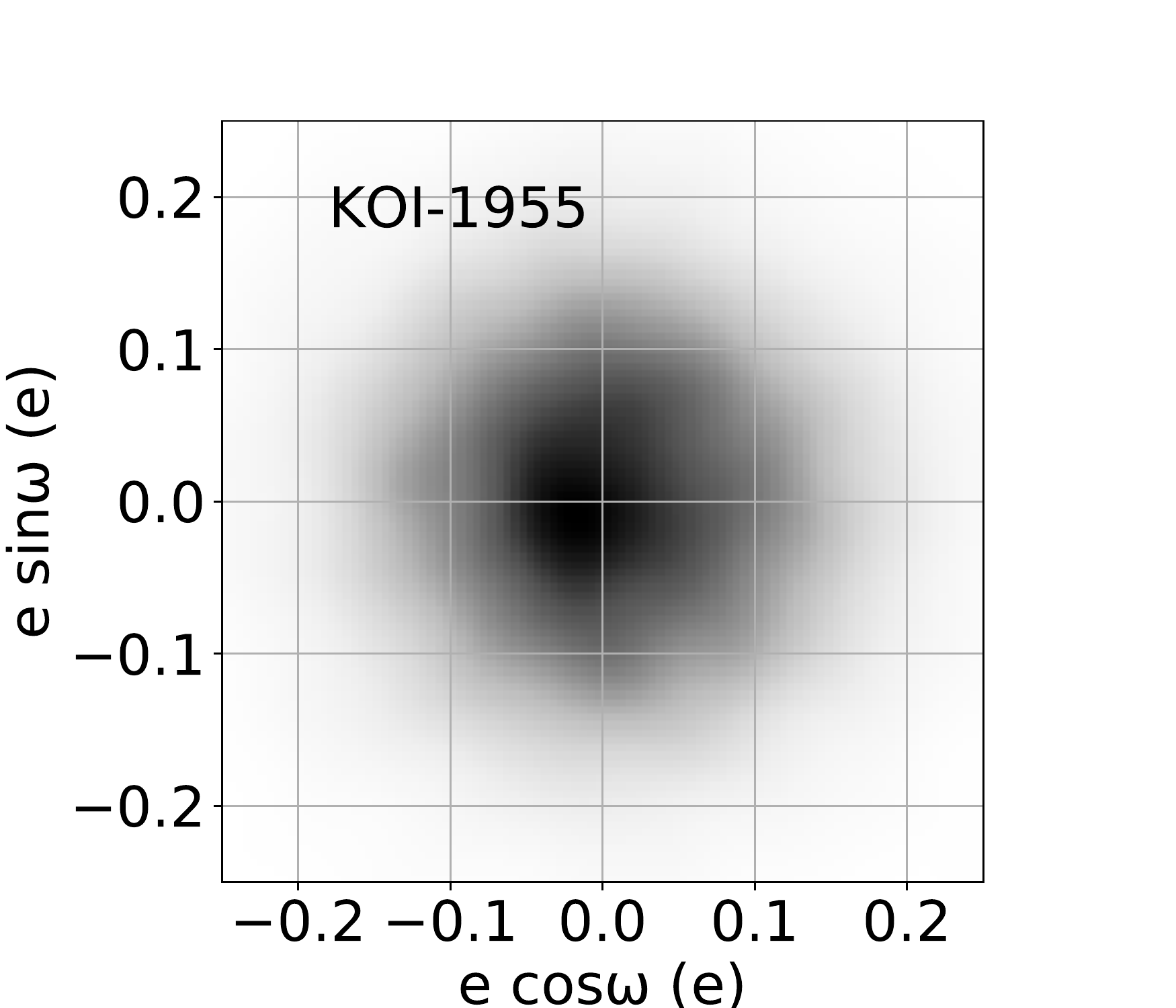}
\includegraphics [height = 1.1 in]{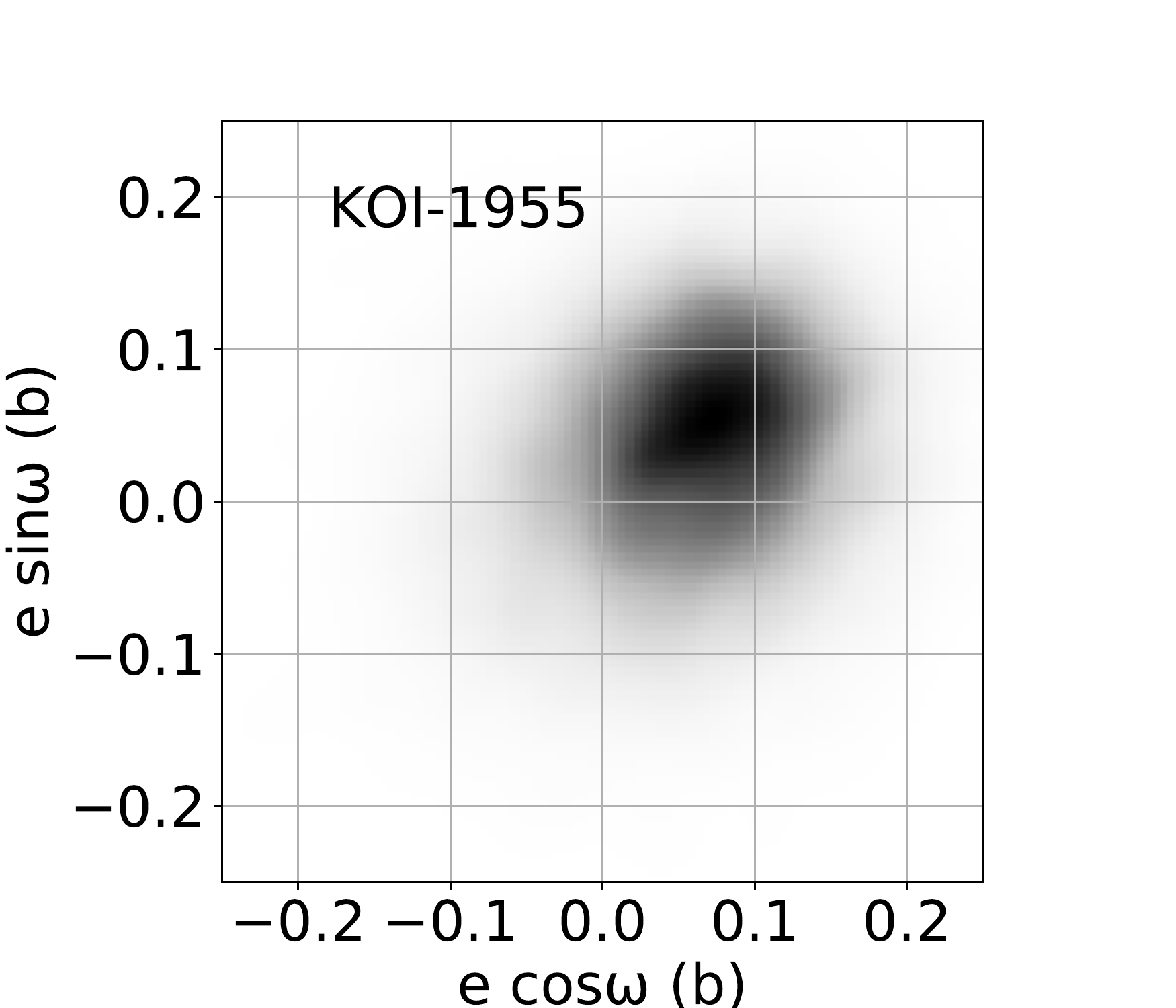}
\includegraphics [height = 1.1 in]{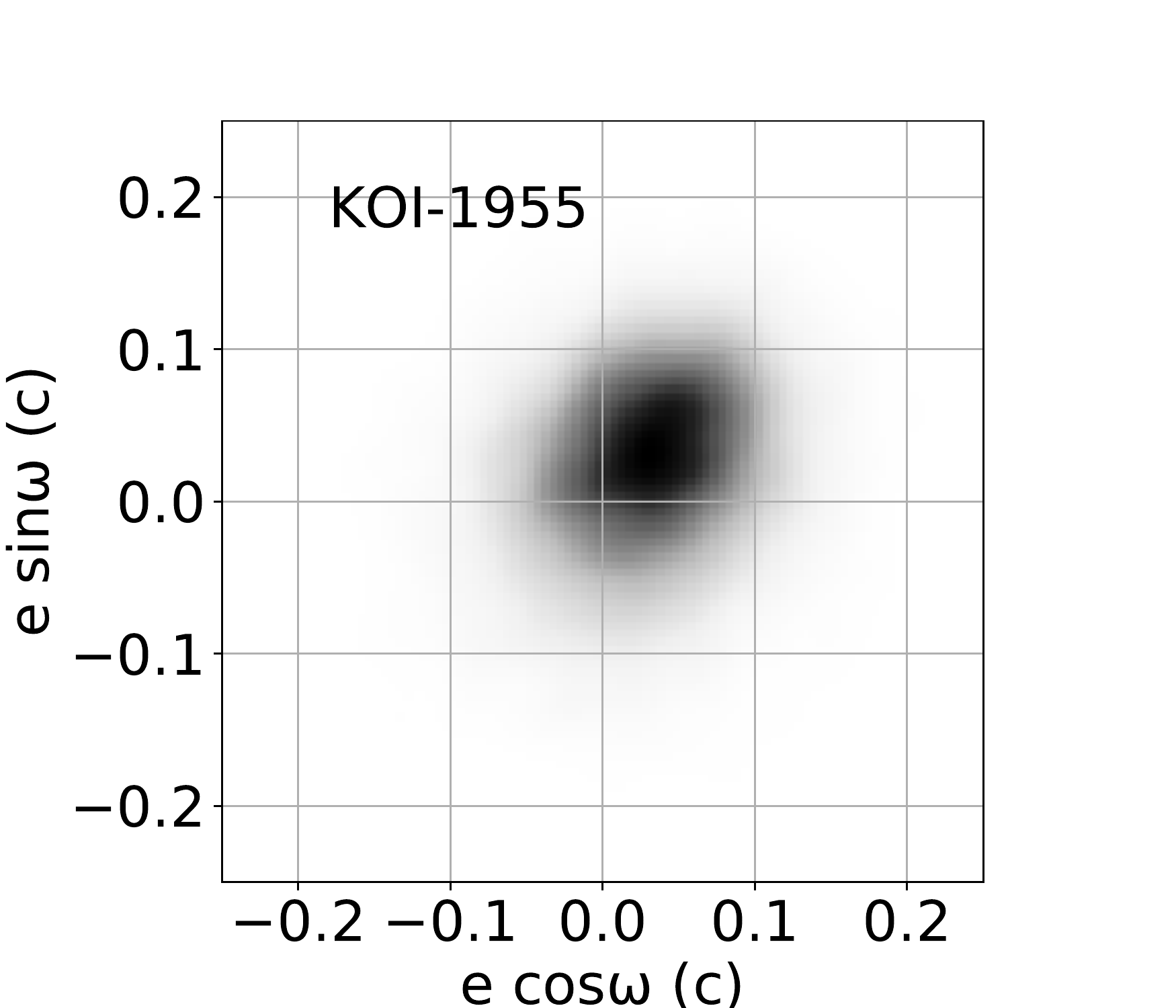}
\includegraphics [height = 1.1 in]{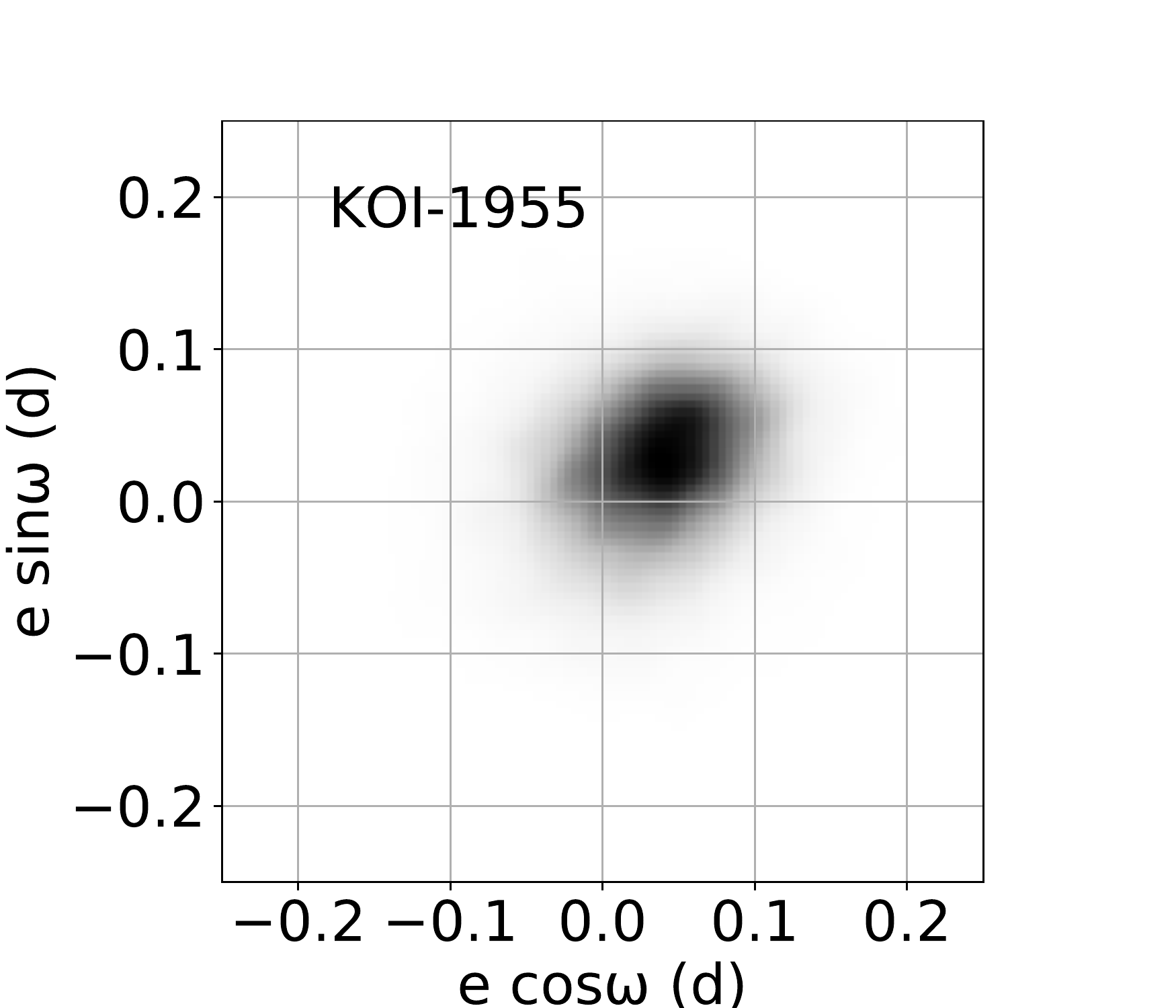} \\
\includegraphics [height = 1.1 in]{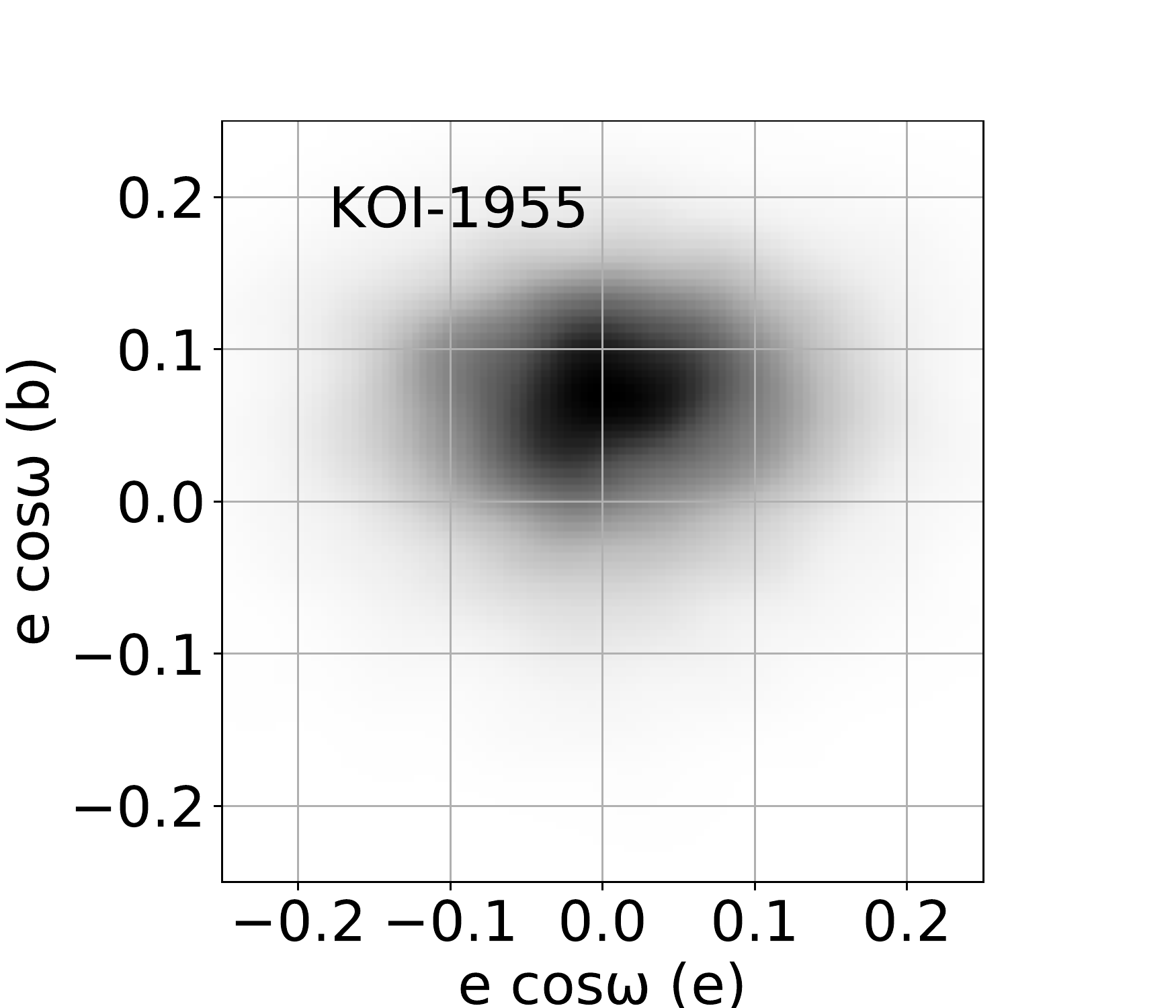}
\includegraphics [height = 1.1 in]{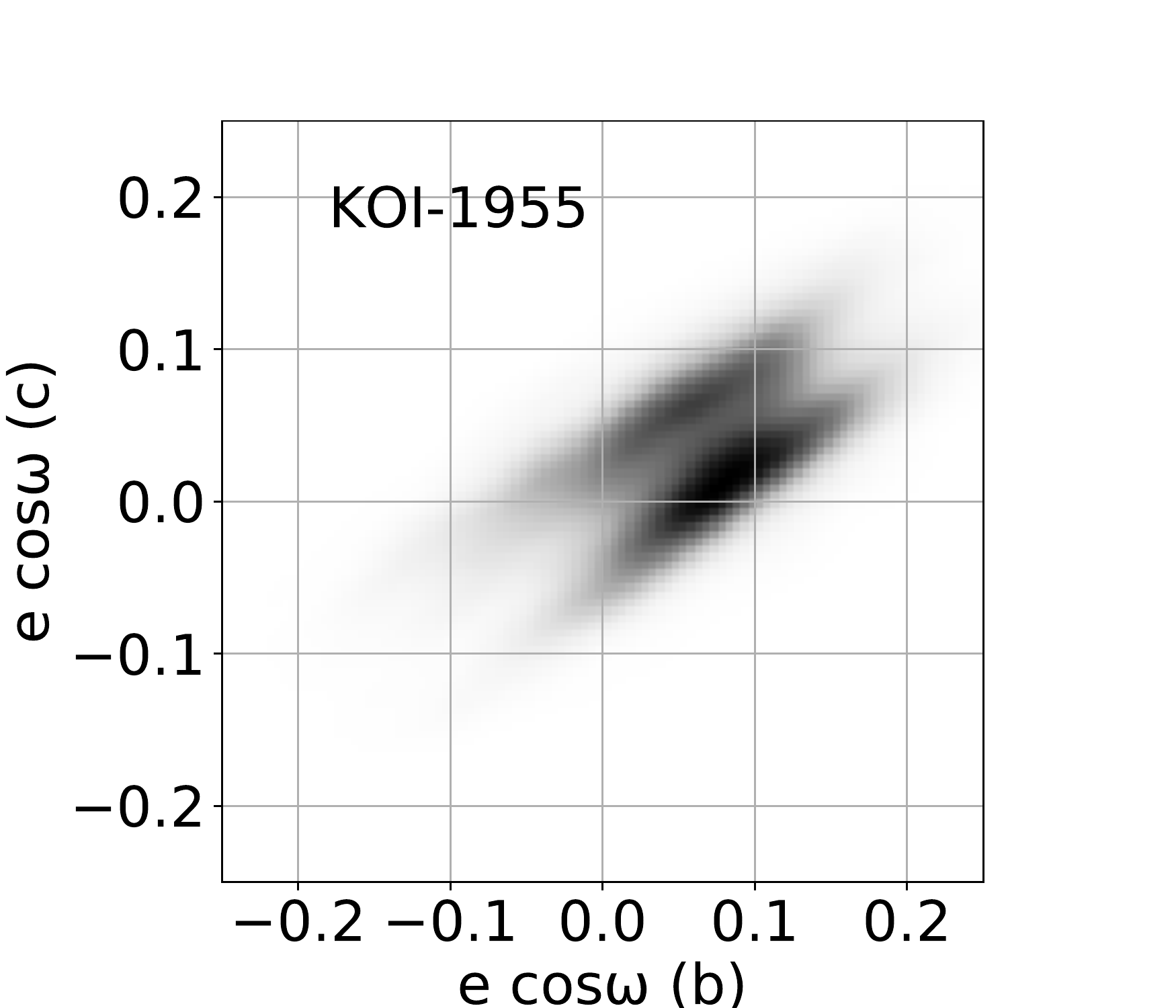}
\includegraphics [height = 1.1 in]{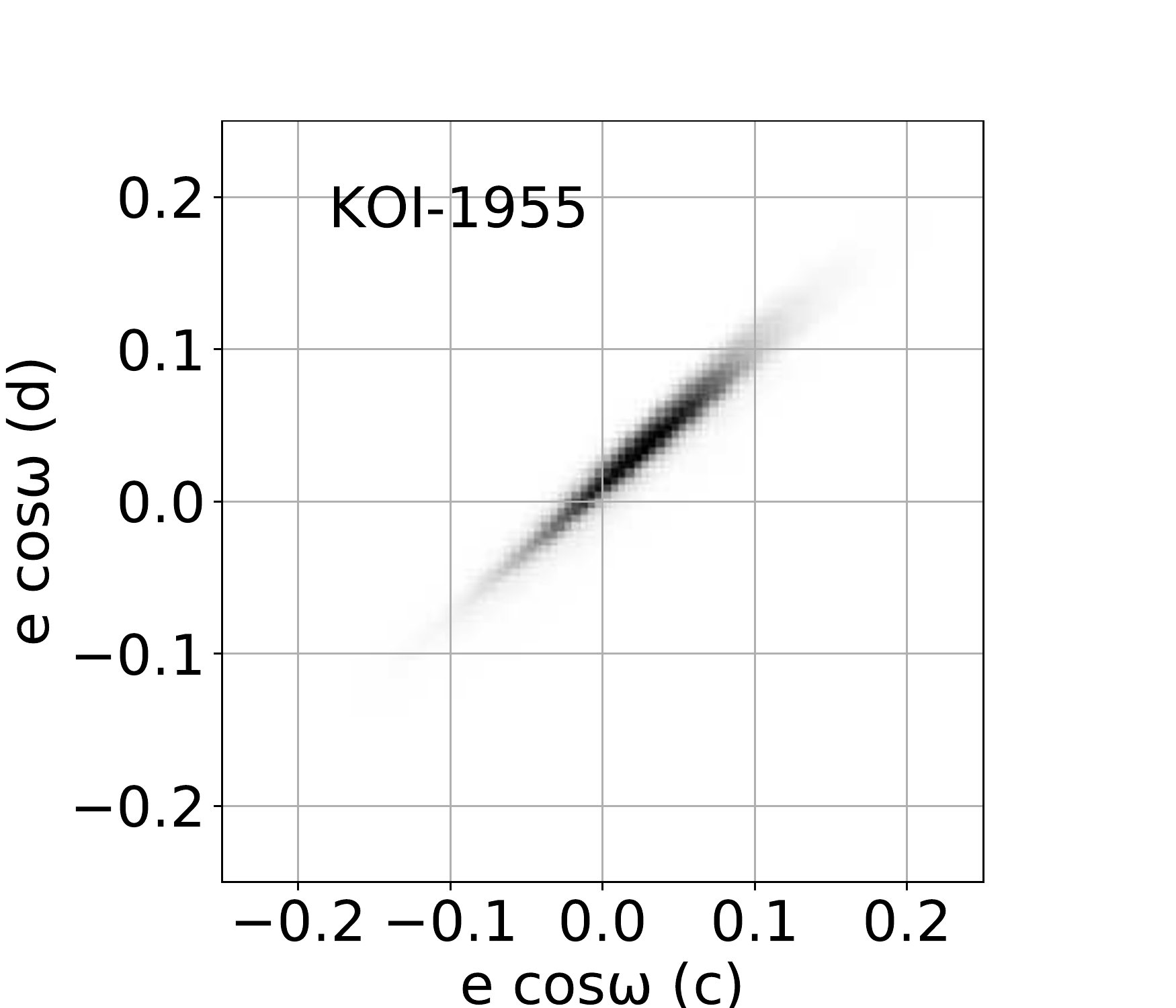}
\includegraphics [height = 1.1 in]{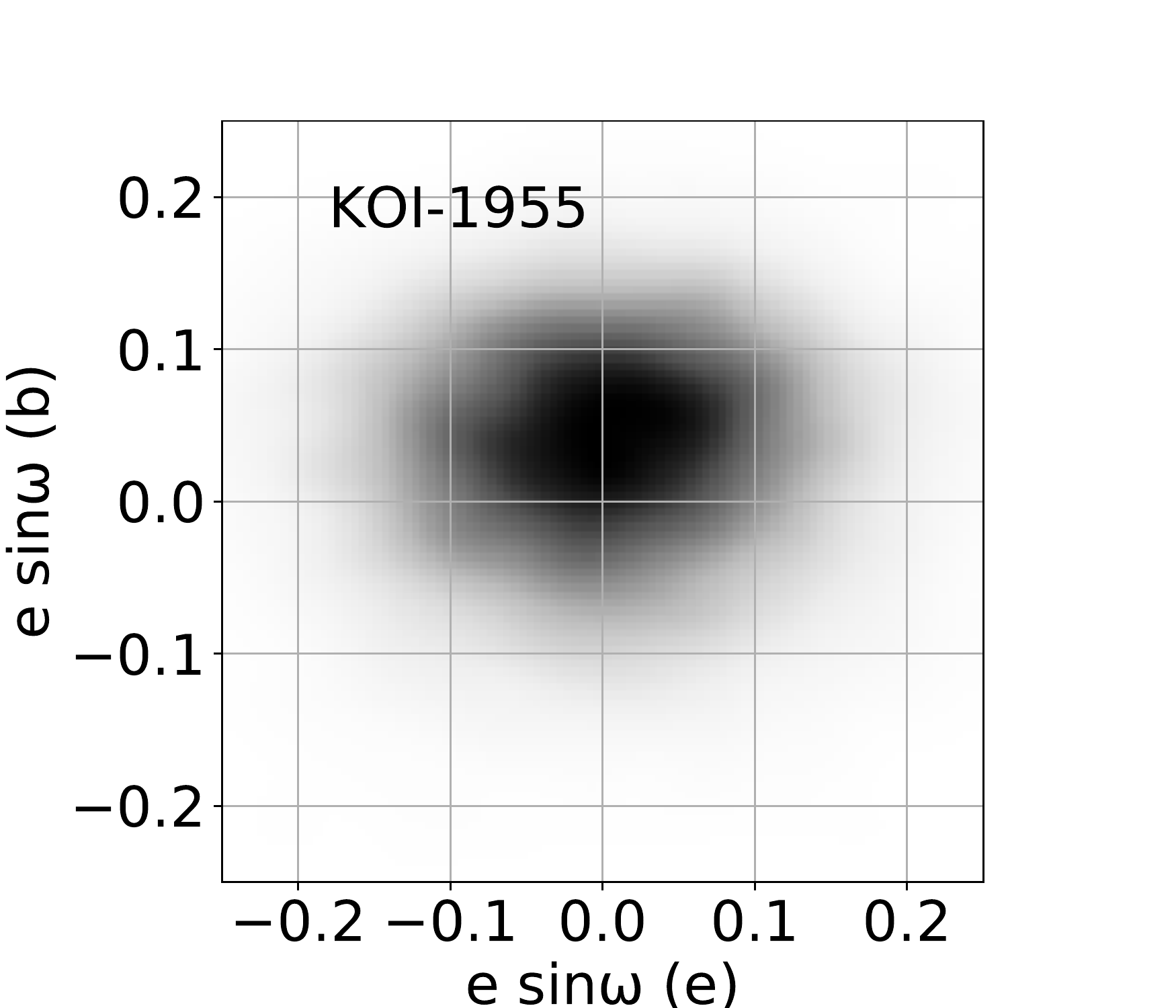} \\
\includegraphics [height = 1.1 in]{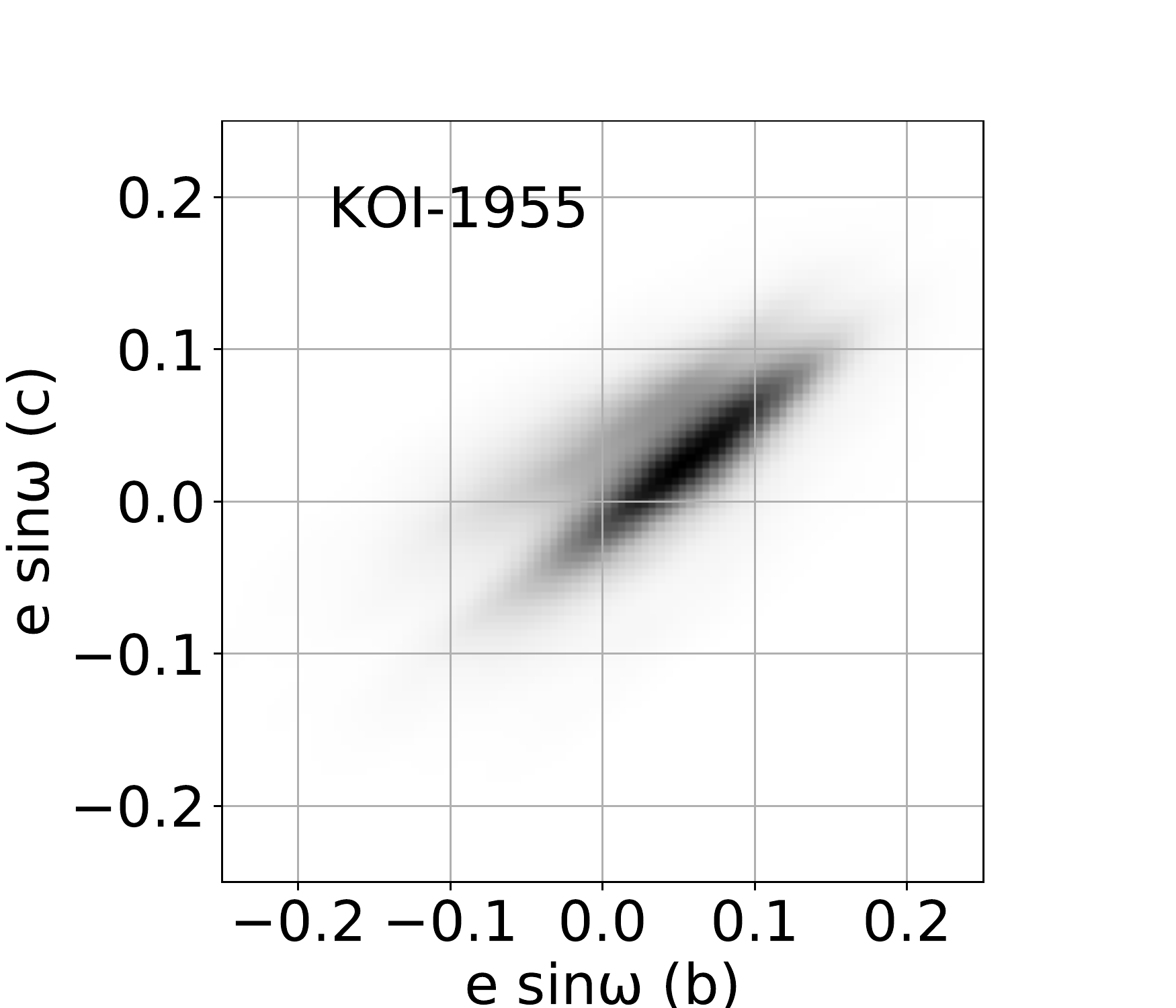}
\includegraphics [height = 1.1 in]{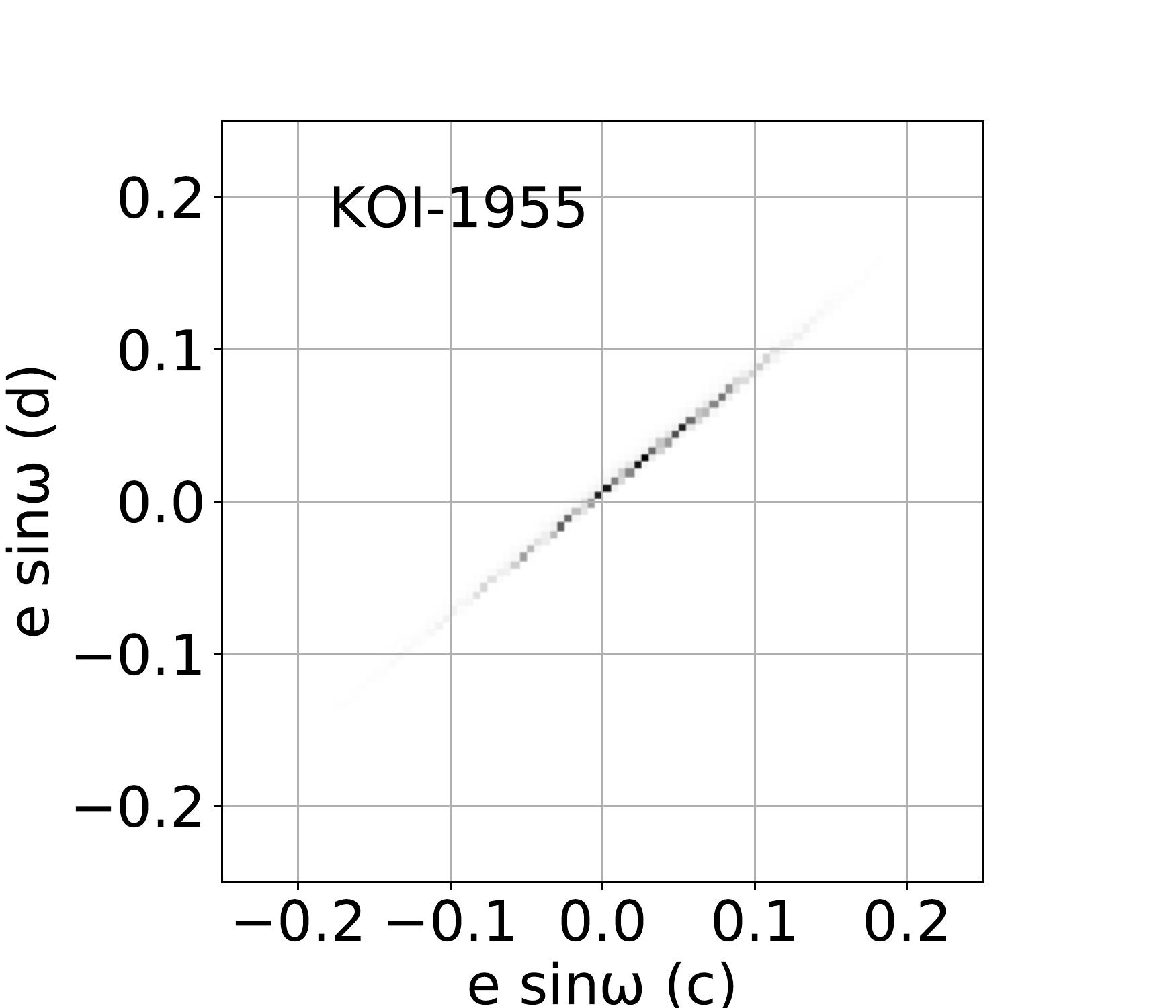}
\caption{Two-dimensional kernel density estimators on joint posteriors of eccentricity vector components: four-planet systems (Part 4 of 4). 
\label{fig:ecc4d} }
\end{center}
\end{figure}

The remainder of the candidates have eccentricity constraints that improve little on our prior. In systems with moderately detected TTVs, individual eccentricities are poorly constrained. Nevertheless, TTVs are sensitive to relative eccentricities in interacting pairs \citep{Lith2012}, whereby adding the same eccentricity vector to both planetary orbits has little effect on the TTVs. We are left with tightly correlated eccentricity vector components between interacting pairs, unless the TTVs contain enough information to break this eccentricity-eccentricity degeneracy. There are many examples of these correlations in Figures~\ref{fig:ecc2a}--\ref{fig:ecc4d}, and in general the solutions are stable since the orbits are apsidally aligned and locked \citep{Jontof-Hutter2016}. Nevertheless, we expect eccentricities in compact transiting systems to be small from tidal eccentricity damping. \citet{Hansen2015} show that eccentricity damping in a planet near the host propagates to secularly coupled neighbors, leading to circular orbits even beyond the range of efficient tidal damping from star-planet interactions. Yet the possibility of aligned and eccentric orbits still cannot be ruled out, since the aligned component damps the slowest (\citealt{Mardling2007,VanLaerhoven2012}). 

Despite these limitations on constraining eccentricity from TTVs, there is still useful information in our eccentricity posteriors when only relative eccentricities are constrained. In some cases, joint posteriors of eccentricity vector components exclude the possibility of zero eccentricity in both interacting planets, and hence some eccentricity in the system is required. See for example, KOI-2086 in Figure~\ref{fig:ecc3f} where there is no likelihood in joint posteriors of eccentricity vector components both equalling zero. In such cases, we detect nonzero orbital eccentricities even though the data cannot determine which planetary orbit must be eccentric. We identified planet pairs that require some relative eccentricity where for adjacent planet pairs $e\cos\omega$=0 and $e'\cos\omega'$=0 are excluded from the joint posterior, or where for adjacent planet pairs $e\sin\omega$=0 and $e'\sin\omega'$=0 are excluded from the joint posterior (i.e., where the joint distributions in Figures~\ref{fig:ecc2a}--\ref{fig:ecc4d} for $e\cos\omega$=0 and $e'\cos\omega'$ or $e\sin\omega$=0 and $e'\sin\omega'$ do not include the graph origin). We counted the number of samples in the parameter space where $e\cos\omega$ < 0 and $e'\cos\omega'$ < 0, $e\cos\omega$ < 0 and $e'\cos\omega'$ > 0,  $e\cos\omega$ > 0 and $e'\cos\omega'$ >0, and where $e\cos\omega$ > 0 and $e'\cos\omega'$ <0 and repeated the same for pairs in $e\sin\omega$ and $e'\sin\omega'$. Planet pairs where none of our 10,000 samples are in one of these eight `quadrants' were considered detections of relative eccentricity. This criterion works well to identify the pairs where individual eccentricities are not well constrained, but relative eccentricities are constrained. However, where some individual eccentricities are strongly constrained, like in the three-planet KOI-377 (Kepler-9) system, this test mistakenly places the noninteracting inner pair in the `detections' column (see joint posteriors for KOI-377 in Figure~\ref{fig:ecc3b}). 

In the tables in Appendix C, we list planet pairs that show evidence of relative eccentricity (excluding pairs with planets where individual eccentricities are strongly constrained). Two of the systems with nonzero eccentricity are known to be in Laplace-like resonant chains, KOI-2086 (Kepler-60, \citealt{Gozdziewski2016, Jontof-Hutter2016}) and KOI-730 (Kepler-223, \citealt{Mills2016}), and are likely in pairwise mean motion resonances. We also find nonzero eccentricity at KOI-738, (Kepler-29) which is likely in the second-order 9:7 mean motion resonance \citep{Migaszewski2017}. 

Other examples of nonzero relative eccentricities, with no evidence of mean motion resonances include KOI-277 (Kepler-36), KOI-1783, KOI-3503, KOI-115.01 and KOI-115.02 (Kepler-105 b and Kepler-105 c), KOI-314.01 and KOI-314.02 (Kepler-138 c and Kepler-138 d), KOI-886.01 and KOI-886.02 (Kepler-54 b and Kepler-54 c), and KOI-250.01 and KOI-250.02 (Kepler-26 b and Kepler-26 c). In the remaining systems, zero eccentricity cannot be excluded. This includes KOI-1599 (Kepler-1659) which is known to be in resonance \citep{Panichi2019}.

The planet pairs with poorly constrained eccentricities and poorly constrained relative eccentricities are listed in Table ~\ref{tbl:Ecc-nondetections} in Appendix C. We note that there are no detections of relative eccentricity among planets orbiting at less than 5.7 days, while detectable interactions were expected for two pairs with an inner planet orbiting at less than 5.7 days. This is consistent with reduced tidal damping at greater orbital distances. However, it is also consistent with stronger TTV signals at greater orbital periods. 

\section{Stellar Parameters and Planet Characterization}
Our adopted stellar parameters for planet characterization are in Table~\ref{tbl-StellarProperties}. We rely on \citet{Fulton2018} for stellar effective temperatures, masses and radii as our first choice of stellar parameters. Several of our systems are missing data in that catalog. Our second choice was to retrieve stellar radii from \citet{Berger2020}. 
\begin{table}[ht!]
 \tiny
  \begin{center}
    \begin{tabular}{|c|c|c|c|c|}
      \hline
      KOI &  T$_{eff}$  (K) & M$_{\star}$ (M$_{\odot}$) &  R$_{\star}$ (R$_{\odot}$) & source \\
\hline
85  & 6220   $^{+  60 }_{- 60  }$  & 1.249    $^{+0.010  }_{- 0.030   }$  & 1.443   $^{+0.033   }_{- 0.032  }$  &  1 \\ 
115  & 5933   $^{+  60 }_{- 60  }$  & 0.991    $^{+0.032  }_{- 0.033   }$  & 1.026   $^{+0.025   }_{- 0.024  }$  &  1 \\ 
137  & 5441   $^{+  60 }_{- 60  }$  & 0.983    $^{+0.026  }_{- 0.024   }$  & 0.904   $^{+0.023   }_{- 0.022  }$  &  1 \\ 
152  & 6389   $^{+  60 }_{- 60  }$  & 1.244    $^{+0.027  }_{- 0.042   }$  & 1.316   $^{+0.038   }_{- 0.037  }$  &  1 \\ 
156  & 4473   $^{+  70 }_{- 61  }$  & 0.731    $^{+0.029  }_{- 0.022   }$  & 0.725   $^{+0.011   }_{- 0.012  }$  &  2 \\ 
168  & 5823   $^{+  60 }_{- 60  }$  & 0.990    $^{+0.024  }_{- 0.023   }$  & 1.491   $^{+0.039   }_{- 0.038  }$  &  1 \\ 
222  & 4542   $^{+  75 }_{- 72  }$  & 0.697    $^{+0.029  }_{- 0.022   }$  & 0.698   $^{+0.013   }_{- 0.013  }$  &  2 \\ 
244  & 6285   $^{+  60 }_{- 60  }$  & 1.148    $^{+0.035  }_{- 0.033   }$  & 1.342   $^{+0.029   }_{- 0.028  }$  &  1 \\ 
248  & 4096   $^{+  61 }_{- 75  }$  & 0.607    $^{+0.014  }_{- 0.014   }$  & 0.618   $^{+0.019   }_{- 0.020  }$  &  2 \\ 
250  & 4124   $^{+  43 }_{- 68  }$  & 0.593    $^{+0.015  }_{- 0.016   }$  & 0.595   $^{+0.020   }_{- 0.026  }$  &  2 \\ 
255  & 4066   $^{+  75 }_{- 78  }$  & 0.617    $^{+0.013  }_{- 0.013   }$  & 0.634   $^{+0.018   }_{- 0.018  }$  &  2 \\ 
277  & 5979   $^{+  60 }_{- 60  }$  & 1.034    $^{+0.022  }_{- 0.022   }$  & 1.634   $^{+0.042   }_{- 0.040  }$  &  1 \\ 
314  & 3975   $^{+  64 }_{- 62  }$  & 0.535    $^{+0.012  }_{- 0.012   }$  & 0.535   $^{+0.013   }_{- 0.014  }$  &  2 \\ 
377  & 5788   $^{+  60 }_{- 60  }$  & 1.024    $^{+0.024  }_{- 0.029   }$  & 0.971   $^{+0.025   }_{- 0.024  }$  &  1 \\ 
401  & 5455   $^{+  60 }_{- 60  }$  & 1.013    $^{+0.031  }_{- 0.033   }$  & 0.975   $^{+0.026   }_{- 0.026  }$  &  1 \\ 
430  & 4267   $^{+  78 }_{- 67  }$  & 0.636    $^{+0.018  }_{- 0.018   }$  & 0.626   $^{+0.013   }_{- 0.012  }$  &  2 \\ 
457  & 5108   $^{+  60 }_{- 60  }$  & 0.853    $^{+0.009  }_{- 0.011   }$  & 0.751   $^{+0.020   }_{- 0.019  }$  &  1 \\ 
520   &  5106   $^{+   60  }_{-  60  }$    & 0.860   $^{+   0.016  }_{-  0.024  }$   &  0.787  $^{+    0.023  }_{-  0.022   }$   &  1  \\
523  & 5732   $^{+  60 }_{- 60  }$  & 0.921    $^{+0.025  }_{- 0.023   }$  & 1.324   $^{+0.053   }_{- 0.051  }$  &  1 \\ 
567  & 5530   $^{+  60 }_{- 60  }$  & 0.872    $^{+0.024  }_{- 0.028   }$  & 0.821   $^{+0.023   }_{- 0.022  }$  &  1 \\ 
620  & 5674   $^{+  60 }_{- 60  }$  & 0.929    $^{+0.016  }_{- 0.013   }$  & 0.820   $^{+0.027   }_{- 0.026  }$  &  1 \\ 
654  & 5785   $^{+  60 }_{- 60  }$  & 0.972    $^{+0.031  }_{- 0.034   }$  & 0.974   $^{+0.025   }_{- 0.024  }$  &  1 \\ 
730  & 5697   $^{+  60 }_{- 60  }$  & 1.041    $^{+0.032  }_{- 0.031   }$  & 1.574   $^{+0.106   }_{- 0.096  }$  &  1 \\ 
738  & 5378   $^{+  60 }_{- 60  }$  & 0.761    $^{+0.024  }_{- 0.028   }$  & 0.732   $^{+0.033   }_{- 0.031  }$  &  1 \\ 
750   &  5048    $^{+  80 }_{-   76   }$   &  0.819    $^{+  0.030  }_{-  0.023   }$   &  0.897    $^{+  0.028  }_{-  0.026   }$ &  2  \\
806  & 5464   $^{+  60 }_{- 60  }$  & 0.945    $^{+0.016  }_{- 0.022   }$  & 0.819   $^{+0.036   }_{- 0.034  }$  &  1 \\ 
877  & 4331   $^{+  84 }_{- 83  }$  & 0.647    $^{+0.020  }_{- 0.020   }$  & 0.633   $^{+0.012   }_{- 0.011  }$  &  2 \\ 
886  & 3854   $^{+  69 }_{- 80  }$  & 0.518    $^{+0.012  }_{- 0.013   }$  & 0.522   $^{+0.014   }_{- 0.013  }$  &  2 \\ 
898  & 4163   $^{+  74 }_{- 42  }$  & 0.592    $^{+0.017  }_{- 0.021   }$  & 0.584   $^{+0.027   }_{- 0.026  }$  &  2 \\ 
934  & 5503   $^{+  60 }_{- 60  }$  & 0.849    $^{+0.034  }_{- 0.027   }$  & 0.902   $^{+0.059   }_{- 0.054  }$  &  1 \\ 
1070  & 5600   $^{+  114 }_{- 111  }$  & 0.973    $^{+0.069  }_{- 0.067   }$  & 1.094   $^{+0.055   }_{- 0.051  }$  &  2 \\ 
1279  & 5705   $^{+  60 }_{- 60  }$  & 0.927    $^{+0.032  }_{- 0.029   }$  & 1.051   $^{+0.029   }_{- 0.028  }$  &  1 \\ 
1338  & 5680   $^{+  60 }_{- 60  }$  & 0.870    $^{+0.032  }_{- 0.028   }$  & 0.967   $^{+0.030   }_{- 0.029  }$  &  1 \\ 
1353   &  5989  $^{+   60}_{-   60 }$  &   1.061  $^{+   0.011}_{-   0.035  }$  &  0.994  $^{+   0.026 }_{-  0.025  }$   &  1  \\
1574  & 5885   $^{+  127 }_{- 119  }$  & 1.016    $^{+0.090  }_{- 0.072   }$  & 1.422   $^{+0.043   }_{- 0.040  }$  &  2 \\ 
1576  & 5559   $^{+  60 }_{- 60  }$  & 1.007    $^{+0.014  }_{- 0.027   }$  & 0.919   $^{+0.023   }_{- 0.023  }$  &  1 \\ 
1598  & 5451   $^{+  60 }_{- 60  }$  & 0.854    $^{+0.034  }_{- 0.032   }$  & 0.839   $^{+0.022   }_{- 0.022  }$  &  1 \\ 
1599 & 5823  $^{+  158 }_{-   193 }$ &1.02 $^{+  0.115 }_{- 0.127   }$ & 0.972 $^{+  0.291  }_{-  0.104 }$  & 3 \\
1783  & 5923   $^{+  60 }_{- 60  }$  & 1.076    $^{+0.036  }_{- 0.032   }$  & 1.143   $^{+0.031   }_{- 0.030  }$  &  1 \\ 
1831  & 5233   $^{+  60 }_{- 60  }$  & 0.901    $^{+0.017  }_{- 0.027   }$  & 0.835   $^{+0.023   }_{- 0.022  }$  &  1 \\ 
1833  & 4413   $^{+  75 }_{- 69  }$  & 0.681    $^{+0.025  }_{- 0.024   }$  & 0.667   $^{+0.012   }_{- 0.012  }$  &  2 \\ 
1955  & 6272   $^{+  60 }_{- 60  }$  & 1.213    $^{+0.031  }_{- 0.018   }$  & 1.363   $^{+0.034   }_{- 0.033  }$  &  1 \\ 
2028  & 5213   $^{+  97 }_{- 91  }$  & 0.827    $^{+0.047  }_{- 0.053   }$  & 0.805   $^{+0.035   }_{- 0.032  }$  &  2 \\ 
2086  & 5834   $^{+  60 }_{- 60  }$  & 1.000    $^{+0.026  }_{- 0.025   }$  & 1.433   $^{+0.042   }_{- 0.041  }$  &  1 \\ 
2092  & 5585   $^{+  60 }_{- 60  }$  & 0.827    $^{+0.030  }_{- 0.029   }$  & 0.835   $^{+0.068   }_{- 0.062  }$  &  1 \\ 
2113  & 5183   $^{+  98 }_{- 93  }$  & 0.763    $^{+0.045  }_{- 0.040   }$  & 0.770   $^{+0.037   }_{- 0.033  }$  &  2 \\ 
2174  & 4356   $^{+  66 }_{- 66  }$  & 0.602    $^{+0.015  }_{- 0.015   }$  & 0.572   $^{+0.013   }_{- 0.013  }$  &  3 \\ 
2195  & 6121   $^{+  60 }_{- 60  }$  & 1.037    $^{+0.029  }_{- 0.029   }$  & 1.249   $^{+0.053   }_{- 0.050  }$  &  1 \\ 
2414  & 5541   $^{+  60 }_{- 60  }$  & 0.882    $^{+0.015  }_{- 0.014   }$  & 1.617   $^{+0.042   }_{- 0.041  }$  &  1 \\ 
3503   &  6001 $^{+   60}_{-   60  }$    &  0.858 $^{+   0.022 }_{-  0.019  }$    &  1.255 $^{+   0.034 }_{-  0.033  }$     &  1  \\
\hline
     \end{tabular}
    \caption{Adopted stellar effective temperatures, masses and radii used for planetary characterization, with the source listed in the last columns. Source 1: \citet{Fulton2018} .  Source 2: \citet{Berger2020}. Source 3:  \textit{Kepler} DR25 as listed on https://exoplanetarchive.ipac.caltech.edu  \citep{Thompson2018}. 
    }\label{tbl-StellarProperties}
  \end{center}
\end{table}

We drew samples from posterior summary statistics of stellar parameters and combined them with our samples of planetary dynamical masses and published planetary radii to characterize credible intervals of planetary masses, densities, equilibrium blackbody temperatures, and atmospheric transmission annuli. We summarize our posteriors of planetary parameters with sample medians and bounds enclosing 68.3\% of samples with equal weight in the tails. In some cases, parameters drawn from other studies have substantial skewness. This is particularly a problem for derived quantities based on skewed parameters like bulk density or atmospheric scale height that are sensitive to planetary radii. We considered posteriors with quoted negative error bars within 10\% of the positive error bar as Gaussian distributions and adopted the average of the published uncertainties as the uncertainty. For moderately skewed distributions over a variable $x$ where we rely on skewed posteriors, we adopt the following approximate likelihood:

\begin{equation}
p(x|x_{mode},\kappa^{+}, \kappa^{-}) \simeq  \frac{1}{\sqrt{2\pi\kappa_m^2}}    \exp    \left[ -\frac{1}{2} \left(\frac{\log(1+\gamma (x - x_{mode} ))} {\log \beta}\right)^2 \right]
\label{barlow}
\end{equation}
where for non-Gaussian uncertainties $\kappa^{+}$ and $\kappa^{-}$, $\beta = \kappa^{+}/\kappa^{-}$ and $\gamma = \frac{\kappa^{+}-\kappa^{-}}{\kappa^{+}\kappa^{-}}$  \citep{Barlow2004}. 

While the denominator in the exponential function in Equation~\ref{barlow} approaches zero in the limit of symmetric uncertainties, it does provide a smooth distribution with skew from just three parameters and closely approximates a Gaussian where $\beta = 1.1$ or 0.9. Equation~\ref{barlow} requires positive definite quantities, which is satisfied for our parameters of interest. For moderate asymmetries in the uncertainties, the function gives excellent agreement with the 68.3\% central region of a log-normal distribution. One disadvantage of this function is that it is not normalized; if $\beta = 1.1$ or 0.9, the integrated likelihoods sum to 1.003, and the normalization error rises to $\approx$ 17$\%$ if $\beta = 2$. 

To calculate planetary bulk density, we rely on \citet{Berger2018} for planetary radii where available. However, where planetary radii in \citet{Berger2018} have significantly different upper and lower uncertainties such that $\beta > 2$ or  $\beta < 1/2$, where the approximation of equation~\ref{barlow} loses accuracy, we adopt relevant parameters from other sources if they are skewed less, since excessively skewed radius distributions cause an excess of posterior likelihood at extreme bulk density or atmospheric scale height. For example, \citet{Berger2018} estimate the radius of KOI-620.03 (Kepler-51 c), which has a grazing transit and large TTVs, as R$_{p} = 5.44 ^{+29.43}_{-1.15}$ R$_{\oplus}$. In this case, the poorly constrained upper bound coupled with a stronger lower bound could lead to an overly optimistic estimate of the transmission annulus. Other large catalogs of \textit{Kepler} planetary parameters like \textit{Kepler} DR 24, DR 25, or \citet{Fulton2018} have similar difficulty with either skewed or imprecise errors for KOI-620.03. Hence, for this system, we adopt the transit parameters of \citet{LibbyRoberts2020}. 

Elsewhere, when our default source has skewed uncertainties, such that  $\beta > 2$ or  $\beta < 1/2$, we adopt the parameters of \textit{Kepler} DR 25 if less skewed. This included the planet radii of KOIs 85.03, 115.02, 115.03, 152.04, 255.02, 314.01, 430.01, 567.01, 567.03, 654.01, 730.02, 730.03,  750.01, 750.02, 877.03, 1338.02, 1598.01, 1598.02, 1598.03, 1831.01, 1831.04, 1833.02, 1833.03, 1955.01, 1955.03, 1955.04, 2086.01, 2086.02, 2086.03, 2195.01, 2195.02 and 2195.03.
 
Among our selected systems, KOIs 1599, 2092, and 2174 are missing from \citet{Berger2018}. For KOI-1599, we adopt the planet radii of \citet{Panichi2019}: $R_p$ = 1.9$\pm 0.2$ $R_{\oplus}$ for KOI-1599.01 and 1.9 $\pm 0.3$  $R_{\oplus}$ for KOI-1599.02. Although the dynamical masses of the planets are constrained to an uncertainty $\sim$3\% \citep{Panichi2019}, the mass of the star KOI-1599 is poorly constrained, leading to weak constraints on the planetary bulk properties. For KOI-2092.03, we adopt our own measurement of the radius from analysis of the light curve. 

In cases where the planetary mass upper limits are weak, we imposed a maximum bulk density of 10 g cm$^{-3}$. Our results are in Tables~\ref{tbl-massresults2}--\ref{tbl-massresults4}. For some individual systems, other authors have used short-cadence transit-timing data \citep{Jontof-Hutter2016}, or performed photodynamical models to self-consistently account for eccentricities and the host density \citep{Almenara2018} or have adopted different priors in mass or eccentricity than we have here \citep{Hadden2017}. In several cases, our adoption of post-Gaia stellar parameters improves upon the published precision in measured planetary parameters, but not always. Hence, we report both our results and the results of other studies in Tables~\ref{tbl-massresults2}--\ref{tbl-massresults4}. We attempt to highlight which is the preferred source for planet parameters although for many candidates this cannot be done in an objective and uniform way. For example, the radius of Kepler-105 following Gaia is $\sim$ 39\% larger than the size adopted by \citet{Jontof-Hutter2016}, but also has larger uncertainties. As another example, in the case of Kepler-9, TTV data are supplemented with RV data which further constrain planetary masses \citep{Borsato2019}. However, the high S/N TTVs enable precise measurements of dynamical masses, orbital eccentricities, and the bulk density of the host. The Gaia stellar parameters agree closely with the photodynamical model of \citet{Freudenthal2018}. 

For KOI-137 (Kepler-18), updated planetary radii are smaller and inconsistent with those adopted by \citet{coch11}. We also find lower masses for KOI-137.01 and KOI-137.02 than \citet{coch11} did. On the other hand, \citet{coch11} combined RV and TTV data in their mass measurements, and while they did not have the complete transit-timing data set, the amplitude, phase and superperiod of the near-resonant TTV signal of the outer two planets were well constrained by the data at that time. The TTVs do little to constrain the mass of the inner planet KOI-137.03, and the RV measurement of \citet{coch11} remains preferred.

\citet{Panichi2018} report the size of KOI-806.03 (Kepler-30 b) as 3.75 $\pm 0.18$ $R_{\oplus}$, consistent with the earlier measurement of 3.9 $\pm 0.2$ $R_{\oplus}$ by \citet{Sanchis2012} but significantly greater than the radius measurement we adopt from \citet{Berger2018} (1.84 $^{+ 0.13 }_{- 0.17 }$ $R_{\oplus}$). Some of the disagreement may be attributed to the stellar radius measurement. \citet{Sanchis2012} report R$_{\star}$ = 0.95 $\pm 0.12$ R$_{\odot}$. \citet{Fulton2018} revise this to R$_{\star}$ = 0.819 $^{+0.036}_{-0.034}$  R$_{\odot}$. The revised planet radius increases the inferred density to 8.50 $^{+ 1.50 }_{- 1.60 }$ g cm$^{-3}$, which is consistent with a rocky composition.  

\citet{Berger2020} report a larger mass for KOI-886 (Kepler-54) than adopted by \citet{Hadden2017}. In addition, our revised planet sizes are inconsistent with earlier estimates. For KOI-1574.01 (Kepler-87 b) and KOI-1574.02 (Kepler-87 c), our revised radii are smaller than those adopted by \citet{ofir14}, leading to a slightly higher density for the extreme low-density planet KOI-1574.02 (Kepler-87 c). On the other hand, our planetary radii for KOI-1576 (Kepler-307), taken from \citet{Berger2018}, are less precise than those adopted by \citet{Jontof-Hutter2016} and agree more closely with \citet{Hadden2017}. 

For several candidates, we find higher masses and consequently higher densities than those found by \citet{Hadden2017}. In that study, the default prior on planetary masses was uniform on a logarithmic scale and hence weighted towards lower masses. The alternative prior adopted by \citet{Hadden2017} was uniform in planet masses and hence closer to our adopted prior, although in their study, this was coupled with a logarithmic prior in eccentricity. This provided a conservative test for the ``robustness" of the inferred parameters, but also effectively added more weight to low eccentricity solutions leading to tighter constraints on mass. Hence in Tables~\ref{tbl-massresults2}--\ref{tbl-massresults4}, there are several cases where we report upper limits only, where \citet{Hadden2017} report tighter constraints under their high mass prior. Among the TTV candidates that have not been subsequently studied by other authors, we compared the results of \citet{Hadden2017} following their high mass prior with our own results.

All of the planets with a strong expectation of nonresonant and near-resonant TTVs have strongly detected TTVs and the measured transit times impose useful constraints on interacting planets of those systems. All of these systems have TTVs detected and cataloged by \citet{Holczer2016}, and have been characterized in other studies. Of the systems with an expectation of near-resonant TTVs only, the TTVs of KOIs 255, 430, 750, and 1833 were not detected by \citet{Holczer2016} but we have found useful mass upper limits for KOI-255.01, KOI-430.01, KOI-750.01 and KOI-1833.03. Despite an expectation of TTVs due to near-resonance, the transit times of KOIs 1338, 2195 and 2414 provide weak constraints on planetary masses, most likely because their TTV superperiods far exceed the Kepler baseline. 

Among systems with an expectation of strong nonresonant TTVs that are not identified in the catalog of \citet{Holczer2016}, we have found dynamical mass constraints on KOI-85.01, KOIs 654.01 and KOI-654.02, and KOI-1070.02. The existing transit data do not usefully constrain the masses of the planets at KOI-2174. For the systems where we had an expectation of either weakly detected TTVs or a nondetection, several have mass constraints that have not been characterized in other studies. KOIs 222.01, 255.01, and 1279.01 have weak constraints whereby the mass upper limits are lower than those imposed by our prior on bulk density although the densities are still conspicuously high. KOIs 401.03, 413.02, 567.02, 934.02, 2113.01, and 2113.02 all have strong and useful mass constraints imposed by the TTVs. 

\begin{figure}
\begin{center}
\figurenum{17}
\includegraphics [height = 1.1 in]{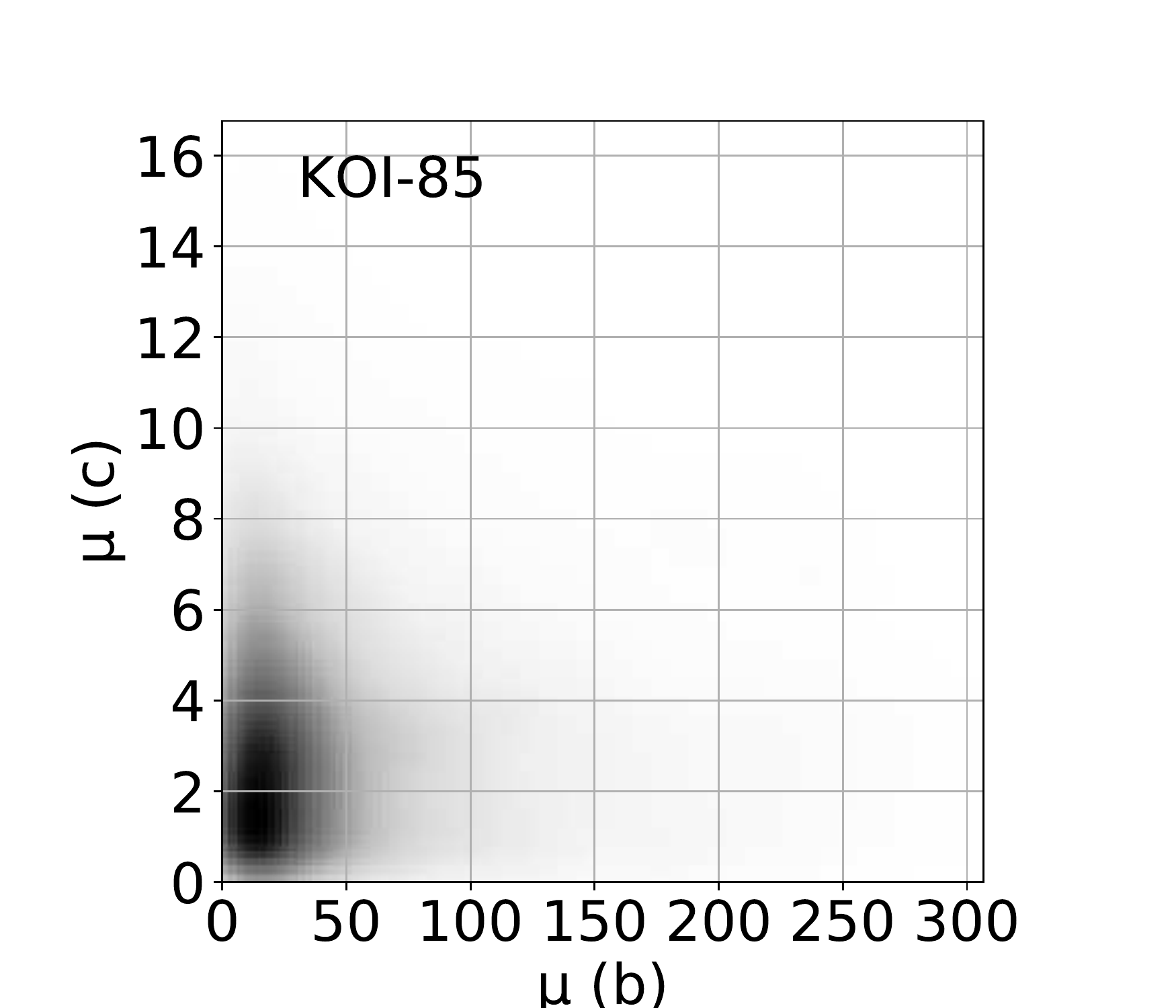}
\includegraphics [height = 1.1 in]{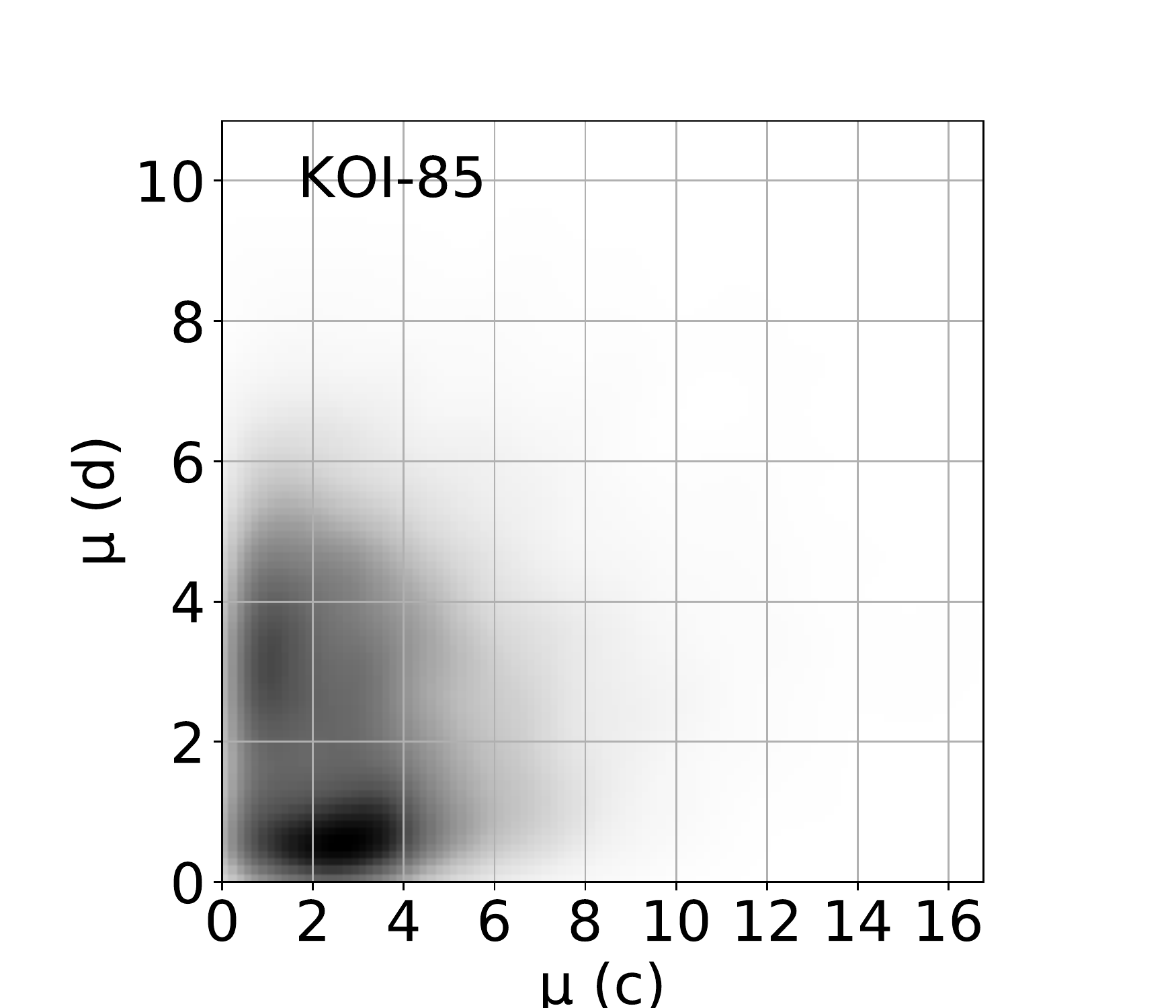}
\includegraphics [height = 1.1 in]{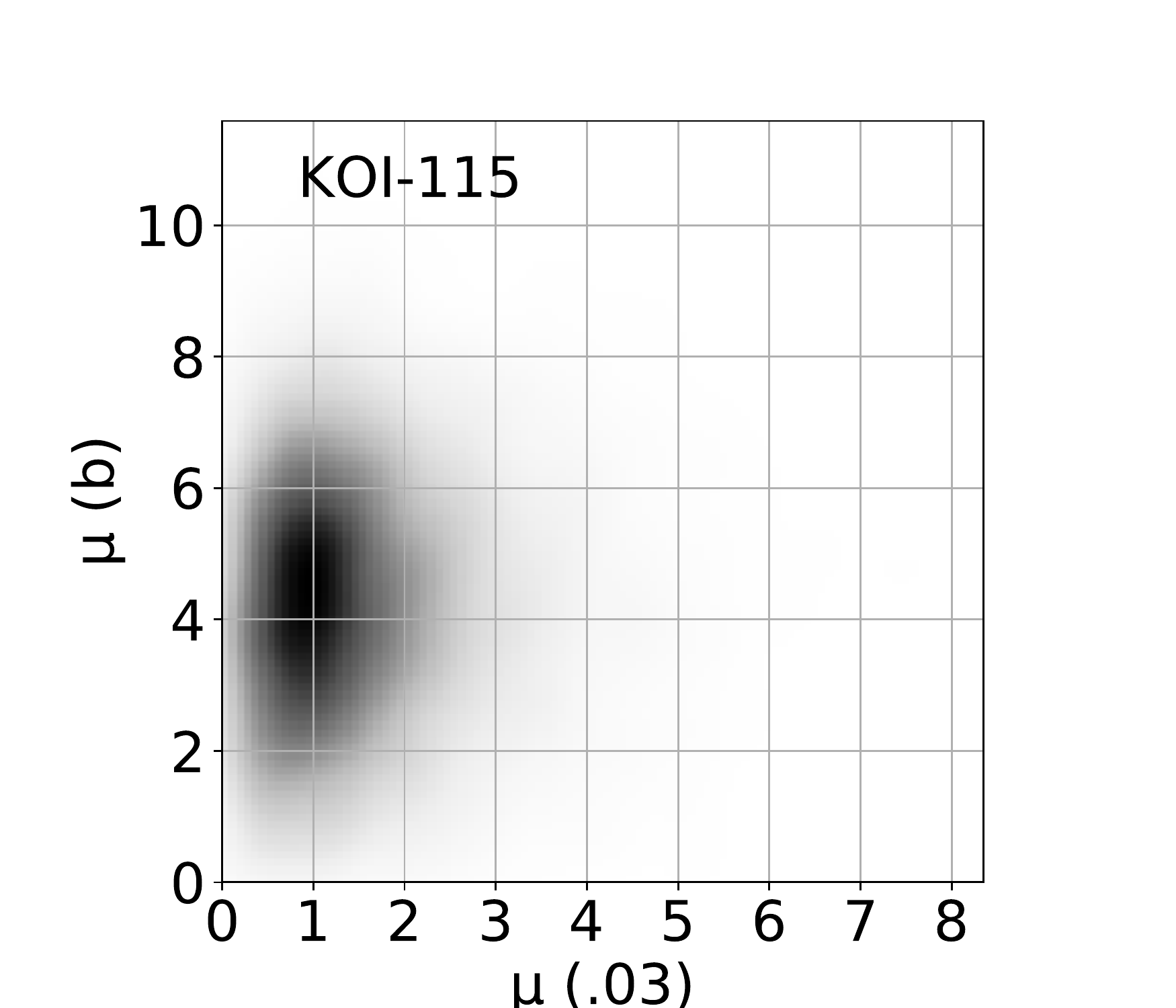}
\includegraphics [height = 1.1 in]{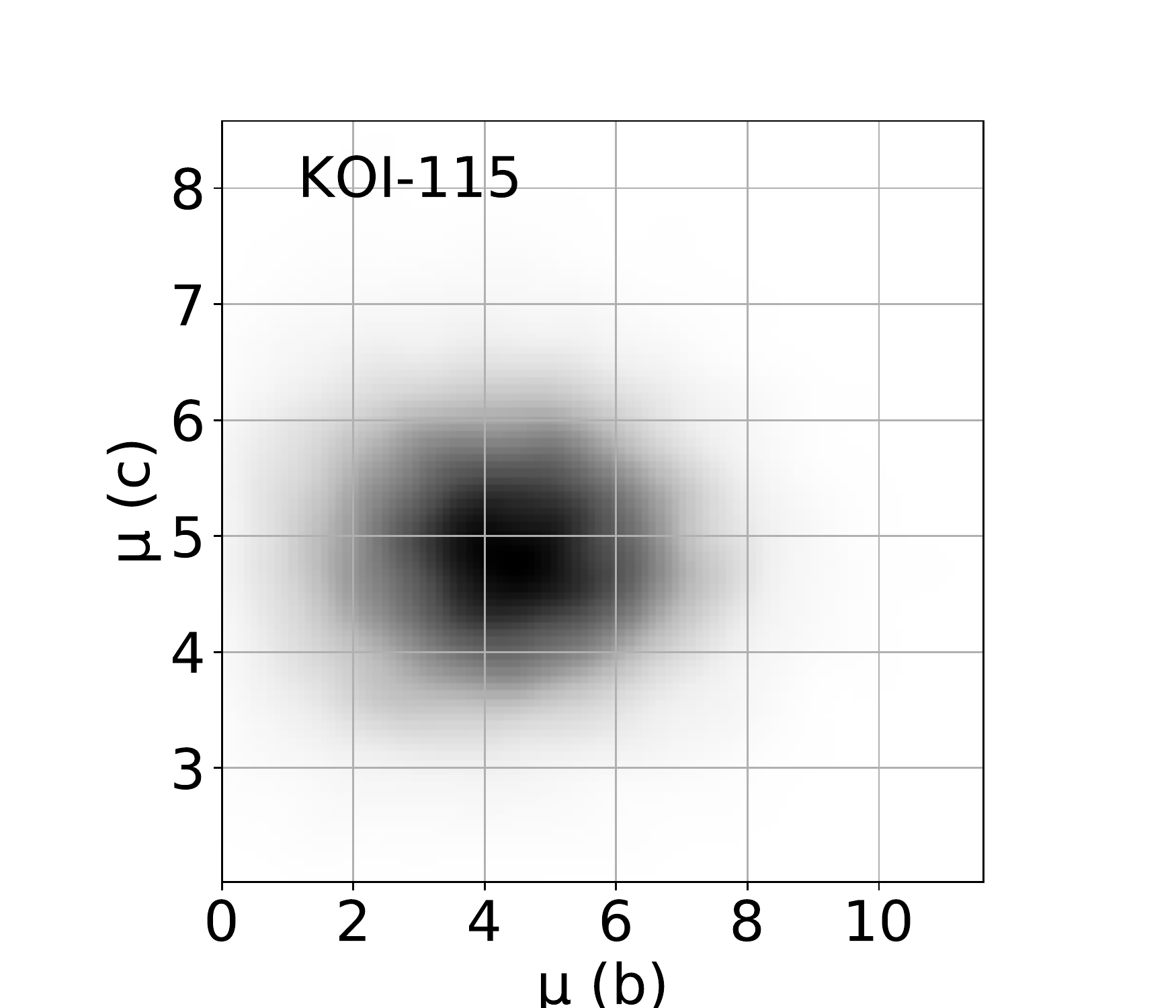} \\
\includegraphics [height = 1.1 in]{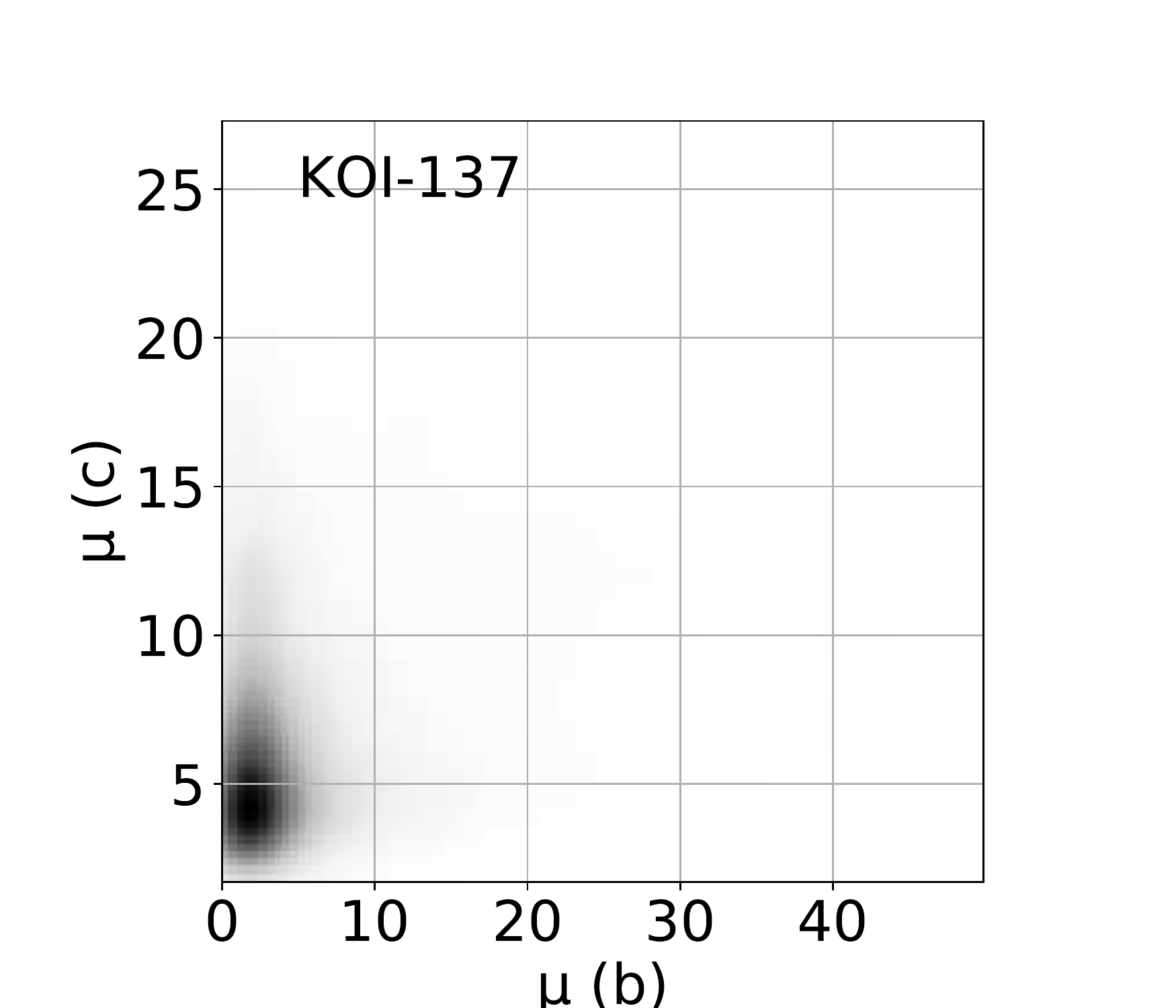}
\includegraphics [height = 1.1 in]{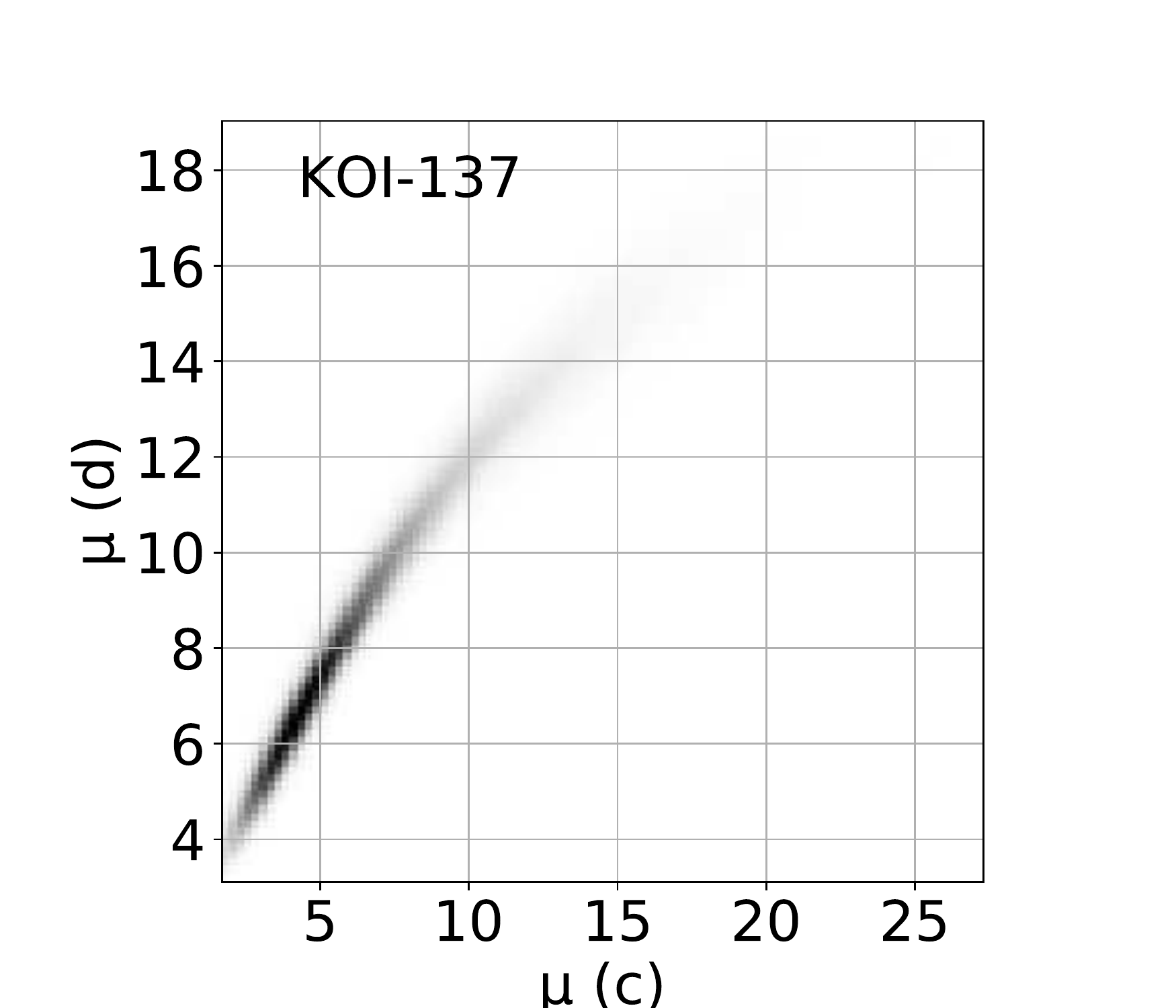}
\includegraphics [height = 1.1 in]{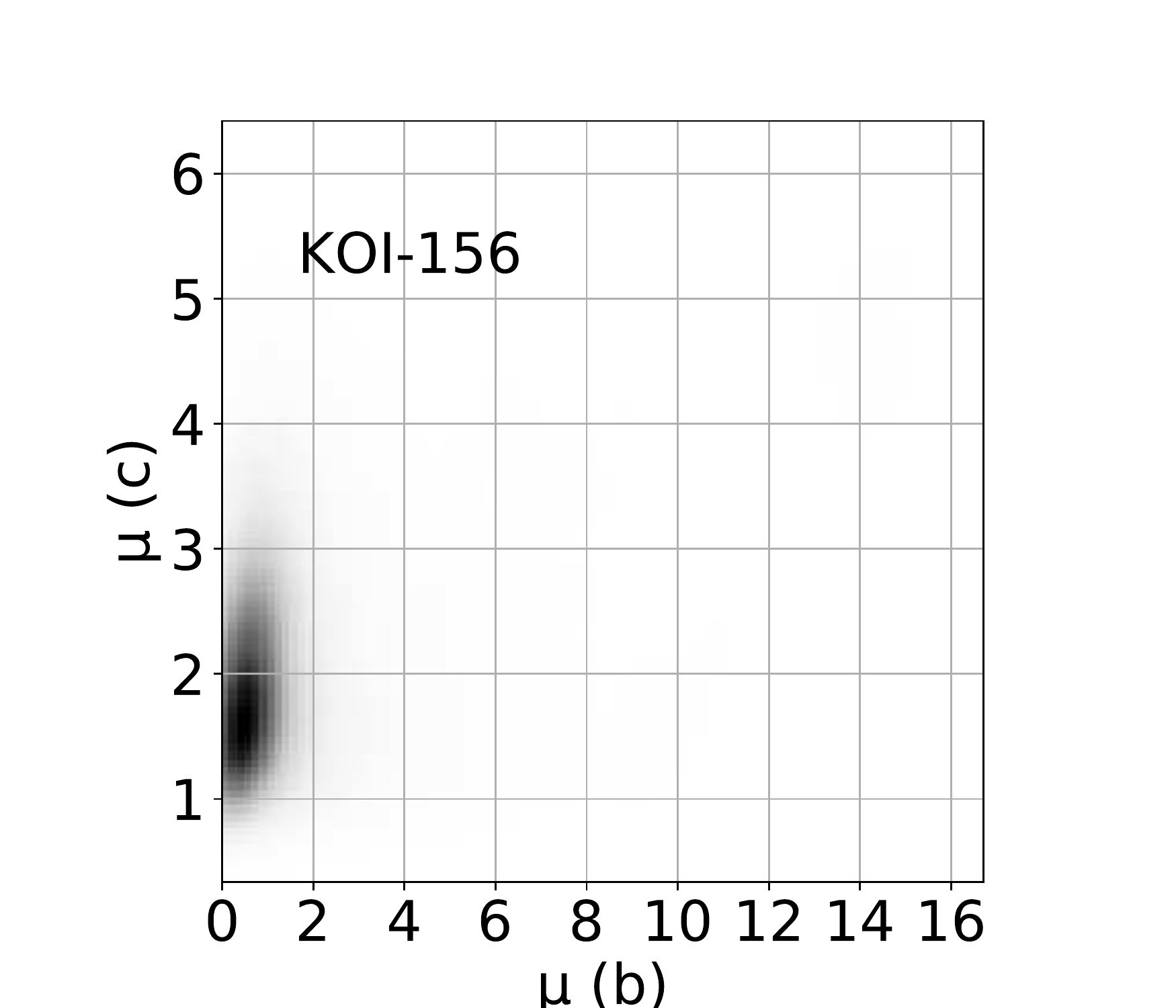}
\includegraphics [height = 1.1 in]{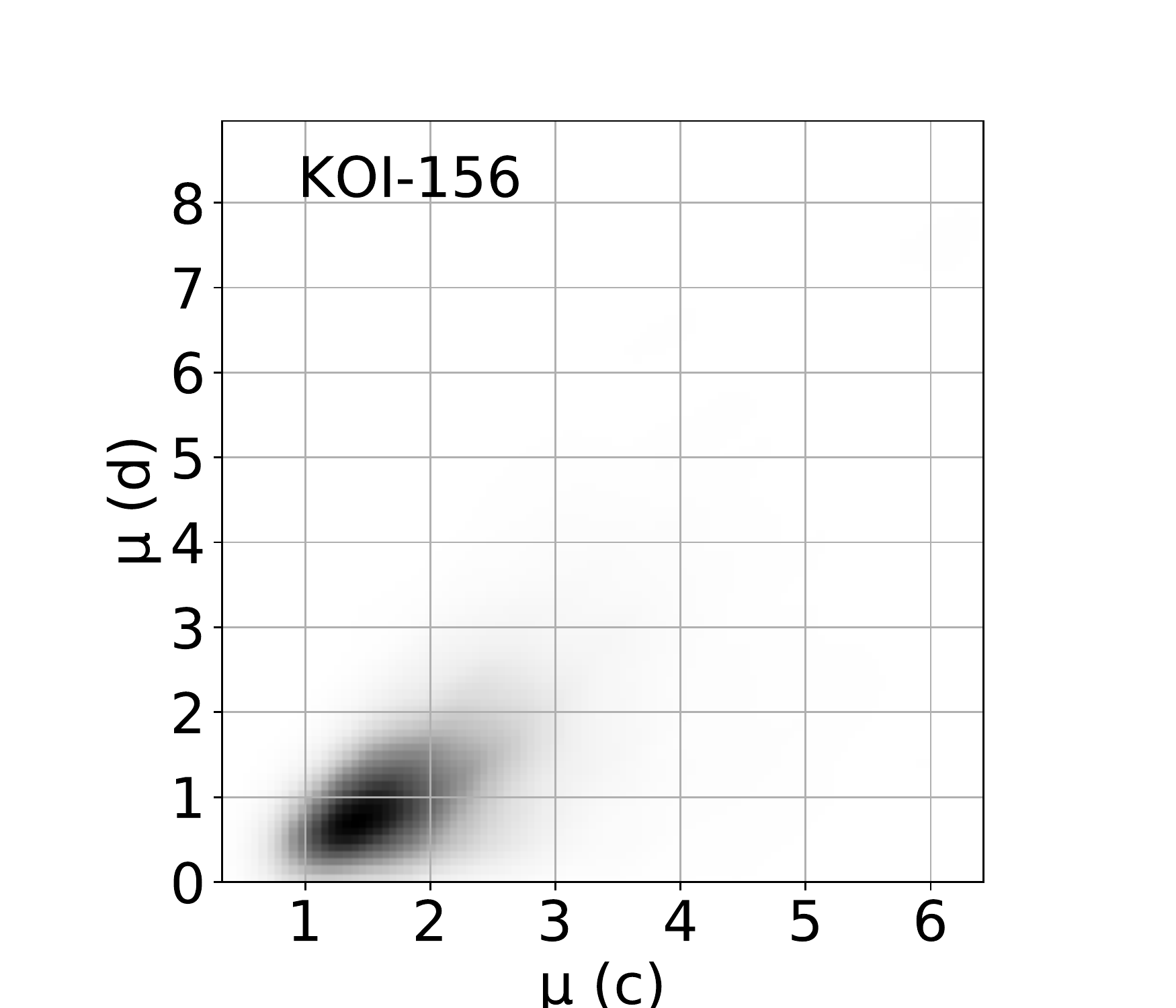} \\
\includegraphics [height = 1.1 in]{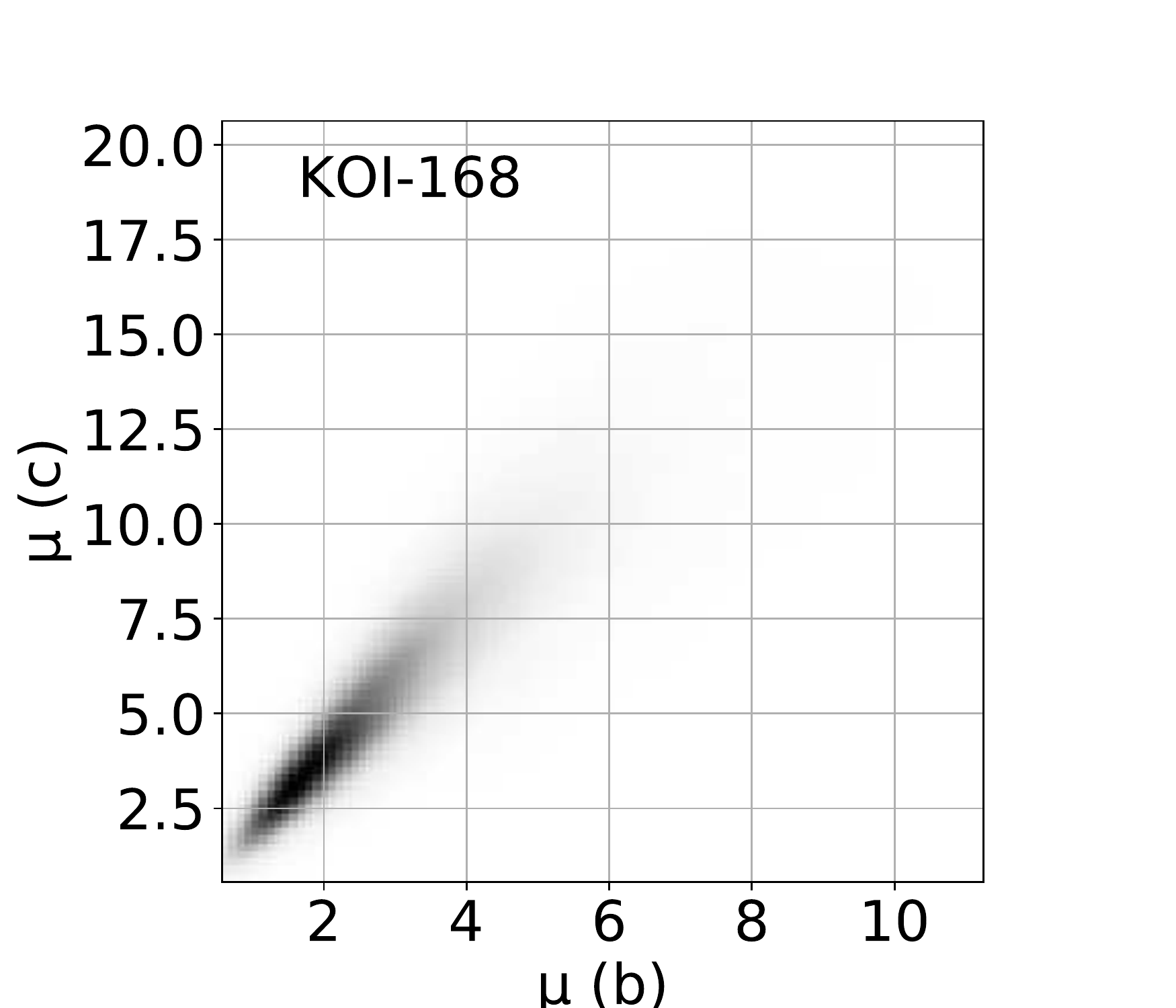}
\includegraphics [height = 1.1 in]{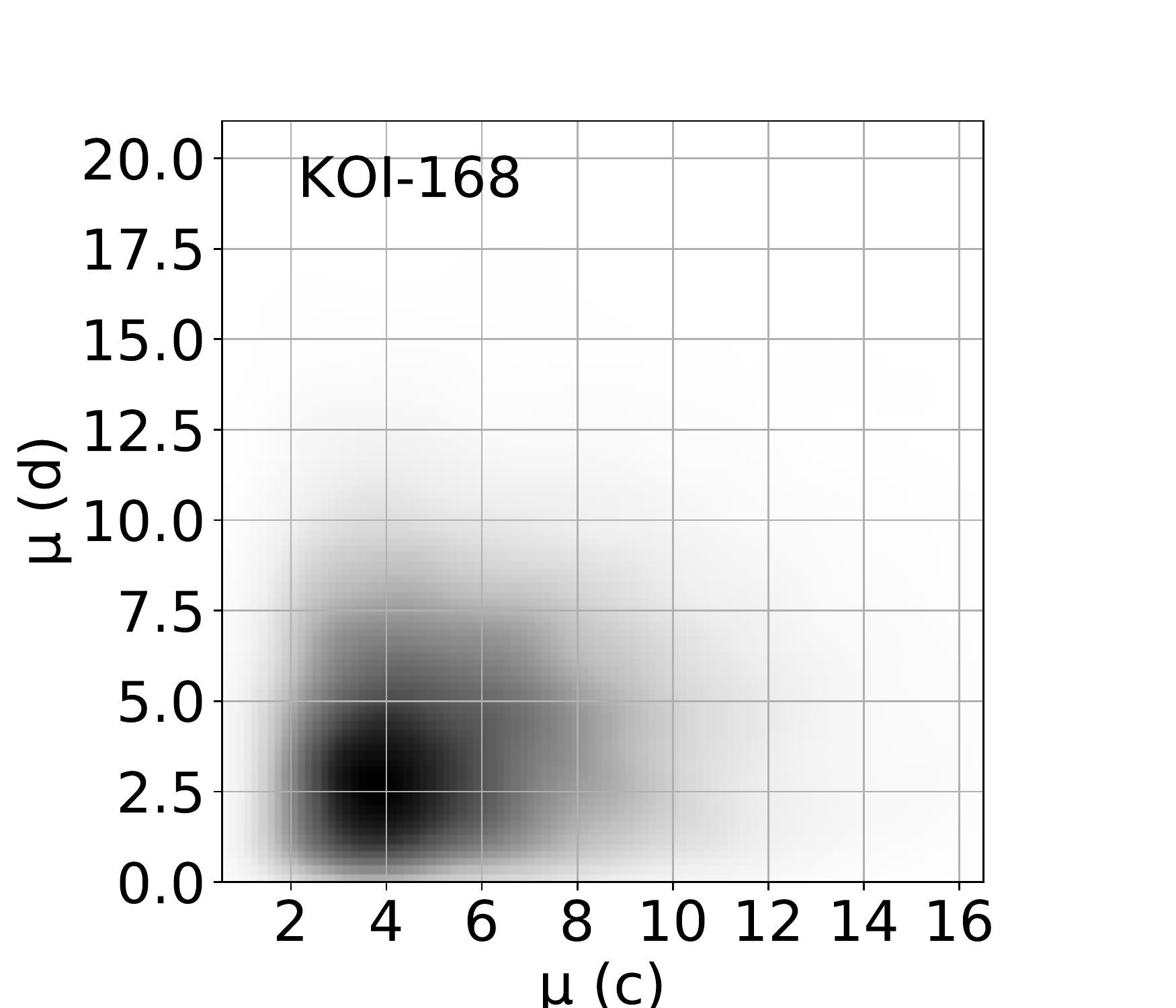}
\includegraphics [height = 1.1 in]{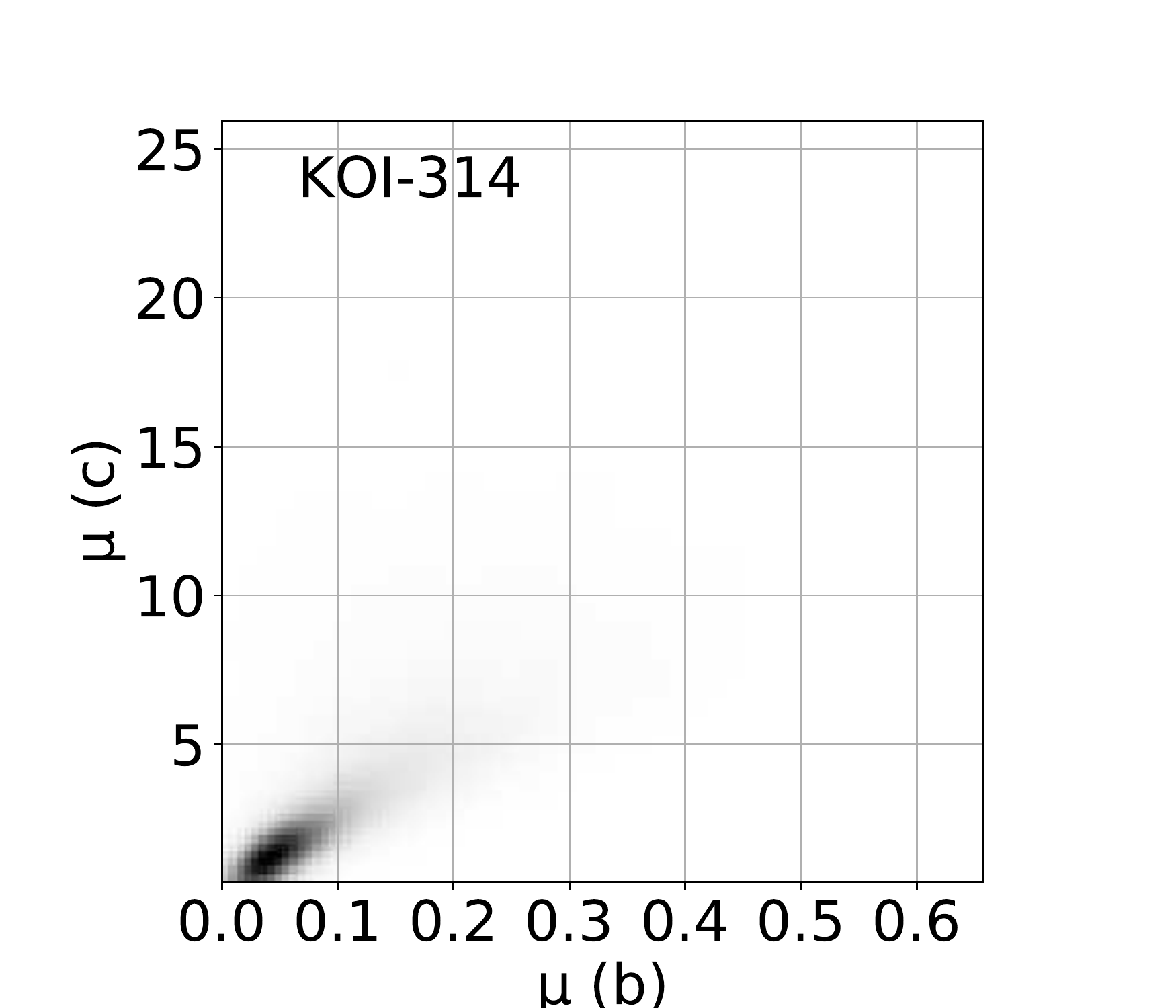}
\includegraphics [height = 1.1 in]{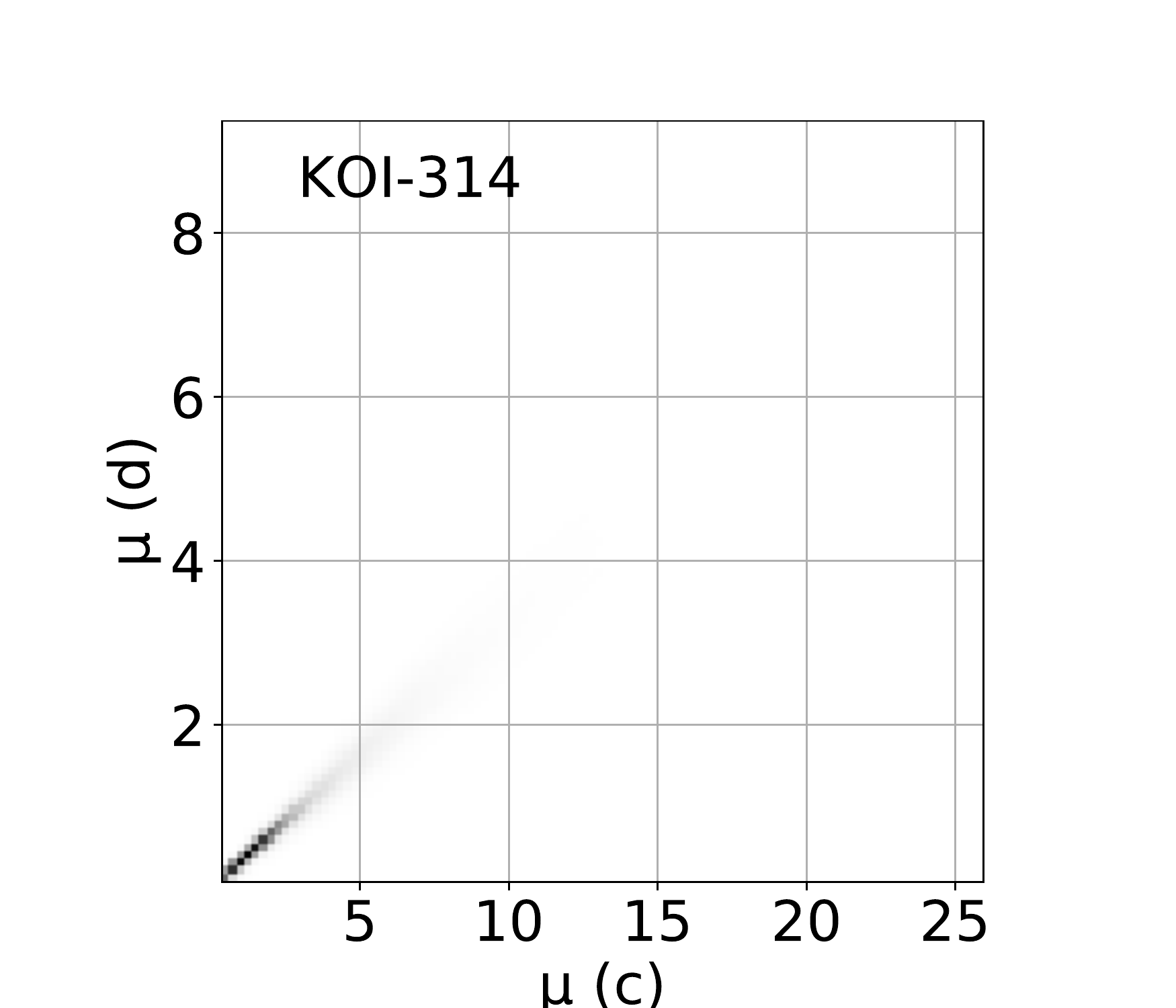} \\
\includegraphics [height = 1.1 in]{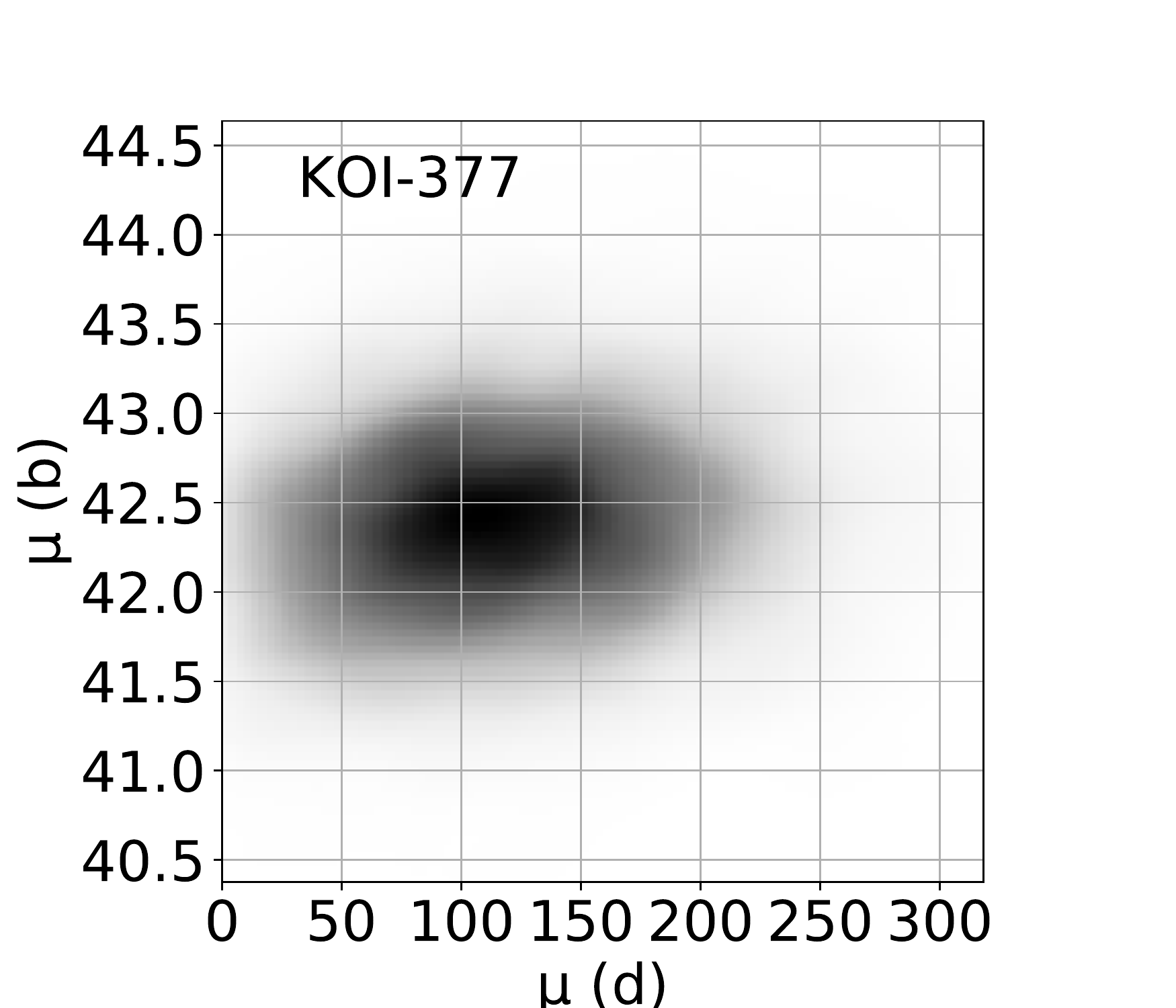}
\includegraphics [height = 1.1 in]{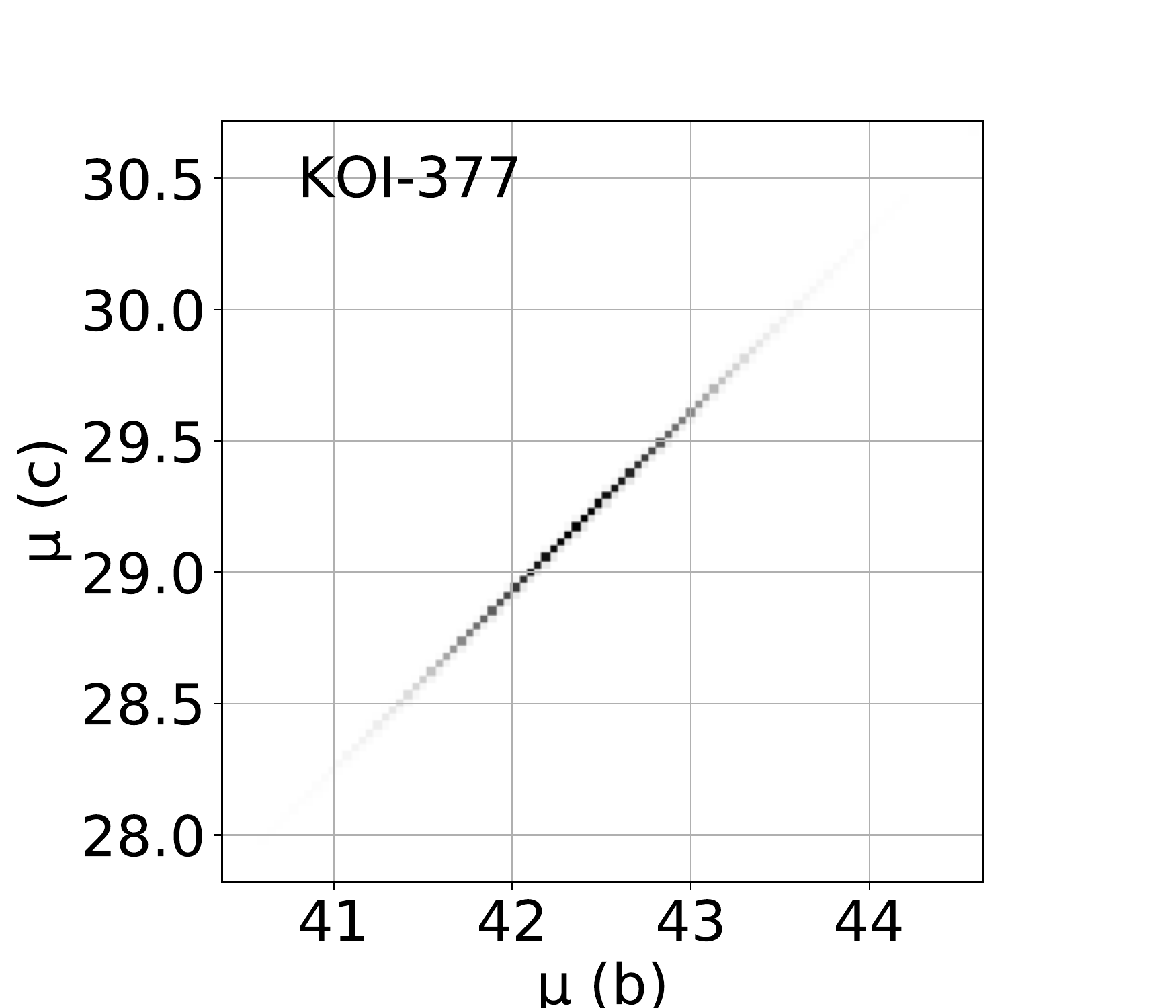}
\includegraphics [height = 1.1 in]{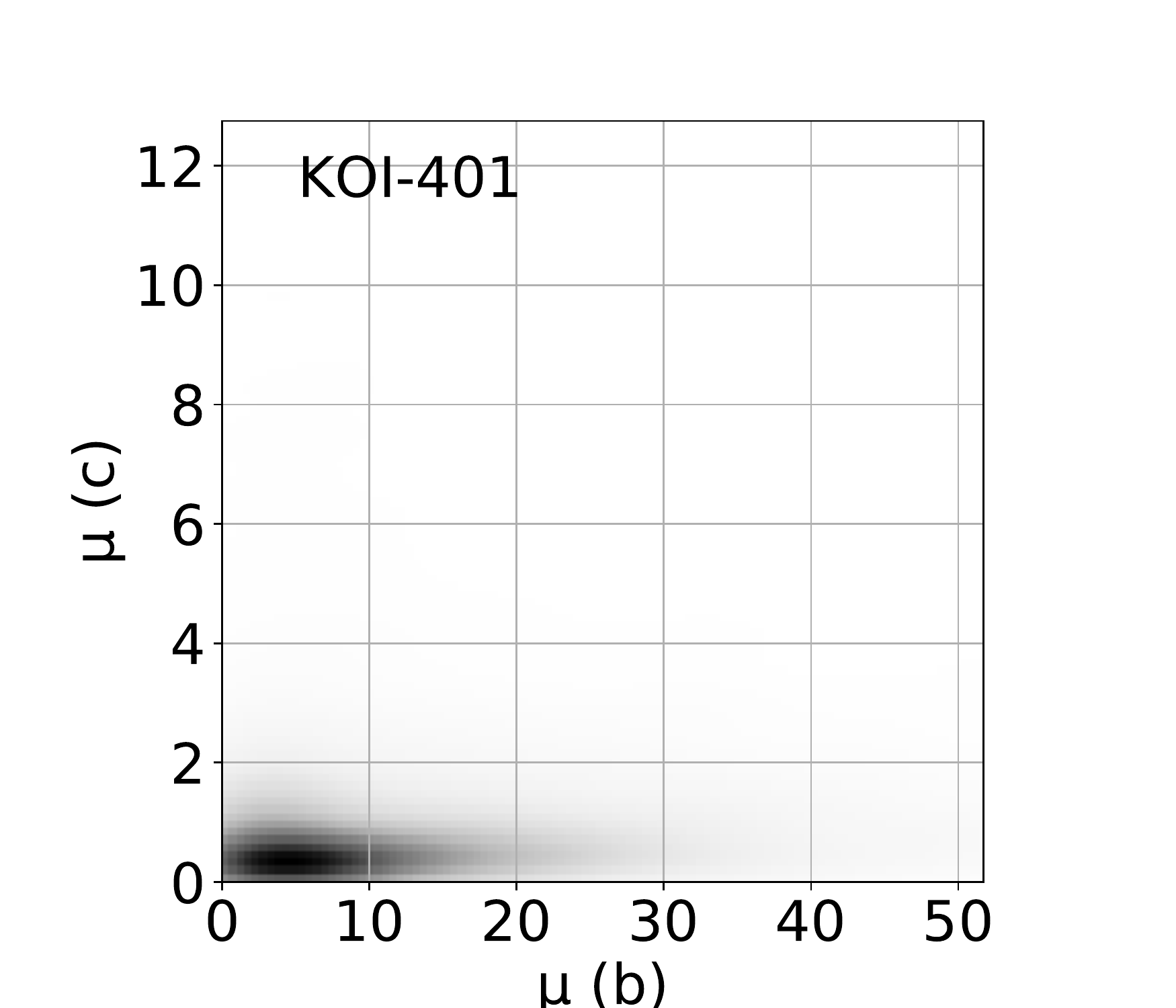}
\includegraphics [height = 1.1 in]{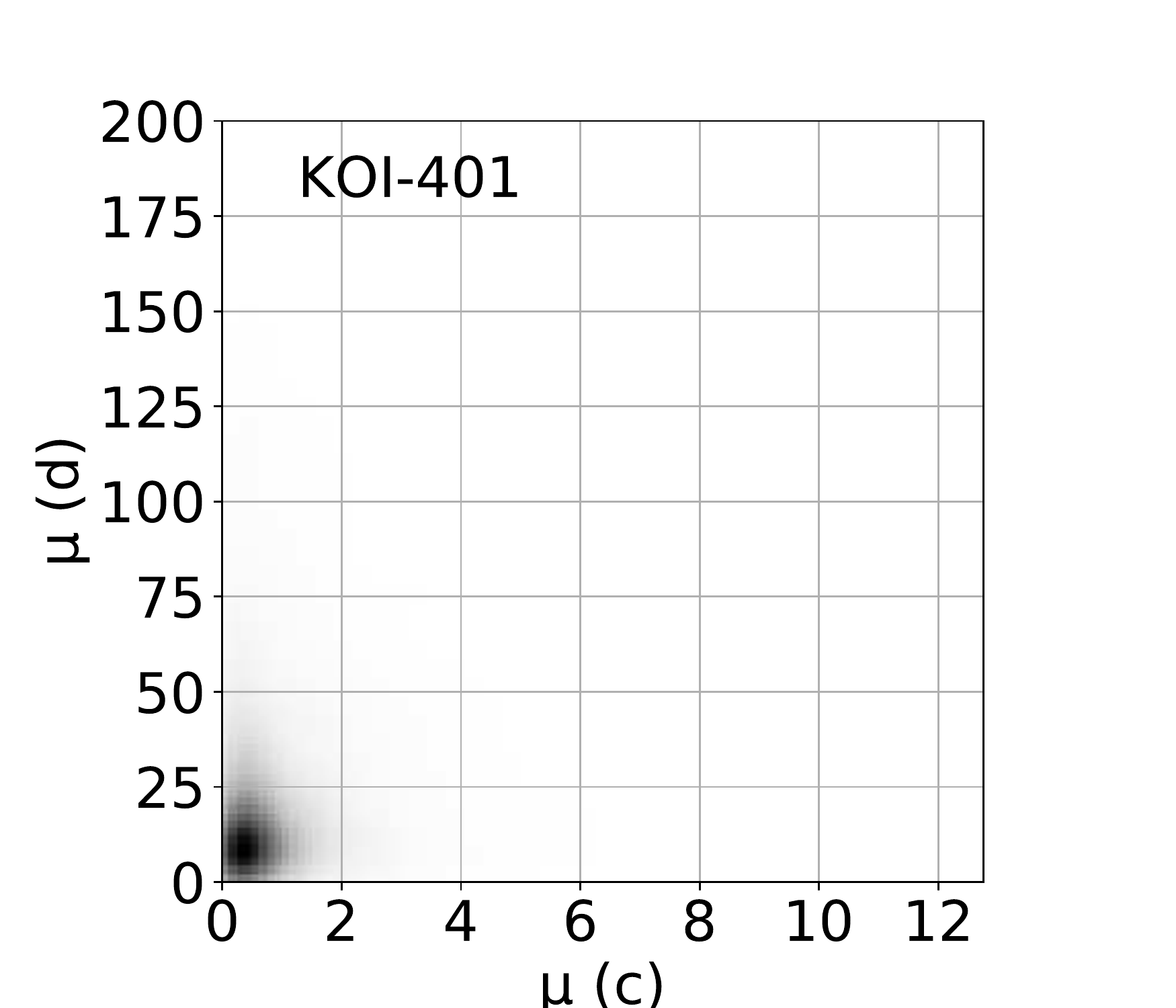} \\
\includegraphics [height = 1.1 in]{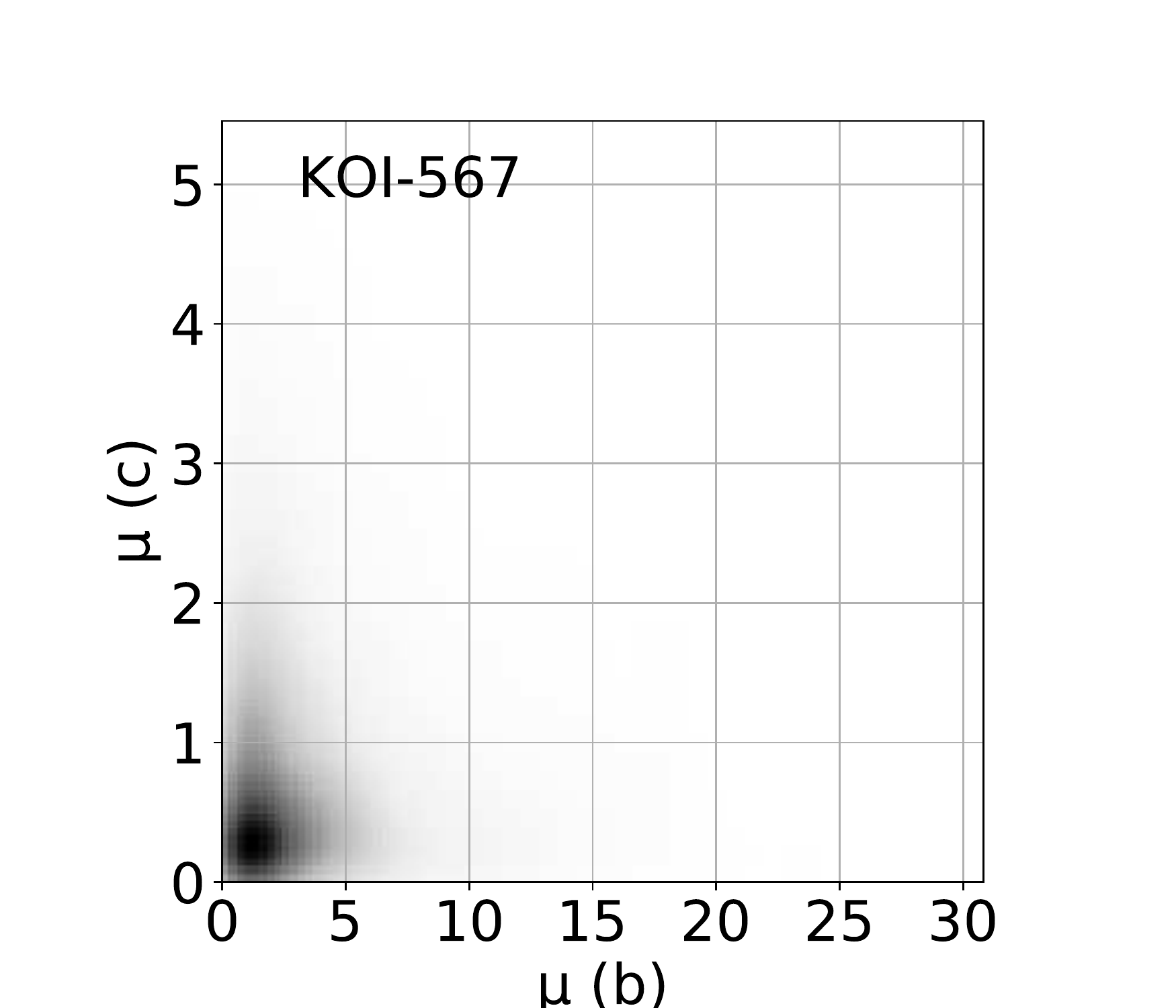}
\includegraphics [height = 1.1 in]{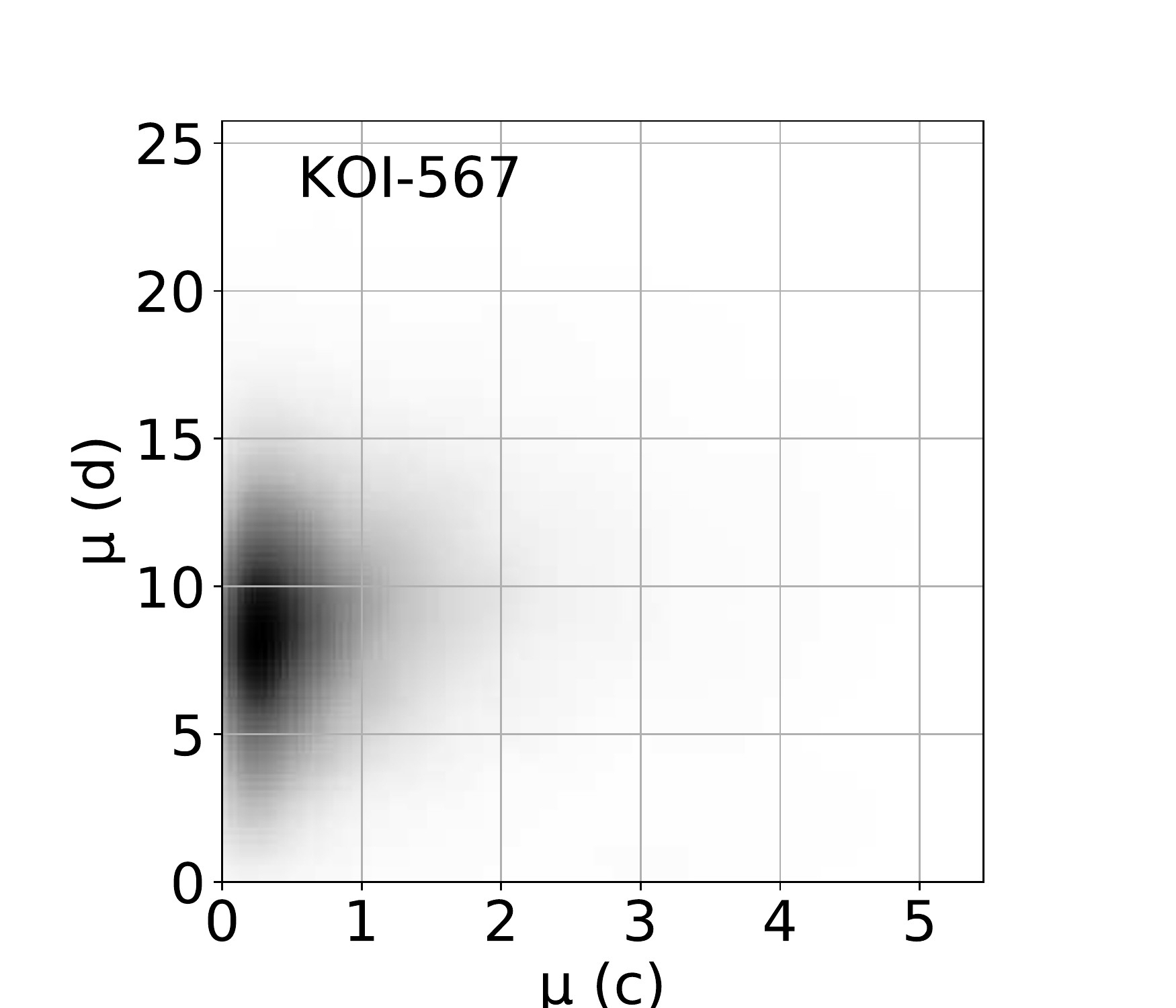} 
\includegraphics [height = 1.1 in]{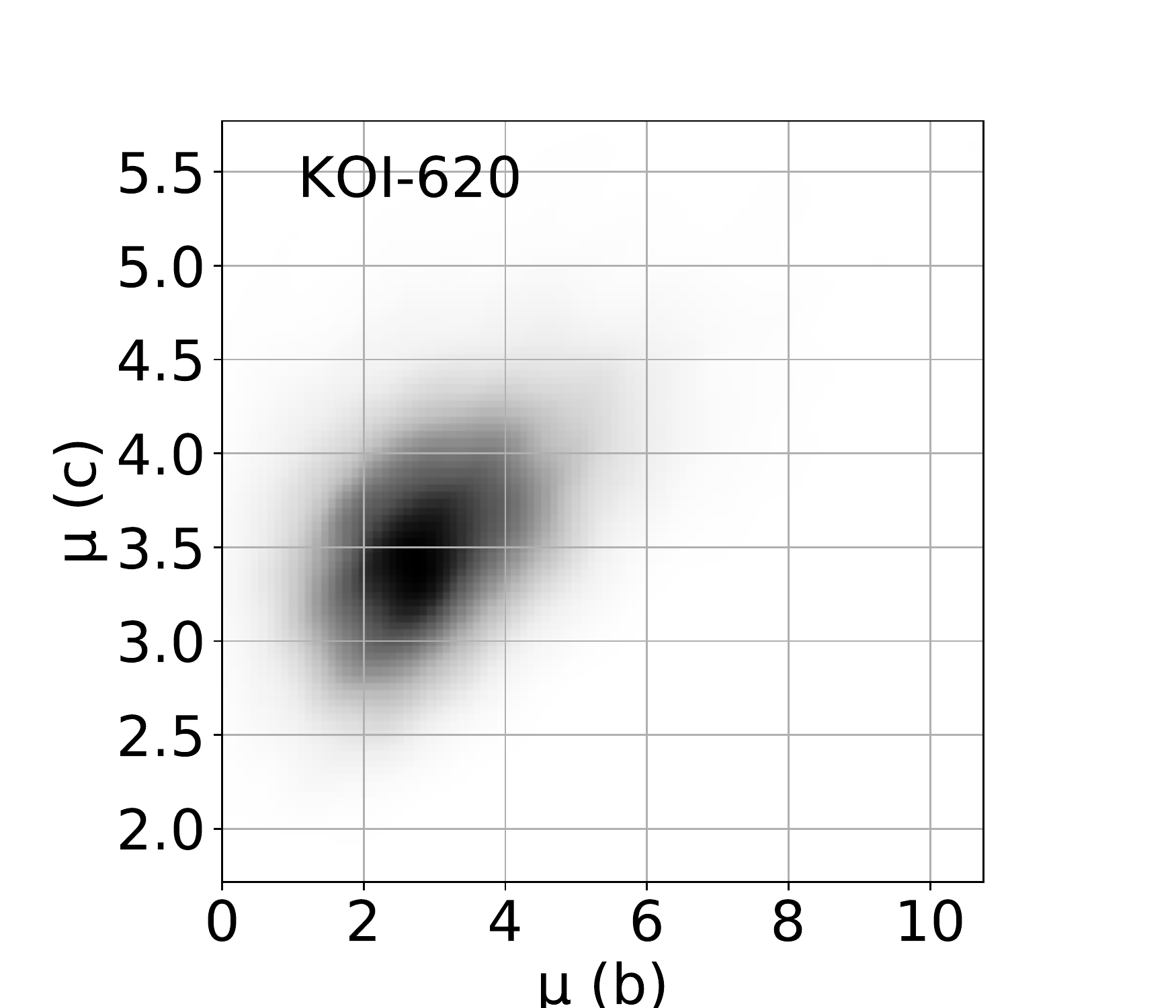}
\includegraphics [height = 1.1 in]{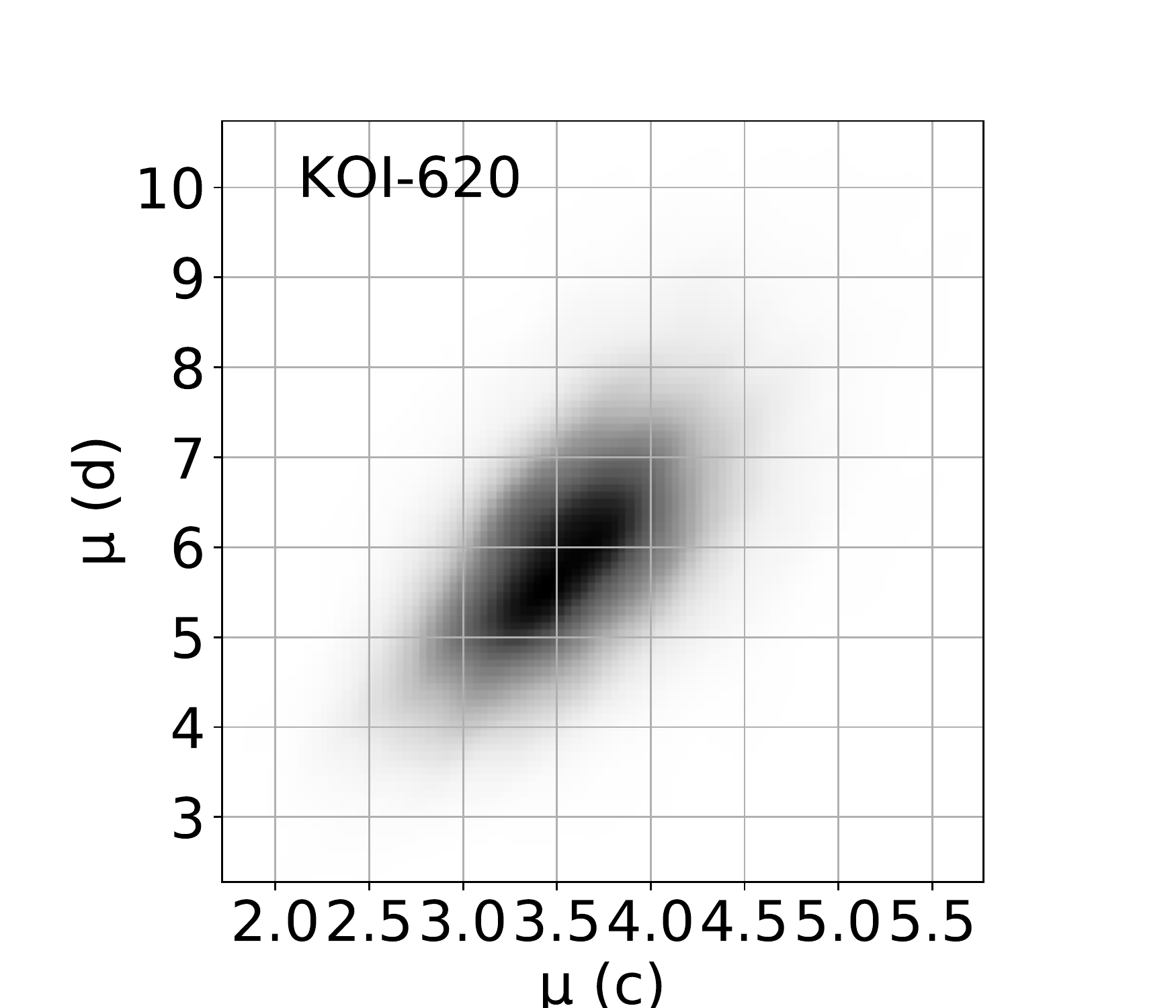} \\
\includegraphics [height = 1.1 in]{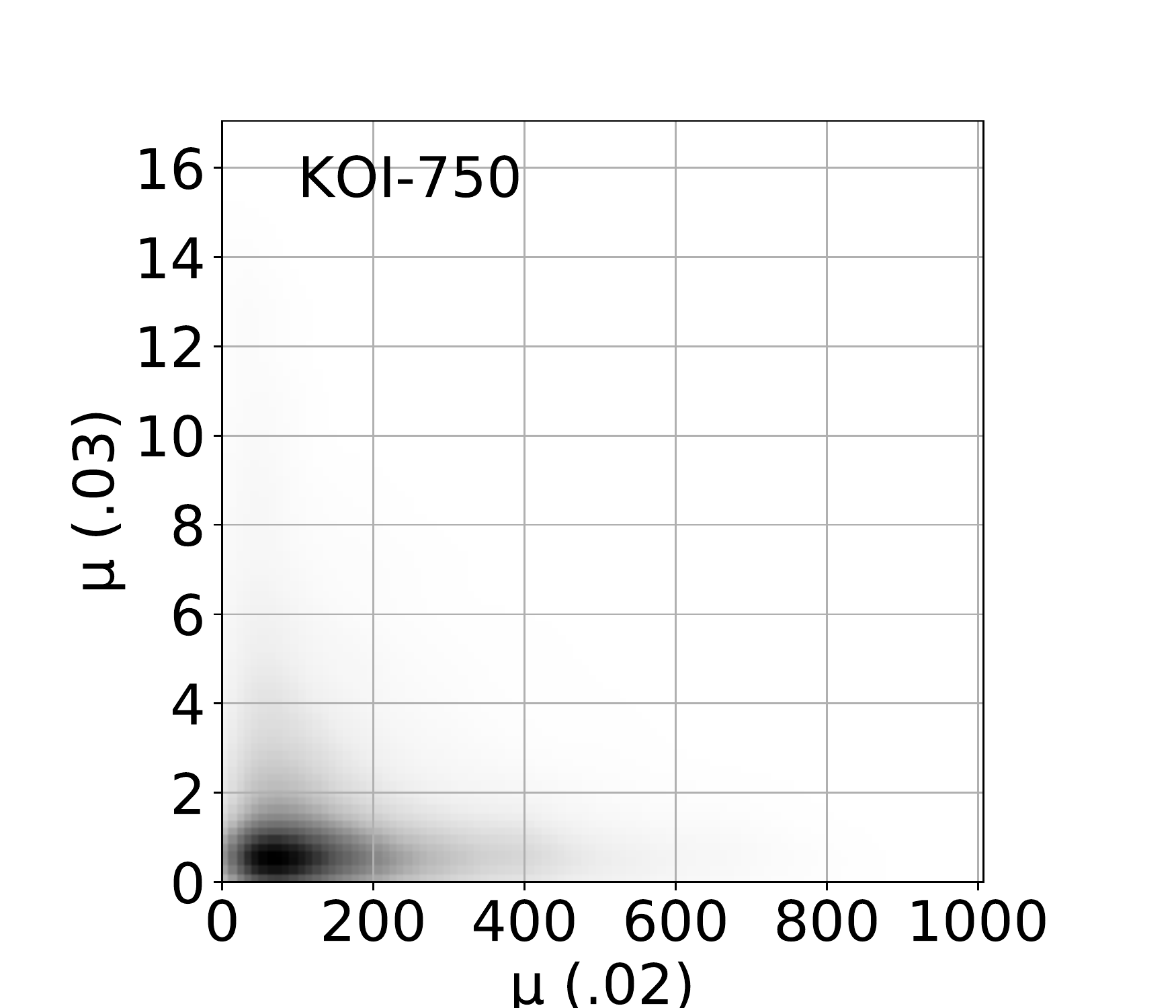}
\includegraphics [height = 1.1 in]{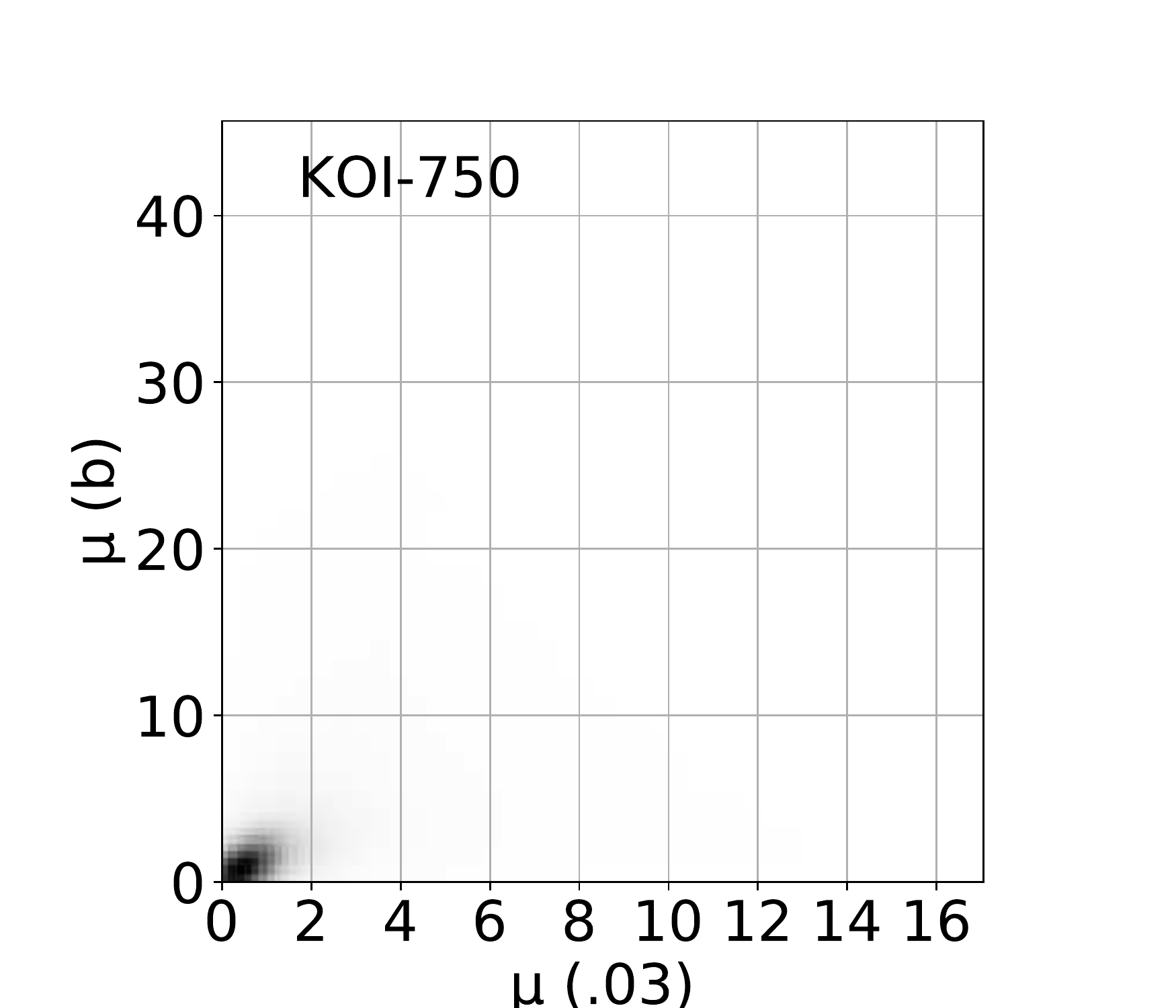} 
\includegraphics [height = 1.1 in]{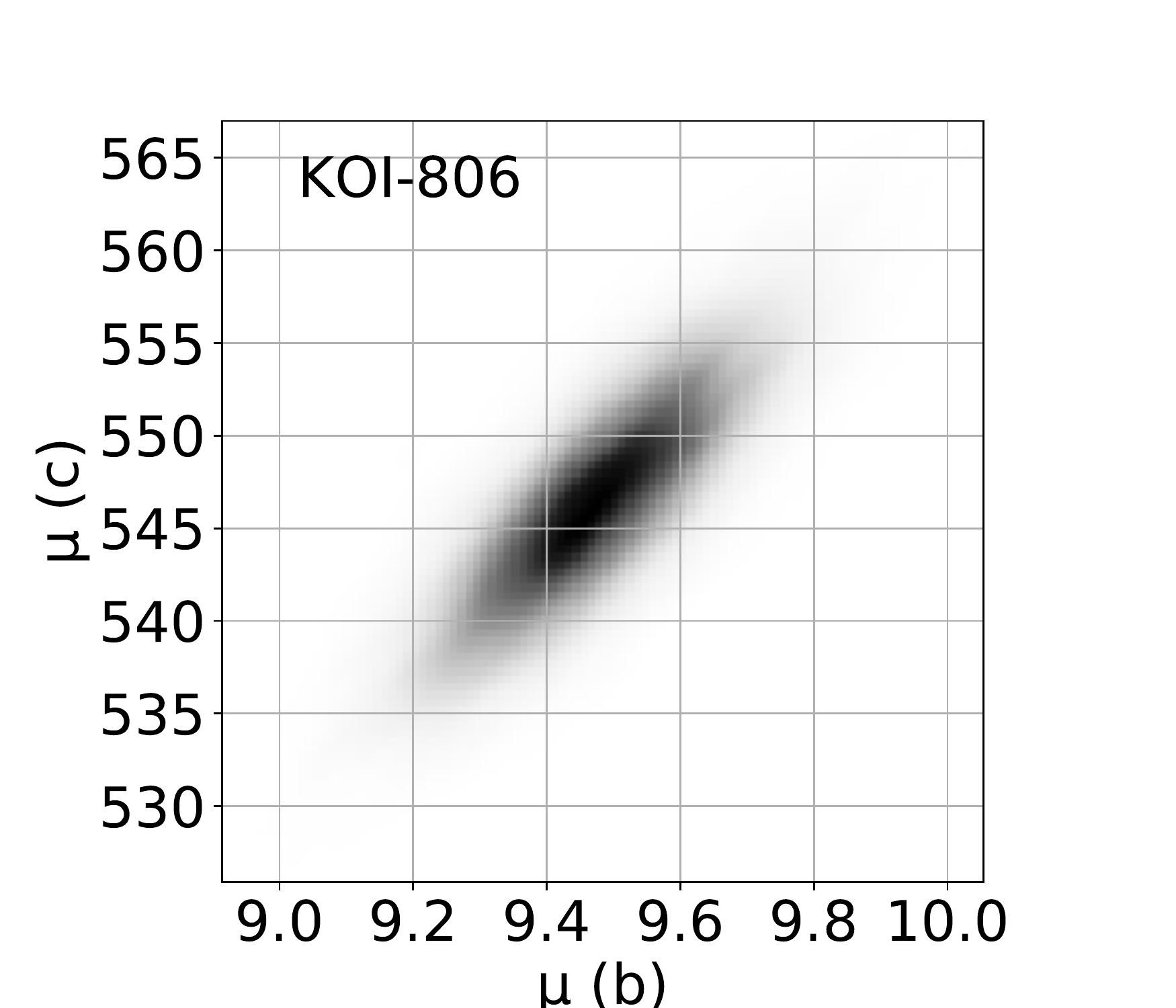}
\includegraphics [height = 1.1 in]{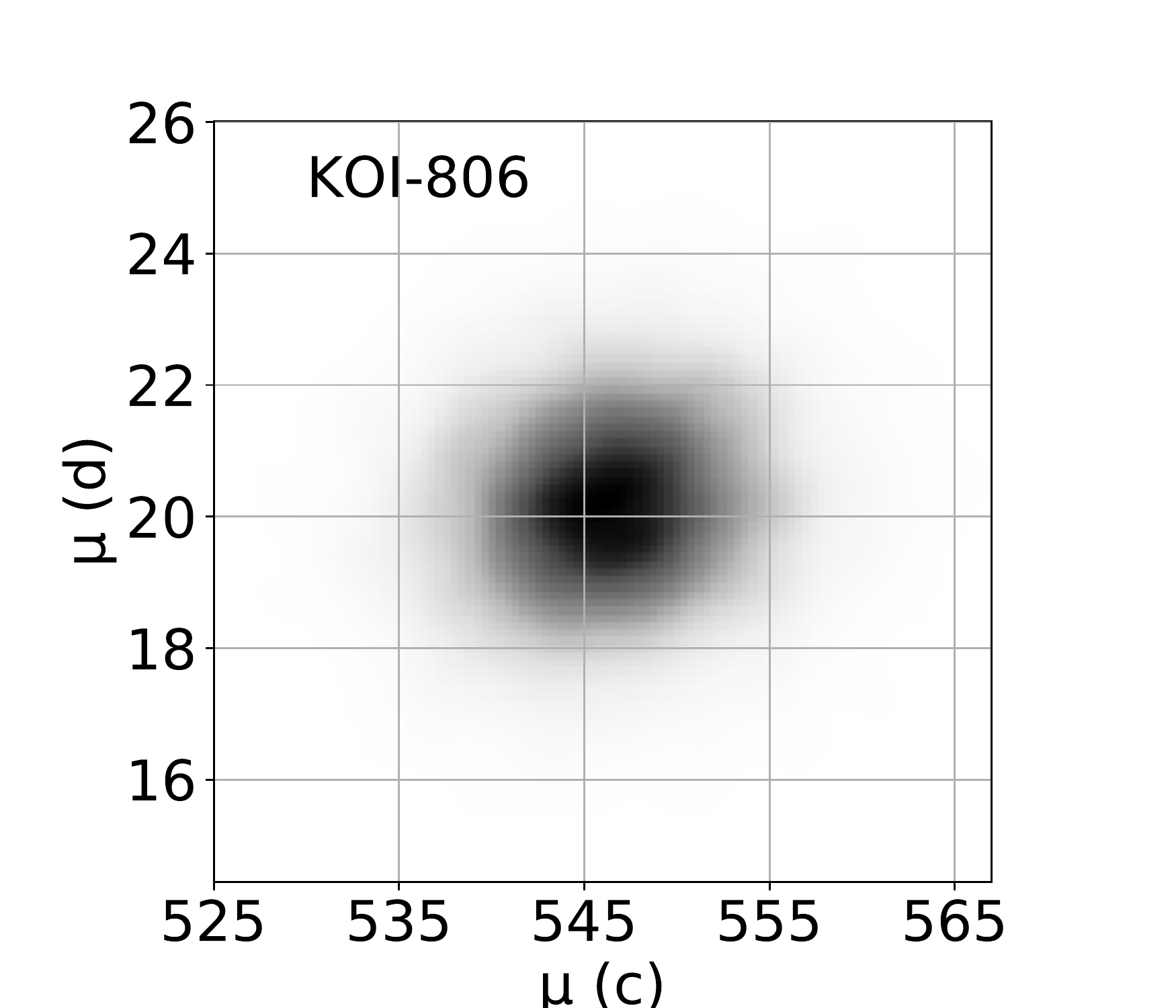} \\
\includegraphics [height = 1.1 in]{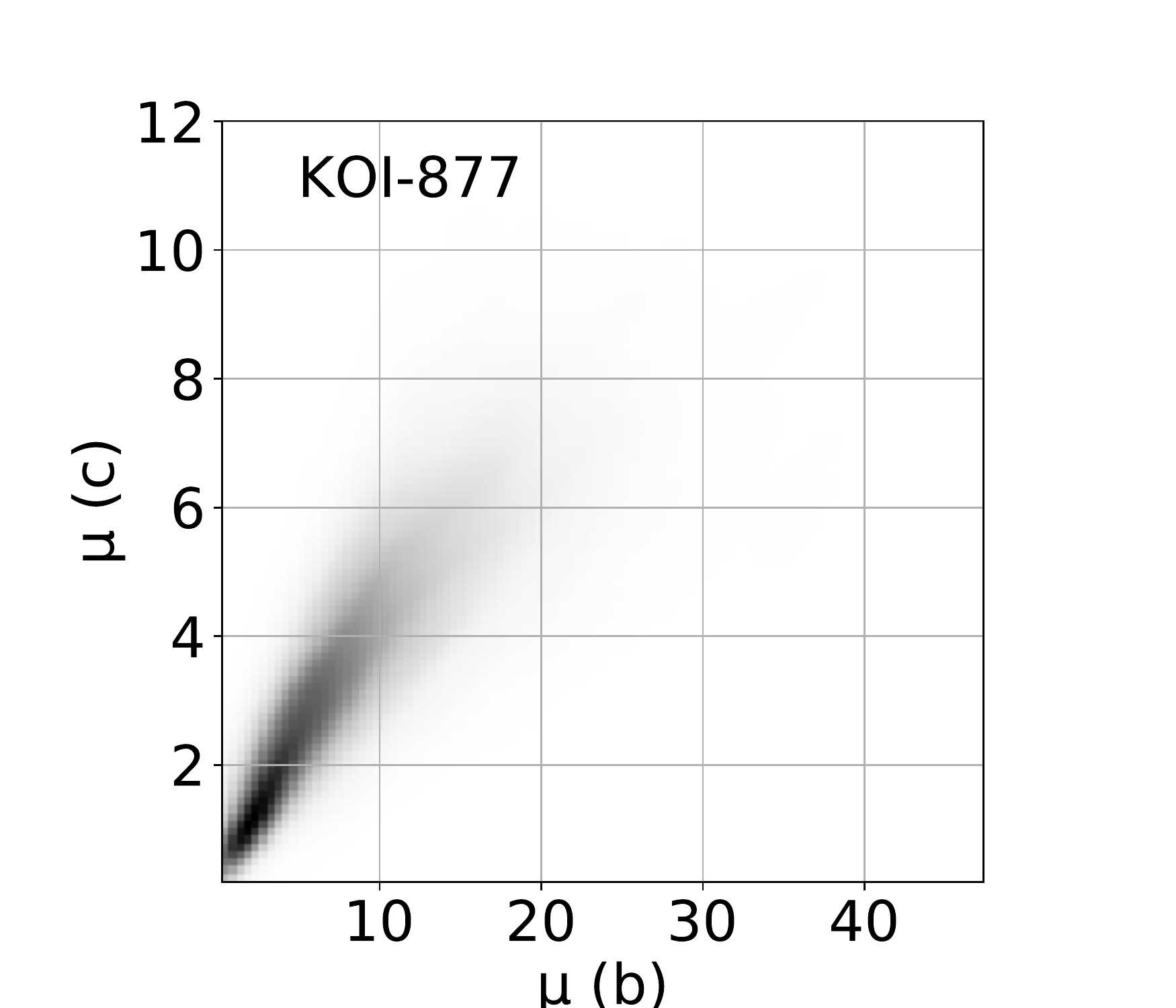}
\includegraphics [height = 1.1 in]{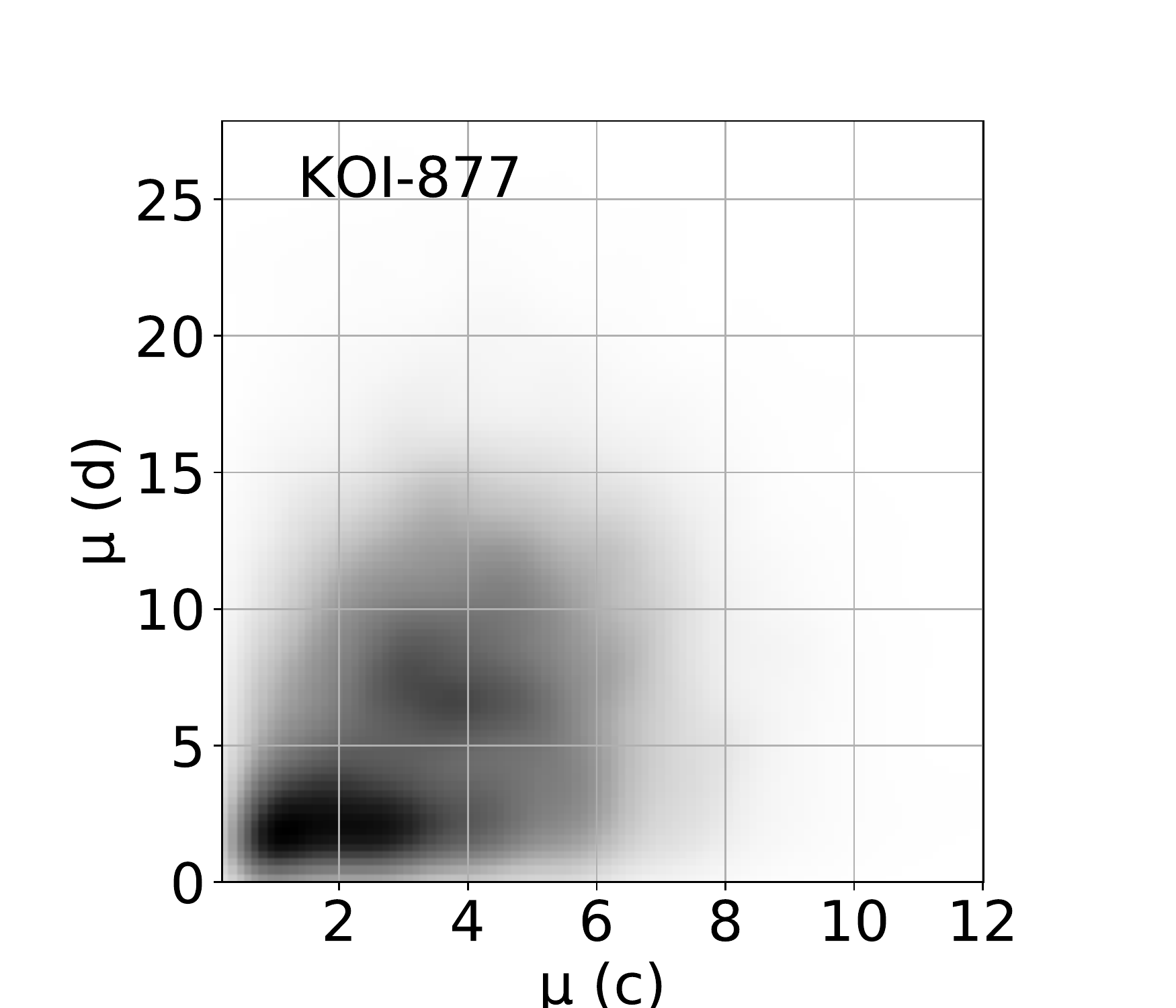}
\includegraphics [height = 1.1 in]{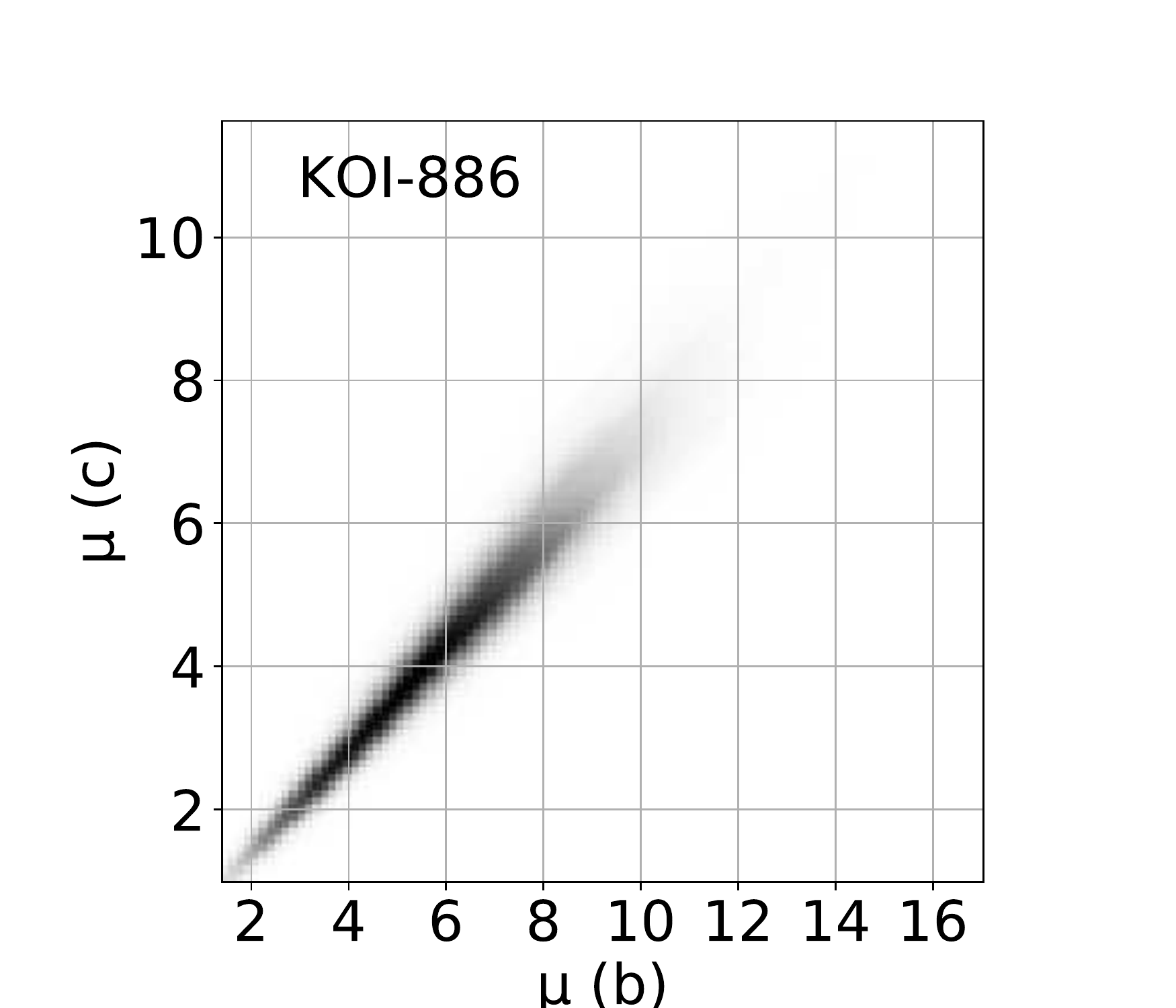}
\includegraphics [height = 1.1 in]{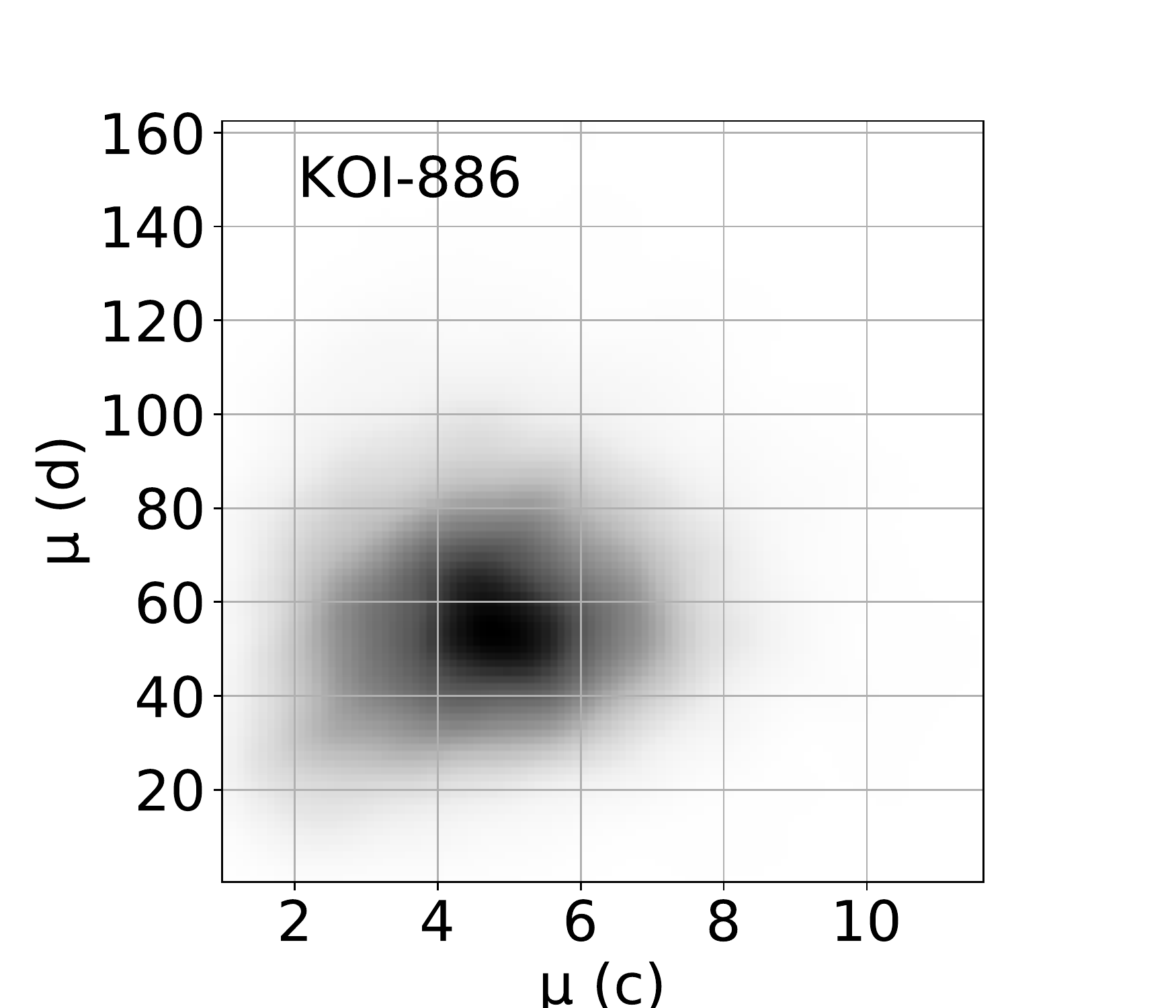} 
\caption{Two-dimensional kernel density estimators on joint posteriors of dynamical masses, $\mu$, scaled by factor $\frac{M_{\odot}}{M_{\oplus}}$: three-planet systems. 
\label{fig:mu3a} }
\end{center}
\end{figure}

Most of the candidates with mass upper limits that are tighter than that imposed by our prior, $\rho < 10$ g cm$^{-3}$, were expected to have strong TTV signals (see Tables~\ref{tbl-SystemSelection1}--\ref{tbl-SystemSelection2}). The exceptions to this are planets that have poorly constrained masses. They include KOI-314.01 (Kepler-138 c) which likely has a high bulk density. Different authors have found discrepant stellar parameters for KOI-314, which in turn casts some doubt on the likely bulk densities of the planets despite the dynamical mass constraints \citep{Jontof-Hutter2019}. KOI-314.01 and KOI-314.02 have a similar size but a very tightly constrained mass ratio of $\sim$3 (see Figure~\ref{fig:mu3a}). Thus the density upper limit on KOI-314.01 leads to the expectation that KOI-314.02 is rich in volatiles.  

KOI-934.03 has weak constraints on its mass upper limit following a moderately weak expected resonant TTV score of 5.3 (see Table~\ref{tbl-SystemSelection2}). 

KOI-1070.03 has an expectation of strongly detected nonresonant TTVs due its close neighbor KOI-1070.02  ($\frac{P_{.02}}{P_{.03}}$ = 1.16 ) and its long orbital period ($P_{.03}$ = 92.8 days). Nevertheless, its mass upper limit is imposed by our prior, not the data.

KOI-1338.03 and KOI-1338.02 have a TTV period longer than the Kepler baseline and weak upper limits on their masses. Nevertheless, the TTVs provide a strong nonzero lower bound on the mass of KOI-1338.03. Among our 10,000 samples, the lowest dynamical mass is 0.99 $\frac{M_{\oplus}}{M_{\odot}}$. 

KOI-2414.01 (Kepler-384 b) and KOI-2414.02 (Kepler-384 c) have expected resonant TTVs (see Table~\ref{tbl-SystemSelection2}), although in both cases, our mass upper limits are imposed by the prior in density. The TTV period is far longer than the Kepler baseline, and a wide range of masses are consistent with the data. In both cases, the prospects for improved mass constraints with future transit times and RV data are high. Similarly, KOI-2092.03 (Kepler-359 d) has a TTV baseline longer than the Kepler period and a weak upper limit on its mass.

Among planets that have not been characterized in prior studies, we note the newly characterized small planet KOI-654.02 (Kepler-200 c) that has size $R_{p} = $1.86 $^{+ 0.32 }_{- 0.35 }$  $R_{\oplus}$ and inferred density $\rho_{p} = $ 3.36 $^{+ 3.89 }_{- 1.68 }$ g cm$^{-3}$. Although nominally not dense enough to be rocky, there is a nonnegligible likelihood that the planet is dense enough to be rocky. Unfortunately, tighter dynamical mass constraints may not become available for some time: the future TTVs diverge slowly, and the expected RV signal from the each planet at KOI-654 is just $\sim$1 m s$^{-1}$. We also note that at KOI-3503, future transit times should improve the mass constraints on two Earth-sized planets. 

\subsection{Estimated Atmospheric Transmission Annuli}
We estimated the 68.3\% central credible intervals on the planetary atmospheric scale heights $H$, where 
\begin{equation}
H = \frac{kT_{eq}R_{p}^{2}}{\mu m_{H} G m_{p}}. 
\end{equation}
Here, $k$ is the Boltzmann constant, $\mu$ is the mean molecular weight in atomic mass units, $m_{H}$ is the mass of atomic hydrogen, $G$ is the gravitational constant and  $T_{eq}$ is equilibrium black body temperature given a Bond albedo of 0.3 and a circular orbit. We assume an atmospheric composition of 75\% H$_{2}$ and 25\% He by mass. We estimated the atmospheric transmission annuli ($\delta_{ann}$) of our candidates assuming that the annulus probes five scale heights:
\begin{equation}
\delta_{ann} \approx \frac{10H\delta}{R_{p}} 
\end{equation}
where $\delta$ is the transit depth.

We consider the planetary masses where the 15.9th percentile is more than half of the median of our posterior samples to be strong detections. These are prime candidates for initial atmospheric characterization with the James Webb Space Telescope (JWST, \citealt{BatalhaNatasha2019}). For these exoplanets, we convolve our posterior samples of mass, radius, and equilibrium blackbody temperature to derive posteriors on the estimated planetary atmospheric scale height and the transmission annulus in a cloud-free atmosphere. For the weaker detections, we report mass upper limits only corresponding to the 97.7th percentile of mass posterior samples, or the upper limit imposed by a maximum bulk density of 10 g cm$^{-3}$ on our samples. All else being equal, a mass upper limit corresponds to a lower bound on the atmospheric scale height. Here we estimate the minimum transmission annulus from the 2.27th percentile of our samples; that is, 97.7\% of our posterior samples give a transmission annulus above this lower bound. 

Our results reveal promising exoplanets for atmospheric transmission observations, over a wide range of hosts and equilibrium temperatures. Tables~\ref{tbl-massresults2}---\ref{tbl-massresults4} list constraints on planetary parameters including bulk density, equilibrium blackbody temperature, and estimated atmospheric scale height. We compare our mass estimates to those of prior studies for the same candidates where available, and we estimate derived quantities including scale height and transmission annulus given the information available in the prior studies, with additional parameters drawn from Kepler DR 25 as needed. 

Among our candidate sample, several planets have high estimated atmospheric scale heights, and expected transmission annuli are above 100 ppm, including 35 strong mass detections and 16 planets with upper limits only on the masses.

\begin{table}[ht!]
\tiny
  \begin{center}
    \begin{tabular}{|c|c|c|c|c|c|c|c|c|}
      \hline
      name &  P (days) & R$_p$ (R$_{\oplus}$) & M$_p$ (M$_{\oplus}$) & Study (preferred in italics) & $\rho_{p}$ (g/cm$^3$)& T$_{eq}$ (K) & Sc. H. ($R_{\oplus}$) & Annulus (ppm) \\               
\hline 
  222.01 (\object{Kepler-120} b) & 6.312 & 2.55 $^{+ 0.10 }_{- 0.07 }$ & < 15.58  & \textit{This Study} & < 5.12  &  618 $^{+   14 }_{-  14 }$ & > 0.014  & > 69 \\ 
 222.02  (Kepler-120 c) & 12.795 & 2.04 $^{+ 0.08 }_{- 0.07 }$  & < 13.61   & \textit{This Study} & < 9.02  &  489 $^{+   11 }_{-  11 }$ & > 0.008  & > 31 \\ 
 244.01  (\object{Kepler-25} c)  & 12.720 & 5.20 $\pm 0.09$  & 24.60 $\pm 5.7$  & \textit{\citet{mar14}} & 0.90 $\pm 0.21$  &   942 $^{+   16 }_{-  16 }$  & 0.05 $^{+ 0.02 }_{- 0.01 }$ & 135 $^{+ 42 }_{- 25 }$ \\ 
 244.01  (Kepler-25 c)  & 12.721 & 5.09 $^{+ 0.21 }_{- 0.41 }$  & 3.01 $^{+ 3.02 }_{- 1.46 }$ $  \left(^{+ 9.24 }_{- 2.06 }\right)$  & \textit{This Study} & 0.17 $^{+ 0.19 }_{- 0.09 }$ &   961 $^{+   14 }_{-  14 }$  & 0.360 $^{+ 0.369 }_{- 0.186 }$ & 915 $^{+ 881 }_{- 465 }$ \\ 
244.02  (Kepler-25 b)  & 6.239 & 2.71 $ \pm $ 0.05 & 9.6 $\pm 4.2$  & \textit{ \citet{mar14}} & 2.5 $\pm 1.0$   &  1195 $^{+   20 }_{-  20 }$ &  0.048 $^{+ 0.036 }_{- 0.015 }$ & 72 $^{+ 53 }_{- 22 }$ \\ 
 244.02  (Kepler-25 b)  & 6.238 & 2.71 $^{+ 0.12 }_{- 0.11 }$ & < 7.02  & This Study & < 1.91  &  1219 $^{+   18 }_{-  19 }$ & > 0.068  & > 100 \\ 
 255.01  (\object{Kepler-505} b)  & 27.522 & 3.11 $^{+ 0.09 }_{- 0.11 }$  & < 41.26   & \textit{This Study} & < 7.72  &  341 $^{+   10 }_{-  10 }$ & > 0.004  & > 31 \\ 
 255.02 & 13.603 & 0.75 $^{+ 0.05 }_{- 0.06 }$ & < 0.92  & \textit{This Study} & < 10.00  &  431 $^{+   13 }_{-  12 }$ & > 0.016  & > 37 \\ 
  277.01  (\object{Kepler-36} c)  & 16.219 & 3.68 $^{+ 0.056 }_{- 0.055 }$  & 7.84 $^{+ 0.33 }_{- 0.36 }$  &  \textit{ \citet{Deck2012}} & 0.87 $^{+ 0.06 }_{- 0.05 }$ &   928 $^{+   13 }_{-  13 }$  & 0.085 $^{+ 0.005 }_{- 0.005 }$ & 116 $^{+ 6 }_{- 5 }$ \\ 
  277.01  (Kepler-36 c)  & 16.219 & 3.69 $^{+ 0.17 }_{- 0.15 }$  & 7.71 $^{+ 0.25 }_{- 0.23 }$ $  \left(^{+ 0.49 }_{- 0.47 }\right)$  & This Study & 0.85 $^{+ 0.12 }_{- 0.11 }$ &   946 $^{+   16 }_{-  15 }$  & 0.088 $^{+ 0.009 }_{- 0.008 }$ & 120 $^{+  7 }_{-  7 }$ \\ 
   277.02  (Kepler-36 b)  & 13.868 & 1.49 $^{+ 0.035 }_{- 0.035 }$  & 4.32 $^{+ 0.19 }_{- 0.20 }$   &  \textit{\citet{Deck2012}} & 7.19 $^{+ 0.62 }_{- 0.57 }$ & 978 $^{+    14 }_{-  14 }$ & 0.026 $^{+ 0.002 }_{- 0.002 }$ & 15 $^{+ 1 }_{- 1 }$ \\ 
 277.02  (Kepler-36 b)  & 13.868 & 1.51 $^{+ 0.08 }_{- 0.07 }$  & 4.13 $^{+ 0.13 }_{- 0.13 }$   $ \left(^{+ 0.27 }_{- 0.25 }\right)$  & This Study & 6.43 $^{+ 1.04 }_{- 0.96 }$ & 998 $^{+    16 }_{-  16 }$ & 0.030 $^{+ 0.003 }_{- 0.003 }$ & 16 $^{+  1 }_{-  1 }$ \\ 
 430.01  (\object{Kepler-551} b)  & 12.376 & 2.53 $^{+ 0.12 }_{- 0.11 }$  & < 8.49   & \textit{This Study} & < 3.02  &  490 $^{+   11 }_{-  11 }$ & > 0.019  & > 129 \\ 
 430.02 & 9.341 & 0.89 $^{+ 0.09 }_{- 0.06 }$ & < 1.47  & \textit{This Study} & < 10.00  &  538 $^{+   13 }_{-  12 }$ & > 0.017  & > 37 \\ 
457.01  (\object{Kepler-161} b)  & 4.922 & 2.03 $^{+ 0.10 }_{- 0.08 }$ & < 7.99  & \textit{This Study} & < 5.32  &  844 $^{+   15 }_{-  15 }$ & > 0.023  & > 86 \\ 
 457.02  (Kepler-161 c)  & 7.064 & 2.24 $^{+ 0.10 }_{- 0.23 }$  & < 5.19   & \textit{This Study} & < 10.00  &  748 $^{+   13 }_{-  13 }$ & > 0.021  & > 103 \\ 
 523.01  (\object{Kepler-177} c)  & 49.409 & 8.72 $^{+ 0.36 }_{- 0.34 }$  & 14.6 $^{+ 2.7 }_{- 2.6 }$   &  \textit{ \citet{Vissapragada2020}} & 0.121 $^{+ 0.026 }_{- 0.022 }$ &   575 $^{+   13 }_{-  13 }$  & 0.159 $^{+ 0.04 }_{- 0.03 }$ & 578 $^{+ 137 }_{- 95 }$ \\ 
 523.01  (Kepler-177 c)  & 49.410 & 9.13 $^{+ 0.50 }_{- 0.49 }$  & 13.54 $^{+ 2.52 }_{- 2.39 }$ $  \left(^{+ 5.27 }_{- 4.58 }\right)$  & This Study & 0.10 $^{+ 0.03 }_{- 0.02 }$ &   575 $^{+   13 }_{-  13 }$  & 0.186 $^{+ 0.047 }_{- 0.035 }$ & 650 $^{+ 145 }_{- 110 }$ \\ 
 523.02  (\object{Kepler-176} b)  & 36.856 & 3.50 $^{+ 0.19 }_{- 0.15 }$  & 5.83 $^{+ 0.85 }_{- 0.65 }$     &  \textit{ \citet{Vissapragada2020} } & 0.750 $^{+ 0.117 }_{- 0.089 }$ & 634 $^{+    15 }_{-  15 }$ & 0.070 $^{+ 0.015 }_{- 0.012 }$ & 143 $^{+ 26 }_{- 20 }$ \\  
 523.02  (Kepler-176 b)  & 36.856 & 3.62 $^{+ 0.24 }_{- 0.20 }$  & 5.32 $^{+ 0.78 }_{- 0.75 }$   $ \left(^{+ 1.60 }_{- 1.45 }\right)$  & This Study & 0.59 $^{+ 0.15 }_{- 0.13 }$ & 634 $^{+    14 }_{-  14 }$ & 0.085 $^{+ 0.018 }_{- 0.014 }$ & 164 $^{+ 29 }_{- 23 }$ \\ 
  654.01  (\object{Kepler-200} b)  & 8.603 & 1.75 $^{+ 0.23 }_{- 0.13 }$ & < 3.95  & \textit{This Study} & < 4.00  &  883 $^{+   15 }_{-  15 }$ & > 0.037  & > 72 \\ 
 654.02  (Kepler-200 c)  & 10.217 & 1.86 $^{+ 0.32 }_{- 0.35 }$  & 3.48 $^{+ 1.90 }_{- 1.37 }$ $  \left(^{+ 3.76 }_{- 2.22 }\right)$  & \textit{This Study} & 3.39 $^{+ 3.91 }_{- 1.70 }$ &   834 $^{+   14 }_{-  14 }$  & 0.039 $^{+ 0.032 }_{- 0.016 }$ & 52 $^{+ 35 }_{- 18 }$ \\ 
738.01  (\object{Kepler-29} b)  & 10.338 & 2.55 $^{+ 0.12 }_{- 0.12 }$  & 5.0 $^{+ 1.5 }_{- 1.5 }$     &  \textit{ \citet{Vissapragada2020} } & 1.66 $^{+ 0.50 }_{- 0.51 }$ & 697 $^{+    18 }_{-  18 }$ & 0.047 $^{+ 0.022 }_{- 0.012 }$ & 222 $^{+ 97 }_{- 53 }$ \\ 
 738.01  (Kepler-29 b)  & 10.338 & 2.56 $^{+ 0.16 }_{- 0.14 }$  & 3.50 $^{+ 1.09 }_{- 1.10 }$   $ \left(^{+ 2.24 }_{- 2.23 }\right)$  & \textit{This Study} & 1.12 $^{+ 0.43 }_{- 0.38 }$ & 698 $^{+    17 }_{-  17 }$ & 0.070 $^{+ 0.033 }_{- 0.018 }$ & 322 $^{+ 147 }_{- 78 }$ \\ 
 738.02  (Kepler-29 c)  & 13.286 & 2.34 $^{+ 0.12 }_{- 0.12 }$  & 4.5 $^{+ 1.2 }_{- 1.2 }$   & \textit{  \citet{Vissapragada2020} }& 1.91 $^{+ 0.52 }_{- 0.54 }$ &   641 $^{+   17 }_{-  17 }$  & 0.041 $^{+ 0.016 }_{- 0.009 }$ & 182 $^{+ 68 }_{- 39 }$ \\ 
738.02   (Kepler-29 c) & 13.288 & 2.39 $^{+ 0.15 }_{- 0.13 }$  & 3.10 $^{+ 0.95 }_{- 0.97 }$ $  \left(^{+ 1.93 }_{- 1.97 }\right)$  & \textit{This Study} & 1.21 $^{+ 0.47 }_{- 0.40 }$ &   642 $^{+   16 }_{-  16 }$  & 0.064 $^{+ 0.030 }_{- 0.016 }$ & 274 $^{+ 124 }_{- 67 }$ \\ 
   1279.01  (\object{Kepler-804} b)  & 14.375 & 1.94 $^{+ 0.10 }_{- 0.09 }$  & < 11.35   & \textit{This Study} & < 9.05  &  768 $^{+   14 }_{-  14 }$ & > 0.013  & > 23 \\ 
 1279.02  (Kepler-804 c)  & 9.651 & 1.07 $^{+ 0.07 }_{- 0.06 }$ & < 2.57  & \textit{This Study} & < 10.00  &  877 $^{+   16 }_{-  16 }$ & > 0.023  & > 21 \\ 
 1599.01  (\object{Kepler-1659} c)  & 20.4415 & 1.9 $ \pm 0.3$ & 4.6$\pm 0.3$ &    \textit{\citet{Panichi2019}}  &  3.7 $^{+   1.5 }_{-  1.0 }$ &  660 $^{+  20 }_{- 19 }$ & 0.027 $^{+ 0.007 }_{- 0.006 }$ & 52 $^{+ 8 }_{- 7 }$ \\  
  1599.01  (Kepler-1659 c)  & 20.407 & 1.90 $^{+ 0.20 }_{- 0.20 }$  & 4.48 $^{+ 0.86 }_{- 0.74 }$ $  \left(^{+ 1.85 }_{- 1.40 }\right)$  & \textit{This Study} & 3.63 $^{+ 1.60 }_{- 1.09 }$ &   705 $^{+   82 }_{-  65 }$  & 0.030 $^{+ 0.010 }_{- 0.008 }$ & 66 $^{+ 18 }_{- 14 }$ \\ 
    1599.02  (Kepler-1659 b)  & 13.609 & 1.9 $ \pm $ 0.2 &  9.0$\pm 0.3$ &  \textit{\citet{Panichi2019}} &   7.2 $^{+   2.8 }_{-  2.6 }$  &  755 $^{+  23 }_{- 23 }$ &  0.016 $^{+ 0.006 }_{- 0.005 }$ & 32 $^{+ 6 }_{- 6 }$ \\  
 1599.02  (Kepler-1659 b)  & 13.618 & 1.90 $^{+ 0.30 }_{- 0.30 }$  & 5.84 $^{+ 1.27 }_{- 1.17 }$   $ \left(^{+ 2.58 }_{- 2.40 }\right)$  & \textit{This Study} & 4.81 $^{+ 3.55 }_{- 1.87 }$ & 805 $^{+    93 }_{-  73 }$ & 0.026 $^{+ 0.012 }_{- 0.008 }$ & 56 $^{+ 18 }_{- 12 }$ \\ 
1783.01  (\object{Kepler-1662} b)  & 134.4622 & 8.85 $^{+ 0.25 }_{- 0.24 }$  & 71.2 $^{+ 11.0 }_{- 9.0 }$    &  \textit{\citet{Vissapragada2020}} & 0.563 $^{+ 0.089 }_{- 0.073 }$ & 385 $^{+     7 }_{-   7 }$ & 0.022 $^{+ 0.003 }_{- 0.003 }$ & 99 $^{+ 15 }_{- 13 }$ \\ 
 1783.01  (Kepler-1662 b)  & 134.457 & 9.18 $^{+ 0.43 }_{- 0.41 }$  & 98.52 $^{+ 32.87 }_{- 24.70 }$   $ \left(^{+ 83.88 }_{- 46.20 }\right)$  & This Study & 0.71 $^{+ 0.27 }_{- 0.20 }$ & 385 $^{+     7 }_{-   7 }$ & 0.017 $^{+ 0.006 }_{- 0.005 }$ & 76 $^{+ 26 }_{- 19 }$ \\ 
   1783.02  (Kepler-1662 c)  & 284.213 & 5.43 $^{+ 0.52 }_{- 0.30 }$  & 15.0 $^{+ 4.0 }_{- 2.9 }$   &  \textit{\citet{Vissapragada2020}} & 0.51 $^{+ 0.15 }_{- 0.10 }$ &   300 $^{+    5 }_{-   5 }$  & 0.032 $^{+ 0.009 }_{- 0.007 }$ & 92 $^{+ 24 }_{- 20 }$ \\     
    1783.02  (Kepler-1662 c)  & 284.246 & 5.55 $^{+ 0.56 }_{- 0.36 }$  & 18.69 $^{+ 5.75 }_{- 4.71 }$ $  \left(^{+ 11.84 }_{- 8.68 }\right)$  & This Study & 0.54 $^{+ 0.23 }_{- 0.17 }$ &   300 $^{+    5 }_{-   5 }$  & 0.028 $^{+ 0.011 }_{- 0.007 }$ & 80 $^{+ 28 }_{- 19 }$ \\ 
 2113.01  (\object{Kepler-417} c)  & 15.943 & 2.49 $^{+ 0.25 }_{- 0.22 }$  & < 5.18   & \textit{This Study} & < 1.91  &  620 $^{+   29 }_{-  29 }$ & > 0.039  & > 170 \\ 
 2113.02  (Kepler-417 b)  & 12.331 & 2.22 $^{+ 0.22 }_{- 0.19 }$ & < 4.17  & \textit{This Study} & < 2.35  &  675 $^{+   32 }_{-  31 }$ & > 0.040  & > 161 \\ 
 2414.01  (\object{Kepler-384} b)  & 22.595 & 1.95 $^{+ 0.12 }_{- 0.09 }$ & < 18.55  & \textit{This Study} & < 10.00  &  802 $^{+   14 }_{-  13 }$ & > 0.011  & >  7 \\ 
 2414.02  (Kepler-384 c)  & 45.351 & 2.40 $^{+ 0.38 }_{- 0.35 }$  & < 53.07   & \textit{This Study} & < 10.00  &  636 $^{+   11 }_{-  11 }$ & > 0.006  & >  3 \\ 
 3503.01* (\object{KIC 6368175}) & 21.185 & 1.19 $^{+ 0.24 }_{- 0.19 }$ & < 9.20  & \textit{This Study} & < 10.00  &  785 $^{+   14 }_{-  13 }$ & > 0.013  & >  5 \\ 
 3503.02* & 31.828 & 1.21 $^{+ 0.28 }_{- 0.09 }$  & < 10.36   & \textit{This Study} & < 10.00  &  686 $^{+   12 }_{-  11 }$ & > 0.011  & >  8 \\ 
\hline
      \end{tabular}
    \caption{Results for two-planet systems: Planetary sizes,  masses, bulk densities, equilibrium blackbody temperatures, atmospheric scale heights, and estimated atmospheric transmission annuli. Uncertainties enclose the 68.3\% confidence intervals (with 95.5\% confidence intervals in parentheses to inform skewness). We include the Kepler numbers of candidates that are previously confirmed or validated in the first column. KOI names with an asterisk indicate newly confirmed planets.}\label{tbl-massresults2}
  \end{center}
\end{table}

\begin{table}[ht!]
\tiny
  \begin{center}
    \begin{tabular}{|c|c|c|c|c|c|c|c|c|}
      \hline
           name &  P (days) & R$_p$ (R$_{\oplus}$) & M$_p$ (M$_{\oplus}$) & Study (preferred in italics) & $\rho_{p}$ (g/cm$^3$)& T$_{eq}$ (K) & Sc. H. ($R_{\oplus}$) & Annulus (ppm) \\               
      \hline      
   85.01 (\object{Kepler-65} c) & 5.860 & 2.61 $^{+ 0.12 }_{- 0.11 }$ & < 11.88   & \textit{This Study}  & < 3.82  &  1270 $^{+   21 }_{-  20 }$ & > 0.038  & > 47 \\ 
 85.02 (Kepler-65 b) & 2.155 & 1.45 $^{+ 0.07 }_{- 0.06 }$   & < 7.51  & \textit{This Study} & < 10.00  &  1772 $^{+   29 }_{-  28 }$ & > 0.032  & > 20 \\ 
 85.03 (\object{Kepler-54} d)  & 8.134 & 1.58 $^{+ 0.06 }_{- 0.10 }$  & < 6.95   & \textit{This Study}  & < 10.00  &  1138 $^{+   18 }_{-  18 }$ & > 0.021  & > 15 \\ 
115.01 (\object{Kepler-105} b)  & 5.412 & 2.80 $^{+ 0.12 }_{- 0.24 }$  & 4.17 $^{+ 1.91 }_{- 1.97 }$ $  \left(^{+ 3.70 }_{- 3.55 }\right)$  & \textit{This Study} & 1.36 $^{+ 1.01 }_{- 0.65 }$ &   1082 $^{+   18 }_{-  18 }$  & 0.089 $^{+ 0.081 }_{- 0.033 }$ & 208 $^{+ 183 }_{- 70 }$ \\  
 115.01 (Kepler-105 b)  & 5.412 & 2.22 $^{+ 0.11 }_{- 0.11 }$  & 3.88 $^{+ 1.92 }_{- 1.85 }$   & \citet{Jontof-Hutter2016} & 1.98 $^{+ 1.05 }_{- 0.93 }$ &   996 $^{+   31 }_{-  31 }$  & 0.06 $^{+ 0.05 }_{- 0.02 }$ & 174 $^{+ 143 }_{- 57}$ \\   
 115.02 (Kepler-105 c)  & 7.126 & 1.79 $^{+ 0.26 }_{- 0.23 }$  & 4.77 $^{+ 0.92 }_{- 0.89 }$ $  \left(^{+ 1.83 }_{- 1.74 }\right)$  & \textit{This Study} & 4.32 $^{+ 2.51 }_{- 1.55 }$ &   986 $^{+   16 }_{-  16 }$  & 0.036 $^{+ 0.014 }_{- 0.010 }$ & 38 $^{+ 11 }_{-  7 }$ \\ 
115.02 (Kepler-105 c)  & 7.126 & 1.31 $ \pm $ 0.07 & 4.60 $^{+ 0.92 }_{- 0.85 }$   & \citet{Jontof-Hutter2016}  &  < 10.00   &  908 $^{+   28 }_{-  28 }$ & > 0.014  & > 19 \\  
115.03 & 3.437 & 0.55 $^{+ 0.08 }_{- 0.07 }$   & < 0.67  & \textit{This Study} & < 10.00  &  1259 $^{+   20 }_{-  21 }$ & > 0.051  & > 16 \\ 
 115.03 & 3.436 & 0.73 $ \pm $ 0.04 & < 1.27  & \citet{Jontof-Hutter2016} & < 10.00  &  1159 $^{+   36 }_{-  35 }$ & > 0.031  & > 8 \\ 
 137.01 (\object{Kepler-18} c)  & 7.642 & 4.29 $^{+ 0.18 }_{- 0.17 }$  & 5.78 $^{+ 5.04 }_{- 2.33 }$ $  \left(^{+ 10.95 }_{- 3.49 }\right)$  & \textit{This Study} & 0.40 $^{+ 0.36 }_{- 0.16 }$ &   831 $^{+   14 }_{-  14 }$  & 0.140 $^{+ 0.097 }_{- 0.065 }$ & 738 $^{+ 506 }_{- 344 }$ \\ 
137.01 (Kepler-18 c)  & 7.642 & 5.49 $^{+ 0.26 }_{- 0.26 }$  & 17.3 $^{+ 1.9 }_{- 1.9 }$  & \textit{\citet{coch11}} & 0.59 $\pm 0.07$ &   905 $^{+   28 }_{-  27 }$  & 0.08 $^{+ 0.01 }_{- 0.01 }$ &  354 $^{+ 48 }_{- 41 }$ \\ 
137.02 (Kepler-18 d)  & 14.859 & 5.20 $^{+ 0.22 }_{- 0.22 }$  & 8.30 $^{+ 4.17 }_{- 2.70 }$ $  \left(^{+ 7.36 }_{- 4.27 }\right)$  & \textit{This Study} & 0.33 $^{+ 0.17 }_{- 0.11 }$ &   666 $^{+   12 }_{-  11 }$  & 0.114 $^{+ 0.057 }_{- 0.038 }$ & 720 $^{+ 352 }_{- 241 }$ \\ 
137.02 (Kepler-18 d)  & 14.859 & 6.98 $^{+ 0.33 }_{- 0.33 }$  & 16.41 $\pm 1.40$   &  \textit{\citet{coch11}} & 0.27 $^{+ 0.03 }_{- 0.03 }$ &   725 $^{+   22 }_{-  22 }$  & 0.11 $^{+ 0.02 }_{- 0.01 }$ & 532 $^{+ 59 }_{- 50 }$ \\ 
 137.03 (Kepler-18 b)  & 3.505 & 2.00 $ \pm $ 0.1 & 6.9 $\pm 3.4$  & \textit{ \citet{coch11}} &  4.9$\pm 2.4$  &  1174 $^{+   36 }_{-  36 }$ & 0.04$^{+  0.03 }_{- 0.01 }$  & 44$^{+  39 }_{- 15 }$  \\ 
 137.03 (Kepler-18 b)  & 3.505 & 1.78 $^{+ 0.19 }_{- 0.17 }$   & < 13.08  & This Study & < 10.00  &  1077 $^{+   18 }_{-  19 }$ & > 0.016  & > 27 \\ 
 156.01 (\object{Kepler-114} c)  & 8.042 & 1.72 $^{+ 0.08 }_{- 0.08 }$  & 1.43 $^{+ 0.71 }_{- 0.47 }$ $  \left(^{+ 1.84 }_{- 0.77 }\right)$  & \textit{This Study} & 1.56 $^{+ 0.84 }_{- 0.54 }$ &   656 $^{+   15 }_{-  15 }$  & 0.071 $^{+ 0.037 }_{- 0.024 }$ & 246 $^{+ 124 }_{- 83 }$ \\ 
   156.01 (Kepler-114 c)  & 8.041 & 1.3 $^{+ 0.1 }_{- 0.1 }$  & 2.4 $^{+ 0.6 }_{- 0.7 }$   & \citet{Hadden2017} & 5.5 $^{+ 2.0 }_{- 1.5 }$ &   501 $^{+   18 }_{-  18 }$  & 0.019 $^{+ 0.009 }_{- 0.005 }$ & 88 $^{+ 40 }_{- 19 }$ \\    
   156.02 (Kepler-114 b)  & 5.188 & 1.33 $^{+ 0.06 }_{- 0.06 }$   & < 4.19  & \textit{This Study} & < 10.00  &  760 $^{+   18 }_{-  17 }$ & > 0.017  & > 46 \\
 156.02 (Kepler-114 b)  & 5.189 & 1.0 $ \pm $ 0.1 &   2.9 $^{+ 2.8 }_{- 1.9 }$  &  \citet{Hadden2017}   & 13.9$^{+  16.4 }_{-  9.4 }$  & 580 $^{+   22 }_{-  22 }$ &  0.013 $^{+   0.007 }_{-  0.001 }$  &  49 $^{+   26 }_{-  10 }$  \\ 
  156.03 (Kepler-114 d)  & 11.777 & 2.76 $^{+ 0.22 }_{- 0.14 }$  & < 2.91   & \textit{This Study}  & < 0.73  &  578 $^{+   13 }_{-  14 }$ & > 0.083  & > 424 \\
156.03 (Kepler-114 d)  & 11.776 & 2.0 $ \pm $ 0.1 & 0.8 $^{+ 0.8 }_{- 0.8 }$   &  \citet{Hadden2017} &  0.4 $^{+ 0.6 }_{- 0.4 }$   &    442 $^{+ 17 }_{- 17 }$  &   --- & ---  \\ 
168.01  (\object{Kepler-23} c) & 10.738 & 3.08 $^{+ 0.14 }_{- 0.13 }$  & 5.33 $^{+ 3.75 }_{- 2.42 }$ $  \left(^{+ 8.11 }_{- 3.66 }\right)$  & \textit{This Study} & 1.02 $^{+ 0.75 }_{- 0.47 }$ &   1018 $^{+   18 }_{-  17 }$  & 0.095 $^{+ 0.081 }_{- 0.040 }$ & 129 $^{+ 108 }_{- 53 }$ \\ 
 168.01  (Kepler-23 c) & 10.742 & 3.2 $^{+ 0.2 }_{- 0.2 }$  & 9.1$^{+ 3.4 }_{- 3.9 }$   & \citet{Hadden2017} & 1.3 $^{+ 0.8 }_{- 0.4 }$ &  1037 $^{+   31 }_{-  29 }$  & 0.066 $^{+ 0.059 }_{- 0.020 }$ & 86 $^{+ 74 }_{- 25 }$ \\ 
  168.02  (Kepler-23 d) & 15.278 & 2.17 $^{+ 0.11 }_{- 0.10 }$  & < 11.75   & \textit{This Study}  & < 6.46  &  905 $^{+   15 }_{-  15 }$ & > 0.019  & > 18 \\ 
 168.02  (Kepler-23 d) & 15.274 & 2.3 $ \pm $ 0.1 &  4.9 $^{+ 3.6 }_{- 3.5 }$  &  \citet{Hadden2017}  & 2.1 $^{+ 1.8 }_{- 1.5 }$  &  922 $^{+   27 }_{-  26 }$ & >0.02  &  >37 \\  
 168.03  (Kepler-23 b) & 7.108 & 1.73 $^{+ 0.09 }_{- 0.08 }$  & 2.79 $^{+ 2.04 }_{- 1.22 }$   $ \left(^{+ 4.68 }_{- 1.80 }\right)$  & \textit{This Study} & 2.88 $^{+ 2.18 }_{- 1.29 }$ & 1168 $^{+    20 }_{-  20 }$ & 0.068 $^{+ 0.054 }_{- 0.029 }$ & 51 $^{+ 40 }_{- 22 }$ \\ 
  168.03  (Kepler-23 b) & 7.107 & 1.8 $\pm 0.1$  & 4.7 $^{+ 1.9 }_{- 1.9 }$  & \citet{Hadden2017} & 4.2 $^{+ 2.3 }_{- 1.6 }$ & 1190 $^{+    35 }_{-  33 }$ & --- & --- \\  
314.01 (\object{Kepler-138} c)  & 13.782 & 1.67 $ \pm $ 0.15 & 5.2$\pm 1.2$   & \textit{\citet{Almenara2018}} &  6.1$^{+ 2.6 }_{- 1.9 }$ &  429 $\pm 20$ &  0.012 $^{+ 0.005 }_{- 0.003 }$  & 53 $^{+ 17 }_{- 11 }$ \\ 
 314.01 (Kepler-138 c)  & 13.781 & 1.42 $^{+ 0.09 }_{- 0.13 }$ & < 5.22   & This Study & < 10.00  &  412 $^{+   10 }_{-  10 }$ & > 0.008  & > 45 \\ 
 314.02 (Kepler-138 d)  & 23.093 & 1.68 $ \pm $ 0.15 & 1.17$\pm 0.30$   &  \textit{\citet{Almenara2018}}  & 1.3$^{+ 0.6 }_{- 0.4 }$   &  361 $^{+   17 }_{-  27 }$ & 0.05 $^{+ 0.02 }_{- 0.01 }$  & 164 $^{+ 60 }_{- 37 }$ \\ 
  314.02 (Kepler-138 d)  & 23.087 & 1.32 $^{+ 0.04 }_{- 0.04 }$  & < 2.41   &  This Study  & < 5.86  &  347 $^{+    8 }_{-   8 }$ & > 0.013  & > 59 \\ 
   314.03 (Kepler-138 b)  & 10.313 & 0.70 $ \pm $ 0.07 &  0.19 $\pm 0.05$ &  \textit{\citet{Almenara2018}} & 2.9$^{+ 1.4 }_{- 1.0 }$  &  473 $^{+   22 }_{-  22 }$ & 0.07 $^{+ 0.03 }_{- 0.02 }$  & 129 $^{+ 50 }_{- 30 }$ \\ 
 314.03 (Kepler-138 b)  & 10.313 & 0.63 $^{+ 0.03 }_{- 0.03 }$   & < 0.19  & This Study & < 4.23  &  454 $^{+   11 }_{-  10 }$ & > 0.050  & > 111 \\
   377.01 (\object{Kepler-9} b)   &  19.243   & 7.91  $ ^{+ 0.24   }_{-  0.24 } $  & 44.36   $^{+ 0.44 }_{-  0.44 } $  &  \textit{\citet{Freudenthal2018}} & 0.49   $^{+ 0.05 }_{-  0.04  } $  &  656.09   $^{+ 10.54 }_{-  10.46 } $  & 0.04889   $^{+ 0.00314 }_{-  0.00299 } $  & 411.52   $^{+ 14.76 }_{-  14.45 }  $\\
   377.01 (Kepler-9 b)  & 19.248 & 7.97 $^{+ 0.35 }_{- 0.33 }$  & 43.15 $^{+ 1.17 }_{- 1.40 }$ $  \left(^{+ 2.16 }_{- 2.94 }\right)$  & This Study & 0.47 $^{+ 0.07 }_{- 0.06 }$ &   669 $^{+   11 }_{-  11 }$  & 0.052 $^{+ 0.005 }_{- 0.005 }$ & 436 $^{+ 26 }_{- 25 }$ \\ 
 377.02 (Kepler-9 c)  & 38.969 & 7.76  $^{+ 0.23   }_{-  0.23 }$ &   30.56  $^{+ 0.92 }_{-  0.91 }$ &  \textit{\citet{Freudenthal2018}} &   0.36 $ ^{+ 0.04 }_{-  0.03  } $  &  518.56  $^{+ 8.11 }_{-  8.21 }$  & 0.05396  $^{+ 0.00381 }_{-  0.00353 }$ &  428.29  $^{+ 19.75 }_{-  18.79 }$\\
377.02 (Kepler-9 c)  & 38.944 & 8.21 $^{+ 0.36 }_{- 0.35 }$  & 29.71 $^{+ 0.80 }_{- 0.94 }$ $  \left(^{+ 1.49 }_{- 2.02 }\right)$  & \textit{This Study} & 0.29 $^{+ 0.04 }_{- 0.04 }$ &   529 $^{+    9 }_{-   9 }$  & 0.064 $^{+ 0.006 }_{- 0.006 }$ & 451 $^{+ 27 }_{- 25 }$ \\ 
 377.03 (Kepler-9 d)  & 1.593 & 1.53 $^{+ 0.09 }_{- 0.07 }$   & < 9.27  & \textit{This Study} & < 10.00  &  1537 $^{+   26 }_{-  27 }$ & > 0.026  & > 36 \\
401.01 (\object{Kepler-149} b)  & 29.199 & 4.37 $^{+ 0.21 }_{- 0.19 }$   & < 66.53  & \textit{This Study} & < 4.59  &  551 $^{+   10 }_{-  10 }$ & > 0.008  & > 39 \\ 
 401.02 (Kepler-149 d)  & 160.018 & 4.51 $^{+ 0.34 }_{- 0.60 }$  & < 89.35   & \textit{This Study}  & < 10.00  &  312 $^{+    6 }_{-   6 }$ & > 0.003  & > 12 \\ 
 401.03 (Kepler-149 c)  & 55.321 & 1.75 $^{+ 0.10 }_{- 0.10 }$ & < 4.24   & \textit{This Study}  & < 4.46  &  445 $^{+    8 }_{-   8 }$ & > 0.017  & > 31 \\ 
567.01 (\object{Kepler-184} b)  & 10.688 & 2.59 $^{+ 0.13 }_{- 0.26 }$   & < 12.27  & \textit{This Study} & < 8.63  &  735 $^{+   13 }_{-  14 }$ & > 0.015  & > 55 \\ 
 567.02 (Kepler-184 c)  & 20.302 & 2.12 $^{+ 0.27 }_{- 0.22 }$ & < 2.96   & \textit{This Study}  & < 1.88  &  593 $^{+   11 }_{-  11 }$ & > 0.046  & > 120 \\ 
 567.03 (Kepler-184 d)  & 29.022 & 2.55 $^{+ 0.66 }_{- 0.29 }$  & 7.68 $^{+ 3.36 }_{- 3.39 }$ $  \left(^{+ 6.97 }_{- 6.55 }\right)$  & \textit{This Study} & 1.67 $^{+ 1.58 }_{- 0.90 }$ &   527 $^{+   10 }_{-  10 }$  & 0.031 $^{+ 0.029 }_{- 0.013 }$ & 67 $^{+ 54 }_{- 23 }$ \\ 
 620.01 (\object{Kepler-51} b)  & 45.154 & 6.89 $^{+ 0.14  }_{- 0.14}$ & 3.69 $^{+ 1.86  }_{- 1.59}$ & \textit{ \citet{LibbyRoberts2020}} & 0.064 $^{+ 0.024  }_{- 0.024}$ &  473 $^{+ 6  }_{- 6}$ & 0.30  $^{+ 0.22  }_{- 0.10 }$ & 2725$^{+ 1961  }_{- 876}$ \\
 620.01 (Kepler-51 b)  & 45.153 & 6.65 $^{+ 0.15 }_{- 0.12 }$  & 2.70 $^{+ 1.42 }_{- 1.11 }$   $ \left(^{+ 3.07 }_{- 2.09 }\right)$  & \textit{This Study} & 0.05 $^{+ 0.03 }_{- 0.02 }$ & 460 $^{+     9 }_{-   9 }$ & 0.403 $^{+ 0.280 }_{- 0.140 }$ & 3738 $^{+ 2605 }_{- 1289 }$ \\ 
 620.02 (Kepler-51 d)  & 130.194  &  9.46$^{+  0.16  }_{- 0.16 }$   &  5.70$^{+  1.12  }_{- 1.12 }$ & \textit{ \citet{LibbyRoberts2020}}  &  0.038 $^{+ 0.006  }_{- 0.006 }$  &   332 $^{+ 4  }_{- 4  }$  &  0.27 $^{+ 0.07  }_{- 0.05 }$   &  3463 $^{+ 858  }_{- 565 }$  \\ 
  620.02 (Kepler-51 d)  & 130.184 & 9.08 $^{+ 0.17 }_{- 0.15 }$  & 5.51 $^{+ 1.16 }_{- 1.10 }$ $  \left(^{+ 2.47 }_{- 2.06 }\right)$  & \textit{This Study} & 0.04 $^{+ 0.01 }_{- 0.01 }$ &   323 $^{+    6 }_{-   6 }$  & 0.257 $^{+ 0.065 }_{- 0.046 }$ & 3263 $^{+ 827 }_{- 576 }$ \\ 
 620.03 (Kepler-51 c)  & 85.312 &  8.98 $^{+ 2.84  }_{-  2.84}$  & 4.43 $^{+ 0.54  }_{-  0.54 }$  &\textit{ \citet{LibbyRoberts2020}} & 0.034 $^{+ 0.069 }_{- 0.019}$ &  382.27 $^{+ 4.81 }_{- 4.71 }$ & 0.37 $^{+ 0.28 }_{- 0.20 }$ & 4899$^{+ 1726 }_{-1637 }$ \\
 620.03 (Kepler-51 c)  & 85.318 & 5.51 $^{+ 0.13 }_{- 0.13 }$  & 3.33 $^{+ 0.53 }_{- 0.49 }$ $  \left(^{+ 1.08 }_{- 0.95 }\right)$  &This Study  & 0.11 $^{+ 0.02 }_{- 0.02 }$ &   372 $^{+    7 }_{-   7 }$  & 0.180 $^{+ 0.033 }_{- 0.026 }$ & 627 $^{+ 110 }_{- 88 }$ \\ 
 750.01 (\object{Kepler-662} b)  & 21.678 & 2.95 $^{+ 0.12 }_{- 0.23 }$  & < 16.69   & \textit{This Study}  & < 5.38  &  558 $^{+   13 }_{-  12 }$ & > 0.012  & > 40 \\ 
 750.02  & 5.044 & 1.18 $^{+ 0.04 }_{- 0.10 }$   & < 2.93  & \textit{This Study} & < 10.00  &  907 $^{+   20 }_{-  20 }$ & > 0.022  & > 33 \\ 
 750.03  & 14.515 & 1.65 $^{+ 0.33 }_{- 0.31 }$ & < 6.71   & \textit{This Study}  & < 10.00  &  638 $^{+   14 }_{-  14 }$ & > 0.013  & > 19 \\ 
 806.01 (\object{Kepler-30} d)  & 142.642 & 8.79 $\pm 0.13$  & 23.7 $\pm 1.3$    & \citet{Panichi2017} & 0.19 $^{+ 0.01 }_{- 0.01 }$ &   324 $^{+    21 }_{-   22 }$  & 0.056 $^{+ 0.005 }_{- 0.005 }$ & 670 $^{+ 60 }_{- 57 }$ \\  
 806.01 (Kepler-30 d)  & 143.494 & 8.58 $^{+ 0.62 }_{- 0.47 }$  & 18.79 $^{+ 1.37 }_{- 1.32 }$ $  \left(^{+ 2.79 }_{- 2.73 }\right)$  & This Study & 0.16 $^{+ 0.03 }_{- 0.03 }$ &   302 $^{+    7 }_{-   7 }$  & 0.065 $^{+ 0.010 }_{- 0.009 }$ & 783 $^{+ 83 }_{- 73 }$ \\ 
 806.02 (Kepler-30 c)  & 60.325 & 11.98 $\pm 0.28$  & 536 $\pm 5$   & \citet{Panichi2017} & 1.71 $\pm 0.13$ &   432 $^{+   27 }_{-  29 }$  & 0.006 $^{+ 0.001 }_{- 0.001 }$ & 103 $^{+ 7 }_{- 7 }$ \\  
  806.02 (Kepler-30 c)  & 60.319 & 11.75 $^{+ 0.66 }_{- 0.63 }$  & 512.22 $^{+ 10.51 }_{- 13.58 }$ $  \left(^{+ 18.92 }_{- 32.05 }\right)$  & This Study & 1.74 $^{+ 0.33 }_{- 0.27 }$ &   402 $^{+   10 }_{-  10 }$  & 0.006 $^{+ 0.001 }_{- 0.001 }$ & 99 $^{+  7 }_{-  6 }$ \\ 
 806.03 (Kepler-30 b)  & 29.355 & 1.84 $^{+ 0.13 }_{- 0.17 }$  & 8.74 $^{+ 0.29 }_{- 1.38 }$   $ \left(^{+ 0.50 }_{- 4.45 }\right)$  & \textit{This Study} & 8.53 $^{+ 1.47 }_{- 1.62 }$ & 513 $^{+    12 }_{-  12 }$ & 0.010 $^{+ 0.001 }_{- 0.001 }$ & 22 $^{+  2 }_{-  1 }$ \\  
 806.03 (Kepler-30 b)  & 29.219 & 3.75  $\pm 0.18$ & 9.2 $\pm 0.1$  & \citet{Panichi2017} & 0.96 $\pm 0.15$ & 550 $^{+    35 }_{-  36 }$ & 0.044 $^{+ 0.005 }_{- 0.005 }$ & 58 $^{+ 7 }_{- 7 }$ \\ 
  \hline
     \end{tabular}
    \caption{Results for three-planet systems (part 1 of 2): Planetary sizes,  masses, bulk densities, equilibrium blackbody temperatures, atmospheric scale heights, and estimated atmospheric transmission annuli of three-planet systems in our sample. In this table we include out three-planet systems. Uncertainties enclose the 68.3\% confidence intervals (with 95.5\% confidence intervals in parentheses to inform skewness). We include the Kepler numbers of candidates that are previously confirmed or validated in the first column. KOI names with an asterisk indicate newly confirmed planets.}\label{tbl-massresults3}
  \end{center}
\end{table}
 
  \clearpage
\newpage

 \begin{table}[ht!]
\tiny
  \begin{center}
    \begin{tabular}{|c|c|c|c|c|c|c|c|c|}
      \hline
           name &  P (days) & R$_p$ (R$_{\oplus}$) & M$_p$ (M$_{\oplus}$) & Study (preferred in italics) & $\rho_{p}$ (g/cm$^3$)& T$_{eq}$ (K) & Sc. H. ($R_{\oplus}$) & Annulus (ppm) \\               
 \hline
   877.01 (\object{Kepler-81} b)  & 5.955 & 2.4 $ \pm $ 0.1 &  8.2  $^{+ 3.4 }_{- 3.8 }$  & \citet{Hadden2017} &  2.9  $^{+ 1.6 }_{- 1.3 }$  &  643 $^{+   20 }_{-  20 }$ &  0.025 $^{+   0.024 }_{-  0.008 }$  & 149 $^{+   137 }_{-  46 }$  \\ 
   877.01 (Kepler-81 b)  & 5.955 & 2.37 $^{+ 0.11 }_{- 0.12 }$   & < 16.70  & This Study & < 7.83  &  635 $^{+   14 }_{-  15 }$ & > 0.010  & > 65 \\
 877.02 (Kepler-81 c)  & 12.040 & 2.3 $ \pm $ 0.1 & 3.9  $^{+ 1.0 }_{- 1.2 }$    & \citet{Hadden2017}  & 1.7  $^{+ 0.6 }_{- 0.5 }$  &  508 $^{+   16 }_{-  15 }$ & 0.039 $^{+   0.020 }_{-  0.009 }$ & 209 $^{+   108 }_{-  46 }$   \\ 
 877.02 (Kepler-81 c)  & 12.042 & 2.22 $^{+ 0.11 }_{- 0.11 }$ & < 5.34   &  This Study  & < 2.86  &  502 $^{+   11 }_{-  11 }$ & > 0.024  & > 136 \\  
  877.03 (Kepler-81 d)  & 20.836 & 1.42 $^{+ 0.06 }_{- 0.07 }$  & < 6.02   & This Study  & < 10.00  &  418 $^{+   10 }_{-  10 }$ & > 0.008  & > 24 \\
  886.01 (\object{Kepler-54} b)  & 8.009 & 1.92 $^{+ 0.06 }_{- 0.06 }$  & 3.26 $^{+ 1.25 }_{- 1.24 }$   $ \left(^{+ 2.52 }_{- 2.12 }\right)$  & \textit{This Study} & 2.54 $^{+ 1.03 }_{- 0.97 }$ & 472 $^{+    11 }_{-  11 }$ & 0.028 $^{+ 0.017 }_{- 0.008 }$ & 196 $^{+ 123 }_{- 55 }$ \\ 
  886.01 (Kepler-54 b)  & 8.011& 2.3 $^{+ 0.4 }_{- 0.3 }$  & 2.6 $^{+ 0.8 }_{- 0.8 }$   & \citet{Hadden2017} & 1.1 $^{+ 0.5 }_{- 0.5}$ & 452 $^{+    30 }_{-  31 }$ & 0.049 $^{+ 0.031 }_{- 0.019 }$ & 281 $^{+ 138 }_{- 81 }$ \\   
 886.02 (Kepler-54 c)  & 12.068 & 1.44 $^{+ 0.08 }_{- 0.05 }$  & 2.34 $^{+ 0.95 }_{- 0.93 }$ $  \left(^{+ 1.90 }_{- 1.56 }\right)$  & \textit{This Study} & 4.01 $^{+ 1.79 }_{- 1.60 }$ &   412 $^{+   10 }_{-  10 }$  & 0.020 $^{+ 0.013 }_{- 0.006 }$ & 105 $^{+ 68 }_{- 30 }$ \\ 
   886.02 (Kepler-54 c)  & 12.071 & 1.3 $^{+ 0.1 }_{- 0.1 }$  &  2.3$^{+ 0.7 }_{- 0.7 }$  & \citet{Hadden2017} & 5.6 $^{+ 2.9 }_{- 1.9 }$ &   394$^{+   27 }_{-  28 }$  & 0.015 $^{+ 0.007 }_{- 0.004 }$ & 90 $^{+ 40 }_{- 24 }$ \\  
 886.03 (Kepler-54 d)  & 20.996 & 1.46 $^{+ 0.07 }_{- 0.06 }$  & < 7.64   & \textit{This Study}  & < 10.00  &  342 $^{+    8 }_{-   8 }$ & > 0.006  & > 29 \\ 
    934.01 (\object{Kepler-254} b)  & 5.827 & 3.59 $^{+ 0.28 }_{- 0.25 }$   & < 102.27  & \textit{This Study} & < 10.00  &  939 $^{+   32 }_{-  32 }$ & > 0.007  & > 29 \\ 
 934.02  (Kepler-254 c) & 12.407 & 2.26 $^{+ 0.20 }_{- 0.17 }$ & < 5.38   & \textit{This Study}  & < 2.54  &  730 $^{+   25 }_{-  25 }$ & > 0.037  & > 99 \\ 
 934.03 (Kepler-254 d)  & 18.748 & 2.68 $^{+ 0.20 }_{- 0.24 }$  & < 26.46   & \textit{This Study}  & < 10.00  &  636 $^{+   22 }_{-  21 }$ & > 0.008  & > 26 \\ 
 1070.01 (\object{Kepler-266} b)  & 6.618 & 2.58 $^{+ 0.19 }_{- 0.16 }$   & < 48.19  & \textit{This Study} & < 10.00  &  975 $^{+   39 }_{-  38 }$ & > 0.009  & > 17 \\ 
 1070.02 (Kepler-266 c)  & 107.742 & 4.12 $^{+ 0.30 }_{- 0.25 }$  & < 10.31   & \textit{This Study}  & < 0.84  &  385 $^{+   16 }_{-  15 }$ & > 0.032  & > 111 \\ 
 1070.03 & 92.803 & 2.11 $^{+ 0.21 }_{- 0.35 }$ & < 4.00   & \textit{This Study}  & < 10.00  &  404 $^{+   17 }_{-  16 }$ & > 0.013  & > 27 \\ 
1338.01 (\object{Kepler-822} b)  & 3.223 & 1.51 $^{+ 0.08 }_{- 0.08 }$   & < 8.51  & \textit{This Study} & < 10.00  &  1219 $^{+   24 }_{-  23 }$ & > 0.021  & > 29 \\ 
 1338.02 & 42.041 & 1.66 $^{+ 0.45 }_{- 0.24 }$  & < 28.79   & \textit{This Study}  & < 10.00  &  518 $^{+   10 }_{-  10 }$ & > 0.006  & >  6 \\ 
 1338.03 & 21.016 & 1.44 $^{+ 0.36 }_{- 0.23 }$ & < 19.50   & \textit{This Study}  & < 10.00  &  653 $^{+   13 }_{-  13 }$ & > 0.009  & >  6 \\ 
 1353.01 (\object{Kepler-289} c) & 125.870 & 11.21 $^{+ 0.50 }_{- 0.47 }$  & 115.25 $^{+ 12.36 }_{- 12.22 }$ $  \left(^{+ 23.94 }_{- 24.51 }\right)$  & \textit{This Study} & 0.45 $^{+ 0.08 }_{- 0.07 }$ &   376 $^{+    7 }_{-   7 }$  & 0.022 $^{+ 0.003 }_{- 0.003 }$ & 239 $^{+ 33 }_{- 26 }$ \\ 
   1353.01 (Kepler-289 c) & 125.852  &11.59 $^{+ 0.19 }_{-   0.19 }$   & 132  $^{+17 }_{-17}$   & \citet{Schmitt2014b} &  0.47 $^{+ 0.06 }_{- 0.06}$ &  372 $^{+ 4.37 }_{- 4.73}$ &  0.020 $^{+ 0.003}_{- 0.002  }$ & 213 $^{+  33 }_{- 24}$    \\
 1353.02 (Kepler-289 b) & 34.543 & 2.42 $^{+ 0.42 }_{- 0.36 }$   & < 16.61  & \textit{This Study} & < 7.95  &  579 $^{+   11 }_{-  10 }$ & > 0.010  & > 19 \\ 
  1353.02 (Kepler-289 b) & 34.545 &  2.15 $^{+ 0.10 }_{- 0.10 }$  & 7.3 $^{+ 6.8 }_{- 6.8 }$   & \citet{Schmitt2014b} & <10  &  573$^{+7}_{- 7}$ & ---  & --- \\ 
(PH 3c) Kepler-289 d & 65.959 & 2.68 $^{+ 0.17 }_{- 0.17 }$  & 2.86 $^{+ 0.51 }_{- 0.44 }$ $  \left(^{+ 1.07 }_{- 0.85 }\right)$  & \textit{This Study} & 0.82 $^{+ 0.24 }_{- 0.18 }$ &   467 $^{+    9 }_{-   9 }$  & 0.062 $^{+ 0.015 }_{- 0.012 }$ & 154 $^{+ 31 }_{- 26 }$ \\ 
(PH 3c) Kepler-289 d & 66.063 & 2.68 $^{+ 0.17 }_{- 0.17}$ & 4.0 $^{+ 0.9 }_{- 0.9}$ &  \citet{Schmitt2014b} & 1.2 $^{+ 0.3 }_{- 0.3}$  & 461  $^{+6 }_{- 6}$ &  0.04  $^{+0.01 }_{-  0.01}$  & 110 $^{+ 32 }_{- 22}$ \\
 1574.01 (\object{Kepler-87} b) & 114.756 & 10.67 $^{+ 0.52 }_{- 0.51 }$  & 302.27 $^{+ 125.51 }_{- 90.64 }$ $  \left(^{+ 262.59 }_{- 149.75 }\right)$  & \textit{This Study} & 1.38 $^{+ 0.61 }_{- 0.44 }$ &   433 $^{+   14 }_{-  13 }$  & 0.009 $^{+ 0.004 }_{- 0.003 }$ & 39 $^{+ 17 }_{- 12 }$ \\ 
 1574.01 (Kepler-87 b)  & 114.757 & 13.49 $ \pm $ 0.55 &  326.1 $\pm 8.8$  & \citet{ofir14}  &  0.728$\pm 0.026$  &  483 $^{+   8 }_{-  8 }$ & 0.014$\pm 0.001$  &  51$\pm 3$  \\  
 1574.02 (Kepler-87 c)  & 189.802 & 5.62 $^{+ 0.48 }_{- 0.84 }$  & 6.57 $^{+ 1.61 }_{- 1.37 }$ $  \left(^{+ 3.34 }_{- 2.53 }\right)$  & \textit{This Study} & 0.28 $^{+ 0.32 }_{- 0.09 }$ &   365 $^{+   12 }_{-  11 }$  & 0.073 $^{+ 0.030 }_{- 0.030 }$ & 150 $^{+ 49 }_{- 42 }$ \\ 
  1574.02 (Kepler-87 c)  & 191.232 & 6.14 $\pm 0.29$  & 6.50 $\pm 0.80$   & \citet{ofir14} & 0.153 $\pm 0.019$ &   407 $\pm 7$  & 0.125 $^{+ 0.022 }_{- 0.018 }$ & 231 $^{+ 36}_{- 29 }$ \\  
 1574.03 & 5.834 & 1.72 $^{+ 0.10 }_{- 0.11 }$   & < 12.24  & \textit{This Study} & < 10.00  &  1169 $^{+   38 }_{-  36 }$ & > 0.018  & > 13 \\ 
 1576.01 (\object{Kepler-307} b)  & 10.421 & 2.83 $^{+ 0.27 }_{- 0.21 }$  & 7.77 $^{+ 0.88 }_{- 0.85 }$   $ \left(^{+ 1.85 }_{- 1.63 }\right)$  & \textit{This Study} & 1.74 $^{+ 0.54 }_{- 0.43 }$ & 772 $^{+    13 }_{-  13 }$ & 0.044 $^{+ 0.010 }_{- 0.008 }$ & 126 $^{+ 20 }_{- 16 }$ \\   
 1576.01 (Kepler-307 b)  & 10.421 & 2.43 $\pm 0.09$  & 7.48 $^{+ 0.91 }_{- 0.87 }$    & \citet{Jontof-Hutter2016} & 2.87 $^{+ 0.50 }_{- 0.43 }$ & 711 $^{+    17 }_{-  17 }$ & 0.030 $^{+ 0.005 }_{- 0.005 }$ & 101 $^{+ 14 }_{- 12 }$ \\ 
  1576.02 (Kepler-307 c)  & 13.075 & 2.55 $^{+ 0.28 }_{- 0.21 }$  & 4.11 $^{+ 0.66 }_{- 0.59 }$ $  \left(^{+ 1.39 }_{- 1.14 }\right)$  & \textit{This Study} & 1.26 $^{+ 0.46 }_{- 0.36 }$ &   716 $^{+   13 }_{-  12 }$  & 0.063 $^{+ 0.018 }_{- 0.013 }$ & 160 $^{+ 33 }_{- 25 }$ \\ 
  1576.02 (Kepler-307 c)  & 13.073 & 2.20 $\pm 0.07$  & 3.64 $^{+ 0.65 }_{- 0.58 }$   & \citet{Jontof-Hutter2016} & 1.74 $\pm 0.30 $ &   659 $^{+   16 }_{-  16 }$  & 0.046 $^{+ 0.009 }_{- 0.007 }$ & 137 $^{+ 26 }_{- 21 }$ \\  
 1576.03 & 23.341 & 1.06 $^{+ 0.07 }_{- 0.24 }$  & < 1.30   & \textit{This Study}  & < 10.00  &  590 $^{+   11 }_{-  10 }$ & > 0.019  & > 49 \\ 
 1598.01 (\object{Kepler-310} c)  & 56.476 & 3.22 $^{+ 0.24 }_{- 0.14 }$ & < 15.30   & \textit{This Study}  & < 2.42  &  422 $^{+    8 }_{-   8 }$ & > 0.016  & > 51 \\ 
 1598.01 (Kepler-310 c)  & 56.476 & 2.9 $^{+ 0.3 }_{- 0.2 }$  &  5.9 $^{+ 4.1 }_{- 4.0 }$   & \citet{Hadden2017}    &  1.1 $^{+ 0.9 }_{- 0.8 }$   &  414 $^{+    17 }_{-   17 }$ &   0.022 $^{+ 0.013 }_{- 0.007 }$ &    82 $^{+ 44 }_{- 22 }$  \\ 
 1598.02 (Kepler-310 d)  & 92.876 & 2.64 $^{+ 0.21 }_{- 0.11 }$  & 7.34 $^{+ 3.16 }_{- 2.61 }$ $  \left(^{+ 7.07 }_{- 5.02 }\right)$  & \textit{This Study} & 1.95 $^{+ 0.96 }_{- 0.74 }$ &   357 $^{+    7 }_{-   7 }$  & 0.019 $^{+ 0.011 }_{- 0.006 }$ & 46 $^{+ 25 }_{- 14 }$ \\ 
  1598.02 (Kepler-310 d)   &  92.874  &  2.1 $^{+ 0.2 }_{- 0.1 }$   &  8.4  $^{+ 2.9  }_{- 2.8 }$  &  \citet{Hadden2017} & 4.3  $^{+ 1.9  }_{-  1.6 }$  &  350 $^{+ 15 }_{-   14  }$  &  0.010  $^{+  0.006  }_{-  0.003 }$  &  30  $^{+ 16 }_{-  8}$ \\
 1598.03 (Kepler-310 b)  & 13.930 & 1.40 $^{+ 0.11 }_{- 0.06 }$   & < 7.88  & \textit{This Study} & < 10.00  &  672 $^{+   13 }_{-  12 }$ & > 0.012  & > 14 \\ 
 1833.01 (\object{Kepler-968} b)  & 3.693 & 2.08 $^{+ 0.37 }_{- 0.25 }$   & < 17.66  & \textit{This Study} & < 10.00  &  783 $^{+   29 }_{-  29 }$ & > 0.011  & > 40 \\ 
 1833.02 & 7.685 & 2.58 $^{+ 0.23 }_{- 0.22 }$  & < 11.02   & \textit{This Study}  & < 4.50  &  613 $^{+   23 }_{-  23 }$ & > 0.017  & > 83 \\ 
 1833.03 (Kepler-968 c)  & 5.707 & 1.51 $^{+ 0.13 }_{- 0.13 }$ & < 2.63   & \textit{This Study}  & < 4.39  &  677 $^{+   25 }_{-  24 }$ & > 0.030  & > 120 \\ 
2086.01 (\object{Kepler-60} b)  & 7.133 & 1.71 $\pm 0.13$  & 4.19 $^{+ 0.56 }_{- 0.52 }$   &   \citet{Jontof-Hutter2016} & 4.62 $^{+ 1.40 }_{- 1.10 }$ & 1078 $^{+    50 }_{-  50 }$ & 0.039 $^{+ 0.009 }_{- 0.007 }$ & 33 $^{+ 6 }_{- 5 }$ \\ 
 2086.01 (Kepler-60 b)  & 7.133 & 1.64 $^{+ 0.26 }_{- 0.32 }$  & 3.92 $^{+ 0.48 }_{- 0.72 }$   $ \left(^{+ 0.96 }_{- 2.81 }\right)$  & \textit{This Study} & 5.80 $^{+ 4.20 }_{- 2.23 }$ & 1145 $^{+    21 }_{-  21 }$ & 0.038 $^{+ 0.014 }_{- 0.009 }$ & 36 $^{+  7 }_{-  5 }$ \\  
  2086.02 (Kepler-60 c)  & 8.919 & 1.90 $\pm 0.15$  & 3.85 $\pm 0.81$  &  \citet{Jontof-Hutter2016}  & 3.06 $^{+ 1.14 }_{- 0.86 }$ &   1000 $^{+   47 }_{-  46 }$  & 0.050 $^{+ 0.016 }_{- 0.011 }$ & 47 $^{+ 14 }_{- 9 }$ \\ 
 2086.02 (Kepler-60 c)  & 8.919 & 1.84 $^{+ 0.29 }_{- 0.36 }$  & 3.62 $^{+ 0.77 }_{- 0.85 }$ $  \left(^{+ 1.53 }_{- 2.37 }\right)$  & \textit{This Study} & 3.72 $^{+ 4.09 }_{- 1.49 }$ &   1062 $^{+   19 }_{-  20 }$  & 0.047 $^{+ 0.022 }_{- 0.017 }$ & 49 $^{+ 16 }_{- 10 }$ \\ 
2086.03 (Kepler-60 d)  & 11.898 & 1.99 $ \pm $ 0.16 &   4.60 $^{+ 0.92 }_{- 0.85 }$   &  \citet{Jontof-Hutter2016}    &  2.91 $^{+ 1.03 }_{- 0.78 }$ &  908 $^{+   43 }_{-  42 }$ &  0.045 $^{+ 0.013 }_{- 0.010 }$  & 29 $^{+ 8 }_{- 6 }$  \\ 
 2086.03 (Kepler-60 d)  & 11.898 & 1.69 $^{+ 0.27 }_{- 0.33 }$  & 3.86 $^{+ 0.80 }_{- 0.85 }$ $  \left(^{+ 1.61 }_{- 2.73 }\right)$  & \textit{This Study} & 5.24 $^{+ 4.76 }_{- 2.11 }$ &   965 $^{+   18 }_{-  17 }$  & 0.034 $^{+ 0.014 }_{- 0.009 }$ & 32 $^{+  9 }_{-  6 }$ \\ 
 2092.01 (\object{Kepler-359} c)  & 57.711 & 5.14 $^{+ 1.54 }_{- 0.66 }$ & < 9.90   & \textit{This Study}  & < 0.39  &  430 $^{+   17 }_{-  18 }$ & > 0.064  & > 208 \\ 
   2092.01 (Kepler-359 c)  & 57.693 & 4.8 $^{+ 1.0 }_{- 0.9 }$ & 5.1 $^{+ 2.3 }_{- 2.1 }$   &  \citet{Hadden2017}   & 0.2$^{+ 0.2 }_{- 0.1 }$  &  540 $^{+   66 }_{-  52 }$ & 0.129 $^{+ 0.131 }_{- 0.061 }$   & 430 $^{+ 369 }_{- 161 }$  \\  
 2092.02 (Kepler-359 b)  & 25.563 & 4.00 $^{+ 1.19 }_{- 0.51 }$   & < 125.15  & \textit{This Study} & < 9.26  &  564 $^{+   23 }_{-  23 }$ & > 0.005  & > 12 \\ 
 2092.03 (Kepler-359 d)  & 77.085 & 3.68 $^{+ 0.24 }_{- 1.17 }$  & < 7.74   & \textit{This Study}  & < 10.00  &  390 $^{+   16 }_{-  16 }$ & > 0.008  & > 54 \\ 
2092.03 (Kepler-359 d)  & 77.083 & 4.6 $ \pm $ 0.9 &  4.4$^{+ 2.4 }_{- 1.3 }$  &  \citet{Hadden2017}   &  0.3$^{+ 0.2 }_{- 0.2 }$  &  490 $^{+   60 }_{-  48 }$ &  0.103 $^{+ 0.070 }_{- 0.044 }$   &  209 $^{+ 110 }_{- 73 }$  \\ 
 2195.01 (\object{Kepler-372} c)  & 20.053 & 2.63 $^{+ 0.44 }_{- 0.29 }$ & < 32.08   & \textit{This Study}  & < 10.00  &  790 $^{+   18 }_{-  18 }$ & > 0.009  & > 11 \\ 
 2195.02 (Kepler-372 d)  & 30.091 & 2.25 $^{+ 0.38 }_{- 0.25 }$  & < 30.95   & \textit{This Study}  & < 10.00  &  689 $^{+   16 }_{-  16 }$ & > 0.008  & >  7 \\ 
 2195.03 (Kepler-372 b)  & 6.850 & 1.83 $^{+ 0.30 }_{- 0.21 }$   & < 27.28  & \textit{This Study} & < 10.00  &  1129 $^{+   26 }_{-  27 }$ & > 0.013  & >  7 \\ 
 \hline
     \end{tabular}
    \caption{Results for three-planet systems (part 2 of 2): Planetary sizes,  masses, bulk densities, equilibrium blackbody temperatures, atmospheric scale heights, and estimated atmospheric transmission annuli. Uncertainties enclose the 68.3\% confidence intervals (with 95.5\% confidence intervals in parentheses to inform skewness). We include the Kepler numbers of candidates that are previously confirmed or validated in the first column. KOI names with an asterisk indicate newly confirmed planets.}\label{tbl-massresults3c}
  \end{center}
\end{table}
 
 \clearpage
\newpage

\begin{table}[ht!]
\tiny
  \begin{center}
    \begin{tabular}{|c|c|c|c|c|c|c|c|c|}
      \hline
           name &  P (days) & R$_p$ (R$_{\oplus}$) & M$_p$ (M$_{\oplus}$) & Study (preferred in italics) & $\rho_{p}$ (g/cm$^3$)& T$_{eq}$ (K) & Sc. H. ($R_{\oplus}$) & Annulus (ppm) \\               
\hline
152.01 (\object{Kepler-79} d) &  52.090 &  7.15  $^{+ 0.20   }_{-  0.20 }$ &  5.27  $^{+ 0.91 }_{-  0.90 }$ & \textit{\citet{Chachan2020}}  &  0.08  $^{+ 0.02 }_{-  0.01  }$   &  579  $^{+ 13 }_{-  13 }$ &  0.296  $^{+ 0.064 }_{-  0.046 }$ &  1198  $^{+ 252 }_{-  180 }$ \\
 152.01 (Kepler-79 d)  & 52.092 & 7.24 $^{+ 0.34 }_{- 0.32 }$  & 6.34 $^{+ 1.87 }_{- 1.62 }$   $ \left(^{+ 4.17 }_{- 2.95 }\right)$  & This Study & 0.09 $^{+ 0.03 }_{- 0.03 }$ & 600 $^{+    11 }_{-  11 }$ & 0.262 $^{+ 0.096 }_{- 0.065 }$ & 1046 $^{+ 366 }_{- 245 }$ \\ 
 152.02 (Kepler-79 c)   &  27.403 & 3.76  $^{+ 0.11   }_{-  0.11 }$ &  4.59  $^{+ 1.30 }_{-  1.33 }$ & \textit{\citet{Chachan2020}}  &  0.47  $^{+ 0.14 }_{-  0.14  }$   &  717 $ ^{+ 17 }_{-  16 }$ &  0.117  $^{+ 0.048 }_{-  0.027 }$ &  243 $^{+ 99}_{-  54 }$\\
 152.02 (Kepler-79 c)  & 27.404 & 3.68 $^{+ 0.18 }_{- 0.17 }$  & 5.86 $^{+ 2.49 }_{- 1.98 }$   $ \left(^{+ 5.36 }_{- 3.38 }\right)$  &This Study & 0.65 $^{+ 0.30 }_{- 0.23 }$ & 742 $^{+    14 }_{-  13 }$ & 0.090 $^{+ 0.048 }_{- 0.028 }$ & 183 $^{+ 95 }_{- 55 }$ \\ 
 152.03 (Kepler-79 b)   &  13.485 & 3.51  $^{+ 0.10   }_{-  0.10 }$ &  8.63  $^{+ 3.84 }_{-  2.83 }$ & \textit{\citet{Chachan2020}}  &  1.10  $^{+ 0.50 }_{-  0.37  }$   &  909  $^{+ 21 }_{-  20 } $&  0.069  $^{+ 0.034}_{-  0.021 }$ &  127 $^{+ 63 }_{-  39 }$ \\
 152.03 (Kepler-79 b)  & 13.484 & 3.44 $^{+ 0.17 }_{- 0.16 }$  & 8.56 $^{+ 6.22 }_{- 3.64 }$   $ \left(^{+ 17.44 }_{- 5.71 }\right)$  & This Study & 1.17 $^{+ 0.89 }_{- 0.51 }$ & 940 $^{+    17 }_{-  17 }$ & 0.069 $^{+ 0.051 }_{- 0.029 }$ & 130 $^{+ 96 }_{- 55 }$ \\ 
 152.04 (Kepler-79 e)   &  81.067 & 3.53  $^{+ 0.16   }_{-  0.16 }$ &  3.90  $^{+ 0.70 }_{-  0.61 }$ & \textit{\citet{Chachan2020}} &  0.49 $ ^{+ 0.12 }_{-  0.09  }$   &  500  $^{+ 11 }_{-  11 }$ &  0.084  $^{+ 0.018}_{-  0.014 }$ &  114  $^{+ 23 }_{-  19 }$ \\
 152.04 (Kepler-79 e)  & 81.072 & 3.01 $^{+ 0.56 }_{- 0.46 }$  & 3.86 $^{+ 0.98 }_{- 0.89 }$   $ \left(^{+ 2.08 }_{- 1.61 }\right)$  & This Study & 0.71 $^{+ 0.53 }_{- 0.30 }$ & 517 $^{+    10 }_{-   9 }$ & 0.069 $^{+ 0.035 }_{- 0.023 }$ & 105 $^{+ 37 }_{- 26 }$ \\ 
 248.01 (\object{Kepler-49} b)  & 7.203 & 2.61 $^{+ 0.08 }_{- 0.07 }$  & 7.03 $^{+ 9.14 }_{- 2.89 }$   $ \left(^{+ 13.87 }_{- 3.93 }\right)$  & \textit{This Study} & 2.21 $^{+ 2.81 }_{- 0.93 }$ & 535 $^{+    15 }_{-  15 }$ & 0.027 $^{+ 0.020 }_{- 0.015 }$ & 185 $^{+ 131 }_{- 104 }$ \\ 
  248.01 (Kepler-49 b)   & 7.204  &  2.35  $ \pm $ 0.09   &  5.09 $^{+  2.11}_{-  1.9 }$ & \citet{Jontof-Hutter2016}  &  2.17  $^{+  0.95 }_{-  0.92 }$  &  499 $^{+  25 }_{-  24 }$  & 0.03  $^{+ 0.02 }_{-  0.01 }$ &  211 $^{+ 149}_{- 62} $ \\
 248.02 (Kepler-49 c)  & 10.913 & 2.47 $^{+ 0.23 }_{- 0.15 }$  & 4.62 $^{+ 7.47 }_{- 2.12 }$   $ \left(^{+ 11.39 }_{- 2.79 }\right)$  & \textit{This Study} & 1.61 $^{+ 2.48 }_{- 0.80 }$ & 466 $^{+    14 }_{-  13 }$ & 0.034 $^{+ 0.031 }_{- 0.021 }$ & 186 $^{+ 161 }_{- 115 }$ \\ 
 248.02 (Kepler-49 c)  & 10.912 & 2.06  $ \pm $ 0.09 & 3.28 $^{+  1.45}_{-  1.32 }$   & \citet{Jontof-Hutter2016} & 2.05  $^{+  1.00 }_{-  0.94 }$   &  435 $^{+   22 }_{- 22 }$ & 0.03$^{+ 0.02 }_{-  0.01 }$  & 198 $^{+ 148}_{- 63} $  \\
  248.03 (Kepler-49 d)  & 2.577 &  1.82 $^{+ 0.06 }_{- 0.06 }$  & < 12.39  & \textit{This Study} & < 10.00  &  754 $^{+   22 }_{-  22 }$ & > 0.011  & > 51 \\ 
 248.04 (Kepler-49 e)  & 18.597 & 1.77 $^{+ 0.08 }_{- 0.06 }$  & < 6.20  & \textit{This Study} & < 6.09  &  391 $^{+   11 }_{-  11 }$ & > 0.010  & > 48 \\ 
 250.01 (\object{Kepler-26} b)  & 12.283 & 3.19 $^{+ 0.10 }_{- 0.09 }$  & 4.29 $^{+ 0.78 }_{- 0.72 }$   $ \left(^{+ 1.60 }_{- 1.47 }\right)$  & \textit{This Study} & 0.73 $^{+ 0.16 }_{- 0.13 }$ & 444 $^{+    17 }_{-  17 }$ & 0.056 $^{+ 0.013 }_{- 0.010 }$ & 486 $^{+ 111 }_{- 83 }$ \\ 
  250.01 (Kepler-26 b)  & 12.2796 & 2.78 $^{+ 0.11 }_{- 0.11 }$  & 5.12 $^{+ 0.65 }_{- 0.61 }$     & \citet{Jontof-Hutter2016} & 1.26 $^{+ 0.21 }_{- 0.19 }$ & 423 $^{+    15 }_{-  15 }$ & 0.034 $^{+ 0.006 }_{- 0.005 }$ & 339 $^{+ 53 }_{- 41 }$ \\
 250.02 (Kepler-26 c)  & 17.251 & 2.98 $^{+ 0.28 }_{- 0.23 }$  & 5.64 $^{+ 0.93 }_{- 0.89 }$   $ \left(^{+ 1.86 }_{- 1.81 }\right)$  & \textit{This Study} & 1.12 $^{+ 0.37 }_{- 0.30 }$ & 397 $^{+    14 }_{-  16 }$ & 0.034 $^{+ 0.010 }_{- 0.007 }$ & 229 $^{+ 53 }_{- 38 }$ \\ 
  250.02 (Kepler-26 c)  & 17.2559 & 2.72 $^{+ 0.12 }_{- 0.12 }$  & 6.20 $^{+ 0.65 }_{- 0.65 }$   & \citet{Jontof-Hutter2016} & 1.61 $^{+ 0.27 }_{- 0.22 }$ & 378 $^{+    14 }_{-  14 }$ & 0.024 $^{+ 0.004 }_{- 0.003 }$ & 180 $^{+ 23 }_{- 20 }$ \\ 
 250.03 (Kepler-26 d)  & 3.544 &  1.36 $^{+ 0.05 }_{- 0.04 }$  & < 5.82  & \textit{This Study} & < 10.00  &  672 $^{+   25 }_{-  26 }$ & > 0.013  & > 45 \\ 
 250.04 (Kepler-26 e)  & 46.831 & 2.40 $^{+ 0.08 }_{- 0.10 }$  & < 24.94  & \textit{This Study} & < 10.00  &  285 $^{+   10 }_{-  11 }$ & > 0.003  & > 22 \\ 
 520.01 (\object{Kepler-176} c)  & 12.758 & 2.48 $^{+ 0.11 }_{- 0.20 }$  & < 13.37  & This Study & < 8.18  &  629 $^{+   12 }_{-  12 }$ & > 0.012  & > 49 \\ 
 520.01 (Kepler-176 c)   &  12.759   &  2.34 $^{+ 0.11 }_{-   0.20 } $ &  5.03 $^{+ 3.14  }_{-  2.25  }$  &  \citet{Hadden2017}  & 2.32  $^{+1.75 }_{-   1.10 }$ & 736 $^{+ 22  }_{-  19 }$ &  0.040$^{+  0.034   }_{- 0.016 }$  & 155 $^{+ 125  }_{-  61  }$\\
 520.02 (Kepler-176 b)  & 5.433 &  1.55 $^{+ 0.23 }_{- 0.16 }$  & < 15.01  & This Study & < 10.00  &  836 $^{+   16 }_{-  16 }$ & > 0.012  & > 20 \\ 
 520.03 (Kepler-176 d)  & 25.754 & 2.27 $^{+ 0.09 }_{- 0.16 }$  & 3.42 $^{+ 3.02 }_{- 1.51 }$   $ \left(^{+ 6.58 }_{- 2.21 }\right)$  & This Study & 2.02 $^{+ 1.91 }_{- 0.97 }$ & 498 $^{+     9 }_{-  10 }$ & 0.034 $^{+ 0.029 }_{- 0.016 }$ & 118 $^{+ 96 }_{- 55 }$ \\ 
 520.03 (Kepler-176 d)    &   25.753   &   2.17 $^{+ 0.09  }_{-  0.16  }$  &   5.90 $^{+ 2.21  }_{-  2.17  }$  & \citet{Hadden2017}   &   3.31 $^{+ 1.72 }_{-   1.30  }$  &   583$^{+ 17  }_{-  15 }$  &   0.024 $^{+ 0.014  }_{-  0.007  }$  &   81 $^{+ 48  }_{-  23 }$ \\
 520.04 (Kepler-176 e)  & 51.167 & 1.45 $^{+ 0.27 }_{- 0.22 }$  & < 8.69  & This Study & < 10.00  &  396 $^{+    7 }_{-   8 }$ & > 0.007  & > 11 \\ 
  730.01 (\object{Kepler-223} d)   & 14.789 & 5.24 $^{+ 0.26 }_{- 0.45 }$  & 8.0 $^{+ 1.5 }_{- 1.3 }$    &\textit{ \citet{Mills2016}}  & 0.31 $^{+ 0.12 }_{- 0.07 }$ & 923 $^{+    39 }_{-  63 }$ & 0.14 $^{+ 0.04 }_{- 0.04 }$ & 223 $^{+ 52 }_{- 45 }$ \\ 
 730.01 (Kepler-223 d)   & 14.788 & 4.54 $^{+ 0.38 }_{- 0.33 }$  & 7.65 $^{+ 1.79 }_{- 1.81 }$   $ \left(^{+ 3.61 }_{- 3.59 }\right)$  & \textit{This Study} & 0.43 $^{+ 0.16 }_{- 0.13 }$ & 916 $^{+    33 }_{-  30 }$ & 0.134 $^{+ 0.050 }_{- 0.032 }$ & 228 $^{+ 75 }_{- 47 }$ \\ 
 730.02 (Kepler-223 c)   & 9.846 & 3.44 $^{+ 0.20 }_{- 0.30 }$  & 5.1 $^{+ 1.7 }_{- 1.1 }$    &\textit{ \citet{Mills2016} } & 0.71 $^{+ 0.33 }_{- 0.20 }$ & 1057 $^{+    44 }_{-  69 }$ & 0.104 $^{+ 0.036 }_{- 0.029 }$ & 139 $^{+ 43 }_{- 34 }$ \\
 730.02 (Kepler-223 c)   & 9.848 & 2.68 $^{+ 0.45 }_{- 0.45 }$  & 7.64 $^{+ 2.94 }_{- 2.92 }$   $ \left(^{+ 6.53 }_{- 5.30 }\right)$  & \textit{This Study} & 2.15 $^{+ 1.92 }_{- 1.04 }$ & 1048 $^{+    37 }_{-  34 }$ & 0.052 $^{+ 0.039 }_{- 0.020 }$ & 85 $^{+ 54 }_{- 27 }$ \\ 
 730.03 (Kepler-223 e)   & 19.726 & 4.60 $^{+ 0.27 }_{- 0.41 }$  & 4.8 $^{+ 1.4 }_{- 1.2 }$  &\textit{ \citet{Mills2016}}  &  0.28 $^{+ 0.12 }_{- 0.08 }$ &  839 $^{+   35 }_{-  57 }$  & 0.166 $^{+ 0.064 }_{- 0.048 }$ & 219 $^{+ 78 }_{- 56 }$  \\
 730.03 (Kepler-223 e)   & 19.728 & 3.57 $^{+ 0.60 }_{- 0.59 }$  & < 6.14  & \textit{This Study} & < 1.39  &  832 $^{+   29 }_{-  27 }$ & > 0.065  & > 132 \\ 
 730.04 (Kepler-223 b)   & 7.384 & 2.99 $^{+ 0.18 }_{- 0.27 }$  & 7.4 $^{+ 1.3 }_{- 1.1 }$    &\textit{ \citet{Mills2016}} & 1.54 $^{+ 0.63 }_{- 0.35 }$ & 1164 $^{+    50 }_{-  79 }$ & 0.065 $^{+ 0.016 }_{- 0.015 }$ & 79 $^{+ 17 }_{- 15 }$ \\ 
 730.04 (Kepler-223 b)   & 7.384 & 3.18 $^{+ 0.53 }_{- 0.27 }$  & 5.80 $^{+ 1.62 }_{- 1.53 }$   $ \left(^{+ 3.66 }_{- 3.14 }\right)$  & \textit{This Study} & 0.79 $^{+ 0.42 }_{- 0.31 }$ & 1153 $^{+    40 }_{-  39 }$ & 0.125 $^{+ 0.060 }_{- 0.036 }$ & 125 $^{+ 49 }_{- 30 }$ \\ 
   1831.01 (\object{Kepler-324} c) & 51.832 & 2.60 $^{+ 0.19 }_{- 0.16 }$  & < 2.73  & \textit{This Study} & < 0.85  &  413 $^{+    8 }_{-   8 }$ & > 0.054  & > 226 \\ 
1831.01 (Kepler-324 c)  & 51.810 & 2.6$^{+ 0.6 }_{- 0.2 }$  & <0.7 &  \citet{Hadden2017} & < 0.2  &  413 $^{+    17 }_{-   16 }$ &  -- & --- \\ 
 1831.02 (Kepler-324 b)  & 4.386 &  1.18 $^{+ 0.06 }_{- 0.07 }$  & < 3.95  & \textit{This Study} & < 10.00  &  941 $^{+   18 }_{-  17 }$ & > 0.021  & > 32 \\ 
 1831.03* & 34.169 & 1.13 $^{+ 0.05 }_{- 0.05 }$  & 2.23 $^{+ 0.50 }_{- 0.84 }$   $ \left(^{+ 0.94 }_{- 1.40 }\right)$  & \textit{This Study} & 9.95 $^{+ 0.05 }_{- 4.62 }$ & 475 $^{+     9 }_{-   9 }$ & 0.013 $^{+ 0.010 }_{- 0.001 }$ & 22 $^{+ 15 }_{-  3 }$ \\ 
   1831.03* & 34.207 & 1.2 $^{+ 0.1 }_{- 0.1 }$  & 3.5 $^{+ 2.8 }_{- 2.0 }$    &  \citet{Hadden2017}  & 10.3 $^{+ 11.0 }_{- 6.0 }$ & 474 $^{+     20 }_{-   18 }$ & 0.009 $^{+ 0.008 }_{- 0.001 }$ & 15 $^{+ 11 }_{- 3 }$ \\  
 1831.04 & 13.981 & 1.97 $^{+ 0.15 }_{- 0.12 }$  & < 22.04  & \textit{This Study} & < 10.00  &  640 $^{+   12 }_{-  12 }$ & > 0.008  & > 12 \\ 
 1955.01 (\object{Kepler-342} b)  & 15.170 & 2.47 $^{+ 0.41 }_{- 0.34 }$  & < 40.03  & \textit{This Study} & < 10.00  &  901 $^{+   14 }_{-  14 }$ & > 0.010  & >  8 \\ 
 1955.02 (Kepler-342 d)  & 39.462 & 2.83 $^{+ 0.13 }_{- 0.19 }$  & < 20.29  & \textit{This Study} & < 6.26  &  655 $^{+   11 }_{-  10 }$ & > 0.012  & > 11 \\ 
 1955.03 (Kepler-342 e)  & 1.644 &  1.18 $^{+ 0.19 }_{- 0.16 }$  & < 7.44  & \textit{This Study} & < 10.00  &  1891 $^{+   30 }_{-  29 }$ & > 0.034  & > 10 \\ 
 1955.04 (Kepler-342 c)  & 26.237 & 2.76 $^{+ 0.45 }_{- 0.38 }$  & < 43.06  & \textit{This Study} & < 10.00  &  751 $^{+   12 }_{-  12 }$ & > 0.008  & >  5 \\
 2174.01 (\object{KIC 8261920}) & 6.695 & 2.44 $^{+ 0.10 }_{- 0.12 }$  & < 3.91  & \textit{This Study} & < 1.61  &  607 $^{+   21 }_{-  23 }$ & > 0.046  & > 151 \\ 
 2174.02 & 33.136 & 2.05 $^{+ 0.08 }_{- 0.10 }$  & < 18.39  & \textit{This Study} & < 10.00  &  356 $^{+   13 }_{-  13 }$ & > 0.005  & > 18 \\ 
 2174.03 & 7.725 & 1.71 $^{+ 0.07 }_{- 0.09 }$  & < 2.24  & \textit{This Study} & < 2.78  &  577 $^{+   21 }_{-  21 }$ & > 0.037  & > 85 \\ 
 2174.04 & 3.016 &  0.96 $^{+ 0.04 }_{- 0.05 }$  & < 1.91  & \textit{This Study} & < 10.00  &  791 $^{+   28 }_{-  29 }$ & > 0.022  & > 43 \\ 
 \hline
     \end{tabular}
    \caption{Results for four-planet systems: Planetary sizes,  masses, bulk densities, equilibrium blackbody temperatures, atmospheric scale heights, and estimated atmospheric transmission annuli of four-planet systems in our sample. Uncertainties enclose the 68.3\% confidence intervals (with 95.5\% confidence intervals in parentheses to inform skewness). We include the Kepler numbers of candidates that are previously confirmed or validated in the first column. KOI names with an asterisk indicate newly confirmed planets.}\label{tbl-massresults4}
  \end{center}
\end{table}

\begin{figure}[ht!]
\begin{center}
\includegraphics [height = 2.2 in]{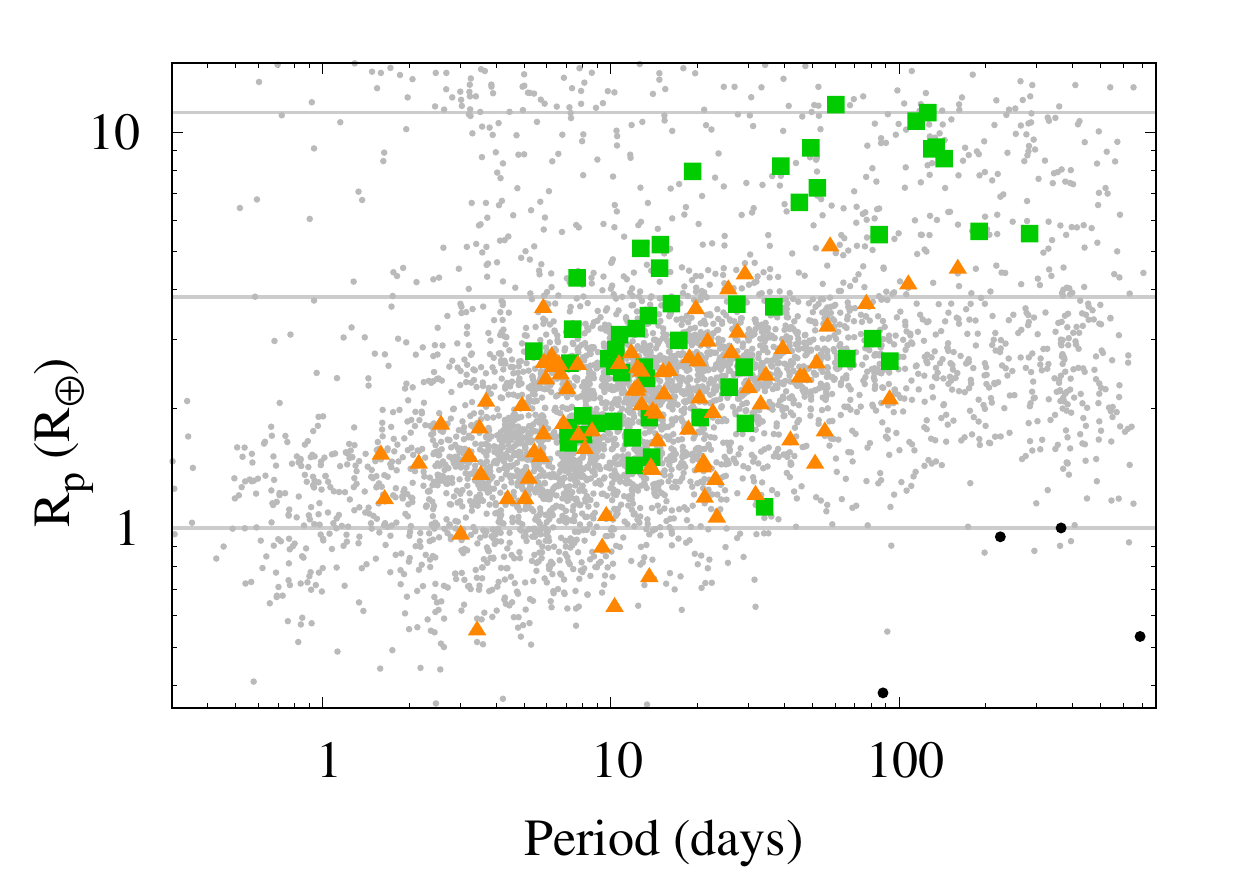}
\includegraphics [height = 2.2 in]{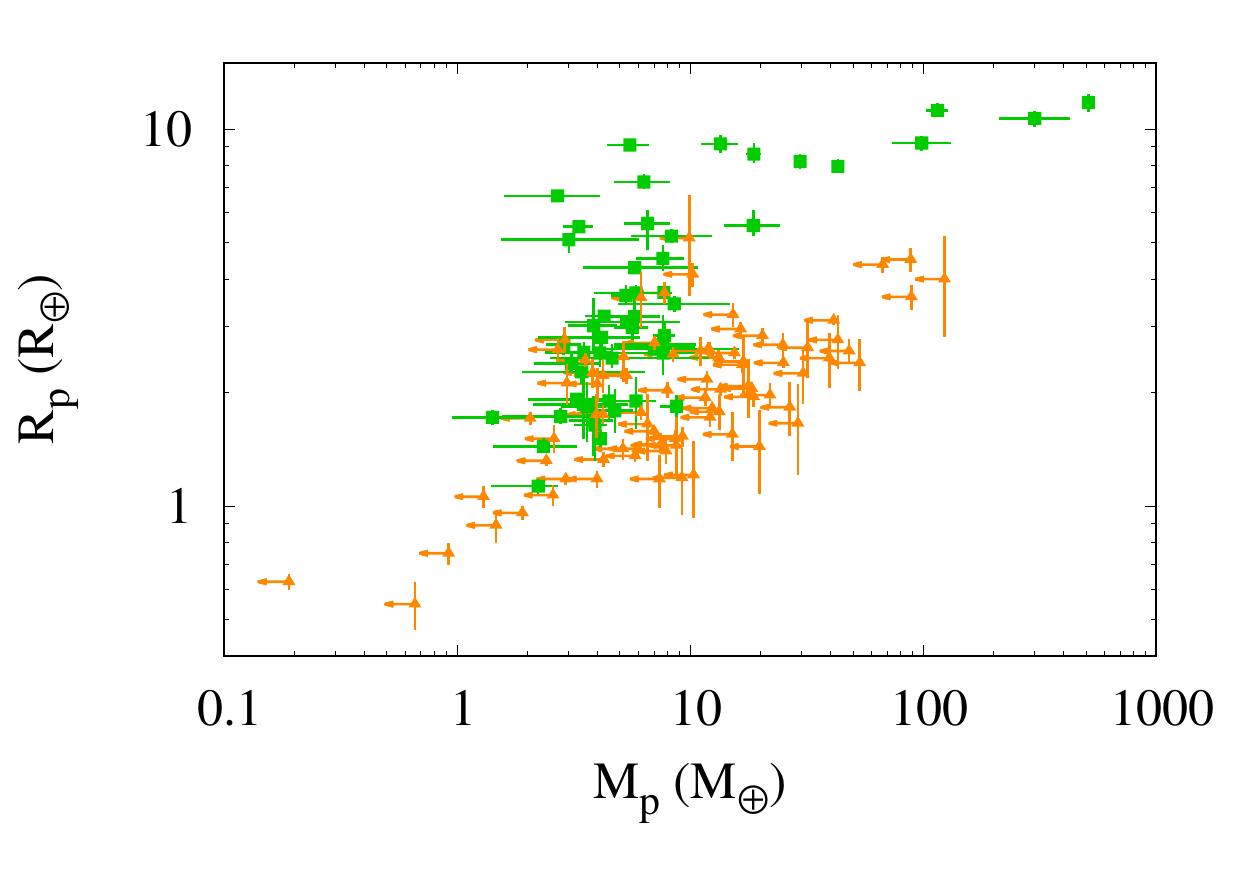}
\caption{Left panel: Period and nominal radii of exoplanets in the Kepler field characterized with TTVs, with strong detections (green squares) and planets with mass upper limits only (orange triangles) overlaid on the nominal period-radius sample from the Kepler mission (grey points). Black points mark Solar System planets, and gray lines mark the sizes of Earth, Neptune and Jupiter. Right panel: Mass-radius diagram of our sample, with the same color scheme as on the left. Error bars mark 68.3\% confidence intervals centered on the median. Upper limits marked with arrows signify 97.7th percentile boundaries. 
}
\label{fig:P-R} 
\end{center}
\end{figure}

Figure~\ref{fig:P-R} shows the period-radius distribution and mass-radius distribution of our sample. Although the detections and nondetections are well mixed in the relevant ranges of period, radius, and mass, we see that the upper limits are, on average, at shorter orbital periods and among smaller planets, consistent with the detection biases of TTVs (\citealt{Steffen2016, Mills2017b}). The majority of our planet sample have sub-Neptune masses.

\begin{figure}[ht!]
\begin{center}
\includegraphics [height = 2.1 in]{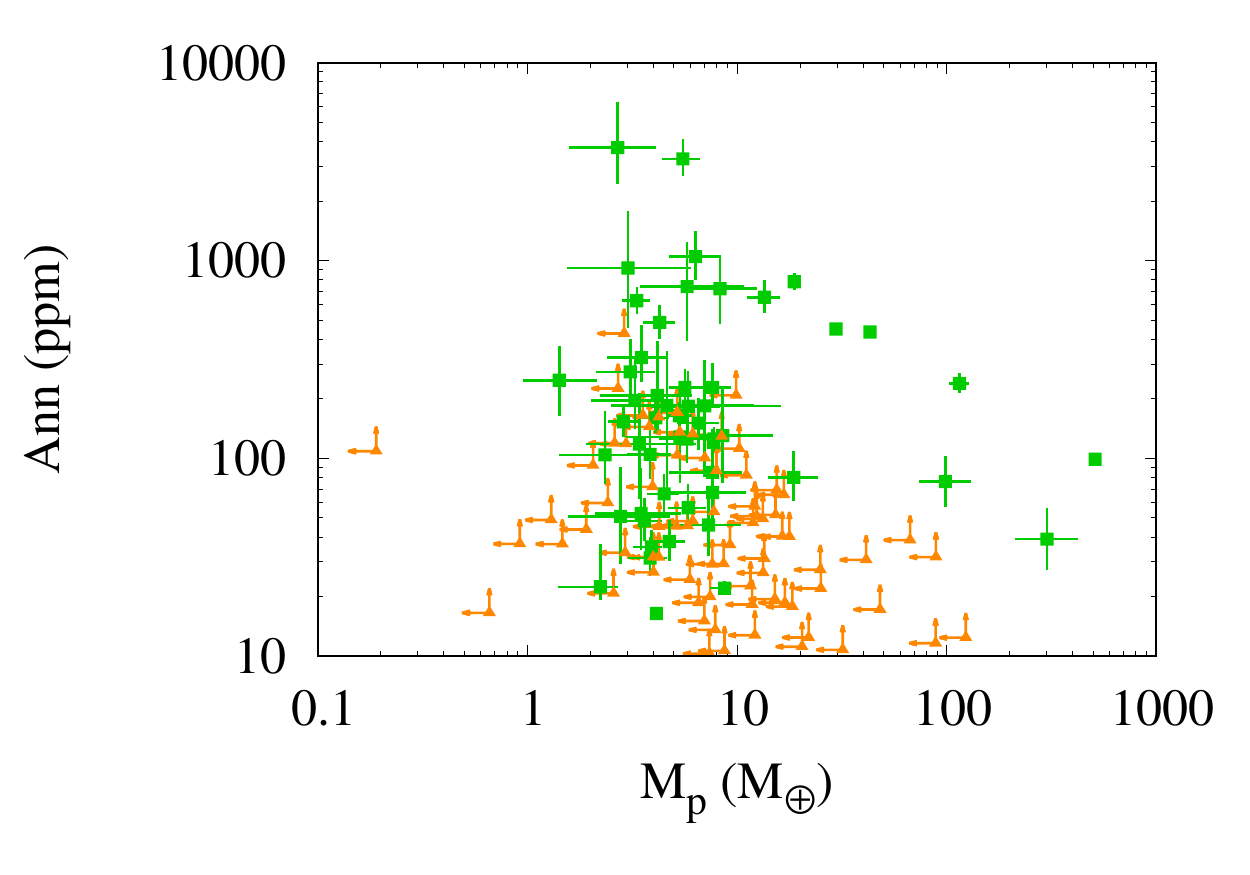}
\includegraphics [height = 2.1 in]{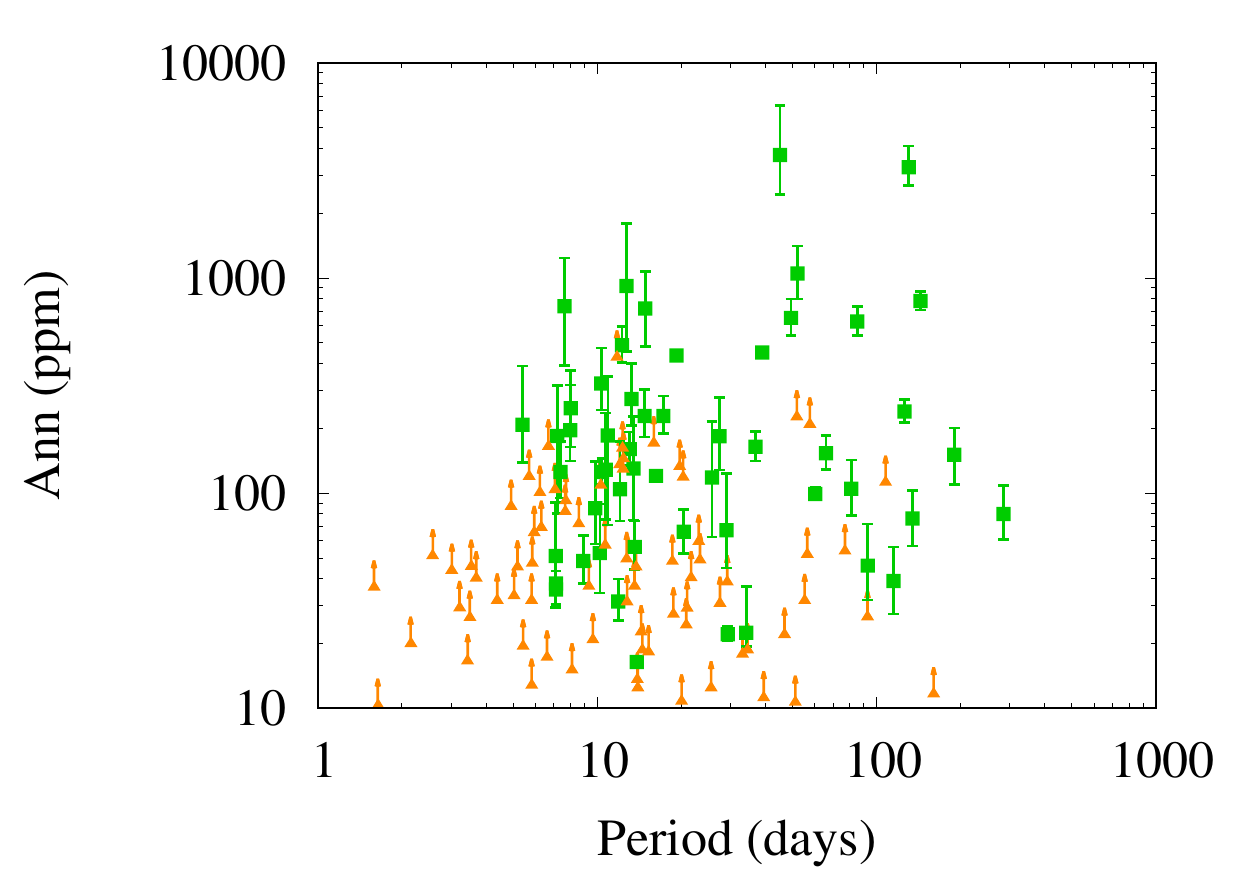}
\includegraphics [height = 2.1 in]{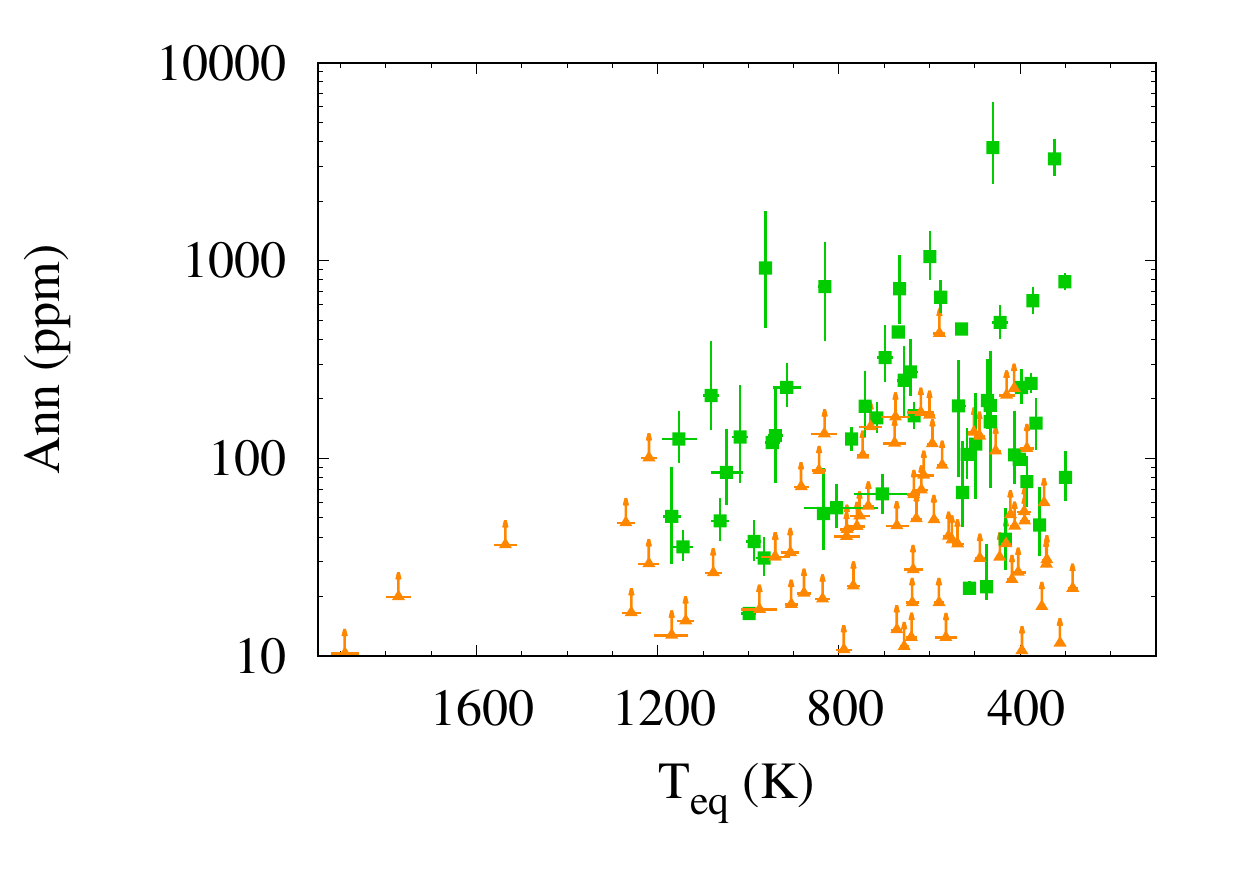}
\includegraphics [height = 2.1 in]{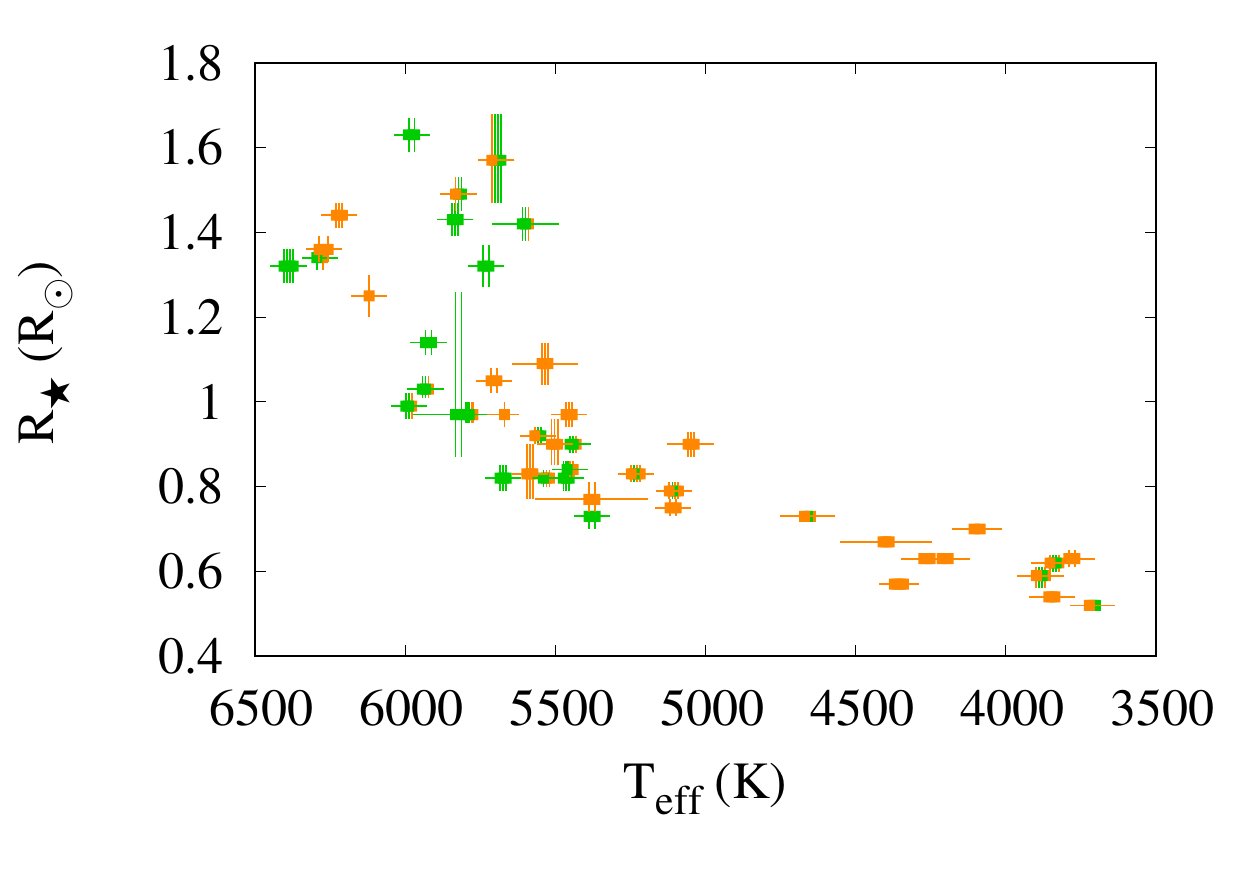}
\caption{Top Left: Planetary masses and transmission annuli above 10 ppm from of our sample, with strong TTV detections in green and planets with mass upper limits only in orange. Error bars mark 68.3\% confidence intervals. Upper limits in mass marked with arrows signify 97.7th percentile boundaries. Lower limits in annulus depth are marked with arrows from the 2.3th percentile boundaries (97.7 percent of samples are above the data point).Top Right: Orbital period and transmission annuli  for our transiting exoplanet sample, with the same color scheme. Bottom Left: Tranmission annulus and equilibrium blackbody temperature. Bottom Right: Host radii and effective temperatures among planets with an estimated transmission annulus above 10 ppm. The multiplicity of the included planets is indicated by the number of errorbars, with colors indicating either strong TTV detections or mass upper limits only.
}
\label{fig:Ann} 
\end{center}
\end{figure}

Figure~\ref{fig:Ann} highlights the planets among our sample with estimated transmission annuli above 10 ppm, alongside their orbital periods, masses, equilibrium temperatures, and some of their host properties. The highest transmission annuli occur where TTV mass determinations are at their most sensitive, from $\sim$10 to 200 days. This group is dominated by sub-Neptune-mass planets over a wide range of equilibrium blackbody temperatures that reach as low as $\sim$300 K. The hosts have a wide range of sizes and effective temperatures, although the strong planet mass detections are primarily at hotter stars. 

\subsection{Radial Velocity Candidates}
 
We estimated the RV signal expected from our mass determinations or upper limits, and we list the strongest candidates in Table~\ref{tbl-rv} in Appendix D. Of the planets with strongly detected masses, we identify planets with a nominal RV signal above 2 m s$^{-1}$. For candidates with mass upper limits only, we estimate the expected upper limit in RV amplitude. Where these upper limits exceed 2 m/s, RV may impose tighter mass upper limits than the existing transit timing data, assuming a precision of 2 m s$^{-1}$  for Kepler magnitudes < 13.5 \citep{mar14}, and we mark these with an asterisk. In addition to these, we note that \textit{ExoFOP Kepler} lists the Kepler magnitude of KOI-2414 as 13.6, making it a strong candidate for combined RV and TTV follow-up. RV may also be useful where dynamical mass posteriors of interacting planets are strongly correlated, even if only one of the planet pair is detectable in RV. Such targets include KOI-222, KOI-877, KOI-2195, KOI-2414, and KOI-3503.

\section{Conclusion}
We have identified a sample of planets with an expectation of TTV mass detections, and we include the nondetections in our analysis to reduce the detection bias in population studies with TTV masses. Our sample of systems for TTV analysis is based on planet sizes and orbital periods and not any prior identification of periodicities in the transit times. With forward-modeling of our TTV posteriors, we project future transit times with uncertainties for our planetary sample up to 2026 and have further identified a target list for follow-up transit photometry that can be efficiently planned. This is a compromise between systems with TTVs that diverge slowly enough for additional data to be of value, but not so much that a transit observation cannot be planned for a specific future date. We have found that the uncertainty on projected transit times is most sensitive to the coherence time of near-resonant signals, with TTV periodicities longer than the Kepler baseline causing the most rapid divergence in predicted times. However, we also found that the expectation of TTV signals of both resonant and nonresonant periodicities is moderately correlated with smaller projected uncertainties in transit times. 

We verify the planetary nature of four candidates: KOI-1831.03, KOI-1833.02, KOI-3503.01, and KOI- 3503.02, where TTV models improve at the $3-\sigma$ level with free masses compared to massless models for the previously unverified candidates.

We have identified where eccentricities are well constrained and where joint posteriors imply non-zero relative eccentricity among planet pairs. 

To characterize our planet sample, we have used stellar parameters that in many cases improve upon data that were available in prior TTV studies of Kepler's multitransiting systems.

Finally, we have identified candidate systems for atmospheric transmission spectroscopy by estimating the cloud-free transmission annulus for TTV systems to provide an unbiased target list for atmospheric retrieval with JWST, over a wide range of stellar parameters and planetary orbits. While the Transiting Exoplanet Survey Satellite (TESS) will provide typically brighter targets than our sample, the Kepler sample will likely provide an important contribution to atmospheric studies among Neptunes and sub-Neptunes in the range of orbital distances probed by TTV analyses.

We thank our anonymous referee for insightful comments that improved this manuscript.

D.J. and E.B.F. acknowledge support from NASA Exoplanet Research Program award \#NNX15AE21G.  
E.B.F. acknowledges the support of the Ambrose Monell Foundation and the Institute for Advanced Study.  
This work was supported by a grant from the Simons Foundation/SFARI (675601, E.B.F.). D.J. acknowledges the support of the University of the Pacific. D.J., E.F., and A.W. acknowledge support by funding from the Pennsylvania State University's Office of Science Engagement and Center for Exoplanets and Habitable Worlds.  The Center for Exoplanets and Habitable Worlds is supported by the Pennsylvania State University, the Eberly College of Science, and the Pennsylvania Space Grant Consortium. Portions of this research were conducted with Advanced CyberInfrastructure computational resources provided by The Institute for Computational and Data Sciences at The Pennsylvania State University (http://icds.psu.edu). D.J. and J.L. acknowledge the NASA Astrophysics Data Analysis Program award \#NNX17AL57G. This study benefited from collaborations and/or information exchanged within NASA's Nexus for Exoplanet System Science (NExSS) research coordination network sponsored by NASA's Science Mission Directorate. 
\newpage
\section*{Appendix A: Projected TTV plots}

\clearpage

\clearpage
\newpage
\section*{Appendix B: Joint posterior plots}
\begin{figure}[!h]
\begin{center}
\figurenum{16}
\includegraphics [height = 1.1 in]{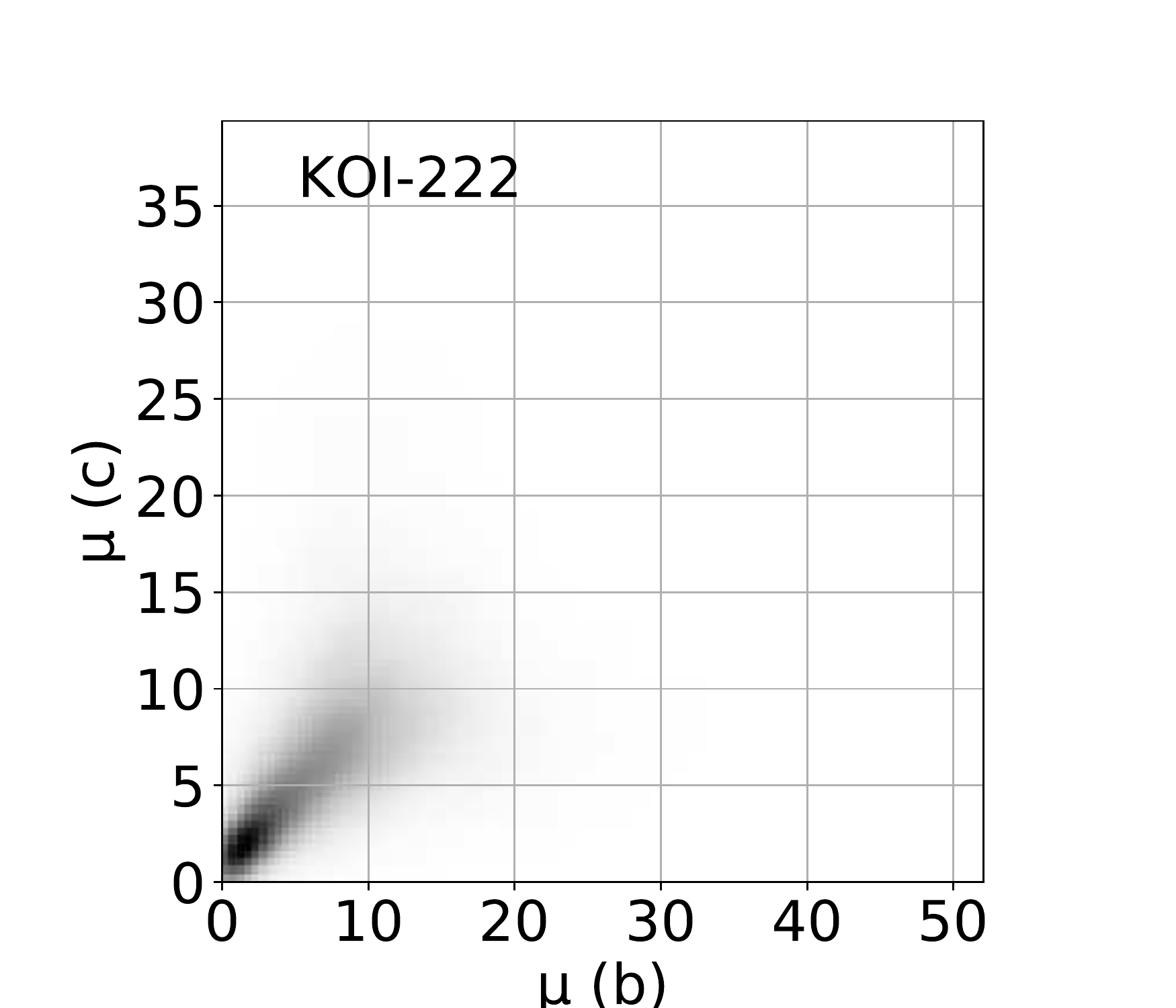}
\includegraphics [height = 1.1 in]{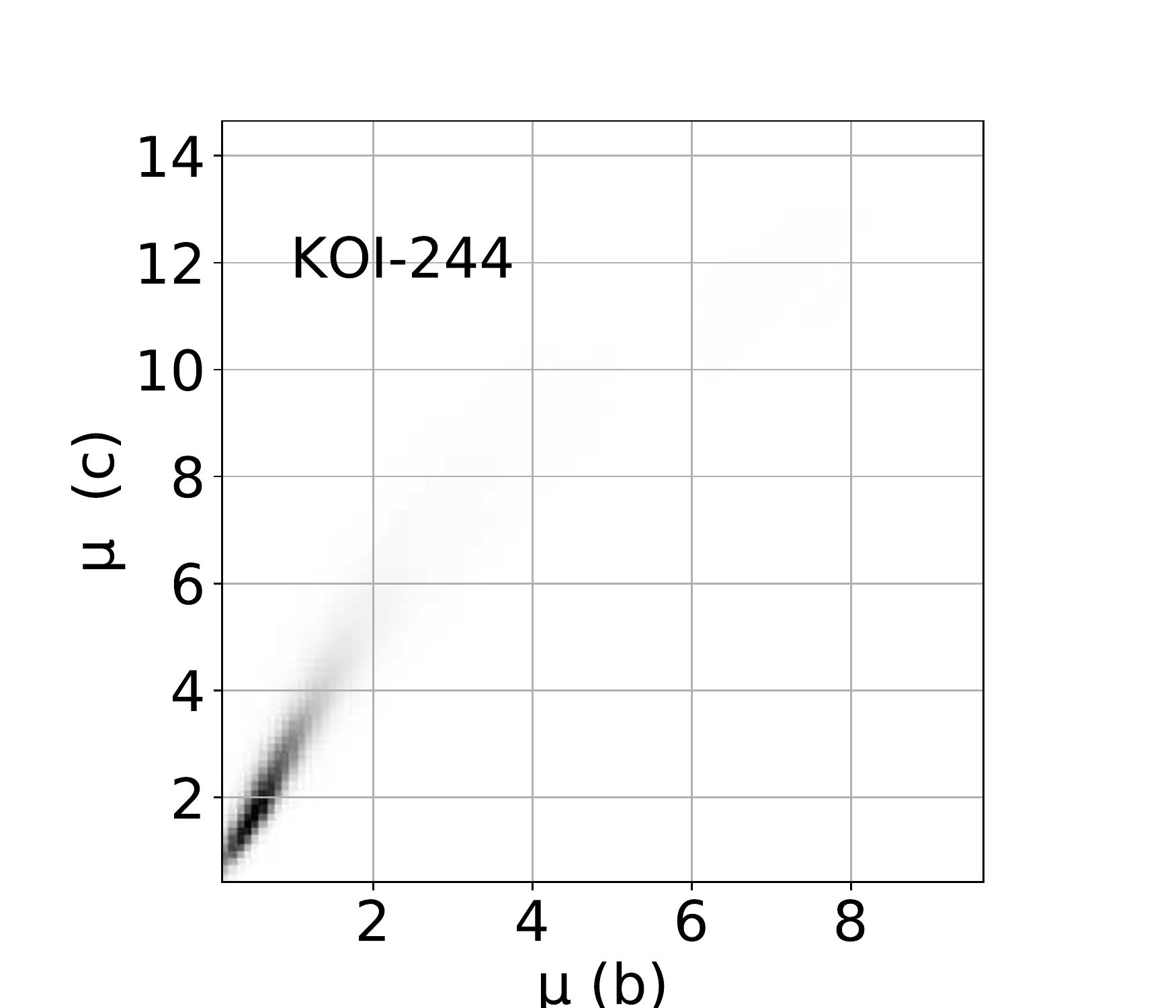}
\includegraphics [height = 1.1 in]{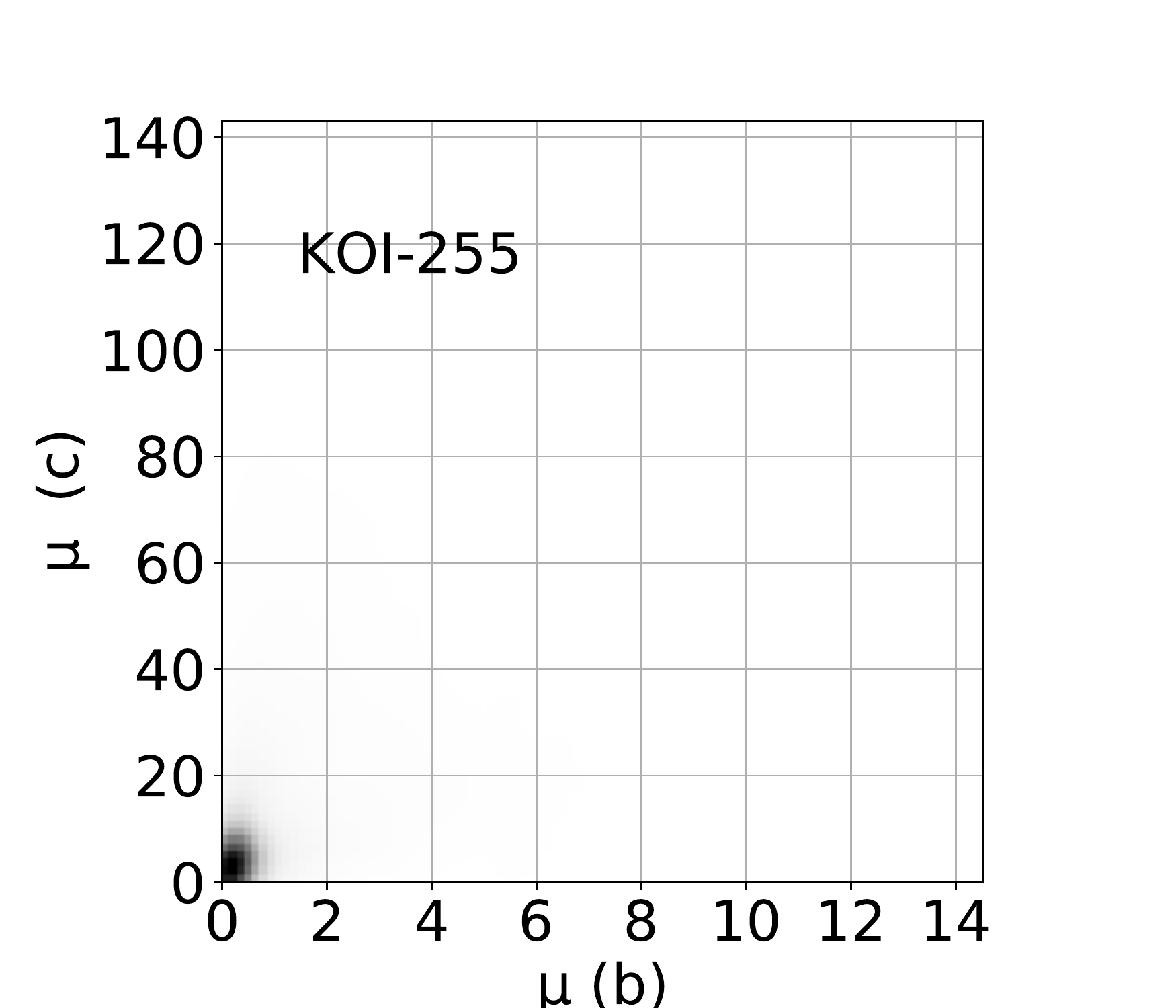}
\includegraphics [height = 1.1 in]{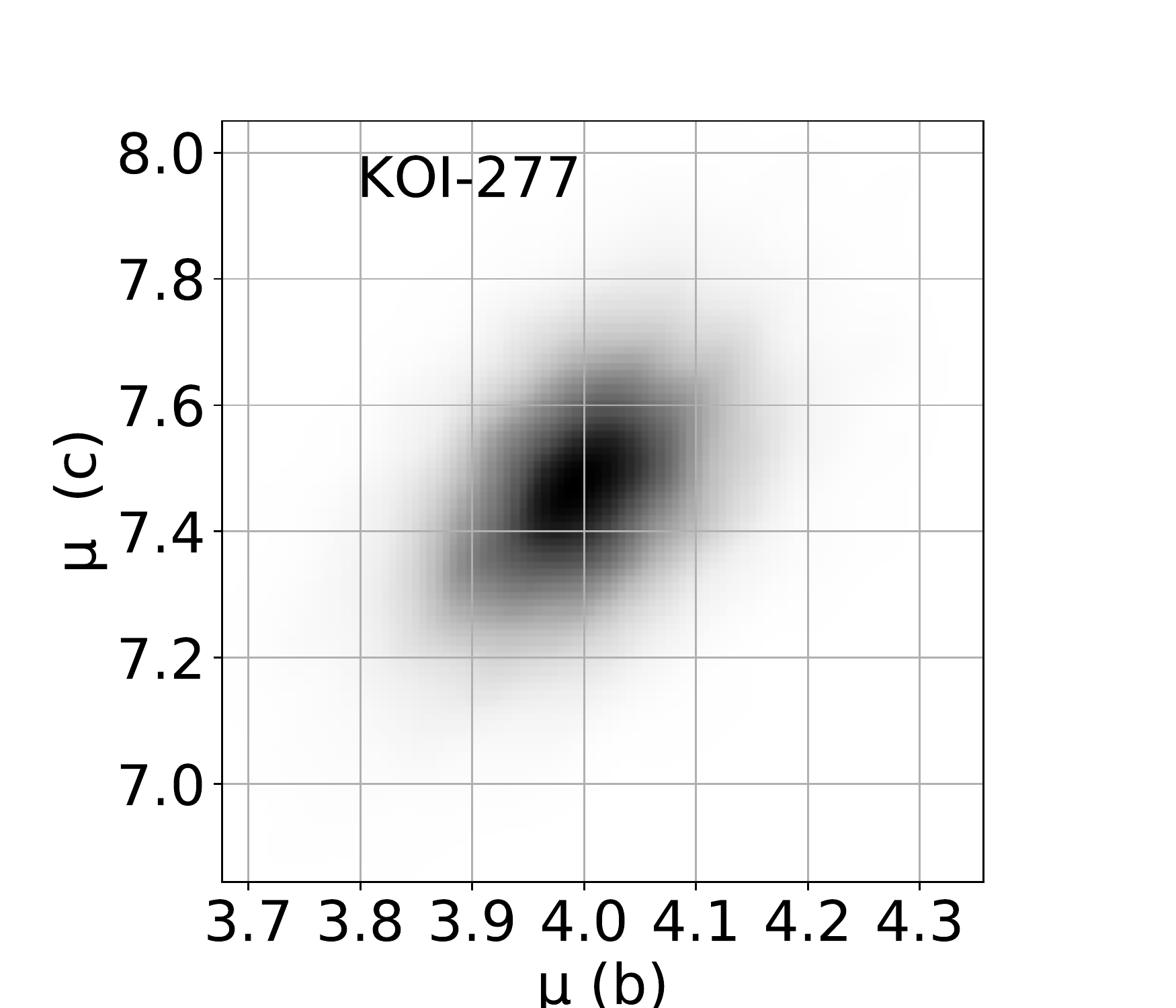} \\
\includegraphics [height = 1.1 in]{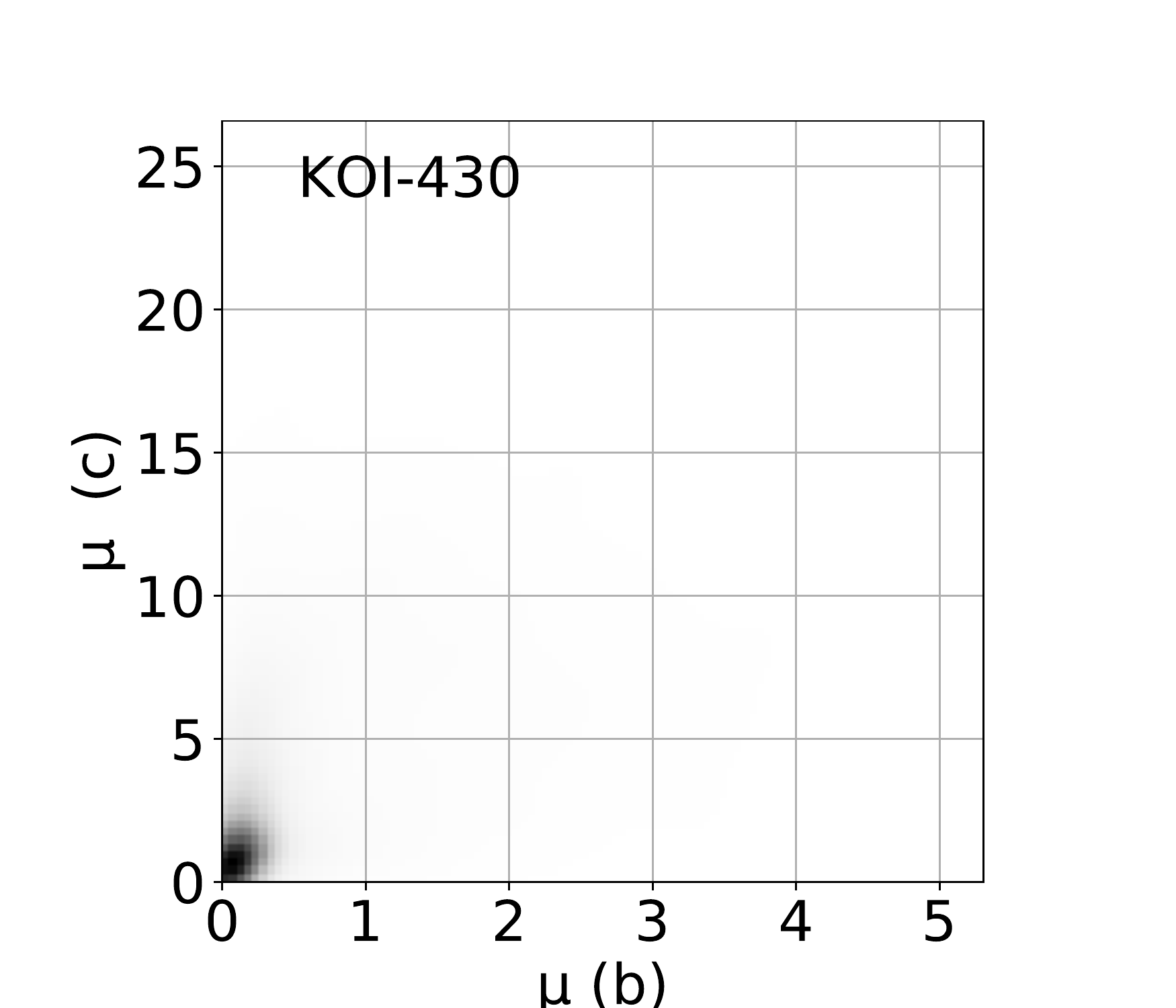}
\includegraphics [height = 1.1 in]{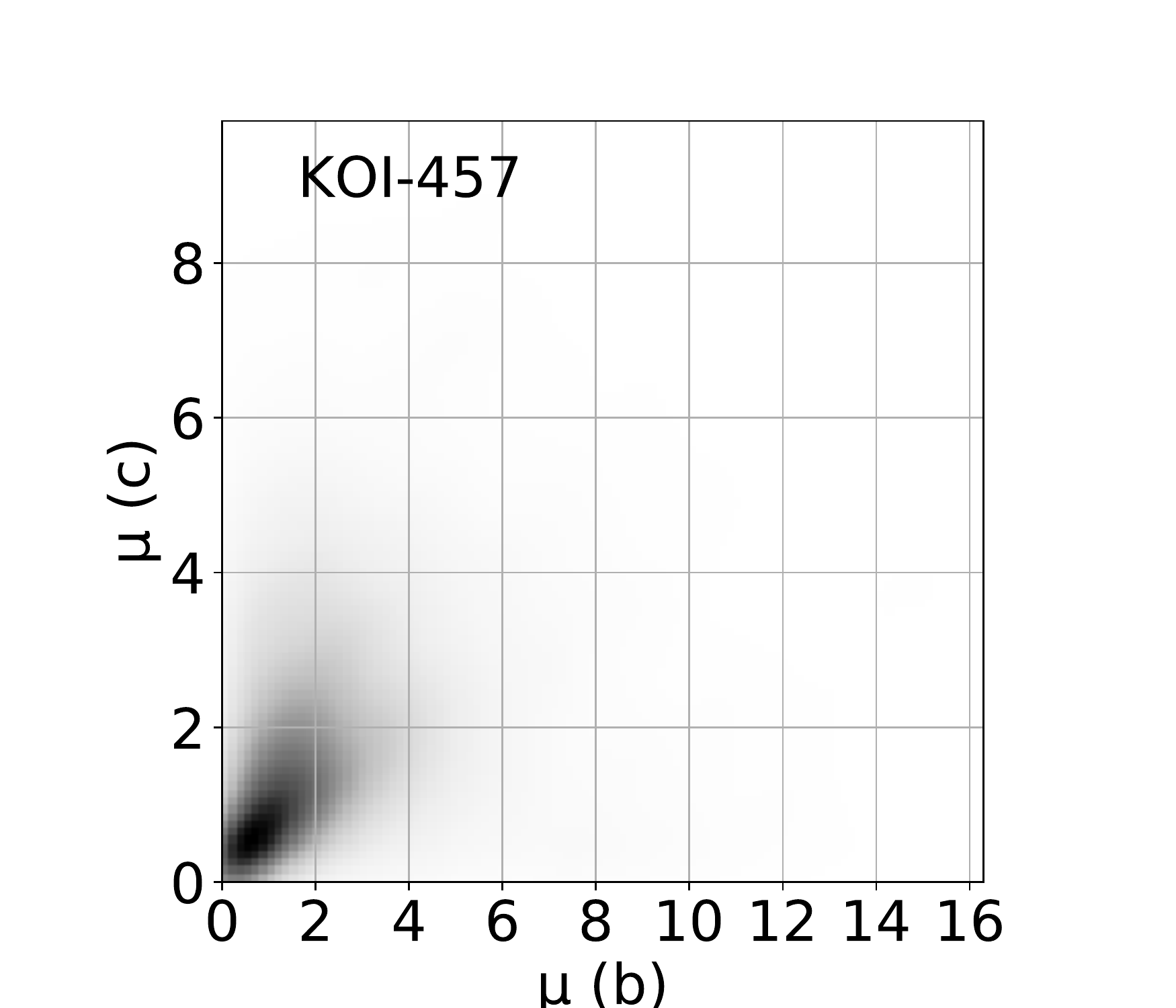}
\includegraphics [height = 1.1 in]{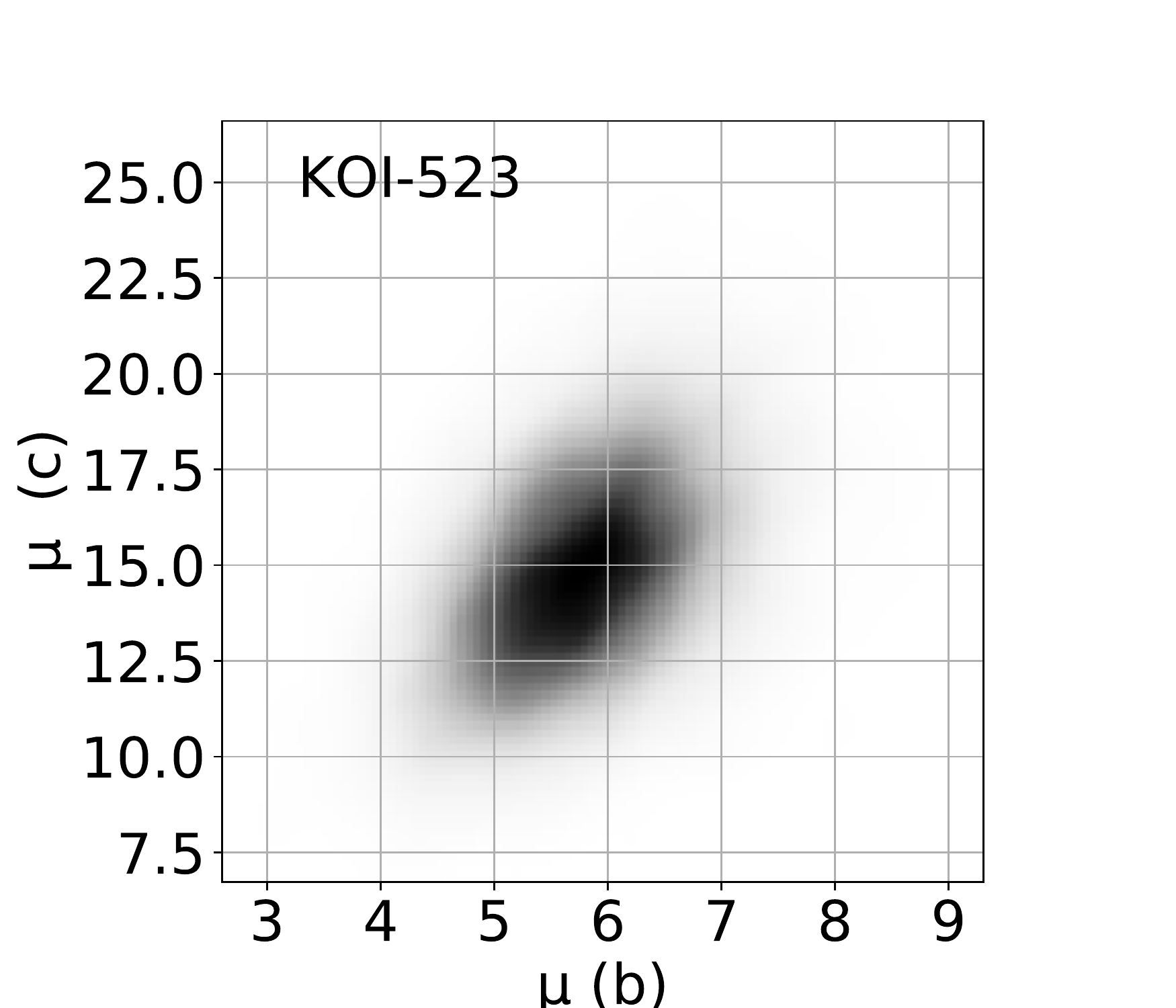}
\includegraphics [height = 1.1 in]{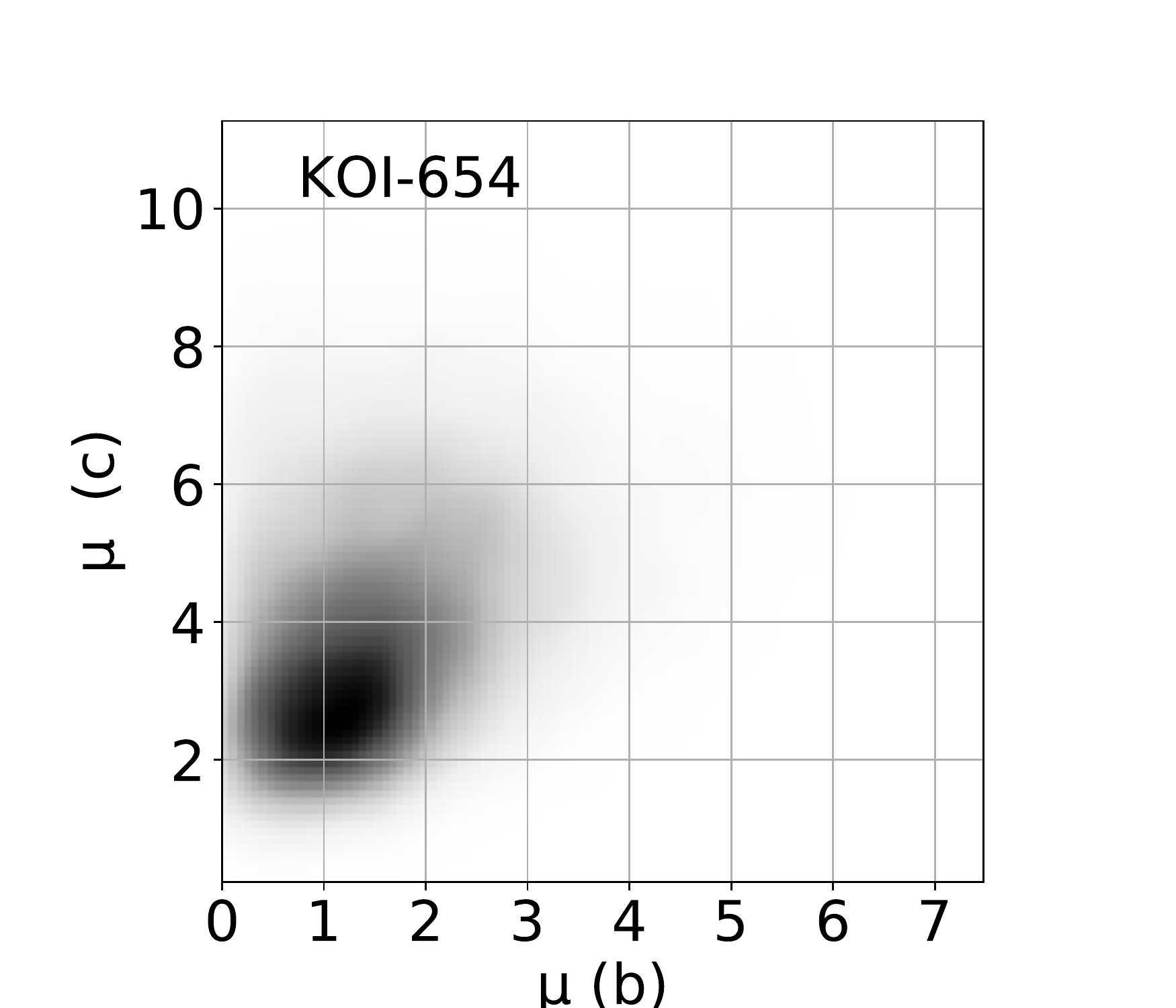} \\
\includegraphics [height = 1.1 in]{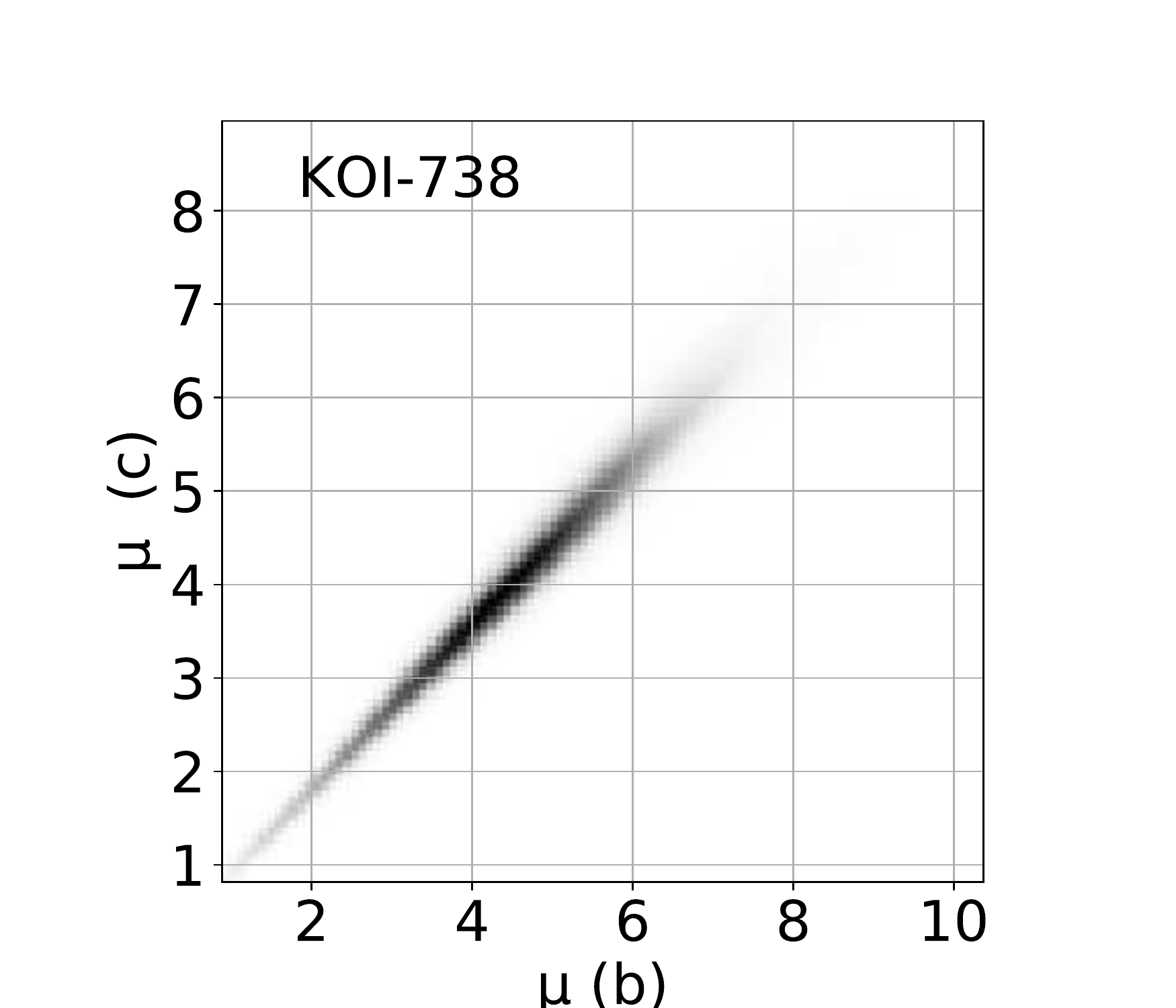}
\includegraphics [height = 1.1 in]{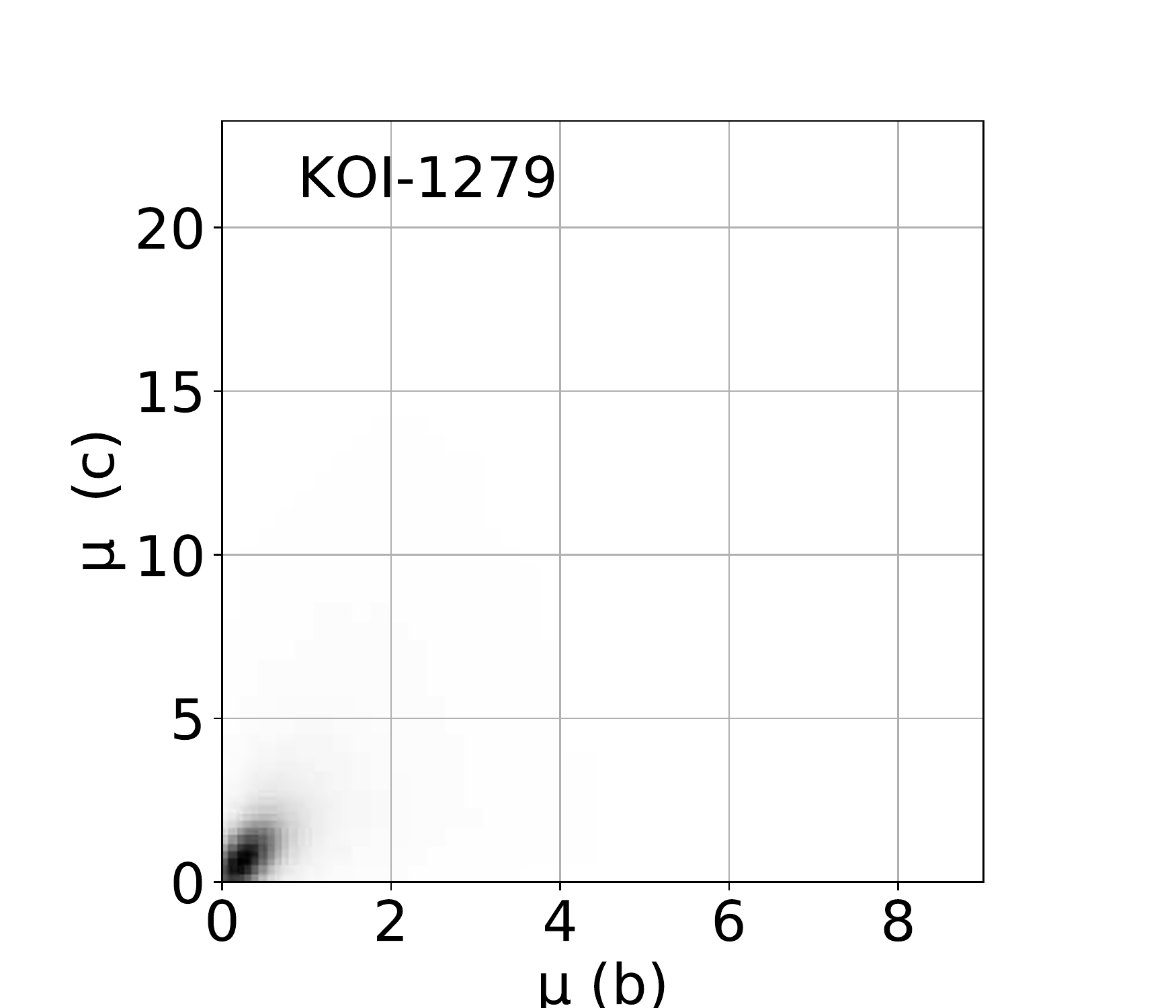}
\includegraphics [height = 1.1 in]{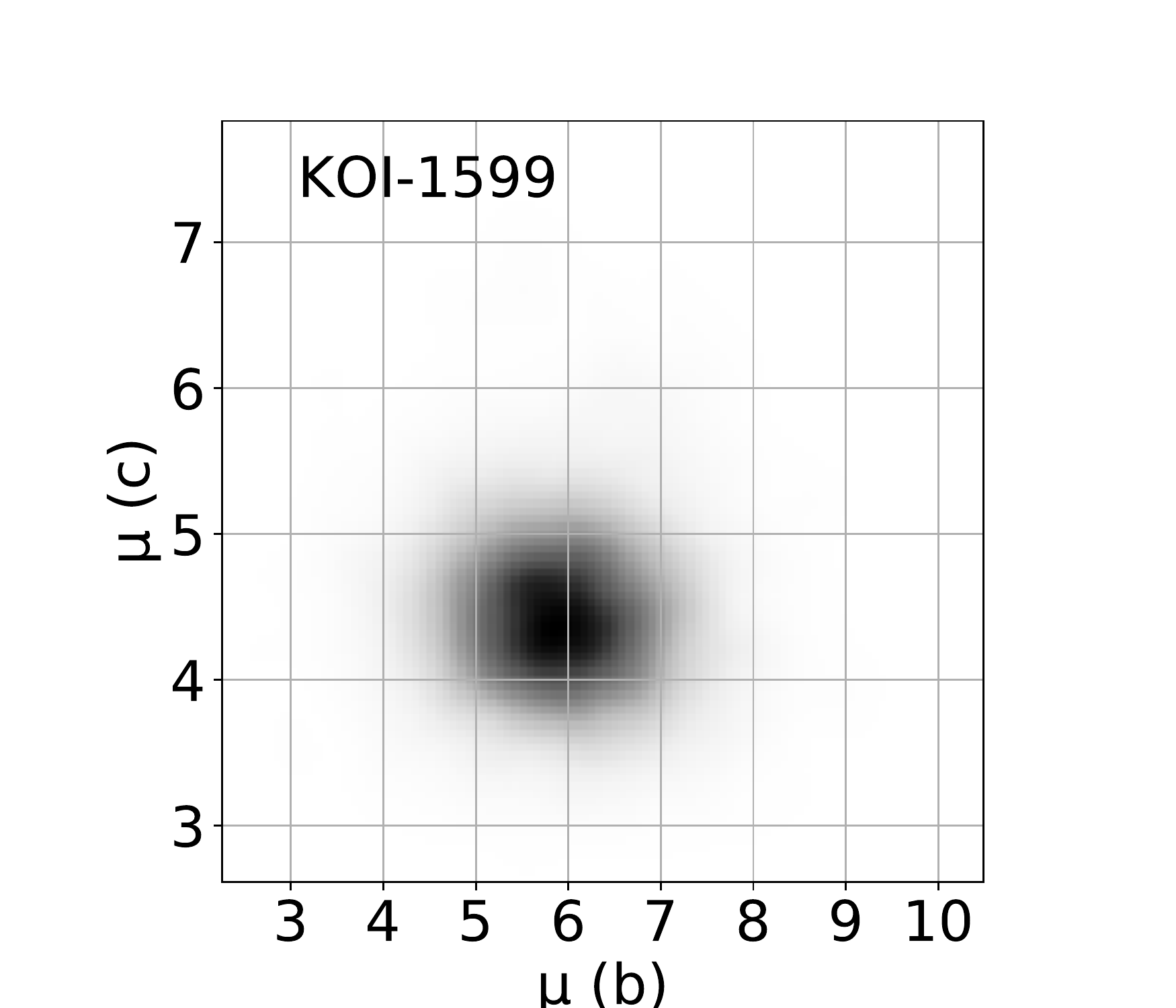}
\includegraphics [height = 1.1 in]{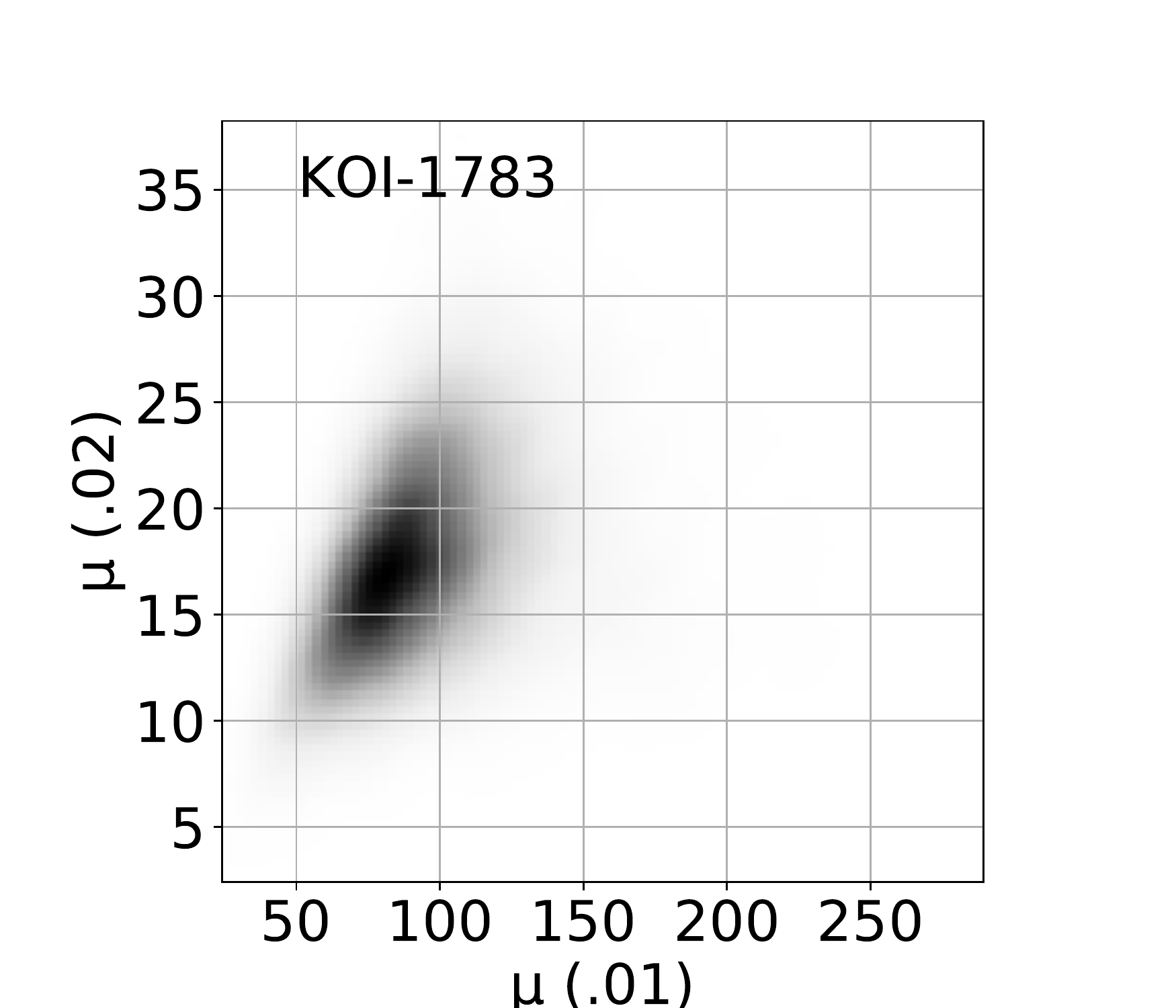} \\
\includegraphics [height = 1.1 in]{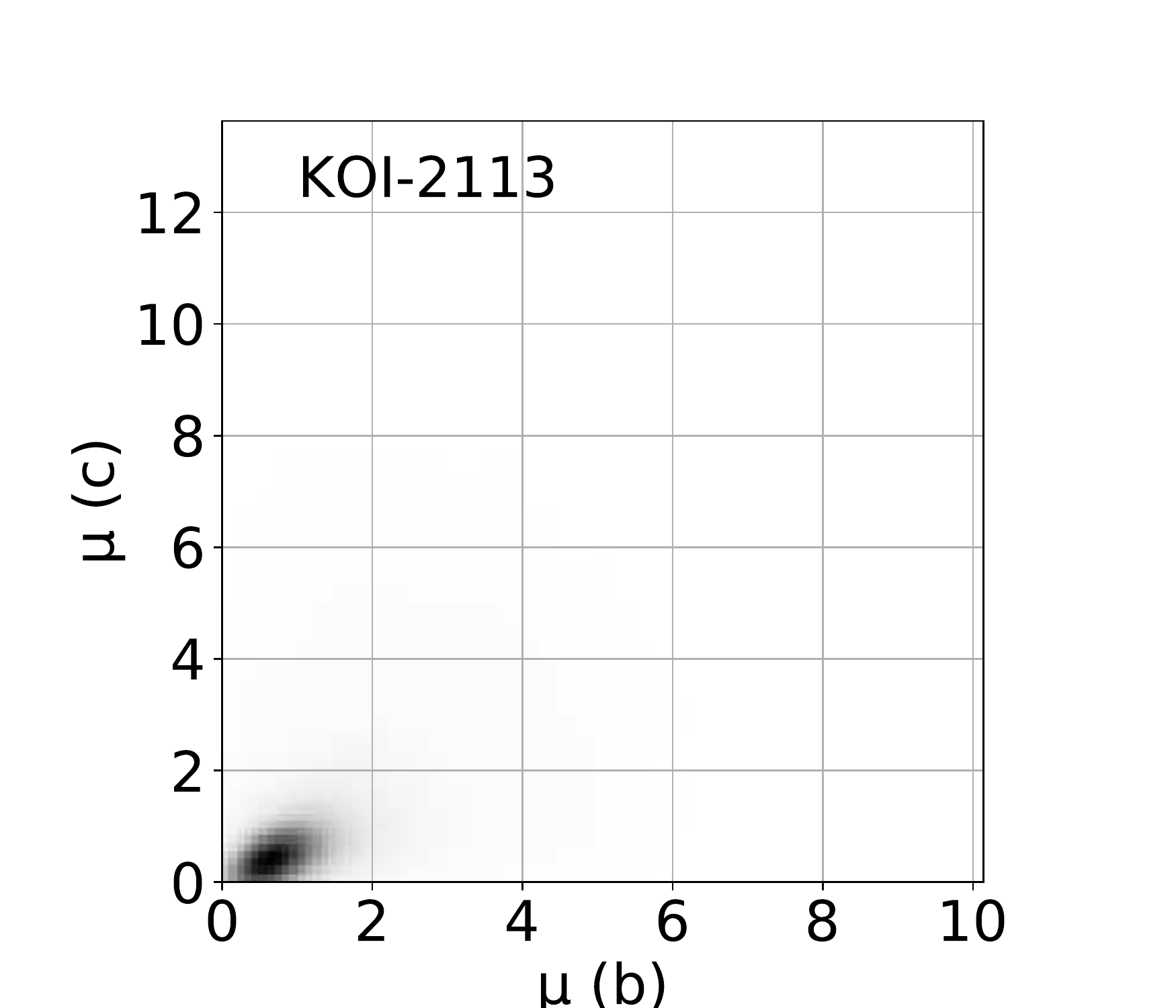}
\includegraphics [height = 1.1 in]{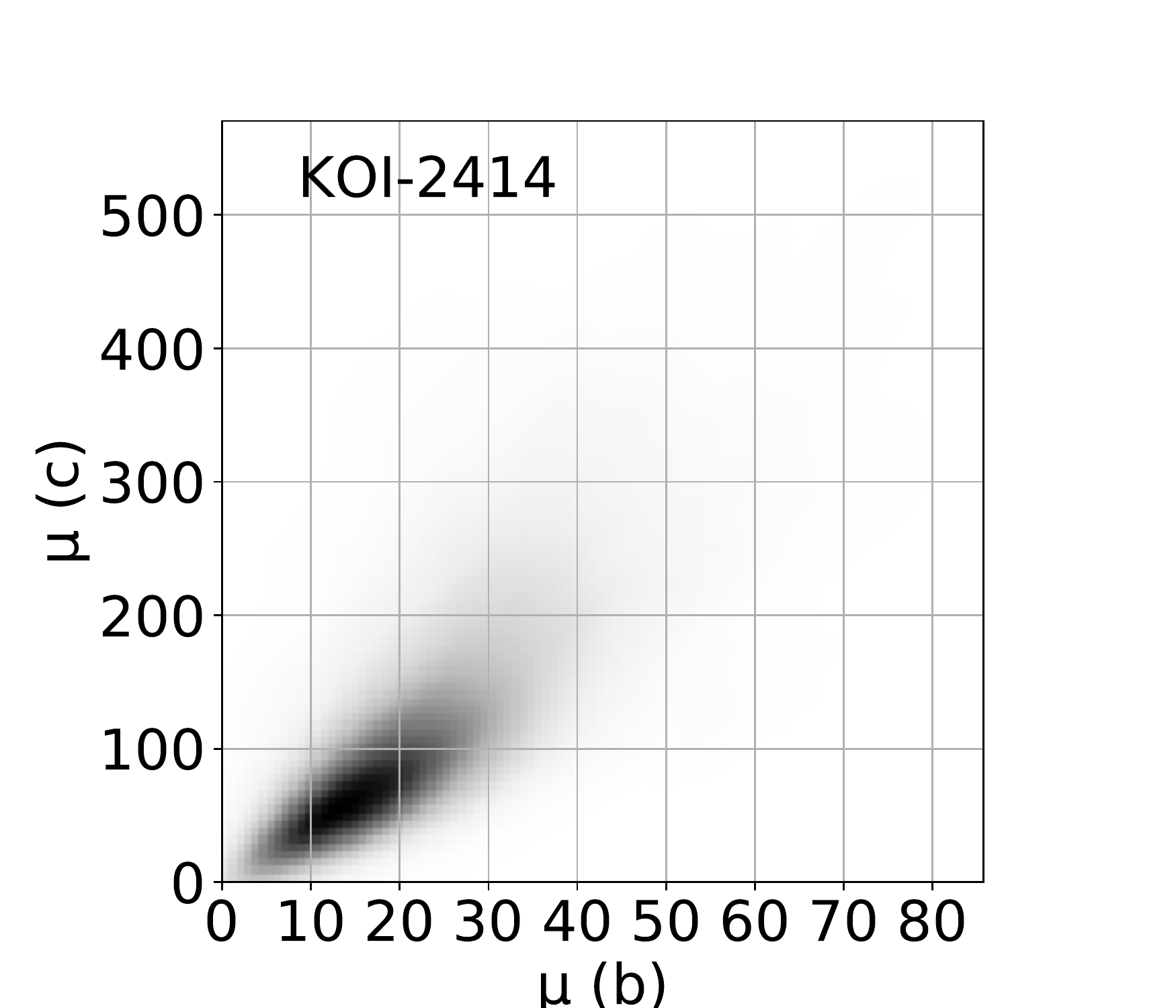}
\caption{Two-dimensional kernel density estimators on joint posteriors of dynamical masses, $\mu$, scaled by factor $\frac{M_{\odot}}{M_{\oplus}}$: two-planet systems. 
\label{fig:mu2a}}
\end{center}
\end{figure}

\begin{figure}
\begin{center}
\figurenum{18}
\includegraphics [height = 1.1 in]{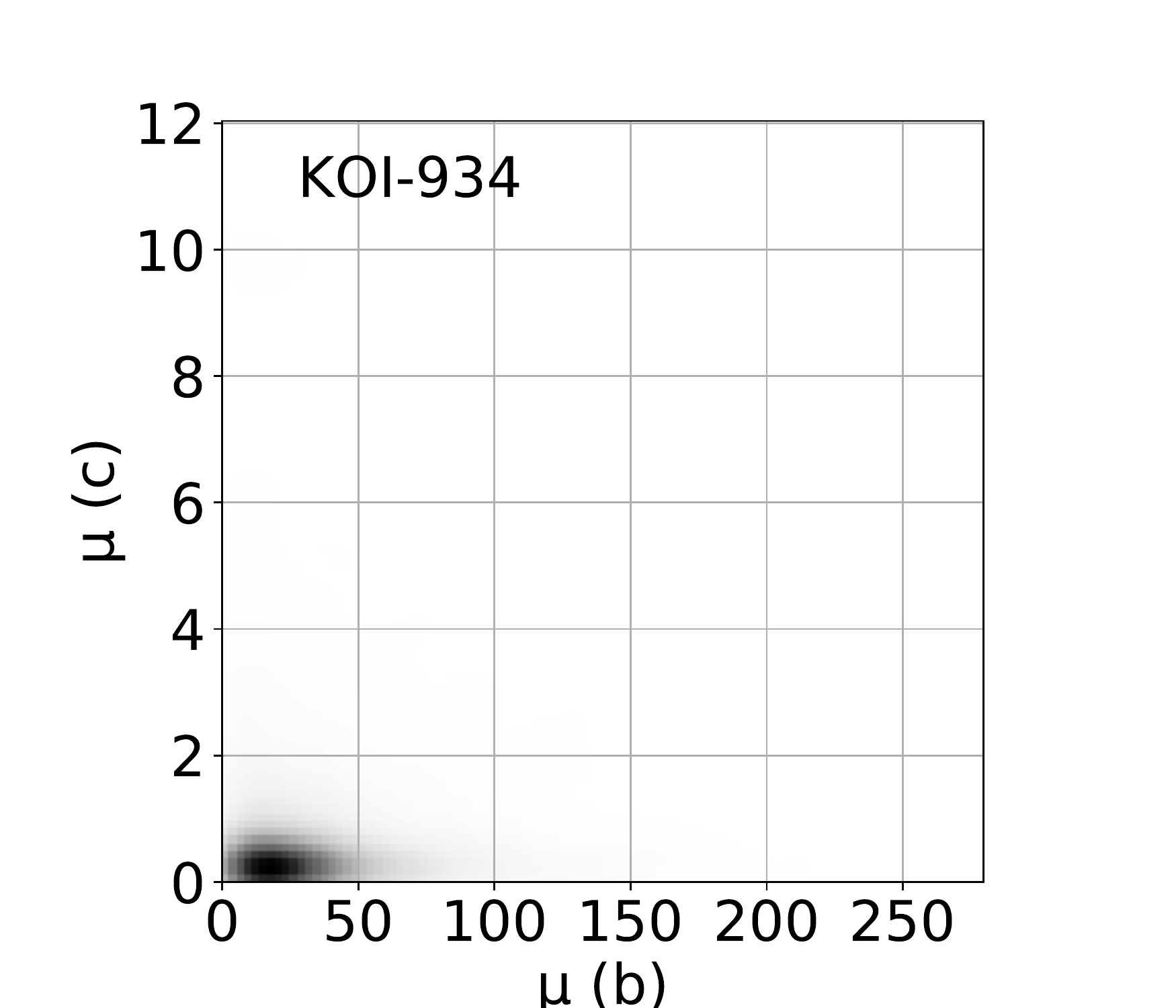}
\includegraphics [height = 1.1 in]{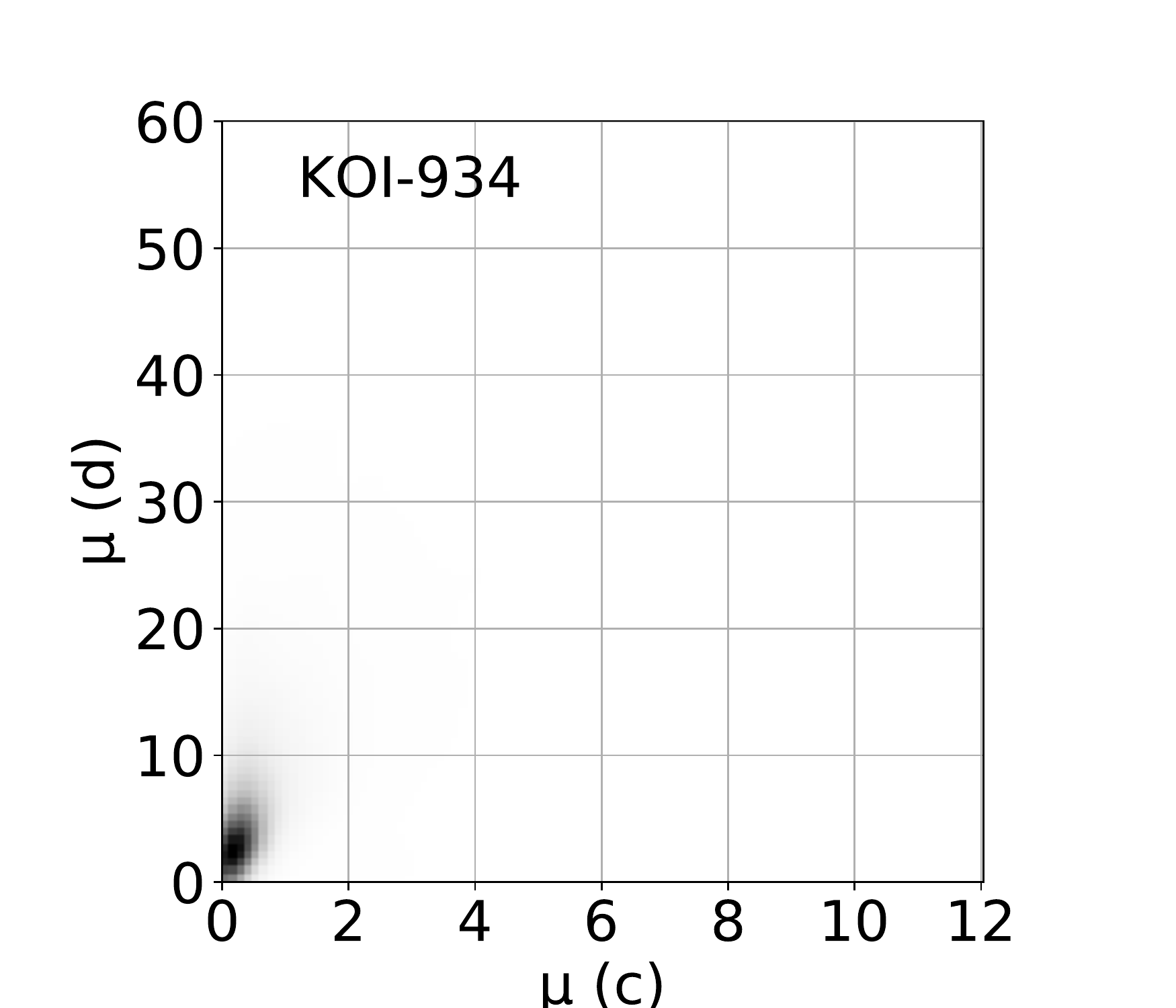}
\includegraphics [height = 1.1 in]{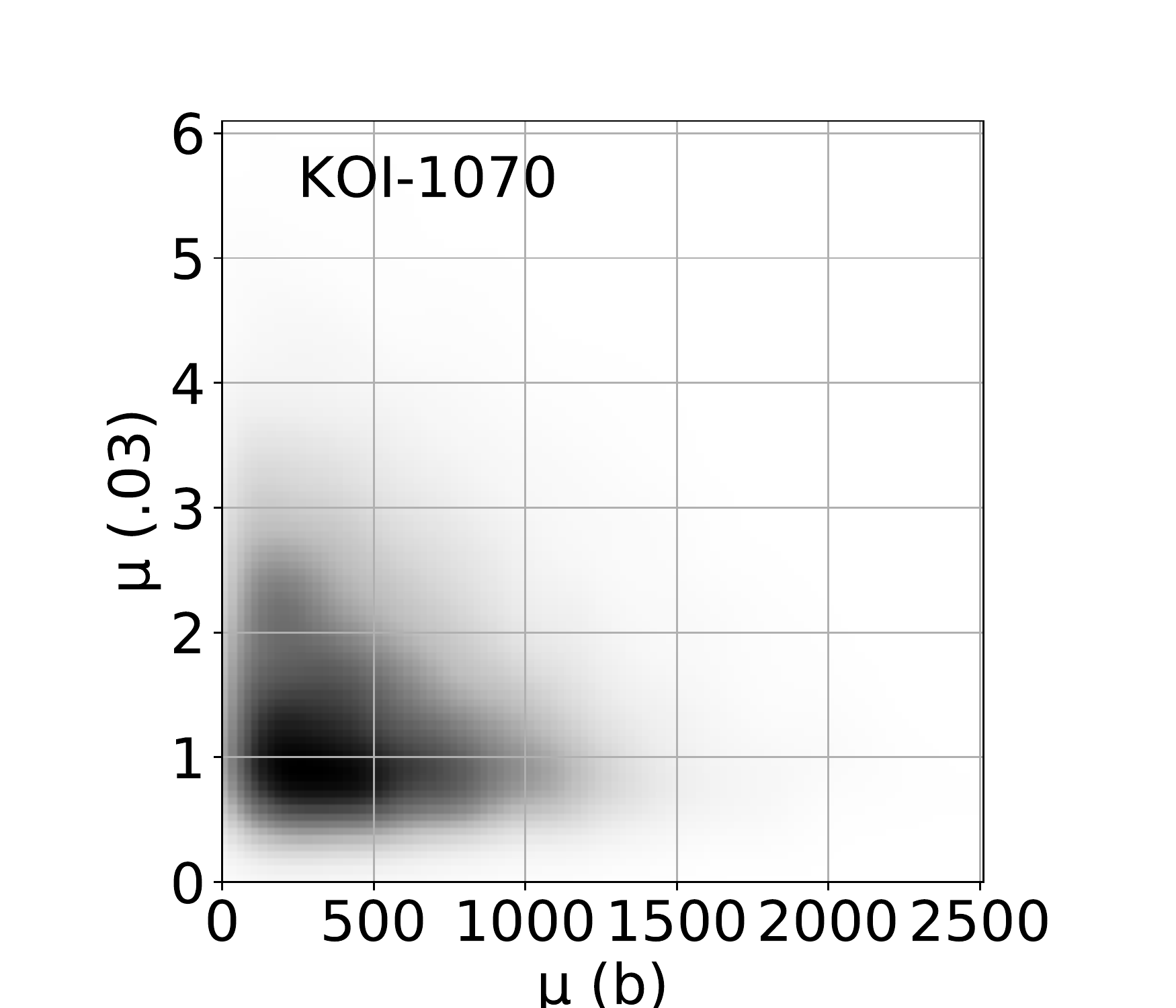}
\includegraphics [height = 1.1 in]{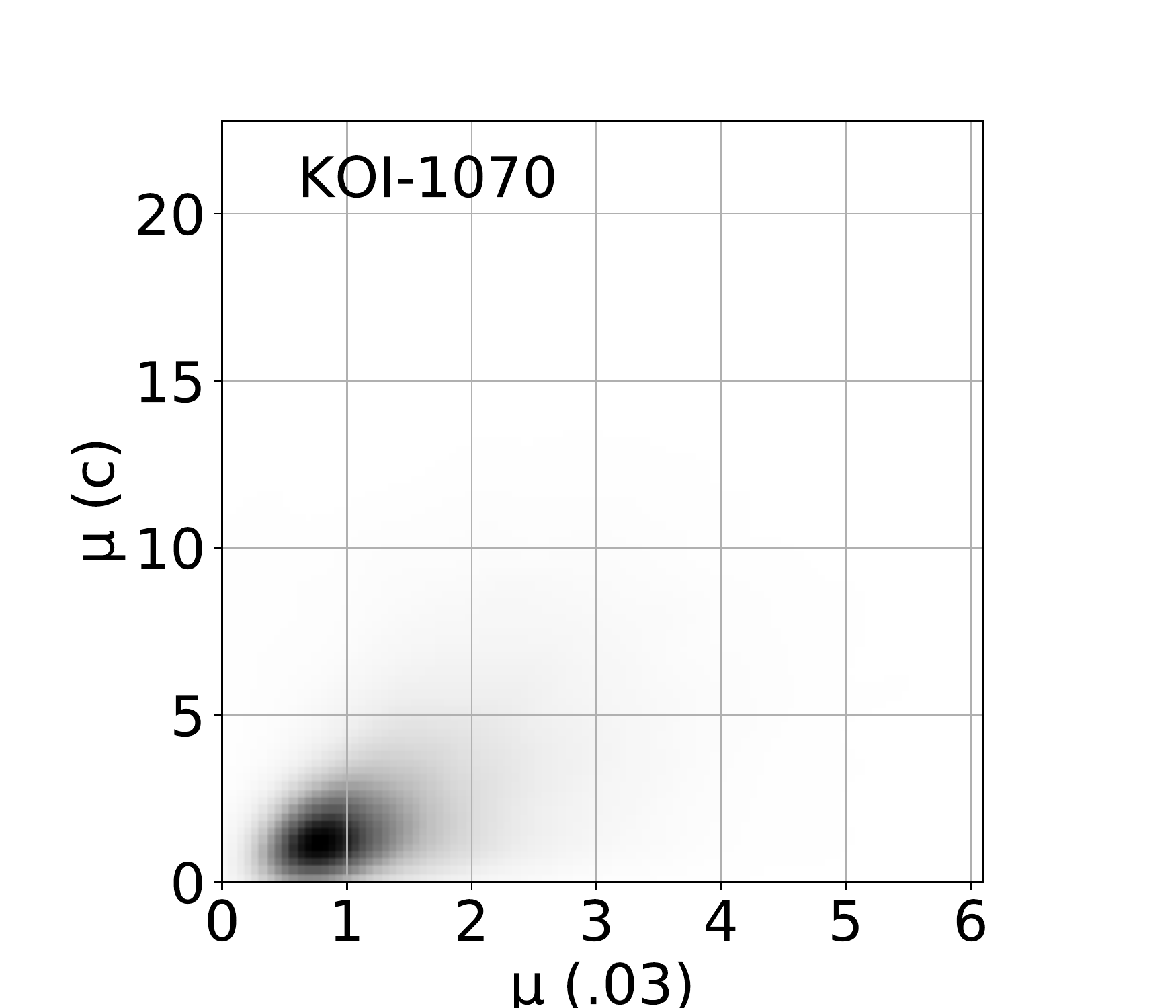} \\
\includegraphics [height = 1.1 in]{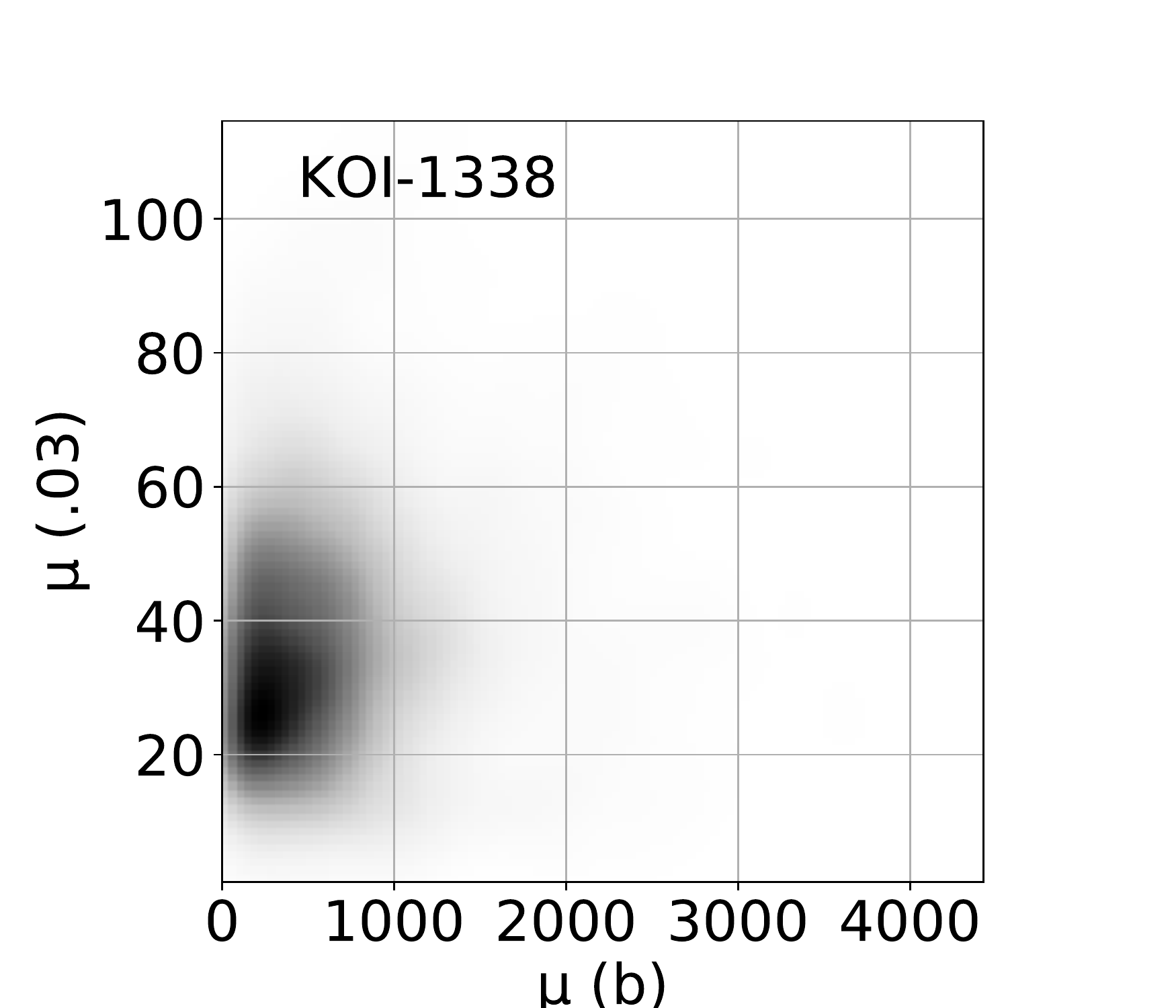}
\includegraphics [height = 1.1 in]{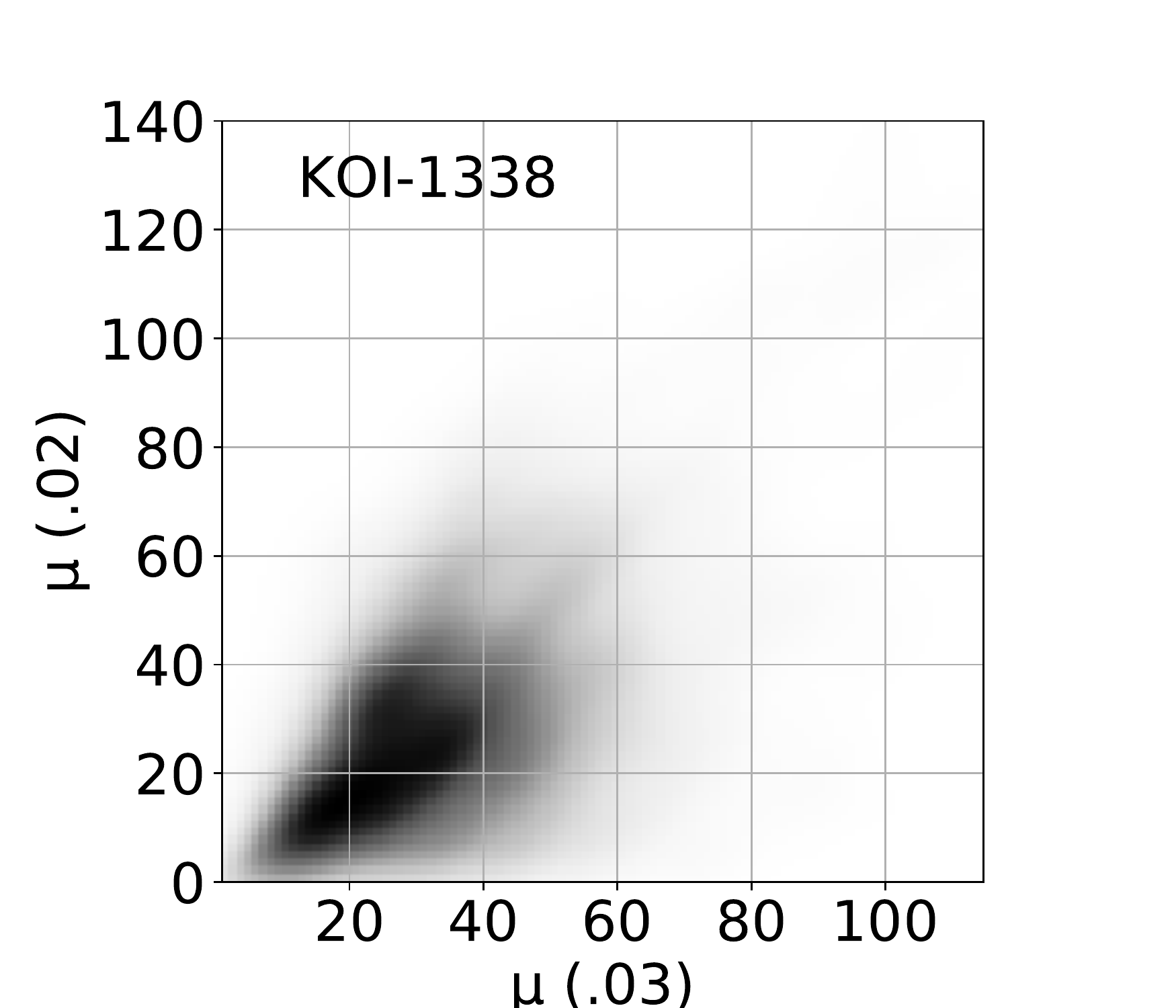}
\includegraphics [height = 1.1 in]{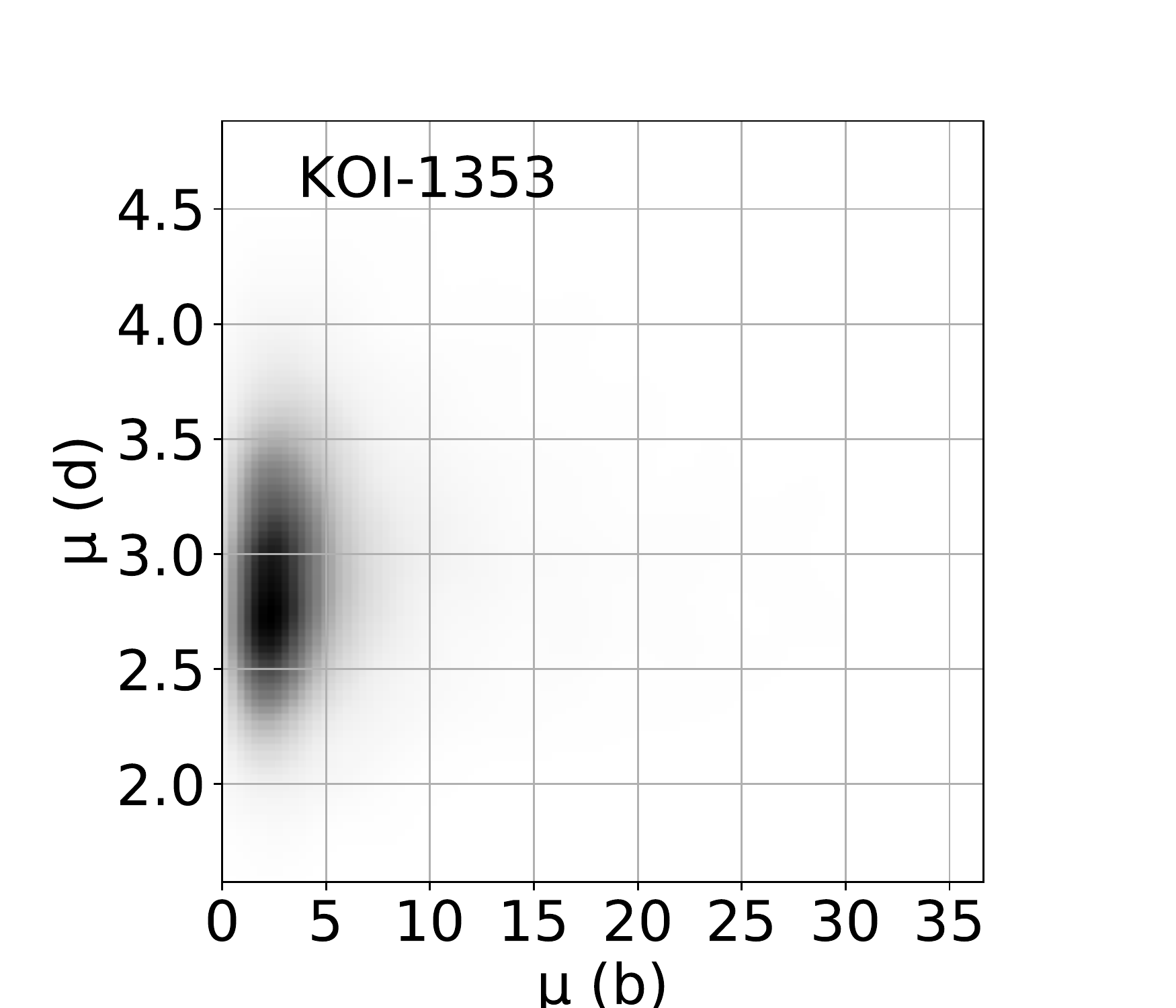}
\includegraphics [height = 1.1 in]{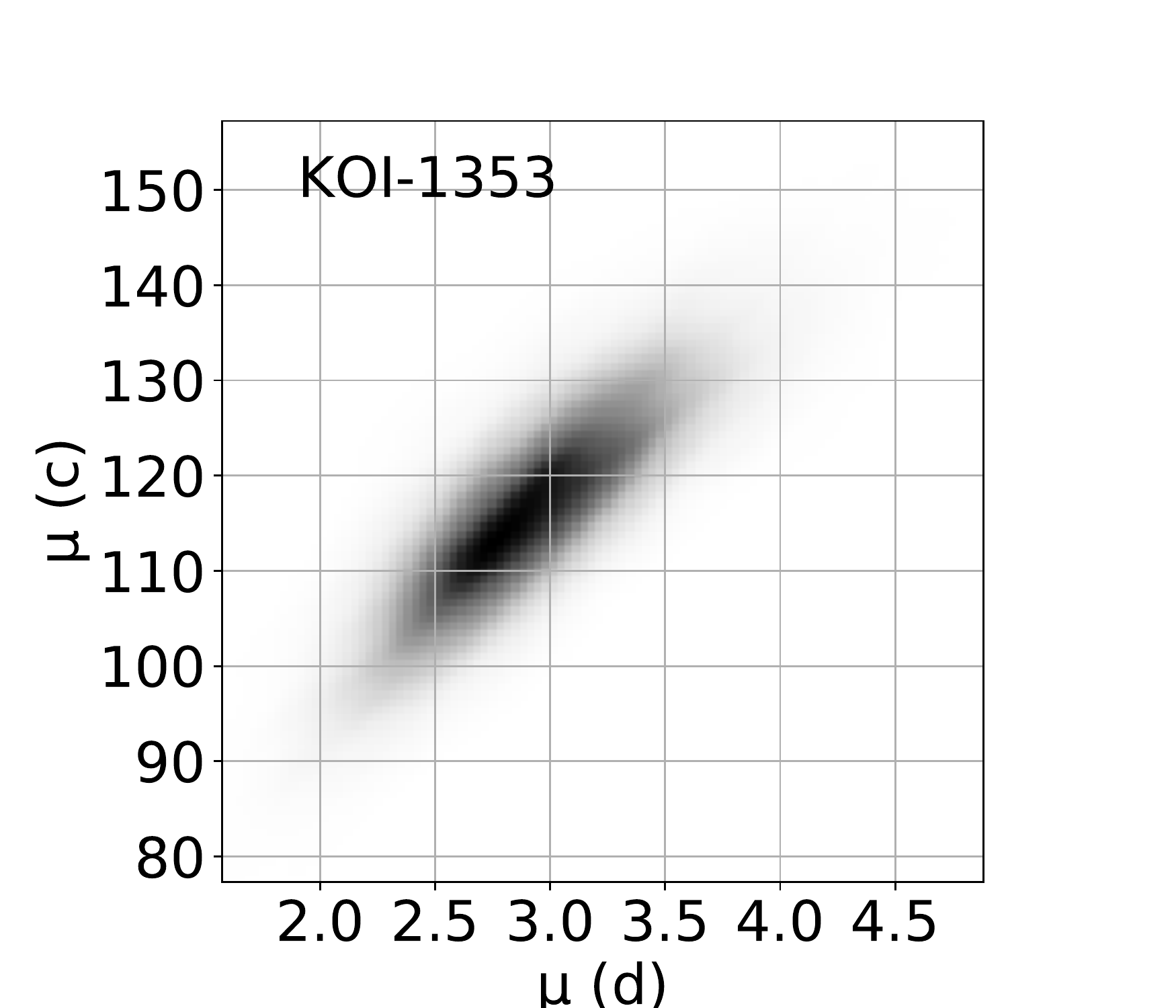} \\
\includegraphics [height = 1.1 in]{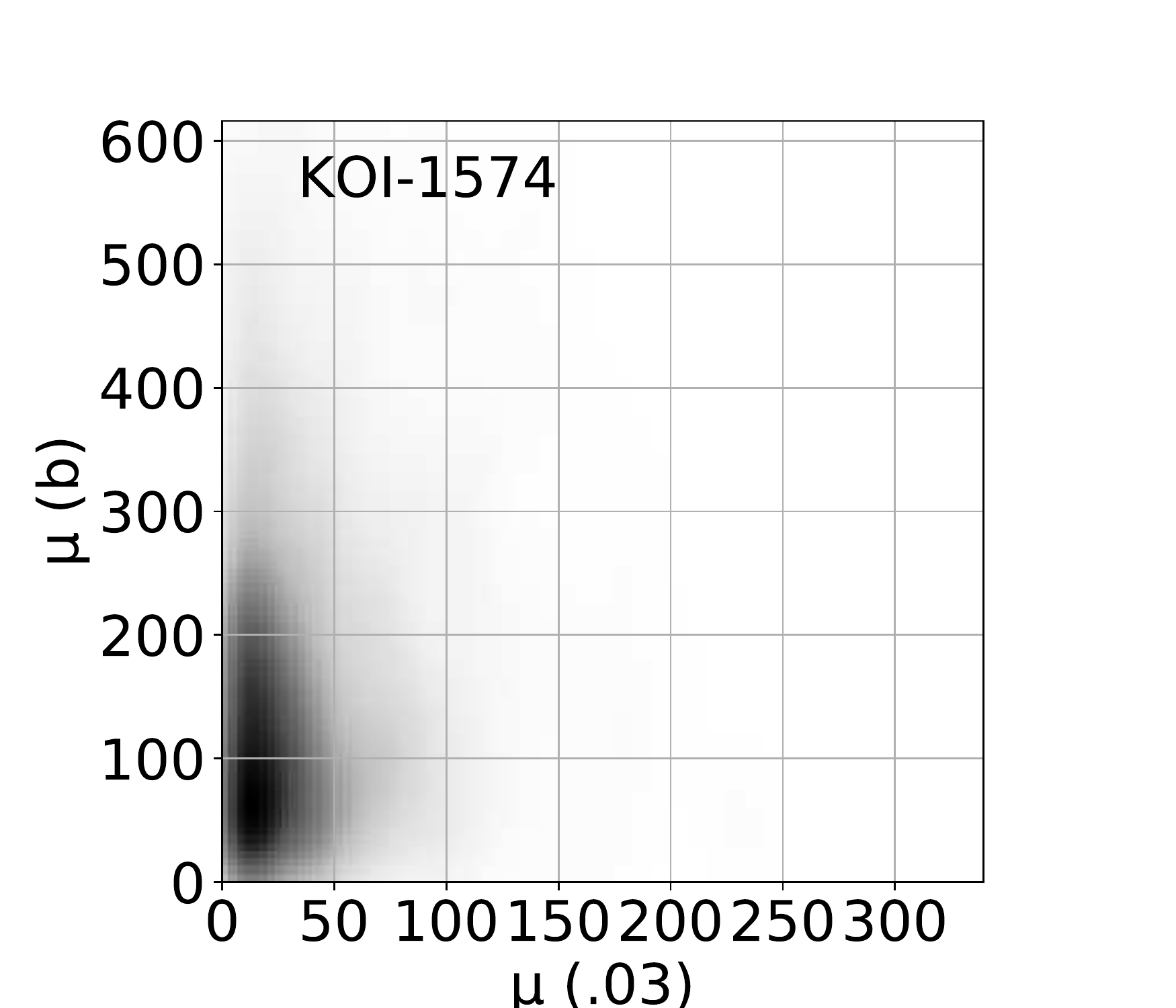}
\includegraphics [height = 1.1 in]{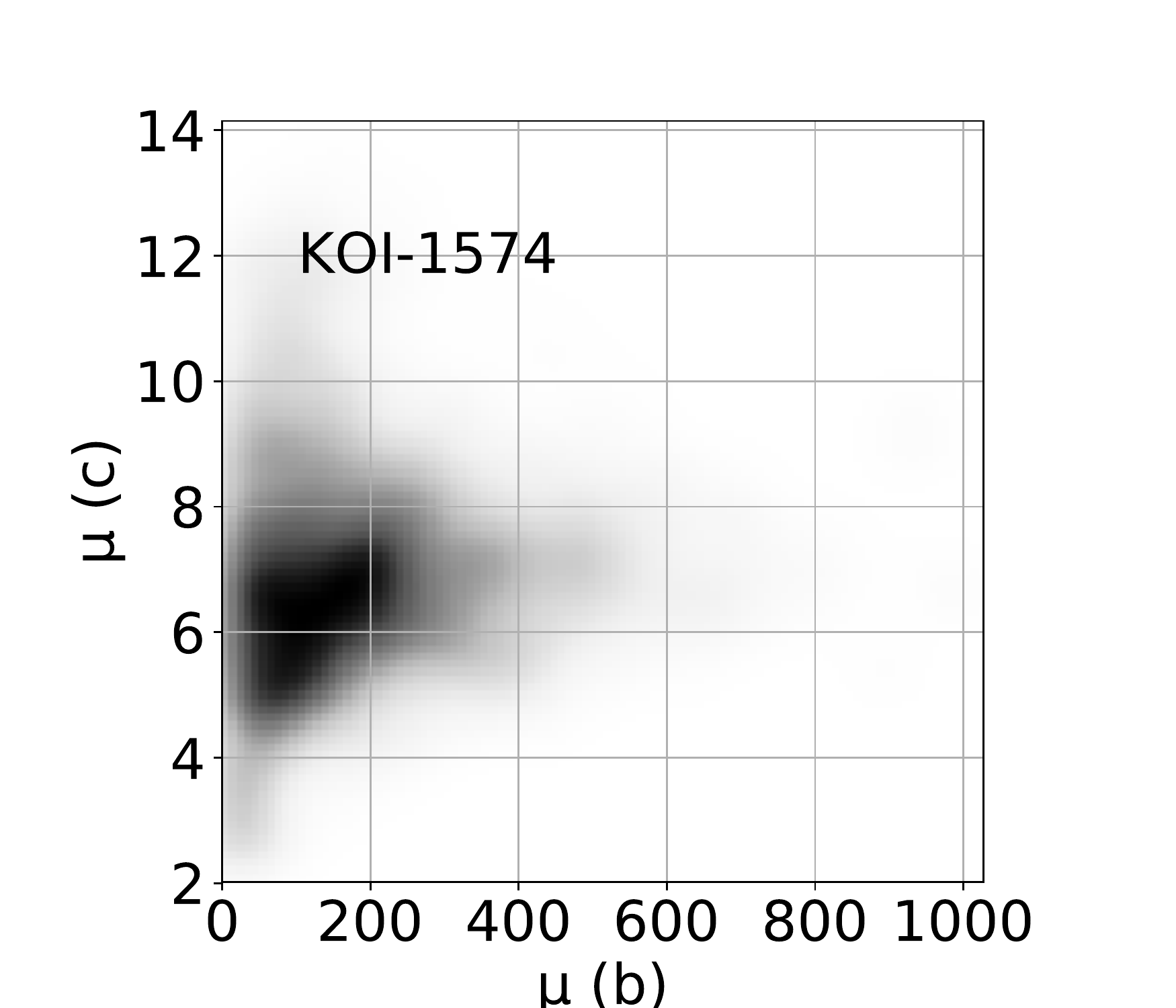}
\includegraphics [height = 1.1 in]{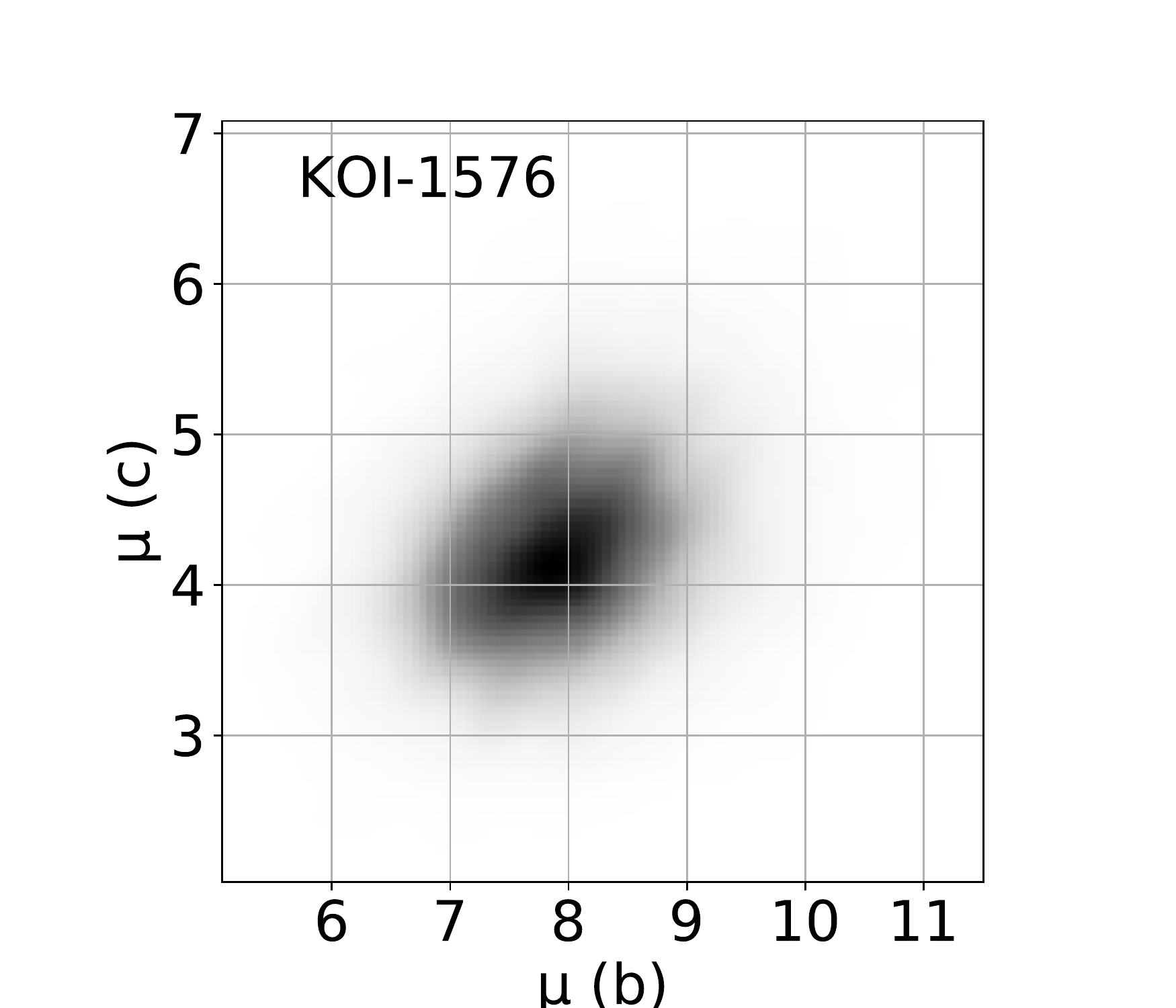}
\includegraphics [height = 1.1 in]{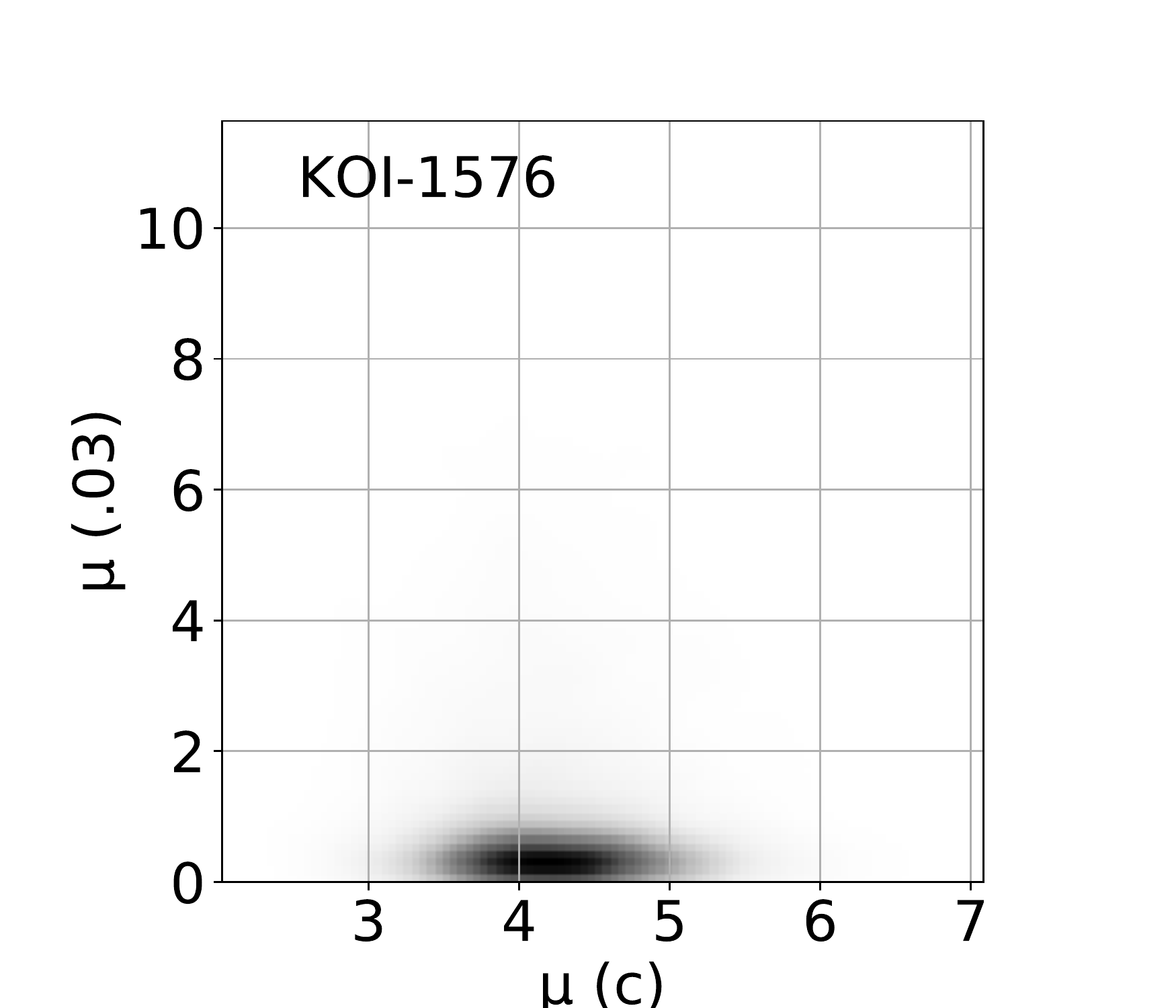} \\
\includegraphics [height = 1.1 in]{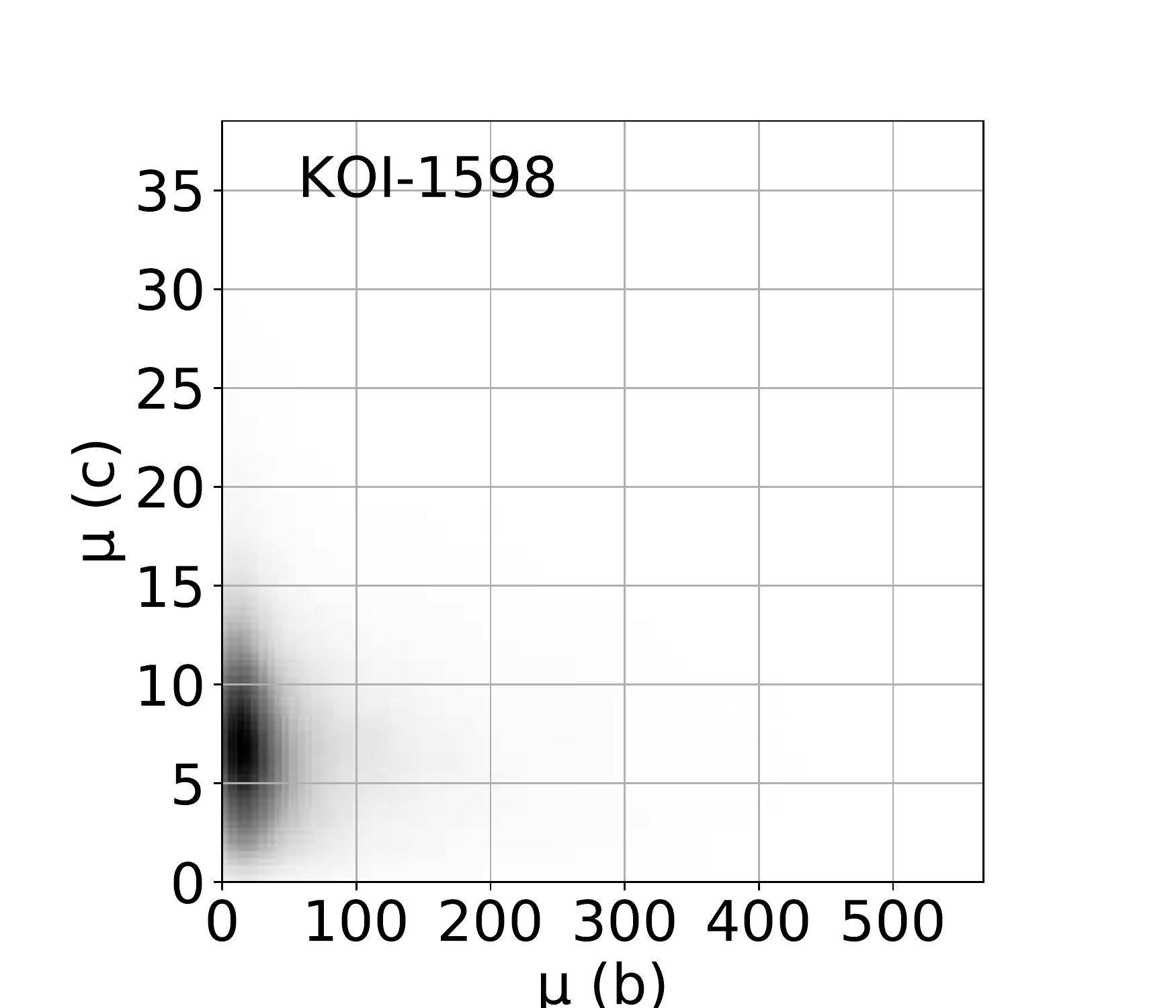}
\includegraphics [height = 1.1 in]{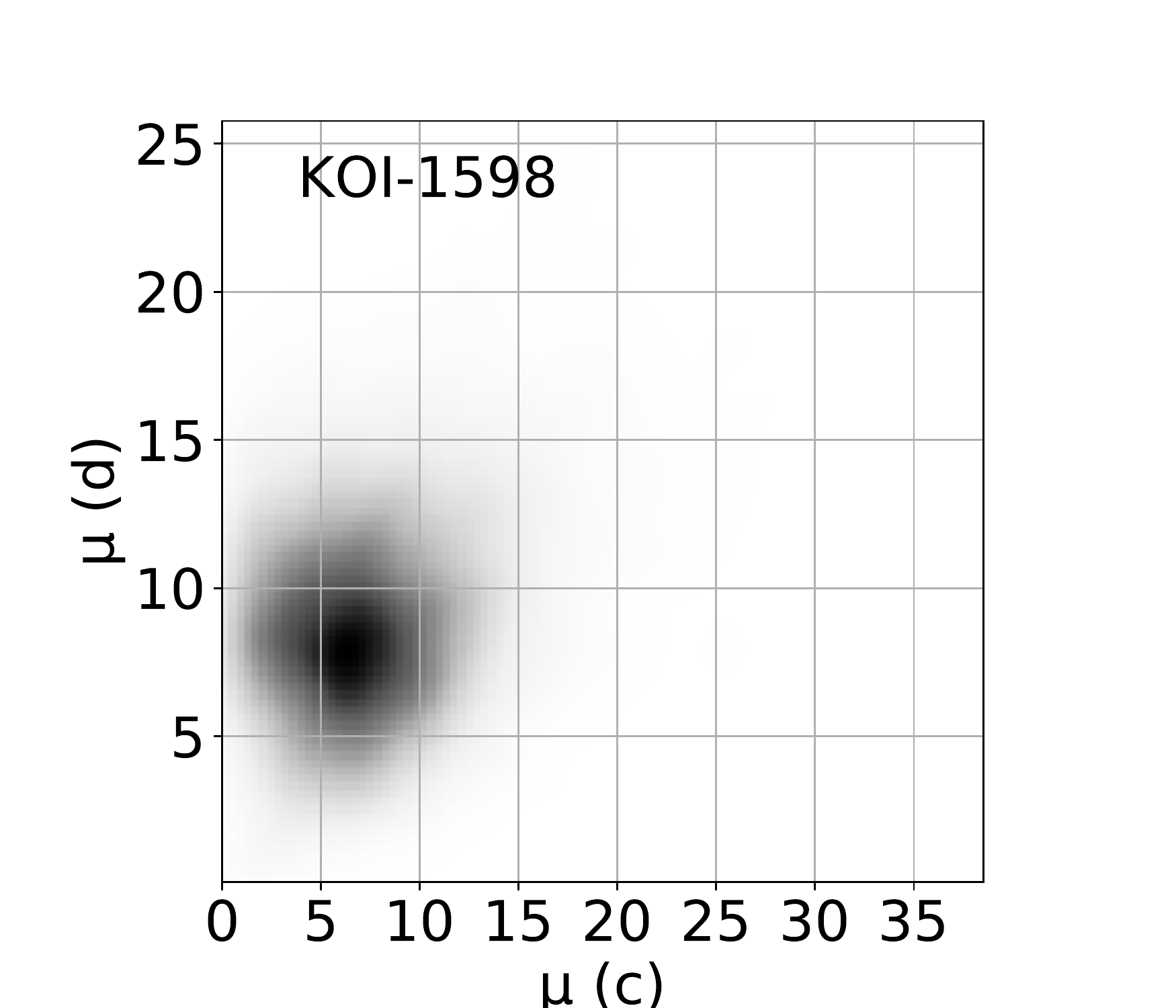}
\includegraphics [height = 1.1 in]{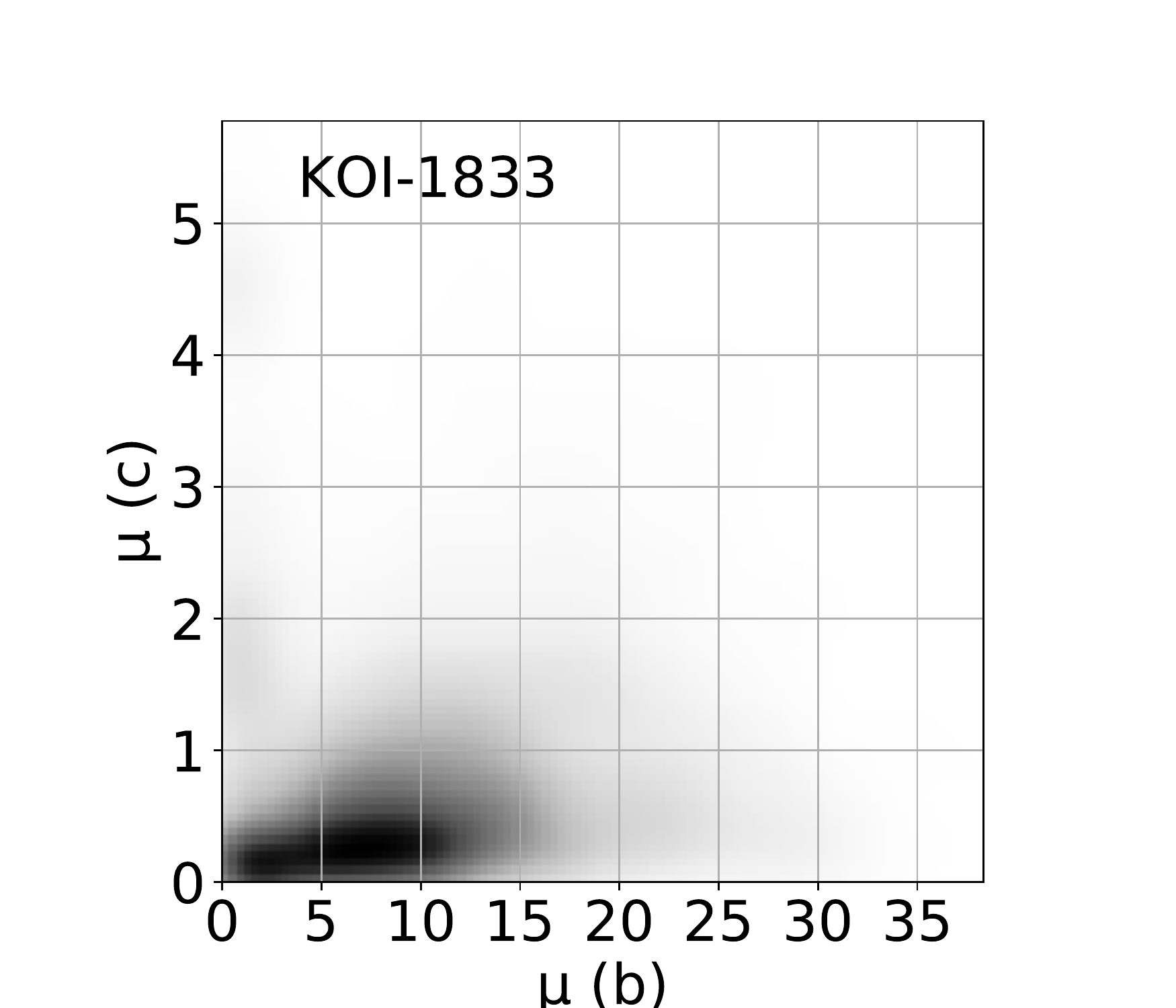}
\includegraphics [height = 1.1 in]{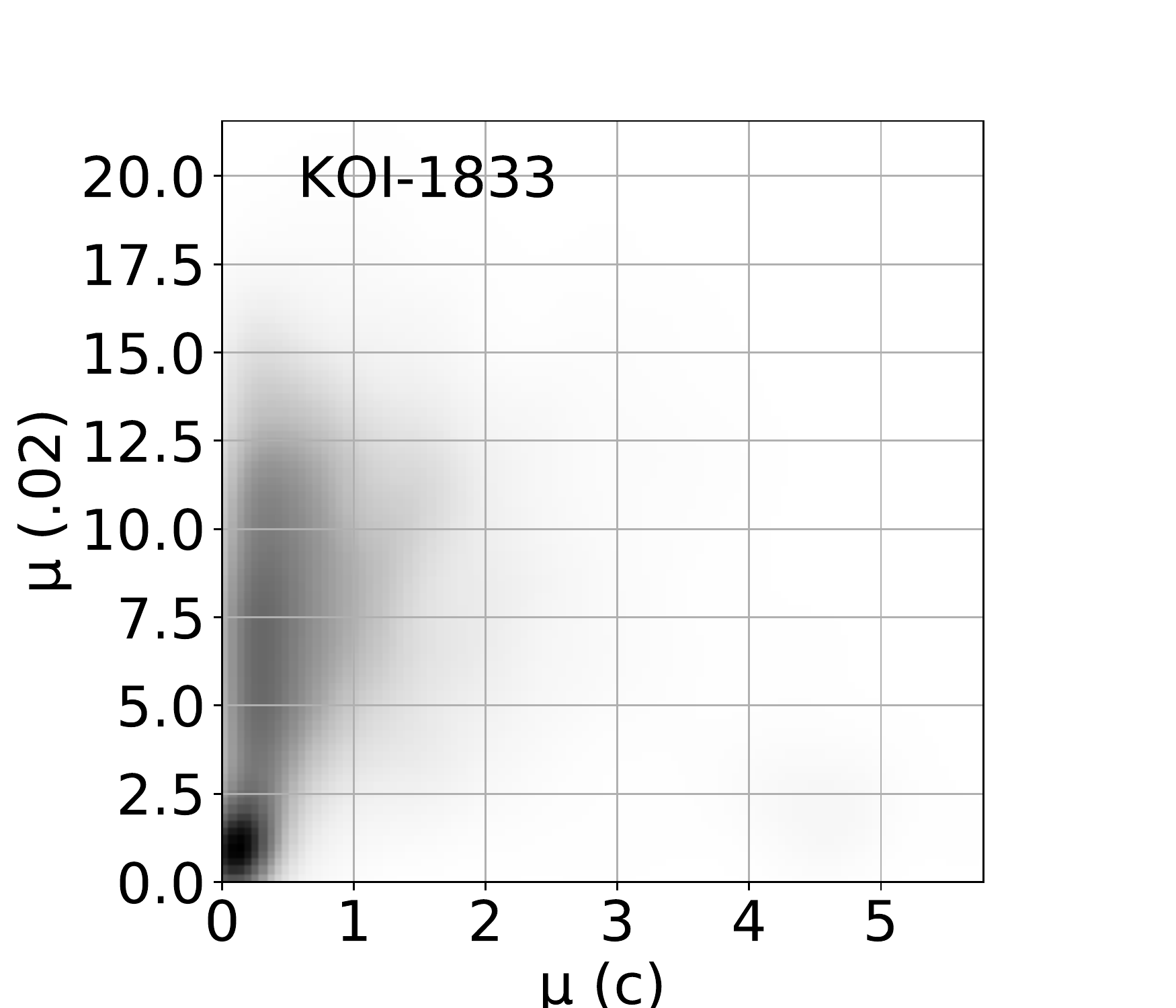} \\
\includegraphics [height = 1.1 in]{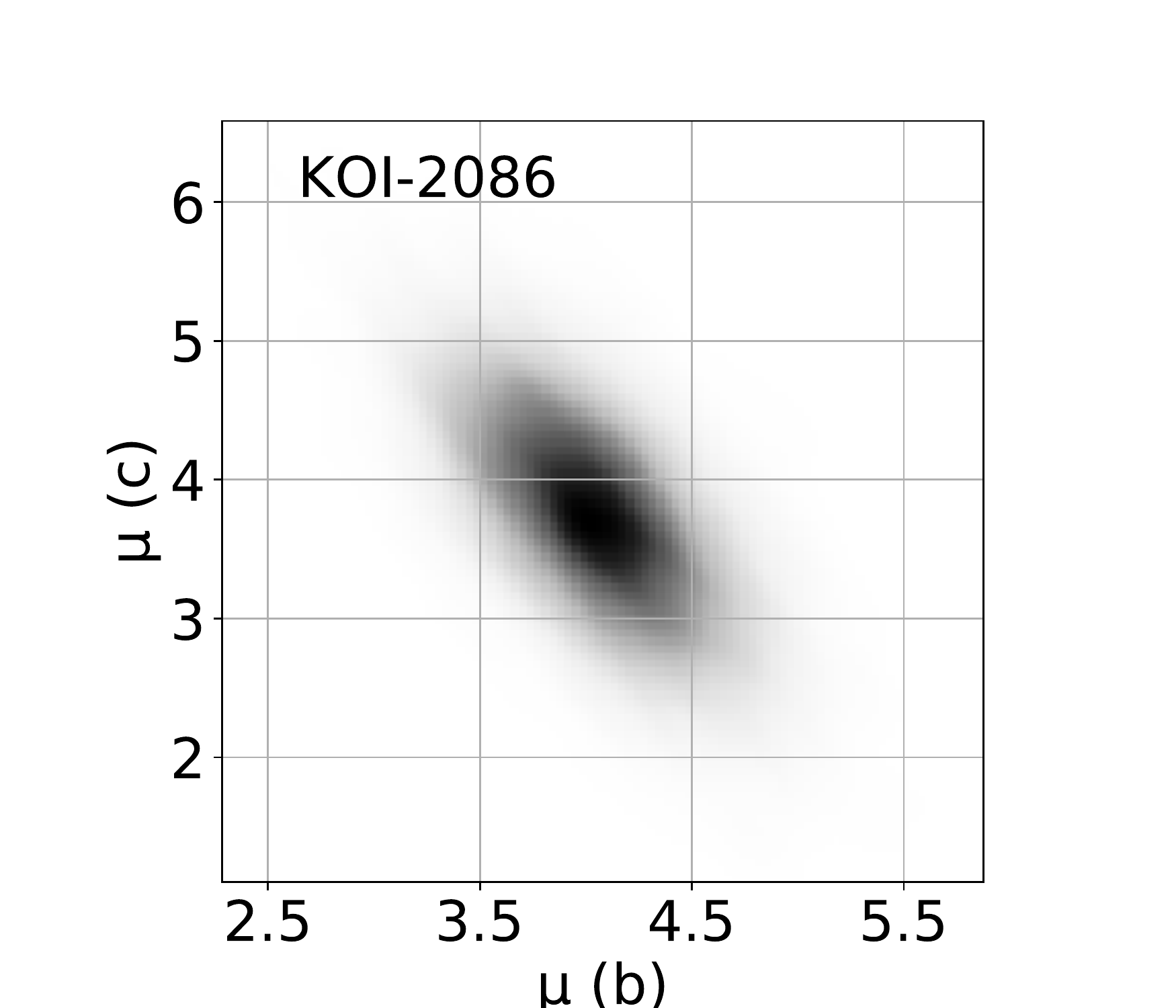}
\includegraphics [height = 1.1 in]{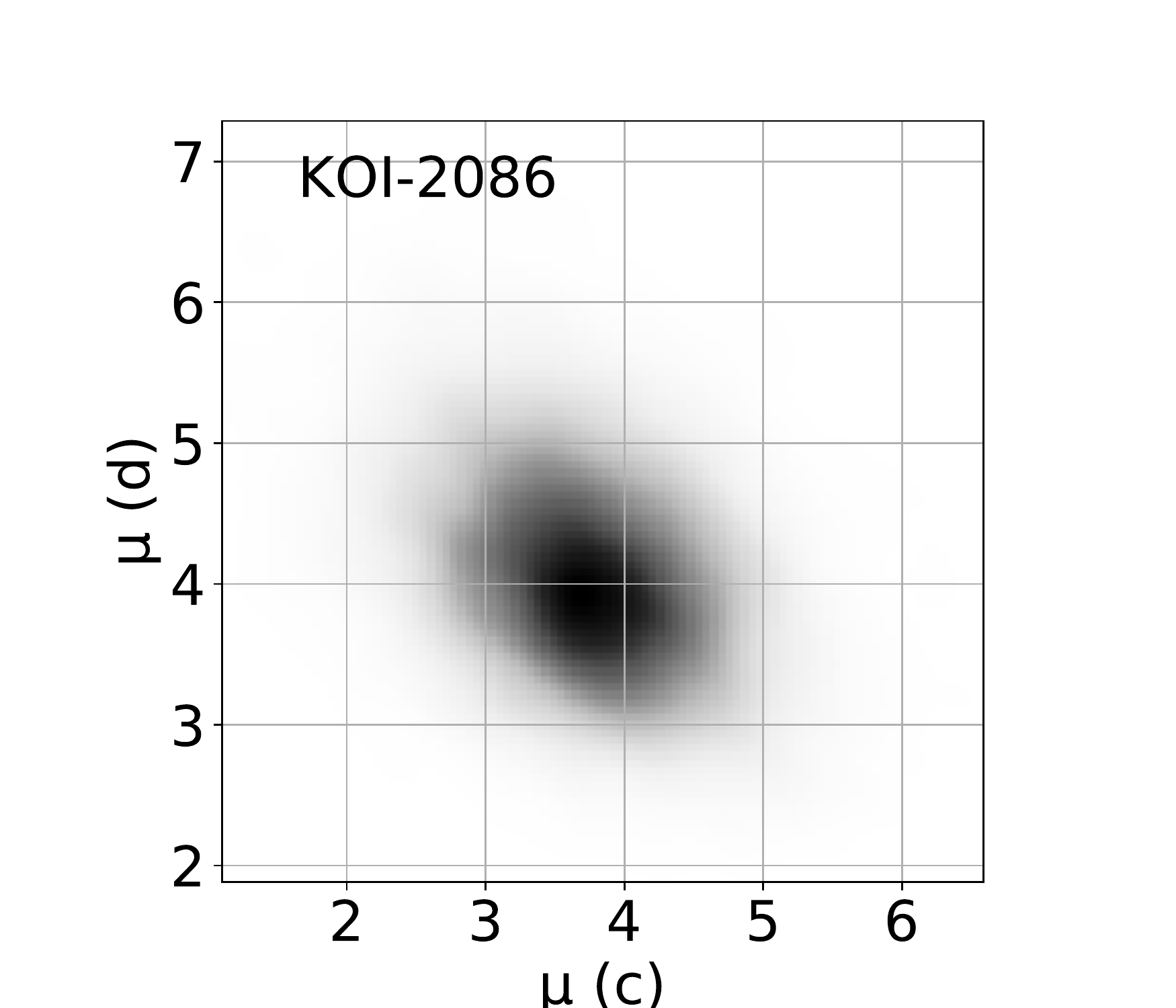} 
\includegraphics [height = 1.1 in]{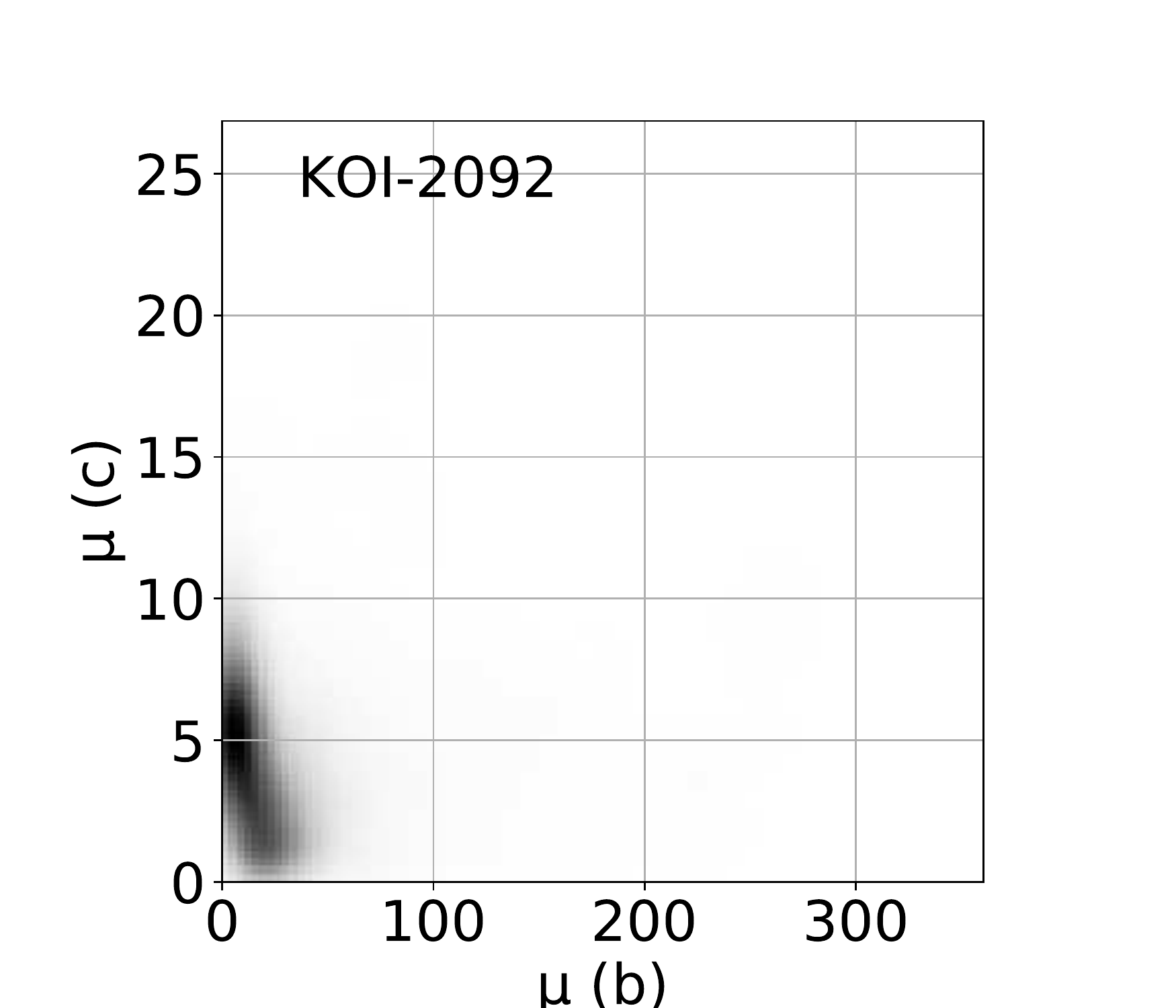}
\includegraphics [height = 1.1 in]{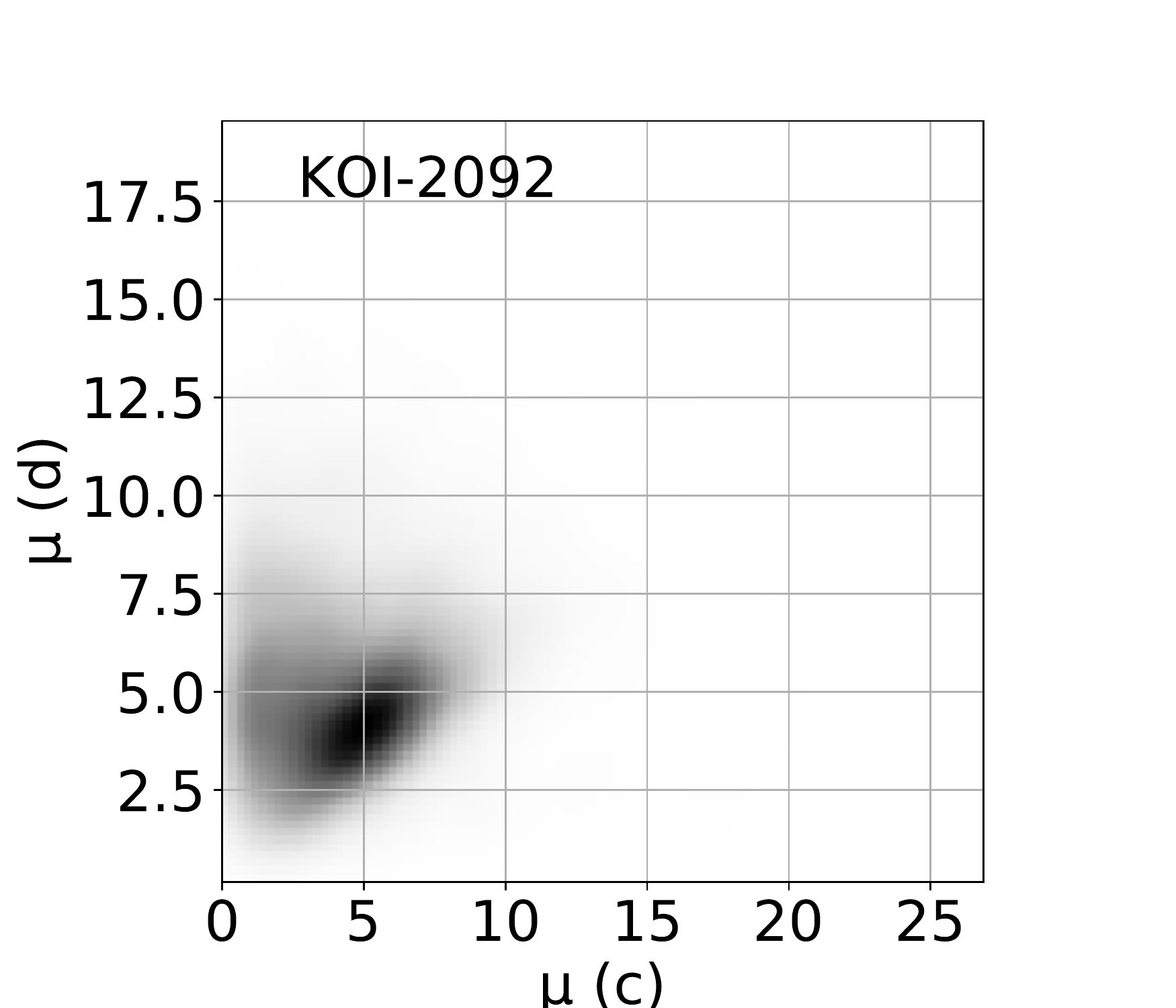} \\
\includegraphics [height = 1.1 in]{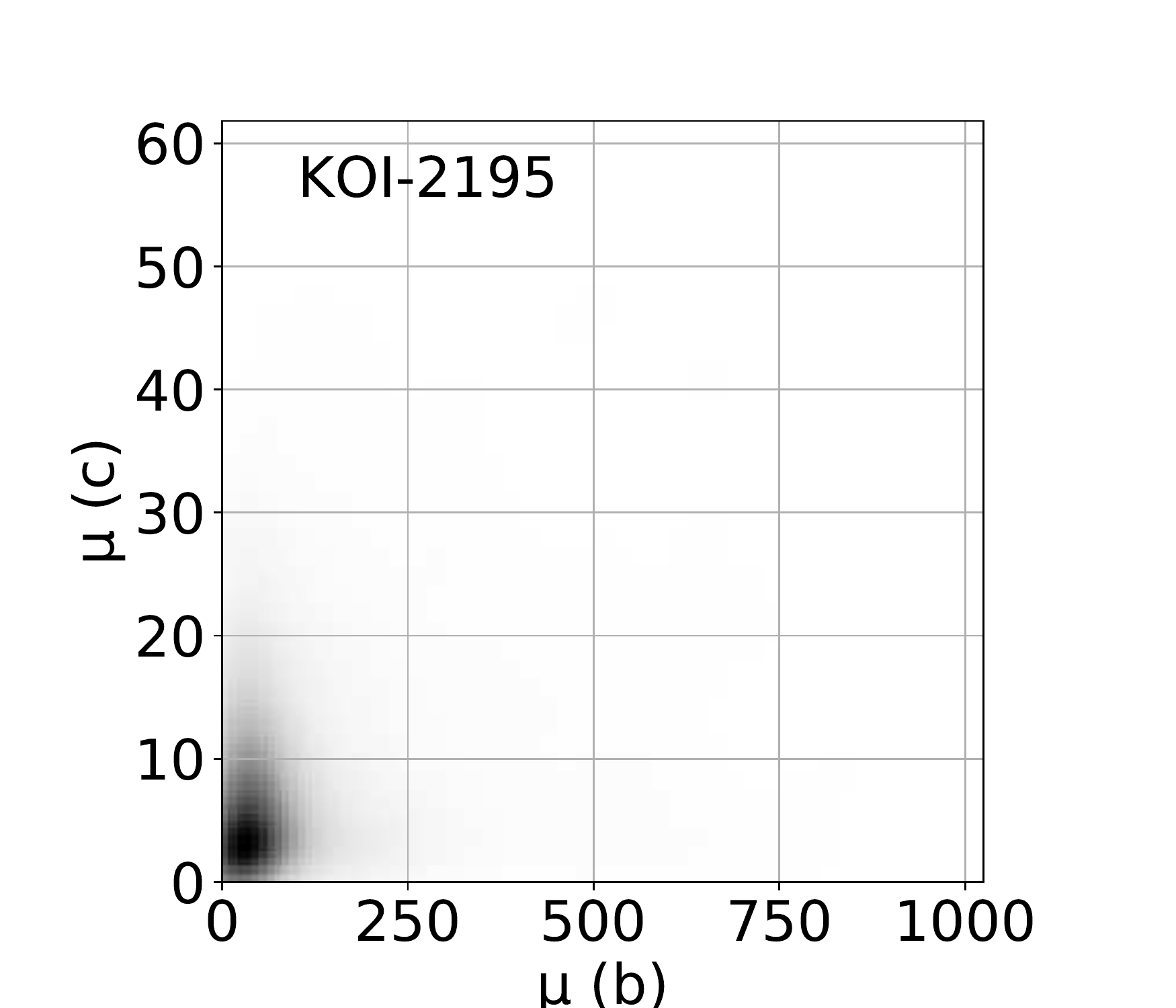}
\includegraphics [height = 1.1 in]{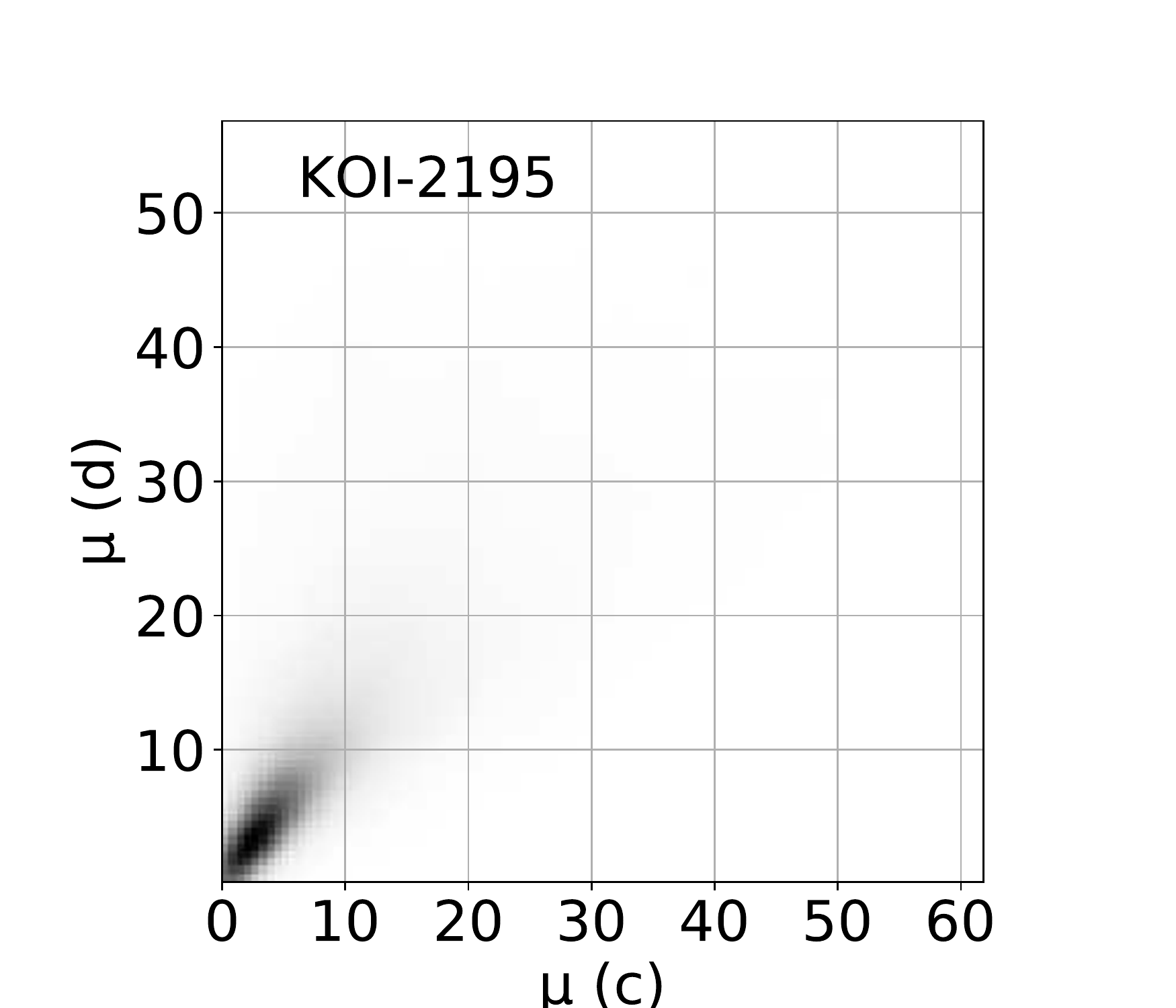} 
\caption{Two-dimensional kernel density estimators on joint posteriors of dynamical masses, $\mu$, scaled by factor $\frac{M_{\odot}}{M_{\oplus}}$: three-planet systems (cont'd). 
\label{fig:mu3b} }
\end{center}
\end{figure}

\begin{figure}
\begin{center}
\figurenum{19}
\includegraphics [height = 1.1 in]{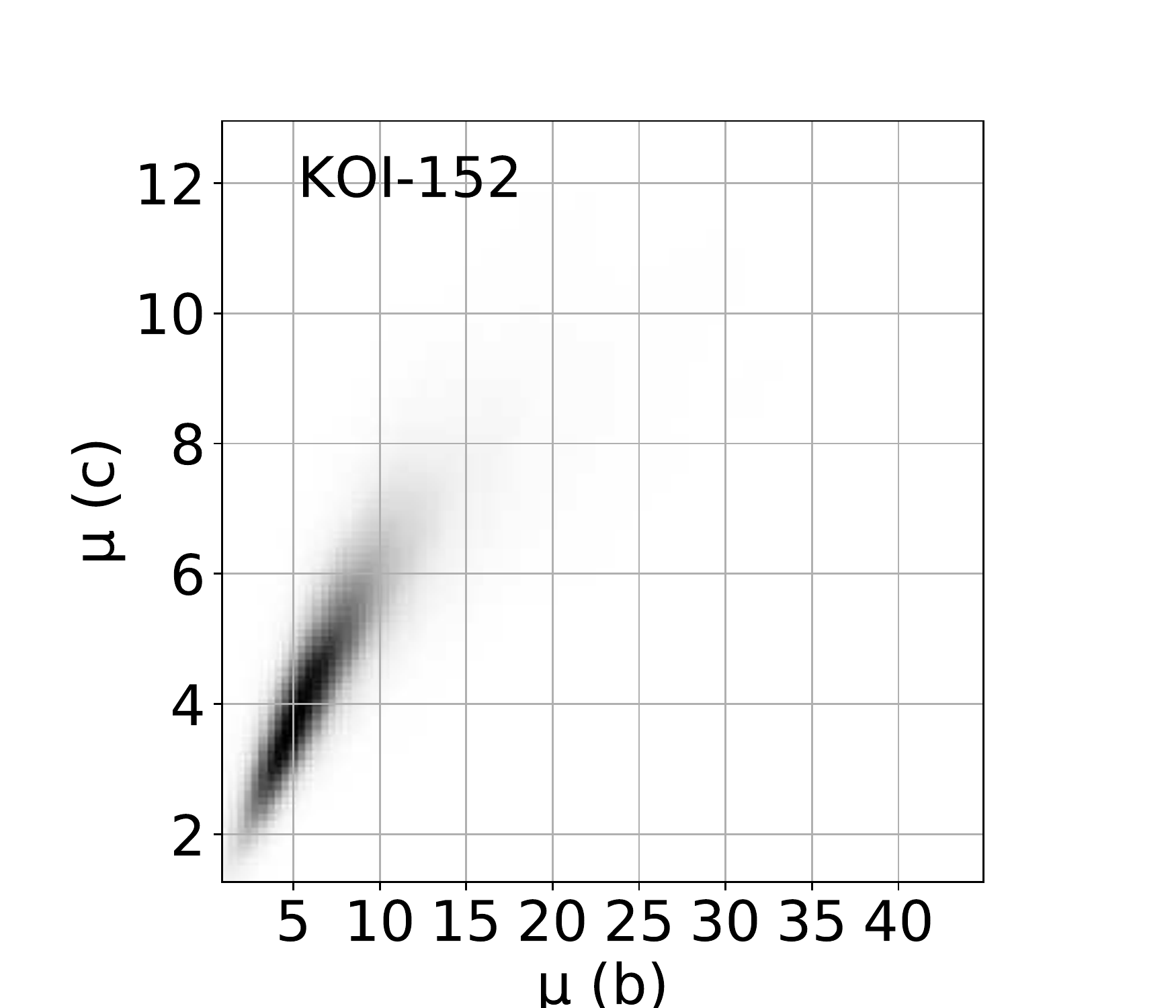}
\includegraphics [height = 1.1 in]{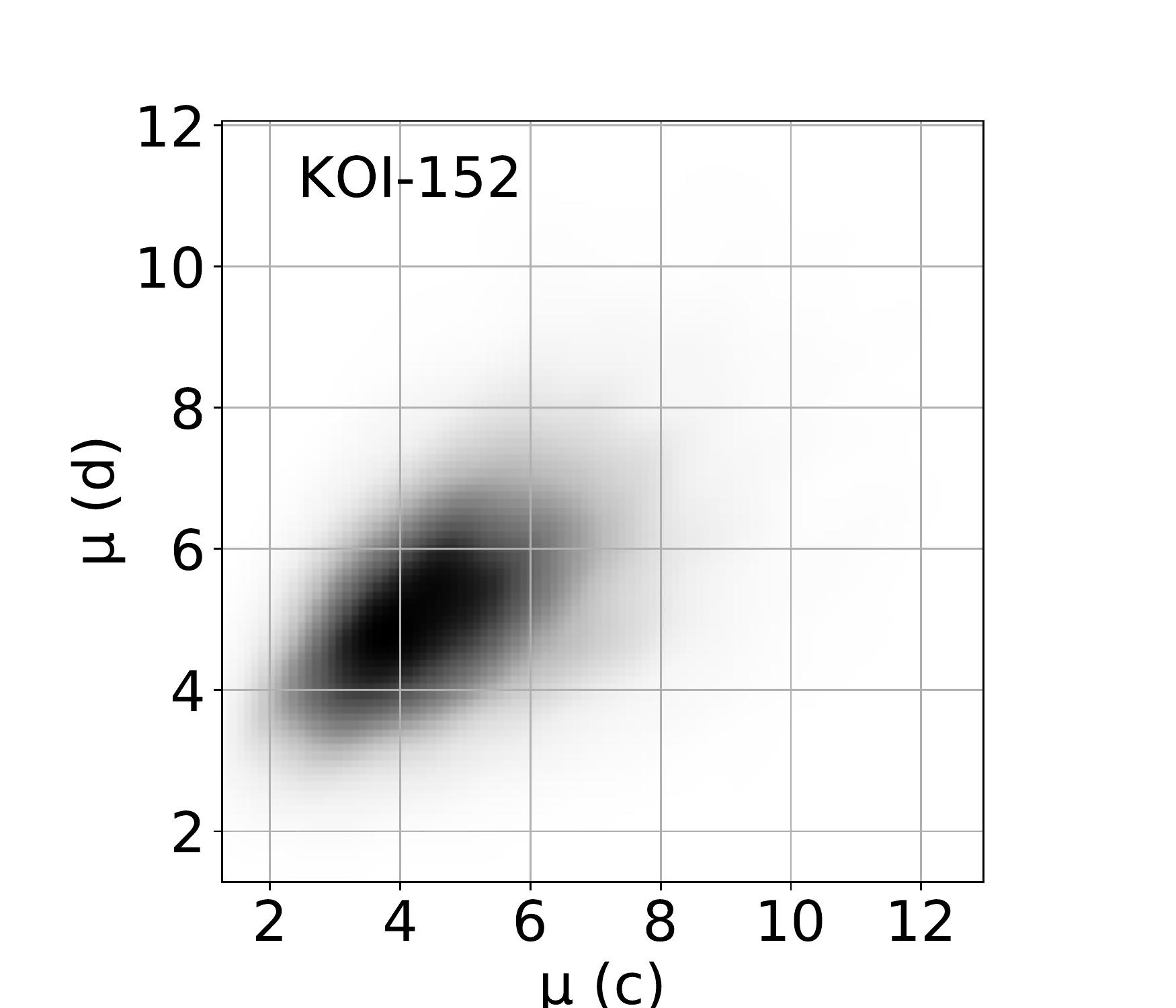}
\includegraphics [height = 1.1 in]{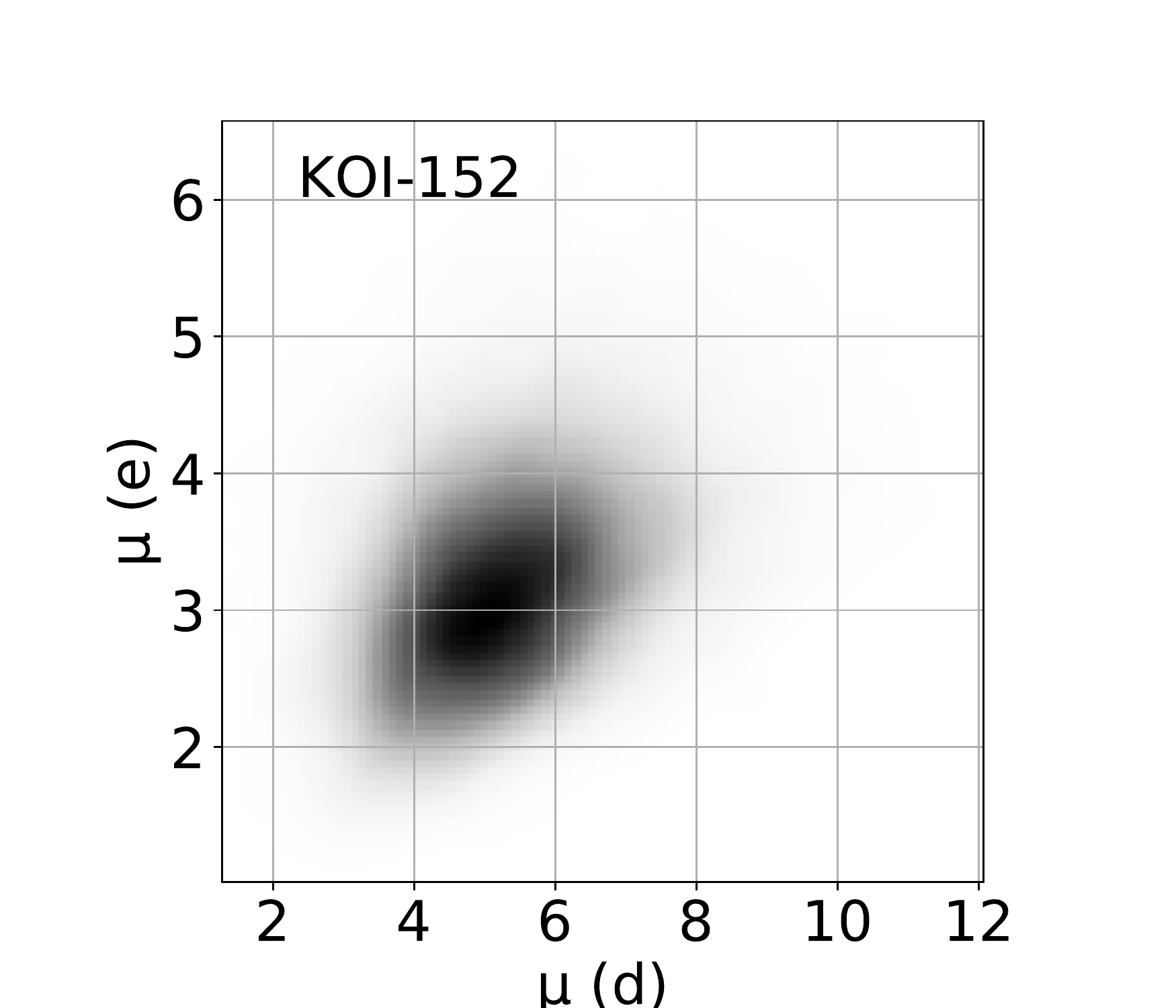}
\includegraphics [height = 1.1 in]{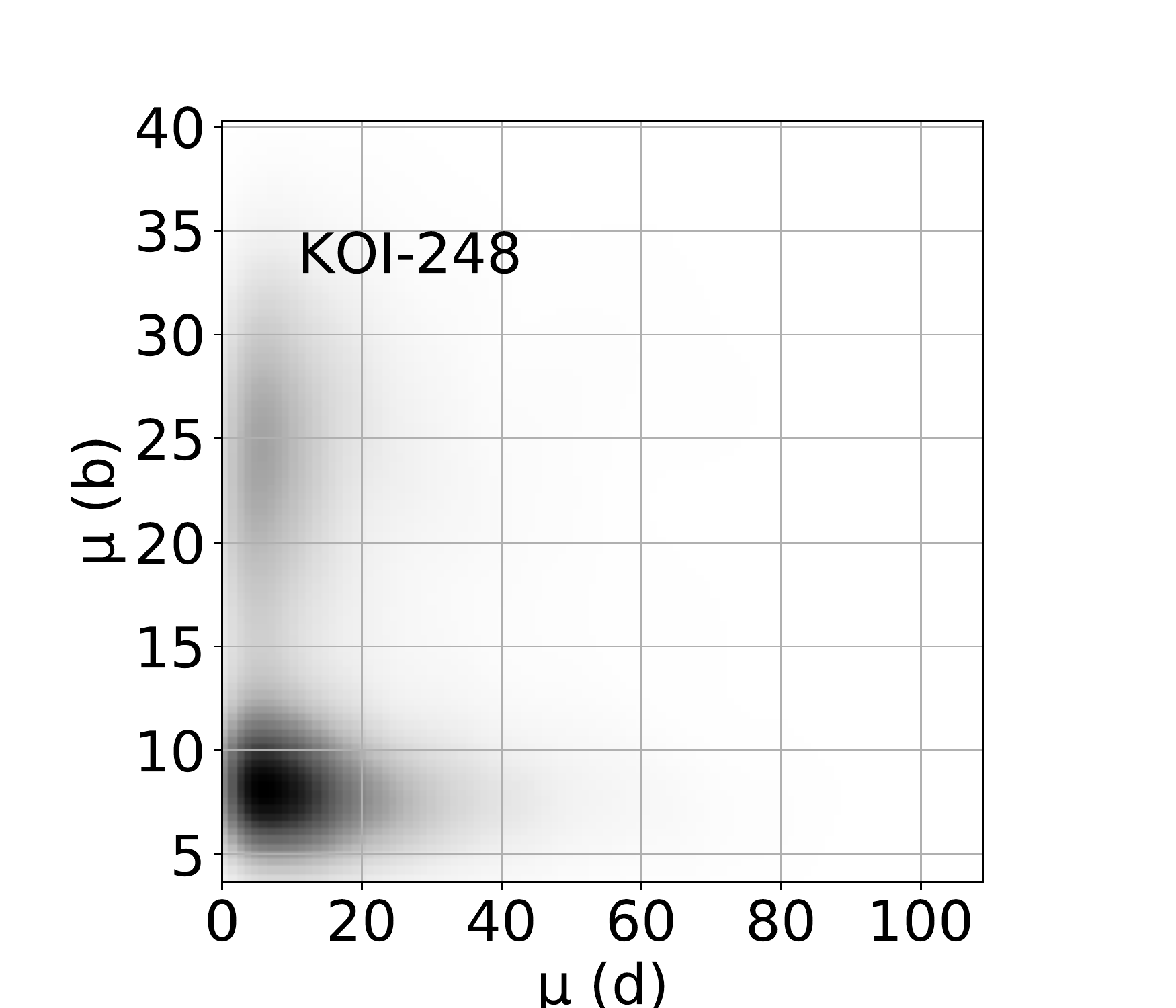} \\
\includegraphics [height = 1.1 in]{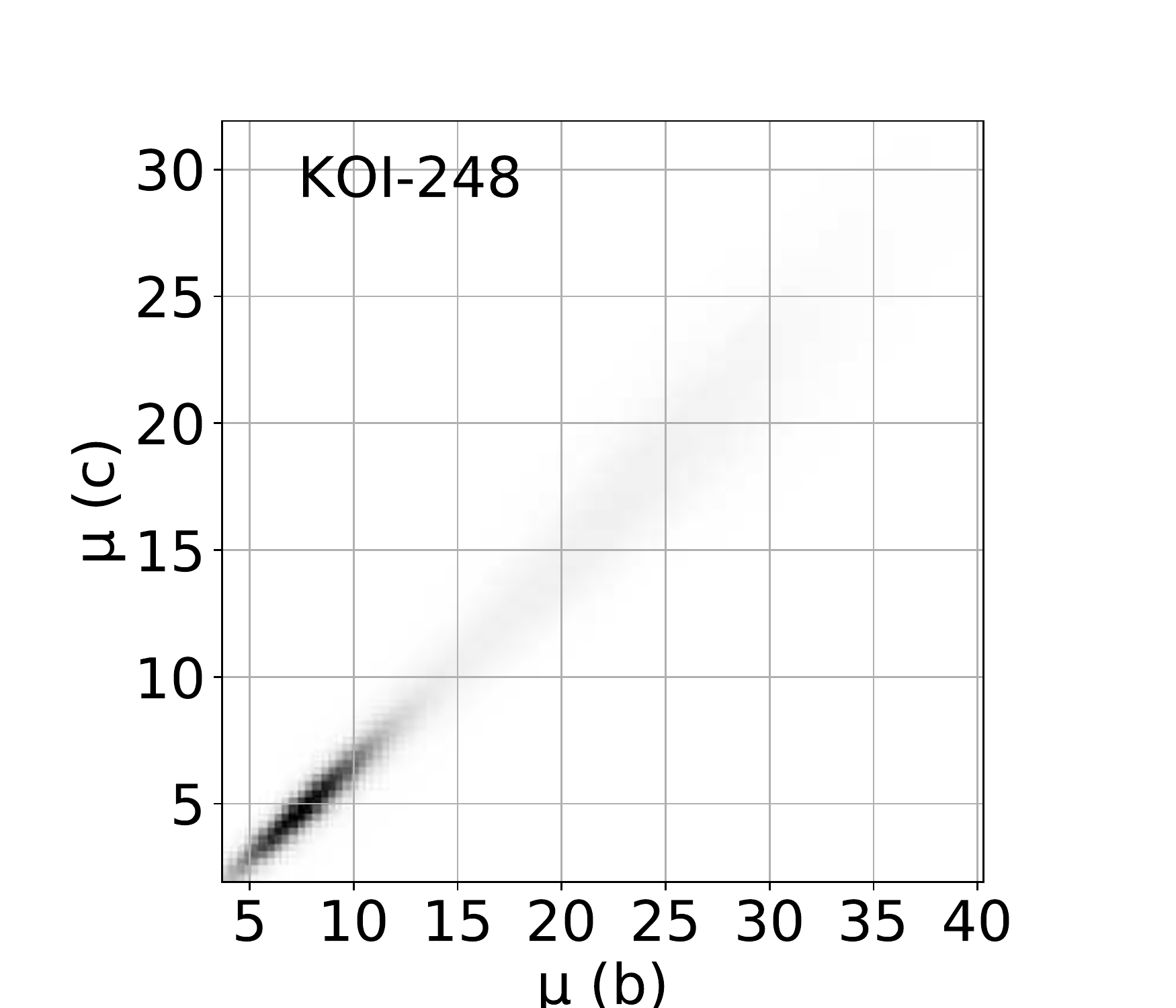}
\includegraphics [height = 1.1 in]{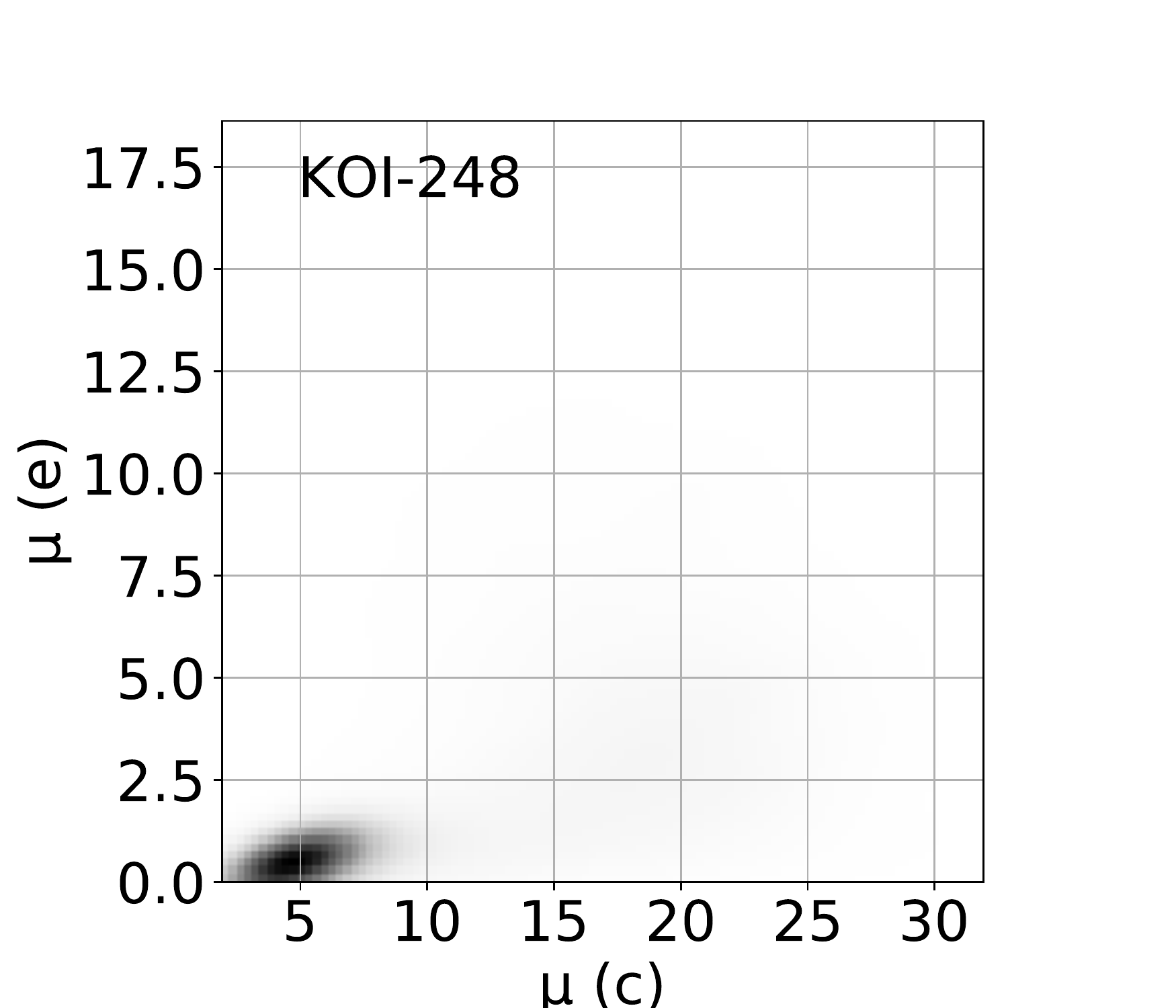}
\includegraphics [height = 1.1 in]{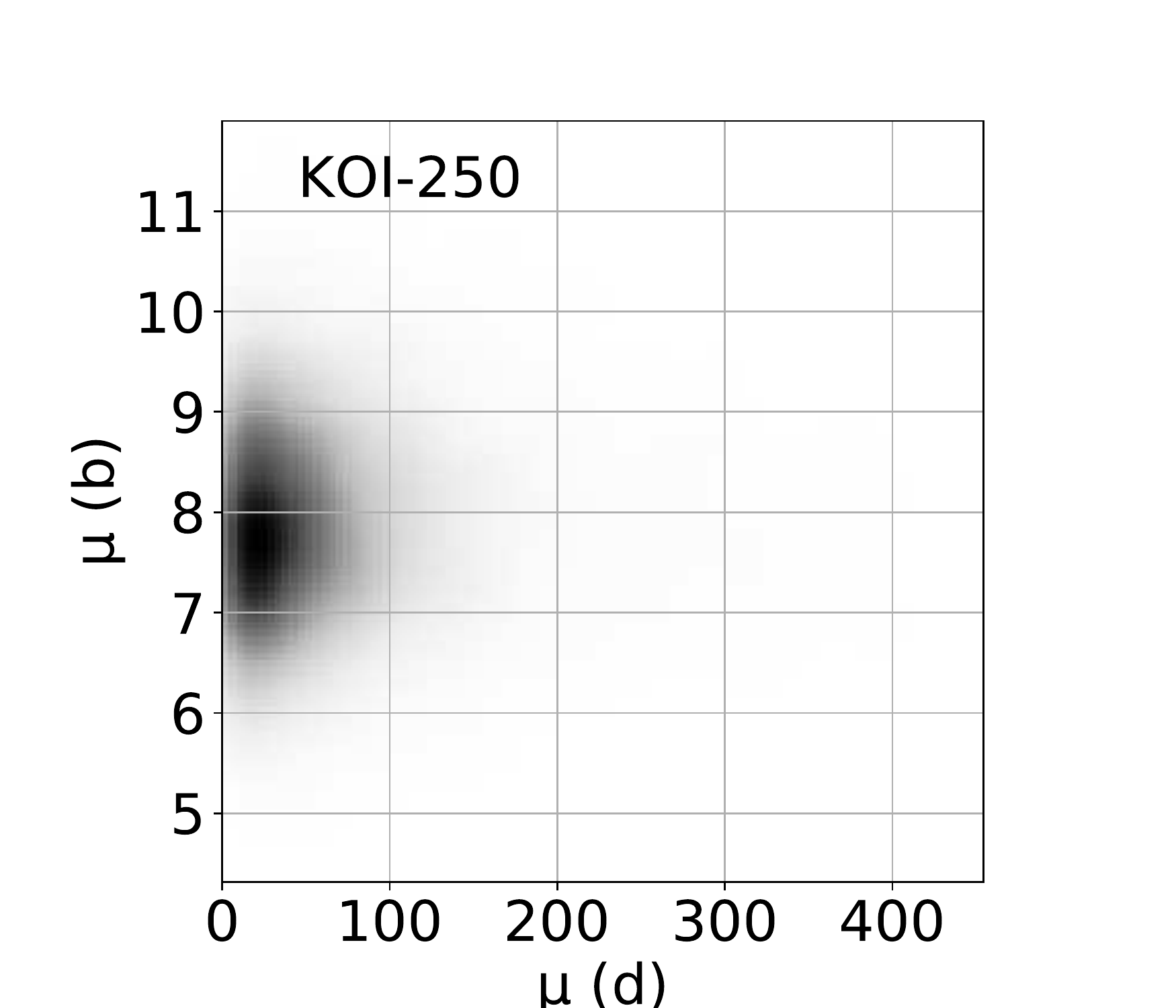}
\includegraphics [height = 1.1 in]{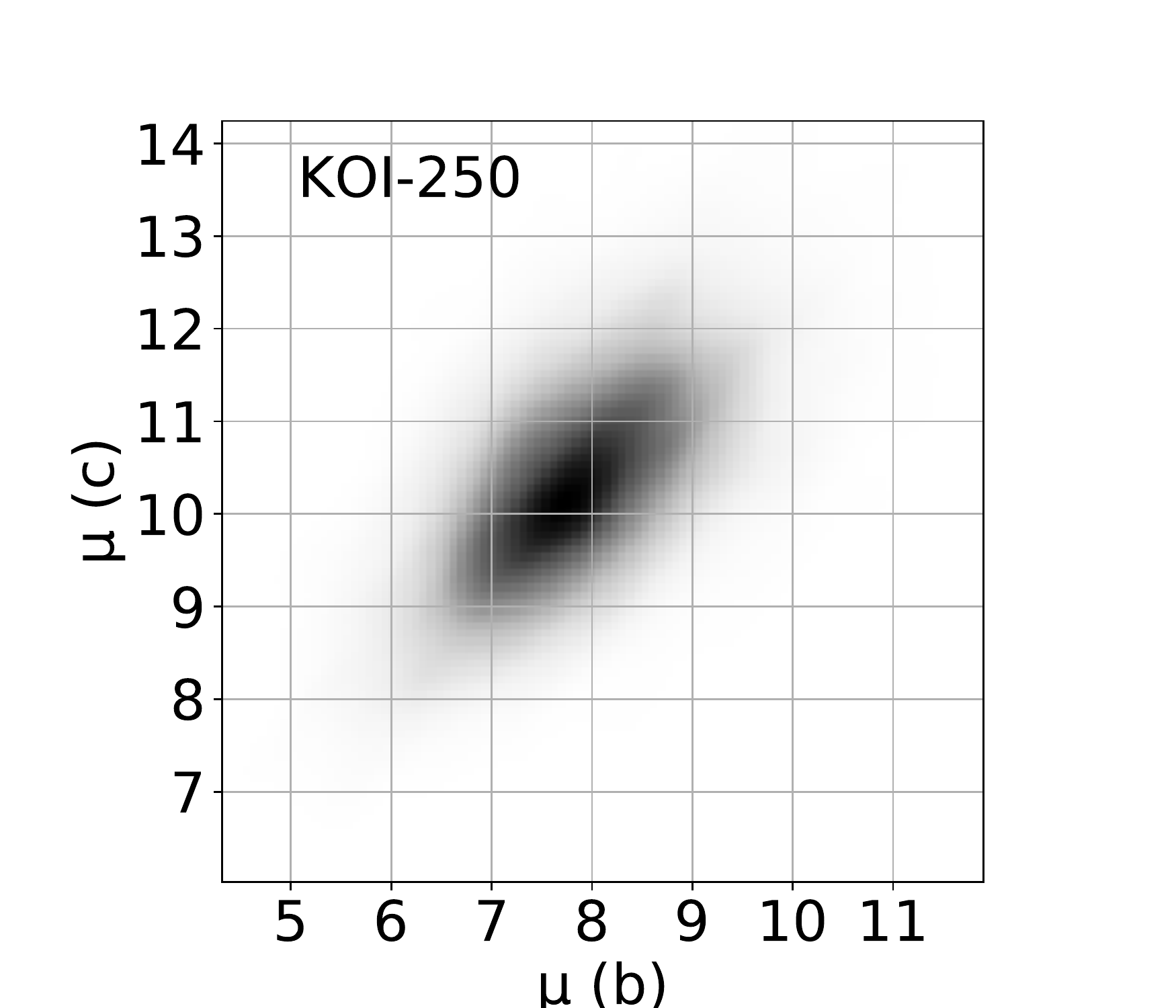} \\
\includegraphics [height = 1.1 in]{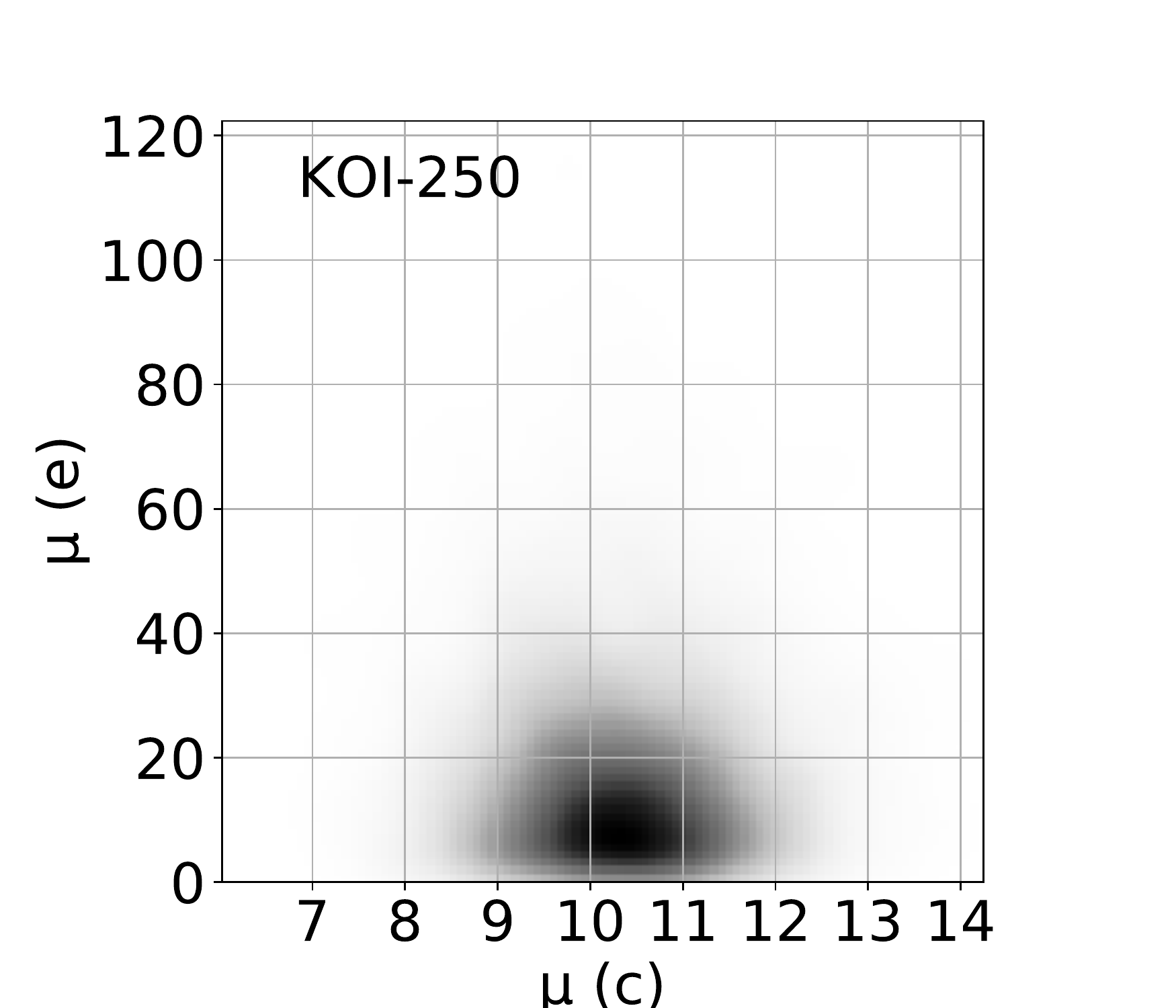}
\includegraphics [height = 1.1 in]{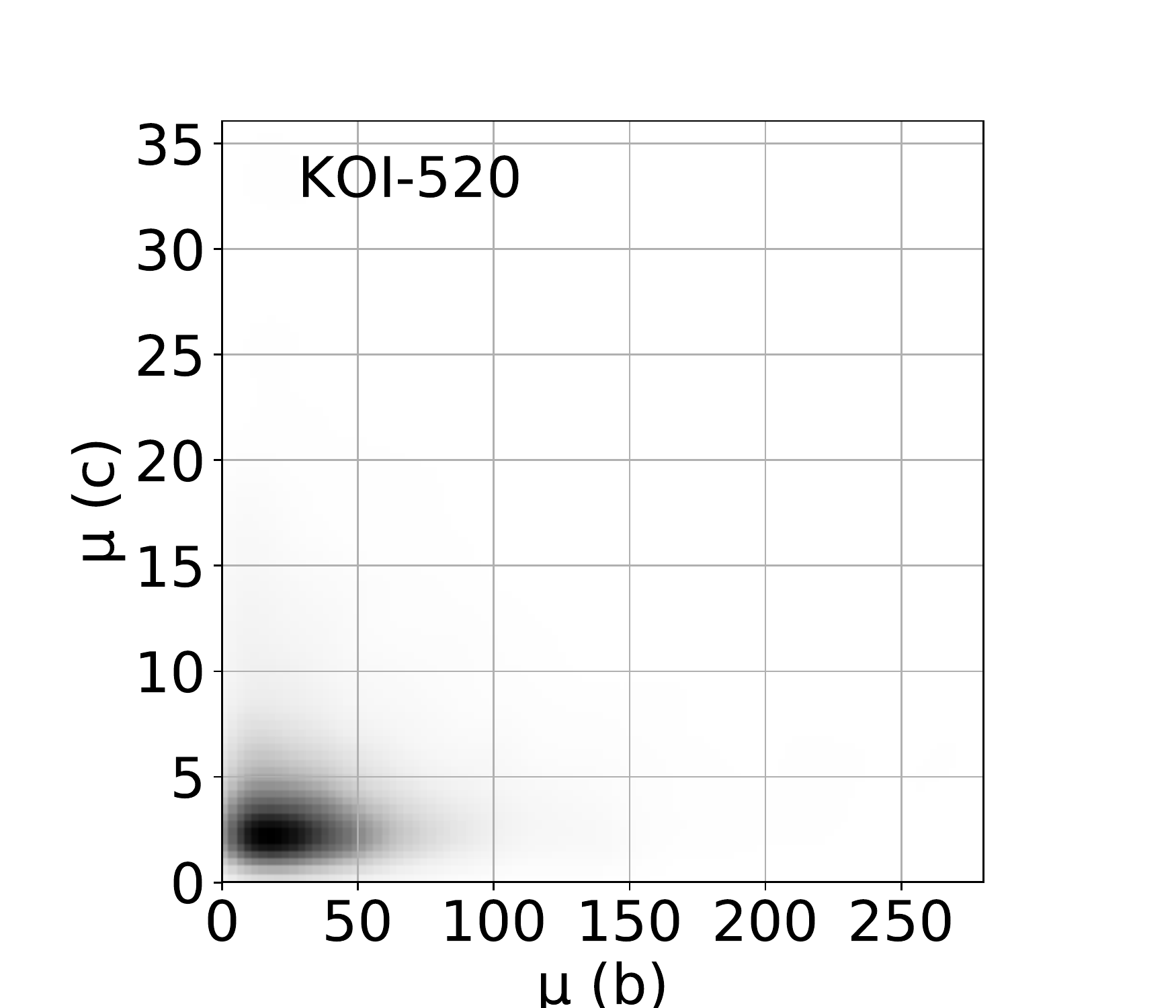}
\includegraphics [height = 1.1 in]{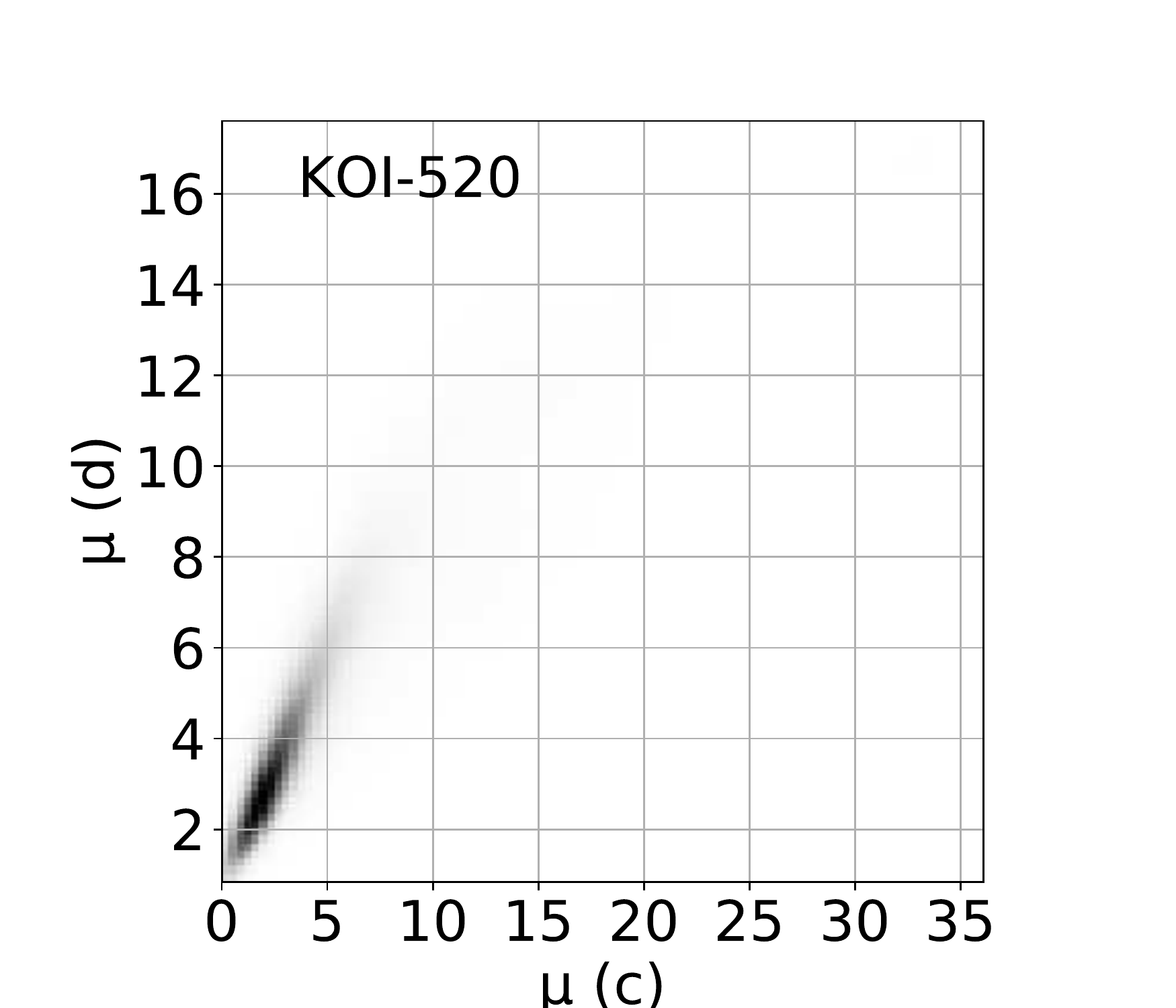}
\includegraphics [height = 1.1 in]{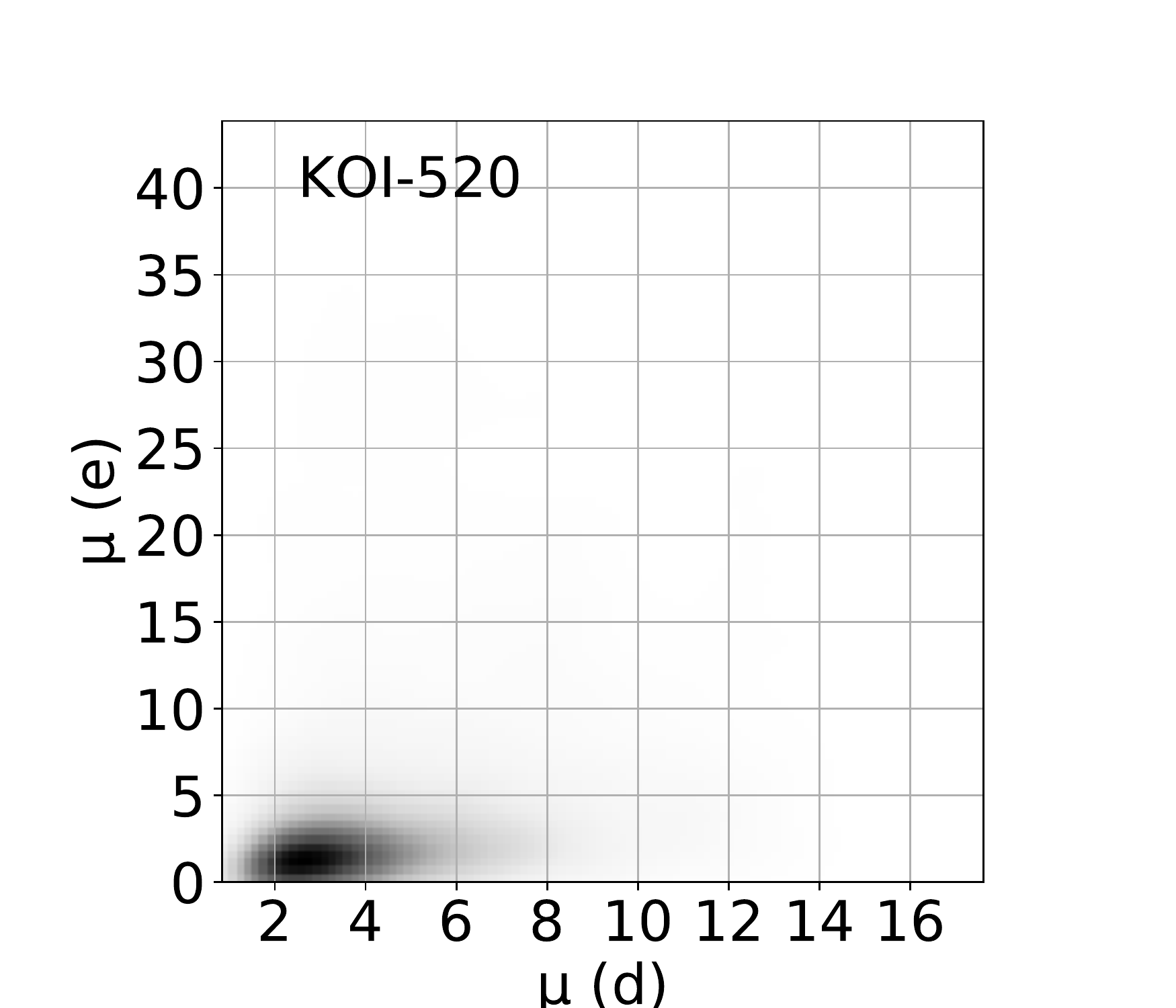} \\
\includegraphics [height = 1.1 in]{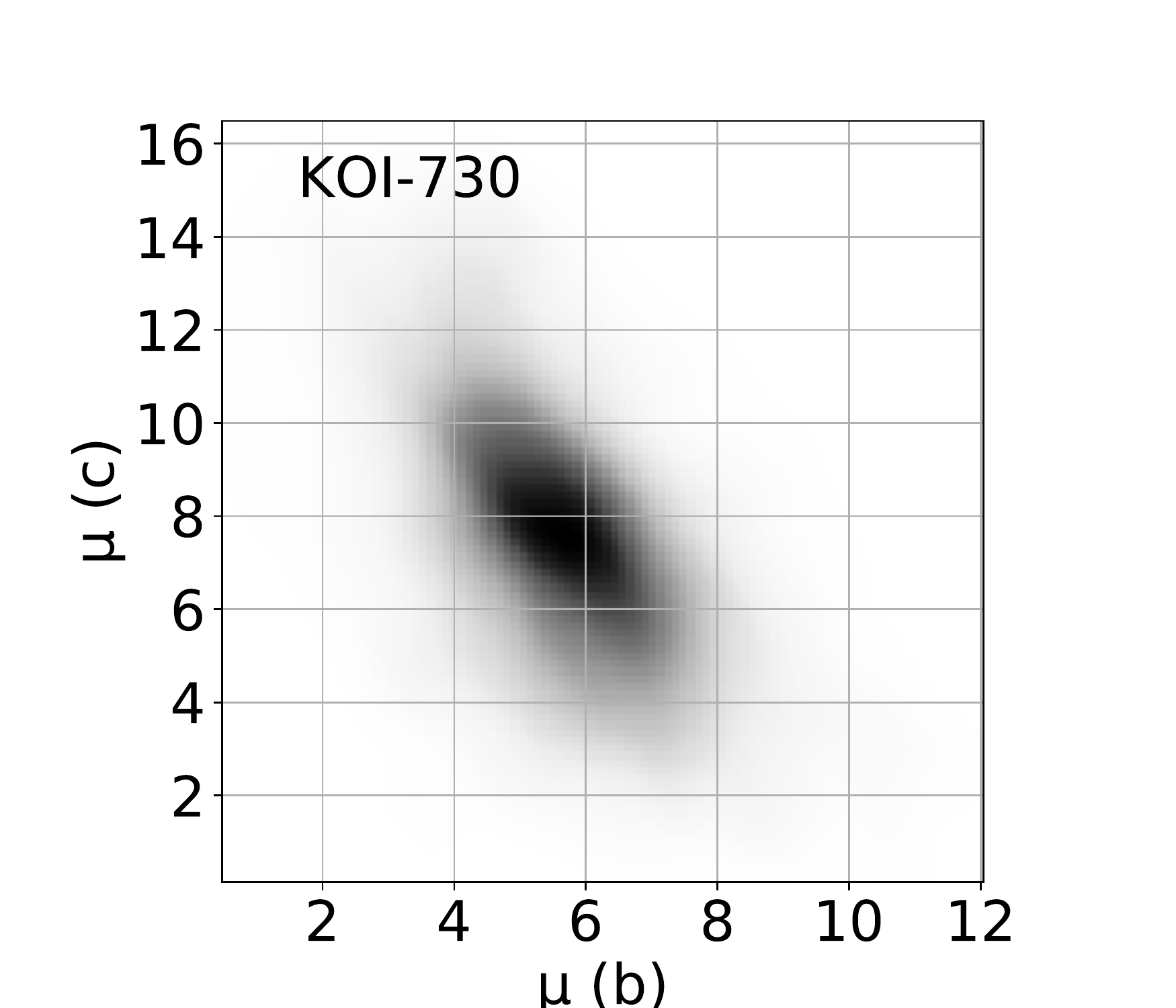}
\includegraphics [height = 1.1 in]{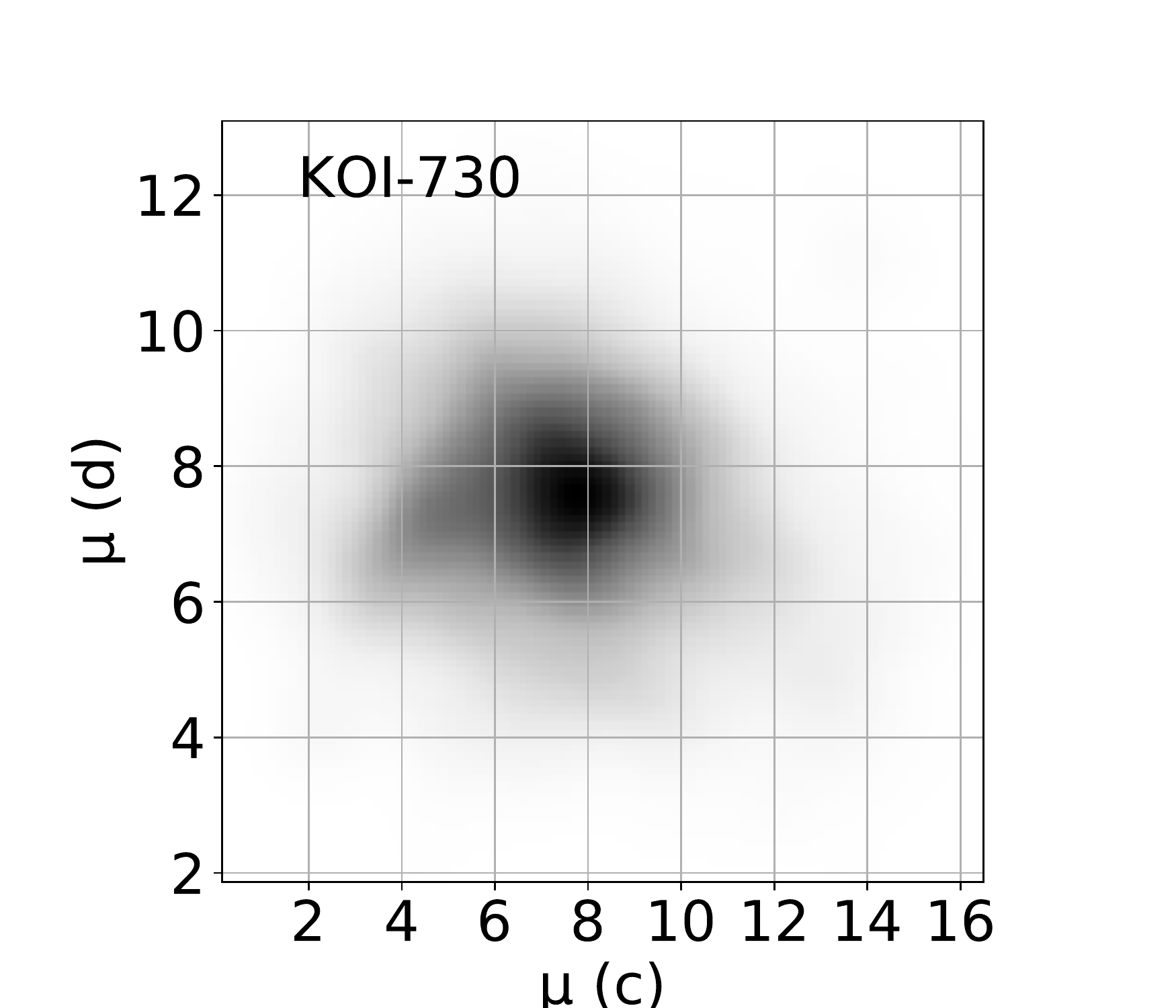}
\includegraphics [height = 1.1 in]{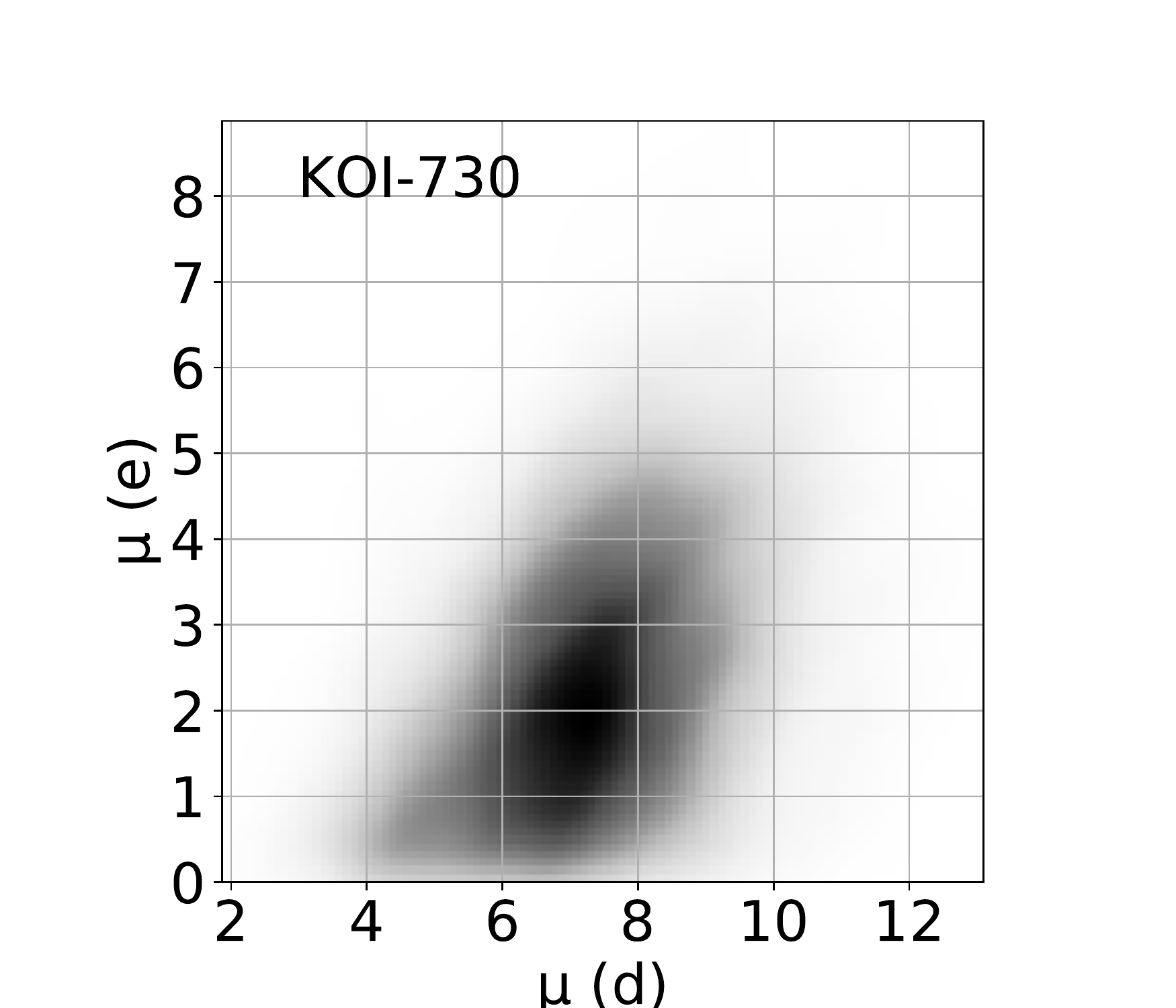}
\includegraphics [height = 1.1 in]{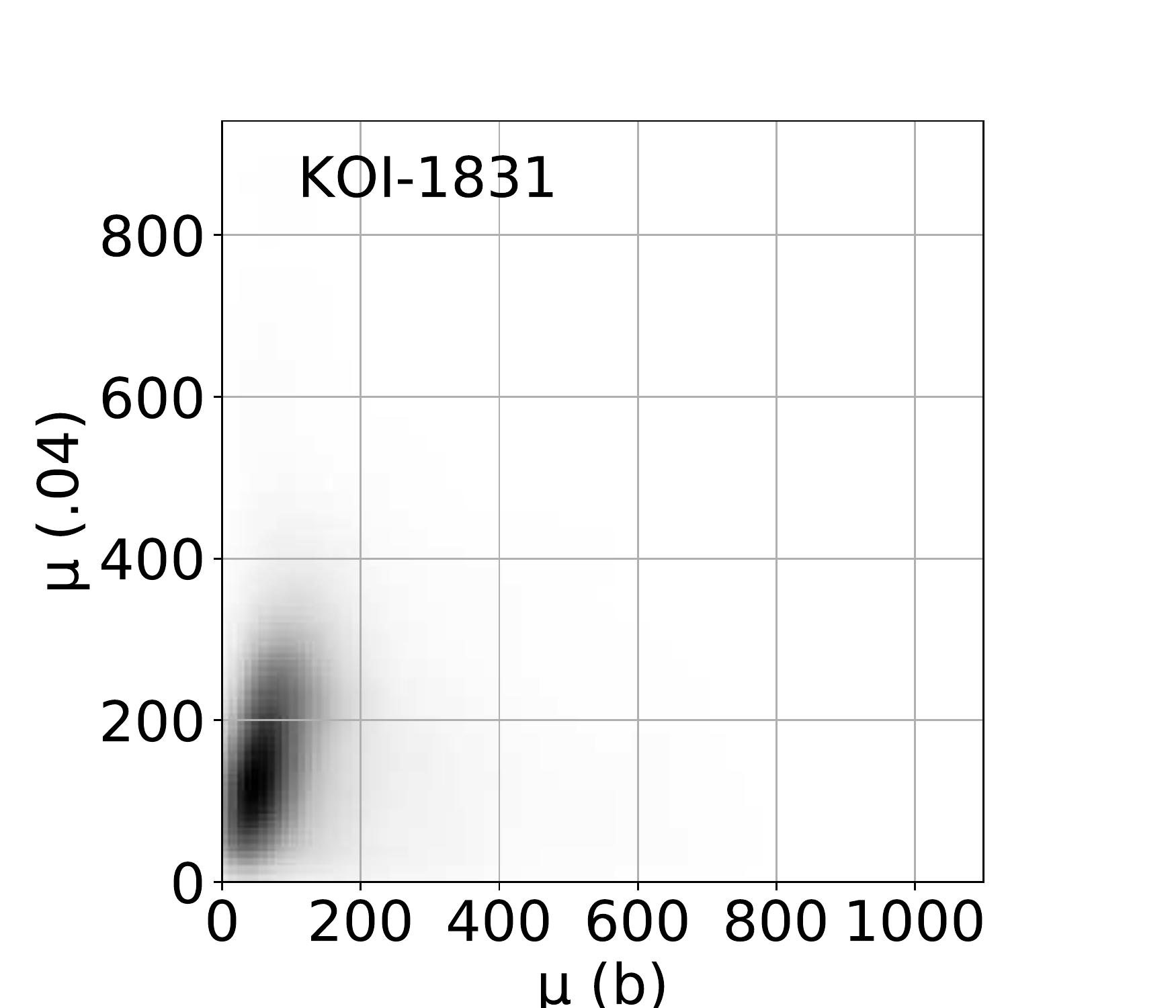} \\
\includegraphics [height = 1.1 in]{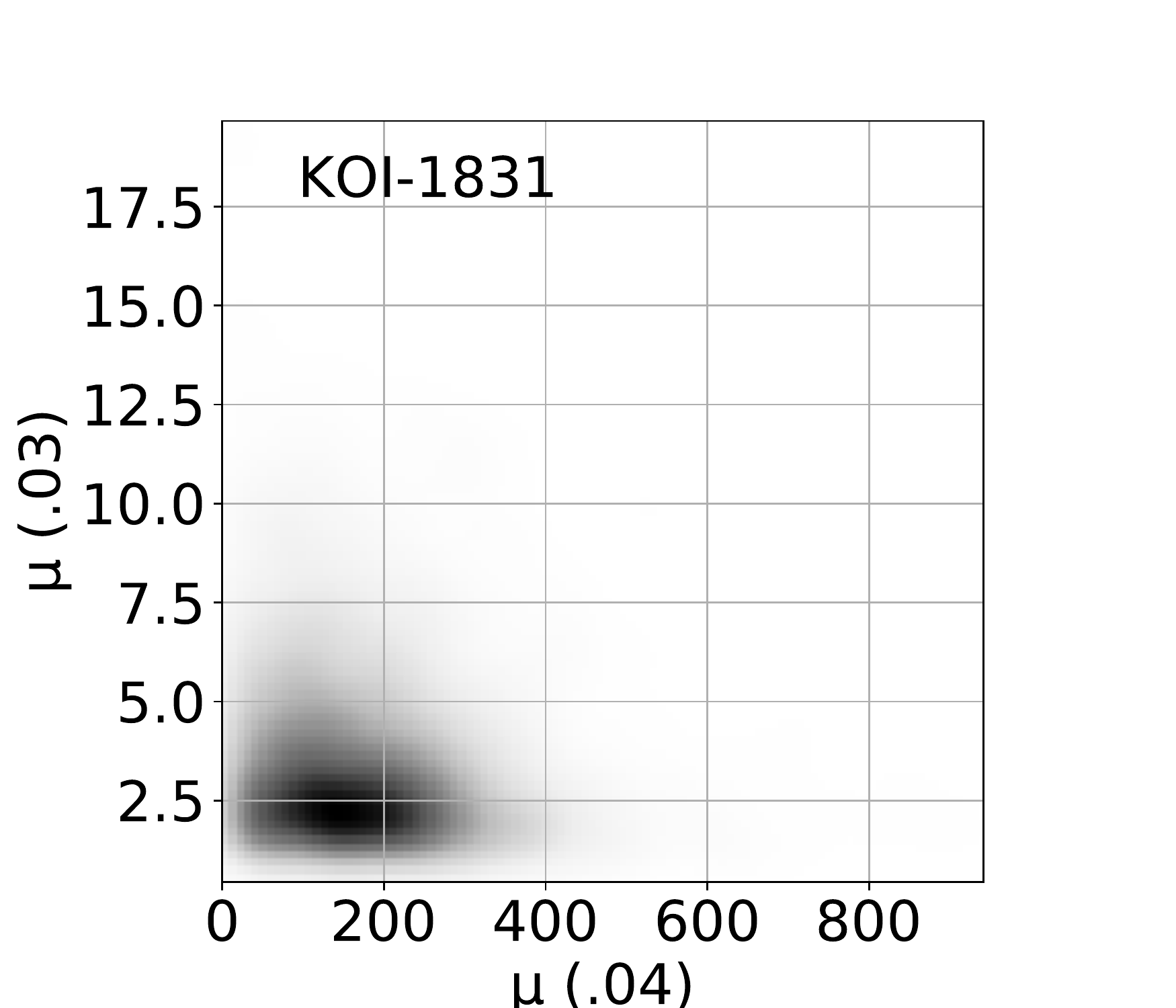}
\includegraphics [height = 1.1 in]{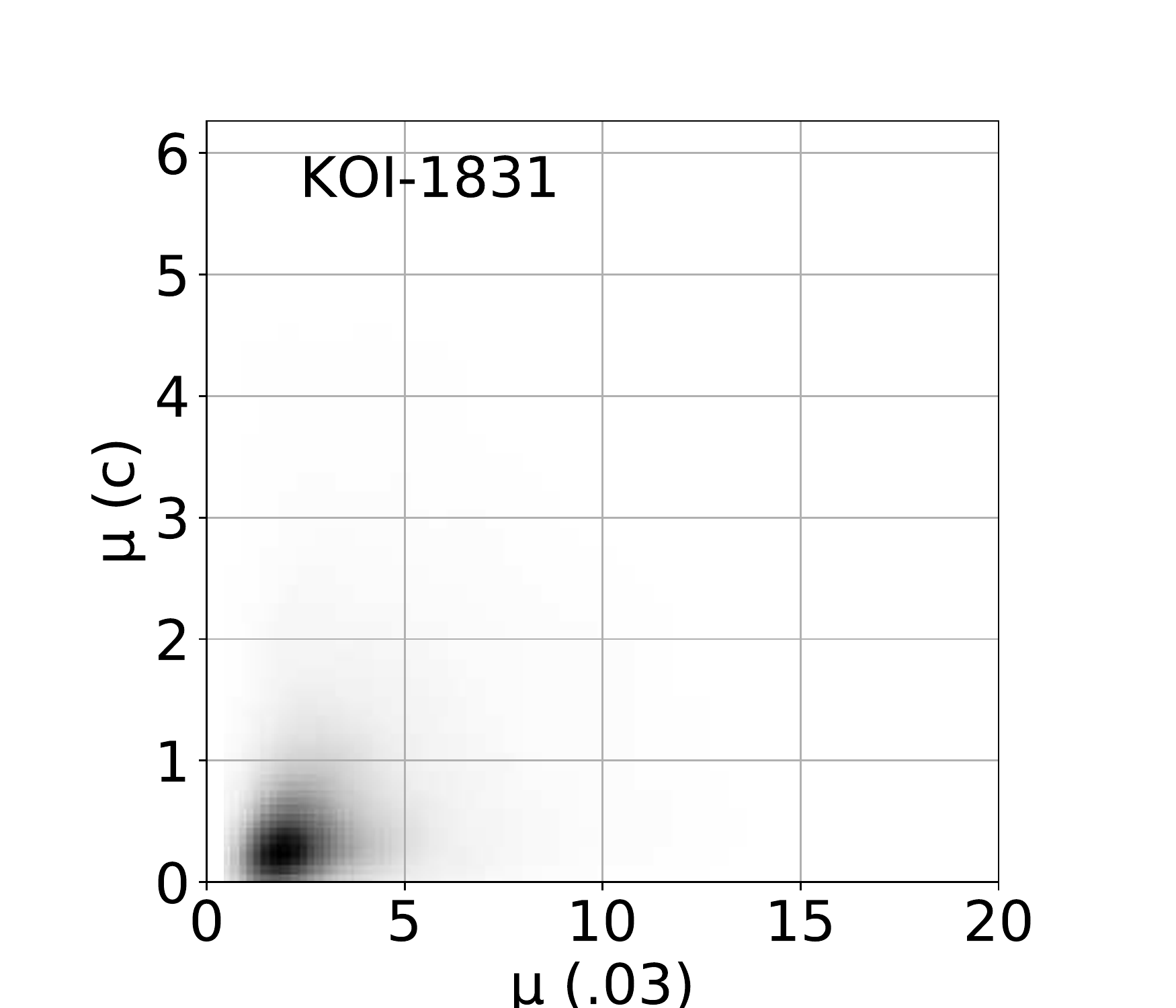} 
\includegraphics [height = 1.1 in]{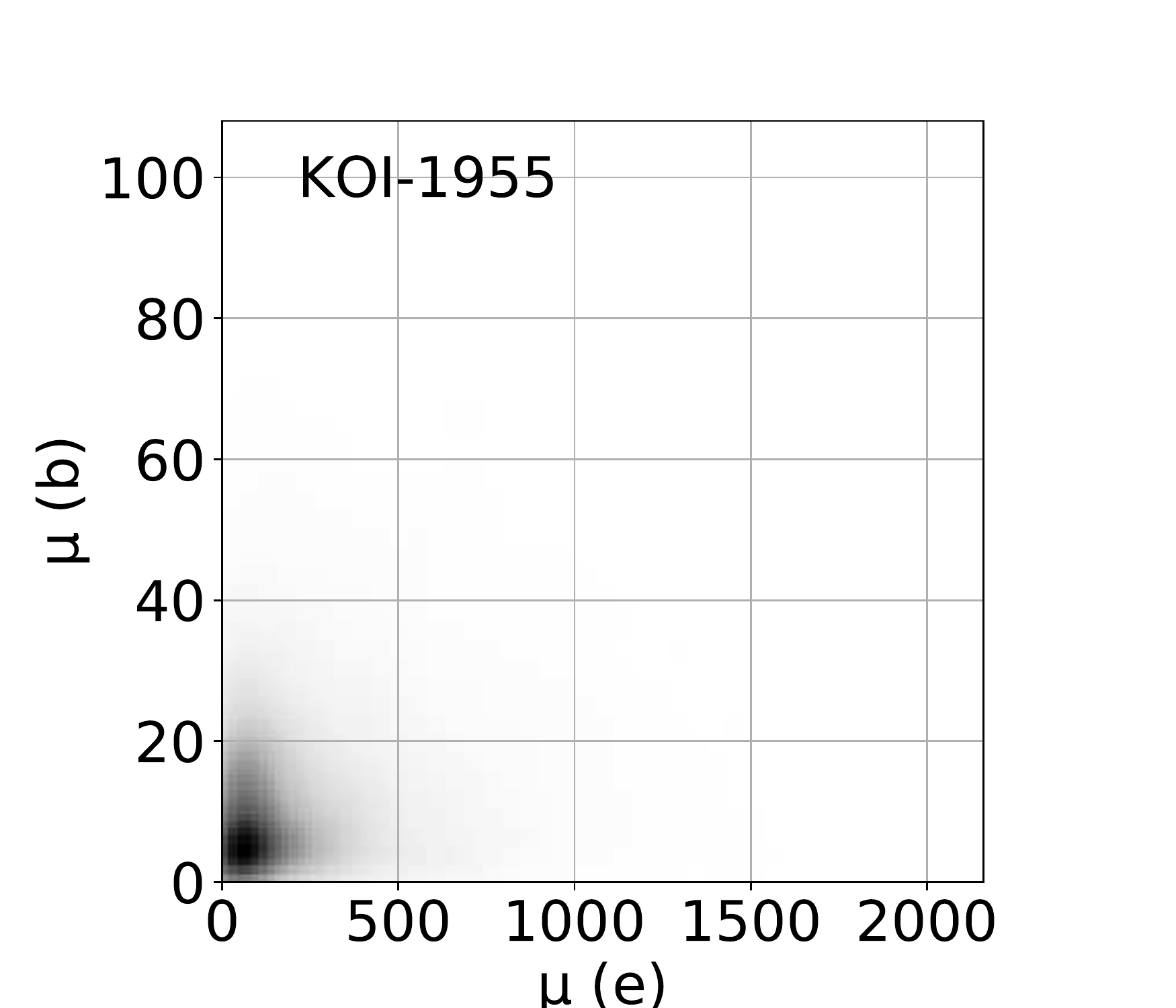}
\includegraphics [height = 1.1 in]{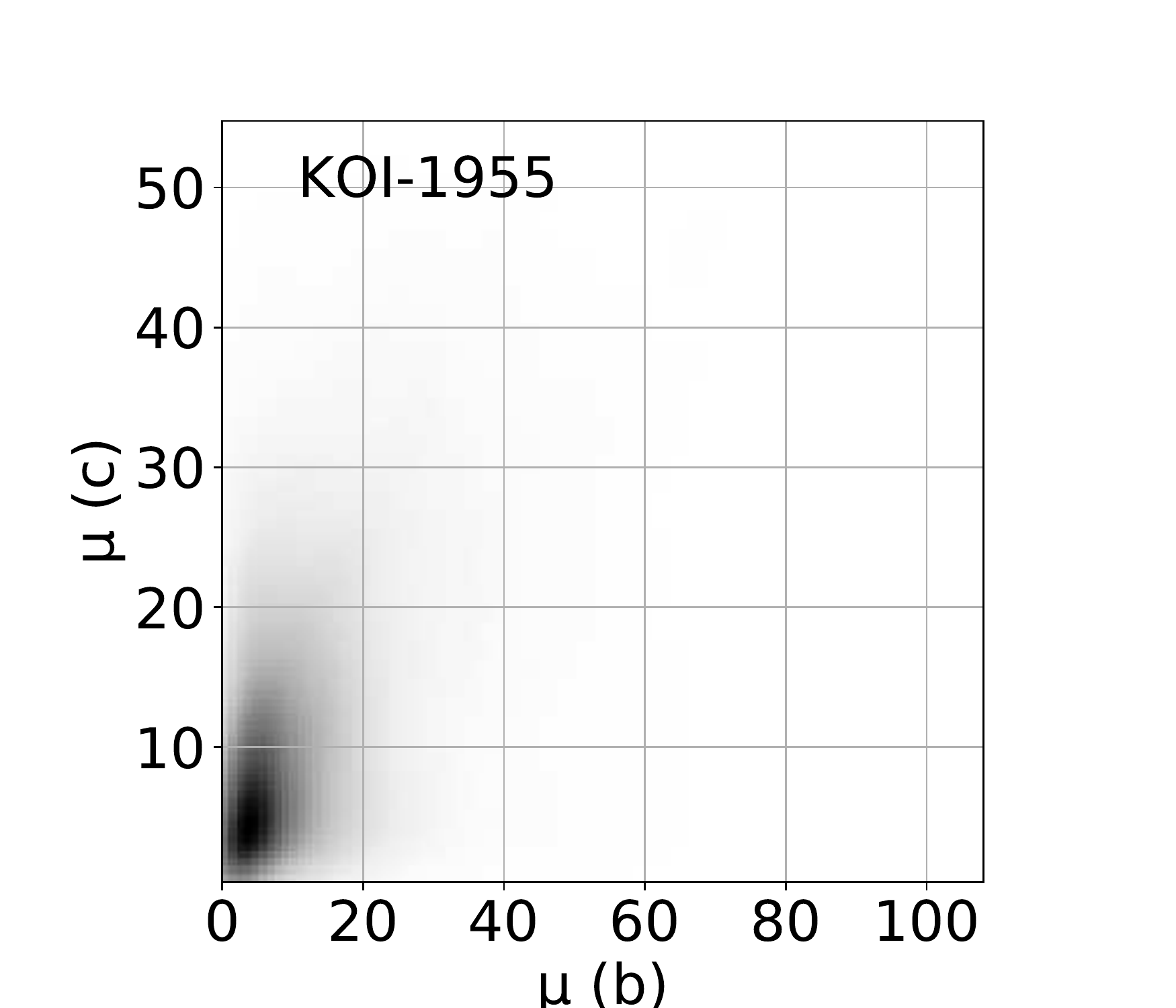} \\
\includegraphics [height = 1.1 in]{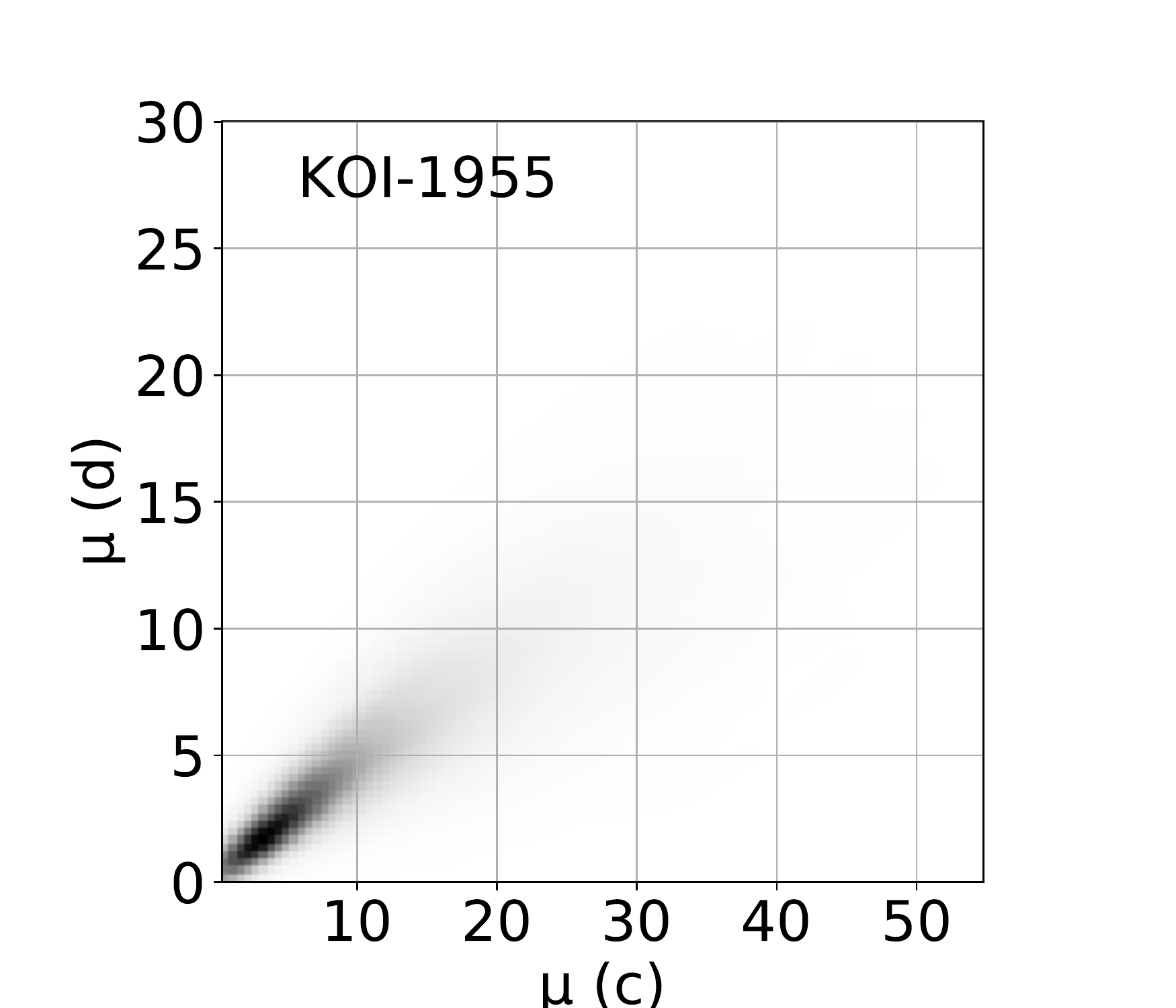}
\includegraphics [height = 1.1 in]{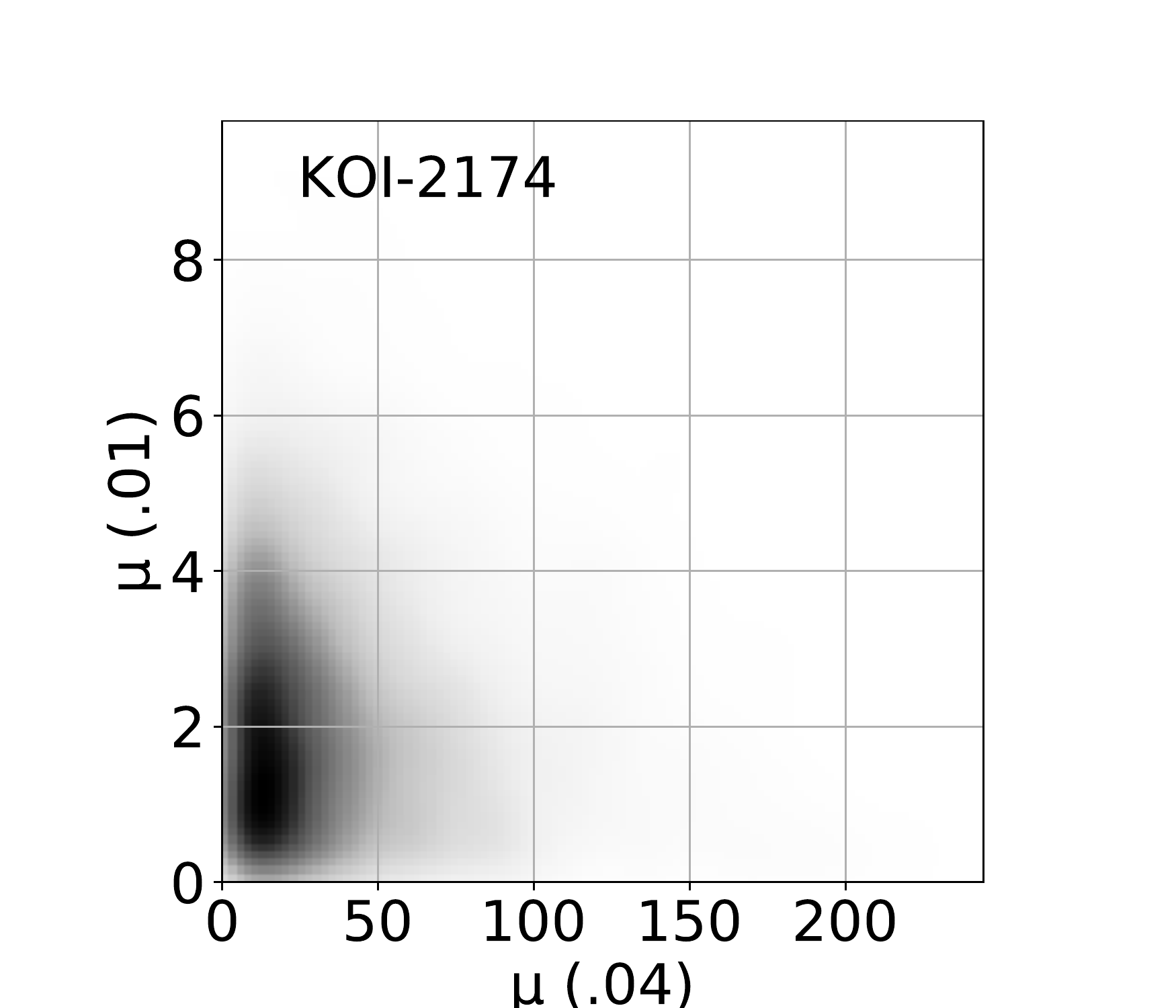} 
\includegraphics [height = 1.1 in]{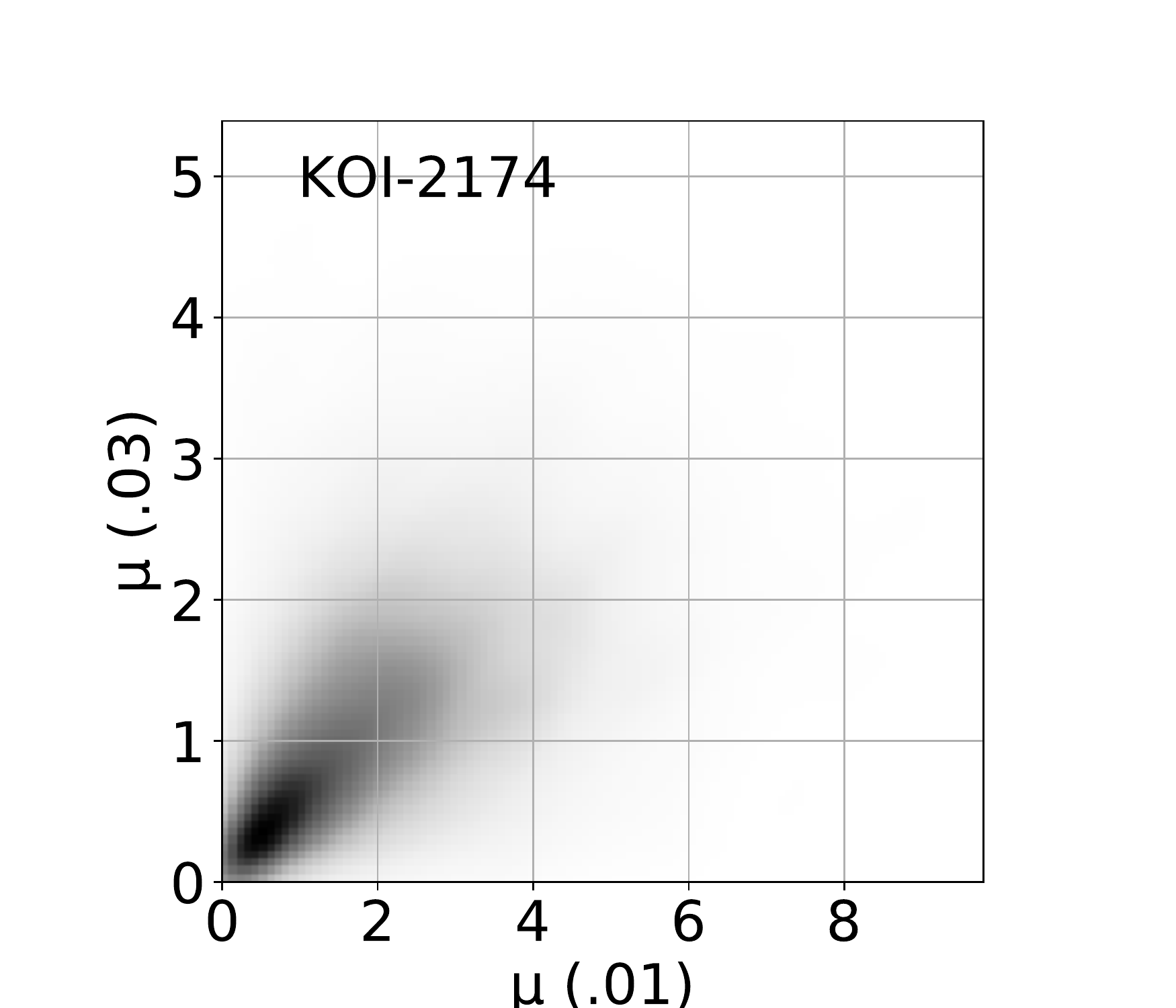}
\includegraphics [height = 1.1 in]{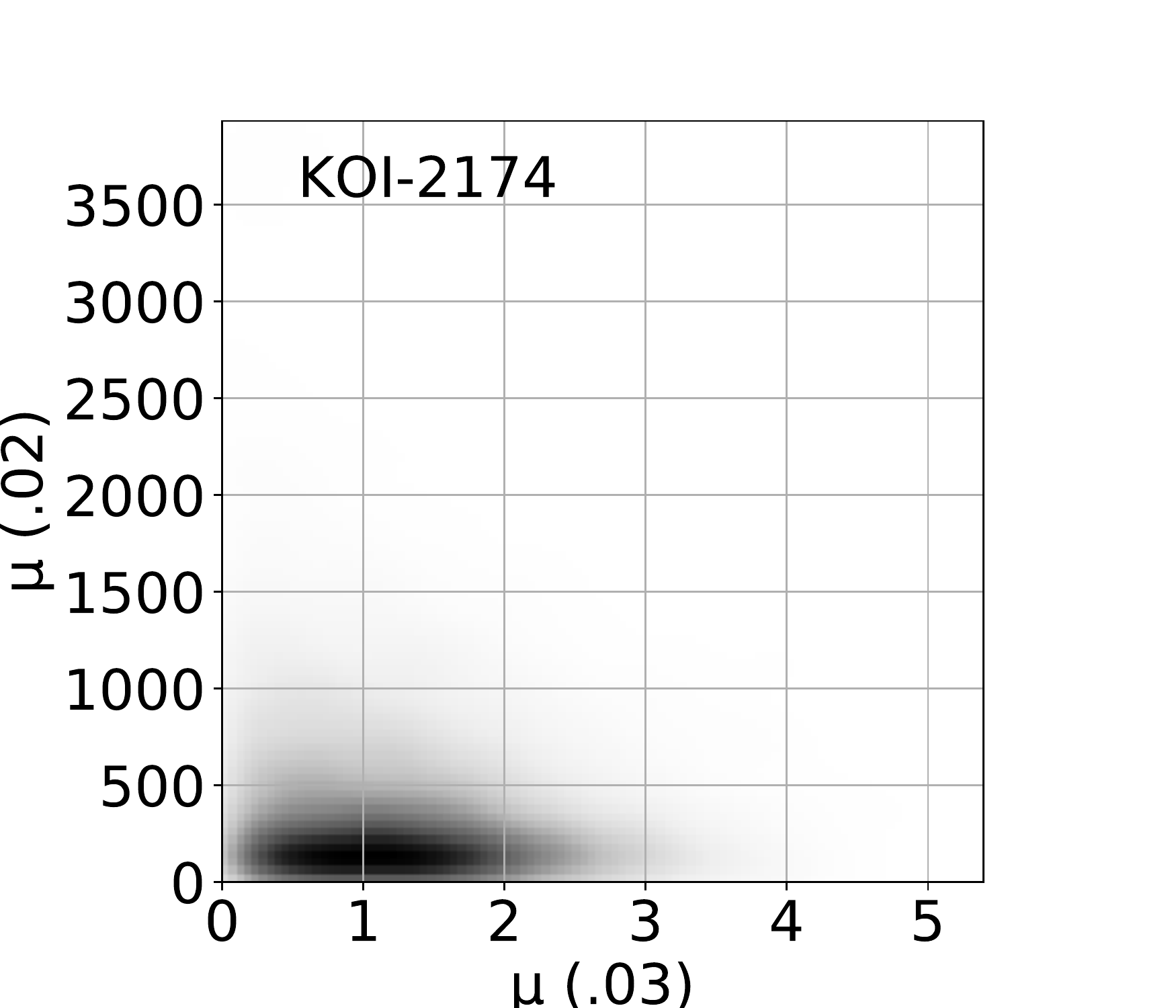} 
\caption{Two-dimensional kernel density estimators on joint posteriors of dynamical masses, $\mu$, scaled by factor $\frac{M_{\odot}}{M_{\oplus}}$: four-planet systems. 
\label{fig:mu4a} }
\end{center}
\end{figure}

\clearpage

\section*{Appendix C: Eccentricity constraints for individual candidates}
\begin{table}[ht!]
\tiny
  \begin{center}
    \begin{tabular}{|c|c||c|c||}
      \hline
      KOI &   P (days) &  e$\sin\omega$   &  e$\cos\omega$  \\
\hline
222.01 & 6.312 & -0.003 $^{+ 0.008 }_{- 0.018 }$  & -0.008 $^{+ 0.036 }_{- 0.010 }$     \\ 
 222.02 & 12.795 & 0.004 $^{+ 0.020 }_{- 0.046 }$  & -0.041 $^{+ 0.098 }_{- 0.029 }$     \\ 
 244.01 & 12.721 & 0.017 $^{+ 0.070 }_{- 0.044 }$  & 0.009 $^{+ 0.039 }_{- 0.030 }$     \\ 
 244.02 & 6.238 & 0.057 $^{+ 0.055 }_{- 0.038 }$  & 0.018 $^{+ 0.031 }_{- 0.017 }$     \\ 
 255.01 & 27.522 & 0.007 $^{+ 0.069 }_{- 0.085 }$  & 0.003 $^{+ 0.038 }_{- 0.039 }$     \\ 
 255.02 & 13.603 & 0.002 $^{+ 0.031 }_{- 0.032 }$  & 0.004 $^{+ 0.020 }_{- 0.033 }$     \\ 
 277.01 & 16.219 & -0.039 $^{+ 0.031 }_{- 0.030 }$  & 0.054 $^{+ 0.021 }_{- 0.023 }$     \\ 
 277.02 & 13.868 & -0.025 $^{+ 0.034 }_{- 0.033 }$  & 0.051 $^{+ 0.023 }_{- 0.025 }$     \\ 
 430.01 & 12.376 & 0.004 $^{+ 0.065 }_{- 0.068 }$  & 0.005 $^{+ 0.066 }_{- 0.065 }$     \\ 
 430.02 & 9.341 & 0.003 $^{+ 0.075 }_{- 0.079 }$  & -0.004 $^{+ 0.079 }_{- 0.077 }$     \\ 
 523.01 & 49.410 & -0.013 $^{+ 0.057 }_{- 0.060 }$  & -0.026 $^{+ 0.064 }_{- 0.064 }$     \\ 
 523.02 & 36.856 & -0.012 $^{+ 0.066 }_{- 0.069 }$  & -0.024 $^{+ 0.074 }_{- 0.074 }$     \\ 
 654.01 & 8.603 & -0.000 $^{+ 0.074 }_{- 0.074 }$  & 0.038 $^{+ 0.073 }_{- 0.075 }$     \\ 
 654.02 & 10.217 & 0.004 $^{+ 0.068 }_{- 0.067 }$  & 0.027 $^{+ 0.068 }_{- 0.069 }$     \\ 
 738.01 & 10.338 & -0.029 $^{+ 0.073 }_{- 0.070 }$  & -0.005 $^{+ 0.073 }_{- 0.069 }$     \\ 
 738.02 & 13.288 & -0.021 $^{+ 0.063 }_{- 0.061 }$  & 0.009 $^{+ 0.064 }_{- 0.060 }$     \\ 
 1279.01 & 14.375 & 0.002 $^{+ 0.065 }_{- 0.064 }$  & 0.014 $^{+ 0.065 }_{- 0.065 }$     \\ 
 1279.02 & 9.651 & -0.004 $^{+ 0.078 }_{- 0.077 }$  & -0.007 $^{+ 0.077 }_{- 0.079 }$     \\ 
 1599.01 & 20.407 & -0.063 $^{+ 0.082 }_{- 0.087 }$  & 0.052 $^{+ 0.057 }_{- 0.063 }$     \\ 
 1599.02 & 13.618 & 0.048 $^{+ 0.090 }_{- 0.086 }$  & -0.039 $^{+ 0.083 }_{- 0.085 }$     \\ 
 1783.01 & 134.457 & -0.038 $^{+ 0.012 }_{- 0.021 }$  & 0.008 $^{+ 0.008 }_{- 0.005 }$     \\ 
 1783.02 & 284.246 & -0.010 $^{+ 0.027 }_{- 0.032 }$  & 0.019 $^{+ 0.018 }_{- 0.015 }$     \\ 
 2113.01 & 15.943 & -0.020 $^{+ 0.070 }_{- 0.067 }$  & -0.008 $^{+ 0.066 }_{- 0.066 }$     \\ 
 2113.02 & 12.331 & 0.018 $^{+ 0.077 }_{- 0.078 }$  & 0.004 $^{+ 0.077 }_{- 0.077 }$     \\ 
 2414.01 & 22.595 & 0.078 $^{+ 0.058 }_{- 0.043 }$  & 0.047 $^{+ 0.046 }_{- 0.057 }$     \\ 
 2414.02 & 45.351 & -0.072 $^{+ 0.034 }_{- 0.037 }$  & -0.028 $^{+ 0.020 }_{- 0.008 }$     \\ 
 3503.01 & 21.191 & 0.014 $^{+ 0.076 }_{- 0.076 }$  & -0.143 $^{+ 0.095 }_{- 0.086 }$     \\ 
 3503.02 & 31.810 & -0.018 $^{+ 0.071 }_{- 0.070 }$  & 0.012 $^{+ 0.064 }_{- 0.061 }$     \\ 
 \hline
     \end{tabular}
    \caption{Eccentricity constraints for 2-planet systems, with a Gaussian prior of width 0.1 on eccentricity vector components.}\label{tbl-ecc2}
  \end{center}
\end{table}

\begin{table}[ht!]
\tiny
  \begin{center}
    \begin{tabular}{|c|c||c|c||}
      \hline
      KOI &   P (days) &  e$\sin\omega$   &  e$\cos\omega$  \\
\hline
 85.01 & 5.860 & -0.003 $^{+ 0.059 }_{- 0.055 }$  & 0.012 $^{+ 0.058 }_{- 0.051 }$     \\ 
 85.02 & 2.155 & -0.032 $^{+ 0.059 }_{- 0.067 }$  & -0.018 $^{+ 0.047 }_{- 0.063 }$     \\ 
 85.03 & 8.133 & -0.010 $^{+ 0.052 }_{- 0.049 }$  & 0.006 $^{+ 0.052 }_{- 0.051 }$     \\ 
 115.01 & 5.412 & -0.019 $^{+ 0.046 }_{- 0.044 }$  & -0.042 $^{+ 0.046 }_{- 0.042 }$     \\ 
 115.02 & 7.126 & -0.028 $^{+ 0.040 }_{- 0.038 }$  & -0.036 $^{+ 0.039 }_{- 0.036 }$     \\ 
 115.03 & 3.437 & -0.049 $^{+ 0.055 }_{- 0.055 }$  & -0.015 $^{+ 0.063 }_{- 0.054 }$     \\ 
 137.01 & 7.642 & -0.034 $^{+ 0.023 }_{- 0.028 }$  & 0.002 $^{+ 0.005 }_{- 0.004 }$     \\ 
 137.02 & 14.859 & -0.031 $^{+ 0.025 }_{- 0.033 }$  & -0.005 $^{+ 0.017 }_{- 0.017 }$     \\ 
 137.03 & 3.505 & 0.013 $^{+ 0.059 }_{- 0.034 }$  & -0.007 $^{+ 0.052 }_{- 0.042 }$     \\ 
 156.01 & 8.042 & -0.046 $^{+ 0.058 }_{- 0.058 }$  & 0.020 $^{+ 0.053 }_{- 0.054 }$     \\ 
 156.02 & 5.188 & 0.049 $^{+ 0.072 }_{- 0.071 }$  & -0.009 $^{+ 0.074 }_{- 0.075 }$     \\ 
 156.03 & 11.777 & -0.008 $^{+ 0.048 }_{- 0.048 }$  & -0.001 $^{+ 0.046 }_{- 0.047 }$     \\ 
 168.01 & 10.738 & -0.003 $^{+ 0.057 }_{- 0.055 }$  & 0.017 $^{+ 0.055 }_{- 0.054 }$     \\ 
 168.02 & 15.278 & 0.000 $^{+ 0.048 }_{- 0.047 }$  & 0.015 $^{+ 0.047 }_{- 0.049 }$     \\ 
 168.03 & 7.108 & 0.007 $^{+ 0.071 }_{- 0.067 }$  & -0.011 $^{+ 0.069 }_{- 0.071 }$     \\ 
 314.01 & 13.781 & -0.021 $^{+ 0.054 }_{- 0.053 }$  & -0.006 $^{+ 0.054 }_{- 0.058 }$     \\ 
 314.02 & 23.087 & 0.022 $^{+ 0.042 }_{- 0.040 }$  & 0.050 $^{+ 0.052 }_{- 0.046 }$     \\ 
 314.03 & 10.313 & -0.022 $^{+ 0.062 }_{- 0.061 }$  & -0.001 $^{+ 0.062 }_{- 0.068 }$     \\ 
 377.01 & 19.248 & 0.013 $^{+ 0.000 }_{- 0.000 }$  & 0.060 $^{+ 0.001 }_{- 0.001 }$     \\ 
 377.02 & 38.944 & 0.002 $^{+ 0.000 }_{- 0.000 }$  & -0.064 $^{+ 0.000 }_{- 0.000 }$     \\ 
 377.03 & 1.593 & 0.014 $^{+ 0.100 }_{- 0.102 }$  & 0.070 $^{+ 0.102 }_{- 0.103 }$     \\ 
 401.01 & 29.199 & 0.013 $^{+ 0.036 }_{- 0.033 }$  & -0.015 $^{+ 0.038 }_{- 0.045 }$     \\ 
 401.02 & 160.018 & -0.001 $^{+ 0.072 }_{- 0.062 }$  & 0.010 $^{+ 0.080 }_{- 0.082 }$     \\ 
 401.03 & 55.321 & -0.034 $^{+ 0.058 }_{- 0.069 }$  & 0.005 $^{+ 0.050 }_{- 0.058 }$     \\ 
 567.01 & 10.688 & 0.041 $^{+ 0.077 }_{- 0.074 }$  & -0.018 $^{+ 0.065 }_{- 0.073 }$     \\ 
 567.02 & 20.302 & 0.012 $^{+ 0.062 }_{- 0.064 }$  & 0.020 $^{+ 0.072 }_{- 0.080 }$     \\ 
 567.03 & 29.022 & 0.009 $^{+ 0.053 }_{- 0.053 }$  & 0.021 $^{+ 0.059 }_{- 0.067 }$     \\ 
 620.01 & 45.153 & -0.056 $^{+ 0.019 }_{- 0.022 }$  & -0.019 $^{+ 0.009 }_{- 0.010 }$     \\ 
 620.02 & 130.184 & -0.034 $^{+ 0.024 }_{- 0.024 }$  & 0.014 $^{+ 0.011 }_{- 0.011 }$     \\ 
 620.03 & 85.318 & -0.044 $^{+ 0.028 }_{- 0.028 }$  & 0.025 $^{+ 0.014 }_{- 0.014 }$     \\ 
 750.01 & 21.678 & 0.002 $^{+ 0.091 }_{- 0.052 }$  & 0.031 $^{+ 0.052 }_{- 0.054 }$     \\ 
 750.02 & 5.044 & -0.031 $^{+ 0.103 }_{- 0.095 }$  & 0.002 $^{+ 0.070 }_{- 0.070 }$     \\ 
 750.03 & 14.515 & -0.008 $^{+ 0.109 }_{- 0.061 }$  & 0.028 $^{+ 0.064 }_{- 0.064 }$     \\ 
 806.01 & 143.494 & -0.008 $^{+ 0.001 }_{- 0.001 }$  & -0.026 $^{+ 0.001 }_{- 0.001 }$     \\ 
 806.02 & 60.319 & -0.008 $^{+ 0.000 }_{- 0.000 }$  & 0.009 $^{+ 0.001 }_{- 0.001 }$     \\ 
 806.03 & 29.355 & 0.033 $^{+ 0.000 }_{- 0.000 }$  & -0.008 $^{+ 0.001 }_{- 0.001 }$     \\ 
 877.01 & 5.955 & 0.011 $^{+ 0.012 }_{- 0.010 }$  & -0.009 $^{+ 0.030 }_{- 0.008 }$     \\ 
 877.02 & 12.042 & 0.026 $^{+ 0.018 }_{- 0.029 }$  & -0.043 $^{+ 0.070 }_{- 0.028 }$     \\ 
 877.03 & 20.836 & -0.047 $^{+ 0.070 }_{- 0.031 }$  & -0.056 $^{+ 0.083 }_{- 0.025 }$     \\ 
 886.01 & 8.009 & -0.038 $^{+ 0.065 }_{- 0.066 }$  & -0.004 $^{+ 0.063 }_{- 0.061 }$     \\ 
 886.02 & 12.068 & -0.051 $^{+ 0.052 }_{- 0.054 }$  & 0.001 $^{+ 0.051 }_{- 0.050 }$     \\ 
 886.03 & 20.996 & 0.016 $^{+ 0.039 }_{- 0.042 }$  & 0.020 $^{+ 0.041 }_{- 0.039 }$     \\
 934.01 & 5.827 & -0.041 $^{+ 0.048 }_{- 0.058 }$  & 0.010 $^{+ 0.041 }_{- 0.051 }$     \\ 
 934.02 & 12.407 & 0.009 $^{+ 0.062 }_{- 0.070 }$  & 0.027 $^{+ 0.069 }_{- 0.074 }$     \\ 
 934.03 & 18.748 & 0.018 $^{+ 0.050 }_{- 0.057 }$  & 0.037 $^{+ 0.058 }_{- 0.055 }$     \\ 
 1070.01 & 6.618 & -0.010 $^{+ 0.102 }_{- 0.101 }$  & 0.003 $^{+ 0.094 }_{- 0.097 }$     \\ 
 1070.02 & 107.723 & 0.003 $^{+ 0.054 }_{- 0.053 }$  & 0.003 $^{+ 0.054 }_{- 0.053 }$     \\ 
 1070.03 & 92.783 & 0.002 $^{+ 0.058 }_{- 0.057 }$  & 0.002 $^{+ 0.058 }_{- 0.057 }$     \\ 
 1338.01 & 3.223 & -0.003 $^{+ 0.098 }_{- 0.088 }$  & -0.011 $^{+ 0.100 }_{- 0.099 }$     \\ 
 1338.02 & 42.041 & -0.020 $^{+ 0.089 }_{- 0.076 }$  & -0.051 $^{+ 0.083 }_{- 0.059 }$     \\ 
 1338.03 & 21.016 & 0.017 $^{+ 0.056 }_{- 0.057 }$  & 0.038 $^{+ 0.067 }_{- 0.068 }$     \\ 
 1353.01 & 125.870 & -0.001 $^{+ 0.008 }_{- 0.011 }$  & 0.009 $^{+ 0.005 }_{- 0.005 }$     \\ 
 1353.02 & 34.543 & -0.058 $^{+ 0.037 }_{- 0.041 }$  & -0.004 $^{+ 0.029 }_{- 0.027 }$     \\ 
Kep-289d & 65.959 & -0.027 $^{+ 0.007 }_{- 0.009 }$  & -0.003 $^{+ 0.001 }_{- 0.001 }$     \\ 
 1574.01 & 114.756 & 0.001 $^{+ 0.057 }_{- 0.060 }$  & -0.003 $^{+ 0.065 }_{- 0.061 }$     \\ 
 1574.02 & 189.802 & -0.004 $^{+ 0.040 }_{- 0.047 }$  & -0.006 $^{+ 0.054 }_{- 0.048 }$     \\ 
 1574.03 & 5.834 & -0.011 $^{+ 0.103 }_{- 0.096 }$  & 0.008 $^{+ 0.099 }_{- 0.107 }$     \\ 
 1576.01 & 10.421 & -0.032 $^{+ 0.050 }_{- 0.053 }$  & 0.015 $^{+ 0.037 }_{- 0.034 }$     \\ 
 1576.02 & 13.075 & -0.023 $^{+ 0.045 }_{- 0.048 }$  & 0.008 $^{+ 0.033 }_{- 0.030 }$     \\ 
 1576.03 & 23.341 & -0.064 $^{+ 0.142 }_{- 0.098 }$  & -0.012 $^{+ 0.039 }_{- 0.092 }$     \\ 
 1598.01 & 56.476 & 0.026 $^{+ 0.054 }_{- 0.047 }$  & -0.004 $^{+ 0.042 }_{- 0.046 }$     \\ 
 1598.02 & 92.876 & -0.000 $^{+ 0.047 }_{- 0.047 }$  & -0.004 $^{+ 0.039 }_{- 0.040 }$     \\ 
 1598.03 & 13.930 & 0.004 $^{+ 0.077 }_{- 0.075 }$  & 0.006 $^{+ 0.075 }_{- 0.072 }$     \\ 
 1833.01 & 3.693 & -0.008 $^{+ 0.018 }_{- 0.022 }$  & -0.019 $^{+ 0.016 }_{- 0.024 }$     \\ 
 1833.02 & 7.685 & 0.014 $^{+ 0.019 }_{- 0.028 }$  & -0.037 $^{+ 0.019 }_{- 0.027 }$     \\ 
 1833.03 & 5.707 & 0.009 $^{+ 0.023 }_{- 0.027 }$  & -0.038 $^{+ 0.021 }_{- 0.027 }$     \\ 
  2086.01 & 7.133 & 0.010 $^{+ 0.061 }_{- 0.060 }$  & 0.026 $^{+ 0.067 }_{- 0.069 }$     \\ 
 2086.02 & 8.919 & 0.035 $^{+ 0.054 }_{- 0.053 }$  & -0.000 $^{+ 0.062 }_{- 0.063 }$     \\ 
 2086.03 & 11.898 & 0.003 $^{+ 0.048 }_{- 0.047 }$  & 0.023 $^{+ 0.052 }_{- 0.053 }$     \\ 
 2092.01 & 57.711 & 0.068 $^{+ 0.064 }_{- 0.100 }$  & -0.008 $^{+ 0.056 }_{- 0.059 }$     \\ 
 2092.02 & 25.563 & 0.052 $^{+ 0.069 }_{- 0.094 }$  & 0.047 $^{+ 0.083 }_{- 0.050 }$     \\ 
 2092.03 & 77.085 & 0.028 $^{+ 0.065 }_{- 0.079 }$  & -0.055 $^{+ 0.057 }_{- 0.049 }$     \\ 
 2195.01 & 20.053 & 0.007 $^{+ 0.059 }_{- 0.052 }$  & 0.030 $^{+ 0.086 }_{- 0.059 }$     \\ 
 2195.02 & 30.091 & -0.064 $^{+ 0.060 }_{- 0.061 }$  & -0.074 $^{+ 0.066 }_{- 0.052 }$     \\ 
 2195.03 & 6.850 & -0.005 $^{+ 0.041 }_{- 0.056 }$  & -0.003 $^{+ 0.048 }_{- 0.057 }$     \\ 
 \hline
     \end{tabular}
    \caption{Eccentricity constraints for 3-planet systems, with a Gaussian prior of width 0.1 on eccentricity vector components.}\label{tbl-ecc3}
  \end{center}
\end{table}

\clearpage

\newpage

\begin{table}[ht!]
\tiny
  \begin{center}
    \begin{tabular}{|c|c||c|c|}
      \hline
      KOI &   P (days) &  e$\sin\omega$   &  e$\cos\omega$  \\
\hline
 152.01 & 52.092 & 0.009 $^{+ 0.021 }_{- 0.022 }$  & -0.009 $^{+ 0.034 }_{- 0.032 }$     \\ 
 152.02 & 27.404 & -0.028 $^{+ 0.016 }_{- 0.020 }$  & -0.024 $^{+ 0.012 }_{- 0.018 }$     \\ 
 152.03 & 13.484 & -0.001 $^{+ 0.005 }_{- 0.004 }$  & -0.020 $^{+ 0.008 }_{- 0.014 }$     \\ 
 152.04 & 81.072 & -0.012 $^{+ 0.019 }_{- 0.019 }$  & -0.011 $^{+ 0.027 }_{- 0.026 }$     \\ 
 248.01 & 7.203 & 0.043 $^{+ 0.044 }_{- 0.049 }$  & -0.053 $^{+ 0.056 }_{- 0.047 }$     \\ 
 248.02 & 10.913 & 0.035 $^{+ 0.036 }_{- 0.043 }$  & -0.043 $^{+ 0.042 }_{- 0.037 }$     \\ 
 248.03 & 2.577 & -0.053 $^{+ 0.085 }_{- 0.093 }$  & 0.028 $^{+ 0.076 }_{- 0.077 }$     \\ 
 248.04 & 18.597 & -0.045 $^{+ 0.088 }_{- 0.062 }$  & 0.090 $^{+ 0.072 }_{- 0.137 }$     \\ 
 250.01 & 12.283 & -0.038 $^{+ 0.041 }_{- 0.045 }$  & 0.058 $^{+ 0.043 }_{- 0.043 }$     \\ 
 250.02 & 17.251 & -0.041 $^{+ 0.034 }_{- 0.037 }$  & 0.040 $^{+ 0.035 }_{- 0.036 }$     \\ 
 250.03 & 3.544 & -0.038 $^{+ 0.086 }_{- 0.078 }$  & 0.040 $^{+ 0.076 }_{- 0.076 }$     \\ 
 250.04 & 46.831 & -0.026 $^{+ 0.058 }_{- 0.062 }$  & -0.037 $^{+ 0.068 }_{- 0.060 }$     \\ 
 520.01 & 12.758 & 0.015 $^{+ 0.016 }_{- 0.010 }$  & 0.009 $^{+ 0.016 }_{- 0.010 }$     \\ 
 520.02 & 5.433 & 0.042 $^{+ 0.058 }_{- 0.046 }$  & -0.030 $^{+ 0.035 }_{- 0.042 }$     \\ 
 520.03 & 25.754 & 0.005 $^{+ 0.019 }_{- 0.014 }$  & -0.001 $^{+ 0.033 }_{- 0.024 }$     \\ 
 520.04 & 51.167 & -0.023 $^{+ 0.067 }_{- 0.069 }$  & -0.031 $^{+ 0.043 }_{- 0.023 }$     \\ 
 730.01 & 14.788 & -0.010 $^{+ 0.033 }_{- 0.038 }$  & 0.006 $^{+ 0.035 }_{- 0.034 }$     \\ 
 730.02 & 9.848 & 0.062 $^{+ 0.062 }_{- 0.068 }$  & -0.002 $^{+ 0.055 }_{- 0.056 }$     \\ 
 730.03 & 19.728 & 0.027 $^{+ 0.037 }_{- 0.041 }$  & 0.016 $^{+ 0.048 }_{- 0.043 }$     \\ 
 730.04 & 7.384 & -0.002 $^{+ 0.050 }_{- 0.067 }$  & 0.039 $^{+ 0.040 }_{- 0.041 }$     \\ 
   1831.01 & 51.832 & 0.010 $^{+ 0.046 }_{- 0.042 }$  & 0.011 $^{+ 0.039 }_{- 0.040 }$     \\ 
 1831.02 & 4.386 & 0.045 $^{+ 0.064 }_{- 0.062 }$  & 0.060 $^{+ 0.068 }_{- 0.092 }$     \\ 
 1831.03 & 34.169 & -0.007 $^{+ 0.057 }_{- 0.052 }$  & -0.012 $^{+ 0.045 }_{- 0.046 }$     \\ 
 1831.04 & 13.981 & 0.039 $^{+ 0.054 }_{- 0.053 }$  & -0.021 $^{+ 0.051 }_{- 0.045 }$     \\ 
 1955.01 & 15.170 & 0.034 $^{+ 0.069 }_{- 0.080 }$  & 0.062 $^{+ 0.067 }_{- 0.078 }$     \\ 
 1955.02 & 39.462 & 0.025 $^{+ 0.044 }_{- 0.049 }$  & 0.036 $^{+ 0.048 }_{- 0.052 }$     \\ 
 1955.03 & 1.644 & -0.001 $^{+ 0.097 }_{- 0.100 }$  & 0.000 $^{+ 0.103 }_{- 0.098 }$     \\ 
 1955.04 & 26.237 & 0.024 $^{+ 0.055 }_{- 0.061 }$  & 0.029 $^{+ 0.053 }_{- 0.056 }$     \\ 
 2174.01 & 6.695 & 0.035 $^{+ 0.066 }_{- 0.079 }$  & -0.009 $^{+ 0.049 }_{- 0.053 }$     \\ 
 2174.02 & 33.136 & 0.008 $^{+ 0.050 }_{- 0.061 }$  & -0.008 $^{+ 0.059 }_{- 0.056 }$     \\ 
 2174.03 & 7.725 & 0.031 $^{+ 0.061 }_{- 0.072 }$  & -0.008 $^{+ 0.046 }_{- 0.050 }$     \\ 
 2174.04 & 3.016 & 0.004 $^{+ 0.072 }_{- 0.073 }$  & 0.019 $^{+ 0.066 }_{- 0.067 }$     \\ 
 \hline
     \end{tabular}
    \caption{Eccentricity constraints for 4-planet systems, with a Gaussian prior of width 0.1 on eccentricity vector components.}\label{tbl-ecc4}
  \end{center}
\end{table}

\begin{table}[ht!]
 \tiny
       \begin{center}
       \vspace{-0.15 in}
    \begin{tabular}{|c|c|c|}
      \hline
      KOI  &  P (days)   &     P' (days) \\
\hline
KOI-152 &  13.4844 & 27.4024 \\
KOI-152 &  27.4024 &  52.0908 \\
KOI-152 & 52.0908 & 81.0702 \\
KOI-168 & 7.1079 & 10.7414  \\
KOI-168 &  6.3121 & 12.7951  \\
KOI-244 & 6.2383 & 12.7205  \\
KOI-250 * & 12.2825 & 17.2510 \\
KOI-277 * &  13.8680 & 16.2185 \\
KOI-314 * & 10.3108 & 13.7809 \\
KOI-314 * & 13.7809 & 23.0890 \\
KOI-520 & 12.758 & 25.754 \\
KOI-523 * & 36.8581 & 49.4103 \\
KOI-620 * & 45.1540 &  85.3159 \\
KOI-620 * &  85.3159 & 130.1818 \\
KOI-730 * &  7.3841 & 9.8476 \\
KOI-730 * & 9.8476 & 14.7880 \\
KOI-738 * & 10.3385 & 13.2880 \\
KOI-806 * & 29.3547 & 60.3189 \\
KOI-806 & 60.3189 & 143.5056 \\
KOI0-877 * & 5.9547 & 12.042 \\
KOI-886  & 8.0096 & 12.0699 \\
KOI-1353 &  34.545 & 65.959 \\
KOI-1353* & 65.959 & 125.870 \\
KOI-1576 * & 10.4220 & 13.0723 \\
KOI-1783 * & 134.4623 &   284.2931 \\
KOI-1831 * &  34.1671 &  51.8335 \\
KOI-1833 * &   5.7090 &  7.6850 \\
KOI-2086 * & 7.1333 & 8.9188 \\
KOI-2086 * &  8.9188 &  11.8977 \\
KOI-2195 * & 20.0570 &    30.0981 \\
KOI-3503  & 21.1876 &    31.8245 \\
 \hline
     \end{tabular}
    \caption{Detections of non-zero eccentricity among planet pairs following joint posteriors of eccentricity vector components. The left column identifies the system and the remaining two columns in each planet identifies the planet candidates of interacting pairs by the orbital periods of inner (P) and outer (P$'$) planets in each pair. Pairs with a prior expectation of detectable interactions, with a resonant or non-resonant TTV score above 5 in Tables~\ref{tbl-SystemSelection1} or Tables~\ref{tbl-SystemSelection2}  are marked with an asterisk.}\label{tbl:Ecc-detections}
  \end{center}
\end{table}

 \begin{table}[ht!]
 \tiny
       \begin{center}
       \vspace{-0.15 in}
    \begin{tabular}{|c|c|c|}
      \hline
   KOI  &  P (days)   &  P' (days) \\
\hline
 KOI-85  & 2.1549 & 5.8597 \\
 KOI-85 * &  5.8597 & 8.13135 \\
 KOI-115  & 3.4363 &  5.4120  \\
  KOI-115 * & 5.4120 & 7.1261 \\
  KOI-137 & 3.5047 & 7.6412 \\ 
  KOI-137 * & 7.6412 & 14.8594 \\
 KOI-156 & 5.1881 &  8.0414  \\
 KOI-156 * &  8.0414 &  11.7767 \\
 KOI-168 &  10.7414 & 15.2725 \\
 KOI-248 &  2.5766 & 7.2038 \\ 
 KOI-248 * &  7.2038 &  10.9116 \\ 
 KOI-248 & 10.9116 &  18.5973 \\
 KOI-250 & 3.5438 & 12.2825 \\
 KOI-250 & 17.2510 & 46.8314  \\
 KOI-255 * & 13.6026  & 27.5230   \\
 KOI-401 &  29.1983 &  55.3301 \\
 KOI-430 * & 9.3398 &  12.3765 \\
KOI-457 & 4.9218 & 7.0633 \\ 
 KOI-520 & 5.433 & 12.758 \\
 KOI-520 & 25.754 & 51.168 \\
 KOI-567 &  10.6877 & 20.3075 \\
 KOI-567 & 20.3075 & 29.0223 \\
 KOI-654 * & 8.5994 & 10.2211 \\
 KOI-730 * & 14.7880 & 19.7276 \\
 KOI-877 & 12.0428 & 20.8374 \\
 KOI-886  & 12.0699 & 20.9961 \\
 KOI-934  & 5.8266 & 12.3978 \\
 KOI-934 * & 12.3978 &  18.7528 \\
  KOI-1070 & 6.6183 & 92.7879 \\
KOI-1070 * &  92.7879  & 107.7301 \\
KOI-1279 & 9.6508 & 14.3750 \\
KOI-1338 & 3.2229 &  21.0193 \\
KOI-1338 * &  21.0193  & 42.0435 \\
KOI-1574 & 5.8339 &  114.7553 \\
KOI-1574 * & 114.7553 & 190.9922 \\
 KOI-1576 & 13.0723 & 23.3419 \\
KOI-1598  & 13.9305 & 56.4689 \\
KOI-1598 * & 56.4689 & 92.8801 \\
KOI-1599 * & 13.6180 &  20.4067 \\
KOI-1831 & 4.3852 &  13.9822 \\
KOI-1831 &  13.9822 &  34.1671 \\
KOI-1833 * & 3.6927 &   5.7090 \\
KOI-2092 * & 25.5636 &  57.6983 \\
KOI-2092 * &  57.6983 &  77.0851 \\
KOI-2113 & 12.3313 &  15.9423 \\
KOI-2174 & 3.0160 &  6.6951 \\
KOI-2174 * &  6.6951 &   7.7269 \\
KOI-2174 &   7.7269 &  33.1364 \\
KOI-2195 & 6.8497 & 20.0570 \\ 
KOI-2414 * &  22.5992 &   45.3492 \\
 \hline
     \end{tabular}
    \caption{Eccentricity `non-detections' among planet pairs following joint posteriors of eccentricity vector components. The left column in each panel identifies the system and the remaining two columns in each planet identifies the planet candidates of interacting pairs by the orbital periods of inner (P) and outer (P$'$) planets in each pair. Pairs with a prior expectation of detectable interactions, with a resonant or non-resonant TTV score above 5 in Tables~\ref{tbl-SystemSelection1} or Tables~\ref{tbl-SystemSelection2}  are marked with an asterisk.}\label{tbl:Ecc-nondetections}
  \end{center}
\end{table}

\clearpage
\section*{Appendix D: Estimated Radial Velocity Semiamplitudes}
\begin{table}[ht!]
\tiny
  \begin{center}
    \begin{tabular}{|c|c|c||c|c|c||c|c|c|}
      \hline
     KOI &   P (days) &  RV amplitude (m/s)   &   KOI &   P (days) &  RV amplitude (m/s) &   KOI &   P (days) &  RV amplitude (m/s) \\
\hline
222.01 & 6.312 & < 7.19  &   85.01* & 5.860 & < 3.74 &  152.03 & 13.484 & 2.02 $^{+ 1.47 }_{- 0.86 }$ \\ 
 222.02 & 12.795 & < 5.08  &  85.02* & 2.155 & < 116.40  & 248.01 & 7.203 & < 8.28  \\ 
 244.02 & 6.238 & < 2.23  &  85.03* & 8.134 & < 2.35  &  248.02 & 10.913 & < 5.52  \\ 
 255.01 & 27.522 & < 10.18 &  137.02 & 14.859 & 2.19 $^{+ 1.10 }_{- 0.71 }$ &    250.02 & 17.251 & 2.34 $^{+ 0.30 }_{- 0.27 }$ \\ 
 430.01 & 12.376 & < 3.05  & 137.03 & 3.505 & < 10.79  &  250.03 & 3.544 & < 90.10  \\ 
  523.01 & 49.410 & 2.49 $^{+ 0.46 }_{- 0.43 }$ &    156.02 & 5.188 & < 3.00  &  730.02 & 9.848 & 2.23 $^{+ 0.85 }_{- 0.85 }$ \\ 
 1279.01 & 14.375 & < 3.33  &  168.02* & 15.278 & < 3.05  &     1831.02 & 4.386 & < 194.83  \\ 
 1783.01 & 134.457 & 11.69 $^{+ 3.87 }_{- 2.92 }$ &  314.01* & 13.781 & < 2.83 &  1831.04 & 13.981 & < 119.48  \\ 
2414.01 & 22.595 & < 10.76  &  377.01 & 19.248 & 10.17 $^{+ 0.16 }_{- 0.17 }$ & 1955.01* & 15.170 & < 13.58  \\ 
 2414.02 & 45.351 & < 53.35  & 377.02 & 38.944 & 5.54 $^{+ 0.09 }_{- 0.09 }$ &  1955.02* & 39.462 & < 3.34  \\ 
  3503.01 & 21.191 & < 28.95  &  377.03 & 1.593 & < 143.67  &  1955.03* & 1.644 & < 581.62  \\ 
 3503.02 & 31.810 & < 37.32  &  401.01 & 29.199 & < 13.88 &  1955.04* & 26.237 & < 9.17  \\ 
 & & & 401.02 & 160.018 & < 11.42  &  2174.02 & 33.136 & < 265.56   \\ 
 & & &  567.01 & 10.688 & < 4.03  &  2174.04 & 3.016 & < 50.92   \\ 
 & & &  806.01 & 143.494 & 2.40 $^{+ 0.17 }_{- 0.16 }$ & & & \\ 
 & & &  806.02 & 60.319 & 87.18 $^{+ 1.07 }_{- 1.16 }$ & & &  \\  
 & & &  877.01 & 5.955 & < 7.76 & & &  \\ 
 & & &  877.03 & 20.836 & < 3.53 & & &  \\ 
 & & &   886.03 & 20.996 & < 22.38 & & &  \\ 
 & & &  934.01 & 5.827 & < 70.75 & & &  \\ 
 & & &  934.03 & 18.748 & < 7.96 & & & \\ 
  & & &  1070.01 & 6.618 & < 549.70 & & &  \\ 
 & & &  1338.01 & 3.223 & < 897.01 & & &  \\ 
 & & &  1338.02 & 42.041 & < 14.38 & & &  \\ 
 & & &  1338.03 & 21.016 & < 17.90 & & &  \\ 
 & & &  1353.01 & 125.870 & 14.81 $^{+ 1.41 }_{- 1.39 }$ & & & \\ 
 & & &  1353.02 & 34.543 & < 3.47 & & & \\  
 & & &  1574.01 & 114.756 & 39.07 $^{+ 15.02 }_{- 11.32 }$ & & & \\ 
 & & &  1574.03 & 5.834 & < 44.94 & & & \\ 
 & & &  1576.01 & 10.421 & 2.31 $^{+ 0.25 }_{- 0.24 }$ & & & \\ 
 & & &  1598.01 & 56.476 & < 2.83 & & & \\ 
 & & &  1598.03 & 13.930 & < 66.72 & & & \\ 
 & & &  1833.01 & 3.693 & < 10.78 & & & \\ 
 & & &  1833.02 & 7.685 & < 4.84 & & &  \\ 
 & & &  2092.02 & 25.563 & < 36.04 & & & \\ 
 & & &  2195.01 & 20.053 & < 8.32 & & & \\ 
 & & &  2195.02 & 30.091 & < 7.99 & & & \\ 
 & & &  2195.03 & 6.850 & < 205.09 & & & \\ 
 \hline
     \end{tabular}
    \caption{Expected radial velocity amplitudes for signals or upper limits above 2 m/s for two-planet systems (left), three-planet systems (middle) and four-planet systems (right). Intervals for strongly detected planetary masses include the median, 16th and 84th percentile, while upper limits only include either 97.7th percentile upper limits or limits imposed by our bulk density prior, $\rho < 10$ g/cm$^3$. Where hosts have Kepler magnitude $<13.5$, candidates are marked with an asterisk.}\label{tbl-rv}
  \end{center}
\end{table}
\clearpage

\section*{Appendix E: Measured Long-cadence Transit Times for Systems}
For most systems, we adopt the transit times of \citet{rowe15a}. For Kepler-289 d (Planet Hunters PH3 c), we use the times of \citet{Schmitt2014b}. For planets that are likely interacting that are not in the TTV catalog of \citet{rowe15a}, we measured center of transit times using the photometric transit modeling \textsc{transitfit5} package \citep{Rowe2016}.  The model assumes a noninteracting Keplerian orbit  with the analytic quadratic limb-darkening model of \citet{man02}.  An initial model fit was performed assuming no transiting timing variations. The model parameters for the template are stellar mean density, photometric zero-point, mean orbital period, center of transit time, impact parameter, and scaled planetary radius. Model parameters were initialized using the parameters reported in the Kepler DR 25 catalog \citep{Thompson2018}. The template is then fitted to each individual transit with only the center of transit time allowed to vary. The individual midtransit times are then used to {\it deTTV} the photometric light curve by adjusting observation times to produce a perfectly periodic transit curve.  Observation time adjustments are based on a linear interpolation of measured transit times relative to the orbital period (observed versus calculated).  The transit-template model is then updated by the fitted model against the deTTVed light curve. Uncertainties are based on the variance of model fit residuals that assume independent photometric measurements with Gaussian uncertainties. 

Our measured transit times are available in their entirety in electronic format. We list the first few rows of Table~\ref{tbl-koi520tt} (KOI-520),  Table~\ref{tbl-koi750tt} (KOI-750), Table~\ref{tbl-koi1353tt} (KOI-1353), and Table~\ref{tbl-koi3503tt} (KOI-3503) in this appendix to show formatting.

\begin{table}[ht!]
\tiny
  \begin{center}
    \begin{tabular}{|c|c||c|c||c|c||c|c|}
    \hline
65.91529371 & 0.01382856 & 65.03270462 & 0.00750664 & 69.44980075 & 0.00850947 & 105.31678628 & 0.01912911 \\
71.37416188 & 0.01384941 & 77.78823900 & 0.00676071 & 95.20250164 & 0.00595128 & 156.44698831 & 0.03683910  \\
76.76455745 & 0.02016537 & 90.54315823 & 0.00549775 & 120.94866571 & 0.00845493 & 207.65312842 & 0.01806724  \\
82.23254021 & 0.01524773 & 103.31082904 & 0.00590469 & 146.70983328 & 0.01080931 & 258.79064173 & 0.02446249  \\
     \hline
     \end{tabular}
    \caption{Transit times for KOI-520. Odd columns give the transit times (BJD-2,454,900), from left to right in order of orbital period. Even columns give measurement uncertainty in transit times (in days). The table is available in its entirety in electronic format.}\label{tbl-koi520tt}
  \end{center}
\end{table}

\begin{table}[ht!]
\tiny
  \begin{center}
    \begin{tabular}{|c|c||c|c||c|c||c|c|}
    \hline
65.52383433  & 0.03519860  & 71.83372710 &  0.03185410  & 82.84541108 &  0.01379358  \\
70.54157712  & 0.07151604  & 86.32901433 &  0.03501218  & 104.54478278 &  0.01084324  \\
75.61044521  & 0.04587991  & 129.90179411 &  0.03641057  & 126.22898672 &  0.00691597  \\
80.69592394  & 0.05461808  & 144.40835160 &  0.03535571  & 169.58693056 &  0.01012767  \\
     \hline
     \end{tabular}
    \caption{Transit times for KOI-750. Odd columns give the transit times (BJD-2,454,900), from left to right in order of orbital period. Even columns give measurement uncertainty in transit times (in days). The table is available in its entirety in electronic format.}\label{tbl-koi750tt}
  \end{center}
\end{table}

\begin{table}[ht!]
\tiny
  \begin{center}
    \begin{tabular}{|c|c||c|c||c|c|}
    \hline
 65.69734128  & 0.00680384  & 75.6406 &  0.0060  & 169.66132408 &  0.00056087  \\
134.75905456  & 0.01156442  & 141.7767 &  0.0041  & 295.52314401 &  0.00068940 \\ 
169.31253567  & 0.01319349  & 207.8402 &  0.0051  & 421.38803663 &  0.00061263  \\
203.86562825  & 0.01119769  & 273.9573 &  0.0038  & 547.25387748 &  0.00069121  \\
     \hline
     \end{tabular}
    \caption{Transit times for KOI-1353. Odd columns give the transit times (BJD-2,454,900), from left to right in order of orbital period. Even columns give measurement uncertainty in transit times (in days). We adopted the transit times of Kepler-289 d from \citet{Schmitt2014b}.  The table is available in its entirety in electronic format.}\label{tbl-koi1353tt}
  \end{center}
\end{table}

\begin{table}[ht!]
\tiny
  \begin{center}
    \begin{tabular}{|c|c||c|c|}
    \hline
81.84788981  & 0.03723880  & 77.50612207  & 0.04062840 \\
102.99217572  & 0.04443407  & 109.34471530  & 0.03969897 \\ 
145.30910821  & 0.04715805  & 141.21054970  & 0.04784740 \\
166.52880612  & 0.04548758  & 173.01620158  & 0.04274378 \\
     \hline
     \end{tabular}
    \caption{Transit times for KOI-3503. Odd columns give the transit times (BJD-2,454,900), from left to right in order of orbital period. Even columns give measurement uncertainty in transit times (in days). The table is available in its entirety in electronic format.}\label{tbl-koi3503tt}
  \end{center}
\end{table}

\suppressAffiliationsfalse
\allauthors

\end{document}